\RequirePackage{lhchiggs}
\documentclass{cernyrep}
\usepackage{rotating}
\usepackage{color}
\usepackage{subfigure}
\usepackage{lscape}
\usepackage{floatpag}
\floatpagestyle{plain}
\usepackage{axodraw}
\usepackage{lineno}
\begin{document}

  

\thispagestyle{empty}
{
\setlength{\unitlength}{1mm}
\begin{picture}(0.001,0.001)

\put(18,-100){\LARGE\bfseries
            Handbook of LHC Higgs cross sections:}
\put(38,-110){\LARGE\bfseries
             2. Differential Distributions}
\put(9,-145){\Large\bfseries
             Report of the LHC Higgs Cross Section Working Group}
\put(117,-185){\Large Editors:}
\put(135,-185){\Large S.~Dittmaier}
\put(135,-191){\Large C.~Mariotti}
\put(135,-197){\Large G.~Passarino}
\put(135,-203){\Large R.~Tanaka}

\end{picture}
}


\pagenumbering{roman}




\newpage

\leftline{\bf Conveners}


\noindent \emph{Gluon-fusion process:}
         M.~Grazzini,\,  	 
         F.~Petriello,\,
         J.~Qian,\,  	 
         F.~St\"ockli 	    	 
\vspace{0.1cm}

\noindent \emph{Vector-boson-fusion process:} 	
         A.~Denner,\, 	
         S.~Farrington,\, 	
         C.~Hackstein,\,
         C.~Oleari,\,
         D.~Rebuzzi 
\vspace{0.1cm}

\noindent{$\PW\PH/\PZ\PH$ \emph{production mode}:} 	
         S.~Dittmaier,\, 	
         R.~Harlander,\,
         C.~Matteuzzi,\, 	
         J.~Olsen,\, 	
         G.~Piacquadio
\vspace{0.1cm}

\noindent{$\PQt\PQt\PH$ \emph{process}:} 	
         C.~Neu,\, 	  	
         C.~Potter,\, 	
         L.~Reina,\, 	
         M.~Spira
\vspace{0.1cm}

\noindent\emph{MSSM neutral Higgs:} 	
         M.~Spira,\, 	
         M.~Vazquez Acosta,\, 	  	
         M.~Warsinsky,\, 	
         G.~Weiglein 
\vspace{0.1cm}

\noindent\emph{MSSM charged Higgs:} 	
         M.~Flechl,\, 	
         S.~Heinemeyer,\,
         M.~Kr\"amer,\, 	
         S.~Lehti
\vspace{0.1cm}

\noindent\emph{PDF:} 	
         S.~Forte,\, 	
         J.~Huston,\, 	
         K.~Mazumdar,\, 	  	
         R.~Thorne
\vspace{0.1cm}

\noindent\emph{Branching ratios:} 	
         A.~Denner,\, 	
         S.~Heinemeyer,\,
         I.~Puljak,\, 	  	
         D.~Rebuzzi
\vspace{0.1cm}

\noindent\emph{NLO MC:}
         M.~Felcini,\, 	  	
         F.~Krauss,\,
         F.~Maltoni,\, 	
         P.~Nason,\,
         J.~Yu	
\vspace{0.1cm}

\noindent\emph{Higgs pseudo-observables:} 	
         M.~D\"uhrssen,\, 	
         M.~Felcini,\, 	  	
         G.~Passarino 

\noindent\emph{$\gamma\gamma$ decay:} 	
         S.~Gascon-Shotkin,\, 	
         M.~Kado

\noindent\emph{$\PZ\PZ^*$ decay:} 	
         N.~De Filippis,\, 
         S.~Paganis 

\noindent\emph{$\PW\PW^*$ decay:} 	
         J.~Cuevas,\, 
         T.~Dai 

\noindent\emph{$\tau\tau$ decay:} 	
        A.~Nikitenko,\,
	M.~Schumacher 

\noindent\emph{$\PQb\bar\PQb$ decay:} 	
        C.~Matteuzzi,\, 
        J.~Olsen,\, 
	C.~Potter 

\noindent\emph{$\PH^\pm$ decay:} 	
	M.~Flechl,\,
        S.~Lehti


\vfill

\newpage
\vspace*{10cm}
\begin{center} 
 \bf {Abstract}
\end{center}
\vspace{0.5cm}
This Report summarises the results of the second year's activities of 
the LHC Higgs Cross Section Working Group.
The main goal of the working group was to present the state of the art of
Higgs Physics at the LHC, integrating all new results that have appeared in the
last few years.
The first working group report 
{\it Handbook of LHC Higgs Cross Sections: 1.~Inclusive Observables} (CERN-2011-002)
focuses on predictions (central values and errors) for
total Higgs production cross sections and Higgs branching ratios
in the Standard Model and its minimal supersymmetric extension, covering also related 
issues such as Monte Carlo generators, parton distribution functions, and
pseudo-observables.
This second Report represents the next natural step towards realistic predictions
upon providing results on cross sections with benchmark cuts, differential distributions, 
details of specific decay channels, and further recent developments.

\newpage
\vspace*{10cm}
\begin{center}
We, the authors, would like to dedicate this Report to the memory of \\
Robert Brout and Simon van der Meer.
\end{center}

\newpage
\begin{flushleft}

S.~Dittmaier$^{1}$,\,
C.~Mariotti$^{2,3}$,\,
G.~Passarino$^{2,4}$\,and\,
R.~Tanaka$^{5}$\,(eds.);\\


 S.~Alekhin$^{6,7}$,\,
 J.~Alwall$^{8}$,\,
 E.~A.~Bagnaschi$^{9,10}$,\, 
 A.~Banfi$^{11}$,\, 
 J.~Bl\"umlein$^{6}$,\,
 S.~Bolognesi$^{12}$
% ...

\end{flushleft}
 
 \begin{itemize}
  \item[$^{1}$] 
   Physikalisches Institut, Albert-Ludwigs-Universit\"at Freiburg,
   D-79104 Freiburg, Germany
 
  \item[$^{2}$] 
   INFN, Sezione di Torino, Via P. Giuria 1, 10125 Torino, Italy

  \item[$^{3}$] 
   CERN, CH-1211 Geneva 23, Switzerland 
 
  \item[$^{4}$] 
   Dipartimento di Fisica Teorica, Universit\`a di Torino,  Via P. Giuria 1, 10125 Torino, Italy

  \item[$^{5}$] 
   Laboratoire de l'Acc\'el\'erateur Lin\'eaire, CNRS/IN2P3, F-91898 Orsay CEDEX, France

  \item[$^{6}$] 
   DESY, Zeuthen, Platanenallee 6, D-15738 Zeuthen, Germany
  
  \item[$^{7}$] 
   Institute for High Energy Physics, 142281 Protvino,
   Moscow region, Russia 

  \item[$^{8}$] 
   Theoretical Physics Department, Fermi National Accelerator Laboratory\\
   MS 106, Batavia, IL 60510-0500, USA

  \item[$^{9}$] 
   Dipartimento di Fisica, Universit\`a degli Studi di Milano and INFN,\\ 
   Sezione di Milano, Via Celoria 16, I-20133 Milan, Italy
 
  \item[$^{10}$] 
   Laboratoire de Physique Th\'eorique et des Hautes Energies, \\
   4 Place Jussieu, F-75252 Paris CEDEX 05, France

  \item[$^{11}$]
   Zurich, ETH CH-8093 Zurich, Schafmattstrasse 16, 8093, Zurich, Switzerland 

  \item[$^{12}$] 
   Johns Hopkins University, Baltimore, Maryland. 410-516-8000, USA

  \item[$^{13}$]
   University of Florida, 215 Williamson Hall, P.O. Box 118440 Gainesville, FL 32611 

  \item[$^{14}$] 
   Departamento de F\'isica, Facultad de Ciencias Exactas y Naturales\\ 
   Universidad de Buenos Aires, 
   Pabellon I, Ciudad Universitaria (1428) \\
   Capital Federal, Argentina 

  \item[$^{15}$] 
   Department of Physics, University of Oxford, Denys Wilkinson Building, \\
   Keble Road, Oxford OX1 3RH, UK
 
  \item[$^{16}$] 
   Physics Dept., Blackett Laboratory, Imperial College London, \\
   Prince Consort Rd, London SW7 2BW, UK

  \item[$^{17}$] 
   Department of Physics, Brookhaven National Laboratory, Upton, NY 11973, USA 

  \item[$^{18}$]
   Dipartimento Interateneo di Fisica dell'Universit\`a e del Politecnico di Bari,\\
   INFN Sezione di Bari, Via Orabona 4, 70125 Bari, Italy

  \item[$^{19}$]
   Universit\`a degli Studi di "Roma Tre" Dipartimento di Fisica\\
   Via Vasca Navale 84, 00146 Rome, Italy

  \item[$^{20}$] 
   Institut f\"ur Theoretische Physik und Astrophysik, Universit\"at W\"urzburg, \\ 
   Emil-Hilb-Weg 22, D-97074 W\"urzburg, Germany
 
  \item[$^{21}$]
   School of Physics, University of Melbourne, Victoria, Australia

  \item[$^{22}$]
   Univversit\`a di Roma 'Tor Vergata' Dipartimento di Fisica\\
   Via della Ricerca Scientifica, 1 I-00133 Rome, Italy

  \item[$^{23}$] 
   Theoretical Physics Department, Fermi National Accelerator Laboratory \\
   MS 106, Batavia, IL 60510-0500, USA
 
  \item[$^{24}$] 
   Instituto de F\'isica de Cantabria (IFCA), CSIC-Universidad de Cantabria, \\
   Santander, Spain
  
  \item[$^{25}$] 
   CEA - Centre d'Etudes de Saclay, France

  \item[$^{26}$] 
   Institute of Physics, University of Debrecen H-4010 Debrecen P.O.Box 105, Hungary

  \item[$^{27}$] 
   University of Nova Gorica, Laboratory for Astroparticle Physics\\
   SI-5000 Nova Gorica, Slovenia

  \item[$^{28}$]
   Universit\'e Claude Bernard Lyon 1, CNRS-IN2P3,  Institut de Physique Nucleaire de Lyon (IPNL)\\
   4, rue Enrico Fermi F-69622 Villeurbanne, CEDEX, France

 \item[$^{29}$] 
   DESY, Notkestrasse 85, D-22607 Hamburg, Germany
  
  \item[$^{30}$]
   Institute for Theoretical Physics University of Zurich\\
   Winterthurerstrasse 190, CH-8057 Zurich, Switzerland 

  \item[$^{31}$] 
   LAPTH, Universit\'e de Savoie and CNRS, Annecy-le-Vieux, France

  \item[$^{32}$] 
   Institut f\"ur Theoretische Physik und Institut f\"ur Experimentelle Teilchenphysik, \\ 
   Karlsruhe Institut of Technology, D-76131 Karlsruhe, Germany

  \item[$^{33}$] 
   Bergische Universit\"at Wuppertal, D-42097 Wuppertal, Germany  
 
  \item[$^{34}$] 
   Helsinki Institute of Physics, P.O. Box 64, FIN-00014 University of Helsinki, Finland
 
  \item[$^{35}$] 
   SLAC National Accelerator Laboratory, Menlo Park, CA  94025 USA

  \item[$^{36}$] 
   Department of Physics and Astronomy, Michigan State University, \\
   East Lansing, MI 48824, USA
  
  \item[$^{37}$] 
   Department of Physics, Univ. of Texas at Arlington, SH108, University of Texas,\\
   Arlington, TX 76019, USA 
  
  \item[$^{38}$] 
   Niels Bohr Institute (NBI) University of Copenhagen\\
   Blegdamsvej 17 DK-2100 Copenhagen DENMARK

  \item[$^{39}$] 
   Institute of Nuclear Research of the Hungarian Academy of Sciences, Hungary

  \item[$^{40}$]
   Department of Physics, Royal Holloway, University of London Egham TW20 0EX, UK

  \item[$^{41}$] 
   University of Split, FESB, R. Boskovica bb, 21 000 Split, Croatia
 
  \item[$^{42}$] 
   Institut f\"ur Theoretische Teilchenphysik und Kosmologie, RWTH Aachen University, \\
   D-52056 Aachen, Germany 
 
  \item[$^{43}$] 
   Institute for Particle Physics Phenomenology, Department of Physics, \\
   University of Durham, Durham DH1 3LE, UK

  \item[$^{44}$] 
   National Central University, Chung-Li, Taiwan 

  \item[$^{45}$] 
   School of Physics, and State Key Laboratory of Nuclear Physics and Technology,\\
   Peking University, China

  \item[$^{46}$] 
   Centre for Cosmology, Particle Physics and Phenomenology (CP3), \\ 
   Universit\'e Catholique de Louvain, B-1348 Louvain-la-Neuve, Belgium
  
  \item[$^{47}$]
   Univ. of Wisconsin Dept. of Physics, High Energy Physics\\
   2506 Sterling Hall 1150 University Ave, Madison, WI 53706  

  \item[$^{48}$]
   Southern Methodist University Dept. of Physics, Dallas, TX 75275, USA
 
  \item[$^{49}$] 
   Universit\`a di Milano-Bicocca and INFN, Sezione di Milano-Bicocca,\\
   Piazza della Scienza 3, 20126 Milan, Italy
  
  \item[$^{50}$] 
   University of Virginia, Charlottesville, VA 22906, USA 
 
  \item[$^{51}$] 
   Department of Physics, Princeton University, Princeton, NJ 08542, USA
 
  \item[$^{52}$]
   Dept. of Physics and Astronomy University of Sheffield, Sheffield S3 7RH, UK

  \item[$^{53}$] 
   NCSR Demokritos, Institute of Nuclear Physics, Athens, Greece

  \item[$^{54}$] 
   High Energy Physics Division, Argonne National Laboratory, Argonne, IL 60439, USA 
 
  \item[$^{55}$] 
   Department of Physics \& Astronomy, Northwestern University, Evanston, IL 60208, USA

  \item[$^{56}$] 
   Department of Physics, University of Oregon, Eugene, OR 97403-1274, USA 

  \item[$^{57}$] 
   University of Liverpool Dept. of Physics Oliver Lodge Lab, \\
   Oxford St. Liverpool L69 3BX UK

  \item[$^{58}$]
   Physikalisches Institut, Universit\"at Heidelberg, Heidelberg, Germany
 
  \item[$^{59}$] 
   Universit\`a di Pavia and INFN, Sezione di Pavia\\ 
   Via A. Bassi, 6, 27100 Pavia, Italy
 
  \item[$^{60}$] 
   Physics Department, Florida State University\\
   Tallahassee, FL 32306-4350, USA

  \item[$^{61}$]
   Joint Institute for Nuclear Research (JINR), Joliot-Curie st., 6, \\
   Dubna, 141980 Moscow region, Russia

  \item[$^{62}$] 
   Paul Scherrer Institut, CH--5232 Villigen PSI, Switzerland
 
  \item[$^{63}$]
   Massachusetts Institute of Technology 77 Massachusetts Avenue\\
   Cambridge, MA 02139-4307 USA

  \item[$^{64}$]
   Cambridge University Dept. of Physics Cavendish Laboratory\\
   Cambridge CB3 0HE, UK  

  \item[$^{65}$] 
   Max-Planck-Institut f\"ur Physik, Werner-Heisenberg-Institut,\\
   F\"ohringer Ring 6, D-80805 M\"unchen, Germany

 \item[$^{66}$] 
   Department of Physics and Astronomy, University College London, \\
   Gower Street, London WC1E 6BT, UK 

  \item[$^{67}$] 
   INFN, Sezione di Firenze and Dipartimento di Fisica e Astronomia, Universit\`a di Firenze,\\
   I-50019 Sesto Fiorentino, Florence, Italy 

  \item[$^{68}$]
   ITPP, Lausanne CH-1015 Lausanne, PHB - Ecublens, 1015, Lausanne, Switzerland 
  
  \item[$^{69}$]
   Universit\`a di Napoli Federico II Dipartimento di Scienze Fisiche\\
   via Cintia I-80126 Napoli, Italy

  \item[$^{70}$] 
   School of Physics, University of the Witwatersrand, Private Bag 3, \\
   Wits 2050, Johannesburg, South Africa
   
  \item[$^{71}$]
   Department of Physics, University of California at San Diego, \\
   La Jolla, CA 92093, USA

  \item[$^{72}$] 
   Department of Physics, SUNY at Buffalo, Buffalo, NY 14260-1500, USA
 
  \item[$\dagger$]
   On leave of absence from INFN, Sezione di Firenze, Italy 


  
  


  






 \end{itemize}

\newpage
\begin{center}
 {\bf Prologue}
\end{center}
\vspace{0.5cm}
The implementation of spontaneous symmetry breaking in the framework
of gauge theories in the 1960s triggered the breakthrough in the
construction of the standard electroweak theory, as it still persists
today. The idea of driving the spontaneous breakdown of a gauge
symmetry by a self-interacting scalar field, which thereby lends mass
to gauge bosons, is known as the {\it Higgs mechanism} and goes back
to the early work of 
\Brefs{Englert:1964et,Higgs:1964ia,Higgs:1964pj,Guralnik:1964eu,Higgs:1966ev,Kibble:1967sv}. 
The postulate of a
new scalar neutral boson, known as the {\it Higgs particle}, comes as
a phenomenological imprint of this mechanism. Since the birth of this
idea, the Higgs boson has successfully escaped detection in spite
of tremendous search activities at the high-energy colliders
LEP and Tevatron, leaving open the crucial question whether the
Higgs mechanism is just a theoretical idea or a `true model'
for electroweak symmetry breaking. 


The experiments at the Large Hadron Collider (LHC) have made an impressive step forward
in answering this question, by closing down the space available for the long sought Higgs 
and supersymmetric particles to hide in, putting the Standard Model of particle physics 
through increasingly gruelling tests.
Results based on the analysis of considerably more data than those presented at the Summer 
Conferences are sufficient to make significant progress in the search for the Higgs boson, 
but not enough to make any conclusive statement on the existence or non-existence of the Higgs.  
The outcome of the Higgs search at the LHC will either carve our present understanding of 
electroweak interactions in stone or will be the beginning of a theoretical revolution.


\newpage
\mbox{}

\newpage
\tableofcontents

\newpage
\pagenumbering{arabic}
\setcounter{footnote}{0}


\section{Introduction\footnote{S.~Dittmaier, C.~Mariotti, G.~Passarino and R.~Tanaka.}}

\newcommand{\ggF}{\ensuremath{\Pg\Pg \to \PH}}
\newcommand{\VBF}{\ensuremath{\PQq\PQq' \to \PQq\PQq'\PH}}
\newcommand{\VH}{\ensuremath{\PQq\PAQq \to \PW\PH/\PZ\PH}}
\newcommand{\ttH}{\ensuremath{\PQq\PAQq/\Pg\Pg \to \PQt\PAQt\PH}}


The quest for the origin of electroweak symmetry breaking is one
of the major physics goals of the Large Hadron Collider (LHC) at
CERN. After the successful start of $\Pp\Pp$ collisions in 2009 and 2010,
the LHC machine has been operated at the centre-of-mass energy of
$7$\UTeV\ in 2011, and data corresponding to a luminosity of $5.7\Ufb^{-1}$ have been
delivered. 
The LHC is expected to operate at $\sqrt{s} = 7$ or $8\UTeV$ in 2012
and a long shutdown (2013--2014) is scheduled to prepare for the run
at the design centre-of-mass energy of $14\UTeV$. 


At the LHC, the most important Standard Model (SM) Higgs-boson production processes are:
the gluon-fusion process (\ggF), where a $\Pg\Pg$ pair annihilates into the Higgs boson
through a loop with heavy-quark dominance;
vector-boson fusion (\VBF), where vector bosons are radiated off quarks
and couple to produce a Higgs boson;
vector-boson associated production (\VH), where the Higgs boson is radiated off a gauge boson;
top-quark pair associated production (\ttH), where the Higgs boson is radiated of a top quark.

ATLAS and CMS, with data currently analysed, are able to exclude a substantial region 
of the possible Higgs-boson mass range. The results that were presented in December 2011
show that the region of Higgs masses between approximately $116$ and $127\UGeV$ is not
excluded, and the excess of events observed
for hypothesised Higgs-boson masses at this low end of the explored range makes the
observed limits weaker than expected. To ascertain the origin of this excess, more
data are required.
With much more data accumulated in 2012, one may eventually reach the discovery of the 
Higgs boson. For this, predictions with the highest precision for
Higgs-boson production and decay rates and
associated uncertainty estimates are crucial.  
If there is a discovery then theoretical predictions for Higgs-boson property measurements
will become even more important. This is the reason why the LHC Higgs Cross Section Working 
Group has been created in 2010 as the joint forum of the experimental
collaborations (ATLAS, CMS, and LHCb) and the theory communities. 

In the LHC Higgs Cross Section Working Group, there are $16$ subgroups.
The first four address different Higgs-boson production modes: \ggF, \VBF, \VH, and 
\ttH\ processes. Two more subgroups are focusing on MSSM neutral- and MSSM charged-Higgs production.
In addition, six new subgroups were created in 2010 to study the Higgs-boson decay modes:
$\PH \to \PGg\PGg, \PWp\PWm, \PZ\PZ, \PGtprp\PGtprm, \PQb\PAQb$, and $\PH^{\pm} $.
Four subgroups discuss common issues across the various production modes:
Higgs-boson decay branching ratios (BR) in the SM and in the Minimal Supersymmetric Standard Model
(MSSM),
parton distribution functions (PDFs),
next-to-leading order (NLO) Monte Carlo (MC) generators
for both Higgs signal and SM backgrounds and, finally, the definition of Higgs
pseudo-observables, in particular, the heavy-Higgs-boson lineshape.

\begin{figure}
  \begin{center}
    \includegraphics[width=0.55\textwidth]{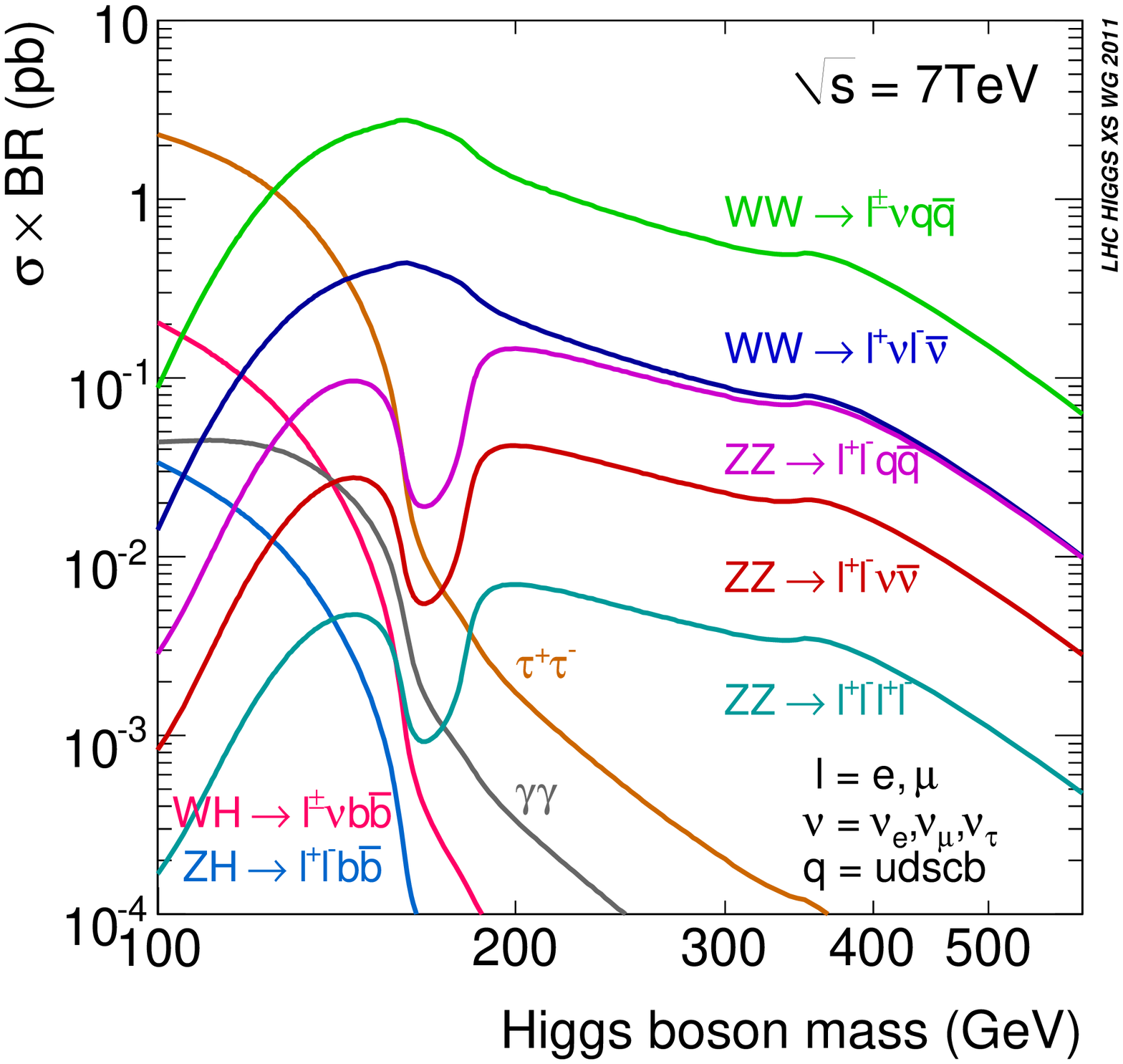}
        \caption{The SM Higgs-boson production cross sections
                      multiplied by decay branching ratios in $\Pp\Pp$ collisions 
                      at \mbox{$\sqrt{s}=7\UTeV$}  as a function of Higgs-boson mass.
                      All production modes are summed in the channels of 
                      $\PH \to \PGtprp\PGtprm$, $\PGg\PGg$, or
                      $\PW\PW/\PZ\PZ(\to 4~\mathrm{fermions})$.
                      In the $\PH\to \PQb\PAQb$ channel, only the vector-boson associated 
                      production is considered.}
            \label{fig:sm-crossbr}
  \end{center}
\end{figure}

In the first Report~\cite{Dittmaier:2011ti}, the state-of-the-art inclusive Higgs-boson 
production cross sections and decay branching ratios have been compiled. 
The major part of the Report was devoted to discussing the computation of cross 
sections and branching ratios for the SM and MSSM Higgs bosons.
The related theoretical uncertainties due to QCD scale and PDF were discussed. 
The Higgs-boson production cross sections are calculated with varying
precision in the perturbative expansion.
For total cross sections,
the calculations are performed up to the next-to-next-to-leading-order
(NNLO) QCD correction for the \ggF, \VBF, and \VH\ processes,
while up to NLO for \ttH\ process.
In most cases, the NLO electroweak (EW) corrections have been applied assuming factorisation 
with the QCD corrections.
The Higgs-boson decay branching ratios take into account the recently 
calculated higher-order NLO QCD and EW corrections in each Higgs-boson decay mode.
The resulting SM Higgs-boson production cross sections times branching
ratios are shown in \refF{fig:sm-crossbr}.
For these calculations, the common SM input parameter set has been used as given in
\Bref{LHCHiggsCrossSectionWorkingGroup:SMinput}.
The coherent theory results to the experimental collaborations facilitated the first LHC 
combination of the Higgs-boson search results, as described in
\Bref{LHCHiggsCombination:2011}.

The present Report, in particular, covers updates on inclusive observables. 
%
%
%
%
The goal of this second Report is to extend the previous study of inclusive cross sections 
to differential distributions. The experimental analysis must impose cuts on the final 
state in order to extract the signal from background: a precise determination of the 
corresponding signal acceptance is therefore necessary. 

Various studies are performed in different Higgs-boson production modes (\ggF, \VBF, \VH, 
and \ttH\ processes); the benchmark cuts for these processes have been defined, and the
differential distributions have been compared at various levels of theoretical accuracy, \ie 
at NLO/NNLO and with MC generators:
\begin{itemize}
\item
In addition, many search modes for the Higgs boson are carried out in the
exclusive mode, \ie by separating the events according to number of jets or the
transverse momentum ($\pT$) of the Higgs boson.
A particularly important channel is $\PH \to \PW\PW \to \Pl\nu\Pl\nu$
in \ggF\ process, where the events are classified in $\PH{+}0,1,2$-jet multiplicity bins to 
improve the signal-to-noise ratio. There are large logarithms associated with the ratio of 
the Higgs-boson mass over the defining $\pT$ of the jet: the theoretical error assignment 
in the exclusive jet bins has been extensively discussed, and a precise prescription is given 
in this Report. 
\item
The $\pT$ of the Higgs boson is a particularly interesting quantity, as it can be used as 
the discriminant variable against the SM backgrounds. 
Possible large logarithms that can occur when cuts are imposed should be studied carefully:
for instance, the Higgs-boson transverse-momentum spectrum in \ggF\ process has been
studied at NLO accuracy and supplemented with next-to-next-to-leading logarithmic (NNLL)
resummation of small-$\pT$ logarithms. A systematic study of the uncertainties of the shape 
of the resummed Higgs-boson $\pT$ spectrum has also been carried out.
\looseness-1
\item
The differential distributions of the SM backgrounds (in particular the irreducible
backgrounds to Higgs-boson searches) have been studied extensively in this Report. 
In the searches at LHC, most of the backgrounds in the signal regions are derived from
measurements in control regions (so-called ``data-driven'' methods);
the extrapolation to the signal region relies on MC simulations, and the related theoretical 
uncertainty is usually estimated by comparing different MC generators and by varying the 
QCD scales. Whenever possible, not only the normalisation, but also the parametrisation
of the background shape should be taken from the control region.
However, there are backgrounds for which one must rely on theoretical predictions;
the main example is represented by di-boson backgrounds $\PQq\PAQq/\Pg\Pg \to \PW\PW/\PZ\PZ$ 
where data control regions cannot be used, due to the limited size of the data sample 
that is available.
%
\item
In addition, the interference between the Higgs-boson signal in $\Pg\Pg \to \PH \to \PW\PW/\PZ\PZ$
and the gluon-induced continuum $\PW\PW/\PZ\PZ$ production may not be negligible after 
introducing experimental cuts. It will be important to compute the interference between 
signal and background and to try to access this at NLO level. The NLO Monte Carlo's will be 
used to simulate this background and to determine how the $K$-factor is changing with the 
chosen kinematic cuts.
\end{itemize}

This Report also discusses issues common to different Higgs-boson search channels. 
\begin{itemize}
\item
The Higgs-boson BRs are the important ingredients for Higgs physics.
Their most precise estimate with the state-of-the-art calculations for SM
and MSSM is presented and the associated uncertainties are discussed. 
\item
PDFs are crucial for the prediction of Higgs-boson production processes, hence PDFs and 
their uncertainties are of particular significance. At present, these PDFs are obtained from 
fits to data from deep-inelastic scattering, Drell--Yan processes, and jet production from a wide
variety of different experiments. Upon arrival of new LHC data, significant improvements 
are expected for the PDF predictions. Different groups have produced publicly available 
PDFs using different data sets and analysis frameworks, and updates are reported. 
\item
NLO MCs are now widely used at LHC; the main progress is represented by a consistent 
inclusion of exact NLO corrections matched to the parton-shower (PS) simulations.
At present, all the main Higgs-boson production channels (\ggF, \VBF, \VH, and \ttH)
are simulated with NLO+PS, together with most important SM backgrounds, like 
$\PQq\PAQq/\Pg\Pg \to \PW\PW/\PZ\PZ, \PW\PQb\PAQb/\PZ\PQb\PAQb, \PQt\PAQt$, etc.  
Tuning of NLO+PS generators is an important issue, and particularly relevant
is the $\pT$ of the Higgs boson. 
Estimates of uncertainties in NLO+PS simulations due to QCD scale uncertainties or
different matching procedures are also reported, and uncertainties due to hadronisation 
and underlying events are discussed.
\item
The current searches for a heavy Higgs boson assume on-shell (stable)
Higgs-boson production. The production cross section is then sampled over
a Breit--Wigner distribution (either fixed-width or running-width scheme), as implemented 
in the MC simulations. Recent studies have shown that the effects due to off-shell 
Higgs-boson production and decay and to interference of the signal with the SM backgrounds 
may become sizable for Higgs-boson masses $\MH > 300\UGeV$; the Higgs-boson lineshape is expected 
to be altered as well. Thus concrete theoretical predictions for the heavy-Higgs-boson lineshape is
discussed in this Report. 
\end{itemize}



Several models beyond the SM are also discussed in this Report: in the MSSM the Higgs-boson 
sector contains, two scalar doublets, accommodating five physical Higgs bosons
of the light and heavy CP-even $\PSh$ and $\PH$, the CP-odd $\PSA$, and the charged 
Higgs bosons $\PSHpm$; BRs and various kinematical distributions are discussed.
A model which contains a 4th generation of heavy fermions, consisting of an up- and a 
down-type quark $(\PQtpr,\PQbpr)$, a charged lepton $(\Pl')$, and a massive 
neutrino $(\PGnl')$ has been studied: a large effect of the higher-order electroweak
corrections has been found.


\enlargethispage{1em}
This Report is based upon the outcome of a series of workshops
throughout 2010--2011, joint effort for Higgs-boson cross sections between
ATLAS, CMS, LHCb collaborations, and the theory community.
These results are recommended as the common inputs to the experimental
collaborations, and for the Higgs combinations at LHC
\footnote{Any updates will be bade available at the TWiki page: 
https://twiki.cern.ch/twiki/bin/view/LHCPhysics/CrossSections}.




\newpage
\providecommand{\HDECAY}{{\sc HDECAY}}
\providecommand{\HIGLU}{{\sc HIGLU}}
\providecommand{\Prophecy}{{\sc Prophecy4f}}
\providecommand{\CPsuperH}{{\sc CPsuperH}}
\providecommand{\FeynHiggs}{{\sc FeynHiggs}}

\providecommand{\MH}{M_\mathrm{H}}
\providecommand{\Hbb}{\PH \to \PQb \PQb}
\providecommand{\Htautau}{\PH \to \PGtp\PGtm}
\providecommand{\Hmumu}{\PH \to \PGmp\PGmm}
\providecommand{\Hss}{\PH \to \PQs \PQs}
\providecommand{\Hcc}{\PH \to \PQc \PAQc}
\providecommand{\Htt}{\PH \to \PQt \PAQt}
\providecommand{\Hgg}{\PH \to \Pg\Pg}
\providecommand{\Hgaga}{\PH \to \PGg\PGg}
\providecommand{\HZga}{\PH \to \PZ\PGg}
\providecommand{\HWW}{\PH \to \PW\PW}
\providecommand{\HZZ}{\PH \to \PZ\PZ}
\providecommand{\phibb}{\PH \to \PQb \PAQb}
\providecommand{\phitautau}{\PH \to \PGtp \PGtm}
\providecommand{\phimumu}{\PH \to \PGmp \PGmm}
\providecommand{\phiss}{\PH \to \PQs \PAQs}
\providecommand{\phicc}{\PH \to \PQc \PAQc}
\providecommand{\phitt}{\PH \to \PQt \PAQt}
\providecommand{\phigg}{\PH \to \Pg\Pg}
\providecommand{\phigaga}{\PH \to \PGg\PGg}
\providecommand{\phiZga}{\PH \to \PZ\PGg}
\providecommand{\phiVV}{\PH \to \PV^{(*)}\PV^{(*)}}
\providecommand{\phiWW}{\PH \to \PW^{(*)}\PW^{(*)}}
\providecommand{\phiZZ}{\PH \to \PZ^{(*)}\PZ^{(*)}}

\providecommand{\br}{{\mathrm{BR}}}
\providecommand{\Gatot}{\Gamma_{\mathrm H}}
\providecommand{\lsim}
{\;\raisebox{-.3em}{$\stackrel{\displaystyle <}{\sim}$}\;}
\providecommand{\gsim}
{\;\raisebox{-.3em}{$\stackrel{\displaystyle >}{\sim}$}\;}
\providecommand{\Pqb}{\bar{\Pq}}
\providecommand{\Pfb}{\bar{\Pf}}
\providecommand{\PV}{\mathrm{V}}
\providecommand{\tb}{\tan\beta}
\providecommand{\Mh}{M_\mathrm{h}}
\providecommand{\MA}{M_\mathrm{A}}

\providecommand{\mhmaxx}{\ensuremath{m_{\Ph}^{\rm max}}}
\providecommand{\eqn}[1]{Eq.\,(\ref{#1})}
\providecommand{\fig}[1]{Fig.\,\ref{#1}}

\providecommand{\zehomi}[1]{$\cdot 10^{-#1}$}
\providecommand{\zehoze}{}
\providecommand{\zehopl}[1]{$\cdot 10^{#1}$}


\section{Branching ratios\footnote{%
    A.~Denner, S.~Heinemeyer, I.~Puljak, D.~Rebuzzi (eds.);
    S.~Dittmaier, M.~M\"uhlleitner, A.~M\"uck, M.~Spira, M.M.~Weber and
    G.~Weiglein.}}
\label{se:BRs}

For a correct interpretation of experimental data precise calculations
not only of the various production cross sections, but also for the
relevant decay widths are essential, including their respective
uncertainties. Concerning the SM Higgs boson in
\Bref{Dittmaier:2011ti} a first precise estimate
of the branching ratios was presented. In \refS{sec:SMHiggsBR} we
update this prediction and supplement it with an estimate of the various
uncertainties. For the lightest Higgs boson in the MSSM in
\Bref{Dittmaier:2011ti} preliminary results for
$\br(\phitautau)$ ($\phi = \Ph,\PH,\PA$) were given. 
In \refS{sec:MSSMHiggsBR} we present a prediction for all relevant
decay channels evaluated in the \mhmaxx\ scenario~\cite{Carena:2002qg}.


\subsection{SM Higgs branching ratios with uncertainties}
\label{sec:SMHiggsBR}

In this section we present an update of the BR calculation as well as
results for the uncertainties of the decay widths and BRs for a SM
Higgs boson.  
Neglecting these uncertainties would yield in the case
of negative search results too large excluded regions of the parameter
space. In case of a Higgs-boson signal these uncertainties are crucial
to perform a reliable and accurate determination of $\MH$ and the
Higgs-boson couplings~\cite{Aad:2009wy,Ball:2007zza,Duhrssen:2004cv}.
The uncertainties arise from two sources, the missing
higher-order corrections yield the ``theoretical'' uncertainties,
while the experimental errors on the SM input parameters, such as the
quark masses or the strong coupling constant, give rise to the
``parametric'' uncertainties.  Both types of uncertainty have to be
taken into account and combined for a reliable estimate.  We
investigate all relevant channels for the SM Higgs boson, $\Htt$,
$\Hbb$, $\Hcc$, $\Htautau$, $\Hmumu$, $\Hgg$, $\Hgaga$, $\HZga$,
$\HWW$ and $\HZZ$ (including detailed results also for the various
four-fermion final states). We present results for the total width,
$\Gatot$, as well as for various BRs. These results have also been
published in \Bref{Denner:2011mq}.

\subsubsection{Programs and Strategy for Branching Ratio Calculations}
\label{sec:Programs}

The branching ratios of the Higgs boson in the SM have been
determined using the programs {\HDECAY}
\cite{Djouadi:1997yw,Spira:1997dg,hdecay2} and {\Prophecy}
\cite{Bredenstein:2006rh,Bredenstein:2006ha,Prophecy4f}. In a first
step, all partial widths have been calculated as accurately as possible.
Then the branching ratios have been derived from this full set of
partial widths. Since the widths are calculated for on-shell Higgs
bosons, the results have to be used with care for a heavy Higgs boson
($\MH\gsim 500\UGeV$).

\begin{itemize}
\item {\HDECAY} calculates the decay widths and branching ratios
of the Higgs boson(s) in the SM and the MSSM. For the SM it includes
all kinematically allowed channels and all relevant higher-order
QCD corrections to decays into quark pairs and into gluons.
Below the thresholds for two-particle decays, the corresponding three-particle 
decays are used, \eg  below the
$\PQt\PAQt$ threshold the branching ratio for $\Htt$ is calculated
from the three-body decay $\PH\to\PQt\PQb\PW$ including finite-width effects.
More details are given below.

\item {\Prophecy} is a Monte Carlo event generator for $\PH \to
  \PW\PW/\PZ\PZ \to 4\Pf$ (leptonic, semi-leptonic, and hadronic) final
  states. It provides the leading-order (LO) and next-to-leading-order
  (NLO) partial widths for any possible 4-fermion final state. It
  includes the complete NLO QCD and electroweak corrections and all
  interferences at LO and NLO. In other words, it takes into account
  both the corrections to the decays into intermediate $\PW\PW$ and
  $\PZ\PZ$ states as well as their interference for final states that
  allow for both. The dominant two-loop contributions in the
  heavy-Higgs-mass limit proportional to $\GF^2 \MH^4$ are included
  according to \Brefs{Ghinculov:1995bz,Frink:1996sv}.  Since the
  calculation is consistently performed with off-shell gauge bosons
  without any on-shell approximation, it is valid above, near, and
  below the gauge-boson pair thresholds. Like all other light quarks
  and leptons, bottom quarks are treated as massless.  Using the
  LO/NLO gauge-boson widths in the LO/NLO calculation ensures that the
  effective branching ratios of the $\PW$ and $\PZ$ bosons obtained by
  summing over all decay channels add up to one.

\item Electroweak NLO corrections to the decays
$\PH\to\PGg\PGg$ and $\PH\to \Pg\Pg$ have been calculated in
\Brefs{Aglietti:2004ki,Aglietti:2004nj,Aglietti:2006ne,Degrassi:2004mx,
Degrassi:2005mc,Actis:2008ug,Actis:2008ts}.  They are implemented in
{\HDECAY} in form of grids based on the calculations of
\Brefs{Actis:2008ug,Actis:2008ts}.
\end{itemize}

The results presented below have been obtained as follows. The Higgs total
width resulting from {\HDECAY} has been modified according to the
prescription
\begin{equation}
\Gamma_{\PH} = \Gamma^{\mathrm{HD}} - \Gamma^{\mathrm{HD}}_{\PZ\PZ} 
            - \Gamma^{\mathrm{HD}}_{\PW\PW} + \Gamma^{\mathrm{Proph.}}_{4\Pf}~,
\end{equation}
where $\Gamma_{\PH}$ is the total Higgs width, $\Gamma^{\mathrm{HD}}$
the Higgs width obtained from {\HDECAY},
$\Gamma^{\mathrm{HD}}_{\PZ\PZ}$ and $\Gamma^{\mathrm{HD}}_{\PW\PW}$
stand for the partial widths to $\PZ\PZ$ and $\PW\PW$ calculated with
{\HDECAY}, while $\Gamma^{\mathrm{Proph.}}_{4\Pf}$ represents the
partial width of $\PH\to 4\Pf$ calculated with {\Prophecy}.  The
latter can be split into the decays into $\PZ\PZ$, $\PW\PW$, and the
interference,
\begin{equation}
\Gamma^{\mathrm{Proph.}}_{4\Pf}=\Gamma_{{\PH}\to \PW^*\PW^*\to 4\Pf}
+ \Gamma_{{\PH}\to \PZ^*\PZ^*\to 4\Pf}
+ \Gamma_{\mathrm{\PW\PW/\PZ\PZ-int.}}\,.
\end{equation}

\subsubsection{The SM input-parameter set}
\label{sec:Param}

The production cross sections and decay branching ratios of the Higgs
bosons depend on a large number of SM parameters.  For our
calculations, the input-parameter set as defined in Appendix A of
\Bref{Dittmaier:2011ti} has been used.  


As input values for the gauge-boson masses we use the pole masses
 $\MZ=91.15349\UGeV$ and $\MW= 80.36951\UGeV$, derived from the PDG
values given in Appendix~A of \Bref{Dittmaier:2011ti}. The gauge-boson
widths have been calculated at NLO from
the other input parameters resulting in $\Gamma_{\PZ}=2.49581\UGeV$ and
$\Gamma_{\PW}=2.08856\UGeV$.

It should be noted that for our numerical analysis we have used the
one-loop pole masses for the charm and bottom quarks and their
uncertainties, since these values do not exhibit a significant
dependence on the value of the strong coupling constant $\alphas$ in
contrast to the $\MSbar$ masses \cite{Narison:1994ag}.

\subsubsection{Procedure for determining uncertainties}
\label{sec:Procedure}

We included two types of uncertainty: Parametric uncertainties (PU),
which originate from uncertainties in input parameters, and theoretical
uncertainties (THU), which arise from unknown contributions to the
theoretical predictions, typically missing higher orders.
Here we describe the way these uncertainties have been determined.

\paragraph{Parametric uncertainties}

In order to determine the parametric uncertainties of the Higgs-decay
branching ratios we took into account the uncertainties of the input
parameters $\alphas$, $\Mc$, $\Mb$, and $\Mt$. The considered
variation of these input parameters is given in \refT{tab:inputpu}.
\begin{table}\small
\renewcommand{\arraystretch}{1.2}
\setlength{\arraycolsep}{1.5ex}
\caption{Input parameters and their relative uncertainties, as used for the
uncertainty estimation of the branching ratios. The masses of the
central values correspond to the 1-loop pole masses, while the last
column contains the corresponding $\MSbar$ mass values.}
\label{tab:inputpu}
$$\begin{array}{cccc}
\hline
\text{Parameter} & \text{Central value} & \text{Uncertainty}
& \text{$\MSbar$ masses $\Mq(\Mq)$} \\
\hline
  \alphas (\MZ)& 0.119 & \pm0.002\\
\Mc & 1.42\UGeV & \pm0.03\UGeV & 1.28\UGeV \\
\Mb & 4.49\UGeV & \pm0.06\UGeV & 4.16\UGeV \\
\Mt & 172.5\UGeV & \pm2.5\UGeV & 165.4\UGeV \\
\hline
\end{array}$$
\end{table}
The variation in $\alphas$ corresponds to three times the error given in
\Brefs{Bethke:2009jm,Nakamura:2010zzi}. The uncertainties for $\Mb$
and $\Mc$ are a compromise between the errors of
\Bref{Nakamura:2010zzi} and the errors from the most precise
evaluations \cite{Kuhn:2007vp,Chetyrkin:2010ic,signer}.  For $\Mc$ our
error corresponds roughly to the one obtained in
\Bref{Dehnadi:2011gc}.  Finally, the assumed error for $\Mt$ is about
twice the error from the most recent combination of CDF and D\O{}
\cite{:1900yx}.

We did not consider parametric uncertainties resulting from experimental
errors on $\GF$, $\MZ$, $\MW$, and the lepton masses, because their
impact is below one per mille. We also did not include uncertainties for
the light quarks $\PQu, \PQd, \PQs$ as the corresponding branching ratios are very
small and the impact on other branching ratios is negligible. Since we
used $\GF$ to fix the electromagnetic coupling $\alpha$,
uncertainties in the hadronic vacuum polarisation do not matter.

Given the uncertainties in the parameters, the parametric uncertainties
have been determined as follows. For each parameter
$p=\alphas,\Mc,\Mb,\Mt$ we have calculated the Higgs branching ratios
for $p$, $p+\Delta p$ and $p-\Delta p$, while all other parameters have
been left at their central values. The error on each branching ratio
has then been determined by
\begin{eqnarray}
\Delta^p_+ \br &=&  \max \{\br(p+\Delta p),\br(p),\br(p-\Delta p)\} -
\br(p),\nonumber\\
\Delta^p_- \br &=&  \br(p) -\min \{\br(p+\Delta p),\br(p),\br(p-\Delta p)\}.
\end{eqnarray}
Note that this definition leads to asymmetric errors.
The total parametric errors have been obtained by adding the parametric
errors from the four parameter variations in quadrature. This procedure
ensures that the branching ratios add up to unity for all parameter
variations individually.

The uncertainties of the partial and total decay widths have been
obtained in an analogous way,
\begin{eqnarray}
\Delta^p_+ \Gamma &=&  \max \{\Gamma(p+\Delta
p),\Gamma(p),\Gamma(p-\Delta p)\} - \Gamma(p),\nonumber\\
\Delta^p_- \Gamma &=&  \Gamma(p) -\min \{\Gamma(p+\Delta
p),\Gamma(p),\Gamma(p-\Delta p)\},
\end{eqnarray}
where $\Gamma$ denotes the partial decay width for each considered decay
channel or the total width, respectively.  The total parametric errors
have been derived by adding the individual parametric errors in
quadrature.

\paragraph{Theoretical uncertainties}

The second type of uncertainty for the Higgs branching ratios
results from approximations in the theoretical calculations, the
dominant effects being due to missing higher orders. Since the decay
widths have been calculated with \HDECAY\ and \Prophecy\ the missing
contributions in these codes are relevant. For QCD corrections the
uncertainties have been estimated by the scale dependence of the
widths resulting from a variation of the scale up and down by a factor
$2$ or from the size of known omitted corrections. For electroweak
corrections the missing higher orders have been estimated based on the
known structure and size of the NLO corrections. For cases where
\HDECAY\ takes into account the known NLO corrections only
approximatively the accuracy of these approximations has been used.
The estimated relative theoretical uncertainties for the partial
widths resulting from missing higher-order corrections are summarised
in \refT{tab:uncertainty}. The corresponding uncertainty for the total
width is obtained by adding the uncertainties for the partial widths
linearly.
\begin{table}
\caption{Estimated theoretical uncertainties from missing higher orders.}
\label{tab:uncertainty}%
\renewcommand{\arraystretch}{1.2}%
\setlength{\tabcolsep}{1.5ex}%
\centerline{
\begin{tabular}{lllll}
\hline
\text{Partial width} & \text{QCD} & \text{electroweak} & \text{total} \\
\hline
 $\PH \to \PQb\PAQb/\PQc\PAQc$ &    $\sim 0.1\%$
&     $\sim 1$--$2\%$ for $\MH \lsim 135\UGeV$     &      $\sim 2 \%$\\
$\PH\to \PGtp \PGtm/\PGmp\PGmm$ & & $\sim1$--$2\%$  for $\MH \lsim 135\UGeV$ &       $\sim 2 \%$ \\
$\PH \to \PQt\PAQt$ & $\lsim 5\%$& 
      ${\lsim  2}$--${5\%}$   for $\MH < 500 \UGeV$     &       $\sim5\%$ \\
& &      $\sim 0.1 (\frac{\MH}{1\UTeV})^4$ for $\MH > 500\UGeV$  &       $\sim5$--$10\%$ \\      
$\PH \to \Pg\Pg$ & ${\sim 3\%}$   & 
$\sim 1\%$   &   $\sim3\%$\\
$\PH \to \PGg \PGg$  & ${<1\%}$ & $<1\%$    &  $\sim1\%$ \\
$\PH \to \PZ \PGg$  & ${<1\%}$ & $\sim 5\%$    &  $\sim5\%$ \\
$\PH \to \PW\PW/\PZ\PZ\to4\Pf$ & $<0.5\%$ &   $\sim 0.5\%$ for $\MH < 500\UGeV$ &  $\sim0.5\%$\\
$\phantom{\PH\to    \to 4\Pf}$             && $\sim 0.17 (\frac{\MH}{1\UTeV})^4$ for $\MH > 500\UGeV$
&        $\sim0.5$--$15\%$                          \\ 
\hline
\end{tabular}
}
\end{table}

Specifically, the uncertainties of the partial widths calculated with
\HDECAY\ are obtained as follows: For the decays $\PH \to \PQb\PAQb,
\PQc\PAQc$, \HDECAY\ includes the complete massless QCD corrections
up to and including NNNNLO, with a corresponding scale dependence of
about $0.1\%$
\cite{Gorishnii:1990zu,Gorishnii:1991zr,Kataev:1993be,Surguladze:1994gc,
Larin:1995sq,Chetyrkin:1995pd,Chetyrkin:1996sr,Baikov:2005rw}. The NLO
electroweak corrections
\cite{Fleischer:1980ub,Bardin:1990zj,Dabelstein:1991ky,Kniehl:1991ze}
are included in the approximation for small Higgs masses
\cite{Djouadi:1991uf} which has an accuracy of about $1{-}2\%$ for $\MH
< 135\UGeV$. The same applies to the electroweak corrections to $\PH \to
\PGtp \PGtm$.  For Higgs decays into top quarks \HDECAY\ includes the
complete NLO QCD corrections
\cite{Braaten:1980yq,Sakai:1980fa,Inami:1980qp,Gorishnii:1983cu,
Drees:1989du,Drees:1990dq,Drees:1991dq} interpolated to the
large-Higgs-mass results at NNNNLO far above the threshold
\cite{Gorishnii:1990zu,Gorishnii:1991zr,Kataev:1993be,Surguladze:1994gc,
Larin:1995sq,Chetyrkin:1995pd,Chetyrkin:1996sr,Baikov:2005rw}.  The
corresponding scale dependence is below $5\%$.  Only the NLO electroweak
corrections due to the self-interaction of the Higgs boson are included,
and the neglected electroweak corrections amount to about $2{-}5\%$ for
$\MH < 500\UGeV$, where $5\%$ refers to the region near the
$\PQt\PAQt$ threshold and $2\%$ to Higgs masses far above.  For $\MH
> 500\UGeV$ higher-order heavy-Higgs corrections
\cite{Ghinculov:1994se,Ghinculov:1995err,Durand:1994pk,Durand:1994err,
Durand:1994pw,Borodulin:1996br} serve as error estimate, resulting in an
uncertainty of about $0.1\times(\MH/1\UTeV)^4$ for $\MH > 500\UGeV$.
For $\PH \to \Pg\Pg$, \HDECAY\ uses the NLO
\cite{Inami:1982xt,Djouadi:1991tka,Spira:1995rr}, NNLO
\cite{Chetyrkin:1997iv}, and NNNLO  \cite{Baikov:2006ch} QCD corrections
in the limit of heavy top quarks.  The uncertainty from the scale
dependence at NNNLO is about $3\%$.  The NLO electroweak corrections are
included via an interpolation based on a grid from \Bref{Actis:2008ts};
the uncertainty from missing higher-order electroweak corrections is
estimated to be $1\%$.  For the decay $\PH\to\PGg \PGg$, \HDECAY\
includes the full NLO QCD corrections
\cite{Zheng:1990qa,Djouadi:1990aj,Dawson:1992cy,Djouadi:1993ji,
Melnikov:1993tj,Inoue:1994jq,Spira:1995rr} and a grid from
\Bref{Actis:2008ug,Actis:2008ts} for the NLO electroweak corrections.
Missing higher orders are estimated to be below $1\%$.  The contribution
of the $\PH \rightarrow \PGg \Pe^{+}\Pe^{-}$ decay via virtual photon
conversion, evaluated in \Bref{Firan:2007tp} is not taken into account
in the following results. Its correct treatment and its inclusion in
HDECAY are in progress.\footnote{The contribution of $\PH \rightarrow
\PGg \Pe^{+}\Pe^{-}$ is part of the QED corrections to $\PH \rightarrow
\PGg \PGg$ which are expected to be small in total.} The partial decay
width $\PH\to \PZ \PGg$ is included in \HDECAY\ at LO including the virtual
$\PW$, top, bottom, and $\PGt$ loop contributions. The QCD corrections are
small in the intermediate-Higgs-mass range \cite{Spira:1991tj} and can
thus safely be neglected. The associated theoretical uncertainty ranges
at the level below one per cent. The electroweak corrections to this
decay mode are unknown and thus imply a theoretical uncertainty of about
$5\%$ in the intermediate-Higgs-mass range.

The decays $\PH \to \PW\PW/\PZ\PZ\to4\Pf$ are based on \Prophecy, which
includes the complete NLO QCD and electroweak corrections with all
interferences and leading two-loop heavy-Higgs corrections.  For small
Higgs-boson masses the missing higher-order corrections are estimated to
roughly $0.5\%$. For $\MH > 500\UGeV$ higher-order heavy-Higgs
corrections dominate the error leading to an uncertainty of about
$0.17\times(\MH/1\UTeV)^4$.

Based on the error estimates for the partial widths in
\refT{tab:uncertainty}, the theoretical uncertainties for the
branching ratios are determined as follows. For the partial widths
$\PH \to \PQb\PAQb,\PQc\PAQc,\PGtp \PGtm, \Pg\Pg, \PGg \PGg$
the total uncertainty given in \refT{tab:uncertainty} is used.  For
$\PH \to \PQt\PAQt$ and $\PH\to \PW\PW/\PZ\PZ\to4\Pf$, the total
uncertainty is used for $\MH<500\UGeV$, while for higher Higgs masses
the QCD and electroweak uncertainties are added linearly.  Then the
shifts of all branching ratios are calculated resulting from the
scaling of an individual partial width by the corresponding relative
error (since each branching ratio depends on all partial widths,
scaling a single partial width modifies all branching ratios). This is
done by scaling each partial width separately while fixing all others
to their central values, resulting in individual theoretical
uncertainties of each branching ratio.  However, since the errors for
all $\PH\to \PW\PW/\PZ\PZ\to4\Pf$ decays are correlated for
$\MH>500\UGeV$ or small below, we only consider the simultaneous
scaling of all 4-fermion partial widths. The thus obtained individual
theoretical uncertainties for the branching ratios are combined
linearly to obtain the total theoretical uncertainties.

\vspace{1ex}

Finally, the total uncertainties are obtained by adding linearly the
total parametric uncertainties and the total theoretical uncertainties.

\subsubsection{Results}
\label{sec:Results}

\begin{figure}
\includegraphics[width=0.48\textwidth]{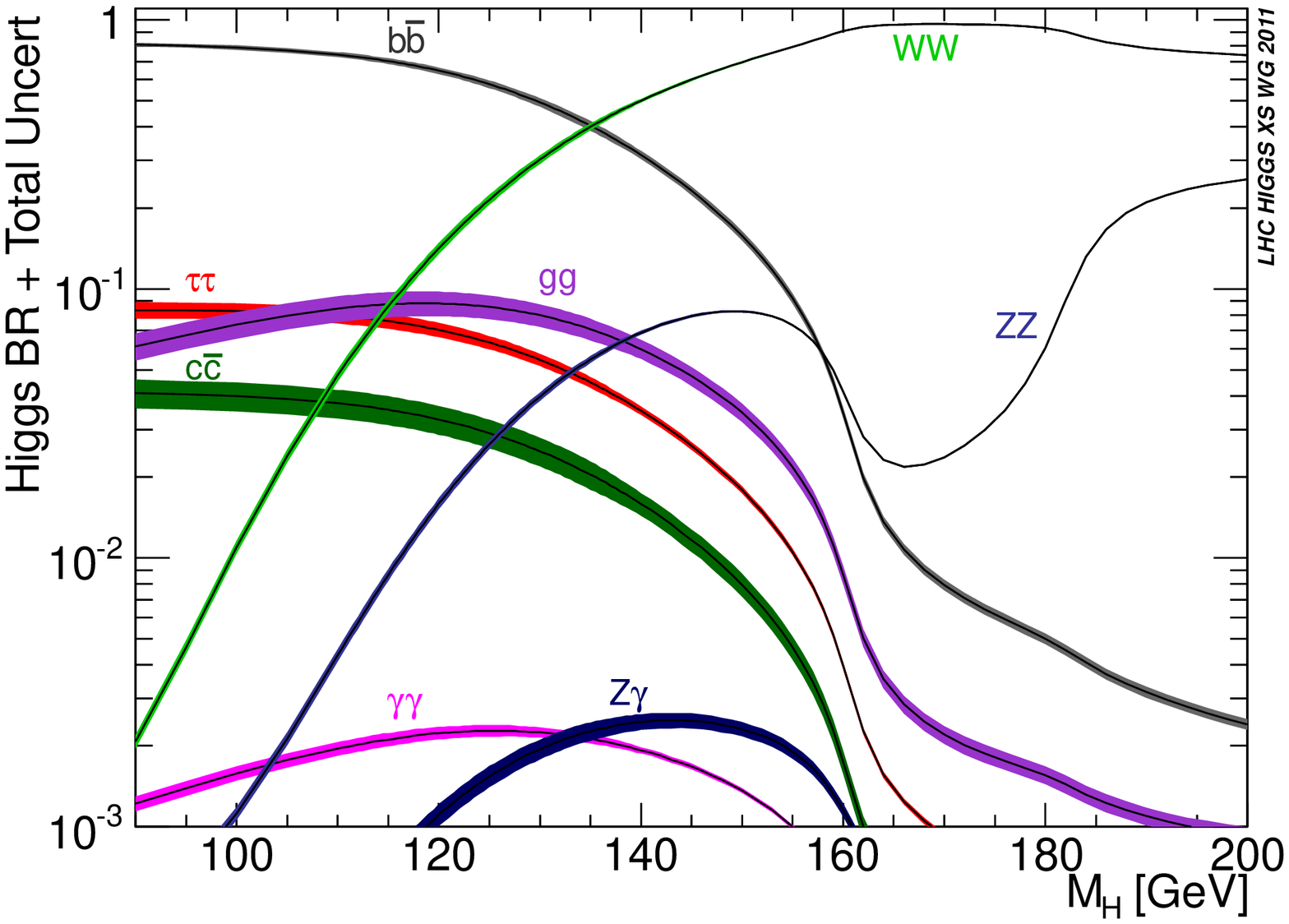}
\includegraphics[width=0.48\textwidth]{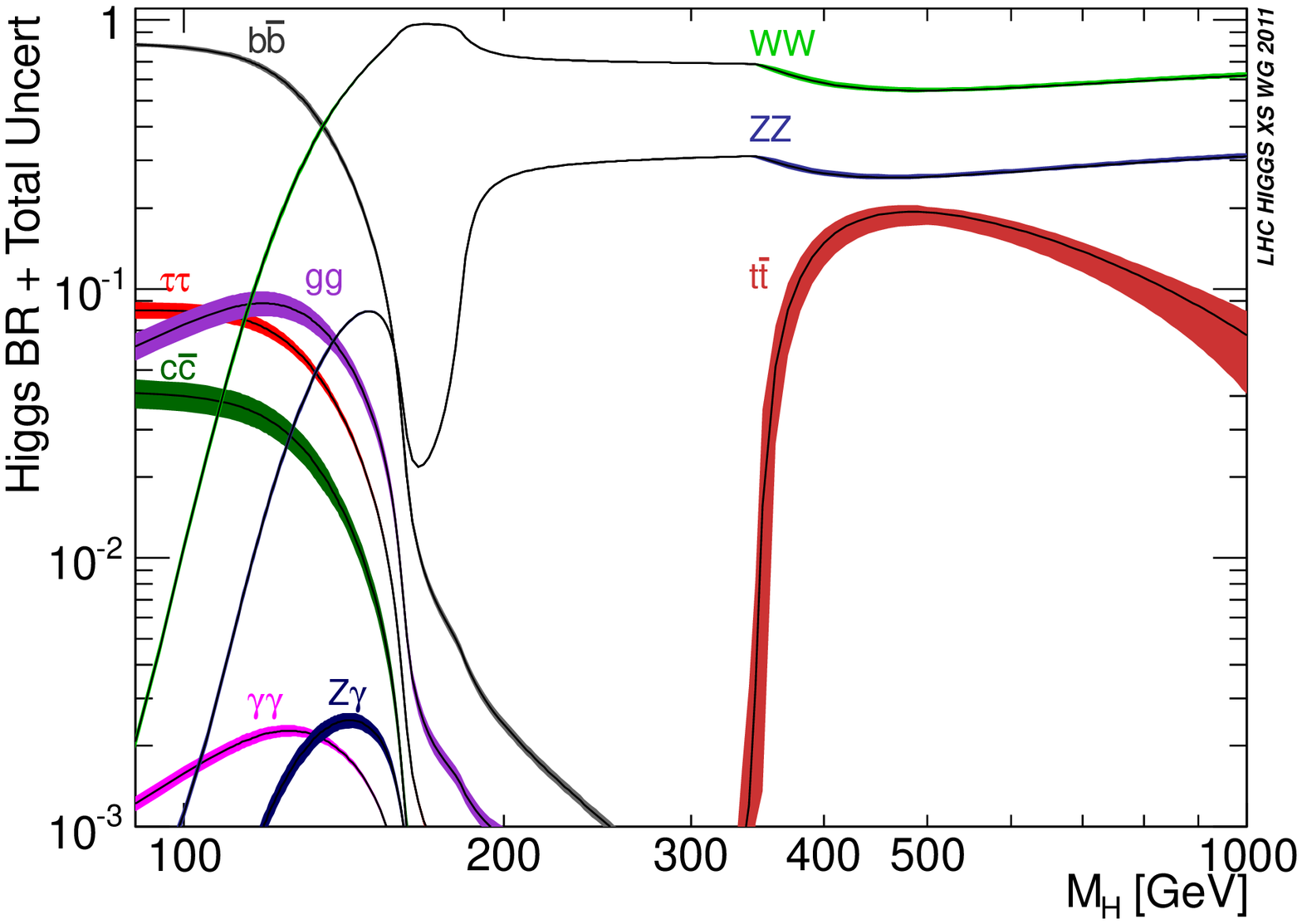}
\vspace*{-0.3cm}
\caption{Higgs branching ratios and their uncertainties for the low
  mass range (left) and for the full mass range (right).}
\label{fig:BRTotUnc}
\end{figure}

\begin{figure}
\includegraphics[width=0.48\textwidth]{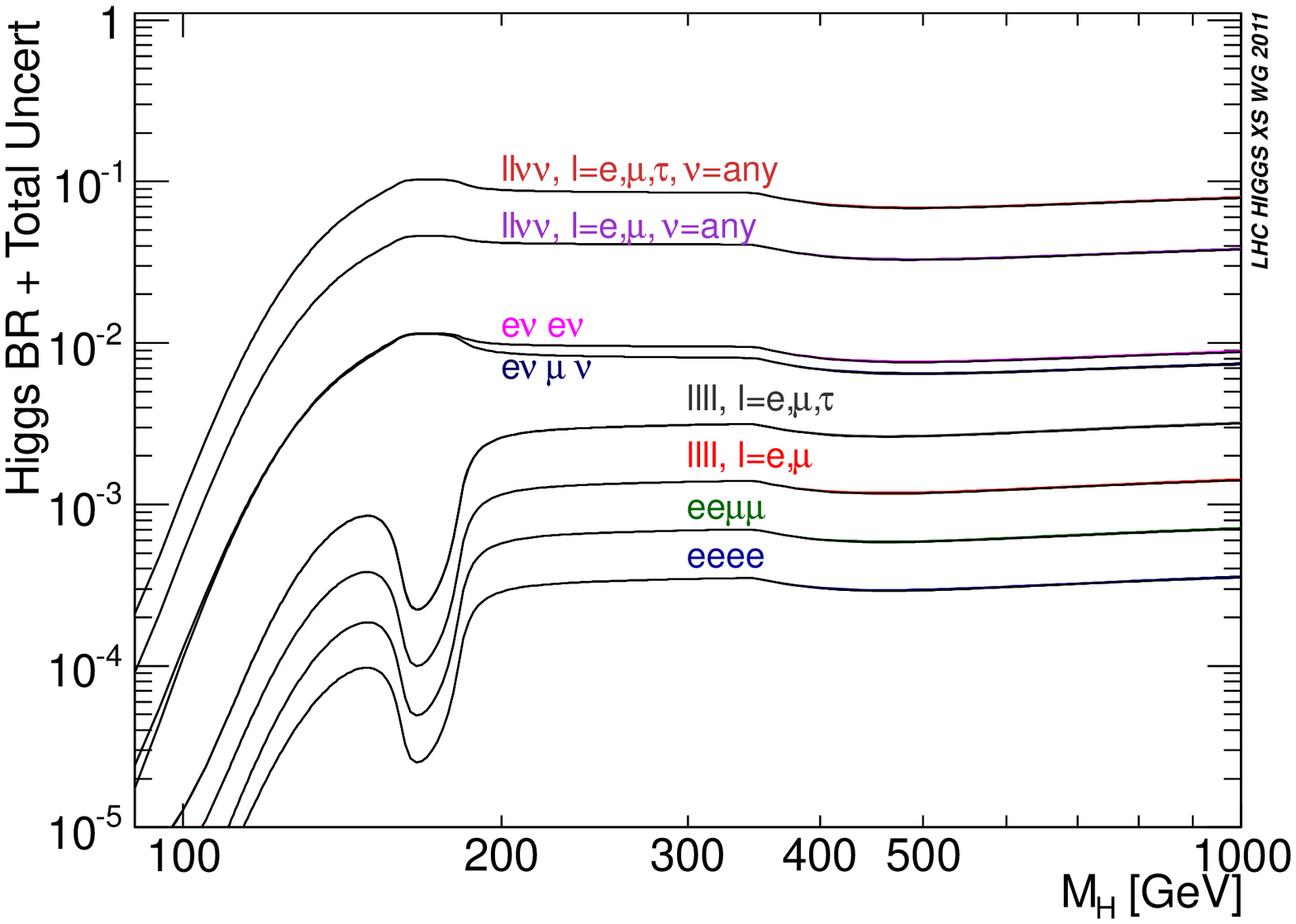}
\includegraphics[width=0.48\textwidth]{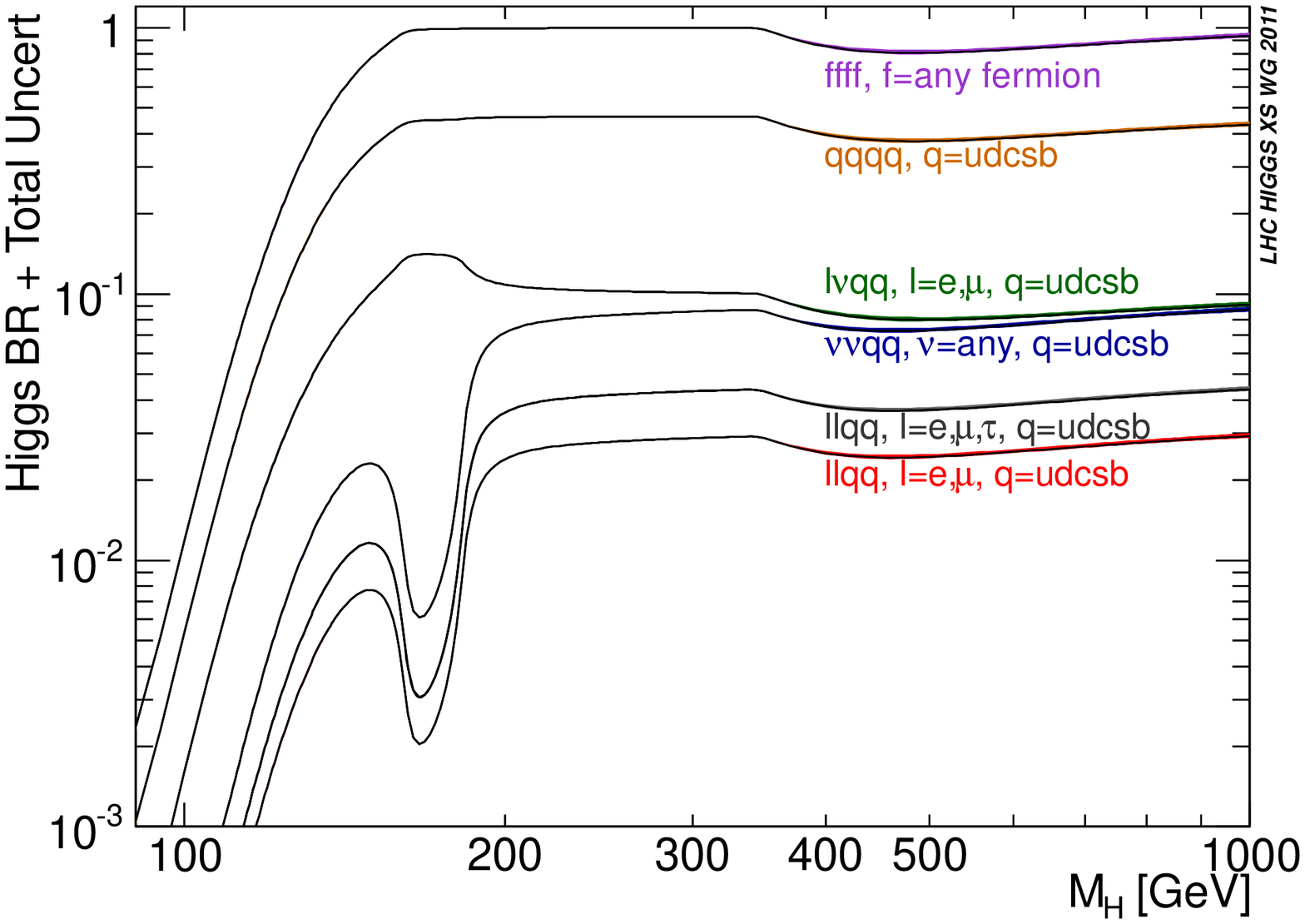}
\vspace*{-0.3cm}
\caption{Higgs branching ratios for the different $\PH \to 4\Pl$ and
  $\PH \to 2\Pl2\PGn$ final states (left) and for $\PH \to 4\Pq$, $\PH
  \to 4\Pf$ and $\PH \to 2\Pq2\Pl, 2\Pq \Pl \PGn, 2\Pq 2 \PGn$ final
  states (right) and their uncertainties for the full mass range.}
\label{fig:BRTotUnc_VV}
\end{figure}

In this section the results of the SM Higgs branching ratios,
calculated according to the procedure described above, are shown and
discussed.  \refF{fig:BRTotUnc} shows the SM Higgs branching ratios in
the low mass range, $100 \UGeV < \MH < 200 \UGeV$, and in the
``full'' mass range, $100 \UGeV < \MH < 1000 \UGeV$, as solid
lines. The (coloured) bands around the lines show the respective
uncertainties, estimated considering both the theoretical and the
parametric uncertainty sources (as discussed in \refS{sec:Procedure}).
More detailed results on the decays $\HWW$ and $\HZZ$ with the
subsequent decay to $4\Pf$ are presented in \refFs{fig:BRTotUnc_VV}.
The largest ``visible'' uncertainties are found for the channels
$\Htautau$, $\Hgg$, $\Hcc$, and $\Htt$, see below.

In the following we list the branching ratios for the Higgs
two-body fermionic and bosonic final states, together with their 
uncertainties, estimated
as discussed in \refS{sec:Procedure}.
Detailed results for four representative Higgs-boson masses are
given in \refT{BR1}. Here we show the BR, the PU separately for the
four parameters as given in \refT{tab:inputpu}, the total PU, the
theoretical uncertainty THU as well as the total uncertainty on the
Higgs branching ratios. 
The THU are most relevant for the $\Hgg$, $\HZga$, and $\Htt$ branching
ratios, reaching $\cal O$($10\%$). For the $\Hbb$, $\Hcc$, and $\Htautau$
branching ratios they remain below a few per cent. 
PU are relevant mostly for the $\Hcc$ and $\Hgg$ branching ratios,
reaching up to
$\cal O$($10\%$) and $\cal O$($5\%$), respectively. 
They are mainly induced by the parametric uncertainties in $\alphas$
and $\Mc$. 
The PU resulting from $\Mb$ affect the $\br(\Hbb)$ at the level of $3\%$,
and the PU from $\Mt$ influences in particular the $\br(\Htt)$ near the
$\PQt\PAQt$ threshold.
For the $\Hgaga$ channel the total uncertainty can reach up to about
$5\%$ in the relevant mass range.
Both THU and PU on the important channels $\HZZ$ and $\HWW$ 
remain at the level of $1\%$ over the full mass range, giving rise to a total
uncertainty below $3\%$ for $\MH > 135 \UGeV$. 
In \refTs{BR-2fa}--\ref{BR-2gbe} (see Appendix) we list the branching ratios for the Higgs
two-body fermionic and bosonic final states, together with their total
uncertainties, for various Higgs-boson masses.%
\footnote{The value $0.0\%$ means that the uncertainty is below
  $0.05\%$.}
\refTs{BR-2gba}--\ref{BR-2gbe}
also contain the total Higgs width $\Gamma_{\PH}$ in the last column.

\begin{table}\footnotesize
\renewcommand{\arraystretch}{1.3}
\setlength{\tabcolsep}{1.0ex}
\caption{SM Higgs branching ratios and their relative parametric (PU),
  theoretical (THU) and total uncertainties for a selection of Higgs
  masses. For PU, all the single contributions are shown. For these
  four columns, the upper percentage value (with its sign) refers 
  to the positive variation of the parameter, while the lower one
  refers to the negative variation of  the parameter. 
}
\label{BR1}
\begin{center}
\begin{tabular}{lccrrrrrrr}
\hline
Channel & $\MH$ [GeV]  &  BR & $\Delta \Mc$ & $\Delta \Mb$ & $\Delta \Mt$ & $\Delta \alphas$ & PU  & THU & Total \\
\hline
& 120  & 6.48\zehomi{1}&$^{-0.2\%}_{+0.2\%}$  & $^{+1.1\%}_{-1.2\%}$  & $^{+0.0\%}_{-0.0\%}$  & $^{-1.0\%}_{+0.9\%}$  & $^{+1.5\%}_{-1.5\%}$  &$^{+1.3\%}_{-1.3\%}$   & $^{+2.8\%}_{-2.8\%}$   \\ 
& 150  & 1.57\zehomi{1}&$^{-0.1\%}_{+0.1\%}$  & $^{+2.7\%}_{-2.7\%}$  & $^{+0.1\%}_{-0.1\%}$  & $^{-2.2\%}_{+2.1\%}$  & $^{+3.4\%}_{-3.5\%}$  &$^{+0.6\%}_{-0.6\%}$   & $^{+4.0\%}_{-4.0\%}$   \\[-1ex]
\raisebox{1.5ex}{$\Hbb$}
& 200  & 2.40\zehomi{3}&$^{-0.0\%}_{+0.0\%}$  & $^{+3.2\%}_{-3.2\%}$  & $^{+0.0\%}_{-0.1\%}$  & $^{-2.5\%}_{+2.5\%}$  & $^{+4.1\%}_{-4.1\%}$  &$^{+0.5\%}_{-0.5\%}$   & $^{+4.6\%}_{-4.6\%}$   \\ 
& 500  & 1.09\zehomi{4}&$^{-0.0\%}_{+0.0\%}$  & $^{+3.2\%}_{-3.2\%}$  & $^{+0.1\%}_{-0.1\%}$  & $^{-2.8\%}_{+2.8\%}$  & $^{+4.3\%}_{-4.3\%}$  &$^{+3.0\%}_{-1.1\%}$   & $^{+7.2\%}_{-5.4\%}$   \\ 
\hline

& 120  & 7.04\zehomi{2}        &$^{-0.2\%}_{+0.2\%}$  & $^{-2.0\%}_{+2.1\%}$  & $^{+0.1\%}_{-0.1\%}$  & $^{+1.4\%}_{-1.3\%}$  & $^{+2.5\%}_{-2.4\%}$ &$^{+3.6\%}_{-3.6\%}$  & $^{+6.1\%}_{-6.0\%}$       \\ 
& 150  & 1.79\zehomi{2}        &$^{-0.1\%}_{+0.1\%}$  & $^{-0.5\%}_{+0.5\%}$  & $^{+0.1\%}_{-0.1\%}$  & $^{+0.3\%}_{-0.3\%}$  & $^{+0.6\%}_{-0.6\%}$ &$^{+2.5\%}_{-2.5\%}$  & $^{+3.0\%}_{-3.1\%}$       \\[-1ex] 
\raisebox{1.5ex}{$\Htautau$} 
& 200  & 2.87\zehomi{4}        &$^{-0.0\%}_{+0.0\%}$  & $^{-0.0\%}_{+0.0\%}$  & $^{+0.0\%}_{-0.1\%}$  & $^{+0.0\%}_{-0.0\%}$  & $^{+0.0\%}_{-0.1\%}$ &$^{+2.5\%}_{-2.5\%}$  & $^{+2.5\%}_{-2.6\%}$       \\ 
& 500  & 1.53\zehomi{5}        &$^{-0.0\%}_{+0.0\%}$  & $^{-0.0\%}_{+0.0\%}$  & $^{+0.1\%}_{-0.1\%}$  & $^{-0.1\%}_{+0.0\%}$  & $^{+0.1\%}_{-0.1\%}$ &$^{+5.0\%}_{-3.1\%}$  & $^{+5.0\%}_{-3.2\%}$       \\ 
\hline

& 120  & 2.44\zehomi{4}        &$^{-0.2\%}_{+0.2\%}$  & $^{-2.0\%}_{+2.1\%}$  & $^{+0.1\%}_{-0.1\%}$  & $^{+1.4\%}_{-1.3\%}$  & $^{+2.5\%}_{-2.5\%}$ &$^{+3.9\%}_{-3.9\%}$  & $^{+6.4\%}_{-6.3\%}$       \\ 
& 150  & 6.19\zehomi{5}        &$^{-0.0\%}_{+0.0\%}$  & $^{-0.5\%}_{+0.5\%}$  & $^{+0.1\%}_{-0.1\%}$  & $^{+0.3\%}_{-0.3\%}$  & $^{+0.6\%}_{-0.6\%}$ &$^{+2.5\%}_{-2.5\%}$  & $^{+3.1\%}_{-3.2\%}$       \\[-1ex] 
\raisebox{1.5ex}{$\Hmumu$}
& 200  & 9.96\zehomi{7}        &$^{-0.0\%}_{-0.0\%}$  & $^{-0.0\%}_{+0.0\%}$  & $^{+0.1\%}_{-0.1\%}$  & $^{+0.0\%}_{-0.0\%}$  & $^{+0.1\%}_{-0.1\%}$ &$^{+2.5\%}_{-2.5\%}$  & $^{+2.6\%}_{-2.6\%}$       \\ 
& 500  & 5.31\zehomi{8}        &$^{-0.0\%}_{+0.0\%}$  & $^{-0.0\%}_{+0.0\%}$  & $^{+0.1\%}_{-0.1\%}$  & $^{-0.0\%}_{+0.0\%}$  & $^{+0.1\%}_{-0.1\%}$ &$^{+5.0\%}_{-3.1\%}$  & $^{+5.1\%}_{-3.1\%}$       \\ 
\hline

& 120  & 3.27\zehomi{2}        &$^{+6.0\%}_{-5.8\%}$  & $^{-2.1\%}_{+2.2\%}$  & $^{+0.1\%}_{-0.1\%}$  & $^{-5.8\%}_{+5.6\%}$  & $^{+8.5\%}_{-8.5\%}$  &$^{+3.8\%}_{-3.7\%}$ & $^{+12.2\%}_{-12.2\%}$      \\ 
& 150  & 7.93\zehomi{3}        &$^{+6.2\%}_{-6.0\%}$  & $^{-0.6\%}_{+0.6\%}$  & $^{+0.1\%}_{-0.1\%}$  & $^{-6.9\%}_{+6.8\%}$  & $^{+9.2\%}_{-9.2\%}$  &$^{+0.6\%}_{-0.6\%}$ & $^{+9.7\%}_{-9.7\%}$        \\[-1ex] 
\raisebox{1.5ex}{$\Hcc$} 
& 200  & 1.21\zehomi{4}        &$^{+6.2\%}_{-6.1\%}$  & $^{-0.2\%}_{+0.1\%}$  & $^{+0.1\%}_{-0.2\%}$  & $^{-7.2\%}_{+7.2\%}$  & $^{+9.5\%}_{-9.5\%}$  &$^{+0.5\%}_{-0.5\%}$ & $^{+10.0\%}_{-10.0\%}$      \\ 
& 500  & 5.47\zehomi{6}        &$^{+6.2\%}_{-6.0\%}$  & $^{-0.1\%}_{+0.1\%}$  & $^{+0.1\%}_{-0.1\%}$  & $^{-7.6\%}_{+7.6\%}$  & $^{+9.8\%}_{-9.7\%}$  &$^{+3.0\%}_{-1.1\%}$ & $^{+12.8\%}_{-10.7\%}$      \\ 
\hline

& 350  & 1.56\zehomi{2}        &$^{+0.0\%}_{+0.0\%}$  & $^{-0.0\%}_{+0.0\%}$  & $^{+120.9\%} _{~-78.6\%}$  & $^{+0.9\%}_{-0.9\%}$ & $^{+120.9\%} _{~-78.6\%}$  &$^{~+6.9\%}_{-12.7\%}$  & $^{+127.8\%} _{~-91.3\%}$      \\ 
& 360  & 5.14\zehomi{2}        &$^{-0.0\%}_{-0.0\%}$  & $^{-0.0\%}_{+0.0\%}$  & $^{-36.2\%}_{+35.6\%}$   & $^{+0.7\%}_{-0.7\%}$ & $^{+35.6\%}  _{-36.2\%}$  &$^{~+6.6\%}_{-12.2\%}$  & $^{+42.2\%}  _{-48.4\%}$      \\[-1ex] 
\raisebox{1.5ex}{$\Htt$}
& 400  & 1.48\zehomi{1}        &$^{+0.0\%}_{+0.0\%}$  & $^{-0.0\%}_{+0.0\%}$  & $^{-6.8\%} _{+6.2\%}$    & $^{+0.4\%}_{-0.3\%}$ & $^{+6.2\%}   _{-6.8\%}$   &$^{~+5.9\%}_{-11.1\%}$  & $^{+12.2\%}  _{-17.8\%}$      \\ 
& 500  & 1.92\zehomi{1}        &$^{-0.0\%}_{+0.0\%}$  & $^{-0.0\%}_{+0.0\%}$  & $^{-0.3\%} _{+0.1\%}$    & $^{+0.1\%}_{-0.2\%}$ & $^{+0.1\%}   _{-0.3\%}$   &$^{+4.5\%}_{-9.5\%}$   & $^{+4.6\%}   _{-9.8\%}$       \\ 
\hline

& 120  & 8.82\zehomi{2}        &$^{-0.2\%}_{+0.2\%}$  & $^{-2.2\%}_{+2.2\%}$  & $^{-0.2\%}_{+0.2\%}$   & $^{+5.7\%}_{-5.4\%}$ & $^{+6.1\%}_{-5.8\%}$  &$^{+4.5\%}_{-4.5\%}$  & $^{+10.6\%}_{-10.3\%}$   \\ 
& 150  & 3.46\zehomi{2}        &$^{-0.1\%}_{+0.1\%}$  & $^{-0.7\%}_{+0.6\%}$  & $^{-0.3\%}_{+0.3\%}$   & $^{+4.4\%}_{-4.2\%}$ & $^{+4.4\%}_{-4.3\%}$  &$^{+3.5\%}_{-3.5\%}$  & $^{+7.9\%} _{-7.8\%}$    \\[-1ex] 
\raisebox{1.5ex}{$\Hgg$}
& 200  & 9.26\zehomi{4}        &$^{-0.0\%}_{-0.0\%}$  & $^{-0.1\%}_{+0.1\%}$  & $^{-0.6\%}_{+0.6\%}$   & $^{+3.9\%}_{-3.8\%}$ & $^{+3.9\%}_{-3.9\%}$  &$^{+3.7\%}_{-3.7\%}$  & $^{+7.6\%} _{-7.6\%}$    \\ 
& 500  & 6.04\zehomi{4}        &$^{-0.0\%}_{+0.0\%}$  & $^{-0.0\%}_{+0.0\%}$  & $^{+1.6\%} _{-1.6\%}$  & $^{+3.4\%}_{-3.3\%}$ & $^{+3.7\%}_{-3.7\%}$  &$^{+6.2\%}_{-4.3\%}$  & $^{+9.9\%} _{-7.9\%}$    \\ 
\hline

& 120  & 2.23\zehomi{3}        &$^{-0.2\%}_{+0.2\%}$  & $^{-2.0\%}_{+2.1\%}$  & $^{+0.0\%} _{+0.0\%}$  & $^{+1.4\%}    _{-1.3\%}$   & $^{+2.5\%}_{-2.4\%}$   &$^{+2.9\%}_{-2.9\%}$  & $^{+5.4\%} _{-5.3\%}$   \\ 
& 150  & 1.37\zehomi{3}        &$^{+0.0\%} _{+0.1\%}$ & $^{-0.5\%}_{+0.5\%}$  & $^{+0.1\%} _{-0.0\%}$ & $^{+0.3\%}    _{-0.3\%}$   & $^{+0.6\%}_{-0.6\%}$   &$^{+1.6\%}_{-1.5\%}$  & $^{+2.1\%} _{-2.1\%}$    \\[-1ex] 
\raisebox{1.5ex}{$\Hgaga$}
& 200  & 5.51\zehomi{5}        &$^{-0.0\%}_{-0.0\%}$  & $^{-0.0\%}_{+0.0\%}$  & $^{+0.1\%} _{-0.1\%}$ & $^{+0.0\%}    _{-0.0\%}$   & $^{+0.1\%}_{-0.1\%}$   &$^{+1.5\%}_{-1.5\%}$  & $^{+1.6\%} _{-1.6\%}$    \\ 
& 500  & 3.12\zehomi{7}        &$^{-0.0\%}_{+0.0\%}$  & $^{-0.0\%}_{+0.0\%}$  & $^{+8.0\%} _{-6.5\%}$ & $^{-0.7\%}   _{+0.7\%}$    & $^{+8.0\%}_{-6.6\%}$   &$^{+4.0\%}_{-2.1\%}$  & $^{+11.9\%}_{~-8.7\%}$    \\ 
\hline

& 120  & 1.11\zehomi{3}        &$^{-0.3\%}_{+0.2\%}$  & $^{-2.1\%}_{+2.1\%}$  & $^{+0.0\%} _{-0.1\%}$ & $^{+1.4\%}    _{-1.4\%}$   & $^{+2.5\%}_{-2.5\%}$   &$^{+6.9\%}_{-6.8\%}$  & $^{+9.4\%} _{-9.3\%}$    \\ 
& 150  & 2.31\zehomi{3}        &$^{-0.1\%}_{+0.0\%}$  & $^{-0.6\%}_{+0.5\%}$  & $^{+0.0\%} _{-0.1\%}$ & $^{+0.2\%}    _{-0.3\%}$   & $^{+0.5\%}_{-0.6\%}$   &$^{+5.5\%}_{-5.5\%}$  & $^{+6.0\%} _{-6.2\%}$    \\[-1ex] 
\raisebox{1.5ex}{$\HZga$}
& 200  & 1.75\zehomi{4}        &$^{-0.0\%}_{-0.0\%}$  & $^{-0.0\%}_{+0.0\%}$  & $^{+0.0\%} _{-0.1\%}$ & $^{+0.0\%}    _{-0.0\%}$   & $^{+0.0\%}_{-0.1\%}$   &$^{+5.5\%}_{-5.5\%}$  & $^{+5.5\%} _{-5.6\%}$    \\ 
& 500  & 7.58\zehomi{6}        &$^{-0.0\%}_{+0.0\%}$  & $^{-0.0\%}_{+0.0\%}$  & $^{+0.8\%} _{-0.6\%}$ & $^{-0.0\%}   _{+0.0\%}$    & $^{+0.8\%}_{-0.6\%}$   &$^{+8.0\%}_{-6.1\%}$  & $^{+8.7\%} _{-6.7\%}$    \\ 
\hline

& 120  & 1.41\zehomi{1}        &$^{-0.2\%}_{+0.2\%}$  & $^{-2.0\%}_{+2.1\%}$  & $^{-0.0\%}_{+0.0\%}$  & $^{+1.4\%}    _{-1.4\%}$   & $^{+2.5\%}_{-2.5\%}$   &$^{+2.2\%}_{-2.2\%}$  & $^{+4.8\%} _{-4.7\%}$    \\ 
& 150  & 6.96\zehomi{1}        &$^{-0.1\%}_{+0.1\%}$  & $^{-0.5\%}_{+0.5\%}$  & $^{-0.0\%}_{+0.0\%}$  & $^{+0.3\%}    _{-0.3\%}$   & $^{+0.6\%}_{-0.6\%}$   &$^{+0.3\%}_{-0.3\%}$  & $^{+0.9\%} _{-0.8\%}$    \\[-1ex] 
\raisebox{1.5ex}{$\HWW$}
& 200  & 7.41\zehomi{1}        &$^{-0.0\%}_{-0.0\%}$  & $^{-0.0\%}_{+0.0\%}$  & $^{-0.0\%}_{+0.0\%}$  & $^{+0.0\%}    _{-0.0\%}$   & $^{+0.0\%}_{-0.0\%}$   &$^{+0.0\%}_{-0.0\%}$  & $^{+0.0\%} _{-0.0\%}$    \\ 
& 500  & 5.46\zehomi{1}        &$^{-0.0\%}_{+0.0\%}$  & $^{-0.0\%}_{+0.0\%}$  & $^{+0.1\%} _{-0.0\%}$ & $^{-0.0\%}   _{+0.0\%}$    & $^{+0.1\%}_{-0.1\%}$   &$^{+2.3\%}_{-1.1\%}$  & $^{+2.4\%} _{-1.1\%}$    \\ 
\hline

& 120  & 1.59\zehomi{2}        &$^{-0.2\%}_{+0.2\%}$  & $^{-2.0\%}_{+2.1\%}$  & $^{-0.0\%}_{+0.0\%}$  & $^{+1.4\%}    _{-1.4\%}$   & $^{+2.5\%}_{-2.5\%}$   &$^{+2.2\%}_{-2.2\%}$  & $^{+4.8\%} _{-4.7\%}$    \\ 
& 150  & 8.25\zehomi{2}        &$^{-0.1\%}_{+0.1\%}$  & $^{-0.5\%}_{+0.5\%}$  & $^{+0.0\%} _{+0.0\%}$  & $^{+0.3\%}    _{-0.3\%}$   & $^{+0.6\%}_{-0.6\%}$   &$^{+0.3\%}_{-0.3\%}$  & $^{+0.9\%} _{-0.8\%}$   \\[-1ex] 
\raisebox{1.5ex}{$\HZZ$}
& 200  & 2.55\zehomi{1}        &$^{-0.0\%}_{+0.0\%}$  & $^{-0.0\%}_{+0.0\%}$  & $^{+0.0\%} _{-0.0\%}$ & $^{+0.0\%}    _{-0.0\%}$   & $^{+0.0\%}_{-0.0\%}$   &$^{+0.0\%}_{-0.0\%}$  & $^{+0.0\%} _{-0.0\%}$    \\ 
& 500  & 2.61\zehomi{1}        &$^{+0.0\%}_{-0.0\%}$  & $^{-0.0\%}_{+0.0\%}$  & $^{+0.0\%} _{+0.0\%}$  & $^{-0.0\%}   _{+0.0\%}$    & $^{+0.1\%}_{-0.0\%}$   &$^{+2.3\%}_{-1.1\%}$  & $^{+2.3\%} _{-1.1\%}$   \\ 
\hline

\end{tabular}\end{center}
\end{table}

Finally, \refTs{BR-4f1a}--\ref{BR-4f1e} and
\refTs{BR-4f2a}--\ref{BR-4f2e}, list the branching ratios for the most
relevant Higgs decays into four-fermion final states. The right column
in these tables shows the total relative uncertainties on these
branching ratios in per cent.  These are practically equal for all
the $\PH \to 4\Pf$ branching ratios and the same as those for $\PH \to
\PW\PW/\PZ\PZ$.
It should be noted that the charge-conjugate state is not included for
$\PH \to \Pl\PGn \Pq\Pqb$.

\begin{figure}[htb!]
\includegraphics[width=1.00\textwidth]{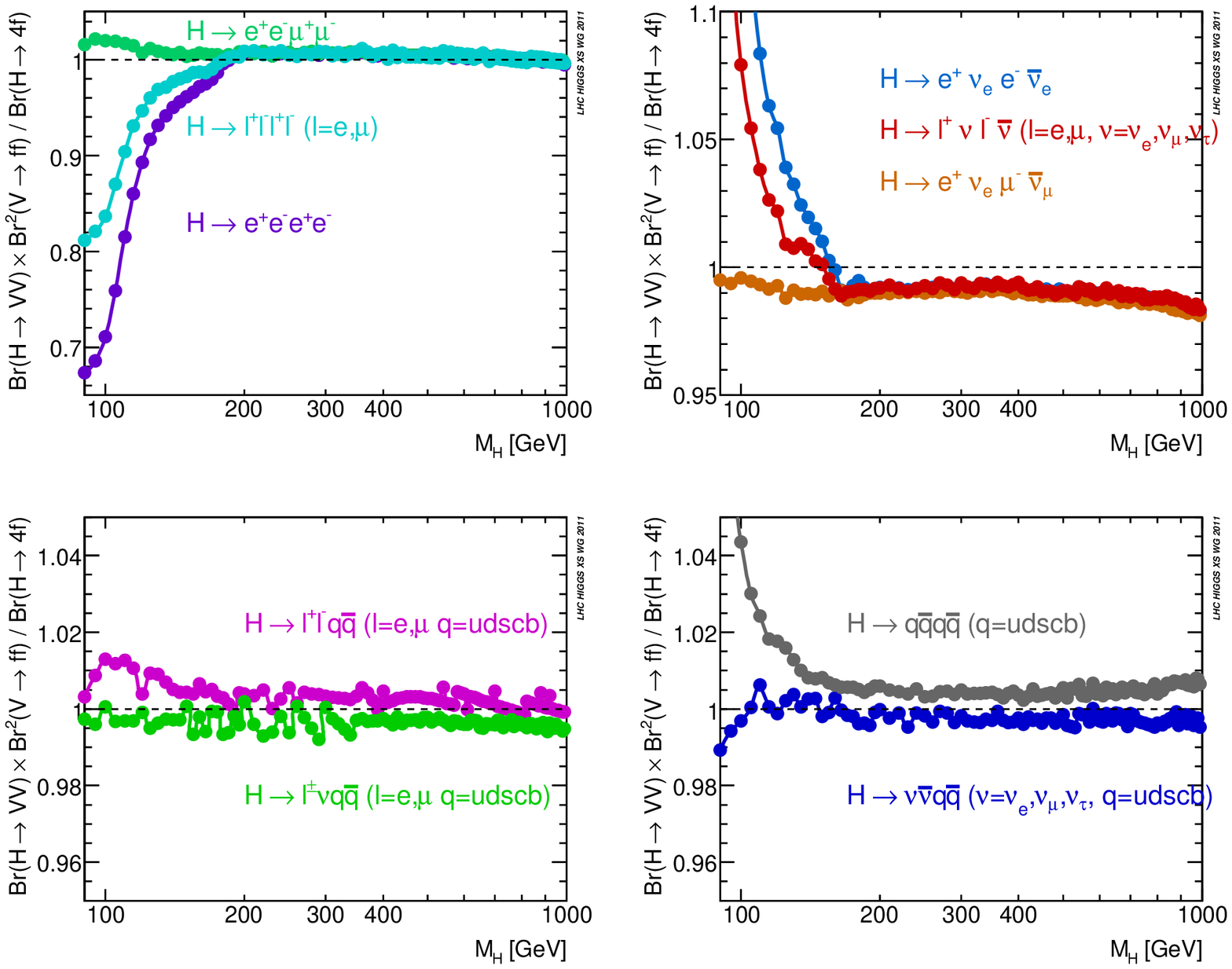}
\vspace*{-0.1cm}
\caption{The ratio between formula \refE{approx} and \Prophecy\ for 
$\PH \to \PZ\PZ \to \Pl\PAl\Pl\PAl$ (top-left),
$\PH \to \PW\PW/\PZ\PZ \to \Pl\PGn\Pl\PGn$ (top-right),
$\PH \to \PZ\PZ \to \Pl\PAl\PQq\PAQq, \PH \to \PW\PW \to \Pl\PGn\PQq\PAQq$ (bottom-left),
and  $\PH \to \PZ\PZ \to \PGn\PAGn\PQq\PAQq, \PH \to \PW\PW/\PZ\PZ \to \PQq\PAQq\PQq\PAQq$ (bottom-right).
}
\label{fig:BR-ratio}
\end{figure}

We would like to remark that, when possible, the branching ratios for
Higgs into four fermions, explicitly calculated and listed in 
\refTs{BR-4f1a}--\ref{BR-4f2e}, should be preferred over
the option of calculating
\begin{equation}
\label{approx}
\br(\PH \to \PV\PV) \times \br(\PV \to \Pf\Pfb) 
                    \times \br(\PV \to \Pf\Pfb) 
                    \times \mbox{(statistical factor)}
\end{equation}
where $\PV = \PW, \PZ$,
and $\br(\PH \to \PV\PV)$ is estimated by \Prophecy, 
while $\br(\PV \to \Pf\Pfb)$ are from Particle Data Group (PDG).
The formula \refE{approx} is based on the narrow-Higgs-width 
approximation and supposes the $\PW$ and $\PZ$ gauge
bosons to be on shell and thus neglects, in particular, all
interferences between different four-fermion final states.
This approximation is generally not accurate enough for Higgs masses below
and near the $\PW\PW/\PZ\PZ$ thresholds.
For precision Higgs physics, it is strongly recommended
to use $\PH \to 4\Pf$ BRs whenever possible.
The ratio of above approximation \refE{approx} over \Prophecy\
prediction is shown in \refF{fig:BR-ratio}. 
For $\PH \to \Pep\Pem\Pep\Pem$ or $\PGmp\PGmm\PGmp\PGmm$ there is a
large enhancement in the BR due to interference for $\MH < 200$\UGeV\ (i.e.\ below $\PW\PW/\PZ\PZ$ thresholds).
Approximation \refE{approx} underestimates the BR for $\PH \to \Pep\Pem\Pep\Pem$
or $\PGmp\PGmm\PGmp\PGmm$ by $11\%$ at $\MH = 120\UGeV$.
For $\PH \to \Pep\PGne\Pem\PAGne$ or $\PH \to \PGmp\PGnGm\PGmm\PAGnGm$ 
there is an interference effect for $\MH < 200$\UGeV. 
Approximation \refE{approx} overestimates the BR for
$\PH \to \Pep\PGne\Pem\PAGne$ or $\PH \to \PGmp\PGnGm\PGmm\PAGnGm$ 
by $5.4\%$ at $\MH = 120\UGeV$.
Above the $\PW\PW/\PZ\PZ$ threshold, the approximation agrees with \Prophecy\ at the level of $1\%$.
For $\PH \to \PZ\PZ \to \Pl\PAl\PQq\PAQq, \PH \to \PW\PW \to \Plpm\PGn\PQq\PAQq$,
and $\PH \to \PZ\PZ \to \PGn\PAGn\PQq\PAQq$ the agreement is at the $1\%$ level.
For $\PH \to \PW\PW/\PZ\PZ \to \PQq\PAQq\PQq\PAQq$ there is an interference
effect for $\MH < 200\UGeV$.



A comparison of our results with previous calculations can be found in
\Bref{Denner:2011mq}.

\clearpage

\subsection{MSSM Higgs branching ratios}
\label{sec:MSSMHiggsBR}

In the MSSM the evaluation of cross sections and of branching ratios
have several common issues as outlined in \refS{sec:mssmvssm}.
It was discussed that {\em before} any branching-ratio calculation can be
performed in a first step the Higgs-boson masses, couplings, and mixings
have to be evaluated from the underlying set of (soft SUSY-breaking)
parameters. A brief comparison of the dedicated codes that provide this kind of
calculations
(\FeynHiggs~\cite{Heinemeyer:1998yj,Heinemeyer:1998np,Degrassi:2002fi,Frank:2006yh}
and \CPsuperH~\cite{Lee:2003nta,Lee:2007gn}) was been given in
\Bref{Dittmaier:2011ti}, where it was concluded
that in the case of real parameters more corrections are included into
\FeynHiggs. Consequently, \FeynHiggs\ was chosen for the corresponding
evaluations in this Report. The results for Higgs-boson masses and
couplings can be provided to other codes
(especially \HDECAY~\cite{Djouadi:1997yw,Spira:1997dg,hdecay2}) via the SUSY
Les Houches Accord~\cite{Skands:2003cj,Allanach:2008qq}. 

In the following subsections we describe how the relevant codes for the
calculation of partial decay widths, \FeynHiggs\ and \HDECAY, are
combined to give the most precise result for the Higgs-boson branching
ratios in the MSSM. Numerical results are shown within the \mhmaxx\
scenario~\cite{Carena:2002qg}, where it should be stressed that it would
be desirable to 
interpret the model-independent results of various Higgs-boson searches
at the LHC also in other benchmark models, see for instance
\Brefs{Carena:2002qg,AbdusSalam:2011fc}. 
We restrict the evaluation to $\tb = 1 \ldots 60$ and 
$\MA = 90 \UGeV \ldots 500 \UGeV$.


\subsubsection{Combination of calculations}


After the calculation of Higgs-boson masses and mixings from the
original SUSY input the branching-ratio calculation has to be
performed. This can be done with the codes \CPsuperH\ and
\FeynHiggs\ for real or complex parameters, or
\HDECAY\ for real parameters. The higher-order corrections included
in the calculation of the various decay channels differ in the three
codes. 

Here we concentrate on the MSSM with real parameters. We combine the
results from \HDECAY\ and \FeynHiggs\ on various decay channels to
obtain the most accurate result for the branching ratios currently
available. In a first
step, all partial widths have been calculated as accurately as
possible. Then the branching ratios have been derived from this full
set of partial widths. 
Concretely, we used \FeynHiggs\ for the evaluation of the
Higgs-boson masses and couplings from the original input
parameters, including corrections up to the two-loop level. 
\FeynHiggs\ results are furthermore used for the channels 
($\phi = \Ph,\PH,\PA$),

\begin{itemize}

\item
$\Gamma(\phitautau)$: 
a full one-loop calculation of the decay width is
included~\cite{Williams:2011bu}. 

\item
$\Gamma(\phimumu)$: 
a full one-loop calculation of the decay width is
included~\cite{Williams:2011bu}. 

\item
$\Gamma(\phiVV)$, $V = \PW^\pm, \PZ$:
results for a SM Higgs boson, taken from
\Prophecy~\cite{Bredenstein:2006rh,Bredenstein:2006ha,Prophecy4f} and
based on a full one-loop calculation, are dressed with effective
couplings for the respective coupling of the MSSM Higgs boson to SM
gauge bosons, see \refS{sec:effcoup}. It should be noted that this does
not correspond to a full one-loop calculation in the MSSM, and the
approximation may be insufficient for very low values of $\MA$.

\end{itemize}

The results for the Higgs-boson masses and couplings are passed to
\HDECAY\ via the SUSY Les Houches Accord~\cite{Skands:2003cj,Allanach:2008qq}.
Using these results the following channels have been evaluated by
\HDECAY, 

\begin{itemize}

\item
$\Gamma(\phibb)$: SM QCD corrections are included up to the four-loop
level
\cite{Gorishnii:1990zu,Gorishnii:1991zr,Kataev:1993be,Surguladze:1994gc,
Larin:1995sq,Chetyrkin:1995pd,Chetyrkin:1996sr,Baikov:2005rw}. The full
SUSY-QCD corrections
\cite{Dabelstein:1995js,Coarasa:1995yg,Eberl:1999he,Guasch:2003cv}
matched to the resummed bottom Yukawa coupling with respect to
$\Delta_{\PQb}$ terms have been included.  The $\Delta_{\PQb}$ terms are included
up to the leading two-loop
corrections~\cite{Noth:2008tw,Noth:2010jy,Mihaila:2010mp} within the
resummed Yukawa coupling. 

\item
$\Gamma(\phitt)$: SM QCD corrections are included up to NLO
\cite{Braaten:1980yq,Sakai:1980fa,Inami:1980qp,Gorishnii:1983cu,
Drees:1989du,Drees:1990dq,Drees:1991dq} interpolated to the massless
four-loop result
\cite{Gorishnii:1990zu,Gorishnii:1991zr,Kataev:1993be,Surguladze:1994gc,
Larin:1995sq,Chetyrkin:1995pd,Chetyrkin:1996sr,Baikov:2005rw} far above
the threshold, while no MSSM specific corrections are taken into
account.

\item
$\Gamma(\phicc)$: SM QCD corrections are included up to the four-loop
level
\cite{Gorishnii:1990zu,Gorishnii:1991zr,Kataev:1993be,Surguladze:1994gc,
Larin:1995sq,Chetyrkin:1995pd,Chetyrkin:1996sr,Baikov:2005rw}, while no
MSSM specific corrections are taken into account.

\item
$\Gamma(\phigg)$: SM QCD corrections are included up to four-loop order
in the limit of heavy top quarks
\cite{Inami:1982xt,Djouadi:1991tka,Spira:1995rr,
Chetyrkin:1997iv,Baikov:2006ch}. The squark-loop contributions including
the NLO QCD corrections \cite{Muhlleitner:2006wx} in the limit of heavy
squarks but no genuine MSSM corrections are taken into account.

\item
$\Gamma(\phigaga)$: The full SM QCD corrections are included up to NLO
\cite{Zheng:1990qa,Djouadi:1990aj,Dawson:1992cy,Djouadi:1993ji,
Melnikov:1993tj,Inoue:1994jq,Spira:1995rr}.  The additional charged
Higgs, squark, and chargino loop contributions including the full NLO QCD
corrections to the squark loops \cite{Muhlleitner:2006wx} are taken into 
account, but no genuine MSSM corrections.

\item
$\Gamma(\phiZga)$: No higher-order corrections are included. The
additional charged Higgs, squark, and chargino loop contributions are
taken into account at LO.

\end{itemize}

Other decay channels such as $\phiss$ and the decays to lighter fermions
are not included since they are neither relevant for LHC searches, nor
do they contribute significantly to the total decay width.
With future releases of \FeynHiggs, \HDECAY, and other codes the
evaluation of individual channels might change to another code.

The total decay width is calculated as,
\begin{align}
\Gamma_\phi &= 
  \Gamma^{\mathrm{FH}}_{\phitautau} 
+ \Gamma^{\mathrm{FH}}_{\phimumu} 
+ \Gamma^{\mathrm{FH/P4f}}_{\phiWW} 
+ \Gamma^{\mathrm{FH/P4f}}_{\phiZZ} \nonumber\\
&\quad 
+ \Gamma^{\mathrm{HD}}_{\phibb}
+ \Gamma^{\mathrm{HD}}_{\phitt}
+ \Gamma^{\mathrm{HD}}_{\phicc}
+ \Gamma^{\mathrm{HD}}_{\phigg}
+ \Gamma^{\mathrm{HD}}_{\phigaga}
+ \Gamma^{\mathrm{HD}}_{\phiZga}~,
\end{align}
followed by a corresponding evaluation of the respective branching
ratio. Decays to strange quarks or other lighter fermions have been neglected. 
Due to the somewhat different calculation compared to the SM case in
\refS{sec:Programs} no full decoupling of the decay widths and
branching ratios of the light MSSM Higgs to the respective SM values can
be expected.


\subsubsection{Results in the \boldmath{$m_h^{\rm max}$} scenario}

\begin{figure}[htb!]
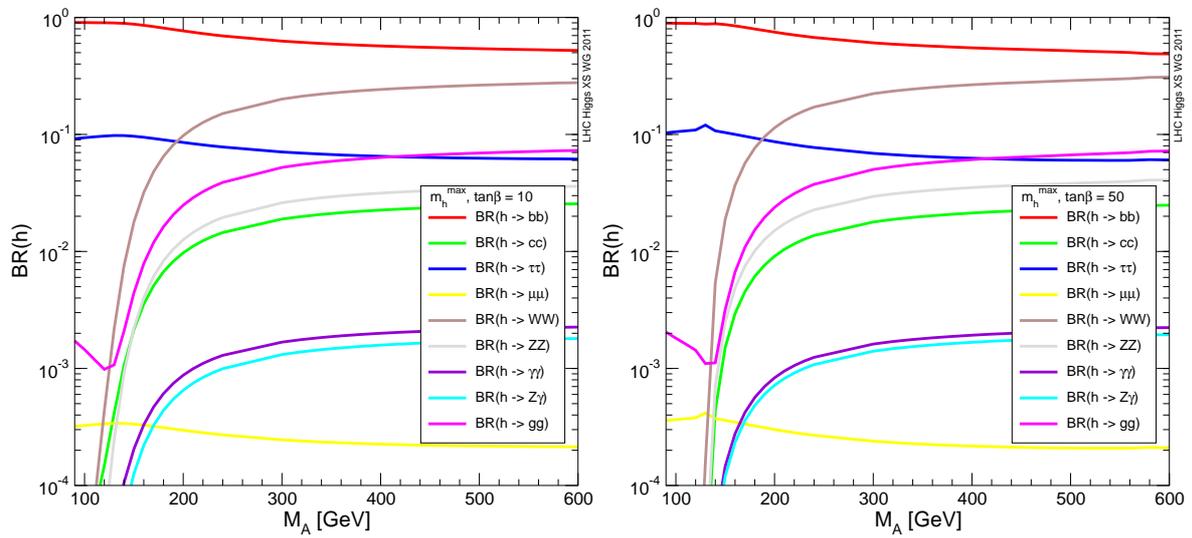

\includegraphics[width=0.48\textwidth]{YRHXS2_BR/YRHXS2_BR_Fig6.eps}
\includegraphics[width=0.48\textwidth]{YRHXS2_BR/YRHXS2_BR_Fig7.eps}
\vspace*{-0.1cm}
\caption{Higgs branching ratios for the light MSSM Higgs boson
for the relevant final states. The parameters are chosen according to the \mhmaxx\
scenario, see \Eref{YRHXS_MSSM_neutral_eq:mhmax}, with $\tb = 10 (50)$ in the
left (right) plot.
}
\label{fig:BR-h-mhmax}
\end{figure}

The procedure outlined in the previous subsection can be applied to
arbitrary points in the MSSM parameter space. Here we show
representative results in the \mhmaxx\ scenario.
In \refF{fig:BR-h-mhmax} the branching ratios for the light MSSM Higgs boson
are shown as a function of $\MA$ for $\tb = 10 (50)$ in the left (right)
plot. For low $\MA$ a strong variation of the branching ratios is visible,
while for large $\MA$ the SM limit is reached.


More detailed sample results for the branching ratio of the lightest MSSM
Higgs boson to $\PQb\PAQb$, $\PGtp\PGtm$, $\PGmp\PGmm$, $\PGg\PGg$,
$\PW^{(*)}\PW^{(*)}$, and $\PZ^{(*)}\PZ^{(*)}$ are shown in 
\refTs{tab:BR-hbb}--\ref{tab:BR-hZZ}. 
As before they have been obtained in the \mhmaxx\ scenario
with $\MA$ and $\tb$ varied (as indicated in the tables). 

\begin{table}
  \vspace{-\headsep}
  \caption{MSSM Higgs branching ratio BR($\Ph \to \PQb\PAQb$) in the 
$m_{\PQb}^{\rm max}$ scenario as a function of $\MA$ [GeV] and $\tb$.
The format in each cell is $M_{\Ph}$ [GeV], BR.
}
  \label{tab:BR-hbb}
  \centering
{
  \setlength{\tabcolsep}{4.5pt}
  \small
  \begin{tabular}{lrcrcrcrcrc}\hline
{\small $\MA$} & \multicolumn{2}{c}{$\tb = 20$}
      & \multicolumn{2}{c}{$\tb = 30$}
            & \multicolumn{2}{c}{$\tb = 40$}
            & \multicolumn{2}{c}{$\tb = 50$}
            & \multicolumn{2}{c}{$\tb = 60$} \\
\hline
 90 & 89.6 & 9.02 \zehomi{1} & 89.8 & 9.00 \zehomi{1} & 
      89.9 & 8.97 \zehomi{1} & 89.9 & 8.94 \zehomi{1} & 
      89.9 & 8.91 \zehomi{1} \\

100 & 99.4 & 9.01 \zehomi{1} & 99.7 & 8.98 \zehomi{1} & 
      99.9 & 8.95 \zehomi{1} & 99.9 & 8.92 \zehomi{1} & 
      99.9 & 8.89 \zehomi{1} \\

110 & 109.0 & 8.99 \zehomi{1} & 109.6 & 8.96 \zehomi{1} & 
      109.8 & 8.93 \zehomi{1} & 109.8 & 8.90 \zehomi{1} & 
      109.9 & 8.87 \zehomi{1} \\

120 & 118.2 & 8.98 \zehomi{1} & 119.1 & 8.95 \zehomi{1} & 
      119.5 & 8.92 \zehomi{1} & 119.7 & 8.89 \zehomi{1} & 
      119.7 & 8.86 \zehomi{1} \\

130 & 125.2 & 8.95 \zehomi{1} & 126.9 & 8.91 \zehomi{1} & 
      127.8 & 8.86 \zehomi{1} & 128.4 & 8.78 \zehomi{1} & 
      128.9 & 8.62 \zehomi{1} \\

140 & 128.1 & 8.89 \zehomi{1} & 129.3 & 8.87 \zehomi{1} & 
      129.8 & 8.86 \zehomi{1} & 130.2 & 8.84 \zehomi{1} & 
      130.5 & 8.84 \zehomi{1} \\

150 & 128.9 & 8.75 \zehomi{1} & 129.7 & 8.73 \zehomi{1} & 
      130.1 & 8.71 \zehomi{1} & 130.4 & 8.69 \zehomi{1} & 
      130.6 & 8.68 \zehomi{1} \\

160 & 129.2 & 8.54 \zehomi{1} & 129.9 & 8.52 \zehomi{1} & 
      130.2 & 8.49 \zehomi{1} & 130.4 & 8.47 \zehomi{1} & 
      130.6 & 8.45 \zehomi{1} \\

170 & 129.4 & 8.30 \zehomi{1} & 130.0 & 8.27 \zehomi{1} & 
      130.2 & 8.25 \zehomi{1} & 130.4 & 8.22 \zehomi{1} & 
      130.6 & 8.20 \zehomi{1} \\

180 & 129.5 & 8.06 \zehomi{1} & 130.0 & 8.03 \zehomi{1} & 
      130.2 & 8.00 \zehomi{1} & 130.4 & 7.97 \zehomi{1} & 
      130.6 & 7.94 \zehomi{1} \\

190 & 129.6 & 7.82 \zehomi{1} & 130.0 & 7.78 \zehomi{1} & 
      130.2 & 7.75 \zehomi{1} & 130.4 & 7.72 \zehomi{1} & 
      130.6 & 7.69 \zehomi{1} \\

200 & 129.7 & 7.60 \zehomi{1} & 130.1 & 7.56 \zehomi{1} & 
      130.2 & 7.52 \zehomi{1} & 130.4 & 7.49 \zehomi{1} & 
      130.6 & 7.46 \zehomi{1} \\

210 & 129.7 & 7.39 \zehomi{1} & 130.1 & 7.35 \zehomi{1} & 
      130.2 & 7.31 \zehomi{1} & 130.4 & 7.28 \zehomi{1} & 
      130.6 & 7.24 \zehomi{1} \\

220 & 129.7 & 7.20 \zehomi{1} & 130.1 & 7.15 \zehomi{1} & 
      130.3 & 7.12 \zehomi{1} & 130.4 & 7.08 \zehomi{1} & 
      130.6 & 7.05 \zehomi{1} \\

230 & 129.8 & 7.02 \zehomi{1} & 130.1 & 6.98 \zehomi{1} & 
      130.3 & 6.94 \zehomi{1} & 130.4 & 6.91 \zehomi{1} & 
      130.6 & 6.87 \zehomi{1} \\

240 & 129.8 & 6.86 \zehomi{1} & 130.1 & 6.82 \zehomi{1} & 
      130.3 & 6.78 \zehomi{1} & 130.4 & 6.75 \zehomi{1} & 
      130.6 & 6.71 \zehomi{1} \\

250 & 129.8 & 6.72 \zehomi{1} & 130.1 & 6.68 \zehomi{1} & 
      130.3 & 6.64 \zehomi{1} & 130.4 & 6.60 \zehomi{1} & 
      130.6 & 6.56 \zehomi{1} \\

260 & 129.8 & 6.59 \zehomi{1} & 130.1 & 6.55 \zehomi{1} & 
      130.3 & 6.51 \zehomi{1} & 130.4 & 6.47 \zehomi{1} & 
      130.6 & 6.43 \zehomi{1} \\

270 & 129.8 & 6.48 \zehomi{1} & 130.1 & 6.43 \zehomi{1} & 
      130.3 & 6.39 \zehomi{1} & 130.4 & 6.36 \zehomi{1} & 
      130.6 & 6.32 \zehomi{1} \\

280 & 129.8 & 6.37 \zehomi{1} & 130.1 & 6.33 \zehomi{1} & 
      130.3 & 6.29 \zehomi{1} & 130.4 & 6.25 \zehomi{1} & 
      130.6 & 6.21 \zehomi{1} \\

290 & 129.8 & 6.28 \zehomi{1} & 130.1 & 6.23 \zehomi{1} & 
      130.3 & 6.19 \zehomi{1} & 130.4 & 6.16 \zehomi{1} & 
      130.6 & 6.11 \zehomi{1} \\

300 & 129.8 & 6.19 \zehomi{1} & 130.1 & 6.14 \zehomi{1} & 
      130.3 & 6.11 \zehomi{1} & 130.4 & 6.07 \zehomi{1} & 
      130.6 & 6.02 \zehomi{1} \\

\hline
  \end{tabular}
}
\end{table}

\begin{table}
  \vspace{-.5\headsep}
  \caption{MSSM Higgs branching ratio BR($\Ph \to \PGtp\PGtm$) in the 
$m_{\Ph}^{\rm max}$ scenario as a function of $\MA$ [GeV] and $\tb$.
The format in each cell is $M_{\Ph}$ [GeV], BR.
}
  \label{tab:BR-htautau}
  \centering
{
  \setlength{\tabcolsep}{4.5pt}
  \small
  \begin{tabular}{lrcrcrcrcrc}\hline
{\small $\MA$} & \multicolumn{2}{c}{$\tb = 20$}
      & \multicolumn{2}{c}{$\tb = 30$}
            & \multicolumn{2}{c}{$\tb = 40$}
            & \multicolumn{2}{c}{$\tb = 50$}
            & \multicolumn{2}{c}{$\tb = 60$} \\
\hline
 90 & 89.6 & 9.50 \zehomi{2} & 89.8 & 9.77 \zehomi{2} & 
      89.9 & 1.01 \zehomi{1} & 89.9 & 1.03 \zehomi{1} & 
      89.9 & 1.06 \zehomi{1} \\

100 & 99.4 & 9.68 \zehomi{2} & 99.7 & 9.96 \zehomi{2} & 
      99.9 & 1.02 \zehomi{1} & 99.9 & 1.05 \zehomi{1} & 
      99.9 & 1.08 \zehomi{1} \\

110 & 109.0 & 9.84 \zehomi{2} & 109.6 & 1.01 \zehomi{1} & 
      109.8 & 1.04 \zehomi{1} & 109.8 & 1.07 \zehomi{1} & 
      109.9 & 1.1 \zehomi{1} \\

120 & 118.2 & 1.00 \zehomi{1} & 119.1 & 1.03 \zehomi{1} & 
      119.5 & 1.06 \zehomi{1} & 119.7 & 1.09 \zehomi{1} & 
      119.7 & 1.12 \zehomi{1} \\

130 & 125.2 & 1.02 \zehomi{1} & 126.9 & 1.06 \zehomi{1} & 
      127.8 & 1.12 \zehomi{1} & 128.4 & 1.20 \zehomi{1} & 
      128.9 & 1.36 \zehomi{1} \\

140 & 128.1 & 1.01 \zehomi{1} & 129.3 & 1.04 \zehomi{1} & 
      129.8 & 1.06 \zehomi{1} & 130.2 & 1.07 \zehomi{1} & 
      130.5 & 1.08 \zehomi{1} \\

150 & 128.9 & 9.83 \zehomi{2} & 129.7 & 1.00 \zehomi{1} & 
      130.1 & 1.02 \zehomi{1} & 130.4 & 1.04 \zehomi{1} & 
      130.6 & 1.05 \zehomi{1} \\

160 & 129.2 & 9.54 \zehomi{2} & 129.9 & 9.70 \zehomi{2} & 
      130.2 & 9.86 \zehomi{2} & 130.4 & 1.00 \zehomi{1} & 
      130.6 & 1.02 \zehomi{1} \\

170 & 129.4 & 9.25 \zehomi{2} & 130.0 & 9.38 \zehomi{2} & 
      130.2 & 9.52 \zehomi{2} & 130.4 & 9.65 \zehomi{2} & 
      130.6 & 9.79 \zehomi{2} \\

180 & 129.5 & 8.95 \zehomi{2} & 130.0 & 9.06 \zehomi{2} & 
      130.2 & 9.18 \zehomi{2} & 130.4 & 9.31 \zehomi{2} & 
      130.6 & 9.43 \zehomi{2} \\

190 & 129.6 & 8.67 \zehomi{2} & 130.0 & 8.76 \zehomi{2} & 
      130.2 & 8.86 \zehomi{2} & 130.4 & 8.97 \zehomi{2} & 
      130.6 & 9.08 \zehomi{2} \\

200 & 129.7 & 8.42 \zehomi{2} & 130.1 & 8.48 \zehomi{2} & 
      130.2 & 8.58 \zehomi{2} & 130.4 & 8.67 \zehomi{2} & 
      130.6 & 8.77 \zehomi{2} \\

210 & 129.7 & 8.18 \zehomi{2} & 130.1 & 8.23 \zehomi{2} & 
      130.2 & 8.31 \zehomi{2} & 130.4 & 8.40 \zehomi{2} & 
      130.6 & 8.49 \zehomi{2} \\

220 & 129.7 & 7.96 \zehomi{2} & 130.1 & 8.00 \zehomi{2} & 
      130.3 & 8.08 \zehomi{2} & 130.4 & 8.15 \zehomi{2} & 
      130.6 & 8.23 \zehomi{2} \\

230 & 129.8 & 7.77 \zehomi{2} & 130.1 & 7.80 \zehomi{2} & 
      130.3 & 7.86 \zehomi{2} & 130.4 & 7.93 \zehomi{2} & 
      130.6 & 8.01 \zehomi{2} \\

240 & 129.8 & 7.59 \zehomi{2} & 130.1 & 7.62 \zehomi{2} & 
      130.3 & 7.67 \zehomi{2} & 130.4 & 7.74 \zehomi{2} & 
      130.6 & 7.81 \zehomi{2} \\

250 & 129.8 & 7.43 \zehomi{2} & 130.1 & 7.45 \zehomi{2} & 
      130.3 & 7.50 \zehomi{2} & 130.4 & 7.56 \zehomi{2} & 
      130.6 & 7.62 \zehomi{2} \\

260 & 129.8 & 7.28 \zehomi{2} & 130.1 & 7.30 \zehomi{2} & 
      130.3 & 7.34 \zehomi{2} & 130.4 & 7.40 \zehomi{2} & 
      130.6 & 7.45 \zehomi{2} \\

270 & 129.8 & 7.15 \zehomi{2} & 130.1 & 7.16 \zehomi{2} & 
      130.3 & 7.20 \zehomi{2} & 130.4 & 7.25 \zehomi{2} & 
      130.6 & 7.30 \zehomi{2} \\

280 & 129.8 & 7.03 \zehomi{2} & 130.1 & 7.04 \zehomi{2} & 
      130.3 & 7.07 \zehomi{2} & 130.4 & 7.12 \zehomi{2} & 
      130.6 & 7.17 \zehomi{2} \\

290 & 129.8 & 6.92 \zehomi{2} & 130.1 & 6.92 \zehomi{2} & 
      130.3 & 6.96 \zehomi{2} & 130.4 & 7.00 \zehomi{2} & 
      130.6 & 7.05 \zehomi{2} \\

300 & 129.8 & 6.82 \zehomi{2} & 130.1 & 6.82 \zehomi{2} & 
      130.3 & 6.85 \zehomi{2} & 130.4 & 6.90 \zehomi{2} & 
      130.6 & 6.94 \zehomi{2} \\

\hline
  \end{tabular}
}
\end{table}

\begin{table}
  \vspace{-\headsep}
  \caption{MSSM Higgs branching ratio BR($\Ph \to \PGmp\PGmm$) in the 
$m_{\Ph}^{\rm max}$ scenario as a function of $\MA$ [GeV] and $\tb$.
The format in each cell is $M_{\Ph}$ [GeV], BR.
}
  \label{tab:BR-hmumu}
  \centering
{
  \setlength{\tabcolsep}{4.5pt}
  \small
  \begin{tabular}{lrcrcrcrcrc}\hline
{\small $\MA$} & \multicolumn{2}{c}{$\tb = 20$}
      & \multicolumn{2}{c}{$\tb = 30$}
            & \multicolumn{2}{c}{$\tb = 40$}
            & \multicolumn{2}{c}{$\tb = 50$}
            & \multicolumn{2}{c}{$\tb = 60$} \\
\hline
 90 & 89.6 & 3.30 \zehomi{4} & 89.8 & 3.39 \zehomi{4} & 
      89.9 & 3.49 \zehomi{4} & 89.9 & 3.59 \zehomi{4} & 
      89.9 & 3.68 \zehomi{4} \\

100 & 99.4 & 3.36 \zehomi{4} & 99.7 & 3.46 \zehomi{4} & 
      99.9 & 3.56 \zehomi{4} & 99.9 & 3.66 \zehomi{4} & 
      99.9 & 3.76 \zehomi{4} \\
 
110 & 109.0 & 3.42 \zehomi{4} & 109.6 & 3.52 \zehomi{4} & 
      109.8 & 3.62 \zehomi{4} & 109.8 & 3.72 \zehomi{4} & 
      109.9 & 3.82 \zehomi{4} \\
 
120 & 118.2 & 3.47 \zehomi{4} & 119.1 & 3.58 \zehomi{4} & 
      119.5 & 3.68 \zehomi{4} & 119.7 & 3.78 \zehomi{4} & 
      119.7 & 3.89 \zehomi{4} \\
 
130 & 125.2 & 3.54 \zehomi{4} & 126.9 & 3.69 \zehomi{4} & 
      127.8 & 3.87 \zehomi{4} & 128.4 & 4.17 \zehomi{4} & 
      128.9 & 4.72 \zehomi{4} \\
 
140 & 128.1 & 3.5 \zehomi{4} & 129.3 & 3.59 \zehomi{4} & 
      129.8 & 3.67 \zehomi{4} & 130.2 & 3.72 \zehomi{4} & 
      130.5 & 3.75 \zehomi{4} \\
 
150 & 128.9 & 3.41 \zehomi{4} & 129.7 & 3.48 \zehomi{4} & 
      130.1 & 3.54 \zehomi{4} & 130.4 & 3.60 \zehomi{4} & 
      130.6 & 3.65 \zehomi{4} \\
 
160 & 129.2 & 3.31 \zehomi{4} & 129.9 & 3.36 \zehomi{4} & 
      130.2 & 3.42 \zehomi{4} & 130.4 & 3.47 \zehomi{4} & 
      130.6 & 3.53 \zehomi{4} \\
 
170 & 129.4 & 3.21 \zehomi{4} & 130.0 & 3.25 \zehomi{4} & 
      130.2 & 3.30 \zehomi{4} & 130.4 & 3.35 \zehomi{4} & 
      130.6 & 3.40 \zehomi{4} \\
 
180 & 129.5 & 3.11 \zehomi{4} & 130.0 & 3.14 \zehomi{4} & 
      130.2 & 3.18 \zehomi{4} & 130.4 & 3.23 \zehomi{4} & 
      130.6 & 3.27 \zehomi{4} \\
 
190 & 129.6 & 3.01 \zehomi{4} & 130.0 & 3.04 \zehomi{4} & 
      130.2 & 3.07 \zehomi{4} & 130.4 & 3.11 \zehomi{4} & 
      130.6 & 3.15 \zehomi{4} \\
 
200 & 129.7 & 2.92 \zehomi{4} & 130.1 & 2.94 \zehomi{4} & 
      130.2 & 2.97 \zehomi{4} & 130.4 & 3.01 \zehomi{4} & 
      130.6 & 3.04 \zehomi{4} \\
 
210 & 129.7 & 2.84 \zehomi{4} & 130.1 & 2.86 \zehomi{4} & 
      130.2 & 2.88 \zehomi{4} & 130.4 & 2.91 \zehomi{4} & 
      130.6 & 2.94 \zehomi{4} \\
 
220 & 129.7 & 2.76 \zehomi{4} & 130.1 & 2.78 \zehomi{4} & 
      130.3 & 2.80 \zehomi{4} & 130.4 & 2.83 \zehomi{4} & 
      130.6 & 2.86 \zehomi{4} \\
 
230 & 129.8 & 2.69 \zehomi{4} & 130.1 & 2.70 \zehomi{4} & 
      130.3 & 2.73 \zehomi{4} & 130.4 & 2.75 \zehomi{4} & 
      130.6 & 2.78 \zehomi{4} \\
 
240 & 129.8 & 2.63 \zehomi{4} & 130.1 & 2.64 \zehomi{4} & 
      130.3 & 2.66 \zehomi{4} & 130.4 & 2.68 \zehomi{4} & 
      130.6 & 2.71 \zehomi{4} \\
 
250 & 129.8 & 2.58 \zehomi{4} & 130.1 & 2.58 \zehomi{4} & 
      130.3 & 2.60 \zehomi{4} & 130.4 & 2.62 \zehomi{4} & 
      130.6 & 2.64 \zehomi{4} \\
 
260 & 129.8 & 2.53 \zehomi{4} & 130.1 & 2.53 \zehomi{4} & 
      130.3 & 2.55 \zehomi{4} & 130.4 & 2.56 \zehomi{4} & 
      130.6 & 2.59 \zehomi{4} \\
 
270 & 129.8 & 2.48 \zehomi{4} & 130.1 & 2.48 \zehomi{4} & 
      130.3 & 2.50 \zehomi{4} & 130.4 & 2.51 \zehomi{4} & 
      130.6 & 2.53 \zehomi{4} \\
 
280 & 129.8 & 2.44 \zehomi{4} & 130.1 & 2.44 \zehomi{4} & 
      130.3 & 2.45 \zehomi{4} & 130.4 & 2.47 \zehomi{4} & 
      130.6 & 2.49 \zehomi{4} \\
 
290 & 129.8 & 2.40 \zehomi{4} & 130.1 & 2.40 \zehomi{4} & 
      130.3 & 2.41 \zehomi{4} & 130.4 & 2.43 \zehomi{4} & 
      130.6 & 2.45 \zehomi{4} \\
 
300 & 129.8 & 2.37 \zehomi{4} & 130.1 & 2.37 \zehomi{4} & 
      130.3 & 2.38 \zehomi{4} & 130.4 & 2.39 \zehomi{4} & 
      130.6 & 2.41 \zehomi{4} \\
 
\hline
  \end{tabular}
}
\end{table}

\begin{table}
  \vspace{-.5\headsep}
  \caption{MSSM Higgs branching ratio BR($\Ph \to \PGg\PGg$) in the 
$m_{\Ph}^{\rm max}$ scenario as a function of $\MA$ [GeV] and $\tb$.
The format in each cell is $M_{\Ph}$ [GeV], BR.
}
  \label{tab:BR-hgaga}
  \centering
{
  \setlength{\tabcolsep}{4.5pt}
  \small
  \begin{tabular}{lrcrcrcrcrc}\hline
{\small $\MA$} & \multicolumn{2}{c}{$\tb = 20$}
      & \multicolumn{2}{c}{$\tb = 30$}
            & \multicolumn{2}{c}{$\tb = 40$}
            & \multicolumn{2}{c}{$\tb = 50$}
            & \multicolumn{2}{c}{$\tb = 60$} \\
\hline
 90 & 89.6 & 9.76 \zehomi{7} & 89.8 & 6.73 \zehomi{7} & 
      89.9 & 5.83 \zehomi{7} & 89.9 & 5.48 \zehomi{7} & 
      89.9 & 5.34 \zehomi{7} \\
 
100 & 99.4 & 1.19 \zehomi{6} & 99.7 & 7.24 \zehomi{7} & 
      99.9 & 5.90 \zehomi{7} & 99.9 & 5.36 \zehomi{7} & 
      99.9 & 5.12 \zehomi{7} \\
 
110 & 109.0 & 1.70 \zehomi{6} & 109.6 & 8.76 \zehomi{7} & 
      109.8 & 6.48 \zehomi{7} & 109.8 & 5.56 \zehomi{7} & 
      109.9 & 5.13 \zehomi{7} \\
 
120 & 118.2 & 3.43 \zehomi{6} & 119.1 & 1.45 \zehomi{6} & 
      119.5 & 9.04 \zehomi{7} & 119.7 & 6.91 \zehomi{7} & 
      119.7 & 5.89 \zehomi{7} \\
 
130 & 125.2 & 1.37 \zehomi{5} & 126.9 & 6.88 \zehomi{6} & 
      127.8 & 4.24 \zehomi{6} & 128.4 & 2.89 \zehomi{6} & 
      128.9 & 2.07 \zehomi{6} \\
 
140 & 128.1 & 6.25 \zehomi{5} & 129.3 & 5.22 \zehomi{5} & 
      129.8 & 4.79 \zehomi{5} & 130.2 & 4.54 \zehomi{5} & 
      130.5 & 4.36 \zehomi{5} \\
 
150 & 128.9 & 1.60 \zehomi{4} & 129.7 & 1.51 \zehomi{4} & 
      130.1 & 1.47 \zehomi{4} & 130.4 & 1.45 \zehomi{4} & 
      130.6 & 1.44 \zehomi{4} \\
 
160 & 129.2 & 2.88 \zehomi{4} & 129.9 & 2.80 \zehomi{4} & 
      130.2 & 2.77 \zehomi{4} & 130.4 & 2.77 \zehomi{4} & 
      130.6 & 2.77 \zehomi{4} \\
 
170 & 129.4 & 4.28 \zehomi{4} & 130.0 & 4.21 \zehomi{4} & 
      130.2 & 4.20 \zehomi{4} & 130.4 & 4.21 \zehomi{4} & 
      130.6 & 4.22 \zehomi{4} \\
 
180 & 129.5 & 5.70 \zehomi{4} & 130.0 & 5.63 \zehomi{4} & 
      130.2 & 5.63 \zehomi{4} & 130.4 & 5.65 \zehomi{4} & 
      130.6 & 5.68 \zehomi{4} \\
 
190 & 129.6 & 7.06 \zehomi{4} & 130.0 & 7.00 \zehomi{4} & 
      130.2 & 7.00 \zehomi{4} & 130.4 & 7.03 \zehomi{4} & 
      130.6 & 7.07 \zehomi{4} \\
 
200 & 129.7 & 8.33 \zehomi{4} & 130.1 & 8.28 \zehomi{4} & 
      130.2 & 8.29 \zehomi{4} & 130.4 & 8.32 \zehomi{4} & 
      130.6 & 8.37 \zehomi{4} \\
 
210 & 129.7 & 9.50 \zehomi{4} & 130.1 & 9.45 \zehomi{4} & 
      130.2 & 9.47 \zehomi{4} & 130.4 & 9.51 \zehomi{4} & 
      130.6 & 9.56 \zehomi{4} \\
 
220 & 129.7 & 1.06 \zehomi{3} & 130.1 & 1.05 \zehomi{3} & 
      130.3 & 1.05 \zehomi{3} & 130.4 & 1.06 \zehomi{3} & 
      130.6 & 1.06 \zehomi{3} \\
 
230 & 129.8 & 1.15 \zehomi{3} & 130.1 & 1.15 \zehomi{3} & 
      130.3 & 1.15 \zehomi{3} & 130.4 & 1.16 \zehomi{3} & 
      130.6 & 1.16 \zehomi{3} \\
 
240 & 129.8 & 1.24 \zehomi{3} & 130.1 & 1.24 \zehomi{3} & 
      130.3 & 1.24 \zehomi{3} & 130.4 & 1.24 \zehomi{3} & 
      130.6 & 1.25 \zehomi{3} \\
 
250 & 129.8 & 1.32 \zehomi{3} & 130.1 & 1.32 \zehomi{3} & 
      130.3 & 1.32 \zehomi{3} & 130.4 & 1.32 \zehomi{3} & 
      130.6 & 1.33 \zehomi{3} \\
 
260 & 129.8 & 1.39 \zehomi{3} & 130.1 & 1.39 \zehomi{3} & 
      130.3 & 1.39 \zehomi{3} & 130.4 & 1.39 \zehomi{3} & 
      130.6 & 1.4 \zehomi{3} \\
 
270 & 129.8 & 1.46 \zehomi{3} & 130.1 & 1.45 \zehomi{3} & 
      130.3 & 1.45 \zehomi{3} & 130.4 & 1.46 \zehomi{3} & 
      130.6 & 1.47 \zehomi{3} \\
 
280 & 129.8 & 1.51 \zehomi{3} & 130.1 & 1.51 \zehomi{3} & 
      130.3 & 1.51 \zehomi{3} & 130.4 & 1.52 \zehomi{3} & 
      130.6 & 1.52 \zehomi{3} \\
 
290 & 129.8 & 1.57 \zehomi{3} & 130.1 & 1.56 \zehomi{3} & 
      130.3 & 1.56 \zehomi{3} & 130.4 & 1.57 \zehomi{3} & 
      130.6 & 1.58 \zehomi{3} \\
 
300 & 129.8 & 1.61 \zehomi{3} & 130.1 & 1.61 \zehomi{3} & 
      130.3 & 1.61 \zehomi{3} & 130.4 & 1.62 \zehomi{3} & 
      130.6 & 1.62 \zehomi{3} \\
 
\hline
  \end{tabular}
}
\end{table}

\begin{table}
  \vspace{-\headsep}
  \caption{MSSM Higgs branching ratio BR($\Ph \to \PW^{(*)}\PW^{(*)}$) in the 
$m_{\Ph}^{\rm max}$ scenario as a function of $\MA$ [GeV] and $\tb$.
The format in each cell is $M_{\Ph}$ [GeV], BR.
}
  \label{tab:BR-hWW}
  \centering
{
  \setlength{\tabcolsep}{4.5pt}
  \small
  \begin{tabular}{lrcrcrcrcrc}\hline
      {\small $\MA$} & \multicolumn{2}{c}{$\tb = 20$}
      & \multicolumn{2}{c}{$\tb = 30$}
            & \multicolumn{2}{c}{$\tb = 40$}
            & \multicolumn{2}{c}{$\tb = 50$}
            & \multicolumn{2}{c}{$\tb = 60$} \\
\hline
 90 & 89.6 & 6.95 \zehomi{8} & 89.8 & 1.54 \zehomi{8} & 
      89.9 & 5.21 \zehomi{9} & 89.9 & 2.28 \zehomi{9} & 
      89.9 & 1.1 \zehomi{9} \\
 
100 & 99.4 & 9.14 \zehomi{7} & 99.7 & 2.04 \zehomi{7} & 
      99.9 & 6.91 \zehomi{8} & 99.9 & 2.96 \zehomi{8} & 
      99.9 & 1.49 \zehomi{8} \\
 
110 & 109.0 & 8.19 \zehomi{6} & 109.6 & 1.89 \zehomi{6} & 
      109.8 & 6.41 \zehomi{7} & 109.8 & 2.75 \zehomi{7} & 
      109.9 & 1.37 \zehomi{7} \\
 
120 & 118.2 & 7.36 \zehomi{5} & 119.1 & 1.93 \zehomi{5} & 
      119.5 & 6.85 \zehomi{6} & 119.7 & 2.96 \zehomi{6} & 
      119.7 & 1.45 \zehomi{6} \\
 
130 & 125.2 & 8.81 \zehomi{4} & 126.9 & 4.46 \zehomi{4} & 
      127.8 & 2.58 \zehomi{4} & 128.4 & 1.60 \zehomi{4} & 
      128.9 & 1.02 \zehomi{4} \\
 
140 & 128.1 & 6.29 \zehomi{3} & 129.3 & 5.79 \zehomi{3} & 
      129.8 & 5.55 \zehomi{3} & 130.2 & 5.41 \zehomi{3} & 
      130.5 & 5.31 \zehomi{3} \\
 
150 & 128.9 & 1.84 \zehomi{2} & 129.7 & 1.85 \zehomi{2} & 
      130.1 & 1.87 \zehomi{2} & 130.4 & 1.88 \zehomi{2} & 
      130.6 & 1.9 \zehomi{2} \\
 
160 & 129.2 & 3.48 \zehomi{2} & 129.9 & 3.57 \zehomi{2} & 
      130.2 & 3.63 \zehomi{2} & 130.4 & 3.68 \zehomi{2} & 
      130.6 & 3.74 \zehomi{2} \\
 
170 & 129.4 & 5.31 \zehomi{2} & 130.0 & 5.47 \zehomi{2} & 
      130.2 & 5.58 \zehomi{2} & 130.4 & 5.67 \zehomi{2} & 
      130.6 & 5.77 \zehomi{2} \\
 
180 & 129.5 & 7.18 \zehomi{2} & 130.0 & 7.40 \zehomi{2} & 
      130.2 & 7.54 \zehomi{2} & 130.4 & 7.66 \zehomi{2} & 
      130.6 & 7.79 \zehomi{2} \\
 
190 & 129.6 & 9.00 \zehomi{2} & 130.0 & 9.26 \zehomi{2} & 
      130.2 & 9.43 \zehomi{2} & 130.4 & 9.59 \zehomi{2} & 
      130.6 & 9.75 \zehomi{2} \\
 
200 & 129.7 & 1.07 \zehomi{1} & 130.1 & 1.10 \zehomi{1} & 
      130.2 & 1.12 \zehomi{1} & 130.4 & 1.14 \zehomi{1} & 
      130.6 & 1.16 \zehomi{1} \\
 
210 & 129.7 & 1.23 \zehomi{1} & 130.1 & 1.26 \zehomi{1} & 
      130.2 & 1.28 \zehomi{1} & 130.4 & 1.30 \zehomi{1} & 
      130.6 & 1.33 \zehomi{1} \\
 
220 & 129.7 & 1.37 \zehomi{1} & 130.1 & 1.41 \zehomi{1} & 
      130.3 & 1.43 \zehomi{1} & 130.4 & 1.45 \zehomi{1} & 
      130.6 & 1.48 \zehomi{1} \\
 
230 & 129.8 & 1.50 \zehomi{1} & 130.1 & 1.54 \zehomi{1} & 
      130.3 & 1.57 \zehomi{1} & 130.4 & 1.59 \zehomi{1} & 
      130.6 & 1.62 \zehomi{1} \\
 
240 & 129.8 & 1.62 \zehomi{1} & 130.1 & 1.66 \zehomi{1} & 
      130.3 & 1.69 \zehomi{1} & 130.4 & 1.71 \zehomi{1} & 
      130.6 & 1.74 \zehomi{1} \\
 
250 & 129.8 & 1.73 \zehomi{1} & 130.1 & 1.77 \zehomi{1} & 
      130.3 & 1.80 \zehomi{1} & 130.4 & 1.82 \zehomi{1} & 
      130.6 & 1.85 \zehomi{1} \\
 
260 & 129.8 & 1.83 \zehomi{1} & 130.1 & 1.87 \zehomi{1} & 
      130.3 & 1.89 \zehomi{1} & 130.4 & 1.92 \zehomi{1} & 
      130.6 & 1.95 \zehomi{1} \\
 
270 & 129.8 & 1.91 \zehomi{1} & 130.1 & 1.95 \zehomi{1} & 
      130.3 & 1.98 \zehomi{1} & 130.4 & 2.01 \zehomi{1} & 
      130.6 & 2.04 \zehomi{1} \\
 
280 & 129.8 & 1.99 \zehomi{1} & 130.1 & 2.03 \zehomi{1} & 
      130.3 & 2.06 \zehomi{1} & 130.4 & 2.09 \zehomi{1} & 
      130.6 & 2.12 \zehomi{1} \\
 
290 & 129.8 & 2.06 \zehomi{1} & 130.1 & 2.11 \zehomi{1} & 
      130.3 & 2.14 \zehomi{1} & 130.4 & 2.17 \zehomi{1} & 
      130.6 & 2.20 \zehomi{1} \\
 
300 & 129.8 & 2.13 \zehomi{1} & 130.1 & 2.17 \zehomi{1} & 
      130.3 & 2.20 \zehomi{1} & 130.4 & 2.23 \zehomi{1} & 
      130.6 & 2.26 \zehomi{1} \\
 
\hline
  \end{tabular}
}
\end{table}

\begin{table}
   \vspace{-.5\headsep}
   \caption{MSSM Higgs branching ratio BR($\Ph \to \PZ^{(*)}\PZ^{(*)}$) in the 
$m_{\Ph}^{\rm max}$ scenario as a function of $\MA$ [GeV] and $\tb$.
The format in each cell is $M_{\Ph}$ [GeV], BR.
}
  \label{tab:BR-hZZ}
  \centering
{
   \setlength{\tabcolsep}{4.5pt}
   \small
   \begin{tabular}{lrcrcrcrcrc}\hline
{\small $\MA$} & \multicolumn{2}{c}{$\tb = 20$}
      & \multicolumn{2}{c}{$\tb = 30$}
            & \multicolumn{2}{c}{$\tb = 40$}
            & \multicolumn{2}{c}{$\tb = 50$}
            & \multicolumn{2}{c}{$\tb = 60$} \\
\hline
 90 & 89.6 & 1.79 \zehomi{9} & 89.8 & 1.79 \zehomi{9} & 
      89.9 & 1.79 \zehomi{9} & 89.9 & 1.79 \zehomi{9} & 
      89.9 & 1.79 \zehomi{9} \\
 
100 & 99.4 & 5.91 \zehomi{8} & 99.7 & 1.35 \zehomi{8} & 
      99.9 & 4.60 \zehomi{9} & 99.9 & 1.99 \zehomi{9} & 
      99.9 & 1.99 \zehomi{9} \\
 
110 & 109.0 & 7.34 \zehomi{7} & 109.6 & 1.70 \zehomi{7} & 
      109.8 & 5.82 \zehomi{8} & 109.8 & 2.49 \zehomi{8} & 
      109.9 & 1.24 \zehomi{8} \\
 
120 & 118.2 & 7.95 \zehomi{6} & 119.1 & 2.13 \zehomi{6} & 
      119.5 & 7.61 \zehomi{7} & 119.7 & 3.30 \zehomi{7} & 
      119.7 & 1.62 \zehomi{7} \\
 
130 & 125.2 & 1.09 \zehomi{4} & 126.9 & 5.65 \zehomi{5} & 
      127.8 & 3.31 \zehomi{5} & 128.4 & 2.07 \zehomi{5} & 
      128.9 & 1.32 \zehomi{5} \\
 
140 & 128.1 & 8.11 \zehomi{4} & 129.3 & 7.56 \zehomi{4} & 
      129.8 & 7.30 \zehomi{4} & 130.2 & 7.14 \zehomi{4} & 
      130.5 & 7.03 \zehomi{4} \\
 
150 & 128.9 & 2.39 \zehomi{3} & 129.7 & 2.43 \zehomi{3} & 
      130.1 & 2.46 \zehomi{3} & 130.4 & 2.49 \zehomi{3} & 
      130.6 & 2.52 \zehomi{3} \\
 
160 & 129.2 & 4.55 \zehomi{3} & 129.9 & 4.69 \zehomi{3} & 
      130.2 & 4.79 \zehomi{3} & 130.4 & 4.87 \zehomi{3} & 
      130.6 & 4.96 \zehomi{3} \\
 
170 & 129.4 & 6.96 \zehomi{3} & 130.0 & 7.20 \zehomi{3} & 
      130.2 & 7.36 \zehomi{3} & 130.4 & 7.05 \zehomi{3} & 
      130.6 & 7.64 \zehomi{3} \\
 
180 & 129.5 & 9.42 \zehomi{3} & 130.0 & 9.74 \zehomi{3} & 
      130.2 & 9.96 \zehomi{3} & 130.4 & 1.01 \zehomi{2} & 
      130.6 & 1.03 \zehomi{2} \\
 
190 & 129.6 & 1.18 \zehomi{2} & 130.0 & 1.22 \zehomi{2} & 
      130.2 & 1.25 \zehomi{2} & 130.4 & 1.27 \zehomi{2} & 
      130.6 & 1.29 \zehomi{2} \\
 
200 & 129.7 & 1.40 \zehomi{2} & 130.1 & 1.45 \zehomi{2} & 
      130.2 & 1.48 \zehomi{2} & 130.4 & 1.51 \zehomi{2} & 
      130.6 & 1.53 \zehomi{2} \\
 
210 & 129.7 & 1.61 \zehomi{2} & 130.1 & 1.66 \zehomi{2} & 
      130.2 & 1.70 \zehomi{2} & 130.4 & 1.73 \zehomi{2} & 
      130.6 & 1.76 \zehomi{2} \\
 
220 & 129.7 & 1.80 \zehomi{2} & 130.1 & 1.86 \zehomi{2} & 
      130.3 & 1.89 \zehomi{2} & 130.4 & 1.92 \zehomi{2} & 
      130.6 & 1.96 \zehomi{2} \\
 
230 & 129.8 & 1.98 \zehomi{2} & 130.1 & 2.03 \zehomi{2} & 
      130.3 & 2.07 \zehomi{2} & 130.4 & 2.10 \zehomi{2} & 
      130.6 & 2.14 \zehomi{2} \\
 
240 & 129.8 & 2.13 \zehomi{2} & 130.1 & 2.19 \zehomi{2} & 
      130.3 & 2.23 \zehomi{2} & 130.4 & 2.27 \zehomi{2} & 
      130.6 & 2.31 \zehomi{2} \\
 
250 & 129.8 & 2.27 \zehomi{2} & 130.1 & 2.33 \zehomi{2} & 
      130.3 & 2.37 \zehomi{2} & 130.4 & 2.41 \zehomi{2} & 
      130.6 & 2.45 \zehomi{2} \\
 
260 & 129.8 & 2.4 \zehomi{2} & 130.1 & 2.46 \zehomi{2} & 
      130.3 & 2.5 \zehomi{2} & 130.4 & 2.54 \zehomi{2} & 
      130.6 & 2.59 \zehomi{2} \\
 
270 & 129.8 & 2.51 \zehomi{2} & 130.1 & 2.58 \zehomi{2} & 
      130.3 & 2.62 \zehomi{2} & 130.4 & 2.66 \zehomi{2} & 
      130.6 & 2.71 \zehomi{2} \\
 
280 & 129.8 & 2.62 \zehomi{2} & 130.1 & 2.68 \zehomi{2} & 
      130.3 & 2.73 \zehomi{2} & 130.4 & 2.77 \zehomi{2} & 
      130.6 & 2.81 \zehomi{2} \\
 
290 & 129.8 & 2.71 \zehomi{2} & 130.1 & 2.78 \zehomi{2} & 
      130.3 & 2.82 \zehomi{2} & 130.4 & 2.86 \zehomi{2} & 
      130.6 & 2.91 \zehomi{2} \\
 
300 & 129.8 & 2.80 \zehomi{2} & 130.1 & 2.86 \zehomi{2} & 
      130.3 & 2.91 \zehomi{2} & 130.4 & 2.95 \zehomi{2} & 
      130.6 & 3.00 \zehomi{2}  \\
 
\hline
  \end{tabular}
}
\end{table}

\clearpage

\newpage
\section{Parton distribution functions\protect\footnote{S.~Forte, J.~Huston and R.~S.~Thorne  (eds); S.~Alekhin, J.~Bl\"umlein, A.M.~Cooper-Sarkar, S.~Glazov, P.~Jimenez-Delgado, S.~Moch, P.~Nadolsky, V.~Radescu, J.~Rojo, A.~Sapronov and W.J.~Stirling.}}
\label{se:PDFs}

\subsection{PDF set updates}

Several of the PDF sets which were discussed in the previous 
Yellow Report~\cite{Dittmaier:2011ti} have been updated since. 
NNPDF2.1~\cite{Ball:2011mu} is an updated 
version of NNPDF2.0~\cite{Ball:2010de} which uses the FONLL general mass 
VFN scheme~\cite{Forte:2010ta}  
instead of a zero-mass scheme. There are also now NNPDF 
NNLO  and LO sets~\cite{Ball:2011uy}, which are based on the same data
and methodology as NNPDF2.1. 
The HERA PDF group now use HERAPDF1.5~\cite{CooperSarkar:2010wm}, 
which contains more data than 
HERAPDF1.0~\cite{:2009wt}, a wider examination of parameter dependence, and an NNLO set 
with uncertainty. This set is available in LHAPDF, however it is partially
based on preliminary HERA-II structure-function data. 


The current PDF4LHC prescription~\cite{Botje:2011sn,PDF4LHCwebpage}
for calculating a central value and uncertainty for a given process should
undergo the simple modification of   
using the most up-to-date relevant set from the relevant 
group, \ie  NNPDF2.0 should be
replaced with NNPDF2.1~\cite{Ball:2011mu} and CTEQ6.6~\cite{Nadolsky:2008zw} 
with CT10~\cite{Lai:2010vv}. At NNLO the existing prescription should be 
used, but with the uncertainty envelope calculated using the up-to-date sets 
noted above.

\subsection{Correlations}

The main aim of this section is to examine the correlations between 
different Higgs production processes and/or backgrounds to these processes. 
The PDF uncertainty analysis may be extended to define a \emph{correlation}
between the uncertainties of two variables, say $X(\vec{a})$ and
$Y(\vec{a}).$  As for the case of PDFs, the physical concept of PDF
correlations can be determined both from PDF determinations based
on the Hessian approach and on the Monte Carlo approach. 
For convenience and commonality, all physical processes were 
calculated using the {\sc MCFM NLO} program 
(versions 5.8 and 6.0)~\cite{Campbell:2000bg,Campbell:2002tg}
with a common set of input files for all groups. 

We present the results for the PDF correlations at NLO 
for Higgs production via gluon--gluon fusion, vector-boson fusion, 
in association with $\PW$ or with a $\PQt \PAQt$ pair at masses $\MH=120\UGeV$,
$\MH=160\UGeV$, $\MH=200\UGeV$, $\MH=300\UGeV$, $\MH=500\UGeV$. We also include 
a wide variety of background processes and other standard production 
mechanisms, \ie  $\PW$, $\PW\PW$, $\PW\PZ$, $\PW\PGg$, $\PW\PQb\PAQb$, $\PQt \PAQt$, 
$\PQt \PAQb$ and the $t$-channel $\PQt (\to \PAQb) + \PQq$, where $\PW$ denotes the 
average of $\PW^+$ and $\PW^-$. 
   
For MSTW2008~\cite{Martin:2009iq}, CT10~\cite{Lai:2010vv}, 
GJR08~\cite{Gluck:2007ck,Gluck:2008gs}, and ABKM09~\cite{Alekhin:2009ni}
 PDFs the correlations of any two quantities $X$ and $Y$ 
are calculated using the 
standard  formula~\cite{Pumplin:2001ct}
\begin{equation}
\rho\left( X,Y\right) \equiv \cos{\varphi}=
\frac{\sum_i (X_i-X_0) (Y_i-Y_0)}{\sqrt{ \sum_i  (X_i-X_0)^2 \sum_i  (Y_i-Y_0)^2}}.
\end{equation}
The index in the sum runs over the number of free parameters and
$X_0, Y_0$ correspond to the values obtained by the central PDF value. 
When positive and negative direction PDF eigenvectors are used this is equivalent to 
(see \eg \Bref{Nadolsky:2008zw})
\begin{eqnarray}
\cos{\varphi} &=& \frac{\vec{\Delta} X \cdot \vec{\Delta} Y}{\Delta X \Delta Y}= \frac{1}{4\Delta X\Delta Y}\sum_{i=1}^{N} \left( X_i^{(+)}-X_i^{(-)}\right) \left( Y_i^{(+)}-Y_i^{(-)}\right), \cr
\Delta X &=& \left| \vec{\Delta} X \right| = \frac{1}{2} \sqrt{\sum_{i=1}^{N} \left( X_i^{(+)}-X_i^{(-)} \right)^2},  
\end{eqnarray}
where the sum is over the $N$ pairs of positive and negative 
PDF eigenvectors. For MSTW2008 and CT10 the 
eigenvectors by default contain only PDF parameters, and $\alphas$ variation may 
be considered separately. For GJR08  and ABKM09, $\alphas$ is one of the 
parameters used directly in calculating the correlation coefficient,
with the central value and variation determined by the fit. 

Due to the specific error 
calculation prescription for HERAPDF1.5 which includes
parametrisation and model errors, the correlations can not be 
calculated in exactly the same way. Rather, a
formula for uncertainty propagation can be used 
in which correlations can be expressed
via relative errors of compounds and their combination:
\begin{align}
\biggl[\sigma\left(\frac{X}{\sigma(X)}+\frac{Y}{\sigma(Y)}\right)\biggr]^2 = 2+2\cos{\varphi},
\end{align}
where $\sigma(O)$ is the PDF error of observable $O$ calculated 
using the HERAPDF prescription. 

The correlations for the NNPDF prescription 
are calculated as discussed in \Bref{Demartin:2010er}, namely
\begin{equation}
\rho\left( X,Y\right)=\frac{\left\langle XY \right\rangle_{\rm rep}-
\left\langle X \right\rangle_{\rm rep}\left\langle Y \right\rangle_{\rm rep}}{\sigma_{X}
\sigma_{Y}}
\end{equation}
where the averages are performed over the $N_{\rm rep}=100$ replicas of the NNPDF2.1 set.

For all sets the correlations are for the appropriate value of $\alphas$ for 
the relevant PDF set.

\begin{table}
\begin{center}
\caption[]{The up-to-date PDF4LHC average for the
correlations between all signal processes
with other signal and background processes for
Higgs production
considered here. The processes have
been classified in correlation classes, as discussed in the text. 
\label{tab:signal-correlations}} 
\small
\vspace{0.2cm}
\begin{center}
{
\begin{tabular}{ccccc}
\hline 
$\MH=120\UGeV$ & $\Pg\Pg\PH$ & VBF & $\PW\PH$ & $\PQt\PAQt \PH$  \\
\hline 
 $\Pg\Pg\PH$            
&$    1 $
&$  -0.6$
&$  -0.2$
&$  -0.2$
  \\
 VBF            
&$  -0.6$
&$    1 $
&$   0.6$
&$  -0.4$
  \\
 $\PW\PH$             
&$  -0.2$
&$   0.6$
&$    1 $
&$  -0.2$
  \\
 $\PQt\PAQt H$            
&$  -0.2$
&$  -0.4$
&$  -0.2$
&$    1 $
  \\
 $\PW$              
&$  -0.2$
&$   0.6$
&$   0.8$
&$  -0.6$
  \\
 $\PW\PW$             
&$  -0.4$
&$   0.8$
&$    1 $
&$  -0.2$
  \\
 $\PW\PZ$             
&$  -0.2$
&$   0.4$
&$   0.8$
&$  -0.4$
  \\
  $\PW\PGg$             
&$   0  $
&$   0.6$
&$   0.8$
&$  -0.6$
  \\
 $\PW\PQb\PAQb$            
&$  -0.2$
&$   0.6$
&$    1 $
&$  -0.2$
  \\
  $\PQt\PAQt$             
&$   0.2$
&$  -0.4$
&$  -0.4$
&$    1 $
  \\
 $\PQt\PAQb$             
&$  -0.4$
&$   0.6$
&$    1 $
&$  -0.2$
  \\
 $\PQt(\to \PAQb)\PQq$            
&$   0.4$
&$   0  $
&$   0  $
&$   0  $
  \\
\hline 
\end{tabular}
}
\quad
{
\begin{tabular}{ccccc}
\hline 
$\MH=160 \UGeV$ & $\Pg\Pg\PH$ & VBF & $\PW\PH$ & $\PQt\PAQt \PH$  \\
\hline 
 $\Pg\Pg\PH$            
&$    1 $
&$  -0.6$
&$  -0.4$
&$   0.2$
  \\
 VBF            
&$  -0.6$
&$    1 $
&$   0.6$
&$  -0.2$
  \\
 $\PW\PH$             
&$  -0.4$
&$   0.6$
&$    1 $
&$   0  $
  \\
 $\PQt\PAQt \PH$            
&$   0.2$
&$  -0.2$
&$   0  $
&$    1 $
  \\
 $\PW$              
&$  -0.4$
&$   0.4$
&$   0.6$
&$  -0.4$
  \\
 $\PW\PW$             
&$  -0.4$
&$   0.6$
&$   0.8$
&$  -0.2$
  \\
 $\PW\PZ$             
&$  -0.4$
&$   0.4$
&$   0.8$
&$  -0.2$
  \\
  $\PW\PGg$             
&$  -0.4$
&$   0.6$
&$   0.6$
&$  -0.6$
  \\
 $\PW\PQb\PAQb$            
&$  -0.2$
&$   0.6$
&$   0.8$
&$  -0.2$
  \\
  $\PQt\PAQt$             
&$   0.4$
&$  -0.4$
&$  -0.2$
&$   0.8$
  \\
 $\PQt\PAQb$             
&$  -0.4$
&$   0.6$
&$    1 $
&$   0  $
  \\
 $\PQt(\to \PAQb)\PQq$            
&$   0.6$
&$   0  $
&$   0  $
&$   0  $
  \\
\hline 
\end{tabular}
}
\end{center}
\begin{center}
{
\begin{tabular}{ccccc}
\hline 
$\MH=200 \UGeV$ & $\Pg\Pg\PH$ & VBF & $\PW\PH$ & $\PQt\PAQt \PH$  \\
\hline 
 $\Pg\Pg\PH$            
&$    1 $
&$  -0.6$
&$  -0.4$
&$   0.4$
  \\
 VBF            
&$  -0.6$
&$    1 $
&$   0.6$
&$  -0.2$
  \\
 $\PW\PH$             
&$  -0.4$
&$   0.6$
&$    1 $
&$   0  $
  \\
 $\PQt\PAQt \PH$            
&$   0.4$
&$  -0.2$
&$   0  $
&$    1 $
  \\
 $\PW$              
&$  -0.6$
&$   0.4$
&$   0.6$
&$  -0.4$
  \\
 $\PW\PW$             
&$  -0.4$
&$   0.6$
&$   0.8$
&$  -0.2$
  \\
 $\PW\PZ$             
&$  -0.4$
&$   0.4$
&$   0.8$
&$  -0.2$
  \\
  $\PW\PGg$             
&$  -0.4$
&$   0.4$
&$   0.6$
&$  -0.6$
  \\
 $\PW\PQb \PAQb$            
&$  -0.2$
&$   0.6$
&$   0.8$
&$  -0.2$
  \\
  $\PQt\PAQt$             
&$   0.6$
&$  -0.4$
&$  -0.2$
&$   0.8$
  \\
 $\PQt\PAQb$             
&$  -0.4$
&$   0.6$
&$   0.8$
&$   0  $
  \\
 $\PQt(\to \PAQb)\PQq$            
&$   0.6$
&$  -0.2$
&$   0  $
&$   0  $
  \\
\hline 
\end{tabular}
}
 \quad 
{
\begin{tabular}{ccccc}
\hline 
$\MH=300 \UGeV$ & $\Pg\Pg\PH$ & VBF & $\PW\PH$ & $\PQt\PAQt \PH$  \\
\hline 
 $\Pg\Pg\PH$            
&$    1 $
&$  -0.4$
&$  -0.2$
&$   0.6$
  \\
 VBF            
&$  -0.4$
&$    1 $
&$   0.4$
&$  -0.2$
  \\
 $\PW\PH$             
&$  -0.2$
&$   0.4$
&$    1 $
&$   0.2$
  \\
 $\PQt\PAQt \PH$            
&$   0.6$
&$  -0.2$
&$   0.2$
&$    1 $
  \\
 $\PW$              
&$  -0.6$
&$   0.4$
&$   0.4$
&$  -0.6$
  \\
 $\PW\PW$             
&$  -0.4$
&$   0.6$
&$   0.8$
&$  -0.2$
  \\
 $\PW\PZ$             
&$  -0.6$
&$   0.4$
&$   0.6$
&$  -0.4$
  \\
  $\PW\PGg$             
&$  -0.6$
&$   0.4$
&$   0.4$
&$  -0.6$
  \\
 $\PW\PQb \PAQb$            
&$  -0.2$
&$   0.4$
&$   0.8$
&$  -0.2$
  \\
  $\PQt\PAQt$             
&$    1 $
&$  -0.4$
&$   0  $
&$   0.8$
  \\
 $\PQt\PAQb$             
&$  -0.4$
&$   0.4$
&$   0.8$
&$  -0.2$
  \\
 $\PQt(\to \PAQb)\PQq$            
&$   0.4$
&$  -0.2$
&$   0  $
&$  -0.2$
  \\
\hline 
\end{tabular}
}
\end{center}
\begin{center}
{
\begin{tabular}{ccccc}
\hline 
$\MH=500 \UGeV$ & $\Pg\Pg\PH$ & VBF & $\PW\PH$ & $\PQt\PAQt \PH$  \\
\hline 
 $\Pg\Pg\PH$            
&$    1 $
&$  -0.4$
&$   0  $
&$   0.8$
  \\
 VBF            
&$  -0.4$
&$    1 $
&$   0.4$
&$  -0.2$
  \\
 $\PW\PH$             
&$   0  $
&$   0.4$
&$    1 $
&$   0  $
  \\
 $\PQt\PAQt \PH$            
&$   0.8$
&$  -0.2$
&$   0  $
&$    1 $
  \\
 $\PW$              
&$  -0.6$
&$   0.4$
&$   0.2$
&$  -0.6$
  \\
 $\PW\PW$             
&$  -0.4$
&$   0.6$
&$   0.6$
&$  -0.4$
  \\
 $\PW\PZ$             
&$  -0.6$
&$   0.4$
&$   0.6$
&$  -0.4$
  \\
  $\PW\PGg$             
&$  -0.6$
&$   0.4$
&$   0.2$
&$  -0.6$
  \\
 $\PW\PQb \PAQb$            
&$  -0.4$
&$   0.4$
&$   0.6$
&$  -0.4$
  \\
  $\PQt\PAQt$             
&$    1 $
&$  -0.4$
&$   0  $
&$   0.8$
  \\
 $\PQt\PAQb$             
&$  -0.4$
&$   0.4$
&$   0.8$
&$  -0.2$
  \\
 $\PQt(\to \PAQb)\PQq$            
&$   0.2$
&$  -0.2$
&$   0  $
&$  -0.2$
  \\
\hline 
\end{tabular}
}
  \\
\end{center}
\end{center}
\end{table}

\begin{table}
\begin{center}
\caption[]{The same as  \refT{tab:signal-correlations}
for the correlations between background processes.
\label{tab:back-correlations}}
\small
\vspace{0.2cm}
\begin{center}
{
\small
\begin{tabular}{ccccccccc}
\hline 
&  $\PW$                  &  $\PW\PW$                 &  $\PW\PZ$                 &  $\PW\PGg$            &  $\PW\PQb \PAQb$         
&  $\Pt \PAQt$           &  $\PQt\PAQb$            &  $\PQt(\to \PAQb)\PQq$      \\
\hline 
 $\PW$                 
&$    1 $
&$   0.8$
&$   0.8$
&$    1 $
&$   0.6$
&$  -0.6$
&$   0.6$
&$  -0.2$
  \\
 $\PW\PW$                
&$   0.8$
&$    1 $
&$   0.8$
&$   0.8$
&$   0.8$
&$  -0.4$
&$   0.8$
&$   0  $
  \\
 $\PW\PZ$                
&$   0.8$
&$   0.8$
&$    1 $
&$   0.8$
&$   0.8$
&$  -0.4$
&$   0.8$
&$   0  $
  \\
 $\PW\PGg$           
&$    1 $
&$   0.8$
&$   0.8$
&$    1 $
&$   0.6$
&$  -0.6$
&$   0.8$
&$   0  $
  \\
 $\PW\PQb \PAQb$         
&$   0.6$
&$   0.8$
&$   0.8$
&$   0.6$
&$    1 $
&$  -0.2$
&$   0.6$
&$   0  $
  \\
 $\PQt \PAQt$          
&$  -0.6$
&$  -0.4$
&$  -0.4$
&$  -0.6$
&$  -0.2$
&$    1 $
&$  -0.4$
&$   0.2$
  \\
 $\PQt\PAQb$           
&$   0.6$
&$   0.8$
&$   0.8$
&$   0.8$
&$   0.6$
&$  -0.4$
&$    1 $
&$   0.2$
  \\
 $\PQt(\to \PAQb)\PQq$    
&$  -0.2$
&$   0  $
&$   0  $
&$   0  $
&$   0  $
&$   0.2$
&$   0.2$
&$    1 $
  \\
\hline 
\end{tabular}
}
  \\
\end{center}
\end{center} 
\end{table}

\subsubsection{Results for the correlation study}

Our main result is the computation of the correlation
between physical processes relevant to Higgs production,
either as signal or as background.
It is summarised in 
\refTs{tab:signal-correlations} and 
\ref{tab:back-correlations}, where we show the  PDF4LHC average
for each of the correlations between signal and
background processes considered. These tables classify each correlation 
in classes with $\Delta\rho=0.2$, that is, if the correlation
is  $1> \rho > 0.9$ the processes is assigned correlation 1,
if $0.9> \rho > 0.7$ the processes is assigned correlation $0.8$ and so on.
The class width is typical of the spread of the results from the 
PDF sets, which are in generally very good, but not perfect agreement. 
The average is obtained using the most up-to-date PDF sets (CT10,
MSTW2008, NNPDF2.1) in the PDF4LHC 
recommendation, and it is  appropriate for use in conjunction 
with the cross-section results obtained in \Bref{Dittmaier:2011ti}, and with 
the background processes listed; the change of correlations due to 
updating the prescription is insignificant in comparison  to 
the class width.

We have also compared this PDF4LHC average to 
the average using all six PDF sets. In general there is rather little change. 
There are quite a few cases where the 
average moves into a neighbouring class but in many cases due to a very small 
change taking the average just over a boundary between two classes. There is 
only a move into the next-to-neighbouring class, \ie  a change of more than $0.2$, in a very small number of 
cases. For the VBF--$\PW\PGg$ correlation at $\MH=120,160\UGeV$
it reduces from 
$0.6$ to $0.2$, for the VBF--$\PW$ correlation at $\MH=500\UGeV$ it reduces from $0.4$
to $0$, for the $\PW\PGg$--$\PQt\PAQb$ correlation it reduces from $0.8$ to 
$0.4$, for $\PW\PZ$--$\PQt\PAQt\PH$ at $\MH=120\UGeV$ it increases from $-0.4$ to $0.0$, 
and for $\PW\PQb \PAQb$--$\PQt \PAQt \PH$ at $\MH=200\UGeV$ it increases from $-0.2$ 
to $0.2$.

\begin{figure}
\centerline{  
\includegraphics[width=0.49\textwidth]{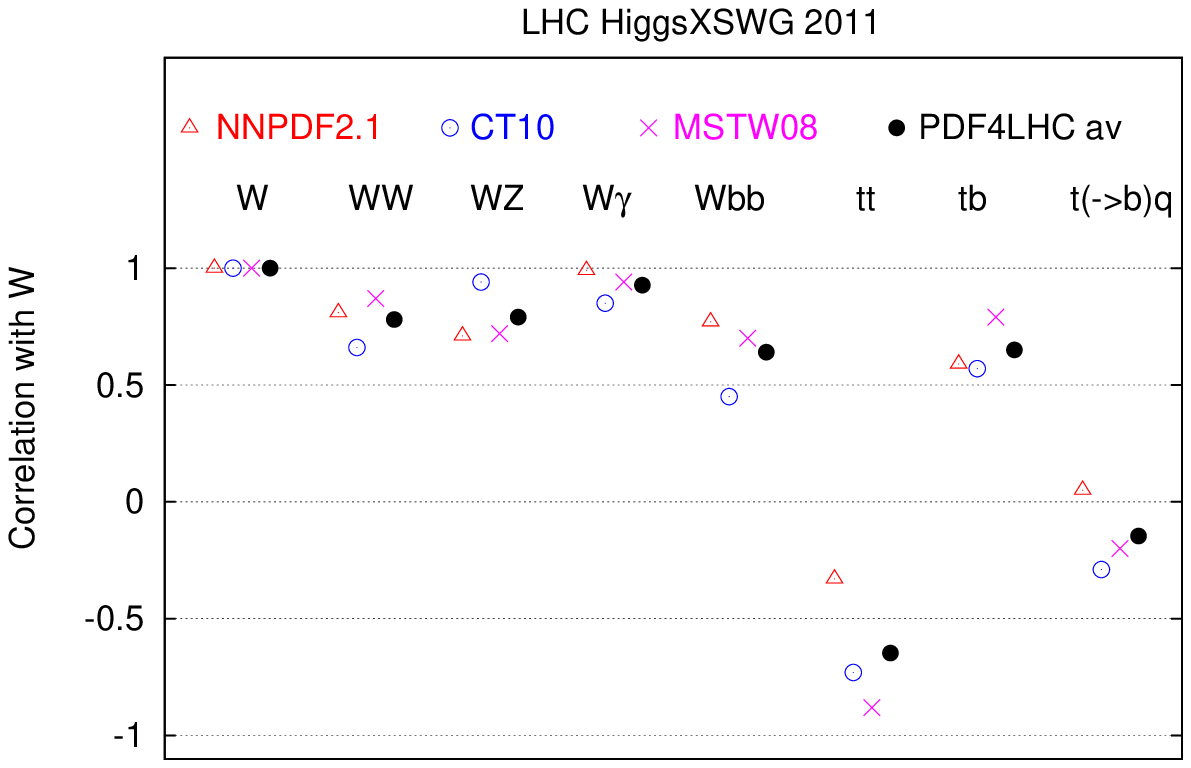}
\includegraphics[width=0.49\textwidth]{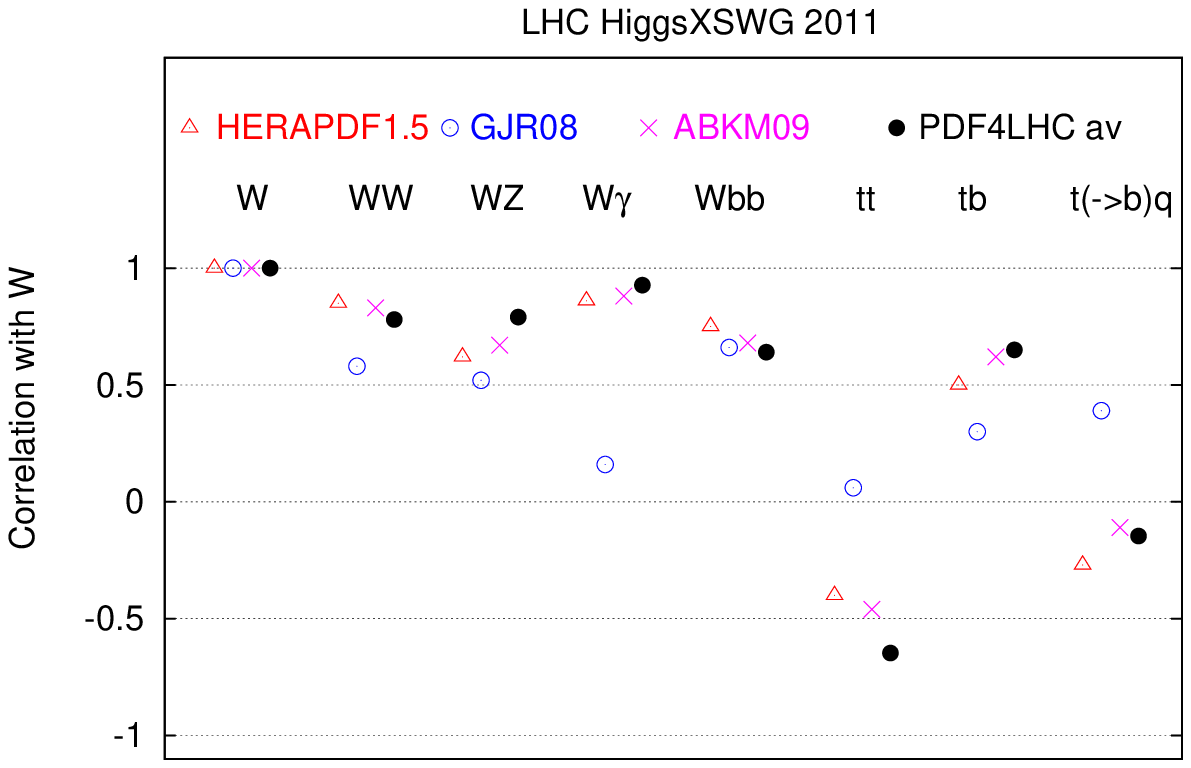}
}
\vspace*{-1em}
\caption{The correlations between $\PW$ production
and the other background processes considered. We show
the results for NNPDF2.1, CT10 and MSTW2008 in the left plot,
and HERAPDF, JR and ABKM in the right plot. In both cases we
show the up-to-date PDF4LHC average result.
\label{fig:corr-back-w}} 
\end{figure}

\begin{figure}
\centerline{  
\includegraphics[width=0.49\textwidth]{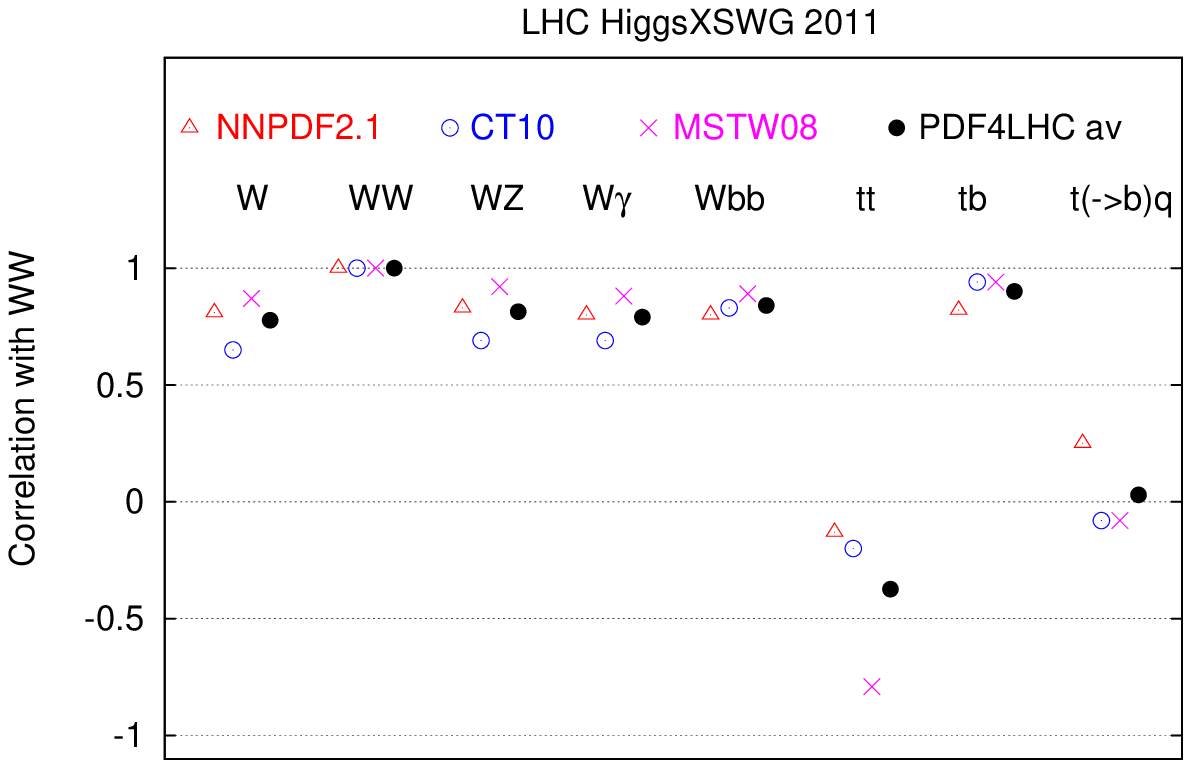}
\includegraphics[width=0.49\textwidth]{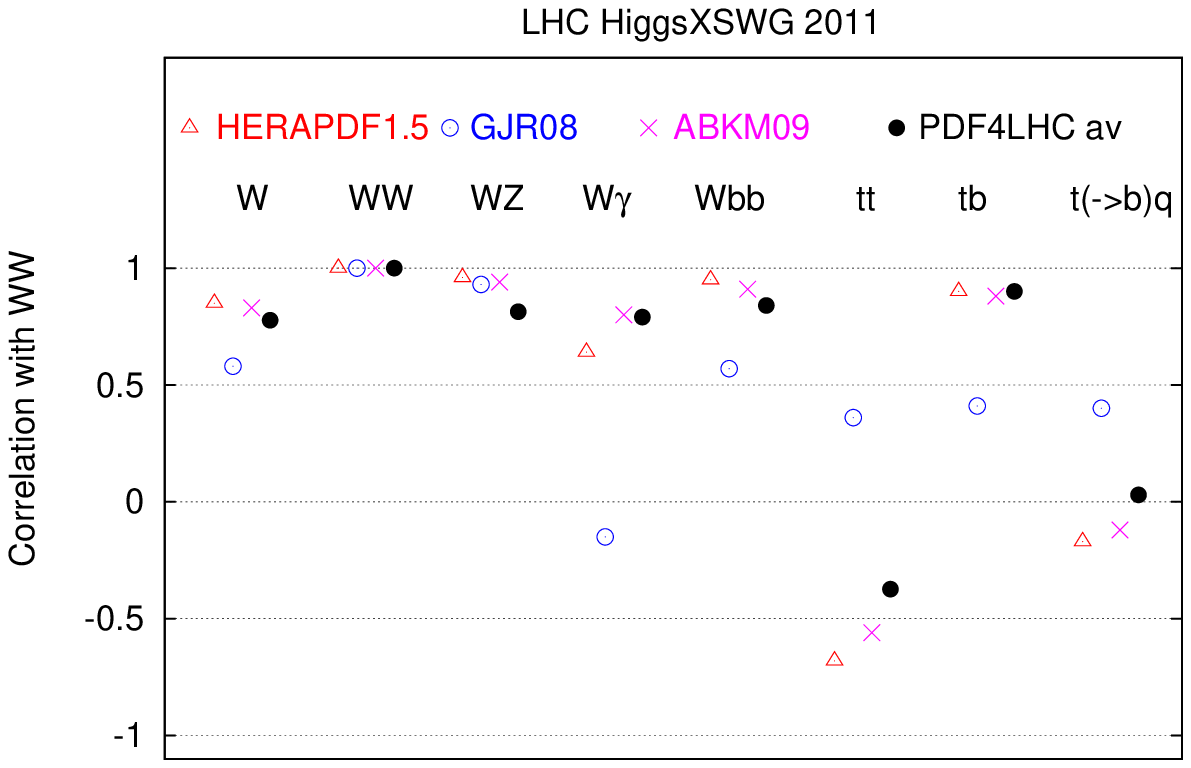}
}
\vspace*{-1em}
\caption{The same as \refF{fig:corr-back-w} for 
 $\PW\PW$ production. } 
\end{figure}

\begin{figure}
\centerline{  
\includegraphics[width=0.49\textwidth]{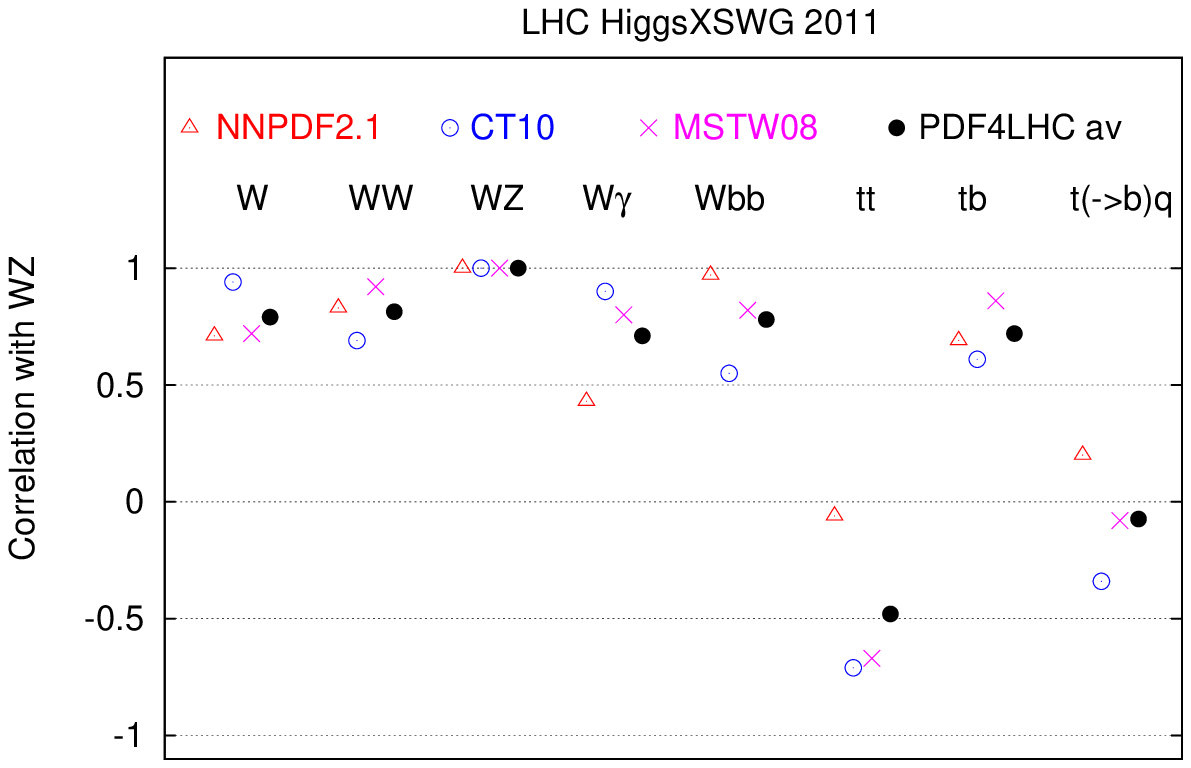}
\includegraphics[width=0.49\textwidth]{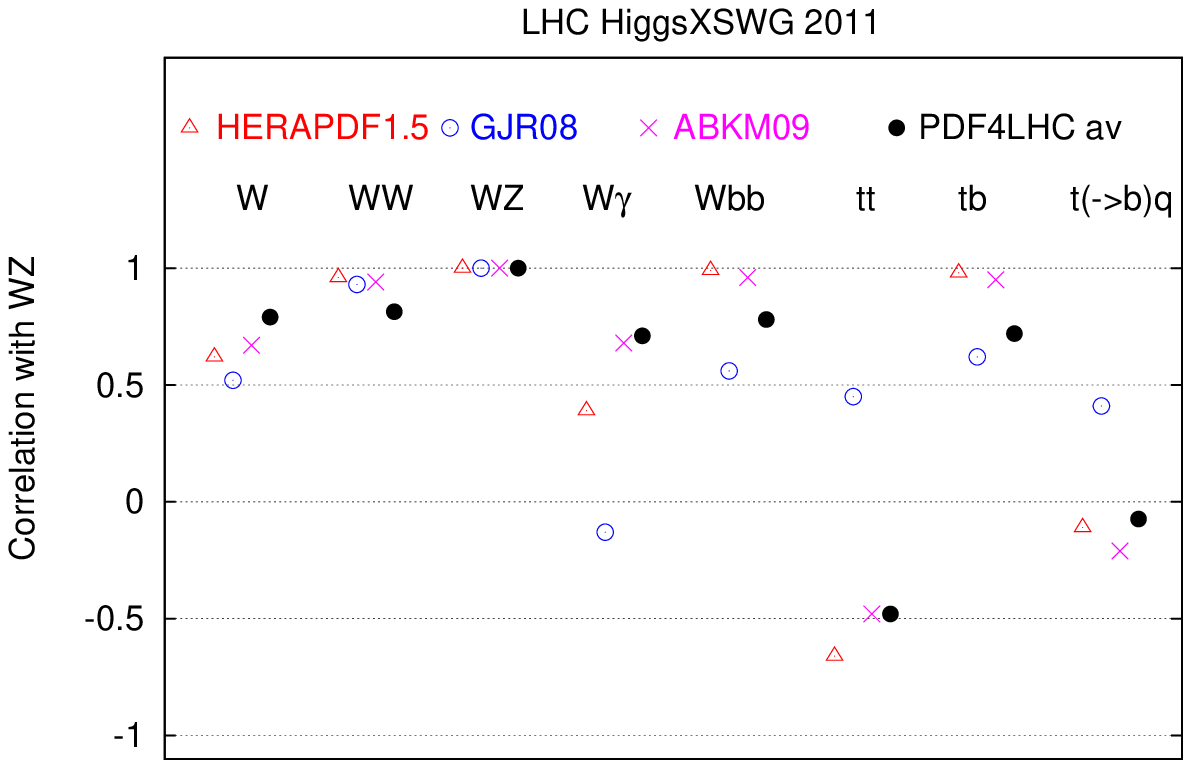}
}
\vspace*{-1em}
\caption{The same as \refF{fig:corr-back-w} for 
 $\PW\PZ$ production.
\label{fig:corr-back-wz}} 
\end{figure}

\begin{figure}
\centerline{  
\includegraphics[width=0.49\textwidth]{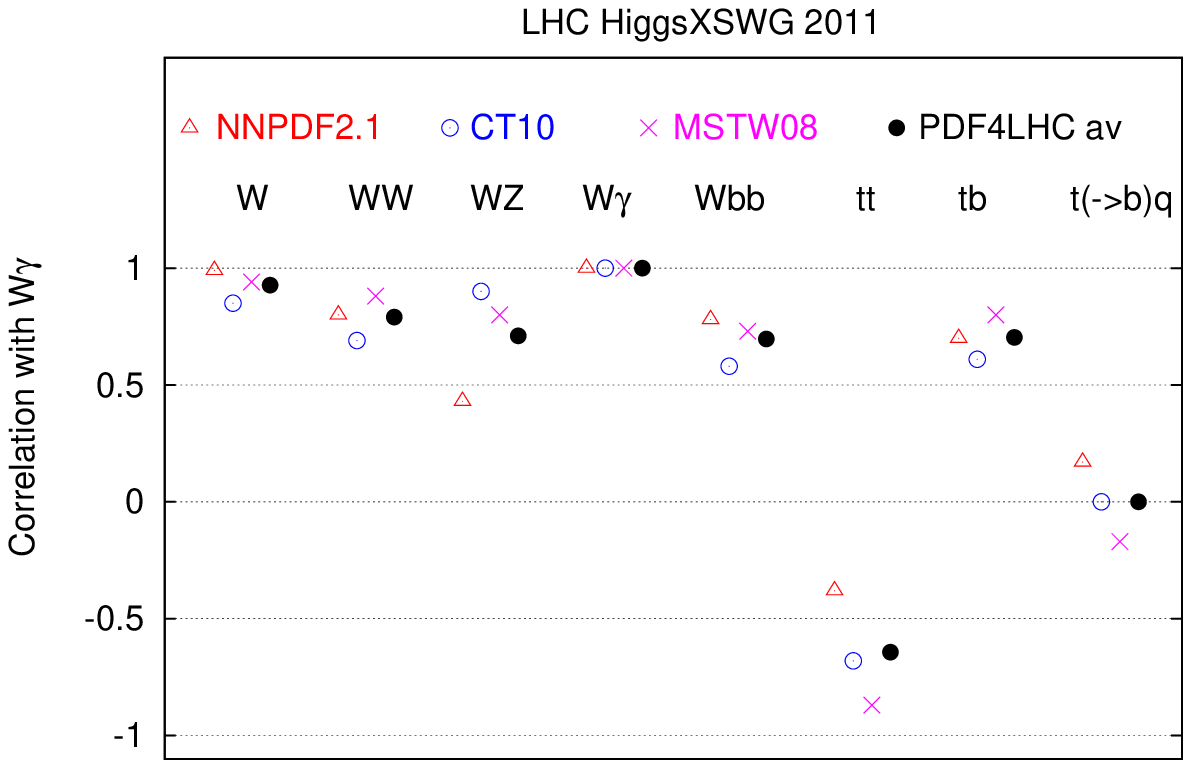}
\includegraphics[width=0.49\textwidth]{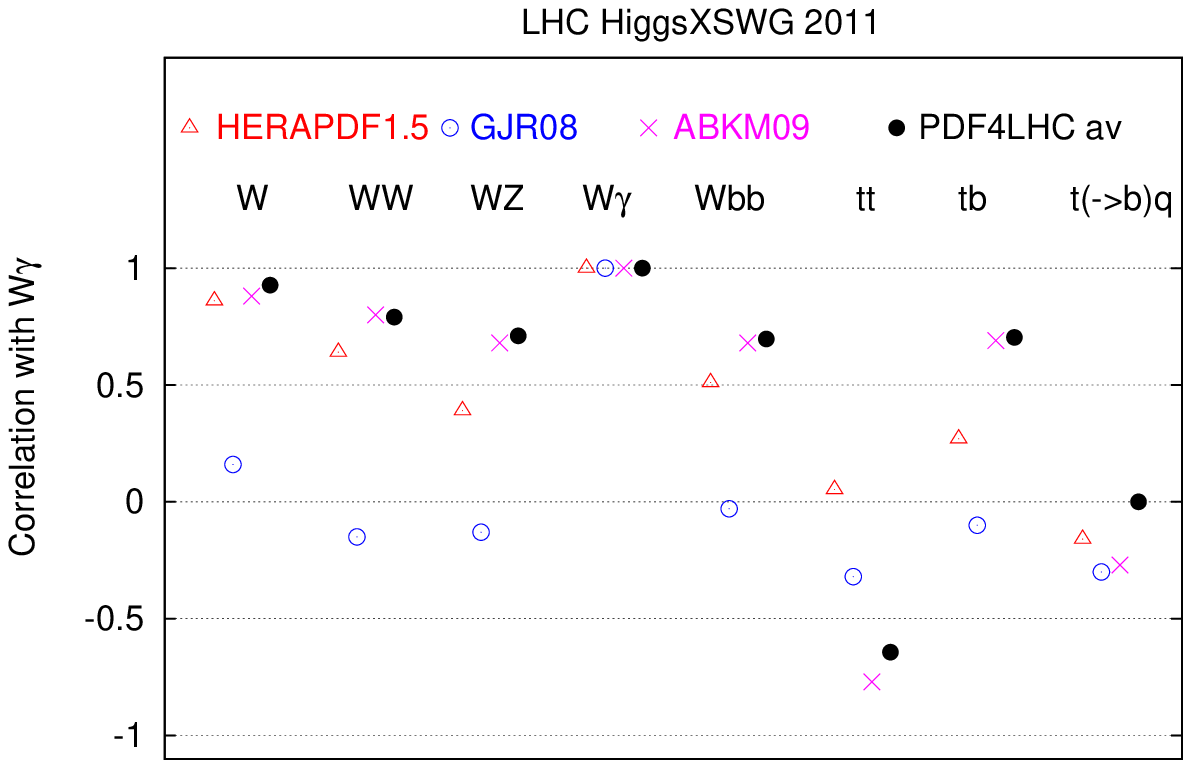}
}
\vspace*{-1em}
\caption{The same as \refF{fig:corr-back-w} for 
 $\PW\PGg$ production.
\label{fig:corr-back-wg}} 
\end{figure}

\begin{figure}
\centerline{  
\includegraphics[width=0.49\textwidth]{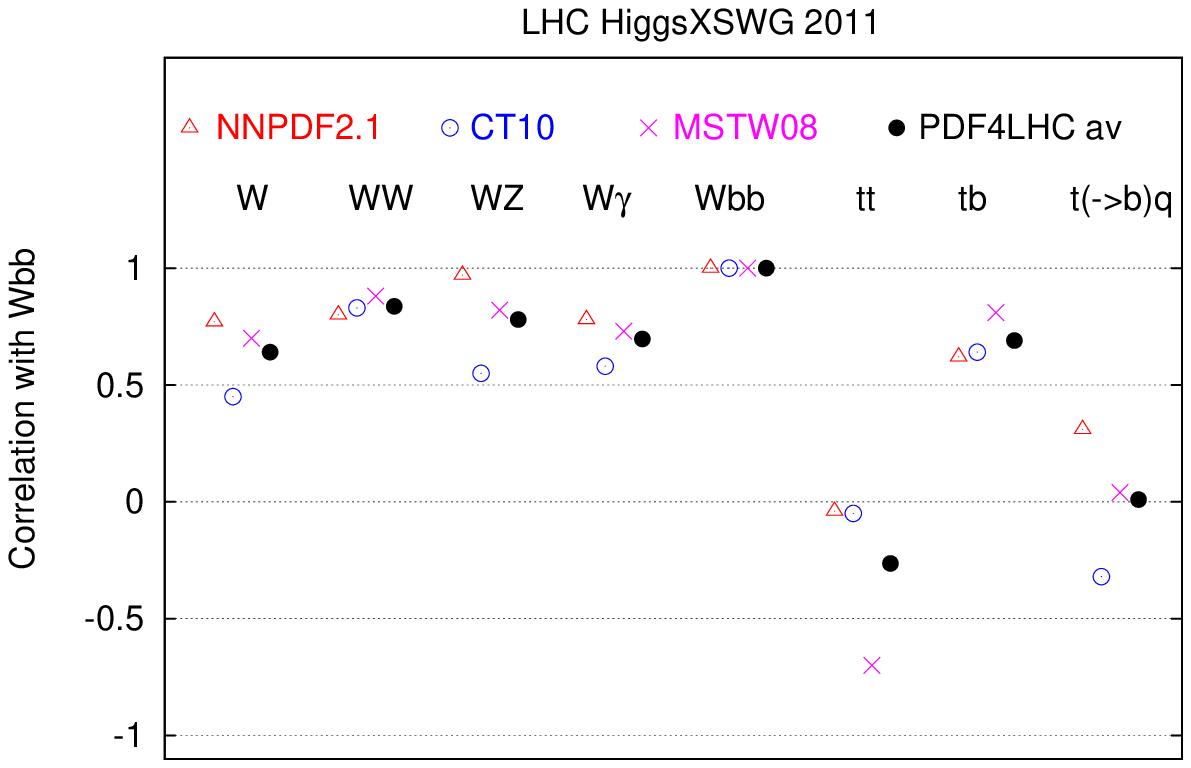}
\includegraphics[width=0.49\textwidth]{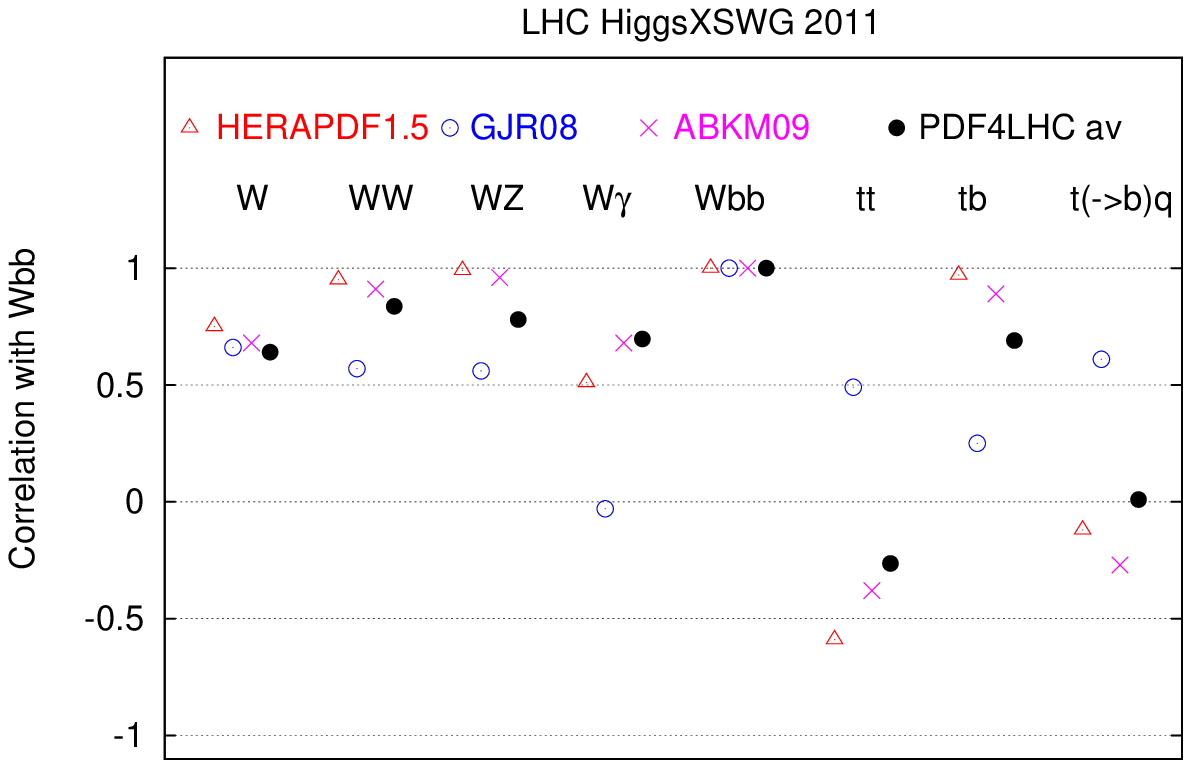}
}
\vspace*{-1em}
\caption{The same as \refF{fig:corr-back-w} for 
 $\PW\PQb\PAQb$ production.
\label{fig:corr-back-wqq}} 
\end{figure}

\begin{figure}
\centerline{  
\includegraphics[width=0.49\textwidth]{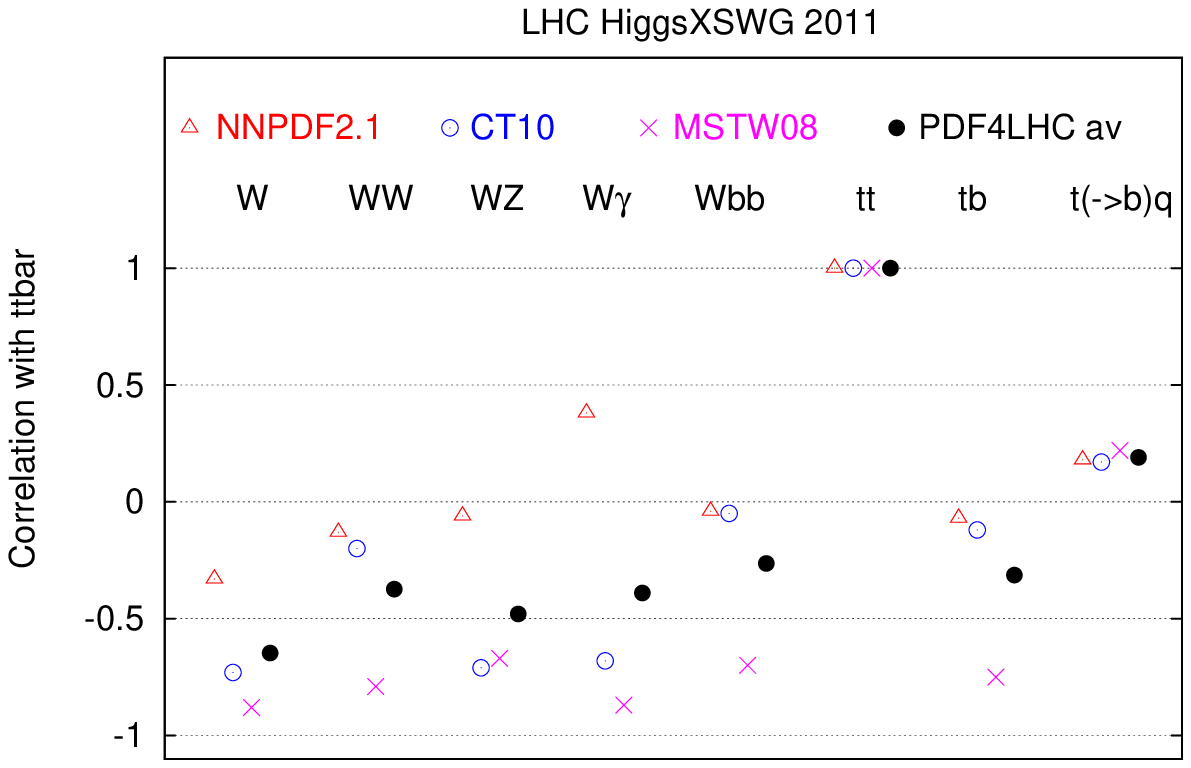}
\includegraphics[width=0.49\textwidth]{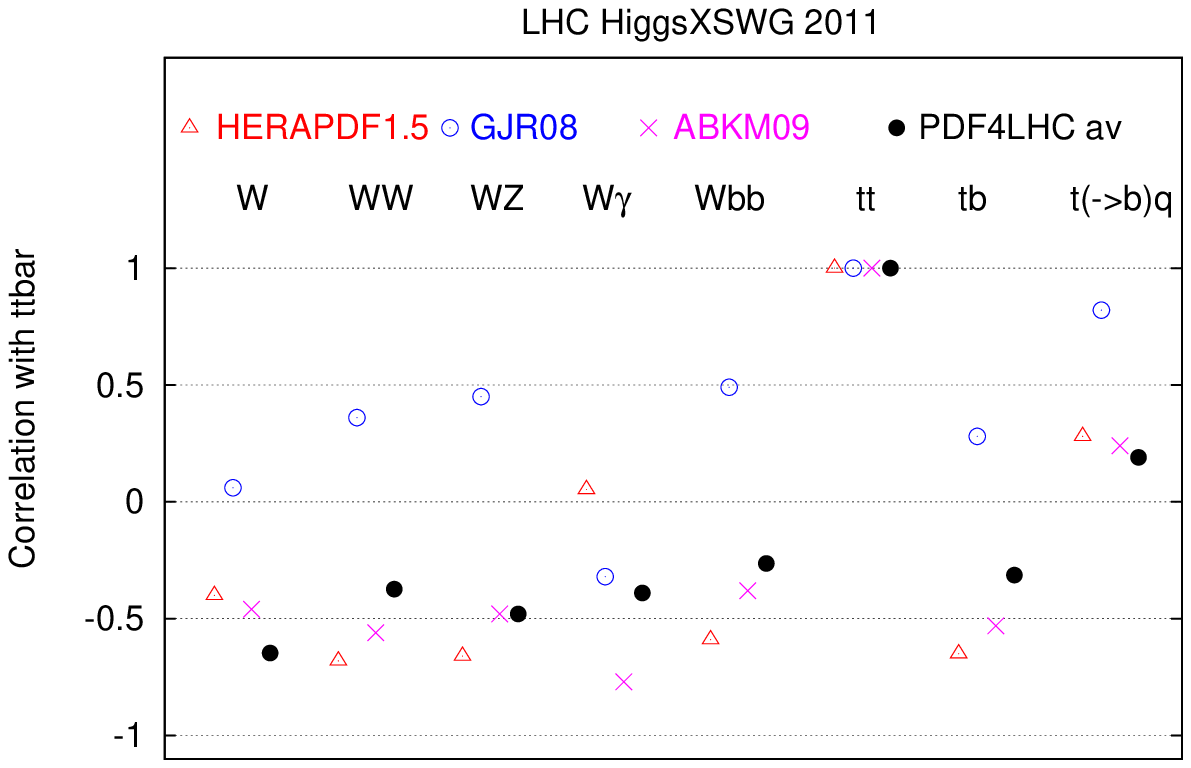}
}
\vspace*{-1em}
\caption{The same as \refF{fig:corr-back-w} for 
 $\PQt\PAQt$ production.
\label{fig:corr-back-tt}} 
\end{figure}

\begin{figure}
\centerline{  
\includegraphics[width=0.49\textwidth]{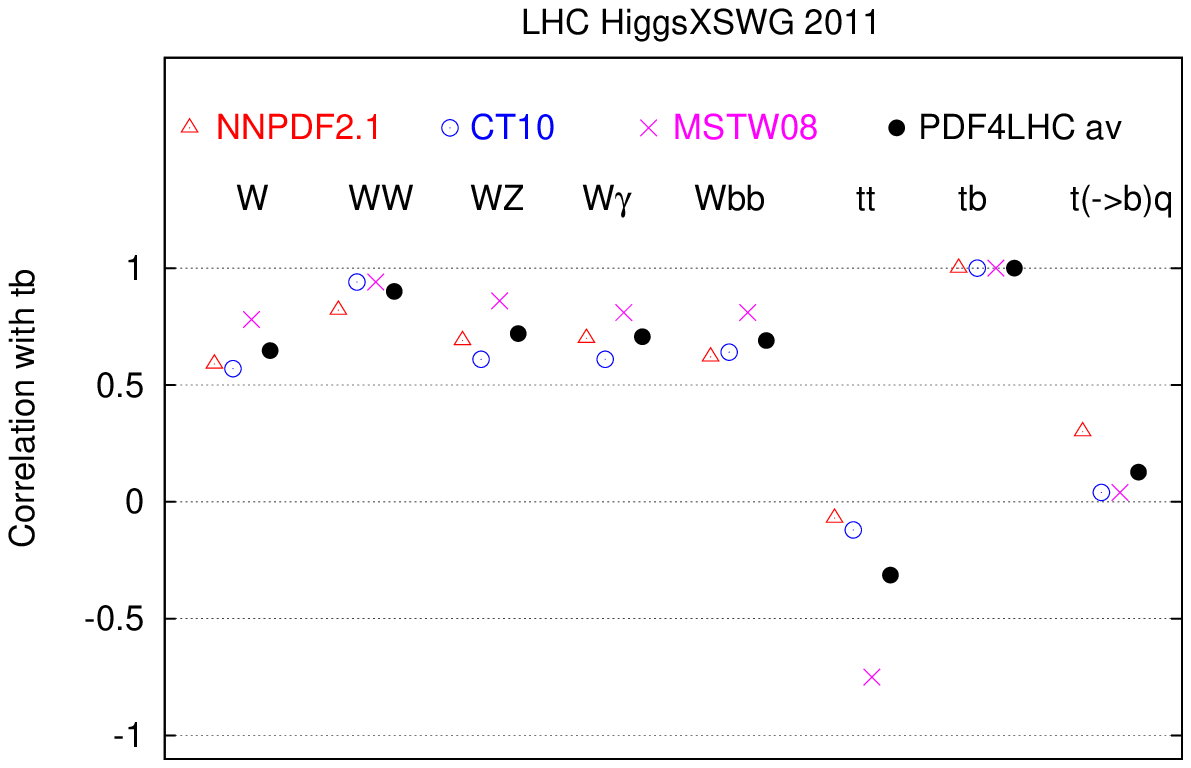}
\includegraphics[width=0.49\textwidth]{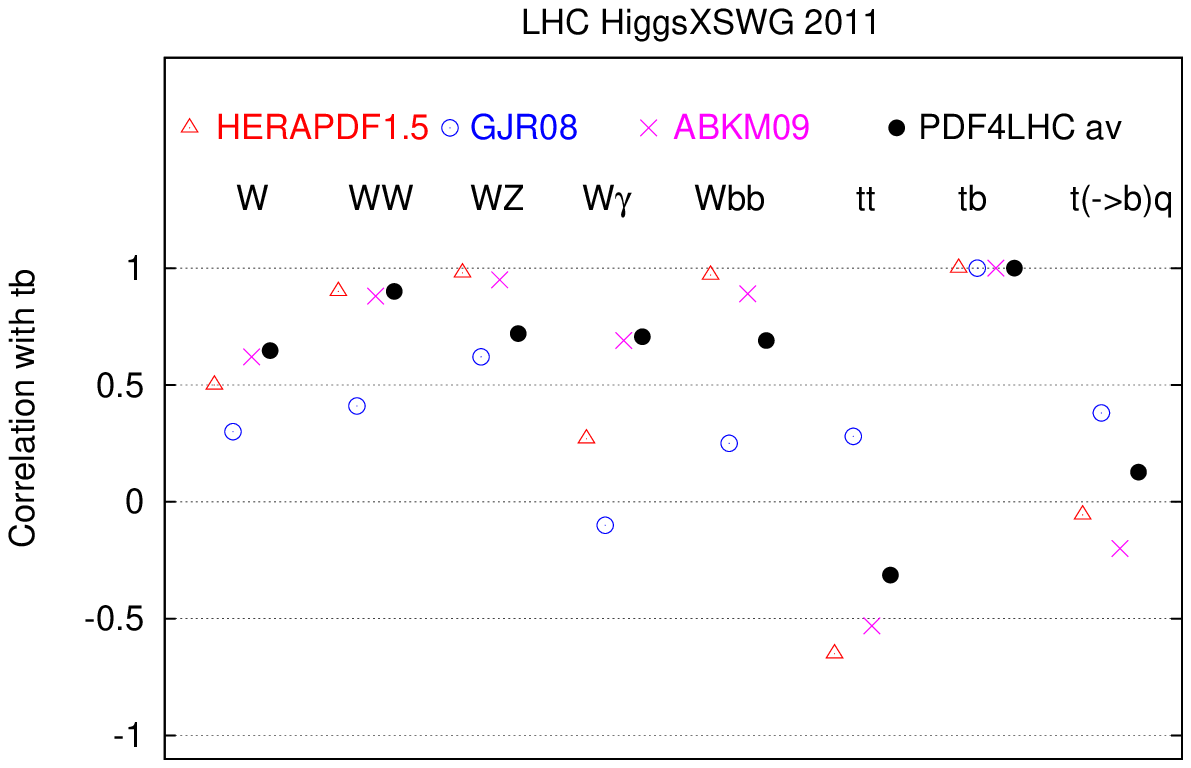}
}
\vspace*{-1em}
\caption{The same as \refF{fig:corr-back-w} for 
 $\PQt\PAQb$ production.
\label{fig:corr-back-tb}} 
\end{figure}

\begin{figure}
\centerline{  
\includegraphics[width=0.49\textwidth]{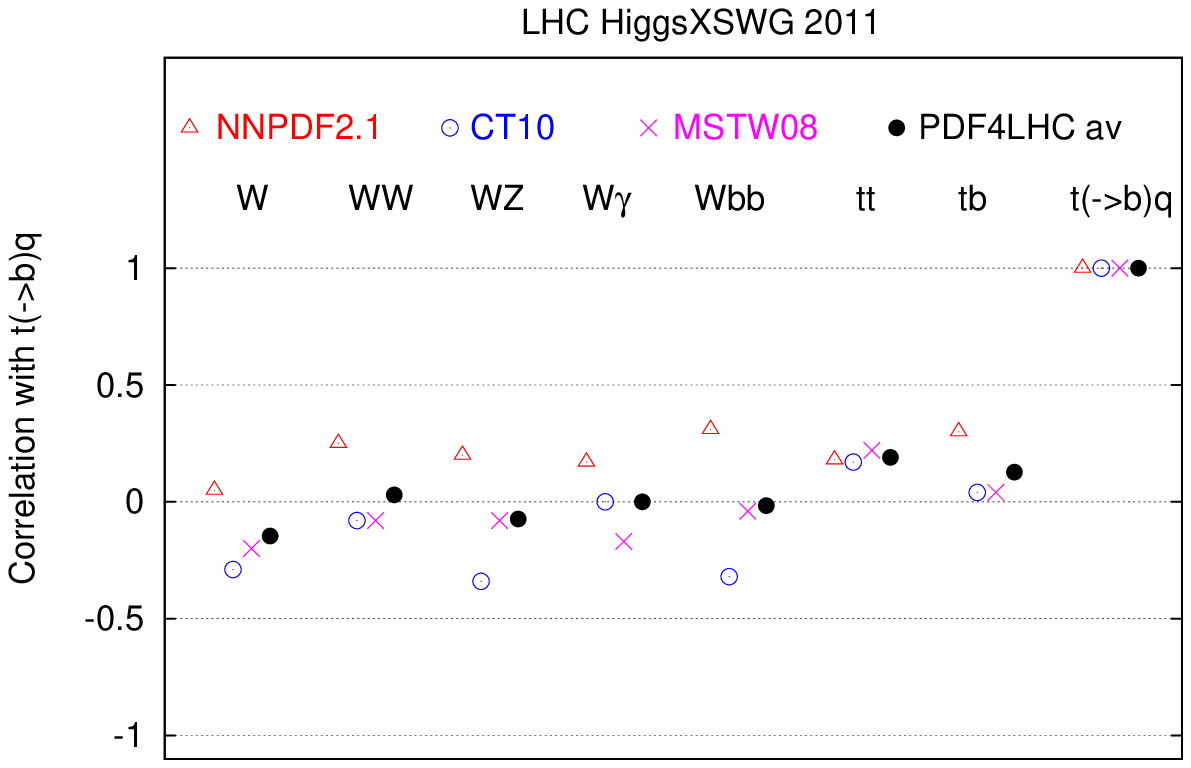}
\includegraphics[width=0.49\textwidth]{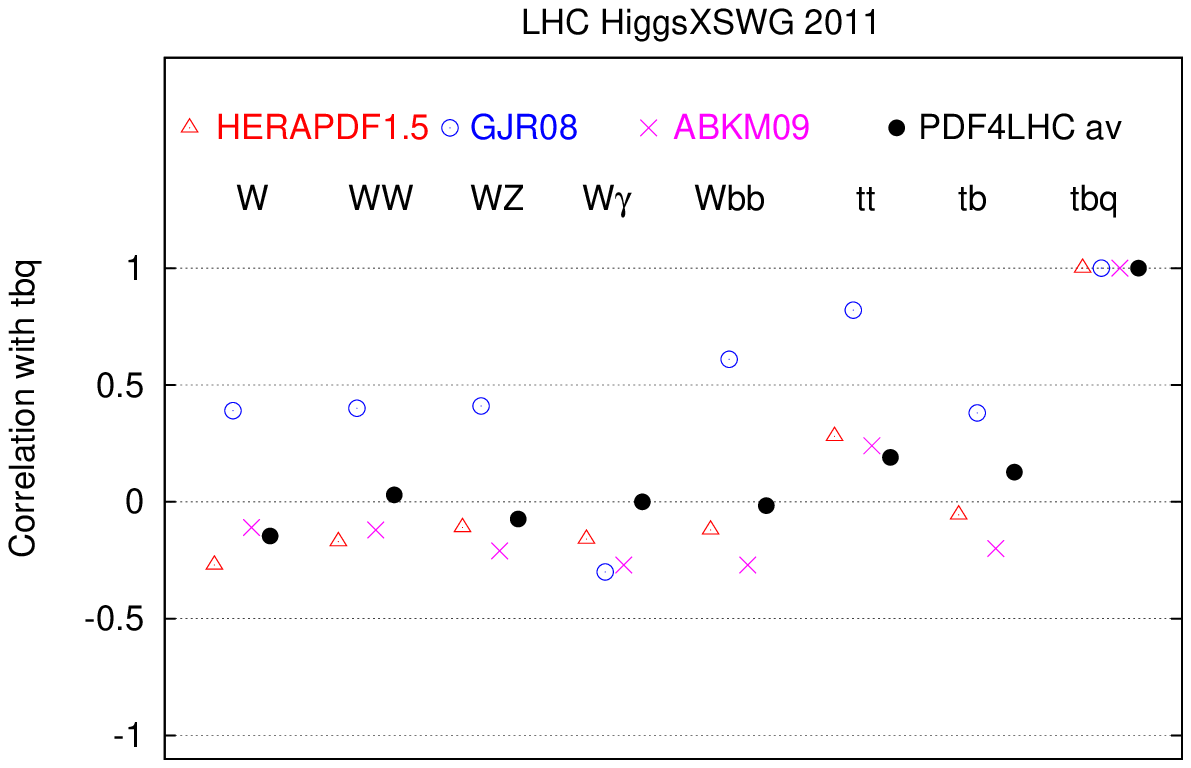}
}
\vspace*{-1em}
\caption{The same as \refF{fig:corr-back-w} for 
 $t$-channel 
$\PQt(\to \PAQb)\PQq$ production.
\label{fig:corr-back-tbq}} 
\end{figure}

\begin{figure}
\centerline{  
\includegraphics[width=0.99\textwidth]{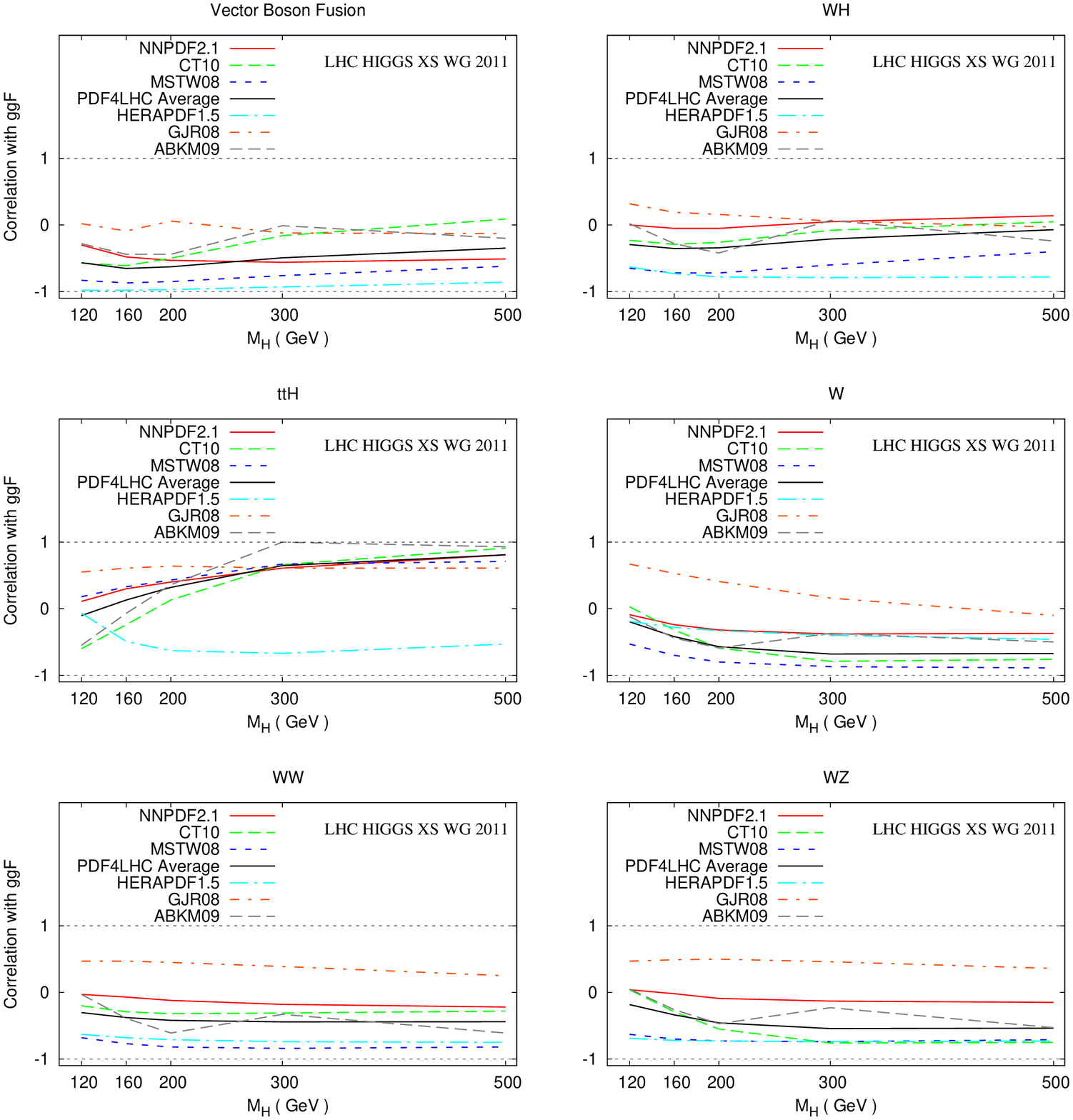}
}
\vspace*{-.5em}
\caption{Correlation between the gluon-fusion $\Pg\Pg\to \PH$ process
and other signal and background processes as a function of $\MH$.
We show the results for the individual PDF sets as well
as the up-to-date PDF4LHC average.
\label{fig:corr-ggh1}} 
\end{figure}

\begin{figure}
\centerline{  
\includegraphics[width=0.99\textwidth]{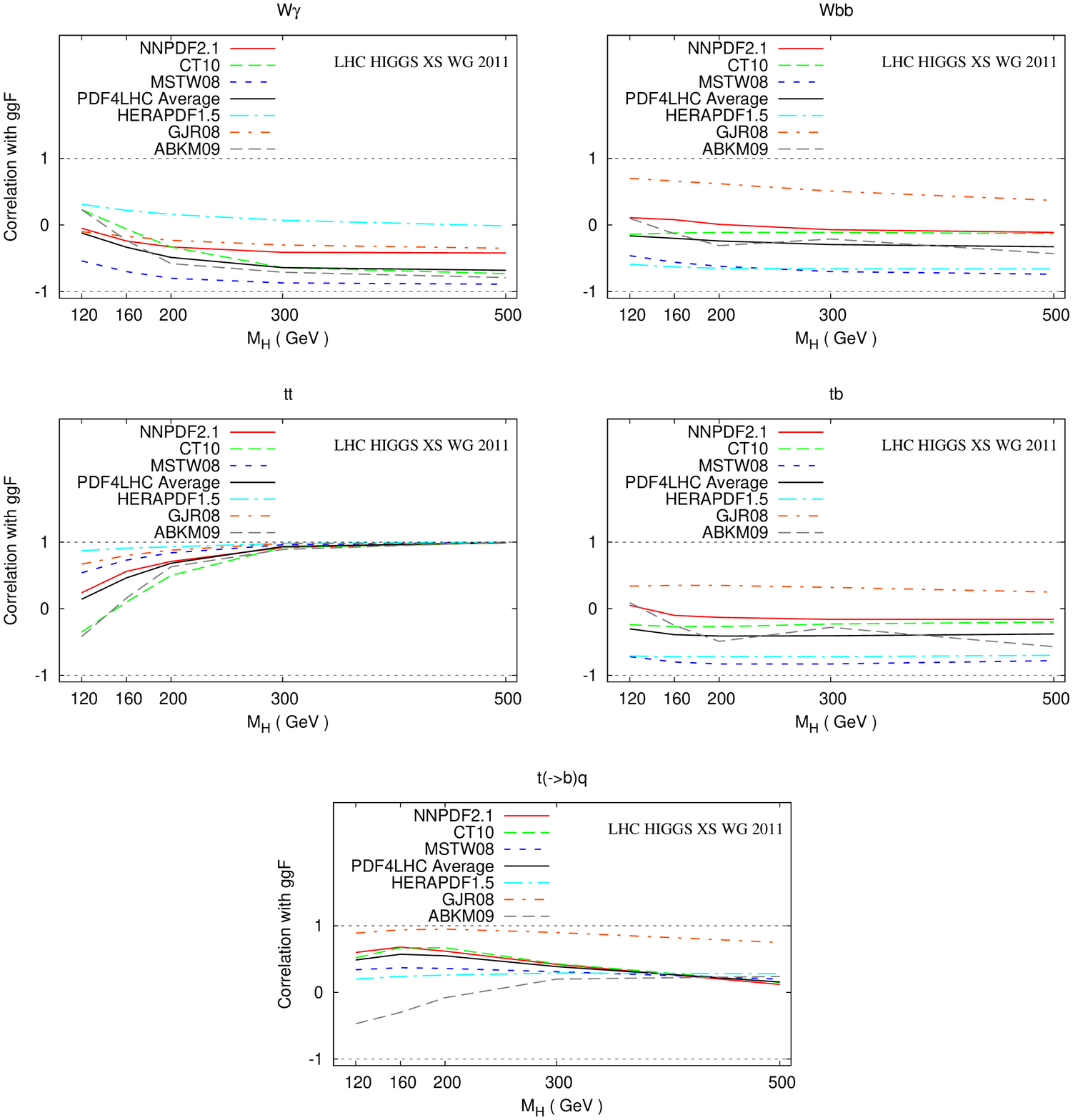}
}
\vspace*{-.5em}
\caption{Correlation between the gluon fusion $\Pg\Pg\to \PH$ process
and other signal and background processes as a function of $\MH$.
We show the results for the individual PDF sets as well
as the up-to-date PDF4LHC average.
\label{fig:corr-ggh2}} 
\end{figure}

\begin{figure}
\centerline{  
\includegraphics[width=0.99\textwidth]{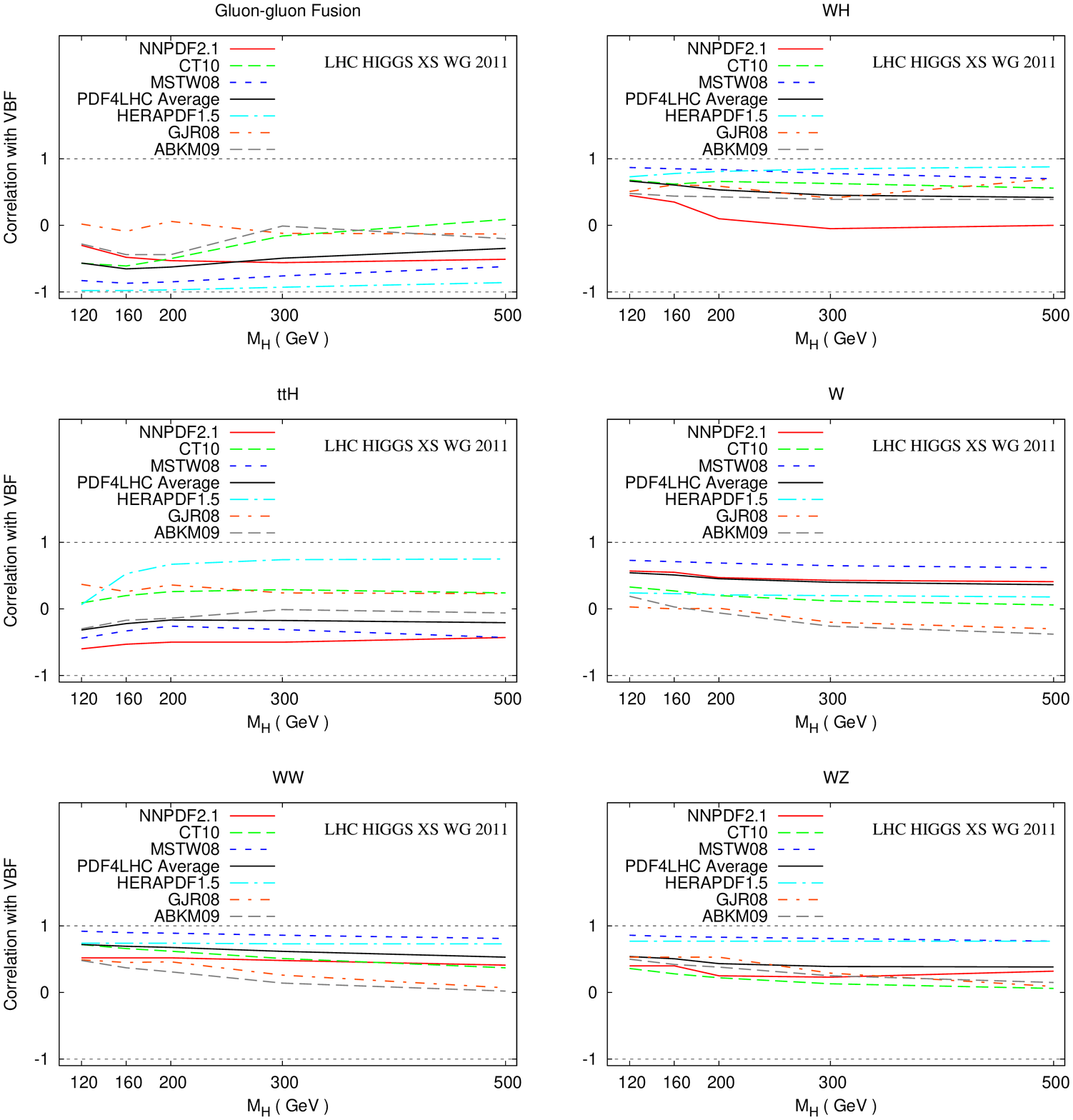}
}
\vspace*{-.5em}
\caption{The same as \refF{fig:corr-ggh1} for the vector
boson fusion process.
\label{fig:corr-vbf1}} 
\end{figure}

\begin{figure}
\centerline{  
\includegraphics[width=0.99\textwidth]{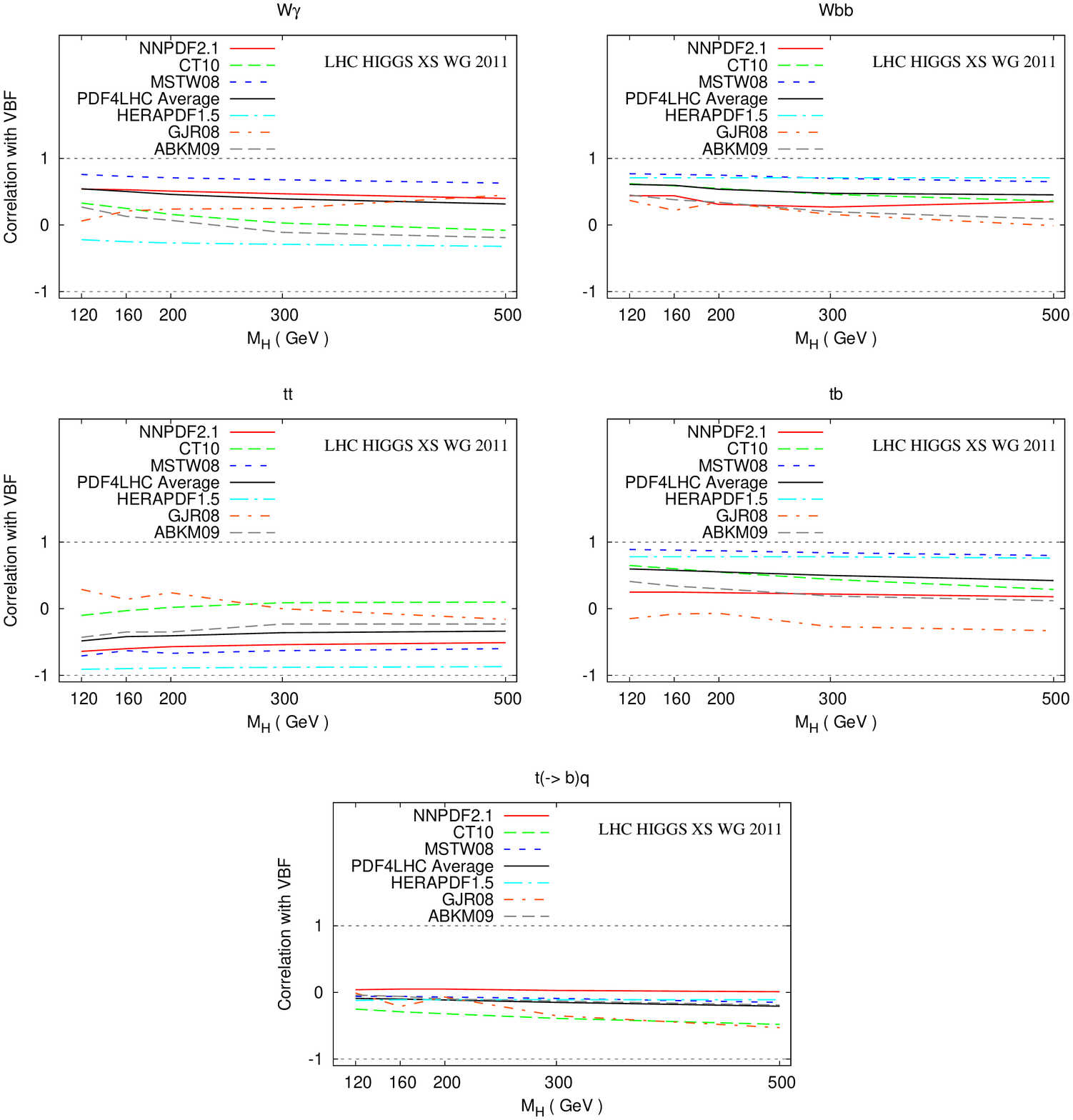}
}
\vspace*{-.5em}
\caption{The same as \refF{fig:corr-ggh2} for the vector
boson fusion process.
\label{fig:corr-vbf2}} 
\end{figure}

A more complete list of processes, with results for each 
individual PDF set, may be found in the tables on the webpage at the
LHC Higgs Cross Section Working Group TWiki~\cite{Twiki}. Note, however, that there 
is a high degree of redundancy in the approximate correlations of 
many processes. For example $\PW$ production is very similar to $\PZ$ production, 
both depending on partons (quarks in this case) at very similar hard 
scales and $x$ values. 
Similarly for $\PW\PW$ and $\PZ\PZ$, and the subprocesses $\Pg\Pg\to \PW\PW(\PZ\PZ)$ and 
Higgs production via gluon--gluon fusion for $\MH=200\UGeV$.  

\begin{figure}
\centerline{  
\includegraphics[width=0.99\textwidth]{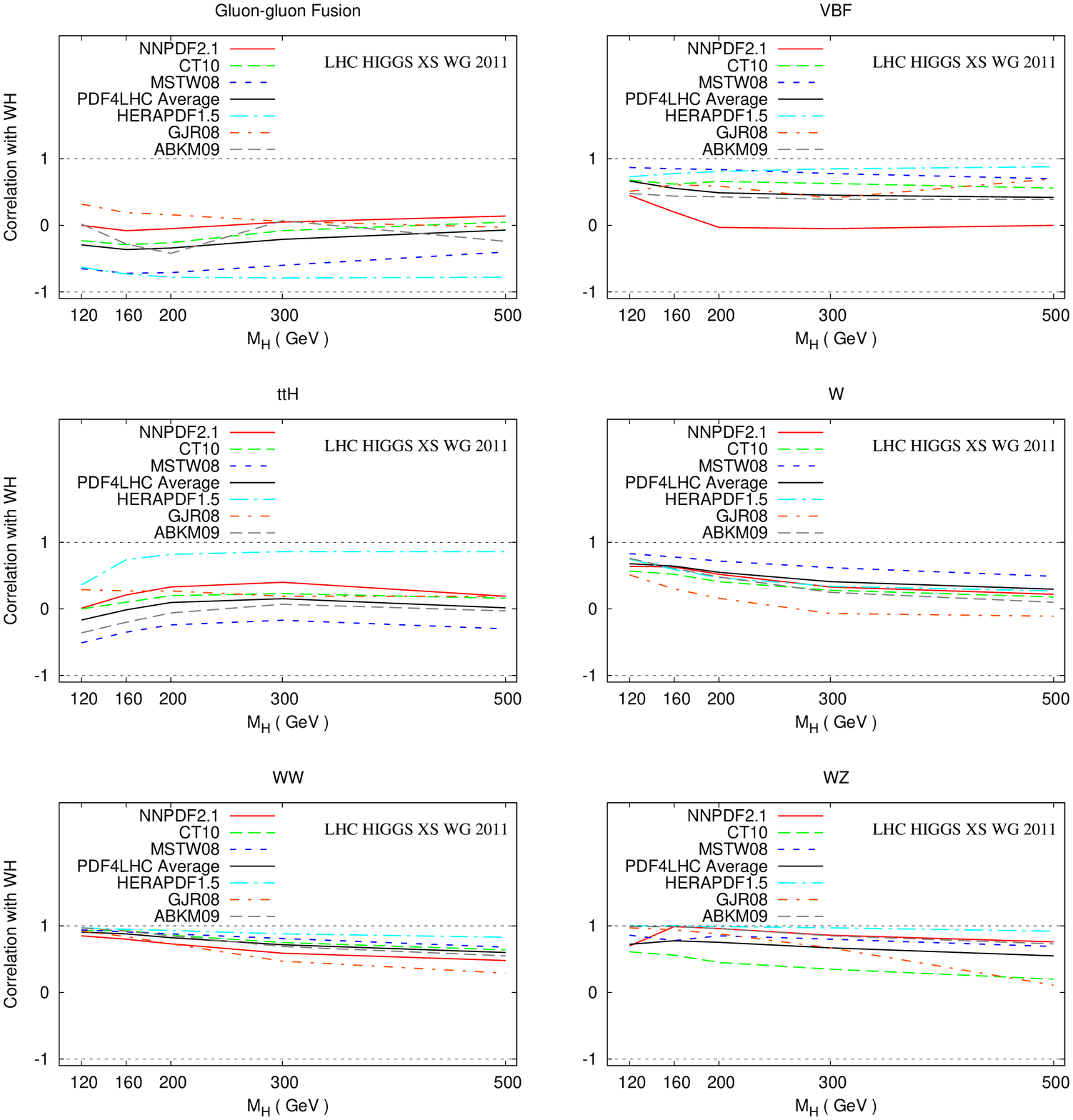}
}
\vspace*{-.5em}
\caption{The same as \refF{fig:corr-ggh1} for the $\PW\PH$ production process.
\label{fig:corr-wh1}} 
\end{figure}

\begin{figure}
\centerline{  
\includegraphics[width=0.99\textwidth]{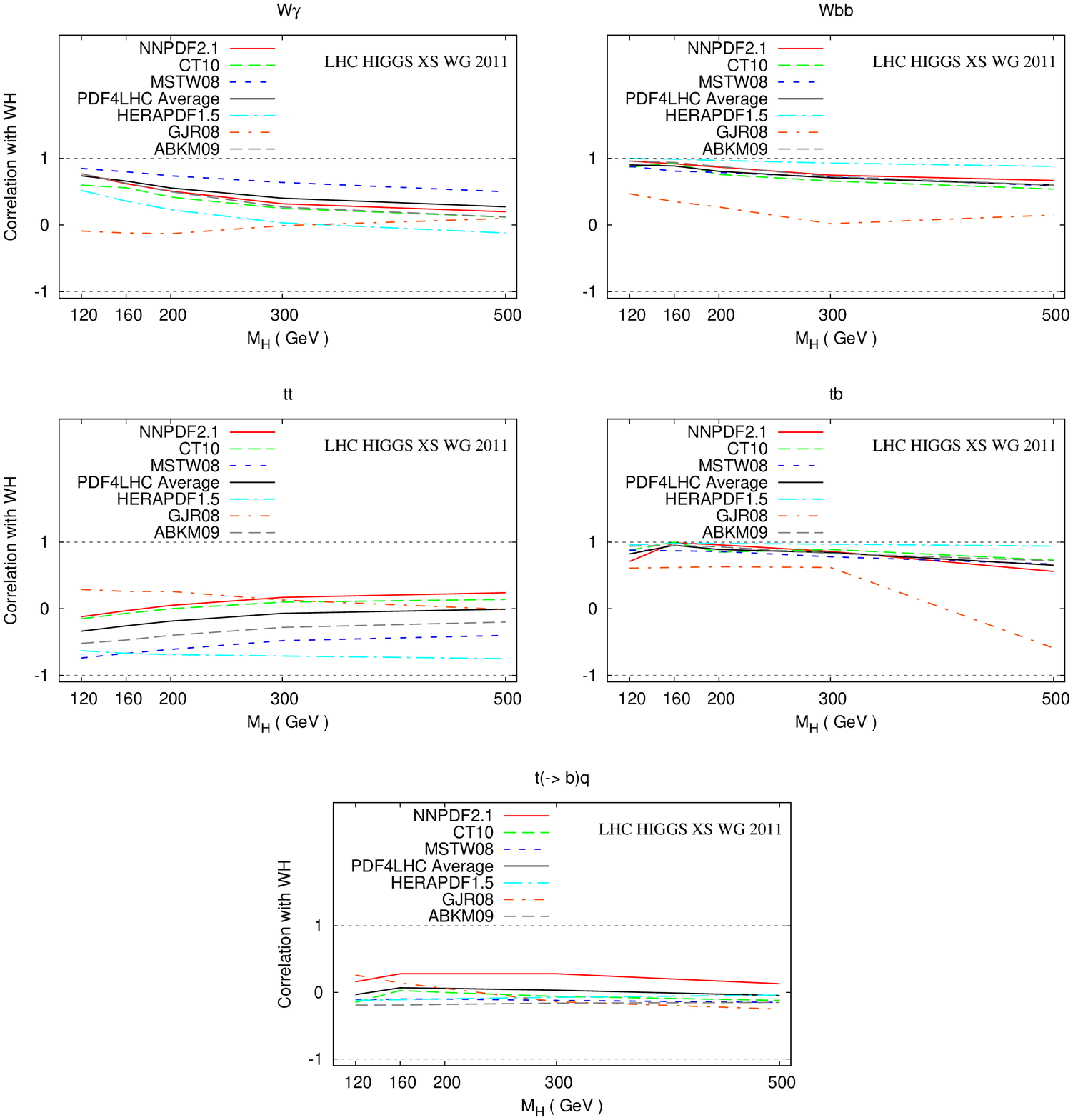}
}
\vspace*{-.5em}
\caption{The same as \refF{fig:corr-ggh2} for the $\PW\PH$ production process.
\label{fig:corr-wh2}} 
\end{figure}

\begin{figure}
\centerline{  
\includegraphics[width=0.99\textwidth]{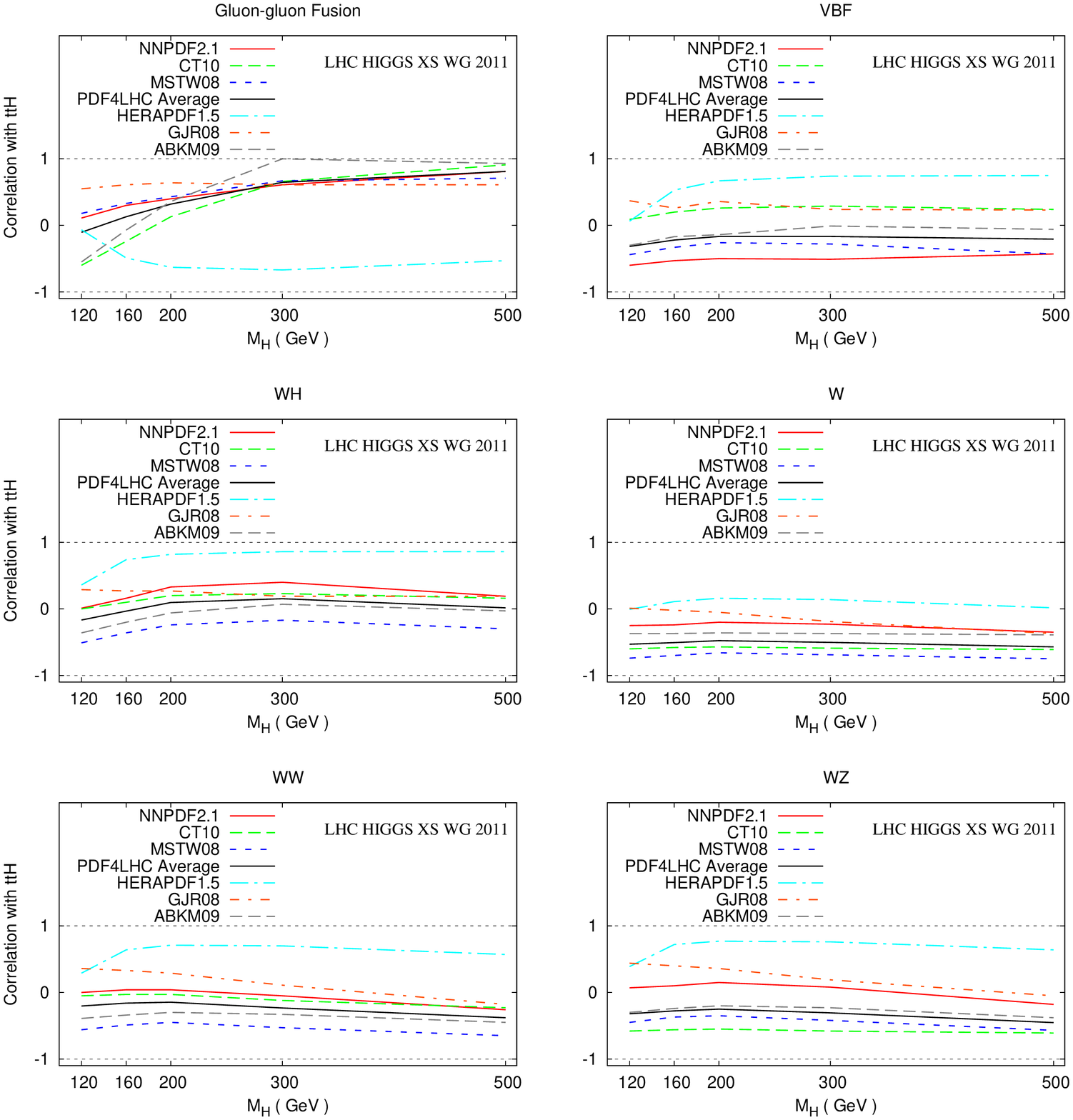}
}
\vspace*{-.5em}
\caption{The same as \refF{fig:corr-ggh1} for the $\PQt\PAQt\PH$ production process.
\label{fig:corr-tth1}} 
\end{figure}

\begin{figure}
\centerline{  
\includegraphics[width=0.99\textwidth]{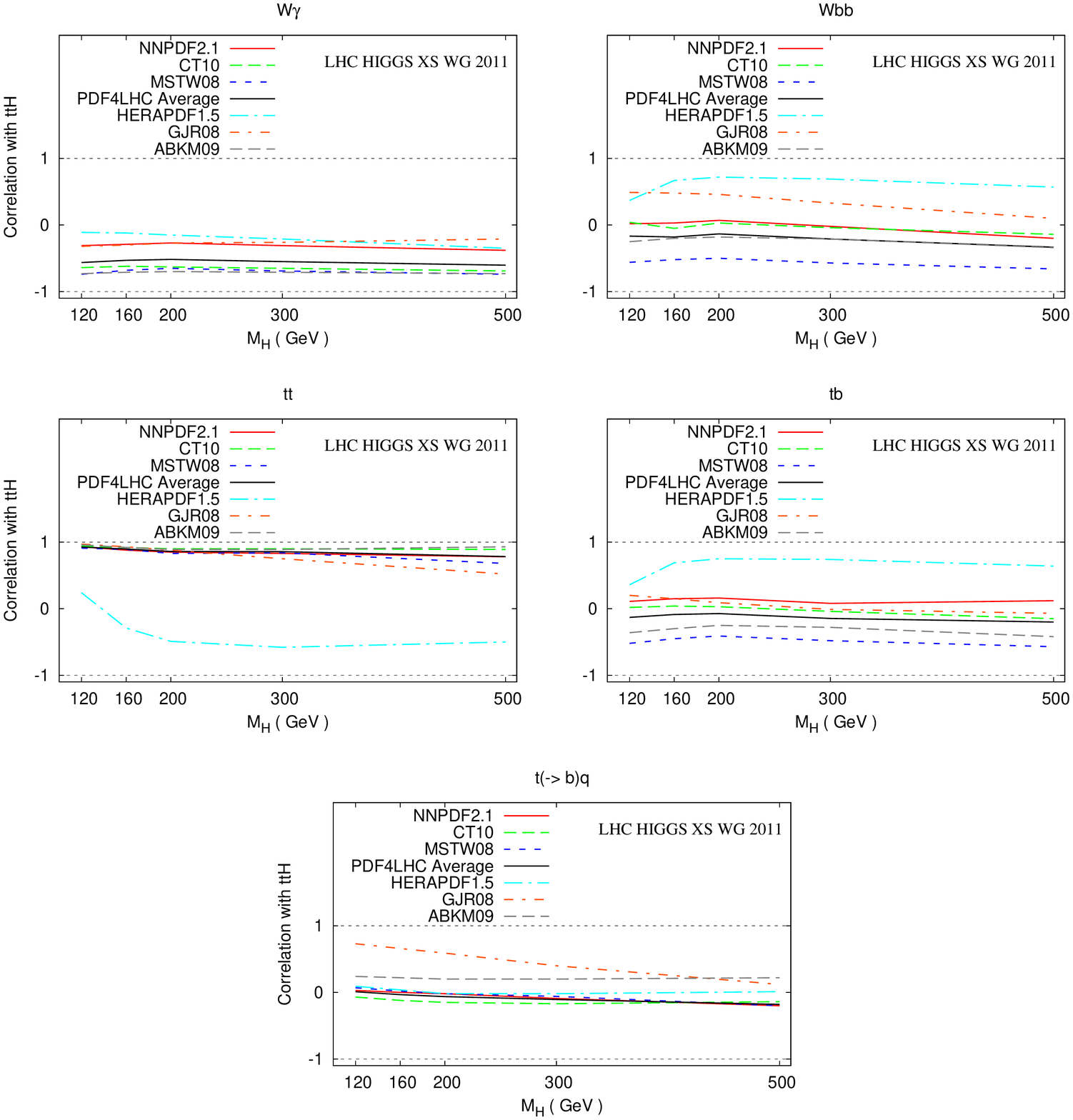}
}
\vspace*{-.5em}
\caption{The same as \refF{fig:corr-ggh2} for the $\PQt\PAQt\PH$ production process.
\label{fig:corr-tth2}} 
\end{figure}

\begin{sloppypar}
More detailed results are presented for the MSTW2008, NNPDF2.1, CT10, 
HERAPDF1.5, GJR08, and ABKM09 PDFs in 
\refFs{fig:corr-back-w}--\ref{fig:corr-tth2}. 
The result using each individual PDF sets is compared to the 
(updated) PDF4LHC average. There is usually a fairly narrow clustering of the 
individual results about the average, with a small number of cases 
where there is one, or perhaps two outliers. The sets with the largest 
parametrisations 
for the PDFs generally tend to give smaller magnitude correlations or 
anticorrelations, 
but this is not always the case, \eg  NNPDF2.1 gives the largest 
anti-correlation for VBF--$\PQt\PAQt\PH$.
There are some unusual features, \eg  for HERAPDF1.5 and high values of $\MH$, 
the $\PQt \PAQt\PH$ 
correlations with quantities depending on the high-$x$ gluon, \eg  
$\Pg\Pg\PH$ and $\PQt\PAQt$ is opposite to the other sets and the correlations with 
quantities depending on high-$x$ quarks and antiquarks, \eg  VBF and $\PW\PW$, is
stronger. This is possibly related to the large high-$x$ 
antiquark distribution in 
HERAPDF which contributes to $\PQt \PAQt\PH$ but not $\Pg\Pg\PH$ or very much 
to $\PQt\PAQt$. GJR08 has a tendency to obtain more correlation between 
some gluon dominated processes, \eg  $\Pg\Pg\PH$ and $\Pt \PAQt$ and quark 
dominated processes, \eg  $\PW$  and $\PW\PZ$, perhaps because the dynamical 
generation of PDFs couples the gluon and quark more strongly.     
\end{sloppypar}

We can now also see the origin of 
the cases where the averages move two classes.
For VBF--$\PW\PGg$ at lower masses GJR08, ABKM09, and HERAPDF1.5 all lie lower than 
the (updated) PDF4LHC average, but not too different to CT10. For VBF--$\PW$ at 
$\MH=500\UGeV$, ABKM09, and GJR08 give a small anticorrelation, while others 
give a correlation, though it is only large in the case of MSTW2008. For 
$\PW\PGg$--$\PQt\PAQb$ the change is due to slightly lower 
correlations for GJR08. However, although the change is two 
classes, in practice it is barely more than $0.2$.  
For $\PW\PZ$--$\PQt\PAQt\PH$ at $\MH=120\UGeV$ both HERAPDF1.5 and GJR08 have a larger 
correlation. This increases with $\MH$ for HERAPDF1.5 as noted, whereas 
GJR08 heads closer to the average, but the move at $\MH=120\UGeV$ is only 
marginally two classes, and is only one class for other masses. For 
$\PW\PQb \PAQb$--$\PQt \PAQt \PH$ at $\MH=200\UGeV$ the situation is similar, and 
MSTW2008 gives easily the biggest pull in the direction of anticorrelation. 

\subsubsection{Additional correlation studies}

The inclusion of the $\alphas$ uncertainty on the correlations, compared to 
PDF only variation, was also 
studied for some PDF sets, \eg  MSTW2008 using the 
approximation that the PDF sets for 
the upper and lower $\alphas$ values~\cite{Martin:2009bu} simply form another pair of orthogonal
eigenvectors (show to be true in the quadratic approximation~\cite{Lai:2010nw}). 
This increases the correlation between some processes, \ie  $\PW$ 
production 
and Higgs via gluon fusion, because the former increases with $\alphas$ due 
to increased evolution of quarks while the latter increases due to direct 
dependence on $\alphas$. Similarly for \eg  Higgs production via gluon 
fusion and $\PQt\PAQt$ production since each depends directly on $\alphas$.
This may also contribute to some of the stronger correlations 
seen using GJR08 in some similar cases.   
In a handful of processes it reduces correlation, \eg  $\PW$ and $\PW\PQb\PAQb$ 
since the latter has a much stronger $\alphas$ dependence. In most cases the 
change compared to PDF-only correlation for a given PDF sets small, and it is not an obvious contributing factor to the cases where the 
(updated) PDF4LHC average is noticeably different to the average using six sets, 
except possibly to $\PW\PQb \PAQb$--$\PQt \PAQt \PH$ at $\MH=200\UGeV$, where 
$\alphas$ does increase correlation.   
  
A small number of correlations were also calculated at NNLO for MSTW2008 PDFs, 
\ie  $\PW$, $\PZ$ and $\Pg\Pg \to \PH$ for the same range of $\MH$. The correlations 
when taking into account PDF uncertainty alone were almost identical to those 
at NLO, variations being less than $0.05$. When $\alphas$ uncertainty 
was included the correlations changed a little more due to $\Pg\Pg \to \PH$
having more direct dependence on $\alphas$, but this is a relatively minor 
effect. Certainly the results in  
\refTs{tab:signal-correlations} 
and \ref{tab:back-correlations}, though calculated at NLO,  can be 
used with confidence at NNLO.

\clearpage

\newpage
\section{NLO parton shower\footnote{%
    M. Felcini, F. Krauss, F. Maltoni, P. Nason and J. Yu. (eds.); J. Alwall, 
E. Bagnaschi, G. Degrassi, M. Grazzini, K. Hamilton, S. H\"oche, C. Jackson, H. Kim, Q. Li, F. Petriello, M. Sch\"onherr, F. Siegert,
P. Slavich, P. Torrielli and A. Vicini.}}

\newcommand\figfact{0.38}  
\newcommand\figfacto{0.34}  

\newcommand\Rups{R^s}
\newcommand\Rupf{R^f}

\newcommand\mT{\ensuremath{m_{\mathrm{T}}}} 

\providecommand\pperpmin{p_\perp^{\rm min}}

\providecommand\fortran{Fortran\xspace}
\providecommand{\pythia}{{\sc Pythia}}
\providecommand{\madgraph}{{\sc MadGraph}}
\providecommand\phojet{Phojet\xspace}
\providecommand{\pythiasix}{{\sc Pythia~6}}
\providecommand{\pythiaeight}{{\sc Pythia~8}}
\providecommand{\herwig}{{\sc Herwig}}
\providecommand{\herwigsix}{{\sc Herwig~6}}
\providecommand{\herwigpp}{{\sc Herwig}\raisebox{0.1ex}{\small $++$}}
\providecommand{\sherpa}{{\sc Sherpa}}
\providecommand{\alpgen}{{\sc AlpGen}}
\providecommand\helac{HELAC\xspace}
\providecommand\charybdis{Charybdis\xspace}
\providecommand\jimmy{Jimmy\xspace}
\providecommand\rivet{Rivet\xspace}
\providecommand\rivetgun{Rivetgun\xspace}
\providecommand\professor{Professor\xspace}
\providecommand\fastjet{FastJet\xspace}
\providecommand\hepmc{HepMC\xspace}
\providecommand\agile{AGILe\xspace}
\providecommand\hztool{HZTool\xspace}
\providecommand\numpy{NumPy\xspace}
\providecommand\scipy{SciPy\xspace}
\providecommand\minuit{Minuit\xspace}
\providecommand\pyminuit{PyMinuit\xspace}
\providecommand\python{Python\xspace}
\providecommand\pt{p_T\xspace}
\providecommand\as{\alpha_{\rm \scriptscriptstyle s}\xspace}
\providecommand\muR{\mu_{\rm \scriptscriptstyle R}\xspace}
\providecommand\muF{\mu_{\rm \scriptscriptstyle F}\xspace}
\providecommand\kT{\ensuremath{k_{\mathrm{T}}}} 
\providecommand\HqT{{\sc HqT}\xspace}
\providecommand\PYTHIA{\pythia}
\providecommand\HERWIG{\herwig}
\providecommand\SHERPA{\sherpa}
\providecommand\POWHEG{{\sc POWHEG}\xspace}
\providecommand\POWHEGPYTHIA{{\sc POWHEG+\pythia}\xspace}
\providecommand\POWHEGHERWIG{{\sc POWHEG+\herwig}\xspace}
\providecommand\POWHEGBOX{{\sc POWHEG BOX}\xspace}
\providecommand\MCatNLO{{\sc MC@NLO}\xspace}
\providecommand\aMCatNLO{{\sc aMC@NLO}\xspace}
\providecommand\ARIADNE{{\sc A\scalebox{0.9}{RIADNE}}\xspace}
\providecommand\ADICIC{{\sc A\scalebox{0.9}{DICIC}++}\xspace}
\providecommand\AMEGIC{{\sc A\scalebox{0.9}{MEGIC}++}\xspace}
\providecommand\Comix{{\sc Comix}\xspace}
\providecommand\HNNLO{{\sc H\scalebox{0.9}{NNLO}}\xspace}
\providecommand\MGME{{\sc M\scalebox{0.9}{AD}G\scalebox{0.9}{RAPH/}M\scalebox{0.9}{AD}E\scalebox{0.9}{VENT}}\xspace}

\subsection{Introduction}

Recently, Monte Carlo event generators have profited by a number of theoretical achievements
that have significantly improved the capability of making accurate predictions and simulations
of events taking place at high-energy colliders. A very short review of the state-of-the-art of the field has been
given in the first volume of this Yellow Report~\cite{Dittmaier:2011ti}, and we refer the reader there for the basic principles  and
the most important improvements with respect to a more standard parton-shower (PS) approach. We just recall the
two main results:  the possibility of consistently including exact next-to-leading order (NLO) corrections, i.e.\ to have NLO+PS generators~\cite{Frixione:2002ik,
Frixione:2003ei,Frixione:2005vw,Frixione:2006gn,Frixione:2007zp,Frixione:2008yi,
LatundeDada:2007jg,Nason:2004rx,Nason:2006hfa,Frixione:2007nu,Frixione:2007vw,
Frixione:2007nw,LatundeDada:2006gx,Hamilton:2008pd,Hamilton:2009za,
Alioli:2008gx,Alioli:2008tz,Alioli:2009je,LatundeDada:2008bv,Alioli:2010qp,Hoche:2010kg,Frixione:2010ra,Torrielli:2010aw,Alioli:2011nr} 
and the merging of parton-shower simulations and high-multiplicity tree-level matrix-element (ME) 
generators~\cite{Mangano:2001xp,Catani:2001cc,Lonnblad:2001iq,Krauss:2002up,
Mrenna:2003if,Schalicke:2005nv,Alwall:2007fs,Hoeche:2009rj,Hamilton:2009ne,Siegert:2010mk,Hamilton:2010wh}.
Moreover, in the last year, an impressive acceleration in achieving the full automatisation of NLO 
computations~\cite{Frederix:2009yq,Hirschi:2011pa,Bevilacqua:2011xh,Cullen:2011xs,Cascioli:2011va} 
as well as their interface to parton showers~\cite{Alioli:2010xd,Hoche:2010pf,Frederix:2011zi,Hoeche:2011fd} 
has taken place. Thanks to such new techniques and their corresponding implementations, all the main 
Higgs-production channels (ggF,VBF,VH,ttH) at the LHC are now available in the context of 
NLO+PS~\cite{Alioli:2008tz,Hamilton:2009za,Nason:2009ai,D'Errico:2011um,Frederix:2011zi,Garzelli:2011vp} together 
with some of the most important backgrounds~\cite{Alioli:2008gx,Hamilton:2010mb,D'Errico:2011sd,Oleari:2011ey,Frederix:2011qg,Melia:2011tj,Frederix:2011ss,Frederix:2011ig,Garzelli:2011is}.
In this respect, it is certainly fair to state that state-of-the-art Higgs phenomenology at the LHC can 
now be performed at least at NLO+PS accuracy. 
This also implies that several important issues, such as those related to the estimation of uncertainties 
in NLO+PS simulations due, for instance, to scale uncertainties or different matching procedures, 
can be systematically studied for the first time. The aim of this section is to show how such studies can 
now be performed in the context of the current Higgs-boson searches. We stress that the results of our sample studies, which consider a light SM Higgs boson, can be easily 
extended to other Higgs mass ranges or to the search in scenarios with enhanced (or suppressed) couplings. In addition, in the case of a Higgs observation,
such uncertainties would play a crucial role in the accurate determination of its properties.

In more detail, in this section we address the following topics:
\begin{itemize}
\item Uncertainties in NLO+PS generators.
While in fixed-order computations there is considerable experience in
using scale variation to estimate theoretical error, in the framework of NLO+PS generators this might not be sufficient.
Suggestions for obtaining more realistic error bands will be given, and
areas that will require further work will be identified.
Uncertainties having to do with shower effects beyond the NLO level,
hadronisation and underlying event (UE) will also be considered.
\item Tuning of NLO+PS generators. In the case of Higgs production in gluon
fusion, a next-to-next-to-leading-log (NNLL)
matched to a next-to-next-to-leading-order (NNLO)
calculation exists, and is implemented in the program {\sc HqT}~\cite{hqt}.
One can use the {\sc HqT} results to improve the output of NLO+PS generators
in several ways through reweighting the generated events, tuning the parameters of the NLO+PS
generators or doing both things to achieve better agreement with {\sc HqT}. 
Issues related to these procedures are discussed.
\item
Since the publication of the first volume of this
Yellow Report~\cite{Dittmaier:2011ti}, some improvements on existing
generators have appeared. We present here these new developments.
\end{itemize}
The section is organised as follows. In \refS{NLOPSsec:compHqT} the POWHEG BOX and the
MC@NLO Higgs-production generators are compared with the
{\sc HqT} prediction for the Higgs $\pT$ spectrum.
In \refS{NLOPSsec:Sherpa}
a study of uncertainties in $\Pg\Pg\to \PH\to \PW\PW^{(*)}$ using {\sc Sherpa} is
presented.
In \refS{NLOEx} a study of systematic effects of NLO+PS
tools in their implementation within the ATLAS and CMS event generation
framework.
In \refS{NLOPSsec:guidelines} we discuss
issues related to the use of the {\sc HqT} result to improve the
NLO+PS programs. In particular, we provide recommendations
on how to perform event reweighting,
if needed, and how to set up the parameters of the NLO+PS
generators to achieve better agreement with the {\sc HqT} result.
In \refS{NLOPSsec:ggHfullmt}, we present new generators
for Higgs production in gluon fusion, including full top and
bottom mass dependence, in the \madgraph{} and in the \POWHEGBOX{}
framework.

In \refS{NLOPSsec:aMCatNLO}, we present a new \aMCatNLO{} feature that allows 
performing studies of scale and PDF dependences by reweighting the same set of events.

Finally, some controversial points yet to be resolved have emerged during the preparation of this
report and are summarised in \refS{NLOPSsec:controversy}.

\subsection{\POWHEGBOX{} and \MCatNLO{} comparison with \HqT{}%
\footnote{K.~Hamilton, F.~Maltoni, P.~Nason and P.~Torrielli.}}

\label{NLOPSsec:compHqT}

In general, there are several sources of  uncertainties possibly affecting the NLO+PS results:
\begin{itemize}
\item Factorisation and renormalisation scale uncertainties. Normally, the NLO
calculation underlying the NLO+PS generator has independent factorisation and
renormalisation scales, and results are affected at the NNLO level if these scales are varied.
\item Uncertainties related to the part of the real cross section that is treated with the shower algorithm.
\item Uncertainties related to how the shower algorithm is implemented.
\item Uncertainties related to the PDF's themselves and also  whether the PDF's used in the
shower algorithm are different from those used in the NLO calculation.
\item Further uncertainties common to all shower generators, i.e.\ hadronisation, underlying event, etc.
\end{itemize}
We focus here on the first three items, which are by far the dominant ones.
As a relevant and phenomenologically important observable, we consider the 
transverse-momentum distribution of the Higgs boson, in the process $\Pg\Pg\to \PH$. This observable is not  a fully NLO
one in the $\Pg\Pg\to \PH+X$ NLO cross-section calculation.  For this reason it is very sensitive to both soft and hard effects and
the NLO+PS results displays more marked differences with respect to the pure
NLO calculation. The latter, in fact, is divergent for $\pT\to 0$, with
and infinite negative spike at $\pT=0$, representing the contribution
of the virtual corrections. The NLO+PS approaches, instead, yield a positive
cross section for all $\pT$, with the characteristic Sudakov peak at small $\pT$.
Furthermore, this distribution can be computed
(using the \HqT{} program \cite{hqt}) at a matched NNLL+NNLO accuracy, which can
serve as a benchmark to characterise the output of the generators.

\begin{figure}
\centering
\includegraphics[width=.5\textwidth,height=\figfact\linewidth]{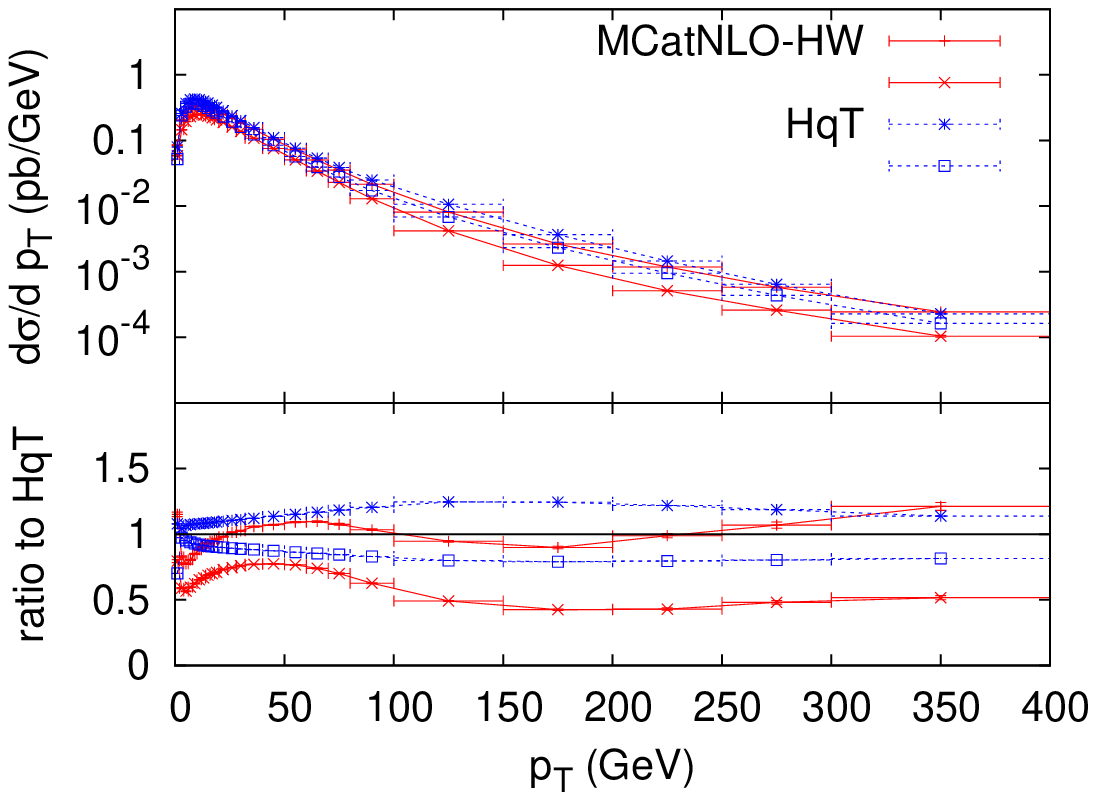}\nolinebreak
\includegraphics[width=0.5\textwidth,height=\figfact\linewidth]{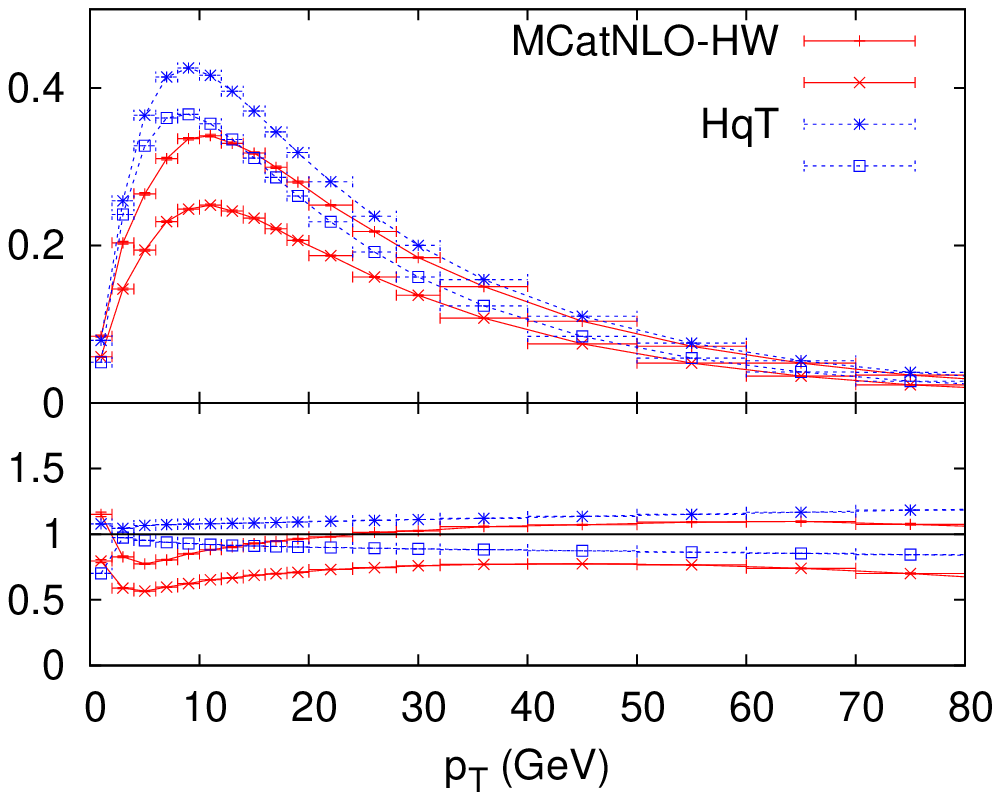}
\\[1em]
\includegraphics[width=.5\textwidth,height=\figfact\linewidth]{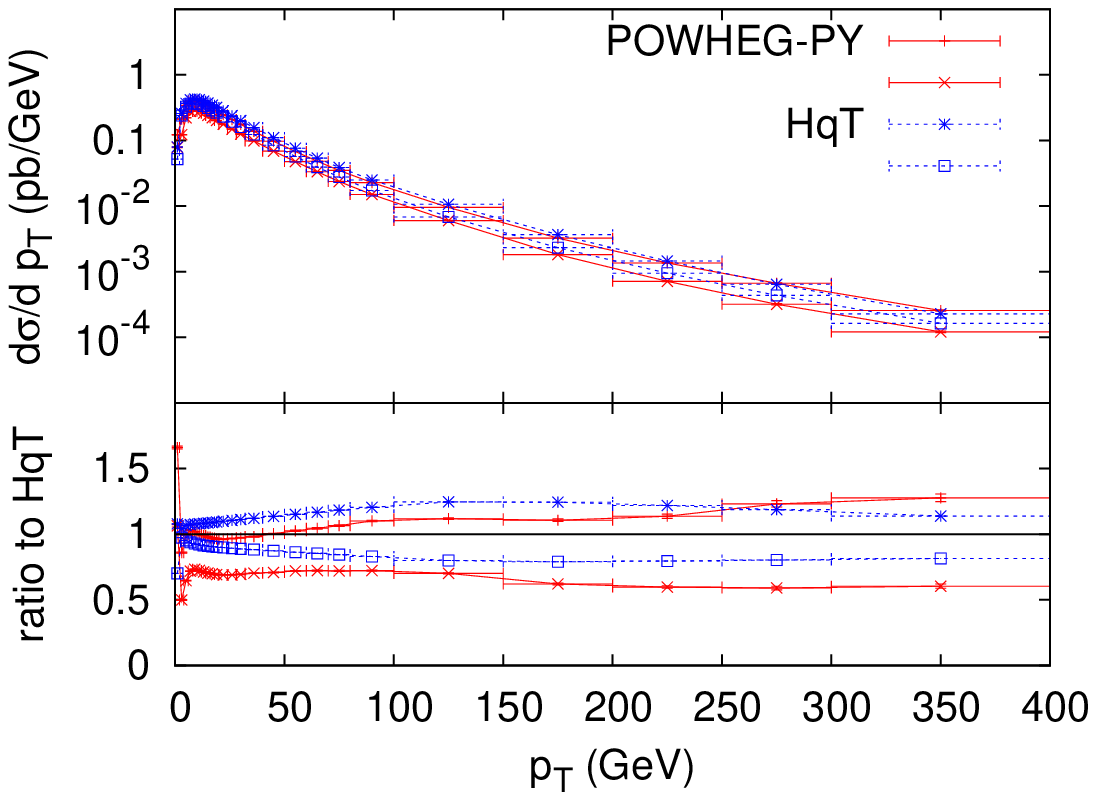}\nolinebreak
\includegraphics[width=.5\textwidth,height=\figfact\linewidth]{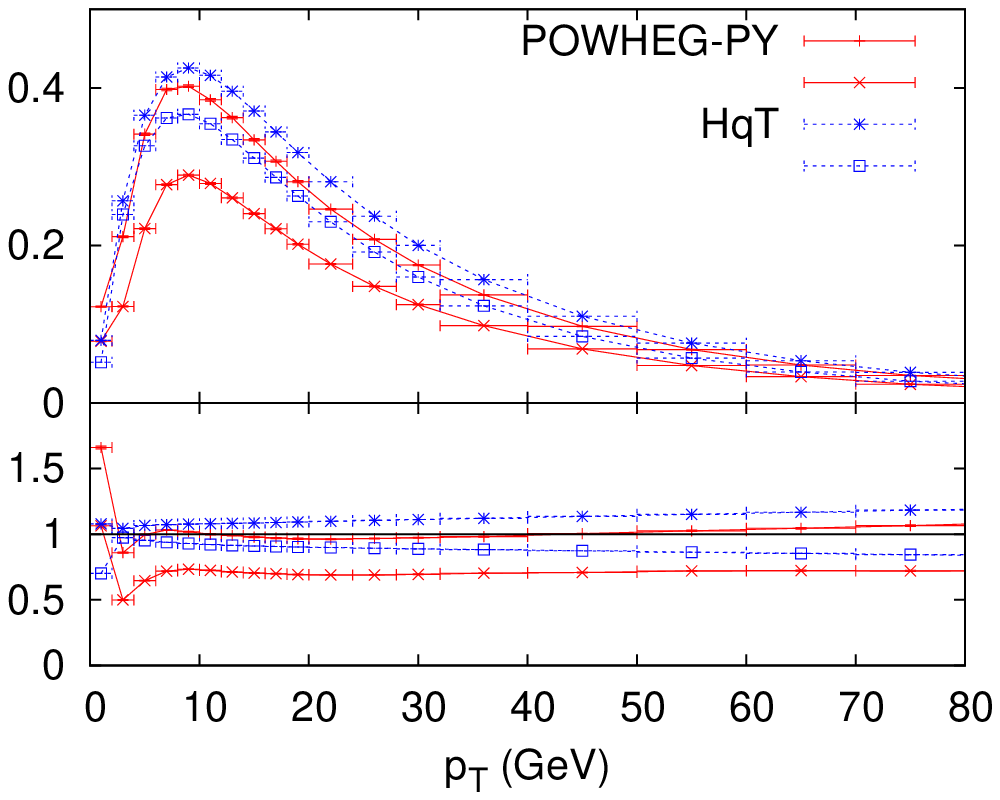}
\caption{Uncertainty bands for the transverse-momentum spectrum of the
  Higgs boson at LHC, $7\UTeV$, for a Higgs mass $\MH=120\UGeV$.  On the
  upper plots,
  the \MCatNLO{}+\HERWIG{} result obtained using the non-default value
  of the reference scale equal to $\MH$. On the lower plots, the
  \POWHEG{}+\PYTHIA{} output, using the non-default $\Rups +\Rupf $
  separation.  The uncertainty bands are obtained by changing $\muR$
  and $\muF$ by a factor of two above and below the central value,
  taken equal to $\MH$, with the restriction $0.5<\muR/\muF<2$.}
\label{figNLOMC:ptMCatNLOMH}
\end{figure}

We begin by showing in \refF{figNLOMC:ptMCatNLOMH} the
comparisons of \MCatNLO{} + \HERWIG{} and \POWHEG{} + \PYTHIA{} with \HqT{}, with
a choice of parameters that yields the best agreement,
in shape, of the $\pT$ distributions. This choice of parameters can be  therefore considered  the
main outcome of this study, i.e.\ it embodies our recommendation for the settings of the two generators.
The settings are as follows:
\begin{itemize}
\item
\MCatNLO{} should be run with the factorisation and renormalisation
scale equal to $\MH$.
\item
\POWHEG{} should be run with the $h$ parameter equal to $\MH/1.2$.
For $\MH=120\UGeV$, this setting is achieved introducing the line
{\tt hfact 100} in the {\tt powheg.input} file.
\end{itemize}
In the figures, the uncertainty band of the \MCatNLO{}+\HERWIG{}
and of \POWHEG{}+\PYTHIA{},
both compared to the \HqT{} uncertainty band, are displayed.
In the lower insert in the figure,
the ratio of all results to the central value of \HqT{} is also displayed.
As stated above, \MCatNLO{}+\HERWIG{}
is run with the central value of the renormalisation scale fixed to $\MH$, and
\POWHEG{} is run with the input line {\tt hfact 100}.
The red, solid curves represent the uncertainty band of the NLO+PS, while
the dotted blue lines represent the band obtained with the \HqT{} program.
The reference scale is chosen equal to $\MH/2$ in \HqT{}, and the scale variations
are performed in the same way as in the NLO+PS generators:
once considers all variations of a factor of two above and below the central scale,
with the restriction $0.5<\muR/\muF<2$.
We have used the MSTW2008 NNLO central set for all curves. This is because \HqT{}
requires NNLO parton densities, and because we want to focus upon differences
that have to do with the calculation itself, rather than the PDF's. The
\HqT{} result has been obtained by running the program with full NNLL+NNLO accuracy,
using the ``switched'' result. The resummation scale $Q$ in \HqT{} has been set
to $\MH/2$.

We notice that both programs are compatible in shape with the \HqT{} prediction.
We also notice that the error band of the two NLO+PS generators is relatively small at
small $\pT$ and becomes larger at larger $\pT$. This should remind us that the
NLO+PS prediction for the high $\pT$ tail is in fact a tree-level-only prediction,
since the production of a Higgs plus a light parton starts at order $\alphas^3$,
its scale variation is of order $\alphas^4$, and its \emph{relative} scale variation
is of order $\alphas^4/\alphas^3$, i.e.\ of order $\alphas$.\footnote{Here we remind the reader that
$\muF^2\frac{d}{d\muF^2} \alphas^3(\muR)=-b_0 3\alphas^4(\muR)$.}
On the other hand the total integral of the curve, i.e.\ the total cross section
(and in fact also the Higgs rapidity distribution, that is obtained by
integrating over all transverse momenta)
are given by a term of order $\alphas^2$ plus
a term of order $\alphas^3$, and their scale variation is also of order $\alphas^4$. Thus, their
relative scale variation is of order $\alphas^4/\alphas^2$, i.e.\ $\alphas^2$.

\begin{figure}
\centering
\includegraphics[width=.5\textwidth,height=\figfact\linewidth]{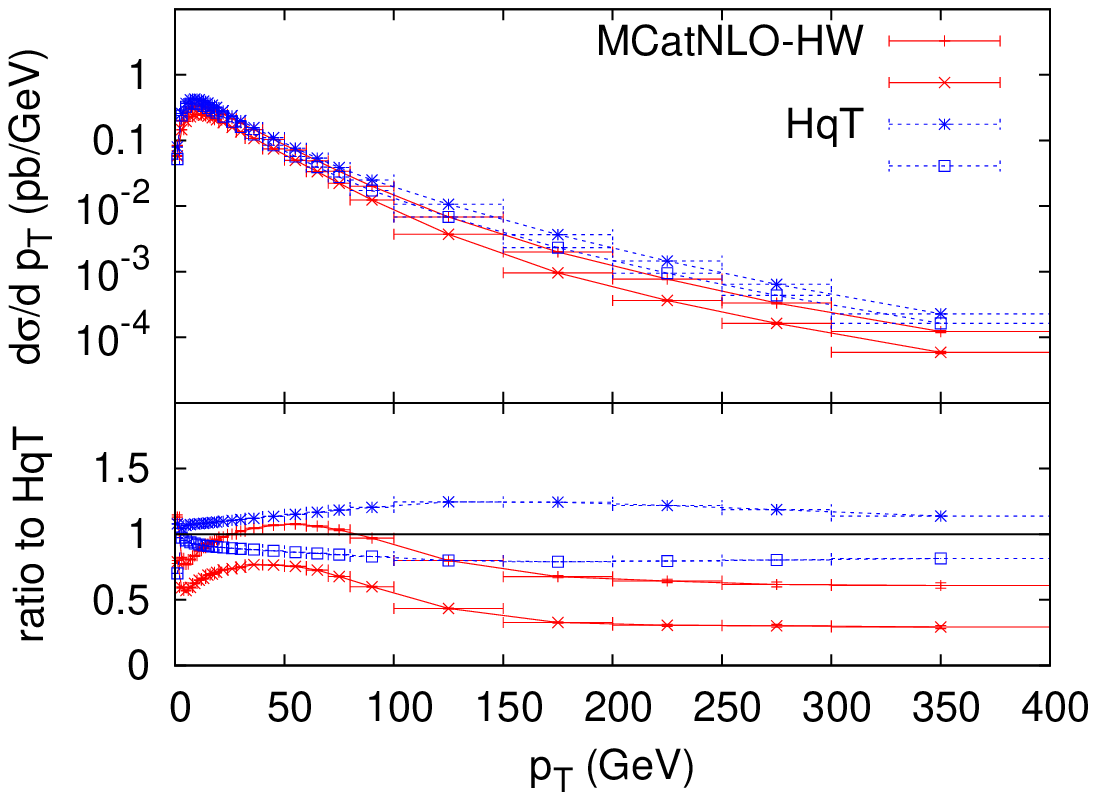}\nolinebreak
\includegraphics[width=.5\textwidth,height=\figfact\linewidth]{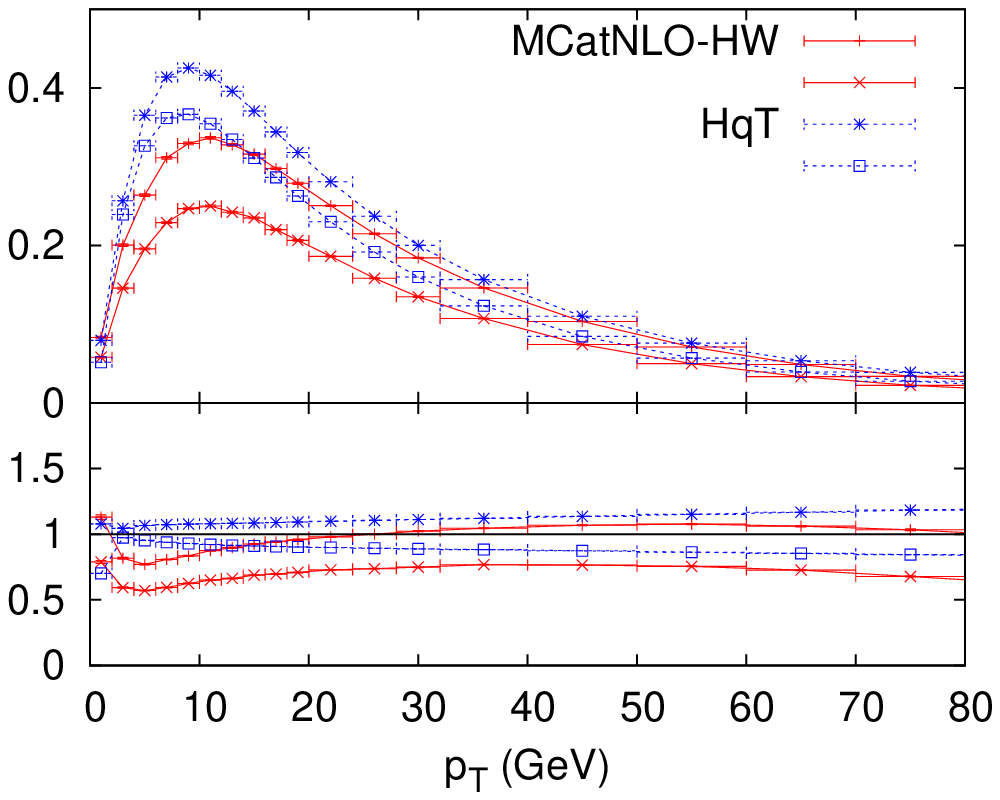}
\\[1em]
\includegraphics[width=.5\textwidth,height=\figfact\linewidth]{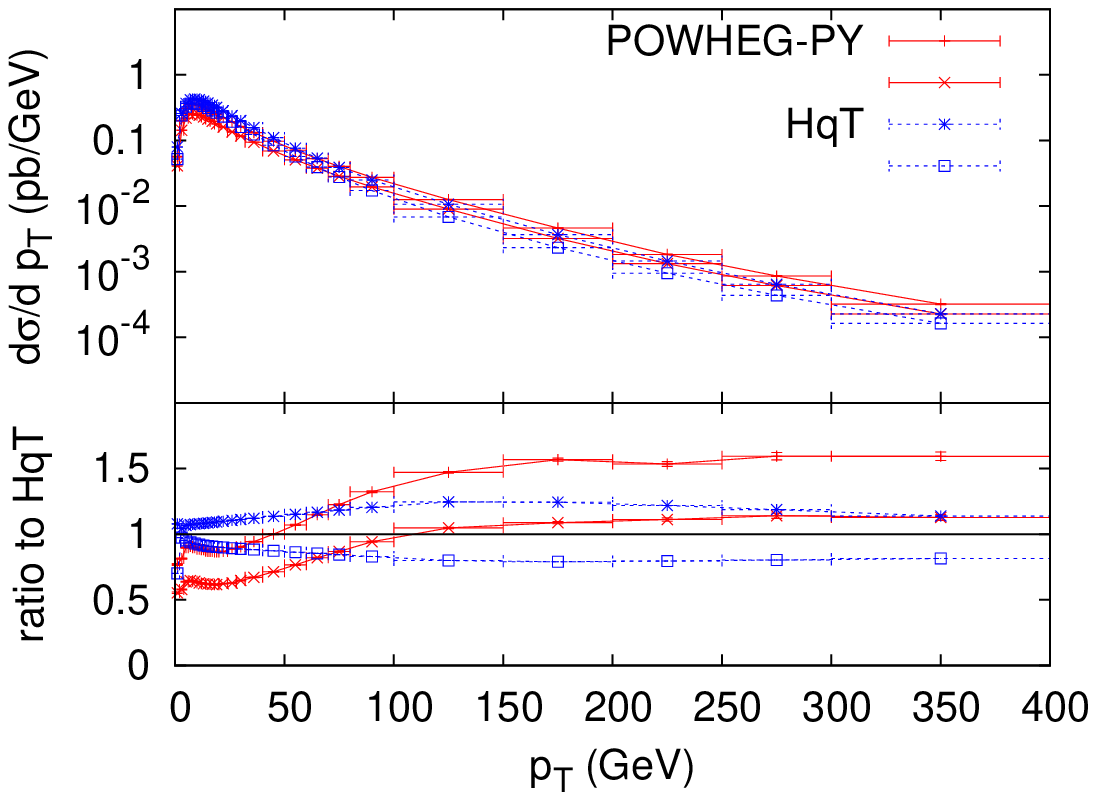}\nolinebreak
\includegraphics[width=.5\textwidth,height=\figfact\linewidth]{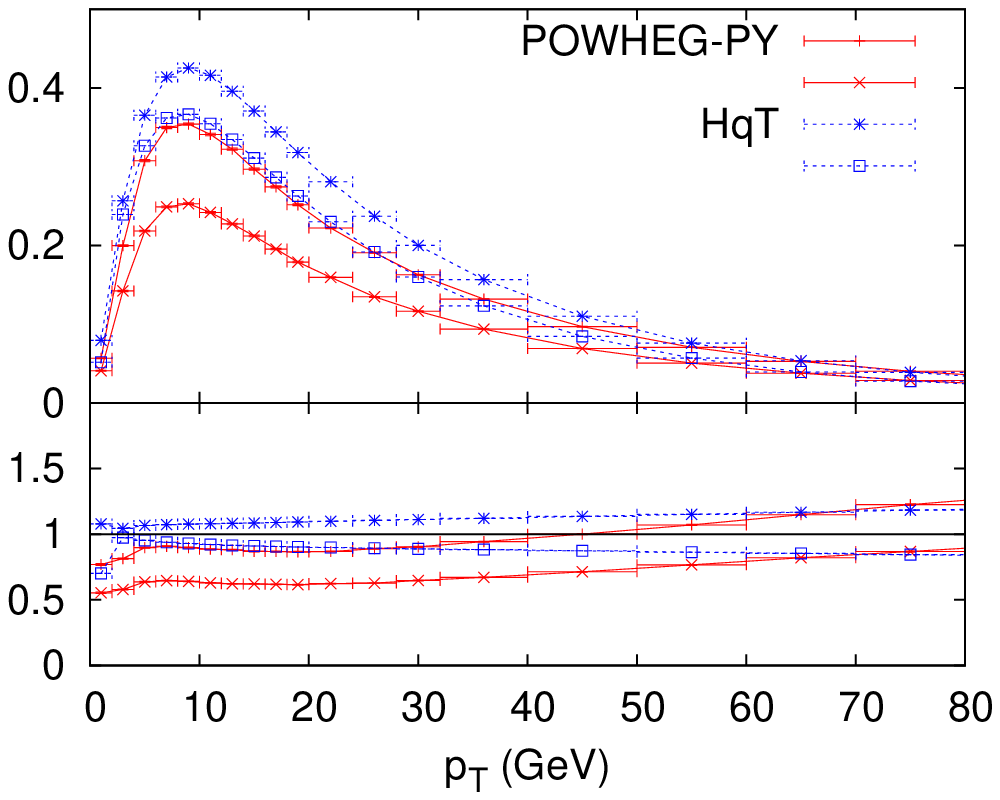}
\caption{The transverse-momentum spectrum of the Higgs in \MCatNLO{}
(upper) and in \POWHEG{}+\PYTHIA{} (lower) compared to the \HqT{} result. In the lower
insert, the same results normalised to the \HqT{} central value are shown.}
\label{figNLOMC:ptHdefault}
\end{figure}

It is instructive to analyse the difference between \MCatNLO{} and \POWHEG{} at their
default value of parameters. This is illustrated in \refF{figNLOMC:ptHdefault}.
The two programs are in reasonable agreement at small transverse momentum, but
display a large difference (about a factor of $3$) in the high transverse
momentum tail. This difference has two causes. One is the different scale
choice in \MCatNLO{}, where by default $\mu=\mT=\sqrt{\MH^2+\pT^2}$, where $\pT$ is the
transverse momentum of the Higgs. That accounts for a factor of $(\alphas(\mT)/\alphas(\MH))^3$,
which is about $1.6$ for the last bin in the plots
(compare the upper plots of \refF{figNLOMC:ptMCatNLOMH} with those of \refF{figNLOMC:ptHdefault}). 
The remaining difference is due to the fact that in \POWHEG{}, used with default parameters, the NLO $K$-factor
multiplies the full transverse-momentum distribution. The \POWHEG{} output is
thus similar to what is obtained with NLO+PS generator, as already observed
in the first volume of this Report.

This point deserves a more detailed explanation, which can be given
along the lines of \Bref{Alioli:2008tz,Nason:2010ap}.
We write below the differential cross section for the hardest emission in 
NLO+PS implementations (see the first volume of this report for details)
\begin{equation}
\label{EqNLOMC:Powheg}
{\rm d}\sigma^{\rm NLO+PS} =
{\rm d}\Phi_B\bar B^{s}(\Phi_B)
\left[\Delta^s(\pperpmin) 
  +  {\rm d}\Phi_{R|B}
      \frac{\Rups (\Phi_R)}{B(\Phi_B)}
      \Delta^s(\pT(\Phi))\right] 
+ {\rm d}\Phi_R \Rupf (\Phi_R),
\end{equation}
where
\begin{equation}\label{eqNLOMC:nlocorr1}
\bar{B}^{s}=B(\Phi_B)+\left[V(\Phi_B)+
\int{\rm d}\Phi_{R|B} \Rups (\Phi_{R|B})\right]\,.
\end{equation}
The sum $\Rups +\Rupf $ yields the real cross section for $\Pg\Pg\to \PH\Pg$,
plus the analogous terms for quark--gluon. Quark--antiquark annihilation is finite
and therefore only contributes to $\Rupf $.

In \MCatNLO{}, the $\Rups $ term is the shower approximation to the real cross
section, and it depends upon the SMC that is being used in conjunction with it.
In \POWHEG{}, one has much freedom in choosing $\Rups $, with the only constraint
$\Rups < R$, in order to avoid negative weights, and $\Rups \to R $
in the small-transverse-momentum limit  (in the sense that $\Rups-R$ should be integrable
in this region).

For the purpose of this review, we call $\mathbb{S}$ events (for shower)
those generated by the first term on the r.h.s.\ of Eq.~(\ref{EqNLOMC:Powheg}), 
i.e.\ those generated using the shower algorithm, and $\mathbb{F}$ (for finite)
events those generated by the $\Rupf$ term.\footnote{In the \MCatNLO{} 
language, these are called $\mathbb{S}$ and $\mathbb{H}$ events, where 
$\mathbb{S}$ stands for standard, and $\mathbb{H}$ for hard.}
The scale dependence typically affects the $\bar{B}$
and the $\Rupf$ terms in a different way.
A scale variation in the square bracket on the r.h.s.\
of Eq.~(\ref{EqNLOMC:Powheg}) is in practice never performed, since
in \MCatNLO{} this can only be achieved by changing
the scale in the Monte Carlo event generator that is being used,
and in \POWHEG{} the most straightforward way to perform it
(i.e.\ varying it naively by a constant factor) would spoil the NLL accuracy
of the Sudakov form factor. We thus assume from now on that the
scales in the square parenthesis are kept fixed. Scale variation
will thus affect $\bar{B}^{s}$ and $\Rupf $.

We observe now that the shape of the transverse momentum of the hardest 
radiation in $\mathbb{S}$ events is not affected by scale variations, given 
that the square bracket on the r.h.s.\ of Eq.~(\ref{EqNLOMC:Powheg}) is not 
affected by it, and that the factor $\bar{B}$
is $\pT$ independent. From this, it immediately
follows that the scale variation of the large-transverse-momentum
tail of the spectrum is of relative order $\alphas^2$, i.e.\ the same relative
order of the inclusive cross section, rather than of relative order $\alphas$,
since $\bar{B}$ is a quantity integrated in the transverse momentum.
Eq.~(\ref{EqNLOMC:Powheg}), in the large-transverse-momentum tail,
becomes
\begin{equation}
\label{EqNLOMC:PowhegLargePt}
{\rm d}\sigma^{\rm NLO+PS} \approx
{\rm d}\Phi_B\bar B^{s}(\Phi_B)
   {\rm d}\Phi_{R|B}
      \frac{\Rups (\Phi_R)}{B(\Phi_B)}
+ {\rm d}\Phi_R \Rupf (\Phi_R).
\end{equation}
From this equation we see that for large transverse momentum, the $\mathbb{S}$ 
event contribution to the cross section is enhanced by a factor $\bar{B}/B$, 
which is in essence the
$K$-factor of the process. We wish to emphasize that this factor does not spoil
the NLO accuracy of the result, since it affects the distribution by terms of
higher order in $\alphas$. Now, in \POWHEG{}, in its default configuration,
$\Rupf $ is only given by the tiny contribution $\PQq\PAQq\to \PH \Pg$, which 
is non-singular, so that $\mathbb{S}$ events dominate the cross section. The 
whole transverse-momentum distribution is thus affected by the $K$-factor, 
yielding a result that is similar to what is obtained in ME+PS calculations, 
where the NLO $K$-factor is applied to the LO distributions.  Notice also that 
changing the form of the central value of the scales again does not
change the transverse-momentum distribution, that can only be affected by
touching the scales in the Sudakov form factor.

\begin{figure}
\centering
\includegraphics[width=.5\textwidth,height=\figfact\linewidth]{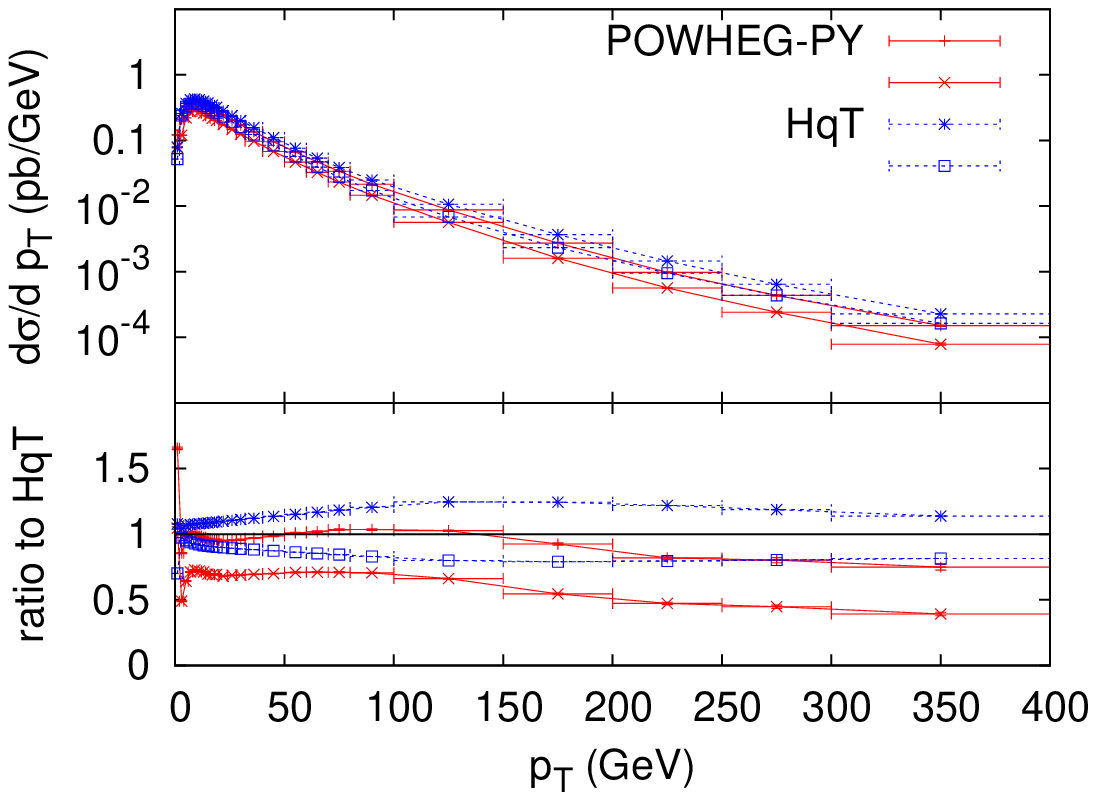}\nolinebreak
\includegraphics[width=.5\textwidth,height=\figfact\linewidth]{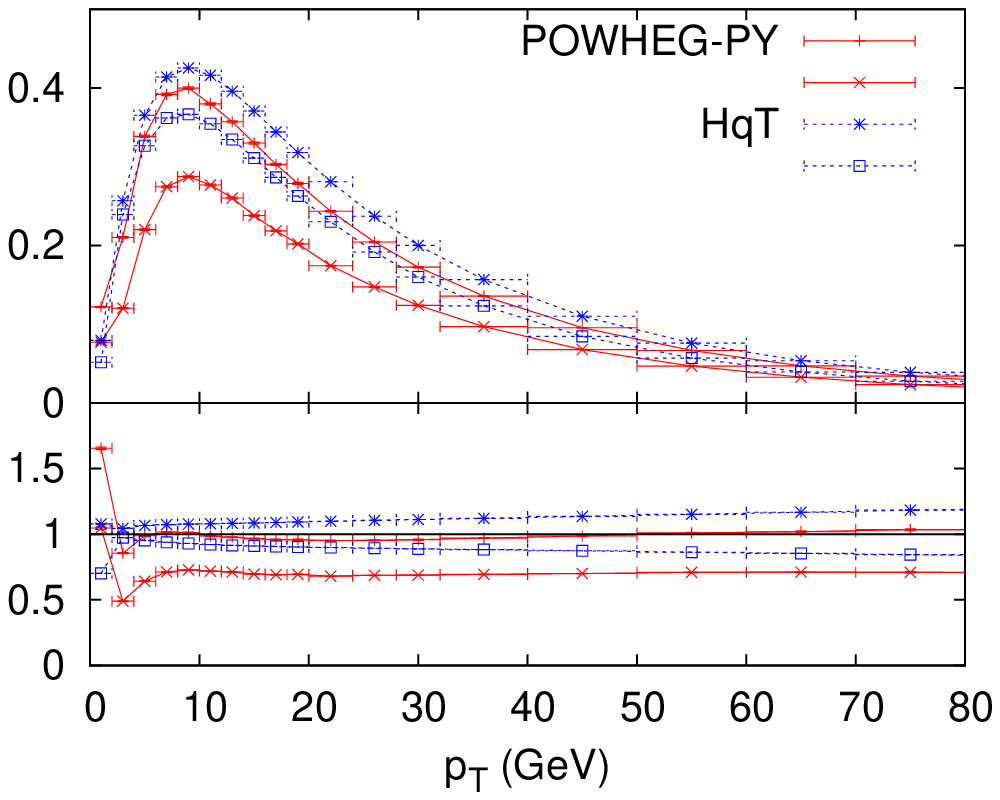}
\caption{The transverse-momentum spectrum of the Higgs in \POWHEG{}+\PYTHIA{},
using the separation of $\mathbb{S}$ and $\mathbb{F}$ events. The central 
scale is chosen equal to $\sqrt{\pT^2+\MH^2}$.}
\label{figNLOMC:pt99POWHEG}
\end{figure}

A simple approach to give a realistic assessment of the uncertainties
in \POWHEG{}, is to also exploit the freedom in the separation $R=\Rups+\Rupf$.
Besides the default value $\Rupf=0$, one can also perform the separation
\begin{equation}\label{eqNLOMC:SFsep}
\Rups=\frac{h^2}{h^2+\pT^2}R\,,\quad\quad \Rupf=\frac{\pT^2}{h^2+\pT^2}R\;.
\end{equation}
In this way, $\mathbb{S}$ and $\mathbb{F}$ events are generated, with the 
former dominating
the region $\pT<h$ and the latter the region $\pT>h$. Notice that by sending $h$
to infinity one recovers the default behaviour. It is interesting to ask what happens
if $h$ is made vanishingly small. It is easy to guess that in this limit the
\POWHEG{} results will end up coinciding with the pure NLO result.
The freedom in the choice of $h$, and also the freedom in changing the form
of the separation in Eq.~(\ref{eqNLOMC:SFsep}) can be exploited to explore further
uncertainties in \POWHEG{}. The Sudakov exponent changes by terms subleading
in $\pT^2$, and so we can explore in this way uncertainties related to the
shape of the Sudakov region.
Furthermore, by suppressing $\Rups$ at large $\pT$ the hard tail of the 
transverse-momentum distribution becomes more sensitive to the scale choice.
The lower plots of \refF{figNLOMC:ptMCatNLOMH} displays the \POWHEG{} result obtained using $h=\MH/1.2$. Notice that in this way the 
large-transverse-momentum tail becomes very similar to the \MCatNLO{} result. The shape of the distribution
at smaller transverse momenta is also altered, and in better agreement
with \HqT{}. If $\mT$ rather than $\MH$ is chosen as reference value for the scale,
we obtain the result of \refF{figNLOMC:pt99POWHEG}, where we see also here
a fall of the cross section at large transverse momentum.

\begin{figure}
\centering
\includegraphics[width=.5\textwidth,height=\figfacto\linewidth]{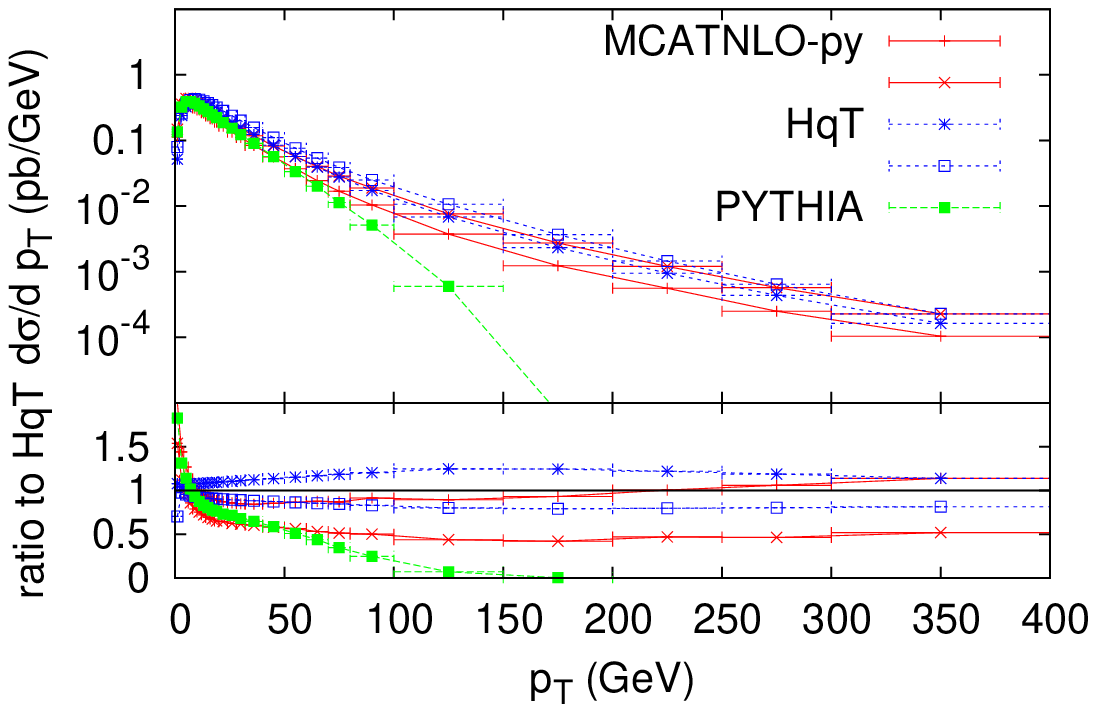}\nolinebreak
\includegraphics[width=.5\textwidth,height=\figfacto\linewidth]{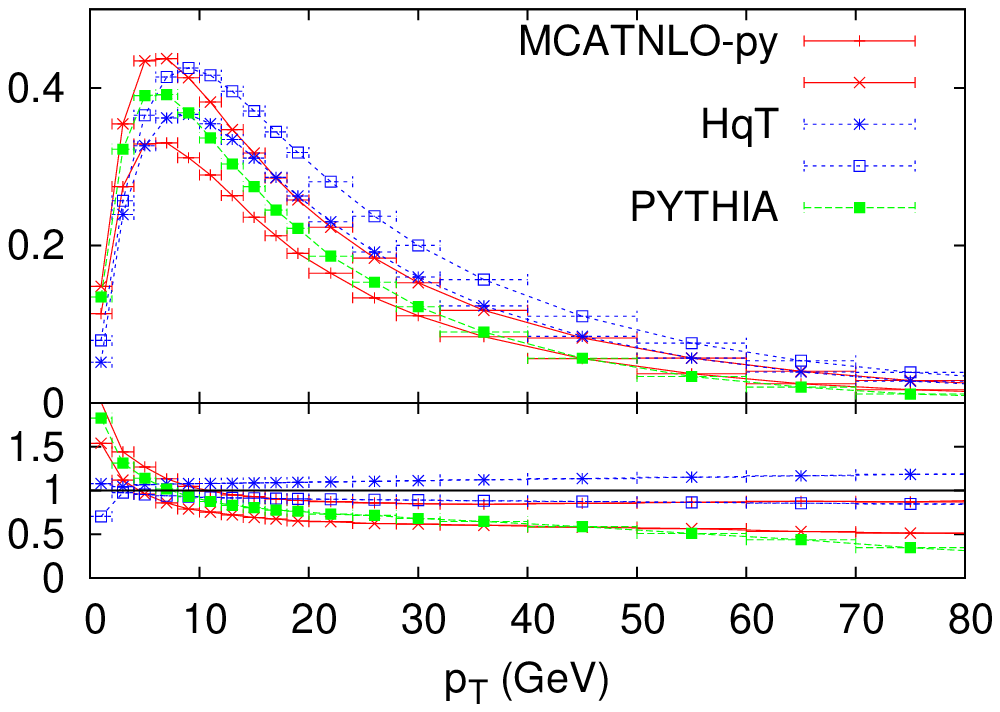}
\\[1em]
\includegraphics[width=.5\textwidth,height=\figfacto\linewidth]{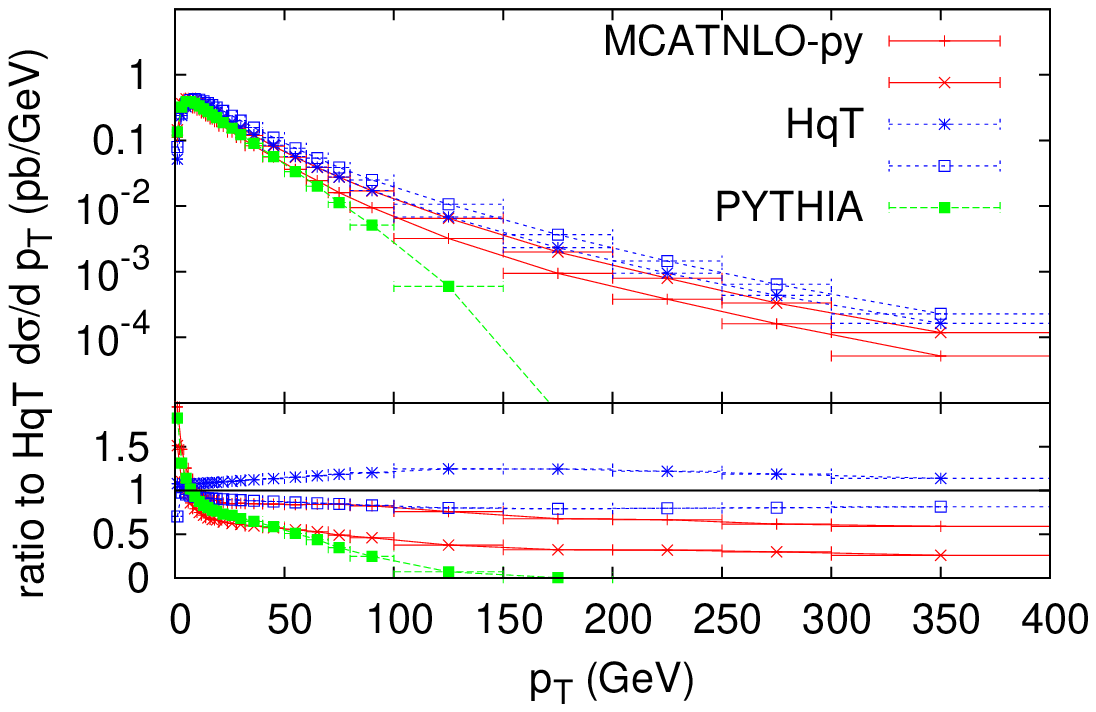}\nolinebreak
\includegraphics[width=.5\textwidth,height=\figfacto\linewidth]{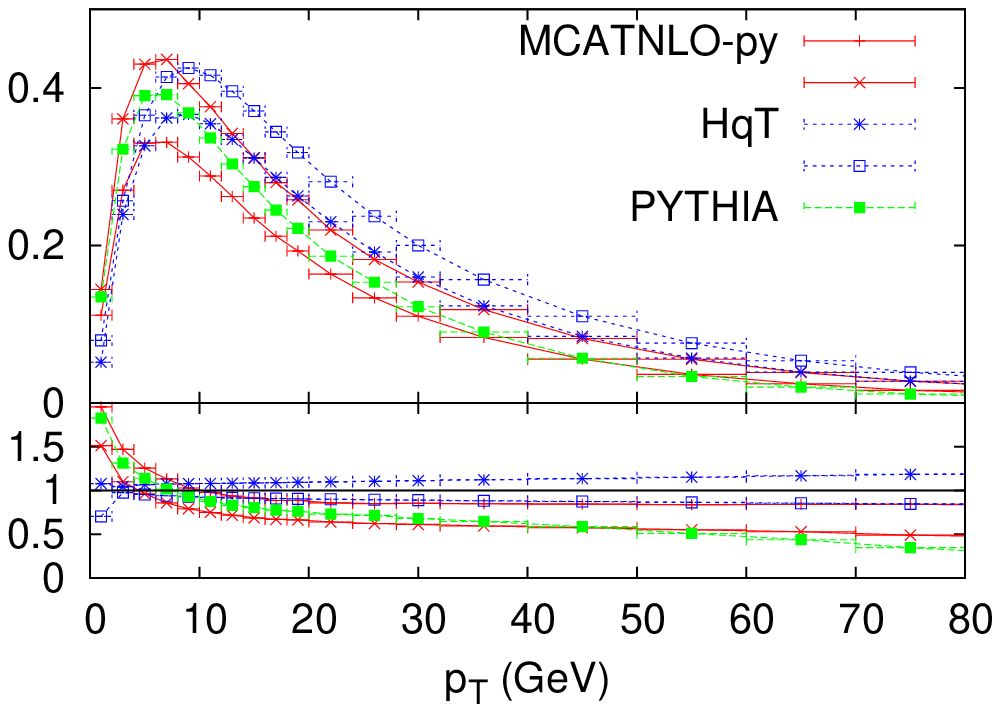}
\caption{The transverse-momentum spectrum of the Higgs in \MCatNLO{}+\PYTHIA{}.
The central scale is chosen
equal to $\MH$ in the upper plot, and to $\sqrt{\pT^2+\MH^2}$ in the lower
plot. The bare \PYTHIA{} result, normalised to the \HqT{} central value total
cross section, is also shown.}
\label{figNLOMC:mcatnlopy}
\end{figure}

The shape of the $\pT$ distribution in \MCatNLO{}+\HERWIG{} is not much affected by
the change of scale for $\pT< 100 \UGeV$. This is due to the fact that
this region is dominated by $\mathbb{S}$ events. It is interesting
to ask whether this region is affected if one changes
the underlying shower Monte Carlo generator.
In \MCatNLO{}, an interface to \PYTHIA{}, using the virtuality-ordered shower,
is also available \cite{Torrielli:2010aw}. The results are
displayed in \refF{figNLOMC:mcatnlopy}.
We observe the following. The large-transverse-momentum tails are consistent with
the \HERWIG{} version. We expect that since this region is dominated by $\mathbb{F}$ events.
The shape of the Sudakov region has changed, showing a behaviour that is more
consistent with the \HqT{} central value, down to scales of about $30\UGeV$. Below
this scale, we observe a $50\%{}$ increase of the cross section as smaller
$\pT$ values are reached. It is clear from the figure that this feature is inherited
from \PYTHIA{}. In fact, the shape from the transverse-momentum spectrum in \MCatNLO{}
is inherited from the shower Monte Carlo.
It is likely that the transverse-momentum-ordered \PYTHIA{} may yield better agreement
with the \HqT{} result.

Summarizing, we find large uncertainties in both \MCatNLO{} and \POWHEG{}
NLO+PS generators for Higgs production in gluon fusion. We have explored here 
uncertainties having to do with scale variation and to the separation of 
$\mathbb{S}$ and $\mathbb{F}$ events. If a higher accuracy result (namely
\HqT{}) was not available, the whole envelope of the uncertainties within
each approach should be considered. These large uncertainties are all
a direct consequence of the large NLO $K$-factor of the $\Pg\Pg\to \PH$
process.  In spite of this fact, we have seen that there are choices
of scales and parameters that bring the NLO+PS results close in shape
to the higher-accuracy calculation of the transverse-momentum spectrum
provided by the \HqT{} program. It is thus advisable to adopt these choices.

\subsection{Uncertainties in \protect{$\mathbf{\Pg\Pg\to\PH\to\PW\PW^{(*)}}$}%
\footnote{M.~Grazzini, H.~Hoeth, F.~Krauss, F.~Petriello,  M.~Sch\"onherr and F.~Siegert.}}

\label{NLOPSsec:Sherpa}

\noindent
In this section a wide variety of different physics effects and uncertainties 
related to key observables in the process 
$\Pg\Pg\to \PH\to \PWp\PWm\to\PGmp\,\PGnGm\,\Pem\,\PAGne$ is presented; 
they include
\begin{itemize} 
\item an appraisal of NLO matching methods, including a comparison with 
  \HNNLO \cite{Grazzini:2008tf} for some observables on the parton level; 
\item perturbative uncertainties, like the impact of scale and PDF
  variations;
\item the impact of parton showering and non-perturbative effects on 
  parton-level results; and
\item non-perturbative uncertainties, and in particular the impact of 
  fragmentation variations and the effect of the underlying event and
  changes in its simulation.
\end{itemize}

\subsubsection{Setup}

In the following sections the \SHERPA\ event generator \cite{Gleisberg:2008ta}
has been used in two modes: Matched samples have been produced according 
to the \POWHEG~\cite{Nason:2004rx,Frixione:2007vw} and
\MCatNLO~\cite{Frixione:2002ik} methods, implemented as described in
\cite{Hoche:2010pf} and \cite{Hoeche:2011fd}, respectively.  In the following
the corresponding samples will be denoted as \SHERPA--\POWHEG\ and
\SHERPA--\MCatNLO, respectively.
Unless stated otherwise, $\MH = 160\,\UGeV$ and the following cuts have been 
applied to leptons ($\Pl = \Pe,\,\PGm$) and jets $j$:
\begin{itemize}
\item leptons: $\pT^{(\Pl)} > 15 \,\UGeV$ and $|\eta^{(\Pl)}| < 2.5$,
\item jets (defined by the anti-$\kT$ algorithm \cite{Cacciari:2008gp} 
  with $R=0.4$): $\pT^{(j)} > 25 \,\UGeV$, $|\eta^{(j)}| < 4.5$.
\end{itemize}
By default, for purely perturbative studies, the central PDF from the MSTW2008 
NLO set~\cite{Martin:2009iq} has been used, while for those plots involving
non-perturbative effects such as hadronisation and the underlying event, the
central set of CT10 NLO~\cite{Lai:2010vv} has been employed, since \SHERPA\
has been tuned to jet data with this set.  In both cases, the PDF set also 
fixes the strong coupling constant and its running behaviour in both matrix
element and parton shower, and in the underlying event.

\subsubsection{Algorithmic dependence: \protect\POWHEG vs.\ \protect\MCatNLO}

Before embarking in this discussion, a number of points need to be stressed:
\begin{enumerate}
\item The findings presented here are very recent, and they are partially at
  odds with previous conclusions.  So they should be understood as
  contributions to an ongoing discussion and hopefully trigger further work;
\item clearly the issue of scales and, in particular, of resummation scales
  tends to be tricky.  By far and large, however, the community agrees that the
  correct choice of resummation scale $Q$ is of the order of the factorisation 
  scale, in the case at hand here, therefore $Q = {\cal O}(\MH)$ \footnote{
    In \HqT\ this scale by default is chosen to be $Q = \MH/2$.
  }.  It is
  notoriously cumbersome to directly translate analytic resummation and scale 
  choices there to parton-shower programs;
\item the version of the \MCatNLO algorithm presented here differs in some
  aspects from the version implemented in the original \MCatNLO program; there, 
  typically the \HERWIG{} parton shower is being used, guaranteeing an upper 
  limitation in the resummed phase space given by the factorisation scale, and
  the FKS method \cite{Frixione:1995ms} is used for the infrared subtraction 
  on the matrix-element level, the difference is compensated by a suitable
  correction term.  In contrast, here, \SHERPA\ is used, where both the parton 
  shower and the infrared subtraction rely on Catani--Seymour subtraction 
  \cite{Catani:1996vz,Schumann:2007mg}, and the phase-space limitation in
  the resummation bit is varied through a suitable parameter \cite{Nagy:2003tz}
  $\alpha\in [0,\,1]$, with $\alpha = 1$ relating to no phase-space 
  restriction.  We will indicate the choice of $\alpha$ by a superscript.
  However, the results shown here should serve as an example only.
\end{enumerate}

\begin{figure}
  \centering
  \includegraphics[width=0.48\textwidth]{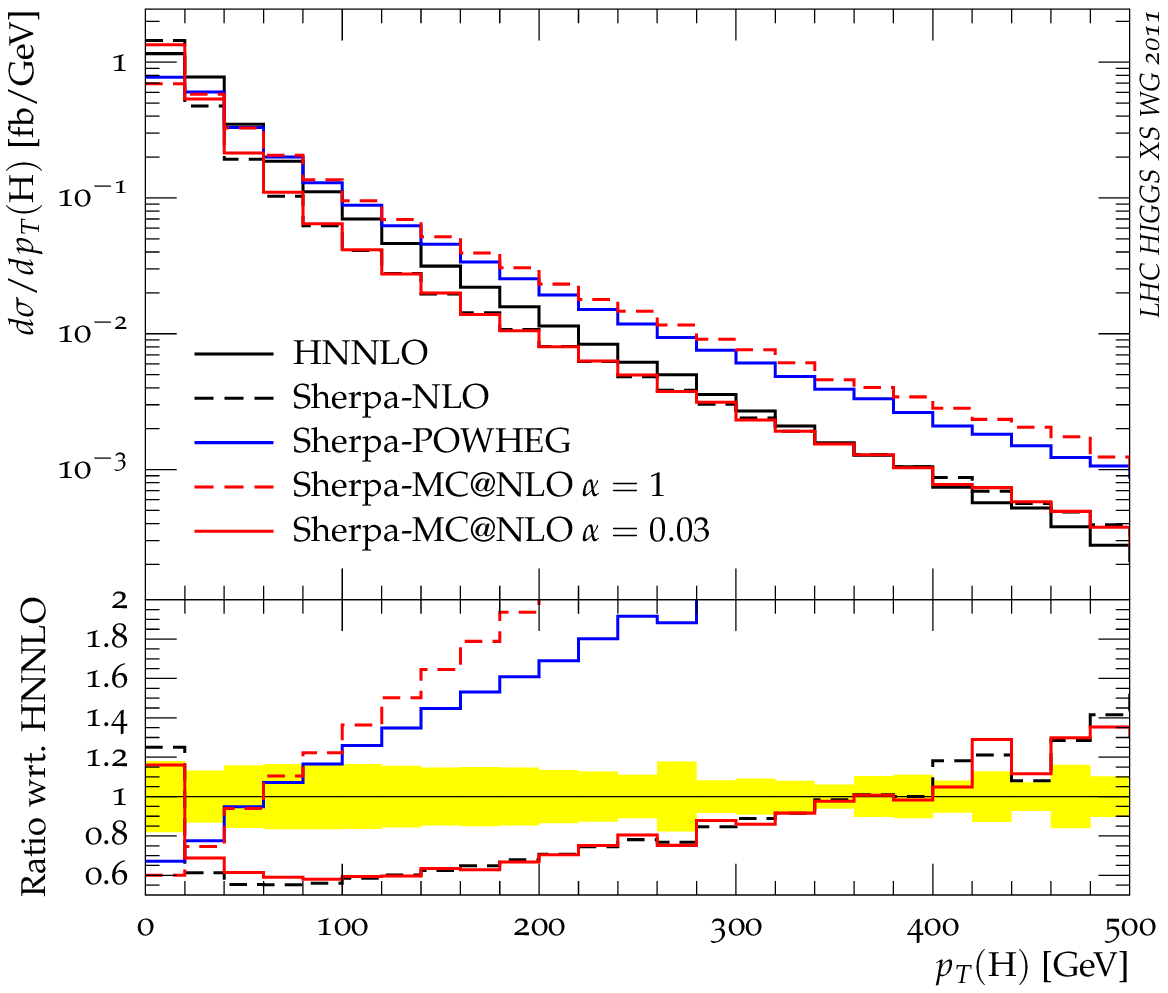}\hfill
  \includegraphics[width=0.48\textwidth]{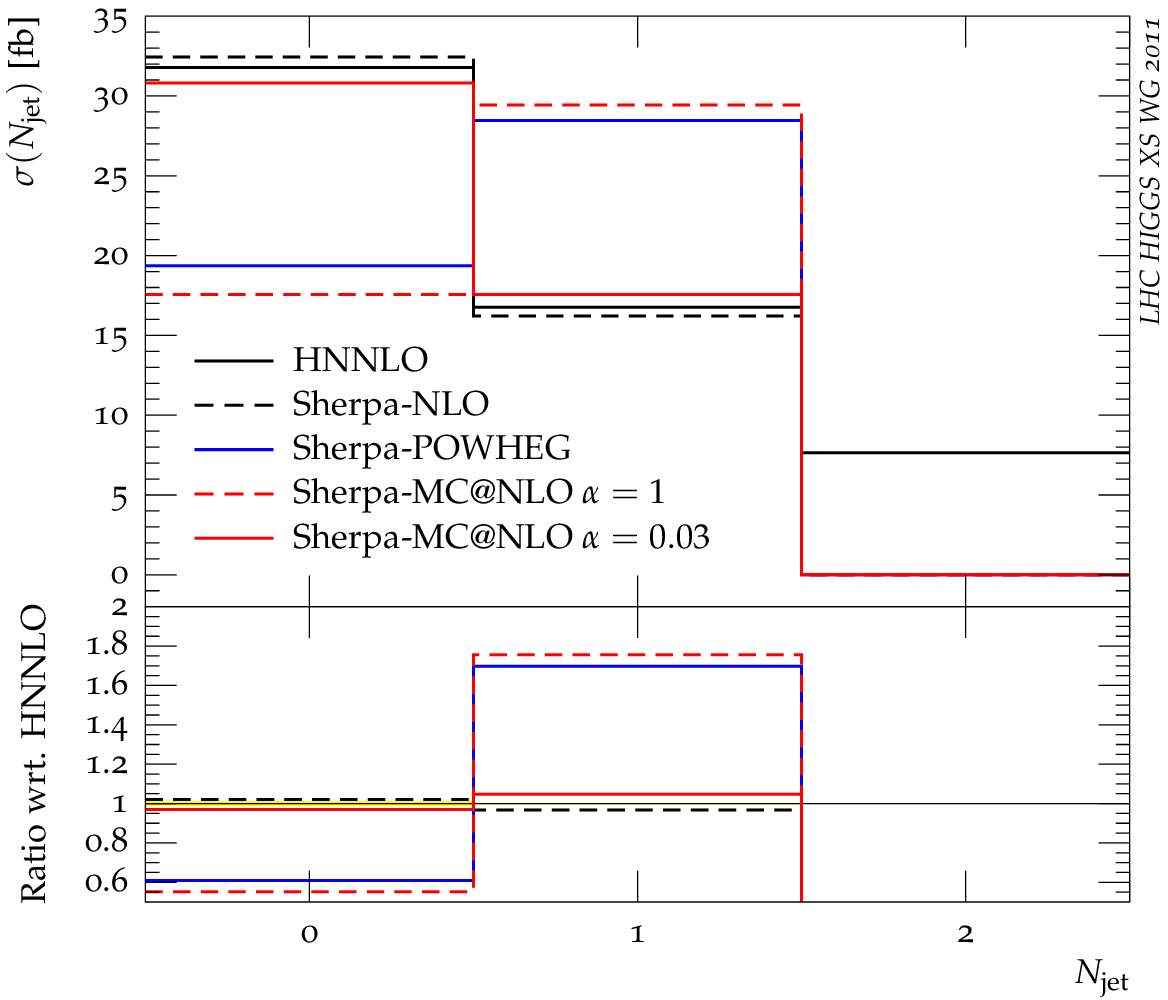}
  \caption{\label{fig:hww:uncert:nlomcs:hnnlo_me}
    The Higgs transverse momentum in all events (left) and the jet 
    multiplicities (right) in $\Pp\Pp\to \PH\to\Pem\,\PGmp\,\PGnGm\,\PAGne$ 
    production.  Here different approaches are compared with each other: two 
    fixed order parton-level calculations at NNLO (\protect\HNNLO, black solid)
    and NLO (black, dashed), and three matched results in the \protect\SHERPA\
    implementations (\protect{\SHERPA-\POWHEG}, blue and 
    \protect{\SHERPA-\MCatNLO} with $\alpha=1$, red dashed and with 
    $\alpha=0.03$, red solid) truncated after the first emission.  The yellow 
    uncertainty band is given by the scale uncertainties of \protect\HNNLO.}
\end{figure}

Starting with a comparison at the matrix-element level, consider 
\refF{fig:hww:uncert:nlomcs:hnnlo_me}, where results from \HNNLO are 
compared with an NLO calculation and the \POWHEG and \MCatNLO implementations 
in \SHERPA\, where the parton shower has been stopped after the first emission.
For the Higgs-boson transverse momentum we find that the \HNNLO result is
significantly softer than both the \SHERPA--\POWHEG and the 
\SHERPA--\MCatNLO$^{(\alpha=1)}$ sample -- both have a significant shape 
distortion w.r.t.\ \HNNLO over the full $\pT(\PH)$ range.  In contrast the 
NLO and the \SHERPA--\MCatNLO$^{(\alpha=0.03)}$ result have a similar shape as 
\HNNLO\ in the mid-$\pT$ region, before developing a harder tail.  The shape 
differences in the low-$\pT$ region, essentially below $\pT(\PH)$ of about 
$20\,\UGeV$, can be attributed to resummation effects.  The picture becomes 
even more interesting when considering jet rates at the matrix element level.  
Here, both the \SHERPA--\POWHEG and the \SHERPA--\MCatNLO$^{(\alpha=1)}$ sample 
have more 1 than 0-jet events, clearly at odds with the three other samples.  
Of course, switching on the parton shower will lead to a migration to higher 
jet bins, as discussed in the next paragraph.  With this in mind, one could 
argue that the 1-jet rates in both the \SHERPA--\POWHEG and the 
\SHERPA--\MCatNLO$^{(\alpha=1)}$ sample seem to be in fair agreement with the 
{\em inclusive} 1-jet rate of \HNNLO\ -- this however does not resolve the 
difference in the 0-jet bin, and ultimately it nicely explains why these two 
samples produce a much harder radiation tail than \HNNLO. 

\begin{figure}
  \centering
  \includegraphics[width=0.48\textwidth]{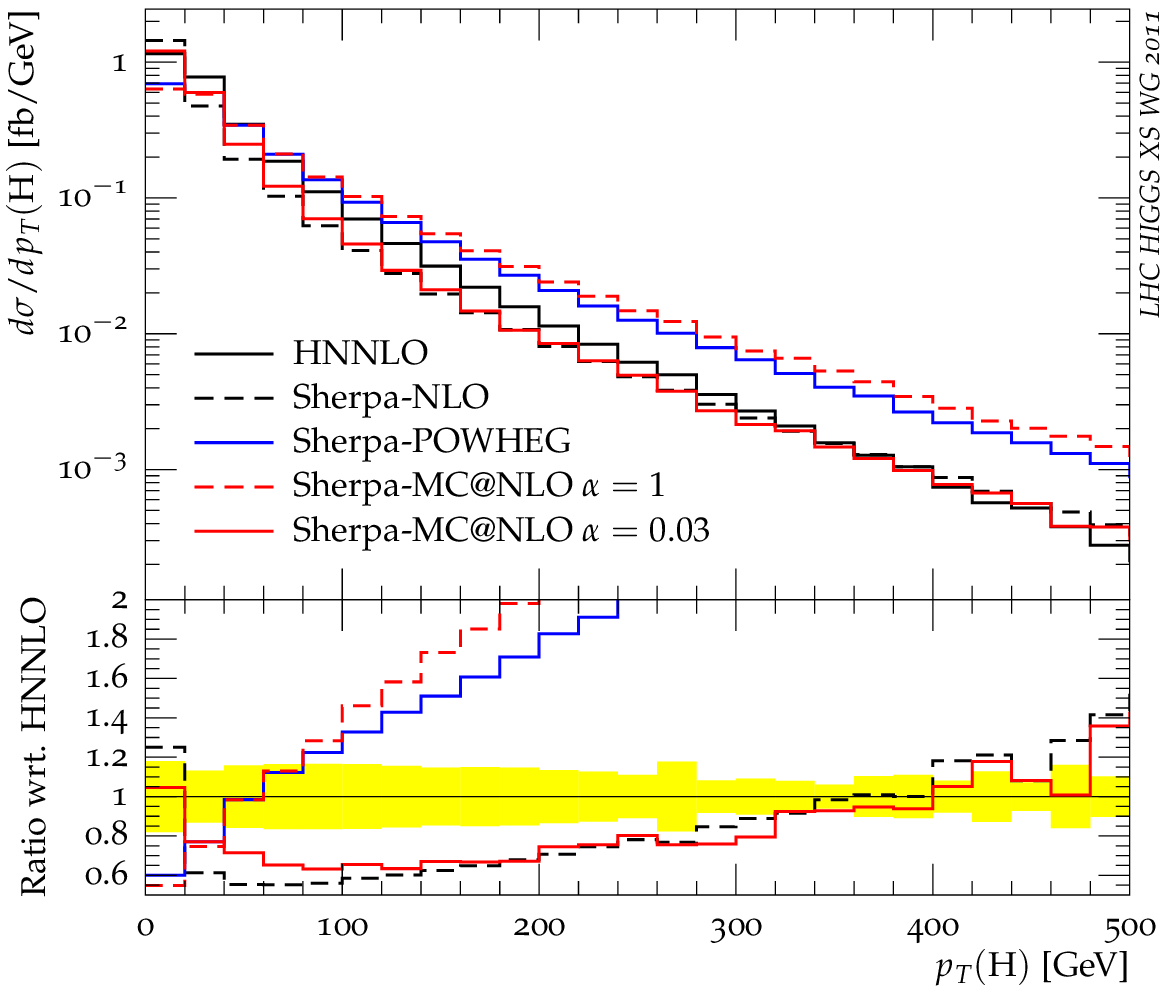}\hfill
  \includegraphics[width=0.48\textwidth]{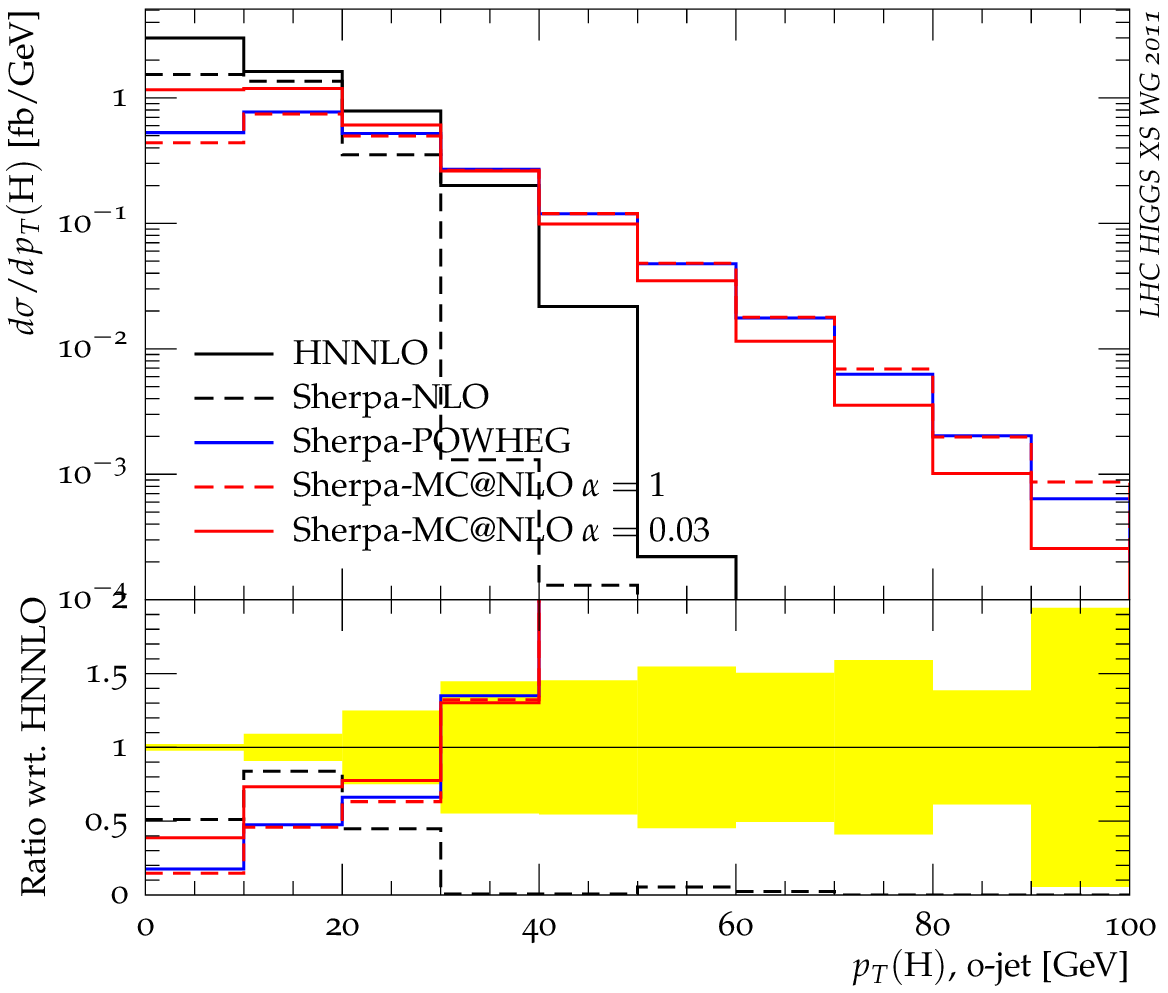}\\[1em]
  \includegraphics[width=0.48\textwidth]{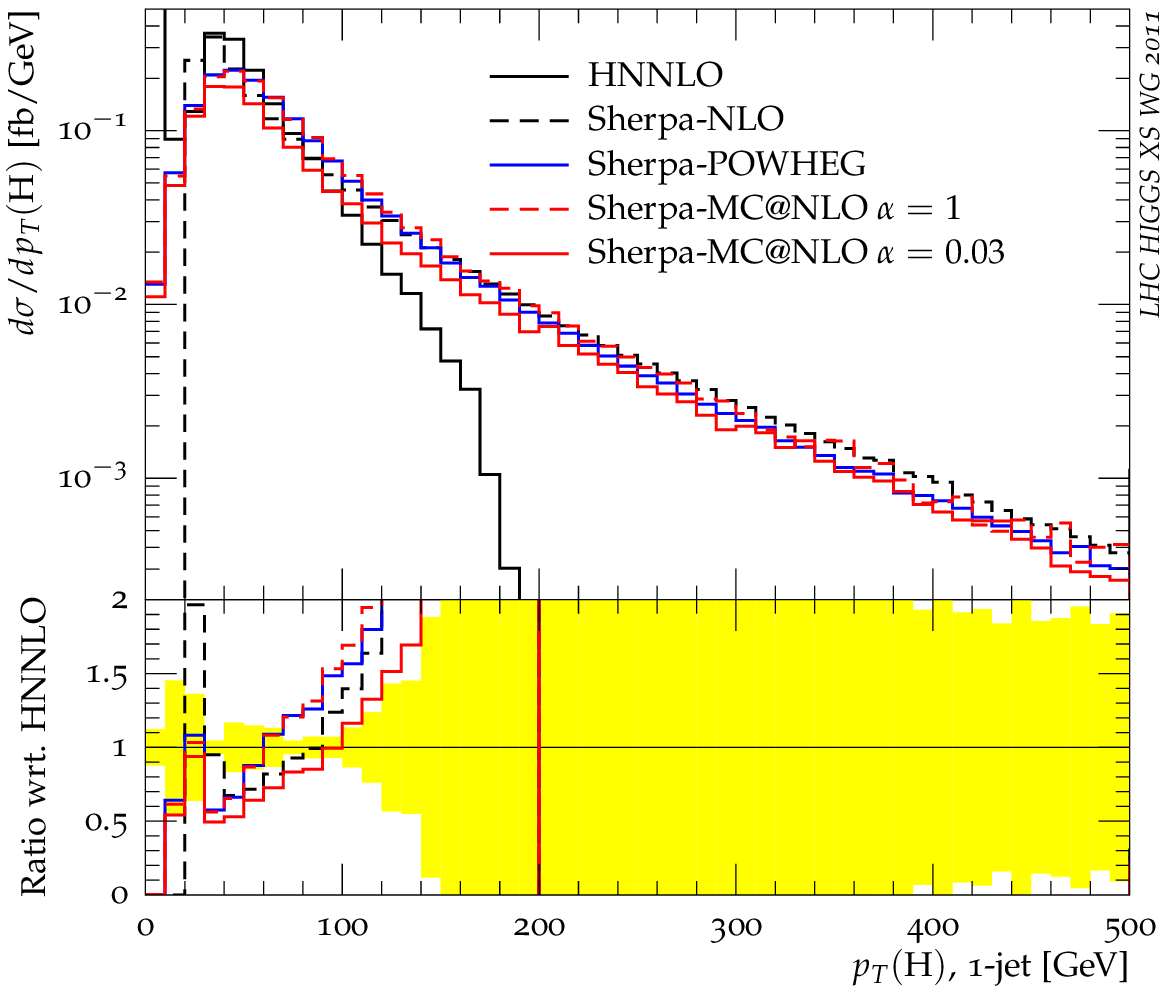}\hfill
  \includegraphics[width=0.48\textwidth]{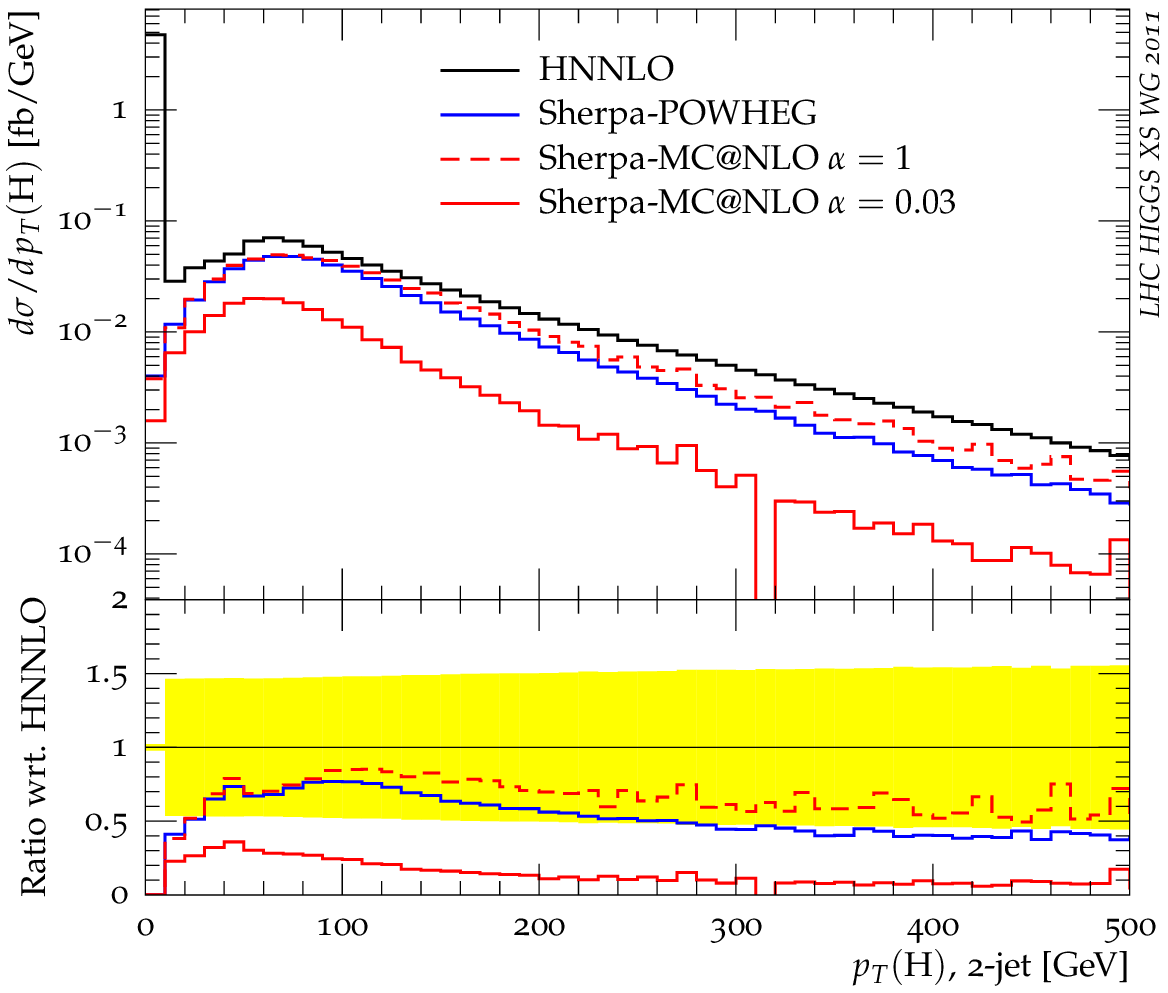}\\[1em]
  \includegraphics[width=0.48\textwidth]{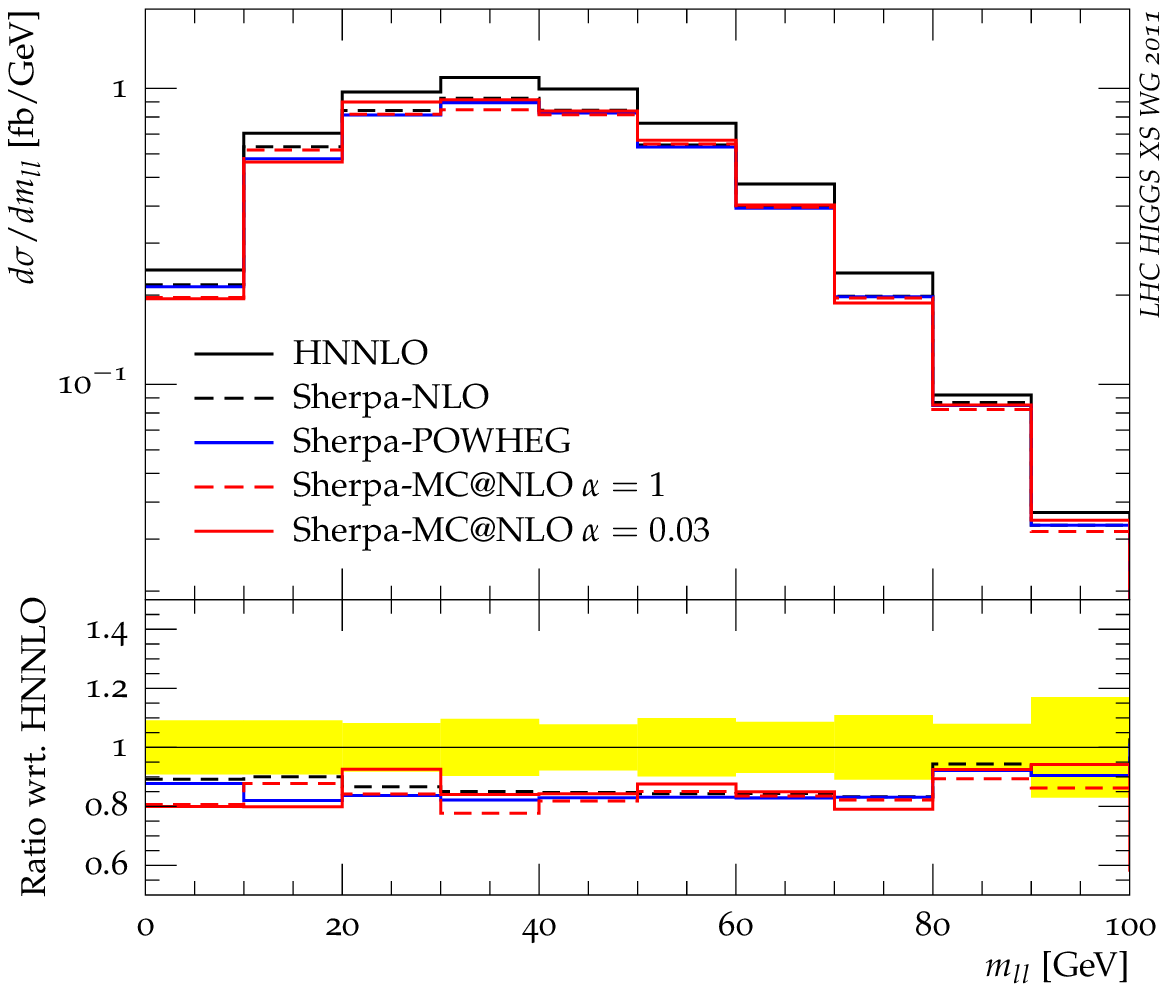}\hfill
  \includegraphics[width=0.48\textwidth]{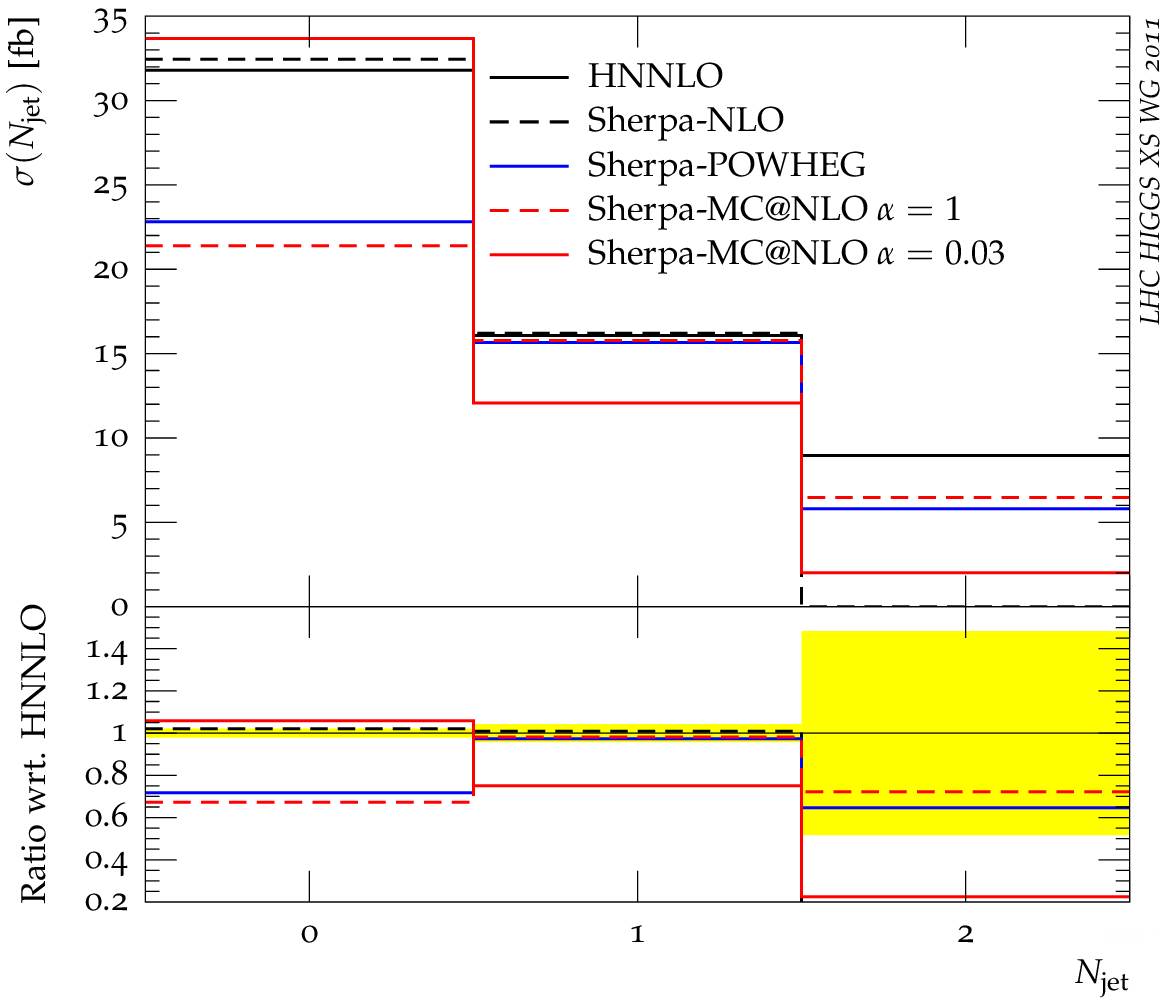}
  \caption{\label{fig:hww:uncert:nlomcs:hnnlo}
    The Higgs transverse momentum in all events, in events with 0, 1, and
    2 and more jets, the invariant lepton mass, and the jet 
    multiplicities in $\Pp\Pp\to \PH\to\Pem\,\PGmp\,\PGnGm\,\PAGne$ production.
    Here different approaches are compared with each other: two 
    fixed-order parton-level calculations at NNLO (\protect\HNNLO, black solid) and
    NLO (black, dashed), and three matched results in the \protect\SHERPA\
    implementations (\protect{\SHERPA--\POWHEG}, blue and 
    \protect{\SHERPA--\MCatNLO} with $\alpha=1$, red dashed and with 
    $\alpha=0.03$, red solid).  The yellow uncertainty band is given by the 
    scale uncertainties of \protect\HNNLO.
  }
\end{figure}

In \refF{fig:hww:uncert:nlomcs:hnnlo} now all plots are at the shower 
level.  We can summarise the findings of this figure as follows: The 
\SHERPA--\POWHEG and \SHERPA--\MCatNLO$^{(\alpha=1)}$ results exhibit fairly
identical trends for most observables, in particular, the Higgs-boson 
transverse momentum in various samples tends to be harder than the pure NLO 
result, \HNNLO, or \SHERPA--\MCatNLO$^{(\alpha=0.03)}$.  Comparing \HNNLO\ with the 
NLO result and \SHERPA--\MCatNLO$^{(\alpha=0.03)}$ we find that in most cases the 
latter two agree fairly well with each other, while there are differences with 
respect to \HNNLO.  In the low-$\pT$ region of the Higgs boson, the 
difference seems to be well described by a global $K$-factor of about $1.3s{-}1.5$, 
while \HNNLO becomes softer in the high-$\pT$ tail, leading to a sizable 
shape difference.  One may suspect that this is due to a different description 
of configurations with two jets, where quantum interferences lead to 
non-trivial correlations of the outgoing particles in phase space, which, of 
course, are correctly accounted for in \HNNLO, while the other results either 
do not include such configurations (the parton-level NLO curve) or rely on the 
spin-averaged and therefore correlation-blind parton shower to describe them.
This is also in agreement with findings in the jet multiplicities, where
the \SHERPA--\MCatNLO$^{(\alpha=0.03)}$ and the NLO result agree with \HNNLO in 
the 0-jet bin, while the \SHERPA--\POWHEG and \SHERPA--\MCatNLO$^{(\alpha=1)}$ 
result undershoot by about $30\%$, reflecting their larger QCD activity.  For 
the 1- and 2-jet bins, however, their agreement with the \HNNLO result 
improves and in fact, as already anticipated from the ME-level results, 
multiplying the \SHERPA--\POWHEG\ result with a global $K$-factor of about $1.5$ 
would bring it to a very good agreement with \HNNLO for this observable.  In 
contrast the \SHERPA--\MCatNLO$^{(\alpha=0.03)}$ result undershoots \HNNLO in the 
1-jet bin by about $25\%$ and in the 2-jet bin by about a factor of $4$.  Clearly 
here support from higher order tree-level matrix elements like in ME+PS or 
MENLOPS-type \cite{Hamilton:2010wh,Hoche:2010kg} approaches would be helpful. 

Where not stated otherwise, in the following sections, all curves relate
to an NLO matching according to the \SHERPA--\MCatNLO prescription with 
$\alpha = 0.03$.

\subsubsection{Perturbative uncertainties}
\begin{figure}
  \centering
  \includegraphics[width=0.48\textwidth]{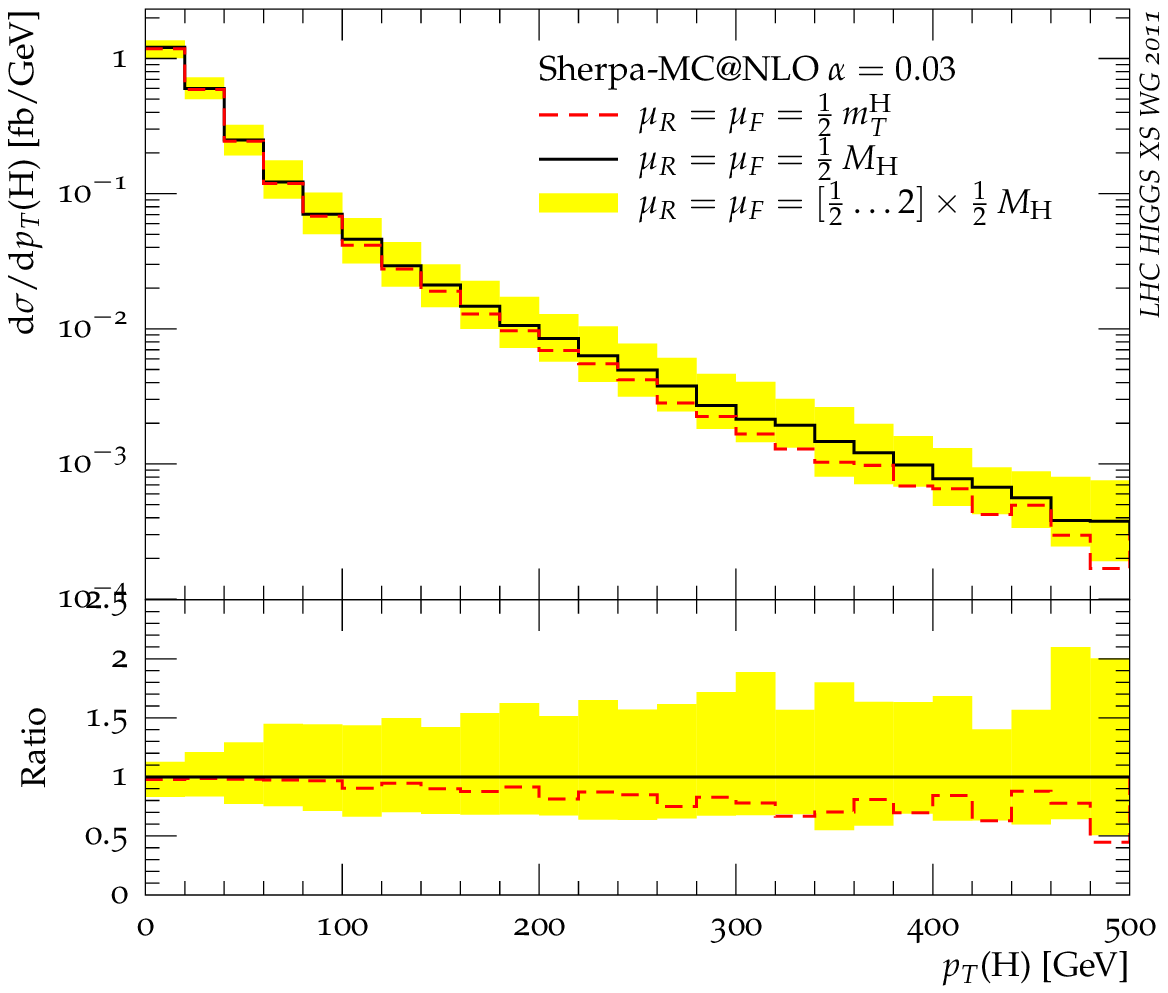}\hfill
  \includegraphics[width=0.48\textwidth]{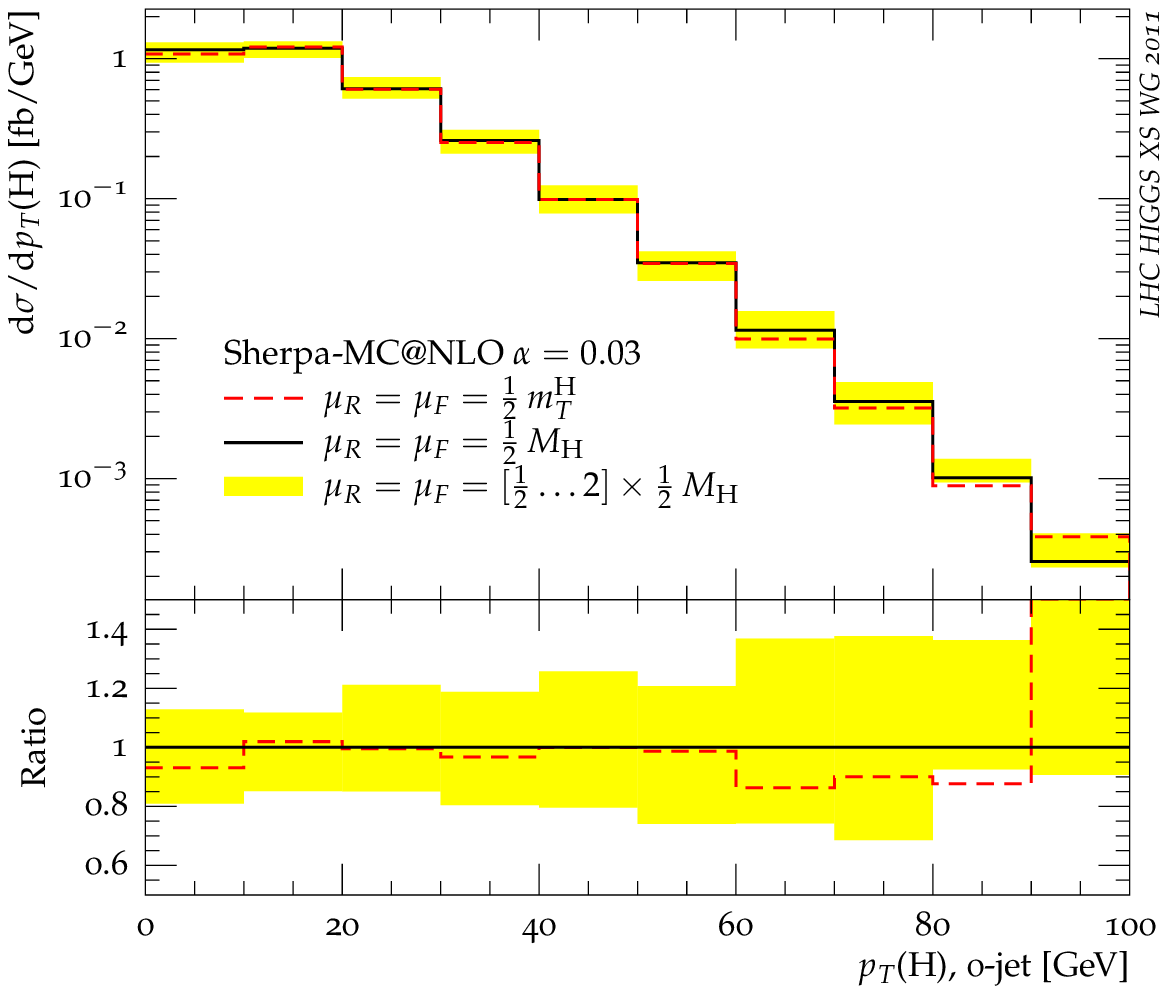}\\[1em]
  \includegraphics[width=0.48\textwidth]{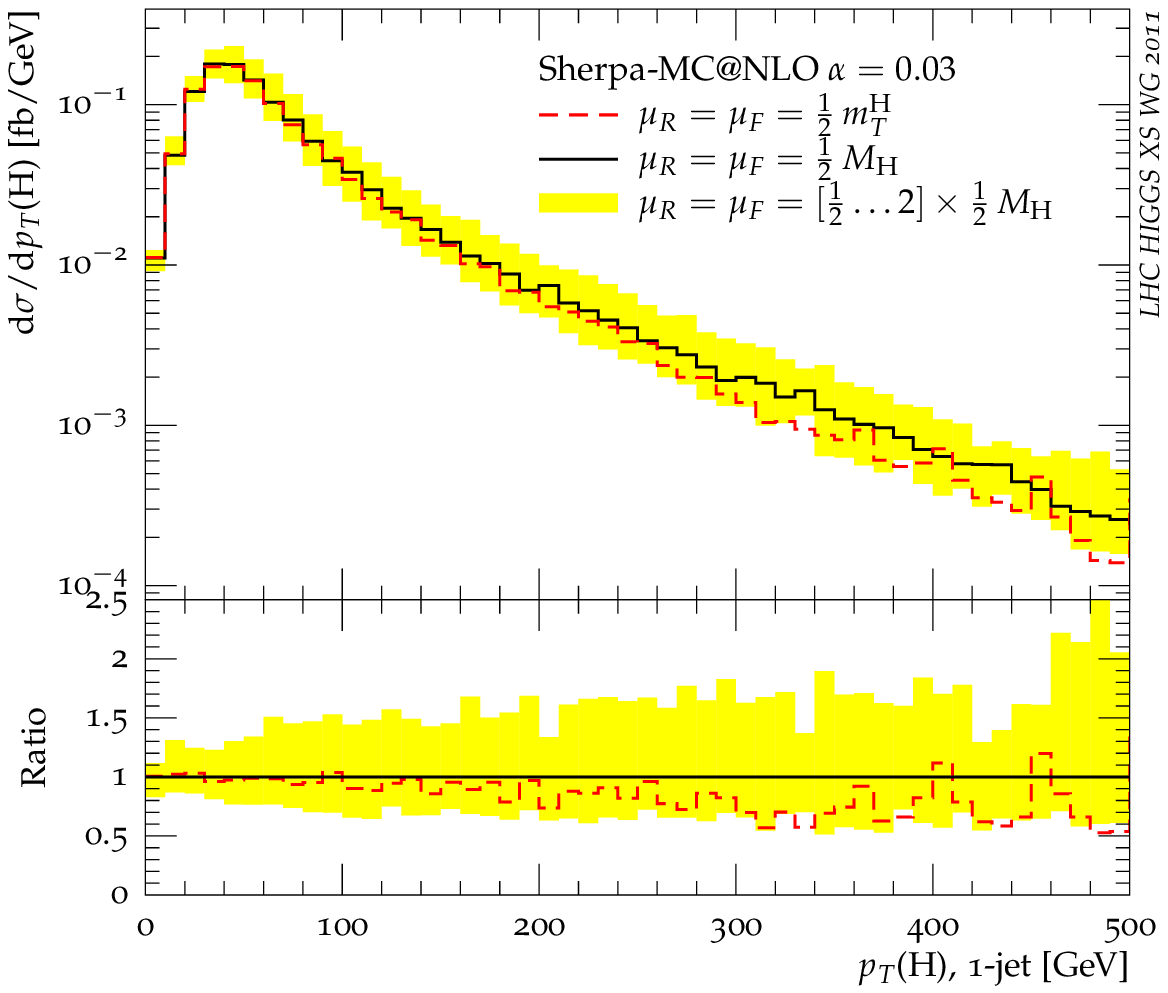}\hfill
  \includegraphics[width=0.48\textwidth]{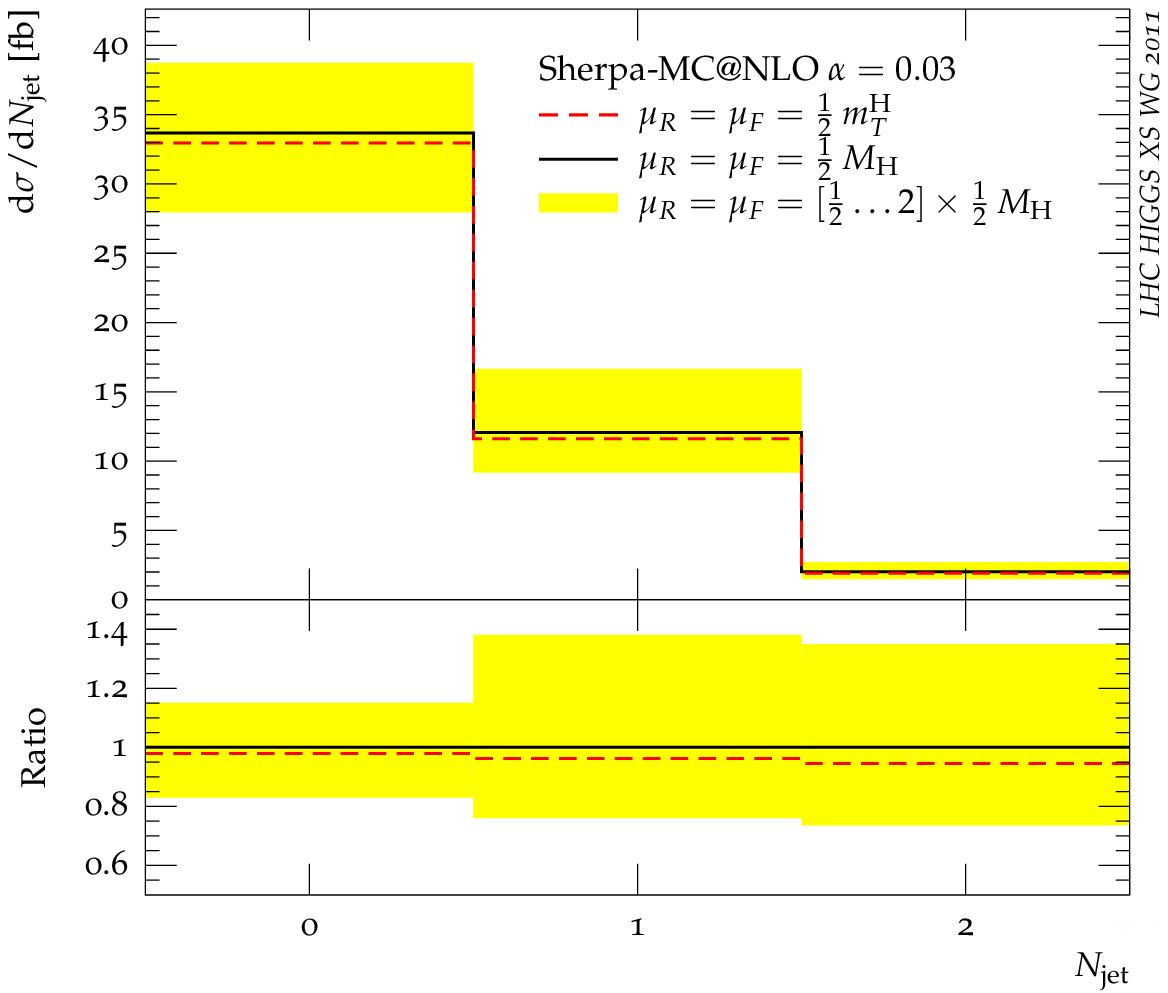}
  \caption{\label{fig:hww:uncert:scale}
    The impact of scale variation by a factor of $2$ and $1/2$ (yellow band) around 
    the central scale $\muR=\muF=\MH/2$ (black solid line) and 
    of a variation of the functional form of the scale to 
    $\muR=\muF=\tfrac{1}{2}\,\mT^{\PH}$ (red dashed line) on the transverse 
    momentum of the Higgs boson in all events, events with no and with one 
    jet, and on jet multiplicities.  All curves are obtained by an NLO 
    matching according to the \protect{\SHERPA--\MCatNLO} prescription with 
    $\alpha = 0.03$.
  }
\end{figure}

The impact of scale variations on the Higgs transverse momentum in all events 
and in events with 0, 1, and 2 or more jets is exhibited in 
\refF{fig:hww:uncert:scale}, where a typical variation by factors of $2$ 
and $1/2$ has been performed around the default scales $\muR = \muF = \MH$.  
For the comparison with the fixed scale we find that distributions that 
essentially are of leading order accuracy, such as $\pT$ distributions of 
the Higgs boson in the 1-jet region or the 1-jet cross section, the scale 
uncertainty is of the order of $40\%$, while for results in the resummation 
region of the parton shower, \ie\ the 0-jet cross section or the Higgs-boson 
transverse momentum in the 0-jet bin, the uncertainties are smaller, at about 
$20\%$.  

In contrast differences between the functional form of the scales choice are
smaller, also exemplified in \refF{fig:hww:uncert:scale}, where the 
central value $\muF = \muR = \MH/2$ has been compared with an alternative 
scale $\muF = \muR = \mT^{\PH}/2 = \sqrt{\MH^2+p_{T,\PH}^2}/2$.  It should be 
noted though that there the two powers of $\alphas$ related to the effective 
$\Pg\Pg\PH$ vertex squared have been evaluated at a scale $\tfrac{1}{2}\,\MH$.
Anyway, with this in mind it is not surprising that differences only start
to become sizable in the large-$\pT$ regions of additional radiation,
where they reach up to about $40\%$.  

\begin{figure}
  \centering
  \includegraphics[width=0.48\textwidth]{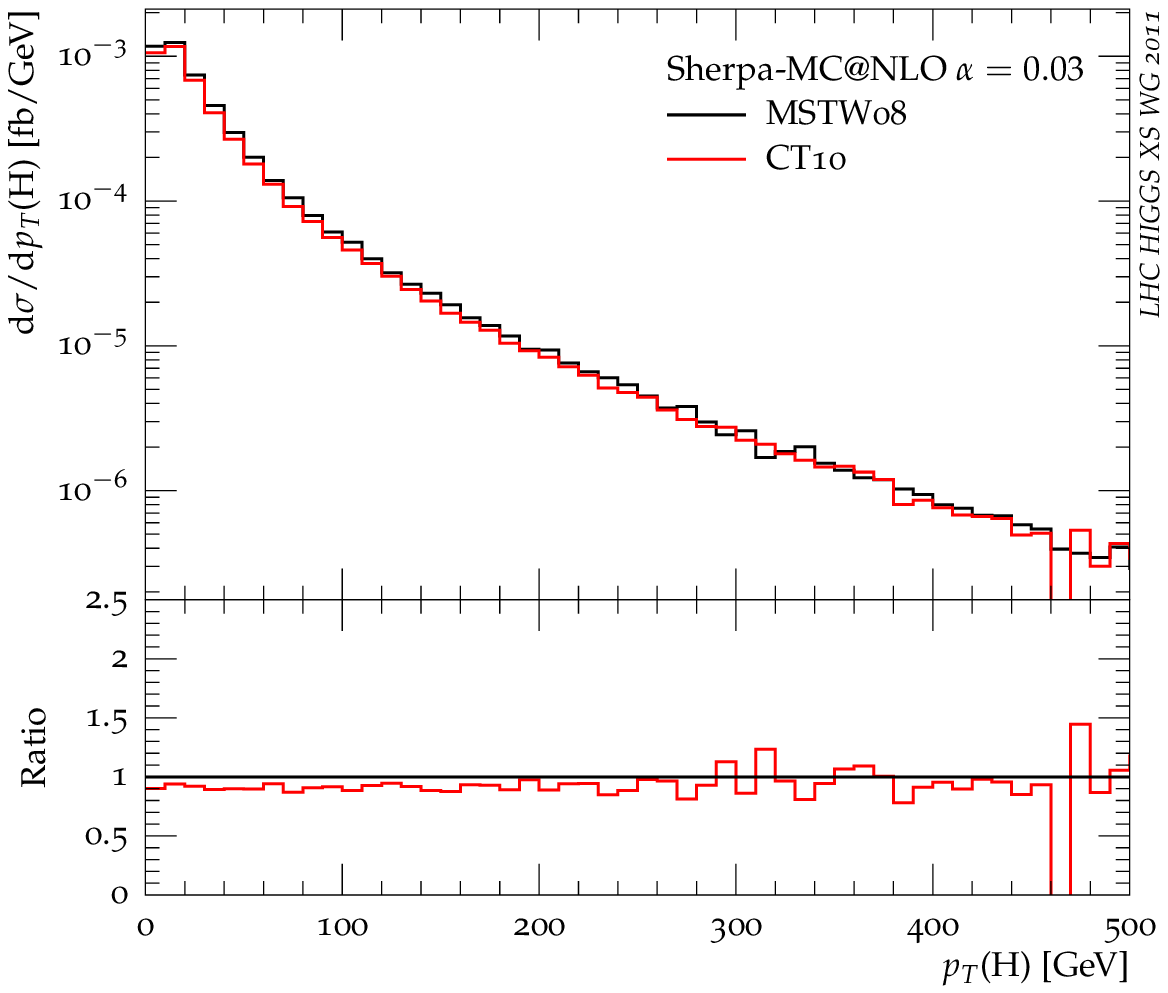}\hfill
  \includegraphics[width=0.48\textwidth]{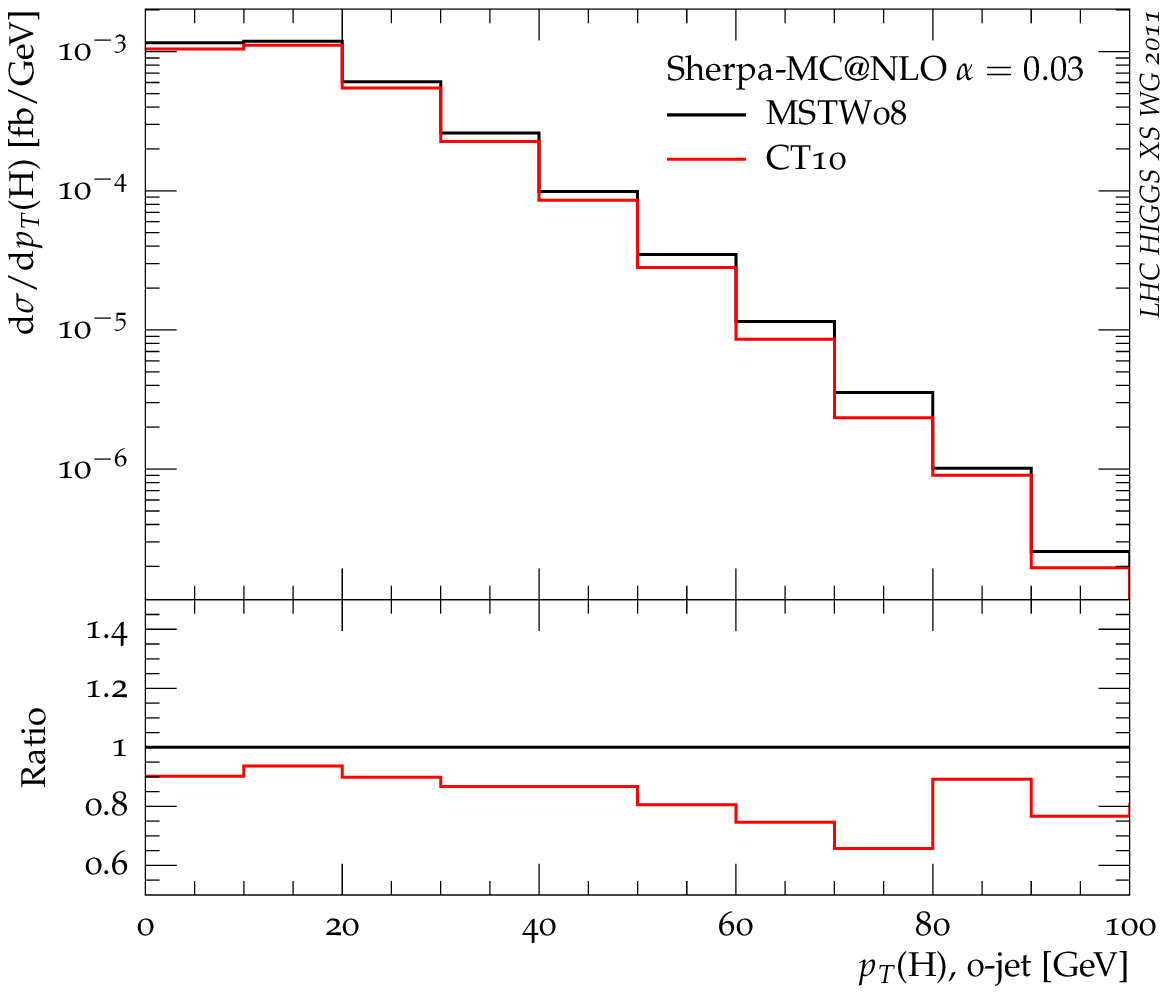}\\[1em]
  \includegraphics[width=0.48\textwidth]{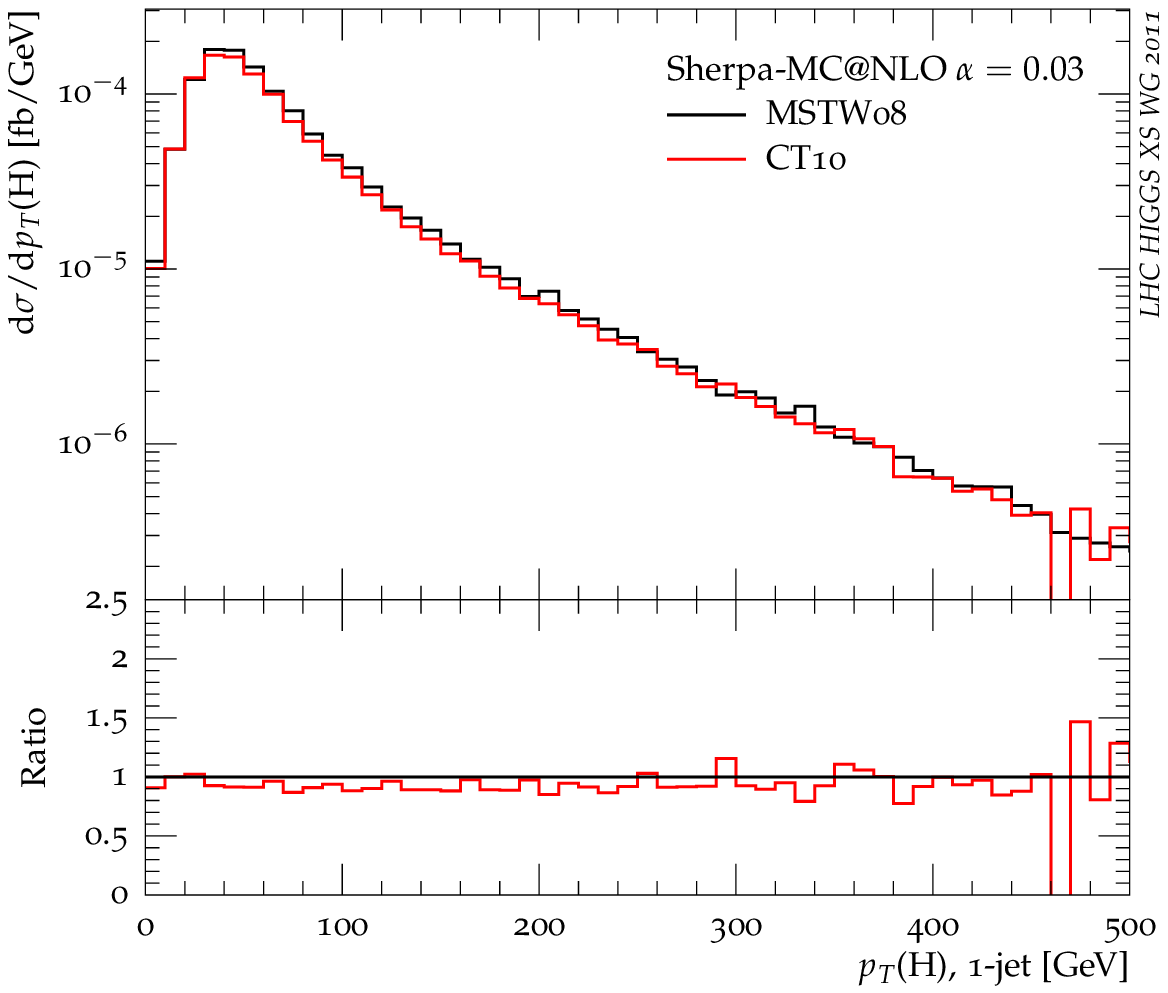}\hfill
  \includegraphics[width=0.48\textwidth]{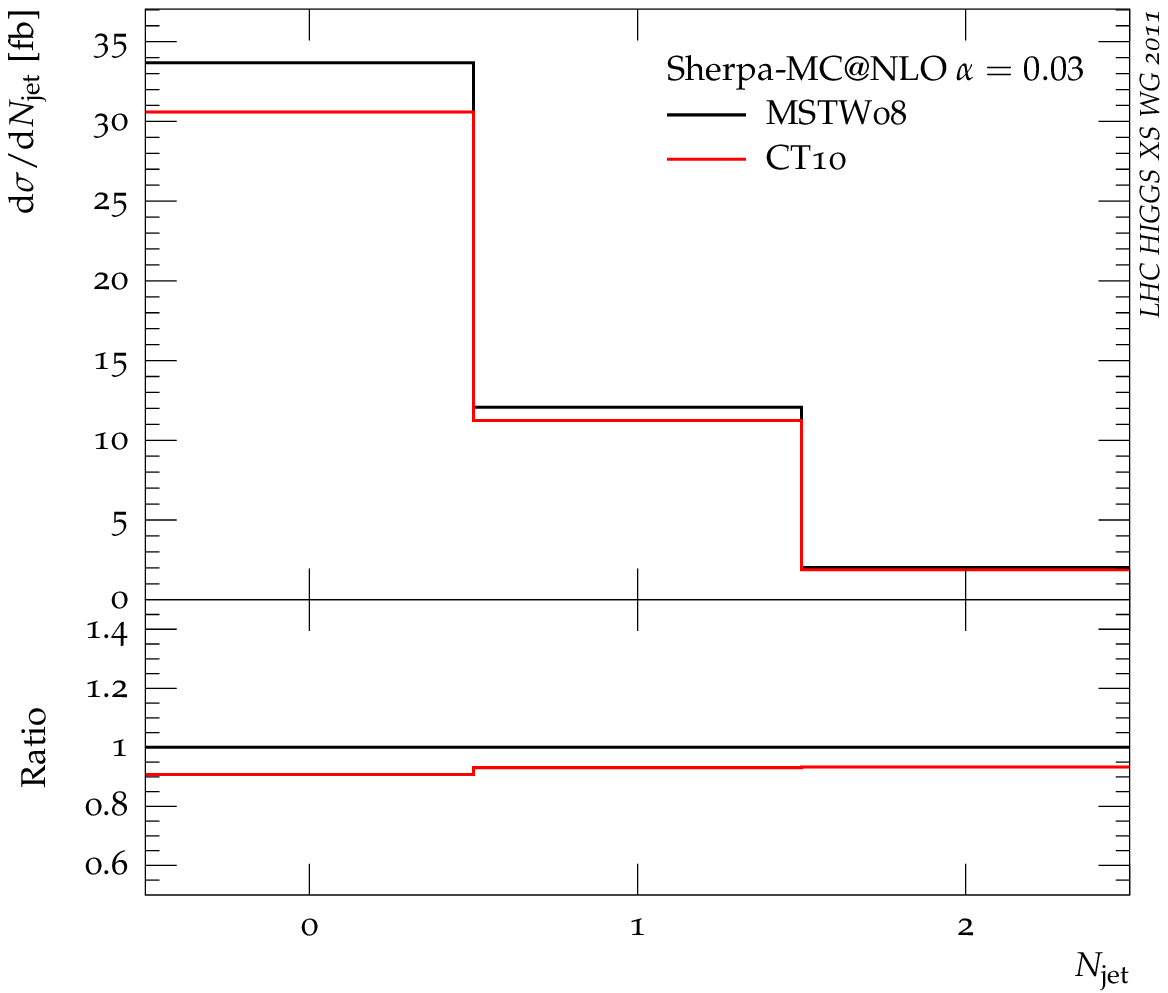}
  \caption{\label{fig:hww:uncert:PDF}
    The impact of different choices of PDFs -- MSTW2008 NLO (black) and 
    CT10 NLO (red) -- for a $\MH = 160\UGeV$ on the transverse momentum of
    the Higgs boson in all events, events with no and with one jet, and on
    jet multiplicities.  All curves are obtained by an NLO matching according 
    to the \protect{\SHERPA-\MCatNLO} prescription with $\alpha = 0.03$.
  }
\end{figure}

Rather than performing a full PDF variation according to the recipe in
\Bref{Alekhin:2011sk}, in \refF{fig:hww:uncert:PDF} results for MSTW2008 
NLO, have been compared to those obtained with the central NLO set of CT10.
By far and large, there is no sizable difference in any relevant shape.
However, a difference of about $10\%$ can be observed in the total normalisation,
which can be traced back to the combined effect of minor differences in both 
the gluon PDF and the value of $\alphas$.  

\subsubsection{Parton to hadron level}

\begin{figure}
  \centering
  \includegraphics[width=0.48\textwidth]{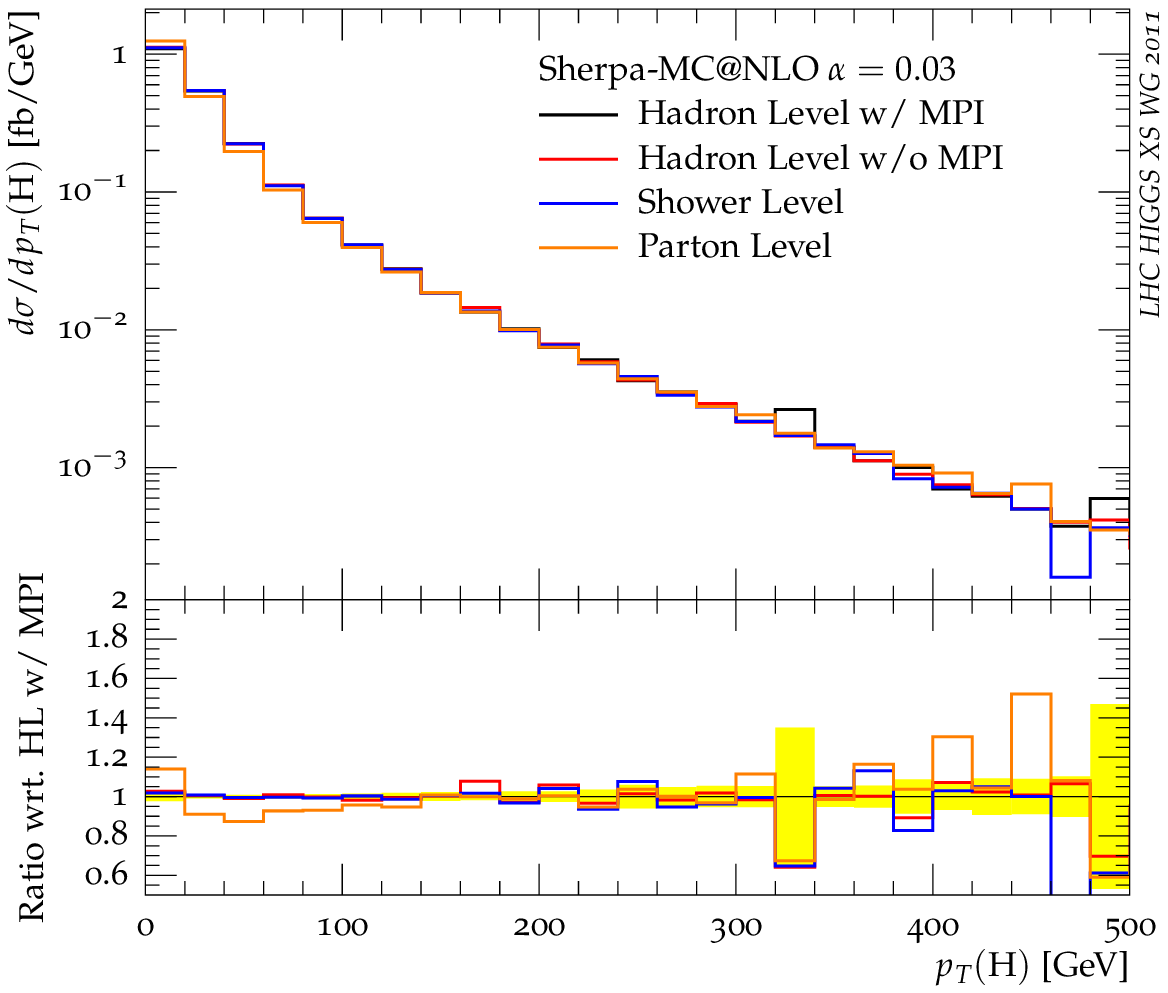}\hfill
  \includegraphics[width=0.48\textwidth]{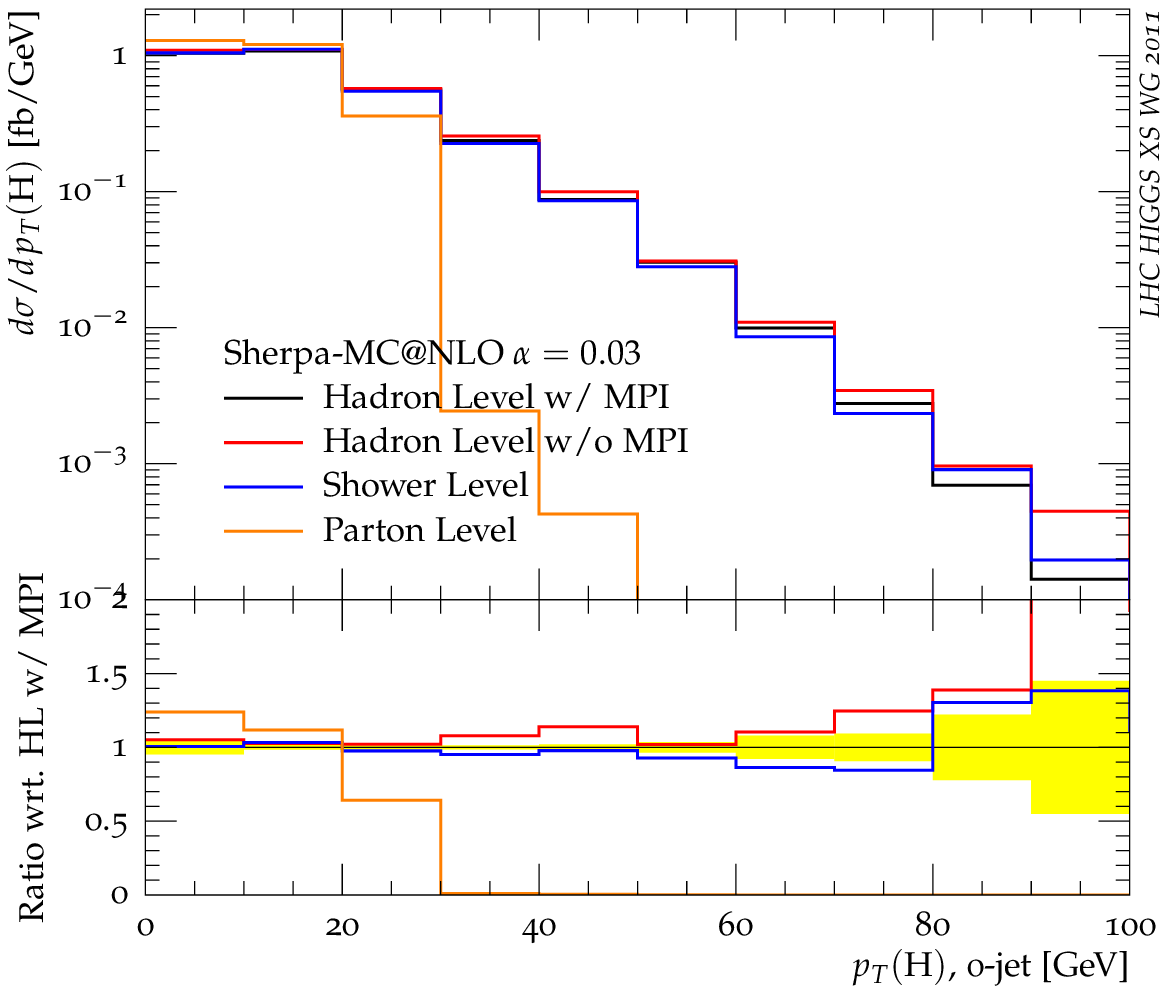}\\[1em]
  \includegraphics[width=0.48\textwidth]{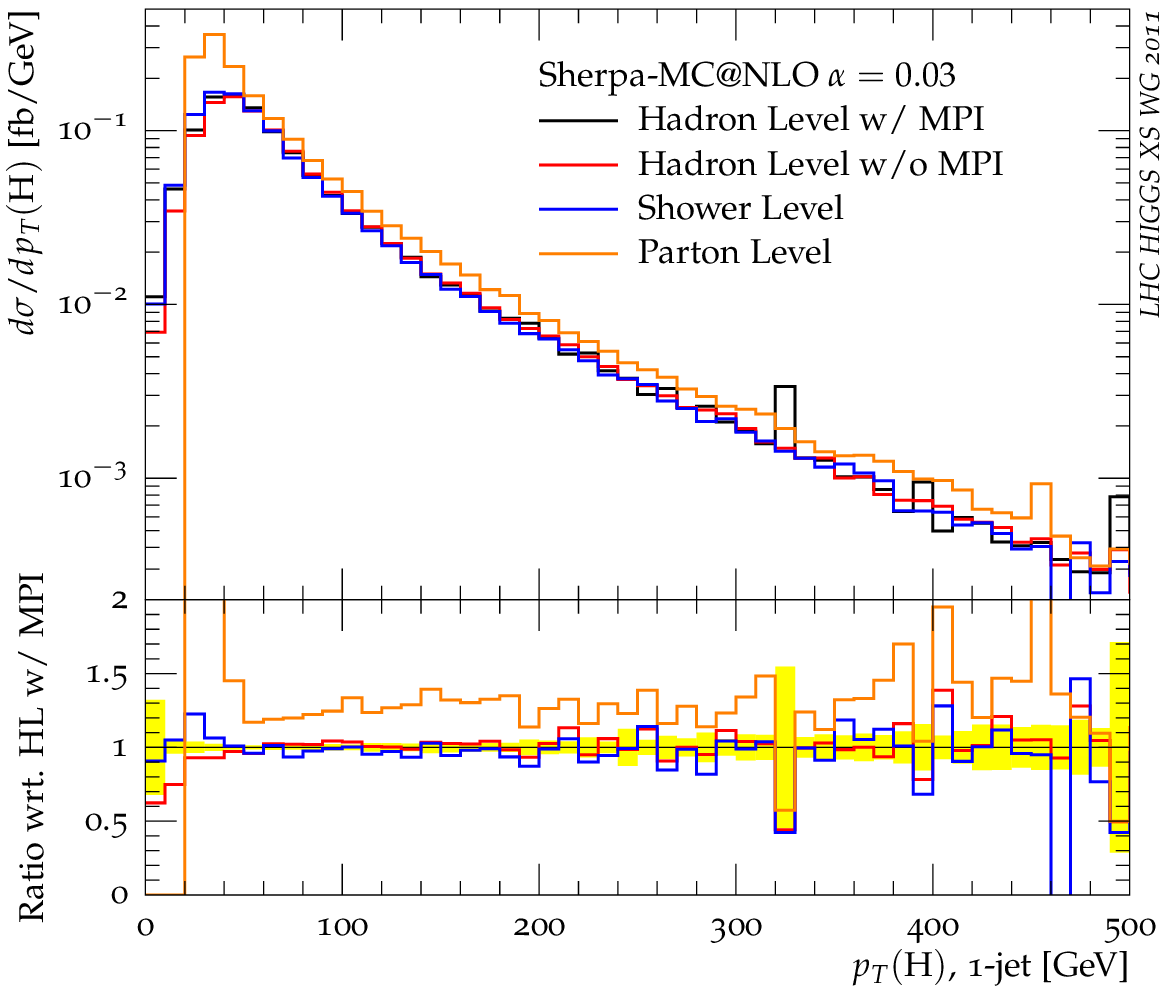}\hfill
  \includegraphics[width=0.48\textwidth]{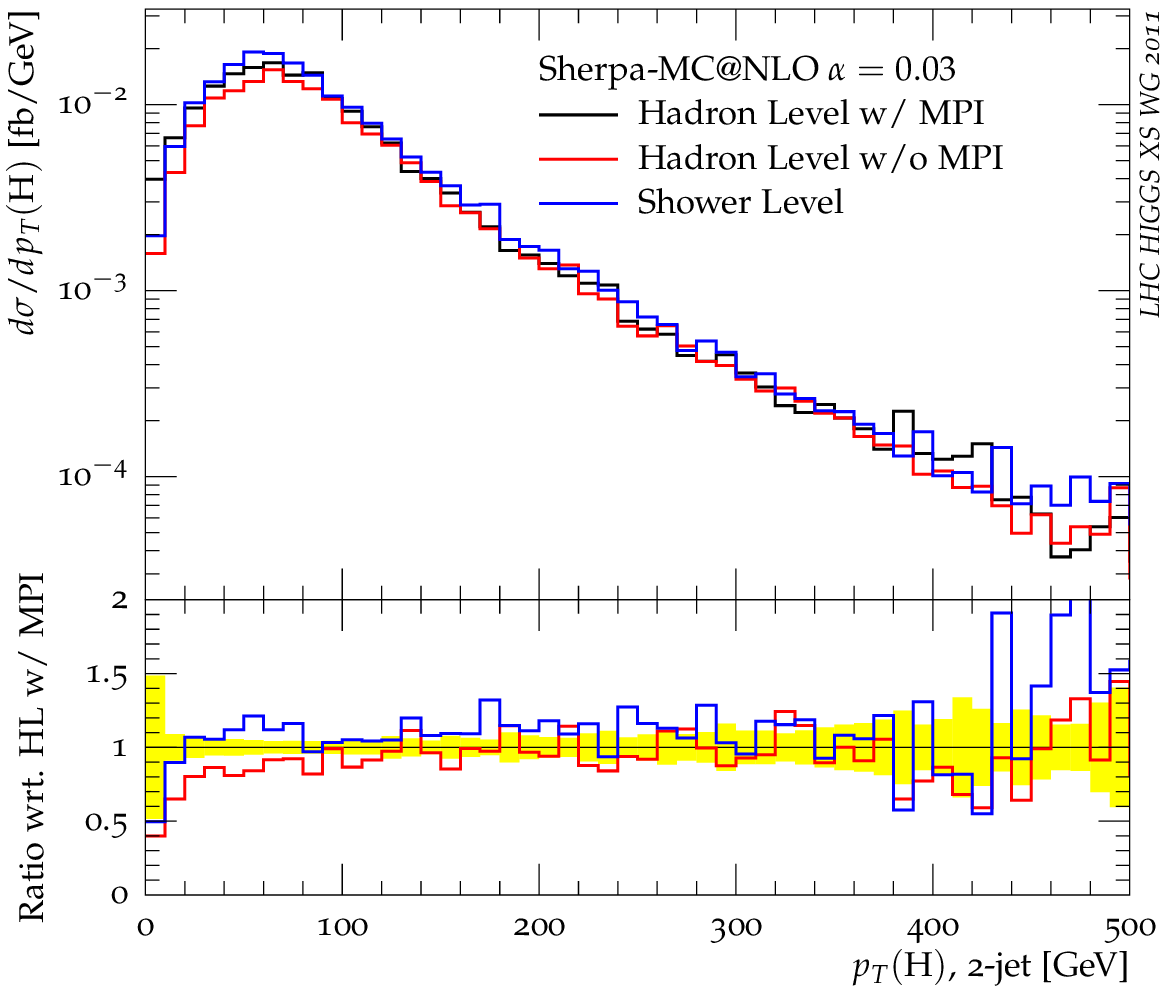}\\[1em]
  \includegraphics[width=0.48\textwidth]{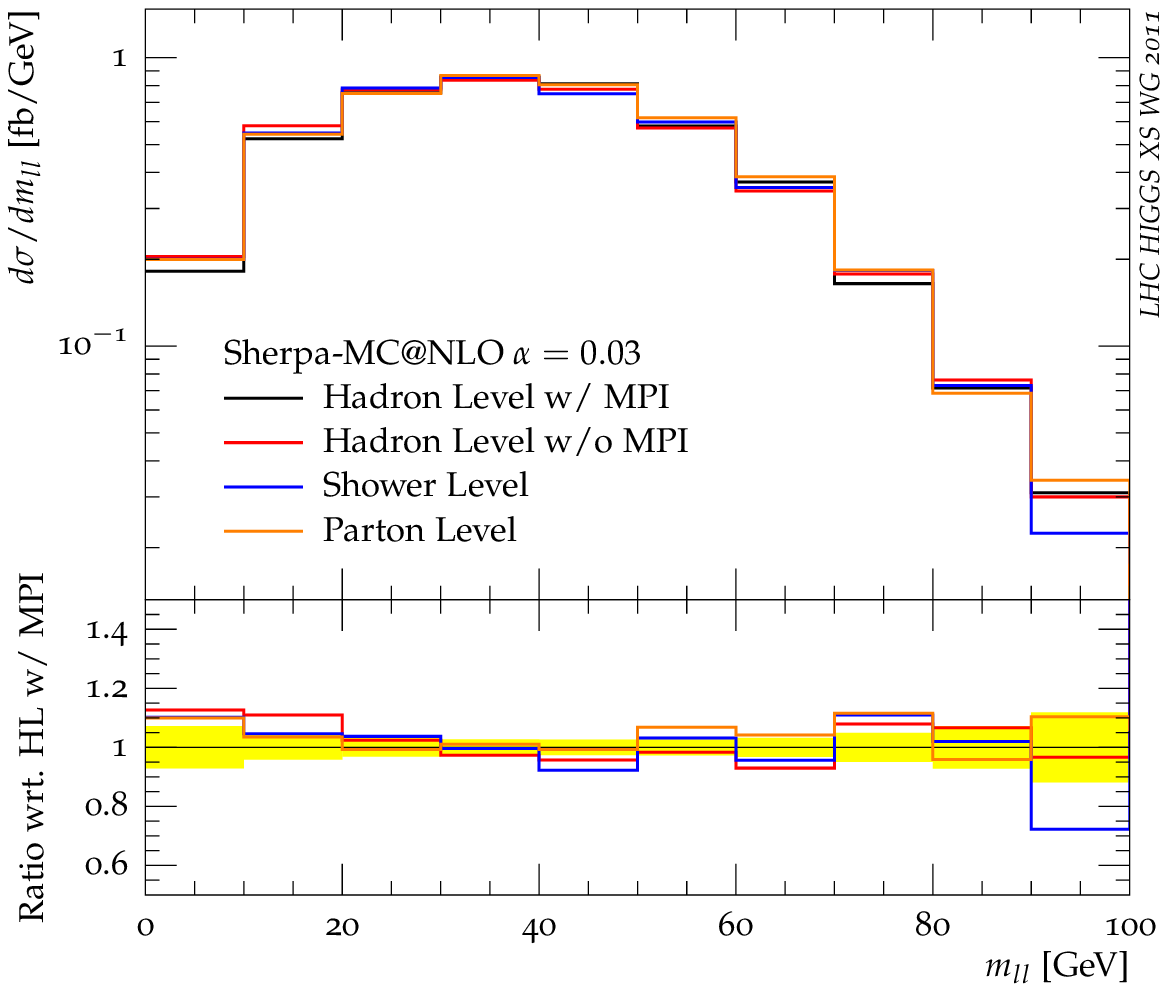}\hfill
  \includegraphics[width=0.48\textwidth]{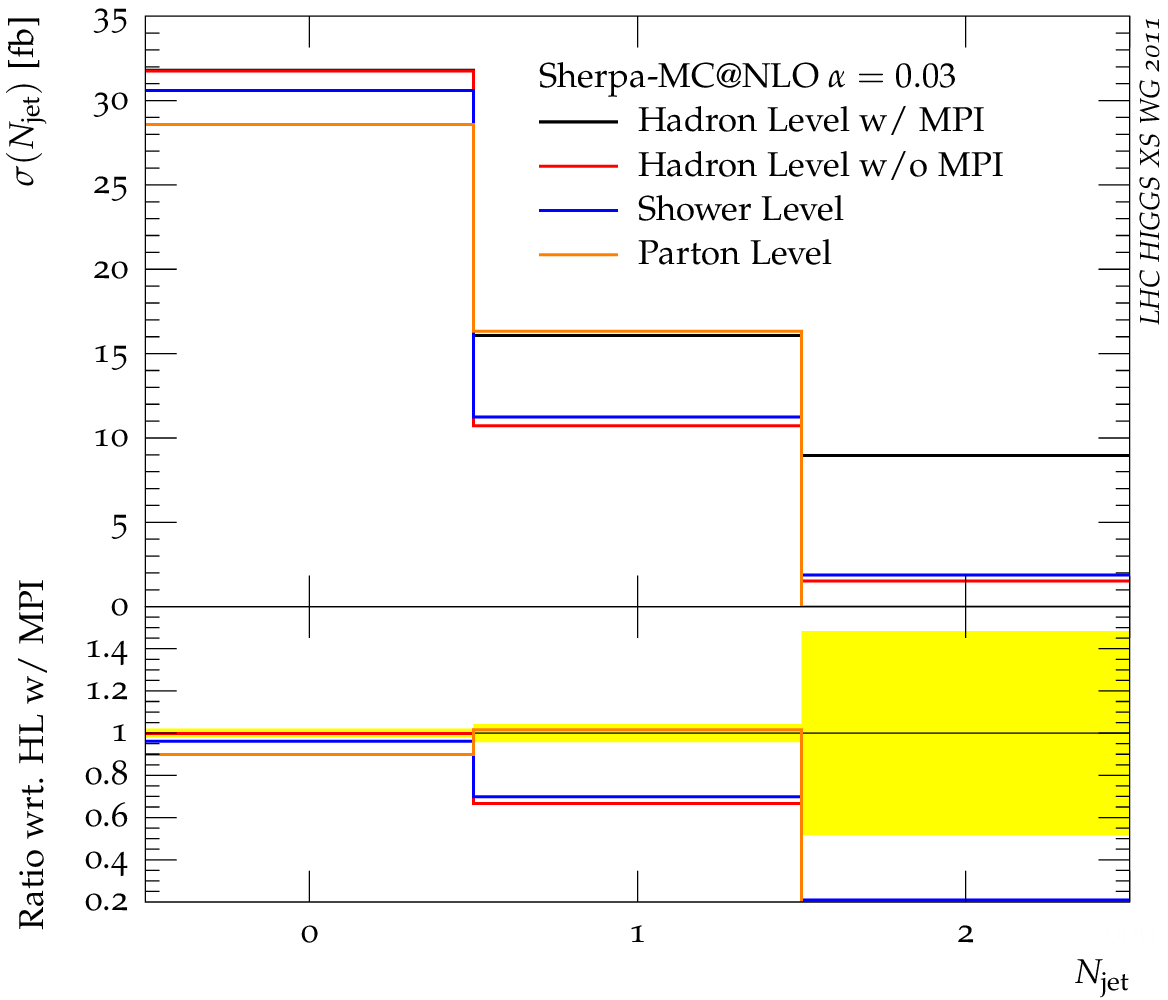}
  \caption{\label{fig:hww:uncert:levels}
    The impact of going from the parton level (orange) over the parton shower
    level (red) to the hadron level without (blue) and with (black) the 
    underlying event.  The error band relates to the statistical fluctuations 
    of the reference result -- the full simulation.  All curves are obtained by 
    an NLO matching according to the \protect{\SHERPA--\MCatNLO} prescription 
    with $\alpha = 0.03$.
  }
\end{figure}

In \refF{fig:hww:uncert:levels} the impact of adding parton showering,
fragmentation, and the underlying event is exemplified for the same 
observables as in the figures before.  In addition, the invariant lepton mass 
and the jet multiplicities are exhibited at the same stages of event 
generation.  All curves are obtained by an NLO matching according to the 
\SHERPA--\MCatNLO prescription with $\alpha = 0.03$.  In the following, the 
CT10 PDF was used to be able to use the corresponding tuning of the 
non-perturbative \SHERPA\ parameters.

In general, the effect of parton showering is notable, resulting in a shift
of the transverse momentum of the Higgs boson away from very low transverse 
momenta to values of about $20{-}80 \UGeV$ with deviations up to $10\%$.  This 
effect is greatly amplified in the exclusive jet multiplicities, where the 
{\em exclusive} 1-jet bin on the parton level feeds down to higher jet 
multiplicities emerging in showering, leading to a reduction of about $50\%$ in 
that bin (which are of course compensated by higher bins such that the net 
effect on the {\em inclusive} 1-jet bin is much smaller).  A similar effect 
can be seen in the Higgs transverse momentum in the exclusive 0-jet bin, where 
additional radiation allows for larger transverse kicks of the Higgs boson 
without actually resulting in jets.  In all cases, however, the additional 
impact of the underlying event is much smaller, with a maximal effect of 
about $15{-}20\%$ in the 2-jet multiplicity and in some regions of the Higgs 
transverse momentum.

\subsubsection{Non-perturbative uncertainties}

In the following, uncertainties due to non-perturbative effects have been 
estimated.  Broadly speaking, two different physics reasons have been 
investigated, namely the impact of fragmentation, which has been assessed
by switching from \SHERPA's default cluster fragmentation~\cite{Winter:2003tt}
to the Lund string fragmentation~\cite{Andersson:1983ia} encoded in \PYTHIA 6.4 
\cite{Sjostrand:2006za} and suitably interfaced\footnote{
  Both fragmentation schemes in the \SHERPA\ framework have been tuned 
  to the same LEP data, yielding a fairly similar quality in the description 
  of data.},
and the impact of variations in the modelling of the underlying event.
There \SHERPA's model, which is based on \cite{Sjostrand:1987su}, has been
modified such that the transverse activity (the plateau region of $N_{\mathrm ch}$ 
in the transverse region) is increased/decreased by $10\%$.  

\begin{figure}
  \centering
  \includegraphics[width=0.48\textwidth]{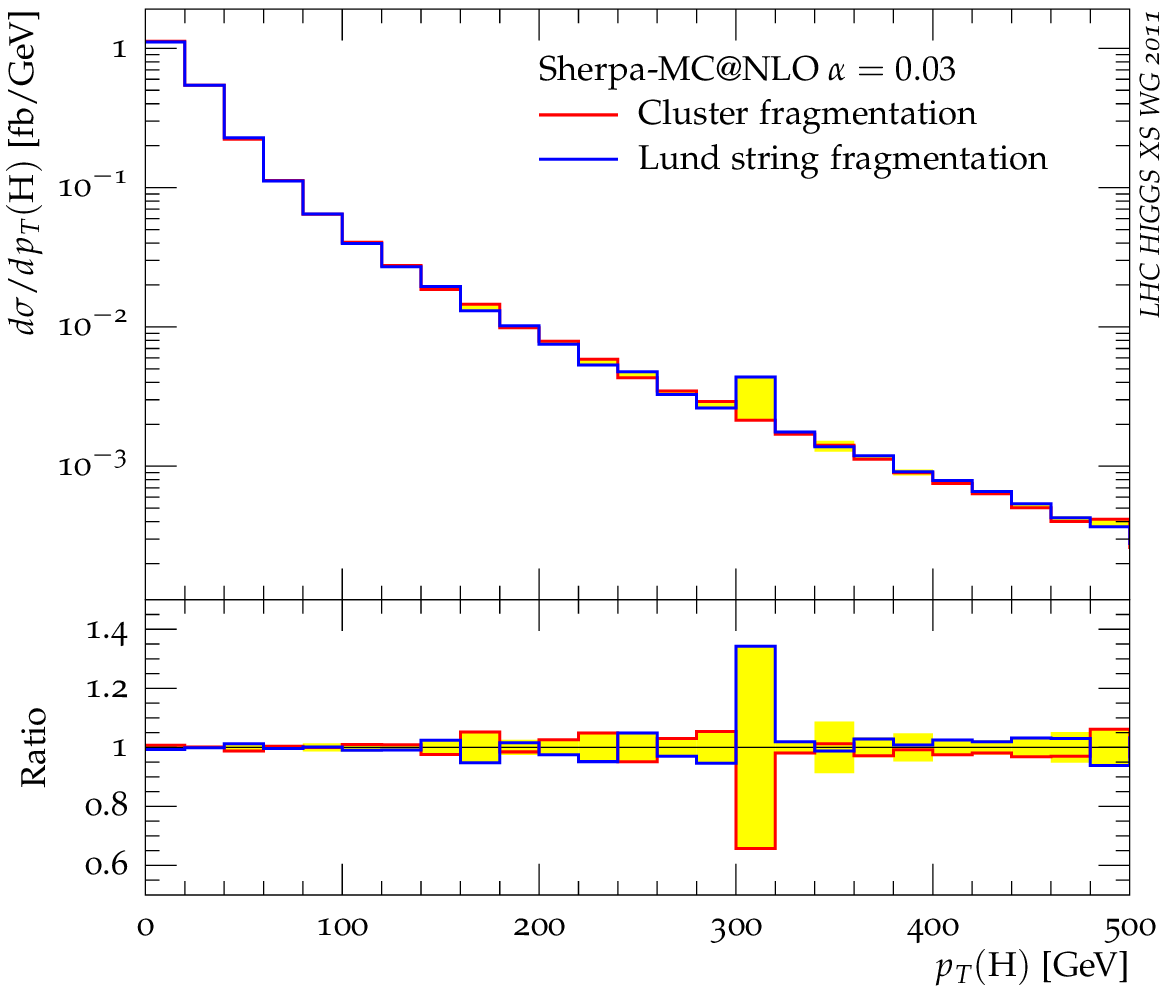}\hfill
  \includegraphics[width=0.48\textwidth]{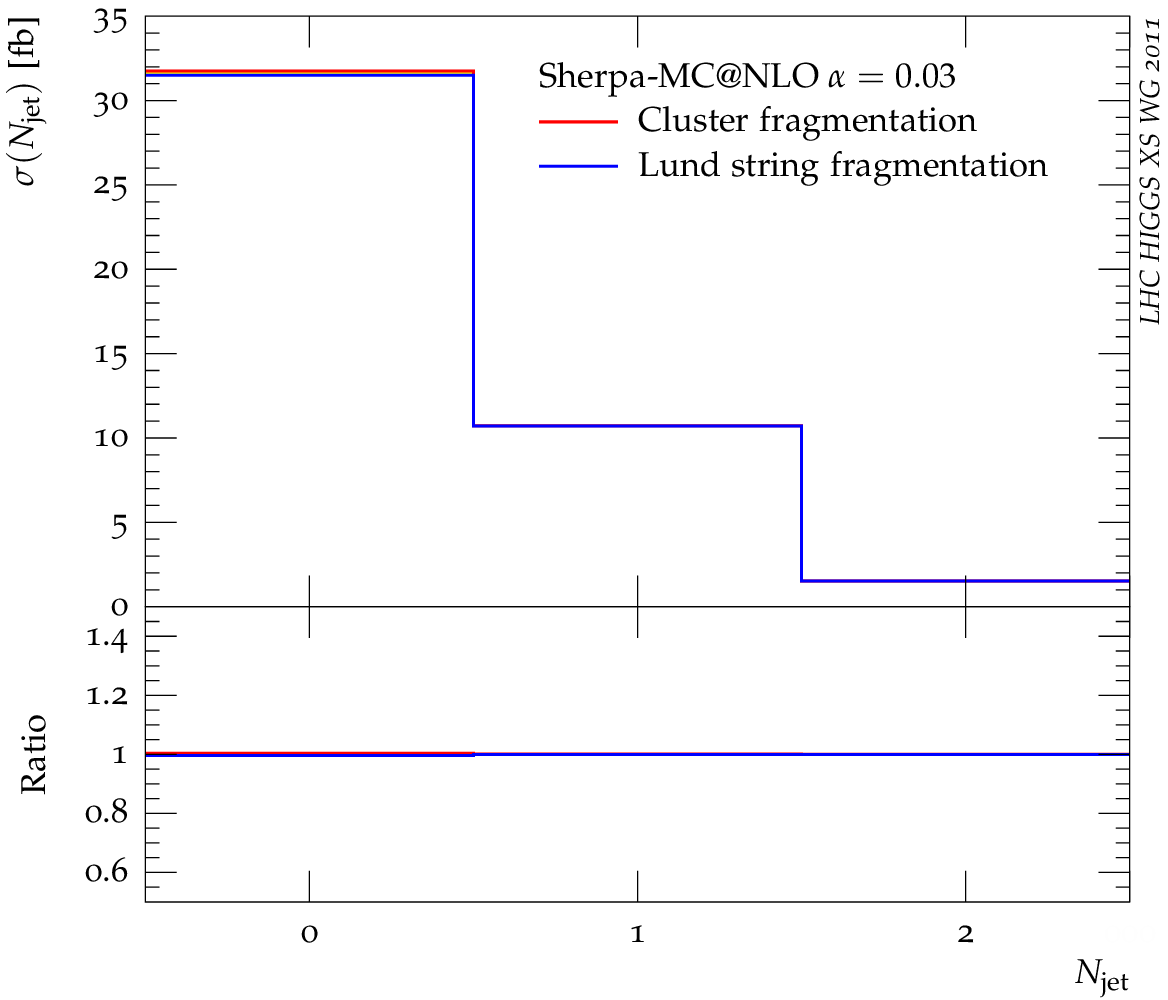}
  \caption{\label{fig:hww:uncert:frag}
    The impact of different fragmentation models -- cluster (black) and 
    string (red) -- for a $\MH = 160\UGeV$ on the transverse momentum of the 
    Higgs boson in all events, and on jet multiplicities.   All curves are 
    obtained by an NLO matching according to the \protect{\SHERPA--\MCatNLO} 
    prescription with $\alpha = 0.03$.  The yellow band indicates a 
    combination of statistical differences and the differences of the two 
    fragmentation schemes.
  }
\end{figure}

We find that in all relevant observables the variation of the fragmentation
model leads to differences which are consistent with statistical fluctuations
in the different Monte Carlo samples.  This is illustrated by the Higgs-boson
transverse momentum and the jet multiplicities, displayed in 
\refF{fig:hww:uncert:frag}.

Similarly, differences due to the underlying event on the Higgs boson transverse
momentum and various jet-related observables are fairly moderate and typically
below $10\%$.  This is especially true for jet multiplicities, where the $10\%$
variation of the underlying event activity translates into differences of the
order of $2{-}3\%$ only.  However, it should be stressed here that the variation 
performed did not necessarily affect the hardness of the parton multiple 
scatterings, \ie the amount of jet production in secondary parton scatters,
but rather increased the number of comparably soft scatters, leading to an
increased soft activity.  In order to obtain a more meaningful handle on jet 
production due to multiple parton scattering dedicated analyses are mandatory
in order to validate and improve the relatively naive model employed here.
\begin{figure}
  \centering
  \includegraphics[width=0.48\textwidth]{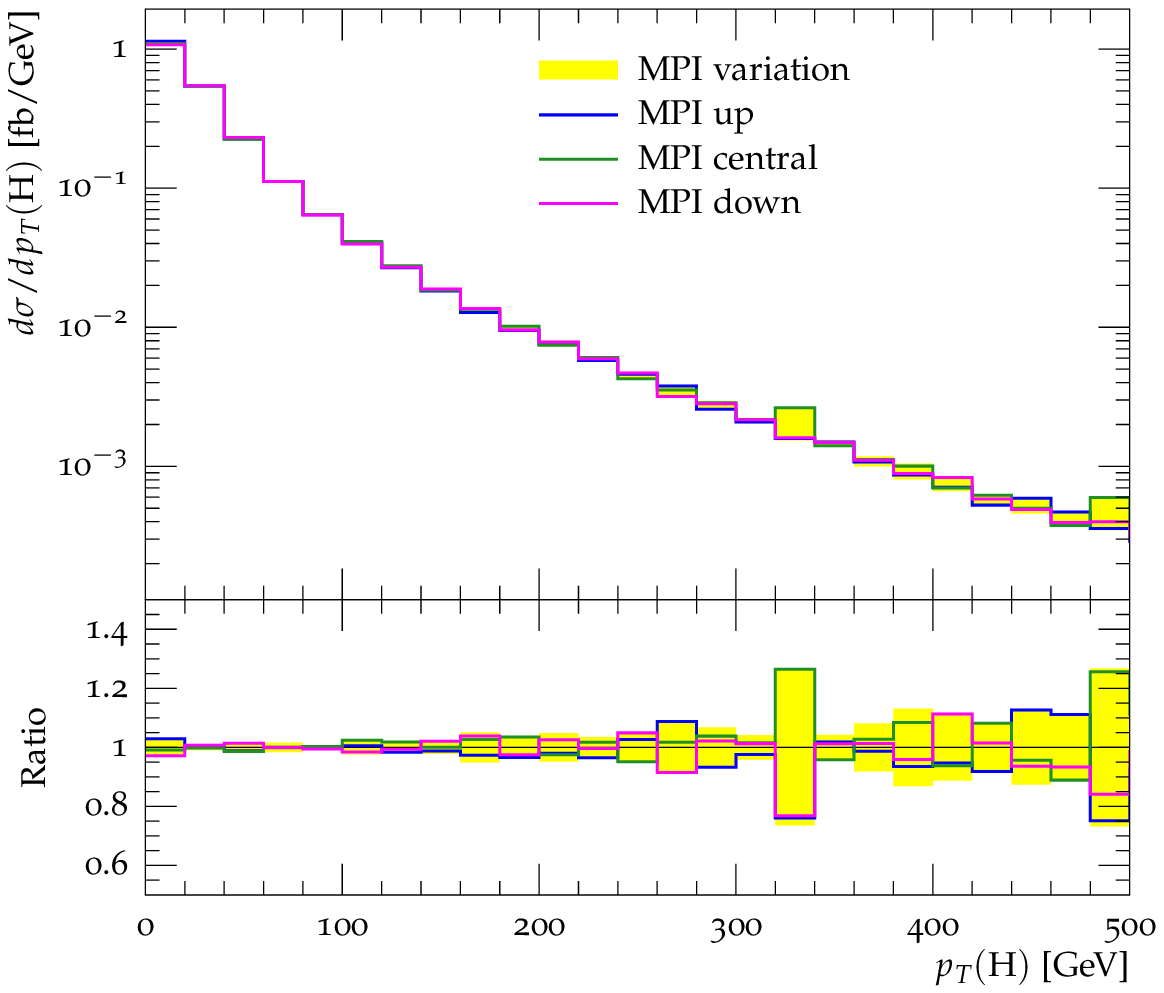}\hfill
  \includegraphics[width=0.48\textwidth]{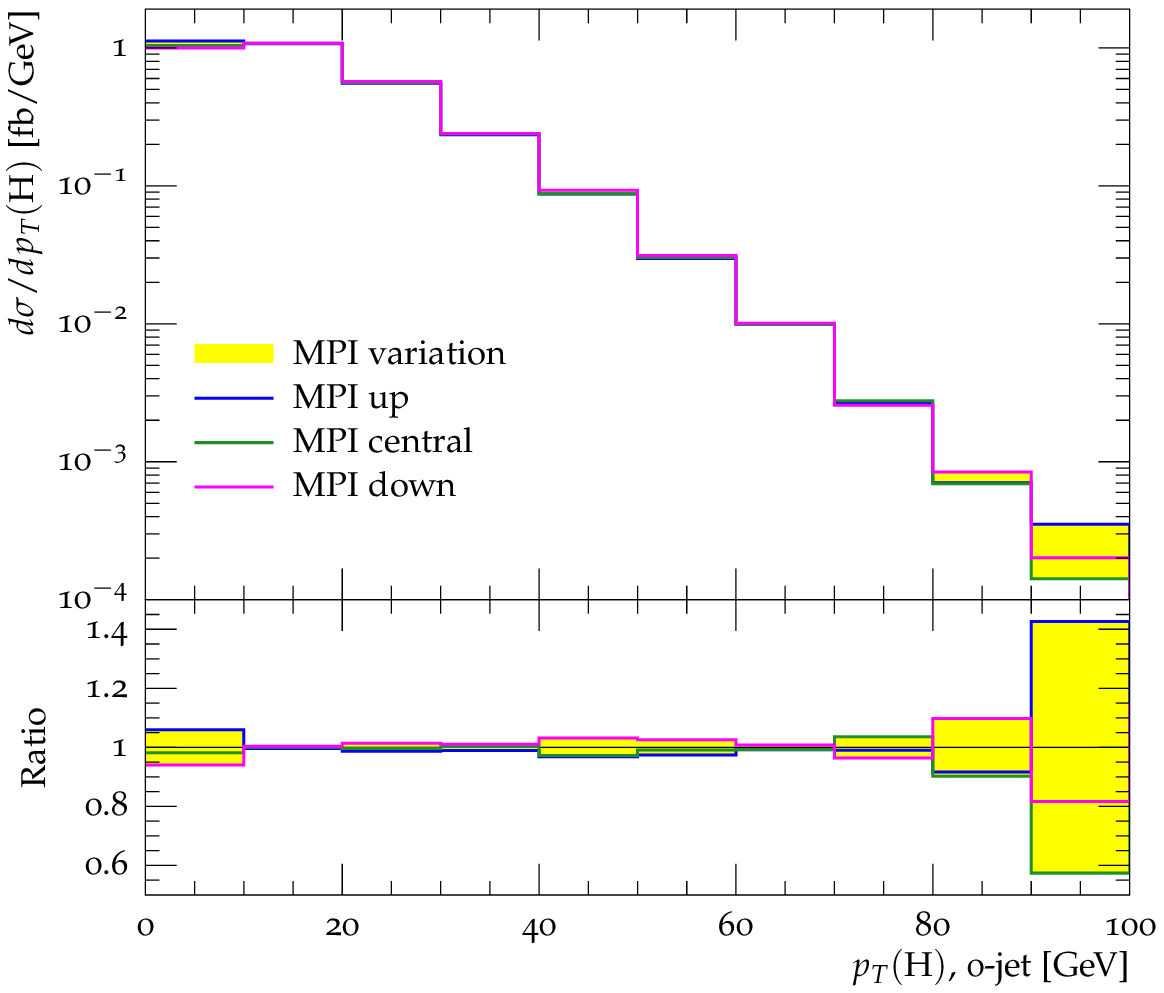}\\[1em]
  \includegraphics[width=0.48\textwidth]{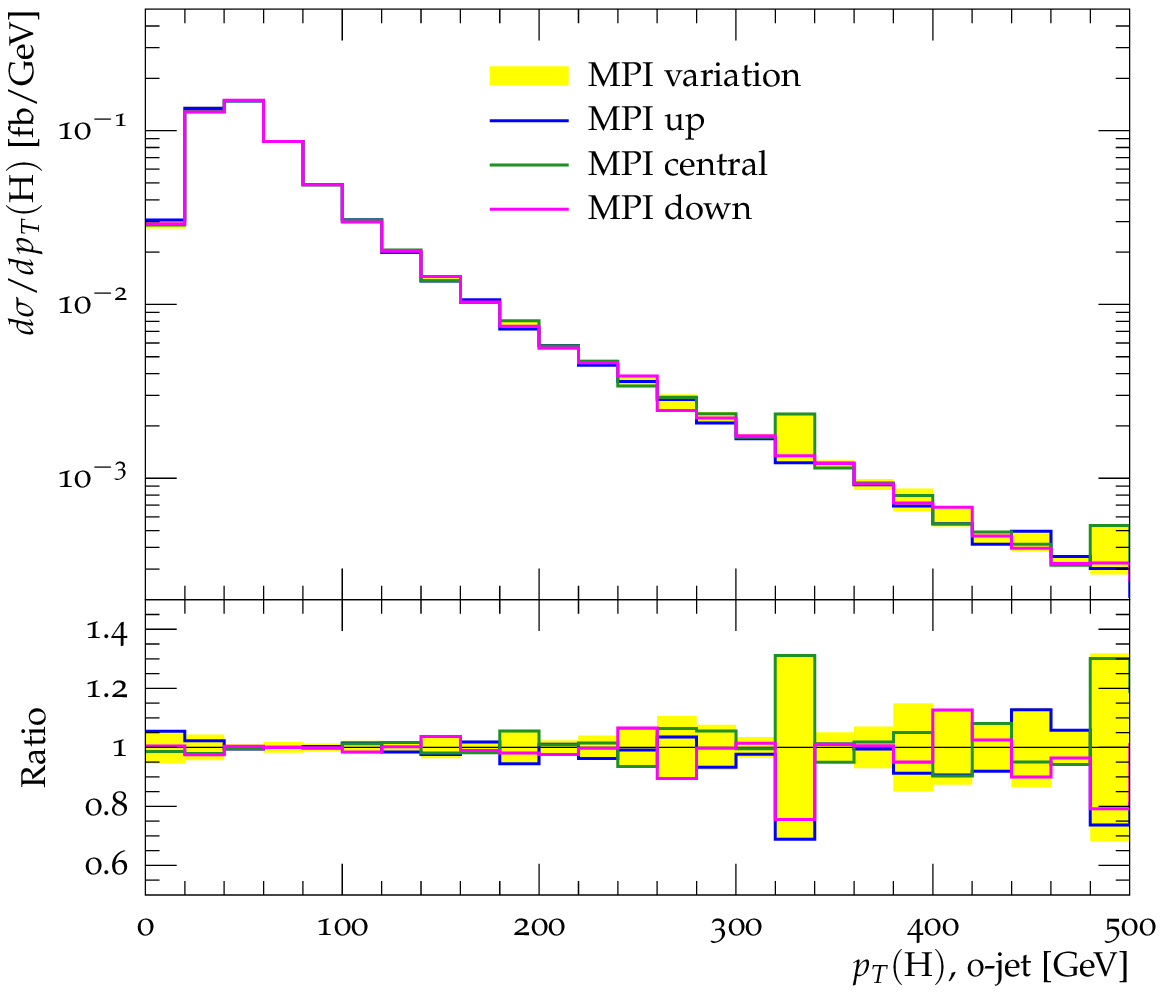}\hfill
  \includegraphics[width=0.48\textwidth]{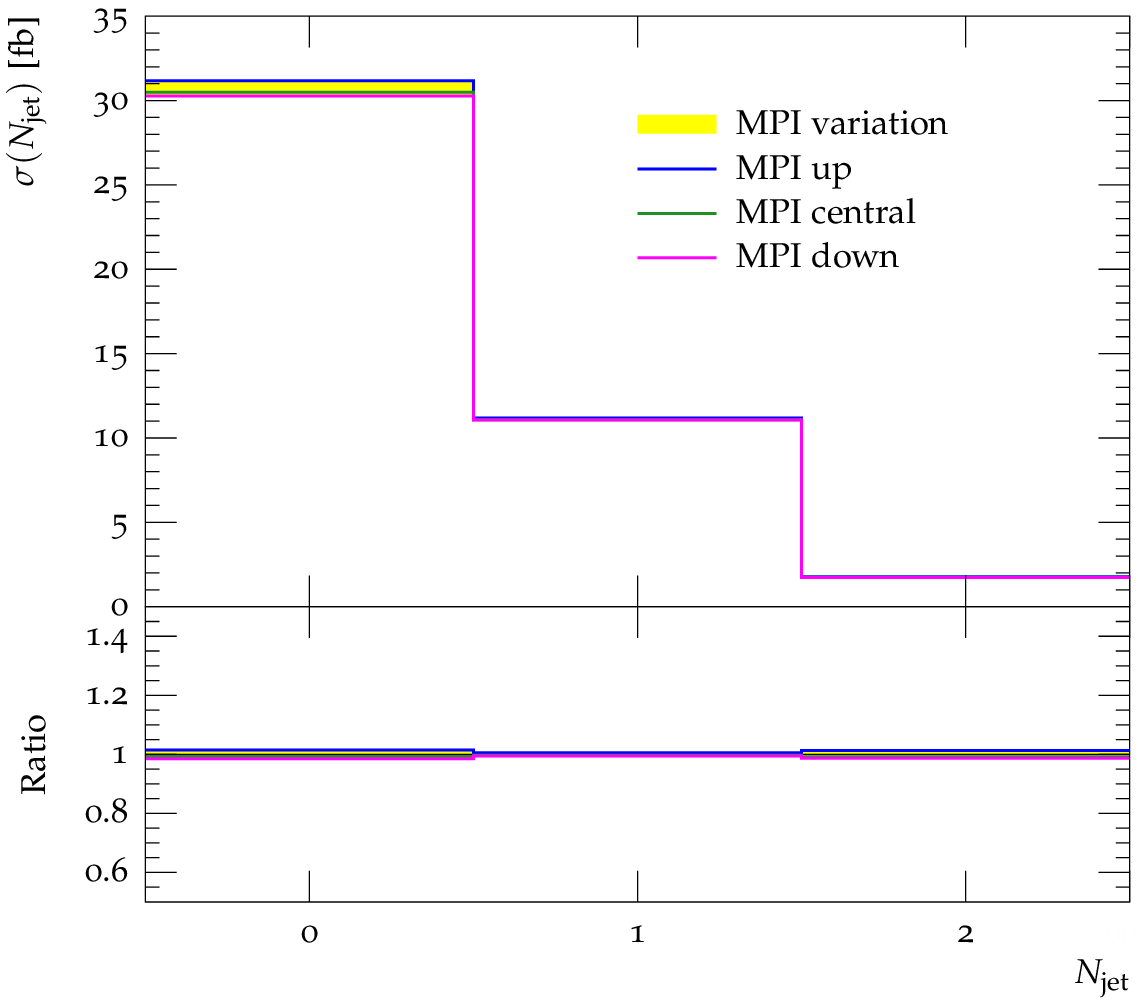}
  \caption{\label{fig:hww:uncert:ue}
    Differences in the Higgs-boson transverse momentum in all events, in events
    with 0, 1, and in jet multiplicities due to variations in the modelling of 
    the underlying event -- central (black) vs.\ increased (blue) and 
    suppressed (purple) activity -- for a $\MH = 160\UGeV$.  All curves are 
    obtained by an NLO matching according to the \protect{\SHERPA--\MCatNLO} 
    prescription with $\alpha = 0.03$.  The yellow band indicates the
    uncertainties related to the modelling of the underlying event.
  }
\end{figure}

\clearpage

\subsection{Systematic studies of NLO MC tools in experimental implementations}\label{NLOEx}
In conjunction with systematic studies of theoretical tools, 
described in the previous sections,
the NLO MC group has estimated the systematic uncertainties of the NLO MC tools in  
their implementation within the event-generation framework of the ATLAS and CMS experiments.
This section describes the results of these studies performed both at the parton level, after hard scatterer 
parton showering, and after a fast detector simulation to account for detector effects.  

The initial plans for the systematic uncertainty studies entail as follows:
\begin{itemize}
\item Higgs production process: gluon fusion, $\Pg\Pg\rightarrow \PH$;
\item decay processes: $\PH\rightarrow \PW\PW \rightarrow \Pl \PGn \Pl \PGn$ and 
$\PH\rightarrow \PZ\PZ \rightarrow 4\Pl$;
\item MC tools: \POWHEGBOX, \MCatNLO, \SHERPA, and \HERWIG ++;
\item parton showering: \PYTHIA\ and \HERWIG;
\item Higgs mass: $130\UGeV$ and $170\UGeV$;
\item PDF: MSTW, CT10, NNPDF, and CTEQ6.6;
\item underlying events: Initial stage to switch off the UE in \PYTHIA\ and switch off the soft UE in \HERWIG\ to focus on matrix-element and parton-shower effects. 
\end{itemize}
In the following sections, we describe the current studies with \POWHEG interfaced with two different parton showering (PS) programs, 
\PYTHIA\ or \HERWIG, as implemented in ATLAS and CMS MC event generation.

\subsubsection{NLO Monte Carlo tools and parameters}
The study presented hereafter is based on the ATLAS implementation of \POWHEGPYTHIA and \POWHEGHERWIG. 
Two different Higgs mass values are considered: $\MH=170\UGeV$ and $\MH=130\UGeV$, with $38\UMeV$
 and $4.9\UMeV$ widths, respectively. 
The parton distribution function (PDF) CTEQ6.6 set is used for this study.
All other parameters are kept at the  default values implemented in \POWHEG~\cite{Nason:2010ap}.
We present here the results for the $\PH\rightarrow \PW\PW \rightarrow \Pl \PGn \Pl \PGn$ final state. 
This is one of the most relevant channels for discovery.
A study for the other golden channel $\PH\rightarrow \PZ\PZ \rightarrow 4\Pl$ is in progress.

\subsubsection{Comparisons of \POWHEGPYTHIA or \POWHEGHERWIG}

The statistics of the $\PH\rightarrow \PW\PW \rightarrow \Pl \PGn \Pl \PGn$ event samples consists  
of  22k events for {\sc POWHEG}+\PYTHIA and 50k event for \POWHEGHERWIG, for each Higgs mass values, $\MH=170\UGeV$ and $130\UGeV$.
The event generator is interfaced to ATLAS fast detector simulation to include detector effects.
This, however, along with the specific final-state selection scheme slowed down the event generation significantly, costing us about a week per sample.
In order to expedite complicated systematic uncertainties, such as that of PDF, it would be desirable to improve the speed of specific selection schemes.
The results presented in this section, however, are using the particle-level information without using the fast simulation to be comparable with studies carried out in other sections.

\begin{figure}
  \centering
  \includegraphics[width=1.0\textwidth]{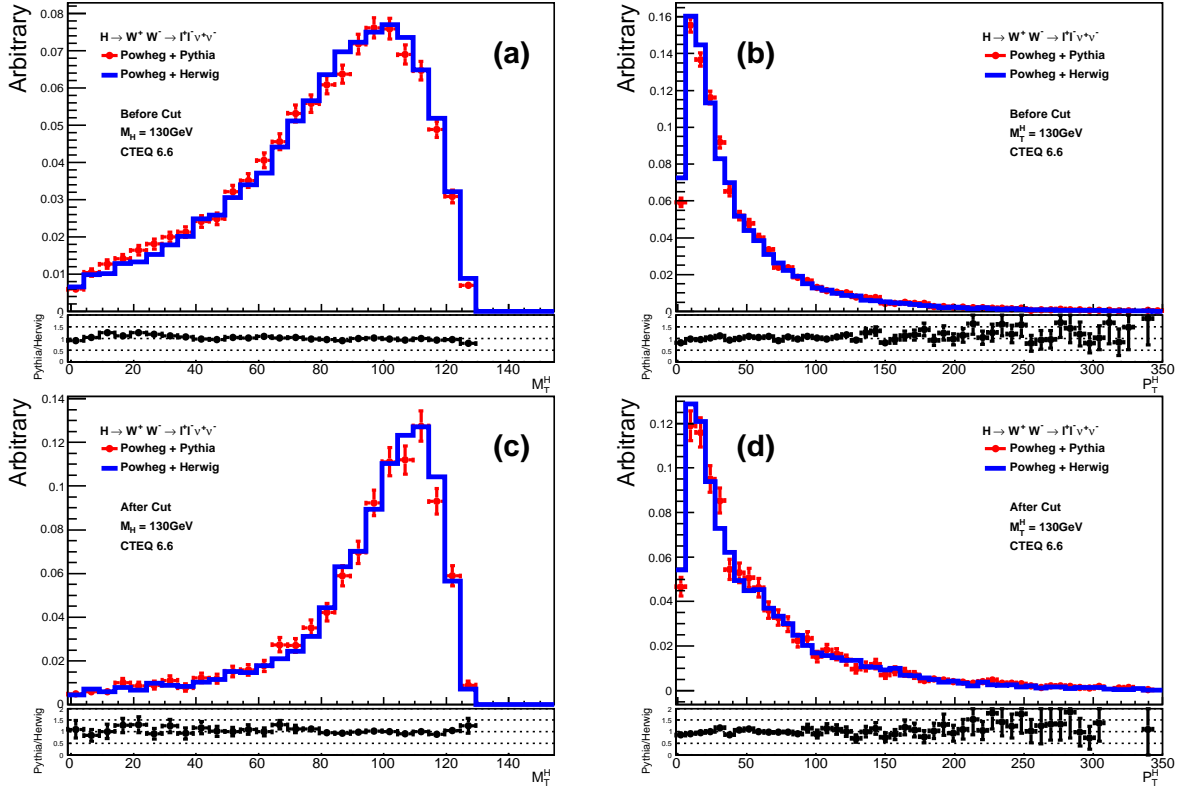}
  \caption{ (a) Higgs transverse mass without cuts, (b) Higgs transverse momentum without cuts, (c) Higgs transverse mass with cuts, and (d) Higgs transverse momentum with cuts for \POWHEGPYTHIA parton showering (red circles) 
and for \POWHEGHERWIG  parton showering (blue histogram) for $\MH=130\UGeV$.  
The plots below each of the histogram are the ratio of \POWHEGHERWIG\ with respect to \POWHEGPYTHIA.}
\label{fig:nlomc:exp:atlas-130}
\end{figure}

\begin{figure}
  \centering
  \includegraphics[width=1.0\textwidth]{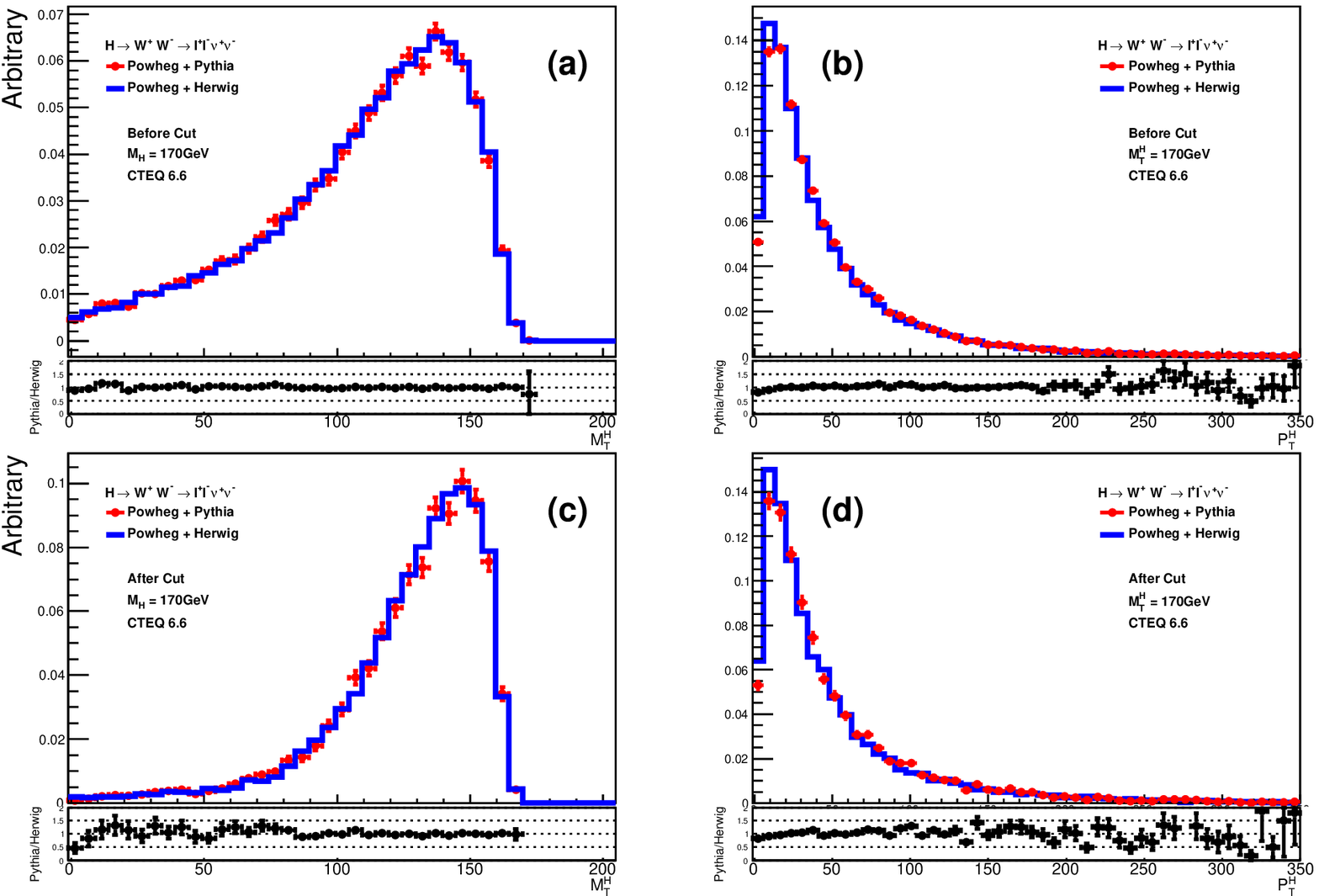}
  \caption{ (a) Higgs transverse mass without cuts, (b) Higgs transverse momentum without cuts, (c) Higgs transverse mass with cuts, and (d) Higgs transverse momentum with cuts for \POWHEGPYTHIA parton showering (red circles) 
and for \POWHEGHERWIG  parton showering (blue histogram) for $\MH=170\UGeV$.  
The plots below each of the histogram are the ratio of \POWHEGHERWIG\ with respect to \POWHEGPYTHIA.}
\label{fig:nlomc:exp:atlas-170}
\end{figure}

\begin{sloppypar}
The comparisons of various Higgs kinematic quantities with {\sc POWHEG}+\PYTHIA\ and {\sc POWHEG}+\HERWIG\ showering have been investigated.  
Most the kinematic variables, such as Higgs mass, W transverse mass and the kinematic variables of the leptons from the decay show no appreciable differences between \POWHEGPYTHIA and \POWHEGHERWIG parton showering.
\refF{fig:nlomc:exp:atlas-130}.(a) for $\MH=130\UGeV$ and \refF{fig:nlomc:exp:atlas-170}.(a) for $\MH=170\UGeV$ show good agreements between {\sc Pythia} and {\sc Herwig} parton showering.
The solid red circles in all plots represent the quantities from \POWHEGPYTHIA while the blue histograms represent those from \POWHEGHERWIG. 
The bin-by-bin ratio of {\sc POWHEG}+\HERWIG\ distributions with respect to \POWHEGPYTHIA  distributions 
is shown in the lower plots, to compare the shapes of the distributions.
However, as can be seen in \refF{fig:nlomc:exp:atlas-130}.(b) for $\MH=130\UGeV$ and \refF{fig:nlomc:exp:atlas-170}.(b) for $\MH=170\UGeV$, a systematic difference between the two PS schemes is observed for this quantity.
A systematic trend that the \POWHEGPYTHIA events display harder Higgs $\pT$ distribution than for \POWHEGHERWIG can be seen clearly from the ratio plots, despite the fact that the statistical uncertainties increases as $\pT$ grows.
The linear fit to the ratio for Higgs $\pT$ shows this trend with the value of the slope at $(9.1\pm 2.6)\times10^{-4}$ and $(6.2\pm 1.6)\times10^{-4}$ for $\MH=130\UGeV$ and $\MH=170\UGeV$, respectively, demonstrating statistically significant systematic effect.
The same trends have been observed in $\pT$ of the $\PWp$ and $\PWm$ from the Higgs decay as well.
\end{sloppypar}

\begin{figure}
  \centering
  \includegraphics[width=1.0\textwidth]{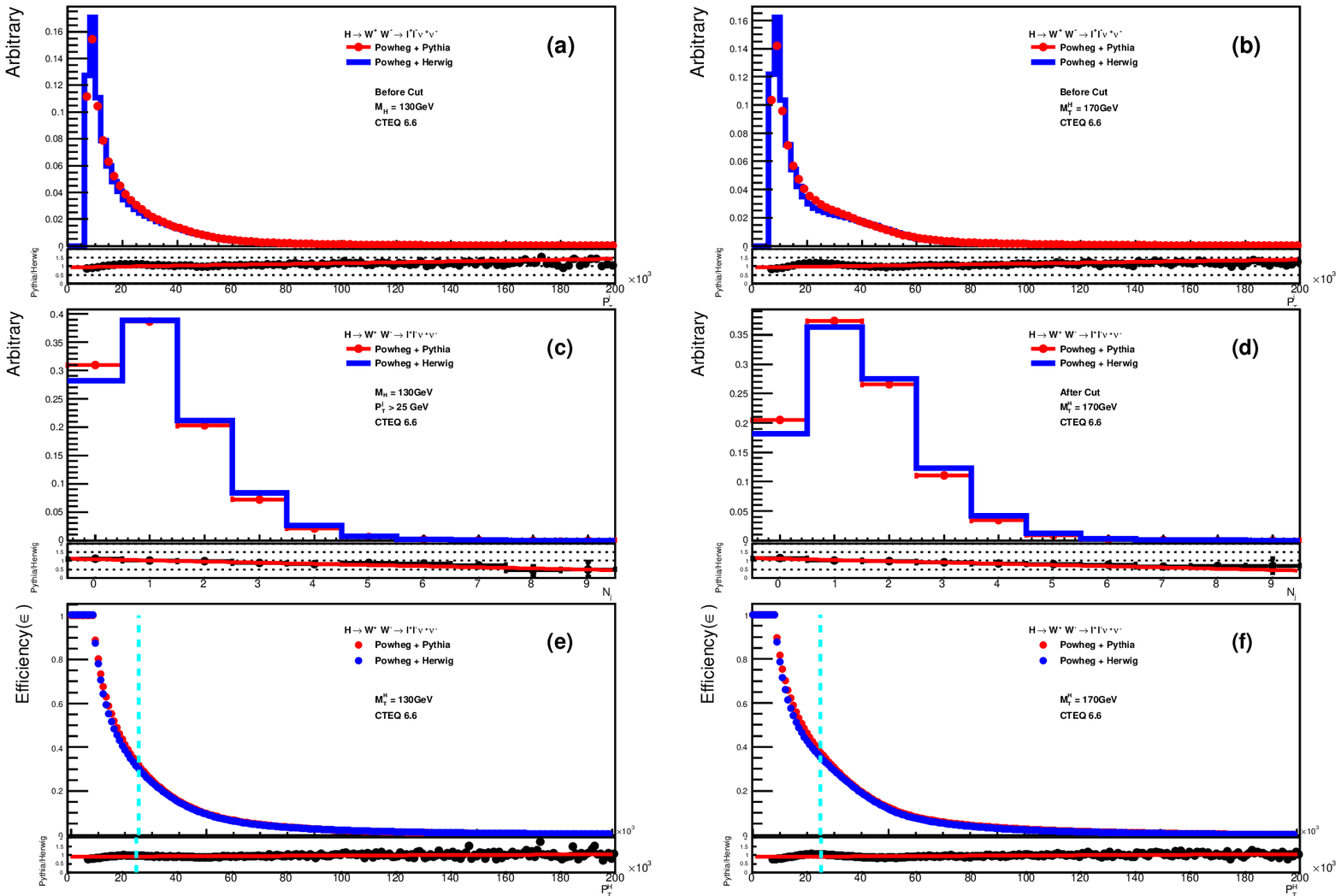}
\caption{$\pT$ distributions of the jets from parton showering for (a) $\MH=130\UGeV$ and (b) $\MH=170\UGeV$, number of associated jets with $\pT>25\UGeV$ for (c) $\MH=130\UGeV$ and (d) $\MH=170\UGeV$, and jet efficiency as a function of jet $\pT$ (e) $\MH=130\UGeV$ and (f) $\MH=130\UGeV$. \POWHEGPYTHIA parton showering (red circles) and for \POWHEGHERWIG\ parton showering (blue histogram).  The plots below each of the histogram are the ratio of \POWHEGHERWIG\ with respect to \POWHEGPYTHIA.}
  \label{fig:nlomc:exp:atlas-11}
\end{figure}

In order to investigate this effect further, we have looked into the $\pT$ distributions of the jets, 
number of associated jets and jet efficiencies  as a function of jet $\pT$ cut values as shown in \refF{fig:nlomc:exp:atlas-11}  for $\MH=170\UGeV$ and $\MH=130\UGeV$. 
It is shown that \PYTHIA\ produces harder jet $\pT$ distributions of the jets since their momentum must balance that of the Higgs.

\subsubsection{Studies with experimental cuts}
In order to ensure the relevance of these results for the experimental searches, 
the cuts applied in this study follow the recommendation from ATLAS
$\PH\rightarrow \PW\PW$
search group~\cite{ATLAS:2011aa},
as follows:
\begin{itemize}
\item exactly two leptons
\item first leading lepton (l1) $\pT> 25\UGeV$, subleading lepton (l2) $\pT > 15\UGeV$;
\item Two leptons have opposite charge $M_{\Pl\Pl} > 15\UGeV$;
\item if l1,l2 have the different flavour, $M_{\Pl\Pl} > 10\UGeV$;
\item if l1,l2 have the same flavours, apply a $Z$ veto, $|M_{\Pl\Pl}-\MZ| > 15\UGeV$;
\item if l1,l2 has  the same flavours $\pT^{\Pl\Pl} > 30\UGeV$;
\item $M_{\Pl\Pl} < 50\UGeV$ for $\MH < 170\UGeV$;
\item $|\Delta\phi_{\Pl\Pl}|< 1.3$ for $\MH< 170\UGeV$; $|\Delta\phi_{\Pl\Pl}|< 1.8$ for $170\UGeV < \MH < 220\UGeV$; 
no $|\Delta\phi_{\Pl\Pl}|$ 
cut for $\MH> 220\UGeV$;
\item $|\eta_{\Pl\Pl}| < 1.3$ for $\MH < 170\UGeV$.
\end{itemize}
\refF{fig:nlomc:exp:atlas-130}.(c) and \refF{fig:nlomc:exp:atlas-170}.(c) show the reconstructed transverse mass of the Higgs from the final-state lepton and neutrino systems before and after the experimental 
selection cuts have been applied, for $\MH=130\UGeV$ and $\MH=170\UGeV$, respectively.
The comparison of the reconstructed transverse-mass distributions show no significant difference, before and after experimental cuts between the two PS schemes.
\refF{fig:nlomc:exp:atlas-130}.(d) and \refF{fig:nlomc:exp:atlas-170}.(d) show the Higgs transverse momentum 
reconstructed from the final-state lepton and neutrino systems, before and after the experimental 
selection cuts have been applied, for $\MH=130\UGeV$ and $\MH=170\UGeV$.
The difference in these distributions between the two PS programs reflects the observation at parton level that  \POWHEGPYTHIA\ displays harder Higgs $\pT$ than that from \POWHEGHERWIG, while the statistical uncertainties at high $\pT$ bins diminish.

\subsubsection{Conclusions and future work}

We have demonstrated that the systematic uncertainties resulting from interfacing \POWHEG to  \PYTHIA\ and  \HERWIG\  programs are small 
in most of the quantities used for the search, independently of the selection cut values and of the Higgs-boson mass. 
On the other hand, the systematic uncertainties on Higgs and jets transverse momenta show sufficiently large 
differences between the two PS algorithms and therefore the associated systematic uncertainties cannot be ignored.
A similar study is in progress for the $\PZ\PZ$ final state. 

\begin{sloppypar}
Thus, we conclude that the main difference between the two PS programs, interfaced to \POWHEG, is observed in the Higgs transverse momentum, reflecting into the experimentally reconstructed transverse momentum of the leptons plus missing transverse energy system.
However, this difference is found to be independent of the selection cuts and of the Higgs mass value within the two masses we have studied.  Additional studies on higher Higgs masses would be needed to  further confirm this conclusion.
\end{sloppypar}

Based on our experience with this study, we emphasize that it is necessary 
to improve the speed of the event generators in order to compute promptly the uncertainties associated to
the PDF set choices, as well as to the description of the underlying event and pile-up effects. As it will be presented in the following, a first
step in this direction has been undertaken by the \aMCatNLO Monte Carlo which provides automatic evaluation of scale and PDF uncertainties by a simple reweighting procedure. Finally, further studies of fast simulation quantities will allow full assessment of the propagation of the theoretical uncertainties to the experimental quantities used for the search.

\subsection{Guidelines for the use of
\HqT{} results to improve NLO+PS programs}\label{NLOPSsec:guidelines}

In several parts of this report regarding specific Higgs signals,
methods to reweight NLO+PS generators have been presented.
In general, reweighting is not a straightforward and free-of-risk procedure, and reaching a
final recommendation would require more studies. Here we summarise a few key 
points that should be kept in mind when setting up a reweighting procedure.

First of all, in general it is always recommended to rescale the total cross section of fully inclusive MC samples obtained via
different techniques, LO+PS, NLO+PS, matching, and so on, to the best available prediction,
at NNLO, fixed order, or resummed. For differential distributions a  general strategy is, of course, not available,
and reweighting should be considered case by case and observable by observable.

An observable that has a key role in the acceptance determination and therefore
in exclusion limits is the transverse-momentum distribution of the Higgs. For this 
observable the best predictions are those provided by the \HqT{} code. It is therefore
natural to suggest to reweight all events from  NLO+PS generators to the \HqT{} distributions.
Our recommendation, in this case, is to reweight at
 the level of showered events,
\emph{without} hadronisation and underlying event. This is especially important in the very-low-$\pT$ region,
where non-perturbative effects are sizeable. One must proceed in this
way, since \HqT{} does not include these non-perturbative effects.
Thus, reweighting the full-hadron-level Monte Carlo output to \HqT{} may lead to washing out 
small-transverse-momentum effects that are determined by hadronisation and the
underlying events.

In the following discussion, we refer to the Higgs $\pT$ spectrum
obtained with the NLO+PS generator by switching off hadronisation and
underlying event as the ``shower distribution''. The same distribution
obtained with the full NLO+PS generator, including hadronisation and
underlying event, as ``hadron distribution''. These definitions work for any
NLO+PS scheme, \MCatNLO{}, or \POWHEG{}. In the \POWHEG{} case, however,
one can also compute the Higgs $\pT$ using what is stored in the LHEF common block, before showering.
We will call this ``\POWHEG{}-level distribution''.

An appropriate reweighting strategy can be the following.
First of all, one determines a reweighting function as the ratio
of the \HqT{} distribution to the shower distribution.
When generating events, one should then apply the reweighting function evaluated
at the 
transverse momentum of the Higgs, as determined at the end of the shower development,
before hadronisation and underlying events are introduced.
The event is then hadronised, and the underlying event is added. The $\pT$ of
the Higgs will be modified at this stage, especially at small $\pT$.
This procedure can only be applied if the shower-level output is
available on an event-by-event basis. This may be the case of the
fortran version of HERWIG, but it is not certainly the case of PYTHIA,
with the underlying event characterised by multiparton interaction
interleaved with the shower development.

An alternative reweighting procedure may be applied in the \POWHEG{}
case, where the Higgs transverse-momentum distribution is
available also at the parton level, before the shower.
It is usually observed that the effect of the shower on this
distribution is quite mild. One then should determine
a reweighting function to be applied as a function of the Higgs
transverse momentum determined at the parton level, such that the
transverse-momentum distribution of the Higgs after full shower
(but before hadronisation) matches the one computed with \HqT{}.
It may be possible to achieve this by an iterative procedure:
one takes the ratio of the shower spectrum over the parton-level spectrum
as the initial reweighting function. One then generates events using
this reweighting function, applied however to the Higgs transverse
momentum at the parton level. This will lead to a residual mismatch
of the shower spectrum with respect to the \HqT{} spectrum,
which can be used to multiplicatively correct the reweighting
function. One keeps iterating the procedure, until convergence is
reached. Events are then generated at the full hadron level, and reweighted
on the basis of the transverse momentum at the \POWHEG{} level.

No studies on the implementation of \HqT{} reweighting according to
the above guidelines have been performed for this report, and this topic
requires further studies.

If the reweighting factor is nearly constant, the reweighting procedure
is considerably simplified, since a nearly constant factor can be
applied safely as a function of the Higgs $\pT$ at the hadron level.
We have noticed here that, with non-default value of parameters,
\POWHEG{} and \MCatNLO{} approach better the \HqT{} result. It is thus advisable
to use this settings before attempting to reweight the distributions.

One simple reweighting option is to use the NLO+PS generators with the
best settings discussed above, and reweight by a constant factor, to match
the NNLO cross section.
In this way, \HqT{} is only used to make a preferred choice of parameter
settings in the NLO+PS generator. A conservative
way to estimate the error band, in this case, would be to use the
NLO+PS scale variation band, multiplied by the
ratio of the NNLO cross section over the NLO+PS cross section
evaluated at the central value of the scales.

The preferred settings are summarised as follows.
In the  \MCatNLO{} case, we recommend to use the reference factorisation and
renormalisation scale equal to $\MH$ rather than $\mT$. In \POWHEG{},
we recommend using the $\Rups/\Rupf$ separation, with the
{\tt powheg.input} variable {\tt hfact} set equal to $\MH/1.2$.
For the scale variation, we recommend to vary $\muR$ and $\muF$
independently by a factor of $2$ above and below the reference scale,
with the constraint $0.5<\muR/\muF<2$.

\subsection{Guidelines to estimate non-perturbative uncertainties}

In this section we propose a mechanism to evaluate the impact of 
non-perturbative uncertainties related to hadronisation and underlying 
event modelling.  In the past, the former typically has been treated by
comparing \pythia\ and \herwig\ results, and analysing, bin by bin, the 
effect of the different fragmentation (\ie\ parton showering) and hadronisation
schemes.  In contrast the latter often is dealt with by merely comparing a
handful of different tunes of the same program, typically \pythia.  Especially
the latter seems to be a fairly unsystematic way, in particular when taking
into account that various tunes rely on very different input data, and some
of the traditional tunes still in use did not even include LHC results (and
typically they therefore fail to describe them very well).  We therefore 
propose the following scheme which essentially relies on systematic variations
around a single tune, with a single PDF:
\begin{itemize}
\item In order to quantify hadronisation uncertainties within a model, use 
  two different tune variations around the central one, defined by producing 
  one charged particle more or less at LEP.  The difference between different
  physics assumptions entering the hadronisation model (\ie cluster vs.\ 
  Lund string hadronisation), but tuned to the same data, still needs to be 
  tested, of course, by running these models\footnote{
    In \SHERPA\ this can be achieved even on top of the {\em same}
    parton showering by contrasting a native cluster hadronisation model with
    the Lund string of \pythia, made available through an interface.  Of course,
    both models have been tuned to the same LEP data.}.
\item For the model- and tune-intrinsic uncertainties of the underlying-event
  simulation we propose to systematically vary the activity, number of charged
  particles, and their summed transverse momentum, in the transverse region.
  There are basically two ways of doing it, one is by increasing or decreasing
  the respective plateaus by $10$\% (this has been done in the study here);
  alternatively, one could obtain tune variations which increase or decrease
  the ``jettiness'' of the underlying event.  Effectively this amounts to
  changing the shape of the various activities in the transverse region.
\end{itemize}
While we appreciate that this way of obtaining systematic uncertainties is
somewhat cumbersome at the moment, we would like to stress that with the
advent of modern tuning tools such as \rivet-\professor \cite{Buckley:2009bj},
this is a perfectly straightforward and controlled procedure.  

At this point it should also be noted that usually different PDFs lead to
different tunes for the underlying event.  Therefore, in order to obtain
a meaningful estimate of uncertainties related to observables which are
susceptible to the underlying event, it is important to also retune its
modelling accordingly.  Just changing the PDF but keeping the underlying event 
model parameters constant will typically lead to overestimated uncertainties.
In a similar fashion, the impact of the strong coupling constant may lead
to the necessity of a retune.  This has recently been intensively discussed
in \Bref{Cooper:2011gk}, a similar discussion for the NLO+PS tools discussed here is still 
missing\footnote{In \SHERPA\ this never was an issue, as the
  strong coupling used throughout the code is consistently given by the PDF.}.

\subsection{Higgs production via gluon fusion: finite-mass effects\footnote{J. Alwall, Q. Li and F. Maltoni.}}
\label{NLOPSsec:ggHfullmt}

Current implementations in NLO+PS codes of  Higgs production process via gluon fusion
~\cite{Frixione:2005gz,Alioli:2008tz,Hamilton:2009za,Hoche:2010pf}
are based on matrix elements evaluated in the Higgs effective theory (ET), i.e.\ in
a theory where the heavy-quark loop(s) are shrunk to effective vertices.  
In several cases  the user is given  the possibility of rescaling the total cross section by a normalisation
factor, defined as the ratio between the exact Born contribution where
the full dependence on the top and bottom masses is kept into account
and the Born contribution evaluated in the ET. While the ET is a very good approximation
for light SM Higgs, fixed-order computations have shown that it fails for heavier Higgs masses,
at large Higgs transverse momentum and for a BSM Higgs with enhanced couplings to $\PQb$ quarks.

Very recently, progress to include finite-mass effects has been achieved on two different fronts:
matched predictions for Higgs production via heavy-quark loops in the SM and beyond have been obtained~\cite{Alwall:2011cy}
and the NLO full-mass-dependent calculation has been implemented in {\sc POWHEG}~\cite{Bagnaschi:2011tu}.

\subsubsection{Higgs production via gluon fusion with loops via LO+PS merging}

In \Bref{Alwall:2011cy} was presented the first fully exclusive simulation
of gluon-fusion inclusive Higgs production based on the (leading-order) exact one-loop
matrix elements for $\PH +0,\,1,\,2$ partons, matched to \pythia\
parton showers using multiple merging schemes implemented in {\sc MadGraph~5}~\cite{Alwall:2011uj}.

In order to take into account the full kinematic dependence of the
heavy-quark loop in Higgs production, the full one-loop amplitudes for
all possible subprocesses contributing to $\PH +0,\,1,\,2$ partons have been calculated. 
To gain in speed and efficiency (the evaluation of multi-parton loop amplitudes is, in general, computationally quite expensive)
a method has been devised to map the integrand in a quick (though approximate) way and to limit
the evaluation of loops  to a small number of points.
Parton-level events for  $\PH +0 ,\,1,\,2$ partons are generated via {\sc MadGraph/MadEvent} in the ET, 
with scale choices optimised  for the subsequent merging procedure. Before passing them to
the PS program, events are reweighted by the ratio of full
one-loop amplitudes over the ET ones, $r= |{\cal M}_{\mathrm{LOOP}}|^2/|{\cal M}_{\mathrm{ET}}|^2$.
The  reweighted parton-level events are unweighted, 
passed through \pythia\ and matched using the $k_{\mathrm{T}}$-MLM or the shower-$k_{\mathrm{T}}$
scheme. Event samples are finally normalised to reference NNLO cross sections.
The reweighting method does not make any approximation and with large enough statistics is exactly equivalent
to the integration of the phase space of the full one-loop amplitudes.

\begin{figure}
\includegraphics[width=0.48\textwidth]{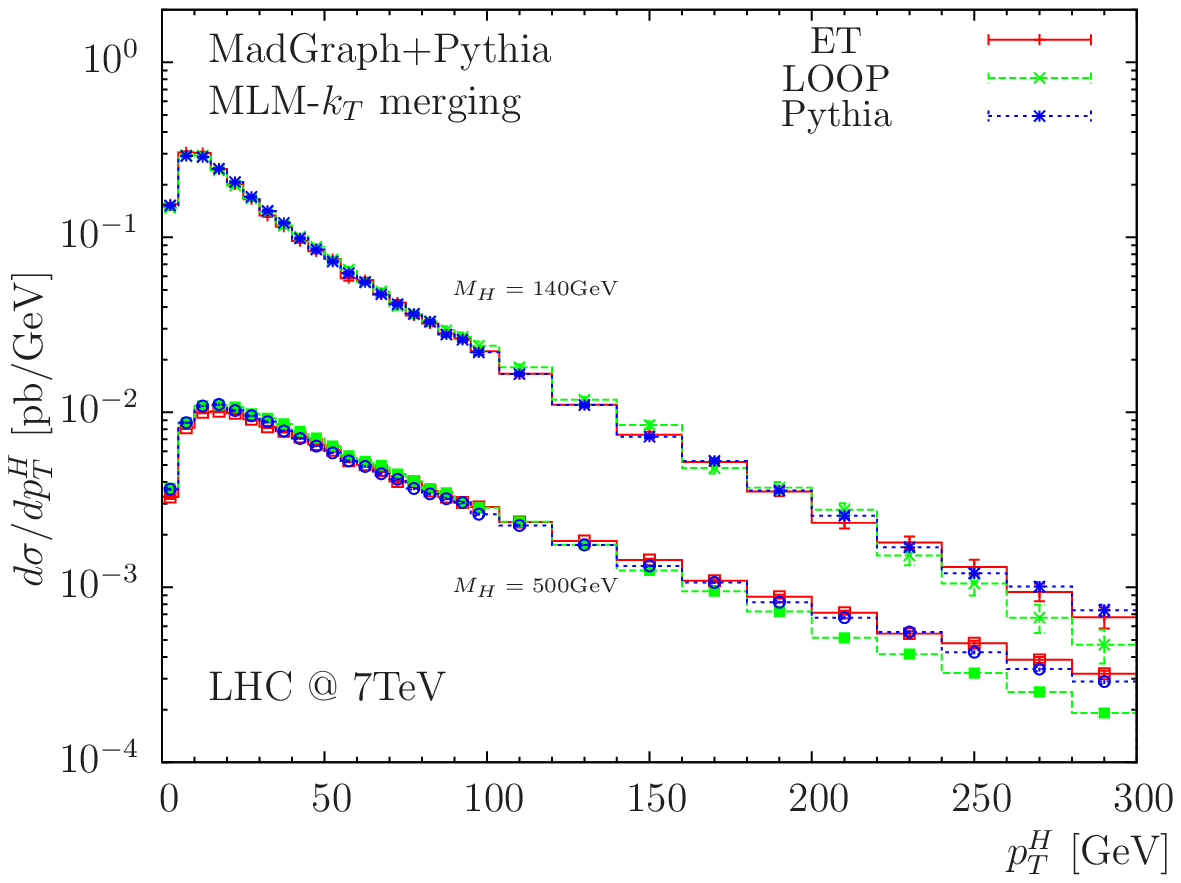}
\includegraphics[width=0.48\textwidth]{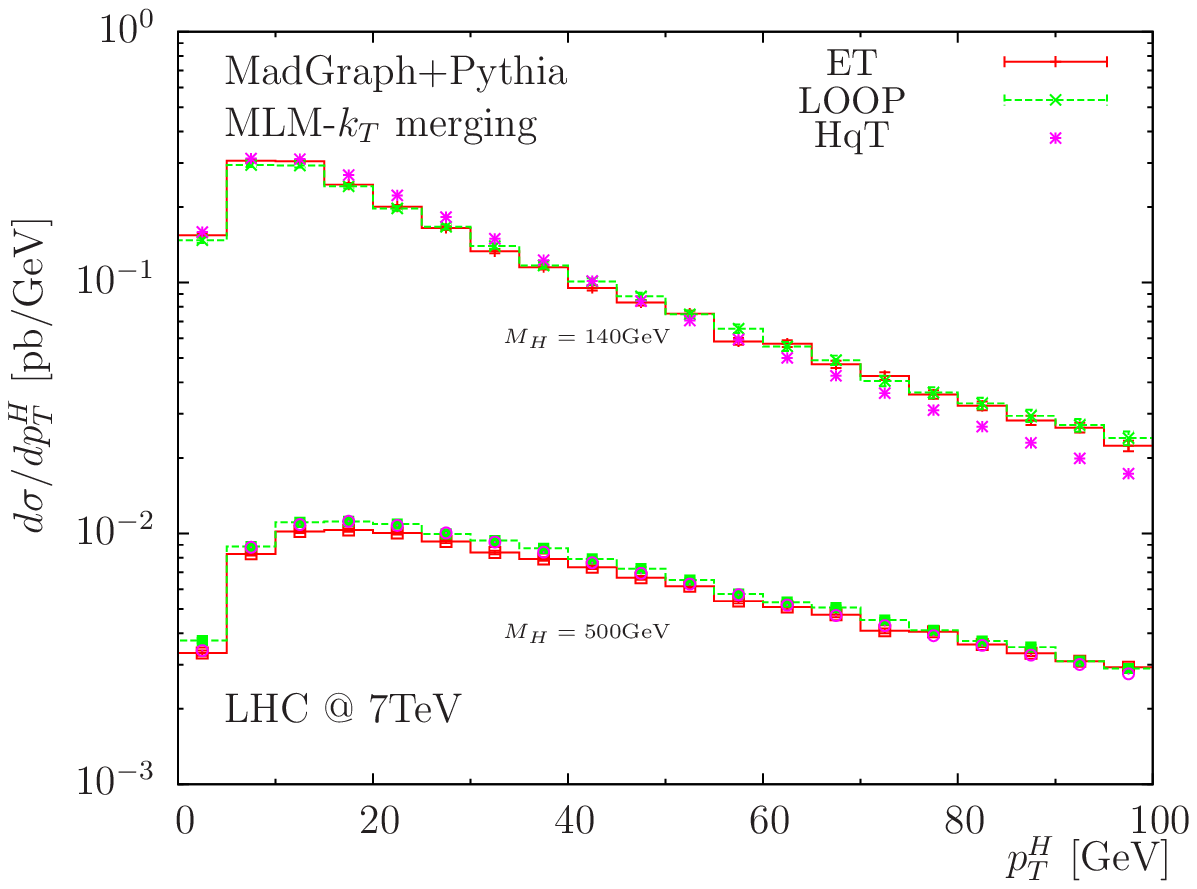}
\caption{SM Higgs  $\pT$ distributions for
$\MH=140\UGeV$ and  $\MH=500\UGeV$ in gluon-fusion production at $7\UTeV$ LHC.  In
the upper plot results in the ET and with full loop dependence
(LOOP) are compared over a large range of $\pT$ values to the default
\pythia\ implementation, which accounts for $2\to2$ matrix-element corrections. In the lower plot the low-$\pT$ range is compared to the NNLO+NNLL results as obtained by \HqT~\cite{Bozzi:2005wk,deFlorian:2011xf}. Curves normalised to the NNLO total cross sections see Ref.~\cite{Alwall:2011cy}.}
\label{hpt1}
\end{figure}

\begin{figure}
\includegraphics[width=0.48\textwidth]{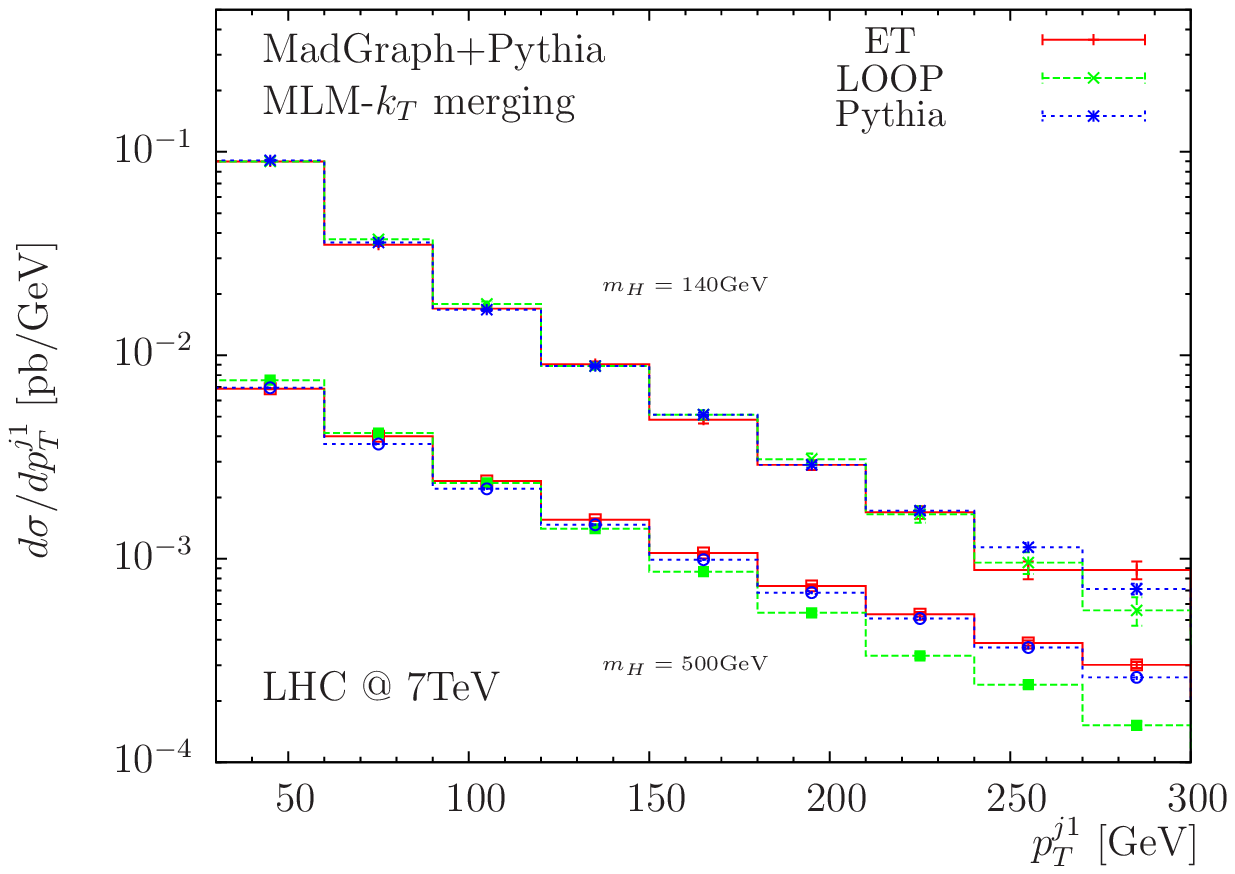}
\includegraphics[width=0.48\textwidth]{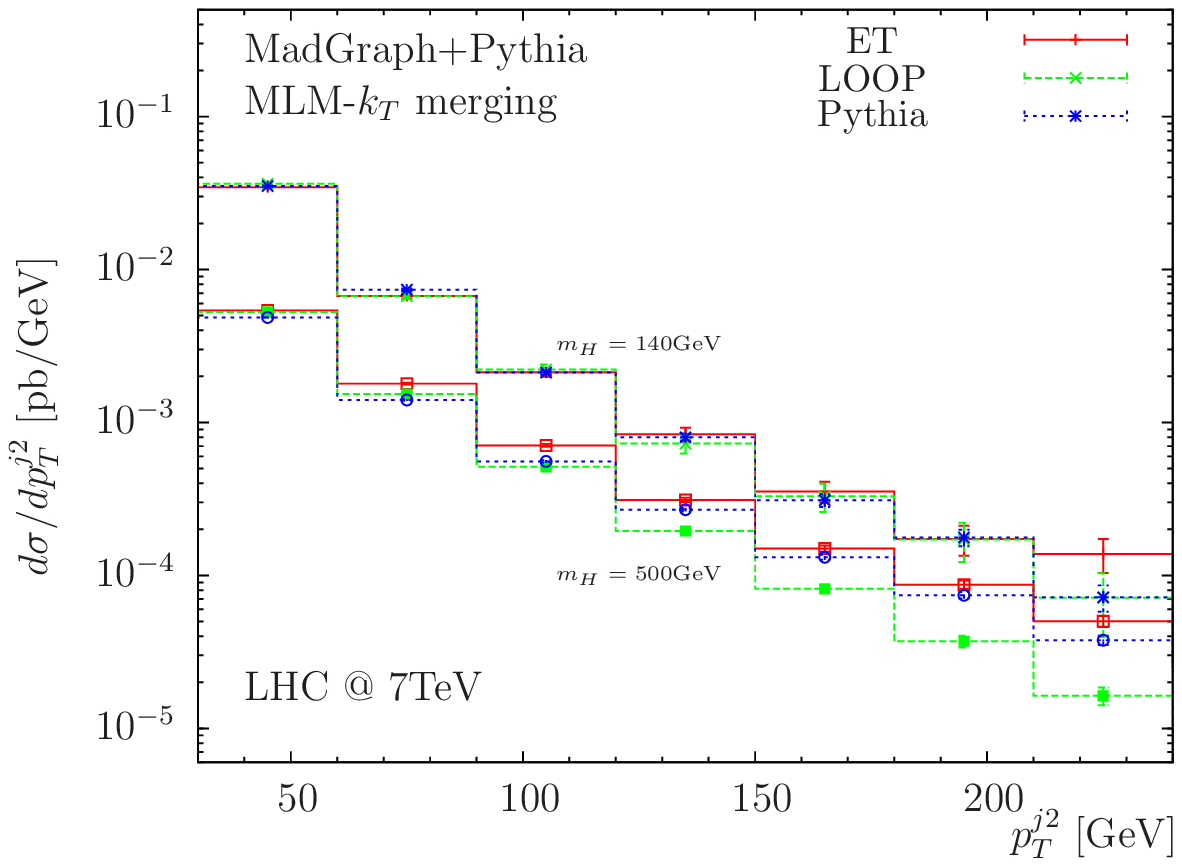}
\caption{Jet $\pT$ distributions for associated jets in gluon-fusion
  production of 
$\MH=140\UGeV$ and  $\MH=500\UGeV$ Higgs bosons at $7\UTeV$ LHC. Plots from \Bref{Alwall:2011cy}.} 
\label{ptj}
\end{figure}

\begin{figure}
\begin{center}
\includegraphics[width=0.48\textwidth]{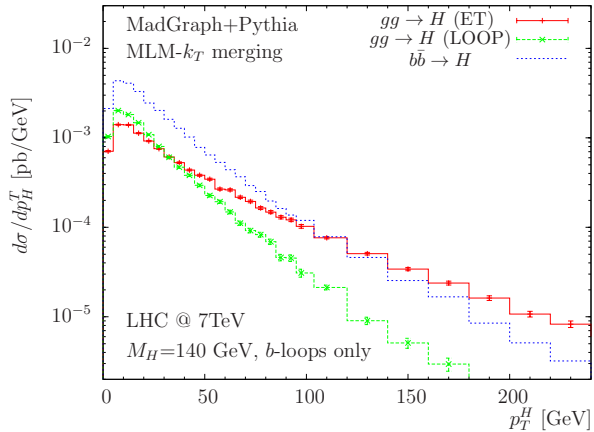}
\end{center}
\caption{$\PQb$-philic Higgs  $\pT$ distribution a the Tevatron and the LHC with $\MH=140\UGeV$.   Results in the ET approximation (red curve) 
and with full loop dependence (green) are shown.  Spectrum of  Higgs produced via $\PQb\PAQb$ fusion in the five-flavour scheme is also shown.  All samples are matrix-element matched with up to two partons in the final state.  Curves normalised to the corresponding NNLO total cross sections see \Bref{Alwall:2011cy}.}
\label{hpt2}
\end{figure}

Representative results for SM Higgs and a $\PQb$-philic Higgs at the LHC at $7\UTeV$ 
are shown in \refFs{hpt1}--\ref{hpt2}. Jets are defined via the $k_{\mathrm{T}}$ algorithm with resolution parameter set to $D = 1$. 
Jets are required to satisfy $|\eta_{j}|< 4.5$ and $\pT^{j}>30\,\UGeV$.
For sake of simplicity, we adopt Yukawa couplings corresponding to the
pole masses, \ie  for the top quark  $\Mt=173\UGeV$ and for the bottom-quark mass $\Mb=4.6\UGeV$. Other quark masses are neglected. 
Throughout our calculation, we adopt the CTEQ6L1 parton distribution functions
(PDFs)~\cite{Pumplin:2002vw} with the core process renormalisation and
factorisation scales $\muR=\muF=M_{\mathrm{T}}^{\PH}\equiv\sqrt{(\pT^{\PH})^2+\MH^2}$.
For the merging performed in {\sc MadGraph/MadEvent}, the  $k_{\mathrm{T}}$-MLM scheme is chosen, with $Q^{\mathrm{ME}}_{\mathrm{min}}=30\UGeV$ and $Q^{\mathrm{jet}}_{\mathrm{min}}=50\UGeV$.

In \refF{hpt1} we show the Higgs $\pT$ distribution for Standard Model Higgs gluon--gluon fusion production at
the LHC with $\MH=140\UGeV$ in a range of $\pT$ relevant for experimental analysis.  
We compare matched results in the ET theory and in the full theory
(LOOP) with \pythia\ with $2\to 2$ matrix element corrections.  We also include the predictions from the analytic computation
at NNLO+NNLL as obtained by {\sc HqT}~\cite{Bozzi:2005wk,deFlorian:2011xf}. The curves are all normalised
to the NNLO+NNLL predictions.  The three Monte Carlo based predictions agree
very well  in all the shown range of $\pT$, suggesting that for this observable,
higher-multiplicity matrix-element corrections (starting from $2\to3$) and
loop effects are not important. This is the case also for jet $\pT$ distributions (not shown) in
the same kinematical range. 
 

In \refFs{hpt1}--\ref{ptj}, the Higgs and jet $\pT$
distributions are shown for Standard Model Higgs gluon--gluon fusion production at
the $7\UTeV$ LHC with $\MH=140$ and $500\UGeV$. 
Monte Carlo based results agree well with each other. As expected, loop effects
show a softening of the Higgs $\pT$, but only at quite high $\pT$. 
We also see that the heavier the Higgs, the more important are the loop
effects.  This is expected, since the heavy Higgs boson can probe the internal structure of 
the top-quark loop already at small $\pT$. The jet $\pT$ distributions do confirm the overall
picture and again indicate loop effects to become relevant only for rather high values
of the $\pT$. The agreement with the NNLO+NNLL predictions at small $\pT$ for
both Higgs masses it is quite remarkable. Key distributions, such
as the $\pT$ of the Higgs, do agree remarkably well with the best available
predictions, for example NNLO+NNLL at small Higgs $\pT$, and offer
improved and easy-to-use predictions for other key observables such as the jet 
rates and distributions. In addition, for heavy Higgs masses and/or large $\pT$, 
loop effects, even though marginal  for phenomenology,  can also be taken 
into account in the same approach, if needed.

In \refF{hpt2},  the $\pT^{\PH}$ distributions for gluon--gluon fusion
production at the $7\UTeV$ LHC of a $\PQb$-philic Higgs with $\MH=140\UGeV$ are shown.
For the sake of illustration, we define a  simplified scenario where the Higgs coupling to the top quark is set to zero and study the Higgs and jet distributions relative to a ``large-$\tan\beta$" scenario with  bottom-quark loops dominating. Note that for simplicity we keep the same normalisation 
as in the Standard Model, i.e., $y_{\PQb}/\sqrt{2} = \Mb/v$ with $\Mb=4.6\UGeV$, as the corresponding cross sections
in enhanced scenarios can be easily obtained by rescaling.  In the $\PQb$-philic Higgs production, the particle running in the loop is nearly massless, and there is no region in $\MH$ or $\pT$ where an effective description is valid. This
also means that a parton-shower generator  alone has no possibility of correctly describing the effects of jet radiation, and genuine loop matrix elements
plus a matched description are needed for achieving reliable simulations. For a $\PQb$-philic Higgs the largest production cross section does not come  from loop induced gluon fusion, but from tree-level $\PQb\PAQb$ fusion. We have therefore included also this production mechanism in \refF{hpt2}. The corresponding histogram is obtained by  merging tree-level matrix elements 
for  $\PH +0,1,2$ partons  (with a $\Ph\PQb\bar\PQb$ vertex)  in the five-flavour scheme to the parton shower.  This provides  a complete  and consistent event 
simulation of inclusive Higgs production in a $\PQb$-philic (or large $\tan \beta$) scenario. 
 
\subsubsection{Finite-quark-mass effects in the gluon-fusion process in POWHEG\footnote{E. Bagnaschi, G. Degrassi, P. Slavich and A. Vicini.}}

Ref.~\cite{Bagnaschi:2011tu} presented an upgraded version of
the code that allows to compute the gluon-fusion cross section in the SM
and in the MSSM.  
The SM simulation is based on matrix elements evaluated retaining the exact top
and bottom mass dependence at NLO QCD and NLO EW.
The MSSM simulation is based on matrix elements in which the exact dependence on
the masses of quarks, squarks, and the Higgs boson has been
retained in the LO amplitude, in the one-loop diagrams with
real-parton emission and in the two-loop diagrams with quarks and
gluons, whereas the approximation of vanishing Higgs mass has been
employed in the two-loop diagrams involving superpartners.
The leading NLO EW effects have also been included in the evaluation of the
matrix elements.
Results obtained with this new version of {\sc POWHEG} are presented
in the SM and in the MSSM  sections of this Report.

The code provides a complete description of on-shell Higgs production,
including NLO QCD corrections matched with a QCD parton shower and
NLO EW corrections.  In the examples discussed in
\Bref{Bagnaschi:2011tu}, the combination {\sc POWHEG+PYTHIA}
has been considered.  In the MSSM case, the relevant parameters of the
MSSM Lagrangian can be passed to the code via a Susy Les Houches
Accord spectrum file.

The code is available from the authors upon request. A release of the
code that includes all the Higgs decay channels is in preparation.

\subsection{Scale and PDF uncertainties in MC@NLO}
\label{NLOPSsec:aMCatNLO}

Computations of (differential) cross sections in hadron--hadron collisions are affected by uncertainties
due to, from the one hand, the unperfected knowledge of parameters in the calculations, both perturbative and non-perturbative nature,
and on the other hand due to truncation errors, i.e.\ to unknown higher-order terms in the perturbative expansion in $\alphas$. 
Among all such uncertainties, scale and PDF ones do have a special status: their variations are typically 
associated with the purely theoretical uncertainty affecting observable predictions.
It is customary to estimate the truncation uncertainty by using the dependence on  renormalisation 
($\muR$) and factorisation ($\muF$) scales. The reason is that such dependence also arises 
because the computation of the cross section is performed only to a given order, and therefore
is thought to give an rough estimate of the possible impact of the unknown terms. 
PDF uncertainties, on the other hand, are evaluated following the directions of the various
PDF groups (see \refS{se:PDFs}) and it amounts to calculating the cross section multiple times 
on a well-definite ensemble of such functions. 
There is, however, something else that make such uncertainties special and particularly
important in complex and therefore time expensive NLO computations: the bulk of the CPU cost of NLO computations can 
be rendered independent of scales and PDFs, as opposed to what happens in the case of other parameters, 
e.g.\ particle masses.  Short-distance cross sections can be written
as linear combinations of scale- and PDF-dependent terms, with coefficients
independent of both scales and PDFs; it is thus possible to compute such
coefficients once and for all, and to combine them at a later stage with
different scales and PDFs at essentially zero cost from the CPU viewpoint.
The crucial point is that this is not  only the feature of the parton-level LO and NLO cross sections,
but also of those performed in the context of MC@NLO.  This implies that from the conceptual point of
view the same procedure for determining scale and PDF uncertainties can be
adopted in MC@NLO as in LO-based Monte Carlo simulation.
This method is automatic, process independent, and being now implemented in {\sc aMCatNLO}~\cite{Frederix:2011ss}, it can
provide scale and PDF uncertainties for an arbitrary process. 
 
In practice parton-level events in the LHE format, soft and hard, are generated as usual through {\sc aMCatNLO} for a central choice of
scales and PDF. After that the weights corresponding to different PDF sets and arbitrary scale choices (that can be decided by the user) 
are determined and stored in ordered arrays associated to each event record.  For any observable, obtained with events after showering and hadronisation and possibly detector simulation,  one can fill a ``central'' histogram with the central weights and as many ``variation''  histograms corresponding to the other weights. This procedure very closely reproduces what is normally done for pure parton-level NLO computations and should be considered
the equivalent one in the context of NLO+PS.

\subsection{On-going debates and open issues}\label{NLOPSsec:controversy}

We summarise here an issue that has recently been reopened for discussion.  
Since this has only quite recently emerged, the present summary is a snapshot 
of the status of the discussion at the time of completion of this Report;
we anticipate a rapid evolution in the following months.  

The debate mainly concerns the origin of the large discrepancies between 
results for the large-transverse-momentum tail of the Higgs-boson $\pT$
distribution obtained with \MCatNLO{} and \POWHEG{} methods, but it also 
addresses differences observed in the rapidity spectrum of the jet in Higgs 
production.

\begin{figure}
\centering
\includegraphics[height=\figfact\linewidth]{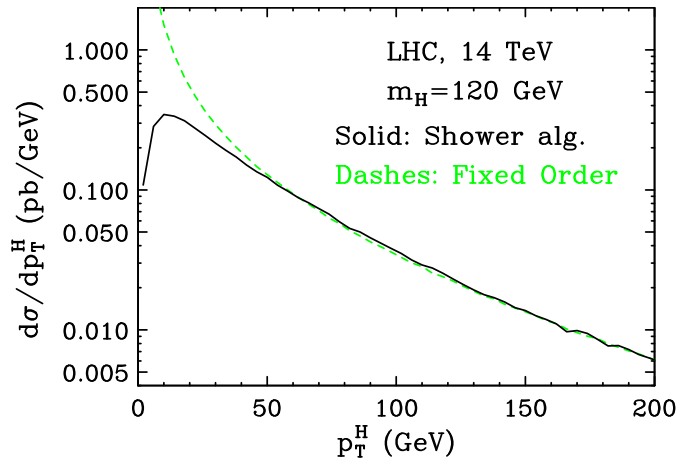}
\caption{
        Illustration of the difference in the Higgs $\pT$ distribution
        computed at fixed order $\alphas^3$ with respect to its generation
        using the shower algorithm.
}
\label{figNLOMC:LOandshower}
\end{figure}

One view for an explanation of these discrepancies has been 
developed in \Brefs{Alioli:2008tz,Nason:2010ap}.  In the following, this 
explanation will be called the ``$K$-factor effect'' and it can be summarised as 
follows: NLO+PS generators produce two kinds of events that are merged in 
their output: $\mathbb{S}$ (for shower) and $\mathbb{F}$ (for finite) events.  
The real-emission cross section is decomposed into these two types of events.  
In the specific case of $\Pg \Pg\to \PH$ the real emission cross section 
$R$ is essentially given by the $\Pg \Pg\to \PH \Pg$ cross section, and 
the  $\Pg \Pq$ and $\PQq\PAQq$ channels can be neglected.  
This cross section is decomposed as follows: $R=\Rups+\Rupf$.  While 
$\Rupf$ has a finite integral, and thus  yields a finite contribution to the real 
cross section, $\Rups$ is divergent,  and its contribution is well defined only 
when suitably combined with the (also divergent) virtual cross section.  
The shower algorithm performs this 
combination, accounting only for leading logarithmic corrections (i.e.\ 
without including all NLO corrections.).  The distribution of events 
generated from $\Rups$ using the shower algorithm is altered with respect to 
the fixed order one, as shown in \refF{figNLOMC:LOandshower}.
The divergence at $\pT\to 0$ of $\Rups$ is turned into a smooth, 
bell--\-shaped curve, characterizing the so--called Sudakov region. The basic 
shower algorithm is such that the integral of this distribution equals the 
Born cross section.  In contrast, the $\Rupf$ contribution can be generated 
with no particular problems, since it is finite.

When promoting the shower algorithm to the NLO level, one must make sure
that the full output yields an NLO accurate cross section.  A contribution
to the NLO cross section comes from the $\Rupf$ term, but also modifications
of the $\Rups$ term are needed. These modifications induce a change in the
transverse-momentum spectrum generated by the $\mathbb{S}$ events, that 
amounts roughly to an overall factor
\begin{equation}\label{eqNLO:SKfact}
K=\frac{\sigma^{({\mathrm{NLO}})}-\sigma^{F}}{\sigma^{({\mathrm{LO}})}}\,,
\end{equation}
so that summing the $\mathbb{S}$ and $\mathbb{F}$ integrals of the 
transverse-momentum spectrum yields the total NLO cross section.

In \MCatNLO{}, the $\Rups$ contribution corresponds to the shower Monte Carlo
approximation to the real-emission cross section. In \POWHEG{}, $\Rups=R$ by
default, although, as we have seen earlier, other choices are possible.
According to \Brefs{Alioli:2008tz,Nason:2010ap}, it is the $K$-factor
of Eq.~\refE{eqNLO:SKfact} that determines the larger high $\pT$ tail of the
Higgs spectrum to be higher in \POWHEG{} than in \MCatNLO{}, since in the former
$\Rupf=0$, and $K$ is in front of the whole distribution. On the other hand,
in \MCatNLO{}, the $K$-factor is only in front  of the $\Rups$ contribution,
that dominates the region of $\pT$ below $\MH$, while $\Rupf$ dominates at
large $\pT$. Since the $K$-factor in Higgs production is particularly large,
this effect is more evident here than in other processes.

We observe that the $K$-factor effect amounts to a correction of 
${\cal O}(\alphas^4)$ to the tail of the transverse-momentum distribution, which 
here is computed at order $\alphas^3$. In the \POWHEGBOX{} implementation of Higgs 
production, the partition $R=\Rups+\Rupf$ can be tuned by setting an input 
parameter, $h$ of Eq.~\eqref{eqNLOMC:SFsep}, and thus the same large variation 
in the transverse-momentum tail can be observed within the \POWHEG{} framework 
alone.

A discording view is expressed in \Bref{Hoeche:2011fd}, which we will refer to 
as ``phase-space effect''.  In this recent publication, a new  
\MCatNLO-inspired implementation in the \SHERPA{} framework is introduced, 
with an $\Rups$ contribution that is a tunable function of one parameter,
called $\alpha$ \cite{Nagy:2003tz}.  This parameter $\alpha$ controls the 
phase space available for radiation, such that $\alpha = 1$ corresponds to 
the full phase space available, and smaller values correspond to restricted
phase spaces.  It should be stressed here that this parameter does not directly
translate into a scale with a clear relation to transverse momenta.  By 
varying this parameter, not only the phase space available to the subtraction 
terms -- the Catani--Seymour dipoles -- but also the phase space available for 
$\mathbb{S}$ events in the parton shower can be varied.  This is because in 
the \SHERPA{} implementation, the same kernels are employed for both 
subtraction and showering, while in the original \MCatNLO implementations 
this is not the case.  There the phase space for soft events is directly 
obtained from the parton shower, and typically related to scales
of the order of the factorisation scale.  The variation in \SHERPA{}, in 
contrast, does not exhibit this feature, which therefore ultimately results 
in large, unphysical variations in the tail of the transverse-momentum 
distribution within the \SHERPA{} implementation of \MCatNLO{}.  However,
this lead the authors of \Bref{Hoeche:2011fd} to argue that the dominant
mechanism of large deviations is not the $K$-factor, as claimed in 
\Brefs{Alioli:2008tz,Nason:2010ap}, but the additional phase space beyond
the factorisation scale.  According to the authors of \Bref{Hoeche:2011fd}, 
this phase-space effect manifests itself in a distortion of the Sudakov form 
factor, affecting $\mathbb{S}$ and, to a lesser extent, $\mathbb{F}$ events.

Phrasing it slightly differently, for $\alpha=1$, the phase space for radiation
in the $\mathbb{S}$ events allows the generation of transverse momenta up to 
the order of the available hadronic energy.  This is also the case for the 
default \POWHEG{} implementation without the additional dampening factor of
Eq.~\eqref{eqNLOMC:SFsep}.  It can be argued that, in addition to the 
in-principle uncontrollable higher-order terms introduced at ${\cal O}(\alphas^4)$ 
through the above discussed $K$-factor, also large $\log^2 (S_{\mathrm had}/\pT^2)$ 
arise to \emph{all} orders in perturbation theory -- instead of terms like
$\log^2 (\muF^2/\pT^2)$ present in standard resummation implemented, e.g.\
in \HqT\footnote{
   Note that for LHC energies, the ratio $S_{\mathrm had}/\muF^2$ is of the 
   same order of magnitude as $\muF^2/\Lambda_\text{QCD}^2$ 
   (for $\muF\sim 100\UGeV$).}.
It is argued that if the $\alpha$ parameter is set to values smaller than $1$, 
a maximum $\pT$ for $\mathbb{S}$ events, $\pT^\text{max}(\alpha)$, is 
introduced. Thus, the $\log^2 (S_{\mathrm{had}}/\pT^2)$ terms become in fact of order 
$\log^2 ({\pT^\text{max}}^2(\alpha)/\pT^2)$.  Values of $\alpha$ such that
$\pT^\text{max}(\alpha)$ becomes of the order of the Higgs mass are thus
justified, and yield differential cross sections that reproduce the NLO results 
for large $\pT$ and that also get rid of the undesired effects in the rapidity
distribution of the radiated jet in the \SHERPA{} implementation of the 
\MCatNLO algorithm.  Too small values of $\alpha$ yield a transverse-momentum 
distribution that approaches the fixed order one, implying that a negative 
dip must arise at low $\pT$ in the $\pT$ distribution, in order for the 
total cross section to remain finite and independent of $\alpha$.

Let us note in passing that in this framework, the notorious dip in the 
rapidity difference of Higgs boson and hardest jet can also be reproduced,
for some values of $\alpha$.  The authors of \Bref{Hoeche:2011fd} argue that
this hints at the dip originating from unaccessible zones in the phase space
for $\mathbb{S}$ events, which are not fully recovered by $\mathbb{F}$ events.

\clearpage


\newpage
\newcommand{\orde}[1]{\mathcal{O}(#1)}
\newcommand{\total}{\mathrm{total}}
\newcommand{\df}{\mathrm{d}}
\newcommand{\jet}{\mathrm{jet}}
\newcommand{\eqr}[1]{Eq.~\eqref{eq:#1}}
\newcommand{\eqs}[2]{Eqs.~\eqref{eq:#1} and \eqref{eq:#2}}
\newcommand{\pb}{\,\mathrm{pb}}
\newcommand{\GeV}{\,\mathrm{GeV}}
\newcommand{\TeV}{\,\mathrm{TeV}}
\newcommand{\abs}[1]{\lvert#1\rvert}
\def\ltap{\raisebox{-.6ex}{\rlap{$\,\sim\,$}}\raisebox{.4ex}{$\,<\,$}}
\def\gtap{\raisebox{-.4ex}{\rlap{$\,\sim\,$}} \raisebox{.4ex}{$\,>\,$}}

\newcommand{\nb}{\,\mathrm{nb}}
\newcommand{\incl}{\mathrm{incl}}
\newcommand{\excl}{\mathrm{excl}}
\newcommand{\Tcm}{{\mathcal{T}_\mathrm{cm}}}
\newcommand{\Tcmc}{\mathcal{T}_\mathrm{cm}^\mathrm{cut}}

\section{The gluon-fusion process\footnote{M.~Grazzini, F.~Petriello (eds.); E.A.~Bagnaschi, A.~Banfi, D.~de~Florian,  G.~Degrassi, G.~Ferrera, G.~P.~Salam, P.~Slavich, I.~W.~Stewart, F.~Stoeckli, F.~J.~Tackmann, D.~Tommasini, A.~Vicini, W.~J.~Waalewijn and G.~Zanderighi.}}


In the first volume of this Handbook, the status of the inclusive cross section for Higgs production in gluon fusion was summarised~\cite{Dittmaier:2011ti}.  Corrections arising from higher-order QCD, electroweak effects, as well as contributions beyond the commonly used effective-theory approximation  were analysed.  Uncertainties arising from missing terms in the perturbative expansion, as well as imprecise knowledge of parton distribution functions, were estimated to range from approximately $15\%$ to $20\%$, with the precise value depending on the Higgs-boson mass.

Our goal in this second volume is to extend the previous study of the gluon-fusion mechanism to include differential cross sections.  The motivation for such investigations is clear; experimental analyses must impose cuts on the final state in order to extract the signal from background.  A precise determination of the corresponding signal acceptance is needed.  Hopefully, it is computed to the same order in the perturbative expansion as the inclusive cross section.  Possible large logarithms that can occur when cuts are imposed should be identified and controlled.  In the case of Higgs production through gluon fusion, numerous tools exist to perform a precise theoretical calculation of the acceptance.  The fully differential Higgs cross section to next-to-next-to-leading order (NNLO) in QCD is available in two parton-level Monte Carlo simulation codes, {\sc FEHiP}~\cite{Anastasiou:2004xq,Anastasiou:2005qj} and {\sc HNNLO}~\cite{Catani:2007vq,Grazzini:2008tf}.  The Higgs transverse-momentum spectrum has been studied with next-to-leading order (NLO) accuracy \cite{deFlorian:1999zd,Ravindran:2002dc,Glosser:2002gm}, and supplemented with next-to-next-to-leading logarithmic (NNLL) resummation of small-$\pT$ logarithms \cite{Bozzi:2005wk,Cao:2009md,deFlorian:2011xf}.  Other calculations and programs for studying Higgs production in the gluon-fusion channel are available, as discussed throughout this section.  All of these results are actively used by the experimental community.

We aim in this section to discuss several issues that arise when studying differential results in the gluon-fusion channel which are relevant to LHC phenomenology.  A short summary of the contents of this section is presented below.
\begin{itemize}

\item The amplitude for Higgs production through gluon fusion begins at one loop.  An exact NNLO calculation of the cross section would therefore require a multi-scale, three-loop calculation.  Instead, an effective-field-theory approach is utilised, which is valid for relatively light Higgs masses.  The validity of this effective-theory approach has been extensively studied for the inclusive cross section, and was reviewed in the first volume of this Handbook~\cite{Dittmaier:2011ti}.  However, the accuracy of the effective-theory approach must also be established for differential distributions.  For example, finite-top-mass corrections of order $(\pT^{\mathrm{H}}/\Mt)^2$ can appear beyond the effective theory.  It is also clear that bottom-quark mass effects on Higgs differential distributions cannot be accurately modeled within the effective theory.  These issues are investigated in \refS{sec:finitemass} of this report.  
A distortion of up to $\orde{10\%}$ of the Higgs transverse-momentum spectrum 
is possible in the low-$\pT$ region,
due to the bottom effects in case of a light Higgs,
while in the high-$\pT$ region the finite top mass can induce 
an even  larger modification of the shape of the distribution~\cite{Bagnaschi:2011tu}.

\item The composition of the background to Higgs production in gluon fusion, followed by the decay $\PH \to \PW^+\PW^-$, differs dramatically depending on how many jets are observed together with the Higgs in the final state.  In the zero-jet bin, the dominant background comes from continuum production of $\PW^+\PW^-$.  In the 1- and 2-jet bins, top-quark pair production becomes large.  The optimisation of the experimental analysis therefore utilises a split into different jet-multiplicity bins.  Such a split induces large logarithms associated with the ratio of the Higgs mass over the defining $\pT$ of the jet.  The most natural method of evaluating the theoretical uncertainty in the zero-jet bin, that of performing naive scale variations, leads to a smaller estimated uncertainty than the error on the inclusive cross section \cite{Anastasiou:2007mz,Grazzini:2008tf,Anastasiou:2009bt,Stewart:2011cf}.  This indicates a possible cancellation between logarithms of the $\pT$ cut and the large corrections to the Higgs cross section, and a potential underestimate of the theoretical uncertainty.  An improved prescription for estimating the perturbative uncertainty in the exclusive jet bins is discussed in \refS{sec:jetveto}.  Evidence is given that the cancellation suggested above does indeed occur, and that the uncertainty in the zero-jet bin is in fact nearly twice as large as the estimated error of the inclusive cross section.  Further evidence for the accidental cancellation between large Sudakov logarithms and corrections to the total cross section is given in \refS{sec:vetosum}, where it is shown that different prescriptions for treating the uncontrolled $\orde{\alphas^3}$ corrections to the zero-jet event fraction lead to widely varying predictions at the LHC. 

\item A significant hindrance in our modeling of the cross section in exclusive jet bins is our inability to directly resum the large logarithms mentioned above.  Instead, we must obtain insight into the all-orders structure of the jet-vetoed cross section using related observables.  An example of such a quantity is the Higgs transverse-momentum spectrum.  Through $\orde{\alphas}$ it is identical to the jet-vetoed cross section, since at this order the Higgs can only recoil against a single parton whose $\pT$ matches that of the Higgs.  The large logarithms which arise when the Higgs transverse momentum becomes small can be resummed.  \refS{sec:hqt}
presents a resummed computation of the $\pT$ spectrum at full NNLL accuracy matched to the NLO result valid at large $\pT$, which is implemented
in an updated version of the code {\sc HqT} \cite{deFlorian:2011xf}.
Although the large logarithmic terms that are resummed are not the same
appearing in the jet-bin cross sections,
a study of the uncertainties of the {\em shape} of the resummed
$\pT$ spectrum may help to quantify those uncertainties.  Another variable used to gain insight into the effect of a jet veto is beam thrust, which is equivalent to a rapidity-weighted version of the standard variable $H_{\mathrm{T}}$.  The resummation of beam thrust~\cite{Stewart:2009yx,Berger:2010xi}, and the reweighting of the Monte Carlo program {\sc MC@NLO} to the NNLL+NNLO prediction for beam thrust in order to gain insight into how well the jet-vetoed cross section is predicted by currently used programs, is described in \refS{sec:jetvetosum}.

\end{itemize}
Unlike the first volume of this report, no explicit numbers are given for use in experimental studies.  The possible distributions of interest in the myriad studies performed at the LHC are too diverse to compile here.  Our goal is to identify and discuss the relevant issues that arise, and to give prescriptions for their solution.  We must also stress that such prescriptions are not set in stone, and are subject to change if theoretical advances occur.  For example, if an exact NNLO calculation of the Higgs cross section with the full dependence on the quark masses becomes available, or the jet-vetoed Higgs cross section is directly resummed, these new results must be incorporated into the relevant experimental investigations.


\noindent
\subsection{Finite-quark-mass effects in the SM gluon-fusion process
  in {\sc POWHEG}\footnote{E.A. Bagnaschi, G. Degrassi, P. Slavich and A. Vicini.}}
\label{sec:finitemass}

\subsubsection{Introduction}
\label{sec:gghtb-intro}
%
The description of the gluon-fusion process can be
approximated, in many cases, by an effective theory (ET) obtained by
taking the limit of infinite mass for the top quark running in the
lowest-order loop that couples the Higgs to the gluons, and neglecting
all the other quark flavors.  This limit greatly simplifies many
calculations, reducing the number of loops that have to be considered
by one.  On the other hand it is important, when possible, to check
whether the effect of an exact treatment of the quark contributions is
significant compared to the actual size of the theoretical
uncertainty.

The validity of the ET approach has been carefully analyzed for the
total cross section. The latter receives very large 
NLO QCD corrections, which have been computed first in the ET
\cite{Dawson:1990zj,Djouadi:1991tka} and then retaining the exact
dependence on the masses of the quarks that run in the loops
\cite{Spira:1995rr,Anastasiou:2006hc,Harlander:2005rq,
  Aglietti:2006tp,Bonciani:2007ex}.  The NLO results computed in the
ET and then rescaled by the exact leading-order (LO) result with full
dependence on the top- and bottom-quark masses provide a description
accurate at the few-per-cent level for $\MH<2\, \Mt$. The deviation
for $2\Mt < \MH < 1\UTeV$ does not exceed the $10\%$ level.
The NNLO QCD corrections to the total
cross section are still large and have been computed in the ET
\cite{Harlander:2002wh, Anastasiou:2002yz, Ravindran:2003um}.  The
finite-top-mass effects at NNLO QCD have been studied in
\Brefs{Marzani:2008az, Harlander:2009bw, Harlander:2009mq,
  Harlander:2009my } and found to be small.  The resummation to all
orders of soft-gluon radiation up to NNLL+NNLO has been studied in
\Bref{Catani:2003zt}, within the ET, and including the exact dependence
of the masses of the top and bottom quarks up to NLL+NLO in \Brefs{deFlorian:2009hc,Dittmaier:2011ti}.
The leading third-order
(NNNLO) QCD terms have been discussed in the ET
\cite{Moch:2005ky}.  The role of electroweak (EW) corrections has been
discussed in
\Brefs{Aglietti:2004nj,Degrassi:2004mx,Degrassi:2005mc,Actis:2008ug,Actis:2008uh,Actis:2008ts,Bonciani:2010ms}
and found to be, for a light Higgs, of the same order of magnitude as
the QCD theoretical uncertainty.  The impact of mixed QCD--EW
corrections has been discussed in \Bref{Anastasiou:2008tj}.  The
residual uncertainty on the total cross section depends mainly on the
uncomputed higher-order QCD effects and on the uncertainties that
affect the parton distribution functions of the proton
\cite{Demartin:2010er,Alekhin:2011sk,Dittmaier:2011ti}.

The Higgs differential distributions have been studied at NNLO
QCD in the ET,
first in the region in which the Higgs $\pT$
is non-vanishing \cite{deFlorian:1999zd,Glosser:2002gm,Ravindran:2002dc}
\footnote{These computations provide an NLO QCD calculation
of the Higgs $\pT$ spectrum.},
and then with proper treatment of the $\pT=0$ contribution \cite{Anastasiou:2004xq,Anastasiou:2005qj,Catani:2007vq}.  The
NLO QCD results including the exact dependence on the top and bottom-quark
masses were first included in the code {\sc HIGLU} based on
\Bref{Spira:1995rr}.  More recently, the same calculation was
repeated in \Brefs{Keung:2009bs,Brein:2010xj,Anastasiou:2009kn}.  The latter
discussed at NLO QCD the role of the bottom quark, for light Higgs-boson
masses.  A description of the Higgs differential
distributions, in the ET,
including transverse-momentum resummation up to NNLL
and matching to NLO and NNLO QCD results
has been provided in
\Brefs{Balazs:2000wv,Balazs:2000sz} and \cite{Cao:2009md}, respectively, and has been implemented in the code
{\sc Resbos}. The Higgs transverse-momentum distribution, in the ET,
including full NNLO QCD results and matched at NNLL QCD with the
resummation of soft-gluon emissions, has been presented in
\Bref{Bozzi:2005wk} and is implemented in the code {\sc HqT}.
The latter allows for a quite accurate estimate of the perturbative
uncertainty on this distribution, which turns out to be of the order
of $\pm 10$\% for light Higgs and for Higgs transverse momentum
$\pT^{\mathrm{H}}< 100 \UGeV$ \cite{deFlorian:2011xf}.

The shower Monte Carlo (SMC) codes matching NLO QCD results with QCD
Parton Shower (PS) \cite{Frixione:2002ik,Alioli:2008tz} consider the
gluon-fusion process only in the ET.
Recently, a step towards the inclusion of the finite-quark-mass effects
in PS was taken in \Bref{Alwall:2011cy}, where parton-level events for Higgs production
accompanied by zero, one, or two partons are generated with matrix elements
obtained in the ET and then, before being passed to the PS, they are re-weighted
by the ratio of the  full one-loop amplitudes over the ET ones.
This procedure is equivalent to generating events directly with
the full one-loop amplitudes, yet it is much faster.

In \Bref{Bagnaschi:2011tu} the implementation
in the {\sc POWHEG} framework
\cite{Nason:2004rx,Frixione:2007vw,Alioli:2010xd} of the NLO QCD
matrix elements for the gluon fusion, including the exact dependence
on the top- and bottom-quark masses, has been presented.  We discuss
here the effect of the exact treatment of the quark masses on the
Higgs transverse-momentum distribution, comparing the results of the
old {\sc POWHEG} release, obtained in the ET, with those of this new
implementation \cite{Bagnaschi:2011tu}. In particular, we consider here the
matching of {\sc POWHEG} with the {\sc PYTHIA} \cite{Sjostrand:2006za}
PS.

\subsubsection{Quark mass effects in the {\sc POWHEG} framework}
\label{sec:gghtb-massffects}

In this section we briefly discuss the implementation of the
gluon-fusion Higgs production process in the {\sc POWHEG
  BOX}\ framework, following closely \Ref~\cite{Alioli:2008tz}. We fix
the notation keeping the discussion at a general level, and refer to
\Bref{Bagnaschi:2011tu} for a detailed description and for the explicit
expressions of the matrix elements.  The generation of the hardest
emission is done in {\sc POWHEG} according to the following formula:
\begin{eqnarray}
\label{eq:gghtb-POWHEG}
d\sigma &=& \bar{B}(\bar{\Phi}_1)\, d \bar{\Phi}_1
 \left\{ \Delta\left(\bar{\Phi}_1,\pT^{\mathrm{min}}\right)
+
\Delta\left(\bar{\Phi}_1, \pT \right)\, \frac{
 R\left(\bar{\Phi}_1,\Phi_{\rm rad} \right)}{ B\left(\bar{
    \Phi}_1\right)} \, d\Phi_{\rm rad} \right\} 
\nonumber\\ 
&+& \sum_q R_{q \bar q}\left(\bar{
  \Phi}_1,\Phi_{\rm rad} \right) d \bar{\Phi}_1 d\Phi_{\rm rad} ~.
\end{eqnarray}
In the equation above the variables $\bar{\Phi}_1 \equiv (M^2,Y)$
denote the invariant mass squared and the rapidity of the Higgs boson,
which describe the kinematics of the Born (i.e., lowest-order) process
$\Pg\Pg\rightarrow\phi$.  The variables $\Phi_{\rm rad}$ describe the
kinematics of the additional final-state parton in the real emission
processes.
The factor $\bar{B}(\bar{\Phi}_1)$ in Eq.~(\ref{eq:gghtb-POWHEG}) is
related to the total cross section computed at NLO in QCD.  It
contains the value of the differential cross section, including real
and virtual radiative corrections, for a given configuration of the
Born final-state variables, integrated over the radiation variables.
The integral of this quantity on $d\bar{\Phi}_1$, without acceptance
cuts, yields the total cross section and is responsible for the
correct NLO QCD normalisation of the result.
The terms within curly brackets in Eq.~(\ref{eq:gghtb-POWHEG})
describe the real emission spectrum of an extra parton: the first term
is the probability of not emitting any parton with transverse momentum
larger than a cutoff $\pT^{\mathrm{min}}$, while the second term is the
probability of not emitting any parton with transverse momentum larger
than a given value $\pT$ times the probability of emitting a parton
with transverse momentum equal to $\pT$. The sum of the two terms
fully describes the probability of having either zero or one
additional parton in the final state. The probability of non-emission
of a parton with transverse momentum $\kT$ larger than $\pT$ is
obtained using the {\sc POWHEG} Sudakov form factor
\begin{equation}
\Delta(\bar \Phi_1,\pT)=
\exp
\left\{
-\int d\Phi_{\rm rad}
\frac{R(\bar{\Phi}_1,\Phi_{\rm rad})}{B(\bar{\Phi}_1)}
\theta(k_{\mathrm{T}}-\pT)
\right\}~,
\label{eq:gghtb-Sudakov}
\end{equation}
where the Born squared matrix element is indicated by
$B(\bar{\Phi}_1)$ and the squared matrix element for the real emission
of an additional parton can be written, considering the subprocesses
$\Pg\Pg\to\phi \Pg$ and $\Pg\PQq\to\phi \PQq$, as
\begin{equation}
R(\bar{\Phi}_1,\Phi_{\rm rad}) =
R_{\Pg\PQq}(\bar{\Phi}_1,\Phi_{\rm rad}) + \sum_{\PQq} \left[
R_{\Pg\PQq}(\bar{\Phi}_1,\Phi_{\rm rad}) + R_{\PQq\Pg}(\bar{\Phi}_1,\Phi_{\rm rad}) \right]~.
\label{eq:gghtb-real}
\end{equation}
Finally, the last term in Eq.~(\ref{eq:gghtb-POWHEG}) describes the effect
of the $\PQq \PAQq \to \phi \Pg$ channel, which has been kept apart in the
generation of the first hard emission, because it does not factorise
into the Born cross section times an emission factor.

\begin{figure}[t]
\begin{center}
\includegraphics[height=76mm,angle=0]{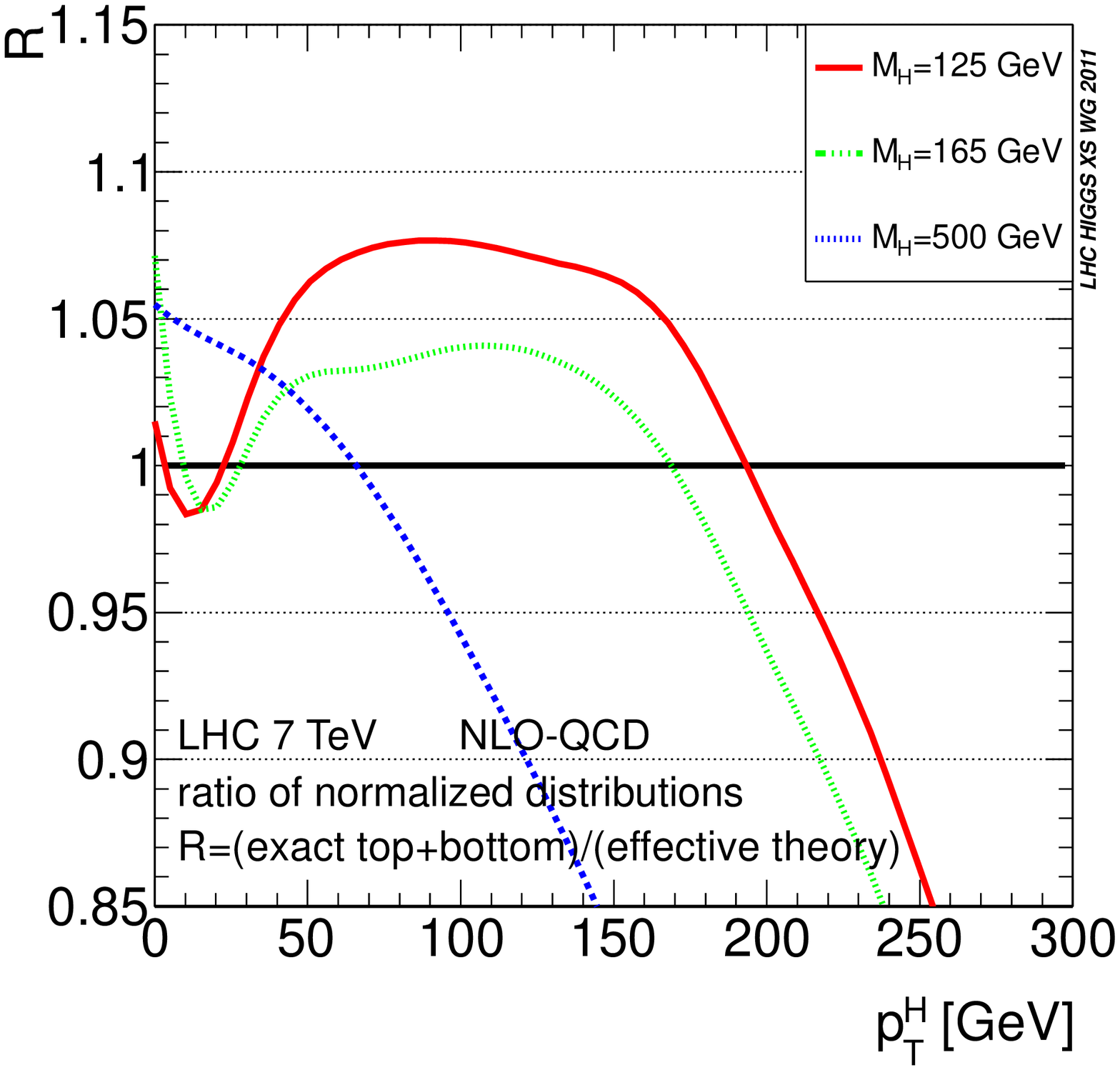}~
\includegraphics[height=76mm,angle=0]{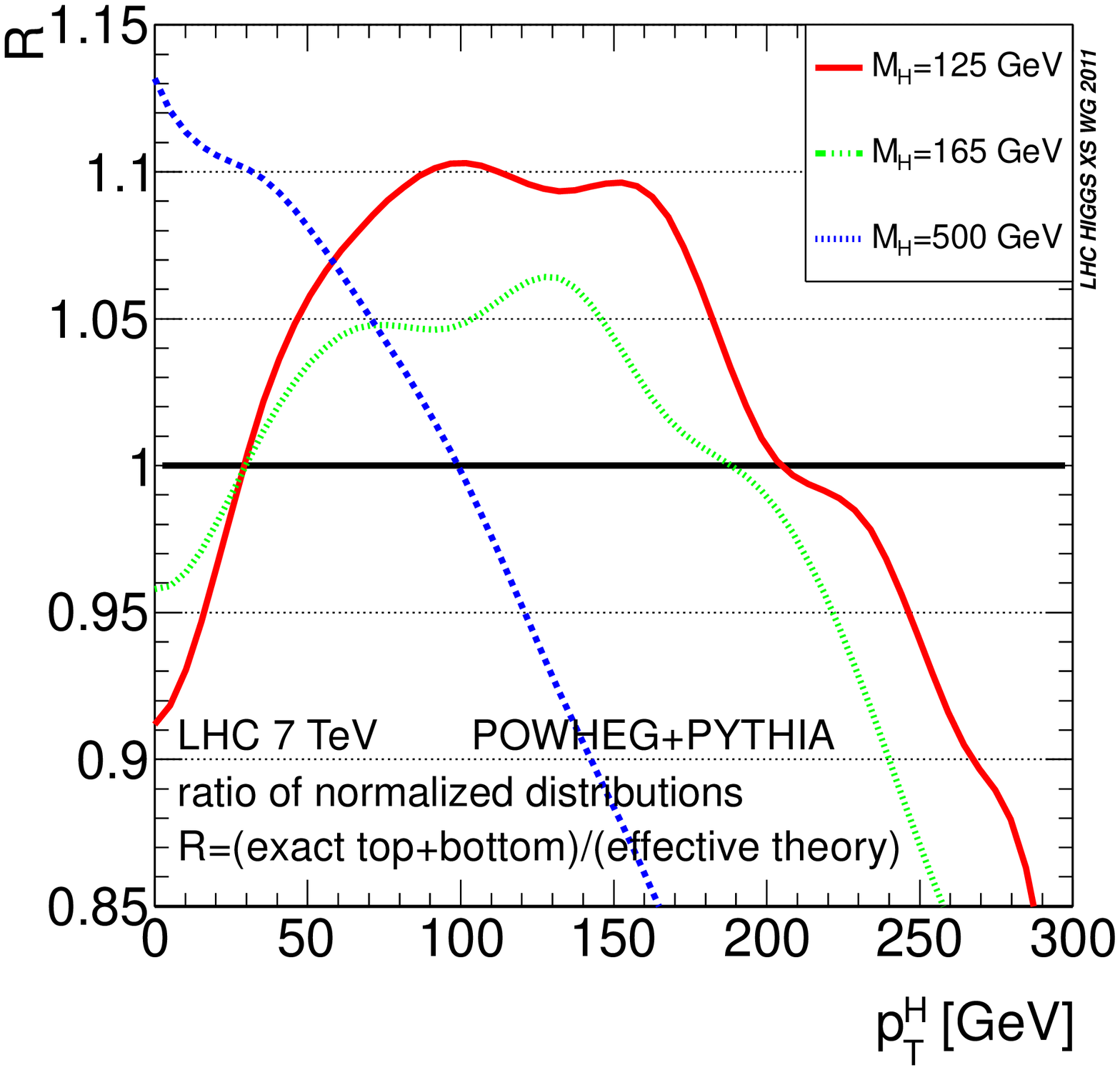}
\end{center}
\vspace{-0.5cm}
\caption{ Ratio, for different values of the SM Higgs-boson mass, of
  the normalised Higgs transverse-momentum distribution computed with
  exact top- and bottom-mass dependence over the one obtained in the
  ET.  Left: ratio of the NLO QCD predictions. Right: ratio of the
  {\sc POWHEG}+{\sc PYTHIA} predictions.}
\label{fig:gghtb-shaperatio}
\end{figure}

The NLO QCD matrix elements used in this implementation have been
computed in \Brefs{Aglietti:2006tp,Bonciani:2007ex}.  We compared
the numerical results for the distributions with those of
the code {\sc FehiPro} \cite{Anastasiou:2006hc}, finding good
agreement.  We also checked that, in the case of a light Higgs and
considering only the top contribution, the ET provides a very good
approximation of the exact result for small values of the Higgs
transverse momentum $\pT^{\mathrm{H}}$, and only when $\pT^{\mathrm{H}}> \Mt$ does a
significant discrepancy appear, due to the fact that the internal
structure of the top-quark loop is resolved.

In order to appreciate the importance of the exact treatment of the
quark masses, we compare in \refF{fig:gghtb-shaperatio} the
normalised distributions computed with the exact top- and bottom-mass
dependence with the corresponding distributions obtained in the
ET. The normalised distributions are defined dividing each
distribution by the corresponding total cross section, and allow a
comparison of the shape of the distributions.  In the left panel of
\refF{fig:gghtb-shaperatio} we plot the ratio of the normalised
$\pT^{\mathrm{H}}$ distributions (exact over ET) computed 
using the real radiation matrix elements that enter in the
NLO QCD calculation,
for different values of the Higgs mass ($\MH=125,\,165,\,500 \UGeV$). The
plot shows that, for a light Higgs, the bottom-quark contribution
induces positive ${\cal O}(10\%)$ corrections in the intermediate
$\pT^{\mathrm{H}}$ range. On the other hand, for a heavy Higgs, the bottom-quark
contribution is negligible, while the exact treatment of the top-quark
contribution tends to enhance the distribution at small $\pT^{\mathrm{H}}$ and
significantly reduce it at large $\pT^{\mathrm{H}}$ (where in any case the cross
section is small).

The matching of the NLO QCD results with a QCD PS is obtained by using
the basic {\sc POWHEG} formula, Eq.~(\ref{eq:gghtb-POWHEG}), for the
first hard emission, and then by vetoing in the PS any emission with a
virtuality larger than the one of the first emission.  The use of the
exact matrix elements in Eq.~(\ref{eq:gghtb-POWHEG}), and in
particular in the {\sc POWHEG} Sudakov form factor,
Eq.\,(\ref{eq:gghtb-Sudakov}), has an important impact on the $\pT^{\mathrm{H}}$
distribution when compared with the distribution obtained in the ET.
Indeed, one should observe that the {\sc POWHEG} Sudakov form factor
is a process-dependent quantity. For a given transverse momentum
$\pT$, in the exponent we find the integral of the ratio of the full
squared matrix elements $R/B$ over all the transverse momenta $k_{\mathrm{T}}$
larger than $\pT$.

The results of the matching of {\sc POWHEG} with the {\sc PYTHIA} PS
are illustrated in the right plot of \refF{fig:gghtb-shaperatio},
where we show the ratio of the normalised $\pT^{\mathrm{H}}$ distribution
computed with exact top- and bottom-mass dependence over the
corresponding distribution computed in the ET, for the same values of
$\MH$ as in the left plot. We shall first discuss the case of a light
Higgs boson, with $\MH=125 \UGeV$. For small $\pT^{\mathrm{H}}$, the Sudakov form
factor with exact top- and bottom-quark mass dependence
$\Delta(\Pt+\Pb,\mathrm{exact})$ is smaller than the corresponding factor
$\Delta(\Pt,\infty)$ with only top in the ET, 
because we have that
$R(\Pt+\Pb,\mathrm{exact})/B(\Pt+\Pb,\mathrm{exact})>R(\Pt,\infty)/B(\Pt,\infty)$.  
The inequality holds for two reasons:
$i)$ the $\pT^{\mathrm{H}}$ distribution is proportional to the squared real matrix element $R$
and, for $\pT^{\mathrm{H}}<200 \UGeV$, $R(\Pt+\Pb,\mathrm{exact})>R(\Pt,\infty)$, 
as  has been discussed in 
\Brefs{Keung:2009bs,Anastasiou:2009kn,Bagnaschi:2011tu};
$ii)$ the bottom quark reduces the LO cross section, with respect 
 to the case with only the top quark in the ET 
\cite{Spira:1995rr}.
Thus, for small
$\pT^{\mathrm{H}}$ the Sudakov form factor suppresses the $\pT^{\mathrm{H}}$ distribution by
almost $10\%$ with respect to the results obtained in the ET.  Since the
emission probability is also proportional to the ratio $R/B$, as can
be read from Eq.~(\ref{eq:gghtb-POWHEG}), starting from $\pT^{\mathrm{H}} \simeq
30 \UGeV$ this factor prevails over the Sudakov factor, and the
distribution with exact dependence on the quark masses becomes larger
than the one in the ET by slightly more than $10\%$ -- as already
observed at NLO QCD in the left plot.  We remark that the effects due
to the exact treatment of the masses of the top and bottom quarks $i)$
are of the same order of magnitude as the QCD perturbative theoretical
uncertainty estimated with {\sc HqT} and $ii)$ have a non-trivial
shape for different values of $\pT^{\mathrm{H}}$. If added to the {\sc HqT}
prediction, these effects would modify in a non-trivial way the
prediction of the central value of the distribution.

A different behaviour is found in the case of a heavy Higgs boson,
with $\MH=500 \UGeV$, because the bottom quark plays a negligible role.
At NLO QCD only the top-quark mass effects are relevant, at large
$\pT^{\mathrm{H}}$, yielding a negative correction.  In turn, the Sudakov form
factor, evaluated for small $\pT^{\mathrm{H}}$, is larger than in the ET (in fact
$R(\Pt+\Pb,\mathrm{exact})/B(\Pt+\Pb,\mathrm{exact})<R(\Pt,\infty)/B(\Pt,\infty)$), yielding what we
would call a Sudakov enhancement.  Also in this case, starting from
$\pT^{\mathrm{H}} \simeq 70 \UGeV$ the effect of the emission probability factor
$R/B$ prevails over the effect of the Sudakov form factor, leading to
negative corrections with respect to the ET case.

\subsubsection{Summary}
\label{sec:gghtb-concl}

An improved release of the code for Higgs-boson production in gluon
fusion in the {\sc POWHEG} framework has been prepared \cite{Bagnaschi:2011tu},
including the complete NLO QCD matrix elements with exact dependence
on the top- and bottom-quark masses.  This code allows a full
simulation of the process, matching the NLO results with the SMC {\sc
  PYTHIA}.

The quark-mass effects on the total cross section and on the Higgs
rapidity distribution are at the level of a few per cent. On the other
hand, the bottom-quark contribution is especially relevant in the
study of the transverse-momentum distribution of a light Higgs
boson. Indeed, the bottom contribution enhances the real emission
amplitude with respect to the result obtained in the ET.  The outcome
is a non-trivial distortion of the shape of the Higgs transverse-momentum 
distribution, at the level of ${\cal O}(10\%)$ of the result
obtained in the ET.  These effects are comparable to the present
estimates of the perturbative QCD theoretical uncertainty (see \refS{sec:hqt}).

\noindent
\subsection{Perturbative uncertainties in exclusive jet bins\footnote{I. W. Stewart and F. J. Tackmann.}}
\label{sec:jetveto}

\subsubsection{Overview}
\label{sec:jetveto:overview}

In this section we discuss the evaluation of perturbative theory uncertainties
in predictions for exclusive jet cross sections, which have a particular number
of jets in the final state. This is relevant for measurements where the data are
divided into exclusive jet bins, and is usually done when the background
composition strongly depends on the number of jets, or when the overall
sensitivity can be increased by optimizing the analysis for the individual jet
bins. The primary example is the $\PH\to \PW\PW$ analysis, which is performed
separately in exclusive $0$-jet, $1$-jet, and $2$-jet channels. Other examples
are vector-boson fusion analyses, which are typically performed in the exclusive
$2$-jet channel, boosted $\PH\to \Pb\bar \Pb$ analyses that include a veto on
additional jets, as well as $\PH\to \PGt\PGt$ and $\PH\to \PGg\PGg$ which
benefit from improved sensitivity when the Higgs recoils against a jet.  When
the measurements are performed in exclusive jet bins, the perturbative
uncertainties in the theoretical predictions must also be evaluated separately
for each individual jet bin~\cite{Anastasiou:2009bt}.  When combining channels
with different jet multiplicities, the correlations between the theoretical
uncertainties can be significant and must be taken into
account~\cite{Stewart:2011cf}.  We will use the notation $\sigma_N$ for an
\emph{exclusive} $N$-jet cross section (with exactly $N$ jets), and the notation
$\sigma_{\ge N}$ for an \emph{inclusive} $N$-jet cross section (with $N$ or more
jets). Three possible methods for evaluating the
uncertainties in exclusive jet cross sections are:
\begin{itemize}
\item[A)] ``Direct Exclusive Scale Variation''.
  Here the uncertainties are evaluated by
  directly varying the renormalisation and factorisation scales in the
  fixed-order predictions for each exclusive jet cross section $\sigma_N$. The
  results are taken as $100\%$ correlated, such that when adding the exclusive jet
  cross sections one recovers the scale variation in the total cross
  section.
\item[B)] ``Combined Inclusive Scale Variation'', as proposed in
  \Bref{Stewart:2011cf}.  Here, the perturbative uncertainties in the
  inclusive $N$-jet cross sections, $\sigma_{\geq N}$, are treated as the
  primary uncertainties that can be evaluated by scale variations in fixed-order
  perturbation theory. These uncertainties are treated as uncorrelated for
  different $N$.  The exclusive $N$-jet cross sections are obtained using
  $\sigma_N = \sigma_{\geq N} - \sigma_{\geq N+1}$. The uncertainties and correlations
  follow from standard error propagation, including the appropriate
  anticorrelations between $\sigma_N$ and $\sigma_{N\pm 1}$ related to the
  division into jet bins.
\item[C)] ``Uncertainties from Resummation with Reweighting.''  Resummed
  calculations for exclusive jet cross sections can provide uncertainty
  estimates that allow one to simultaneously include both types of correlated and
  anticorrelated uncertainties as in methods A and B. The magnitude of the uncertainties
  may also be reduced from the resummation of large logarithms.
\end{itemize}
Method B avoids a potential underestimate of the uncertainties in individual
jet bins due to strong cancellations that can potentially take place in method
A. An explicit demonstration that different treatments of the uncontrolled higher-order $\orde{\alphas^3}$ terms in method A can lead to very different LHC predictions is given in \refS{sec:vetosum}.  Method B produces realistic perturbative uncertainties for exclusive jet cross sections when using fixed-order
predictions for various processes. It is the main topic of this section, which follows
\Bref{Stewart:2011cf}. In method C one can utilise higher-order resummed
predictions for the exclusive jet cross sections, which allow one to obtain
improved central values and further refined uncertainty estimates. This is
discussed in \refS{sec:jetvetosum}. The uncertainties from method B are more consistent
with the information one gains about jet-binning effects from using resummation.

For method B the theoretical motivation from the basic structure of the
perturbative series is outlined in \refS{jetbin_motivation}. An
implementation for the example of the $0$-jet and $1$-jet bins is given in
\refS{jetbin_01jet}.  If theory predictions are required for irreducible
backgrounds in jet bins, then the same method should be used to evaluate the
perturbative uncertainties, for \eg in the case of $\PH\to \PW\PW$, the $\Pq\bar \Pq\to
\PW\PW$ and $\Pg\Pg\to \PW\PW$ direct production channels. 

Note that here we are only discussing the theoretical uncertainties due to
unknown higher-order perturbative corrections, which are commonly estimated
using scale variation. Parametric uncertainties, such as PDF and $\alphas$
uncertainties, must be treated appropriately as common sources for all
investigated channels.

\subsubsection{Theoretical motivation}
\label{jetbin_motivation}

To start, we consider the simplest example of dividing the total cross
section, $\sigma_\total$, into an \emph{exclusive} $0$-jet bin,
$\sigma_0(p^\cut)$, and the remaining \emph{inclusive} $(\geq 1)$-jet bin,
$\sigma_{\geq 1}(p^\cut)$,
\begin{align} \label{eq:pcut}
\sigma_\total = \int_0^{p^\cut}\!d p\, \frac{d\sigma}{d p} + \int_{p^\cut}\!d p\, \frac{d\sigma}{d p}
\equiv  \sigma_0(p^\cut) + \sigma_{\geq 1}(p^\cut)
\,.\end{align}
Here $p$ denotes the kinematic variable which is used to divide up the cross
section into jet bins.  A typical example is $p \equiv \pT^\jet$, defined by the
largest $\pT$ of any jet in the event, such that $\sigma_0(\pT^\cut)$ only
contains events with jets having $\pT < \pT^\cut$, and $\sigma_{\geq
  1}(\pT^\cut)$ contains events with at least one jet with $\pT > \pT^\cut$.
The definition of $\sigma_0$ may also include dependence on rapidity, such as
only considering jets within the rapidity range $\lvert \eta^\jet\rvert <
\eta^\cut$.

The phase-space restriction defining $\sigma_0$ changes its perturbative
structure compared to that of $\sigma_\total$ and in general gives rise to an
additional perturbative uncertainty which we denote by $\Delta_\cut$. This can
be thought of as an uncertainty related to large logarithms of $p^\cut$, or more
generally as an uncertainty associated to computing a less inclusive cross
section, which is theoretically more challenging. In Eq.~\eqref{eq:pcut} both $\sigma_0$
and $\sigma_{\geq 1}$ depend on the phase-space cut, $p^\cut$, and by
construction this dependence cancels in $\sigma_0+\sigma_{\geq 1}$.  Hence, the
additional uncertainty $\Delta_\cut$ induced by $p^\cut$ must be $100\%$
anticorrelated between $\sigma_0(p^\cut)$ and $\sigma_{\geq 1}(p^\cut)$, such
that it cancels in their sum. For example, using a covariance matrix to model
the uncertainties and correlations, the contribution of $\Delta_\cut$ to the
covariance matrix for $\{\sigma_0, \sigma_{\geq 1}\}$ must be of the form
\begin{equation} \label{eq:cutmatrix}
C_\cut = \begin{pmatrix}
   \Delta_\cut^2 &  - \Delta_\cut^2 \\
   -\Delta_\cut^2 & \Delta_\cut^2
\end{pmatrix}\,.
\end{equation}
The questions then are: (1) How can we estimate $\Delta_\cut$ in a simple way,
and (2) how is the perturbative uncertainty $\Delta_\total$ of $\sigma_\total$
related to the uncertainties of $\sigma_0$ and $\sigma_{\geq 1}$?  To answer
these questions, we discuss the perturbative structure of the cross sections in
more detail.

By restricting the cross section to the $0$-jet region, one restricts the
collinear initial-state radiation from the colliding hard partons as well as the
overall soft radiation in the event. This restriction on additional emissions
leads to the appearance of Sudakov double logarithms of the form $L^2 =
\ln^2(p^\cut/Q)$ at each order in perturbation theory, where $Q$ is the hard
scale of the process. For Higgs production from gluon fusion, $Q = \MH$, and the
leading double logarithms appearing at $\orde{\alphas}$ are
\begin{align} \label{eq:sig0dbleL}
\sigma_0(\pT^\cut) &= \sigma_B \Bigl(1 - \frac{3\alphas}{\pi}\, 2\ln^2 \frac{\pT^\cut}{\MH} + \dotsb \Bigr)
\,,\end{align}
where $\sigma_B$ is the Born (tree-level) cross section.

The total cross section only depends on the hard scale $Q$,
which means by choosing the scale $\mu \simeq Q$, the
fixed-order expansion does not contain large logarithms and has the
structure%
\footnote{These expressions for the perturbative series are schematic.  The
  convolution with the parton distribution functions (PDFs) and $\mu$-dependent
  logarithms enter in the coefficients of the series, which are not displayed.
  (The single logarithms related to the PDF evolution are not the logarithms we
  are most interested in discussing.)  }
\begin{equation} \label{eq:sigmatot}
\sigma_\total \simeq \sigma_B\big[ 1 + \alphas + \alphas^2 + \orde{\alphas^3} \big]
\,.\end{equation}
As usual, varying the scale in $\alphas$ (and the PDFs) one obtains an estimate
of the size of the missing higher-order terms in this series, corresponding to
$\Delta_\total$.

The inclusive $1$-jet cross section has the perturbative structure
\begin{align} \label{eq:sigma1}
\sigma_{\geq 1}(p^\cut)
\simeq \sigma_B\bigl[ \alphas (L^2 + L + 1)
+ \alphas^2 (L^4 + L^3 + L^2 + L + 1) + \orde{\alphas^3 L^6} \bigr]
\,,\end{align}
where the logarithms $L = \ln(p^\cut/Q)$. For $p^\cut \muchless Q$ these logarithms
can get large enough to overcome the $\alphas$ suppression. In the limit
$\alphas L^2 \simeq 1$, the fixed-order perturbative expansion breaks down and
the logarithmic terms must be resummed to all orders in $\alphas$ to obtain a
meaningful result. For typical experimental values of $p^\cut$ fixed-order
perturbation theory can still be considered, but the logarithms cause large
corrections at each order and dominate the series. This means varying the scale
in $\alphas$ in Eq.~\eqref{eq:sigma1} tracks the size of the large logarithms and
therefore allows one to get an estimate of the size of missing higher-order
terms caused by $p^\cut$, which corresponds to the uncertainty $\Delta_\cut$.
Therefore, we can approximate $\Delta_\cut = \Delta_{\geq 1}$, where
$\Delta_{\geq 1}$ is obtained from the scale variation for $\sigma_{\geq 1}$.

The exclusive $0$-jet cross section is equal to the
difference between Eqs.~\eqref{eq:sigmatot} and \eqref{eq:sigma1}, and so has the schematic structure
\begin{align} \label{eq:sigma0}
\sigma_0 & (p^\cut) = \sigma_\total - \sigma_{\geq1}(p^\cut)
\nonumber\\
&\simeq \sigma_B \Bigl\{ \bigl[ 1 + \alphas + \alphas^2 + \orde{\alphas^3} \bigr]
- \bigl[\alphas (L^2 \!+ L + 1) + \alphas^2 (L^4 \!+ L^3 \!+ L^2 \!+ L + 1)
+ \orde{\alphas^3 L^6} \bigr] \Bigr\}
\,.\end{align}
In this difference, the large positive corrections in $\sigma_\total$ partly
cancel against the large negative logarithmic corrections in $\sigma_{\geq 1}$.
For example, at $\orde{\alphas}$ there is a value of $L$ for which the
$\alphas$ terms in Eq.~\eqref{eq:sigma0} cancel exactly. At this $p^\cut$ the NLO $0$-jet
cross section has vanishing scale dependence and is equal to the LO cross
section, $\sigma_0(p^\cut)=\sigma_B$.  Due to this cancellation, a standard use
of scale variation in $\sigma_0(p^\cut)$ does not actually probe the size of the
large logarithms, and thus is not suitable to estimate $\Delta_\cut$. This issue
impacts the uncertainties in the experimentally relevant region for $p^\cut$.

For example, for $\Pg\Pg \to \PH$ (with $\sqrt{s} =7\UTeV$, $\MH = 165\UGeV$, $\muF = \muR =
\MH/2$), one finds~\cite{Anastasiou:2004xq, Anastasiou:2005qj, Catani:2007vq, Grazzini:2008tf}
\begin{align}
\sigma_\total &= (3.32 \pb) \bigl[1 + 9.5\,\alphas + 35\,\alphas^2 + \orde{\alphas^3} \bigr]
\,,\nonumber\\
\sigma_{\geq 1}\bigl(\pT^\jet >30\UGeV, \abs{\eta^\jet} < 3.0\bigr)
&= (3.32 \pb) \bigl[4.7\,\alphas + 26\,\alphas^2 + \orde{\alphas^3} \bigr] \,.
\end{align}
In $\sigma_\total$ one can see the impact of the well-known large $K$ factors.
(Using instead $\muF = \muR = \MH$ the $\alphas$ and $\alphas^2$
coefficients in $\sigma_\total$ increase to $11$ and $65$.)  In $\sigma_{\geq 1}$,
one can see the impact of the large logarithms on the perturbative series.
Taking their difference to get $\sigma_0$, one observes a sizeable numerical
cancellation between the two series at each order in $\alphas$.

Since $\Delta_\cut$ and $\Delta_\total$ are by definition uncorrelated, by
associating $\Delta_\cut = \Delta_{\geq 1}$ we are effectively treating
the perturbative series for $\sigma_\total$ and $\sigma_{\geq 1}$ as independent
with uncorrelated perturbative uncertainties. That is, considering
$\{\sigma_\total, \sigma_{\geq 1}\}$, the covariance matrix is diagonal,
\begin{equation} \label{eq:diagmatrix}
\begin{pmatrix}
  \Delta_\total^2 & 0 \\ 0 & \Delta_{\geq1}^2
\end{pmatrix}
\,,\end{equation}
where $\Delta_\total$ and $\Delta_{\geq 1}$ are
evaluated by separate scale variations in the fixed-order predictions for
$\sigma_\total$ and $\sigma_{\geq 1}$.
This is consistent, since for small $p^\cut$ the two series have very different
structures. In particular, there is no reason to believe that the same
cancellations in $\sigma_0$ will persist at every order in perturbation theory
at a given $p^\cut$. It follows that the perturbative uncertainty in
$\sigma_0 = \sigma_\total - \sigma_{\geq 1}$ is given by
$\Delta_\total^2 + \Delta_{\geq 1}^2$, and the resulting
covariance matrix for $\{\sigma_0, \sigma_{\geq 1}\}$ is
\begin{equation} \label{eq:fullmatrix}
C = \begin{pmatrix}
   \Delta_{\geq 1}^2 + \Delta_\total^2 &  - \Delta_{\geq 1}^2 \\
   -\Delta_{\geq 1}^2 & \Delta_{\geq 1}^2
\end{pmatrix}\,.
\end{equation}
The $\Delta_{\geq 1}$
contributions here are equivalent to Eq.~\eqref{eq:cutmatrix} with $\Delta_\cut =
\Delta_{\geq 1}$. Note also that all of $\Delta_\total$
occurs in the uncertainty for $\sigma_0$.  This is reasonable from the point of
view that $\sigma_0$ starts at the same order in $\alphas$ as $\sigma_\total$ and
contains the same leading virtual corrections.

\begin{figure}[t!]
\includegraphics[width=0.5\textwidth]{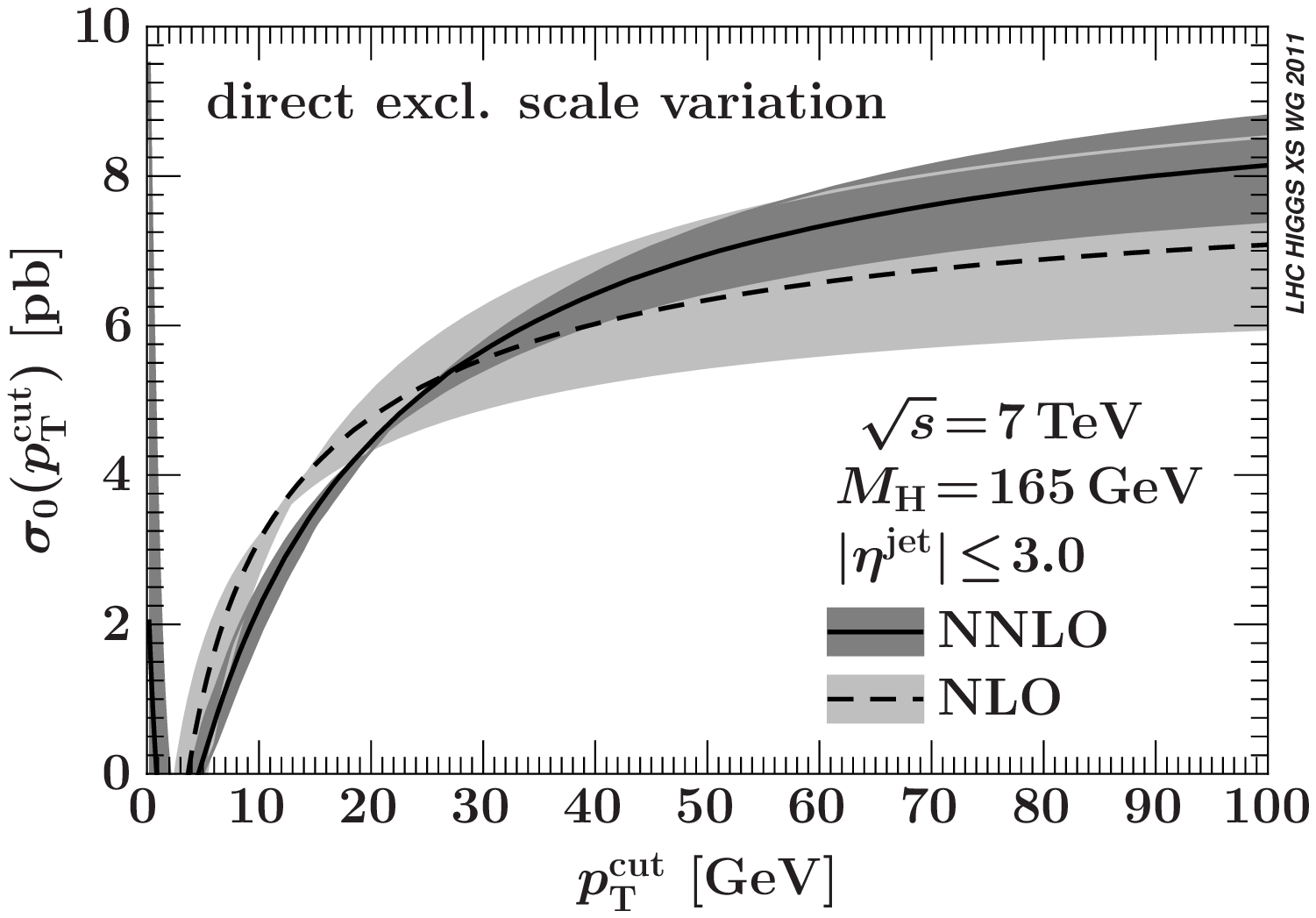}%
\hfill%
\includegraphics[width=0.5\textwidth]{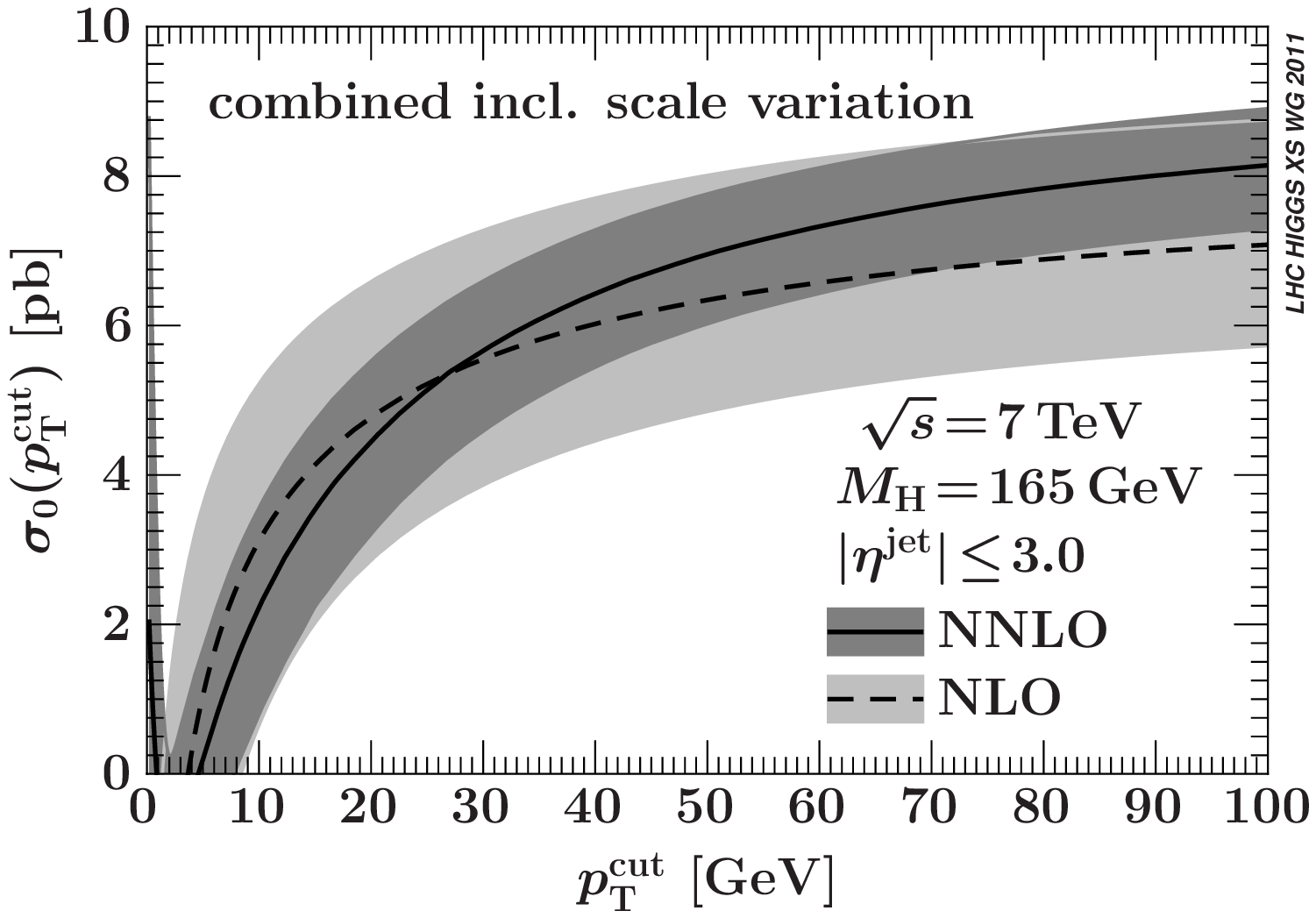}%
\vspace{-0.5ex}
\caption{\label{fig:0jet} Fixed-order perturbative uncertainties for $\Pg\Pg \to \PH +
  0$ jets at NLO and NNLO. On the left, the uncertainties are obtained from the
  direct exclusive scale variation in $\sigma_0(\pT^\cut)$ between $\mu = \MH/4$
  and $\mu = \MH$ (method A). On the right, the uncertainties are obtained by
  independently evaluating the inclusive scale uncertainties in $\sigma_\total$
  and $\sigma_{\geq 1}(p^\cut)$ and combining them in quadrature (method B). The
  plots are taken from \Ref~\cite{Stewart:2011cf}.}
\end{figure}

\begin{figure}[t!]
\includegraphics[width=0.5\textwidth]{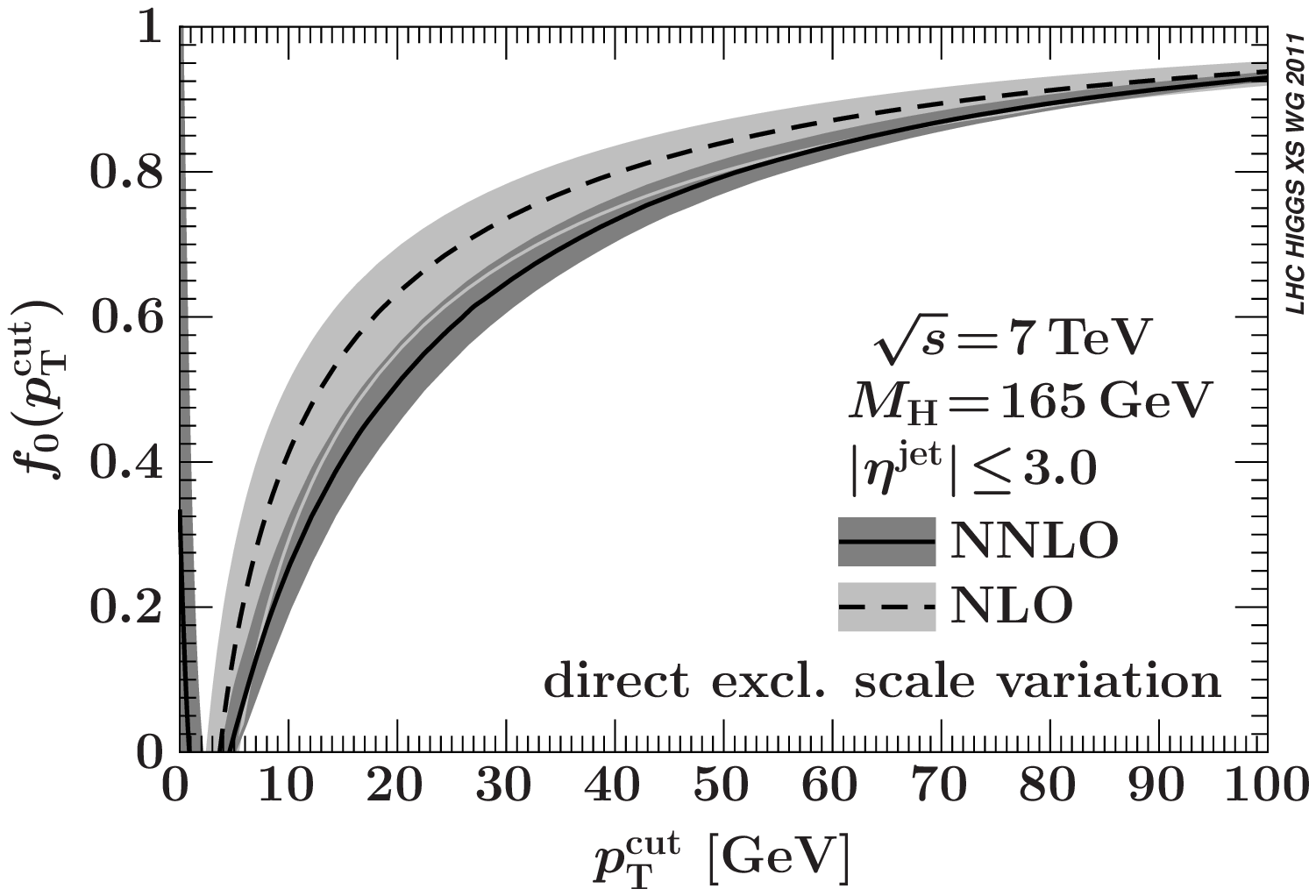}%
\hfill%
\includegraphics[width=0.5\textwidth]{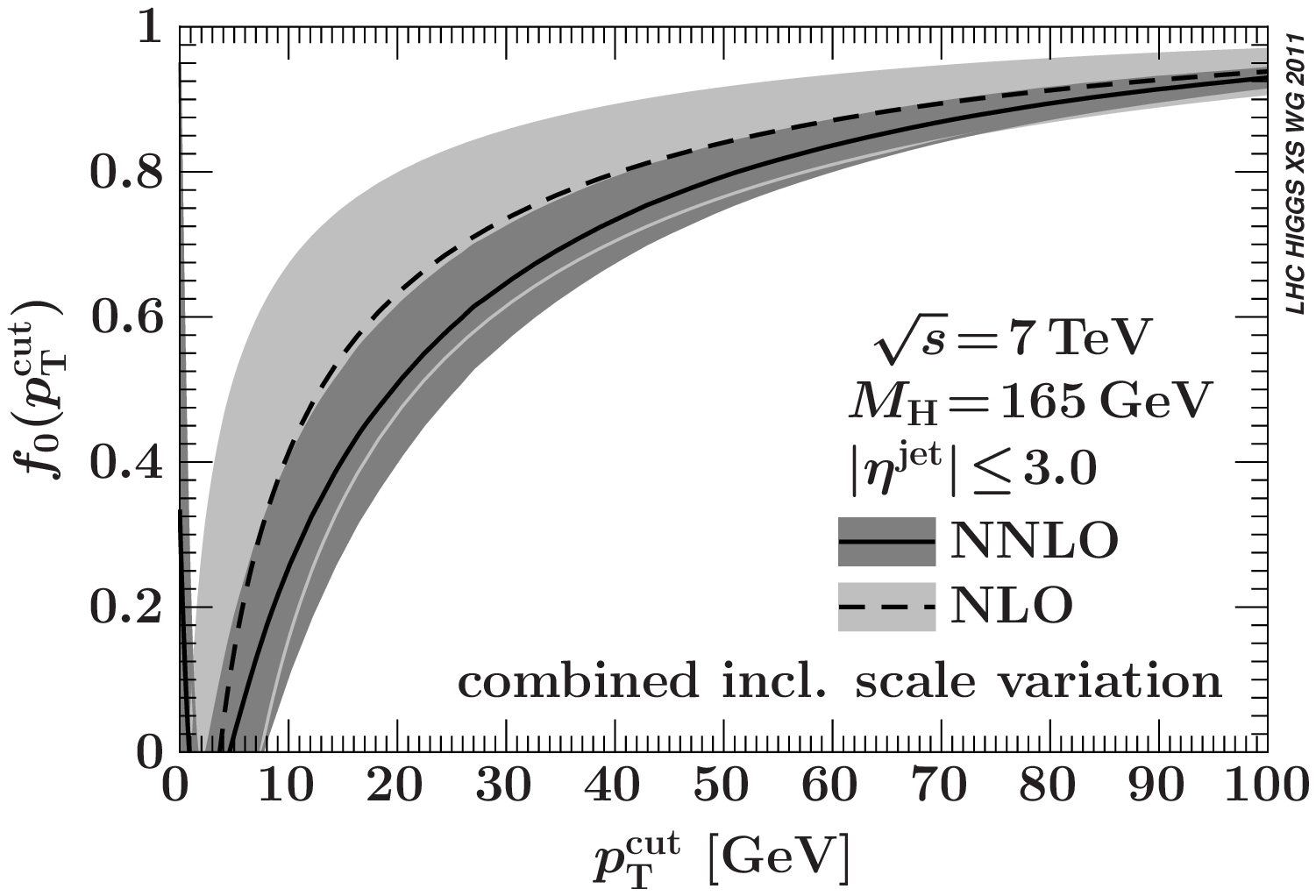}%
\vspace{-0.5ex}
\caption{\label{fig:0jetfrac} Same as \refF{fig:0jet} but for the $0$-jet fraction $f_0(\pT^\cut) = \sigma_0(\pT^\cut)/\sigma_\total$.}
\end{figure}

The limit $\Delta_\cut = \Delta_{\geq 1}$ that Eq.~\eqref{eq:fullmatrix} is based on is of
course not exact. However, the preceding arguments show that it is a more
reasonable starting point than using a common scale variation for the different
jet bins as in method A, since the latter does not account for the additional $p^\cut$ induced
uncertainties. These two methods of evaluating the perturbative uncertainties
are contrasted in \refF{fig:0jet} for $\Pg\Pg \to \PH+0$ jets at NLO (light gray)
and NNLO (dark gray) as a function of $\pT^\cut$ (using $\mu=\MH/2$ for the
central scale choice). The left panel shows the uncertainties from method A obtained from a
direct scale variation by a factor of two in $\sigma_0(\pT^\cut)$.
For small values of $\pT^\cut$ the cancellations that take place in
$\sigma_0(p^\cut)$ cause the error bands to shrink and eventually vanish at
$\pT^\cut\simeq 25\UGeV$, where there is an almost exact cancellation between the
two series in Eq.~\eqref{eq:sigma0}. In contrast, in the right panel the uncertainties are
obtained using the above method B by combining the independent inclusive
uncertainties to obtain the exclusive uncertainty, $\Delta_0^2 =\Delta_\total^2 +
\Delta_{\ge 1}^2$. For large values of $\pT^\cut$ this reproduces the direct
exclusive scale variation, since $\sigma_{\geq 1}(p^\cut)$ becomes small. On the
other hand, for small values of $\pT^\cut$ the uncertainties estimated in this
way are more realistic, because they explicitly estimate the uncertainties due
to the large logarithmic corrections. The features of this plot are quite
generic. In particular, the same pattern of uncertainties is observed for the
Tevatron, when using $\mu=\MH$ as our central scale (with $\mu=2\MH$ and
$\mu=\MH/2$ for the range of scale variation), whether or not we only look at
jets at central rapidities, or when considering the exclusive $1$-jet cross
section.  We also note that using independent variations for $\muF$ and $\muR$
does not change this picture, in particular the $\muF$ variation for fixed
$\muR$ is quite small.  In \refF{fig:0jetfrac} we again compare the two methods, but now for the event
fraction $f_0(\pT^\cut)=\sigma_0(\pT^\cut)/\sigma_\total$. 
At large $\pT^\cut$ the curves approach unity, and the uncertainty asymptotically vanishes.
At small $\pT^\cut$ the
uncertainties in $f_0$ are quite small with method A. In method B the
uncertainties are more realistic, and there is now a significant overlap between
the bands at NLO and NNLO.

The generalisation of the above discussion to more jets and several jet bins is
straightforward.  For the $N$-jet bin we replace $\sigma_\total \to \sigma_{\ge
  N}$, $\sigma_0\to \sigma_N$, and $\sigma_{\ge 1} \to \sigma_{\ge N+1}$, and
take the appropriate $\sigma_B$. If the perturbative series for $\sigma_{\ge N}$
exhibits large $\alphas$ corrections due to its logarithmic series or
otherwise, then the presence of a different series of large logarithms in
$\sigma_{\ge N+1}$ will again lead to cancellations when we consider the
difference $\sigma_N = \sigma_{\geq N} - \sigma_{\geq N+1}$. Hence,
$\Delta_{\geq N+1}$ will again give a better estimate for the extra
$\Delta_\cut$ type uncertainty that arises from separating $\sigma_{\geq N}$
into $\sigma_N$ and $\sigma_{\geq N+1}$. Plots for the $1$-jet bin for Higgs
production from gluon fusion can be found in \Bref{Stewart:2011cf}.

\subsubsection{Example implementation for $H+0$ Jet and $H+1$ jet channels}
\label{jetbin_01jet}

To illustrate the implications for a concrete example we consider the $0$-jet
and $1$-jet bins together with the remaining $(\ge 2)$-jet bin. By construction
only neighboring jet bins are correlated, so the generalisation to more jet bins
is not any more complicated. We denote the total inclusive cross section by
$\sigma_\total$, and the inclusive $1$-jet and $2$-jet cross sections by
$\sigma_{\geq 1}$ and $\sigma_{\geq 2}$. Their respective absolute uncertainties
are $\Delta_\total$, $\Delta_{\geq 1}$, $\Delta_{\geq 2}$, and their relative
uncertainties are given by $\delta_i = \Delta_i/\sigma_i$. The exclusive $0$-jet
cross section, $\sigma_0$, and $1$-jet cross section, $\sigma_1$, satisfy the
relations
\begin{equation} \label{eq:sigma01}
\sigma_0 = \sigma_\total - \sigma_{\geq 1}
\,,\qquad
\sigma_1 = \sigma_\mathrm{\geq 1} - \sigma_{\geq 2}
\,,\qquad
\sigma_\total = \sigma_0 + \sigma_1 + \sigma_{\geq 2}
\,.\end{equation}
Experimentally it is convenient to work with the exclusive $0$-jet and $1$-jet
fractions defined as
\begin{equation} \label{eq:f01}
f_0 = \frac{\sigma_0}{\sigma_\total}
\,,\qquad
f_1 = \frac{\sigma_1}{\sigma_\total}
\,.\end{equation}

Treating the inclusive uncertainties $\Delta_\total$, $\Delta_{\geq 1}$, $\Delta_{\geq 2}$ as uncorrelated, the covariance matrix for the three quantities $\{\sigma_\total, \sigma_0, \sigma_1 \}$ is given by
\begin{equation} \label{eq:Cov}
C =
\begin{pmatrix}
\Delta_\total^2 & \Delta_\total^2 & 0 \\
\Delta_\total^2 & \Delta_\total^2 + \Delta_{\geq 1}^2 & -\Delta_{\geq1}^2 \\
0 &-\Delta_{\geq1}^2 & \Delta_{\geq 1}^2 + \Delta_{\geq 2}^2
\end{pmatrix}
\,.\end{equation}
The relative uncertainties and correlations for $\sigma_0$ and $\sigma_1$ directly follow from Eq.~\eqref{eq:Cov}. Writing them in terms of the relative quantities $f_i$ and $\delta_i$, one gets
\begin{align} \label{eq:deltasigma01}
\delta(\sigma_0)^2 &= \frac{1}{f_0^2}\, \delta_\total^2 + \Bigl(\frac{1}{f_0} - 1\Bigr)^2 \delta_{\geq 1}^2
\,,\nonumber\\
\delta(\sigma_1)^2  &= \Bigl(\frac{1-f_0}{f_1}\Bigr)^2 \delta_{\geq 1}^2 +
\Bigl(\frac{1-f_0}{f_1} - 1\Bigr)^2 \delta_{\geq 2}^2
\,,\nonumber\\
\rho(\sigma_0, \sigma_\total)
&= \biggl[1 + \frac{\delta_{\geq 1}^2}{\delta_\total^2}(1-f_0)^2 \biggr]^{-1/2}
\,,\nonumber\\
\rho(\sigma_1, \sigma_\total) &= 0
\,,\nonumber\\
\rho(\sigma_0, \sigma_1)
&= - \biggl[1 + \frac{\delta_\total^2}{\delta_{\geq 1}^2}\, \frac{1}{(1-f_0)^2}\biggr]^{-1/2}
\biggl[1 + \frac{\delta_{\geq2}^2}{\delta_{\geq 1}^2} \Bigl(1 - \frac{f_1}{1-f_0}\Bigr)^2\biggr]^{-1/2}
\,.\end{align}
Alternatively, we can use $\{\sigma_\total, f_0, f_1\}$ as the three independent
quantities. Their relative uncertainties and correlations following from
Eq.~\eqref{eq:Cov} are then
\begin{align} \label{eq:deltaf01}
\delta(f_0)^2
&= \Bigl(\frac{1}{f_0} - 1 \Bigr)^2\bigl(\delta_\total^2 + \delta_{\geq 1}^2\bigr)
\,,\nonumber\\
\delta(f_1)^2
&= \delta_\total^2 + \Bigl(\frac{1-f_0}{f_1}\Bigr)^2 \delta_{\geq 1}^2 +
\Bigl(\frac{1-f_0}{f_1} - 1\Bigr)^2 \delta_{\geq 2}^2
\,,\nonumber\\
\rho(f_0, \sigma_\total)
&= \biggl[1 + \frac{\delta_{\geq 1}^2}{\delta_\total^2} \biggr]^{-1/2}
\,,\nonumber\\
\rho(f_1, \sigma_\total)
&= -\frac{\delta_\total}{\delta(f_1)}
\,,\nonumber\\
\rho(f_0, f_1)
&= -\biggl(1 + \frac{1-f_0}{f_1}\frac{\delta_{\geq 1}^2}{\delta_\total^2} \biggr)
\Bigl(\frac{1}{f_0} - 1\Bigr) \frac{\delta_\total^2}{\delta(f_0)\delta(f_1)}
\,.\end{align}
The basic steps of the analysis are the same irrespective of the statistical model:
\begin{enumerate}
\item Independently evaluate the inclusive perturbative uncertainties
  $\delta_\total$, $\delta_{\geq 1}$, $\delta_{\geq 2}$ by appropriate scale
  variations in the available fixed-order calculations.
\item Consider the $\delta_i$ uncorrelated and use Eqs.~\eqref{eq:sigma01} or \eqref{eq:f01} to propagate them into the uncertainties of $\sigma_{0,1}$ or $f_{0,1}$.
\end{enumerate}
For example, when using log-likelihoods, the independent uncertainties
$\delta_\total$, $\delta_{\geq 1}$, $\delta_{\geq 2}$ are implemented via three
independent nuisance parameters.

In Step 1, $\delta_\total$ can be taken from the NNLO perturbative uncertainty in
the first Yellow Report~\cite{Dittmaier:2011ti}, while $\delta_{\geq 1}$ is evaluated at
relative NLO and $\delta_{\geq 2}$ at relative LO using any of {\sc HNNLO}~\cite{Catani:2007vq,
Grazzini:2008tf}, {\sc FEHiP}~\cite{Anastasiou:2004xq, Anastasiou:2005qj}, or
{\sc MCFM}~\cite{Campbell:2010cz}. For the central scales $\muF = \muR = \MH/2$ should be used to be
consistent with $\sigma_\total$ from \Bref{Dittmaier:2011ti}.  Numerically
$\sigma_{\geq 1}$ also exhibits better convergence with this choice for the
central scale.  Note that $\sigma_{\geq 2}$ is defined by inverting the cut on
the second jet in $\sigma_1$ and is not independent of $\{\sigma_\total$,
$\sigma_0$, $\sigma_1\}$. Its per-cent uncertainty $\delta_{\ge 2}$ naturally
enters because of the theory correlation model and for
this purpose it should be taken at LO. Also note that $\sigma_{\geq 2}$ is
different from the $2$-jet bin, $\sigma_2$.  If $\sigma_2$ is included one
extends the matrix by an extra row/column, and uses as input that the
uncertainties in $\sigma_\total$, $\sigma_{\ge 1}$, $\sigma_{\ge 2}$, $\sigma_{\ge
  3}$ are independent.

When $\sigma_2$ is used in the analysis with additional vector-boson fusion
cuts, one can determine the uncertainties with the appropriate $\sigma_{\geq 2}$
at NLO and $\sigma_{\geq 3}$ at LO from MCFM~\cite{Campbell:2010cz}. To combine the $\sigma_2$ with
additional vector-boson-fusion cuts with the $0$-jet and $1$-jet gluon-fusion
bins, one treats the VBF uncertainty as uncorrelated with those from gluon
fusion, and assigns a $100\%$ correlation between the perturbative uncertainties
of the two different gluon-fusion $\sigma_{\geq 2}$s that appear in these
computations.

Step 2 above also requires values for $f_0$ and $f_1$ as input. In the
experimental analyses, the central values for $f_0$ and $f_1$ effectively come
from a full Monte Carlo simulation. By only using relative quantities, the
different overall normalisations in the fixed-order calculation and the Monte Carlo
drop out. However, the values for $f_0$ and $f_1$ obtained from the fixed-order
calculation can still be quite different from those in the Monte Carlo, because
the fixed-order results incorporate NNLO $\alphas$ corrections, while the Monte
Carlo includes some resummation as well as detector effects.
A priori the values for $f_{0,1}$ obtained either way could be used as input
for the uncertainty calculation in Step 2. One can cross check
that the two choices lead to similar results for the final uncertainties.

It is useful to illustrate this with a numerical example corresponding to
\refF{fig:0jet}, for which we take $\pT^\cut = 30\UGeV$ and
$\eta^\cut = 3.0$.  For $\MH = 165\UGeV$, we have $\sigma_\total = (8.76
\pm 0.80)\Upb$, $\sigma_{\geq1} = (3.10\pm0.61)\Upb$, and
$\sigma_{\geq 2} = (0.73 \pm 0.42)\Upb$, corresponding to the relative
uncertainties $\delta_\total = 9.1\%$, $\delta_{\geq 1} = 19.9\%$, and
$\delta_{\geq 2} = 57\%$. Using these as inputs in Eq.~\eqref{eq:deltasigma01}
we find for the exclusive jet cross sections $\delta(\sigma_0) = 18\%$,
$\delta(\sigma_1) = 31\%$ with correlations coefficients $\rho(\sigma_0,
\sigma_\total) = 0.79$, $\rho(\sigma_1, \sigma_\total) = 0$, and $\rho(\sigma_0,
\sigma_1) = -0.50$. We see that $\sigma_0$ and $\sigma_1$ have a substantial
negative correlation because of the jet bin boundary they share.  For the
exclusive jet fractions using Eq.~\eqref{eq:deltaf01} we obtain $\delta(f_0) =
12\%$ and $\delta(f_1) = 33\%$ with correlations $\rho(f_0, \sigma_\total) =
0.42$, $\rho(f_1, \sigma_\total) = -0.28$, and $\rho(f_0,f_1) = -0.84$.  The
presence of $\sigma_\total$ in the denominator of the $f_i$'s yields a nonzero
anticorrelation for $f_1$ and $\sigma_\total$, and a decreased correlation for
$f_0$ and $\sigma_\total$ compared to $\sigma_0$ and $\sigma_\total$.  In contrast
to these results, in method A one considers the direct scale variation in the exclusive
$\sigma_{0,1}$ as in the left panel of \refF{fig:0jet}. Due
to the cancellations between the perturbative series, this approach leads to
unrealistically small uncertainties (with the above inputs $\delta(\sigma_0) =
3.2\%$ and $\delta(\sigma_1) = 8.3\%$), which is reflected in the pinching of
the bands in the left plot in \refF{fig:0jet}.

As a final remark, we note that strictly speaking, it is somewhat inconsistent
to use the relative uncertainties for $f_{0,1}$ from fixed order and apply them to the Monte
Carlo predictions for $f_{0,1}$. A more sophisticated treatment would be to
first correct the measurements for detector effects to truth-level using Monte
Carlo. In the second step, the corrected measurements of $\sigma_0$ and
$\sigma_1$ (with truth-level cuts) can be directly compared to the available
calculations, taking into account the theoretical uncertainties as described
above. This also automatically makes a clean separation between experimental and
theoretical uncertainties. Given the importance and impact of the jet selection
cuts, this is very desirable.  It would both strengthen the robustness of the
extracted experimental limits and validate the theoretical description of the
jet selection.

\noindent
\subsection{Perturbative uncertainties in jet-veto
  efficiencies\footnote{A.~Banfi, G.~P.~Salam and G.~Zanderighi.}}
\label{sec:vetosum}


%
In this section we continue our discussion of perturbative uncertainties
in predictions for exclusive cross sections,
concentrating on the 0-jet cross section,
corresponding to
having a Higgs and no jets with a transverse momentum above a given
$\pT^{{\rm cut}}$.
%

We examine here the question of the theoretical uncertainty in the
prediction of the jet-veto efficiency.
In particular, we consider the ambiguity that arises in its
calculation from a ratio of two cross sections that are both known at
next-to-next-to-leading order (NNLO).

In the following, $\sigma_0(\pT^{{\rm cut}})$ will denote the 0-jet cross
section as a function of the jet-transverse-momentum threshold
$\pT^{{\rm cut}}$, whilst $\sigma_\total$ will denote the Higgs total cross
section, without any jet veto. 
It is also useful to consider the ratio of these cross sections,
$f_0(\pT^{{\rm cut}})=\sigma_0(\pT^{{\rm cut}})/\sigma_\total$, which is commonly
referred to as the jet-veto efficiency, or the 0-jet fraction as in \refS{sec:jetveto}.
Knowledge of this efficiency, and its uncertainty, is important in
interpreting measured limits on the Higgs cross section in the 0-jet
bin as a limit on the total Higgs production cross section.

Both $\sigma_0(\pT^{{\rm cut}})$ and $\sigma_\total$ have a fixed-order perturbative
expansion of the form
\begin{subequations}
  \begin{align}
    \label{eq:Sigma-expansion}
    \sigma_0(\pT^{{\rm cut}}) &= \sigma_0^{(0)}(\pT^{{\rm cut}}) + \sigma_0^{(1)}(\pT^{{\rm cut}}) + \sigma_0^{(2)}(\pT^{{\rm cut}}) + \ldots \,,\\
    \sigma_\total &= \sigma^{(0)} + \sigma^{(1)} + \sigma^{(2)} + \ldots \,,
  \end{align}
\end{subequations}
where the superscript $i$ denotes the fact that the given contribution to
the cross section is proportional to $\alphas^i$ relative to the Born
cross section (of order $\alphas^2$ in the present case). Since no
jets are present at the Born level we have $\sigma_0^{(0)}(\pT^{{\rm cut}})
\equiv \sigma^{(0)}$.

The state-of-the-art of fixed-order QCD predictions is NNLO, i.e.\ the
calculation of $\sigma_0(\pT^{{\rm cut}})$ and $\sigma_\total$ with tools like
{\sc FEHiP}~\cite{Anastasiou:2005qj} and
{\sc HNNLO}~\cite{Grazzini:2008tf}.

There is little ambiguity in the definition of the fixed-order results
for the total and jet-vetoed cross sections, with the only freedom
being, as usual, in the choice of the renormalisation and the
factorisation scale.
However, given the expressions of $\sigma_0$ and $\sigma_\total$ at a given
perturbative order, there is some additional freedom in the way one
computes the jet-veto efficiency. For instance, at NNLO the efficiency
can be defined as
\begin{equation}
  f_0^{(a)}(\pT^{{\rm cut}}) \equiv 
  \frac{\sigma_0^{(0)}(\pT^{{\rm cut}})+\sigma_0^{(1)}(\pT^{{\rm cut}})
+\sigma_0^{(2)}(\pT^{{\rm cut}})}{\sigma^{(0)}+\sigma^{(1)}+\sigma^{(2)}}\,.
\label{eq:NNLOa}
\end{equation}
This option is the most widely used and may appear at first sight to
be the most natural, insofar as one keeps as many terms as possible
both in the numerator and denominator.  It corresponds to method A of evaluating the uncertainty 
in the fraction of events in the $0$-jet bin defined in \refS{sec:jetveto:overview}.

However, other prescriptions are possible. For instance, since the
zeroth-order term of $f_0(\pT^{{\rm cut}})$ is equal to $1$, one
can argue that it is really only $1-f_0(\pT^{{\rm cut}})$ that has
a non-trivial perturbative series, given by the ratio of the inclusive
1-jet cross section above $\pT^{{\rm cut}}$,
$\sigma_{\text{1-jet}}^\text{NLO}(\pT^{{\rm cut}})$, to the total cross
section, where
\begin{equation}
  \label{eq:sigma-jet-nlo}
  \sigma_{\text{1-jet}}^\text{NLO}(\pT^{{\rm cut}}) =  \sigma^{(1)} +
  \sigma^{(2)} - \left( \sigma_0^{(1)}(\pT^{{\rm cut}}) + \sigma_0^{(2)}(\pT^{{\rm cut}})
  \right)\,.
\end{equation}
Insofar as the 1-jet cross section is known only to NLO, in taking the
ratio to the total cross section one should also use NLO for the
latter, so that an alternative prescription reads
\begin{equation}
  \label{eq:NNLOb}
  f_0^{(b)}(\pT^{{\rm cut}}) = 1 -
  \frac{\sigma_{\text{1-jet}}^\text{NLO}(\pT^{{\rm cut}})}{\sigma^{(0)}+\sigma^{(1)}}\,.
\end{equation}
Finally, another motivated expression for the jet-veto efficiency is
just the fixed-order expansion up to ${\cal O}(\alphas^2)$ of
Eq.~\eqref{eq:NNLOa}, which can be expressed in terms of the LO and
NLO inclusive jet cross sections above $\pT^{{\rm cut}}$ as follows
\begin{equation}
  \label{eq:NNLOc}
  f_0^{(c)}(\pT^{{\rm cut}}) = 1 -
  \frac{\sigma_{\text{1-jet}}^\text{NLO}(\pT^{{\rm cut}})}{\sigma^{(0)}}+
  \frac{\sigma^{(1)}}{(\sigma^{(0)})^2}\sigma_{\text{1-jet}}^\text{LO}(\pT^{{\rm cut}})\,.
\end{equation}
Prescriptions (a), (b), and (c) differ by terms of relative order
$\alphas^3$ with respect to the Born level, i.e.\ NNNLO.
Therefore, the size of the differences between them is a way to
estimate the associated theoretical uncertainty that goes beyond the
usual variation of scales.

Let us see how these three prescriptions fare in practice in the case
of interest, namely Higgs production at the LHC with 7 \UTeV\
centre-of-mass energy. 
We use MSTW2008NNLO parton distribution
functions~\cite{Martin:2009iq} (even 
for LO and NLO predictions) with $\alphas(\MZ) =0.11707$ and
three-loop running. Furthermore, we use the large-$\Mt$
approximation. We choose a Higgs mass of $145\UGeV$, and default
renormalisation and factorisation scales of $\MH/2$. 
We cluster partons into jets using the anti-$\kT$ jet
algorithm~\cite{Cacciari:2008gp} with $R=0.5$, which is the default
jet definition for CMS.
Switching to $R=0.4$, as used by ATLAS, has a negligible impact
relative to the size of the uncertainties.
We include jets up to infinite rapidity, but have checked that the
effect of a rapidity cut of $4.5/5$, corresponding to the ATLAS/CMS
acceptances, is also much smaller than other uncertainties discussed
here.

\begin{figure}
  \includegraphics[width=.48\textwidth]{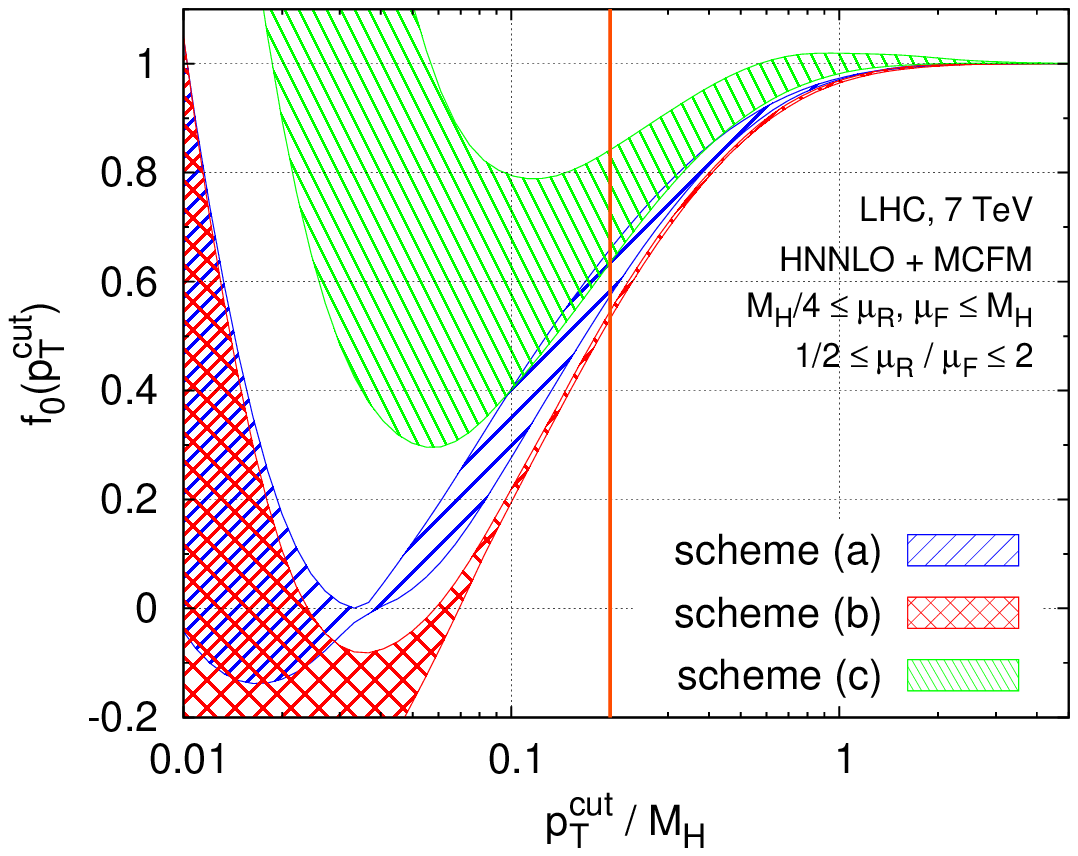}\hfill
  \includegraphics[width=.48\textwidth]{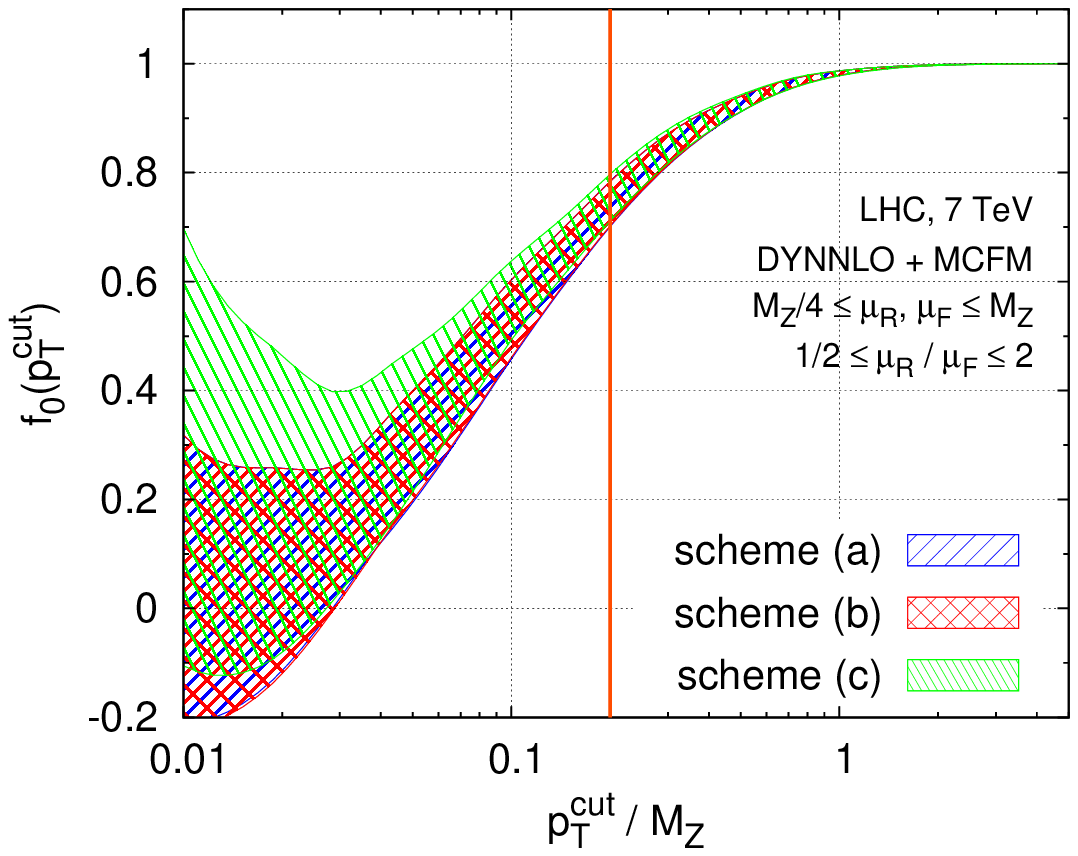}
  \caption{Jet-veto efficiency for Higgs (left) and
    Z-boson production (right) using three different prescriptions for
    the NNLO expansion, see
    Eqs.~(\ref{eq:NNLOa}),~(\ref{eq:NNLOb}),~(\ref{eq:NNLOc}). 
    The bands are obtained by varying renormalisation and
    factorisation scales independently around the central value
    $\MH/2$ ($\MZ/2$) by a factor of two up and down (with the
    constraint $\frac12 < \muR/\muF < 2$). 
    NNLO predictions are obtained by suitably combining the total
    cross sections obtained with {\sc HNNLO}~\cite{Grazzini:2008tf}
    ({\sc DYNNLO}~\cite{Catani:2009sm}) with the 1-jet cross
    section $\sigma_\text{1-jet}(\pT^{{\rm cut}})$ computed with
    {\sc MCFM}~\cite{MCFMweb}.}
\label{fig:ptjv3NNLO}
\end{figure} 

The corresponding results for the jet-veto efficiencies over a wide
range of values of $\pT^{{\rm cut}}/\MH$ are shown in
\refF{fig:ptjv3NNLO} (left). Each of the three prescriptions
Eqs.~\eqref{eq:NNLOa},~\eqref{eq:NNLOb},~\eqref{eq:NNLOc} is presented
together with an associated uncertainty band corresponding to
an independent variation of renormalisation and factorisation scales
$\MH/4 < \muR, \muF < \MH$ (with the constraint $\frac12 <
\muR/\muF < 2$).  The solid red vertical line corresponds to a
reference jet veto of $0.2 \MH\sim 29\UGeV$, which is in the ballpark
of the value used by ATLAS and CMS to split the cross section in 0-,
1-, and 2-jet bins ($25\UGeV$ and $30\UGeV$,
respectively).
Several features can be observed: firstly, the three schemes lead to
substantially different predictions for the jet-veto efficiency,
spanning a range from about 0.50 to 0.85 at the reference jet-veto
value.
Furthermore, the uncertainty bands from the different schemes barely
overlap, indicating that scale uncertainties alone are a poor
indicator of true uncertainties here.
Finally the uncertainty bands' widths are themselves quite different
from one scheme to the next.

The above features are all caused by the poor convergence of the
perturbative series.
In particular, it seems that two classes of effects are at play here.
Firstly, for $\pT^{{\rm cut}} \muchless \MH$, there are large Sudakov logarithms
$\alphas^n\ln^{2n}(\pT^{{\rm cut}}/\MH)$.
These are the terms responsible for the drop in veto efficiency at low
$\pT^{{\rm cut}}$ and the lack of a resummation of these terms to all
orders is responsible for the unphysical increase in veto efficiency
seen at very low $\pT^{{\rm cut}}$ (resummations of related observables are
discussed in \Bref{deFlorian:2011xf,Berger:2010xi}).
The second class of effects stems from the fact that the total cross
section has a very large NLO/LO $K$-factor, $\sim 2$, with substantial
corrections also at NNLO (see \refT{tab:sigma}).
The jet-veto efficiency is closely connected to the 1-jet rate, for
which the NNLO corrections are not currently known.
It
is conceivable that they could be as large, in relative terms, as the
NNLO corrections to the total cross section and
our different schemes for calculating the perturbative efficiency
effectively take that uncertainty into account.

\begin{table}[t]
\begin{center}
\begin{tabular}{cccc}
  \hline
  &LO& NLO& NNLO\\ 
  \hline
  &&&\\[-11pt]
  H [pb] & $3.94^{+1.01}_{-0.73}$ & $8.53^{+1.86}_{-1.34}$ & $10.5^{+0.8}_{-1.0}$ \\[2pt] 
  \hline
  &&&\\[-11pt]
  Z [nb] & $22.84^{+2.07}_{-2.40}$ & $28.6^{+0.8}_{-1.2}$ &
  $28.6^{+0.4}_{-0.4}$ \\[2pt] 
  \hline
\end{tabular}
\end{center}\vspace{-1em}
\caption{Cross sections for Higgs and Z-boson production in 7 \UTeV\ \Pp\Pp
  collisions, at various
  orders in perturbation theory. The central value corresponds to the
  default scale $\muR = \muF = \MH/2$ ($\MZ/2$), 
  the error denotes the scale variation when
  $\muR$ and $\muF$  are varied independently by a factor two around 
  the central value, with the constraint $\frac12 < \muR/\muF < 2$. 
}
\label{tab:sigma}
\end{table}

The reader may wonder whether it is really possible to attribute the
differences between schemes to the poor convergence of the total cross
section. 
One cross-check of this statement is to examine the jet-veto
efficiency for Z-boson production, where, with a central scale choice
$\mu= \MZ/2$, NLO corrections to the total cross section are about
$25\%$, and the NNLO ones are a couple of per cent (see \refT{tab:sigma}).
The results for the jet-veto efficiency are shown in
\refF{fig:ptjv3NNLO} (right). While overall uncertainties remain
non-negligible (presumably due to large Sudakov logarithms), the three
expansion schemes do indeed give almost identical results for $\pT^\cut/\MZ=0.2$.
This supports our interpretation that the poor total-cross-section
convergence is a culprit in causing the differences between the three
schemes in the Higgs case.

To conclude, in determining the jet-veto efficiency for Higgs
production, one should be aware that fixed-order predictions depend
significantly on the precise form of the perturbative expansion used
to calculate the efficiency, Eqs.~(\ref{eq:NNLOa}),~(\ref{eq:NNLOb}),
or (\ref{eq:NNLOc}).
This uncertainty is not reflected in the scale dependence of the
predictions, which is likely due to the accidental cancellations between physically unrelated
large Sudakov effects and large total-cross-section corrections that
are discussed in \refS{sec:jetveto} and \Bref{Stewart:2011cf}.

\def\ltap{\raisebox{-.6ex}{\rlap{$\,\sim\,$}}\raisebox{.4ex}{$\,<\,$}}
\def\gtap{\raisebox{-.4ex}{\rlap{$\,\sim\,$}} \raisebox{.4ex}{$\,>\,$}}
\def\Wcal{{\cal W}}

\subsection{The Higgs $\pT$ spectrum and its uncertainties\footnote{D. de Florian, G.Ferrera, M.Grazzini and D.Tommasini.}}
\label{sec:hqt}

\subsubsection{Introduction}
\label{sec:hqt-intro}

In this section we focus on the
transverse-momentum ($\pT$) spectrum of the SM Higgs boson.
This observable is of direct importance in the experimental search.
A good knowledge of the $\pT$ spectrum can help to define strategies
to improve the statistical significance.
When studying the $\pT$ distribution 
of the Higgs boson in QCD perturbation theory it is convenient to distinguish two regions of
transverse momenta.
In the large-$\pT$ region ($\pT\sim \MH$), where the transverse momentum is
of the order of the Higgs-boson mass $\MH$,
perturbative QCD
calculations based on the truncation of the perturbative series at a
fixed order in $\alphas$ 
are theoretically justified.
In this region, the $\pT$ spectrum is known up to leading order (LO) \cite{Baur:1989cm}
with the full dependence of the masses of the
top and bottom quarks, and up to the next-to-leading
order (NLO) \cite{deFlorian:1999zd,Ravindran:2002dc,Glosser:2002gm}
in the large-$\Mt$ limit.

In the small-$\pT$ region ($\pT\muchless\MH$),
where the bulk of the events is produced, 
the convergence of the fixed-order expansion is spoiled by the presence of 
large logarithmic terms, $\alphas^n\ln^m (\MH^2/\pT^2)$.
To obtain reliable predictions, 
these logarithmically-enhanced terms
have to be systematically
resummed to all
perturbative orders \cite{Dokshitzer:1978yd,Dokshitzer:1978hw,Parisi:1979se,Curci:1979bg,Collins:1981uk,Collins:1981va,Collins:1984kg,Kodaira:1981nh,Catani:1988vd,Catani:2000vq,Bozzi:2005wk,Catani:2010pd}.
It is then important to consistently match the resummed and fixed-order
calculations 
at intermediate values of $\pT$, in order
to obtain accurate QCD predictions for the entire range of transverse momenta.

The resummation of the logarithmically enhanced terms is effectively (approximately) performed
by standard Monte Carlo event generators.
In particular, MC@NLO \cite{Frixione:2002ik} and POWEG \cite{Nason:2004rx} combine soft-gluon resummation through the parton shower
with the LO result valid at large $\pT$, thus achieving a result with formal NLO accuracy.

The numerical program {\sc HqT} \cite{Bozzi:2005wk}
implements
soft-gluon resummation up to NNLL
accuracy \cite{deFlorian:2000pr} combined with fixed-order perturbation theory up to NLO in the large-$\pT$ region \cite{Glosser:2002gm}.
The program is used by the Tevatron and LHC experimental collaborations
to reweight the $\pT$ spectrum of the Monte Carlo event generators used in the analysis and
is thus of direct relevance in the Higgs-boson search.

The program {\sc HqT} is based on the transverse-momentum resummation formalism described
in \Brefs{Catani:2000vq, Bozzi:2005wk, Catani:2010pd}, 
which is valid for 
a generic process in which
a high-mass system of non strongly-interacting particles is produced 
in hadron--hadron collisions.

The phenomenological analysis presented 
in \Bref{Bozzi:2005wk} has been recently extended in \Bref{deFlorian:2011xf}. In particular,
the exact values of the NNLO hard-collinear coefficients ${\cal H}_N^{\PH(2)}$
\Brefs{Catani:2007vq,Catani:2011kr} and of the 
NNLL coefficient $A^{(3)}$ \cite{Becher:2010tm} have been implemented.
The ensuing calculation of the $\pT$ spectrum 
is implemented in the updated version of the numerical code {\sc HqT}, 
which can be downloaded from \Bref{hqt}.

\subsubsection{Transverse-momentum resummation}
\label{sec:theory}

In this section we briefly recall the 
main points of the transverse-momentum resummation approach proposed in  
\Brefs{Catani:2000vq,Bozzi:2005wk,Catani:2010pd},
in the case of a Higgs boson $\PH$  produced
by gluon fusion. As recently pointed out in \Bref{Catani:2010pd},
the gluon fusion $\pT$-resummation formula has a 
structure different from the resummation formula for $\PQq\PAQq$ 
annihilation. The difference originates from the 
collinear correlations that are a specific feature 
of the perturbative evolution of colliding
hadrons into gluon partonic initial states.
These gluon collinear correlations produce, in the small-$\pT$ region,
coherent spin correlations between the helicity
states of the initial-state gluons and definite azimuthal-angle correlations
 between the final-states particles of the observed high-mass system.
Both these kinds of correlations have no analogue for $\PQq\PAQq$ annihilation
processes in the small-$\pT$ region.
In the case of Higgs-boson production, $\PH$ being a scalar particle,
the azimuthal correlations vanish and only gluon spin correlations are
present \cite{Catani:2010pd}.
 
We consider the inclusive hard-scattering process
\begin{equation}
\Pp(p_1) + \Pp(p_2) \;\to\; \PH (\MH,\pT ) + X ,   
\label{first}
\end{equation}
where the protons with momenta
$p_1$ and $p_2$ collide to produce the Higgs boson $\PH$ of mass
$\MH$ and transverse momentum $\pT$,
and $X$ is an arbitrary and undetected final state. 

According to the QCD factorisation theorem
the corresponding transverse-momentum 
differential cross 
section $d\sigma_{\PH}/d\pT^2$ can be written as
\begin{equation}
\label{dcross}
\frac{d\sigma_{\PH}}{d \pT^2}(\pT,\MH,s)= \sum_{a,b}
\int_0^1 dx_1 dx_2 \,f_{a/h_1}(x_1,\muF^2)
\,f_{b/h_2}(x_2,\muF^2) \;
\frac{d{\hat \sigma}_{H,ab}}{d \pT^2}(\pT, \MH,{\hat s};
\alphas(\muR^2),\muR^2,\muF^2) 
\;\;,
\end{equation}
where $f_{a/h}(x,\muF^2)$ ($a=q,{\bar q}, g$)
are the parton densities of the colliding hadron $h$ 
at the factorisation scale $\muF$, 
$d\hat\sigma_{H,ab}/d{\pT^2}$ are the perturbative QCD 
partonic cross sections, 
$s$ ($\hat s = x_1 x_2 s$) 
is the square of the 
hadronic (partonic) centre-of-mass  energy, 
and $\muR$ is the renormalisation 
scale. 

In the region where 
$\pT \sim  \MH$,
the QCD perturbative
series is controlled by a small expansion parameter, 
$\alphas(\MH)$,
and fixed-order calculations are
theoretically justified. In this region, 
the QCD radiative corrections are known up to
NLO \cite{deFlorian:1999zd,Ravindran:2002dc,Glosser:2002gm}. 
In the small-$\pT$ region 
($\pT\muchless\MH$),
the convergence of the fixed-order
perturbative expansion is hampered
by the presence 
of large logarithmic terms, 
$\alphas^n\ln^m (\MH^2/\pT^2)$ (with $1\leq m \leq 2n-1)$.
To obtain reliable predictions these terms have to be resummed to all orders.

The resummation is addressed
at the level of the partonic cross section, which
is decomposed~as
\begin{equation}
\label{resplusfin}
\frac{d{\hat \sigma}_{\PH,ab}}{d\pT^2}=
\frac{d{\hat \sigma}_{\PH,ab}^{(\rm res.)}}{d\pT^2}
+\frac{d{\hat \sigma}_{\PH,ab}^{(\rm fin.)}}{d\pT^2}\; .
\end{equation}
The first term on the right-hand side
includes all the logarithmically-enhanced contributions, at small $\pT$,
and has to be evaluated to all orders in $\alphas$.
The second term
is free of such contributions and
can thus be computed at fixed order in the perturbative expansion. 
To correctly take into account the kinematic constraint of
transverse-momentum conservation, the resummation program has to be carried out
in the impact parameter space $b$. 
Using the Bessel transformation between the conjugate variables 
$\pT$ and  $b$,
the resummed component $d{\hat \sigma}^{({\rm res.})}_{\PH,ac}$
can be expressed as
\begin{equation}
\label{resum}
\frac{d{\hat \sigma}_{\PH,ac}^{(\rm res.)}}{d\pT^2}(\pT,\MH,{\hat s};
\alphas(\muR^2),\muR^2,\muF^2) 
= 
\int_0^\infty db \; \frac{b}{2} \;J_0(b \pT) 
\;\Wcal^{\PH}_{ac}(b,\MH,{\hat s};\alphas(\muR^2),\muR^2,\muF^2) \;,
\end{equation}
where $J_0(x)$ is the $0$th-order Bessel function.
The resummation structure of $\Wcal^{\PH}_{ac}$ can 
be organised in exponential form
considering
the Mellin $N$-moments $\Wcal^{\PH}_N$ of $\Wcal^{\PH}$ with respect to the variable 
$z=\MH^2/{\hat s}$ at fixed 
$\MH$\,\footnote{For the sake of simplicity the resummation
formulae is written only for 
the specific case of 
the diagonal terms in the flavour space. 
In general, the exponential
is replaced by an exponential matrix with respect
to the partonic indices (a 
detailed discussion of the general case can be found in
\Ref~\cite{Bozzi:2005wk}).},
\begin{align}
\label{wtilde}
\Wcal^{\PH}_{N}(b,\MH;\alphas(\muR^2),\muR^2,\muF^2)
&={\cal H}_{N}^{\PH}\left(\MH, 
\alphas(\muR^2);\MH^2/\muR^2,\MH^2/\muF^2,\MH^2/Q^2
\right) \nonumber \\
&\times \exp\{{\cal G}_{N}(\alphas(\muR^2),L;\MH^2/\muR^2,\MH^2/Q^2
)\}
\;\;,
\end{align}
were we have defined the logarithmic expansion parameter 
$L\equiv \ln ({Q^2 b^2}/{b_0^2})$,
and $b_0=2e^{-\gamma_E}$ ($\gamma_E=0.5772...$ 
is the Euler number).

The scale $Q\sim \MH$, appearing in the right-hand side of Eq.~(\ref{wtilde}), 
named resummation scale \cite{Bozzi:2005wk}, 
parameterises the
arbitrariness in the resummation procedure.
The  form factor $\exp\{ {\cal G}_N\}$ is 
{\itshape universal} and contains all
the terms $\alphas^nL^m$ with $1 \leq m \leq 2n$, 
that are logarithmically divergent 
as $b \to \infty$ (or, equivalently, $\pT\to 0$).
The exponent ${\cal G}_N$ 
can  be systematically expanded as
\begin{align}
\label{exponent}
{\cal G}_{N}(\alphas, L;\MH^2/\muR^2,\MH^2/Q^2)&=L 
\;g^{(1)}(\alphas L)+g_N^{(2)}(\alphas L;\MH^2/\muR^2,\MH^2/Q^2)\nonumber\\
&+\frac{\alphas}{\pi} g_N^{(3)}(\alphas L;\MH^2/\muR^2,\MH^2/Q^2)
+{\cal O}(\alphas^n L^{n-2})
\end{align}
where the term $L\, g^{(1)}$ resums the leading logarithmic (LL) 
contributions $\alphas^nL^{n+1}$, the function $g_N^{(2)}$ includes
the NLL contributions $\alphas^nL^{n}$ \cite{Catani:1988vd}, 
$g_N^{(3)}$ controls the NNLL 
terms $\alphas^nL^{n-1}$ \cite{deFlorian:2000pr,Becher:2010tm},
and so forth. The explicit form of the functions
$g^{(1)}$, $g_N^{(2)}$, and $g_N^{(3)}$ can be found in \Bref{Bozzi:2005wk}.

The {\itshape process-dependent} function ${\cal H}_N^{\PH}$ 
does not depend on the impact parameter $b$ and 
includes all the perturbative
terms that behave as constants as $b \to \infty$. 
It can thus be expanded in powers of $\alphas=\alphas(\muR^2)$:
\begin{align}
\label{hexpan}
{\cal H}_N^{\PH}(\MH,\alphas;\MH^2/\muR^2,\MH^2/\muF^2,\MH^2/Q^2)&=
\sigma_{\PH}^{(0)}(\alphas,\MH)
\Bigl[ 1+ \frac{\alphas}{\pi} \,{\cal H}_N^{\PH,(1)}(\MH^2/\muF^2,\MH^2/Q^2) 
\Bigr. \nonumber \\
&\hspace*{-5em}
+ \Bigl.
\left(\frac{\alphas}{\pi}\right)^2 
\,{\cal H}_N^{\PH,(2)}(\MH^2/\muR^2,\MH^2/\muF^2,\MH^2/Q^2)+
{\cal O}(\alphas^3)
\Bigr] \;\;,
\end{align}
where $\sigma_{\PH}^{(0)}(\alphas,\MH)$
is the partonic cross section at the Born level.
The
first order ${\cal H}_{N}^{\PH,(1)}$ \cite{Kauffman:1991cx}
and the second order ${\cal H}_{N}^{\PH,(2)}$ \cite{Catani:2007vq,Catani:2011kr}
coefficients  in Eq.~(\ref{hexpan}),
for the case
of Higgs-boson production in the large-$\Mt$ approximation, are known.

To reduce the impact of unjustified higher-order contributions in 
the large-$\pT$ region,
the logarithmic variable $L$ in Eq.~(\ref{wtilde}), 
which diverges for $b\to 0$, 
is actually replaced  by 
${\widetilde L}\equiv \ln \left({Q^2 b^2}/{b_0^2}+1\right)$ \cite{Bozzi:2005wk, Bozzi:2003jy}.
The variables $L$ and ${\widetilde L}$ are equivalent when $Qb\gg 1$ 
(i.e. at small values of $\pT$), but they 
lead to a different behaviour
of the form factor at small values of $b$. 
An important
consequence of this replacement 
is that, after inclusion of the finite component,  
we exactly recover the fixed-order perturbative value of the total cross section
upon integration of the $\pT$  distribution over $\pT$.

The finite component of the transverse-momentum cross section $d\sigma_{\PH}^{({\rm fin.})}$
(see Eq.~(\ref{resplusfin}))
 does not contain large logarithmic terms
in the small-$\pT$ region,
it can thus be evaluated by truncation of the perturbative series
at a given fixed order.

In summary,
 to carry out the resummation at NLL+LO accuracy, we need the
inclusion of the functions $g^{(1)}$, $g_N^{(2)}$,
${\cal H}_N^{\PH,(1)}$, in Eqs.~(\ref{exponent},\ref{hexpan}),
together with the evaluation of the finite component at LO;
the addition of the functions $g_N^{(3)}$ and ${\cal H}_N^{\PH,(2)}$, together 
with the finite component at NLO (i.e.\ at relative ${\cal O}(\alphas^2)$)
leads to the NNLL+NLO 
accuracy.
We point out that our best theoretical prediction
(NNLL+NLO) includes the {\em full} NNLO 
perturbative contribution in the small-$\pT$ region plus
the NLO correction at large $\pT$.
In particular,  the NNLO  result for the total cross section  
is exactly recovered upon integration
over $\pT$ of the differential cross section $d \sigma_{\PH}/d\pT$ at NNLL+NLO
accuracy.

Finally we recall 
that the resummed form factor 
$\exp \{{\cal G}_N(\alphas(\muR^2),{\widetilde L})\}$
has a singular behaviour, related to the presence of the Landau 
pole in the QCD running coupling, at 
the values of $b$ where $\alphas(\muR^2) {\widetilde L} > \pi/\beta_0$ 
($\beta_0$ is the first-order coefficient of the QCD $\beta$ function).
To perform 
the  inverse Bessel
transformation with respect to the impact parameter $b$ a prescription
is thus necessary.
We follow the regularisation prescription of
\Refs~\cite{Laenen:2000de,Kulesza:2002rh}: 
the singularity is avoided by deforming the 
integration contour in the complex $b$ space.

\subsubsection{Results}
\label{sec:results}

The results we are going to present are obtained with an updated version of the 
numerical code {\sc HqT} \cite{hqt}. 
The new version of this code was improved with respect
to the one used in \Ref~\cite{Bozzi:2005wk}. 
The main differences regard the implementation  
of the exact value of the second-order coefficients  
${\cal H}_{N}^{\PH,(2)}$
computed in \Bref{Catani:2007vq}
and the use of the recently derived 
value of the coefficient $A^{(3)}$ \cite{Becher:2010tm},
which  contributes to NNLL accuracy
(the results in \Bref{Bozzi:2005wk} were obtained by using the 
$A^{(3)}$  value from threshold resummation \cite{Moch:2004pa}). 
Detailed results for the $\pT$ spectrum at the Tevatron and the LHC are presented in \Bref{deFlorian:2011xf}. Here we focus on the uncertainties of the $\pT$ spectrum at the LHC.

The calculation is performed strictly in the large-$\Mt$ approximation.
Effects beyond this approximation were discussed in \refS{sec:finitemass}.

The hadronic $\pT$ cross section at NNLL+NLO accuracy
is computed
by using NNLO parton distributions functions (PDFs)
with $\alphas(\muR^2)$ evaluated at 3-loop order.
We use the MSTW2008 parton densities
unless otherwise stated.

As discussed in \refS{sec:theory}, the 
resummed calculation depends on the factorisation and 
renormalisation scales and on the resummation scale $Q$. 
Our convention to compute factorisation 
and renormalisation scale uncertainties is to consider
independent variations of $\muF$ and $\muR$ by a factor of two around 
the central values $\muF=\muR=\MH$
(i.e. we consider the range $\MH/2< \{\muF,\muR\}< 2\,\MH$), with the constraint
$0.5 < \muF/\muR < 2$. 
Similarly, we follow \Bref{Bozzi:2005wk} and
choose $Q=\MH/2$ as central value of the resummation scale,
considering scale variations in the 
range $\MH/4 < Q < \MH$.

We focus on the uncertainties on the {\em normalised} $\pT$ spectrum (i.e., $1/\sigma \times d\sigma/d\pT$). As mentioned in \refS{sec:hqt-intro}, the
typical procedure of the experimental collaborations is to use the information
on the total cross section \cite{Dittmaier:2011ti} to rescale the best theoretical predictions of Monte Carlo
event generators, whereas the
NNLL+NLO result for the spectrum, obtained with the public program {\sc HqT},
is used to reweight the transverse-momentum
spectrum of the Higgs boson in the simulation.
Such a procedure implies that the important information provided by the resummed NNLL+NLO spectrum
is not its integral, i.e.\ the total cross section, but its {\em shape}.
The sources of uncertainties on the shape of the spectrum are essentially the same
as for the inclusive cross section: the uncertainty from missing higher-order contributions,
estimated through scale variations, PDF uncertainties,
and the uncertainty from the use of the large-$\Mt$ approximation, which is discussed in \Section~\ref{sec:finitemass}.
One additional uncertainty in the $\pT$ spectrum that needs be considered comes from non-perturbative (NP) effects.

It is known that the transverse-momentum
distribution is affected by NP effects, which become important as $\pT$ becomes small.
A customary way of modelling these effects is to introduce an NP transverse-momentum smearing
of the distribution. In the case of resummed calculations in impact-parameter space,
the NP smearing is implemented by multiplying the $b$-space perturbative
form factor by an NP form factor.
The parameters controlling this NP form factor are typically obtained through a comparison to data.
Since the Higgs boson has not been discovered yet,
the way to fix the NP form factor is somewhat arbitrary.
Here we follow the procedure adopted in \Brefs{Bozzi:2005wk,deFlorian:2011xf}, and we multiply the resummed form factor in Eq.~(\ref{resum}) by a gaussian smearing $S_{NP}=\exp\{-g b^2\}$, where the parameter $g$ is taken in the range ($g=1.67{-}5.64\UGeV^2$) suggested by the study of 
\Bref{Kulesza:2003wi}.
The above procedure can give us some insight on the quantitative impact of these NP effects on the Higgs-boson spectrum.

\begin{figure}[htb]
\vspace{5pt}
\begin{center}
\begin{tabular}{cc}
\includegraphics[width=.46\linewidth]{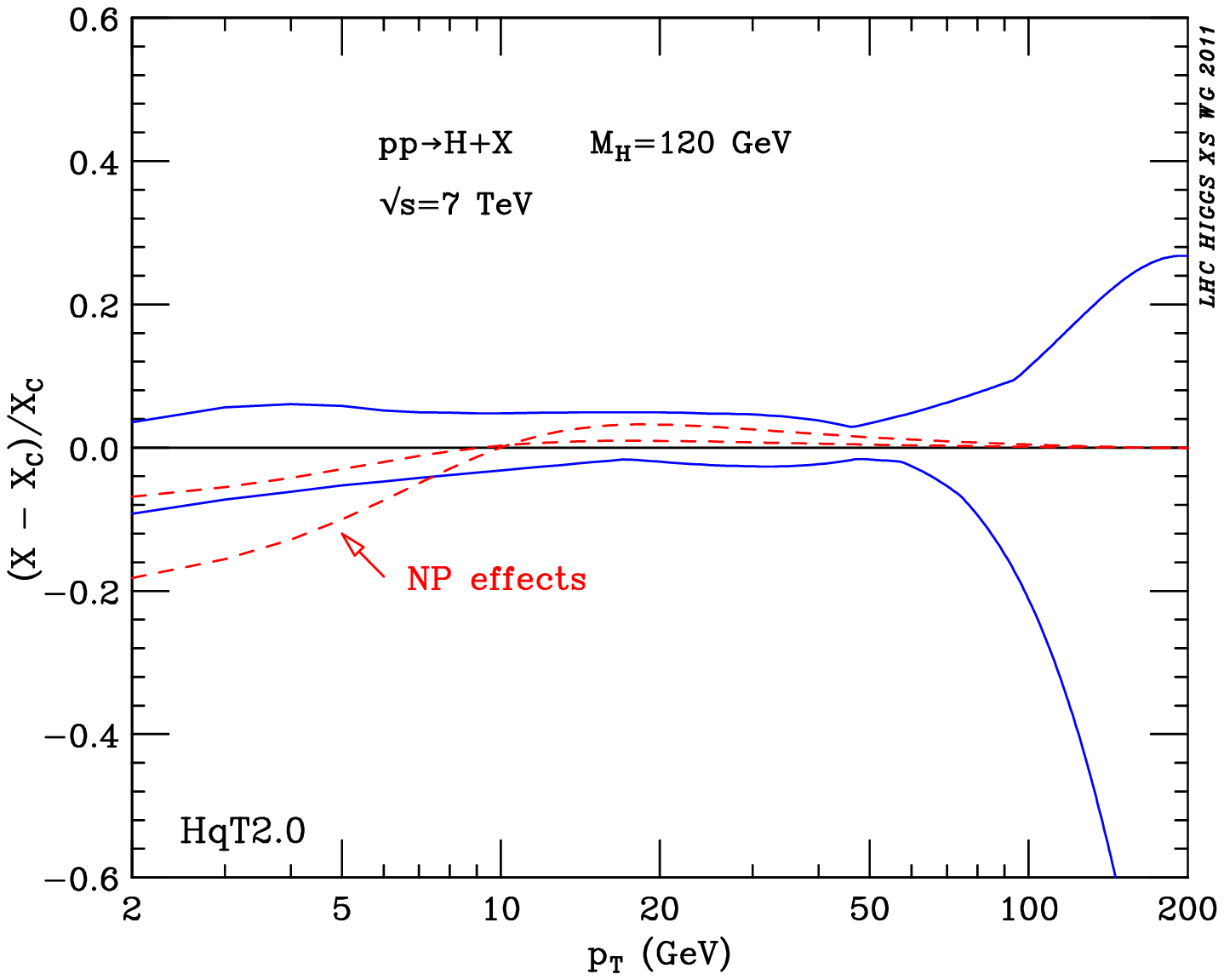} &
\includegraphics[width=.46\linewidth]{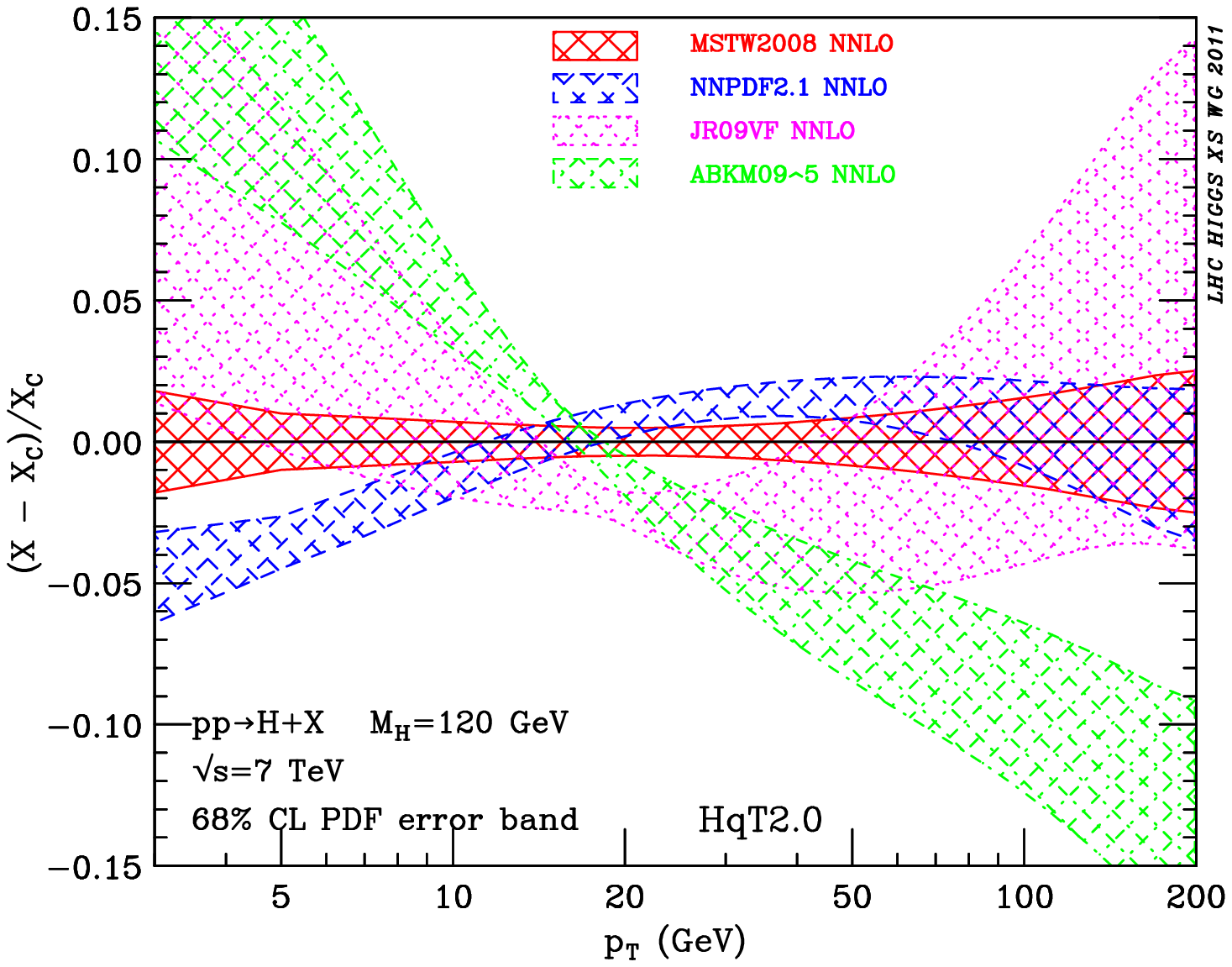}\\
\includegraphics[width=.46\linewidth]{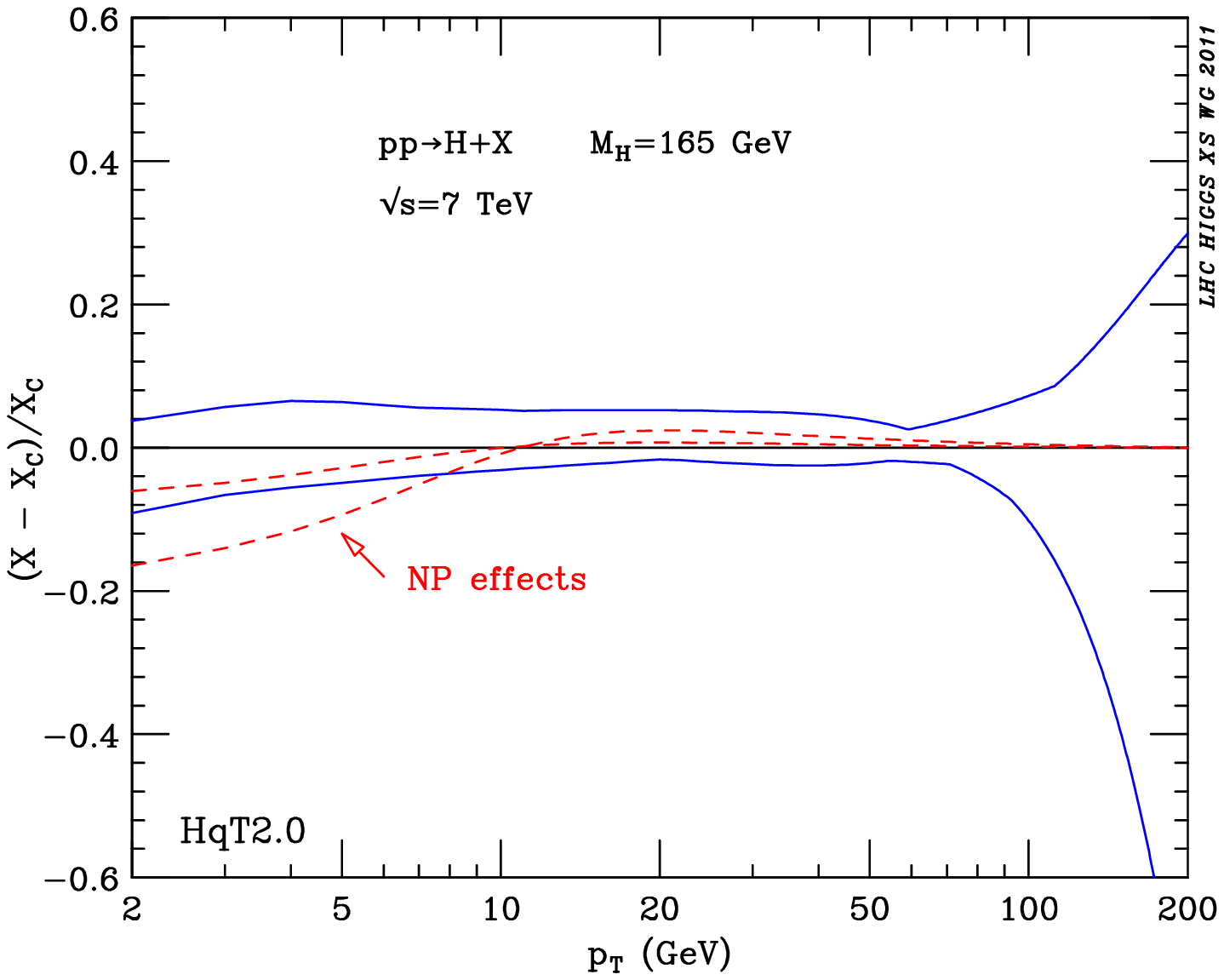} &
\includegraphics[width=.46\linewidth]{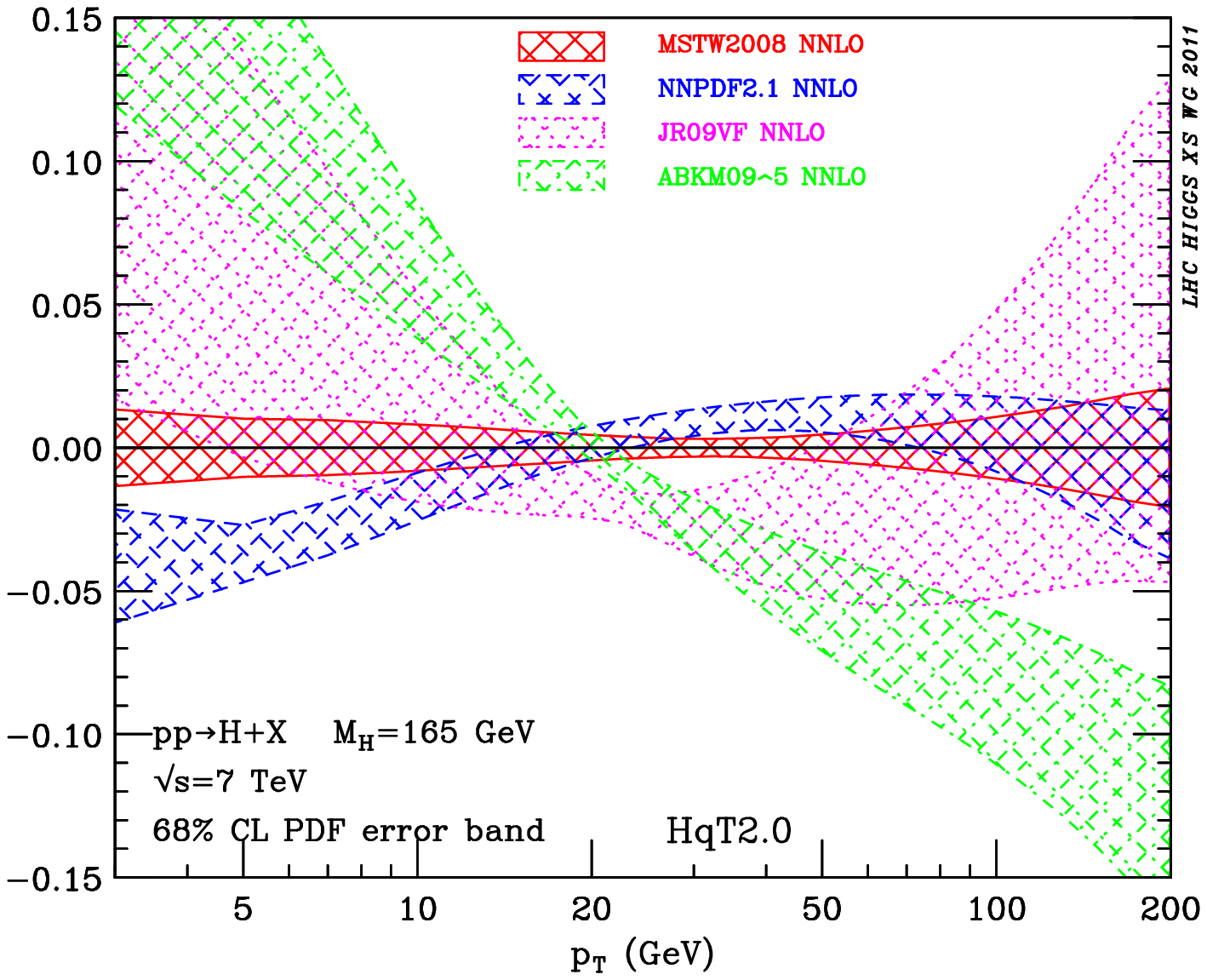}\\
\end{tabular}
\end{center}
\caption{\label{gghqt_fig} Uncertainties in the shape of the $\pT$ spectrum: scale uncertanties compared with NP effects (left panels); PDF uncertainties (right panels).}
\label{fig:comp}
\end{figure}

In \refF{gghqt_fig} (left panels) we compare the NNLL+NLO shape uncertainty as coming from scale variations (solid lines)
to the NP effects (dashed lines) for $\MH=120\UGeV$ and $\MH=165\UGeV$.
Scale variations are performed as follows:
we independently vary $\muF,\muR$ and $Q$
in the ranges
$\MH/2< \{\muF,\muR\} < 2\MH$ and $\MH/4< Q< \MH$,
with the constraints $0.5 < \muF/\muR < 2$ and 
$0.5 < Q/\muR < 2$.
The bands are obtained in practice by normalizing each spectrum to unity, and computing the
relative difference with respect to the central normalised prediction obtained
with the MSTW2008 NNLO set (with $g=0$).
In other words, studying uncertainties on the normalised distribution allows
us to assess the true uncertainty in the shape of the resummed $\pT$ spectrum.

We see that, both for $\MH=120\UGeV$ and for $\MH=165\UGeV$, the scale uncertainty ranges from about $\pm 4\%$ at $\pT\sim 10\UGeV$ to $\pm 5\%$ at $\pT\sim 70\UGeV$.
As $\pT$ increases, the scale uncertainty rapidly increases.
This should not be considered as particularly worrying, since for large
transverse momenta, the resummed result looses predictivity,
and should be replaced by the standard fixed-order result.
The impact of NP effects ranges from about $10\%$ to $20\%$
in the very small-$\pT$ region ($\pT\ltap 10\UGeV$),
is about $3{-}4\%$ for $\pT\sim 20\UGeV$, and quickly decreases as $\pT$ increases.
We conclude that the uncertainty from unknown NP effects is smaller than the scale uncertainty,
and is comparable to the latter only in the very-small-$\pT$ region.

The impact of PDF uncertainties at $68\%$ CL on the shape of the $\pT$ spectrum
is studied in \refFs{gghqt_fig} (right panels).
By evaluating PDF uncertainties with MSTW2008 NNLO PDFs
we see that the uncertainty is at the $\pm (1{-}2)\%$ level.
The use of different PDF sets affects not only the absolute value of the NNLO cross section (see, \eg \Bref{Watt:2011kp}), but also the shape of the $\pT$ spectrum. The predictions obtained with NNPDF~2.1 PDFs are in good agreement with those obtained with the MSTW2008 set, and
the uncertainty bands overlap over a wide range of transverse momenta, both for $\MH=120\UGeV$ and $\MH=165\UGeV$.
On the contrary, the prediction obtained with the ABKM09 NNLO set is softer,
and the uncertainty band does not overlap with the MSTW2008 band.
This behaviour is not completely unexpected: when the Higgs boson is produced
at large transverse momenta, larger values of Bjorken $x$ are probed, where the ABKM gluon is smaller than MSTW2008 one.
The JR09 band shows a good compatibility with the MSTW2008 result
where the uncertainty is, however, rather large.

\subsection{Monte Carlo and resummation for exclusive jet bins%
\footnote{I. W. Stewart, F. St\"ockli, F. J. Tackmann and W. J. Waalewijn.}}

\label{sec:jetvetosum}


\subsubsection{Overview}

In this section we discuss the use of predictions for exclusive jet cross
sections that involve resummation for the jet-veto logarithms,
$\ln(p^\cut/\MH)$, induced by the jet cut parameter $p^\cut$ shown in
Eq.~\eqref{eq:pcut} of \refS{sec:jetveto}.  It is common practice to
account for the $p^\cut$ dependence in exclusive $H+\text{0 jet}$ cross sections
using Monte Carlo programs such as {\sc Pythia}, MC@NLO, or POWHEG, which include a
resummation of at least the leading logarithms (LL). One may also improve the
accuracy of spectrum predictions by reweighting the Monte Carlo to resummed
predictions for the Higgs $q_T$ spectrum (see \refS{sec:hqt}),
which starts to differ from the
jet-veto spectrum at ${\cal O}(\alphas^2)$. The question then remains how to
assess theoretical uncertainties, and three methods (A, B, and C) were outlined
in \refS{sec:jetveto}.

In this section we consider assessing the
perturbative uncertainties when using resummed predictions for variables $p^\cut$
that implement a jet veto, corresponding to method C. An advantage of
using these resummed predictions with method C is that they contain perturbation theory
scale parameters which allow for an evaluation of two components of the
theory error, one which is $100\%$ correlated with the total cross section (as in
method A), and one related to the presence of the jet-bin cut which is
anti-correlated between neighboring jet bins (as in method B). Our discussion of
the correlation matrix obtained from method C follows
\Bref{Stewart:2011cf}.

We consider two choices for the jet-veto variable, the standard $\pT^\jet$
variable with a rapidity cut $|\eta| < \eta^\cut$ (using anti-$k_T$ with $R = 0.5$),
and the beam thrust variable~\cite{Stewart:2009yx}, which is a rapidity-weighted $H_T$, defined as
\begin{equation} \label{eq:Tau_def}
\Tcm = \sum_k | \vec p_{Tk}| e^{-\abs{\eta_k}} = \sum_k (E_k - |p_k^z|)
\,.\end{equation}
The sum here is over all objects in the final state except the Higgs decay products, and
can in principle be considered over particles, topo-clusters, or
jets with a small $R$ parameter. We make use of resummed predictions for
$H+0$ jets from gluon fusion at next-to-next-to-leading logarithmic order with
NNLO fixed-order precision (NNLL+NNLO) from Ref.~\cite{Berger:2010xi}. The
resulting cross section $\sigma_0(\Tcmc)$ has the jet veto implemented by a cut
$\Tcm< \Tcmc$.  This cross section contains a resummation of large logarithms
at two orders beyond standard LL parton-shower programs, and also includes the
correct NNLO corrections for $\sigma_0(\Tcmc)$ for any cut.

A similar resummation for the case of $\pT^\jet$ is not available. Instead, we
use MC@NLO and reweight it to the resummed predictions in $\Tcm$ including
uncertainties and then use the reweighted Monte Carlo sample to obtain 
cross-section predictions for the standard jet veto, $\sigma_0(\pT^\cut)$. We will
refer to this as the reweighted NNLL+NNLO result. Since the Monte Carlo here is
only used to provide a transfer matrix between $\Tcm$ and $\pT^\jet$, and both
variables implement a jet veto, one expects that most of the improvements from
the higher-order resummation are preserved by the reweighting. However, we
caution that this is not equivalent to a complete NNLL+NNLO result for the
$\pT^\cut$ spectrum, since the reweighting may not fully capture effects
associated with the choice of jet algorithm and other effects that enter at this
order for $\pT^\cut$. The dependence on the Monte Carlo transfer matrix also
introduces an additional uncertainty, which should be studied and is not
included in our numerical results below. (We have checked that varying the
fixed-order scale in MC@NLO between $\mu = \MH/4$ and $\mu = 2\MH$ has a very
small effect on the reweighted results.) The transfer matrix is obtained at
the parton level, without hadronisation or underlying event, since we are
reweighting a partonic NNLL+NNLO calculation. In all our results
we consistently use MSTW2008 NNLO PDFs.

In \refS{jetbin_resum} we discuss the determination of perturbative
uncertainties in the resummed calculations (method C), and compare them with those at NNLO
obtained from a direct exclusive scale variation (method A) and from combined
inclusive scale variation (method B) for both $\sigma_0(\pT^\cut)$ and
$\sigma_0(\Tcmc)$. In \refS{jetbin_resumandMC} we compare the predictions
for the 0-jet bin event fraction at different levels of resummation, comparing
results from NNLO, MC@NLO, and the (reweighted) NNLL+NNLO analytic results.

\subsubsection{Uncertainties and correlations from resummation}
\label{jetbin_resum}

The resummed $H+\text{0-jet}$ cross section predictions of
Ref.~\cite{Berger:2010xi} follow from a factorisation theorem for the $0$-jet
cross section~\cite{Stewart:2009yx}, $\sigma_0(\Tcmc) = H\, {\cal I}_{gi}\,{\cal
  I}_{gj}\otimes S f_i f_j$, where $H$ contains hard virtual effects, the ${\cal
  I}$s and $S$ describe the veto-restricted collinear and soft radiation, and
the $f$s are standard parton distributions. Fixed-order perturbation theory is
carried out at three scales, a hard scale $\mu_H^2\sim \MH^2$ in $H$, and beam
and soft scales $\mu_B^2\sim \MH\Tcmc$ and $\mu_S^2\sim (\Tcmc)^2$ for ${\cal
  I}$ and $S$, and are then connected by NNLL renormalisation group evolution
that sums the jet-veto logarithms, which are encoded in ratios of these scales.
The perturbative uncertainties can be assessed by considering two sources: i) an
overall scale variation that simultaneously varies $\{\mu_H,\mu_B,\mu_S\}$ up
and down by a factor of two which we denote by $\Delta_{H0}$, and ii) individual
variations of $\mu_B$ or $\mu_S$ that each hold the other two scales
fixed~\cite{Berger:2010xi}, whose envelope we denote by the uncertainty
$\Delta_{SB}$. Here $\Delta_{H0}$ is dominated by the same sources of
uncertainty as the total cross section $\sigma_\total$, and hence should be
considered $100\%$ correlated with its uncertainty $\Delta_\total$. The uncertainty
$\Delta_{SB}$ is only present due to the jet bin cut, and hence gives the
$\Delta_\cut$ uncertainty discussed in \refS{sec:jetveto} that is
anti-correlated between neighboring jet bins.

If we simultaneously consider the cross sections $\{\sigma_0, \sigma_{\ge 1}\}$ then the full
correlation matrix with method C is
\begin{align} \label{eq:resummatrix}
C &=
\begin{pmatrix}
 \Delta_{SB}^2 &  - \Delta_{SB}^2 \\
-\Delta_{SB}^2 & \Delta_{SB}^2
\end{pmatrix}
+
\begin{pmatrix}
\Delta_{H0}^2 &  \Delta_{H0}\,\Delta_{H\geq 1}  \\
\Delta_{H0}\,\Delta_{H\geq 1} & \Delta_{H\geq 1}^2
\end{pmatrix}
,\end{align}
where $\Delta_{H\geq 1} =\Delta_\total - \Delta_{H0}$ encodes the $100\%$ correlated
component of the uncertainty for the $(\ge 1)$-jet inclusive cross section.
Computing the uncertainty in $\sigma_\total$ gives back $\Delta_\total$.
Eq.~\eqref{eq:resummatrix} can be compared to the corresponding correlation
matrix from method A, which would correspond to taking $\Delta_{SB}\to 0$ and
obtaining the analog of $\Delta_{H0}$ by up/down scale variation without
resummation ($\mu_H=\mu_B=\mu_S$).  It can also be compared to method B, which
would correspond to taking $\Delta_{SB}\to \Delta_{\ge 1}$ and $\Delta_{H\ge
  1}\to 0$, such that $\Delta_{H0}\to \Delta_\total$. Using method C captures
both of the types of uncertainty that appear in methods A and B. Note that the
numerical dominance of $\Delta_{SB}^2$ over $\Delta_{H0}\Delta_{H\ge 1}$ in the
$0$-jet region is another way to justify the
preference for using method B when given a choice between methods A and B.
For example, for $\pT^\cut = 30\UGeV$ and $|\eta^\jet| < 5.0$ we have
$\Delta_{SB}^2 = 0.17$ and $\Delta_{H0}\Delta_{H\ge 1} = 0.02$.
From \eqr{resummatrix} it is straightforward to derive the uncertainties and
correlations in method C when considering the $0$-jet event fraction, $\{\sigma_\total,f_0\}$,
in place of the jet cross sections. We will discuss results for $f_0(\pT^\cut)$
and $f_0(\Tcmc)$ in \refS{jetbin_resumandMC} below.

In \refF{fig:3scales} we show the uncertainties $\Delta_{SB}$ (light green)
and $\Delta_{H0}$ (medium blue) as a function of the jet-veto variable, as well
as the combined uncertainty adding these components in quadrature (dark orange).
From the figure we see
that the $\mu_{H0}$ dominates at large values where the veto is turned off and
we approach the total cross section, and that the jet-cut uncertainty $\Delta_{SB}$
dominates for the small cut values that are typical of experimental analyses
with Higgs jet bins. The same pattern is observed in the left panel which
directly uses the NNLL+NNLO predictions for $\Tcmc$, and the right panel which
shows the result from reweighting these predictions to $\pT^\cut$ as explained above.

\begin{figure}
\includegraphics[width=0.5\textwidth]{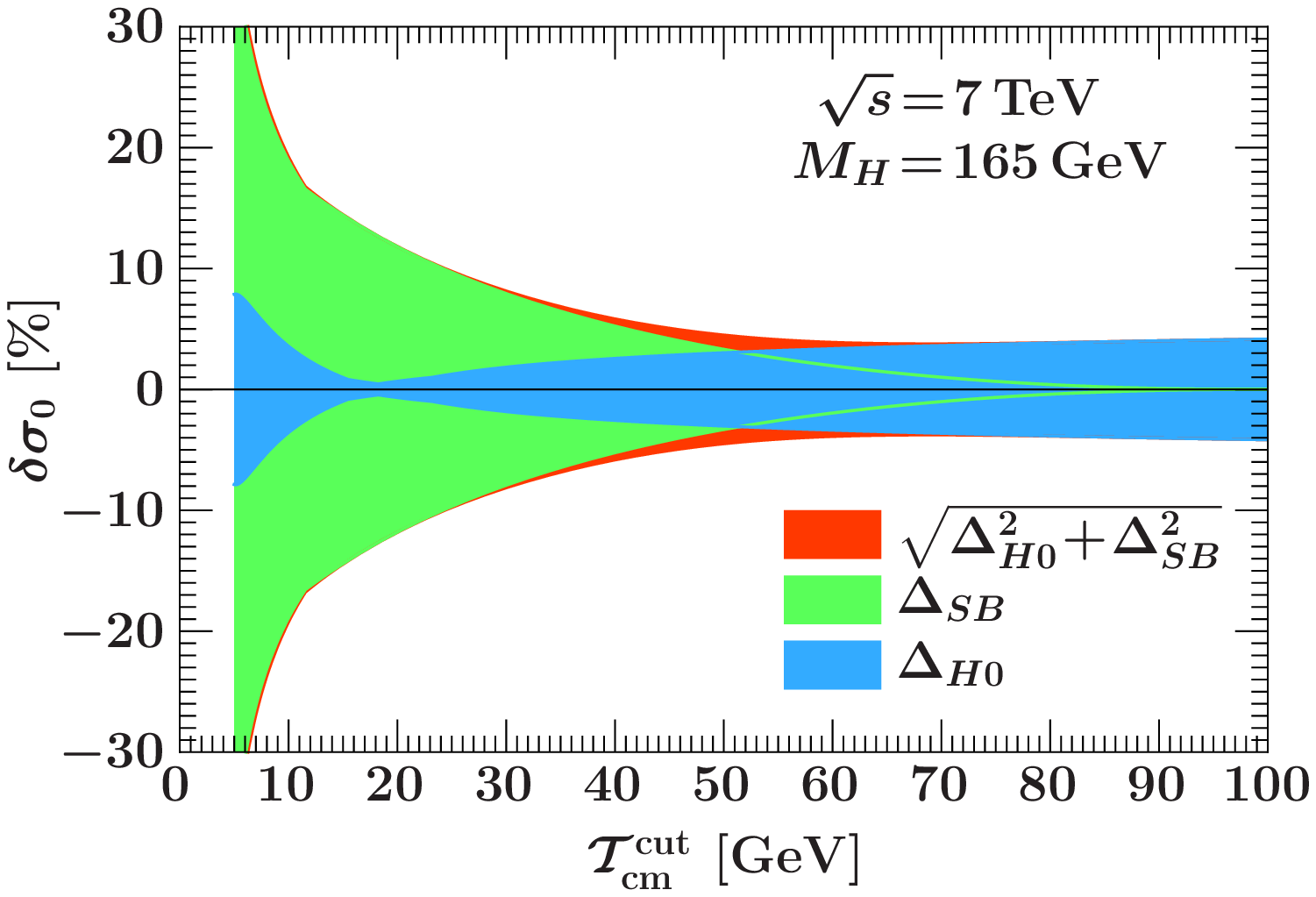}%
\hfill%
\includegraphics[width=0.5\textwidth]{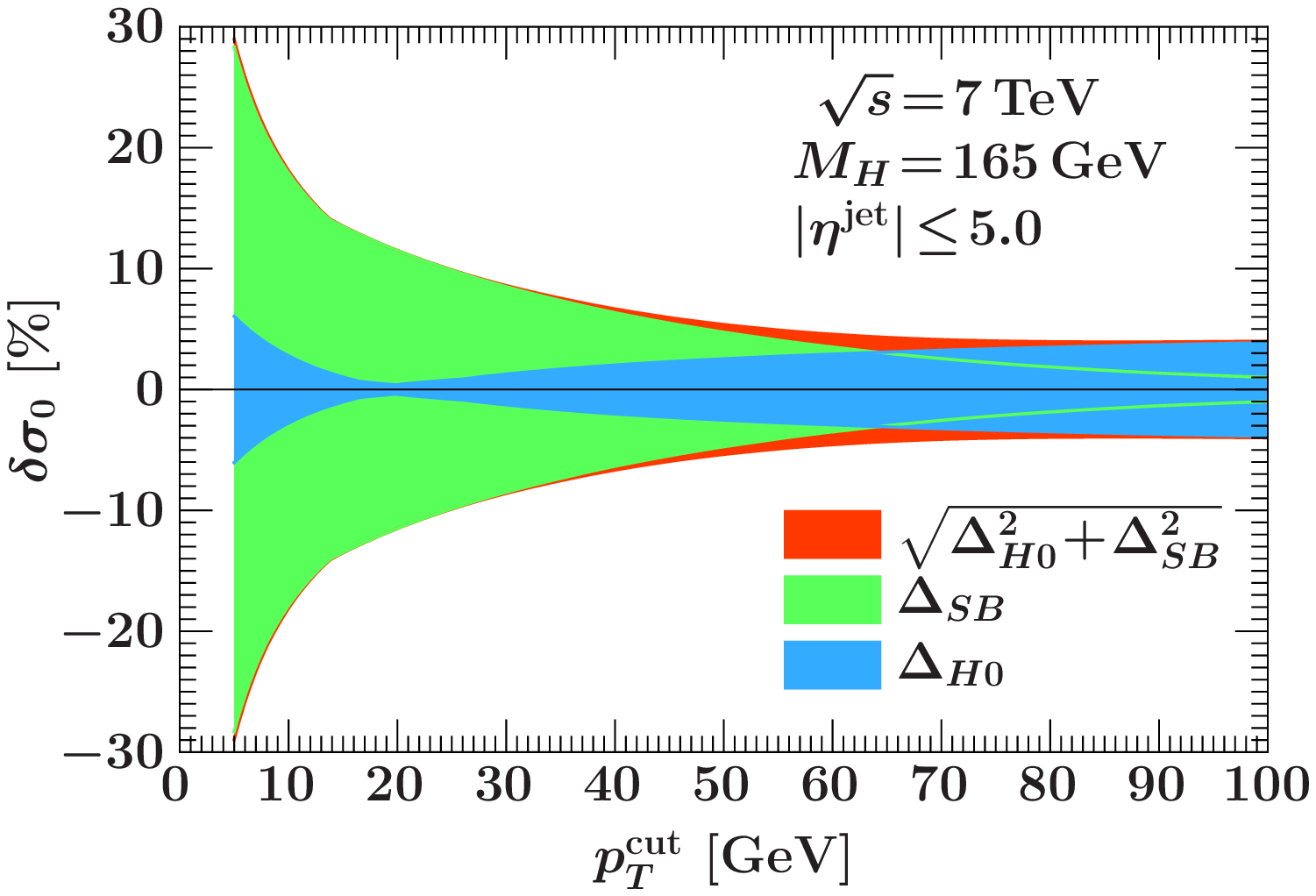}%
\vspace{-0.5ex}
\caption{\label{fig:3scales} Relative uncertainties for the 0-jet bin cross
  section from resummation at NNLL+NNLO for beam thrust $\Tcm$ on the left and
  $\pT^{\rm jet}$ on the right.}
\end{figure}

\begin{figure}
\includegraphics[width=0.5\textwidth]{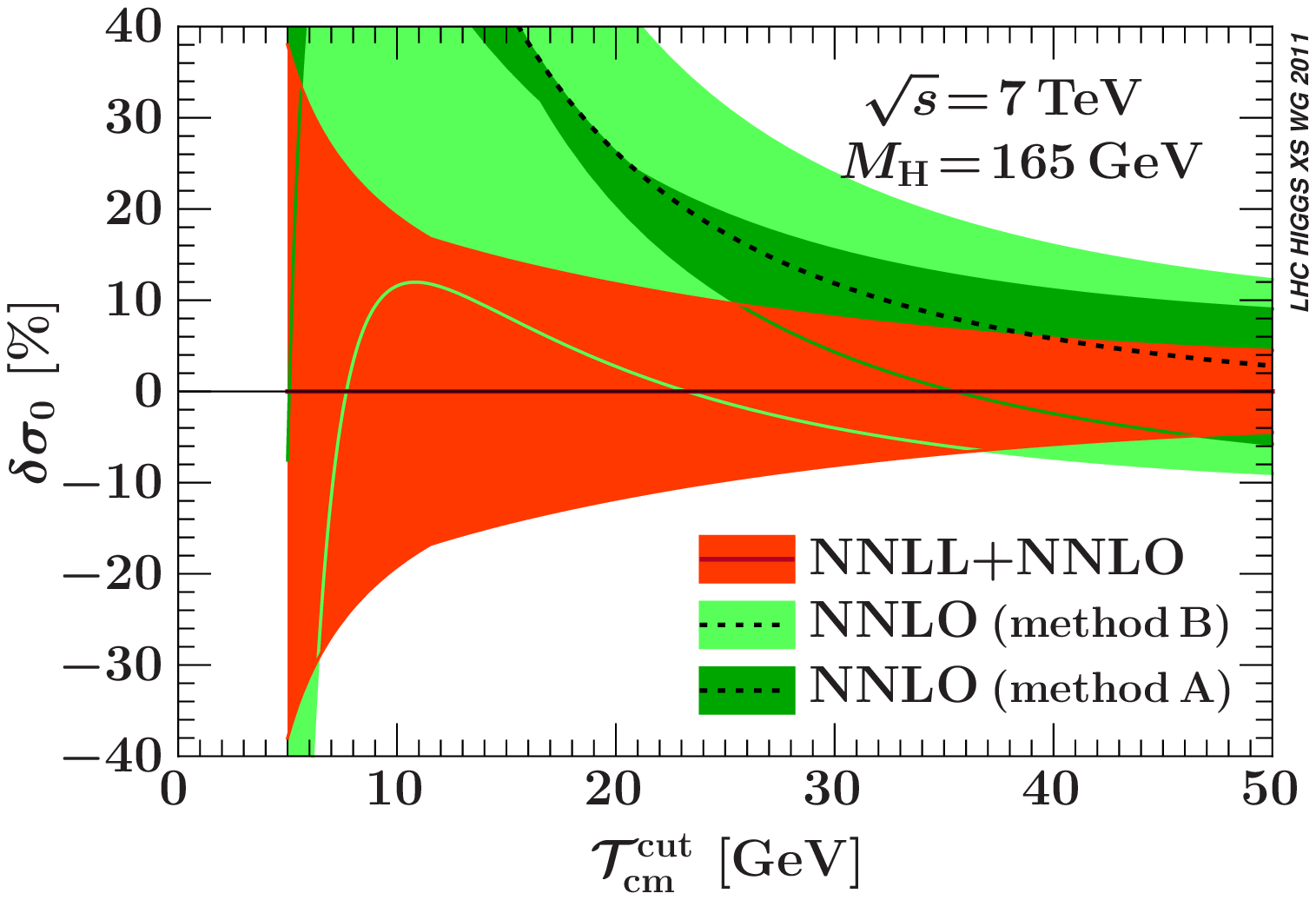}%
\hfill%
\includegraphics[width=0.5\textwidth]{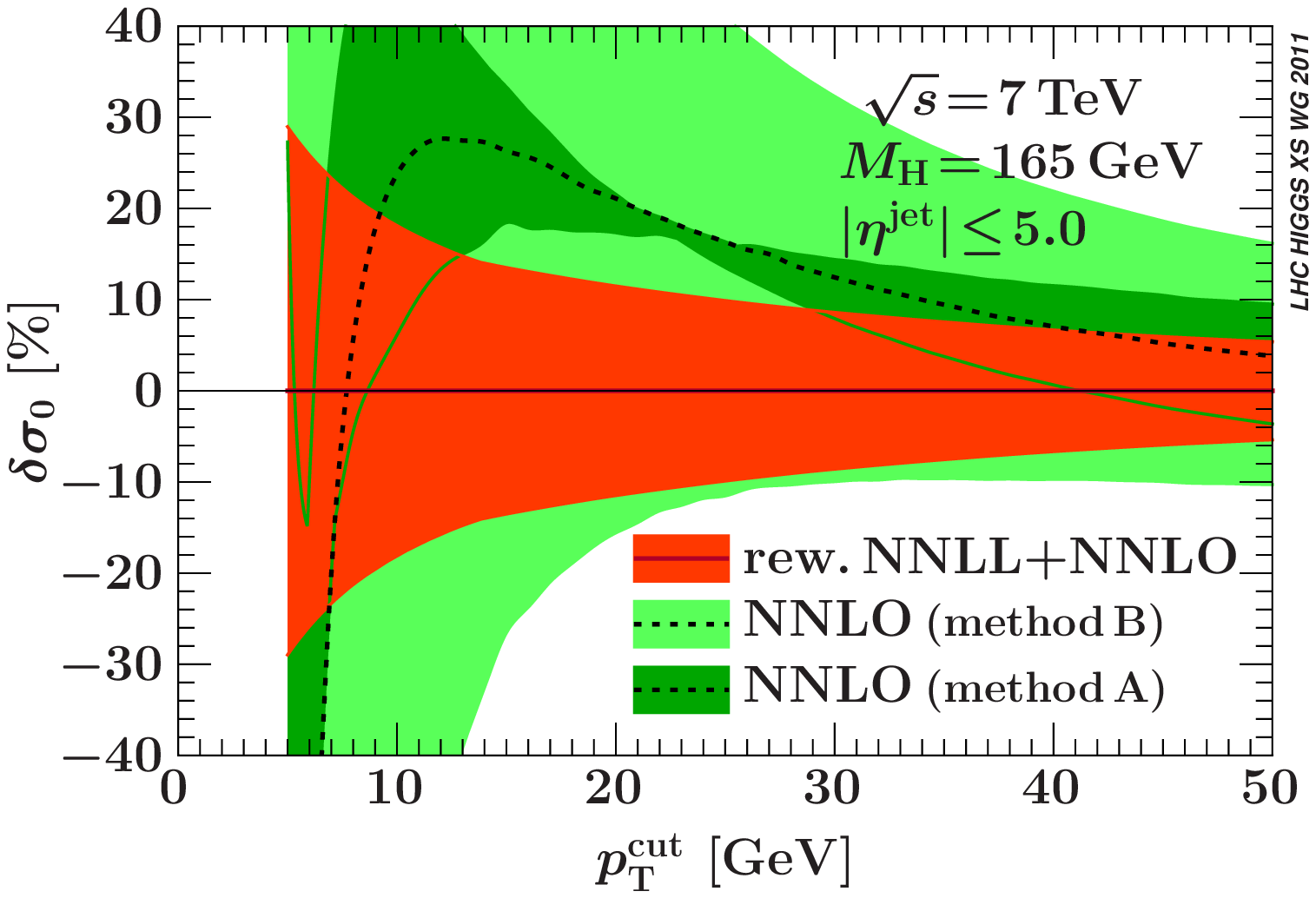}%
\vspace{-0.5ex}
\caption{\label{fig:ABC} Comparison of uncertainties for the
  0-jet bin cross section for beam thrust $\Tcm$ on the left and $\pT^{\rm jet}$ on
  the right. Results are shown at NNLO using methods A and B (direct exclusive
  scale variation and combined inclusive scale variation as discussed in
  \refS{sec:jetveto}), and for the NNLL+NNLO resummed result (method C).
  All curves are normalised relative to the NNLL+NNLO central value.}
\end{figure}

A comparison of the uncertainties for the 0-jet bin cross section from methods A
(medium green), B (light green), and C (dark orange) is shown in
\refF{fig:ABC}, where the results are normalised to the highest-order result
to better show the relative differences and uncertainties.  The NNLO
uncertainties in methods A and B are computed in the manner discussed in
\refS{sec:jetveto}. The uncertainties in method C are the combined
uncertainties from resummation given by $\sqrt{\Delta_{H0}^2+\Delta_{SB}^2}$ in
\refF{fig:3scales}.  In the left panel we use $\Tcmc$ as jet-veto
variable and full results for the NNLO and NNLL+NNLO cross sections, while in
the right panel we use $\pT^\cut$ as jet-veto variable with the full NNLO and
the reweighted NNLL+NNLO results. One observes that the resummation of the large
jet-veto logarithms lowers the cross section in both cases. For cut values
$\gtrsim 25\UGeV$ the relative uncertainties in the resummed result and the
reduction in the resummed central value compared to NNLO are similar for both
jet-veto variables. One can also see that the NNLO uncertainties from method B
are more consistent with the higher-order NNLL+NNLO-resummed results than those
in method A.

From \refF{fig:ABC} we observe that the uncertainties in method C including
resummation (dark orange bands) are reduced by about a factor of two compared
to those in method B (light green bands).
Since the $0$-jet bin plays a crucial role in the $\PH\to \PW\PW$ channel for Higgs
searches, and these improvements will also be reflected in uncertainties for the
$1$-jet bin, the improved theoretical precision obtained with method C has the
potential to be quite important.

\begin{table}[t]
\tabcolsep 10pt
  \centering
  \begin{tabular}{c c c c }
  \hline\hline
  & method A & method B & method C \\
  \hline
  $\delta\sigma_0(\pT^\cut)$         & $3\%$  & $19\%$  & $9\%$ \\
  $\delta\sigma_{\ge 1}(\pT^\cut)$   & $19\%$ & $19\%$  & $14\%$ \\
  $\rho(\sigma_\total,\sigma_0)$       & $1$    & $0.78$  & $0.15$ \\
  $\rho(\sigma_\total,\sigma_{\ge 1})$ & $1$    & $0$     & $0.65$ \\
  $\rho(\sigma_0,\sigma_{\ge 1})$    & $1$    & $-0.63$ & $-0.65$ \\
\hline
  $\delta f_0(\pT^\cut)$             & $6\%$  & $13\%$  & $9\%$ \\
  $\delta f_{\ge 1}(\pT^\cut)$       & $10\%$ & $21\%$  & $11\%$ \\
  $\rho(\sigma_\total,f_0)$            & $-1$   & $0.43$  & $-0.38$ \\
  $\rho(\sigma_\total,f_{\ge 1})$      & $1$    & $-0.43$ & $0.38$ \\
  \hline
  \end{tabular}
  \caption{Example of relative uncertainties $\delta$ and correlations $\rho$ obtained for the
    LHC at $7\,{\rm TeV}$ for $\pT^\cut=30\,{\rm GeV}$ and $\lvert\eta^\jet\rvert < 5.0$.
    (Method A is shown for illustration only and should not be used for the reasons discussed in
    \refS{sec:jetveto}.)
    }
\label{tab:corrcoeffs}
\end{table}
%
To appreciate the effects of the different methods on the correlation matrix we
consider as an example the results for $\pT^\cut = 30\UGeV$ and
$\lvert\eta^\jet\rvert < 5.0$.  The inclusive cross sections are $\sigma_\total
= (8.76 \pm 0.80)\,\mathrm{pb}$ at NNLO, and $\sigma_{\geq1} =
(3.10\pm0.61)\,\mathrm{pb}$ at NLO. The relative uncertainties and correlations
at these cuts for the three methods are shown in \refT{tab:corrcoeffs}.
The numbers for the cross sections are also translated into the equivalent
results for the event fractions, $f_0(\pT^\cut)=\sigma_0(\pT^\cut)/\sigma_\total$
and $f_{\ge 1}(\pT^\cut) = \sigma_{\ge 1}(\pT^\cut)/\sigma_\total$. Note that 
method A should not be used for the reasons discussed in detail in \refS{sec:jetveto},
which are related to the lack of a contribution analogous to $\Delta_{SB}$ in this method, and
the resulting very small and underestimated $\delta\sigma_0$. In methods B and C we
see, as expected, that $\sigma_0$ and $\sigma_{\geq 1}$ have a substantial
anti-correlation due to the jet-bin boundary they share.

In \refS{sec:jetveto}, method B was discussed for
$\{\sigma_\total,\sigma_0,\sigma_1\}$, where we also account for the jet-bin
boundary between $\sigma_1$ and $\sigma_{\ge 2}$. The method C results discussed
here so far are relevant for the jet-bin boundary between $\sigma_0$ and
$\sigma_{\ge 1}$.  To also separate $\sigma_{\ge 1}$ into a $1$-jet bin
$\sigma_1$ and a $\sigma_{\ge 2}$ one can simply use method B for this boundary
by treating $\Delta_{\ge 2}$ as uncorrelated with the total uncertainty
$\Delta_{\ge 1}$ from method C. Once it becomes available one can also use a
resummed prediction with uncertainties for this boundary with method C.

\subsubsection{Comparison of NNLO, MC@NLO, and resummation at NNLL+NNLO}
\label{jetbin_resumandMC}

In this section we compare the results for the $0$-jet event fraction $f_0$ from
different theoretical methods including various levels of logarithmic
resummation. We use the event fraction for this comparison since it is the
quantity used in experimental analyses and what is typically provided by the
Monte Carlo. We compare three different results using both $\pT^{\rm jet}$ and
beam thrust as jet-veto variables:
\begin{enumerate}
\item Fixed-order perturbation theory at NNLO without resummation, where the
  uncertainties are evaluated using method B.
\item MC@NLO, which includes the LL (and partial NLL) resummation provided by
  the parton shower. For the uncertainties we use the relative NNLO
  uncertainties evaluated using method B.
\item Resummation at NNLL+NNLO, with the uncertainties provided by the
  resummation (method C). For beam thrust we directly compare to the full result
  at this order. For $\pT^{\rm \jet}$ we use the resummed beam-thrust result
  reweighted to $\pT^{\rm jet}$ using Monte Carlo as explained at the beginning
  of this section.
\end{enumerate}

\begin{figure}
\includegraphics[width=0.5\textwidth]{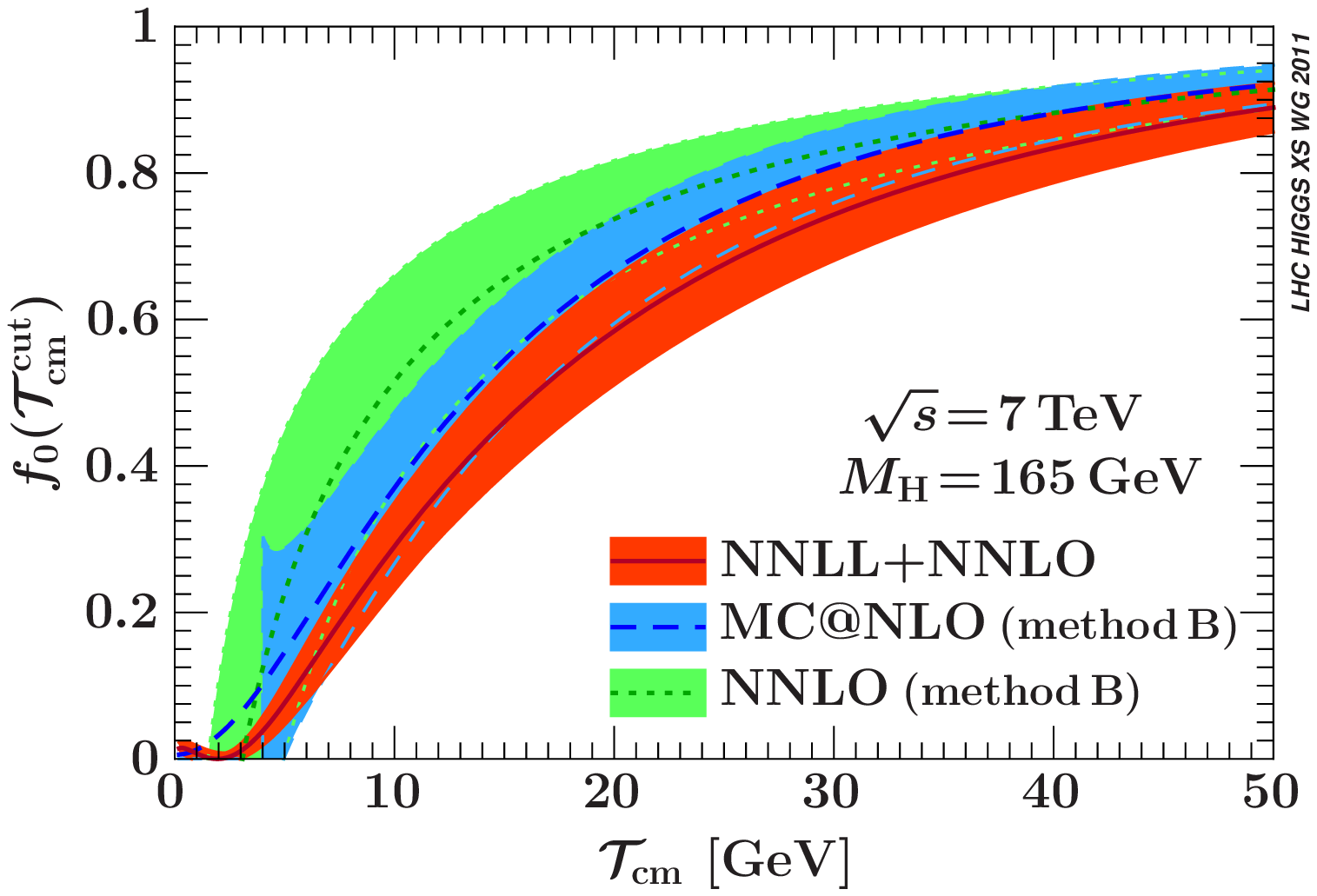}%
\hfill%
\includegraphics[width=0.5\textwidth]{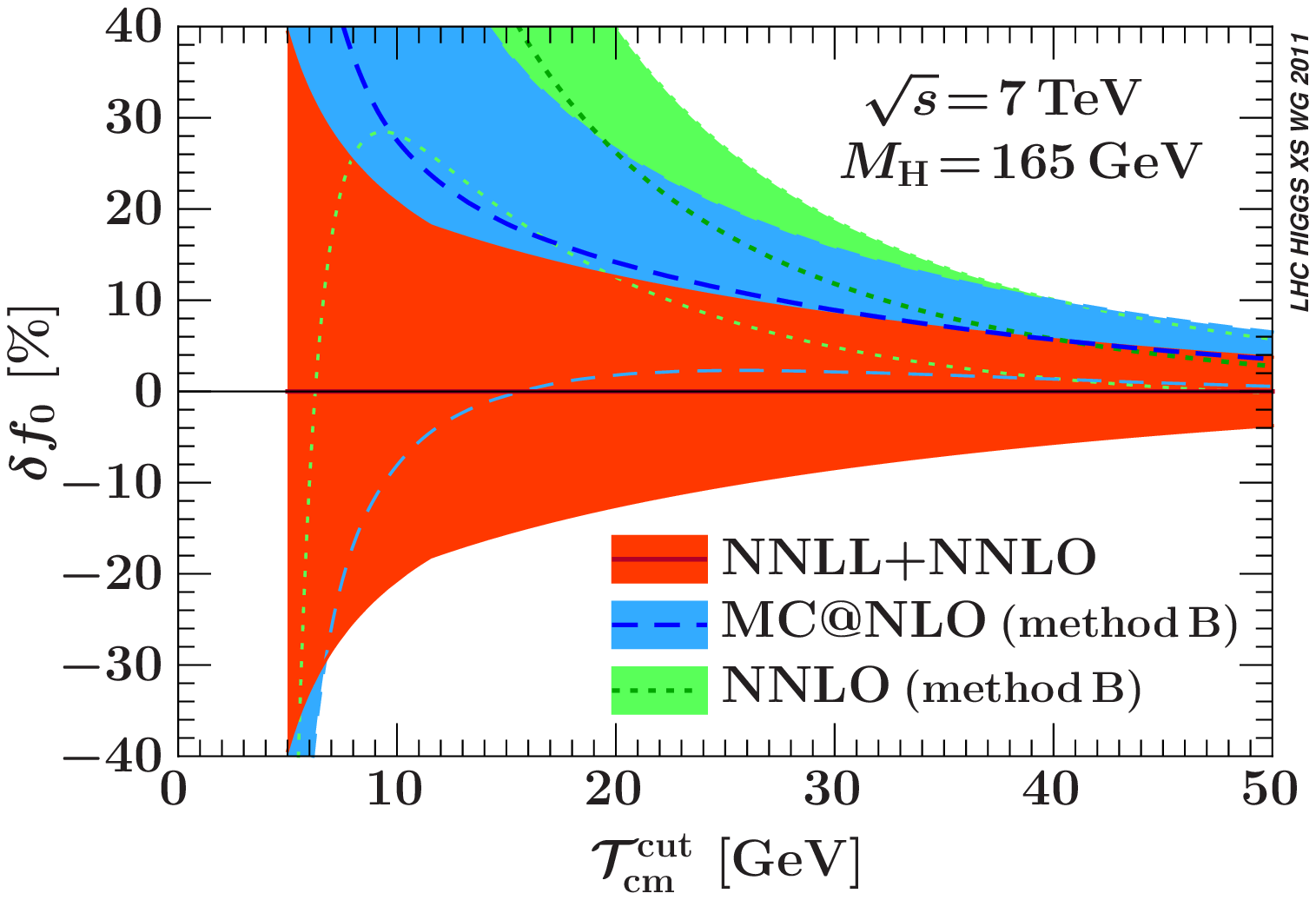}%
\vspace{-0.5ex}
\caption{\label{fig:f0Taucm} Comparison of the $0$-jet fraction for different levels of resummation using beam thrust as the jet-veto variable. Results are shown at NNLO (using method B uncertainties), MC@NLO (using the relative NNLO uncertainty from method B), and for the analytic NNLL+NNLO resummed result.}
\end{figure}

\begin{figure}
\includegraphics[width=0.5\textwidth]{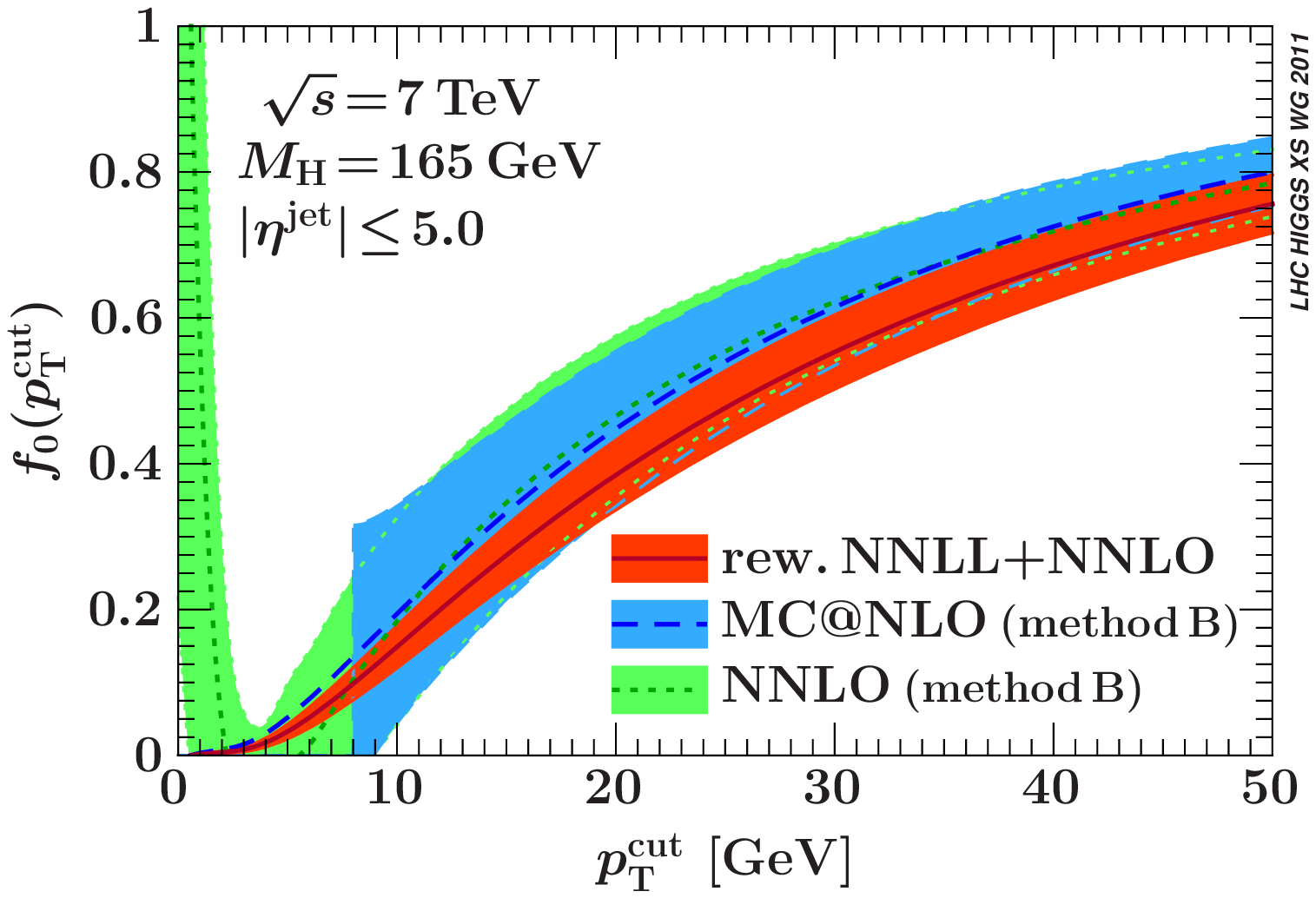}%
\hfill%
\includegraphics[width=0.5\textwidth]{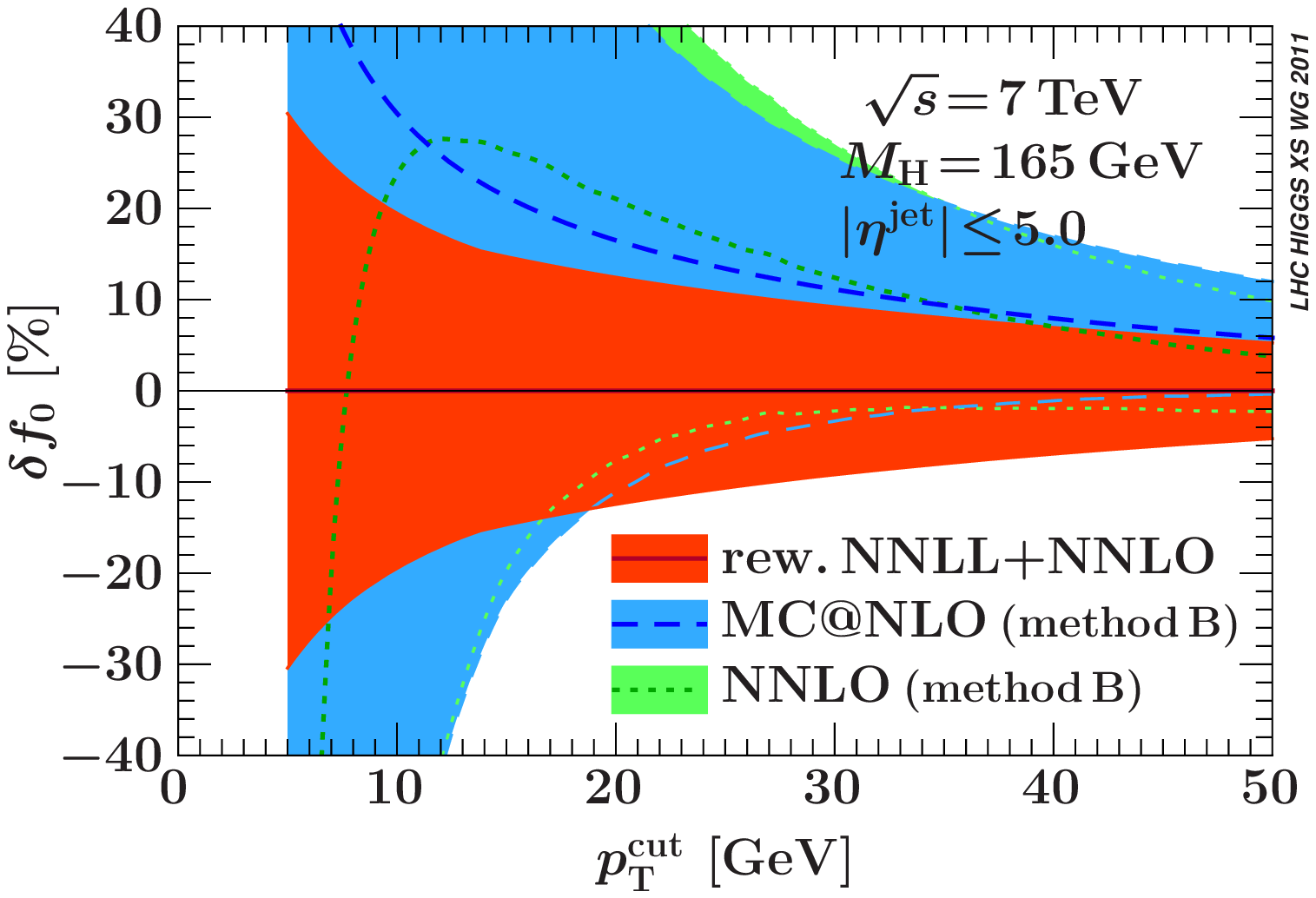}%
\vspace{-0.5ex}
\caption{\label{fig:f0pTjet} Comparison of the $0$-jet fraction for different levels of resummation using $\pT^\jet$ with $\abs{\eta^\jet} < 5.0$ as the jet-veto variable. Results are shown at NNLO (using method B uncertainties), MC@NLO (using the relative NNLO uncertainty from method B), and for the reweighted NNLL+NNLO resummed result.}
\end{figure}

The comparison for $f_0(\Tcmc)$ is shown in \refF{fig:f0Taucm} and for
$f_0(\pT^\jet)$ in \refF{fig:f0pTjet}.  The left panels in each case show
the $0$-jet event fractions, and the right panels normalise these same results
to the highest-order curve to illustrate the relative differences and
uncertainties. Since the parton shower includes the resummation of the leading
logarithms, we expect the MC@NLO results to show a behavior similar to the
NNLL-resummed result, which is indeed the case. For both variables, the MC@NLO
central value is near the upper edge of the NNLL+NNLO uncertainty band (dark
orange band).  From \refF{fig:f0pTjet} we see that the uncertainties
assigned to the MC@NLO results via method B (light blue band) include the
NNLL+NNLO central value for $f_0(\pT^\cut)$.\footnote{The blue uncertainty bands
  for the MC@NLO curves are cut off at very small cut values only because the
  NNLO cross section diverges there and so its relative uncertainties are no
  longer meaningful.  This happens for cut values well below the region of
  experimental interest.} We also see that the uncertainties for $f_0(\pT^\cut)$
in method C are reduced by roughly a factor of two compared to MC@NLO with
method B. This is the analog of the observation that we made for
$\sigma_0(\pT^\cut)$ in \refF{fig:ABC} when comparing NNLO method B and
method C uncertainties.

From the $\delta f_0$ plots in \refF{fig:f0Taucm} and \refF{fig:f0pTjet} we
observe that the impact of the resummation on the NNLL+NNLO central value
compared to the NNLO central value without resummation is similar for both
jet-veto variables for cut values $\gtrsim 25\UGeV$. In this region the MC@NLO
central value lies closer to the NNLO than the NNLL+NNLO for both variables.
However, it is hard to draw conclusions on the impact of resummation by only
comparing MC@NLO and NNLO since they each contain a different set of
higher-order corrections beyond NLO.  For smaller cut values $\lesssim 25\UGeV$,
the two variables start to behave differently, and the NNLL+NNLO resummation
has a larger impact relative to NNLO when cutting on $\Tcm$ than when cutting on
$\pT^\cut$.

To understand these features in more detail one has to take into account two different
effects arising from the two ways in which these jet-veto variables differ.
First, the objects are weighted differently according to their rapidity in the
two jet-veto variables. For $\Tcm$ the particles are weighted by $e^{-|\eta|}$,
while for $\pT^\jet$ no weighting in $\eta$ takes place. Second, the way in
which the cut restriction is applied to the objects in the final state is
different. By cutting on $\Tcm$, the restriction is applied to the sum over all
objects (either particles or jets) in the final state, while by cutting on the
leading $\pT^\jet$ the restriction is applied to the maximum value of all
objects (after an initial grouping into jets with small radius). To disentangle
these two effects we consider two additional jet-veto variables: $H_T$ which
inclusively sums over all object $|\pT|$s in the same way as $\Tcm$ does, but
without the rapidity weighting, and also ${\cal T}_{\rm jet}$, the largest
individual beam thrust of any jet, which has the same object treatment as
$\pT^\jet$, but with the beam-thrust rapidity weighting. The effect of the
different rapidity weighting already appears in the LL series for the jet veto,
i.e., at ${\cal O}(\alphas)$ the coefficient of the leading double logarithm is
a factor of two larger for $\pT^\cut$ than for $\Tcmc$, $\sigma_0(\pT^\cut)
\propto 1-6\alphas/\pi \ln^2(\pT^\cut/\MH)+\ldots$ versus $\sigma_0(\Tcmc)
\propto 1-3\alphas/\pi \ln^2(\Tcmc/\MH)+\ldots$. In contrast, the LL series is
the same for $H_T$ and $\pT^\jet$, and for $\Tcm$ and ${\cal T}_\jet$. The
larger logarithms for $\pT^\cut$ than $\Tcmc$ are reflected by the fact that
$\sigma_0(\pT^\cut)$ is noticeably smaller than $\sigma_0(\Tcmc)$ at equal cut
values. For the same type of object restriction the effect of the resummation
follows the pattern expected from the leading logarithms: The resummation has a
larger impact for $\pT^\jet$ than ${\cal T}_{\rm jet}$, and also for $H_T$ than
$\Tcm$.  On the other hand, a cut on the (scalar) sum of objects is always a
stronger restriction than the same cut on the maximum value of all objects,
since the former restricts wide-angle radiation more. As a result the
resummation has more impact on $\Tcm$ than on ${\cal T}_{\rm jet}$, and also on
$H_T$ than on $\pT^\jet$.  From what we observe in \refF{fig:f0Taucm}
and \refF{fig:f0pTjet}, these two competing effects appear to approximately
balance each other for $\pT^\jet$ and $\Tcm$ for cut values $\gtrsim 25\UGeV$.

To summarise, our results provide an important validation of using method B with
relative uncertainties obtained at NNLO and applying these to the event
fractions obtained from NLO Monte Carlos. While we have only compared to MC@NLO,
we expect the same to be true for POWHEG as well. By reweighting the Monte Carlo
to the NNLL+NNLO result the uncertainties in the predictions can be further
reduced to those obtained from the higher-order resummation using method C, and
this provides an important avenue for improving the analysis beyond method B.
It would also be very useful to investigate experimentally the viability of
using ${\cal T}_\jet$ as the jet-veto variable, by only summing over the jet (or
jets) with the largest individual beam thrust, as this would combine the
advantages of a jet-based variable with the theoretical control provided by the
beam-thrust resummation.

\clearpage

\clearpage

\newpage

\providecommand{\muR}{\mathswitch {\mu_{\mathrm{R}}}}
\providecommand{\muF}{\mathswitch {\mu_{\mathrm{F}}}}
\providecommand\HAWK{{\sc HAWK}}
\providecommand\VBFNLO{{\sc VBFNLO}}
\renewcommand\POWHEG{{\sc POWHEG}}
\renewcommand\POWHEGBOX{{\sc POWHEG BOX}}
\renewcommand\HERWIG{{\sc HERWIG}}
\renewcommand\PYTHIA{{\sc PYTHIA}}

\section{VBF production mode\footnote{%
    A.~Denner, S.~Farrington, C.~Hackstein, C.~Oleari, D.~Rebuzzi (eds.);
S.~Dittmaier, A.~M\"uck, S. Palmer and W. Quayle.}}
\label{sec:YRHX2_VBF}

The production of a Standard Model Higgs boson in association with two hard
jets in the forward and backward regions of the detector, frequently quoted
as the ``vector-boson fusion''~(VBF) channel, is a cornerstone in
the Higgs-boson search both in the ATLAS~\cite{Asai:2004ws} and
CMS~\cite{Abdullin:2005yn} experiments at the LHC.  Higgs-boson production in
the VBF channel also plays an important role in the determination of
Higgs-boson couplings at the LHC (see e.g.~\Bref{Duhrssen:2004cv}). 

The production of a Higgs boson~+~2~jets receives two contributions at
hadron colliders. The first type, where the Higgs boson couples to a
weak boson that links two quark lines, is dominated by $t$- and
$u$-channel-like diagrams and represents the genuine VBF channel.  The
hard jet pairs have a strong tendency to be forward--backward directed
in contrast to other jet-production mechanisms, offering a good
background suppression. The second type of diagrams are actually a
contribution to $\PW\PH$ and $\PZ\PH$ production, with the $\PW/\PZ$
decaying into a pair of jets.

In the previous report~\cite{Dittmaier:2011ti}, state-of-the-art
predictions and error estimates for the total cross sections for
$\Pp\Pp\to\PH+ 2\,\mathrm{jets}$ have been compiled. These were based
on an (approximate) NNLO calculation~\cite{Bolzoni:2010xr} of the
total VBF cross section based on the structure-function approach and
on calculations of the electroweak (EW) corrections by the Monte Carlo
programs \HAWK~\cite{Ciccolini:2007jr,Ciccolini:2007ec,HAWK} and
\VBFNLO{}~\cite{webVBFNLO,Figy:2010ct}. The EW corrections
were simply included as a multiplicative correction factor to the
NNLO QCD prediction.

In this contribution we consider predictions for cross sections with
cuts and distributions. This is particularly interesting for VBF, as
cuts on the tagging jets are used to suppress events from
Higgs~+~2~jet production via gluon fusion and other backgrounds and to
enhance the VBF signal.  Since no NNLO calculation for VBF exists so
far for differential distributions or cross sections with cuts, we use
NLO codes. We also provide results for  NLO QCD corrections + 
parton shower, as generated using the \POWHEG{}~\cite{Nason:2004rx} method
and implemented in the \POWHEGBOX{} code~\cite{Alioli:2010xd}.

\subsection{Theoretical framework}
\label{sec:NLO_theory}

The results presented in \refS{sec:VBF_results} have been obtained
with the Monte Carlo programs \HAWK{} and \VBFNLO{}, which
include both QCD and EW NLO corrections, and with \POWHEG{},
a method to interface QCD NLO calculations with parton showers.
Some details on these codes are given in the following.

\subsubsection{\HAWK}
\label{sec:HAWK}

The Monte Carlo event generator \HAWK{}~\cite{Ciccolini:2007jr,
  Ciccolini:2007ec,HAWK} has already been described
in~\Bref{Dittmaier:2011ti}. Since then, it has been extended to the
processes $\Pp\Pp\to\PW\PH\to\PGn_{\Pl}\,\Pl\,\PH$ and
$\Pp\Pp\to\PZ\PH\to\Pl^-\Pl^+\PH/\PGn_{\Pl}\PAGn_{\Pl}\PH$~\cite{whzhhawk}
(see \refS{sec:YRHX2_WHZH_th}). Here we summarise its most important features for the
VBF channel. \HAWK{} includes the complete NLO QCD and EW
corrections and all weak-boson fusion and quark--antiquark annihilation
diagrams, i.e.~$t$-channel and $u$-channel diagrams with VBF-like
vector-boson exchange and $s$-channel Higgs-strahlung diagrams with hadronic
weak-boson decay, as well as all interferences.  

For the results presented below, $s$-channel contributions and
interferences have been switched off in order to allow for a direct
comparison with the results of the other codes. 
The contributions from $s$-channels and interferences are
below $1\%$ once VBF cuts are applied. While \HAWK{} allows for the inclusion of
contributions of $\PQb$-quark parton distribution functions (PDFs) and
final-state $\PQb$ quarks at LO, these contributions have been
switched off as well. With VBF cuts, they can amount to about $2\%$.
Also contributions from photon-induced processes, which are part of the
real EW corrections and at the level of $1{-}2\%$, have not been included, 
since photon PDFs are not supported by the PDF sets used.
\HAWK{} allows for an on-shell Higgs boson or for an off-shell Higgs
boson (with optional decay into a pair of gauge singlets). For the
virtuality of the off-shell Higgs boson the usual Breit--Wigner
distribution
\begin{equation} 
\label{eq:VBF_BW}
\frac{1}{\pi} \frac{\MH \, \GH} {\left(M^2 -
\MH^2\right)^2 + ( \MH \, \GH )^2 }
\end{equation} 
is used. This form results from the gauge-invariant expansion about
the complex Higgs-boson pole and
is properly normalised. As any other parametrisation, it becomes
unreliable for large Higgs-boson masses, where the Higgs-boson width
gets large and contributions far from the pole become sizeable.

\subsubsection{\VBFNLO{}}

\VBFNLO\ \cite{Arnold:2011wj,Arnold:2008rz} is a parton-level Monte
Carlo event generator that can be used to simulate vector-boson
fusion, double and triple vector-boson production in hadronic
collisions at next-to-leading order in the strong coupling constant,
as well as Higgs-boson plus two jet production via gluon fusion at the
one-loop level.  For Higgs-boson production via VBF, both the NLO QCD
and electroweak corrections to the $t$-channel can be included
\cite{Figy:2004pt,Figy:2010ct} in the Standard Model and the (complex)
MSSM\footnote{For more details see \refS{sec:mssm_vbfnlo}.}, and
  the NLO QCD corrections are included for anomalous couplings between
  the Higgs and a pair of vector bosons.

\VBFNLO\ can also simulate the Higgs decays $\PH \rightarrow
\gamma\gamma, \PGmp \PGmm, \PGtp \PGtm, \PQb \overline{\PQb}$ in
the narrow-width approximation, either taking the appropriate
branching ratios from an input SLHA file or calculating them
internally at LO.  The Higgs-boson decays $\PH \rightarrow \PW^{+} \PW^{-}
\rightarrow \Pl^{+} \PGnl \Pl^{-} \PAGn_{\Pl}$ and $\PH \rightarrow
\PZ\PZ \rightarrow \Pl^{+} \Pl^{-} \Pl^{+} \Pl^{-}, \Pl^{+} \Pl^{-} \PGnl
\overline{\PGnl}$ are calculated using a Breit--Wigner distribution
for the Higgs boson and the full LO matrix element for the $\PH \rightarrow
4f$ decay.

For the results presented here, a modified version of \VBFNLO\ was
used that simulated a Higgs-boson decay with a branching ratio of 1
and a Breit--Wigner distribution \refE{eq:VBF_BW} for the virtuality of
the off-shell Higgs boson.

\subsubsection{\POWHEG{}}
The \POWHEG{} method is a prescription for interfacing NLO calculations with
parton-shower generators, like \HERWIG{} and \PYTHIA{}. It was first
presented in~\Bref{Nason:2004rx} and was described in great detail
in~\Bref{Frixione:2007vw}.  In~\Bref{Nason:2009ai}, Higgs-boson production in
VBF has been implemented in the framework of
the \POWHEGBOX{}~\cite{Alioli:2010xd}, a computer framework that implements
in practice the theoretical construction of~\Bref{Frixione:2007vw}.

All the details of the implementation can be found
in~\Bref{Nason:2009ai}. Here we briefly recall that, in the
calculation of the partonic matrix elements, all partons have been
treated as massless.  This gives rise to a different treatment of
quark flavours for diagrams where a $\PZ$ boson or a $\PW$ boson is
exchanged in the $t$-channel.  In fact, for all $\PZ$-exchange
contributions, the $\PQb$-quark is included as an initial and/or
final-state massless parton.  For the (dominant) $\PW$-exchange
contributions, no initial $\PQb$-quark has been considered, since it
would have produced mostly a $\PQt$ quark in the final state, which would
have been misleadingly treated as massless.  The
Cabibbo--Kobayashi--Maskawa~(CKM) matrix $V_{\rm \scriptscriptstyle
  CKM}$ has been taken as
\begin{equation}
\begin{array}{c}
\\
V_{\rm  \scriptscriptstyle CKM}=
\end{array}
\begin{array}{c c}
& \PQd\quad\quad\quad \ \PQs\ \quad \quad\quad \PQb\ \\
\begin{array}{c}
\PQu\\
\PQc\\
\PQt
\end{array} 
&
\left(
\begin{array}{c c c}
$0.9748$ & $0.2225$ & $0.0036$\\
$0.2225$ & $0.9740$ & $0.041$ \\
$0.009$ & $0.0405$ & $0.9992$
\end{array}
\right).
\end{array}
\end{equation}
We point out that, as long as no final-state hadronic flavour is
tagged, this is practically equivalent to the result obtained using
the identity matrix, due to unitarity.

The Higgs-boson virtuality $M^2$ has been generated distributed according to
\begin{equation}
\label{eq:VBF_BW_PWG}
\frac{1}{\pi} \frac{M^2 \,\GH
  / \MH}{\left(M^2 - \MH^2\right)^2 + ( M^2\, \GH / \MH)^2 }\,,
\end{equation}
 with fixed decay width $\GH$. 

As a final remark, we recall that the renormalisation $\muR$ and factorisation
$\muF$ scales have been taken equal to the transverse momentum of the
radiated parton, during the generation of radiation, as the \POWHEG{} method
requires. The transverse momentum of the radiated parton is taken, in the
case of initial-state radiation, as exactly equal to the transverse momentum
of the parton with respect to the beam axis. For final-state radiation one
takes instead
\begin{equation}
\pT^2=2E^2(1-\cos\theta),
\end{equation}
where $E$ is the energy of the radiated parton and $\theta$ the angle it
forms with respect to the final-state parton that has emitted it, both taken
in the partonic centre-of-mass frame.  The scales used in the calculation of
the \POWHEG{} $\bar{B}$ function (i.e.~the function that is used to generate
the underlying Born variables to which the \POWHEGBOX{} attaches the
radiation ones) are instead the ones defined in the forthcoming
\Eref{eq:VBF_ren_fac_scales}.

\subsection{VBF parameters and cuts}
\label{sec:VBF_setup}
The numerical results presented in \refS{sec:VBF_results} have been
computed using the values of the EW parameters given in Appendix A of
\Bref{Dittmaier:2011ti}. The electromagnetic coupling is fixed in the $\GF$ scheme, to be
\beq
 \alpha_{\GF} = \sqrt{2}\GF\MW^2(1-\MW^2/\MZ^2)/\pi = 1/132.4528\ldots.
\eeq
The weak mixing angle is defined in the on-shell scheme,
\begin{equation}
\sin^2\theta_{\mathrm{w}}=1-\MW^2/\MZ^2=0.222645\ldots \,.
\end{equation}
We consider the following set of Higgs-boson masses and corresponding
total widths, as reported in \Brefs{Dittmaier:2011ti,Denner:2011mq}:
\beq\label{eq:vbf_higgsmasses}
\begin{array}{lcccccc}
\MH \ [\mathrm{GeV}] & $120$ & $150$& $200$ & $250$ & $500$ & $600$ \\
\hline
\GH \ [\mathrm{GeV}]  & $0.00348$ & $0.0173$ & $1.43$ & $4.04$ & $68.0$ & $123$ \\
\end{array}
\eeq
The renormalisation and factorisation scales are set equal to the
$\PW$-boson mass,
\begin{equation}
\label{eq:VBF_ren_fac_scales}
\muR = \muF= \MW.
\end{equation}
In the calculation of the NLO differential cross sections, we have
used the MSTW2008~\cite{Martin:2009iq} and
CTEQ6.6 \cite{Nadolsky:2008zw} PDFs.


Jets are constructed according to the anti-$k_\mathrm{T}$ algorithm, with
the rapidity--azimuthal separation $\Delta R=0.5$, using the default
recombination scheme ($E$ scheme).  Jets are subject to the
following cuts
\begin{equation}
\label{eq:VBF_cuts1}
{\pT}_j > 20~{\rm GeV}, \qquad  |y_j| < 4.5\,,
\end{equation}
where $y_j$ denotes the rapidity of the (massive) jet.
Jets are ordered according to their $\pT$ in decreasing
progression.  The jet with highest $\pT$ is called leading jet, the
one with next highest $\pT$ subleading jet, and both are the tagging
jets.  Only events with at least two jets are kept.  They must satisfy
the additional constraints
\begin{equation}
\label{eq:VBF_cuts2}
|y_{j_1} - y_{j_2}| > 4\,, \qquad m_{jj} > 600~{\rm GeV}.
\end{equation}
The Higgs boson is generated off shell, according to \Erefs{eq:VBF_BW}
or~\refE{eq:VBF_BW_PWG}, and there are no cuts applied to its decay products.

For the calculation of EW corrections, real photons are recombined
with jets according to the same  anti-$\kT$ algorithm as used
for jet recombination.



\subsection{Results}
\label{sec:VBF_results}

In the following we present a few results for the LHC at $7\UTeV$ calculated
with \HAWK, \VBFNLO, and \POWHEG, for the Higgs-boson masses listed in
\Eref{eq:vbf_higgsmasses}.

\subsubsection{Total cross sections with VBF cuts}
We have calculated the cross section for VBF at NLO within the cuts
given in \refS{sec:VBF_setup} using the MSTW2008NLO and CTEQ6.6 PDF
sets with and without EW corrections.  The results are shown in
\refTs{tab:vbf_cteq_nlo} and \ref{tab:vbf_mstw_nlo} including the
statistical errors of the Monte Carlo integration. The results of
\HAWK\ and \VBFNLO\ without EW corrections agree within $1\%$.  For the
results with EW corrections there is a slight discrepancy between
\HAWK\ and \VBFNLO\ at the level of $1{-}1.3\%$ for small Higgs-boson
masses. For heavy Higgs-boson masses the difference increases because  
of leading two-loop heavy-Higgs corrections that are included
in \HAWK\ but not in \VBFNLO. These corrections contribute about $3\%$
and $2\%$ for $\MH=600\UGeV$ and $500\UGeV$, respectively.  The EW
corrections calculated with \HAWK\ amount to $-8\%$ for small
Higgs-boson masses, decrease, and change sign with increasing
Higgs-boson mass and reach $+2.4\%$ for $\MH=600\UGeV$.  We also show
the scale uncertainties obtained with \HAWK\ by varying the
factorisation and renormalisation scales in the range
$\MW/2<\mu<2\MW$\footnote{More precisely, we calculated the cross
  sections for the 5 scale combinations $\{\muR,\muF\}=\{\MW,\MW\}$,
  $\{\MW/2,\MW\}$, $\{\MW,\MW/2\}$, $\{2\MW,\MW\}$, and $\{\MW,2\MW\}$,
  and took the maximum and the minimum of the results as the upper and
  lower bound of the variation.} and the PDF uncertainties obtained
from the corresponding $90\%$ C.L. error PDFs according to the CTEQ
prescription with symmetric errors \cite{Nadolsky:2008zw}. We give
these uncertainties only for the cross section with EW corrections as
they are practically the same without EW corrections. Both the PDF
uncertainties and the scale uncertainties are larger than for the
total cross section without cuts (c.f.\ \Bref{Dittmaier:2011ti}).
Note that we do not include the uncertainties from varying $\alphas$
in the PDFs as these are small (below $1\%$) and thus negligible
compared to the PDF uncertainties~\cite{govoni}.

\refTs{tab:vbf_cteq_powheg} and \ref{tab:vbf_mstw_powheg} show the
corresponding results obtained with \POWHEG\ at pure NLO QCD and with
\PYTHIA\ or \HERWIG\ parton showers. The NLO results in the second
columns can be directly compared with the NLO results without EW corrections
of \refTs{tab:vbf_cteq_nlo} and \ref{tab:vbf_mstw_nlo}.  While \POWHEG\
and \HAWK/\VBFNLO\ agree within integration errors for small Higgs-boson
masses the results differ by $13\%$ and $25\%$ for $\MH=500\UGeV$ and
$\MH=600\UGeV$. We checked that this difference 
is due to the different Breit--Wigner distributions, \refE{eq:VBF_BW} and
\refE{eq:VBF_BW_PWG}, used in the codes, in the  treatment of the
unstable Higgs boson. This difference should be viewed as an estimate
of the theoretical uncertainty of the present calculations based on
Breit--Wigner distributions. The proper treatment of a heavy Higgs
boson is described in \refS{sec:po}, but has not yet been implemented
for VBF. Inclusion of the parton showers reduces the cross
sections by $5\%{-}7\%$ with a larger effect for small Higgs-boson masses,
because more events are cut away after showering. The results for \PYTHIA\
and \HERWIG\ parton showers are in good agreement.
\begin{table}
\caption{Higgs-boson NLO cross sections  at $7\UTeV$ with VBF cuts and
  CTEQ6.6 PDF set with and without EW corrections, relative EW 
  corrections and theoretical uncertainties from PDF and scale variations.}%
\label{tab:vbf_cteq_nlo}%
\begin{center}%
\begin{small}%
\tabcolsep5pt
\begin{tabular}{c|cc|cc|c|cc}%
\hline
 & \multicolumn{2}{c|}{ w/ EW corr} & \multicolumn{2}{c|}{ w/o EW corr}
 & \multicolumn{1}{c|}{ EW corr} & \multicolumn{2}{c}{uncert.} 
\\
 \hline
$\MH$ & \HAWK & \VBFNLO & \HAWK & \VBFNLO & \HAWK & PDF & scale \\
\![GeV] &[fb] &[fb] &[fb] &[fb] & [\%] &  [\%] &  [\%] \\
\hline
$120 $ & $ 261.18 \pm 0.43 $ & $  258.27 \pm 0.41 $ & $ 283.91  \pm
0.42 $ & $282.80 \pm 0.19 $ & $ -8.0 \pm 0.2 $ & $ \pm 3.5 $ & $ +0.5 -0.5$ \\
$150 $ & $ 218.40 \pm 0.36 $ & $  216.84 \pm 0.40 $ & $ 236.75  \pm
0.35 $ & $236.68 \pm 0.14 $ & $ -7.8 \pm 0.2 $ & $ \pm 3.5 $ & $ +1.0 -0.5$ \\ 
$200 $ & $ 165.22 \pm 0.24 $ & $  163.50 \pm 0.24 $ & $ 176.46  \pm
0.24 $ & $176.89 \pm 0.10 $ & $ -6.4 \pm 0.2 $ & $ \pm 3.6 $ & $ +0.6 -0.6$ \\
$250 $ & $ 123.81 \pm 0.17 $ & $  122.67 \pm 0.17 $ & $ 133.13  \pm
0.16 $ & $133.15 \pm 0.07 $ & $ -7.0 \pm 0.2 $ & $ \pm 3.8 $ & $ +0.6 -0.5 $ \\
$500 $ & $ 38.10  \pm 0.07 $ & $  37.31  \pm 0.08 $ & $ ~38.38  \pm
0.07 $ & $~38.41  \pm 0.02 $ & $ -0.7 \pm 0.3 $ & $ \pm 4.3$ & $ +0.4 -0.4$ \\
$600 $ & $ 26.34  \pm 0.12 $ & $  25.46  \pm 0.07 $ & $ ~25.70  \pm
0.11 $ & $~25.55  \pm 0.01 $ & $ ~~2.5 \pm 0.7 $ & $ \pm 4.4 $ & $ +0.7 -0.6 $ \\
\hline
\end{tabular}%
\end{small}%
\end{center}%
\end{table}

\begin{table}
\caption{\POWHEG{} Higgs-boson NLO QCD cross sections at $7\UTeV$ with VBF cuts
  and CTEQ6.6 PDF set: fixed NLO results, \POWHEG{}  showered
  by \PYTHIA{}~(PY) and by \HERWIG{}~(HW).}
\label{tab:vbf_cteq_powheg}%
\begin{center}%
\begin{small}%
\begin{tabular}{cccc}%
\hline
$\MH$ & \POWHEG\ NLO & \POWHEG{} + PY & \POWHEG{} + HW\\
$\mbox{[GeV]}$ &[fb] &[fb] &[fb]  \\
\hline
$120 $ & $  282.87 \pm 0.75 $ & $ 262.96 \pm 0.99 $ & $ 262.04 \pm 0.99 $ \\
$150 $ & $  237.30 \pm 0.57 $ & $ 221.54 \pm 0.79 $ & $ 219.95 \pm 0.79 $ \\
$200 $ & $  177.05 \pm 0.38 $ & $ 164.55 \pm 0.55 $ & $ 163.83 \pm 0.55 $ \\
$250 $ & $  132.93 \pm 0.26 $ & $ 124.19 \pm 0.40 $ & $ 123.65 \pm 0.40 $ \\
$500 $ & $  34.04  \pm 0.07 $ & $ 31.92  \pm 0.09 $ & $ 31.78 \pm 0.10 $ \\
$600 $ & $  20.56  \pm 0.03 $ & $ 19.47  \pm 0.06 $ & $ 19.30 \pm 0.06 $ \\
\hline
\end{tabular}%
\end{small}%
\end{center}%
\end{table}

\begin{table}
\caption{Higgs-boson NLO cross sections at $7\UTeV$ with VBF cuts and
  MSTW2008NLO PDF set with and without EW corrections, relative EW
  corrections and theoretical uncertainties from PDF and scale variations.}%
\label{tab:vbf_mstw_nlo}%
\begin{center}%
\begin{small}%
\tabcolsep5pt
\begin{tabular}{c|cc|cc|c|cc}
\hline
 & \multicolumn{2}{c|}{ w/ EW corr} & \multicolumn{2}{c|}{ w/o EW corr}
 & \multicolumn{1}{c|}{ EW corr} & \multicolumn{2}{c}{uncert.} 
\\
 \hline
$\MH$ & \HAWK & \VBFNLO & \HAWK & \VBFNLO & \HAWK & PDF & scale \\
\![GeV] &[fb] &[fb] &[fb] &[fb] & [\%] &  [\%] &  [\%] \\
\hline
$120 $ & $ 259.74 \pm 0.69  $ & $ 256.69 \pm 0.83    $ & $ 282.17 \pm
0.68 $ & $ 281.37 \pm 0.22 $ & $ -8.0 \pm 0.2 $ & $ \pm 5.0 $ & $ +0.6 - 0.5  $ \\
$150 $ & $ 217.58 \pm 0.37  $ & $ 215.46 \pm 0.33    $ & $ 235.73 \pm
0.36 $ & $ 235.46 \pm 0.15 $ & $ -7.7 \pm 0.2 $ & $ \pm 5.1 $ & $ +0.5 -0.5  $ \\
$200 $ & $ 164.18 \pm 0.24  $ & $ 162.10 \pm 0.26    $ & $ 175.29 \pm
0.23 $ & $ 175.23 \pm 0.15 $ & $ -6.3 \pm 0.2 $ & $ \pm 5.0 $ & $ +0.5 -0.5  $ \\
$250 $ & $ 122.73 \pm 0.20  $ & $ 120.96 \pm 0.61    $ & $ 131.90 \pm
0.19 $ & $ 131.86 \pm 0.07 $ & $ -7.0 \pm 0.2 $ & $ \pm 5.2 $ & $ +0.5 -0.7   $ \\
$500 $ & $ 37.26  \pm 0.09  $ & $ 36.57  \pm 0.10    $ & $ 37.57  \pm
0.10 $ & $ 37.61  \pm 0.02 $ & $ -0.8 \pm 0.2 $ & $ \pm 5.2 $ & $ +0.7 -0.4  $ \\
$600 $ & $ 25.71  \pm 0.08  $ & $ 24.83  \pm 0.06    $ & $ 25.21  \pm
0.08 $ & $   25.01  \pm 0.02 $ & $ ~~2.0 \pm 0.2 $ & $ \pm 5.1 $ & $ +0.4 -0.7 $ \\
\hline
\end{tabular}%
\end{small}%
\end{center}%
\end{table}%

\begin{table}
\caption{\POWHEG{} Higgs-boson NLO QCD cross sections at $7\UTeV$ with VBF cuts
  and MSTW2008NLO PDF set: fixed NLO results, \POWHEG{}  showered
  by \PYTHIA{}~(PY) and by \HERWIG{}~(HW).}
\label{tab:vbf_mstw_powheg}
\begin{center}%
\begin{small}%
\begin{tabular}{cccc}%
\hline
$\MH$ & \POWHEG\ NLO & \POWHEG\ + PY & \POWHEG\ + HW\\
$\mbox{[GeV]}$ &[fb] &[fb] &[fb] \\
\hline
$120 $ & $ 281.97 \pm 0.76 $ & $ 262.23 \pm 0.99 $ & $ 260.44 \pm 1.00 $ \\
$150 $ & $ 235.29 \pm 0.67 $ & $ 218.24 \pm 0.79 $ & $ 216.70 \pm 0.79 $ \\
$200 $ & $ 175.38 \pm 0.42 $ & $ 162.80 \pm 0.55 $ & $ 161.45 \pm 0.55 $ \\
$250 $ & $ 131.57 \pm 0.26 $ & $ 123.21 \pm 0.40 $ & $ 122.62 \pm 0.40 $ \\
$500 $ & $ 33.21  \pm 0.06 $ & $ 31.03  \pm 0.09 $ & $ 30.85 \pm 0.09 $ \\
$600 $ & $ 20.03  \pm 0.03 $ & $ 18.74  \pm 0.05 $ & $ 18.67 \pm  0.06 $ \\
\hline
\end{tabular}%
\end{small}%
\end{center}%
\end{table}%

\subsubsection{Differential distributions}
\label{sec:VBF_Differential_distributions}
In this section we present some results for differential distributions
for the setup defined in \refS{sec:VBF_setup} and MSTW2008NLO PDFs.
For each distribution we show the NLO results from \HAWK\ with (blue
solid) and without (green dash-dotted) EW corrections, the results of
\VBFNLO\ with EW corrections (black long-dashed) and the NLO QCD results
from \POWHEG{} (red short-dashed). Each plot contains results for
$\MH=120\UGeV$ (upper set of curves) and for $\MH=600\UGeV$ (lower set
of curves).  In addition we show the relative EW corrections obtained
from \HAWK. These are insensitive to PDFs and could be taken into
account in any QCD-based prediction for the respective distributions
(based on the cuts of \refS{sec:VBF_setup}) via reweighting. The data
files of the histograms are provided at the TWiki page
of the VBF working group%
\footnote{https://twiki.cern.ch/twiki/bin/view/LHCPhysics/VBF}.

\begin{figure}
  \includegraphics[width=0.49\textwidth, bb = 40 30 530 450, clip]{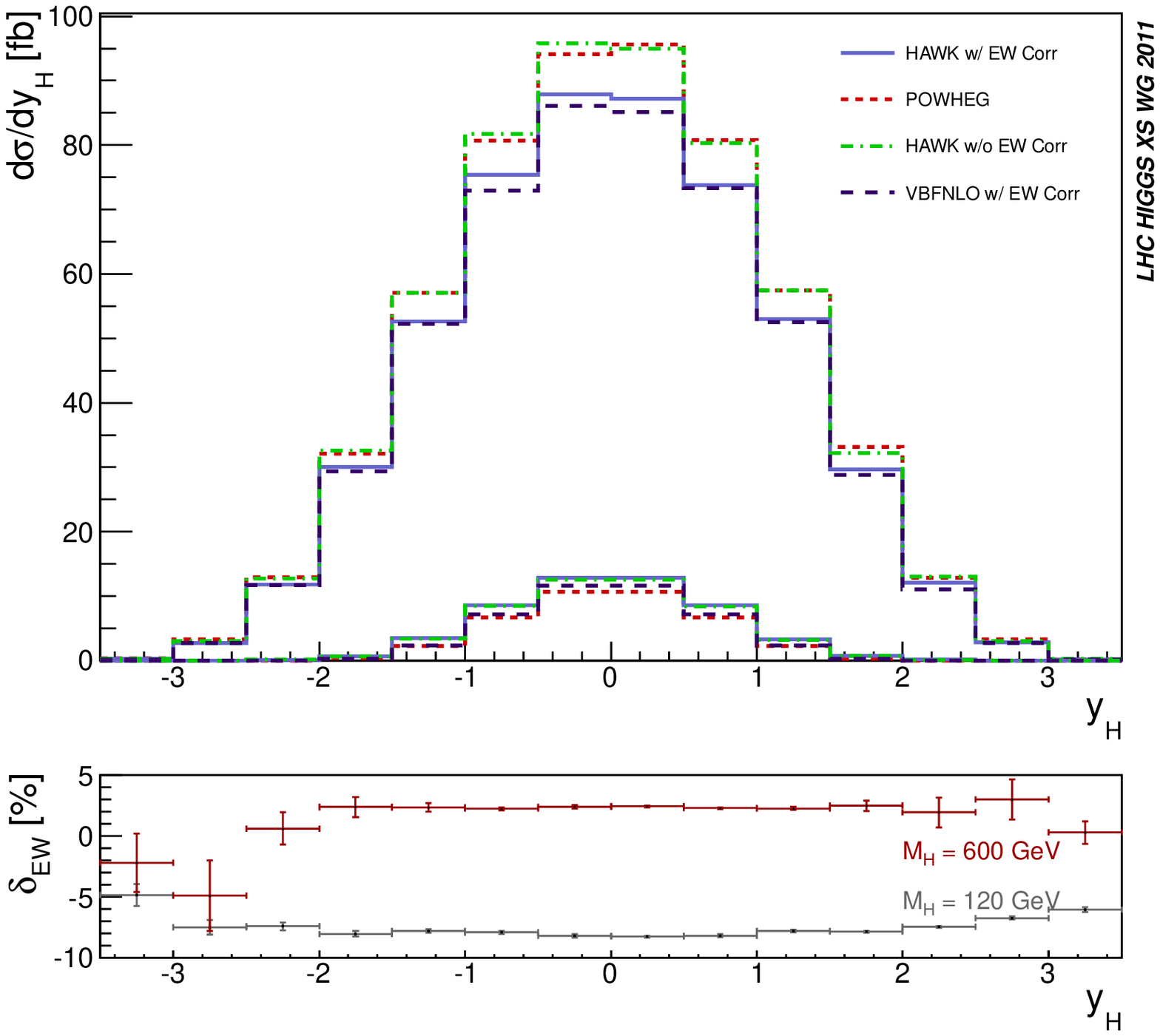}
  \includegraphics[width=0.49\textwidth, bb = 40 30 530 450, clip]{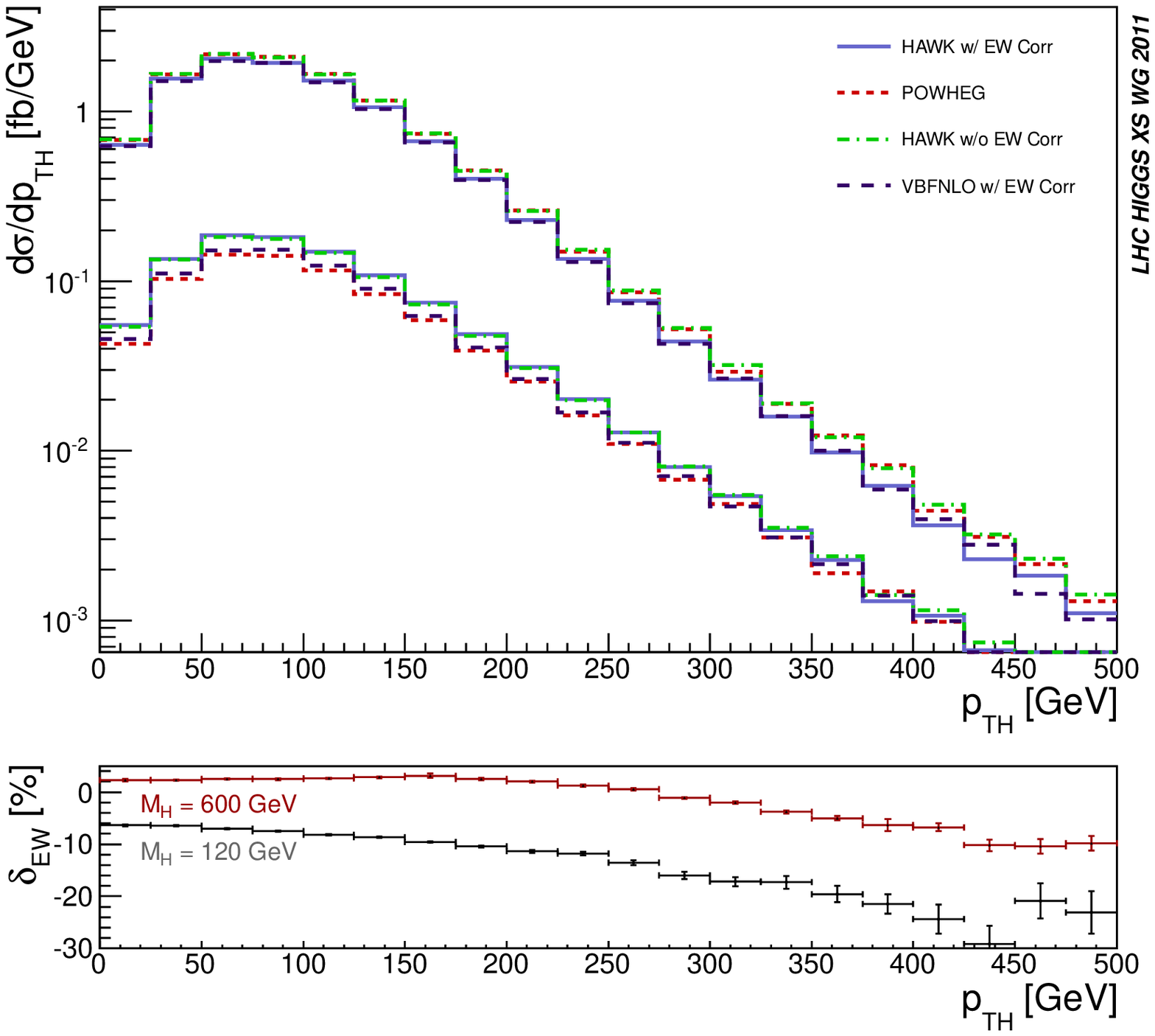}
  \caption{Higgs-boson rapidity (left) and transverse momentum (right). The
    top plots show the comparison between \HAWK{} (with and without EW
    corrections), \POWHEG{} at NLO QCD and \VBFNLO{} (with EW corrections)
    for $\MH = 120$\UGeV{} (upper set of
    curves) and $\MH = 600$\UGeV{} (lower set of curves) for MSTW2008NLO PDF set. The bottom plots
    display the percentage EW corrections, for each of the two mass points.}
\label{fig:YRHXS2_VBF_PTH}
\end{figure}

\refF{fig:YRHXS2_VBF_PTH} displays the rapidity $y_{\PH}$ and the
transverse momentum $\pT{}_{\PH}$ of the Higgs boson. While the
difference in normalisation between \POWHEG\ and \HAWK\ for the high
Higgs-boson mass has been explained for the total cross section, the
shapes of the distributions agree well between the different codes.
While EW corrections are flat for the rapidity distribution, the EW
corrections increase with increasing $\pT{}_{\PH}$.

\begin{figure}
  \includegraphics[width=0.49\textwidth, bb = 40 30 530 450, clip]{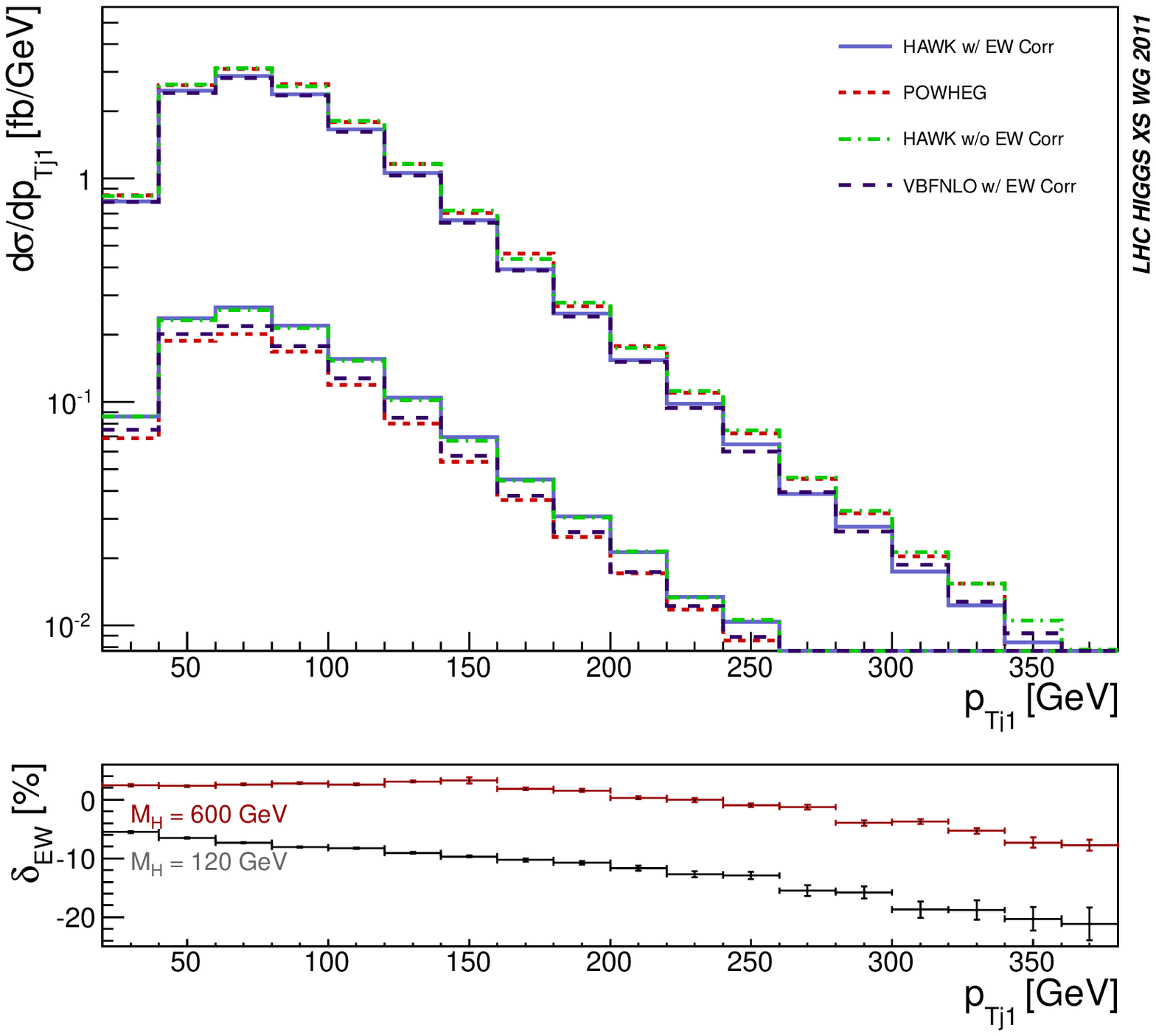}
  \includegraphics[width=0.49\textwidth, bb = 40 30 530 450, clip]{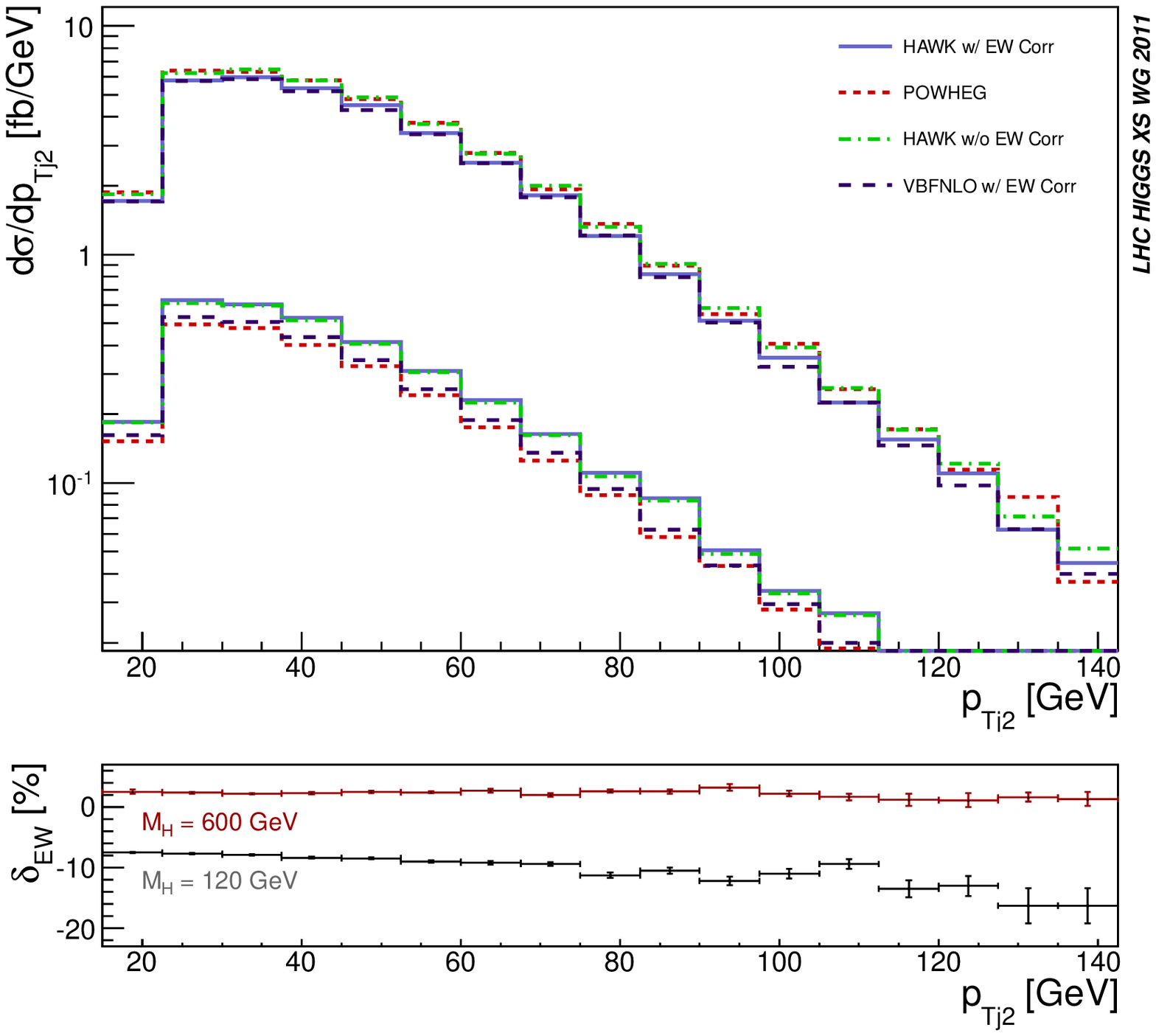}
  \caption{Transverse momentum of the leading (left) and sub-leading jet
    (right).  The top plots show the comparison between \HAWK{} (with and
    without EW corrections), \POWHEG{} at NLO QCD and \VBFNLO{} (with EW corrections)
    for $\MH = 120$\UGeV{}
    (upper set of curves) and $\MH = 600$\UGeV{} (lower set of curves) for MSTW2008NLO PDF set. The
    bottom plots display the percentage EW corrections, for each of the two
    mass points. }
\label{fig:YRHXS2_VBF_PTJ2}
\end{figure}

\begin{figure}
  \includegraphics[width=0.49\textwidth, bb = 40 30 530 450, clip]{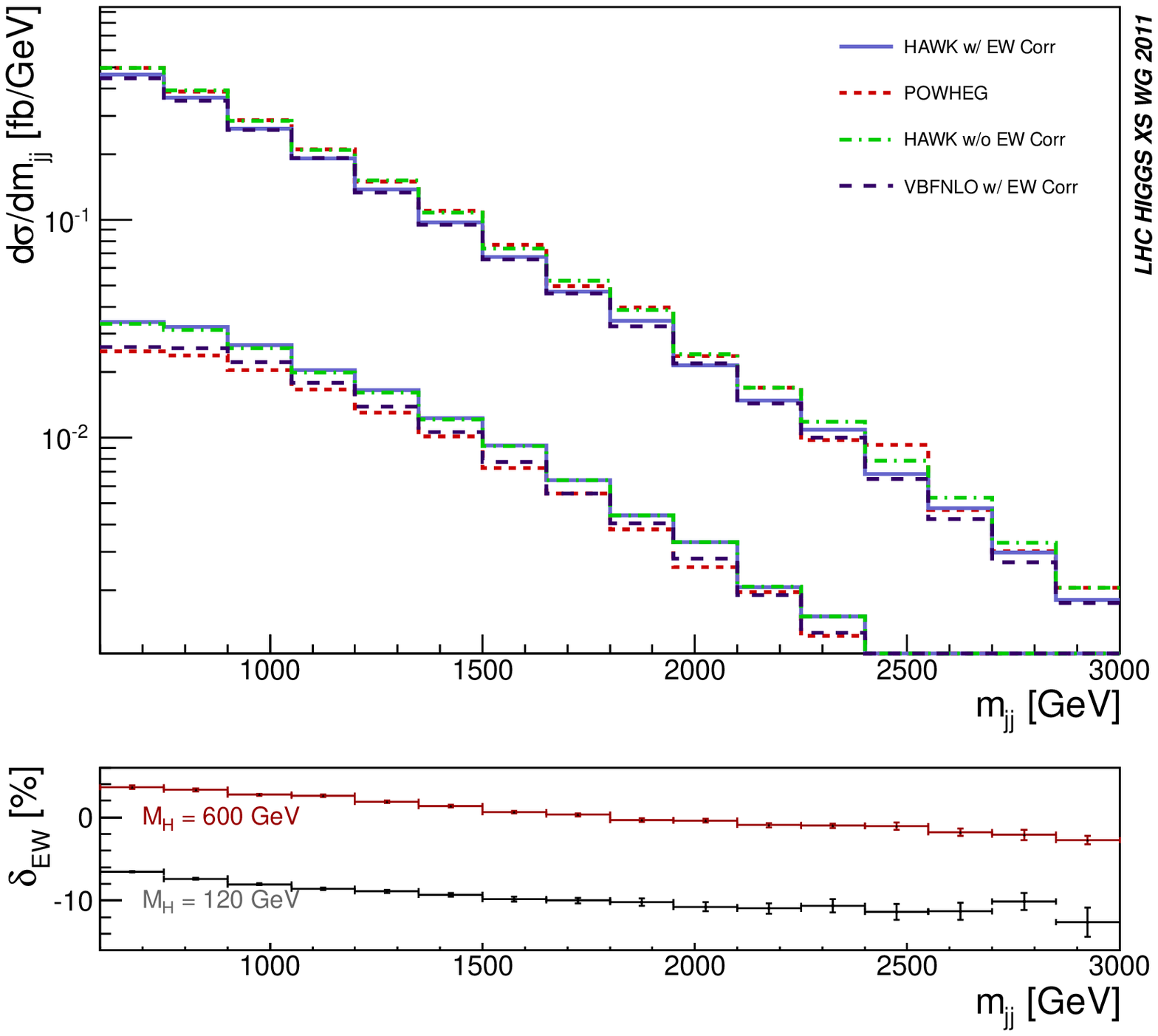}
  \includegraphics[width=0.49\textwidth, bb = 40 30 530 450, clip]{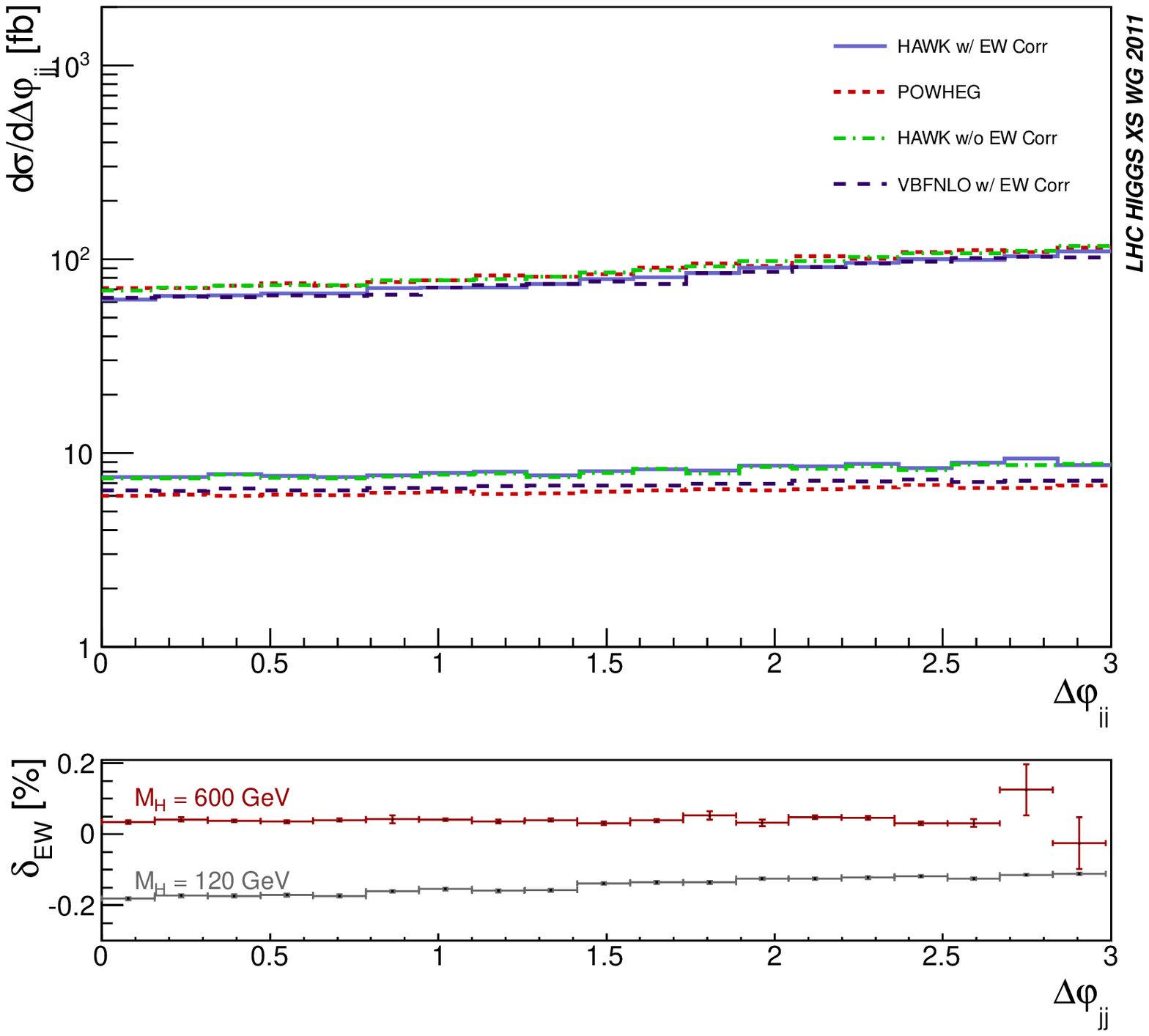}
  \caption{Di-jet invariant mass (left) and azimuthal separation between the two tagging jets (right). 
    The top plots show the comparison between \HAWK{} (with and
    without EW corrections), \POWHEG{} QCD and \VBFNLO{} (with EW corrections)
    for $\MH = 120$\UGeV{}
    (upper set of curves) and $\MH = 600$\UGeV{} (lower set of
    curves) for MSTW2008NLO PDF set. The bottom plots display the percentage EW corrections
    for each of the two mass points. }
\label{fig:YRHXS2_VBF_MJJ}
\end{figure}

For the distributions in the transverse momentum of the leading and
subleading jets, $\pT{}_{j_1}$ and $\pT{}_{j_2}$, shown in
\refF{fig:YRHXS2_VBF_PTJ2} and of the di-jet invariant mass $m_{jj}$,
presented on the left-hand side of \refF{fig:YRHXS2_VBF_MJJ}
similar remarks are applicable.  The shapes of the different
distributions agree well between the different codes.  The EW
corrections reduce the cross section more and more with increasing
energy scale, a generic behaviour of EW corrections that can be
attributed to weak Sudakov logarithms. EW corrections range from a few
to $-20$\%. The distribution in the azimuthal-angle separation between
the two tagging jets $\phi_{jj}$ is shown on the right-hand side of
\refF{fig:YRHXS2_VBF_MJJ}. In particular, for the light Higgs boson,
the electroweak corrections distort the distribution at the level of
several per cent.

\subsubsection{\POWHEG{} differential distributions}
\label{sec:VBF_PWG_results}

In this section, we present a few results obtained by \POWHEG{} interfaced
to \HERWIG{} and \PYTHIA{}, in the configuration described
in \refS{sec:VBF_setup} and for a Higgs-boson mass of $120\UGeV$.  These
results have been generated using the MSTW2008 PDF set. They depend only very
slightly on the value of the Higgs-boson mass and on the PDF set used, so
similar conclusions can be drawn using NNPDF2.0 and CTEQ6.6.  All results are
correct at NLO in QCD. No EW corrections are included.  We have
generated 0.5M events with the
\POWHEGBOX. We have run \PYTHIA{} with the Perugia~0 tuning and
\HERWIG{} in its default configuration, with intrinsic $\pT$-spreading
of $2.5\UGeV$.

\begin{figure}
  \includegraphics[width=0.49\textwidth,bb = 20 20 540 350,
  clip]{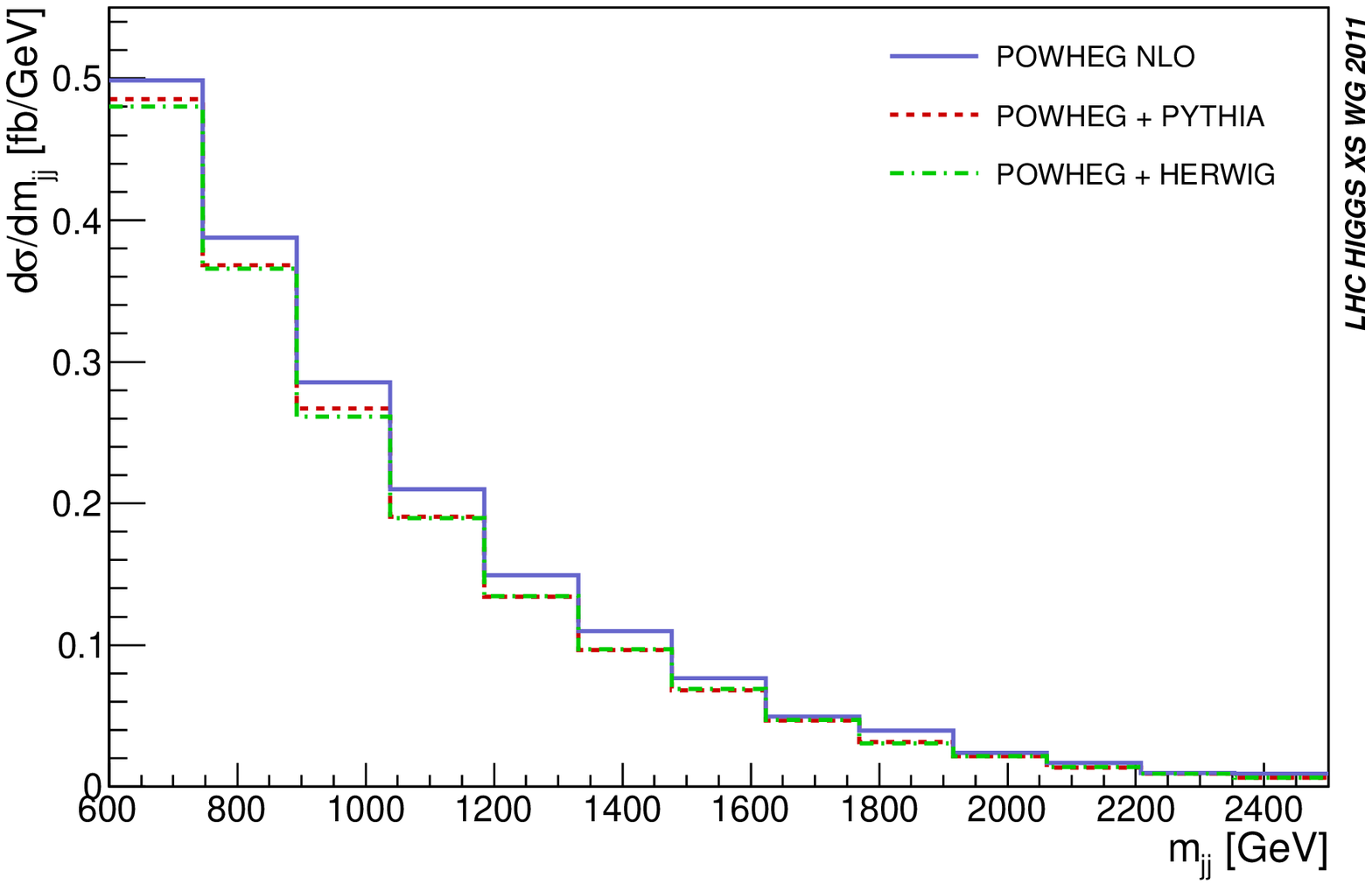}
  \includegraphics[width=0.49\textwidth,bb = 20 20 540 350,
  clip]{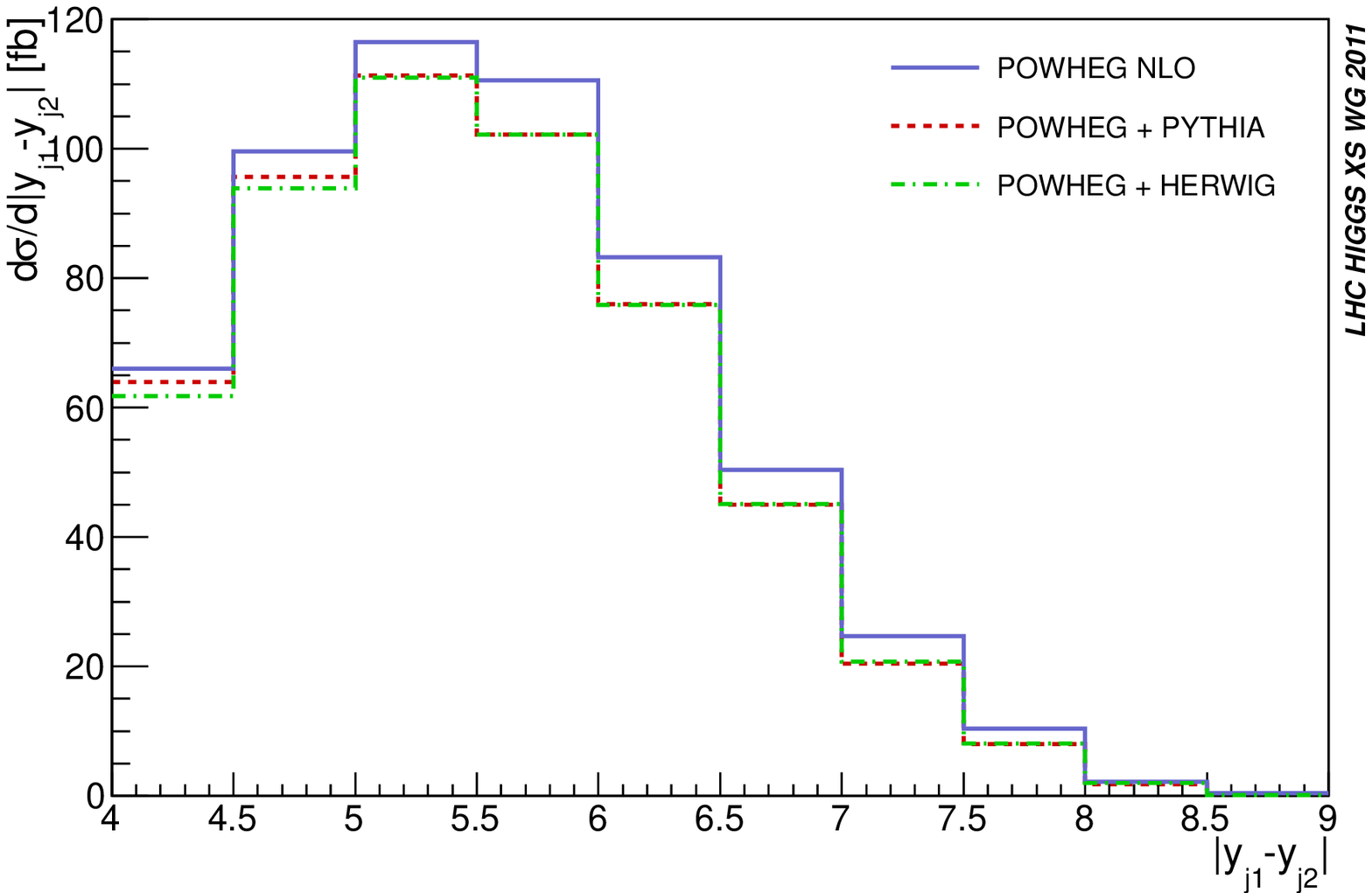} 
    \caption{Invariant
    mass of the two tagging jets, $m_{jj}$, (left) and absolute value
    of the distance in rapidity of the two tagging jets,
    $\abs{y_{j_1}-y_{j_2}}$, (right).  We show the comparison among
    NLO QCD and the \POWHEG{} interfaced to \HERWIG{} and \PYTHIA{}
    results.}
\label{fig:YRHXS2_VBF_PWG1}
\end{figure}
In \refF{fig:YRHXS2_VBF_PWG1} we plot the invariant mass of the two
tagging jets, $m_{jj}$, and the absolute value of the distance in
rapidity of the two tagging jets, $\abs{y_{j_1}-y_{j_2}}$.  In all
distributions presented in this section the solid blue lines
represent the NLO result, and the green and the red curves the results
of \POWHEG{} interfaced to \HERWIG{} and \PYTHIA{}, respectively.  The
effect of the cuts on the showered results produces a reduction of the
total cross section that appears manifest in the two distributions
shown. There is an agreement between the two showered results,
slightly below the NLO curves. This behaviour is generic for several
physical distributions.

\begin{figure}
  \includegraphics[width=0.49\textwidth,bb = 20 20 540 350, clip]{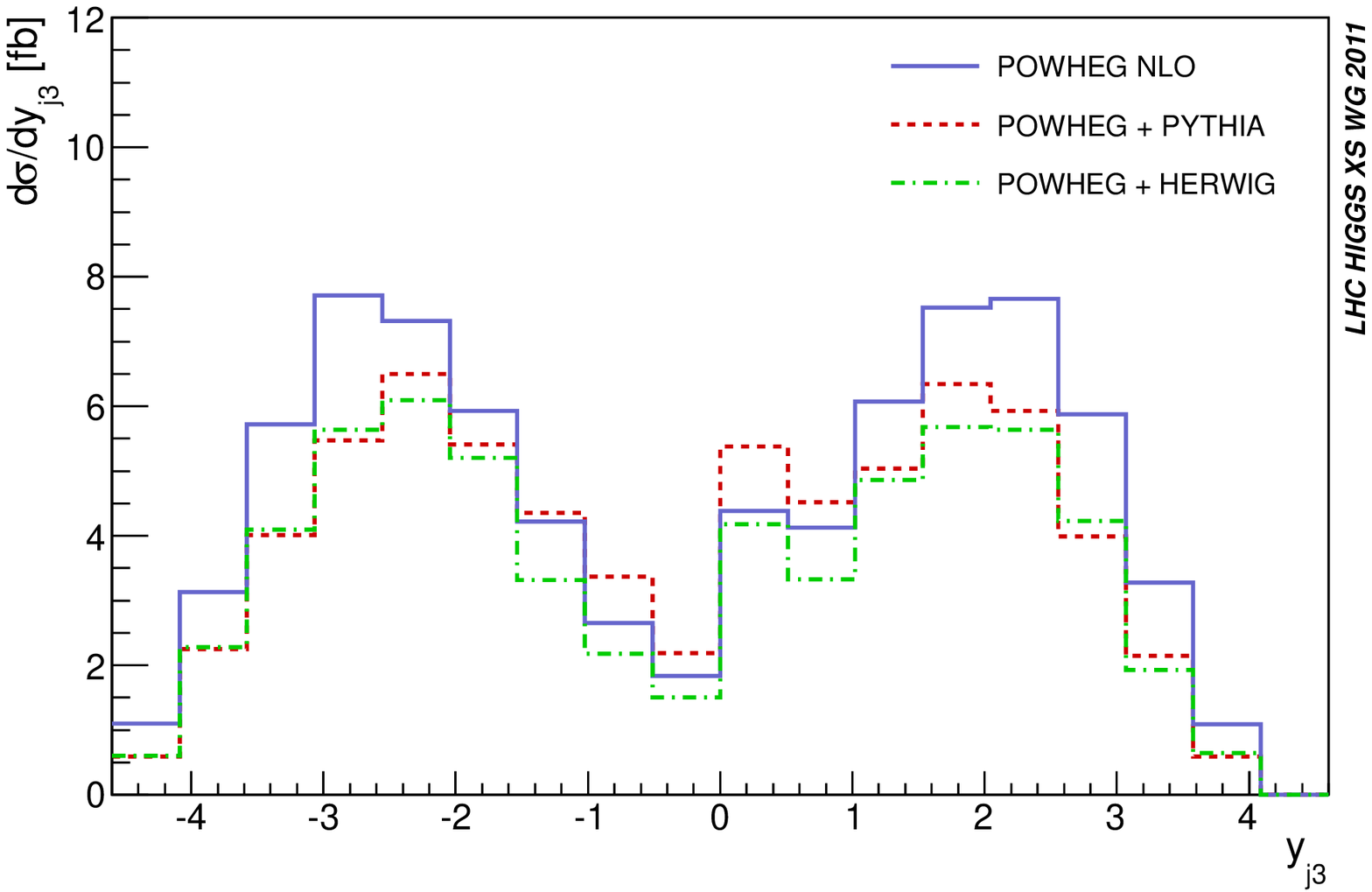}
  \includegraphics[width=0.49\textwidth,bb = 20 20 540 350, clip]{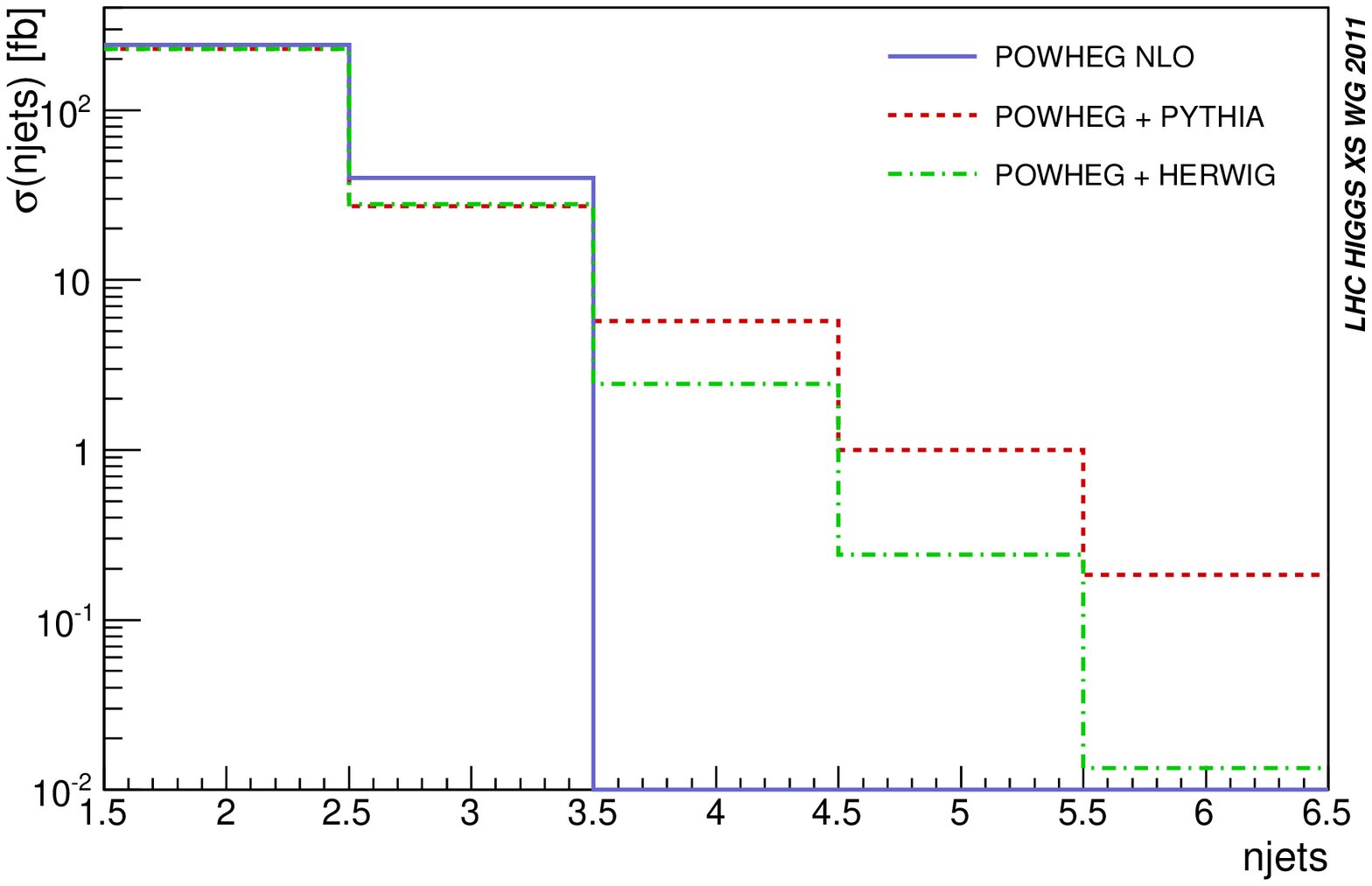}
  \caption{Rapidity distribution of the third
hardest jet, $y_{j_3}$ (left) and exclusive jet-multiplicity (right).   We
  show the comparison among NLO QCD and the \POWHEG{}
  interfaced to \HERWIG{} and \PYTHIA{} results.}
\label{fig:YRHXS2_VBF_PWG2}
\end{figure}
Larger disagreements between the NLO QCD and showered results and between the two
showering programs can be seen in \refF{fig:YRHXS2_VBF_PWG2}.  On the
left-hand side of this figure, we have plotted the rapidity distribution of
the third hardest jet (the one with highest $\pT$ after the two tagging
jets).  The distributions obtained using \POWHEG{} interfaced to \HERWIG{}
and \PYTHIA{} are very similar, but show that fewer events pass
the cuts with respect to the unshowered NLO result.  Nevertheless, they
confirm the behaviour that the third jet  tends to be emitted in the
vicinity of either of the tagging jets.  We recall here that, strictly
speaking, this is a LO distribution, since the third jet in the NLO
calculation comes only from the real-emission contributions.  

We turn now to quantities that are more sensitive to the collinear
and soft physics of the shower.  The jet activity is one such quantity.
In order to quantify it, we plot the exclusive jet-multiplicity distribution
for jets that pass the cuts of \Eref{eq:VBF_cuts1} in the right-hand
plot of \refF{fig:YRHXS2_VBF_PWG2}. The first two tagging jets and the
third jet are well represented by the NLO cross section, which obviously
cannot contribute to events with more than three jets.  From the 4th jet on,
the showers of \HERWIG{} and \PYTHIA{} produce sizable differences (note the
log scale of the plot), the jets from \PYTHIA{} being harder and/or more central
than those from the \HERWIG{} shower.

\begin{figure}
 \includegraphics[width=0.49\textwidth,bb = 20 20 540 350, clip]{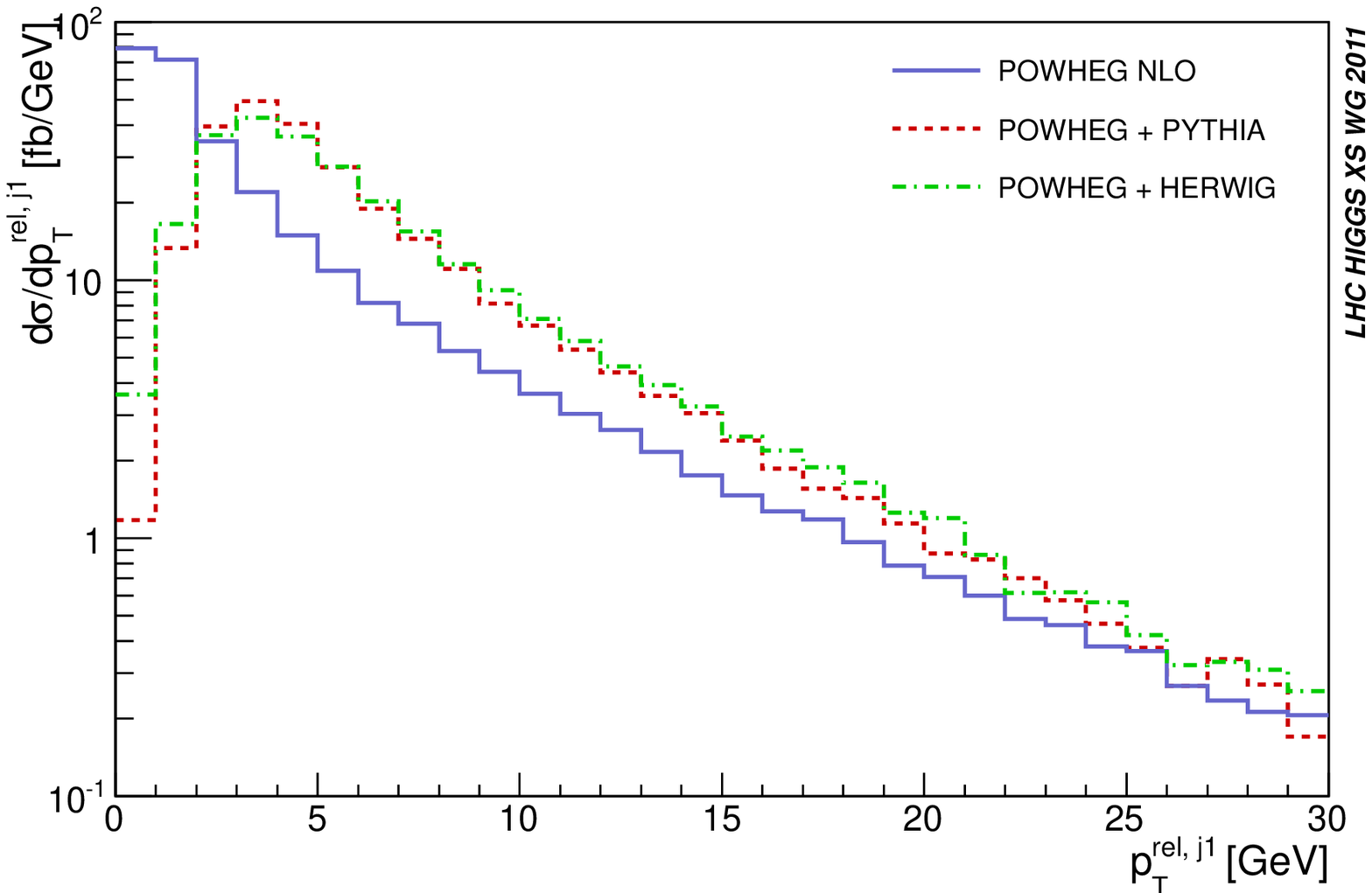} 
 \includegraphics[width=0.49\textwidth,bb = 20 20 540 350, clip]{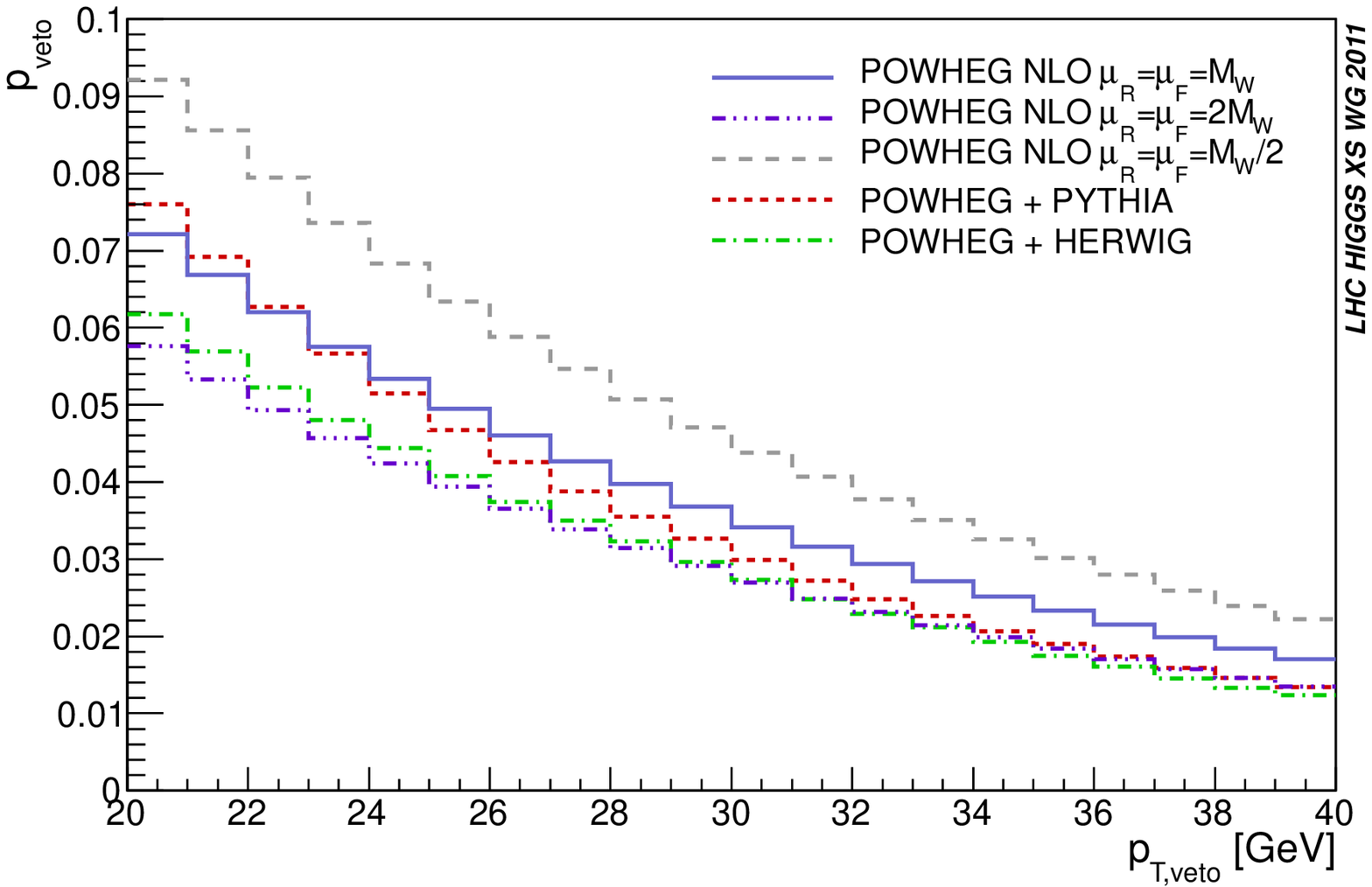}
\vspace*{-.2em}
  \caption{Relative transverse momentum $\pT^{{\rm rel},j_1}$ of all the
    particles clustered inside the first tagging jet, in the reference frame
    where the jet has zero rapidity, defined according
    to \Eref{eq:VBF_def_ptrel} (left) and probability of finding a veto jet
    defined as in \Eref{eq:VBF_P_veto} as a function of $\pT{}_{\rm ,veto}$
    (right).  We show the comparison among NLO QCD and the \POWHEG{}
    interfaced to \HERWIG{} and \PYTHIA{} results. In addition, on the
    right plot, we show the NLO QCD results obtained with
    $\muR=\muF=\MW/2$ and $\muR=\muF=2\MW$.}
\label{fig:YRHXS2_VBF_PWG3}
\end{figure}
Striking differences between the NLO QCD results and \POWHEG{} can be seen in the
relative transverse momentum of all the particles clustered inside one of the
two tagging jets $\pT^{{\rm rel},j}$.  This quantity brings information on
the ``shape'' of the jet. It is defined as follows:
\begin{itemize}
\item[-] For each jet 
we perform a longitudinal boost to
  a frame where the jet has zero rapidity.
\item[-] In this frame, we compute the quantity
\begin{equation}
\label{eq:VBF_def_ptrel}
\pT^{{\rm rel},j}=\sum_{i\in j} \frac{| \vec{k}^i \times
  \vec{p}^{\, j}|}{|\vec{p}^{\, j}|}\,,
\end{equation}
where $k^i$ are the momenta of the particles that belong to the jet
with momentum $p^{j}$.
\end{itemize}
This quantity is thus the sum of the absolute values of the transverse
momenta, taken with respect to the jet axis, of the particles inside the jet,
in the frame specified above.  In the left-hand side of
\refF{fig:YRHXS2_VBF_PWG3} we have plotted the relative transverse
momentum with respect to the first tagging jet, $\pT^{{\rm rel},j_1}$. While
the black NLO curve is diverging as $\pT^{{\rm rel},j_1}$ approaches zero,
the Sudakov form factors damp this region in the \POWHEG{} results.

As a last comparison, we have studied the probability of finding veto jets,
i.e.\ jets that fall within the rapidity interval of the two tagging jets,
\begin{equation}
\min\left(y_{j_1},y_{j_2}\right) < y_j < \max\left(y_{j_1},y_{j_2}\right).
\end{equation}
In fact, for the central-jet-veto proposal, events are discarded if any
additional jet with a transverse momentum above a minimal value, $\pT{}_{\rm
,veto}$, is found between the tagging jets.  The probability, $P_{\rm veto}$,
of finding a veto jet is defined as
\begin{equation}
\label{eq:VBF_P_veto}
P_{\rm veto} = \frac{1}{\sigma_2^{\rm\scriptstyle NLO}} \int_{\pT{}_{\rm ,veto}}^{\infty} 
d\pT^{\,j,{\rm veto}} \frac{d\sigma}{d\pT^{\,j,{\rm veto}}}\,,
\end{equation}
where $\pT^{\,j,{\rm veto}}$ is the transverse momentum of the hardest veto
jet, and $\sigma_2^{\rm \scriptstyle NLO}$ is the total cross section (within
VBF cuts) for $\PH jj$ production at NLO~\footnote{Strictly speaking, $P_{\rm
veto}$ is a LO quantity, since it receives contributions only from the
real-emission diagrams.}.  In the right-hand plot
of \refF{fig:YRHXS2_VBF_PWG3} we have plotted the NLO QCD and \POWHEG{}
interfaced to \HERWIG{} and \PYTHIA{} predictions. For $\MH=120$ \UGeV, we
have $\sigma_2^{\rm \scriptstyle NLO}=282$\Ufb, with the settings described
in \Sref{sec:VBF_setup}.  In addition to the NLO curve obtained with
$\muR=\muF=\MW$, we have added the NLO results computed with
$\muR=\muF=\MW/2$ (upper dashed line) and $\muR=\muF=2\MW$ (lower dash-dotted
line) that show that the \POWHEG{} curves are consistent with the LO band
obtained with a change of the renormalisation and factorisation scale by a
factor of two and that the distance between \POWHEG{}+\PYTHIA{} and
\POWHEG{}+\HERWIG{} is comparable with the scale uncertainty of the
LO result.

\subsubsection{Efficiency of VBF cuts}
The efficiency $\epsilon^{\{\muR,\, \muF\}}$ of the VBF selection cuts
is defined as $\epsilon^{\{\muR,\, \muF\}}=\sigma_{\rm
  cuts}^{\{\muR,\, \muF\}}/\sigma_{\rm inc}^{\{\muR,\, \muF\}}$, where
$\sigma_{\rm cuts}$ is the cross section after all the forward jet
tagging described above and an additional jet-veto cut
$\pT{}_{\rm ,veto}$, and $\sigma_{\rm inc}$ is the inclusive cross
section before cuts, both computed with the renormalisation and
factorisation scales set to $\muR$ and $\muF$, respectively.  The
uncertainty on this ratio is of particular interest to experimental
studies since the theoretical prediction of the efficiency is used to
translate the experimental upper bound on the yield after cuts to a
more useful limit on the inclusive signal cross section.

\begin{figure}
 \includegraphics[width=0.48\textwidth]{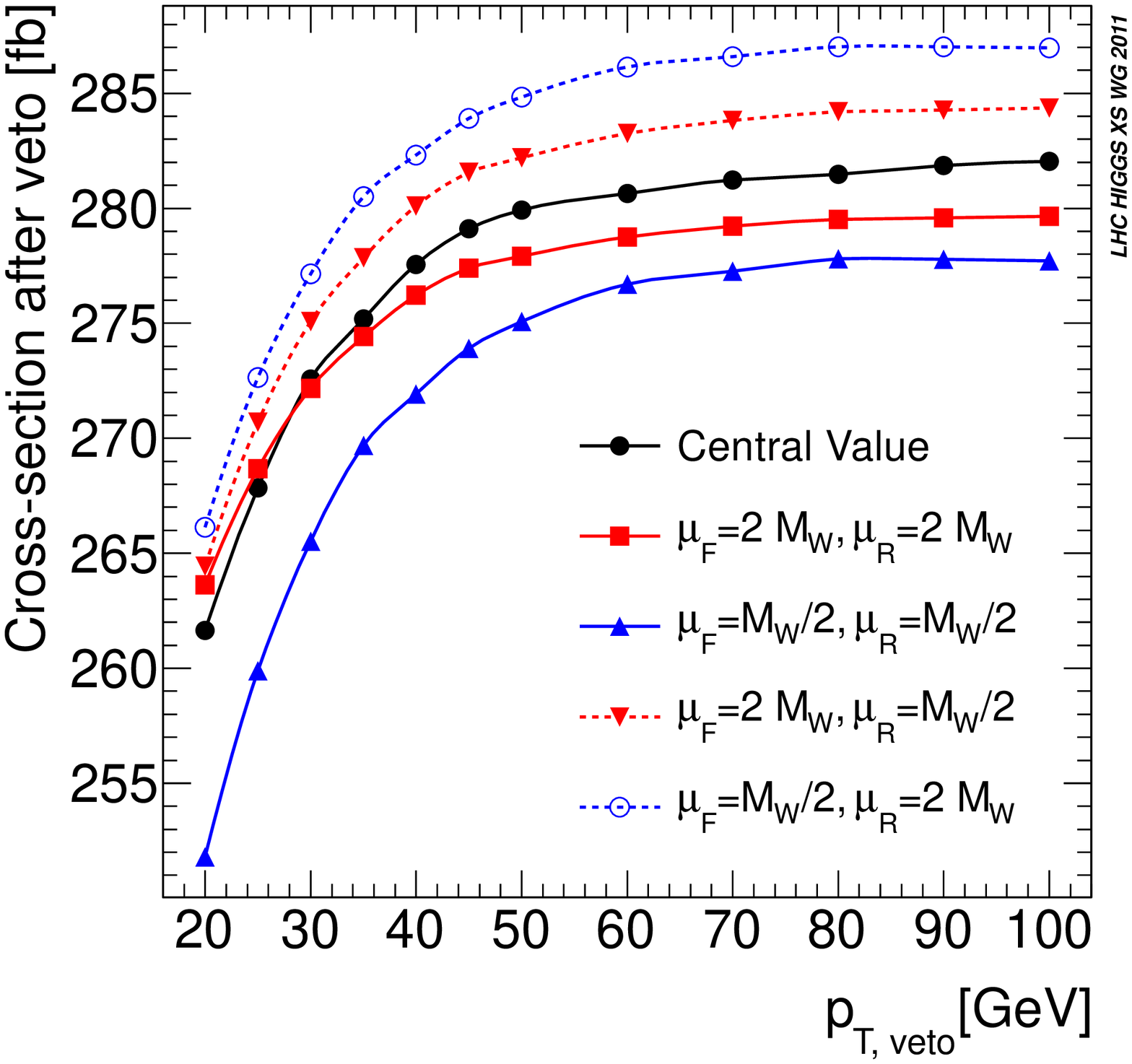}\quad
 \includegraphics[width=0.48\textwidth]{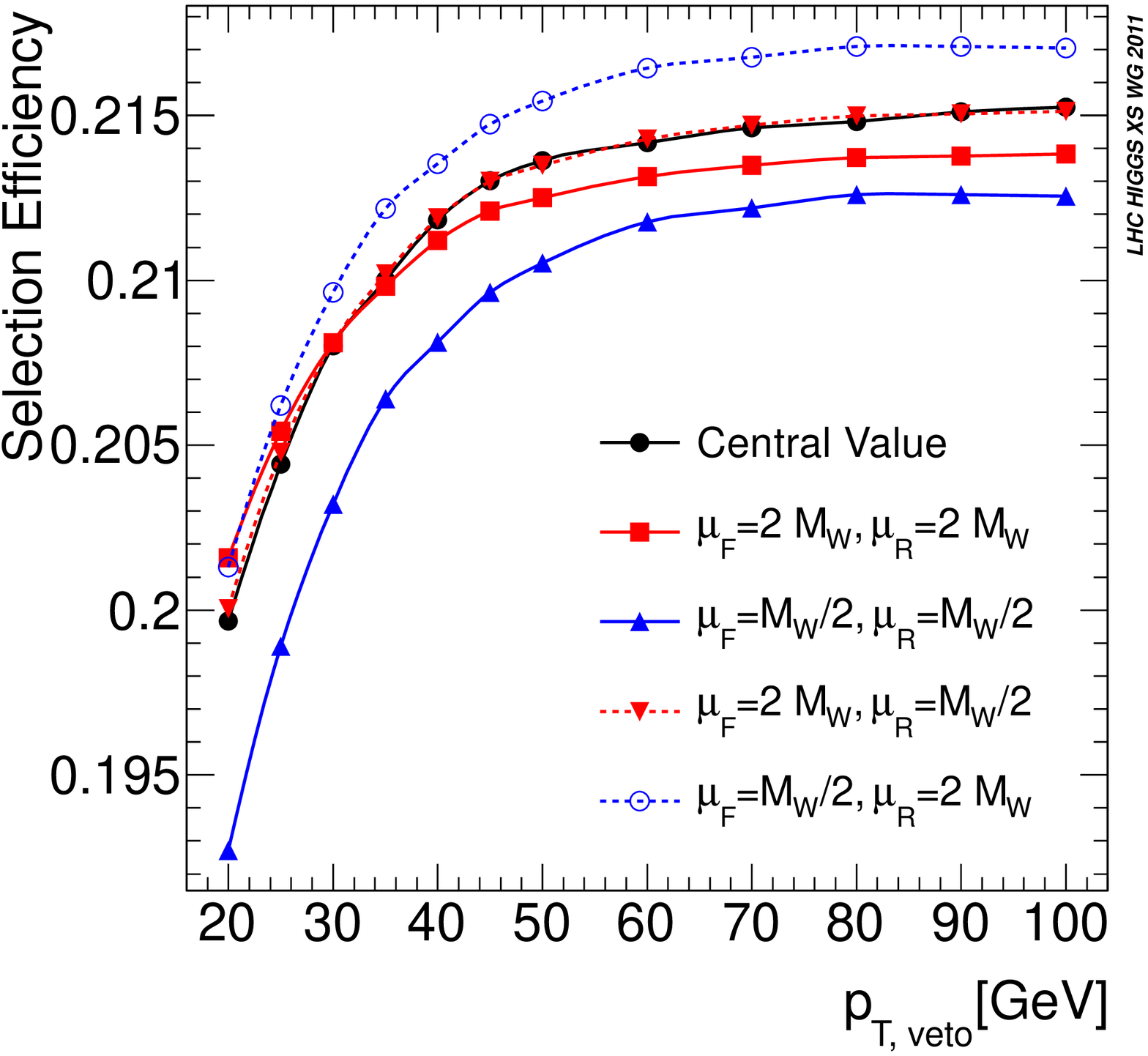}
  \caption{The NLO QCD cross section for $\MH=120\UGeV$ after all cuts
    as a function of 
$\pT{}_{\rm ,veto}$ (left) and the efficiency of the forward jet tagging and
central-jet-veto cuts as a function of $\pT{}_{\rm ,veto}$ (right). Results
obtained with \VBFNLO. }
\label{fig:YRHXS2_VBF_efficiencyplots}
\end{figure}
\refF{fig:YRHXS2_VBF_efficiencyplots} shows the cross section after
cuts (left) and the efficiency of the selection cuts (right) as
functions of $\pT{}_{\rm ,veto}$ for $\MH=120\UGeV$.  The values are
calculated using VBFNLO, at NLO QCD and with the MSTW2008 PDF set.  To
evaluate scale uncertainties, the inclusive cross section and the
exclusive cross sections after cuts are recomputed with four altered
choices of the renormalisation and factorisation scales:
$\{\muR,\muF\}=\{\MW/2,\,\MW/2\}$, $\{2\MW,\,2\MW\}$,
$\{2\MW,\,\MW/2\}$, and $\{\MW/2,\,2\MW\}$.  The cross section and
efficiency curves with these altered scale choices are shown as the
red and blue solid and dashed curves
in~\refF{fig:YRHXS2_VBF_efficiencyplots}.  The scale uncertainty on
the selection efficiency for a given choice of $\pT{}_{\rm ,veto}$ is
taken as the largest deviation of the predicted efficiency obtained
using any of these altered scale choices from the result obtained from
the central-value scale choice $\{\muR,\muF\}=\{\MW,\, \MW\}$.  PDF
uncertainties are estimated using the MSTW2008 $90\%$ C.L.\ error PDF
sets according to the CTEQ prescription with symmetric errors~\cite{Nadolsky:2008zw}.

\begin{table}
\caption{NLO QCD cross sections and efficiencies from \VBFNLO{} for
  the full VBF selection including the jet veto cut of $20\UGeV$ and the
corresponding relative uncertainties from 
the QCD scale and PDFs.}
\label{tab:VBF_afterveto}
\begin{center}%
\begin{small}%
\begin{tabular}{c|ccc|ccc}
\hline
 \multicolumn{1}{c}{$\MH$}  & \multicolumn{3}{c}{Cross section} & \multicolumn{3}{c}{Efficiency}\\
{}[GeV] & [fb] & [\%] & [\%] & [\%] & [\%] & [\%]\\
\hline
$120 $ & $ 261.64 $ & $\pm 3.76 $ & $ \pm 4.91 $ & $
0.200 $ & $\pm  3.485 $ & $ \pm 1.468 $ \\
$150 $ & $ 218.69 $ & $\pm 3.59 $ & $ \pm 4.66 $ & $
0.221 $ & $\pm  3.376 $ & $ \pm 1.196 $ \\
$200 $ & $ 163.34 $ & $\pm 3.92 $ & $ \pm 4.66 $ & $
0.252 $ & $\pm  3.829 $ & $ \pm 1.490 $ \\
$250 $ & $ 123.06 $ & $\pm 4.16 $ & $ \pm 5.11 $ & $
0.279 $ & $\pm  4.145 $ & $ \pm 1.493 $ \\
$500 $ & $ ~~33.55  $ & $\pm 4.37 $ & $ \pm 5.43 $ & $
0.365 $ & $\pm  4.766 $ & $ \pm 1.227 $ \\
$600 $ & $ ~~20.82  $ & $\pm 4.44 $ & $ \pm 5.58 $ & $
0.384 $ & $\pm  4.958 $ & $  \pm 1.002 $ \\
\hline
\end{tabular}%
\end{small}%
\end{center}%
\end{table}%

\refT{tab:VBF_afterveto} shows the central value as well as the
relative scale
and PDF uncertainties for the cross section after cuts and the
selection efficiency, for several values of the Higgs-boson mass and
for a jet-veto cut of $20\UGeV$. For the cross sections, the scale
uncertainties are at the level of $\sim 4\%$, while the PDF
uncertainties are closer to $5\%$.  For the efficiencies, the scale
uncertainties are also close to $4\%$, but the PDF uncertainties are
generally smaller, around $1\%$.


\clearpage

\newpage

\section{$\PW\PH$/$\PZ\PH$ production mode\footnote{%
    S.~Dittmaier, R.V.~Harlander, J.~Olsen, G.~Piacquadio (eds.);
    A.~Denner, G.~Ferrera, M.~Grazzini, S.~Kallweit, A.~M\"uck and F.~Tramontano.}}

\subsection{Theoretical developments}
\label{sec:YRHX2_WHZH_th}

In the previous report~\cite{Dittmaier:2011ti} state-of-the-art predictions
and error estimates for the total cross sections for $\Pp\Pp\to\PW\PH/\PZ\PH$ 
have been compiled, based on (approximate) next-to-next-to-leading-order (NNLO)
QCD and next-to-leading-order (NLO) electroweak (EW) corrections.
In more detail, the QCD corrections ($\sim30\%$) comprise
Drell--Yan-like contributions~\cite{Brein:2003wg}, which respect factorisation
according to $\Pp\Pp\to \PV^*\to \PV\PH$ and represent the dominant
parts, and a smaller remainder, which contributes beyond NLO.
The NLO EW corrections to the total cross sections were evaluated as in
\Bref{Ciccolini:2003jy} and turn out to be about $-(5{-}10)\%$.
In the report~\cite{Dittmaier:2011ti} the Drell--Yan-like NNLO QCD
predictions are dressed with the NLO EW corrections in fully factorised form
as suggested in \Bref{Brein:2004ue},
i.e.\ the EW corrections simply enter as multiplicative correction factor,
which is rather insensitive to QCD scale and parton distribution function (PDF)
uncertainties. For ZH production the one-loop-induced subprocess $\Pg\Pg\to\PZ\PH$,
which is part of the non-Drell--Yan-like NNLO QCD corrections, was taken into account as well.
For the LHC with a centre-of-mass (CM) energy of $7(14)\UTeV$
the QCD scale uncertainties
were assessed to be about $1\%$ and $1{-2}(3{-}4)\%$ for $\PW\PH$
and $\PZ\PH$ production, respectively, while
uncertainties of the PDFs turn out to be about $3{-}4\%$.

After the completion of the report~\cite{Dittmaier:2011ti} theoretical progress
has been made in various directions: 
\begin{itemize}
\item 
On the QCD side, the Drell--Yan-like NNLO corrections to $\PW\PH$ production 
are available now~\cite{Ferrera:2011bk} including the full kinematical information
of the process and leptonic W decays.
\item
For total cross sections the non-Drell--Yan-like remainder at NNLO QCD (apart from
the previously known $\Pg\Pg\to\PZ\PH$ subprocess) has
been calculated recently~\cite{Brein:2011vx}; in particular, this
includes contributions with a top-quark induced gluon--Higgs coupling. As
previously assumed, these effects are at the per-cent level. They
typically increase towards larger Higgs masses and scattering energies,
reaching about $2.5\%(3\%)$ for ZH production at $7\UTeV(14\UTeV)$. 
Since these Yukawa-induced terms arise for the first time at
${\cal O}(\alphas^2)$, they also increase the perturbative uncertainty
which, however, still remains below the error from the PDFs.
\item
On the EW side, the NLO corrections have been generalised to the more complex processes
$\Pp\Pp\to\PW\PH\to\PGn_{\Pl}\Pl\PH$ and
$\Pp\Pp\to\PZ\PH\to\Pl^-\Pl^+\PH/\PGn_{\Pl}\PAGn_{\Pl}\PH$ including the
W/Z decays, also fully supporting differential observables~\cite{whzhhawk}; 
these results are
available as part of the {\HAWK} Monte Carlo 
program~\cite{HAWK}, which was originally designed for the description of Higgs
production via vector-boson fusion including NLO QCD and EW
corrections~\cite{Ciccolini:2007jr}.
\end{itemize}

The following numerical results on differential quantities are, thus, obtained as follows:
\begin{itemize}
\item WH production: 
The fully differential (Drell--Yan-like) NNLO QCD prediction of \Bref{Ferrera:2011bk}
is reweighted with the relative EW correction factor calculated with {\HAWK},
analogously to the previously used procedure for the total cross section.
The reweighting is done bin by bin for each distribution.
\item ZH production: 
Here the complete prediction is obtained with {\HAWK} including NLO QCD and EW
corrections, employing the factorisation of relative EW corrections as well.
\end{itemize}

\subsection{Numerical results}

For the numerical results in this section, we have used the following
setup. The renormalisation and factorisation scales have been identified
and set to
\begin{equation}
\muR=\muF=\MH+\MV,
\end{equation}
where $\MV$ is the \PW/\PZ-boson mass for WH/ZH production. We
have employed the MSTW2008 PDF sets 
at NNLO for WH production and the PDF4LHC prescription to 
calculate the cross section for ZH production at NLO. 
The relative EW corrections have been calculated using the 
central NLO MSTW2008 PDF, but hardly depend
on the PDF and scale choice. We use the $\GF$ scheme to fix the
electromagnetic coupling $\alpha$ and use the values
of $\alphas$ associated to the given PDF set. For the QCD predictions to  
$\PW\PH$ production, we employ the full CKM matrix which
enters via global factors multiplying the different partonic channels. 
For all the \HAWK{} predictions, we neglect mixing of the first two generations
with the third generation. In the EW loop corrections, the CKM matrix
is set to unity, since quark mixing is negligible there.
For $\PZ\PH$ production, bottom quarks in the initial state are only
included in LO within \HAWK{} because of their small impact on the cross sections.

Both the NNLO QCD prediction as well as the \HAWK{} predictions are obtained
for off-shell vector bosons. The vector-boson width can be viewed as a free
parameter of the calculation, and we have chosen 
$\Gamma_{\PW} = 2.08872\UGeV$ and
$\Gamma_{\PZ}$ from the default input. The NNLO QCD calculation~\cite{Ferrera:2011bk} predicts the
integrated \PW-boson production cross section in the presence of the cuts defined
below, so that it had to be multiplied by the branching ratio 
$\mathrm{BR}_{\PW \to \Pl \PGn_{\Pl}}$ for a specific leptonic
final state. Because the EW radiative corrections to the BR are included in
the EW corrections from \HAWK{}, the partial \PW~width has to be used at Born level
in the QCD-improved cross section,
i.e.\ $\mathrm{BR}_{\PW \to \Pl \PGn_{\Pl}}= 
\Gamma^{\mathrm{Born}}_{\PW \to \Pl \PGn_{\Pl}}/\Gamma_{\PW}$, where
$\Gamma^{\mathrm{Born}}_{\PW \to \Pl \PGn_{\Pl}}=\GF \MW^3/(6 \sqrt{2} \pi)$. 
Using any different input value for $\Gamma^{\mathrm{new}}_{\PW}$, all results 
are thus, up to negligible corrections, changed by the ratio 
$\Gamma_{\PW}/\Gamma^{\mathrm{new}}_{\PW}$. In the \HAWK{} prediction, the 
branching ratios for the different leptonic channels are implicitly included by
calculating the full matrix elements. However, up to negligible corrections
the same scaling holds if another numerical value for the input width was used.
In particular, the relative EW corrections hardly depend on 
$\Gamma_{\PW}$.

All results are given for a specific 
leptonic decay mode, e.g.\ for $\PH \Pe^+\Pe^-$ or $\PH \PGmp\PGmm$ production, 
and are not summed over lepton generations. While for charged leptons
the results depend on the prescription for lepton--photon recombination 
(see below), the results for invisible \PZ\ decays, 
of course, do not depend on the neutrino flavour and can be trivially obtained 
by multiplying the $\PH \PGn_{\Pl} \PAGn_{\Pl}$ results by three.

In the calculation of EW corrections, we alternatively apply two versions of
handling photons that become collinear to outgoing charged leptons. The first
option is to assume perfect isolation between charged leptons and photons, an
assumption that is at least approximately fulfilled for (bare) muons. The second
option performs a recombination of photons and nearly collinear charged leptons
and, thus, mimics the inclusive treatment of electrons within electromagnetic
showers in the detector. Specifically, a photon $\PGg$ and a lepton \Pl\ are
recombined for $R_{\Pl\PGg}<0.1$, where
$R_{\Pl\PGg}=\sqrt{(y_{\Pl}-y_{\PGg})^2+\phi^2_{\Pl\PGg}}$
is the usual separation variable in the $y{-}\phi$-plane with $y$ denoting the
rapidity and $\phi_{\Pl\PGg}$ the angle between \Pl\ and $\PGg$ in the
plane perpendicular to the beams. If \Pl\ and $\PGg$ are recombined, we simply
add their four-momenta and treat the resulting object as quasi-lepton. If more
than one charged lepton is present in the final state, the possible
recombination is performed with the lepton delivering the smaller value of
$R_{\Pl\PGg}$. The corresponding EW corrections are labeled
$\delta^\mathrm{bare}$ and $\delta^\mathrm{rec}$, respectively.

After employing the recombination procedure we apply the following cuts on the
charged leptons,
\begin{equation}
p_{\mathrm{T},\Pl} > 20\UGeV, \quad
|y_{\Pl}|< 2.5,
\end{equation}
where $p_{\mathrm{T},\Pl}$ is the transverse momentum of the lepton \Pl. For
channels with at least one neutrino in the final state we require a missing
transverse momentum
\begin{equation}
p_{\mathrm{T,miss}} > 25\UGeV,
\end{equation}
which is defined as the total transverse momentum of the neutrinos in the event.
In addition, we apply the cuts
\begin{equation}
p_{\mathrm{T},\PH} > 200\UGeV, \quad
p_{\mathrm{T},\PW/\PZ} > 190\UGeV
\label{eq:VHpTcuts}
\end{equation}
on the transverse momentum of the Higgs and the weak gauge bosons, respectively.
The corresponding selection of events with boosted Higgs bosons is improving the
signal-to-background ratio in the context of employing the measurement of the
jet substructure in $\PH \to \Pb \bar\Pb$ decays leading to a single fat jet.
The need for background suppression calls for (almost) identical cuts on the
transverse momentum of the vector bosons and the Higgs boson. However, symmetric
cuts induce large radiative corrections in fixed-order calculations
in the corresponding $\pT$ distributions near the cut. Since the
Higgs boson and the vector boson are back-to-back at LO, any
initial-state radiation will either decrease $p_{\mathrm{T},\PH}$ or
$p_{\mathrm{T},\PW/\PZ}$ and the event may not pass the cut anymore. Hence,
the differential cross section near the cut is sensitive to almost collinear
and/or rather soft initial-state radiation. By choosing the above (slightly
asymmetric) cuts this large sensitivity to higher-order corrections can be
removed for the important $p_{\mathrm{T},\PH}$-distribution. Of course, since
the LO distribution for $p_{\mathrm{T},\PW/\PZ}$ is vanishing
for $p_{\mathrm{T},\PW/\PZ}< 200\UGeV$ due to the $p_{\mathrm{T},\PH}$ cut,
the higher-order corrections to the $p_{\mathrm{T},\PW/\PZ}$ distributions
are still large in this region.

In the following plots, we show several relative corrections 
and the absolute cross-section predictions based on 
factorisation for QCD and EW corrections,
\begin{equation}
\sigma=\sigma^\mathrm{QCD}\times
\left( 1 + \delta_{\mathrm{EW}}^\mathrm{rec} \right)
+ \sigma_\gamma\, ,
\end{equation}
where $\sigma^\mathrm{QCD}$ is the best QCD prediction at hand,
$\delta_{\mathrm{EW}}^\mathrm{rec}$ is the relative EW correction with
recombination and $\sigma_\gamma$ is the cross section due to photon-induced
processes which are at the level of $1\%$ and estimated employing the
MRSTQED2004 PDF set for the photon.
In detail, we discuss the distributions in $p_{\mathrm{T},\PH}$,
$p_{\mathrm{T},\PV}$, $p_{\mathrm{T},\Pl}$, and $y_{\PH}$.
More detailed results can be found in \Bref{whzhhawk}.

\begin{figure}
\includegraphics[width=7.5cm]{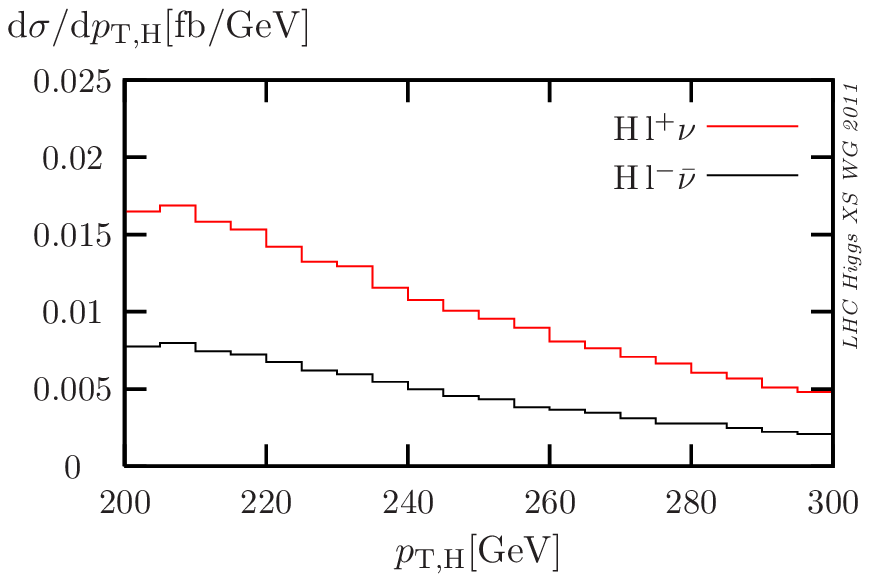}
\hfill
\includegraphics[width=7.5cm]{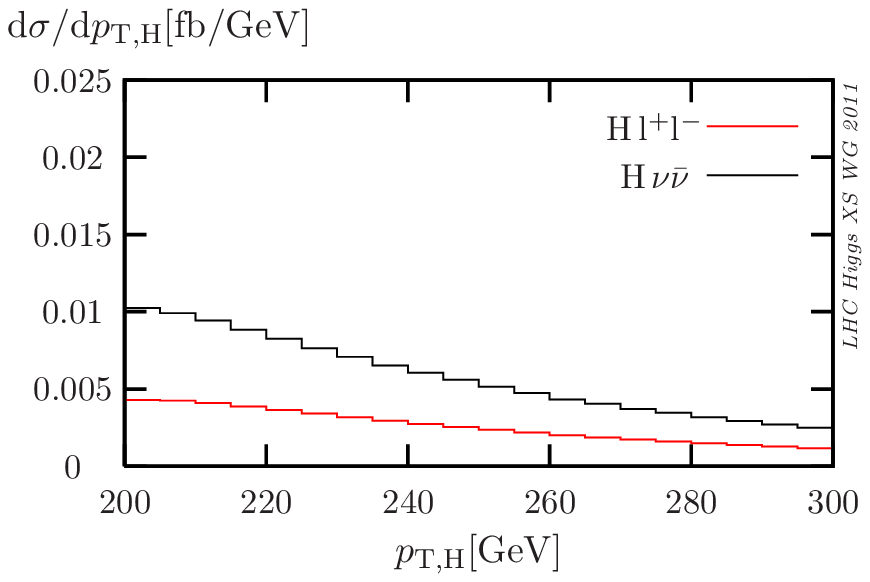}
\\
\includegraphics[width=7.5cm]{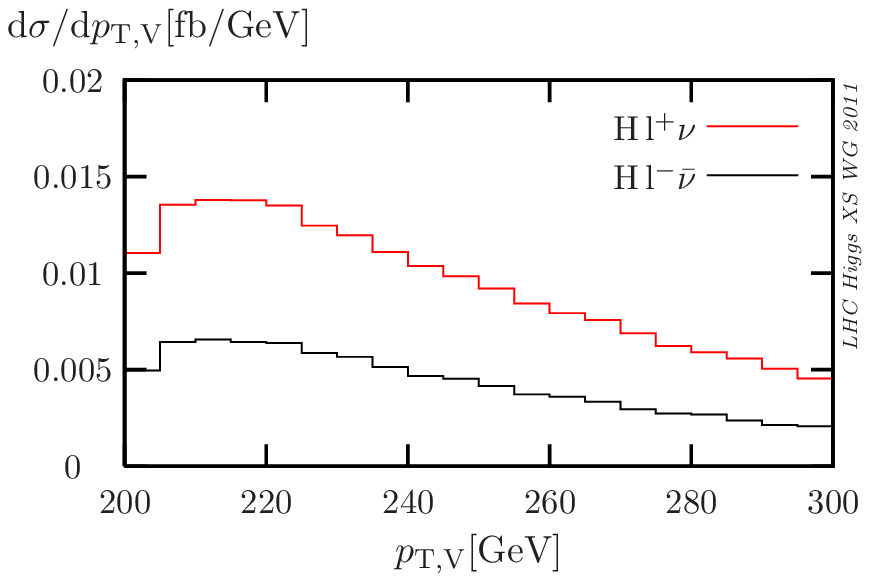}
\hfill
\includegraphics[width=7.5cm]{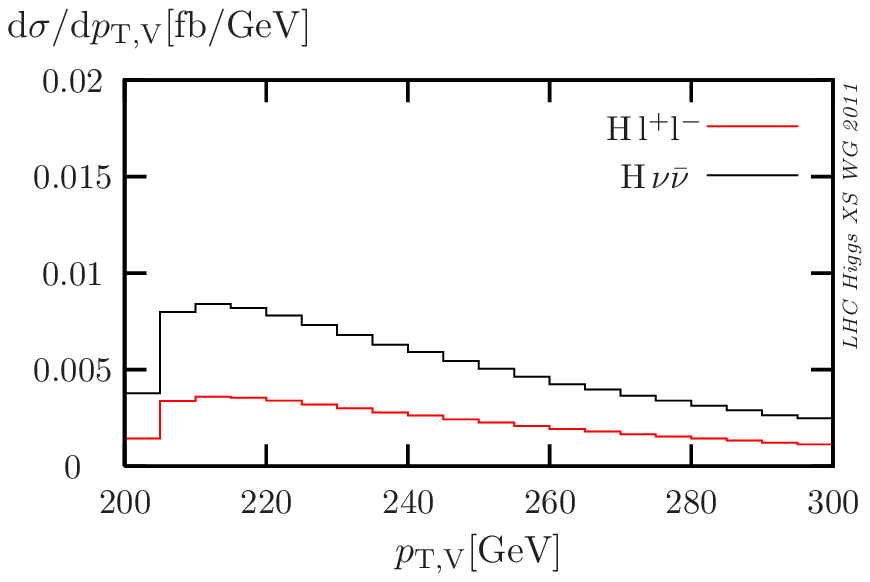}
\\
\includegraphics[width=7.5cm]{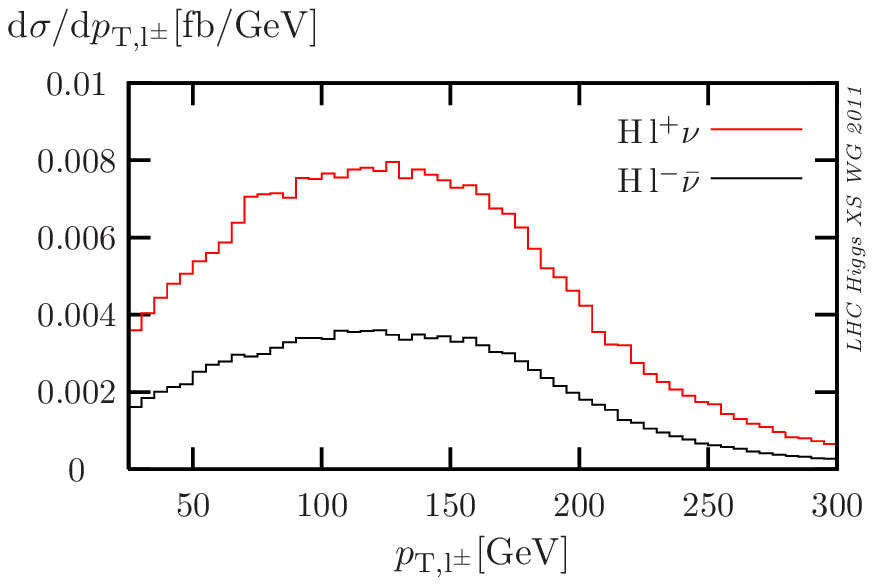}
\hfill
\includegraphics[width=7.5cm]{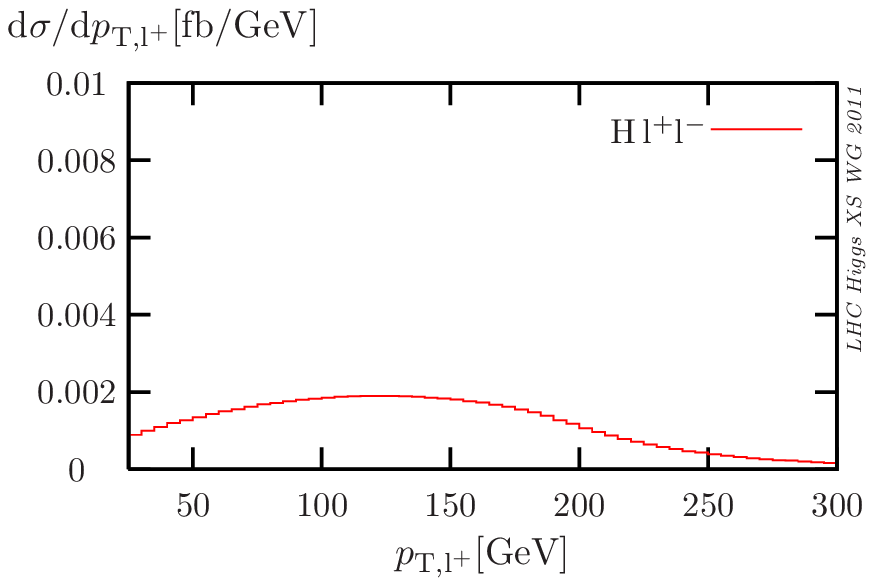}
\\
\includegraphics[width=7.5cm]{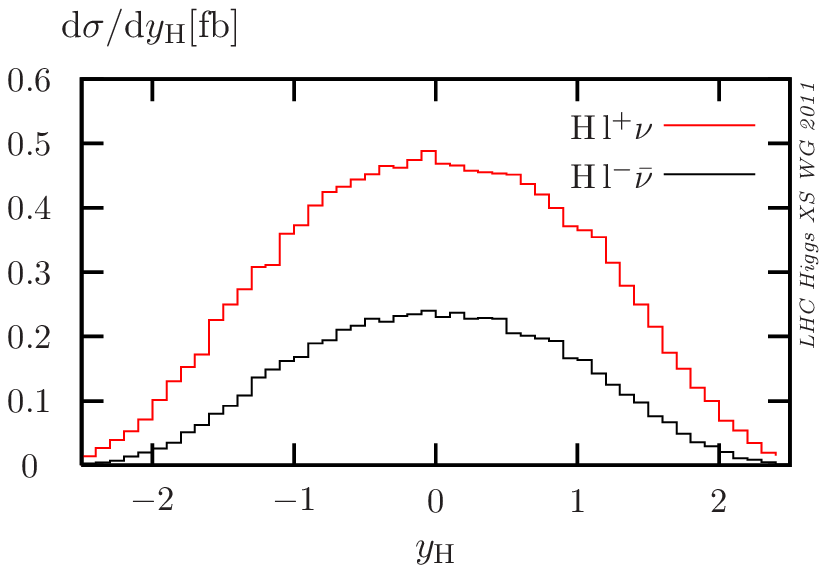}
\hfill
\includegraphics[width=7.5cm]{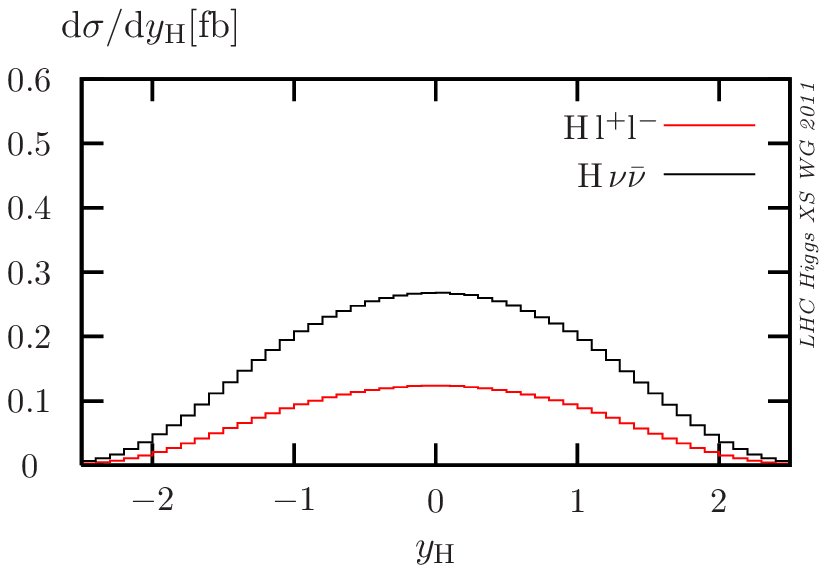}
\caption{\label{fi:abs_pTcut}
Predictions for the
$p_{\mathrm{T},\PH}$, $p_{\mathrm{T},\PV}$,
$p_{\mathrm{T},\Pl}$, and $y_{\PH}$ distributions (top to bottom)
for Higgs strahlung off \PW\ bosons (left) and \PZ\ bosons (right)
for boosted Higgs bosons at
the $7\UTeV$ LHC for $\MH=120\UGeV$.
}
\end{figure}

\begin{figure}
\includegraphics[width=7.5cm]{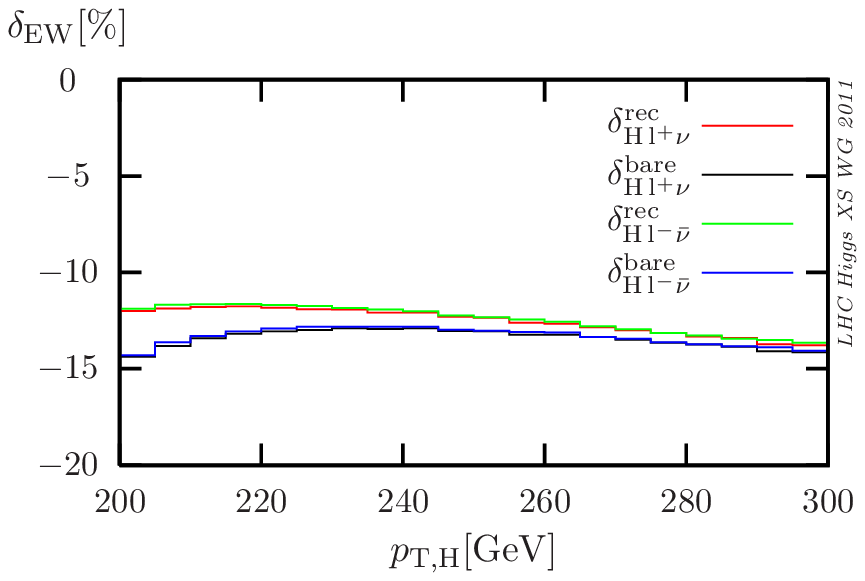}
\hfill
\includegraphics[width=7.5cm]{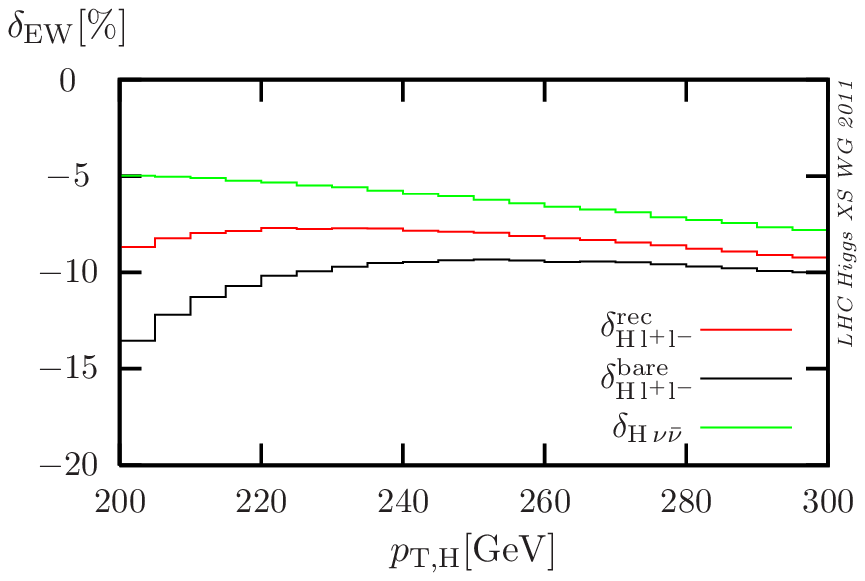}
\\
\includegraphics[width=7.5cm]{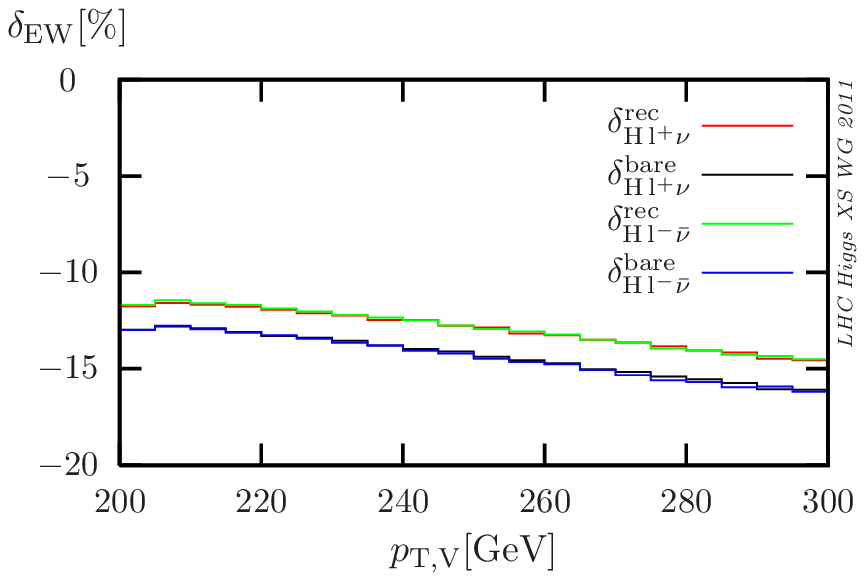}
\hfill
\includegraphics[width=7.5cm]{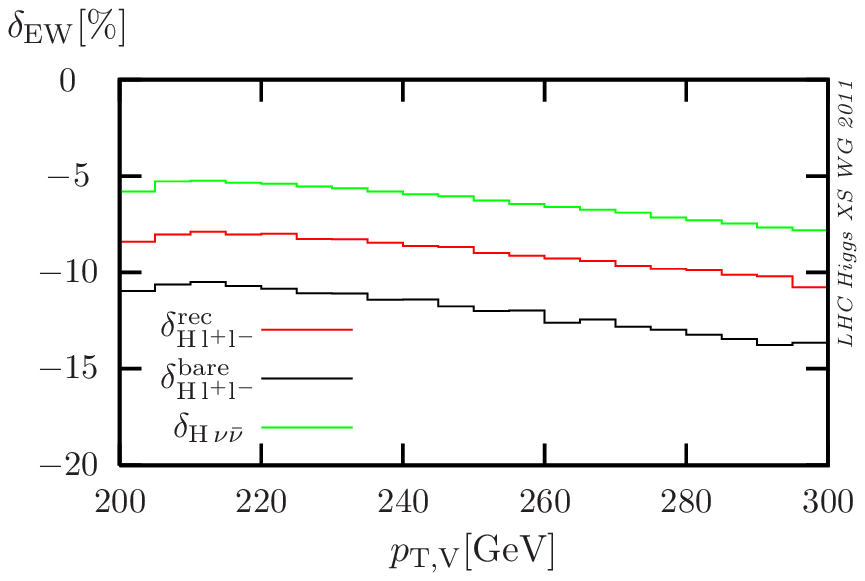}
\\
\includegraphics[width=7.5cm]{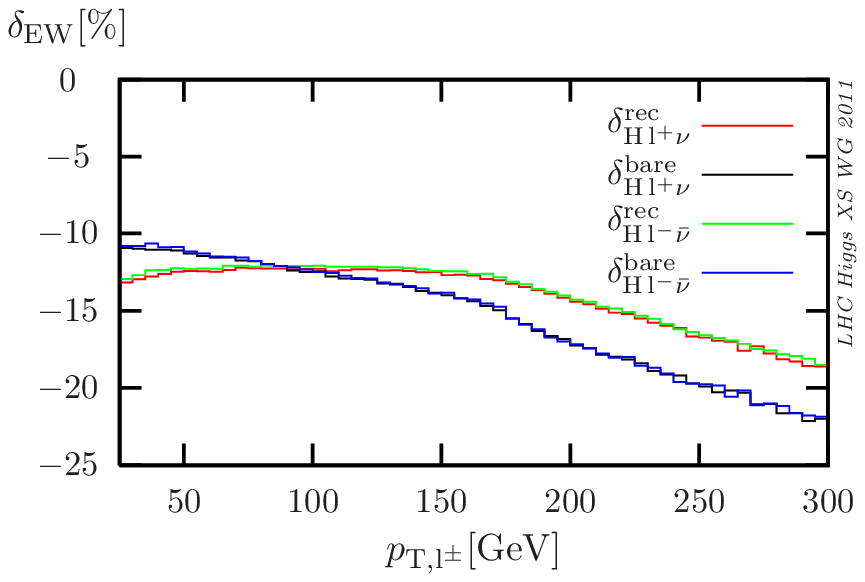}
\hfill
\includegraphics[width=7.5cm]{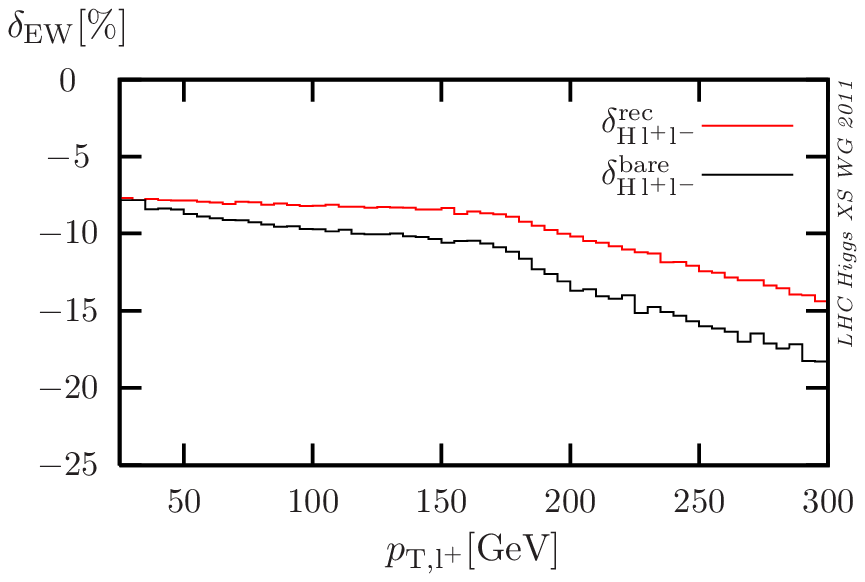}
\\
\includegraphics[width=7.5cm]{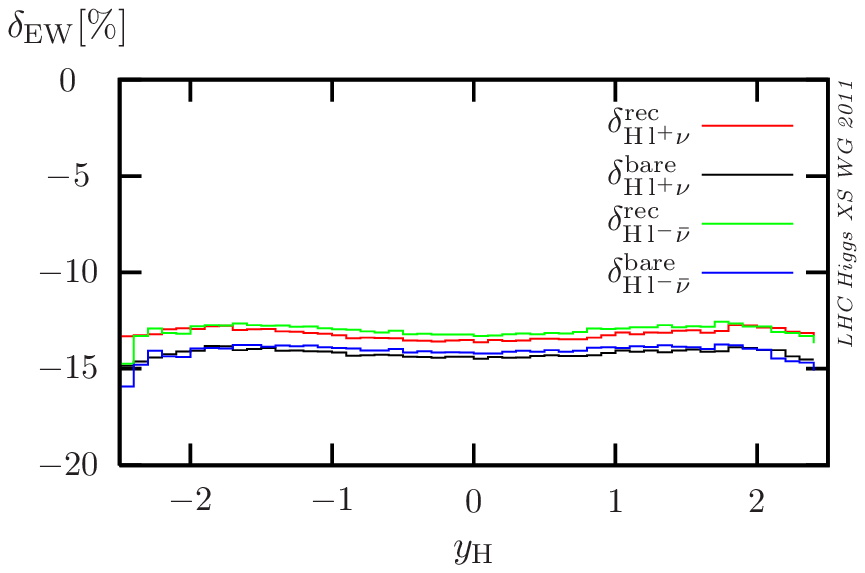}
\hfill
\includegraphics[width=7.5cm]{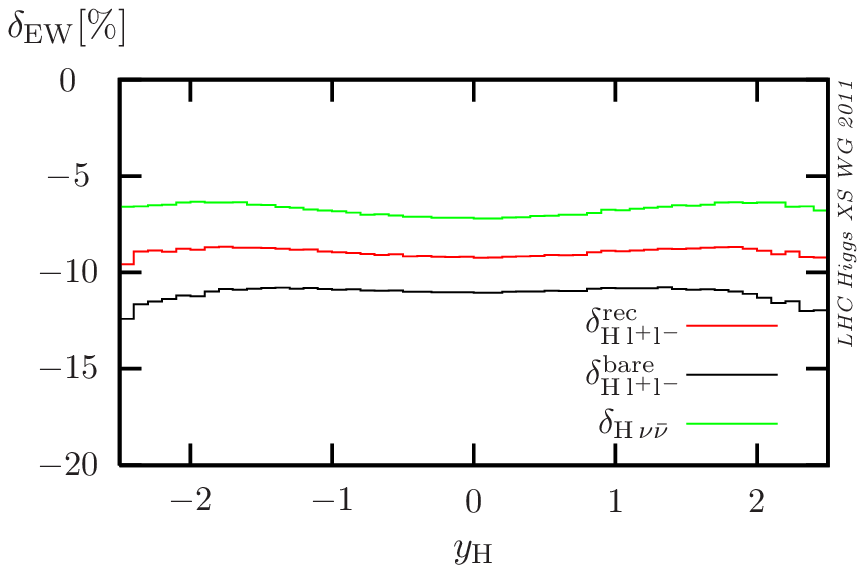}
\caption{\label{fi:rel_pTcut}
Relative EW corrections for the
$p_{\mathrm{T},\PH}$, $p_{\mathrm{T},\PV}$,
$p_{\mathrm{T},\Pl}$, and $y_{\PH}$ distributions (top to bottom)
for Higgs strahlung off \PW\ bosons (left) and \PZ\ bosons (right)
for boosted Higgs bosons at
the $7\UTeV$ LHC for $\MH=120\UGeV$.
}
\end{figure}

\begin{figure}
\includegraphics[width=7.5cm]{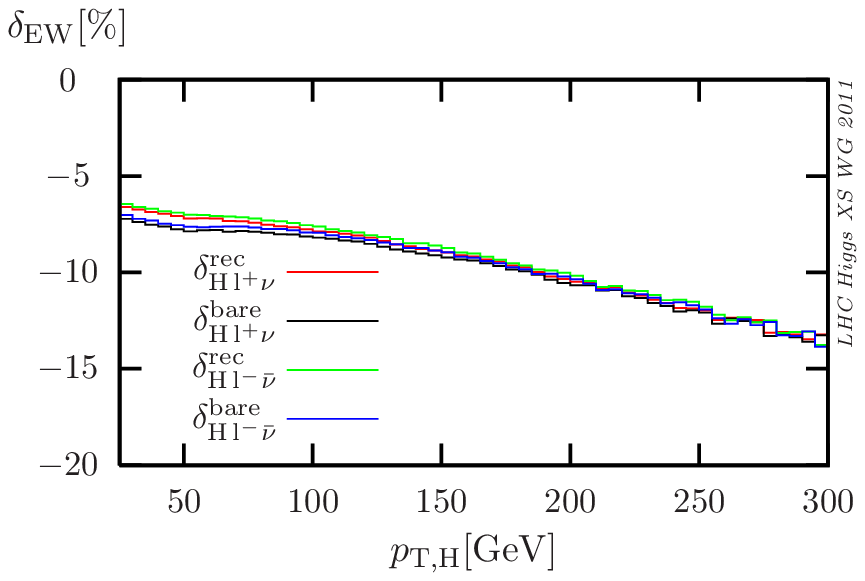}
\hfill
\includegraphics[width=7.5cm]{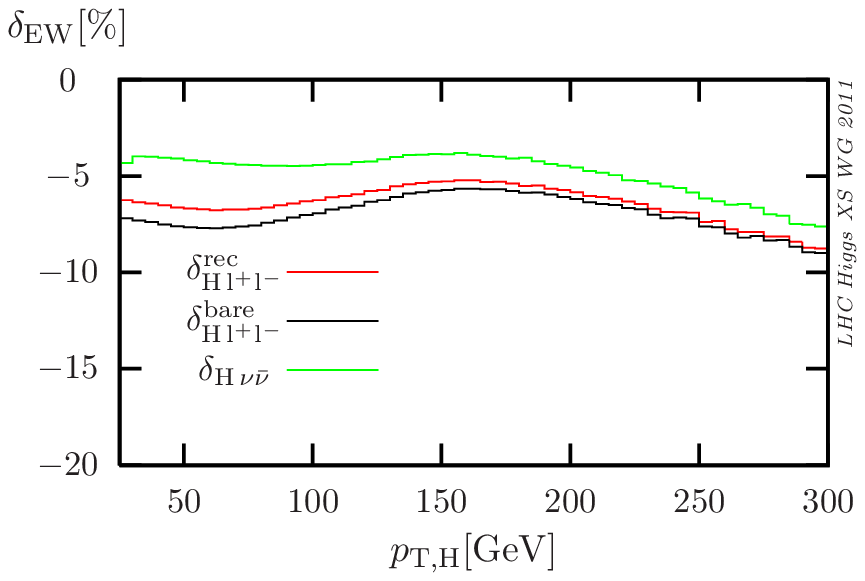}
\\
\includegraphics[width=7.5cm]{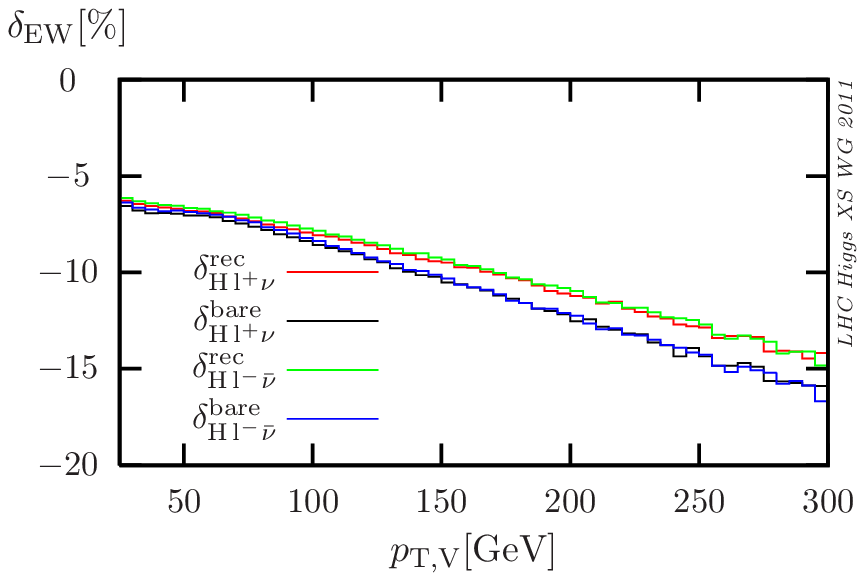}
\hfill
\includegraphics[width=7.5cm]{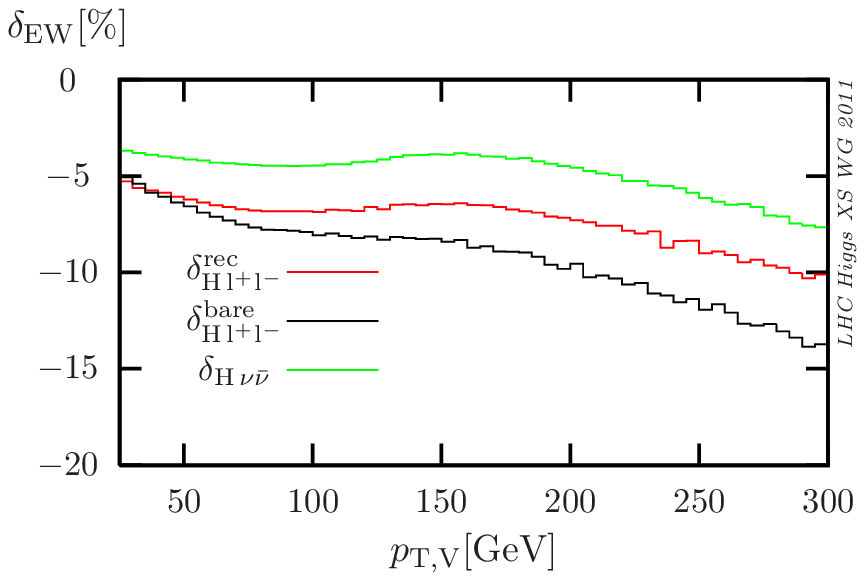}
\\
\includegraphics[width=7.5cm]{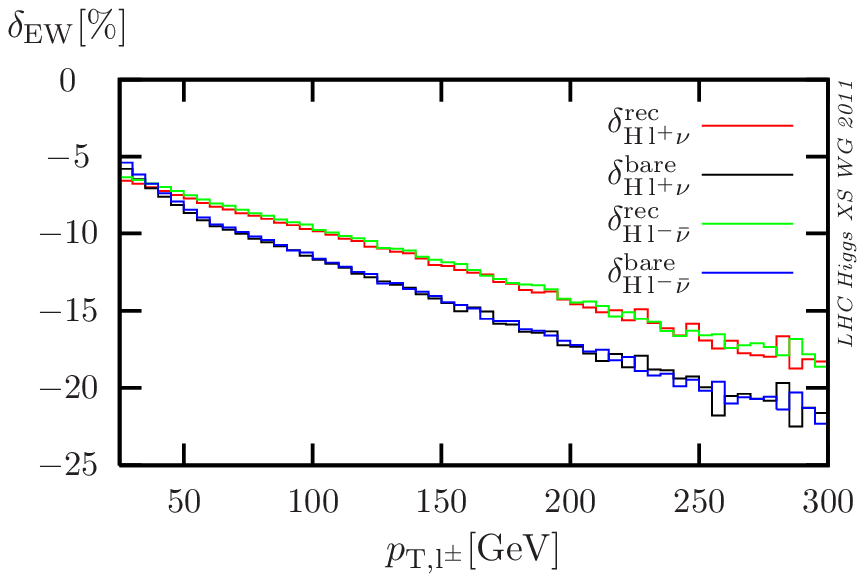}
\hfill
\includegraphics[width=7.5cm]{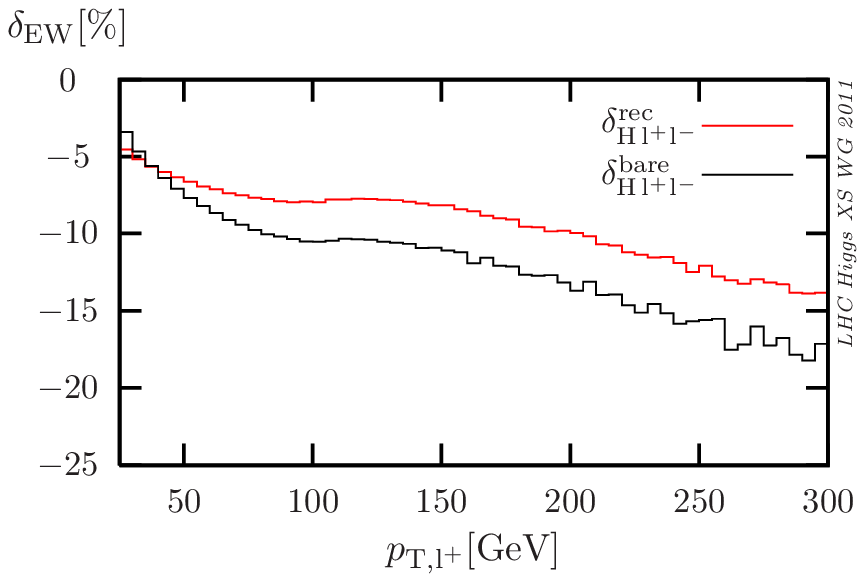}
\\
\includegraphics[width=7.5cm]{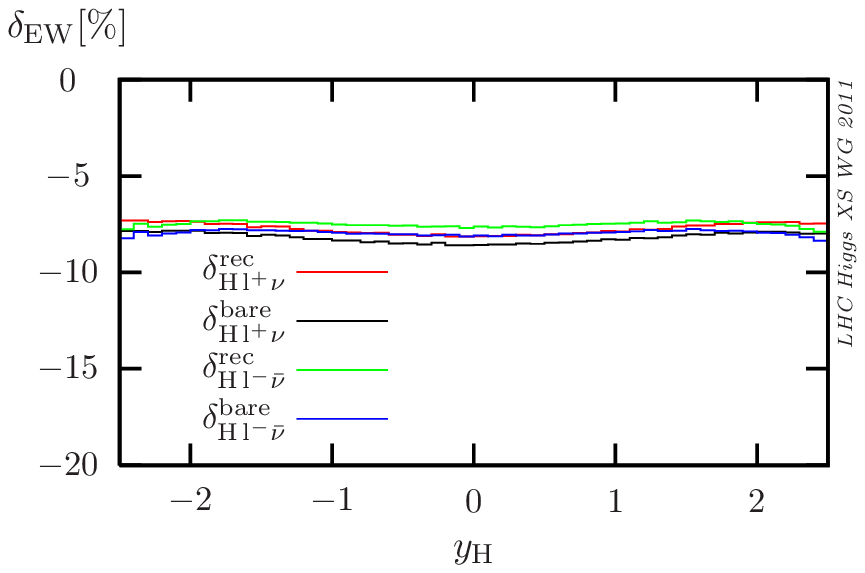}
\hfill
\includegraphics[width=7.5cm]{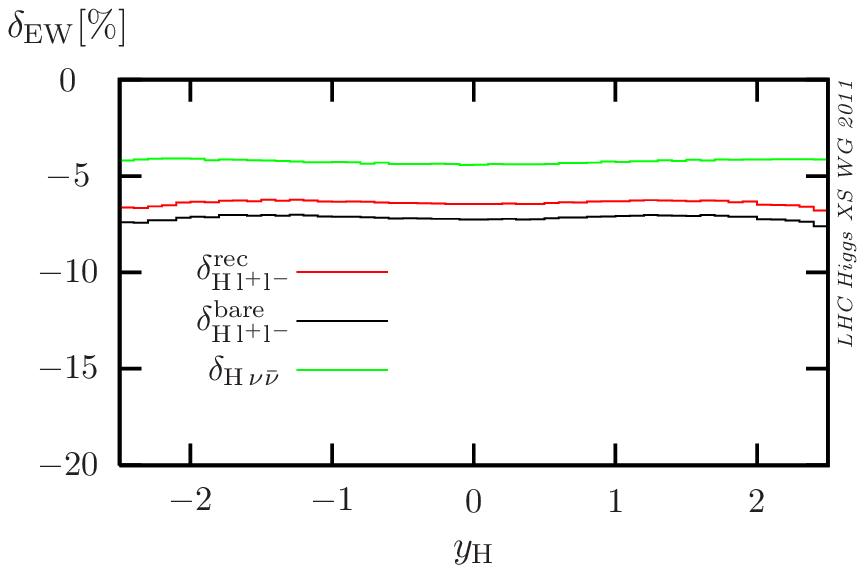}
\caption{\label{fi:rel_basiccuts}
Relative EW corrections for the
$p_{\mathrm{T},\PH}$, $p_{\mathrm{T},\PV}$,
$p_{\mathrm{T},\Pl}$, and $y_{\PH}$ distributions (top to bottom)
for Higgs strahlung off \PW\ bosons (left) and \PZ\ bosons (right)
for the basic cuts at the $7\UTeV$ LHC for $\MH=120\UGeV$.
}
\end{figure}

Figure~\ref{fi:abs_pTcut} shows the 
distributions for the two WH production channels $\PH\Pl^+\PGn$ and
$\PH\Pl^-\PAGn$ and for the ZH production channels $\PH\Pl^+\Pl^-$ and
$\PH\PGn\PAGn$. 
The respective EW corrections are depicted in \refF{fi:rel_pTcut} for
the two different treatments of radiated photons, but the difference
between the two versions, which amounts to $1{-}3\%$, is small. 
The bulk of the EW corrections, which are typically in the range
of $-(10{-}15)\%$, is thus of pure weak origin.
In all $\pT$ distributions the EW corrections show
a tendency to grow more and more negative for larger $\pT$,
signalling the onset of the typical logarithmic high-energy behaviour
(weak Sudakov logarithms). The rapidity distributions receive rather flat
EW corrections, which resemble the ones to the respective integrated
cross sections. Note that the latter are significantly larger in size
than the ones quoted in \Bref{Dittmaier:2011ti} for the total cross sections,
mainly due to the influence of the $\pT$ cuts on the Higgs
and gauge bosons, which enforce the dominance of larger scales in
the process.
This can be clearly seen upon comparing the results with the ones shown
in \refF{fi:rel_basiccuts}, where only the 
basic cuts are applied, but not Eq.~(\ref{eq:VHpTcuts}).
For the basic cuts, the EW corrections are globally smaller in size
by about $5\%$, but otherwise show the same qualitative features.

The relative EW corrections shown here could be taken into account in any
QCD-based prediction for the respective distributions 
(based on the quoted cuts) via reweighting. For this purpose the
data files of the histograms are available at the TWiki page
of the WH/ZH working group%
\footnote{https://twiki.cern.ch/twiki/bin/view/LHCPhysics/WHZH}.
The small photon-induced contributions, which are included in our best prediction
and at the level of $1\%$ for WH production and negligible for ZH production,
are also available and could be simply added.

For definiteness, in \refT{ta:MH120results}, we show the integrated results 
corresponding to the cuts in the boosted setup.

Finally, we estimate the uncertainties resulting from
the remaining spurious QCD scale dependences, missing higher-order contributions, and
uncertainties in the PDFs:
\begin{itemize}
\item
We estimate the scale uncertainties upon varying the renormalisation 
and factorisation scales independently by a factor of two around our default scale choice.
At NNLO for WH production, the integrated cross section for the boosted Higgs analysis varies by
$\Delta_{\mathrm{scale}}=2\%$. 
In the considered distributions, the variation of the scales 
only affects the overall normalisation.
Only in the $p_{\mathrm{T},\PW/\PZ}$ distribution 
near $200\UGeV$, also the scale variation indicates that higher-order corrections are large,
as discussed above. Here, scale variation leads to an error estimate of a few $10\%$ at NLO. 
For ZH production at NLO, the scale variation even leads to an error estimate slightly
below $2\%$. 
\item
Both for WH and ZH production, starting at NNLO new types of higher-order QCD contributions
arise that are not reflected by scale variations at (N)NLO.
Specifically, this comprises the gluon-induced contribution to ZH production
(not taken into account in our NLO prediction here), which
is known to be sizable, and the top-loop-induced NNLO
contributions to WH and ZH production, which have been computed recently at the inclusive  
level~\cite{Brein:2011vx}.
The corresponding uncertainty $\Delta_{\mathrm{HO}}= 7 (1)\%$ for 
ZH (WH) production, which we estimate from the known size of those effects on the total cross sections, 
is also shown in \refT{ta:MH120results}.
The relatively large uncertainty for ZH production will be reduced once the NNLO
QCD corrections are known at the differential level as well.
\item
Concerning PDF uncertainties, as stated above, all
central values for WH correspond to the central MSTW2008 prediction at NNLO. At $68\%$
confidence level (C.L.), the MSTW error sets indicate a PDF error slightly below $2\%$.
In distributions again only the overall normalisation is affected, and the distributions are 
not distorted (not shown in the plots). 
According to the PDF4LHC prescription, we rescale the NNLO uncertainty from MSTW by the additional spread
observed at NLO (which is a factor of $\sim 2.5$) when including the CT10 (rescaled to $68\%$ C.L.) 
and NNPDF~2.1 at $68\%$ C.L.\ in the error estimate. 
While for $\PW^+\PH$ production the error 
band is symmetric, the actual error for $\PW^-\PH$ production covers the region from $-8\%$ to
$+2\%$. For $\PZ\PH$ production, we follow the NLO prescription and use the midpoint of the 
above PDF sets as the central value. The resulting PDF error also amounts to about $\pm 5\%$.
For the $\pT$ distributions, individual PDF sets again only lead to differences in the overall 
normalisation. Only in the rapidity distribution of the Higgs boson, the shape of the distribution is
distorted at the level of a few per cent.
\end{itemize}

\providecommand{\phz}{\phantom{(0)}}
\begin{table}
$$ \begin{array}{ccccc}
\hline
\mathrm{channel} & \;\PH\Pl^+ \PGn_{\Pl} +X \;& \;\PH\Pl^- \PAGn_{\Pl} +X \;& \;\PH\Pl^+ \Pl^- +X \;& \;\PH\PGn_{\Pl} \PAGn_{\Pl} +X \; \\ 
\hline
\sigma/\Ufb                                                  \; & \; 1.384(4)       \; & \; 0.617(2)      \; & \; 0.3467(1)      \; & \; 0.7482(1)           \\ 
\hline                                                                                                                 
\delta_{\ELWK}^{\mathrm{bare}} / \%  \; & \; -14.3\phz	   \; & \; -14.0\phz	   \; & \; -11.0\phz	   \; & \; -6.9\phz		\\ 
\delta_{\ELWK}^{\mathrm{rec}} / \%   \; & \; -13.3\phz	   \; & \; -13.1\phz	   \; & \; -9.0\phz	   \; & \; -6.9\phz		\\ 
\sigma_{\ga}/\Ufb                  \; & \; 0.020	   \; & \; 0.010	   \; & \; 0.0002	   \; & \; 0.0000	 \\ 
\hline                             							     
\Delta_{\mathrm{PDF}} / \%                  \; & \; \pm5\phz	   \; & \; \pm5\phz	   \; & \; \pm5\phz	   \; & \; \pm5\phz		\\ 
\Delta_{\mathrm{scale}} / \%                \; & \; \pm2\phz	   \; & \; \pm2\phz	   \; & \; \pm2\phz	   \; & \; \pm2\phz		\\ 
\Delta_{\mathrm{HO}} / \%                   \; & \; \pm1\phz	   \; & \; \pm1\phz	   \; & \; \pm7\phz	   \; & \; \pm7\phz		\\ 
\hline
\end{array} $$
\caption{\label{ta:MH120results} Integrated cross sections, EW corrections, and estimates 
$\Delta_{\mathrm{PDF/scale}}$ for the PDF and scale uncertainties for the
different Higgs-strahlung channels in the boosted setup for the LHC at $7\UTeV$ for $\MH=120\UGeV$.}
\end{table}

\clearpage

\clearpage

\newpage

\providecommand{\muR}{\mu_{\mathrm{R}}}
\providecommand{\muF}{\mu_{\mathrm{F}}}
\providecommand{\pt}{p_\mathrm{T}}
\providecommand\POWHEG{{\sc POWHEG}}
\providecommand\POWHEGBOX{{\sc POWHEG BOX}}
\providecommand\MCNLO{{\sc MC@NLO}}
\providecommand\HERWIG{{\sc HERWIG}}
\providecommand\HWpp{{\sc HERWIG++}}
\providecommand\PYTHIA{{\sc PYTHIA}}
\providecommand{\helac}{{\sc HELAC}}
\providecommand{\helacdipoles}{{\sc HELAC-Dipoles}}
\providecommand{\helaconeloop}{{\sc HELAC-Oneloop}}
\providecommand{\helacphegas}{{\sc HELAC-PHEGAS}}
\providecommand{\helacnlo}{{\sc HELAC-NLO}}
\providecommand{\amcatnlo}{{\sc aMC@NLO}}
\providecommand{\madloop}{{\sc MadLoop}}
\providecommand{\madfks}{{\sc MadFKS}}
\providecommand{\powheghelac}{{\sc PowHel}}
\providecommand{\fastjet}{{\sc FastJet}}
\providecommand{\bt}{\ensuremath{\bar{{\rm t}}}}

\section{ttH Process\protect\footnote{%
L.~Reina, M.~Spira (eds.);
S.~Dawson, R.~Frederix, M.~V.~Garzelli, A.~Kardos,
C.~G.~Papadopoulos, Z.~Tr\'ocs\'anyi and D.~Wackeroth.}}

The process of Higgs radiation off top quarks $\PQq\PAQq/\Pg\Pg\to
\PH\PQt\PAQt$ becomes particularly relevant for Higgs searches at the
LHC in the light-Higgs-mass region below $\sim 130\UGeV$. Given the
recent Higgs-boson exclusion limits provided by the Tevatron and LHC
experiments, and given the indirect constraints provided by the
electroweak precision fits~\cite{:2010vi}, the light-Higgs-boson mass
region will play a crucial role in confirming or excluding the Higgs
mechanism of electroweak symmetry breaking as minimally implemented in
the Standard Model. Indeed, it will be hard for the LHC running at
$\sqrt{s}=7\UTeV$ to exclude the light-Higgs-boson mass region with
enough statistical significance and more in depth studies will be
necessary in order to extract more information from the available
data. In this context, a careful study of $\PQt\PAQt\PH$
production will become important. Finally, if a light scalar is
discovered, the measurement of the $\PQt\PAQt\PH$ production rate can
provide relevant information on the top--Higgs Yukawa coupling.

The full next-to-leading-order (NLO) QCD corrections to $\PQt\PAQt\PH$
production have been calculated
\cite{Beenakker:2001rj,Beenakker:2002nc,Reina:2001sf,Dawson:2002tg}
resulting in a moderate increase of the total cross section at the LHC
by at most $\sim 20\%$, depending on the value of $\MH$ and on the PDF
set used. More importantly, the QCD NLO corrections reduce the
residual scale dependence of the inclusive cross section from ${\cal
  O}(50\%)$ to a level of ${\cal O}(10\%)$, when the renormalisation
and factorisation scales are varied by a factor of $2$ above and below a
conventional central scale choice (the NLO studies in the literature
use $\mu_0=\Mt+\MH/2$), and clearly show their importance in obtaining
a more stable and reliable theoretical prediction.

Using the NLO codes developed by the authors of
\Brefs{Beenakker:2001rj,Beenakker:2002nc,Reina:2001sf,Dawson:2002tg},
in a previous report~\cite{Dittmaier:2011ti} we studied the inclusive
$\PQt\PAQt\PH$ production at both $\sqrt{s}=7\UTeV$ and $14\UTeV$ and we
provided a breakdown of the estimated theoretical error from
renormalisation- and factorisation-scale dependence, from $\alphas$,
and from the choice of parton distribution functions (PDFs). The total
theoretical errors were also estimated combining the uncertainties
from scale dependence, $\alphas$ dependence, and PDF dependence
according to the recommendation of the LHC Higgs Cross Section Working
Group~\cite{Dittmaier:2011ti}. For low Higgs-boson masses, the
theoretical errors typically amount to $10{-}15\%$ of the corresponding
cross sections.

In this second report we move towards exclusive studies and address
the problem of evaluating the impact of QCD corrections and the
corresponding residual theoretical uncertainty in the presence of
kinematic constraints and selection cuts that realistically model the
experimental measurement.  We focus on three main studies: 1) a study
of the theoretical uncertainty from scale dependence, $\alphas$, and
PDFs on some significant differential distributions for the on-shell
parton-level process $\Pp\Pp\rightarrow \PQt\PAQt\PH$ calculated at
NLO in QCD (see \refS{subsec:ttH_NLO_distributions}); 2) a study
of the effects of interfacing the NLO parton-level calculations with
parton-shower Monte Carlo programs, namely \PYTHIA{} and \HERWIG{},
including a comparison between the \MCNLO{} and \POWHEGBOX{}
frameworks, recently applied to the case of $\PQt\PAQt\PH$ and
$\PQt\PAQt\PA$ production in~\Bref{Frederix:2011zi} and
\Bref{Garzelli:2011vp}, respectively (see
\refS{subsec:ttH_interface_NLO_MC}); 3) a study of the 
background process $\Pp\Pp\rightarrow\PQt\PAQt\PQb\PAQb$ based on the
NLO QCD results presented
in~\Brefs{Bredenstein:2009aj,Bredenstein:2008zb,Bredenstein:2010rs,
  Bevilacqua:2009zn,Binoth:2010ra}, including a comparison between
signal and background ($\Pp\Pp\rightarrow\PQt\PAQt\PQb\PAQb$) at the
parton level, based on the study presented in~\Bref{Binoth:2010ra}
(see \refS{subsec:ttH_ttbb_signal_vs_background}).

\subsection{NLO distributions for $\PQt\PAQt\PH$ associated production}
\label{subsec:ttH_NLO_distributions}
In this section we study the theoretical error on NLO QCD
distributions for $\PQt\PAQt\PH$ on-shell production at a centre-of-mass (CM)
energy of $7\UTeV$. Input parameters are chosen following the
Higgs Cross Section Working Group
recommendation~\cite{Dittmaier:2011ti}. 

We focus on the case of a Higgs boson with $\MH=120\UGeV$ and study
the dependence of the differential cross sections listed above from
the renormalisation and factorisation scale, from $\alphas$, and from
the PDFs. We vary the renormalisation and factorisation scales
together by a factor of $2$ around a central value
$\muR=\muF=\mu_0=\Mt+\MH/2$.  For the $\alphas$ and PDFs
uncertainties, we present in this report results obtained by varying
$\alphas$ and the choice of PDFs within the
CTEQ6.6~\cite{Pumplin:2002vw} set, according to the CTEQ6.6
recommendation.  A complete study including also the
MSTW2008~\cite{Martin:2009iq,Martin:2009bu} and
NNPDF2.1~\cite{Cerutti:2011au} sets of PDFs will be updated on the
$\PQt\PAQt\PH$ TWiki page as soon as available.  Since we consider a
single set of PDFs (CTEQ6.6), we show more conservative $90\%$
C.L.\ errors.  Together with the inclusive results presented in
\Bref{Dittmaier:2011ti}, which included a comparison between CTEQ6.6,
MSTW2008, and NNPDF2.0~\cite{Ball:2010de}, the current study should
provide a good guidance to estimate the theoretical uncertainty on the
differential distributions for the on-shell $\PQt\PAQt\PH$ production.

In this section we consider distributions in transverse momentum (\pT)
and pseudorapidity ($\eta$) of the top/antitop quarks and of the Higgs
boson, and in the $R(\PQt,\PAQt)$, $R(\PQt,\PH)$, $R(\PAQt,\PH)$
variables, where $R$ is the distance in the $(\phi,\eta)$ plane. 

\begin{figure}
\centering
\vspace{0.3truecm}
\includegraphics[width=0.45\textwidth]{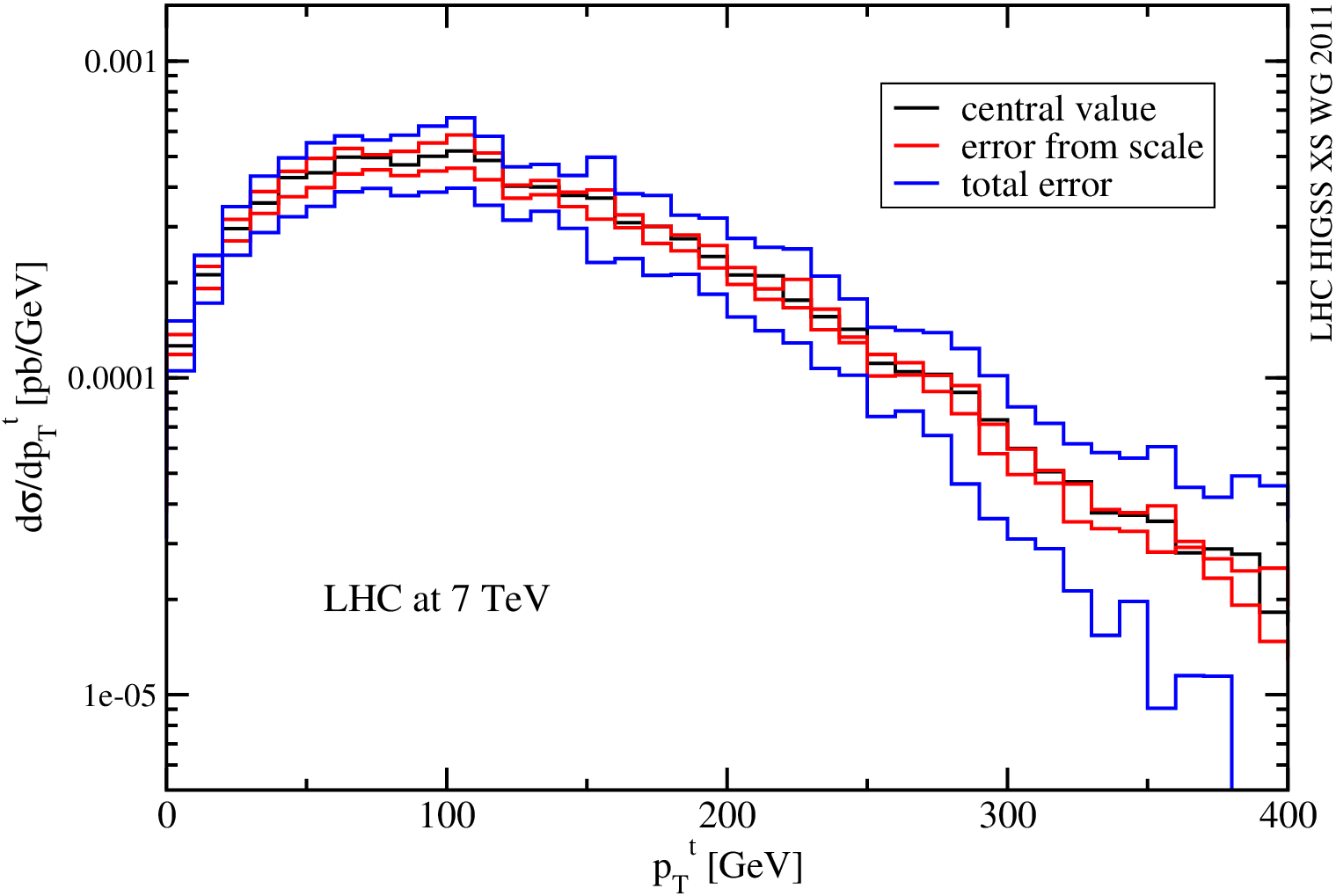}
\hspace{0.3truecm}
\includegraphics[width=0.44\textwidth]{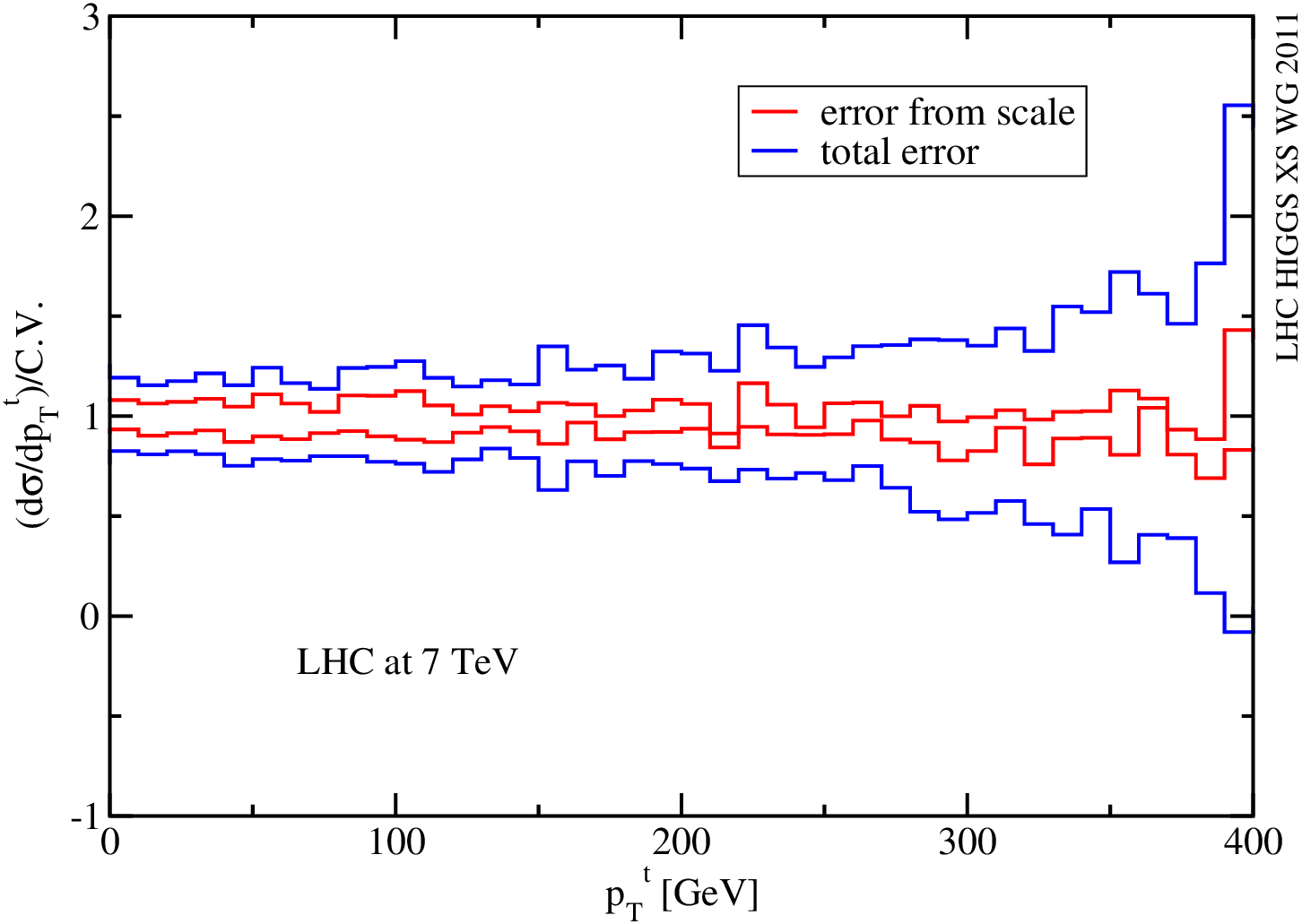}
\caption{Theoretical uncertainty on top(antitop)-quark transverse-momentum
  ($\pT^{\PQt}$) distribution: the l.h.s.\ shows the actual distributions, the
  r.h.s.\ the spread around the central value in per cent. A detailed explanation of the red and
  blue bands is given in the text.}
\label{fig:ttH_pt_t}
\end{figure}

\begin{figure}
\centering
\vspace{0.3truecm}
\includegraphics[width=0.45\textwidth]{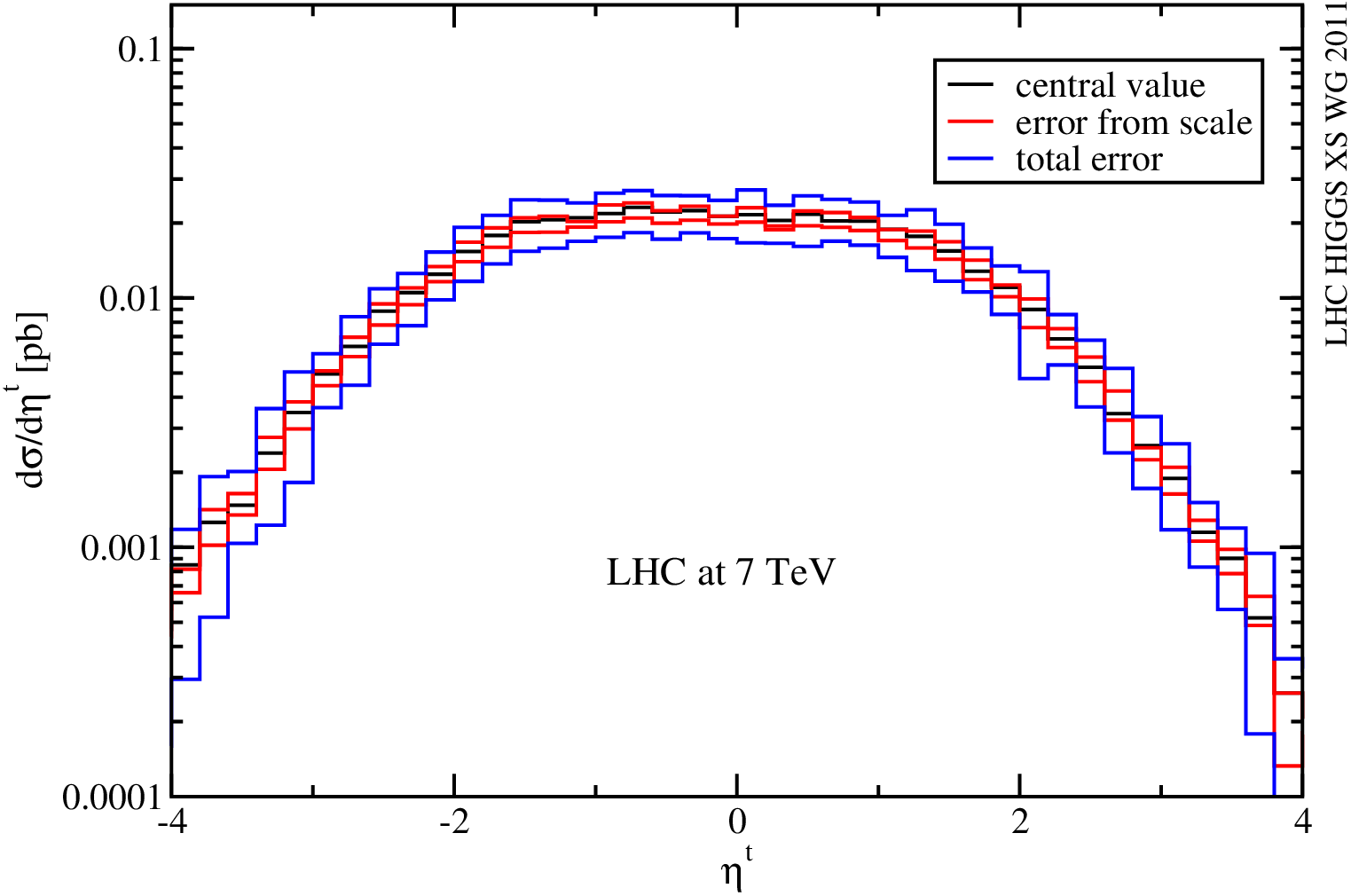} 
\hspace{0.3truecm}
\includegraphics[width=0.44\textwidth]{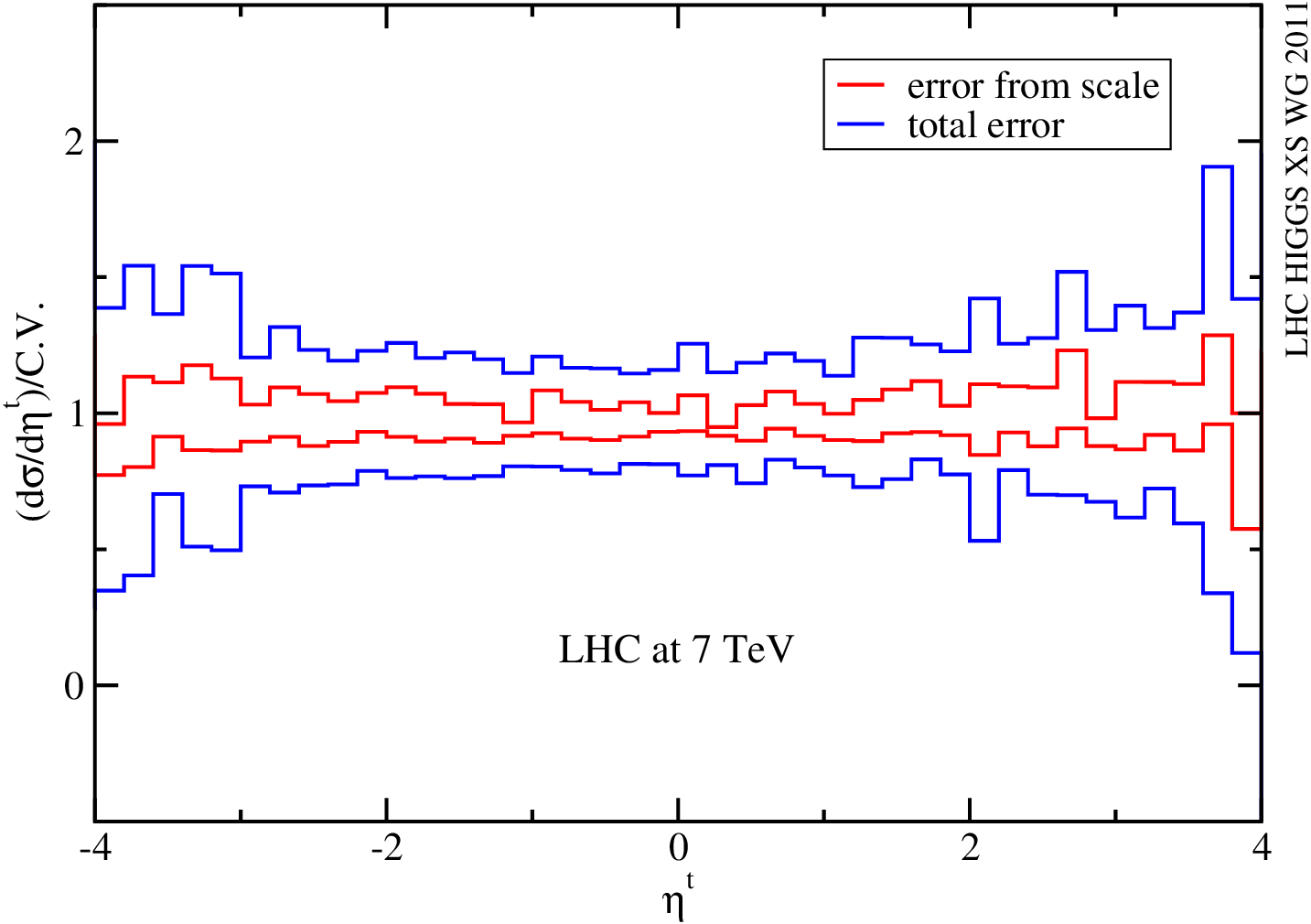}
\caption{Theoretical uncertainty on top(antitop)-quark pseudorapidity ($\eta^{\PQt}$) distribution:
  the l.h.s.\ shows the actual distributions, the r.h.s.\ the 
  spread around the central value in per cent. A detailed explanation of the red and blue bands is given in
  the text.}
\label{fig:ttH_eta_t}
\end{figure}

\begin{figure}
\centering
\vspace{0.3truecm}
\includegraphics[width=0.45\textwidth]{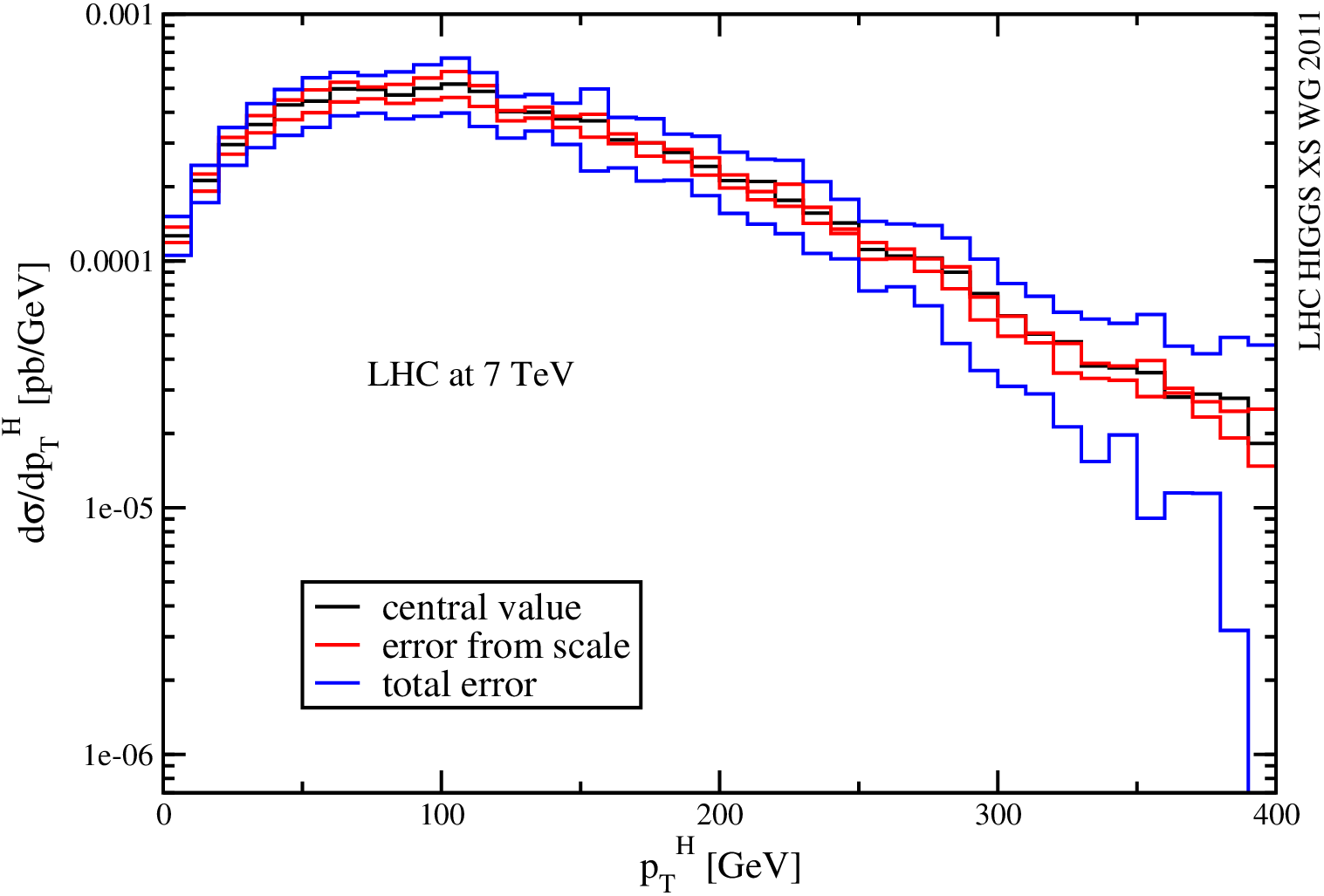}
\hspace{0.3truecm}
\includegraphics[width=0.44\textwidth]{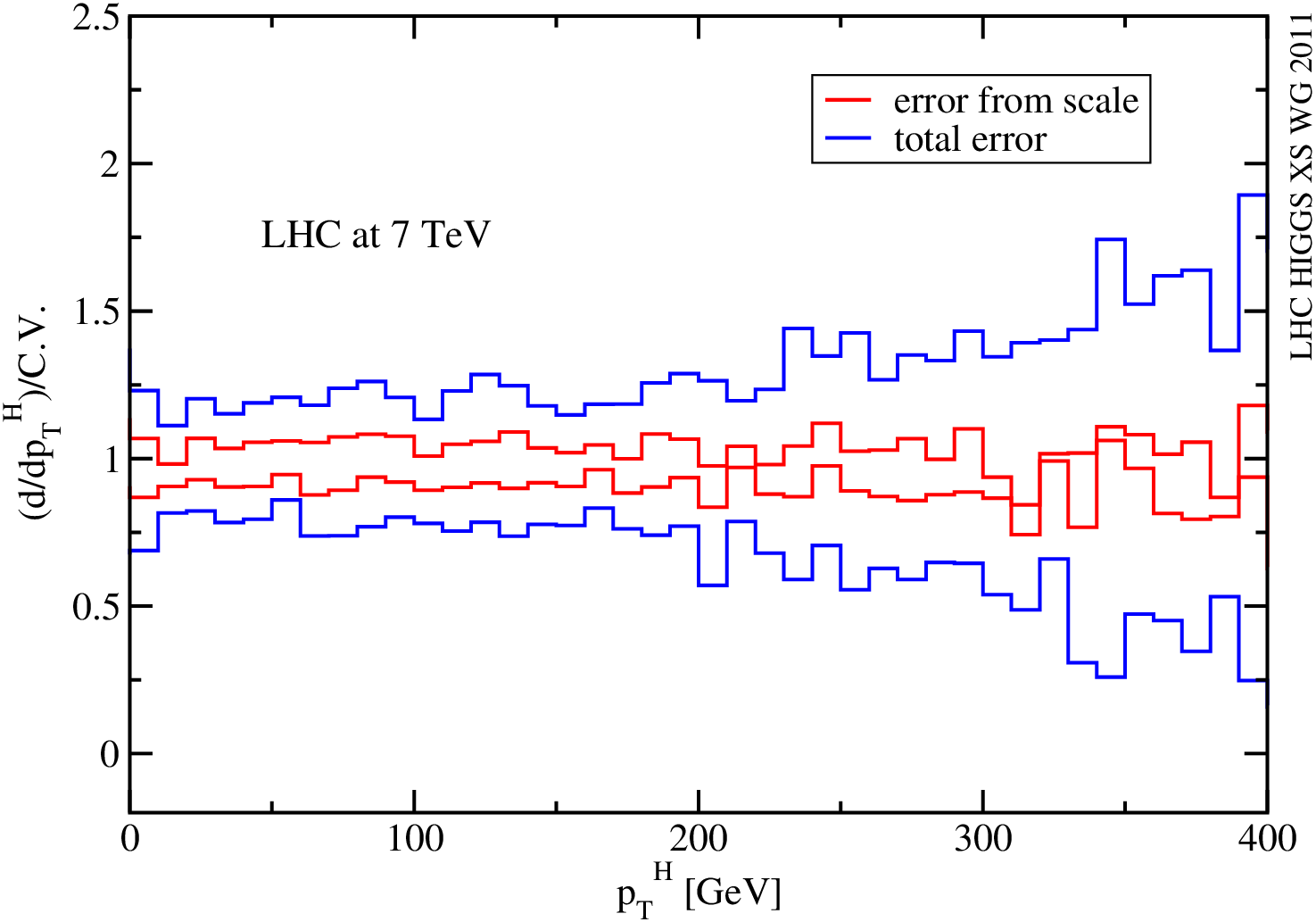}
\caption{Theoretical uncertainty on Higgs-boson transverse momentum
  ($\pT^{\PH}$) distribution: the l.h.s.\ shows the actual distributions, the
  r.h.s.\ the spread around the central value in per cent. A detailed explanation of the red and
  blue bands is given in the text.}
\label{fig:ttH_pt_H}
\end{figure}

\begin{figure}
\centering
\vspace{0.3truecm}
\includegraphics[width=0.45\textwidth]{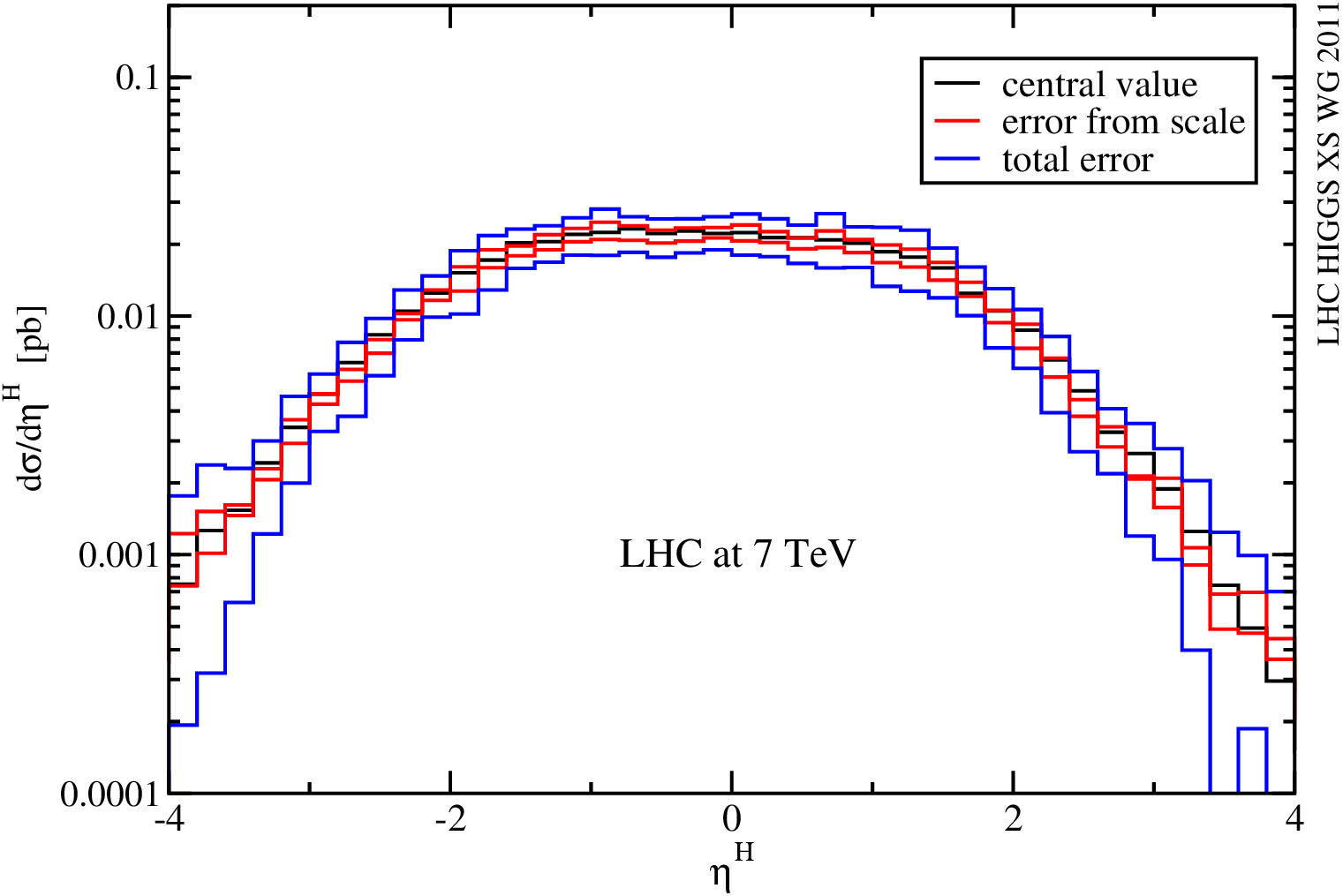} 
\hspace{0.3truecm}
\includegraphics[width=0.44\textwidth]{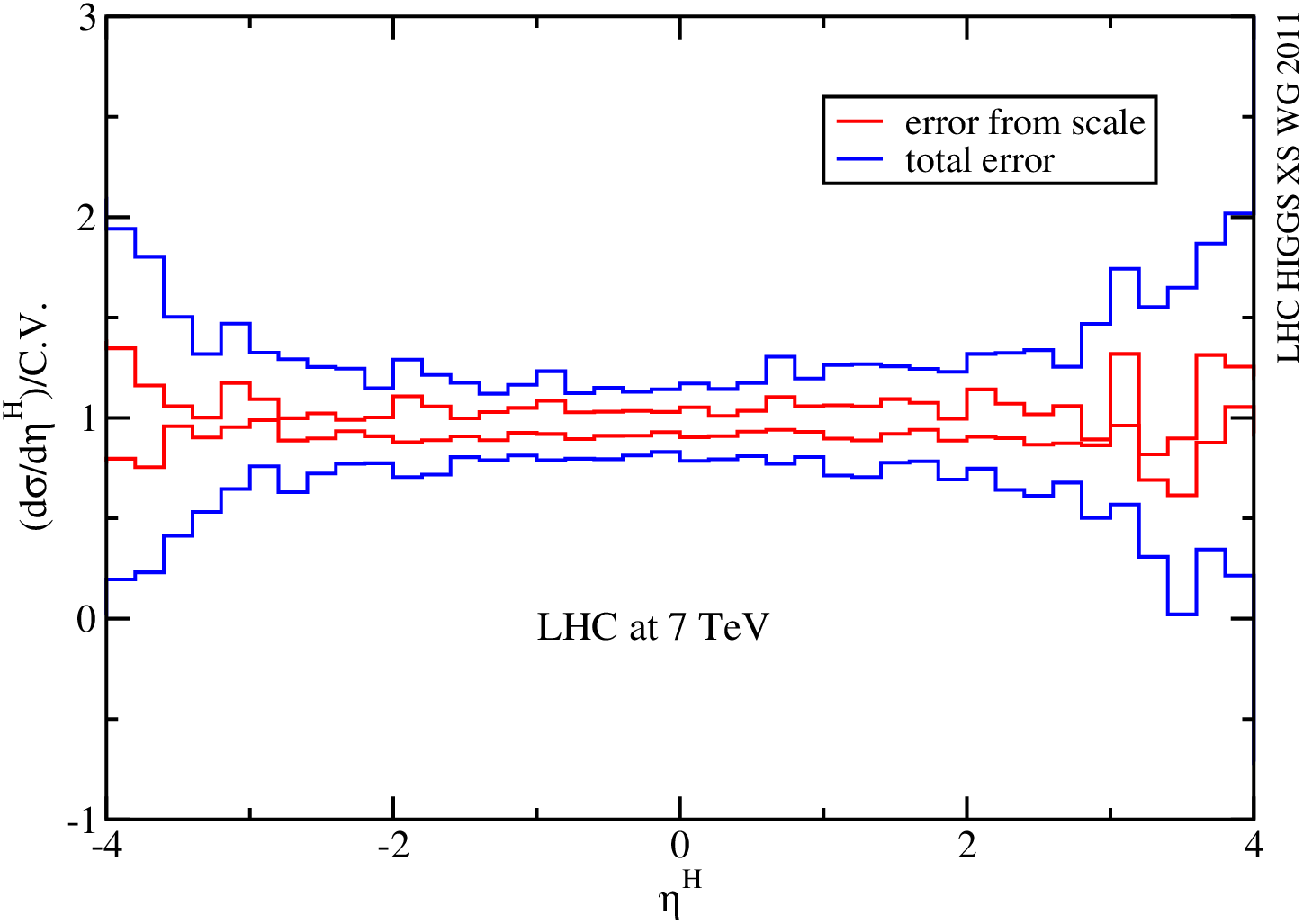}
\caption{Theoretical uncertainty on Higgs-boson pseudorapidity ($\eta^{\PH}$) distribution:
  the l.h.s.\ shows the actual distributions, the r.h.s.\ the 
  spread around the central value in per cent. A detailed explanation of the red and blue bands is given in
  the text.}
\label{fig:ttH_eta_H}
\end{figure}

\begin{figure}
\centering
\vspace{0.3truecm}
\includegraphics[width=0.45\textwidth]{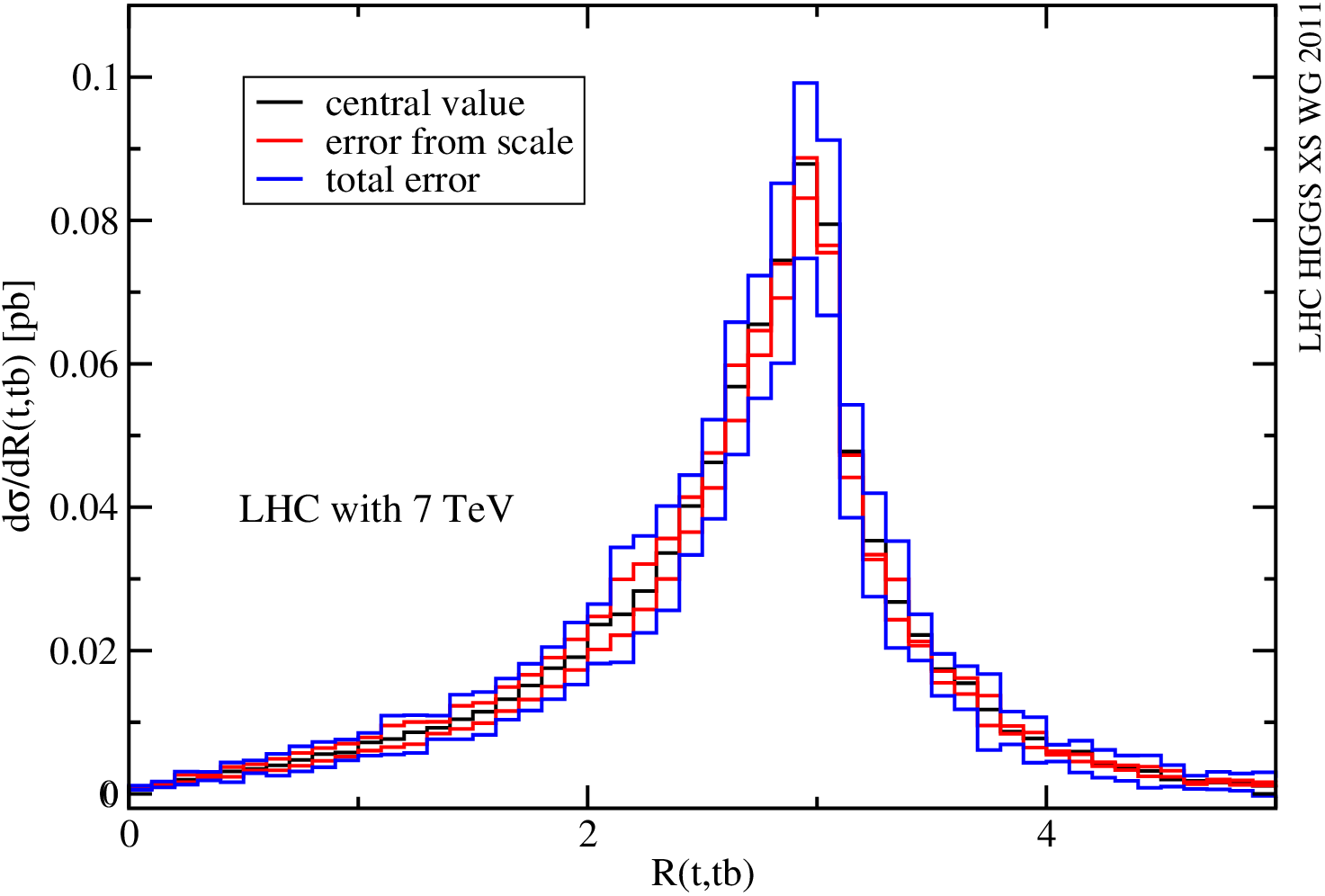}
\hspace{0.3truecm}
\includegraphics[width=0.44\textwidth]{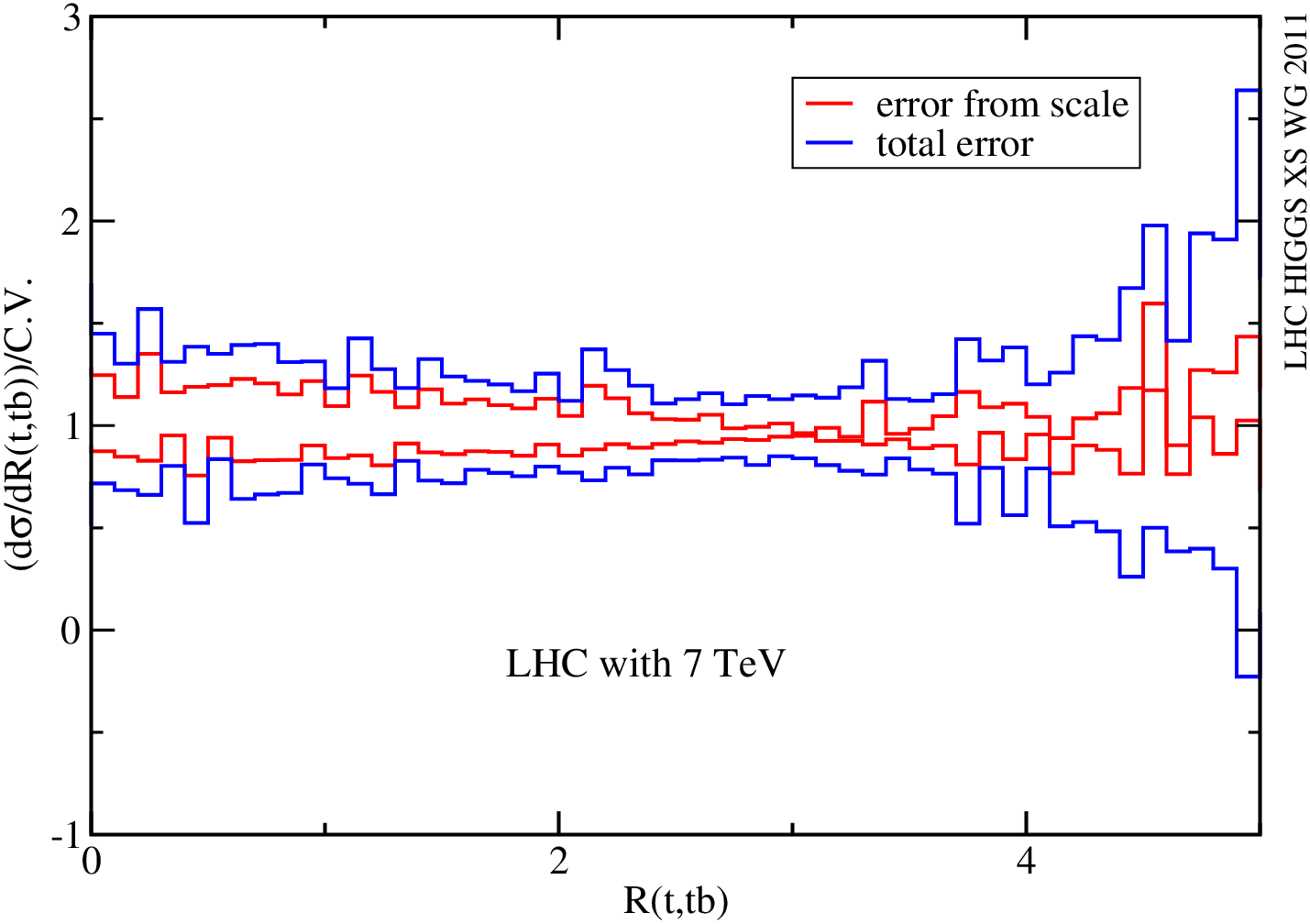}
\caption{Theoretical uncertainty on the top-antitop quark $(\eta,\phi)$ distance
  ($R(\PQt,\PAQt)$) distribution: the l.h.s.\ shows the actual distributions, the
  r.h.s.\ the spread around the central value in per cent. A detailed explanation of the red and
  blue bands is given in the text.}
\label{fig:ttH_R_ttb}
\end{figure}

\begin{figure}
\centering
\vspace{0.3truecm}
\includegraphics[width=0.45\textwidth]{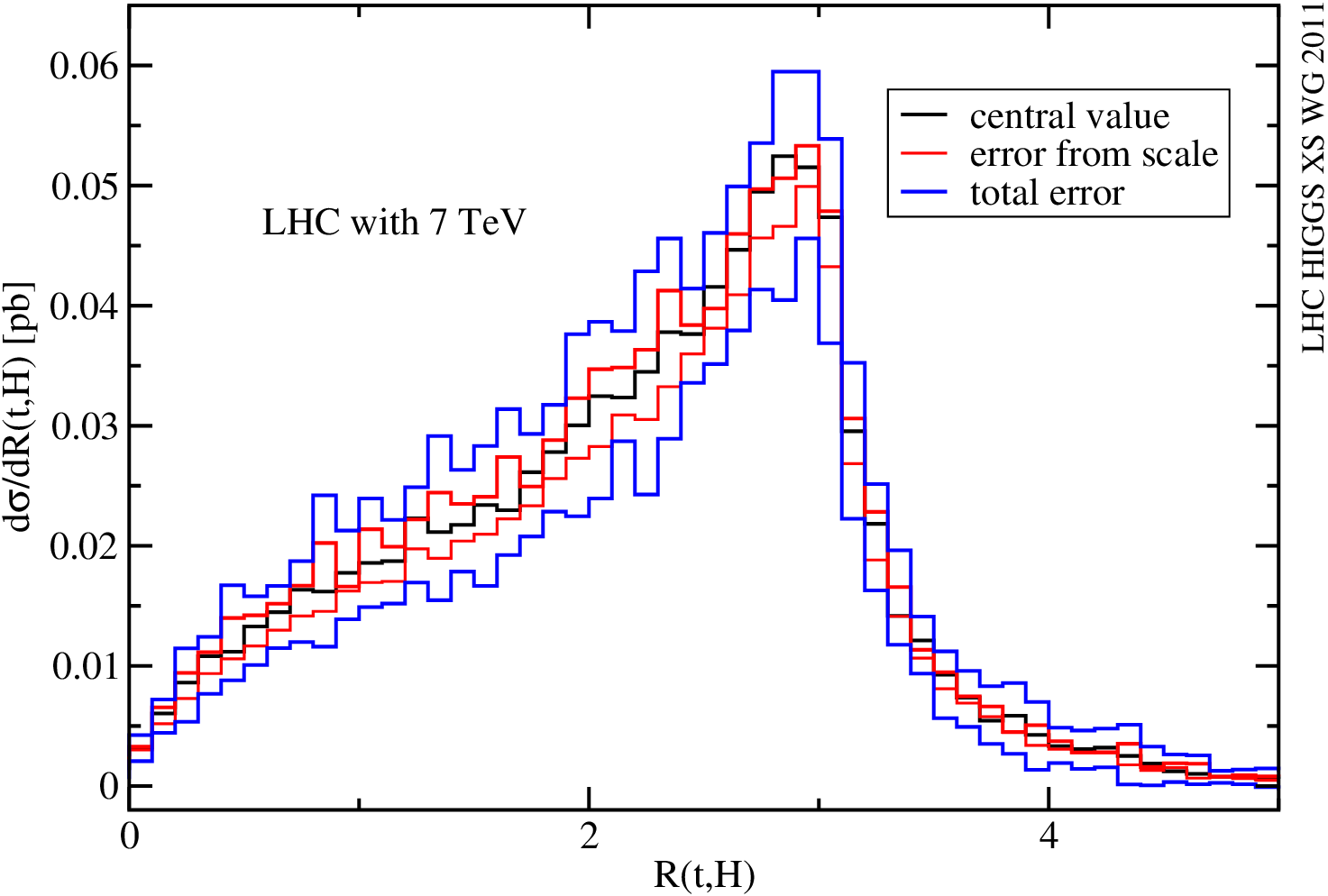} 
\hspace{0.3truecm}
\includegraphics[width=0.44\textwidth]{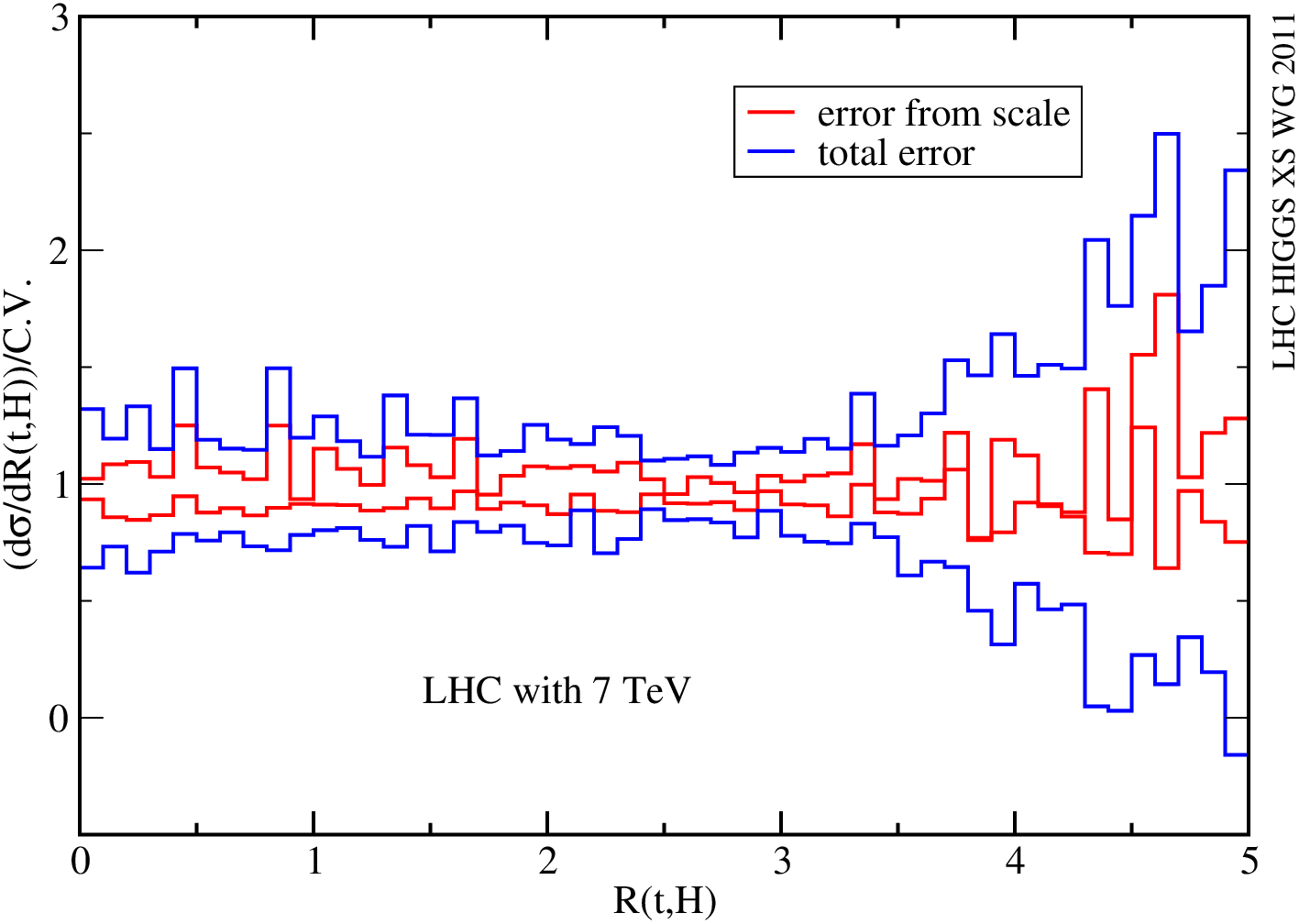}
\caption{Theoretical uncertainty on top-Higgs  $(\eta,\phi)$ distance ($R(\PQt,\PH)$) distribution
  in the no-cut configuration:
  the l.h.s.\ shows the actual distributions, the r.h.s.\ the 
  spread around the central value in per cent. A detailed explanation of the red and blue bands is given in
  the text.}
\label{fig:ttH_R_tH}
\end{figure}

In \refFs{fig:ttH_pt_t}-\ref{fig:ttH_eta_H} we present results for
the differential cross sections in the $\pT$ and $\eta$ of both $\PQt$
($\PAQt$ would be equivalent) and $\PH$, while in
\refFs{fig:ttH_R_ttb} and \ref{fig:ttH_R_tH} we present results for
differential cross sections in $R(\PQt,\PAQt)$, $R(\PQt,\PH)$, and
$R(\PAQt,\PH)$. In the figures, the left-hand-side (l.h.s) plot shows
the actual distributions, while the right-hand-side (r.h.s) plot gives
the spread around the central value in per cent. More specifically, in the l.h.s.\ part of each plot
we show a central distribution obtained for $\muR=\muF=\mu_0$ (black
histogram), a differential band that represents the variation of the
distribution when $\muR=\muF$ is varied from $\mu_0/2$ to $2\mu_0$
(delimited by the upper and lower red histograms), and a differential
band that represents the variation of the distribution when also the
uncertainty from $\alphas$ and PDFs is added (delimited by the upper
and lower blue histograms). Therefore the red band represents an
estimate of the theoretical uncertainty from residual scale
dependence, while the blue band represents an estimate of the total
theoretical error, at NLO in QCD. The theoretical errors from
$\alphas$ and PDF have been determined consistently within the
CTEQ6.6 package and combined in quadratures, before adding the result
linearly to the error from scale dependence, bin by bin.  On the other
hand, in the r.h.s.\ part of each plot we illustrate the error
and to this purpose we just plot the ratio of the histograms that
delimit the scale dependence and ($\alphas$+PDF) uncertainty
differential bands to the corresponding central value, i.e.\ the ratio
of the red and blue histograms to the black one, bin by bin.

All distributions show interesting common features. We notice that
towards the end points of each plot, statistical fluctuations obscure
the main behaviour of the distributions and our estimates become
unreliable. The problem can be easily addressed if needed by using
higher statistics.  All over the statistically significant range of
each plot we see that different kinematic regions bear different
errors, and are more or less sensitive to the residual theoretical
uncertainties. In general, the residual scale dependence ranges
between $10\%$ and $20\%$, while the total error, including the
$\alphas$+PDF residual uncertainty ($90\%$ C.L.), ranges between
$20\%$ and $50\%$.

We point out that more-in-depth-studies would benefit from the inclusion
of the $\PQt\PAQt\PH$ final-state decays, which has recently become
available via the interface with parton-shower Monte Carlo programs
(\PYTHIA{} and \HERWIG{}), as discussed in
\refS{subsec:ttH_interface_NLO_MC}, and we therefore recommend
for future studies that a full study of the theoretical uncertainty be
done in such context. Indeed, all the results presented in
\refS{subsec:ttH_interface_NLO_MC} already include a study of the
PDF error obtained using the MSTW2008 set of PDFs.

\subsection{Interface of NLO $\PQt\PAQt\PH$ and $\PQt\PAQt\PA$
  calculations with parton-shower Monte Carlo programs}
\label{subsec:ttH_interface_NLO_MC}

Recently, $\PQt\PAQt\PH$ production at LHC has been studied by
a\MCNLO~\cite{Frederix:2011zi} and
\powheghelac~\cite{Garzelli:2011vp}, two state-of-the-art independent
frameworks, which allow to combine, through a proper NLO matching
procedure, the computation of hard-scattering processes at NLO
accuracy in QCD, to a parton-shower evolution (resumming at least the
leading logarithmic soft and collinear divergences at all orders in
perturbation theory) down to the hadronisation energy scale.

\powheghelac{} is based on codes included in the \helacnlo{}
package~\cite{Bevilacqua:2011xh,Bevilacqua:2010mx}, used for the
computation of all matrix elements provided as input to the
\POWHEGBOX~\cite{Alioli:2010xd} program, which adopts the FKS
subtraction scheme~\cite{Frixione:1995ms} to factor out the IR
singularities in phase-space integrations and implements the \POWHEG{}
matching scheme~\cite{Nason:2004rx,Frixione:2007vw}.  The a\MCNLO{}
code, on the other hand, is built upon the
Mad{\sc FKS}~\cite{Frederix:2009yq} framework which also uses the FKS
subtraction scheme.  The {\sc MadLoop}{} code~\cite{Hirschi:2011pa} is used
to generate the virtual corrections. To match the results to the
parton shower, the \MCNLO{} method has been
employed~\cite{Frixione:2002ik}.
  
So far, both codes have been used for phenomenological studies at the
hadron level of several different processes interesting for Tevatron
and LHC
physics~\cite{Kardos:2011qa,Frederix:2011zi,Garzelli:2011vp,Frederix:2011qg,Frederix:2011ig,Frederix:2011ss,Garzelli:2011is,Garzelli:2011iu}.
This is the first time their results are compared in full detail,
fixing a common scheme agreed upon by their developers for the study of
$\PQt\PAQt\PH$ production at LHC.  The setup, the set of cuts adopted
for the comparison, as well as the definition of the considered
observables, is presented in the following.
 
We consider two Higgs scenarios:
\begin{enumerate}
\item Standard Model Higgs boson;
\item pseudo-scalar Higgs boson.
\end{enumerate}
For both scenarios the Higgs-boson mass was set to $\MH=120\UGeV$ and
standard Yukawa couplings were assumed. The top mass was assumed to be
$\Mt=172.5\UGeV$.  A dynamical scale, defined as $( M_{\mathrm{T},\PQt}$\,\,$
M_{\mathrm{T},\PAQt} $\,\,$ M_{\mathrm{T},\PH})^{1/3}$, where $M_{\mathrm{T},i}$ is the transverse
mass $\sqrt{ M_i^2 + p_{\mathrm{T},i}^2}$, was used in the generation of the
events at ${\sqrt{s}}= 7\UTeV$. The factorisation and the
renormalisation scales were set equal.  The NLO MSTW2008 PDF set with
$5$ active flavours was used, together with the corresponding $\alphas$
and $68\%$ C.L.\ uncertainty set.  Particle decay, shower, hadronisation,
and hadron decay effects have been simulated by means of the latest
fortran version of the \HERWIG{}
code~\cite{Corcella:2000bw,Corcella:2002jc}, \HERWIG{}~6.5.20.  The
Higgs boson was forced to decay in the $\PQb\PAQb$ channel with a
branching ratio equal to one, $\pi^0$ and $\mu^\pm$ were set stable to
simplify the analyses, whereas all other particles and hadrons
(including B-hadrons) were assumed to be stable or to decay according
to the default implementation of the shower MC.  Multiparticle
interaction effects were neglected.  Jets were reconstructed through
the anti-$k_{\mathrm{T}}$ clustering algorithm~\cite{Cacciari:2008gp}, as
implemented in \fastjet{}~3.0.0, with a recombination radius parameter
$R$ fixed to $0.5$.

The following four sets of cuts were adopted:
\begin{itemize}
\item
[Set 0)] No cut (inclusive analysis);
\item
[set 1)] $p_{\mathrm{T},\PH}$ $> 200\UGeV$, computed after showering and before $\PH$ decay
(boosted analysis);
\item
[Set 2)] (i) $E_{\mathrm{T},\mathrm{min}}^{\mathrm{j}} = 25\UGeV$ and
(ii) $|\eta^{\mathrm{j}}|< 2.5$ for all jets (otherwise the jet is discarded), 
(iii) $\#\mathrm{jets}> 4$ for each event (hadronic-cut analysis) ;
\item
[Set 3)] besides including cuts in set 2), 
(iv) we focused on the di-leptonic channel, asking for at least 
one $\Pl^+$ and one $\Pl^-$ with
(v) $E_{\mathrm{T},\mathrm{min}}^{\mathrm{\Pl^\pm}} = 20\UGeV$ and 
(vi) $|\eta^{\mathrm{\Pl^\pm}}| < 2.5$, whereas the transverse missing energy
of the event was constrained to be 
(vii) $\slashed{E}_{\mathrm{T},\mathrm{min}} > 30\UGeV$. Charged leptons not satisfying
both cut (v) and cut (vi) were discarded in all events
(all-cut analysis). 
\end{itemize}

\begin{figure}
\centering
\includegraphics[bb=100 205 482 563, width=0.49\textwidth]{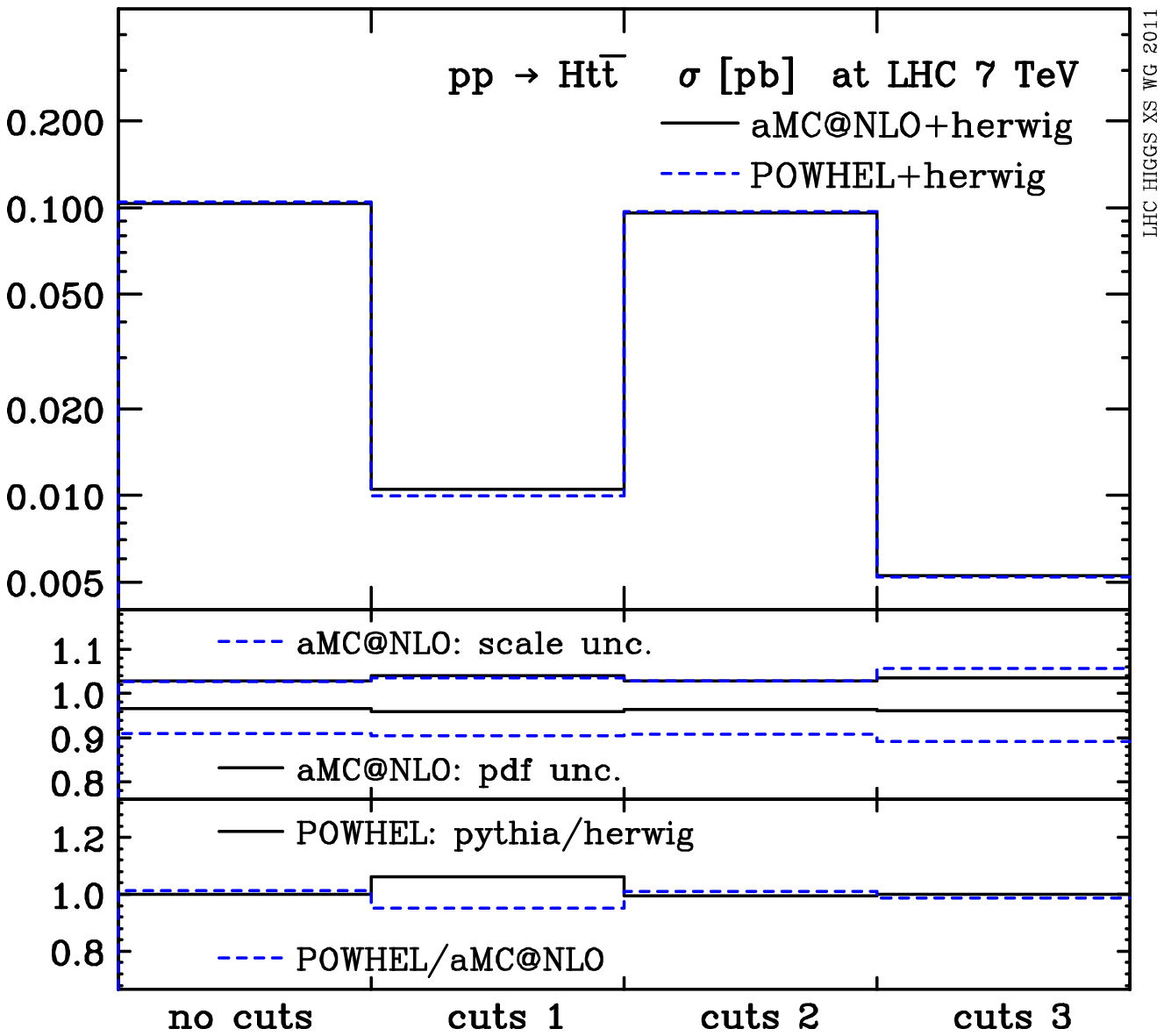}
\includegraphics[bb=100 205 482 563, width=0.49\textwidth]{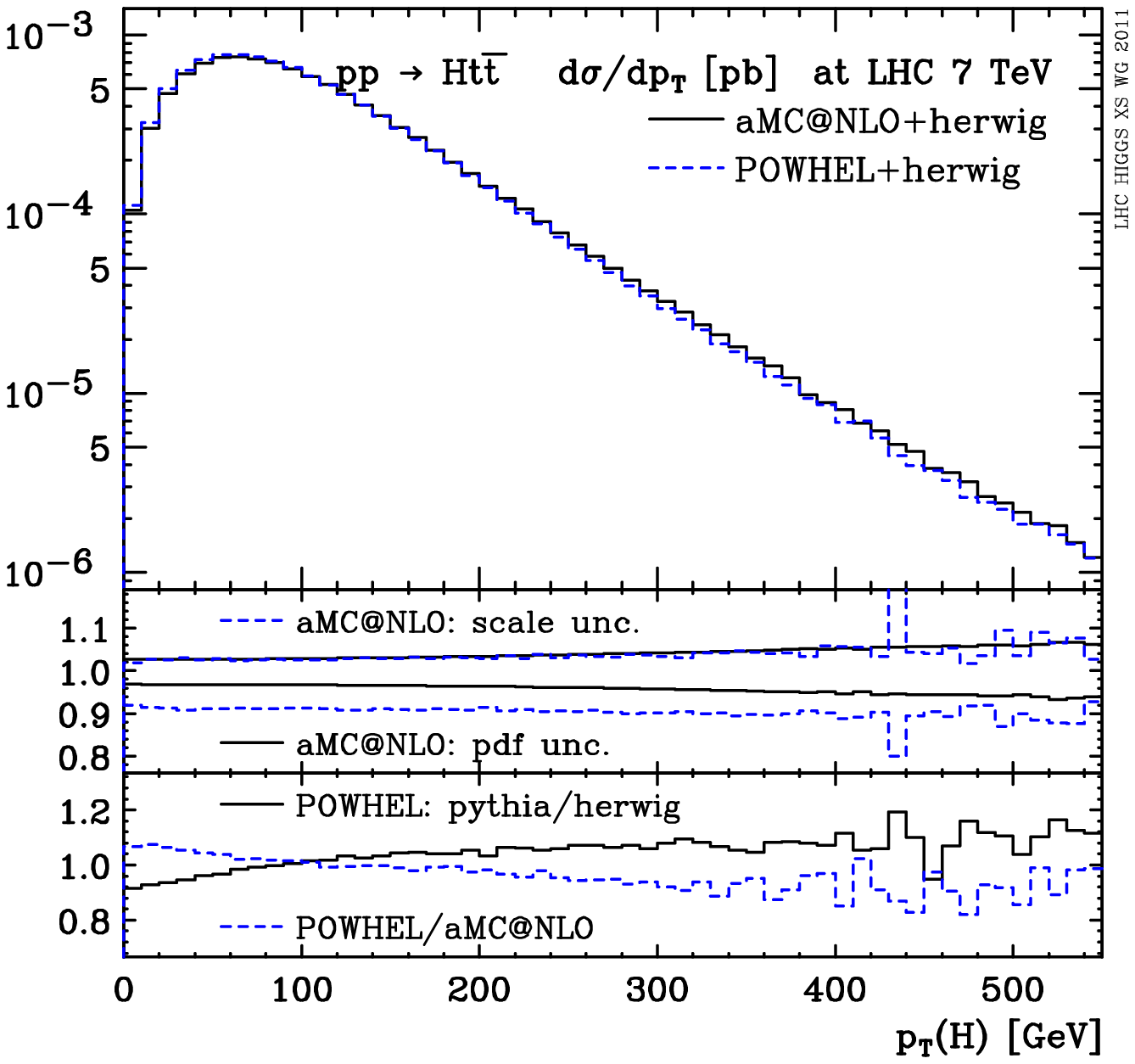}
\caption{Total rates for scalar-Higgs-boson production after the different
  cuts
  defined in the text (\textit{left}) and transverse momentum of
  the Higgs boson (\textit{right}) in the no-cut configuration. In the upper
  inset the scale and PDF uncertainties computed by
  a\MCNLO{} interfaced to \HERWIG{} are shown. The lower inset
  displays the ratio of \powheghelac{} over a\MCNLO{} and the ratio between
  the results computed by interfacing \powheghelac{} to \PYTHIA{} and
  \HERWIG{}. See text for more details.}
\label{fig:ttH_nlomc_fig1}
\end{figure}

\begin{figure}
\centering
\includegraphics[bb = 88 205 482 563, width=0.49\textwidth]{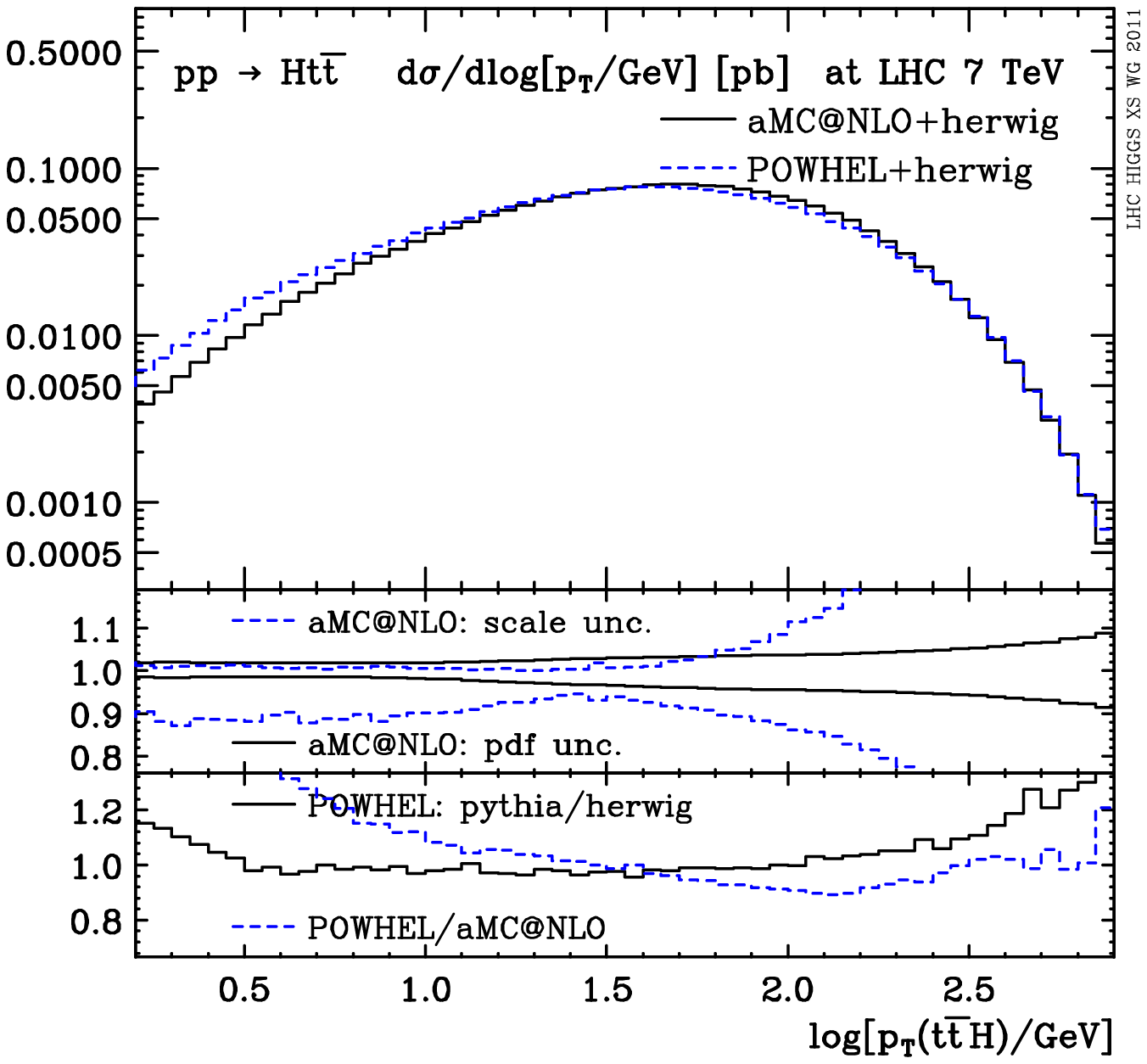}
\includegraphics[bb = 88 205 482 563, width=0.49\textwidth]{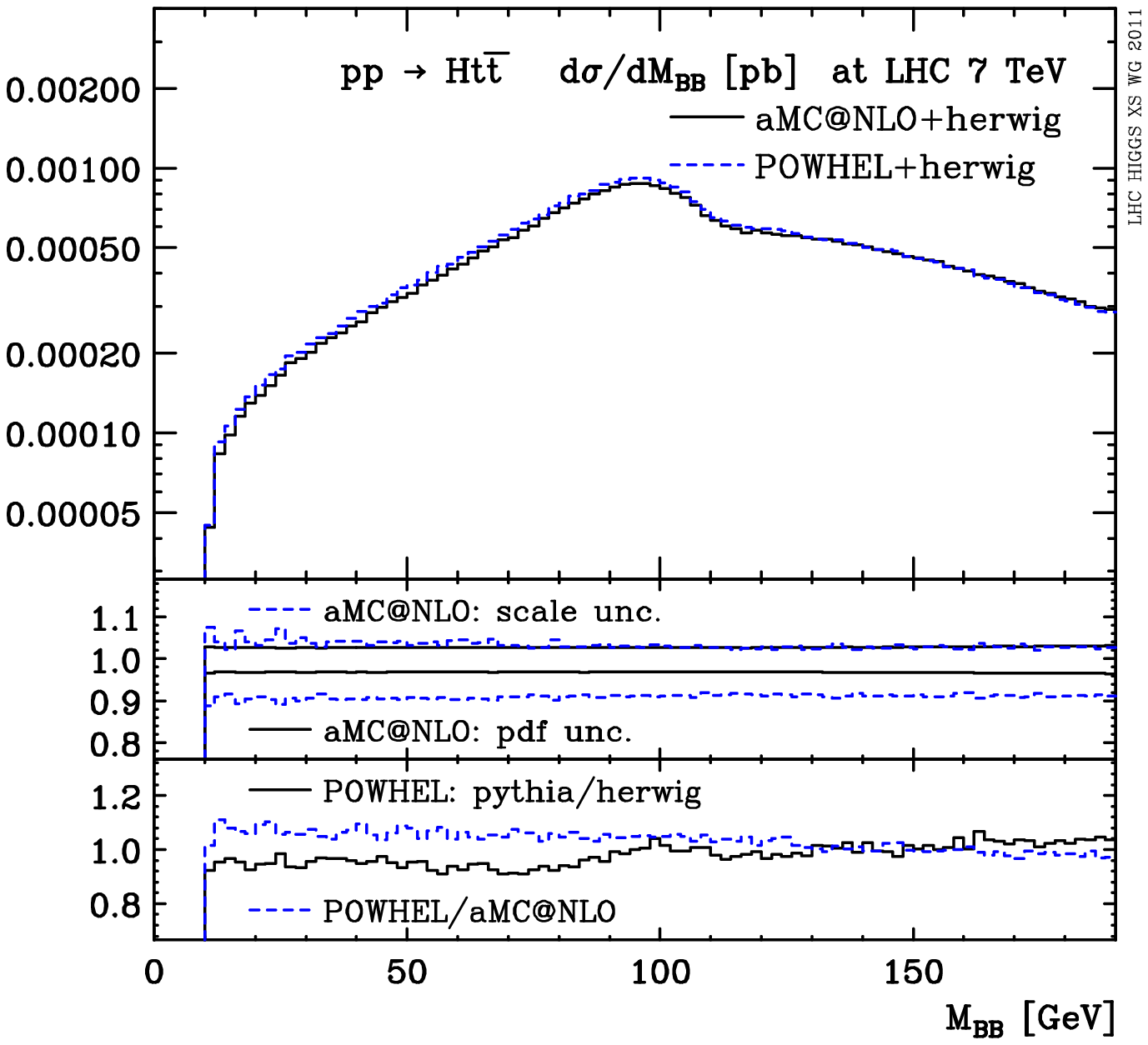}
\caption{Transverse momentum of the Higgs--top--antitop system
  (\textit{left}) and invariant mass of the two hardest lowest-lying B
  hadrons (\textit{right}).  The different regions of the plots are
  defined as in \protect\refF{fig:ttH_nlomc_fig1}. See text for
  more details.}
\label{fig:ttH_nlomc_fig2}
\end{figure}

\begin{figure}
\centering
\includegraphics[bb = 100 205 482 563, width=0.49\textwidth]{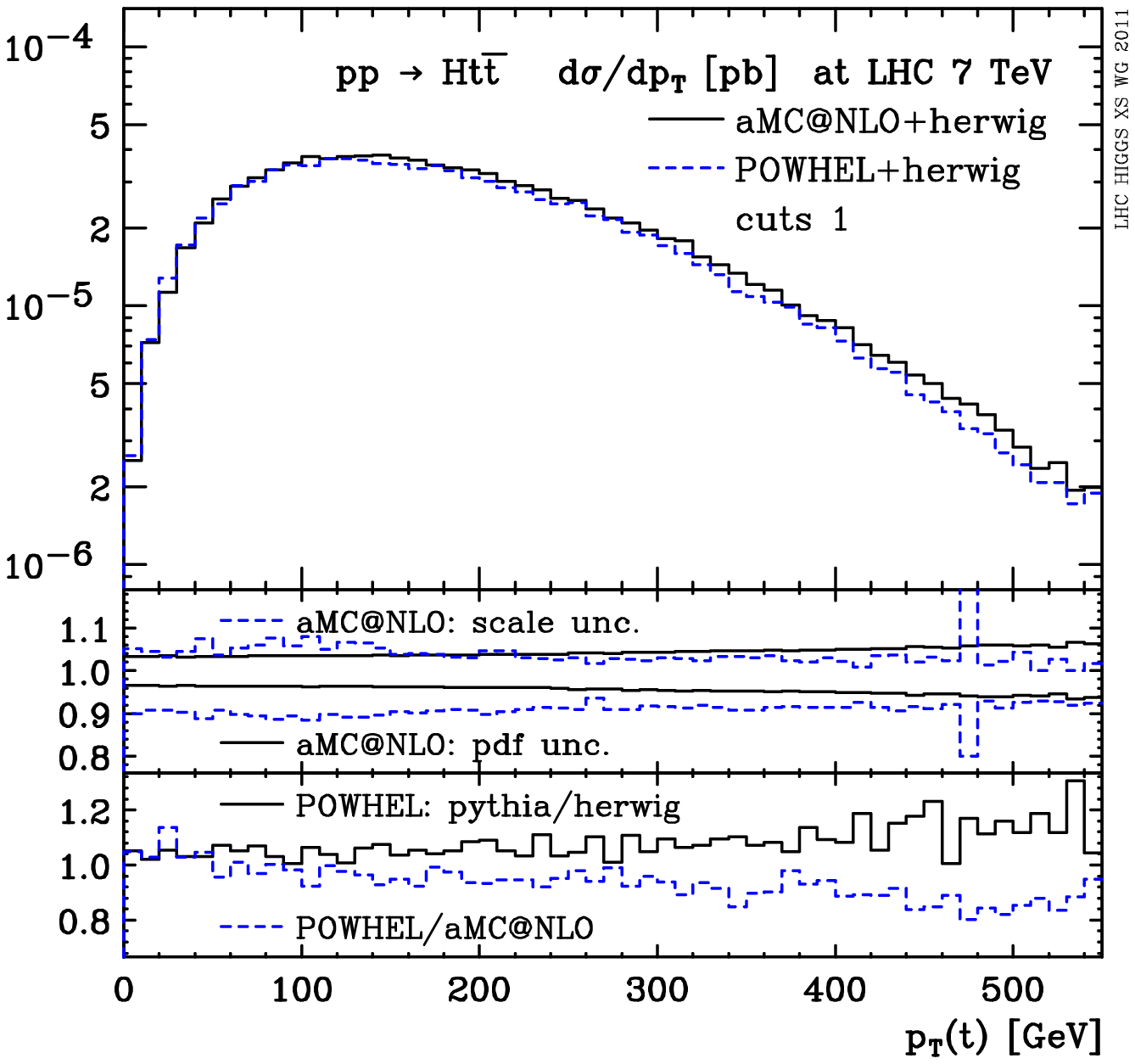}
\includegraphics[bb = 100 205 482 563, width=0.49\textwidth]{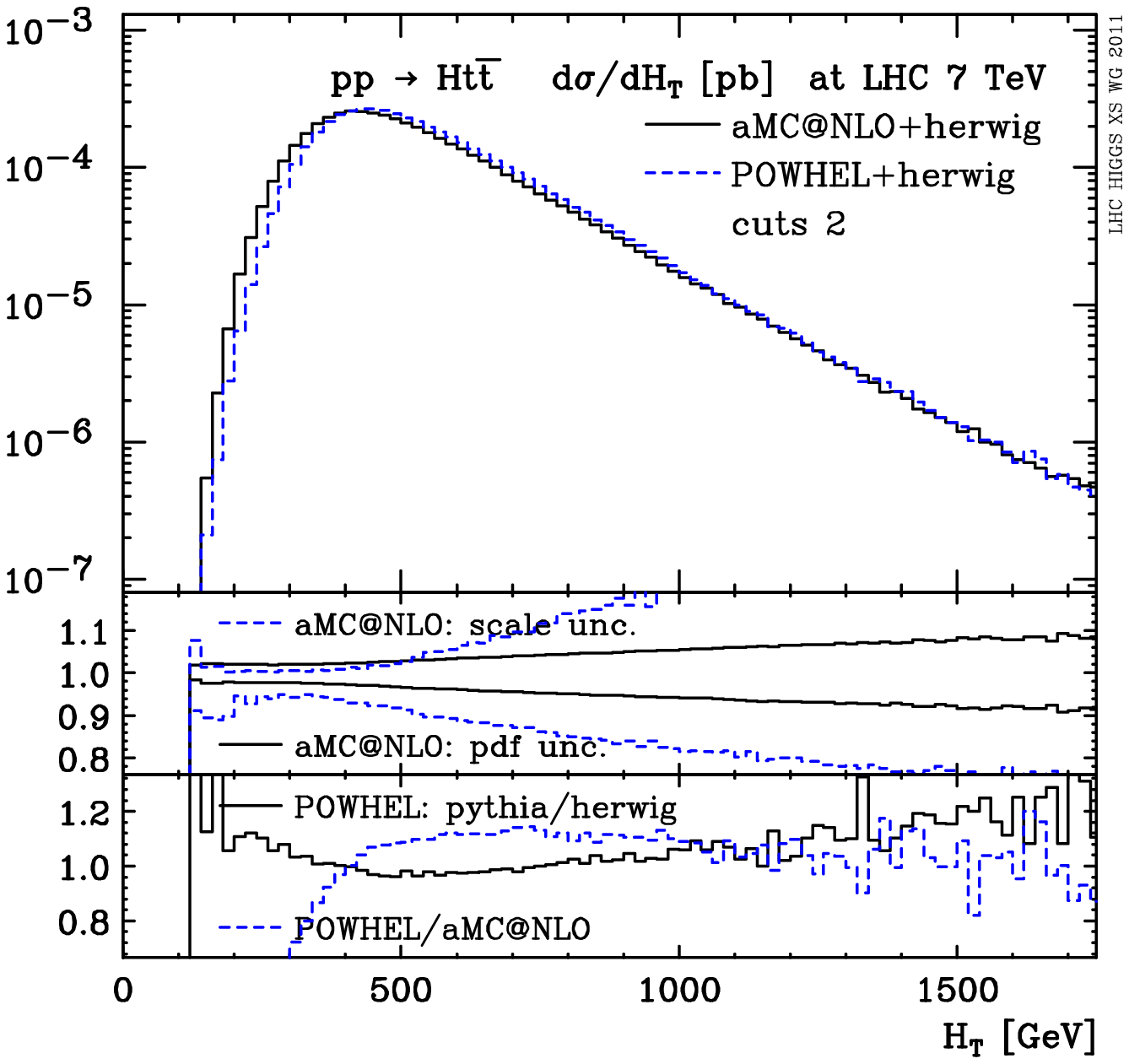}
\caption{Transverse momentum of the top quark in the boosted-Higgs
  scenario (\textit{left}) and scalar sum of transverse energies
  (\textit{right}) under the cuts of set 2). The different regions of
  the plots are defined as in
  Fig.~\protect\ref{fig:ttH_nlomc_fig1}. See text for more details.}
\label{fig:ttH_nlomc_fig3}
\end{figure}

\begin{figure}
\centering
\includegraphics[bb = 100 205 482 563, width=0.49\textwidth]{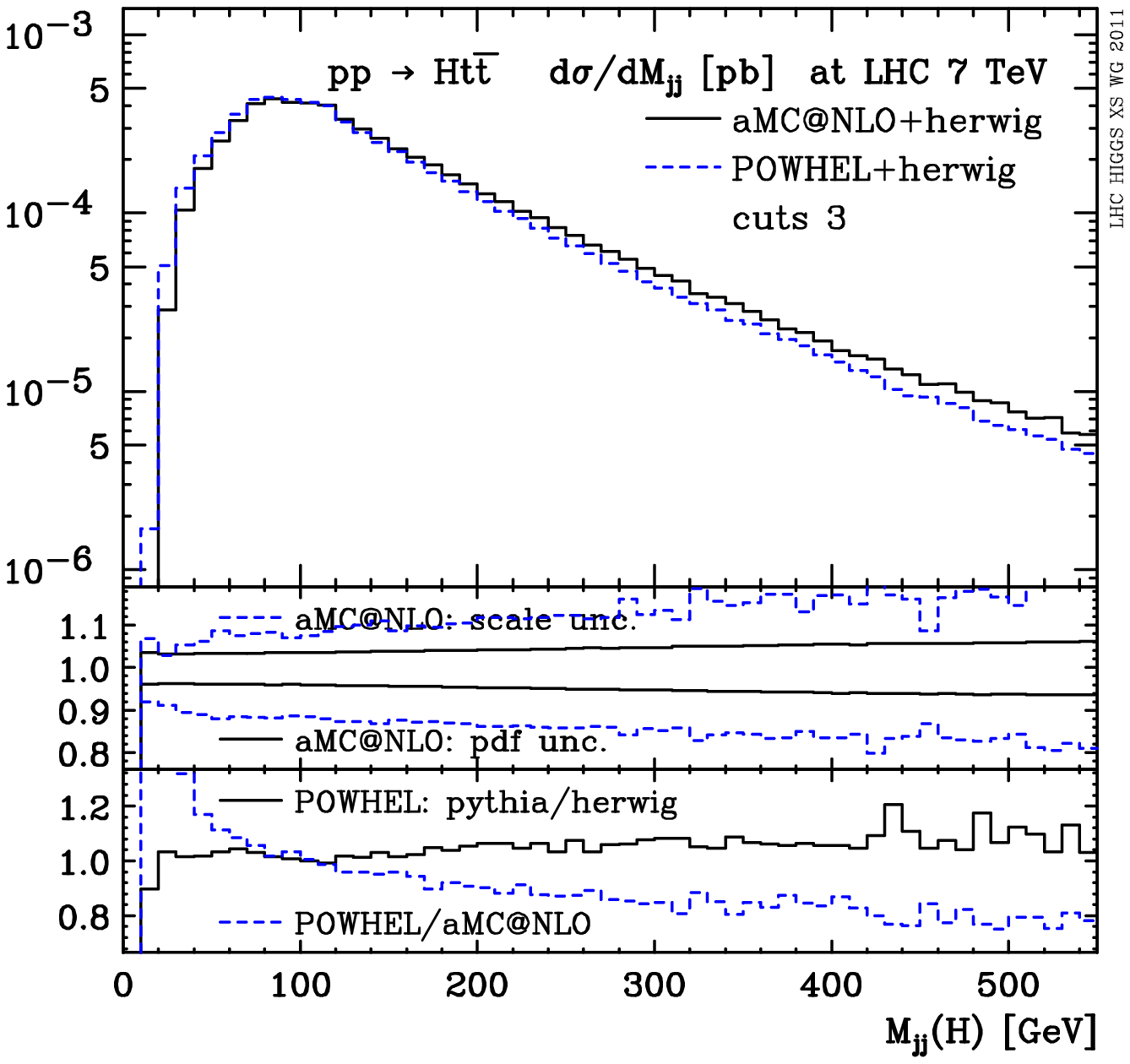}
\includegraphics[bb = 100 205 482 563, width=0.49\textwidth]{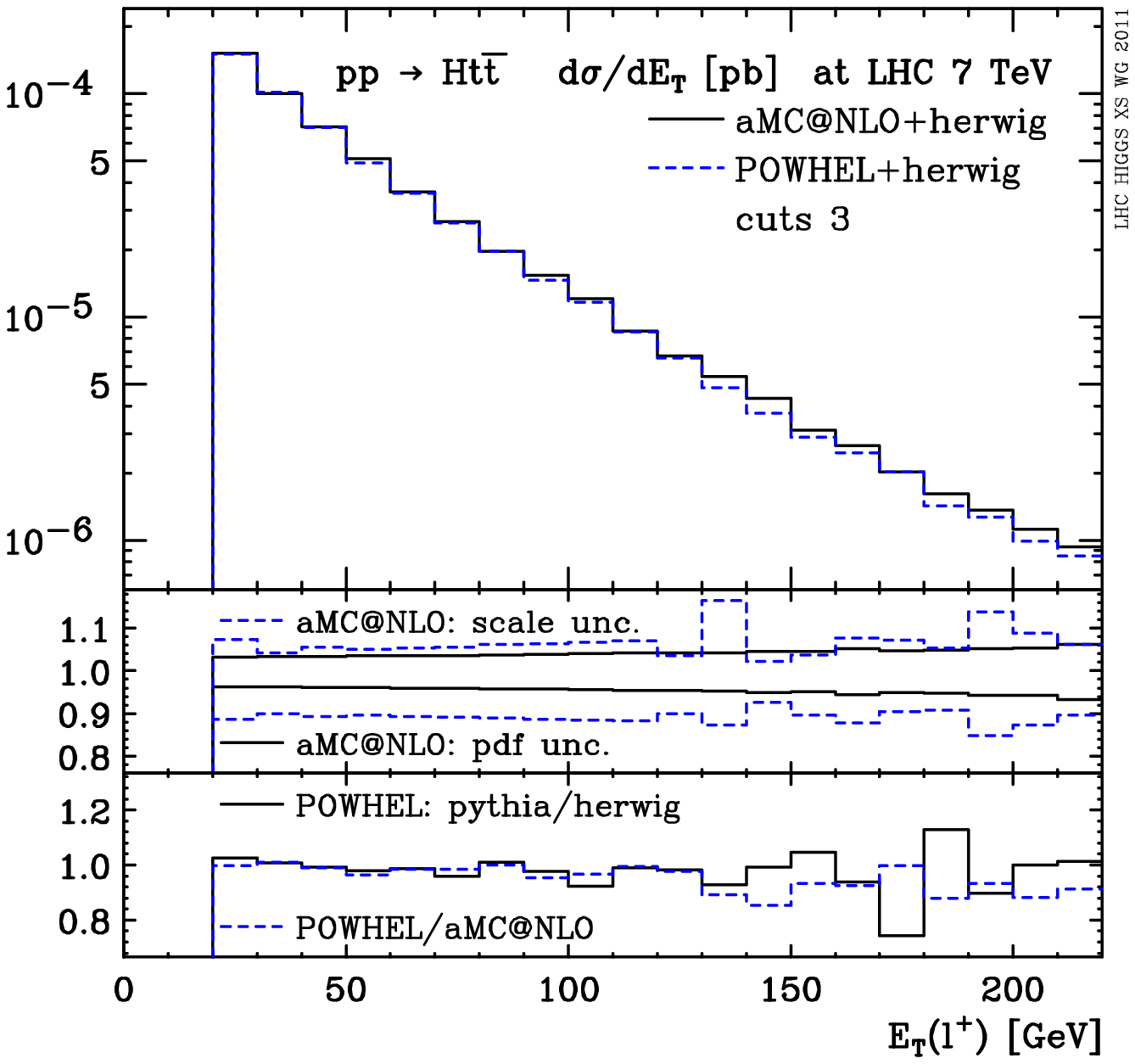}
\caption{Invariant mass of all jet pairs passing the cuts of set 3)
  (\textit{left}) and transverse energy of the hardest
  positively-charged lepton (\textit{right}). The different regions of
  the plots are defined as in
  Fig.~\protect\ref{fig:ttH_nlomc_fig1}. See text for more details.}
\label{fig:ttH_nlomc_fig4}
\end{figure}

\begin{figure}
\centering
\includegraphics[bb = 100 205 482 563, width=0.49\textwidth]{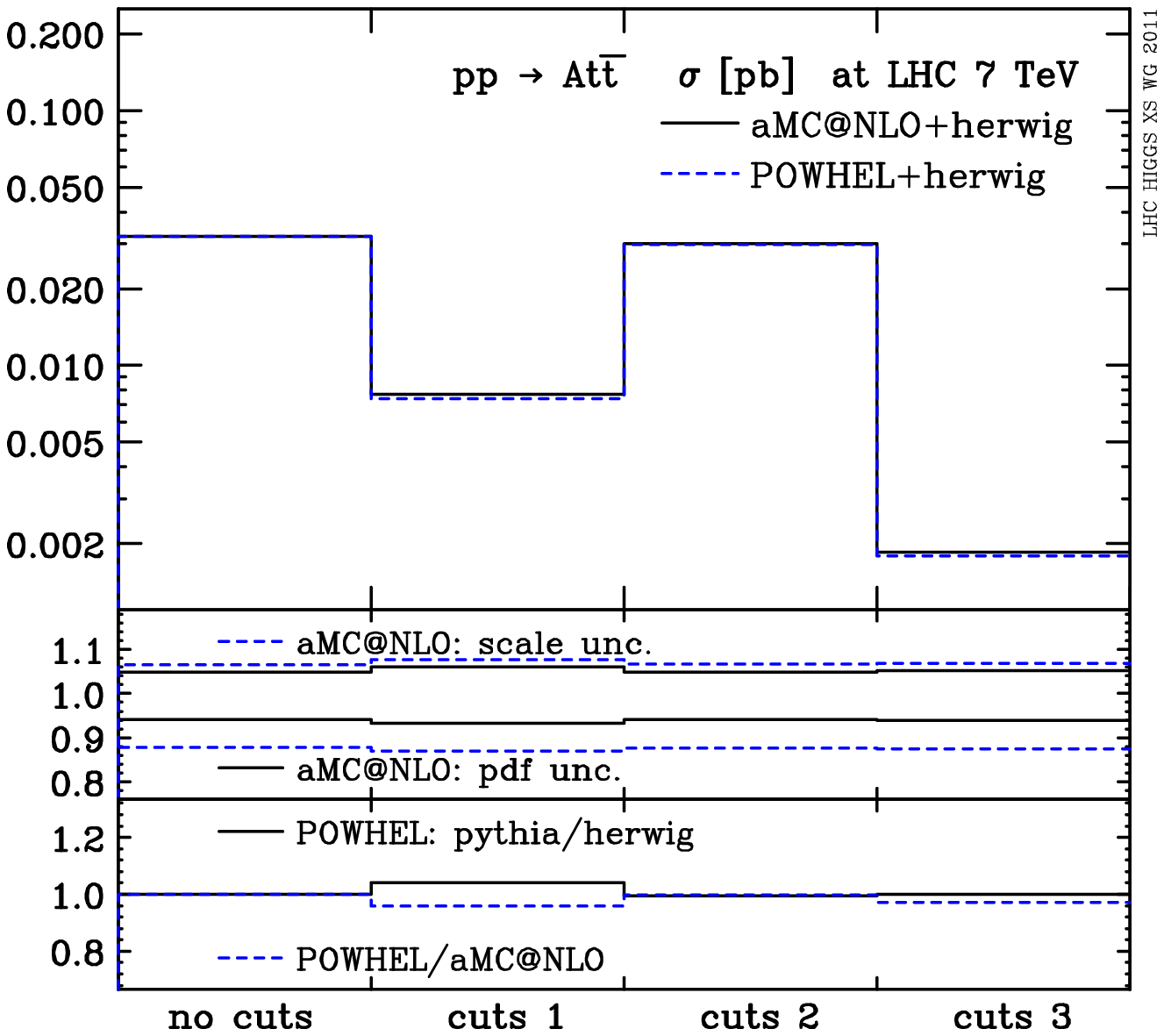}
\includegraphics[bb = 100 205 482 563, width=0.49\textwidth]{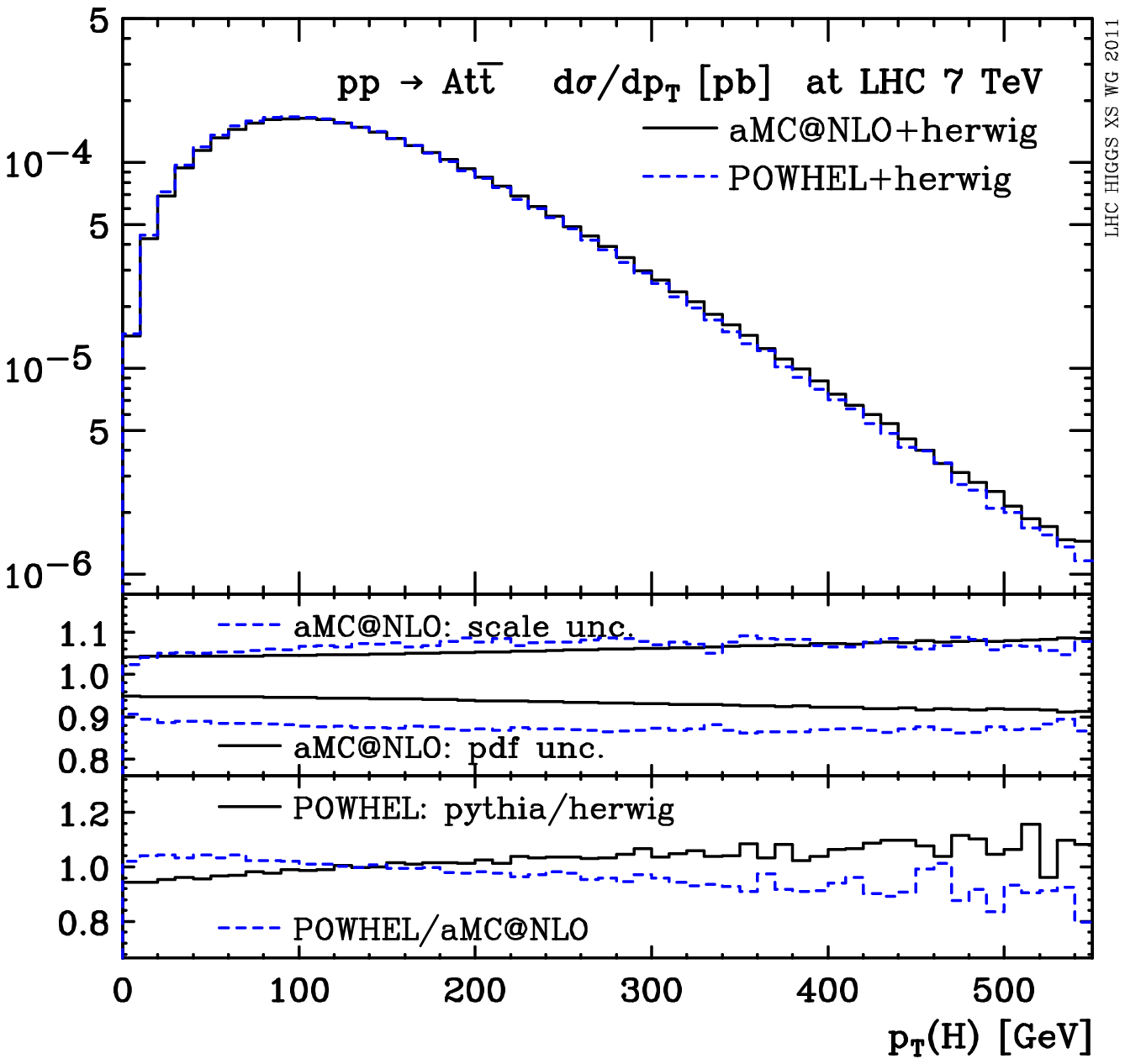}
\caption{Total rates for pseudo-scalar Higgs boson production after
  the different cuts defined in the text (\textit{left}) and
  transverse momentum of the pseudo-scalar Higgs boson
  (\textit{right}) in the no-cut configuration. The different regions
  of the plots are defined as in
  Fig.~\protect\ref{fig:ttH_nlomc_fig1}. See text for more details.}
\label{fig:ttH_nlomc_fig5}
\end{figure}

\begin{figure}
\centering
\includegraphics[bb = 100 205 482 563, width=0.49\textwidth]{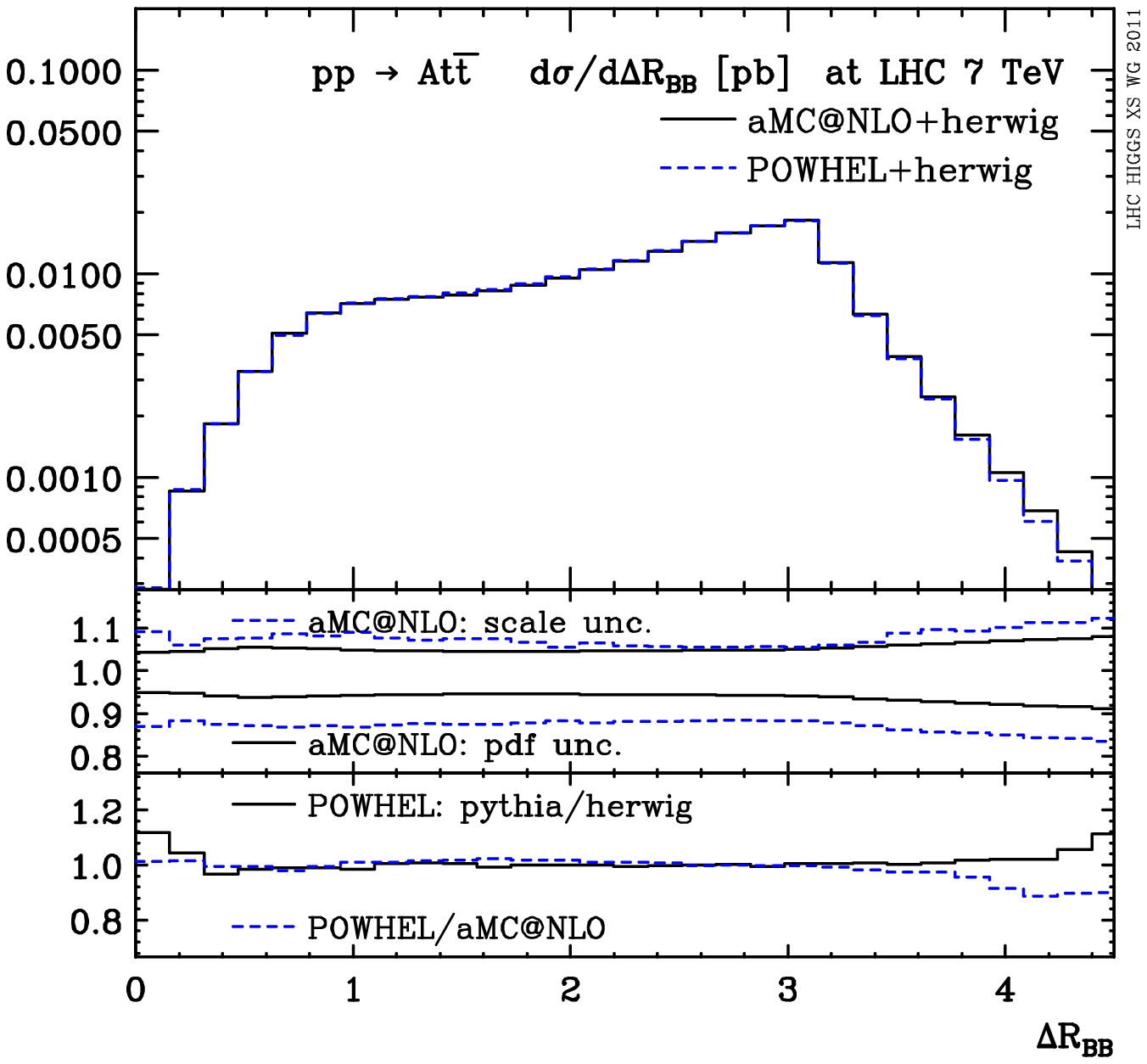}
\includegraphics[bb = 100 205 482 563, width=0.49\textwidth]{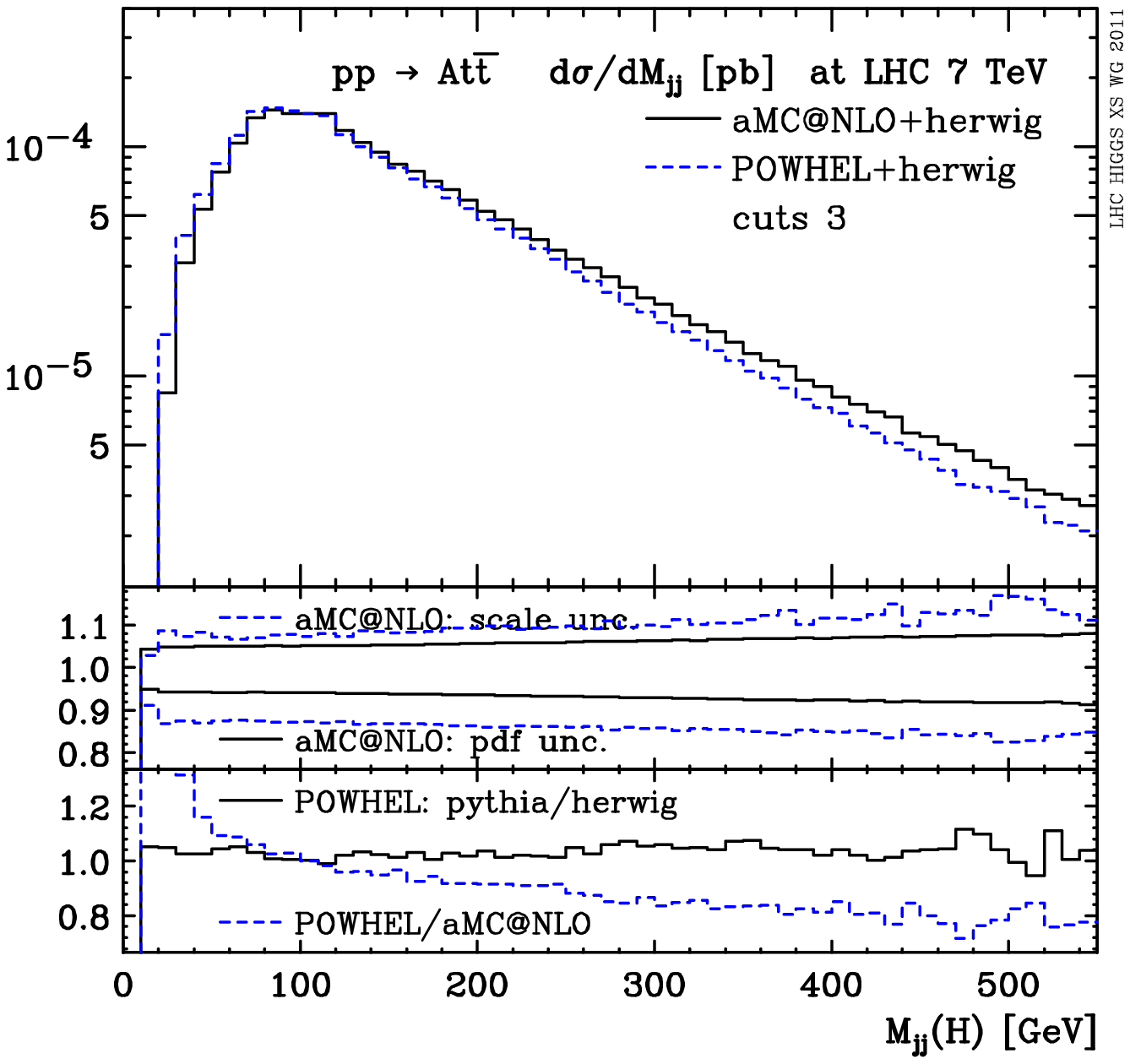}
\caption{Separation in pseudo-rapidity and azimuthal angle of the two
  hardest lowest-lying B hadrons in the events (no-cut analysis) 
  (\textit{left}) and
  invariant mass of all jet pairs passing the set 3) of cuts
  (\textit{right}). The different regions of the plots are defined as in
  Fig.~\protect\ref{fig:ttH_nlomc_fig1}. See text for more details.}
\label{fig:ttH_nlomc_fig6}
\end{figure}

We have studied a set of $\sim 20$ observables\footnote{We can share a
  complete and detailed list of results with other working groups who
  wish to perform the same analysis and make comparisons.} for both
different scenarios (SM and pseudo-scalar Higgs) and found similar
results by \powheghelac{} and a\MCNLO{} simulations. In the following,
we restrict ourselves to a representative set of
distributions. Results for scenario $1$, i.e.\ the scalar Higgs boson,
are presented in \refFs{fig:ttH_nlomc_fig1}--\ref{fig:ttH_nlomc_fig4}, 
whereas results for scenario $2$, i.e.\ the
pseudo-scalar Higgs, are included in
\refFs{fig:ttH_nlomc_fig5} and \ref{fig:ttH_nlomc_fig6}.  In the
upper part of each plot, the predictions of both \powheghelac{} and
a\MCNLO{} interfaced to \HERWIG{} are shown (by blue dashed 
and black solid lines, respectively). The lower part of each plot is
furthermore divided into two regions. In the top region we exhibit the
scale and PDF uncertainties computed by a\MCNLO{} using the procedure
outlined in \Bref{Frederix:2011ss}, while in the bottom region
the ratio of the predictions obtained by \powheghelac{} and a\MCNLO{}
is presented (i.e.\ the ratio of the curves in the main plot), as well
as the ratio of the results of \powheghelac{} interfaced to \PYTHIA{}
and to \HERWIG{}.  The scale dependence is obtained by the independent
variation of factorisation and renormalisation scales around the
default value $\mu_0$ in the range $[\mu_0/2,2 \mu_0]$, with the
restriction that $1/2 < \muR/\muF < 2$; the PDF uncertainty is
obtained by running the 40 MSTW $68\%$ C.L.~sets and combining them
using the Hessian method.  As for \PYTHIA{}, the last Fortran version
available in the web, \PYTHIA{} 6.4.25, has been adopted, in the
Perugia 2011 tune configuration~\cite{Skands:2010ak}, one of the most
updated leading-order tunes that takes into account recent LHC
experimental data, providing a $p_\mathrm{T}$-ordered shower (in the absence
of a tune specifically designed for NLO matched
computations). Furthermore, $\PQt$, $\PH$, and gauge-boson masses, and total
decay widths in \PYTHIA{} have been constrained to the same values as
in \HERWIG{}, and the $\PH$ forced to decay into $\PQb\PAQb$ in all events.

Let us describe the features of the figures in more detail. In the
left panel of \refF{fig:ttH_nlomc_fig1}, the predictions for the total rates
after the various cuts described above are given. For both the
\POWHEG{} as well as the \MCNLO\ method the total rates before
applying cuts are given by the fixed-order NLO results and are in
agreement. The rates after the cuts defined by sets 2) and 3) turned
out to be, as well, very similar in the two approaches.  On the other
hand, there is a $5\%$ difference between the total rate obtained by
\powheghelac{} interfaced to \HERWIG{} and the other predictions
(\powheghelac{} interfaced to \PYTHIA{} and a\MCNLO{} interfaced to
\HERWIG{}) just in case of the boosted-Higgs scenario, identified by
the set 1) of cuts, where only events with a Higgs boson with a
transverse momentum of at least $200\UGeV$ are kept in the
analysis. The origin of this difference can be understood from the
plot on the right-hand side of \refF{fig:ttH_nlomc_fig1}: the transverse
momentum of the Higgs boson as computed by \powheghelac{} + \HERWIG{}
turns out to be slightly softer in comparison to the other two
predictions. The uncertainty coming from scale variations is of the
order of $+ 5\%, - 10\%$, and becomes slightly larger when the cuts of
set 3) (all-cut analysis) are applied. The uncertainties from the PDFs
are smaller, $\pm 5\%$.

In the plot on the left-hand side of \refF{fig:ttH_nlomc_fig2} the
total transverse momentum of the $\PQt\PAQt\PH$ system is shown. This
observable is expected to be very sensitive to the matching procedure
used and, in the low-\pT\ region, very sensitive to the parton
shower. It turns out that the predictions obtained by a\MCNLO{} and
\powheghelac{} are in agreement within expectations, differences being
below $10\%$, except in the very soft region, where the differences
indeed increase.  Like before, both the a\MCNLO{} + \HERWIG{} and the
\powheghelac{} + \PYTHIA{} predictions are marginally harder than the
\powheghelac{} + \HERWIG{} ones. The uncertainties coming from scale
variations are small in the low-\pT\ region, where this observable has
NLO accuracy, while in the large-\pT{} region, i.e.\ \pT{} $>100\UGeV$,
the uncertainty grows and shows the usual large dependence typical of
a LO observable. Note that, even though in the low-\pT{} region this
observable is accurate up to NLO, the results are very sensitive to
large logarithms that are resummed by the parton shower. Therefore,
the scale dependence can not be considered at all an accurate estimate
of the total uncertainties. In the plot on the right-hand side of
\refF{fig:ttH_nlomc_fig2}, the invariant mass of the two hardest
lowest-lying B hadrons is shown. Like before, \powheghelac{} +
\HERWIG{} results are slightly softer than the other two predictions,
however differences amount to less than $10\%$ in the whole range
spanned by this observable and are within the uncertainties coming
from scale dependence and the PDF error sets.

As was already noted above and can be observed from the plot on the
left-hand side of \refF{fig:ttH_nlomc_fig1}, the \powheghelac{} +
\HERWIG{} predictions are $5\%$ smaller than the a\MCNLO{} + \HERWIG{}
and \powheghelac{} + \PYTHIA{} ones. This is again related to the fact
that the \powheghelac{} + \HERWIG{} results are slightly softer than
the other two. When looking into the boosted-Higgs scenario (set 1) of
cuts) (see above), the same features are visible in the plot of the
transverse momentum of the top quark, shown on the left-hand side of
\refF{fig:ttH_nlomc_fig3}.  In the plot on the right-hand side of
\refF{fig:ttH_nlomc_fig3}, on the other hand, the
\HT\ distribution is displayed for events passing the set of cuts 2),
i.e.\ at least $4$ central jets with a minimum transverse energy of
$25\UGeV$. This observable was defined as the scalar sum of transverse
energies \HT=$\sum_j E_{\mathrm{T},j}$ + $\sum_l E_{\mathrm{T},l}$ +
$\slashed{E}_{\mathrm{T}}$, where $j$ runs over all jets passing the cuts,
$l$ runs over all charged leptons and $\slashed{E}_{\mathrm{T}}$ is the
missing transverse energy. For this observable the differences between
a\MCNLO{} and \powheghelac{} turned out to be larger, a\MCNLO{} being
slightly softer this time.

One of the complications of extracting a Higgs signal using the
$\PQt\PAQt\PH$ channel is the combinatorial background: there are many
jets in the signal process and assigning the correct ones to the Higgs
decay is a non-trivial task. For example, if one naively takes all
events with two oppositely charged leptons and at least 4 jets,
i.e.\ the set 3) of cuts (see above), and plots the invariant mass of
all jet pairs satisfying the cuts, there is hardly any peak visible at
the Higgs mass, $\MH=120\UGeV$, as can be seen in the plot on the 
left-hand side of \refF{fig:ttH_nlomc_fig4}. This is one of the
challenges of finding a $\PQt\PAQt\PH$ signal.

As a final comparison between a\MCNLO{} and \powheghelac{} for the
scalar Higgs (scenario 1), we show on the right-hand side of \refF{fig:ttH_nlomc_fig4}
the transverse energy of the hardest positively-charged lepton after
the cuts of set 3). Results are in good agreement and no differences
are visible within statistical fluctuations.

The total rates for the pseudo-scalar Higgs boson, scenario 2, are
presented in the plot on the left-hand side of
\refF{fig:ttH_nlomc_fig5}. Rates without cuts and set 2) and 3) of
cuts, are about a third of the corresponding ones relative to scalar
Higgs-boson production, see \refF{fig:ttH_nlomc_fig1}. For the
boosted-Higgs scenario the difference is much smaller, with rates for
pseudo-scalar Higgs production only about $25\%$ smaller than the rates
for the scalar Higgs. The origin of this effect is that a
pseudo-scalar Higgs boson is in general more boosted, as can clearly
be seen by comparing the plot on the right-hand side of
\refF{fig:ttH_nlomc_fig5} with the one on the right-hand side of
\refF{fig:ttH_nlomc_fig1}.  Therefore, a boosted-Higgs search
following the guidelines of \Bref{Plehn:2009rk}, will work
equally well for a pseudo-scalar boson~\cite{Frederix:2011zi}. The
scale and PDF uncertainties are slightly larger in the case of the
pseudo-scalar, being $+7\%, -13\%$ and $\pm 6\%$ respectively. The
differences among the three predictions obtained by a\MCNLO{} and
\powheghelac{} interfaced to \HERWIG{} and \powheghelac{} interfaced
to \PYTHIA{} are smaller compared to scalar Higgs-boson predictions,
even though, in general, \powheghelac{} + \HERWIG{} results are still
slightly softer than the ones obtained by the other two frameworks.

Finally, in the plot on the left-hand side of \refF{fig:ttH_nlomc_fig6} the
separation for the two hardest lowest-lying B~hadrons in pseudo-rapidity and
azimuthal angle, $\Delta R_{B_1B_2}=\sqrt{(\Delta \eta_{B_1B_2})^2 +
  (\Delta \phi_{B_1B_2})^2}$ is presented. Results for a\MCNLO{} and
\powheghelac{} are in excellent agreement, well below $3\%$ over almost
the entire range. For large separations the difference increase
towards $10\%$ or so. As expected, for large separation also the scale
dependence increases. In the plot on the right-hand side of
\refF{fig:ttH_nlomc_fig6} the invariant mass of all jet pairs
defined for events passing the cuts of set 3) (see above) is
shown. Like for the scalar case, there is hardly any Higgs signal
visible over the continuous background coming from jet pairs that do
not correspond to the Higgs-boson decay (and are instead produced as a
consequence of top and antitop decays).
We point out that all three predictions are the same for the peak around the Higgs mass on 
\refFs{fig:ttH_nlomc_fig2} (left) and \ref{fig:ttH_nlomc_fig4}
(right), making the predictions particularly robust in this most important region.

We conclude that we are confident that the predictions for
$\PQt\PAQt\PH$ and $\PQt\PAQt\PA$ production are under good
theoretical control.  The differences between a\MCNLO{} and
\powheghelac{} interfaced to \HERWIG{} and \powheghelac{} interfaced
to \PYTHIA{} are, in general, below $10\%$ in those regions not
completely dominated by parton-shower effects. In general,
\powheghelac{} + \HERWIG{} gives slightly softer predictions than the
other two frameworks.  Reaching this level of agreement between the
results obtained by different frameworks has required several
interactions among us, since many different details in the independent
setup of the codes and in the precise definition of the observables
can produce sizable differences.\footnote{The codes and/or the event files
ready to-be-showered are available in the \powheghelac{} and a\MCNLO{}
websites, \texttt{http://grid.kfki.hu/twiki/bin/view/DbTheory/TthProd}
and \texttt{http://amcatnlo.cern.ch}, respectively.}

\subsection{The $\Pp\Pp\rightarrow\PQt\PAQt\PQb\PAQb$ background}
\label{subsec:ttH_ttbb_signal_vs_background}

In the low-Higgs-boson-mass region, where a SM Higgs boson mainly
decays to $\PQb\PAQb$, QCD $\PQt\PAQt\PQb\PAQb$ and $\PQt\PAQt jj$
production represent the most important backgrounds. The selection
strategies proposed so far by ATLAS and CMS are based on the full
reconstruction of the $\PQt\PAQt\PQb\PAQb$ signature, starting from a
final state with four $\PQb$ jets and additional light jets. Upon
reconstruction of the top quarks, two $\PQb$ quarks are identified as
originating from the top-quark decays, while the remaining two $\PQb$
quarks represent a Higgs candidate, to be identified via the 
invariant-mass reconstruction of the $\PQb\PAQb$ pair. Simulations indicate
that the presence of other $\PQb$ and light jets in the final state
greatly affects the correct identification of the $\PQb\PAQb$ pair
from the Higgs decay therefore diluting the signal to background
ratio. More recently, the idea of searching for a highly-boosted Higgs
boson (producing a \textit{fat} jet containing the $\PQb\PAQb$ decay
products) has been proposed to enhance the signal to background ratio. 

Whether $\PQt\PAQt\PH$ will provide a discovery channel very much
depends on how well we can study the characteristics of the
$\PQt\PAQt\PQb\PAQb$ and $\PQt\PAQt jj$ backgrounds and find ways to
efficiently discriminate them from the signal.  Requiring three 
$\PQb$ tags would strongly suppress the $\PQt\PAQt jj$ contamination, and
would leave $\PQt\PAQt\PQb\PAQb$ as the dominant background.

\begin{figure}
\centering
\includegraphics[width=0.45\textwidth]{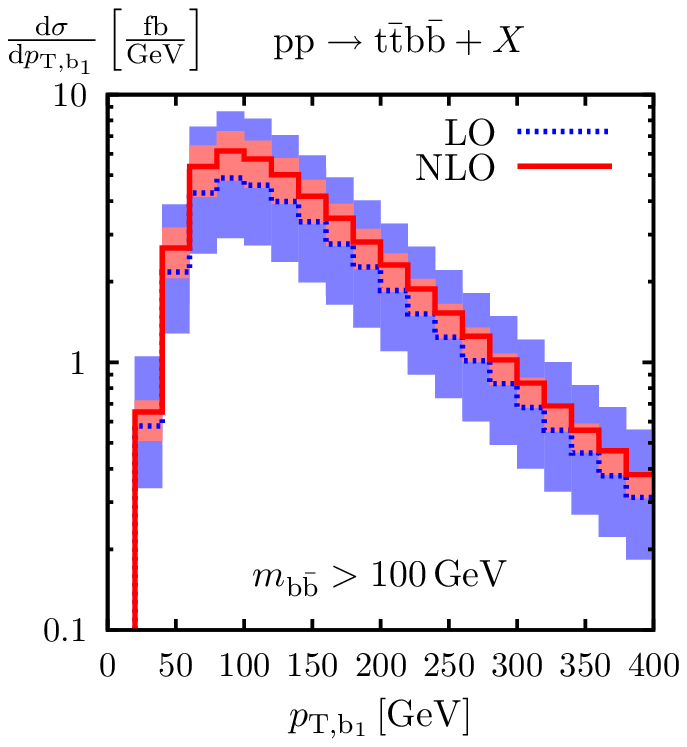}
\hspace*{1em}
\includegraphics[width=0.45\textwidth]{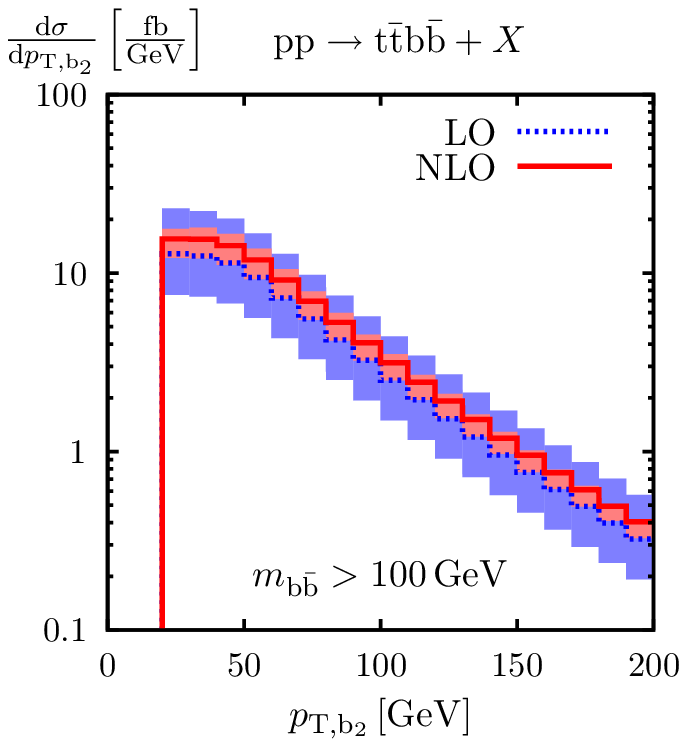}\\[1em]
\includegraphics[width=0.45\textwidth]{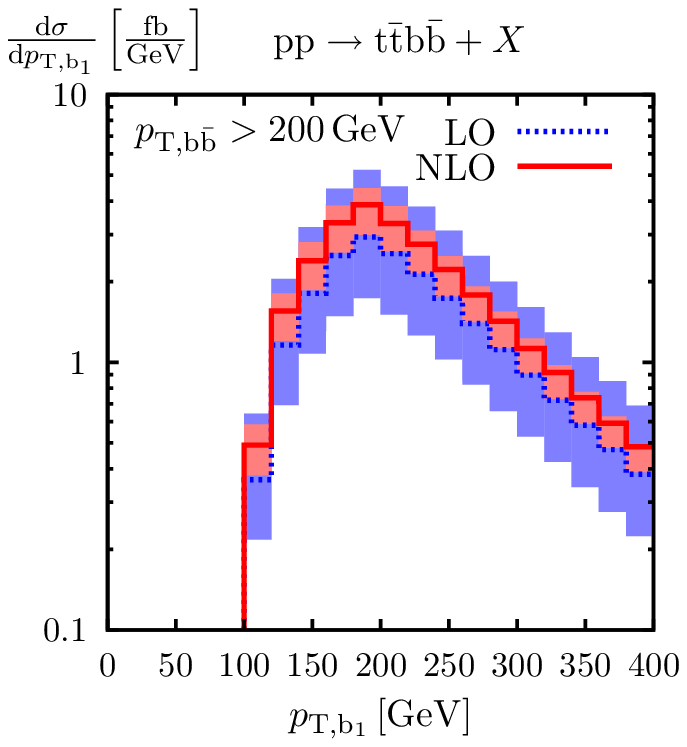}
\hspace*{1em}
\includegraphics[width=0.45\textwidth]{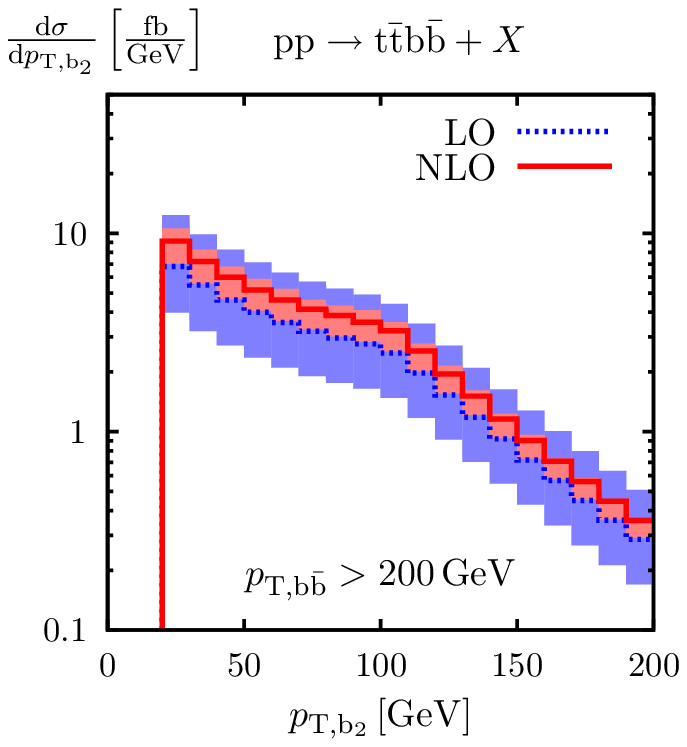}
\caption{Transverse momentum of the harder ($p_{\mathrm{T},\PQb_1}$) and softer
  ($p_{\mathrm{T},\PQb_2}$) $b$ jets at $14\UTeV$ CM energy: absolute LO and NLO prediction for the
  unboosted ($m_{\PQb\PAQb}>100\UGeV$, upper plots) and boosted regimes
  ($p_{T,\PQb\PAQb}>200\UGeV$, lower plots). The uncertainty bands
  correspond to a factor two scale variation. From \Bref{Bredenstein:2010rs}.}
\label{fig:ttbb_ptb1+2_mbb100_ptbb200}
\end{figure}

The NLO QCD corrections to the $\PQt\PAQt\PQb\PAQb$ production
background have been calculated
\cite{Bredenstein:2009aj,Bredenstein:2008zb,Bredenstein:2010rs,
  Bevilacqua:2009zn,Binoth:2010ra} and discussed for the LHC at the 
CM energy of $14\UTeV$.
The corrections enhance and stabilise
the cross section. Traditionally the simulations of
$\PQt\PAQt\PQb\PAQb$ were based on a LO cross section and used
$\muR=\muF=\Mt+m_{\PQb\PAQb}/2$ as the central renormalisation and
factorisation scales. NLO studies have shown that this scale choice
does not provide an adequate description of the QCD dynamics of
$\PQt\PAQt\PQb\PAQb$, since this process is a multiscale process that
involves scales much below $\Mt+m_{\PQb\PAQb}/2$. The theoretical
stability of the cross section is greatly improved by choosing a
dynamical scale like
$\muR^2=\muF^2=\Mt\sqrt{p_{\mathrm{T}\PQb}p_{\mathrm{T}\PAQb}}$. In
this case, the NLO corrections increase the background cross section
within the signal region by about $20{-}30\%$
\cite{Bredenstein:2009aj,Bredenstein:2008zb,Bredenstein:2010rs}. Most
importantly, the scale dependence is significantly reduced to a level
significantly below $30\%$. Examples of $\pT$ distributions for the
hardest and softest $\PQb$ jets are given in
\refF{fig:ttbb_ptb1+2_mbb100_ptbb200}, where the unboosted
($m_{\PQb\PAQb}>100\UGeV$) and boosted
($p_{\mathrm{T},\PQb\PAQb}>200\UGeV$) regimes are compared.

\begin{figure}
\centering
\includegraphics[width=0.45\textwidth]{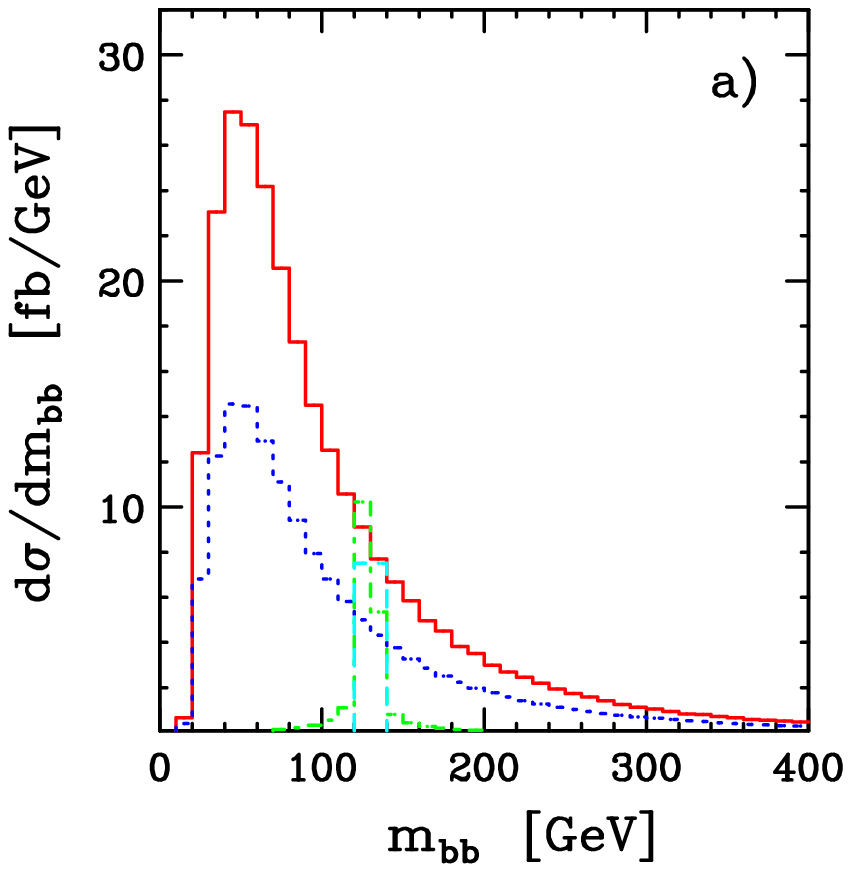}
\hspace*{1em}
\includegraphics[width=0.45\textwidth]{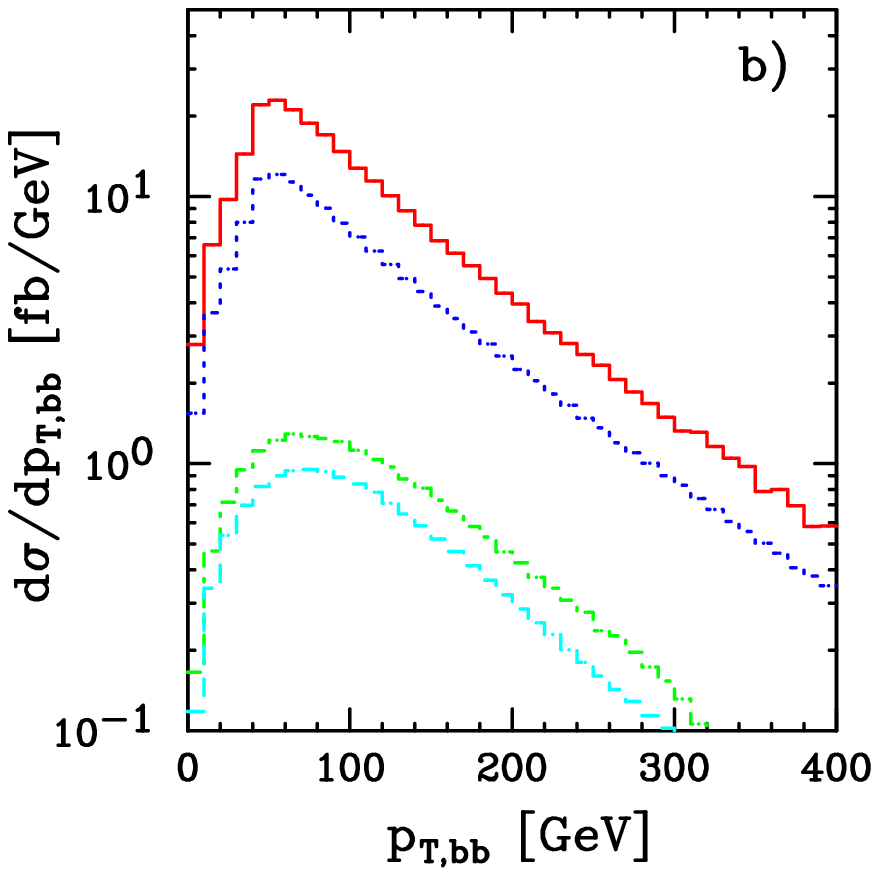}\\[1em]
\includegraphics[width=0.45\textwidth]{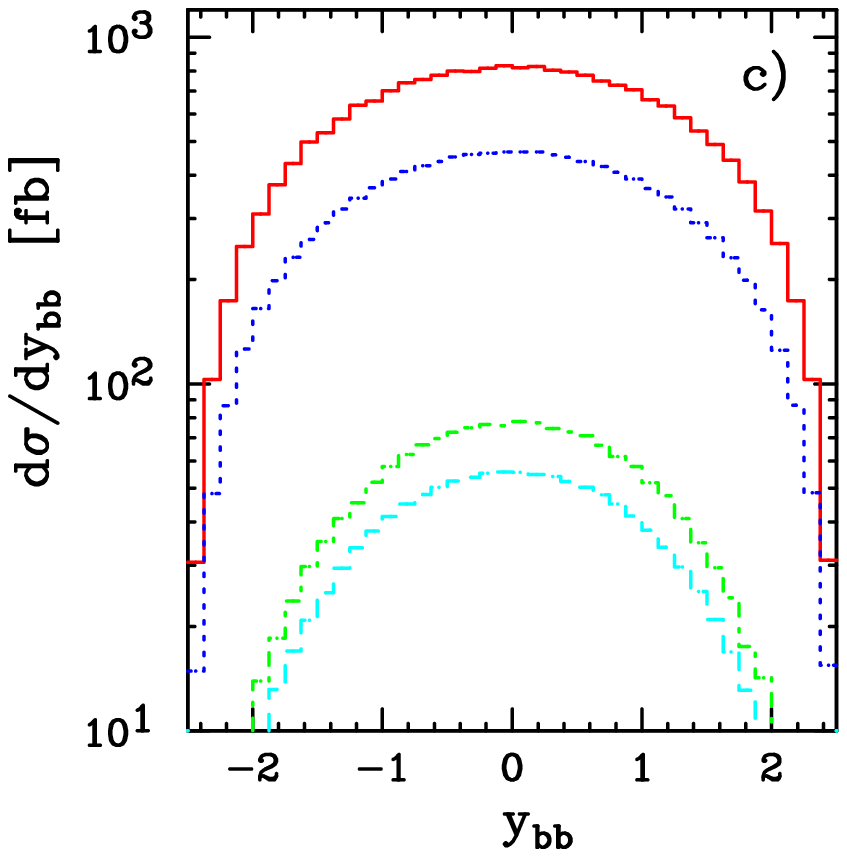}
\hspace*{1em}
\includegraphics[width=0.45\textwidth]{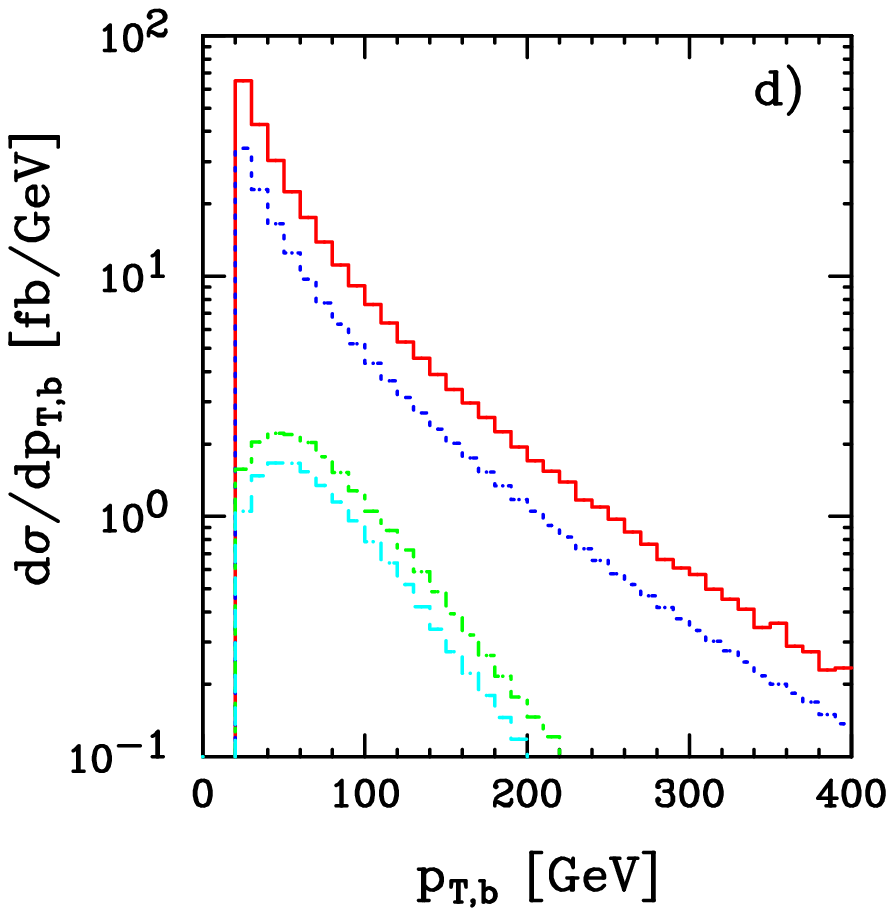}
\caption{Distribution of the invariant mass $m_{\PQb\PAQb}$ of the
  bottom-anti-bottom pair (a), distribution in the transverse momentum
  $p_{\mathrm{T}_{\PQb\PAQb}}$ of the bottom-anti-bottom pair (b),
  distribution in the rapidity $y_{\PQb\PAQb}$ of the
  bottom--anti-bottom pair (c), and distribution in the transverse
  momentum $p_{\mathrm{T},\PQb}$ of the bottom quark (d) for
  $\Pp\Pp\rightarrow \PQt\PAQt\PH \rightarrow \PQt\PAQt\PQb\PAQb +X$
  and $\Pp\Pp\rightarrow \PQt\PAQt\PQb\PAQb+X$ at the LHC for the 
  CM energy of $14\UTeV$.  The red
  solid line refers to the NLO QCD background, the blue dotted line to
  the LO QCD background, while the green dash-dotted and cyan dashed
  line to the NLO and LO signal, respectively. From \Bref{Binoth:2010ra}.}
\label{fig:ttH_ttbb_vs_ttH_nwa}
\end{figure}

In addition a comparison between the signal process $\Pp\Pp\to
\PQt\PAQt\PH\to \PQt\PAQt\PQb\PAQb$ and the $\PQt\PAQt\PQb\PAQb$
background has been obtained in the narrow-width approximation
\cite{Binoth:2010ra}.  In \refF{fig:ttH_ttbb_vs_ttH_nwa}, a few
histograms, namely the invariant mass, the transverse momentum, the
rapidity of the two-$\PQb$-jet system, as well as the transverse
momentum of the single $\PQb$-jet are shown.  In all figures the red solid line
refers to the NLO QCD background, the blue dotted line to the LO QCD
background, while the green dash-dotted and cyan dashed line to the
NLO and LO signal, respectively. Apart from the invariant mass of the
$\PQb\PAQb$ system and the $p_T$ spectrum of the $\PQb$ quark, the
shapes look very similar for signal and background.

This makes it possible to study the signal and background processes
including the final-state Higgs decay into $\PQb\PAQb$ with cuts at
the same time at NLO. However, it should be noted that the final-state
top decays have not been included at NLO so that a full NLO signal and
background analysis including all experimental cuts is not possible so
far. The top-quark decays are expected to affect the final-state
distributions more than the Higgs decays into $\PQb\PAQb$ pairs. The
next natural step will then be to interface the NLO calculation of
$\PQt\PAQt\PQb\PAQb$ production with \PYTHIA{} and \HERWIG{}. This will provide
the ultimate tool to study both signal and background in the presence
of both decays of the Higgs boson and of the top/antitop pair.

Finally, we note that the NLO QCD corrections to 
$\Pp\Pp\to\PQt\PAQt jj$ have been calculated as 
well~\cite{Bevilacqua:2010ve,Bevilacqua:2011hy}, however,
in the boosted-Higgs analysis the major background to the 
$\PQt\PAQt\PH$ signal is due to $\PQt\PAQt\PQb\PAQb$ production~\cite{Plehn:2009rk}.


\clearpage

\clearpage


\newpage


\providecommand{\qTH}{q_{\mathrm{T,H}}}
\providecommand{\YH}{Y_{\mathrm{H}}}
\providecommand{\pTiso}{p_{\mathrm{T},\mathrm{iso}}}

\section{$\PGg\PGg$ decay mode\footnote{%
    S.~Gascon-Shotkin, M.~Kado (eds.); N.~Chanon, L.~Cieri, G.~Davies, 
    D.~D'Enterria, D.~de Florian, S.~Ganjour,
    J.~-Ph.~Guillet, C.~-M.~Kuo, N.~Lorenzo,E.~Pilon and J.~Schaarschmidt.}}

\subsection{Introduction}
Despite its relatively low branching fraction and considerable reducible and irreducible backgrounds from
SM QCD processes, the $\PGg\PGg$ decay mode benefits from a clean signature, provided that a 
sufficiently high-resolution electromagnetic calorimeter is used. The mode $\PH \rightarrow\PGg\PGg$ is generally
considered to be the principal discovery channel at the LHC for a Higgs boson having a mass between 
$100\UGeV$ and $150\UGeV$. 
Furthermore, given the recent combined results from the ATLAS and CMS collaborations from analyses 
corresponding to up to $2.3\Ufb^{-1}$ of integrated luminosity
excluding with $95\%$ confidence a Standard Model Higgs boson in the range $141\UGeV<\MH<476\UGeV$, the 
$\PGg\PGg$ decay mode is the channel whose sensitive range remains at this writing the least constrained.  
The corresponding individual ATLAS~\cite{Collaboration:2011ww,ATLAS-CONF-2011-149} and CMS~\cite{CMS-PAS-HIG-11-021} results have also been made public.

\subsection{Sets of acceptance criteria used}

Three sets of acceptance criteria, shown in \refT{tab:HggAccCriteria}, have been used for the studies presented in 
this section. Two of them ('ATLAS', 'CMS') are based on acceptance
criteria of the ATLAS and CMS
$\PH\rightarrow \PGg\PGg$ searches; the third ('Loose') corresponds roughly to
typical acceptance criteria used in the measurement of Standard Model prompt photon
processes which constitute backgrounds to these searches.
\begin{table}[h]\small
\renewcommand{\arraystretch}{1.2}
\setlength{\arraycolsep}{1.5ex}
\caption{The three sets of acceptance criteria used for studies in this section. }
\label{tab:HggAccCriteria}
\centerline{
$\begin{array}{cccc}
\hline
            & \text{'CMS'} & \text{'ATLAS'} & \text{'Loose'} \\
\hline
E_{\mathrm{T}\PGg 1} \mathrm{[GeV]} & >40  & >40 & >20 \mbox{ or } 23 \\
E_{\mathrm{T}\PGg 2} \mathrm{[GeV]} & >30  & >25 & >20 \\
|\eta_{\PGg}| & <2.5 & <2.37 & <2.5 \\
\mbox{Excluded }|\eta_{\PGg}| & [1.4442,1.566] & [1.37,1.52] &   \\
M_{\PGg\PGg} \mathrm{[GeV]} & [100,160] & [100,160] & >80 \\
\hline
\end{array}$ }
\end{table}

For the studies concerning background processes in the Standard Model, parton-level
isolation requirements have been imposed, requiring a maximum transverse hadronic 
energy of $5\UGeV$ within a solid cone of radius $\Delta R=\sqrt{{\Delta\eta}^2+ {\Delta\phi}^2}=0.4$ or
$0.3$ as noted.

\subsection{Signal modelling and differential $K$-factors}


\subsubsection{Reweighting of the $\pT$ spectrum for the gluon-fusion production process}

\begin{figure}
\includegraphics[width=0.49\textwidth]{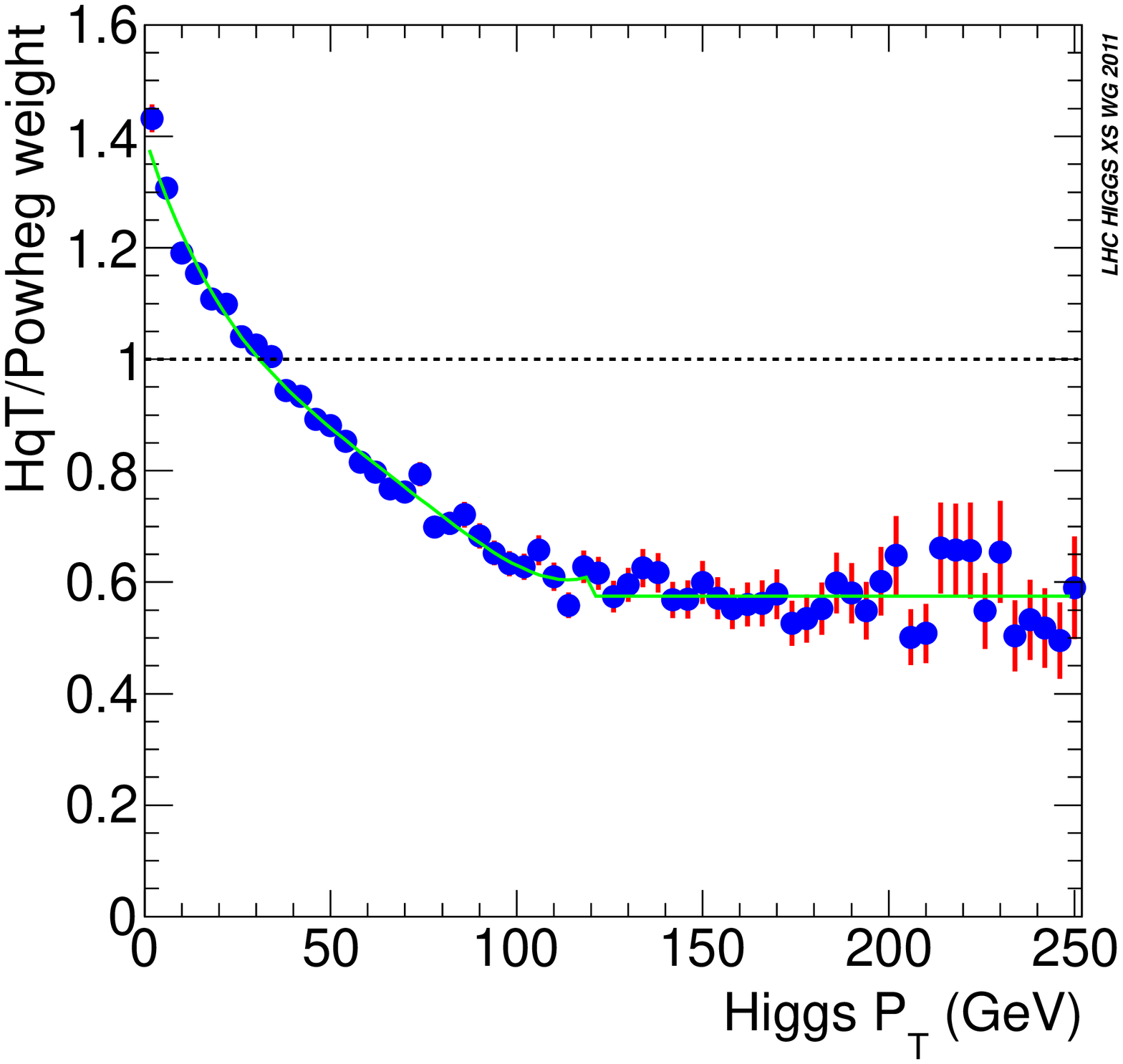}
\includegraphics[width=0.49\textwidth]{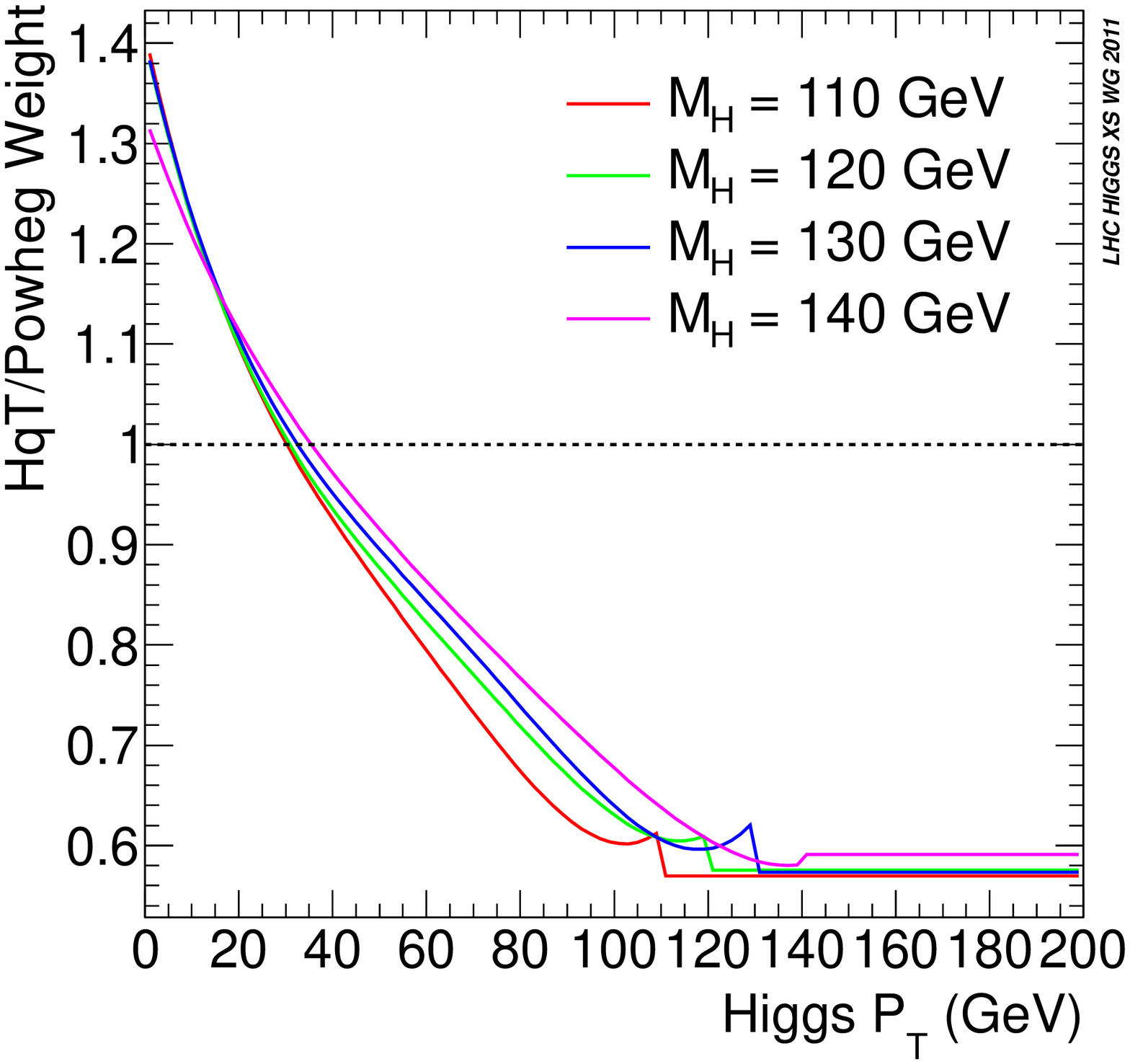}
\caption{(Left) Distribution of relative {\sc HqT}/\POWHEG{} event weights
as a function of Higgs-boson $\pT$ for $\MH=120\UGeV$ with fit superposed. (Right) Fitted event weight
functions for $\MH=110,120,130,140\UGeV$ as a function of Higgs-boson  $\pT$. }
\label{HqtKfactors}
\end{figure}

The effect of reweighting the Higgs-boson $\pT$ spectrum given by the NLO program \POWHEG{} (with parton-shower
simulation, hadronisation, and underlying event from \PYTHIA{}~6.4) to that given by the 
inclusive NNLL program {\sc HqT}~\cite{deFlorian:2011xf} has been evaluated for the $\PGg\PGg$  
final state. Figure~\ref{HqtKfactors} shows the distribution of relative {\sc HqT}/\POWHEG{} event weights
as a function of Higgs-boson $\pT$ for $\MH=120\UGeV$; the distribution is fit with a 4th-degree 
polynomial function for $\pT<\MH$ and a constant function for $\pT>\MH$.  The fitted functions for   
four Higgs-boson masses relevant to the $\PH\rightarrow \PGg\PGg$ search are also shown; there is a slight
to moderate dependence on the Higgs-boson mass.    

\begin{figure}
\includegraphics[width=0.49\textwidth]{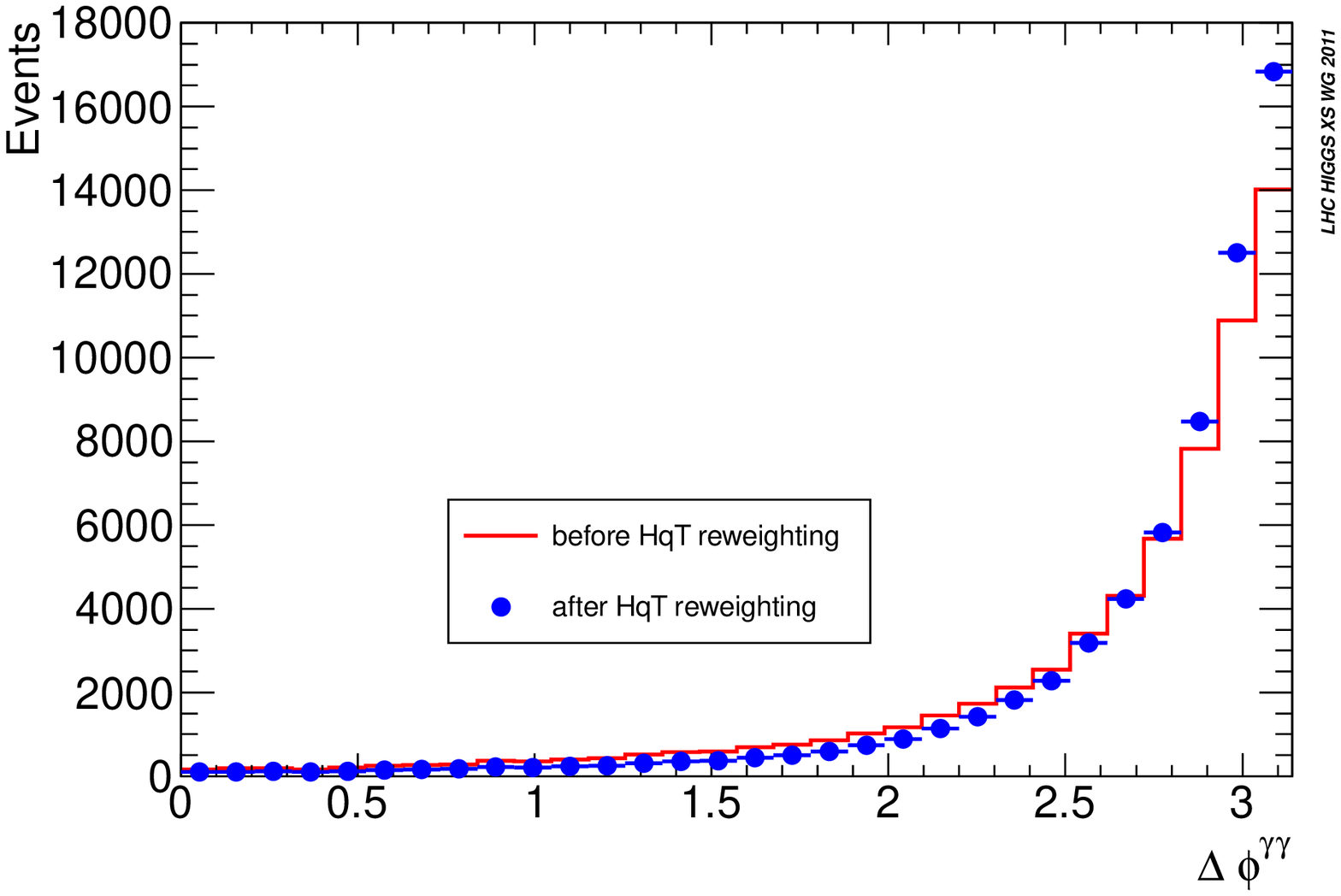}
\includegraphics[width=0.49\textwidth]{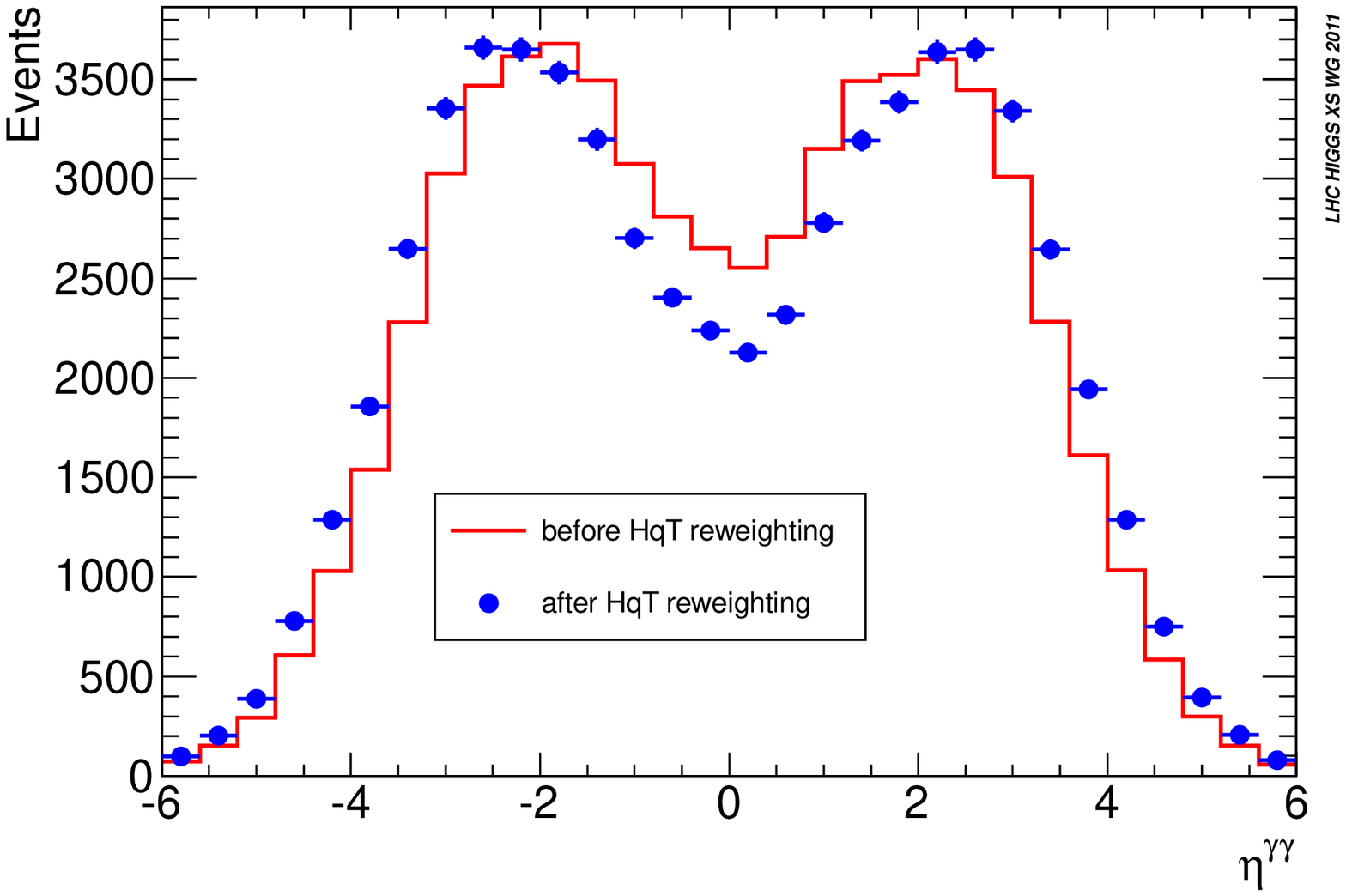}\\
\includegraphics[width=0.49\textwidth]{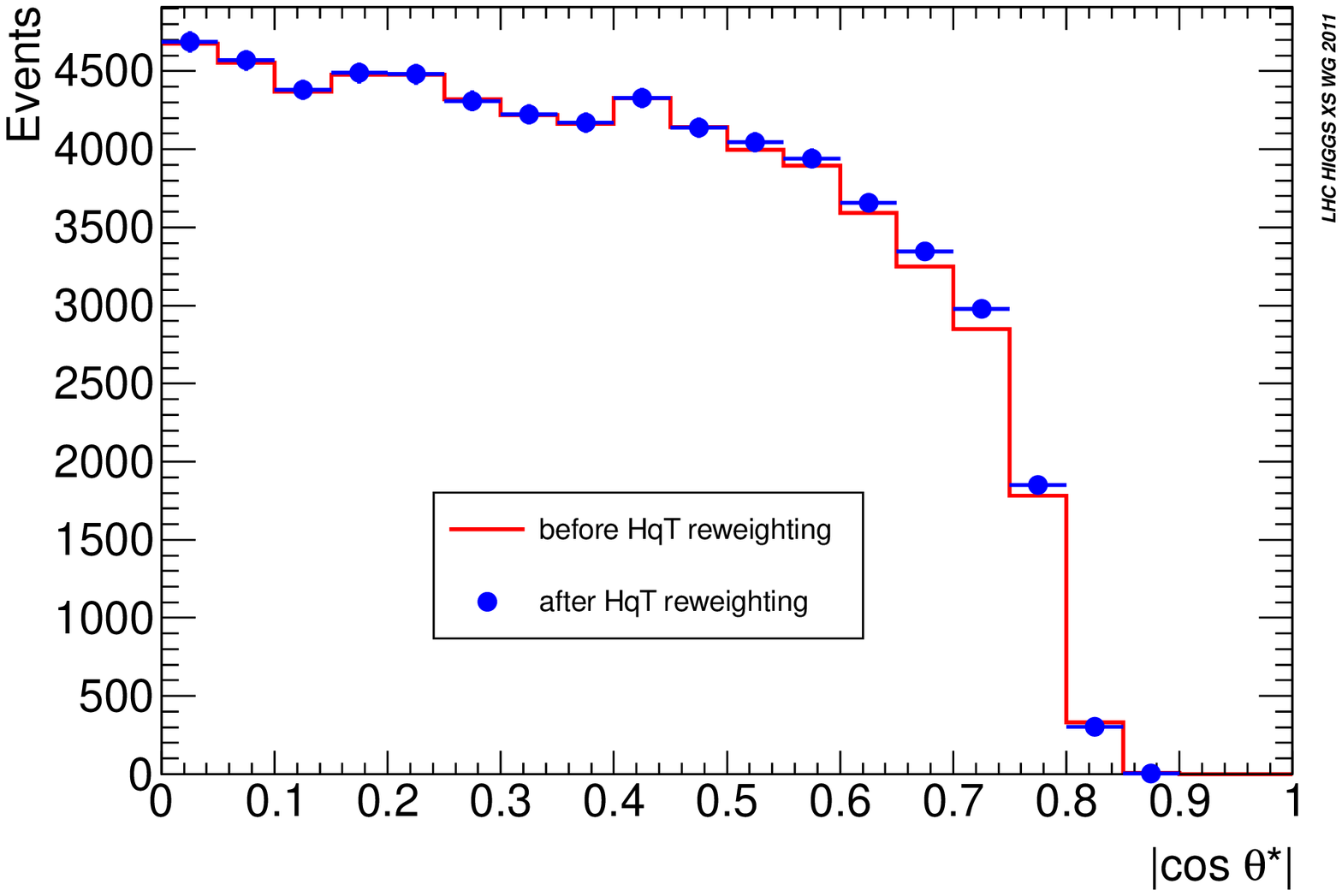}
\caption{Distributions of kinematical observables potentially important for the $\PH\rightarrow \PGg\PGg$ search, for 
$\MH=120\UGeV$, after application of 'CMS' acceptance criteria, before (red histogram) and after (blue dots)
reweighting of the Higgs-boson $\pT$ spectrum to that of {\sc HqT}. Top left:  $\Delta\phi^{\PGg\PGg}$; top right:
$\eta^{\PGg\PGg}$;  bottom: $\cos\theta^*$. }
\label{HggPtReweightedKinematics}
\end{figure}

It is important to evaluate the impact of the $\pT$-reweighting on other observables susceptible to be 
used in the $\PH\rightarrow \PGg\PGg$ search. Figure~\ref{HggPtReweightedKinematics} shows the distributions,
after application of the 'CMS' acceptance criteria, of the kinematical observables $\Delta\phi^{\PGg\PGg}$,
the difference in azimuthal angle between the two photons, 
$\eta^{\PGg\PGg}$, the pseudo-rapidity of the diphoton system, and $\cos\theta^*$, the cosine of the angle between
one of the photons and the beamline in the centre-of-mass frame of the Higgs boson, before and after the
{\sc HqT} reweighting, for $\MH=120\UGeV$.
The $\pT$-reweighting has a significant effect on both the $\Delta\phi^{\PGg\PGg}$ and
$\eta^{\PGg\PGg}$ distributions; the $\Delta\phi^{\PGg\PGg}$ distribution is shifted to higher values,
corresponding to the two photons being more back to back, and that of $\eta^{\PGg\PGg}$ is shifted away 
from small rapidity values towards the forward--backward zones. However, the  $\cos\theta^*$ distribution is
only slightly affected, with a small enhancement around  $\cos\theta^{*}=0.7$ and a corresponding small deficiency at 
the highest values. This has important implications for the treatment of the signal--background interference, which will
be discussed below. 
  
\subsubsection{Doubly-differential $K$-factors for the gluon-fusion production process}
\label{se:Kfac_ggHgaga}

To propagate higher-order effects to kinematical distributions produced by {\sc POWHEG}~\cite{Frixione:2007nw}, one can also perform a 2D reweighting with a 
$K$-factor $K(\qTH,\YH)$ where $\qTH$ is the transverse momentum of the Higgs boson and $\YH$ its 
rapidity \cite{Davatz:2006ut}. In the following we describe such a 2D reweighting procedure using 
{\sc HNNLO} \cite{Catani:2007vq} and {\sc POWHEG}~\cite{TheseNChanon}. This study should be repeated with {\sc HqT}
in place of {\sc HNNLO}.

\begin{figure}
    \includegraphics[height=5.3cm]{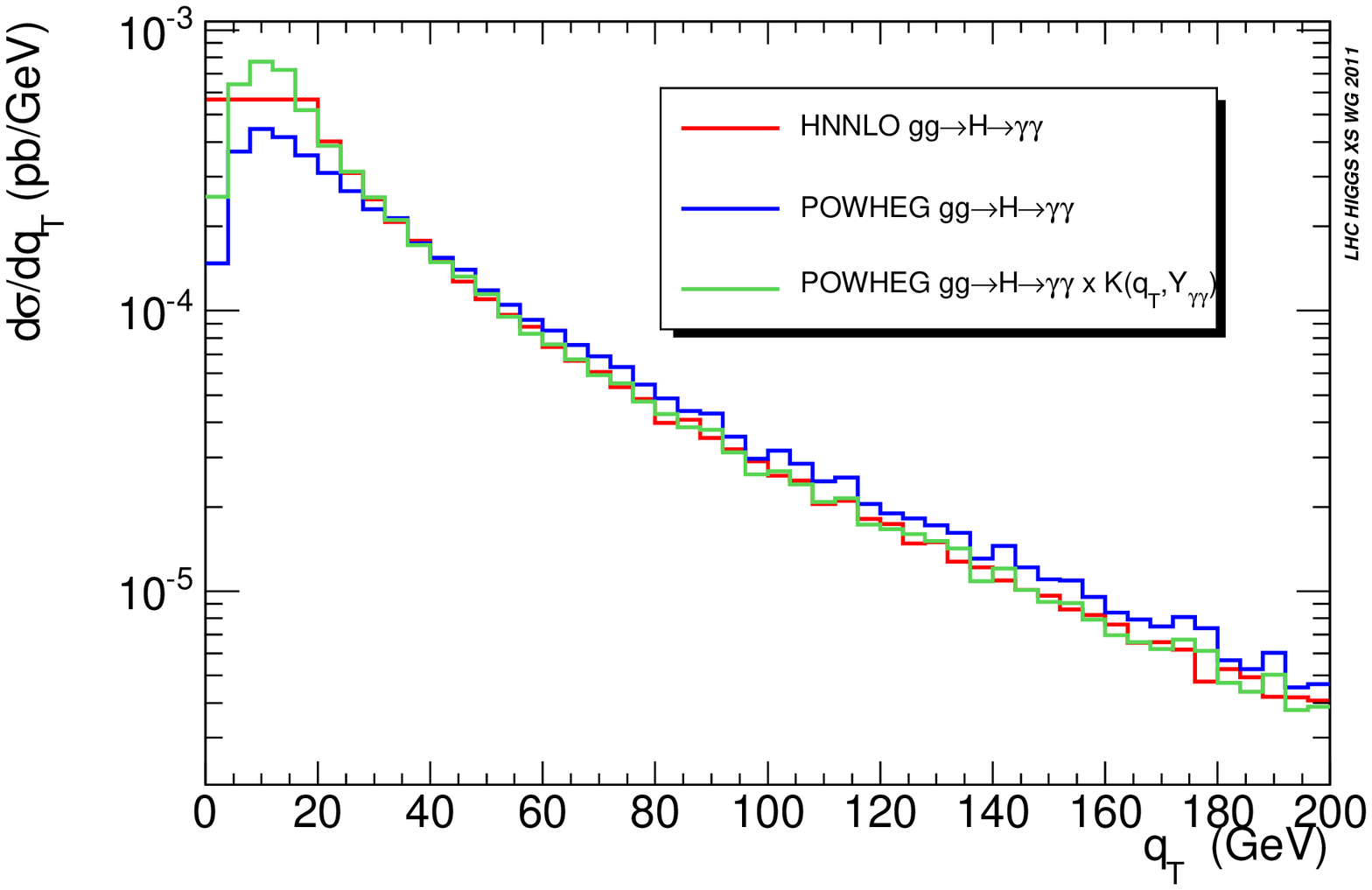}
    \includegraphics[height=5.3cm]{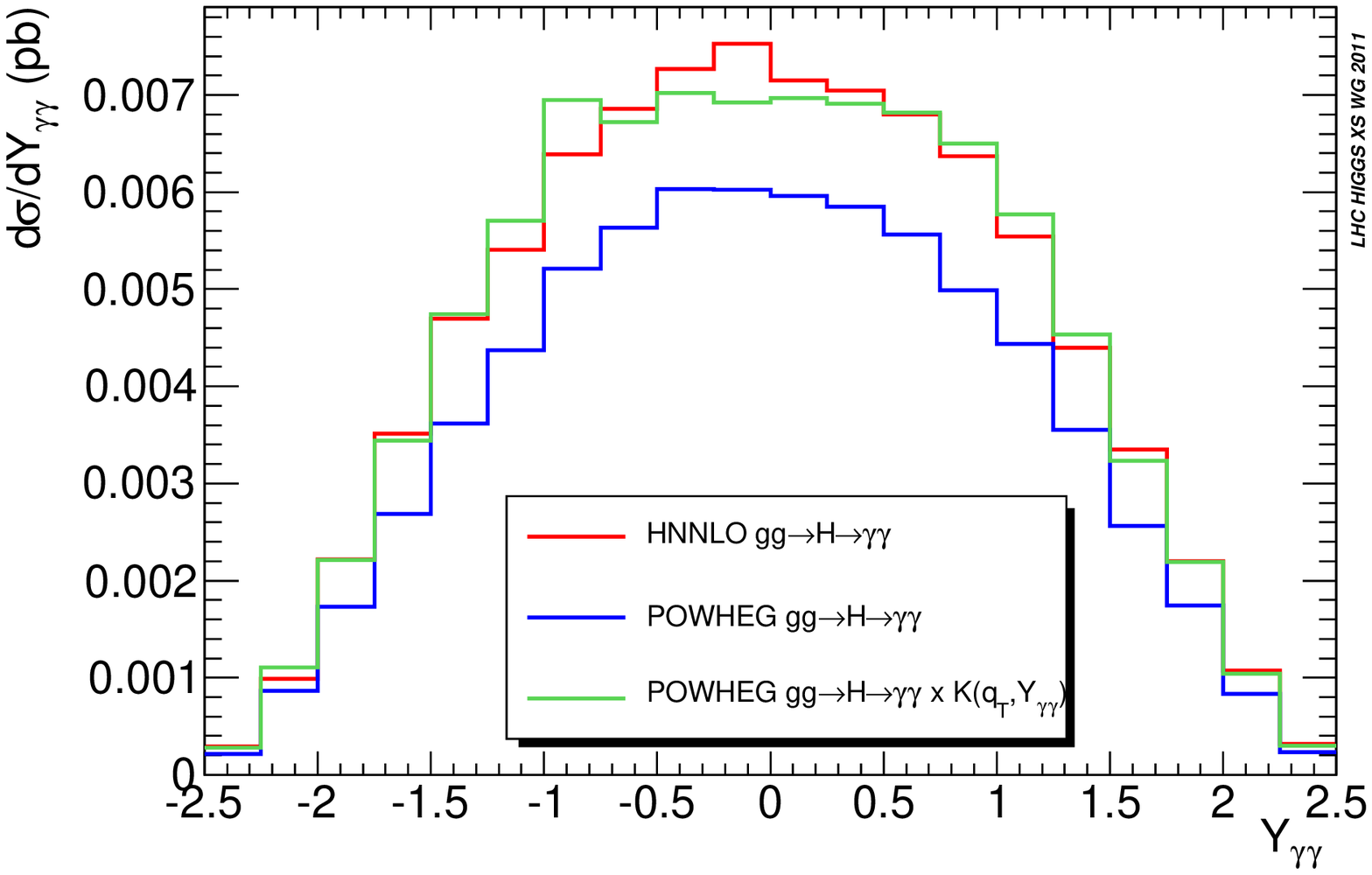}\\
    \includegraphics[height=5.3cm]{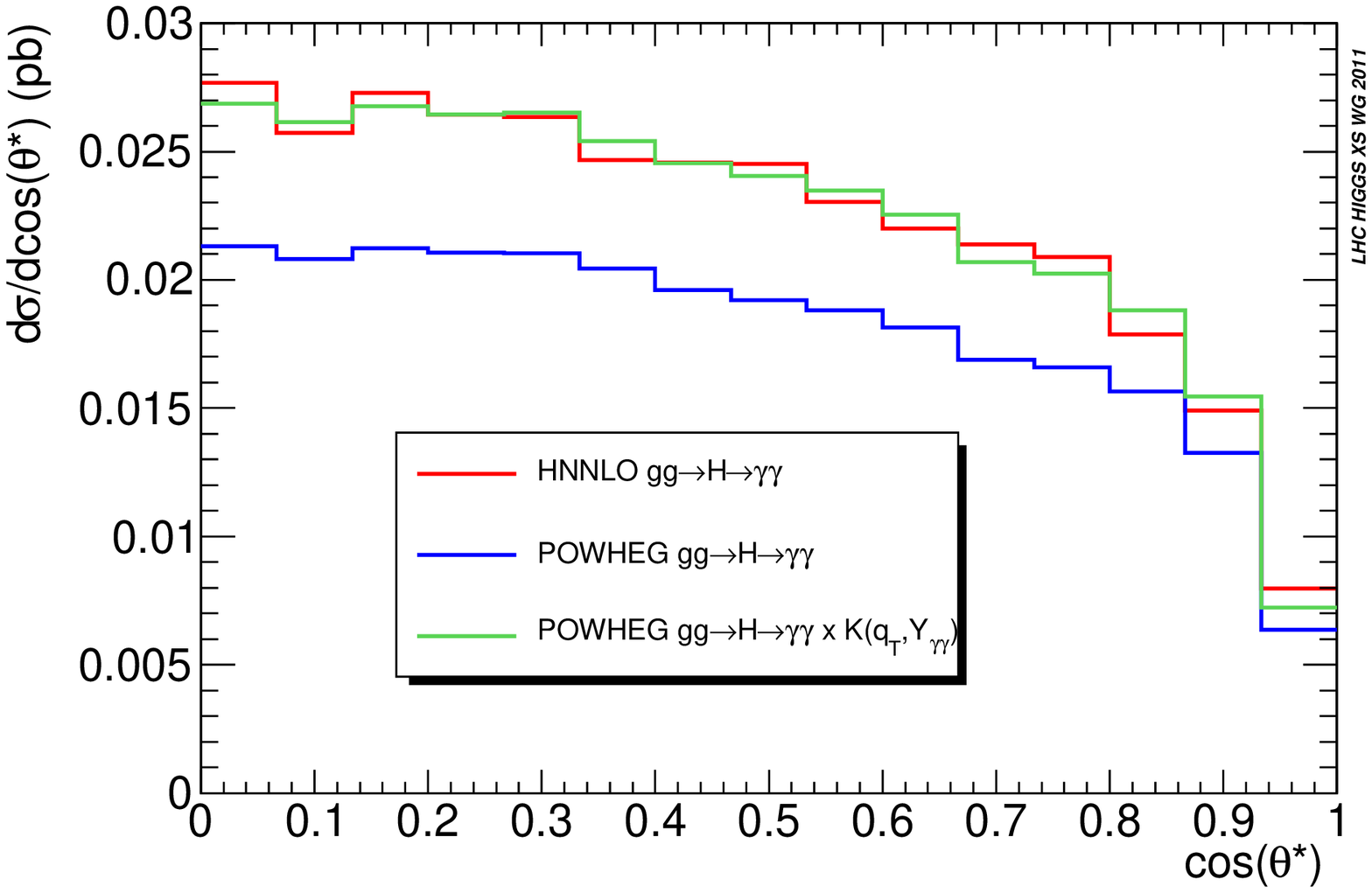}
    \caption{Differential cross sections of the $\Pg\Pg\rightarrow \PH\rightarrow \PGg\PGg$ process for $\MH=120\UGeV$: 
Higgs-boson transverse momentum (top left), Higgs-boson rapidity (top right), and $\cos\theta^{*}$ (bottom) for HNNLO, 
{\sc POWHEG}, and {\sc POWHEG} reweighted with $K(\qTH,\YH)$.}
\label{KfactorHgg2Dapplied}
\end{figure}

\begin{sloppypar}
The $K$-factors $K(\qTH,\YH)$ are computed by applying the 'Loose' kinematical criteria with $E_{\mathrm{T},\PGg 1}>20\UGeV$ 
and $E_{\mathrm{T}, \PGg 2}>20\UGeV$.
An isolation criterion $\sum E_{\mathrm{T}}<5\UGeV$ in a cone
$\Delta R<0.3$ around the photons is applied at parton level in {\sc HNNLO}, while $\sum E_{\mathrm{T}}<7\UGeV$ is used at 
generator level in {\sc POWHEG}. The $K$-factors have been computed by bins of $4\UGeV$ in $\qTH$ and $0.25$ in $\YH$. 
Since the lowest $\qTH$ bins give a divergent cross section at fixed order, they have been merged in 
order to yield a constant $K$-factor in the range $0<\qTH<20\UGeV$ (the reweighted {\sc POWHEG} spectrum profits 
therefore from a leading-log shape resulting from the {\sc PYTHIA} parton shower in this range). Contiguous bins
in the ($\qTH,\YH$) plane are then merged together to smooth out statistical fluctuations 
(they could also be fitted with smooth functions). The $K$-factors thus obtained by this procedure for the 
Higgs-boson 
masses $\MH=110,120,130,140 \UGeV$ are given in \refA{app:2dKfactorsSignalAndBackground} (\refT{tab2dHNNLO}).
The differential cross-section distributions for {\sc HNNLO}, {\sc POWHEG}, and  {\sc POWHEG} after the application
of the $K$-factors are shown in \refF{KfactorHgg2Dapplied}. 
The need for such $K$-factors in the low-$\qTH$, central $\YH$ region is noticeable.
\end{sloppypar}


As expected, the use of the 2D $K$-factors is found to accurately reproduce the transverse momentum and the rapidity of the 
Higgs boson (see \refF{KfactorHgg2Dapplied}) to within $5\%$. It also accurately reproduces angular variables such as 
$\cos\theta^{*}$ to the same level of precision.

\subsubsection{Gluon-fusion signal and background interference}

The $\PGg\PGg$ decay channel is affected by destructive interference between the 
Higgs-boson gluon-fusion production process and the Standard Model continuum $\Pg\Pg\rightarrow\PGg\PGg$
'box' process, which constitutes an irreducible background.  This interference has been calculated at the two-loop
level by Dixon and Siu~\cite{Dixon:2003yb}, and the interference factor $\delta$ can be obtained as a
function of the angle $\theta^*$, where $\theta^*$ is the angle between
one of the photons and the beamline in the centre-of-mass frame of the Higgs boson, the distribution
of which is subject to experimental acceptance criteria, and $y$ is the photon rapidity:  
\begin{equation}
\theta^* = \arccos\Bigl(\tanh\frac{y(\PGg_1)-y(\PGg_2)}{2}\Bigr).
\label{eq:costhetastar}
\end{equation}

\begin{figure}
\centerline{
    \includegraphics[height=6cm]{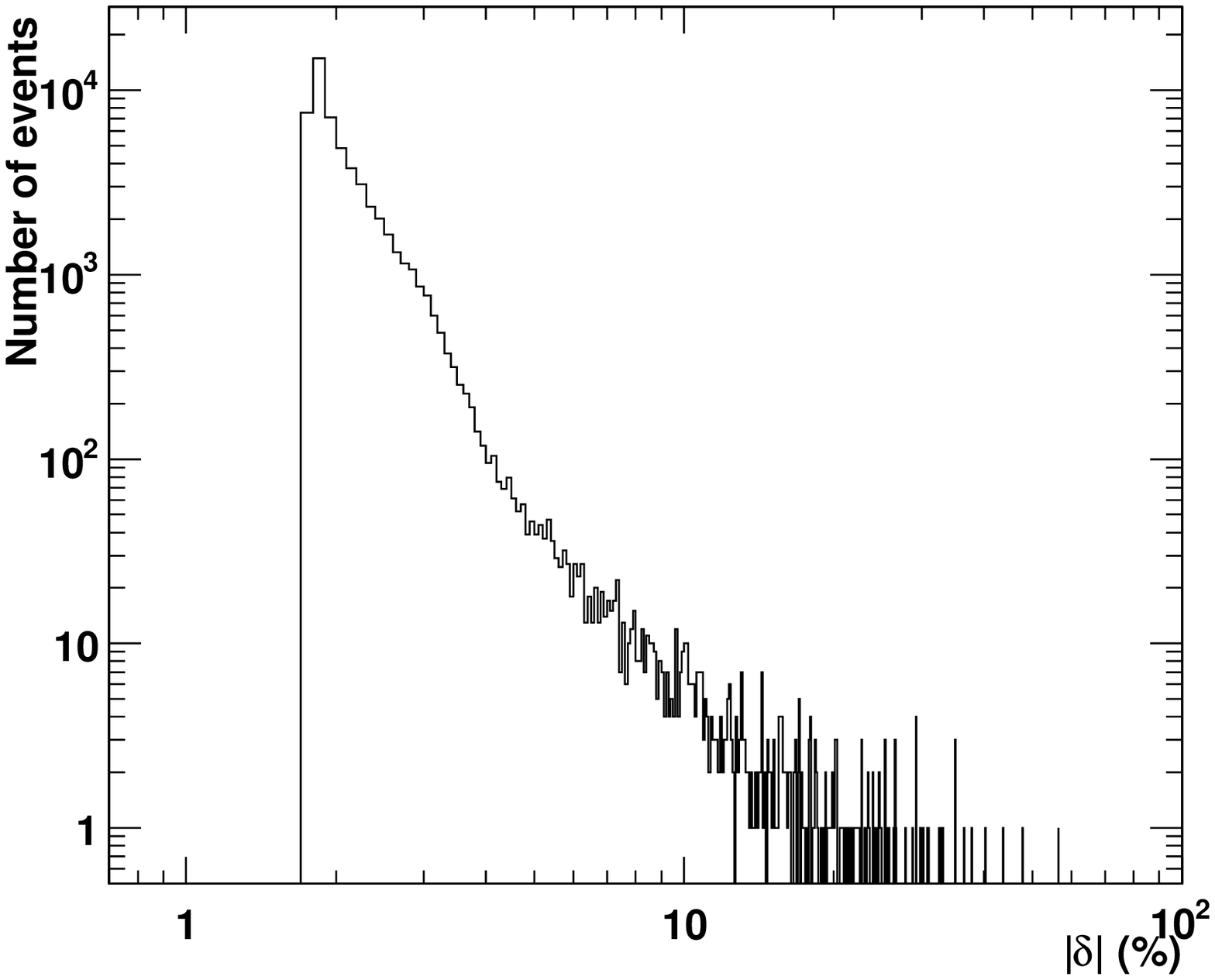}
    \includegraphics[height=6cm]{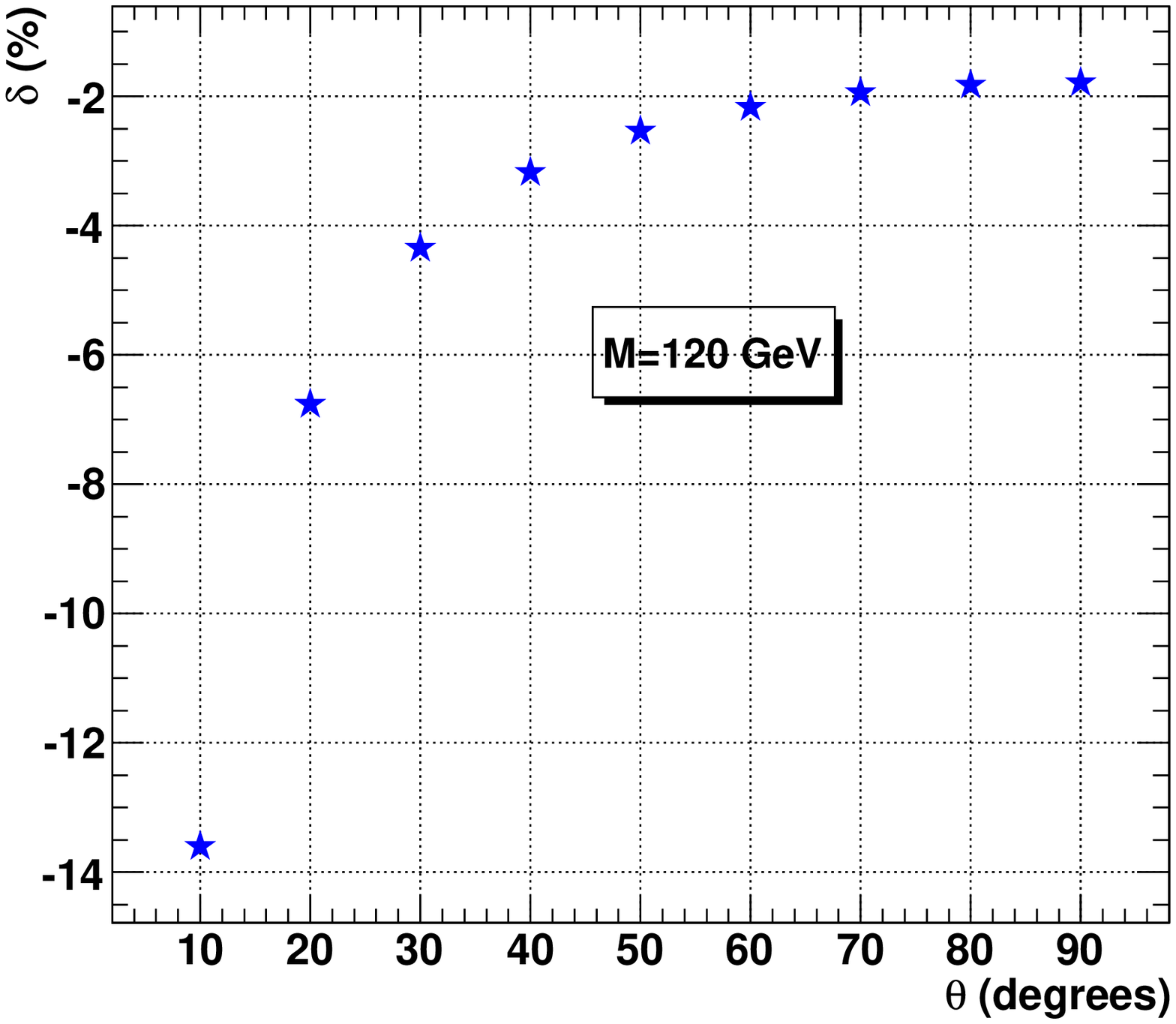}
}
    \caption{Absolute value of the destructive interference factor $\delta$ (left) and its value as
a function of $\theta^*$, for $\MH=120\UGeV$ using the 'ATLAS' acceptance criteria.}
\label{HggInterferenceDelta}
\end{figure}

Average values of $\delta$ are shown in \refT{tab:HggInterference}, where the values of 
$\theta^*$ have been obtained subject to the 'ATLAS' acceptance criteria defined at the beginning
of this section. 
\begin{table}\small
\renewcommand{\arraystretch}{1.2}
\setlength{\arraycolsep}{1.2ex}
\caption{Average values of the destructive interference factor $\delta$ as a function
of Higgs-boson mass, for the 'ATLAS' acceptance criteria.}
\label{tab:HggInterference}
$\begin{array}{cccccccccccc}
\hline
 \text{$\MH$[GeV]}  & 100 & 105 & 110 & 115 & 120 &125 & 130 & 135 & 140 & 145 & 150 \\
\hline
\delta [\%] & -3.16  & -2.83 & -2.59 & -2.42 & -2.31 & -2.28 & -2.36 &-2.54 &-2.87 & -3.40 & -4.33 \\
\hline 
\end{array}$
\end{table}
These are at the level of a few per cent, reaching a minimum of $2.28\%$ for 
$\MH=125\UGeV$ and increasing to $3.16\%$ and $4.33\%$, respectively, for $\MH=100\UGeV$ and
$\MH=150\UGeV$.  However, for very low values of $\theta^*$, $\delta$ can reach far higher
values, as much as $15\%$ or more. \refF{HggInterferenceDelta} shows, for the 'ATLAS' acceptance 
criteria and $\MH=120\UGeV$,
the distribution of $\delta$ values and $\delta$ as a function of $\theta^*$.

At the beginning of this section it was shown that the $\cos\theta^*$ distribution was relatively insensitive
to the $\pT$-reweighting. This means that it probably does not matter which (the $\pT$-reweighting or
the interference correction) is performed first.
It has been suggested~\cite{LanceDixon} that the two steps should be 
performed in both orders, and the difference, for example in the two distributions of the cumulative event weights,
should be taken as a theoretical systematic uncertainty.

In addition, the calculation performed in \Bref{Dixon:2003yb} takes only virtual QCD corrections into account,
and the scattering angle used is that of the beam axis.  To check the stability of the result for those cases 
where the  diphoton system sizeably differs from the framework used for the calculation, the interference term 
is recomputed for  signal events with a transverse momentum of the Higgs boson in excess of $20\UGeV$. The overall 
variation of the  term is of the order of $10\%$. This systematic uncertainty on the $O(3\%)$ correction to 
the $\sigma  \times{}$BR can be neglected when considering the QCD scale, PDF, and $\alpha_s$ systematic 
uncertainties on the  overall signal normalisation.


\subsection{Background extraction and modelling}


\subsubsection{Background modelling biases and systematic uncertainties}

The search for the Higgs boson in the diphoton channel relies, for both 
ATLAS~\cite{Collaboration:2011ww,ATLAS-CONF-2011-149} and 
CMS~\cite{CMS-PAS-HIG-11-021}, on analytic models of the background shape. The scope of these models is to have a reasonable 
fit of the background diphoton invariant-mass distribution in the data to allow an accurate estimate of the 
background in the signal region from the side bands. Various models were investigated, a single or double 
exponential parametrisation or polynomial shapes. These models and in particular the simple exponential 
function are of course not adequate to model the background over any diphoton invariant-mass range, but are a 
good approximation in small ranges. Assessing the ability of the functional form to model correctly the 
background over a given mass range should be determined from large Monte Carlo simulation samples of 
irreducible and reducible backgrounds and checked in the data.

To quantify the potential bias of a given functional form the difference between the fitted function to large 
Monte Carlo samples and its shape in a narrow mass window can be used, or a signal-plus-background model can 
be fitted, and the average number of signal events fitted will give an estimate of the bias. Preliminary 
studies of this type using the {\sc Diphox}~\cite{Binoth:1999qq} and {\sc RESBOS}~\cite{Balazs:1997hv,Balazs:1999yf, Balazs:2006cc,Balazs:2007hr, Nadolsky:2007ba } simulations have been carried 
out~\cite{Collaboration:2011ww,ATLAS-CONF-2011-149} showing that a simple exponential function or a second 
order polynomial when fitted over 
a mass range of $100\UGeV$ to $160\UGeV$ can introduce sizable biases, of the order of $10{-}20\%$ of a Standard Model 
signal. Such bias can be reduced either by using a higher-order polynomial as was done by CMS~\cite{CMS-PAS-HIG-11-021} or 
reducing the fitting mass range. In both cases the cost of reducing a potential bias is a reduction in the 
statistical precision of the determination of the background. The choice of optimal functional form,
the constraints on its parameters, and the fitting range can be studied using the available Monte Carlo programs.
Checks can also be made using the data, but it should be noted that their result can be biased both by 
statistical fluctuations in the search region and the potential presence of a signal.  Such checks were 
carried out in ATLAS~\cite{Collaboration:2011ww,ATLAS-CONF-2011-149} using a double exponential model fitted to the data to generate an ensemble 
of pseudo-experiments which were then fitted using a simple exponential. This check showed a fair agreement 
between the possible bias measured in the Monte Carlo and that measured from the two parametrisations.

In ATLAS~\cite{Collaboration:2011ww,ATLAS-CONF-2011-149}, to account for a potential bias in the statistical treatment of the results of this 
channel, a fraction of the signal fitted is allowed to be assigned to a bias in the background modeling. 
%
%

Detailed Monte Carlo based studies are necessary to accurately estimate the potential bias arising from a given 
choice of parametrisation of the background in the diphoton channel. Although such bias can be accounted for in 
the statistical treatment of the analysis results, it is preferable to keep it small relative to the expected number 
of signal events. Checks on the data are also important to further confirm the choice of background model. Depending 
on the kinematic requirements chosen in ATLAS and CMS, the background modeling systematic uncertainty could be 
correlated between experiments.

\subsubsection{Status of background calculations}

The irreducible background to the $\PH\rightarrow \PGg\PGg$ search is composed of prompt photon
pairs from the quark--antiquark annihilation, gluon-fusion, and gluon--(anti)quark scattering processes.  
One or both photons come either directly from the hard process or from parton fragmentation, in which 
a cascade of successive collinear splittings yields a radiated photon.  The contributions from the 
so-called 'direct' components have been calculated at NLO and implemented in the programs
{\sc Diphox}~\cite{Binoth:1999qq}, {\sc gamma2MC}~\cite{Bern:2002jx}, and {\sc MCFM}~\cite{MCFMweb}. In addition, 
{\sc Diphox} contains a complete implementation
at NLO of single- and double-fragmentation contributions. The calculation implemented in the 
{\sc RESBOS}~\cite{Balazs:1997hv,Balazs:1999yf, Balazs:2006cc,Balazs:2007hr, Nadolsky:2007ba }
program has NNLL resummation accuracy and an effective treatment of LO single fragmentation. 

The direct contribution from the gluon-fusion channel, known as the 'box' contribution, is technically at
NNLO at lowest order, and has been calculated~\cite{Dicus:1987fk}, in addition to many of the higher-order corrections at 
N$^3$LO, and implemented in {\sc gamma2MC} and  {\sc RESBOS}.  These corrections are 
quantitatively of equal importance as the direct contributions, due to the significant gluon luminosity 
at the LHC.


Experimental measurements of differential prompt diphoton pair cross sections at both the 
Tevatron~\cite{Aaltonen:2011vk,Abazov:2010ah } and at
the LHC~\cite{Aad:2011mh, Chatrchyan:2011qt} have exhibited largely satisfactory agreement with the ensemble of theoretical
predictions. The exception has been in the so-called 'collinear' regime, corresponding to low values of 
$\Delta\phi^{\PGg\PGg}$ and $M^{\PGg\PGg}$, high values of $\cos\theta^{*}$, and the characteristic 
'shoulder' in the $\pT$ spectrum of the diphoton system. These disagreements were thought to be due to the absence 
of NNLO contributions in the theoretical predictions from either the direct or fragmentation components.
 
Recently, a fully-differential calculation of the direct components at NNLO using the $q_{\mathrm{T}}$ subtraction 
formalism~\cite{Catani:2007vq} has been performed and implemented in the parton-level program 
{\sc 2gammaNNLO}~\cite{Catani:2011qz}. 
Contributions from fragmentation are not included and are formally eliminated by the application of the 
so-called smooth cone isolation criterion due to Frixione et al.~\cite{Frixione:1998jh}:  For a cone of 
radius $r=\sqrt{{\Delta\eta}^2+{\Delta\phi}^2}< R$ around a photon with transverse momentum $\pT^{\PGg}$,
the total amount of partonic transverse energy $E_{\mathrm{T}}$ must be less than  $E^{\mathrm{max}}_{\mathrm{T}}(r)$, where 
\begin{equation}
E^{\mathrm{max}}_{\mathrm{T}}(r)=\epsilon_{\PGg}\pT^{\PGg}\Bigl(\frac{1-\cos r}{1-\cos R}\Bigr)^n.
\label{eq:Frixione}
\end{equation}

\begin{figure}
\centerline{
\includegraphics[width=0.49\textwidth]{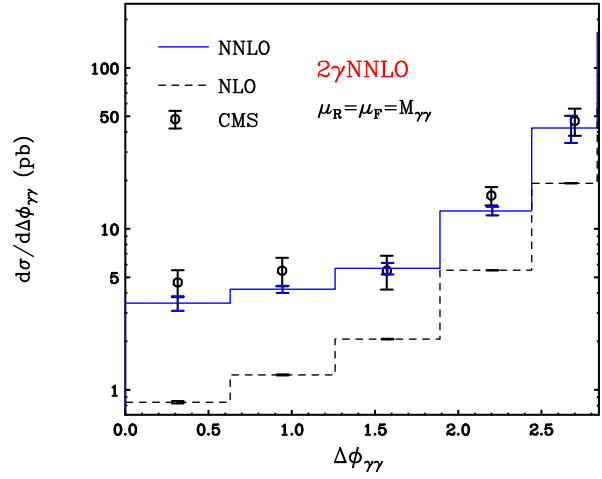}
}
\caption{Differential diphoton cross section as a function of $\Delta\phi_{\PGg\PGg}$ 
at NNLO (blue) and at NLO (dotted black) calculated with a preliminary result from 
the {\sc 2gammaNNLO} program, superimposed
on results from CMS data (points) from 2010~\cite{Chatrchyan:2011qt}.}
\label{2gammaNNLOdsigmadphi}
\end{figure}

In the above definition $\epsilon_{\PGg}$ and $n$ are parameters, with $0<\epsilon_{\PGg}< 1$. 
Figure~\ref{2gammaNNLOdsigmadphi} shows the differential cross section 
$d\sigma/d\Delta\phi_{\PGg\PGg}$ as predicted by {\sc 2gammaNNLO} at both NNLO and NLO (but excluding
the higher-order 'box' diagram corrections), compared
to a recent measurement by the CMS collaboration~\cite{Chatrchyan:2011qt} at $\sqrt{s}=7\UTeV$, 
with acceptance criteria
closely following the 'Loose' selection defined at the beginning of this section. There is
satisfactory agreement between this preliminary theoretical prediction and the CMS data, indicating
that the addition of the direct NNLO contributions alone may correct much of the disagreement, modulo
the fact that the CMS analysis used simple hollow cone isolation requirements and not the Frixione criterion.

\begin{figure}
\includegraphics[width=0.49\textwidth]{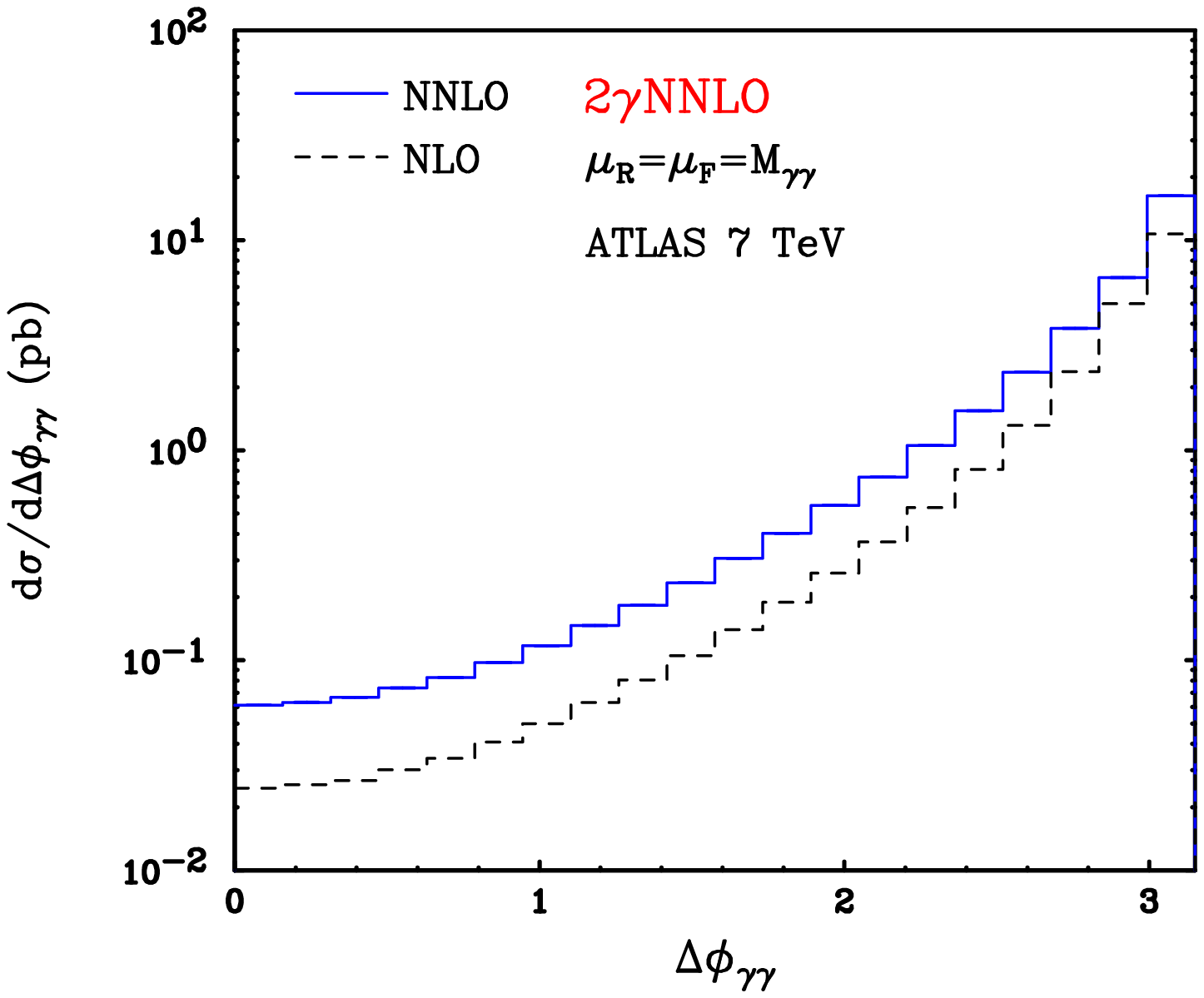}
\includegraphics[width=0.49\textwidth]{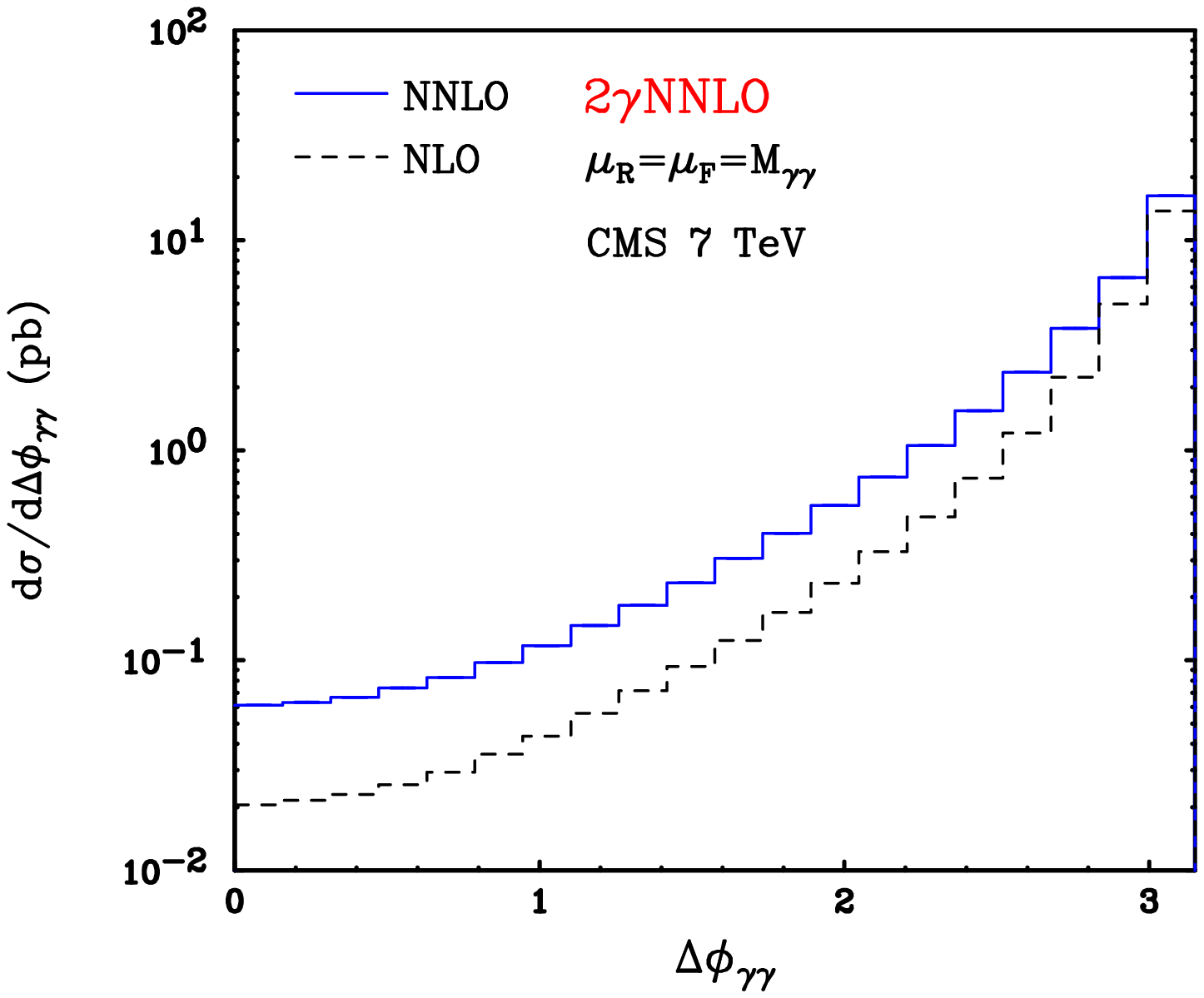}
\caption{Differential diphoton cross section as a function of $\Delta\phi_{\PGg\PGg}$ 
at NNLO (blue) and at NLO (dotted black) calculated with a preliminary result from 
the {\sc 2gammaNNLO} program, for the 'ATLAS' (left) and 'CMS' (right) acceptance criteria.}
\label{2gammaNNLOdsigmadphiHiggs}
\end{figure}

Figure~\ref{2gammaNNLOdsigmadphiHiggs} shows the same differential distribution $d\sigma/d\Delta\phi_{\PGg\PGg}$ 
predicted by {\sc 2gammaNNLO} at both NNLO and NLO, this time for the 'ATLAS' and 'CMS' acceptance criteria defined 
at the beginning of this section.


\subsubsection{Doubly-differential $K$-factors}

Although the ATLAS and CMS $\PH\rightarrow \PGg\PGg$ analyses estimate the background directly from data, it 
is nevertheless useful to benefit from the best possible background estimate from Monte Carlo simulations. 
Furthermore, this is 
needed for meaningful data/Monte Carlo comparisons as well as to train 
classifiers in 
multivariate analyses. For these purposes, we propose in this section 
a differential reweighting of 
parton-shower events to 
NLO calculations. This has been achieved by using parton-shower events obtained with $\PGg\PGg$+jets samples generated with
{\sc Madgraph} \cite{Madgraph} (which contains the Born diagram and up to two supplementary hard jets) 
hadronised with {\sc PYTHIA}~\cite{PYTHIA}, and lowest-order box events generated with {\sc PYTHIA}.
Events have been reweighted to NLO with 
{\sc Diphox}~\cite{Binoth:1999qq} 
(NLO Born and single- and double-fragmentation contributions) and {\sc Gamma2MC}~\cite{Bern:2002jx} 
(NLO box contributions). It should
be noted that the {\sc Madgraph} $\PGg\PGg$+jets process includes the fragmentation contribution at the matrix-element level
as a bremsstrahlung contribution, while {\sc Diphox} includes the full treatment of the fragmentation function at NLO. This 
study should be repeated with {\sc 2gammaNNLO}.

\begin{figure}
    \includegraphics[height=5.3cm]{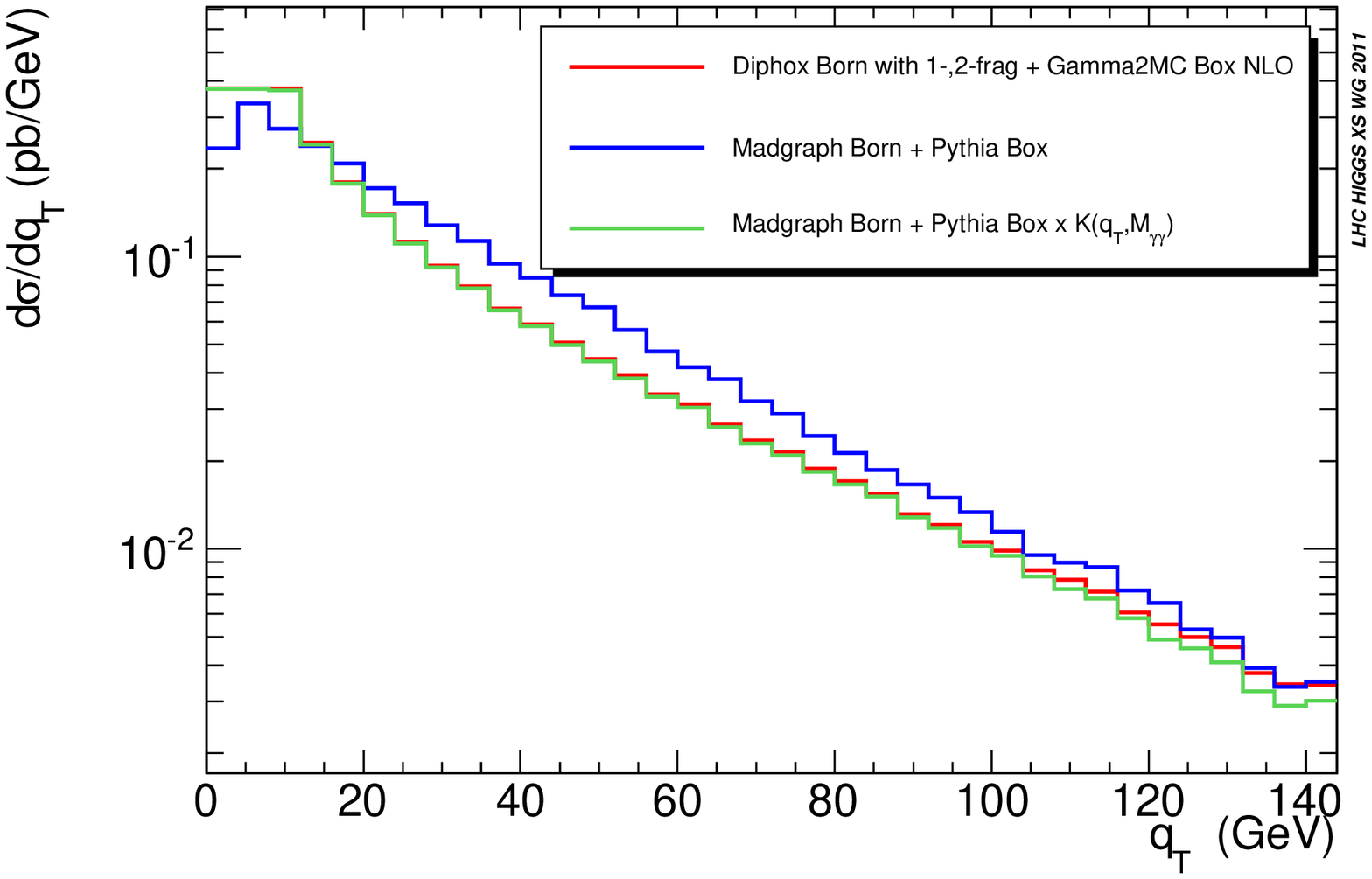}
    \includegraphics[height=5.3cm]{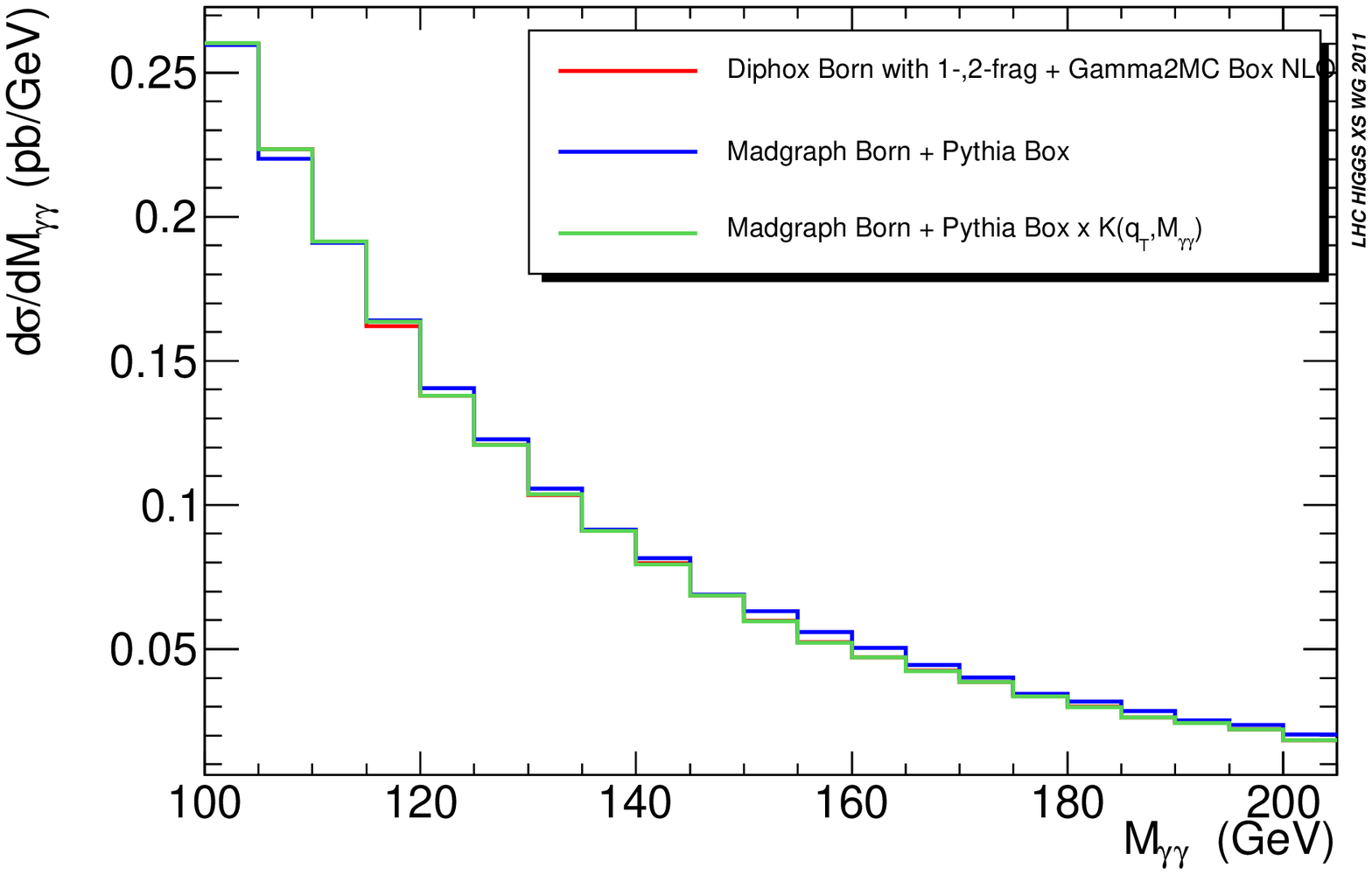}\\
    \includegraphics[height=5.3cm]{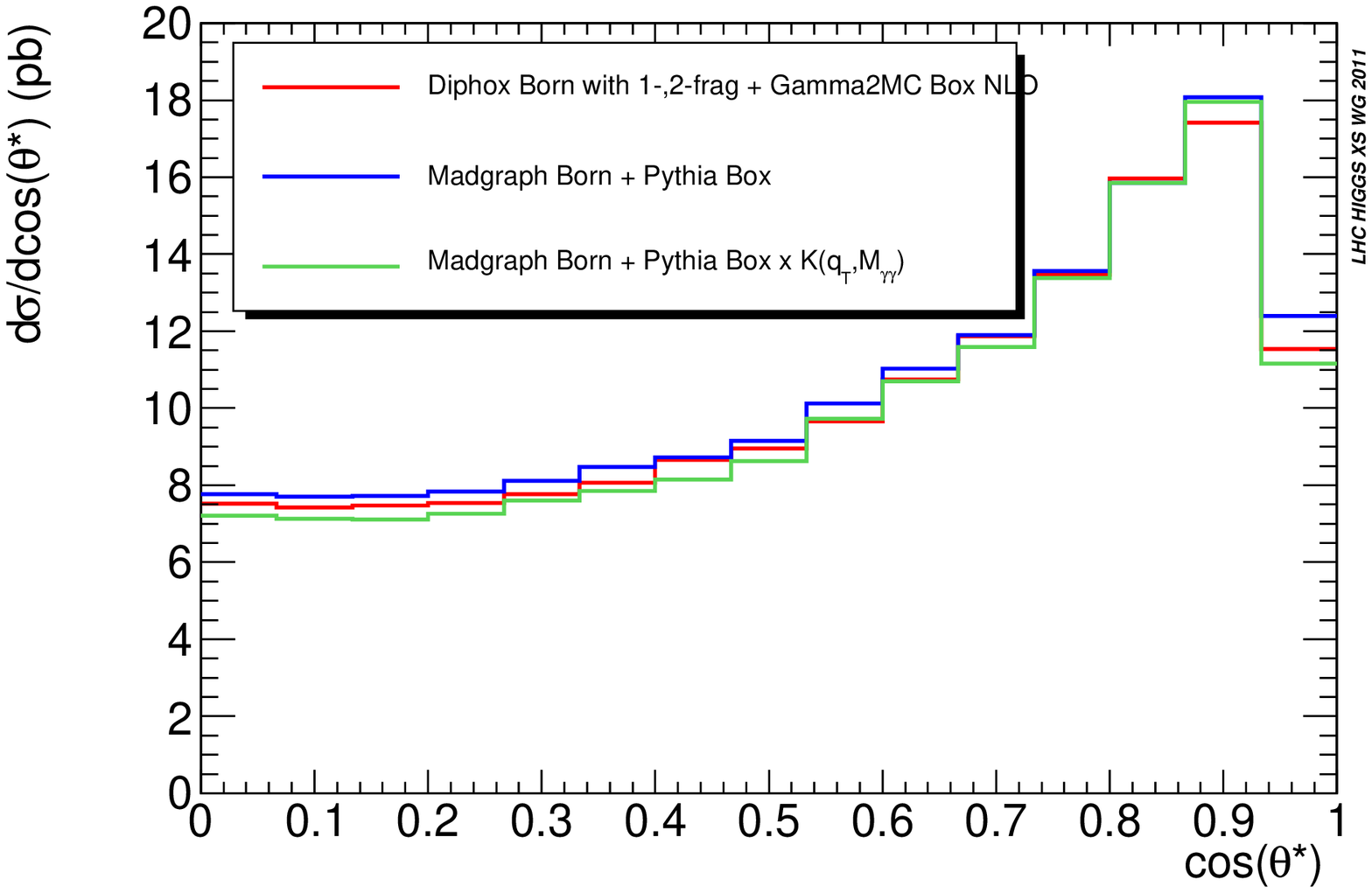}
 \caption{Differential cross sections for 
diphoton production from the sum of the quark--antiquark annihilation, gluon-fusion, and gluon--(anti)quark 
scattering processes:  Diphoton transverse momentum (top left), diphoton invariant mass (top right), and 
$\cos\theta^{*}$ (bottom) for a combination of {\sc Diphox} and {\sc Gamma2MC}, a combination of {\sc Madgraph} 
and {\sc PYTHIA}, and the latter combination reweighted with $K(M_{\PGg\PGg},\qTH)$.}
\label{KfactorPromptPrompt2Dapplied}
\end{figure}

In order to reproduce most of the kinematic features of the NLO processes, it has been found that it is sufficient to 
perform a 2D reweighting with a $K$-factor $K(q_{\mathrm{T},\PGg\PGg}$, $M_{\PGg\PGg}$), where 
$q_{\mathrm{T},\PGg\PGg}$ is the 
transverse momentum of the diphoton system and $M_{\PGg\PGg}$ its invariant mass \cite{TheseNChanon}. The $K$-factors 
$K(q_{\mathrm{T},\PGg\PGg},M_{\PGg\PGg})$ are computed by applying the 'Loose' kinematical cuts with 
$E_{\mathrm{T},\PGg 1}>20\UGeV$ and 
$E_{\mathrm{T}, \PGg 2}>20\UGeV$.
An isolation criterion $\sum E_{\mathrm{T}}<5\UGeV$ in
a cone $\Delta R<0.3$ 
around the photons is applied at parton level and  $\sum E_{\mathrm{T}}<7\UGeV$ at generator level. The $K$-factors
have been 
computed for bins of $4\UGeV$ in $q_{\mathrm{T},\PGg\PGg}$ and $5\UGeV$ in  $M_{\PGg\PGg}$. Contiguous bins in the
($q_{\mathrm{T},\PGg\PGg}$, $M_{\PGg\PGg}$) plane are then merged together to smooth out statistical fluctuations 
(they could be alternatively fitted with smooth functions). The $K$-factors obtained by this procedure are shown in
\refA{app:2dKfactorsSignalAndBackground} (\refT{tab2dPrompt}).
The differential cross-section distributions for a combination of {\sc Diphox} and {\sc Gamma2MC}, and a combination
of {\sc Madgraph} and {\sc PYTHIA} after the application of the $K$-factors are shown in \refF{KfactorPromptPrompt2Dapplied}.
It is interesting to note that the supplementary hard jets in the
{\sc Madgraph} $\PGg\PGg$+jets samples allows
the population of the high-$q_{\mathrm{T},\PGg\PGg}$ and high-$M_{\PGg\PGg}$ regions,
which would have been forbidden by the LO kinematics of the {\sc PYTHIA} Born samples
had they been used.


As expected, the 2D $K$-factors are found to
accurately reproduce the transverse momentum and the invariant-mass spectra of the
diphoton system (see \refF{KfactorPromptPrompt2Dapplied}) in the region where the $\PH\rightarrow \PGg\PGg$ 
searches are performed ($M_{\PGg\PGg}>100\UGeV$).
They also accurately reproduce angular variables such as $\cos\theta^{*}$.

\clearpage

\newpage
\section{WW$^{*}$ decay mode\footnote{%
    S. Diglio, B.~Di Micco, R.~Di~Nardo, A.~Farilla, B.~Mellado and F.~Petrucci (eds.).}}
\subsection{Introduction}

For Higgs-boson masses $\MH\gsim 135\GeV$,
the Higgs boson decays  mainly into a $\PW\PW^{*}$ pair.
The W's decay mainly hadronically,
but this decay topology is experimentally difficult to exploit 
due to the high cross section of multi-jet processes in
$\Pp\Pp$ collisions.  If one of the W's decays to $\Pl\PGn$,
where $\Pl$ indicates an electron or muon, and the other to two quarks,
the background from $\Pp\Pp \to \PW+{}$jets with $\PW \to \Pl\PGn$,
becomes dominant. For high values of the Higgs mass, a strong cut on the
neutrino transverse momentum (measured as missing energy, $E_{\mathrm{T}}^{\rm miss}$, in the detector)
can be used to reject the $\PW+{}$jets
background.
For low values of the Higgs mass only the fully leptonic decay channel
of the W's ($\PW\PW \to \Pl\PGn \Pl\PGn$) can be used for Higgs search with high
sensitivity.

The main background in this channel is $\Pp\Pp \to \PQt \PAQt \to
\PW\PQb \PW \PQb \to \Pl\PGn \Pl\PGn \PQb \PQb$. This background is
characterised by the presence of high missing $E_{\mathrm{T}}$ from the two
neutrinos and high jet multiplicity from the $\PQb$ quarks.

In order to maximise the sensitivity, the analysis is performed in jet
categories, that means the whole  sample is divided in $0$, $1$, and $>1$ jets subsamples.
The use of jet categories is common to other channels like $\PH
\to \PGg \PGg$ and $\PH \to \PGt \PGt$. In the first two subsections  we will discuss the
concrete implementation of the jet categorisation in the $\PH \to \PW\PW $
analysis.
The follwoing sections discuss the treatment of the irreducible SM $\PW\PW$ background.

\subsection{Jet bin categorisation and uncertainties}

The main Higgs  production process at LHC is gluon fusion $\Pg\Pg \to
\PH$ whose inclusive cross section is known at NNLO QCD $+$ NLO EW $+$
higher-order improvements~\cite{Dittmaier:2011ti},
with uncertainties from residual QCD scale dependences and from PDF errors
of $6{-}10\%$  and $8{-}10\%$ for the LHC at $7\UTeV$ centre-of-mass energy.
The $\Pg\Pg \to \PH$  process is characterised by large contributions from extra gluon
 emission. The amount of extra jets produced in the process has been computed using the {\sc POWHEG} MC program
interfaced to {\sc PYTHIA} for the showering with the following results:  $44\%$ of the events have more than one
jet and $13\%$ have
 more than two jets. The jet algorithm used is the anti-$k_{\mathrm{T}}$ 
with a cone $\Delta R = 0.4$, and a cut on the transverse momentum
of the jet of $25\UGeV$ and a pseudorapidity cut of 4.5 are imposed. 
 The CTEQ6.6 NLO PDF set has been used for the computation.
Scale uncertainties on the exclusive jet cross sections result to be smaller than
the scale uncertainty on the inclusive cross sections. This points to
several studies, also reported in this report, showing that the
conventional scale uncertainty variation underestimate the exclusive
jet bin uncertainty. In order to compute more realistic uncertainties a procedure has been set up that 
furnishes more reasonable values. As described in the following, it consists in evaluating the scale
uncertainty on the inclusive multi-jet cross sections, $\sigma_{\ge 0}$, $\sigma_{\ge 1}$, and
$\sigma_{\ge 2}$,  and propagating them uncorrelated to
the exclusive jet bins. These uncertainties produce large correlations
among several channels. In the $\PW\PW$ case, three different channels are
analysed: $\Pep \Pem$, $\Pepm \PGmpm$, and
$\PGmp \PGmm$. The jet bin uncertainties have to be considered fully
correlated among them.

Analysis selection criteria for the several channels are different,
and the scale uncertainties on the inclusive jet cross section can be
different according to the several selections that are considered. 
In order to decouple the largest contribution to the jet counting from
the specific analysis criteria the signal yield in a given jet bin is
defined as follows,
\begin{equation}
N_{0j} = \sigma_{\ge 0}f_{0}A_{0}, \quad N_{1j} =
\sigma_{\ge 0}f_{1} A_{1}, \quad  
N_{2j} = \sigma_{\ge 0}f_{2}A_{2},
\end{equation}
where $f_{0}$, $f_{1}$ and $f_{2}$ are defined as
\begin{equation}
f_0 = \frac{\sigma_{\ge 0} - \sigma_{\ge 1}}{\sigma_{\ge 0}}, \quad f_1
= \frac{\sigma_{\ge 1}-\sigma_{\ge 2}}{\sigma_{\ge 0}}, \quad 
f_2 = \frac{\sigma_{\ge 2}}{\sigma_{\ge 0}}.
\end{equation}
The central values of $f_0$, $f_1$, and $f_2$ are computed using the
{\sc POWHEG} MC after the reweighting of the Higgs $\pT$  to {\sc HqT}~2.0 and the full reconstruction chain of
each experiment, while for the purpose of error propagation they are
evaluated using {\sc HNNLO}. In order to properly correlate the
uncertainties among several measurements and between the ATLAS and CMS
collaborations,
a common set of cuts has been defined in order to compute the
uncertainties on the $\sigma_{\ge n \mathrm{jets}}$ and the $f_{i}$ values used
in the error propagation. The cuts are shown in 
\refT{tab:cuts_agreed}.

\begin{table}
\caption{Common cuts for the $\PH \to \PW\PW^{*}$ analysis for
  the evaluation of jet bin uncertainties on the signal yield. The
  $p_{\rm Tl1}$,  $p_{\rm Tl2}$ are the cuts on the  lepton $p_{\rm
    T}$, $E_{\mathrm{T}}^{\rm miss}$ is the total $\pT$ of the neutrinos,
$m_{\Pl\Pl}$ is the invariant mass of the two leptons, and $p_{\rm T
  jet cut}$ is the cut imposed on the jet $\pT$ for the jet counting.} \label{tab:cuts_agreed}
\centerline{
\begin{tabular}{c}
\hline
$p_{\rm Tl1}$, $p_{\rm Tl2}$ > 20 \UGeV{} , $|\eta_1|$, $|\eta_2| < 2.5$\\
$E_{\mathrm{T}}^{\rm miss} > 30$ \UGeV{} \\
$m_{\Pl\Pl} > 12$ \UGeV{} \\
$p_{\rm T jet cut} < 30$ \UGeV{} with $|\eta| < 3$ and using anti-$k_{\mathrm{T}}$
algorithm with cone size $0.5$\\
\hline
\end{tabular}
}
\end{table}

The scale uncertainties on $\sigma_{\ge 0}$, $\sigma_{\ge 1}$, and
$\sigma_{\ge 2}$ have been computed using {\sc HNNLO} and are shown in 
\refT{tab:scale uncertainties}, the renormalisation and factorisation scales were varied by a factor two
up and down in order to get the error. The errors shown are the maximum
variation obtained in the Higgs mass range $130 < \MH < 300$ \UGeV{}.

\begin{table}
\caption{Inclusive jet uncertainties used for exclusive  jet bin uncertainties
  estimation.} \label{tab:scale uncertainties}
\centerline{  
\begin{tabular}{ccc}
\hline
$\Delta \sigma_{\ge 0}$ & $\Delta \sigma_{\ge 1}$ & $\Delta \sigma_{\ge 2}$ \\
\hline
$10 \%$ & $20 \%$ & $70 \%$ \\
\hline
\end{tabular}
}
\end{table}

The $A_{i}$ coefficients are the acceptances specific for each analysis, they 
are the acceptances due to  the further cuts applied in addition to the cuts 
shown in \refT{tab:cuts_agreed}. The scale uncertainty on $A_{i}$ has been evaluated
studying the fractional variation of events passing all analysis cuts
over the event passing the selection in \refT{tab:cuts_agreed}.
The effects observed are small and have been evaluated only for the
$\PGm\PGm$ channel and two Higgs mass points. 
The cuts used by the ATLAS collaboration for the 0-jet and 1-jet
analysis are shown in \refT{tab:acceptance_cuts}. The analysis is
divided in a low mass analysis, applied for $\MH < 200$ \UGeV{},  and an
high-mass analysis for $\MH > 200$ \UGeV{}. The two analyses differ in
the treatment of the $\PW\PW$ background; details are described in
\refS{sec:background}.

\begin{table}
\caption{Analysis cuts used in the $\PH \to {\PW\PW}^{*} \to \Pl\PGn
  \Pl\PGn$ analysis in the ATLAS experiment, the $\pT^{\Pl\Pl}$
    variable is the sum of the $\pT$'s of the two letpons, the
    $\PZ_{\PGt \PGt}$ veto is a veto applied to the invariant mass of
    the $\PGt \PGt$ system reconstructed in the collinear
    approximation (the neutrino from $\PGt$ decay is supposed to fly in
  the same direction of the charged lepton) of $|m_{\PGt \PGt} -
  \MZ| > 15$ \UGeV{}.} \label{tab:acceptance_cuts}
\centerline{
\begin{tabular}{c|c}
\hline
$\MH <   200 \UGeV$ & $\MH > 200 \UGeV$ \\
\hline
$E_{\mathrm{T}}^{\rm  miss rel} > $ 40 \UGeV{} & $E_{\mathrm{T}}^{\rm miss rel} >40$
\UGeV{}\\
$m_{\Pl\Pl}$ < 50 \UGeV{} & $m_{\Pl\Pl} < 150$ \UGeV{} \\
$\Delta \phi_{\Pl\Pl} < 1.8$ & no $\Delta \phi_{\Pl\Pl}$ cut \\
$|m_{\Pl\Pl} - \MZ| > 15 \UGeV$ & $|m_{\Pl\Pl} - \MZ| > 15 \UGeV$ \\ 
\hline
\multicolumn{2}{l}{0-jet: \quad $\pT^{\Pl\Pl} > 30$ \UGeV{}} \\
\hline
\multicolumn{2}{l}{1-jet: \quad $\pT^{\rm l1} + \pT^{\rm l2} + E_{\mathrm{T}}^{\rm miss} + \pT^{\rm  jet} < 30$ \UGeV{}} \\
\multicolumn{2}{c}{Z$_{\PGt \PGt}$ veto} \\
\hline
\end{tabular}
}
\end{table}
The fractional variation of the $A_0$ and $A_1$ varying the
normalisation and factorisation scale at $\muR = \muF = 2\MH$
and $\muR = \muF = 1/2\MH$ with respect to the central value of
  $\muR = \muF = \MH$ are shown in 
  \refT{tab:acceptances_scale} both for the low-mass and the high-mass analyses.
\begin{table}
\caption{Acceptance variation for the 0 and 1-jet analysis for two
  Higgs mass points. The acceptance are evaluated for the low-mass
  analysis for $\MH=130\UGeV$ and the high-mass  analysis for $\MH = 200\UGeV$.} 
\label{tab:acceptances_scale}
\centerline{
\begin{tabular}{cccc}
\hline
& & $\MH = 130 \UGeV$ & $\MH = 200 \UGeV$ \\
\hline
$A_{0}$ & $\mu_{\mathrm{R}} = \mu_{\mathrm{F}} = 2\MH$ & $- 1$ per mil  & $+ 3\%$\\
$A_{0}$ & $\mu_{\mathrm{R}} = \mu_{\mathrm{F}} = \MH/2$ & $+ 5$ per mil  &  $-1.3\%$\\
\hline
$A_{1}$ & $\mu_{\mathrm{R}} = \mu_{\mathrm{F}} = 2\MH$ & $- 6$ per mil  & $+ 4\%$\\
$A_{1}$ & $\mu_{\mathrm{R}} = \mu_{\mathrm{F}} = \MH/2$ & $+ 4\%$  &  $+3\%$\\
\hline
\end{tabular}
}
\end{table}
The uncertainties obtained are much smaller than the scale
uncertainties on the inclusive jet cross sections.

\subsection{Discussion about the 2-jet bin}
\begin{figure}
\begin{center}
\includegraphics[width=0.7\textwidth]{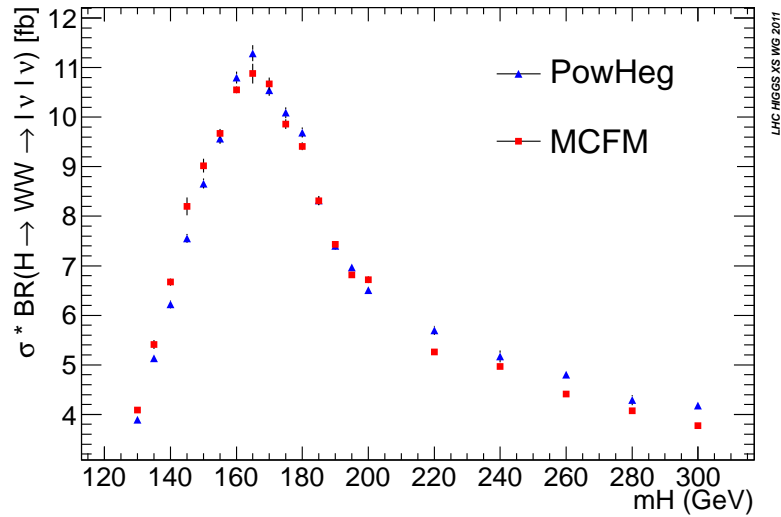}
\end{center}
\caption{The $\Pp\Pp \to \PH+2$ jets cross section evaluated using typical charged leptons
  and jets selection cuts of the $\PW\PW+2$ jets analysis.} \label{fig:plot_POWHEG_MCFM_comparison}
\end{figure}

The error on the 2-jet inclusive bin has been computed using the {\sc HNNLO}
program. It predicts a quite large error on the 2-jet cross section
because the 2-jet bin is computed at LO. The inlcusive 2-jet cross section
$\sigma_{\ge 2}$ is about $30\%$ of the inclusive 1-jet cross section
$\sigma_{\ge 1}$. With further cuts
that are  usually used in the 2-jet channel  this can reach   very small
values. Typically the 2-jet bin is used in vector-boson fusion (VBF)
and VH analyses where the analyses are tuned to maximise the cross sections in the 
VBF and HV channels.
Recently an NLO calculation of the $\Pg\Pg \to \PH + 2$ jets process became
available and implemented in {\sc MCFM}. The scale uncertainties of these
results are smaller than the uncertainty from {\sc HNNLO} that
computes the $\PH+2$~jets cross section at LO.
In order to use the smaller error from {\sc MCFM} we need to compare the
cross section computed with {\sc MCFM} with the respective cross section computed
with the {\sc POWHEG} Monte Carlo after the reweighting to {\sc HqT}~2.0. In fact the signal 
yield is computed using the {\sc POWHEG} MC that 
evaluates the 2-jet cross section only through the
parton shower. It is therefore important to compare the central value
of the 2-jet cross section from {\sc POWHEG} with Higgs $\pT$ reweighting
and with the {\sc MCFM}~v6.0 computation.
The comparison is performed applying the following cuts,
\begin{equation}
\pT^{\rm l1} > 20 \UGeV, \quad \pT^{\rm l2} > 10 \UGeV, \quad |\eta_{\rm l1}| < 2.5,
\quad |\eta_{\rm l2}| < 2.5,
\end{equation}
and at least two anti-$k_{\mathrm{T}}$ jets with $\pT > 25\UGeV$ with a cone
$\Delta R = 0.4$.
The computation has been performed using the CTEQ6.6 PDF set and the
renormalisation and factorisation scales fixed at the value of the
Higgs mass. The total {\sc POWHEG} cross section has been normalised at the
NNLO value. The results are shown in 
\refT{tab:2_jet_MCFM_POWHEG} and plotted in \refF{fig:plot_POWHEG_MCFM_comparison}.
\begin{table}
\caption{Comparison between the {\sc POWHEG} and {\sc MCFM} 2 jet inclusive cross
  section, the error shown are the statistical errors, the ratio of the
  two cross sections doesn't exceed $11\%$ in the whole $\MH$ range.} \label{tab:2_jet_MCFM_POWHEG}
\centerline{
\begin{tabular}{crrc}
\hline
     $\MH$ [GeV]     &        {\sc MCFM}  [fb]       & {\sc POWHEG} [fb] & Ratio: {\sc POWHEG}/{\sc MCFM} \\ 
\hline
$130$& $ 4.09 \pm 0.05 $ & $ 3.89 \pm 0.05 $ &  $  0.95 \pm 0.02 $ \\ 
$135$& $ 5.41 \pm 0.09 $ & $ 5.13 \pm 0.06 $ &  $  0.95 \pm 0.02 $ \\ 
$140$& $ 6.67 \pm 0.07 $ & $ 6.22 \pm 0.08 $ &  $  0.93 \pm 0.01 $ \\ 
$145$& $ 8.20 \pm 0.18 $ & $ 7.55 \pm 0.09 $ &  $  0.92 \pm 0.02 $ \\ 
$150$& $ 9.02 \pm 0.14 $ & $ 8.66 \pm 0.10 $ &  $  0.96 \pm 0.02 $ \\ 
$155$& $ 9.67 \pm 0.09 $ & $ 9.57 \pm 0.11 $ &  $  0.99 \pm 0.01 $ \\ 
$160$& $10.55 \pm 0.07 $ & $ 10.80 \pm 0.12$  & $   1.02 \pm 0.01$  \\ 
$165$& $10.88 \pm 0.20 $ & $ 11.29 \pm 0.16$  & $   1.04 \pm 0.02$  \\ 
$170$& $10.67 \pm 0.13 $ & $ 10.54 \pm 0.11$  & $   0.99 \pm 0.02$  \\ 
$175$& $ 9.86 \pm 0.10 $ & $ 10.09 \pm 0.11$  & $   1.02 \pm 0.01$  \\ 
$180$& $ 9.41 \pm 0.08 $ & $ 9.69 \pm 0.10 $ &  $  1.03 \pm 0.01 $ \\ 
$185$& $ 8.31 \pm 0.09 $ & $ 8.32 \pm 0.09 $ &  $  1.00 \pm 0.01 $ \\ 
$190$& $ 7.43 \pm 0.05 $ & $ 7.40 \pm 0.08 $ &  $  1.00 \pm 0.01 $ \\ 
$195$& $ 6.82 \pm 0.06 $ & $ 6.97 \pm 0.07 $ &  $  1.02 \pm 0.01 $ \\ 
$200$& $ 6.72 \pm 0.07 $ & $ 6.51 \pm 0.07 $ &  $  0.97 \pm 0.01 $ \\ 
$220$& $ 5.26 \pm 0.05 $ & $ 5.70 \pm 0.09 $ &  $  1.08 \pm 0.02 $ \\ 
$240$& $ 4.97 \pm 0.04 $ & $ 5.17 \pm 0.12 $ &  $  1.04 \pm 0.02 $ \\ 
$260$& $ 4.41 \pm 0.02 $ & $ 4.80 \pm 0.07 $ &  $  1.09 \pm 0.01 $ \\ 
$280$& $ 4.07 \pm 0.04 $ & $ 4.29 \pm 0.10 $ &  $  1.05 \pm 0.02 $ \\ 
$300$& $ 3.77 \pm 0.02 $ & $ 4.18 \pm 0.06 $ &  $  1.11 \pm 0.02 $ \\ 
\hline
\end{tabular}
}
\end{table}

Renormalisation and factorisation scale variations from $\MH/4$ to
2 $\MH$  have been evaluated  and tabulated in \refT{tab:2jets_scale_uncertainties}.
\begin{table}
\caption{The $\Pg\Pg \to \PH +2$ jets cross section evaluated for different
  scale choices.} \label{tab:2jets_scale_uncertainties}
\centerline{
\begin{tabular}{cccc}
\hline
      $\MH$ [GeV]     &    $\xi_{\rm R} = \muR/\MH$       & $\xi_{\rm F} = \muF/\MH$ &  $\sigma$ [fb] \\ 
\hline
      & $ 0.25 $&$ 0.25 $&$  5.17  \pm 0.64 $ \\ 
      & $ 0.25 $&$ 0.5  $&$  4.60  \pm 0.24 $ \\ 
      & $ 0.5  $&$ 0.25 $&$  4.88  \pm 0.11 $ \\ 
      & $ 0.5  $&$ 0.5  $&$  5.14  \pm 0.14 $ \\ 
      & $ 0.5  $&$ 1    $&$  4.98  \pm 0.08 $ \\ 
$130$ & $ 1    $&$ 0.5  $&$  4.42  \pm 0.05 $ \\ 
      & $ 1    $&$ 1    $&$  4.17  \pm 0.06 $ \\ 
      & $ 1    $&$ 2    $&$  4.13  \pm 0.05 $ \\ 
      & $ 2    $&$ 1    $&$  3.41  \pm 0.03 $ \\ 
      & $ 2    $&$ 2    $&$  3.34  \pm 0.05 $ \\ 
\hline
      & $ 0.25 $&$ 0.25 $&$  6.47  \pm 0.36 $ \\ 
      & $ 0.25 $&$ 0.5  $&$ 12.46  \pm 0.38 $ \\ 
      & $ 0.5  $&$ 0.25 $&$ 13.27  \pm 0.32 $ \\ 
      & $ 0.5  $&$ 0.5  $&$ 13.03  \pm 0.30 $ \\ 
      & $ 0.5  $&$ 1    $&$ 12.24  \pm 0.36 $ \\ 
$160$ & $ 1    $&$ 0.5  $&$ 10.77  \pm 0.12 $ \\ 
      & $ 1    $&$ 1    $&$ 10.52  \pm 0.13 $ \\ 
      & $ 1    $&$ 2    $&$ 10.28  \pm 0.07 $ \\ 
      & $ 2    $&$ 1    $&$  8.43  \pm 0.09 $ \\ 
      & $ 2    $&$ 2    $&$  8.35  \pm 0.07 $ \\ 
\hline
      & $ 0.25 $&$ 0.25 $&$  6.43  \pm 0.15 $ \\ 
      & $ 0.25 $&$ 0.5  $&$  6.96  \pm 0.11 $ \\ 
      & $ 0.5  $&$ 0.25 $&$  6.84  \pm 0.11 $ \\ 
      & $ 0.5  $&$ 0.5  $&$  6.70  \pm 0.13 $ \\ 
      & $ 0.5  $&$ 1    $&$  6.66  \pm 0.14 $ \\ 
$220$ & $ 1    $&$ 0.5  $&$  5.85  \pm 0.06 $ \\ 
      & $ 1    $&$ 1    $&$  5.56  \pm 0.06 $ \\ 
      & $ 1    $&$ 2    $&$  5.33  \pm 0.04 $ \\ 
      & $ 2    $&$ 1    $&$  4.48  \pm 0.03 $ \\ 
      & $ 2    $&$ 2    $&$  4.59  \pm 0.06 $ \\ 
\hline
 \end{tabular}
}
 \end{table}
The scale uncertainty is computed by taking the maximum spread in
the cross section obtained spanning the values $\MH/2 < \muR, \muF
< 2 \MH$ but keeping $ 1/2 < \muR/\muF < 2$. The obtained
uncertainties are summarised in \refT{tab:scale_uncertainites}.
\begin{table}
\caption{Scale uncertainties on the $\Pg\Pg \to \PH + 2$ jets cross section
  evaluated using the {\sc MCFM} program.} \label{tab:scale_uncertainites}
\centerline{
\begin{tabular}{cc}
\hline
$\MH$ [GeV] & scale uncertainties \\
\hline
$130$ & $23 \%$ \\
$160$ & $24 \%$ \\
$220$ & $20 \%$ \\
\hline
\end{tabular}
}
\end{table}

\subsection{Backgrounds}
\label{sec:background}
The main backgrounds to the $\PH \to \PW\PW $ channel are the top background,
the Drell--Yan process $\Pp\Pp \to \PZ/\PGg^{*}+{}$jets with $\PZ/\PGg^{*} \to \Plp\Plm$, 
the $\Pp\Pp \to \PW+{}$jets, with $\PW\to \Pl\PGn$, with a fake lepton from the jets, 
 and the irreducible $\Pp\Pp \to \PW^{+} \PW^{-}$ background. 
The reducible backgrounds are estimated with data-driven 
technique slightly different between the two collaborations.
The treatment of the irreducible $\PW\PW$ background is mainly affected by
theoretical uncertainties and will be discussed in the following.

For a relatively light Higgs, the $\PW$ bosons in the $\PW\PW^*$ pair produced
in the Higgs decay have opposite spin orientations, since the Higgs has spin
zero. In the weak decay of the $\PW$, due to the $V-A$ nature of the
interaction, the positively charged lepton is preferably emitted in the
direction of the $\PW^{+}$ spin and the negative letpon in the opposite direction
of the $W^{-}$ spin. Therefore the two charged leptons are emitted close to each
other, and their di-lepton invariant mass $m_{\Pl\Pl}$ is small.
This feature is used in the low-mass analysis to define a signal-free control region through
the cut $m_{\Pl\Pl} > 80$ \UGeV{}. The event yield of the $\PW\PW^{*}$ background
is computed in the control region and extrapolated to the signal
region.
The $\PW\PW$ yield in the signal region is therefore
\begin{equation}
N^{\rm WW 0j}_{\rm S.R.} = \alpha_{\rm 0j} N^{\rm WW 0j}_{\rm C.R.}, \quad
N^{\rm WW 1j}_{\rm S.R.} = \alpha_{\rm 1j} N^{\rm WW 1j}_{\rm C.R.}. 
\end{equation}

The value of $\alpha$ is affected both by theoretical and experimental
errors. Because $\alpha$ is defined using only leptonic quantities, the
experimental error is negligible and the theoretical error is
carefully evaluated in \refS{sec:alpha}.

For $\MH > 200\UGeV$ it is not possible to define a signal-free
control region, therefore the $\PW\PW$ yield needs to be determined directly
from the theoretical expectation. In \refS{sec:acceptances} we
discuss the evaluation of the jet bin uncertainties on the expected
$\PW\PW^*$ yield.

\subsection{Theoretical uncertainties on the extrapolation parameters
  $\alpha$ for the 0j and 1j analyses}
\label{sec:alpha}

The $\PW\PW$ background is estimated for $\MH < 200\UGeV$ using event
counts in a well defined control region (C.R.\ in the following).
The control region is defined using cuts on $m_{\Pl\Pl}$ and $\Delta
\phi_{\Pl\Pl}$ variables. 
In \refT{tab:WW_cuts} we show the cuts used by the ATLAS collaborations  to define the signal and the $\PW\PW$ C.R.\
for the different channels.
\begin{table}
\caption{Definition of the signal and control regions.} \label{tab:WW_cuts}
\centerline{
\begin{tabular}{cc}
\hline
$\MH$ [GeV] & $110-200$ \\
\hline
 ATLAS    & $\Delta \phi_{\Pl\Pl} < 1.8$ \\
  S.R.    & $m_{\Pl\Pl} < 50 \UGeV $  \\
\hline
  ATLAS   & $m_{\Pl\Pl} > \MZ + 15 \UGeV (\Pe\Pe,\PGm\PGm)$ \\
  C.R.    & $m_{\Pl\Pl} > 80 \UGeV (\Pe\PGm)$ \\ 
\hline
\end{tabular}
}
\end{table}

\begin{figure}
\begin{center}
\includegraphics[width=0.7\textwidth]{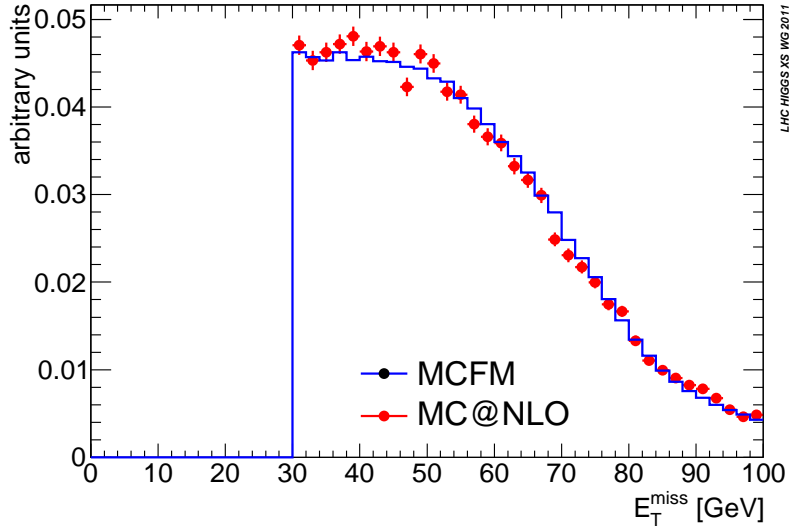}
\end{center}
\caption{Comparison between {\sc MCFM}~v6.0 and MC@NLO: The shape  of the 
missing-$E_{\mathrm{T}}$ distribution is shown
 for events  passing the
  charged-lepton and missing-$E_{\mathrm{T}}$ cuts.} \label{fig:MCFM_MCNLO_comp_etmiss}
\end{figure}

\begin{figure}
\begin{center}
\includegraphics[width=0.49\textwidth]{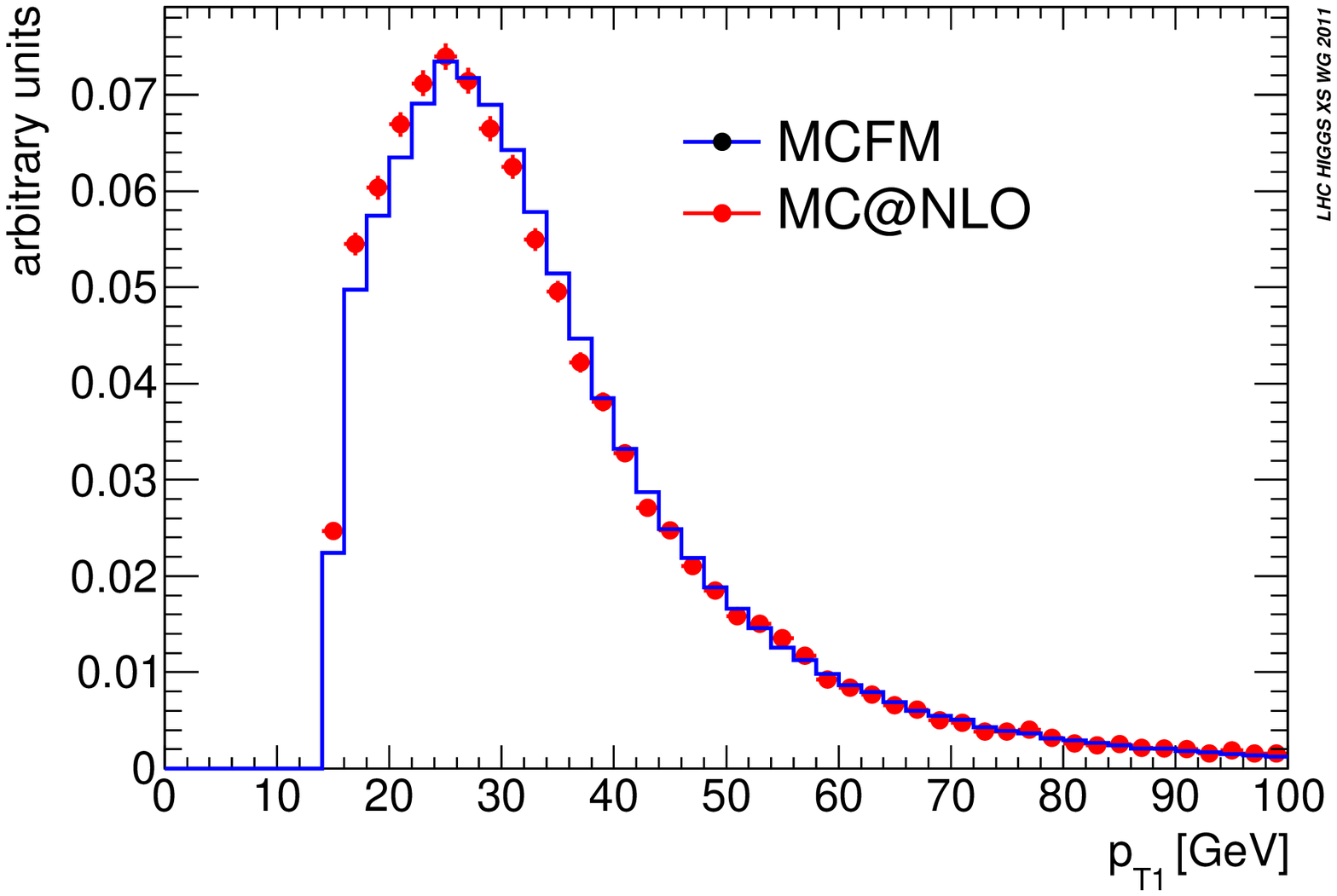}
\includegraphics[width=0.49\textwidth]{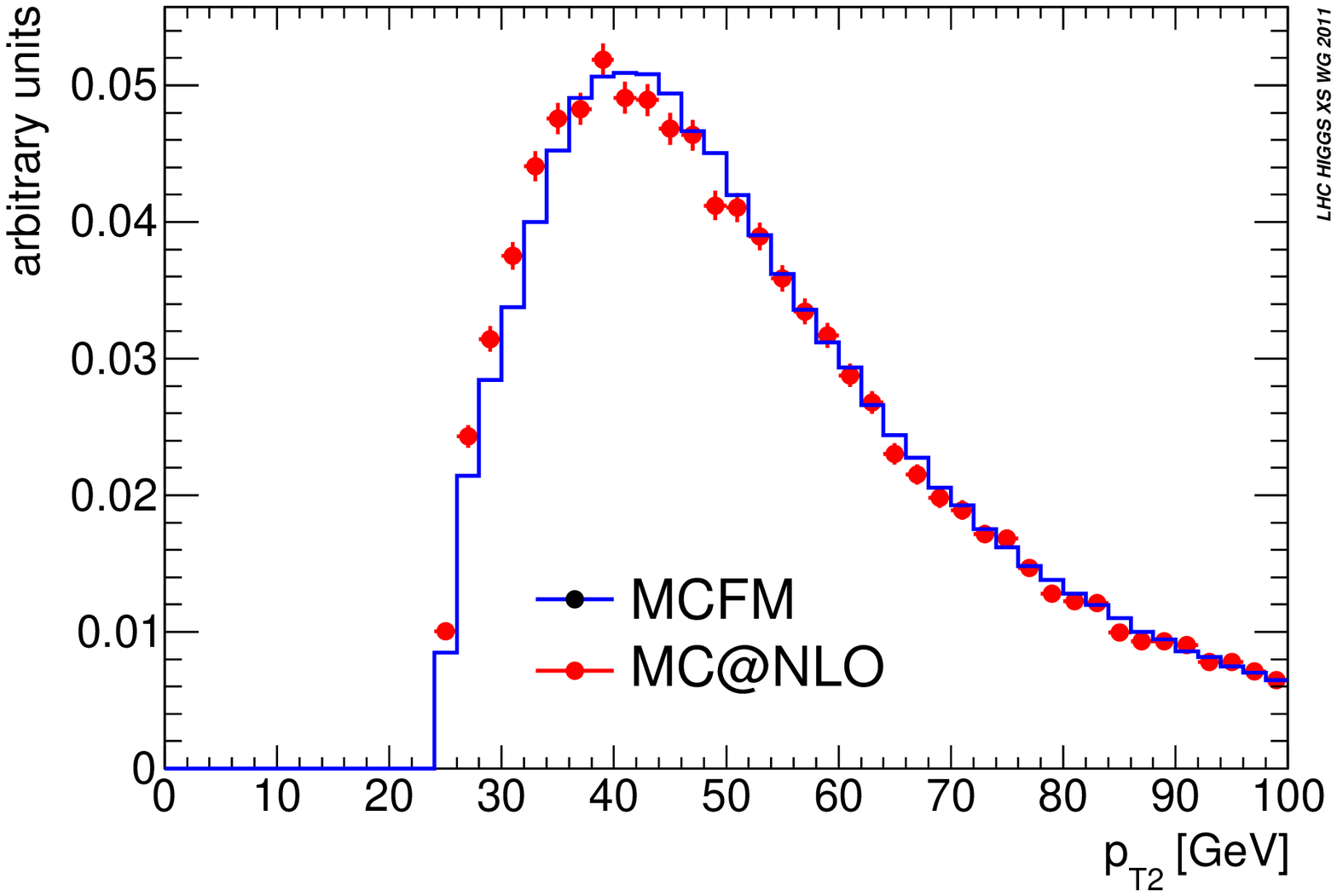}
\end{center}
\caption{Comparison between {\sc MCFM}~v6.0 and MC@NLO: the shapes  of the
  $\pT$ of the softer charged lepton  (left)
  and  of the hardest  charged lepton (right) are shown for events  passing the
  charged-lepton and missing-$E_{\mathrm{T}}$ cuts.} \label{fig:MCFM_MCNLO_comp_1}
\end{figure}

\begin{figure}
\includegraphics[width=0.49\textwidth]{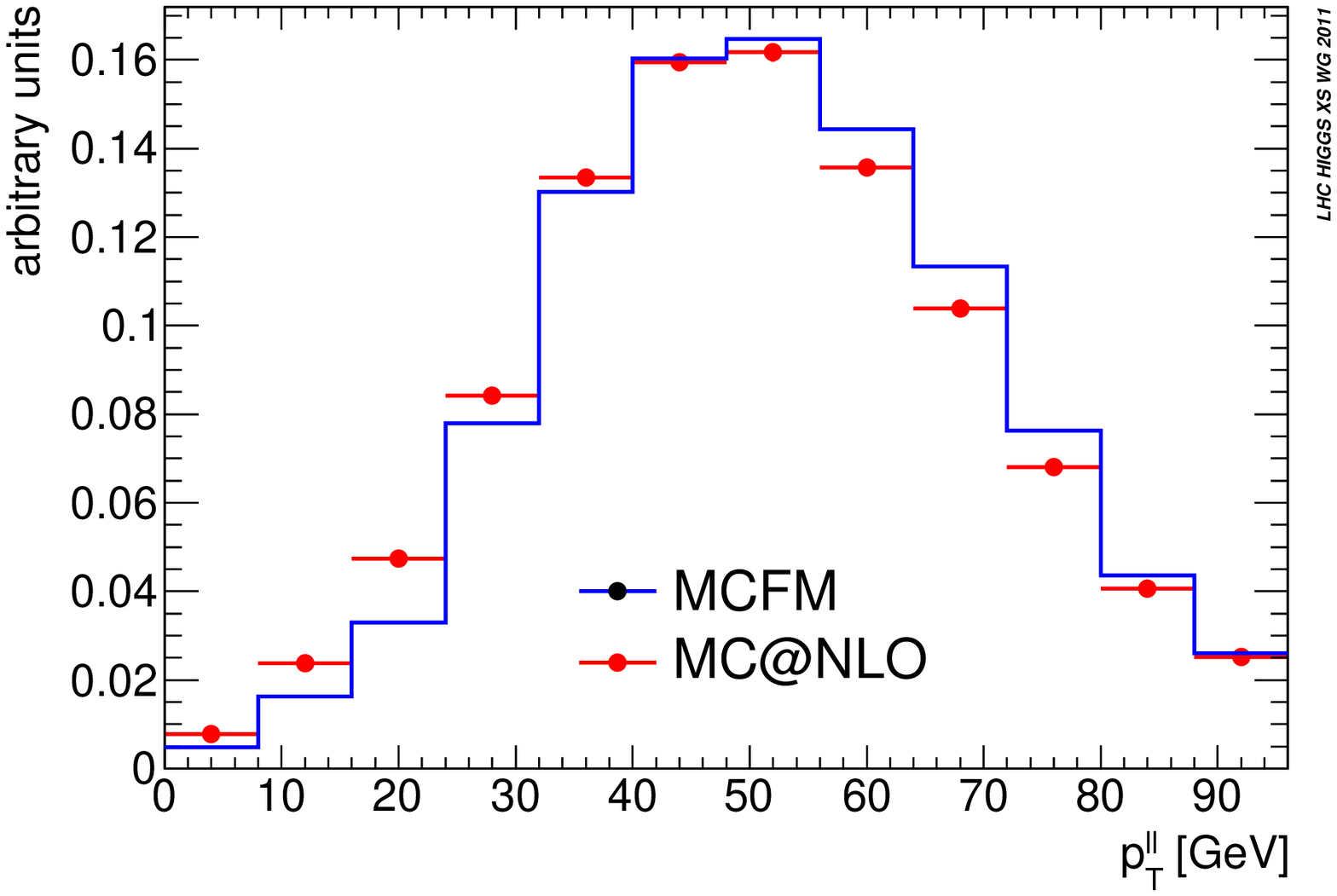}
\includegraphics[width=0.49\textwidth]{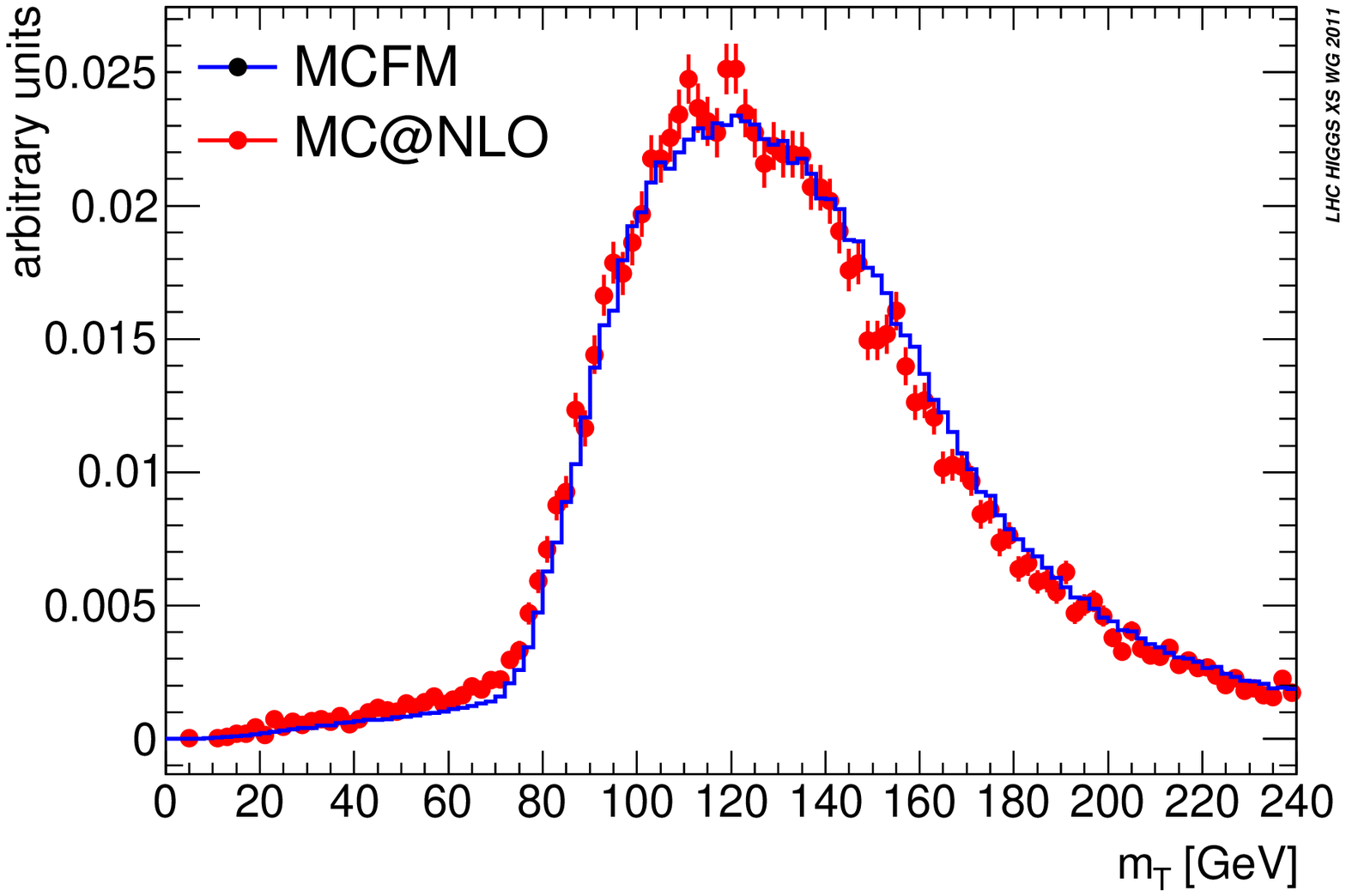}
\caption{Comparison between {\sc MCFM}~v6.0 and MC@NLO: the shapes  of the $\pT$
  of the charged letpon system (left) and the transverse mass of the
  event (right), are shown, the definition of the transverse mass variable is given in the text.} \label{fig:MCFM_MCNLO_comp_2}
\end{figure}

\begin{figure}
\includegraphics[width=0.49\textwidth]{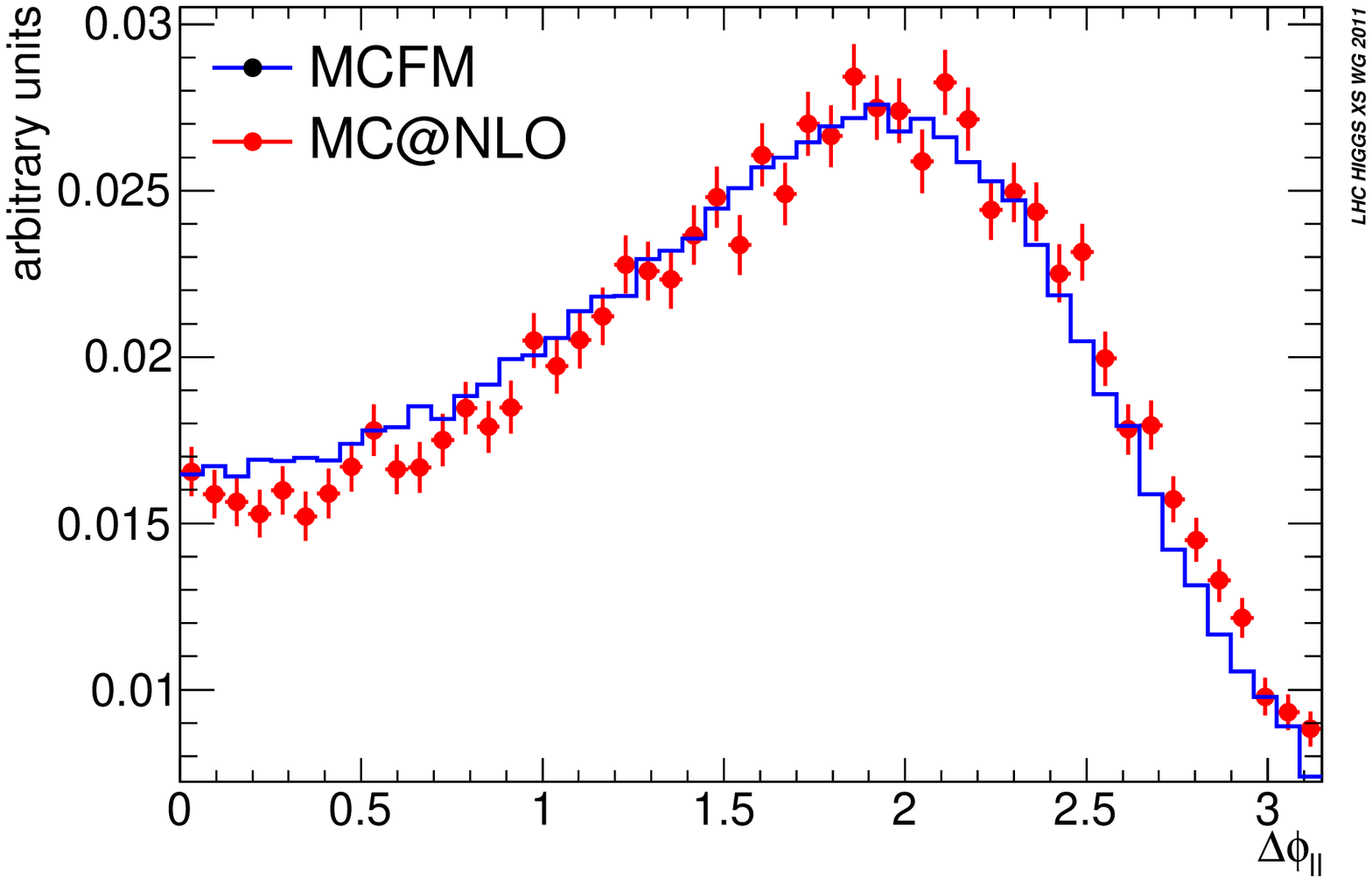}
\includegraphics[width=0.49\textwidth]{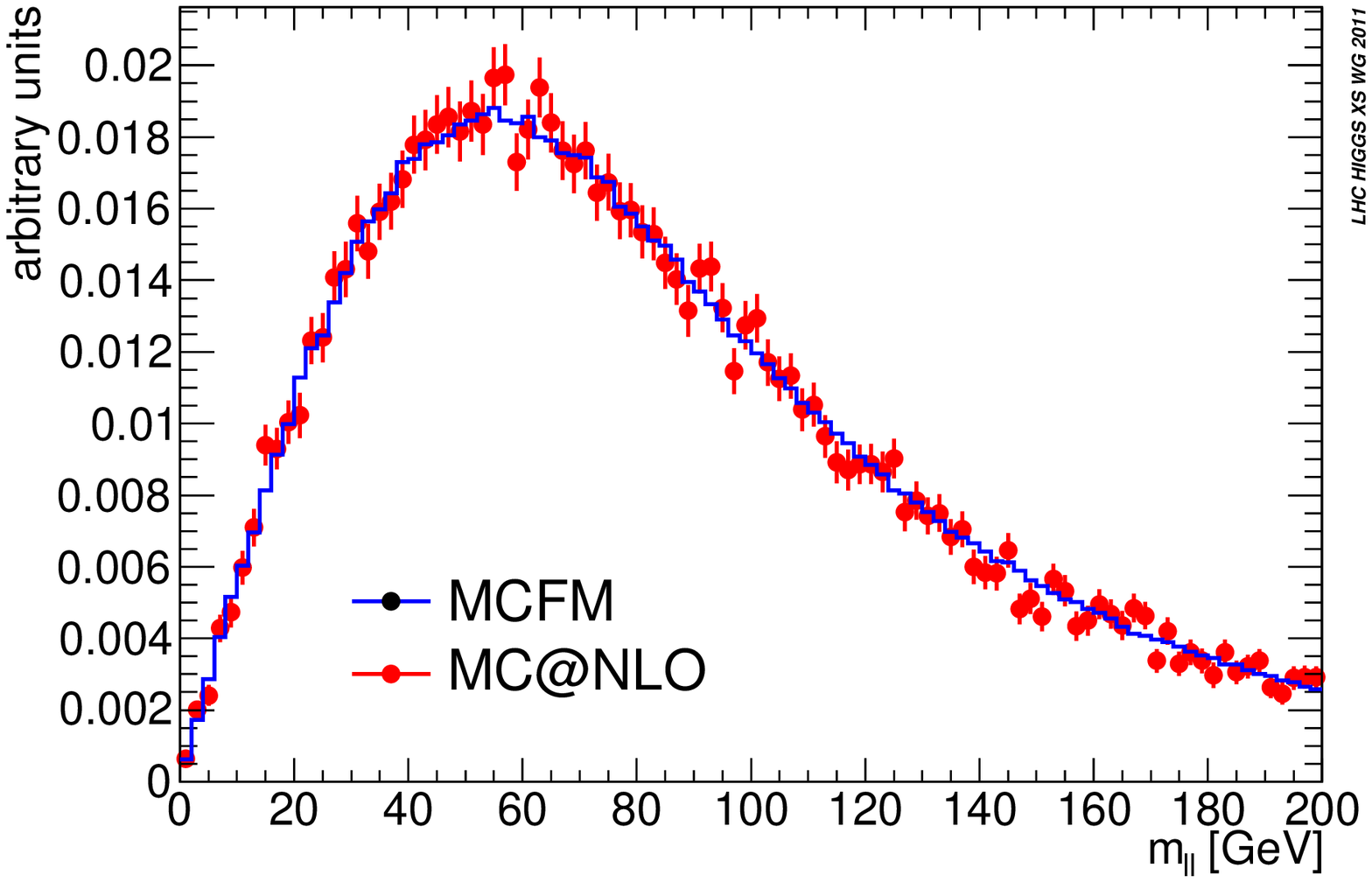}
\caption{Comparison between {\sc MCFM}~v6.0 and MC@NLO shapes  of the angle
  between the two leptons in the transverse plane $\Delta
  \phi_{\Pl\Pl}$ (left) and the invariant mass of the di-lepton system
  $m_{\Pl\Pl}$  (right). The two variables are used for the definition of
  the signal and the $\PW\PW^{*}$ control regions.} \label{fig:MCFM_MCNLO_comp_3}
\end{figure}

\begin{sloppypar}
The amount of $\PW\PW$ background in signal region is determined from the control region through the parameter $\alpha$ defined by
\begin{equation}
\alpha_{\PW\PW} = N^{\PW\PW}_{\rm S.R.}/N^{\PW\PW}_{\rm C.R.},
\end{equation}
where $\alpha$ is evaluated independently for the 0-jet and 1-jet bin.
The Standard Model $\PW\PW^{*}$ yield is obtained using the MC@NLO Monte
Carlo program interfaced to {\sc Herwig} for parton showering. MC@NLO  computes the $\Pp\Pp \to \PW\PW ^{*} \to \Pl\PGn \Pl\PGn$ at NLO including off-shell contributions. Spin correlations
for off-shell $\PW$'s are not treated at matrix-element (ME) level
and a correction is provided after the generation step to take into account 
the spin correlation with some approximation. Furthermore MC@NLO does
not implement all electroweak diagrams contributing to the $\Pp\Pp
\to \Pl\PGn \Pl\PGn$ process. In particular, ``singly-resonant'' processes
are missing in the calculation. A singly-resonant diagram is, e.g.\ 
a diagram where at least a lepton neutrino pair is not connected to the same $\PW$
decay vertex. The full ME calculation for the spin correlation and the inclusion of all
singly-resonant diagrams is implemented in the {\sc MCFM}~v6.0 parton
level Monte Carlo generator. In order to take into account
uncertainties in the modelling of the $\PW\PW$ background, the MC@NLO and
{\sc MCFM}~v6.0 output have been compared. The comparison has been performed 
summing up the contribution of all jet bins in order to integrate out effects
from the simulation of the jet multiplicity, which are not well modelled by parton-level Monte Carlo programs.
The CTEQ6.6 PDF error set has
been used in the comparison and it has been used inclusively in
the number of jets. In \refFs{fig:MCFM_MCNLO_comp_etmiss}, \ref{fig:MCFM_MCNLO_comp_1},
 \ref{fig:MCFM_MCNLO_comp_2}, and \ref{fig:MCFM_MCNLO_comp_3}  we show the
comparison between {\sc MCFM}~v6.0  and MC@NLO on several variables. 
The transverse mass $m_{\rm T}$ is defined by 
\begin{equation}
m_{\rm T} = \sqrt{(E_{\mathrm{T}}^{\Pl \Pl} + E_{\mathrm{T}}^{\rm
    miss})^2-(\vec{p}_{T}^{\Pl \Pl} + \vec{p}_{\mathrm{T}}^{\rm miss})^2},
\end{equation}
where $E_{\mathrm{T}}^{\Pl \Pl} = \sqrt{(\vec{p}_{\mathrm{T}}^{\Pl \Pl})^2 +
  m_{\Pl \Pl}^2}$, $|\vec{p}_{\mathrm{T}}^{\rm miss}| = E_{\mathrm{T}}^{\rm
  miss}$, and $\vec{p}_{T}^{\Pl \Pl}$ is the transverse momentum of
the di-lepton system.
\end{sloppypar}

Small difference between the two calculations are visible in all
variables, the $m_{~ll}$ variable that is used to define the $\PW\PW$
C.R. is in very good agreement. The effect of these discrepancies  
on the $\alpha$ parameter has been evaluated to be:
\begin{equation}
\frac{\alpha(\rm MC@NLO)}{\alpha(\rm MCFM)} = 0.980 \pm 0.015,
\end{equation}
where the error is due to the MC statistics, a conservative error of
$3.5\%$ is used in the analysis.

Further uncertainties on the $\alpha$ parameters are obtained using
different PDF sets in the simulation of the $\PW\PW^{*}$ background.
In particular, the $\alpha$ values have been computed with the
CTEQ6.6 error set and computing the error according the recommended
procedure, the values obtained using the central PDFs of MSTW2008
and NNPDF2.1 has also been computed. The results obtained are shown
in \refT{tab:pdfs}.

\begin{table}
\caption{The $\alpha$ parameters computed using different PDF sets and spread obtained spanning on the CTEQ6.6 error set.} \label{tab:pdfs} 
\centerline{
\begin{tabular}{cccc}
\hline
& CTEQ 6.6 error set & MSTW2008 & NNPDF2.1 \\
\hline
$\alpha^{\rm 0j}_{\PW\PW}$ & $2.5\%$ & $+0.1$ per mil & $+2.7\%$ \\
$\alpha^{\rm 1j}_{\PW\PW}$ & $2.6\%$ & $- 7$  per mil & $+1.2\%$ \\
\hline
\end{tabular}
}
\end{table}

Higher-order corrections can affect the $\pT$ distribution of the
$\PW\PW^{*}$ pair and the $\alpha$ values. Their effect
has been estimated by changing the renormalisation ($\muR$) and the
factorisation ($\muF$) scales in MC@NLO. 
The renormalisation scales are defined as $\muR = \xi_{\rm R} \mu_{0}$
and $\muF = \xi_{\rm F} \mu_0$ where $\mu_0$ is a dynamic scale defined
as
\begin{equation}
 \mu_{0} = \frac{\sqrt{p_{\rm T W1}^2 +
    M_{\rm W1}^2}+\sqrt{p_{\rm T W2}^2+M_{\rm W2}^2}}{2}.
\end{equation}
The nominal scale is obtained with $\xi_{\rm R} = \xi_{\rm F} = 1$ while the
scale uncertainty is obtained by varying $\xi_{\rm R}$ and $\xi_{\rm F}$ in
the range $1/2{-}2$ while keeping $\xi_{\rm R}/\xi_{\rm F}$ in between $1/2$ 
and 2; the maximum spread is taken as scale uncertainties. The
uncertainties obtained are shown in \refT{tab:scale_and_pdfs}
where we summarise also the PDF and modelling uncertainties.
The correlation between the $\alpha$ parameters of the 0-jet and 1-jet
analysis is also evaluated in the calculation and they are found to be
fully correlated.

\begin{table}
\caption{Scale and PDF uncertainties on $\PW\PW$ extrapolation parameters $\alpha$ for the NLO 
$\PQq\PQq,\PQq\Pg \to \PW\PW $ process, the
errors are found to be fully correlated between the 0-jet and 1-jet bin.} \label{tab:scale_and_pdfs}
\centerline{
\begin{tabular}{cc|c|c}
\hline
& scale & PDFs  & modelling \\
\hline
$\alpha^{\rm 0j}_{\PW\PW}$ & $2.5\%$ & $3.7\%$ & $3.5\%$\\
$\alpha^{\rm 1j}_{\PW\PW}$ & $4\%$ & $2.9\%$  & $3.5$ \%\\
\hline
$\alpha^{\rm 0j}_{\PW\PW}$, $\alpha^{\rm 1j}_{\PW\PW}$ correlation & \multicolumn{3}{c}{1} \\
\hline
\end{tabular}
}
\end{table}

\subsection{Contribution from the $\Pg\Pg \to \PW\PW$ process}

MC@NLO MC computes the process $\Pp\Pp \to \PW\PW ^{*}$ at NLO. Up to NLO
only, processes initiated by two incoming quarks contribute to the amplitude of the process. 
At the LHC the process $\Pg\Pg \to \PW\PW $ can deliver a
non-negligible contribution. It is therefore important to estimate the
$\Pg\Pg$ contribution at the LHC. This contribution could be further enhanced
by the analysis cuts that mainly select the Higgs topology that is
induced by a $\Pg\Pg$ process. At low mass the $\Pg\Pg$ contribution is again
normalised through the control region, so that only the uncertainties on the
$\alpha$ parameters are relevant for the analysis.
The PDF uncertainty has been evaluated by the band spanned by the CTEQ6.6 PDF
error set and comparing the central value with those from the MSTW2008 and NNPDF2.1 sets.
Moreover, because the $\Pg\Pg \to \PW\PW $ lowest-order cross section is proportional to $\alphas^2$
there are ambiguities on the set of PDF's that has to be used to evaluate the cross section.
Therefore we include in the PDF uncertainty the values obtained using LO, NLO, and NNLO PDF sets. The 
MSTW family of PDF's has been used for this estimate. The results are shown in \refT{tab:pdf_ggww}.
\begin{table}
\caption{Pdf uncertainties on $\PW\PW$ extrapolation parameters $\alpha$ for the $\Pg\Pg \to \PW\PW $ process.} 
\centerline{
\begin{tabular}{cccccc}
\hline
& CTEQ 6.6 error set & MSTW 2008 & NNPDF2.1 & MSTW LO & MSTW NNLO\\
\hline
$\alpha^{\rm 0j}_{\PW\PW}$ & $2.6\%$  & $-4 $per mil & $-4$ per mil & $+3.6\%$ & $-0.6$ per mil\\
$\alpha^{\rm 1j}_{\PW\PW}$ & $ 2.8\%$ & $-2$ per mil & < $1$ per mil &
$+3.6\%$ & $+0.4$ per mil\\
\hline
\end{tabular}
}
\label{tab:pdf_ggww}
\end{table}

The scale uncertainites have been evaluated by varying
independently by a factor four the renormalisation and factorisation
scales with respect to the nominal value $\mu_{~0} = \MW$ while keeping the ratio between $1/2$ and $2$.
The results are shown in \refT{tab:gg2WW} together with the overall PDF uncertainty.
\begin{table}
\caption{Scale and PDF uncertainties on $\PW\PW$ extrapolation parameters $\alpha$ for the $\Pg\Pg \to \PW\PW $ process.} \label{tab:gg2WW}
\centerline{
\begin{tabular}{ccc}
\hline
& scale & PDFs  \\
\hline
$\alpha^{\rm 0j}_{\PW\PW}$ & $6\%$ & $4.4\%$ \\
$\alpha^{\rm 1j}_{\PW\PW}$ & $9\%$ & $4.6\%$\\
\hline
\end{tabular}
}
\end{table}

The values shown in \refT{tab:gg2WW} have to be applied only to
the $\Pg\Pg \to \PW\PW $ component of the $\PW\PW$ background, the incidence of the
$\Pg\Pg \to \PW\PW $ contribution in the signal region is anyway small given
the low $\Pg\Pg \to \PW\PW $ cross section. The yield of the $\Pg\Pg \to \PW\PW $ is $5\%$
of the total $\PW\PW$ background in the 0-jet channel and $7\%$ in the 1-jet
channel.

\subsection{WW background estimate for the high-mass selection}
\subsubsection{Jet-bin cross sections}
\label{sec:acceptances}
For high values of $\MH$, namely $\MH > 200\UGeV$, the statistics in the $\PW\PW$ control region becomes quite small,
and the control region gets contaminated by
a significant signal fraction. In this region a direct evaluation of the $\PW\PW$ yield in the signal region is recommended.
The  $\PW\PW$ yield in jet bins  is treated in the same way and can be written as follows,
\begin{equation}
N^{\PW\PW}_{\rm 0j} = \sigma_{\ge 0}f_{0}A_{0}, \quad 
N^{\PW\PW}_{\rm 1j} = \sigma_{\ge 0}f_{1} A_{1} \label{eq:cross_multi_jets},
\end{equation}
with $f_0$ and $f_1$ defined as
\begin{equation}
f_0 = \frac{\sigma_{\ge 0} - \sigma_{\ge 1}}{\sigma_{\ge 0}}, \quad 
f_1 = \frac{\sigma_{\ge 1}-\sigma_{\ge 2}}{\sigma_{\ge 0}}.
\end{equation}

The cross sections are computed for jets with $\pT > 25\UGeV$ and $|\eta| < 4.5$ for the ATLAS experiment.
The inclusive-cross-section uncertainty is evaluated using MC@NLO, applying the scale variation prescription as above.
The values of the inclusive cross section obtained and their uncertainties are shown in 
\refT{tab:scale_unc_multijet}; they have been evaluated for a single lepton combination.
\begin{table}
\caption{Inclusive multi-jet cross sections computed with MC@NLO and their scale variation. The values shown are obtained for a single lepton combination. CTEQ6.6 PDFs are used in the computation.} \label{tab:scale_unc_multijet}
\centerline{
\begin{tabular}{cccccccc}
\hline
$\sigma_{\ge 0}$ [fb] & $\Delta \sigma_{\ge 0}$ [\%] & $\sigma_{\ge 1}$ [fb]  & $\Delta \sigma_{\ge 1}$ [\%]  & $\sigma_{\ge 2}$ [fb] & $\Delta \sigma_{\ge 2}$ [\%] & $\sigma_{\ge 3}$ [fb] & $\Delta \sigma_{\ge 3}$ \\
\hline
$532$ & $3.3$ & $159$ & $6.5$ & $41$ &  $8.7$ & $9$ & $11$ \\
\hline
\end{tabular}
}
\end{table}

MC@NLO simulates the $\Pp\Pp \to \PW\PW  +2$~jets process through the NLO $\Pp\Pp \to \PW\PW ^{*} \to \Pl\PGn \Pl\PGn$ plus parton shower. This means that no ME computation for the $\Pp\Pp \to \PW\PW  + 2$~jets is
implemented in the generator. In order to take into account mismodelling of the parton-shower MC we compare the ratio $\sigma_{\ge 2}/\sigma_{\ge 1}$ and $\sigma_{\ge 3}/\sigma_{\ge 2}$ between MC@NLO and ALPGEN which computes $\PW\PW$ production cross section up to three jets. The two $\PW$'s are simulated
on shell by ALPGEN, and spin correlations are not included. 
Moreover, only tree-level diagrams are computed by ALPGEN; therefore the comparison is performed only for jet multiplicity higher than one where ALPGEN provides a ME computation for the jet yield. The comparison is shown in \refT{tab:ALPGEN_MCatNLO_COMPUTATION}. The discrepancy between the two generators on these ratios is added in quadrature to the scale variation for the 2-jets and the 3-jets inclusive cross section. 

\begin{table}
\caption{Multi-jet cross section ratios compared between MC@NLO and ALPGEN.} \label{tab:ALPGEN_MCatNLO_COMPUTATION}
\centerline{
\begin{tabular}{ccc}
\hline
& MC@NLO & ALPGEN \\
\hline
$\sigma_{\ge 2}/\sigma_{\ge 1}$ & $0.256$ & $0.360$ \\ 
$\sigma_{\ge 3}/\sigma_{\ge 2}$ & $0.057$ & $0.112$ \\
\hline
\end{tabular}
}
\end{table}

The inclusive multi-jet uncertainties due to scale and modelling are summarised in \refT{tab:multijet_uncertainties}.

\begin{table}
\caption{Uncertainties on the inclusive jet cross sections due to scale and modelling for jet $\pT > 25\UGeV$ and $|\eta|<4.5$.} \label{tab:multijet_uncertainties}
\centerline{
\begin{tabular}{cccc}
\hline
$\Delta \sigma_{\ge 0}$ [\%] & $\Delta \sigma_{\ge 1}$ [\%] & $\Delta \sigma_{\ge 2}$ [\%] & $\Delta \sigma_{\ge 3}$ [\%]   \\
\hline
 $3$ & $6$ & $42$ & $100$ \\
\hline
\end{tabular}
}
\end{table}

\subsubsection{Theoretical errors on acceptances}
In addition to the scale uncertainties on the jet fractions, $f_{i}$, we need to evaluate also the scale uncertainties on the acceptances $A_{i}$  as defined in Eq.~\ref{eq:cross_multi_jets}.
These include scale variation effects with all other cuts except for the jet counting. 
In order to decouple the jet-bin scale uncertainty from other uncertainties, 
the acceptance needs to be evaluated in a given jet bin.
This is done in MC@NLO by requiring a jet-exclusive bin at the beginning of the selection and then evaluating the ratio
\begin{equation}
N_{\rm cuts~and~jet~bin} / N_{\rm jet~bin}.
\end{equation}
The maximum relative spread of this ratio is used as the fractional acceptance scale uncertainty.
In ATLAS the following cuts are applied in both 0-jet and 1-jet bins for $\MH > 220\UGeV$ 
and the $\Pe\Pe$ and $\PGm\PGm$ channels: $p_{T1} > 25\UGeV$, $p_{T2} > 15\UGeV$, $|\eta| < 2.5$, 
$E_{\rm T}^{\rm miss} > 40\UGeV$,  $\Delta \phi_{\Pl\Pl} < 2.6$, 
$m_{\Pl\Pl} < 140\UGeV$ \&\& $|m_{\Pl\Pl}-\MZ| > 15\UGeV$, while
 in the 0-jet channel $|p_{\rm Tl1} + p_{\rm Tl2}| > 30 \UGeV$ and in the 1-jet channel 
$|p_{\rm Tl1} + p_{\rm Tl2} + p_{\rm T jet} + E^{\rm miss}_{\mathrm{T}}| < 30 \UGeV$ cuts are added. 
The scale uncertainty is $5\%$ in the 0-jet bin and $2\%$ in the 1-jet bin.

In case LO MC are used to compute the acceptance in the analysis, the LO--MC@NLO discrepancy on the given acceptance value has to be added to the scale variation value.

\subsection{PDF errors}
PDF errors are much smaller than scale uncertainties and are less linked to the jet activity associated to the $\PW\PW$ production. Also correlations between jet bins can be neglected.
Therefore in evaluating PDF uncertainties we directly compute the product $f_{i}A_{i}$, which is the relative variation in the event yield in the signal region, by varying PDFs within the CTEQ6.6 error set and comparing
the central values to the ones obtained with MSTW2008 and NNPDF2.1. The uncertainty obtained is $2\%$ for both 0-jet and 1-jet bins.

\clearpage

\newpage

\newcommand{\MZZ}{\mathswitch {M_{{\PZ}{\PZ}}}}
\newcommand{\mzonshell}{\mathswitch {M_{\PZ_{1}}}}
\newcommand{\mzoffshell}{\mathswitch {M_{\PZ_{2}}}}

\section{$\ZZ^*$ decay mode\footnote{%
    N.~De Filippis, S.~Paganis (eds.), S.~Bolognesi, T.~Cheng, 
    R.~K.~Ellis, M.D.~Jorgensen, N.~Kauer, M.~Kovac, 
    C.~Mariotti, P.~Nason, T.C.~Petersen, J.~Price and I.~Puljak.}}
\label{se:ZZ}

\subsection{Introduction}

The $\ZZ$ decay mode of the Standard Model (SM) Higgs boson has a lower branching fraction than the $\WW$ decay over the full range
of the Higgs mass hypotheses inspected; nevertheless the final state with four leptons (electrons and muons) from the $\ZZ$ decay 
is very clean and almost background free so it is considered to be the main discovery channel at the LHC.  
The four-lepton channel is also the most precise final state to 
reconstruct the mass peak of the Higgs boson thanks to the high
resolution of the lepton momentum reconstruction with the $\mathrm{ATLAS}$ and the $\CMS$ detectors. 

Both $\mathrm{ATLAS}$ and $\CMS$ collaborations performed prospective studies in the past and published also results about 
the direct searches for the Higgs boson in $\ZZ$ to four-lepton final state, after collecting data in 2010 and 2011 for 
about $5\,$\ifb\ of integrated luminosity.  
Detailed descriptions of the reference analyses concerning four-lepton final state 
are provided in \Brefs{ATLAS-CONF-2011-162,CMS-PAS-HIG-11-025}.

Complementary final states with two leptons and two jets, two leptons and two neutrinos, two leptons and 
two taus have been also inspected; they are mostly relevant for a Higgs-boson
mass above the $\ZZ$ doubly-resonant peak where
the contribution from reducible and irreducible background processes can be significantly reduced. 
Detailed descriptions of the reference analyses concerning those final states 
are provided in \Brefs{ATLAS:2011af,CMS-PAS-HIG-11-026,Aad:2011ec,CMS-PAS-HIG-11-027,CMS-PAS-HIG-11-028}.
The focus of this section is on the evaluation of the impact of some
theoretical issues on the predictions and on the results for 
the analyses previously mentioned, and on the  estimation of the most
important source of background (the production of $\ZZ$) and its related uncertainties.

\subsection{The $\PH \rightarrow \ZZ$ signal process}

The simulation of the production and the decay of the Higgs boson in the $\ZZ$ final state has been done by using several
Monte Carlo generator programs, both in $\mathrm{ATLAS}$ and in $\CMS$ collaborations. The predictions of different programs in terms of 
cross-section values and distributions for the most important observables have been compared and are
reported in this section. 

The Standard Model input parameters as specified in the Appendix of \Bref{Dittmaier:2011ti} have been used. The following 
set of PDF's, CT10, NNPDF2.0 and MSTW2008, have been considered to estimate the central value of
the geometrical acceptance and the uncertainty on it originating from the PDF parametrisation; 
the PDF4LHC prescription of \Bref{Botje:2011sn} has been followed. In order to estimate the QCD scale uncertainty on the acceptance 
the CT10 PDF has been used as central value and the QCD scale is varied following the prescription of \Bref{Dittmaier:2011ti}.

\subsubsection{Leading-order and next-to-leading-order generators}

The multi-purpose generator program \PYTHIA{} version 6.4~\cite{Sjostrand:2006za} and 
the generator program \POWHEGBOX{} ~\cite{Alioli:2010xd} have been used
for the current Higgs simulation by
$\mathrm{ATLAS}$ and $\CMS$ collaborations; \PYTHIA{} is a leading-order (LO) generator while 
\POWHEGBOX{}~\cite{Alioli:2010xd} implements calculations at the next-to-leading order (NLO).

Events with Higgs bosons produced via gluon fusion and vector-boson fusion mechanisms, and decaying into $\ZZ$ to four-lepton final state are generated
by both the programs; if $\mzonshell$ is the di-lepton combination from $\PZ$ decay that gives the invariant mass closest to the $\PZ$-boson nominal mass and 
$\mzoffshell$ is the other $\PZ^{(*)}$, the distributions of the
transverse momentum and Higgs mass for a Higgs boson with mass $\MH=150 \UGeV$ are compared in \refF{figZZ_1} after applying the 
following cuts defining the acceptance of the selection: $\mzonshell >50 \UGeV$, $\mzoffshell >12 \UGeV$ and $\pT>5 \UGeV$, $|\eta|<2.5$ for 
all the leptons. 
For the vector-boson fusion topology the predictions are different at
low transverse momentum over the full Higgs-boson mass range inspected; in
the case of gluon--gluon fusion production, differences become relevant for very large $\MH$ where anyway the MC's are using an approximation
of the Higgs-boson lineshape that can be substantially different from the
predicted one, as discussed in \refS{sec:po}.

\begin{figure}
  \includegraphics[width=0.49\textwidth]{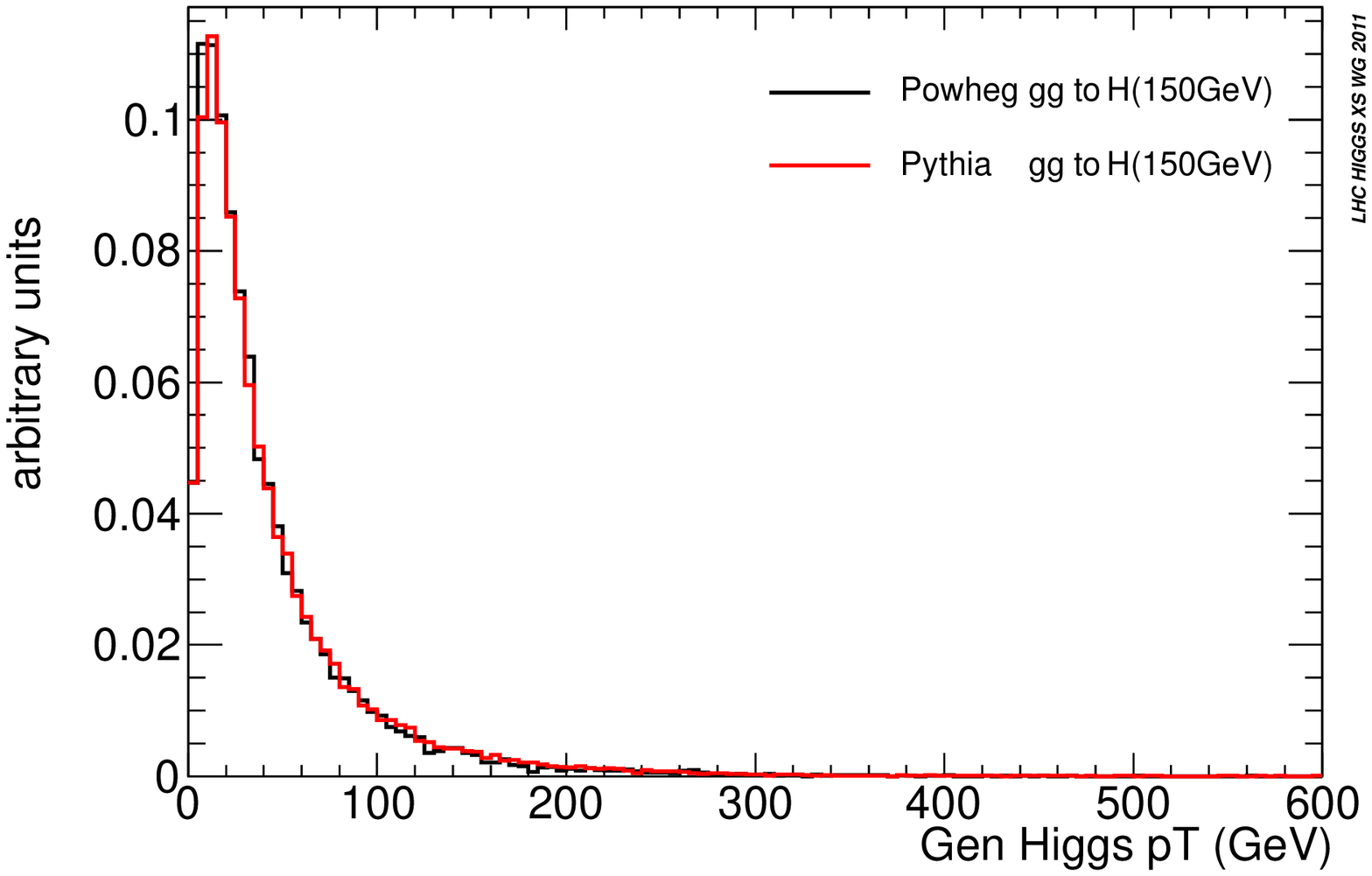} 
  \includegraphics[width=0.49\textwidth]{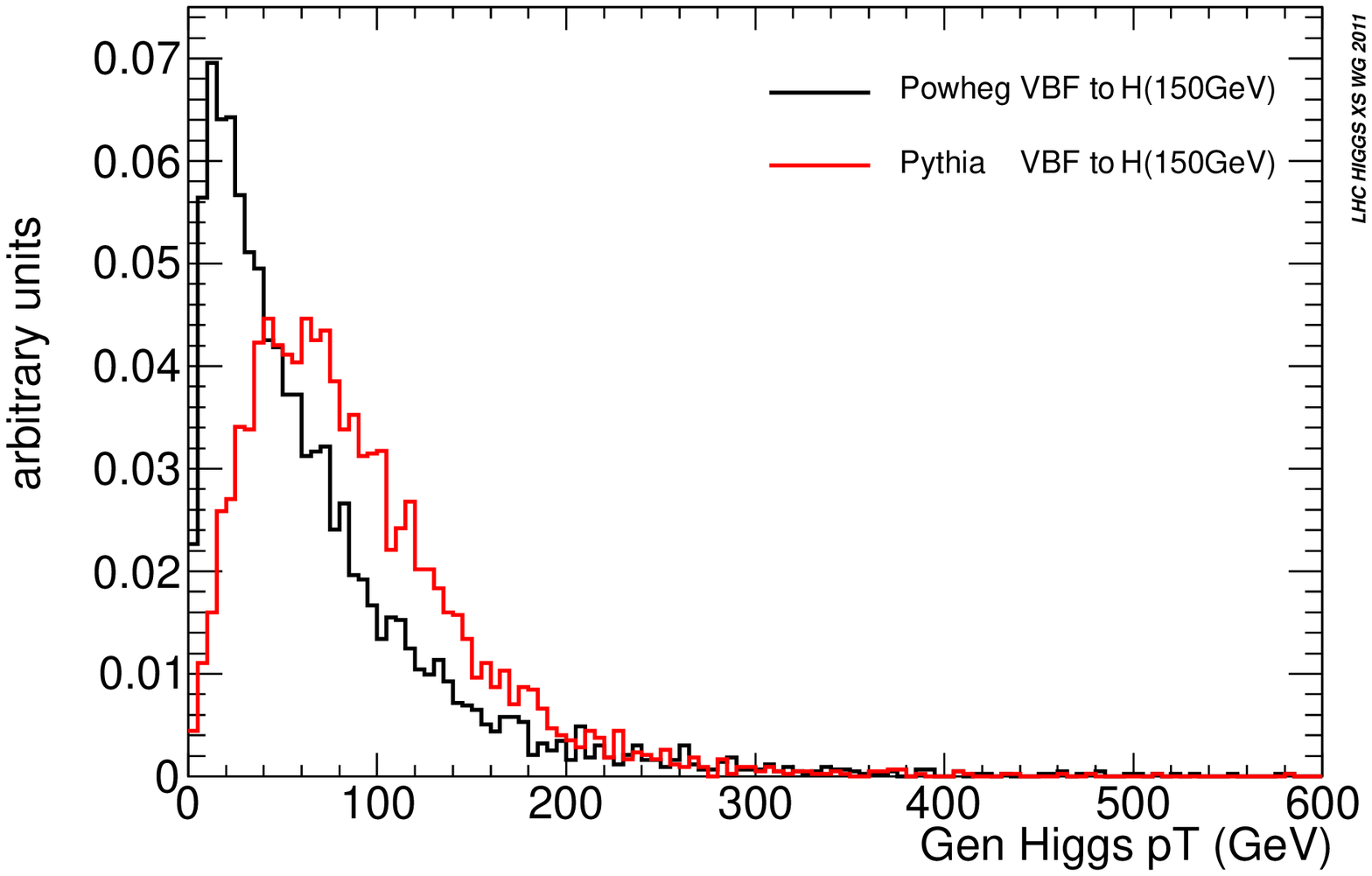} \\
  \includegraphics[width=0.49\textwidth]{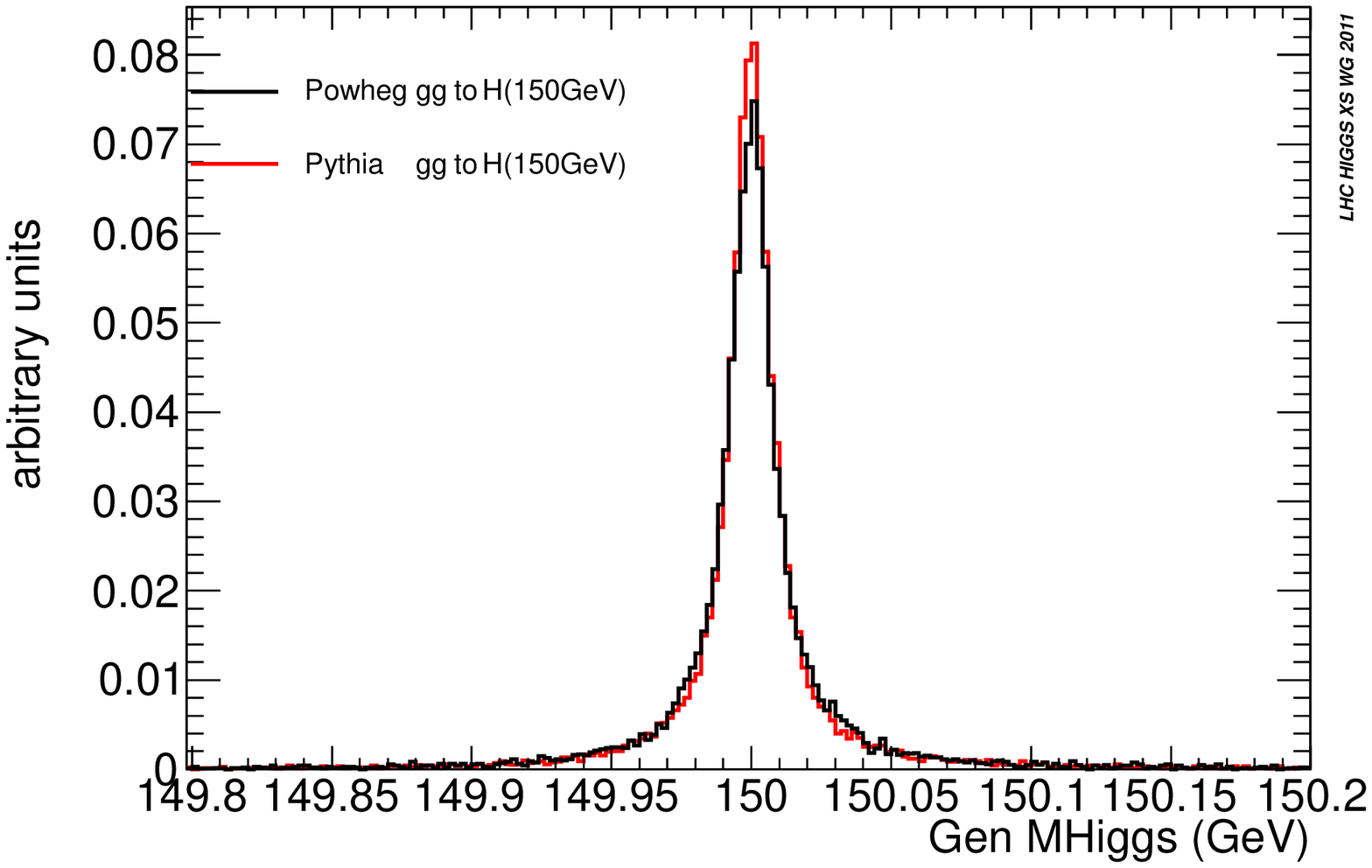} 
  \includegraphics[width=0.49\textwidth]{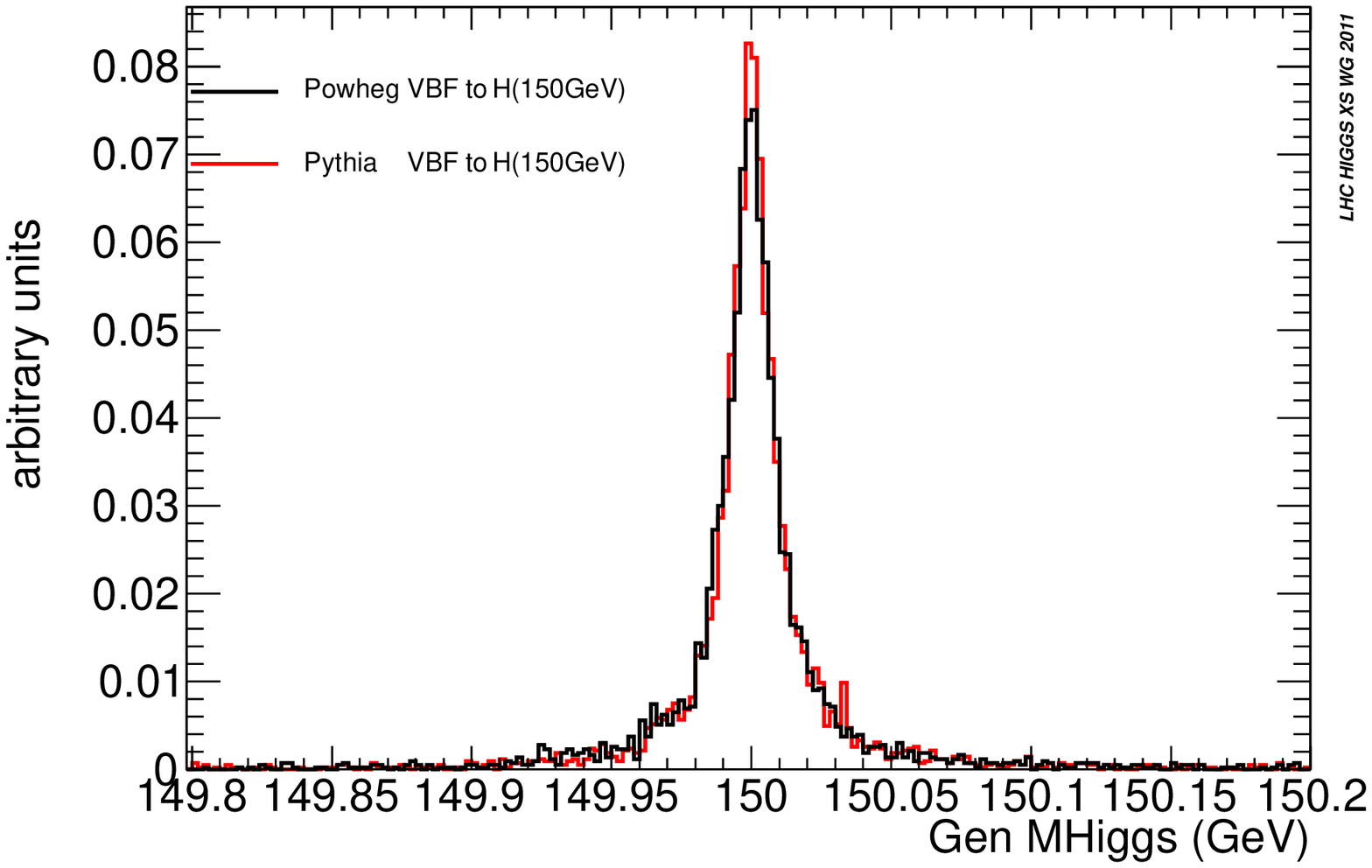}
\caption{Higgs-boson transverse momentum and mass spectrum derived with \PYTHIA~(LO) and
  \POWHEG\ (NLO) generators for the four-lepton final state from the decay of the Higgs boson with mass $\MH=150 \UGeV$, asking $\mzonshell >50 \UGeV$,
$\mzoffshell >12 \UGeV$ and $\pT>5 \UGeV$, $|\eta|<2.5$.  Plots are normalised to the same area.}
\label{figZZ_1}
\end{figure}


\subsubsection{Theoretical uncertainties}

The PDF4LHC prescription is used to estimate the uncertainty on the acceptance for signal events related to the PDF+$\alphas$ choice. 
The results are quoted as the envelope containing all variations for the three sets of PDFs: CT10, MSTW2008, NNPDF.

Concerning the four-lepton analysis the correction factor on the acceptance with respect to the 
central value and the final parametrisation is reported in
\refF{pdfggH} as a function of the Higgs mass. 
A $2\%$ mass-independent error to account for these uncertainties is derived and can be assumed as a conservative estimate over the full mass range. 

The uncertainty on the acceptance and the total Higgs cross section uncertainties from PDF+$\alphas$ have to be considered uncorrelated; 
the level of correlation between the acceptance and cross section for $\MH= 120\UGeV$ and $400\UGeV$ is shown in \refF{pdfggHcorrel}; there is a very little 
correlation for the low Higgs boson mass while a negative correlation seems to develop for very high mass.

\begin{figure}
\begin{center}
\includegraphics[width=0.5\textwidth]{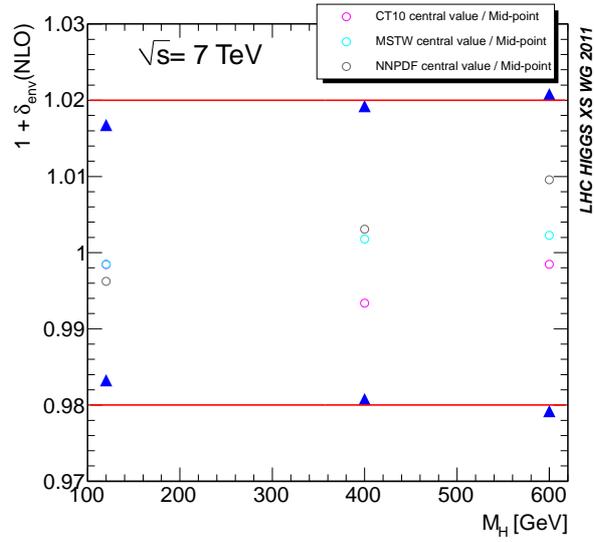} 
\end{center}
\vspace*{-1em}
\caption{The difference between the central value of the acceptance for four-lepton events and the acceptance computed
with plus and minus $1\,\sigma$ of the 
total PDF+$\alphas$ variation (blue markers), following the PDF4LHC prescription; the circles markers correspond to the ratio between central value 
of CT10, MSTW, and NNPDF and the middle value of the envelope.}
\label{pdfggH}
\end{figure}

\begin{figure}
\includegraphics[width=0.49\textwidth]{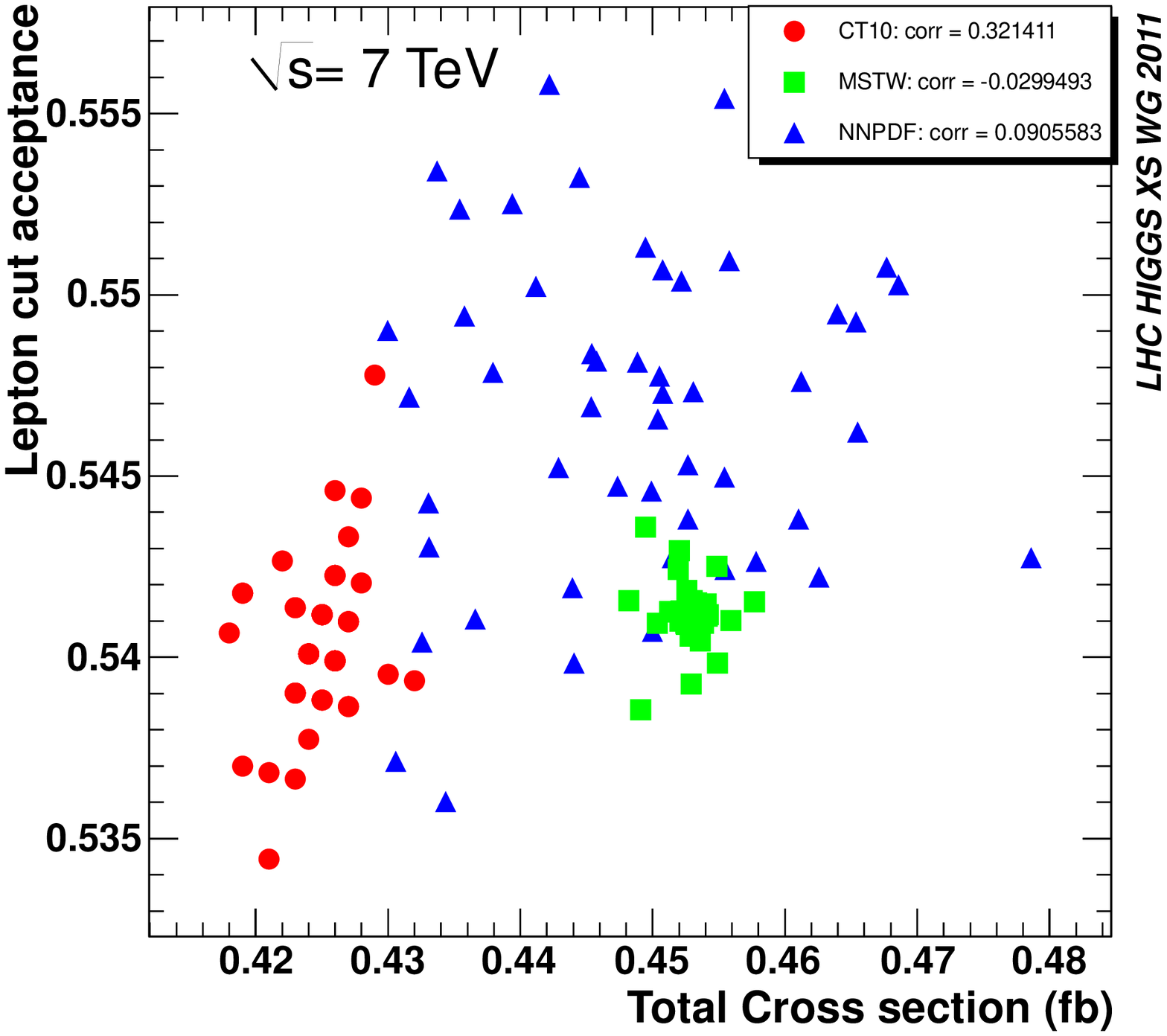} 
\includegraphics[width=0.49\textwidth]{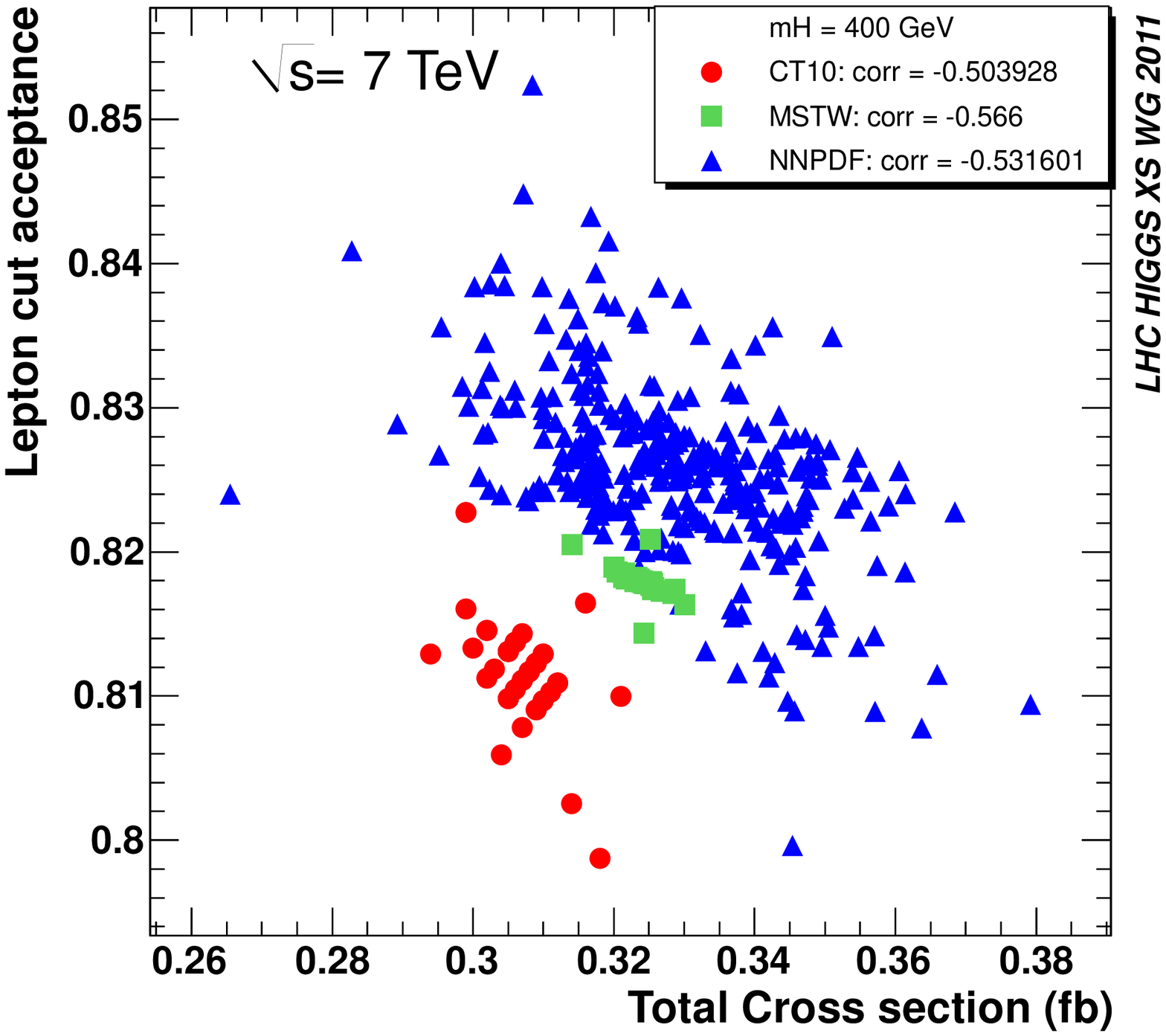} 
\caption{Scatter plots of the signal four-lepton acceptance as a function of the total cross section for signal events 
with $\MH=120  \UGeV$ (left) and $\MH=400  \UGeV$ (right), for different PDF sets. The three PDF sets are shown
in different colours/symbols.}
\label{pdfggHcorrel}
\end{figure}

\subsubsection{Higher-order-corrected Higgs-boson transverse-momentum spectrum}

The total inclusive cross sections for Higgs-boson production have been computed at
NNLO+NNLL~\cite{Dittmaier:2011ti,Anastasiou:2008tj,deFlorian:2009hc,Dawson:1990zj,Spira:1995rr,Harlander:2002wh,Anastasiou:2002yz,Ravindran:2003um,Catani:2003zt,Actis:2008ug}. 
The differential cross section for the transverse momentum is
expected to differ with respect to the one predicted at lower
perturbative order; the Higgs-boson transverse-momentum distribution has been computed 
using the technique of resummation of the large logarithmic 
contributions appearing at transverse momenta much smaller than the
Higgs-boson mass~\cite{deFlorian:2011xf}.

\begin{sloppypar}
In this section the predictions of the tool implementing this
calculation at NNLL, namely {\sc HqT}~\cite{deFlorian:2011xf}, are compared with those from the NLO 
generator program \POWHEGBOX{}. The way to compare the tools is described in \refS{NLOPSsec:compHqT}.
The comparison requires the 
simulation with \POWHEG{} generator of the Higgs production and the usage of parton-shower algorithm.
An alternative way that could be inspected is to compute the weight function $f(\pT)$ to be applied
for each simulated event, with $\pT$ being the Higgs $\pT$ at the \POWHEG{} level (i.e.
before showering) such that the $\pT$ of the Higgs after showering (but
without underlying event and hadronisation) agrees with {\sc HqT}. 
In this section the $\pT$ distributions from \POWHEG{}, after simulating the parton shower, 
the hadronisation, and the underlying event with \PYTHIA{}~6.4 program, are compared with {\sc HqT}. 
\end{sloppypar}

The transverse-momentum ($\pT$) distribution of the Higgs boson produced via gluon fusion is reported in 
\refF{figZZ_1c} (left) as obtained by using the \POWHEG{} generator and 
the {\sc HqT} tool; the two distributions are normalised to the gluon-fusion
cross section reported in YR1~\cite{Dittmaier:2011ti}. The $\pT$ is
significantly affected over the full range and for 
low Higgs-boson masses ($\MH<150  \UGeV$); 
{\sc HqT} gives a larger differential cross section at low $\pT$ ($\pT< 15 \UGeV$) with respect to \POWHEG{}
 (up to $40\%$ increment), while at large $\pT$ the opposite behaviour is
 observed. 
The ratio between the $\pT$ distributions for each $\pT$ bin can be
used as a weight for the \POWHEG{} spectrum that has been used for the current $\PH\rightarrow\ZZ$ analyses in $\CMS$
 and $\mathrm{ATLAS}$. The ratio (or weight) is reported in
 \refF{figZZ_1c} (right) and the best fit curve 
is overimposed to the histogram. 
A table with the weights for several Higgs-boson masses and $\pT$ bins is reported in the 
\refA{app:hqt4powheg} (\refT{tabZZ_1}).


The differences in the Higgs-boson $\pT$ spectrum affects also the $\mzonshell$,
$\mzoffshell$, and four-lepton $\pT$ and $\eta$ distributions, thus the number of accepted events
in the detector. 
The impact of the reweighting of \POWHEG{} events on the acceptance of
the four-lepton analysis has been evaluated. 
The ratio between the acceptance values for \POWHEG{} events with four-electrons, four-muons, 
and two-electrons + two-muons final state 
from $\PH\rightarrow \ZZ$ decay, when turning 
on/off the reweighting with {\sc HqT}, are shown in
\refF{figZZ_1bis} as a function of $\MH$. 
The impact of the reweighting on the acceptance amount to $1{-}2\%$ at most 
at low Higgs-boson masses and to at most $1\%$ at  $\MH > 250 \UGeV$.

All the previous results are estimated with the CT10 PDF~\cite{Martin:2009iq} set and the
input values from \Bref{Dittmaier:2011ti}.

\begin{figure}
  \includegraphics[width=0.49\textwidth]{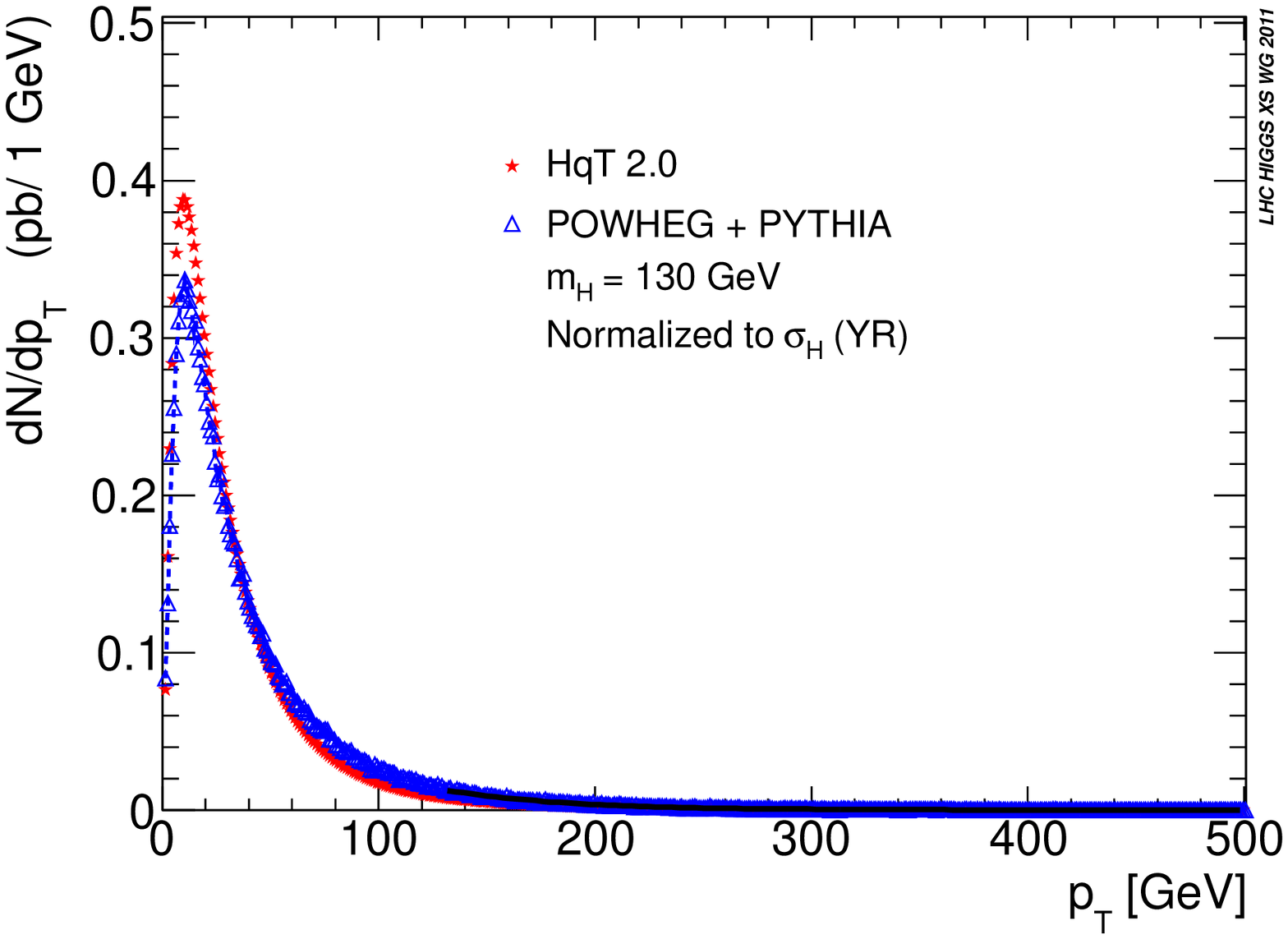} 
  \includegraphics[width=0.49\textwidth]{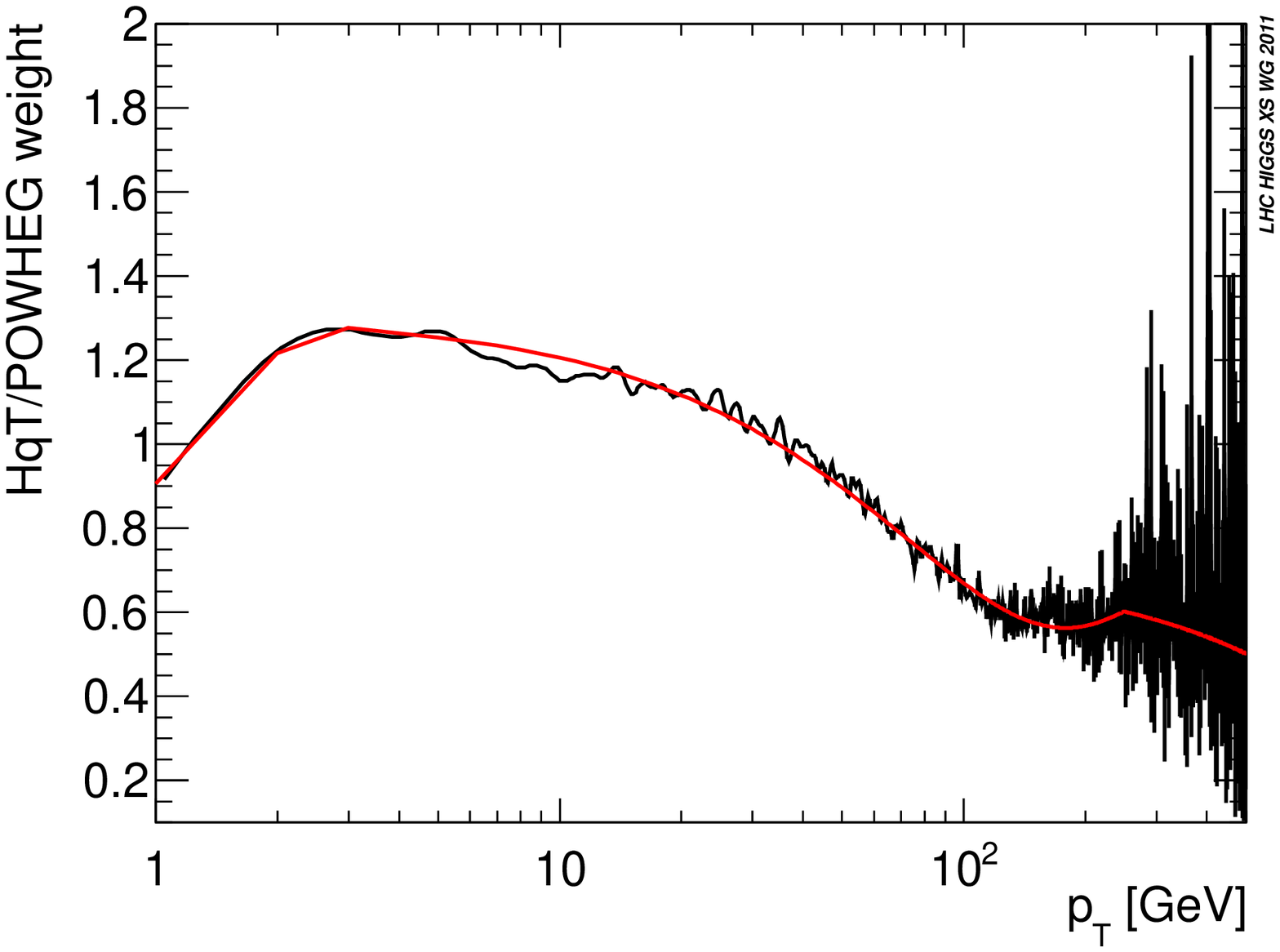} 
\caption{(Left) $\pT$ spectrum of the SM Higgs boson of mass $\MH=130
  \UGeV$ produced via gluon-gluon fusion as obtained by using the 
  \POWHEG{} generator (and {\sc PYTHIA} for parton shower, hadronisation and underlying event) and {\sc HqT} tool. (Right) the ratio {\sc Hqt} vs
  \POWHEG{} as a function of the Higgs-boson $\pT$.}
\label{figZZ_1c}
\end{figure}

\begin{figure}
\begin{center}
  \includegraphics[width=0.6\textwidth]{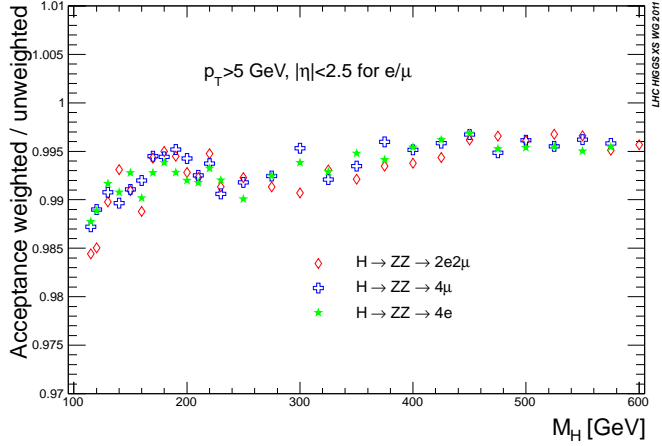} \\
\end{center}
\vspace*{-2em}
\caption{Ratio between the acceptance values for \POWHEG{} events with
four electrons, four muons, and two electrons and two muons final state
from $\PH\rightarrow \ZZ$ decay, when turning on/off the reweighting with
{\sc HqT}.}
\label{figZZ_1bis}
\end{figure}



The predictions of the \POWHEG{} generator at NLO reweighted with and without {\sc HqT} 
are compared with the predictions from {\sc HNNLO}~\cite{Grazzini:2008fs} at NNLO  in \refFs{fig:lo_nlo_nnlo}
and \ref{fig:lo_nlo_nnlo_2}. 

\begin{sloppypar}
The predictions of \POWHEG{} are also reported after Higgs-boson $\pT$
reweighting using {\sc HqT}~\cite{deFlorian:2011xf} in the low-$\pT$ region and 
{\sc HNNLO} in the high-$\pT$ region, as discussed previously and reported in \refT{tabZZ_1}.
As expected, the Higgs-boson $\pT$ is softer in {\sc HNNLO} than {\sc HqT} in the low-$\pT$ region and the effect is more pronounced
in the high-Higgs-boson-mass case (the high-$\pT$ tail matches instead with the reweighted \POWHEG{}, by construction).
It is interesting to notice that, while at low mass the $\pT$ of the leptons has the same behaviour as the
Higgs-boson $\pT$ (i.e., softer for {\sc HNNLO}), at high mass {\sc HNNLO} predicts a stronger $\pT$ for the leptons.
The mass distribution of the $\PZ$ with larger mass has larger tails
in {\sc HNNLO}, while the opposite is true
for the subleading $\PZ$ mass. It is worth noticing that the
Higgs-boson $\pT$ reweighting in \POWHEG{} does not affect
the $\PZ$ invariant-mass distributions. Finally the rapidity of the
Higgs boson is less central in {\sc HNNLO} 
and the same holds for the lepton pseudorapidity, but the difference goes to zero at high mass.
\end{sloppypar}

\begin{figure}
  \includegraphics[width=0.49\textwidth]{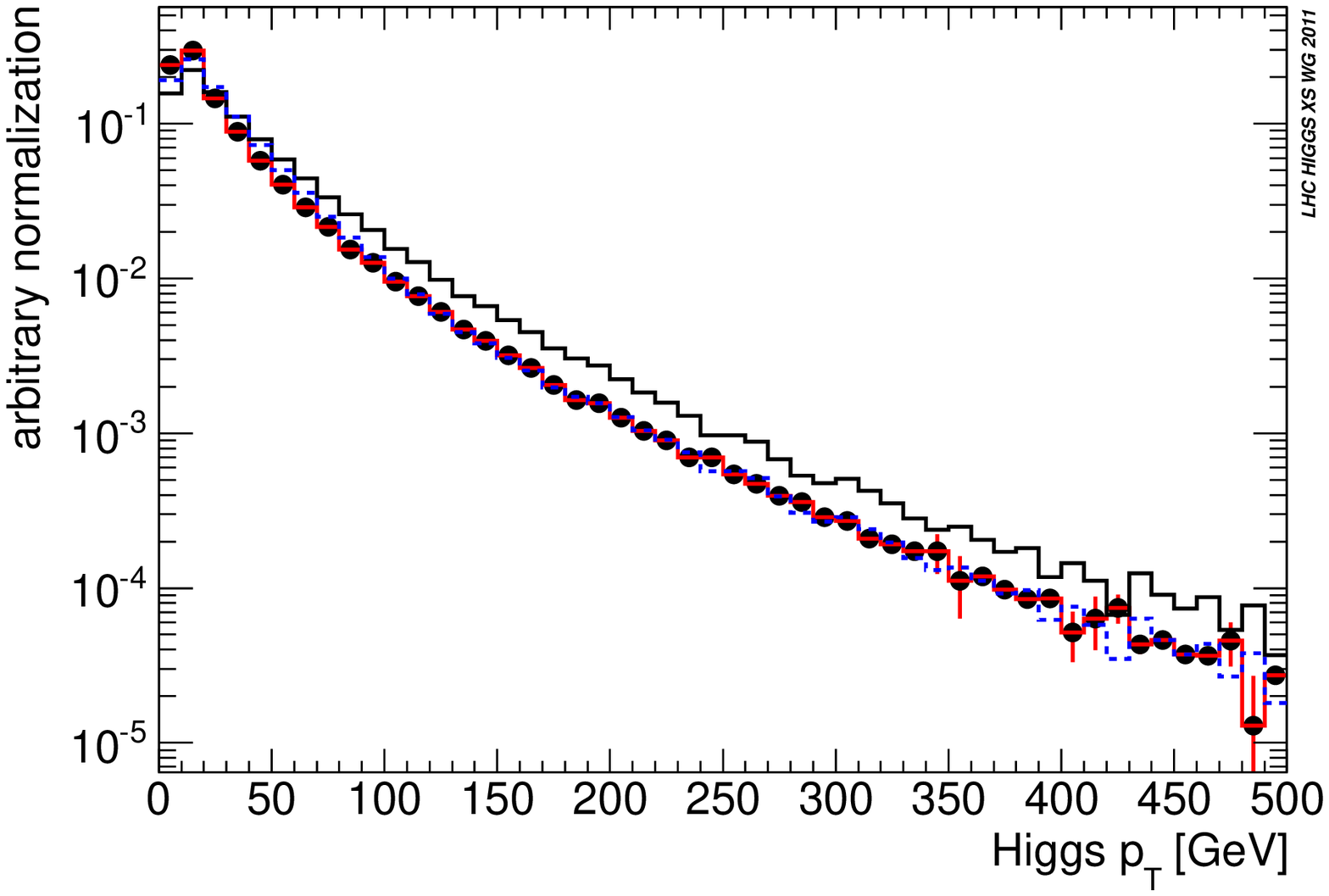}
  \includegraphics[width=0.49\textwidth]{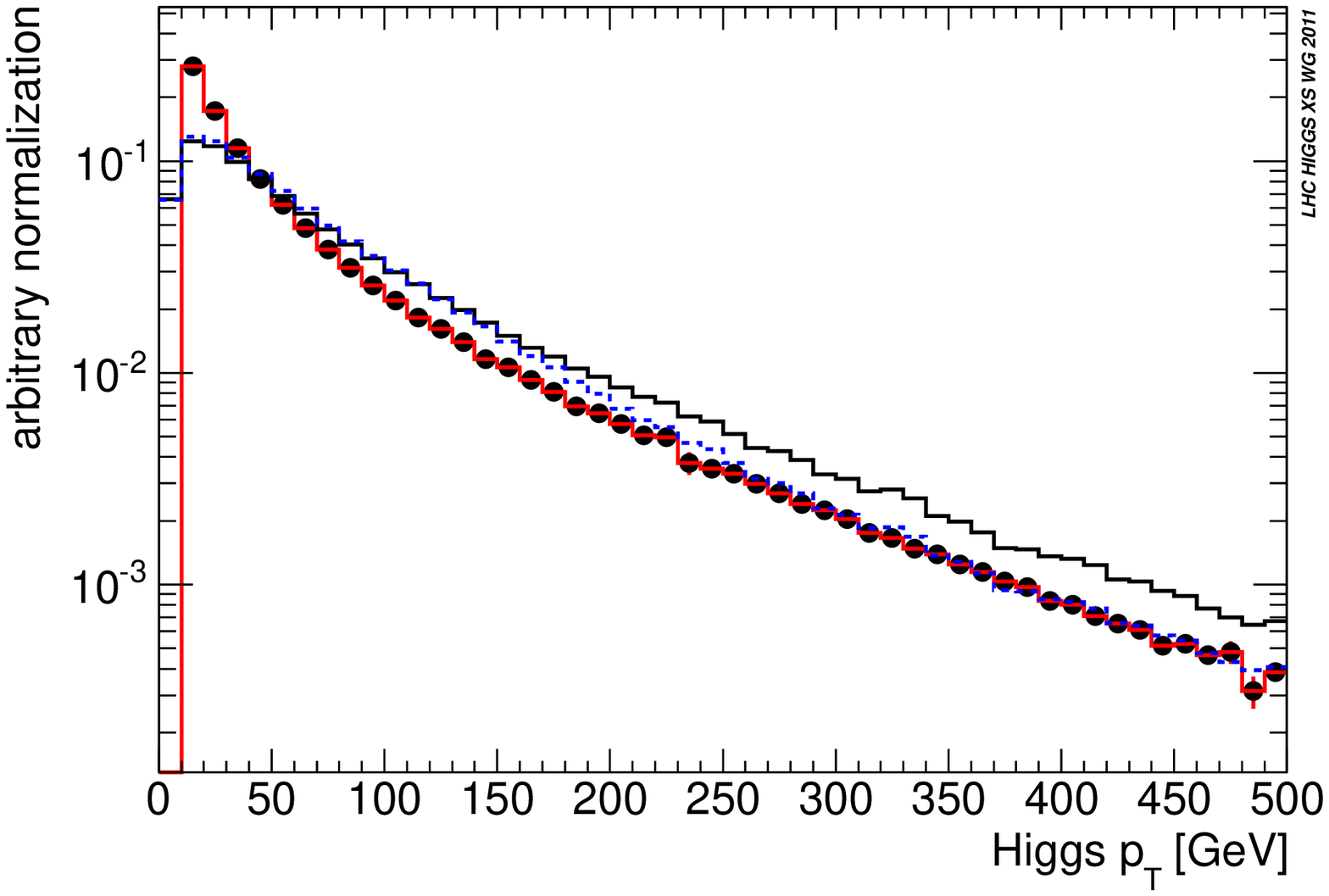} \\
  \includegraphics[width=0.49\textwidth]{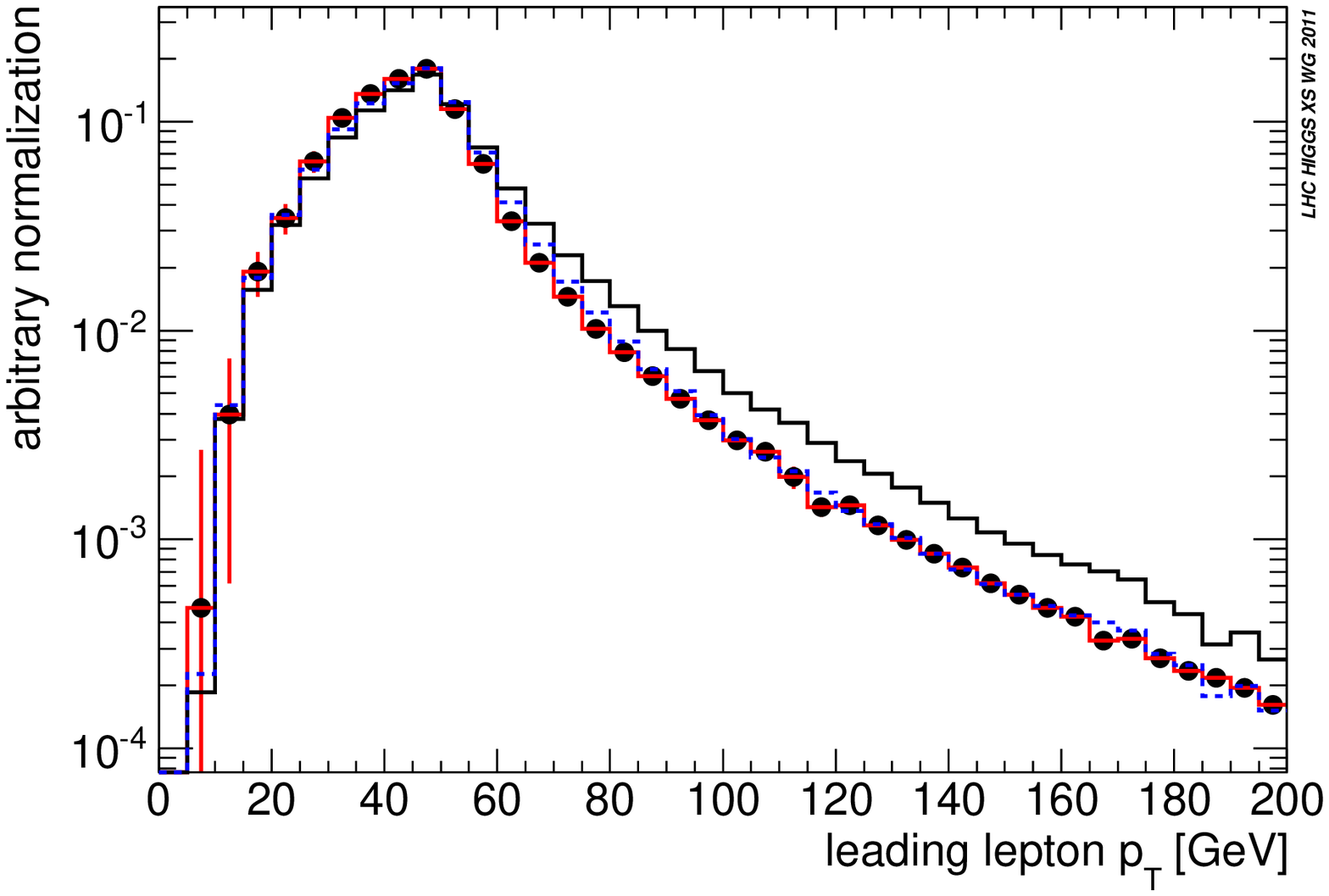}
  \includegraphics[width=0.49\textwidth]{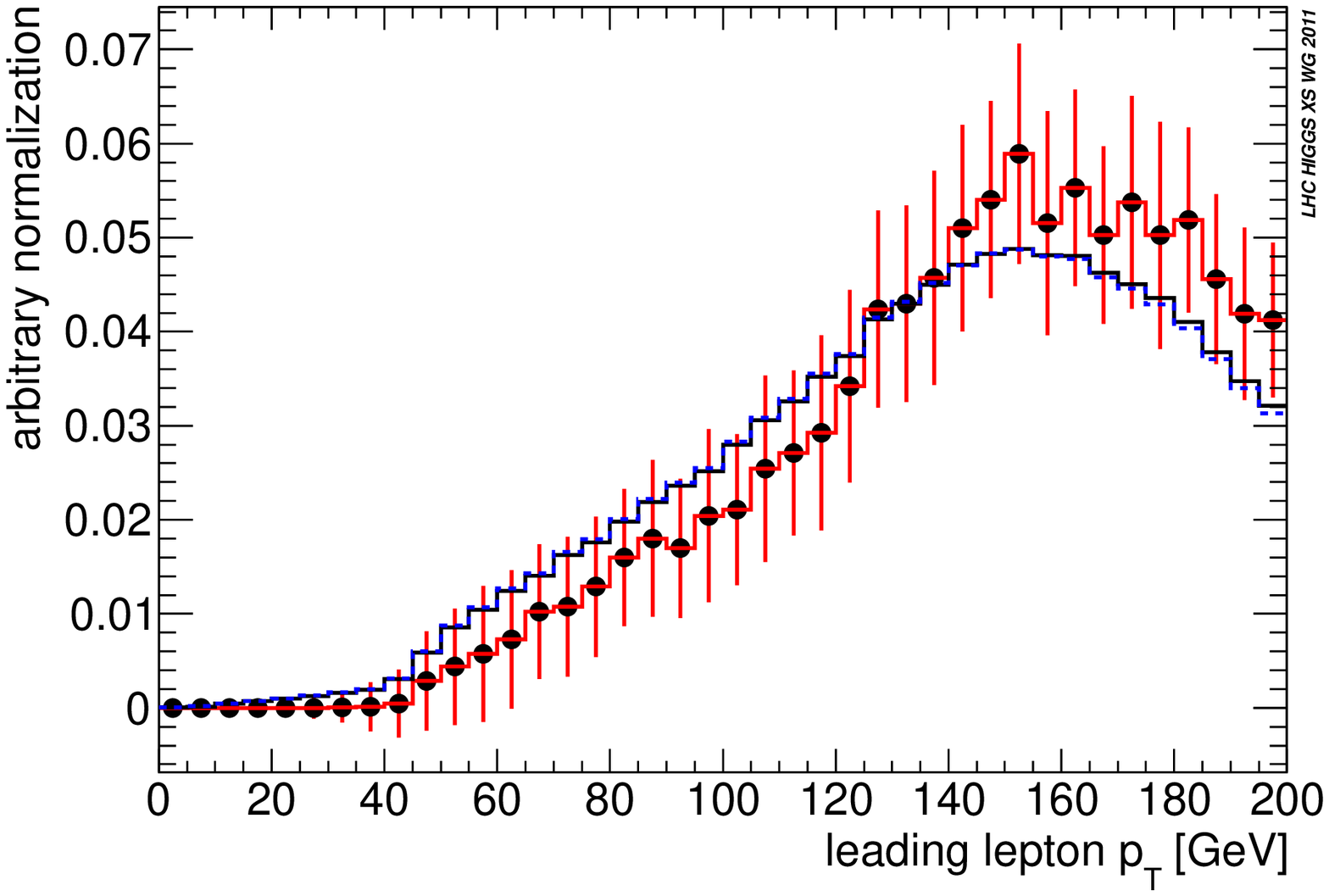} \\
  \includegraphics[width=0.49\textwidth]{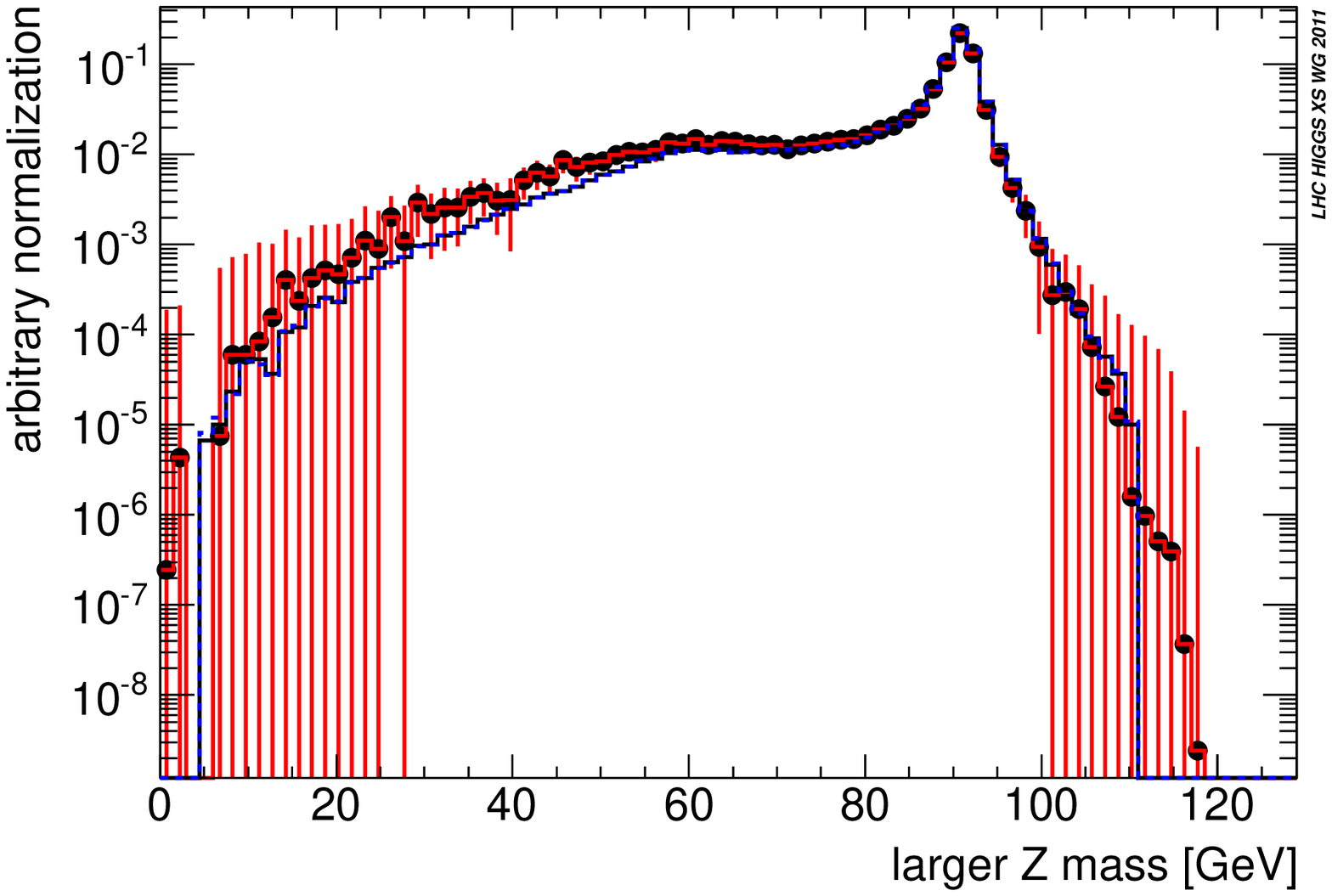}
  \includegraphics[width=0.49\textwidth]{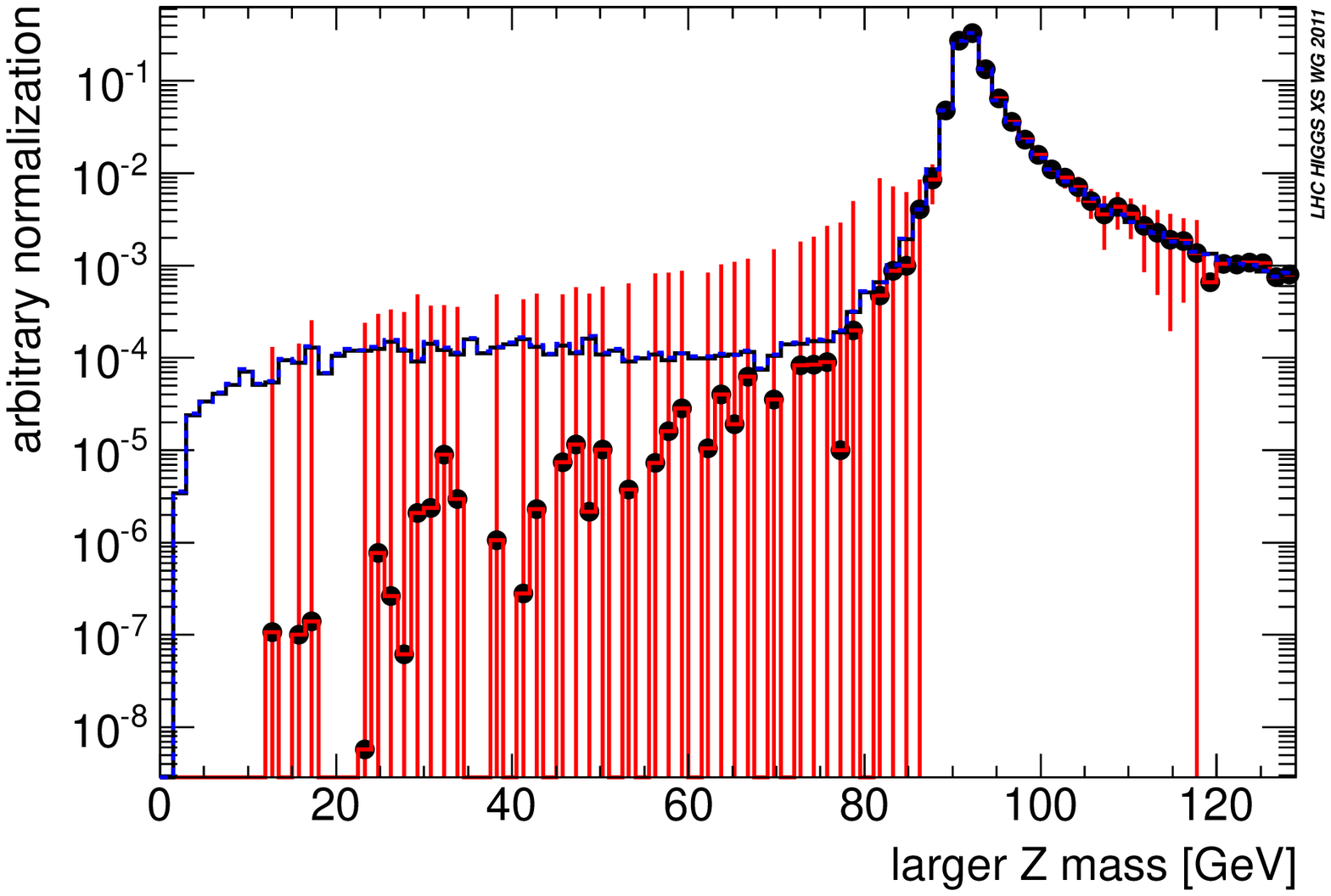} \\
  \includegraphics[width=0.49\textwidth]{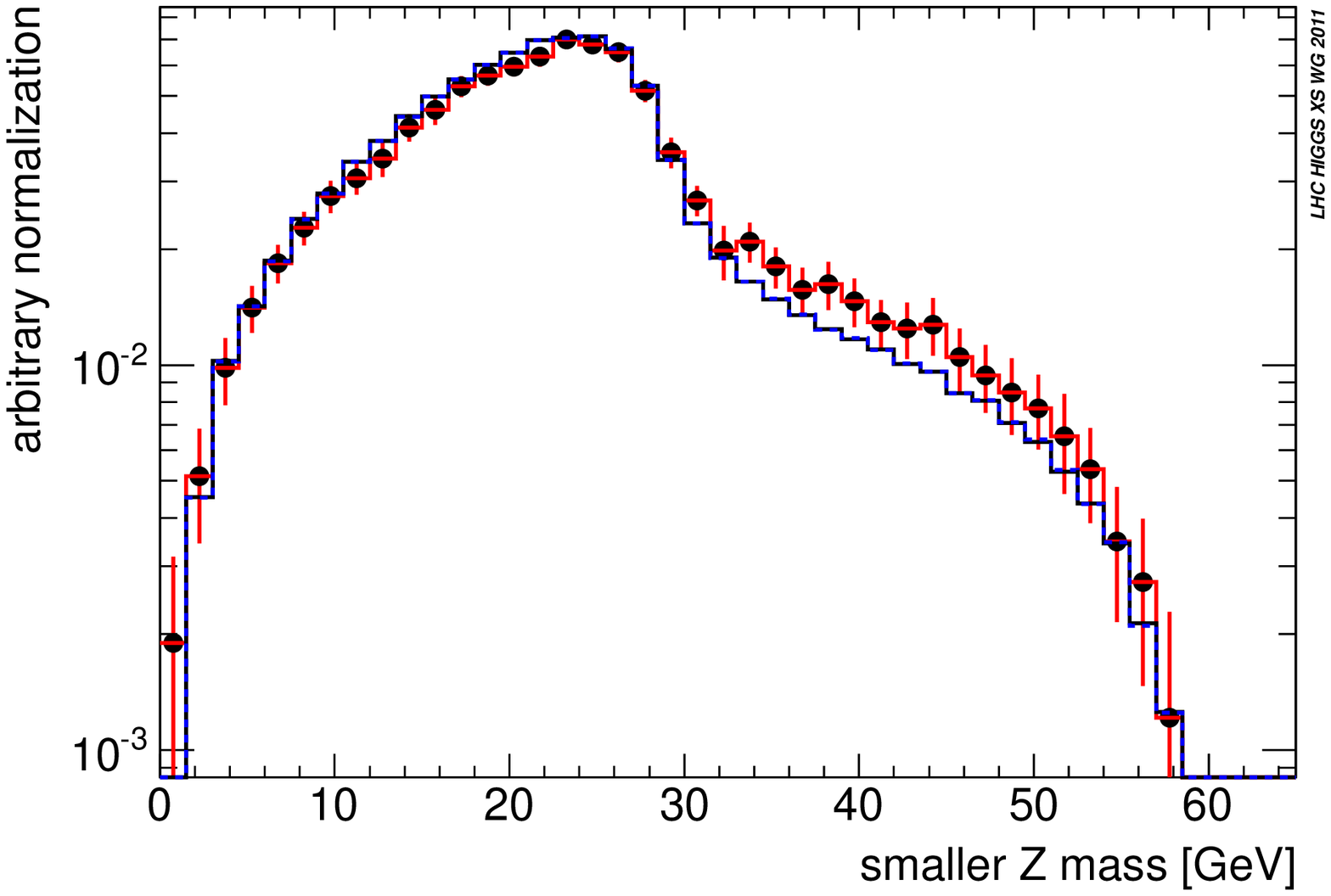}
  \includegraphics[width=0.49\textwidth]{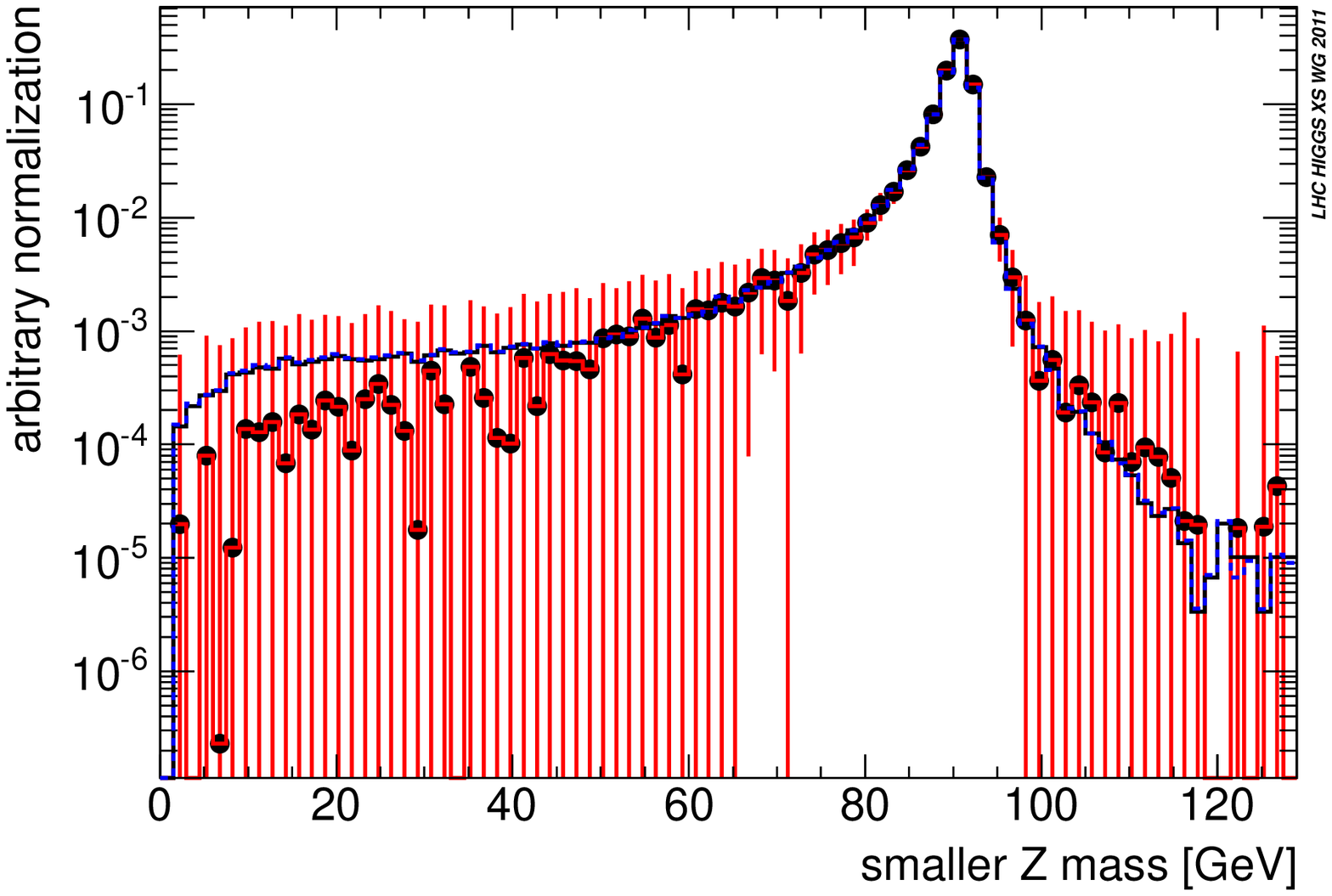}
  \caption{Differential distributions (in arbitrary units) at $7\UTeV$
  for Higgs-boson masses $200\UGeV$ (left) and $500 \UGeV$ (right)
    for three different Monte Carlo generators: \POWHEG{} (black),
  {\sc HNNLO} (red), \POWHEG{} reweighted with {\sc HqT} and {\sc HNNLO} (blue).}
  \label{fig:lo_nlo_nnlo}
\end{figure}

\begin{figure}
  \includegraphics[width=0.49\textwidth]{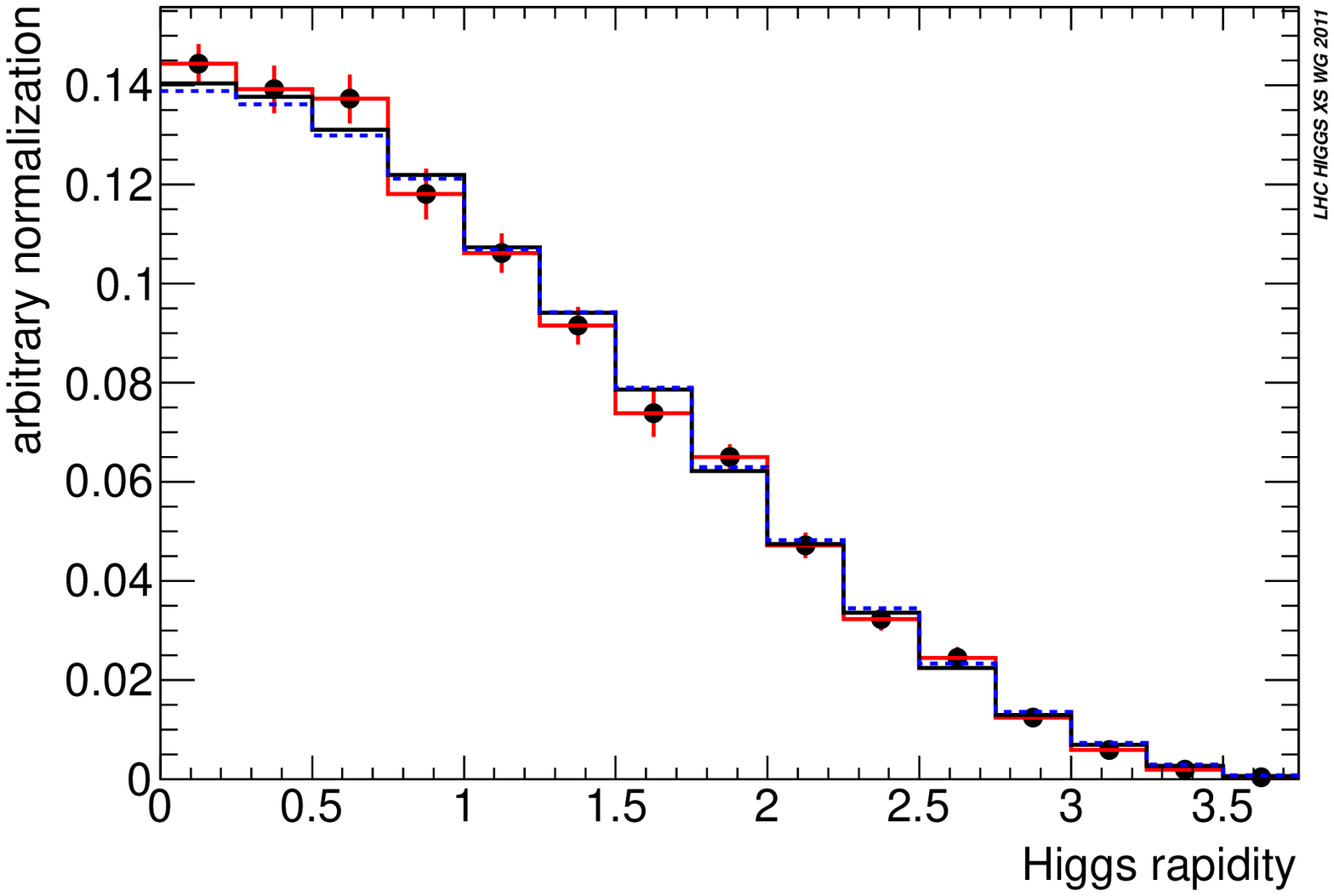}
  \includegraphics[width=0.49\textwidth]{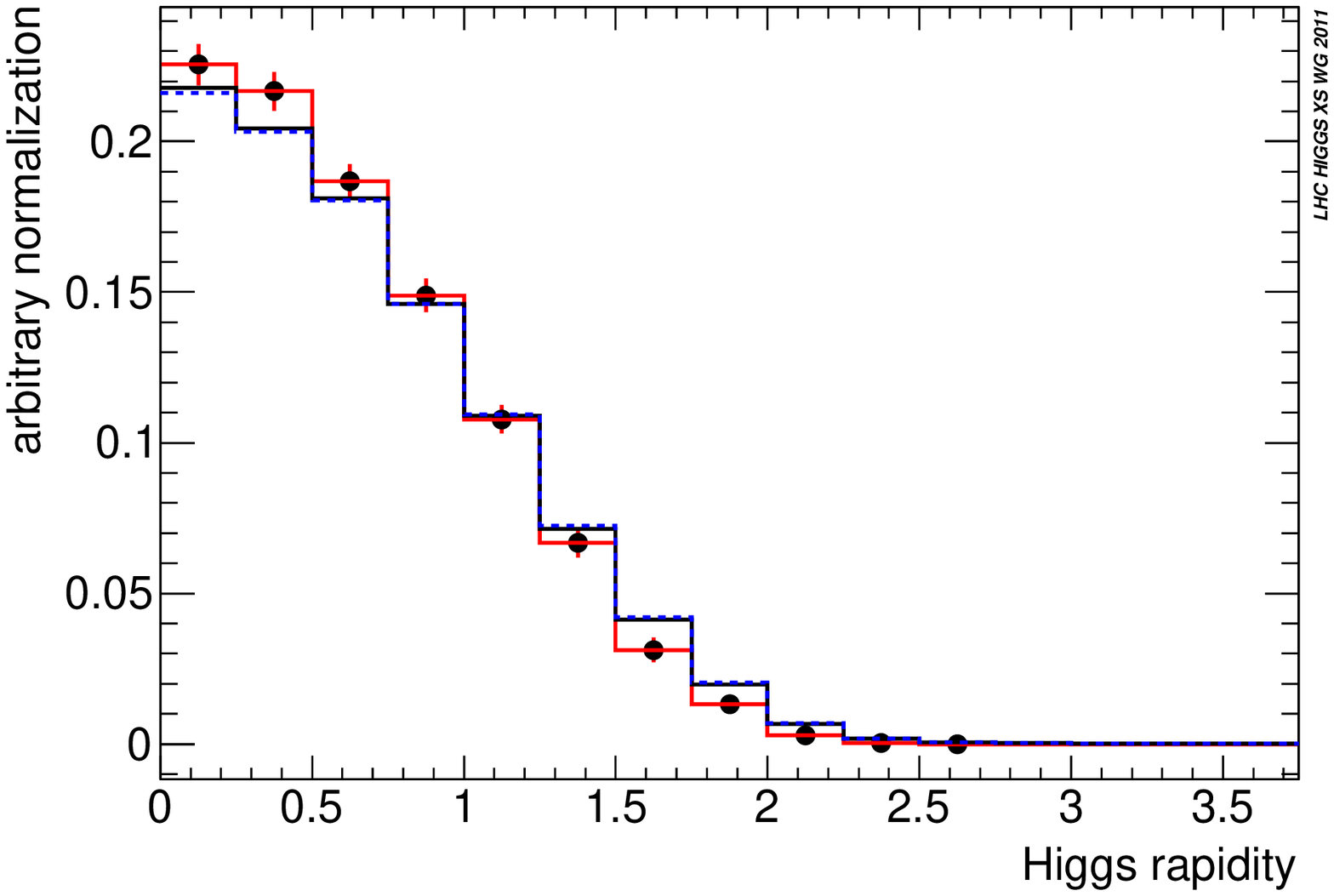} \\
  \includegraphics[width=0.49\textwidth]{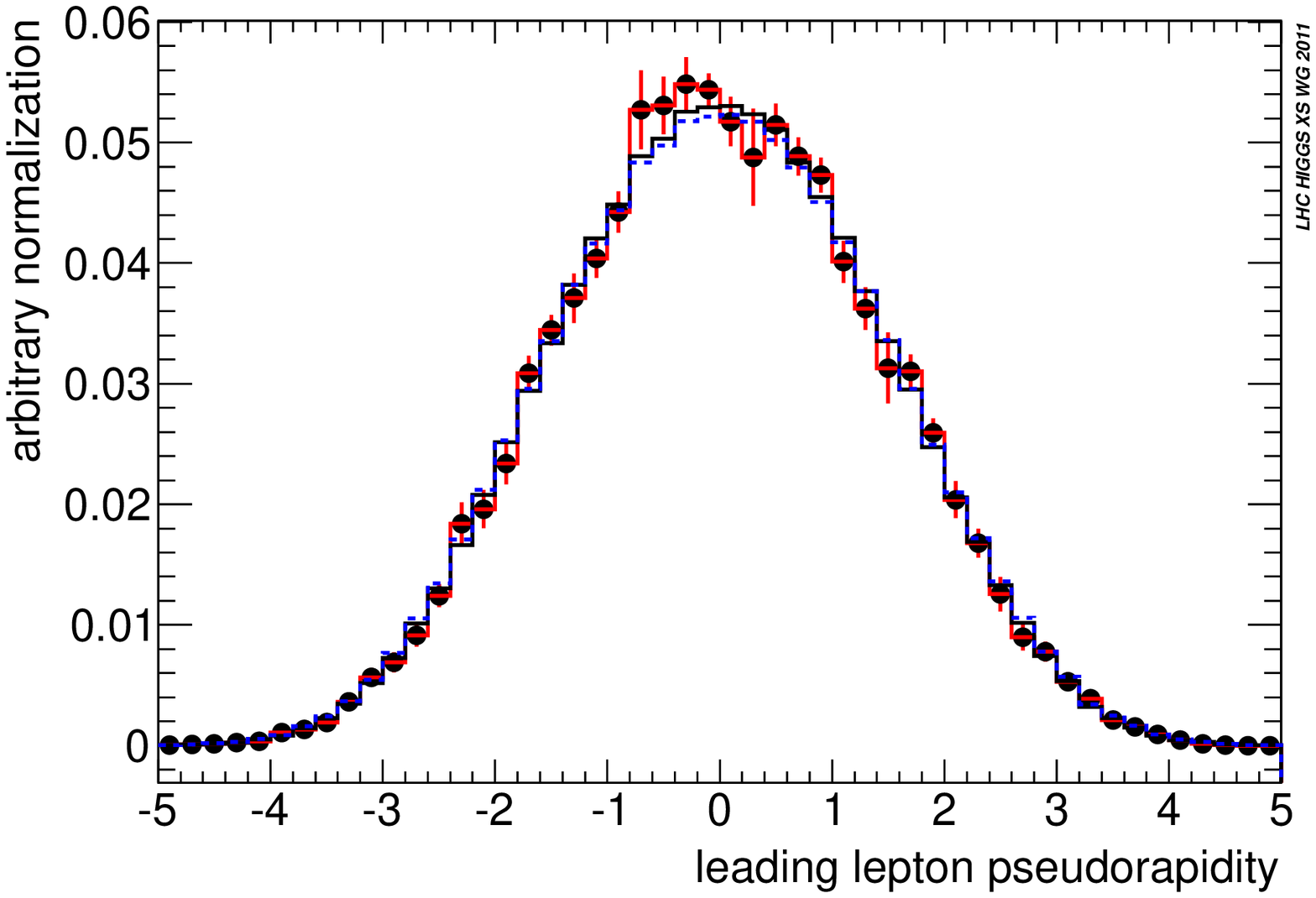}
  \includegraphics[width=0.49\textwidth]{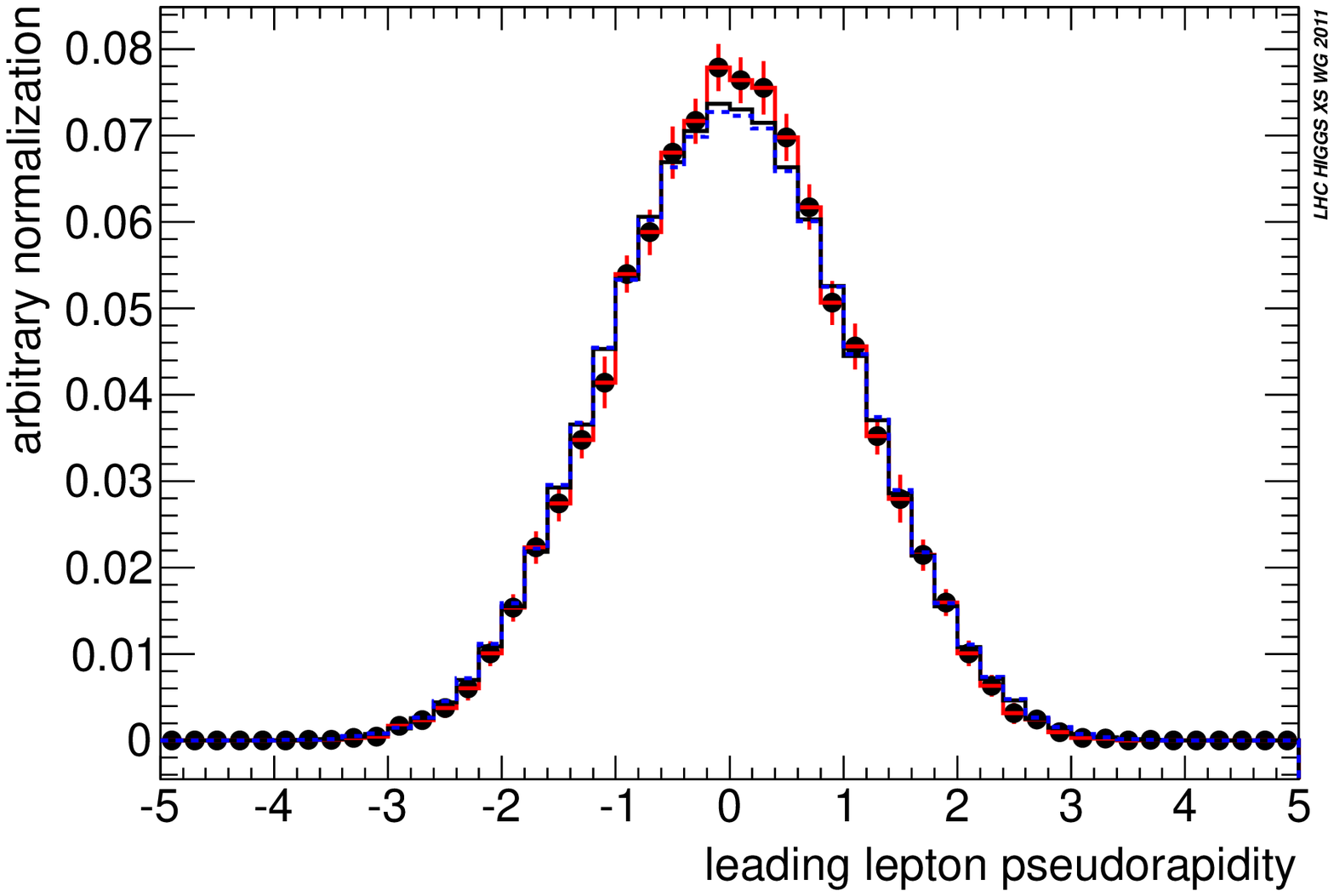}
  \caption{Differential distributions (in arbitrary units) at $7\UTeV$
  for Higgs-boson masses $200\UGeV$ (left) and $500\UGeV$ (right)
    for three different Monte Carlo generators: \POWHEG{} (black),
  {\sc HNNLO} (red), \POWHEG{} reweighted with {\sc HqT} and {\sc HNNLO} (blue).}
  \label{fig:lo_nlo_nnlo_2}
\end{figure}

\subsection{The $\ZZ$ background process}

The understanding of the background processes from the theoretical and
the experimental sides is mandatory to be able to perform a search for
new physics.

The main goal of this section is to compare the predictions from
different generators for the $\ZZ$ process (in particular for the four-lepton final state).
The production of two $\PZ$ bosons represents the irreducible background
for the direct searches for the Higgs boson in the $\PH \rightarrow \ZZ$ channels at LHC. 

The Standard Model input parameters as specified in the Appendix of \Bref{Dittmaier:2011ti} have been used. 
The PDF's used are: CT10, NNPDF2.0, and MSTW2008. The central value of
the cross section and the uncertainty originating from the PDF 
are estimated following the PDF4LHC prescription.
In order to estimate the QCD scale uncertainty the CT10 PDF has
been used as central value and the QCD scale is varied following the prescription of
\Bref{Dittmaier:2011ti}.

Differential distributions and theoretical uncertainties are calculated for a few sets of cuts:
\begin{itemize}
 \item{Cut 1}: $\mzonshell >12  \UGeV$ and $\mzoffshell >12  \UGeV$,
 \item{Cut 2}: $\mzonshell >50  \UGeV$, $\mzoffshell>12  \UGeV$, $\pT(\Pl)>5  \UGeV$, and
 $|\eta(\Pl)|<2.5$,
 \item{Cut 3}: $60 <\mzonshell<120 \UGeV$ and $60 <\mzoffshell<120  \UGeV$. 
\end{itemize}

\subsubsection{$\PQq\PAQq \rightarrow \ZZ$ generators}

\subsubsubsection{{\sc MCFM} predictions}

The {\sc MCFM} program~\cite{MCFMweb,Campbell:1999ah,Campbell:2011bn} v6.1 computes the cross section at LO and NLO
for the process $\PQq\PAQq \rightarrow \Pl\bar{\Pl} \Pl^\prime \bar{\Pl}^\prime$ 
mediated by the exchange of the two bosons $\PZ,\gamma^*$
and their interference, for the doubly-resonant (or $t$-channel) and singly-resonant
(or $s$-channel) diagrams, and for the process $\Pg\Pg\rightarrow \Pl\bar{\Pl} \Pl^\prime \bar{\Pl}^\prime$.

\refF{fig:single_double_mcfm} (left) shows the four-lepton invariant-mass 
distribution obtained with {\sc MCFM} at LO
for  $\ZZ$ production, including or not the singly-resonant (or
$s$-channel) contribution.  The prediction without the singly-resonant
contribution is obtained with {\sc MCFM} v5.8. 
The ``Cut 1'' selection is applied, i.e. ${\mathswitch {m_{\Pl\Pl}}}>12  \UGeV$.

\begin{figure}
  \includegraphics[width=0.49\textwidth, height=0.27\textheight]{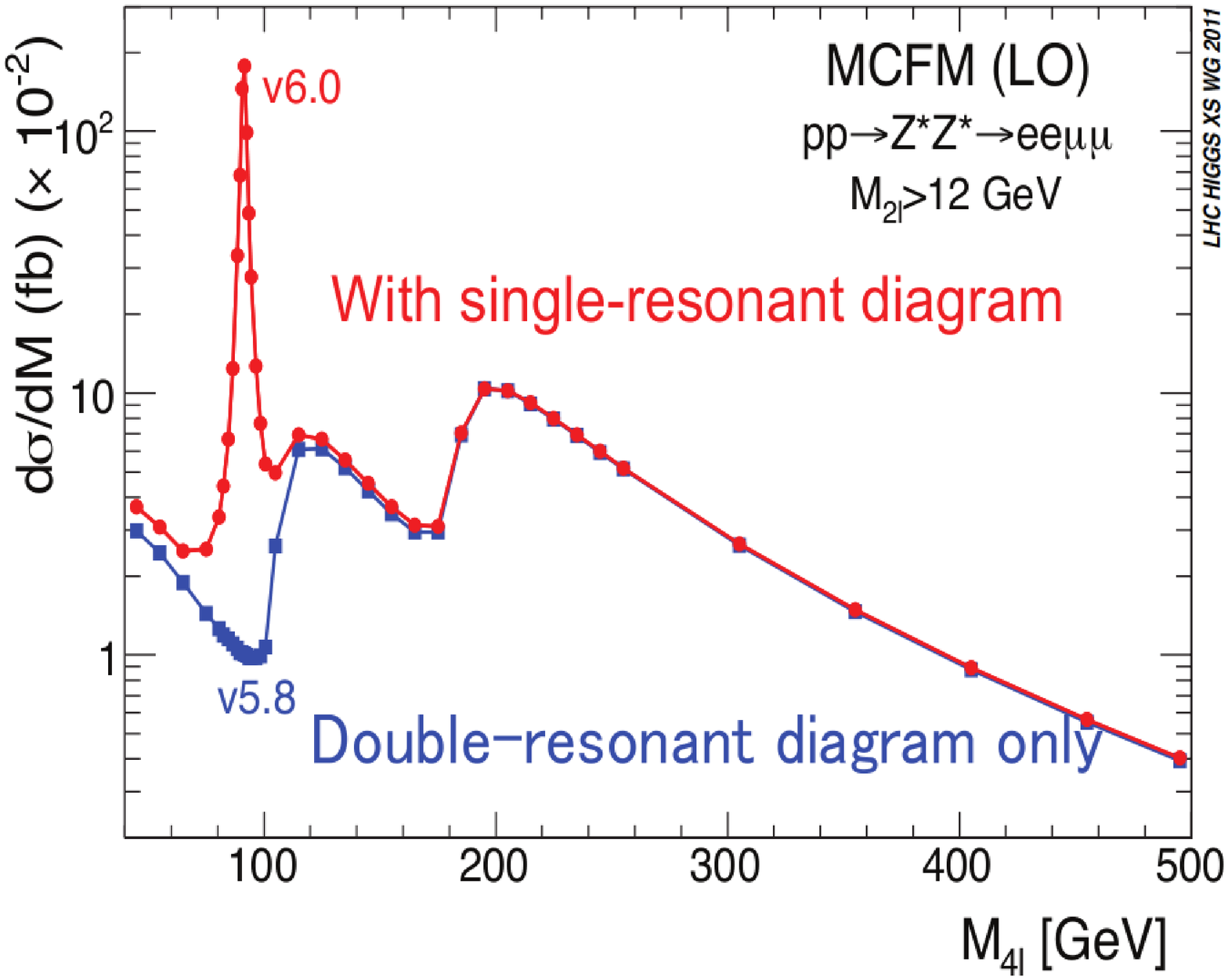}
  \includegraphics[width=0.49\textwidth, height=0.27\textheight]{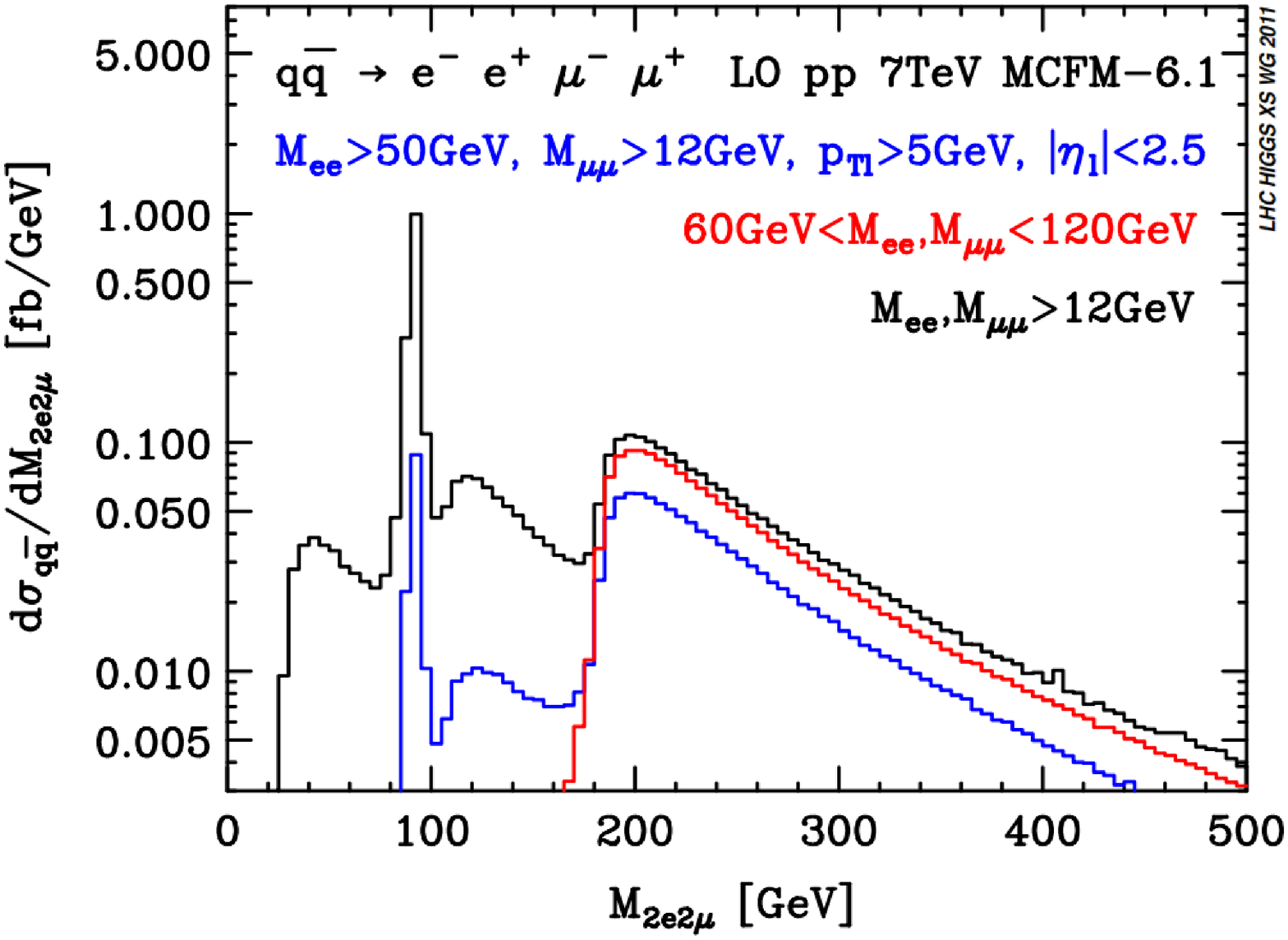}
  \caption{(left) The cross sections for $\PQq\PAQq\rightarrow
    \ZZ^{(*)}\rightarrow{2\Pe 2\PGm}$ as a function
    of ${\mathswitch {m_{2\Pe 2\PGm}}}$ at $7\UTeV$ from {\sc MCFM} for the full calculation
    and without the singly-resonant contribution; (right) LO cross sections for $\PQq\PAQq\rightarrow
    \ZZ^{(*)}\rightarrow {2\Pe 2\PGm}$ as a function of ${\mathswitch {m_{2\Pe 2\PGm}}}$ at $7\UTeV$ from {\sc MCFM} at LO.}  
  \label{fig:single_double_mcfm}
\end{figure}

At LO the cross section as a function of ${\mathswitch {m_{2\Pe 2\PGm}}}$ for the process 
$\PQq\PAQq \rightarrow \ZZ \rightarrow 2\Pe 2\PGm$ is shown in \refF{fig:single_double_mcfm} (right)
and for three different set of cuts.
The black line corresponds to the ``Cut~1'' selection, i.e.\ ${\mathswitch {m_{\Pl\Pl}}} >12
\UGeV$, that is the minimal 
cut requested in the analyses to subtract the contribution to heavy-flavour resonances decaying into
leptons.
The blue line is obtained taking into account the acceptance of the detectors, thus
asking the leptons to have $\pT>5  \UGeV$ and $\eta<2.5$ (``Cut 2'' selection). 
The red line is the differential cross section for the production of
two $\PZ$ bosons, both on shell, i.e.\ asking for $60 \UGeV < m_{\Pl\Pl} <120
\UGeV$ (``Cut 3'' selection).


\refF{fig:ZZ_fig1} (left) shows the cross section at NLO for the full
process $\Pp\Pp \rightarrow \ZZ^{(*)} \rightarrow 2\Pe 2\PGm$ and the
$\Pg\Pg$ contribution separately in the inset, as a function of the
invariant mass of the 4~leptons. The peak at $91\UGeV$ is dominated by the
contribution of the singly-resonant (or $s$-channel) diagrams, while for
${\mathswitch {m_{4\Pl}}} >120  \UGeV$ the doubly-resonant diagrams (or $t$-channel) are
essentially the only contribution.

\begin{figure}
  \includegraphics[width=0.49\textwidth, height=0.3\textheight]{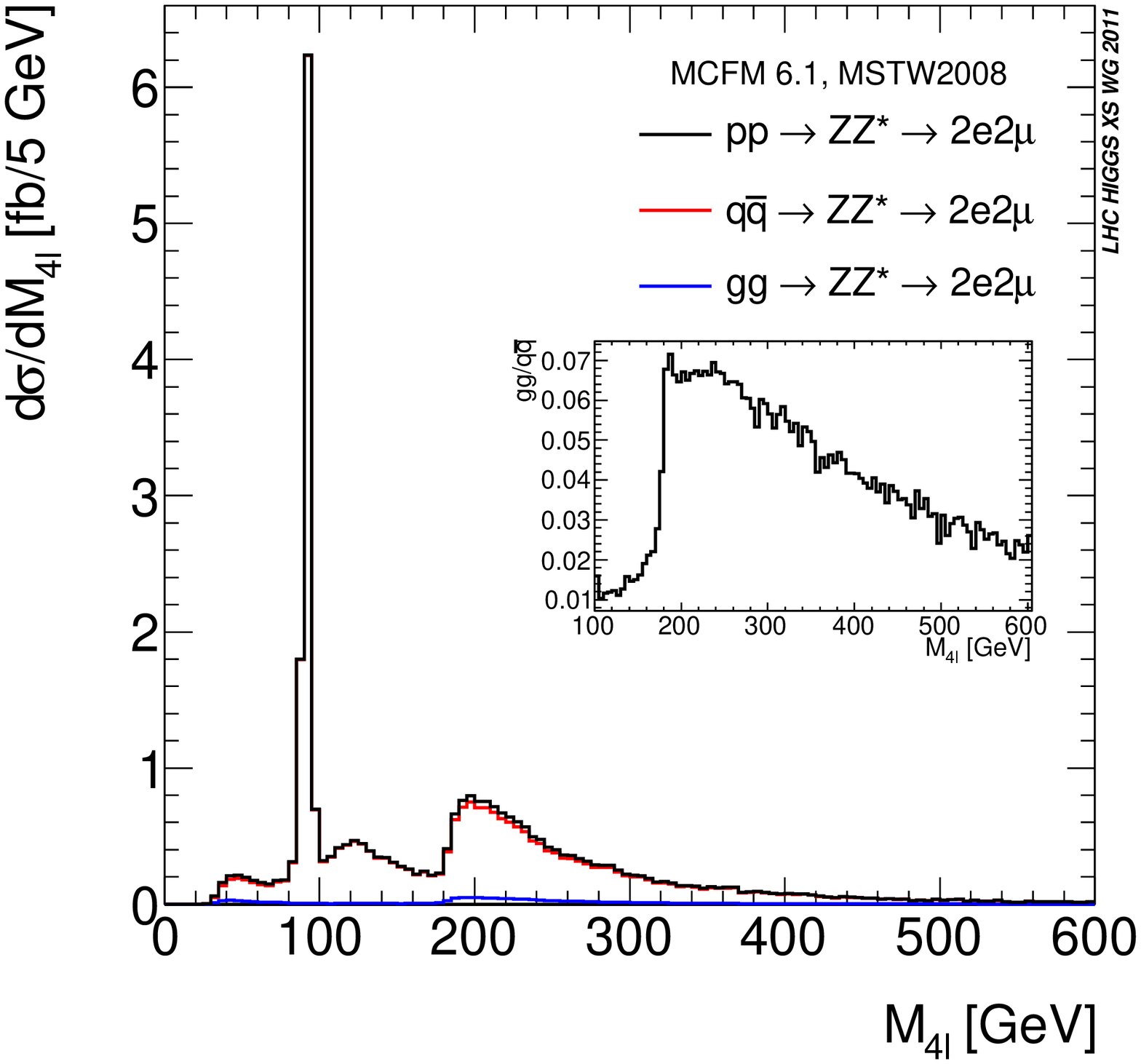}
  \includegraphics[width=0.49\textwidth, height=0.3\textheight]{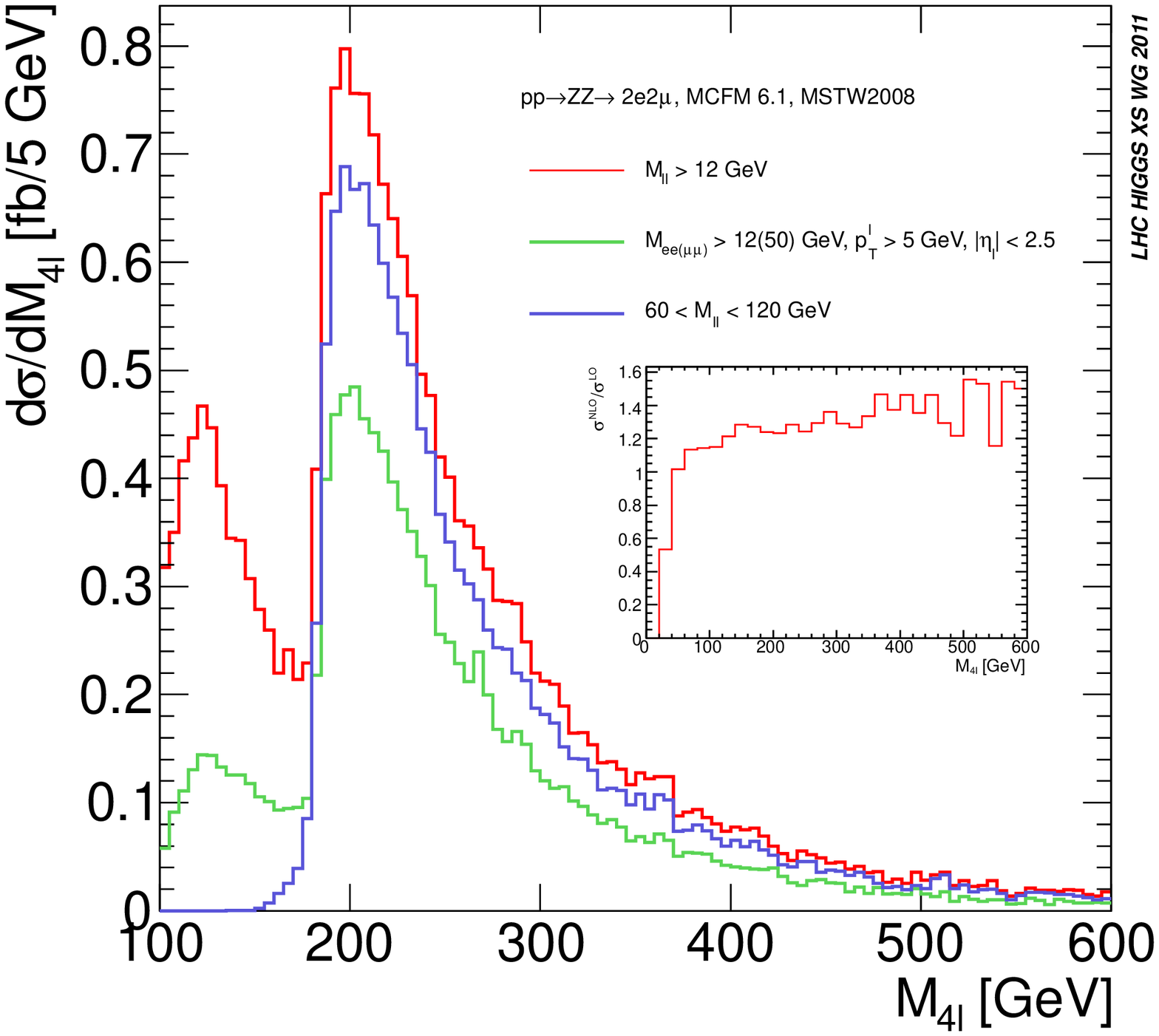}
  \caption{ (Left) The cross sections for $\ZZ^{(*)}$ production as a function
    of ${\mathswitch {m_{2\Pe 2\PGm}}}$ at $7\UTeV$ from {\sc MCFM} at NLO; (right) cross sections for $\ZZ^{(*)}$ production as a function
    of ${\mathswitch {m_{2\Pe 2\PGm}}}$ at $7\UTeV$ from {\sc MCFM} at NLO for three different
    sets of cuts as described in the text.}  
  \label{fig:ZZ_fig1}
\end{figure}

In \refF{fig:ZZ_fig1} (right) the cross section for ${\mathswitch {m_{2\Pe 2\PGm}}}$ from
{\sc MCFM} v6.1 is shown for three different sets of cuts for the
doubly-resonant region only.
In the inset the ratio ``k'' of the NLO and LO predictions is plotted for
the ``Cut 1'' selection. The ``k''-factor depends on 
${\mathswitch {m_{4\Pl}}}$ and increases to about $20\%$ at high masses.



\subsubsubsection{ \POWHEG{} predictions}

In \POWHEGBOX{}~\cite{Alioli:2010xd} a new implementation of the 
vector-boson pair production process at NLO has been
provided~\cite{Melia:2011tj}. The $\PZ/\PGg^*$ interference as well as
singly-resonant contributions are properly included. Interference terms
arising from identical leptons in the final state are considered.

In \refF{fig:cut_pow_mcfm} the four-lepton-invariant-mass
distributions for the final state $2\Pe 2\PGm$ are shown for \POWHEG{},
using two different PDF sets (MSTW and CT10), and compared with {\sc MCFM} v6.1.
The plot on the top-left is obtained asking ${\mathswitch {m_{\Pl\Pl}}}>12  \UGeV$, i.e.\ the
``Cut~1''. The plot on the top-right is asking the second sets of cuts, ``Cut~2''. 
The one on the bottom is for ``Cut~3'' selection, i.e.\ the two $\PZ$ bosons
in the vicinity of their mass shells.
The two programs show a very good agreement. The difference due to the
different PDF sets is discussed in \refS{ZZ_uncert}.
\begin{figure}
  \includegraphics[width=0.49\textwidth]{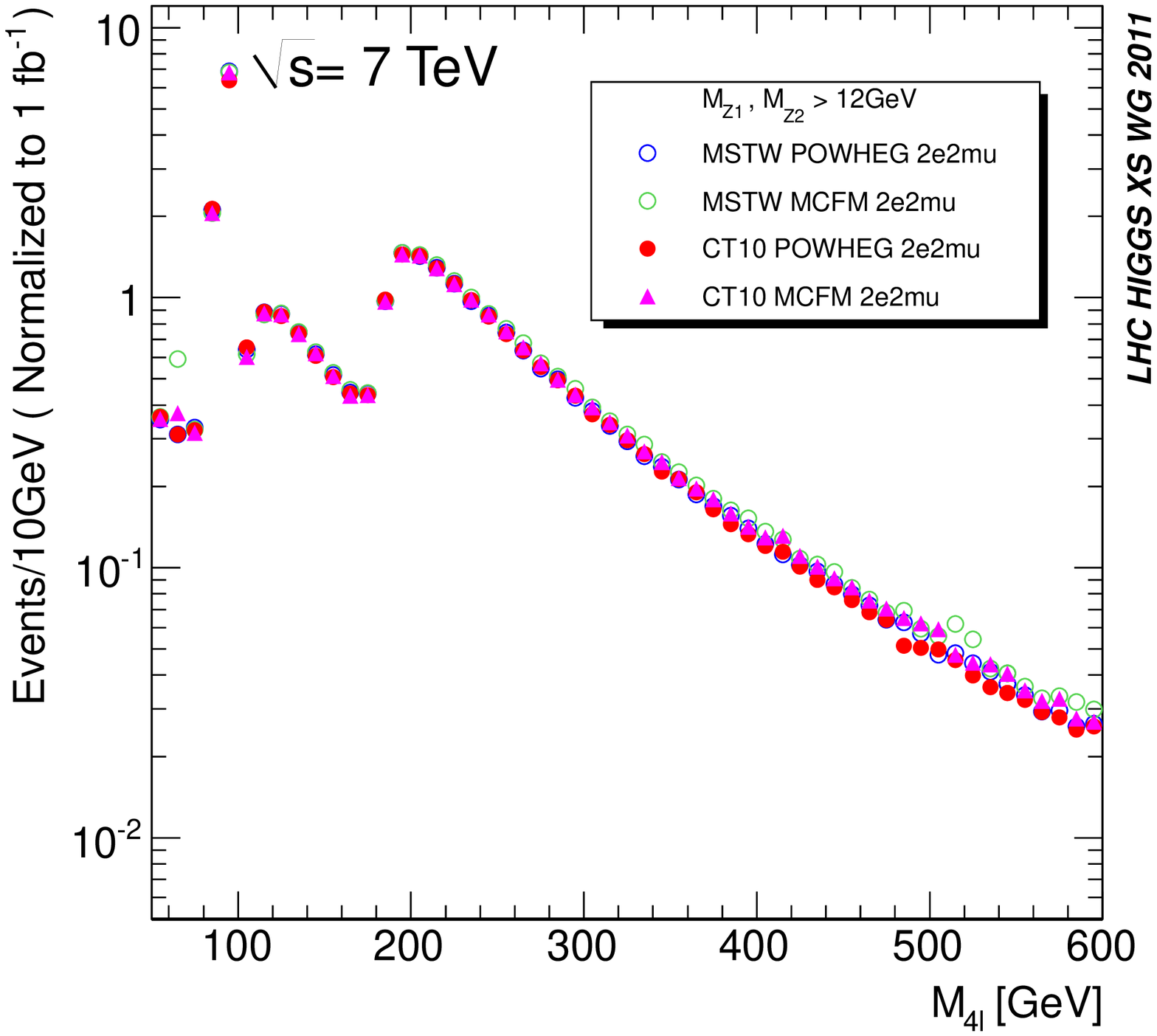} 
  \includegraphics[width=0.49\textwidth]{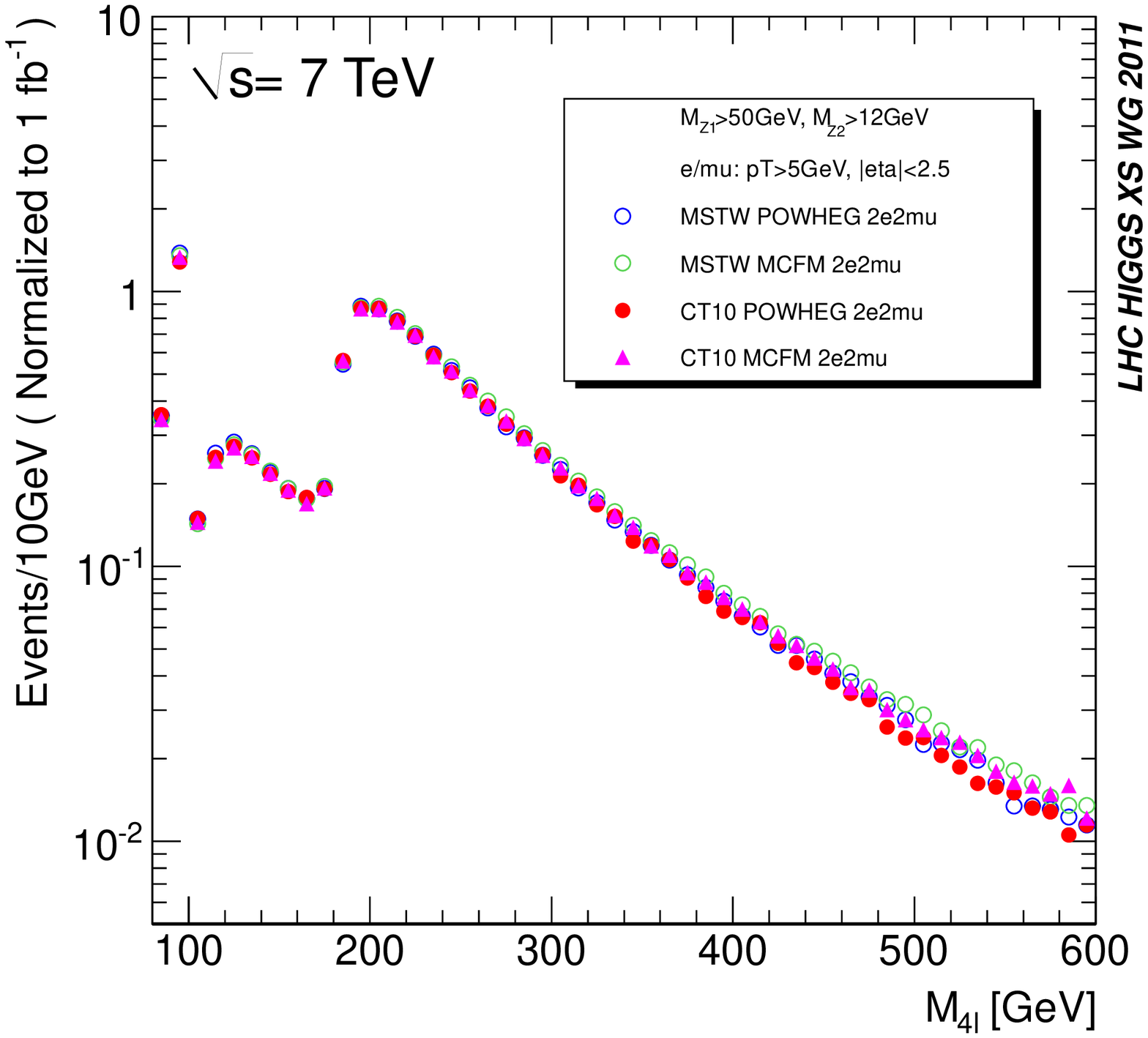}  \\
  \includegraphics[width=0.49\textwidth]{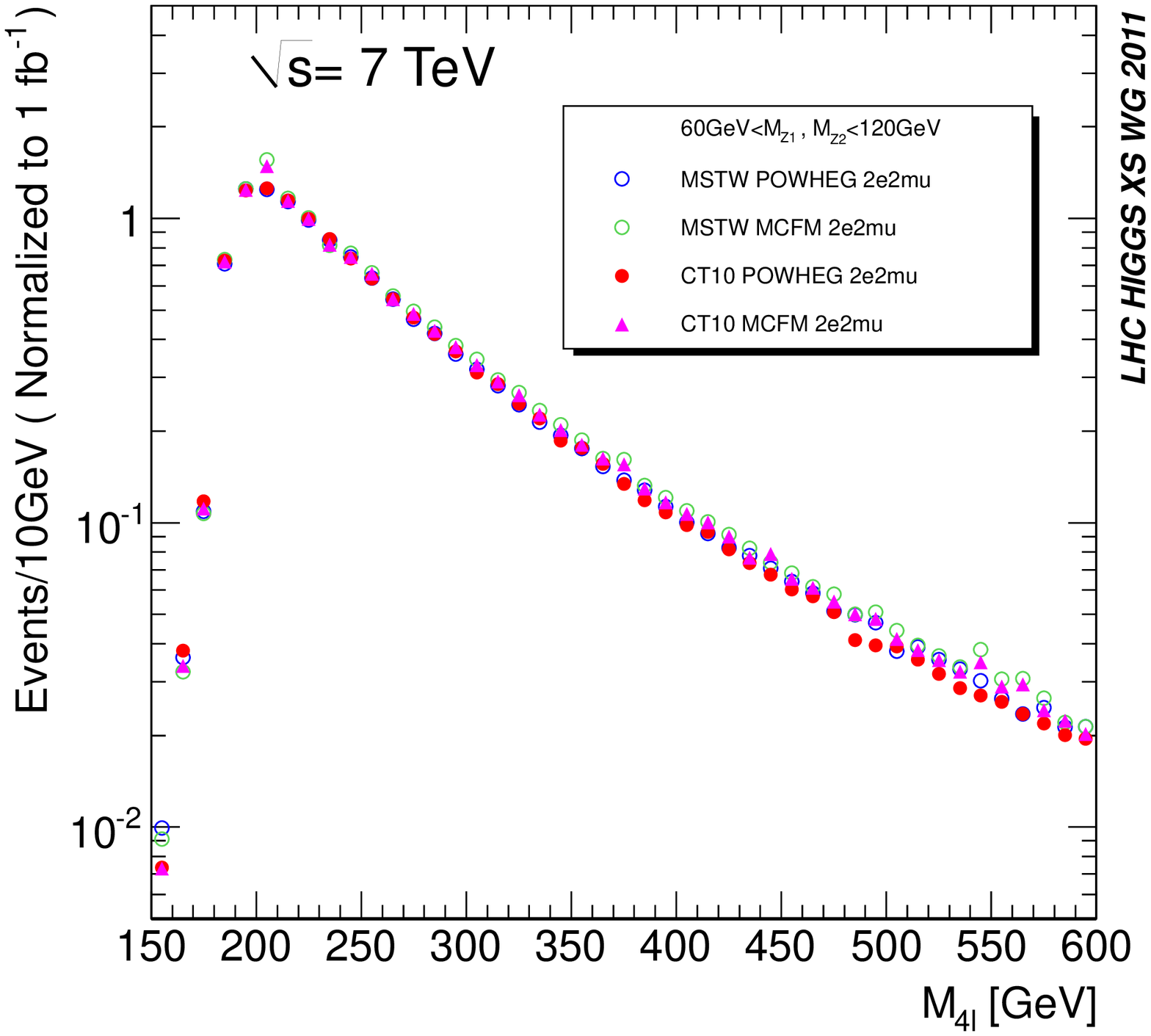} 
  \vspace*{-1em}
  \caption{$ZZ\rightarrow 2\Pe 2\PGm$ invariant-mass distributions at $7  \UTeV$ with \POWHEG{} and {\sc MCFM v6.1}. Two
  different PDF sets are used. The top-left plot is obtained applying only 
${\mathswitch {m_{2l}}}>12  \UGeV$; the top-right plot 
asking $\mzonshell >50  \UGeV$, $\mzoffshell>12  \UGeV$, $\pT(\Pl)>5
    \UGeV$,  $|\eta(\Pl)|<2.5$, and the bottom plot asking 
$60 <\mzonshell<120 \UGeV$ and $60 <\mzoffshell<120  \UGeV$.}
  \label{fig:cut_pow_mcfm}
\end{figure}

In \refF{fig:ZZ_interf} the ${\mathswitch {m_{4\Pl}}}$ distribution is shown for 
the $2\Pe 2\PGm$ and $4\Pe$ final states in order to see the effect of the
interference between identical leptons.
The two top figures are obtained with the ``Cut~1'' selection,  the middle
ones with ``Cut~2'' selection, and the bottom ones with ``Cut~3'' selection. 
The total cross section increases by about $4\%$ because of the interference. 
The effects of using two different PDF sets
is also seen. 

\begin{figure}
  \includegraphics[width=.99\textwidth,height=0.30\textheight]{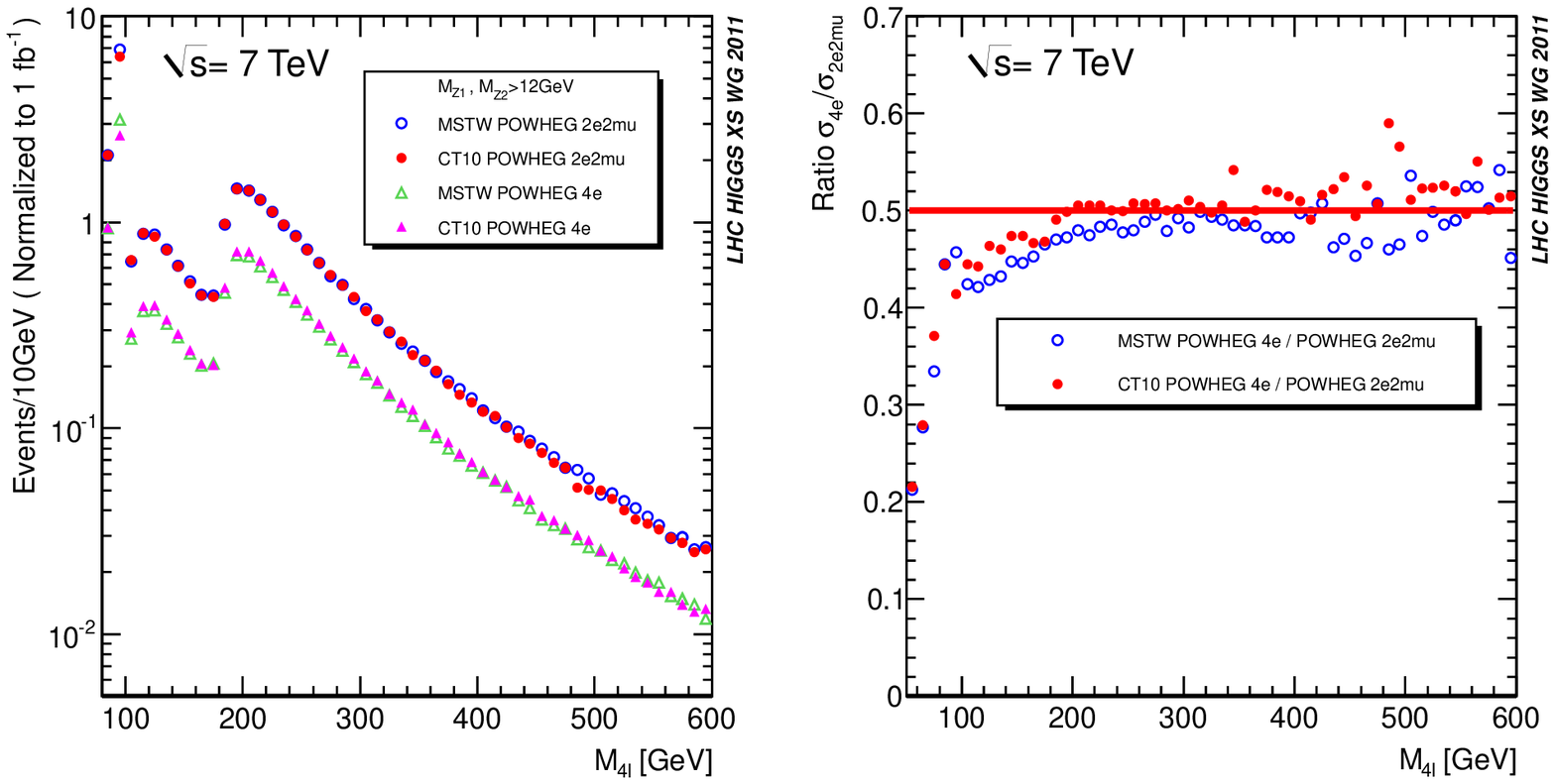} 
  \\[-1em]
  \includegraphics[width=.99\textwidth,height=0.30\textheight]{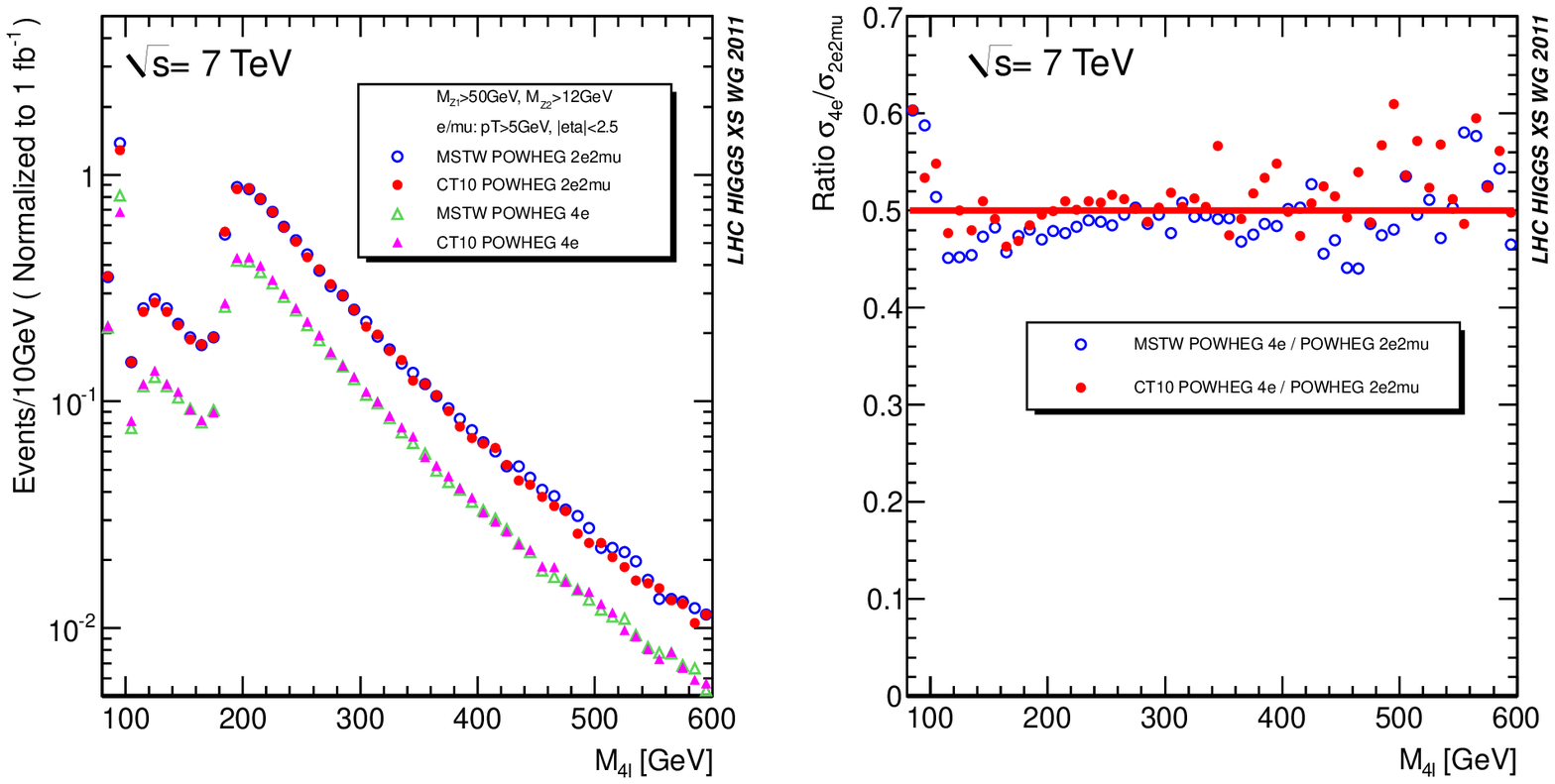} 
  \\[-1em]
  \includegraphics[width=.99\textwidth,height=0.30\textheight]{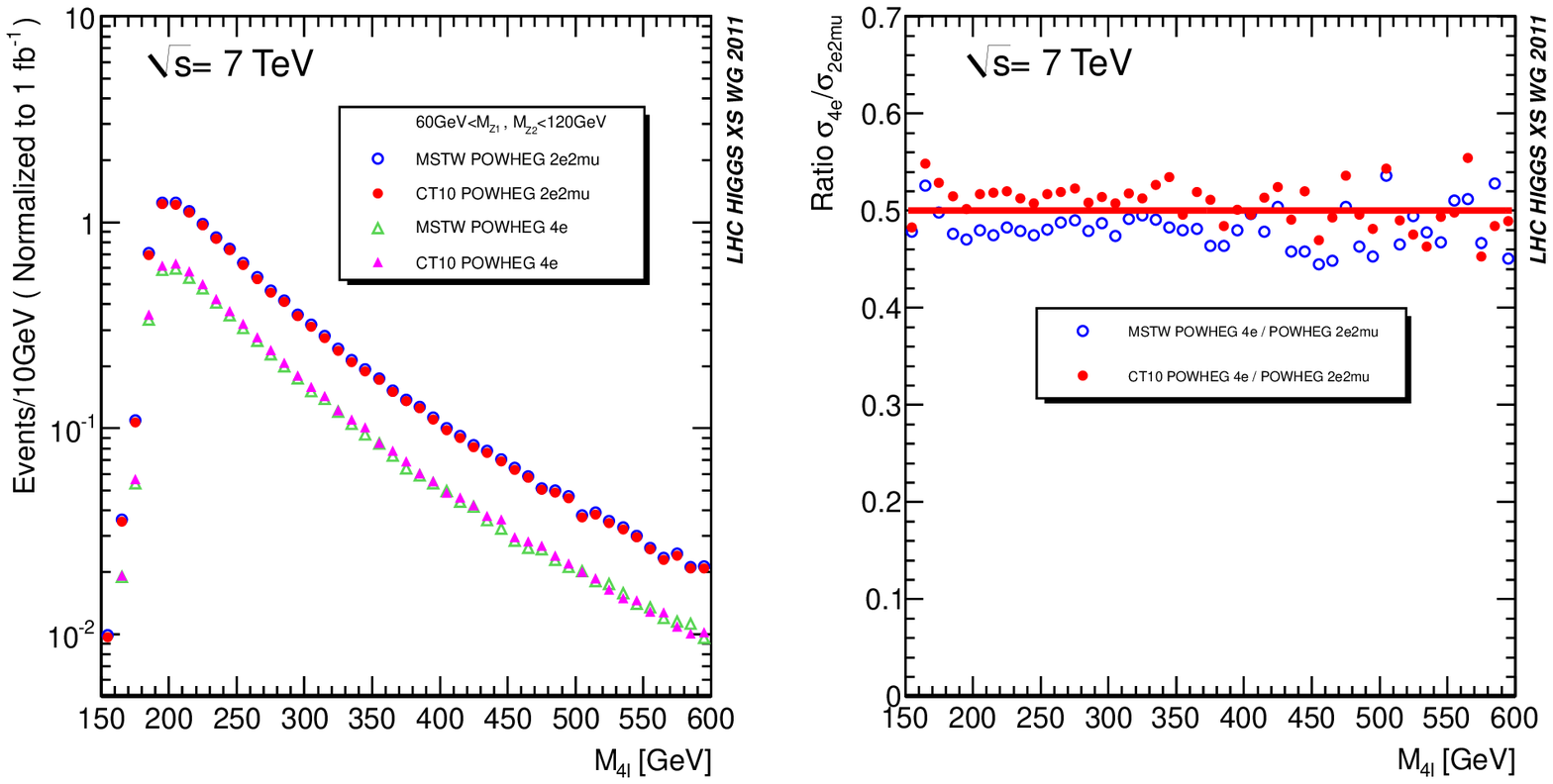} 
  \vspace*{-1em}
  \caption{$\ZZ\rightarrow 2\Pe 2\PGm$ and $\ZZ\rightarrow 4\Pe$ invariant-mass 
  distributions and their ratio at $7\UTeV$ with \POWHEG{}. Two
  different PDF sets are used. The upper plots are obtained applying only
  ${\mathswitch {m_{2\Pl}}}>12  \UGeV$, the middle plots applying $\mzonshell>50  \UGeV$, $\mzoffshell>12  \UGeV$, $\pT(\Pl)>5 
  \UGeV$,  $|\eta(\Pl)|<2.5$, and the bottom plots applying 
  $60 <\mzonshell<120 \UGeV$ and $60 <\mzoffshell<120  \UGeV$.}
  \label{fig:ZZ_interf}
\end{figure}

A full comparison has been also performed between the predictions of both \POWHEG{} at NLO and {\sc PYTHIA} at LO. 
This is shown in \refF{figZZ_5} 
where the distributions are normalised to the the corresponding cross sections. The differences come from the lack of the singly-resonant contribution in {\sc PYTHIA} 
and the LO calculations for the doubly-resonant contribution. 
\begin{figure}
\begin{center}
  \includegraphics[width=0.7\textwidth]{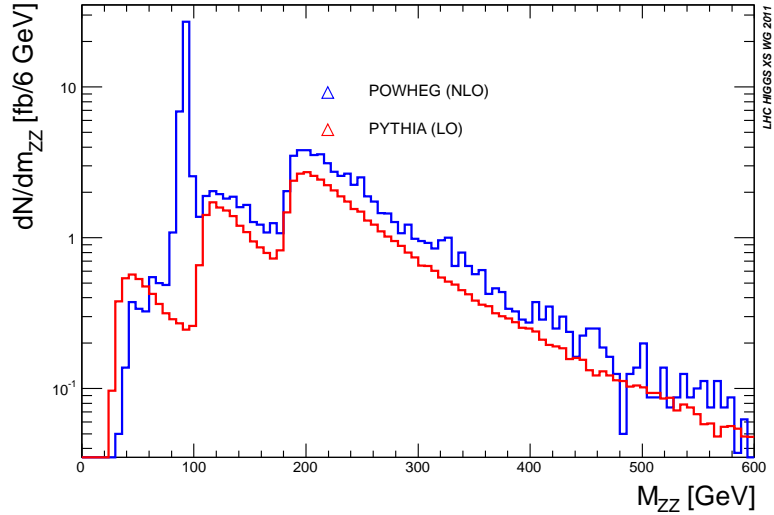} 
  \caption{$\ZZ\rightarrow 4\mu$ invariant-mass distributions at
  $7\UTeV$ as derived by {\sc PYTHIA} and \POWHEG} 
  \label{figZZ_5}
\end{center}
\end{figure}


\subsubsection{$\Pg\Pg\rightarrow \ZZ$ generators}

The gluon-induced $\ZZ$ background, although technically of NNLO compared
to the lowest-order Z-pair production, 
can amount to a substantial fraction of the total irreducible background.
We take into account these diagrams, but a
full NNLO calculation for the process $\PQq\PAQq\rightarrow \ZZ$ is not available.
The contributions of the $\Pg\Pg$ diagrams are included in
{\sc MCFM v6.1} and
the dedicated tool {\sc GG2ZZ}~\cite{Binoth:2008pr}, which computes the 
$\mathrm{\Pg\Pg \rightarrow {\PZ}{\PZ}}$ at LO, which is of order
$\alphas^{2}$, compared to $\alphas^{0}$ for the LO 
$\mathrm{\PQq\PAQq\rightarrow \ZZ}$. 

The two programs provide the functionality to compute the cross section after
applying a cut on the minimally generated invariant mass of the
same-flavour lepton pairs (which can be interpreted as the
$\PZ/\PGg^*$ invariant mass), ${\mathswitch {m_{\Pl\Pl}}}^\mathrm{min}=12\UGeV$.



\subsubsection{{\boldmath $\Pg\Pg\to \PZ\PZ$: Comparison of {\sc gg2ZZ} and {\sc MCFM}}}

In this section, a comparison of results for the 
$\Pg\Pg \to \PZ(\PGg^\ast)\PZ(\PGg^\ast) \to \Pl\bar{\Pl}\Pl'\bar{\Pl'}$ ($\Pl={}$charged lepton) 
continuum background calculated at LO with {\sc gg2ZZ} and {\sc MCFM} is presented.\footnote{%
Off-shell results for the same-flavour process $\Pg\Pg \to \ZZ \to \Pl\bar{\Pl}\Pl\bar{\Pl}$
have been presented in Ref.\ \protect\cite{Zecher:1994kb}.} The {\sc MCFM} calculation of
the $\Pg\Pg$ subprocess is described in \Bref{Campbell:2011bn}.  
{\sc MCFM} also includes a NLO calculation of the $\PQq\PAQq$ subprocess 
\cite{Campbell:1999ah,Campbell:2011bn}. For {\sc gg2ZZ}, first results have been
presented in \Bref{Binoth:2008pr}.  {\sc gg2ZZ} employs the same calculational 
techniques as {\sc gg2WW} \cite{Binoth:2005ua,Binoth:2006mf}.  
Because of the Furry and (generalised) Landau--Yang
theorems, all graphs with $\Pg\Pg V$ triangle quark loop 
(including those with $s$-channel 
$\PZ$ propagator) vanish.  The {\sc MCFM} calculation 
exploits that the top quark decouples in good approximation. The 
$\Pt$ contribution to the quark loop is therefore neglected.  The $\Pb$ contribution 
is included in the massless limit.  In {\sc gg2ZZ}, the $\Pb$ and $\Pt$ contributions are 
included with finite masses.

Parton-level cross sections and $\MZZ$ distributions for $\Pp \Pp$ collisions at 
$\sqrt{s} = 7 \UTeV$ are compared for the three sets of cuts ``Cut 1'', ``Cut 2'' and ``Cut 3''.
In addition, a $\pT(\PZ) > 0.05  \UGeV$ cut is applied to prevent that numerical 
instabilities spoil the amplitude evaluation.  This technical cut reduces the 
cross sections by at most $0.05\%$.
Results are given for a single lepton flavour combination, e.g.\ %
$\Pl=\Pe^-,\Pl'=\mu^-$.  Lepton masses are neglected.
The input-parameter set of \Bref{Dittmaier:2011ti}, App.\ A,   
is used with NLO $\Gamma_{\PZ}$ and $G_\mu$ scheme. 
The renormalisation and factorisation scales are set 
to $\MZ$.  The PDF set MSTW2008 NNLO with 3-loop running for 
$\alphas(\mu^2)$ and $\alphas(\MZ^2)=0.11707$ is used. The fixed-width 
prescription is used for $\PZ$ propagators.

For cut sets ``Cut 1'', ``Cut 2'' and ``Cut 3'', results calculated with {\sc gg2ZZ} and {\sc MCFM} are 
shown in \refT{tab_ggZZ_comparison:1} (cross section) and
\refF{fig_ggZZ_comparison:1} (left, $\MZZ$ distribution).  
The cross sections agree up to $1\%$ or better.
Nevertheless, the residual deviations are significant relative to the MC 
integration errors.  The $\MZZ$ distributions agree, except in the
vicinity of the peaks at $50$ and $200\UGeV$.  In these invariant-mass regions, 
{\sc MCFM}'s differential cross section is slightly larger than {\sc gg2ZZ}'s.
Table~\ref{tab_ggZZ_comparison:2} and \refF{fig_ggZZ_comparison:1} (right) demonstrate that
the observed deviations are consistent with differences of the
calculations in {\sc gg2ZZ} and {\sc MCFM} as discussed above.  In the limit 
$\Mb\to 0$ and $\Mt \to \infty$, {\sc gg2ZZ}'s results are expected to 
agree with {\sc MCFM}'s results within MC integration errors.  
In \refT{tab_ggZZ_comparison:2} and \refF{fig_ggZZ_comparison:1} (right), this is
confirmed for cut set ``Cut 1''.  Note that for cut sets ``Cut 1'' and ``Cut 3'' the finite-mass effects
decrease the cross section, while for cut set ``Cut 2'' the cross section is increased
(see \refT{tab_ggZZ_comparison:1}).\footnote{For cut set ``Cut 2'', one obtains
$0.6297(4)\Ufb$ with {\sc gg2ZZ}-2.0 and $\Mt = 10^5\UGeV$, $\Mb = 10^{-4}\UGeV$, which agrees
with {\sc MCFM}-6.1's result.}
\begin{table}
  \caption{Cross sections in fb for $\Pg\Pg \to \PZ(\PGg^\ast)\PZ(\PGg^\ast) \to \Pl\bar{\Pl}\Pl'\bar{\Pl'}$
  in $\Pp \Pp$ collisions at $\sqrt{s} = 7\UTeV$ and a single lepton flavour combination 
  calculated at LO with {\sc gg2ZZ}-2.0 and {\sc MCFM}-6.1.  Three cut sets are applied:
  $\mzonshell>12 \UGeV$, $\mzoffshell>12\UGeV$ (Cut 1);
  $\mzonshell>50 \UGeV$, $\mzoffshell>12\UGeV$, $\pT(\Pl)>5\UGeV$, $|\eta(\Pl)|<2.5$ (Cut 2);
  $60\UGeV<\mzonshell<120\UGeV$, $60\UGeV<\mzoffshell<120\UGeV$ (Cut 3).}
  \vspace{0cm}
  \label{tab_ggZZ_comparison:1}
  \begin{center}
  \renewcommand{\arraystretch}{1.2}
  \begin{tabular}{lccc}
  \hline
  \multicolumn{1}{c}{} & \multicolumn{3}{c}{$\sigma(\Pg\Pg \to \PZ(\PGg^\ast)\PZ(\PGg^\ast) \to \Pl\bar{\Pl}\Pl'\bar{\Pl'})$ [\Ufb], $\Pp\Pp$, $\sqrt{s} = 7 \UTeV$} \\
  \hline
  \multicolumn{1}{c}{} & {\sc gg2ZZ}-2.0 & {\sc MCFM}-6.1 & $\sigma_\text{{\sc MCFM}}/\sigma_\text{{\sc gg2ZZ}}$ \\
  \hline
  Cut 1 & $1.157(2) $&$ 1.168(2)$ & $1.010(2)$ \\
  Cut 2 & $ 0.6317(4)$ &$ 0.6293(4)$ & $0.9962(8)$ \\
  Cut 3 & $0.8328(3)$ & $0.8343(3)$ & $1.0019(5)$ \\
  \hline
  \end{tabular}\end{center}
\end{table}

\begin{table}
  \caption{Effects of finite $\Pb$ and $\Pt$ masses for the cross section of $\Pg\Pg \to \PZ(\PGg^\ast)\PZ(\PGg^\ast) \to \Pl\bar{\Pl}\Pl'\bar{\Pl'}$ when cut set ``Cut 1'' is applied.  Other details as in \refT{tab_ggZZ_comparison:1}.}
  \vspace{0cm}
  \label{tab_ggZZ_comparison:2}
  \begin{center}
  \renewcommand{\arraystretch}{1.2}
  \begin{tabular}{lcc}
  \hline
  Cut 1: $\mzonshell >12$ GeV, $\mzoffshell >12$$\UGeV$& \multicolumn{1}{c}{$\sigma$ [\Ufb]} & $\sigma_\text{{\sc MCFM}}/\sigma_\text{{\sc gg2ZZ}}$ \\
  \hline
  {\sc gg2ZZ}-2.0 ($\Mt = 172.5$ $\UGeV$, $\Mb = 4.75$ \UGeV) & $1.157(2)$ & $1.010(2)$ \\
  {\sc MCFM}-6.1 (5 massless quark flavours) & 1.168(2) & --- \\
  {\sc gg2ZZ}-2.0 ($\Mt = 10^5$ $\UGeV$, $\Mb = 10^{-4}$ \UGeV) & $1.1677(8)$ & $1.001(2)$ \\
  \hline
  \end{tabular}\end{center}
\end{table}

\begin{figure}
  \includegraphics[width=0.49\textwidth]{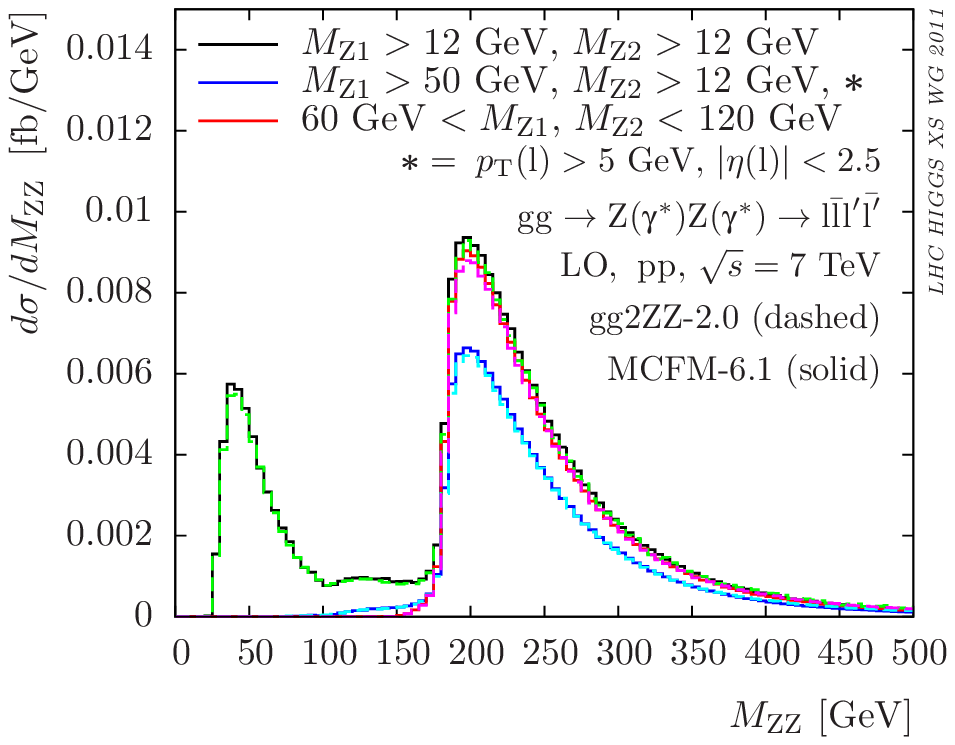}
  \includegraphics[width=0.49\textwidth]{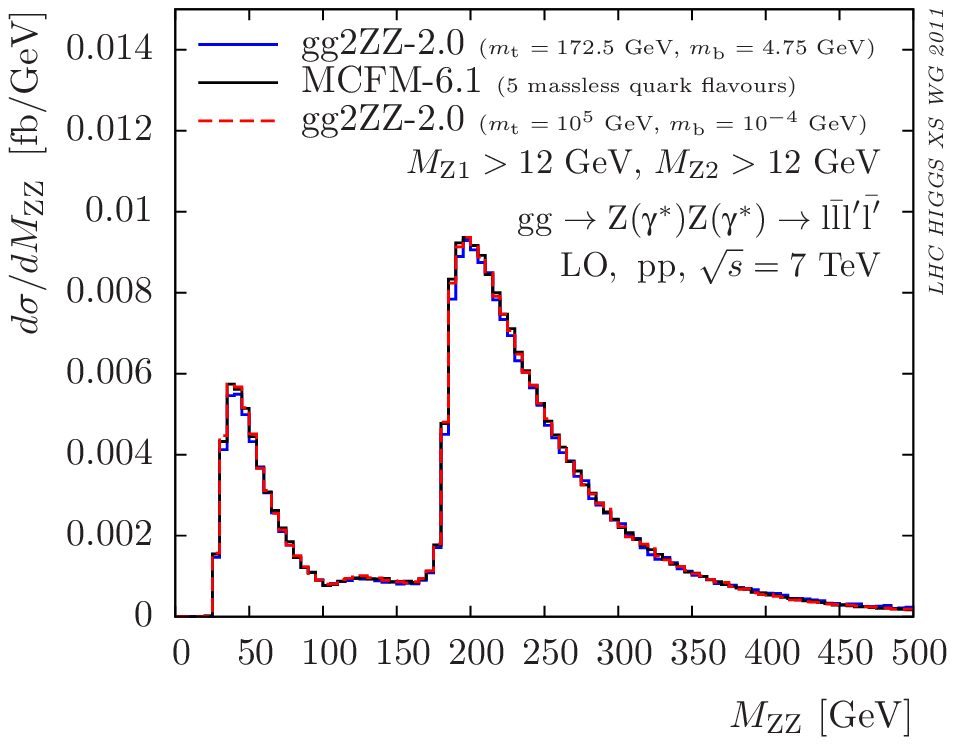}
  \caption{ (left) $\MZZ$ distributions for $\Pg\Pg \to \PZ(\PGg^\ast)\PZ(\PGg^\ast) \to \Pl\bar{\Pl}\Pl'\bar{\Pl'}$ calculated with {\sc MCFM}-6.1 (Cut 1: black, Cut 2: blue, Cut 3: red) and {\sc gg2ZZ}-2.0 (Cut 1: green dashed, Cut 2: cyan dashed, Cut 3: magenta dashed), other details as in \refT{tab_ggZZ_comparison:1}; 
  (right) effects of finite $\Pb$ and $\Pt$ masses for the $\MZZ$ distribution of $\Pg\Pg \to \PZ(\PGg^\ast)\PZ(\PGg^\ast) \to \Pl\bar{\Pl}\Pl'\bar{\Pl'}$ when cut set Cut 1 is applied,  other details as in \refT{tab_ggZZ_comparison:1}.}
  \label{fig_ggZZ_comparison:1}
\end{figure}

\subsubsection{Theoretical uncertainties}
\label{ZZ_uncert}





PDF$+\alphas$ and QCD scale uncertainties for $\Pp\Pp \rightarrow \ZZ
\rightarrow  2\Pe 2\PGm$ at NLO and 
$\Pg\Pg \rightarrow \ZZ \rightarrow  2\Pe 2\PGm$ 
are evaluated using {\sc MCFM} version 6.1.  The following cuts are applied to the leptons: 
${\mathswitch {m_{\Pe\Pe}}}>12\UGeV$ , ${\mathswitch {m_{\PGm\PGm}}}>12 \UGeV$, 
electrons' $\pT>7\UGeV$ and $|\eta|<2.5$, and
muons' $\pT>5\UGeV$ and $|\eta|<2.4$.
No cuts on the minimal $\Delta R$-distance between jets and lepton,
and lepton and lepton pairs are applied. The cross section is
calculated inclusively for the number of jets found. 

\begin{sloppypar}
For the estimation of the PDF$+\alphas$ systematic errors,  the
PDF4LHC prescription~\cite{Botje:2011sn} is applied. 
The upper and lower edges of the envelope are computed with the
three PDF sets: CT10~\cite{Lai:2010vv}, MSTW2008~\cite{Martin:2009iq},
NNPDF~\cite{Ball:2011mu},  which can be parametrised as follow:
\begin{eqnarray}
\mbox{$\ZZ$ @ NLO:} &&\quad \kappa(m_{4\Pl}) = 1 + 0.0035 \sqrt{m_{4\Pl}/\mbox{GeV}-30},
\label{eq:pdf_qq}
\\
\mbox{$\Pg \Pg \to \ZZ$:} &&\quad \kappa(m_{4\Pl}) = 1 + 0.0066 \sqrt{m_{4\Pl}/\mbox{GeV}-10}.
\label{eq:pdf_gg}
\end{eqnarray}
\end{sloppypar}

In \refF{fig:pdf_alphas_2} the difference between the central value of the cross
section and the cross section computed with plus and minus
 $1\sigma$ of the total PDF$+\alphas$ variation, following the
  PDF4LHC prescription,  is
   shown with the filled triangle marker for  
$\PQq\PAQq \rightarrow \ZZ^{(*)}\rightarrow {2\Pe 2\PGm}$ (left) and for 
$\Pg\Pg \rightarrow \ZZ^{(*)}\rightarrow {2\Pe 2\PGm}$ (right) as a function
   of ${\mathswitch {m_{2\Pe 2\PGm}}}$ at $7\UTeV$ from {\sc MCFM}. The red line is the
parametrisation  from Eq.~\refE{eq:pdf_qq} and Eq.~\refE{eq:pdf_gg}. 
The difference between the central value of the cross
section and the cross section computed with the 3 different PDF sets (CT10, MSTW, and NNPDF)
varying them by plus and minus 1~$\sigma$ for 
$\PQq\PAQq \rightarrow \ZZ^{(*)}\rightarrow {2\Pe 2\PGm}$ (left) and for 
$\Pg\Pg \rightarrow ZZ^{(*)}\rightarrow {2\Pe 2\PGm}$ (right) are also

\begin{figure}
  \includegraphics[width=\textwidth]{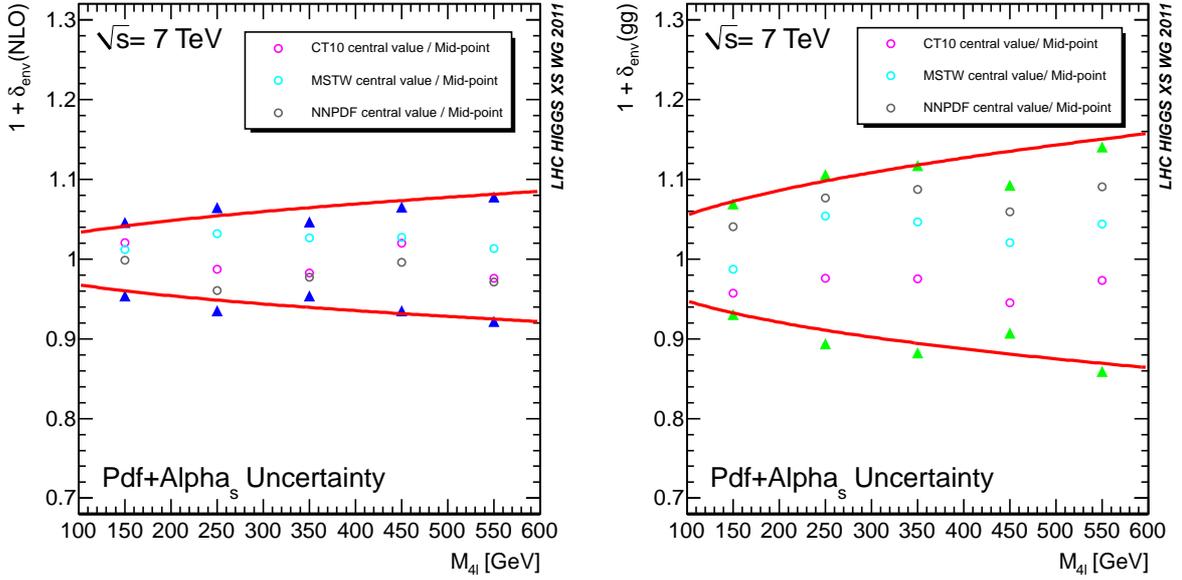}
  \caption{The difference between the central value of the cross
    section and the cross section computed with 3 different PDF sets and the total PDF$+\alphas$ variation (blue markers) varying them by plus and minus $1 \sigma$  for 
$\PQq\PAQq \rightarrow \ZZ^{(*)}\rightarrow {2\Pe 2\PGm}$ (left) and for 
$\Pg\Pg \rightarrow \ZZ^{(*)}\rightarrow {2\Pe 2\PGm}$ (right) as a function
    of ${\mathswitch {m_{2\Pe 2\PGm}}}$ at $7\UTeV$ from {\sc MCFM}.}  
  \label{fig:pdf_alphas_2}
\end{figure}


For the estimation of QCD scale systematic errors, we calculate variations
in the differential cross section $d \sigma / dm_{4\Pl}$ as  the renormalisation and factorisation scales are changed
by a factor of two up and down from their default value  $\muR = \muF = \MZ$. 
The dependence of the systematic QCD scale errors on the four-lepton invariant mass 
can be parametrised as follows:
\begin{eqnarray}
\mbox{$\ZZ$ @ NLO:} &&\qquad \kappa(m_{4\Pl}) = 1.00 + 0.01 \sqrt{(m_{4\Pl}/\mbox{GeV}-20)/13},
\label{eq:qcd_scale_qq}
\\
\mbox{$\Pg \Pg \to \ZZ$:}  &&\qquad  \kappa(m_{4\Pl}) = 1.04 + 0.10 \sqrt{(m_{4\Pl}/\mbox{GeV}+40)/40}.
\label{eq:qcd_scale_gg}
\end{eqnarray}

In \refF{fig:qcd_scale1} the cross section for 
$\PQq\PAQq \rightarrow \ZZ^{(*)}\rightarrow {2\Pe 2\PGm}$ (left) and for 
$\Pg\Pg \rightarrow \ZZ^{(*)}\rightarrow {2\Pe 2\PGm}$ (right) is shown as a function
    of ${\mathswitch {m_{2\Pe 2\PGm}}}$ at $7\UTeV$ from {\sc MCFM} computed with the CT10 PDF
    and varying the QCD scale by a factor of two.
In \refF{fig:qcd_scale} the ratio of the cross section computed
at the different scale and the cross section computed at the central
value of the QCD scale (i.e.\ at $\MZ$) is shown as a function of the
four-lepton invariant mass. The red lines are the parametrisation from
Eq.~\refE{eq:qcd_scale_qq} and Eq.~\refE{eq:qcd_scale_gg} .
\begin{figure}
  \includegraphics[width=\textwidth]{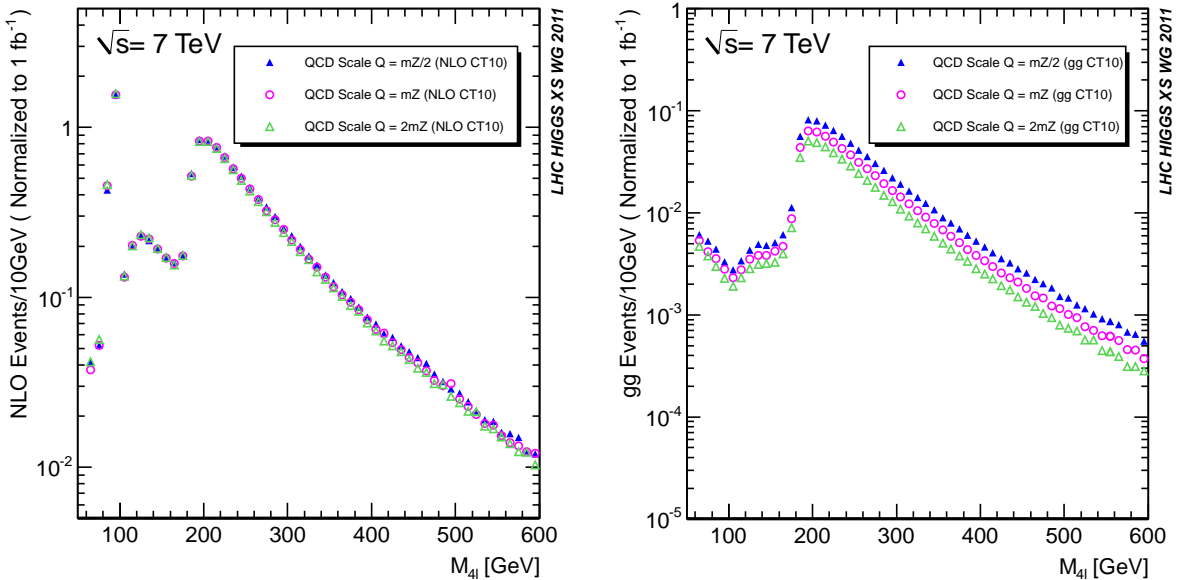}
  \caption{The cross section for 
$\PQq\PAQq \rightarrow \ZZ^{(*)}\rightarrow {2\Pe 2\PGm}$ (left) and for 
$\Pg\Pg \rightarrow \ZZ^{(*)}\rightarrow {2\Pe 2\PGm}$ (right) as a function
    of ${\mathswitch {m_{2\Pe 2\PGm}}}$ at $7\UTeV$ from {\sc MCFM} computed with the CT10 PDF
    and varying the QCD scale by a factor of two.}  
  \label{fig:qcd_scale1}
\end{figure}
\begin{figure}
  \includegraphics[width=\textwidth]{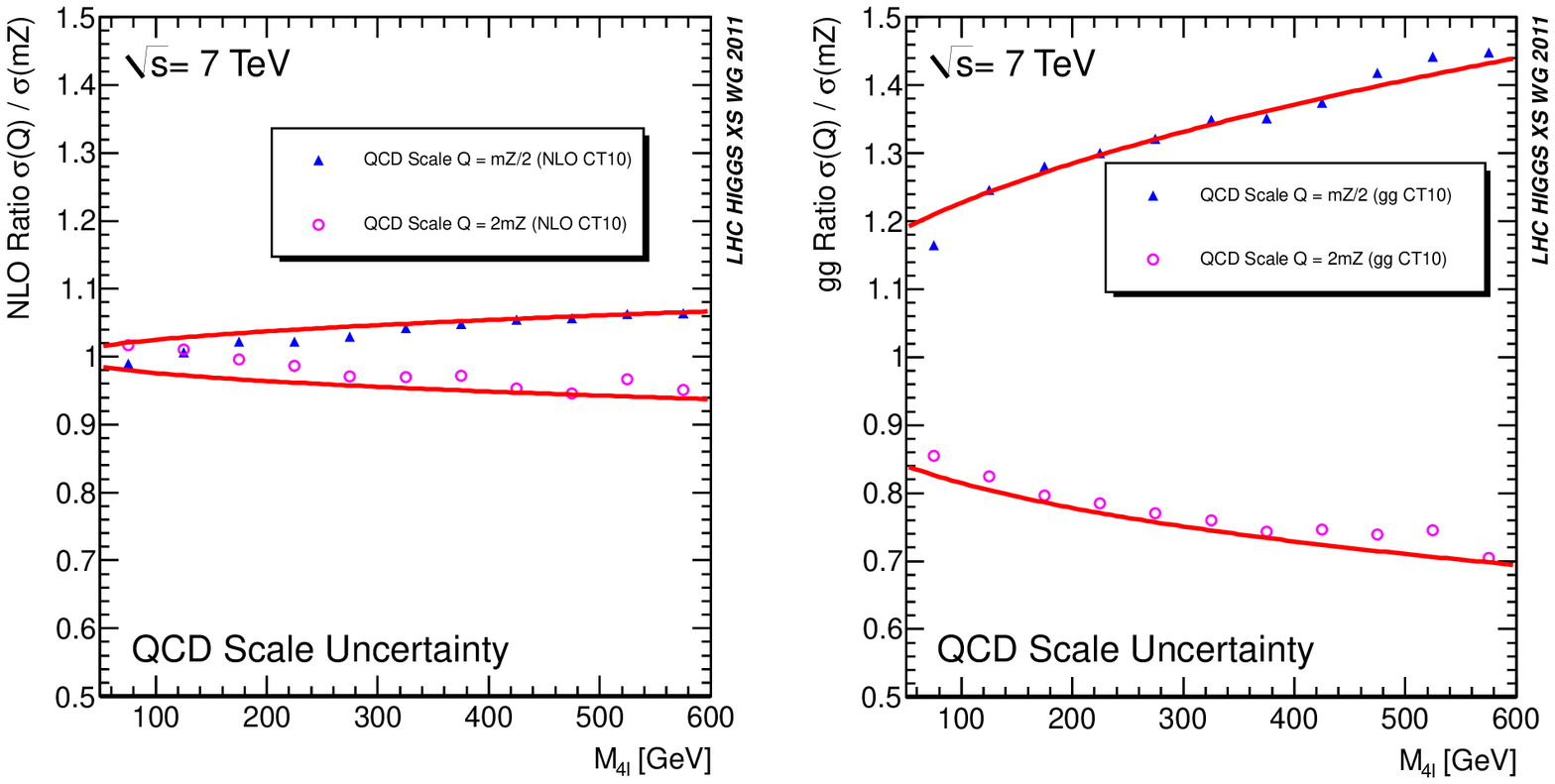}
  \caption{The ratio of the cross sections computed at different
    scales and the cross
    section computed at central value of the QCD scale for $\PQq\PAQq \rightarrow
    \ZZ^{(*)}\rightarrow {2\Pe 2\PGm}$ (left) and $\Pg\Pg \rightarrow
    \ZZ^{(*)}\rightarrow {2\Pe 2\PGm}$ (right)  as a function
    of ${\mathswitch {m_{2\Pe 2\PGm}}}$ at $7\UTeV$ from {\sc MCFM}. The red line is the
    parametrisation described in the text}  
  \label{fig:qcd_scale}
\end{figure}

\subsubsection{Summary}

In \refT{tab_qq_gg_MCFM_ZZ_all} a summary of cross sections for $\ZZ$
production (both quark annihilation and gluon fusion) is 
presented for the three sets of cuts, with uncertainties calculated
following the PDF4LHC prescription. 
In \refT{tab_ZZ_Z_MCFM} calculations are extended for single-\PZ{}
production and the ratio of $\ZZ$ to \PZ, 
which can be used for normalisation of $\ZZ$ production with data, with detailed breakdown of uncertainties.
\begin{table}
\centering
  \caption{Cross sections in fb with uncertainties for $\Pp\Pp \to \PZ(\PGg^\ast)\PZ(\PGg^\ast) \to \Pl\bar{\Pl}\Pl'\bar{\Pl'}$
  in $\Pp\Pp$ collisions at $\sqrt{s} = 7$ \UTeV{} calculated at NLO with {\sc MCFM}-6.1. Three cut sets are applied:
  $\mzonshell>12$ \UGeV, $\mzoffshell>12$ $\UGeV$(Cut 1);
  $\mzonshell>50$ \UGeV, $\mzoffshell>12$ \UGeV, $\pT(\Pl)>5$ \UGeV, $|\eta(\Pl)|<2.5$ (Cut 2);
  $60$ $\UGeV$ $ <\mzonshell<120 $ \UGeV, $60$ $\UGeV$ $<\mzoffshell<120 $ $\UGeV$(Cut 3).}
  \vspace{0cm}
  \label{tab_qq_gg_MCFM_ZZ_all}
\begin{tabular}{ccccc}
\hline
   & $\sigma$ [\Ufb]  &  PDF+$\alphas$ [\%] & QCD Scale [\%] & Total [\%] \\
\hline
{Cut 1} & $15.962$ &$ 3.53$ &$ 2.10$ &$ 5.63$ \\
{Cut 2} & $5.306 $& $3.44 $& $4.23$ & $7.68 $\\
{Cut 3} & $6.965 $& $2.95 $& $3.87$ & $ 6.82$ \\
\hline
\end{tabular}
\end{table}
\begin{table}
\centering
  \caption{Cross sections in fb for $\Pp\Pp \to \PZ(\PGg^\ast)\PZ(\PGg^\ast) \to 4\Pe$ or $4\mu$, $\PZ \to 2\Pe$ or $2\PGm$ and their ratios
  in $\Pp\Pp$ collisions at $\sqrt{s} = 7\UTeV$ calculated at NLO
  with {\sc MCFM}-6.1. For $\ZZ$ cross sections calculations 
the following cuts are applied:
  $\mzonshell >50 \UGeV$, $\mzoffshell >12$ \UGeV, $\pT(\Pl)>5$ \UGeV,
  $|\eta(\Pl)|<2.5$ (Cut 2). 
The set of cuts for $\PZ$ cross sections
  calculations is:  $\MZ>40$ \UGeV, $\pT(\Pl 1)>20$ \UGeV, $\pT(\Pl 2)>10$ \UGeV, $|\eta(\Pl)|<2.5$ }
  \vspace{0cm}
  \label{tab_ZZ_Z_MCFM}
\begin{tabular}{cccc}
\hline
PDF & $\sigma$($\ZZ\rightarrow 4\Pe/4\mu$)~[\Ufb] & $\sigma$($\PZ\rightarrow 2\Pe/2\PGm$)~[$10^5$ \Ufb] & $\sigma(\ZZ)/\sigma(\PZ)$~$[10^{-6}]$ \\
\cline{1-4}
\multicolumn{4}{c}{PDF+$\alphas$} \\
\cline{1-4}
CT10 &$ 5.235 \pm 2.13\% $&$ 5.478 \pm 2.37\% $&$ 9.556 \pm 1.52\% $\\
MSTW2008 &$ 5.408 \pm 1.50\% $&$ 5.494 \pm 1.70\% $&$ 9.844 \pm 0.90\% $\\
NNPDF2.0 &$ 5.290 \pm 1.50\% $&$ 5.259 \pm 1.60\% $&$ 10.06 \pm 0.80\% $\\
Envelope &$ 5.306 \pm 3.44\% $&$ 5.392 \pm 4.02\% $&$ 9.776 \pm 3.73\% $\\
\cline{1-4}
\multicolumn{4}{c}{QCD} \\
\cline{1-4}
$\mu=\MZ/2$ & $+4.23\% $&$ -0.90\% $&$ +5.17\% $\\
$\mu=2\MZ$ & $-3.26\% $&$ +1.20\% $&$ -4.41\% $\\
$|\max|$ & $\phantom{+}4.23\% $&$ \phantom{+}1.20\% $&$ \phantom{+}5.17\% $\\
\cline{1-4}
{Total} & {$5.306 \pm 7.68\%$} & {$5.392 \pm 5.22\%$} & {$9.776 \pm 8.90\%$} \\
\cline{1-4}

\end{tabular}
\end{table}

\subsection{Angular distributions in $\PH \to \ZZ$ decays}
The angular distributions of final decay products from the Higgs-boson decays are strictly related to the scalar nature of the Standard Model Higgs boson.
The angular distributions may be eventually used to separate $\PH \to \ZZ$
events from regular SM di-boson events, and will in the end
serve to verify if any Higgs signature is indeed from a scalar
boson.
The MC program generators used for the $\PH \rightarrow \ZZ$ process are {\sc PYTHIA} and
\POWHEG{} (for which gluon fusion and vector-boson fusion are separated), while
{\sc PYTHIA} is used for the SM di-boson production. The study presented in this section has been 
made with simulated Higgs events with mass of $120\UGeV$.

\subsubsection{Comparison and cross check of angular distributions}

The primary angular distribution to consider is the angle between
the Higgs boson\footnote{For simplicity we refer to the $\ZZ$ system as
the Higgs, even if no Higgs is involved.} direction and a $\PZ$ in
the rest frame of the Higgs, $\theta^*_{\PH,\PZ}$; since the SM Higgs
boson is a scalar, $\cos(\theta^*_{\PH,\PZ})$ it has to be a uniform distribution.
While the dominant production process for the Higgs is gluon fusion,
$\PQq\PAQq$ dominates SM di-boson $\ZZ$ production. As a consequence
of this, the SM di-boson $\ZZ$ system is created in a 
spin state different from the one of the Higgs, and so here
$\cos(\theta^*_{\PH,\PZ})$ is not expected to be uniform.
The $\cos(\theta^*_{\PH,\PZ})$ distribution for $\PH \rightarrow \ZZ$ (both for {\sc PYTHIA} and \POWHEG{}) as well as for SM
di-boson $\ZZ$ production is shown in \refF{fig:angularvar}a; the distribution is indeed
uniform in case of the Higgs for both generators, while it is not
uniform for the SM di-boson $\ZZ$ case. In the same figure the
distribution obtained when the $\ZZ$ invariant mass is required to be around
$120\UGeV$ (which diminishes the separation) is also shown.

Another independent variable sensitive to the spin of the $\ZZ$
system, is the sum of the $\PZ$ masses divided by the Higgs-boson mass, $x_{\PH}$
(quantity related to the $\pT$ of the $\PZ$'s in the Higgs-boson rest frame); that variable is 
expected to be larger for Higgs events than for SM di-boson $\ZZ$ events, but in both cases 
close to unity, due to the lacking phase space, as shown in \refF{fig:angularvar}b.
Some difference has been found between {\sc PYTHIA} and \POWHEG{},
as {\sc PYTHIA} generally yields a larger value of $x_{\PH}$.
\begin{figure}
  \includegraphics[width=0.49\textwidth]{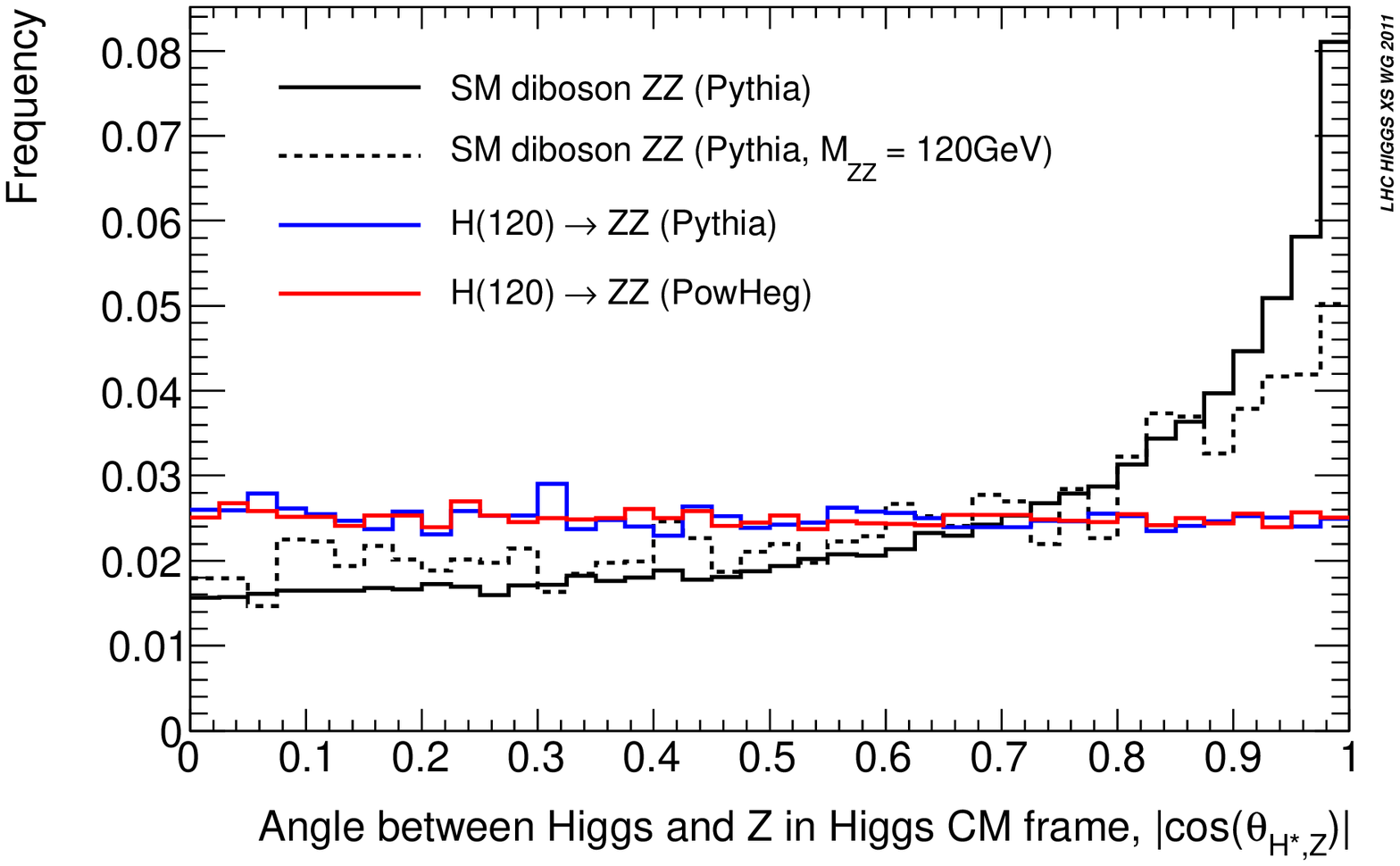}
  \includegraphics[width=0.49\textwidth]{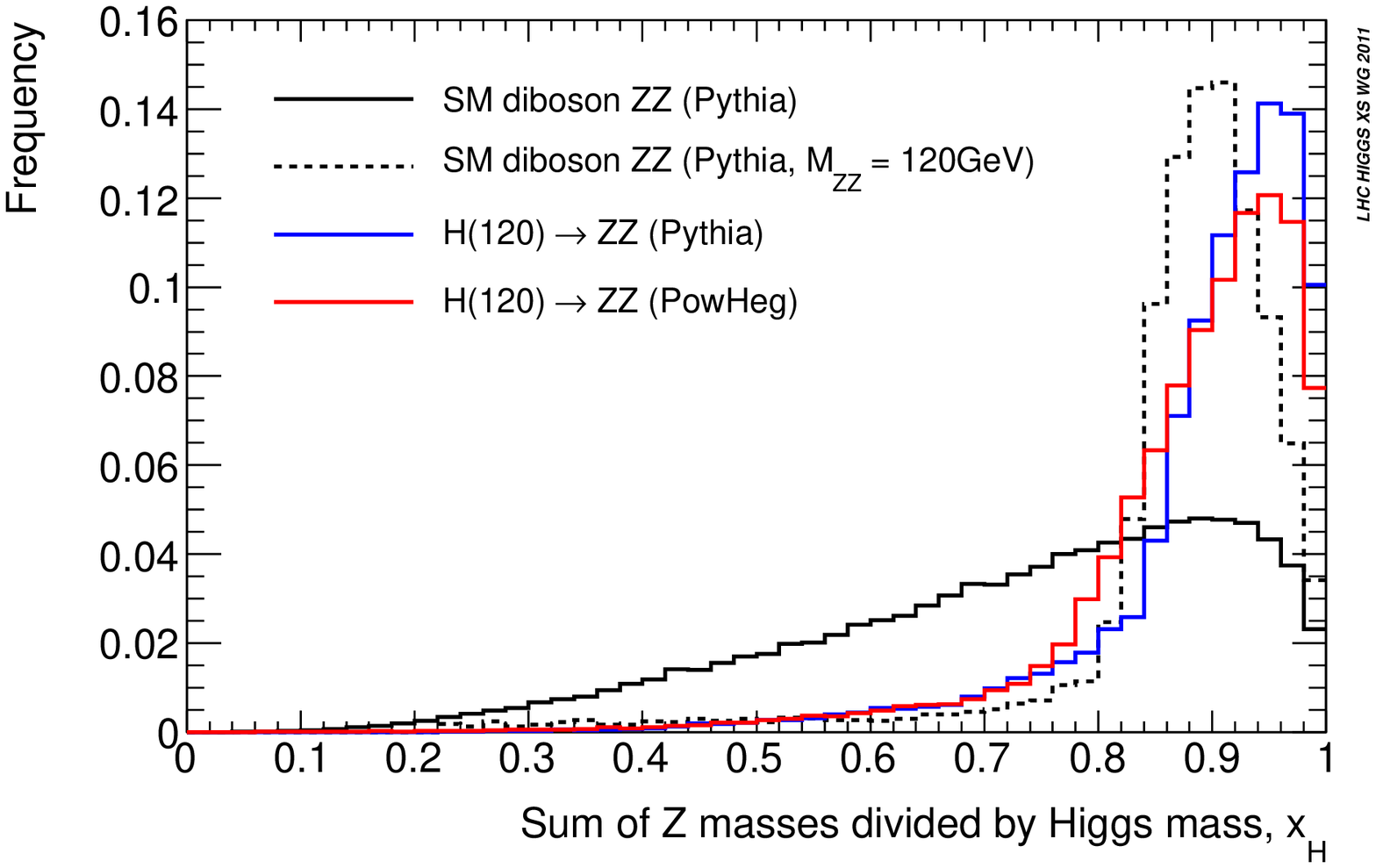}
  \caption{(a) Distribution of $\cos(\theta^*_{\PH,\PZ})$ for Higgs events with $\MH=120  \UGeV$ and $\ZZ$ di-boson events.
  (b) Distribution of $x_{\PH}$ (the sum of the $\PZ$ masses divided by the
  Higgs mass) for the same samples.}
\label{fig:angularvar}
\end{figure}

\subsubsection{Separating $\PH \rightarrow \ZZ$ from $\ZZ$ using angular variables}

In addition to the above two observables, several other variables are also sensitive to the spin of the $\ZZ$ system; the rapidity of that 
system, $y_{\PH}$, and the angle between the $\PZ$ and the leptons in the $\PZ$ rest frame, $\theta^*_{\PZ,\Pl-}$ are different,
yielding five variables in total. Nevertheless most of the variables tend to give less separation once a fiducial selection based on acceptance 
($\pT(\Pl) > 7  \UGeV$, $\eta(\Pl) < 2.5$) is applied.

The final separation between $\PH \rightarrow \ZZ$ and SM di-boson $\ZZ$ derived by combining the angular variables in a neural net 
(Boosted decision tree, BDT), can be found in \refF{fig:angularsep}; the separation is quite poor, and so not much further separation can 
be gained from considering the angular variables.

\begin{figure}
\begin{center}
  \includegraphics[width=0.5\textwidth]{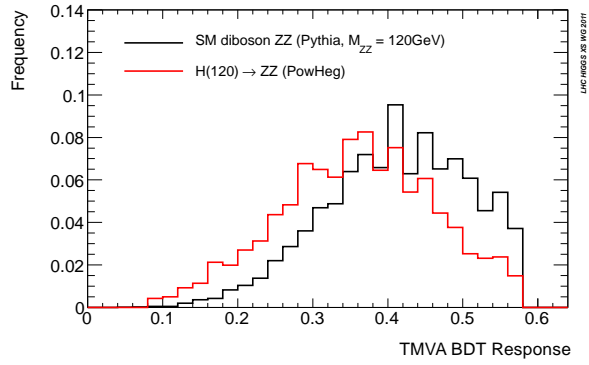}
 \caption{Separation between $\PH \rightarrow \ZZ$ and SM di-boson $\ZZ$ using five
   angular variables combined in a Boosted Decision Tree (BDT).}
\label{fig:angularsep}
\end{center}
\end{figure}

\clearpage


\newpage
\providecommand{\MHpm}{M_{\PSHpm}}
\providecommand{\lsim}
{\;\raisebox{-.3em}{$\stackrel{\displaystyle <}{\sim}$}\;}
\providecommand{\gsim}
{\;\raisebox{-.3em}{$\stackrel{\displaystyle >}{\sim}$}\;}
\providecommand{\orderx}[1]{\ensuremath{{\cal O}(#1)}}
\providecommand{\mhmaxx}{\ensuremath{m_{\rm h}^{\rm max}}}
\providecommand{\MHp}{M_{\PSHpm}}
\providecommand{\gghnnlo}{{\sc ggH@NNLO}}
\providecommand{\bbhnnlo}{{\sc bbH@NNLO}}
\providecommand{\HIGLU}{{\sc HIGLU}}

\section{Neutral-Higgs-boson production in the MSSM\footnote{%
M.~Spira, M.~Vazquez Acosta, M.~Warsinsky, G.~Weiglein (eds.);
E.A.~Bagnaschi, 
M.~Cutajar,
G.~Degrassi, 
R.~Harlander, 
S.~Heinemeyer, 
M.~Kr\"amer, 
A.~Nikitenko,
S.~Palmer, 
M.~Schumacher,
P.~Slavich, 
A.~Vicini, 
T.~Vickey and
M.~Wiesemann.}}
\label{sec:MSSM}

\subsection{The MSSM Higgs sector: general features}

\label{sec:mssmvssm}


Within the Minimal Supersymmetric Standard Model (MSSM) the Higgs-boson
sector contains, contrary to the SM, two scalar doublets, accommodating
five physical Higgs bosons \cite{Fayet:1974pd,Fayet:1977yc,Fayet:1976et,
Dimopoulos:1981zb,Sakai:1981gr,Inoue:1982ej,Inoue:1982pi,Inoue:1983pp}.
These are, at lowest order, the light and heavy CP-even $\PSh$ and
$\PH$, the CP-odd $\PSA$, and the charged Higgs bosons $\PSHpm$. At the
tree level the MSSM Higgs sector can be expressed in terms of the SM
gauge couplings and two further input parameters, conventionally chosen
as $\tanb \equiv v_2/v_1$, the ratio of the two vacuum expectation
values, and either $\MA$ or $\MHp$. Consequently, all other masses and
mixing angles can be predicted.  However, the Higgs sector of the MSSM
is affected by large higher-order corrections, which have to be taken
into account for reliable phenomenological predictions.  The
higher-order corrections arise in particular due to the large top Yukawa
coupling, leading to large loop contributions from the top and stop
sector to the Higgs masses and couplings.  Similarly, for large values
of $\tanb$ effects from the bottom/sbottom sector can also be large.
In particular, the relation between the bottom-quark 
mass and the bottom Yukawa coupling receives
$\tanb$-enhanced contributions entering via
$\Delta_{\PQb}$~\cite{Hall:1993gn,Hempfling:1993kv,
Carena:1994bv,Pierce:1996zz,Carena:1999py,
Guasch:2003cv,Noth:2008tw,Noth:2010jy,Mihaila:2010mp}.  These
corrections are non-vanishing even in the limit of asymptotically large
values of the SUSY mass parameters. For the light Higgs boson these
corrections vanish if the Higgs-boson mass scale, $\MA$ or $\MHp$ also
goes to large values. An analogous contribution, although in general
smaller, exists for the $\PGt$ lepton.  The MSSM Higgs sector is
CP conserving at lowest order. However, CP-violating effects can enter
via the potentially large loop corrections. In this case all three
neutral Higgs bosons mix with each other. As in \Bref{Dittmaier:2011ti}
we will focus here on the CP-conserving case and use $\MA$ as input
parameter.

In a large part of the MSSM parameter space the couplings of the light
CP-even Higgs boson to SM gauge bosons and fermions become SM like. 
This decoupling limit is reached for $\MA \gg \MZ$, however in practice
it is already realised for $\MA \gsim 2 \MZ$. Consequently, in this 
parameter region 
the light CP-even Higgs boson of the MSSM resembles the 
Higgs boson of the SM.
Results for the light CP-even Higgs-boson production cross sections and decay
branching ratios approach the corresponding SM values, provided that the
additional SUSY mass scales are too high to affect the Higgs production
and decay.

However, in general the 
Higgs phenomenology in the MSSM can differ very significantly from the
SM case. Depending on $\MA$ and the SUSY mass scales, the relevant
couplings entering production and decay processes 
of a MSSM Higgs boson can be very different from the corresponding
couplings in the SM case. Consequently, the lower bound on the Higgs
mass in the SM from the searches at LEP cannot be applied directly to the MSSM
case~\cite{Barate:2003sz,Schael:2006cr}; much lighter Higgs 
masses are possible in the MSSM without being in conflict with the
present search limits. Furthermore, the presence of more than one Higgs
boson in the spectrum can give rise to overlapping signals in the Higgs
searches, in particular in parameter regions where the Higgs-boson
widths are large or the experimental resolution does not allow for a
separation of the individual Higgs mass peaks.

Due to the enlarged spectrum in the MSSM, further production and decay
processes are possible compared to the SM case.  In particular, MSSM
Higgs bosons can be produced in association with or in decays of SUSY
particles, and decays of MSSM Higgs bosons into SUSY particles, if
kinematically allowed, can have a large impact on the Higgs branching
ratios. In certain parts of the
parameter space decays of heavy MSSM Higgs bosons into lighter Higgs can
also become important. These decays could potentially yield valuable
information on the Higgs self-couplings. However, in the following we
will mainly focus on SM-type Higgs production processes at the LHC. In
particular, we will investigate Higgs production in gluon fusion, 
weak-boson fusion and in association with bottom quarks. Corresponding
results for the total cross sections were discussed in
\Bref{Dittmaier:2011ti}.

Due to the large number of free parameters in the MSSM, 
it is customary to interpret searches for the Higgs bosons in terms of 
benchmark scenarios where the lowest-order input parameters 
$\tanb$ and $\MA$ (or $\MHp$) are varied, while the other SUSY parameters
entering via radiative corrections are set to certain benchmark values. 
In the following we will focus, in particular, on the $\mhmaxx$ benchmark 
scenario~\cite{Carena:2002qg}. Within the on-shell scheme it is defined
as%
\footnote{
It should be noted that while $M_{\mathrm{SUSY}}$ formally is common for all
three generations of scalar quarks, the only relevant ones are the third
generation squarks. Consequently, limits obtained at the LHC on first
and second generation squarks do not play a relevant role for this scenario.}
\begin{equation}
M_{\mathrm{SUSY}} = 1 \UTeV, \; X_{\PQt} = 2 M_{\mathrm{SUSY}}, \; \mu = 200 \UGeV, \;
M_{\PSg}=800 \UGeV, \; M_2 = 200 \UGeV, \; A_{\PQb} = A_{\PQt},
\label{YRHXS_MSSM_neutral_eq:mhmax}
\end{equation}
where $M_{\mathrm{SUSY}}$ denotes the common soft-SUSY-breaking squark mass of
the third generation, $X_{\PQt}=A_{\PQt}-\mu/\tanb$ the stop mixing parameter,
$A_{\PQt}$ and $A_{\PQb}$ the stop and sbottom trilinear couplings, respectively,
$\mu$ the Higgsino mass parameter, $M_{\PSg}$ the gluino mass, and
$M_2$ the SU(2)-gaugino mass parameter. Adding to the results of
\Bref{Dittmaier:2011ti} for the total inclusive cross sections within
the $\mhmaxx$ benchmark scenario we will present the inclusive results
for the no-mixing scenario in this chapter, too. The no-mixing benchmark
scenario is defined as~\cite{Carena:2002qg}
\begin{equation}
M_{\mathrm{SUSY}} = 2 \UTeV, \; X_{\PQt} = 0 , \; \mu = 200 \UGeV, \;
M_{\PSg}= 1.6 \UTeV, \; M_2 = 200 \UGeV, \; A_{\PQb} = A_{\PQt}\ .
\label{YRHXS_MSSM_neutral_eq:nomixing}
\end{equation}

While in the SM the Higgs-boson mass is a free input
parameter, in the MSSM all masses, mixings, and couplings can be
calculated in terms of the other model parameters. Consequently,
calculations of Higgs production and decay processes in the
MSSM require, as a first step, the evaluation of the Higgs-boson masses
and mixing contributions in terms of $\MA$, $\tanb$, and all other
relevant SUSY parameters that enter via radiative corrections.
Furthermore, the mixing between the different (neutral) Higgs bosons must
be taken into account in order to ensure the correct on-shell properties
of the Higgs fields appearing in the $S$-matrix elements of production or
decay processes.
To perform this kind of evaluations in terms of the MSSM input parameters,
several codes exist. Two codes use electroweak scale parameters as
input, 
\FeynHiggs~\cite{Heinemeyer:1998yj,Heinemeyer:1998np,Degrassi:2002fi,Frank:2006yh}
and \CPsuperH~\cite{Lee:2003nta,Lee:2007gn}, 
while other codes can also work with GUT scale input, 
{\sc SoftSusy}~\cite{Allanach:2001kg},
{\sc Spheno}~\cite{Porod:2003um,Porod:2011nf}, and
{\sc SuSpect}~\cite{Djouadi:2002ze}.
While certain differences in the implemented calculations
exist, all of the above codes incorporate higher-order corrections in
the MSSM Higgs sector up to the two-loop level.
\FeynHiggs\ was chosen for the corresponding calculations in
\Bref{Dittmaier:2011ti}, and following these previous evaluations, we use
\FeynHiggs\ to calculate the
Higgs-boson masses and effective couplings in the MSSM here as well
(where we deviate from this, it will be clearly indicated).
A brief comparison of \FeynHiggs\ and \CPsuperH\ and the respective
differences in the $\mhmaxx$ and no-mixing benchmarks can be found in
\Bref{Dittmaier:2011ti}, while a comparison between \FeynHiggs\ and the
three codes {\sc SoftSusy}, {\sc Spheno}, and {\sc SuSpect} can be found
in \Bref{Allanach:2004rh}. 
The masses, mixings, and couplings obtained in this way can be passed via the
SUSY Les Houches Accord~\cite{Skands:2003cj,Allanach:2008qq} to other
codes for further evaluation.


\subsection{Status of inclusive calculations}

\subsubsection{Summary of existing calculations and implementations}
\label{sec:mssm-calcsum}

\providecommand{\abbrev}{\scalefont{.9}}
\providecommand{\ep}{\epsilon}
\providecommand{\api}{\frac{\alphas}{\pi}}
\providecommand{\eqn}[1]{Eq.\,(\ref{#1})}
\providecommand{\fig}[1]{Fig.\,\ref{#1}}
\providecommand{\tab}[1]{Tab.\,\ref{#1}}
\providecommand{\sct}[1]{Sect.\,\ref{#1}}
\providecommand{\dd}{{\rm d}}
\providecommand{\deriv}[2]{\frac{\dd #1}{\dd #2}}
\providecommand{\order}[1]{{\cal O}(#1)}
\providecommand{\lhc}{{\abbrev LHC}}
\providecommand{\qcd}{{\abbrev QCD}}
\providecommand{\sm}{{\abbrev SM}}
\providecommand{\mssm}{{\abbrev MSSM}}
\providecommand{\susy}{{\abbrev SUSY}}
\providecommand{\bsm}{{\abbrev BSM}}
\providecommand{\pdf}{{\abbrev PDF}}
\providecommand{\lo}{{\abbrev LO}}
\providecommand{\nlo}{{\abbrev NLO}}
\providecommand{\nnlo}{{\abbrev NNLO}}
\providecommand{\dred}{{\abbrev DRED}}
\providecommand{\dreg}{{\abbrev DREG}}
\providecommand{\msbar}{\overline{\mbox{\abbrev MS}}}
\providecommand{\drbar}{\overline{\mbox{\abbrev DR}}}
\providecommand{\bld}[1]{\boldmath{$#1$}}
\providecommand{\mhiggs}{M_\text{H}}
\providecommand{\mtop}{M_\text{t}}

\subsubsubsection*{Gluon fusion}

Gluon fusion has been studied in great detail in the framework of the
SM. In fact, the requirement of higher-order corrections to this process
has been one of the most powerful driving forces for the development of
new theoretical concepts and techniques. The theory prediction currently
used by the experimental collaborations includes the full NLO QCD
corrections
\cite{Graudenz:1992pv,Spira:1995rr,Aglietti:2006tp,Anastasiou:2006hc} as
well as NNLO QCD corrections in the limit of heavy top
quarks~\cite{Harlander:2002wh,Anastasiou:2002yz,Ravindran:2003um},
resummation of soft-collinear logarithms through NNLL in the limit of
heavy top quarks~\cite{Catani:2003zt}, full electroweak
NLO contributions~\cite{Actis:2008ug,Aglietti:2004nj}, and an estimate of the
mixed QCD/EW terms~\cite{Anastasiou:2008tj}.

For the MSSM, however, the corrections to the gluon-fusion process have
been treated much less rigorously until recently. Not even through
${\cal O}(\alphas^3)$, \ie  NLO QCD, has there been a prediction that
takes into account all strongly interacting SUSY particles.

Schematically, the amplitude for Higgs production through gluon fusion
can be written as
\begin{equation}
\begin{split}
{\cal A}(\Pg\Pg\to \Ph) = \sum_{\Pq\in \{\Pt,\Pb\}} 
        \left( a_{\Pq}^{(0)} + a_{\tilde q}^{(0)} + a_{\Pq}^{(1)} + a_{\tilde q}^{(1)} 
             + a_{\tilde \Pq\tilde \Pg}^{(1)}\right)\,.
\end{split}
\end{equation}
The superscripts $(0)$ and $(1)$ denote one- and two-loop contributions,
respectively.  The term $a_{\Pq}^{(0,1)}$ comprises contributions from
Feynman diagrams with only the quark $\Pq$ and gluons running in
loops. Similarly, $a_{\tilde q}^{(0,1)}$ contains only squarks 
$\tilde \Pq$ and gluons, while $a_{\tilde \Pq\tilde \Pg}^{(1)}$ is due to
diagrams containing quarks $\Pq$, squarks $\tilde \Pq$, and gluinos
simultaneously. 

Of course, also the real radiation of quarks and gluons has to be taken
into account: $\Pg\Pg\to \Ph\Pg$, $\Pq\Pg\to \Ph\Pq$, and 
$\PQq\PAQq \to \Ph\Pg$. They are
either quark- or squark-loop mediated, but no mixed quark/squark (or
gluino) terms occur at NLO.

The LO result, $\sum_{\Pq} (a_{\Pq}^{(0)} + a_{\tilde \Pq}^{(0)})$, has been known
since a long time in analytic form. Also the real radiation can be (and
has been) evaluated quite straightforwardly using standard techniques,
see \eg \Brefs{Spira:1995rr,Muhlleitner:2006wx,Bonciani:2007ex}.

The pure quark terms $a_{\Pq}^{(0)} + a_{\Pq}^{(1)}$ correspond to the NLO
SM contribution. The first result that was valid for a
general quark to Higgs mass ratio was presented in
\Bref{Graudenz:1992pv,Spira:1993bb} in the form of one-dimensional
integrals. Meanwhile, more compact expressions in terms of analytic
functions have been
found~\cite{Harlander:2005rq,Aglietti:2006tp,Anastasiou:2006hc}.

These general results justified with hindsight the use of the heavy-top
limit for the gluon-fusion process within the Standard
Model~\cite{Dawson:1990zj,Djouadi:1991tka}: it approximates the exact
QCD correction factor $\sigma^{\NLO}/\sigma^{\LO}$ to within
$2{-}3\%$ for $\Mh<2\Mt$~\cite{Kramer:1996iq,Harlander:2003xy}.  The
question whether this observation carries over to NNLO has been answered
by comparing the asymptotic expansion of the total cross section in
terms of $1/m_{\PQt}$ to the usually adopted expression
\begin{equation}
\begin{split}
\sigma_\text{eff}
&=\left(\frac{\sigma^{\mathrm{NNLO}}}{\sigma^{\LO}}\right)_{\Mt\to\infty}
\sigma^{\LO}(\Mt)\,.
\end{split}
\end{equation}
They agree at the sub-per-cent
level~\cite{Marzani:2008az,Harlander:2009mq,Harlander:2009my,Pak:2009dg,Pak:2011hs}. 
It is thus fair to say that the SM-like 
top-quark induced terms are known through NNLO, while the bottom quark
terms are known through NLO.

Technically, the squark-induced terms are very similar to the
quark-induced ones. Consequently, results for general squark masses are
known also in this case through
NLO~\cite{Aglietti:2006tp,Anastasiou:2006hc,Muhlleitner:2006wx}. In
order to arrive at a consistent result within the MSSM, also mixed
quark/squark/gluino contributions need to be taken into account, however.

For the top sector, this has been done by constructing an effective
Lagrangian where all SUSY particles and the top quark are integrated
out~\cite{Harlander:2004tp,Harlander:2003bb,Degrassi:2008zj}. The SUSY
effects are then reduced to the calculation of the proper Wilson
coefficient, while the actual process diagrams are identical to the ones
in the heavy-top limit in the SM case. The genuine SUSY QCD corrections
in this limit turned out to be of moderate size.

For the bottom sector, this approach is no longer applicable in a
straightforward way, because the bottom quark cannot be assumed
heavy. The problem are the mixed bottom/sbottom/gluino diagrams. A
numerical result for general masses has been presented in
\Brefs{Anastasiou:2008rm,Muhlleitner:2010nm}. A more compact expression can be
obtained through asymptotic expansions in the limit 
$m_{\tilde \Pb},m_{\tilde \Pg} \gg \MH, \Mb$, as applied in
\Brefs{Degrassi:2010eu,Harlander:2010wr}.

In \Bref{Harlander:2010wr}, all contributions of strongly interacting
particles in the MSSM have been combined in a single program {\sc
gghbsusy}. This includes (i) the exact real radiation contributions due
to quark and squark loops for both the top and the bottom sector; (ii)
the exact two-loop virtual terms due to top- and bottom-quark loops;
(iii) the squark-loop and mixed quark/squark/gluino-induced terms in the
limit $M_{\tilde \Pg} = M_{\tilde{\Pq}_1} = M_{\tilde{\Pq}_2} \gg
\{\MH,\Mt,\Mb\}$.  Needless to say that this includes also all the
interference terms.

The program {\sc gghbsusy} links {\sc FeynHiggs} which provides the
on-shell values of the input parameters.


\subsubsubsection*{$\PQb$-quark associated production}

\begin{sloppypar}
The inclusive total cross section for bottom-quark associated
Higgs-boson production, denoted%
\footnote{The notation $(\PQb\PAQb )\PH$ is
  meant to indicate that the $\PQb\PAQb $ pair is not required as part of
  the signature in this process, so that its final-state momenta must be
  integrated over the full phase space.}
$\Pp\Pp/\Pp\bar \Pp \to (\PQb\PAQb )\PH+X$, 
can be calculated in two different schemes. As the mass
of the bottom-quark is large compared to the QCD scale, 
$\Mb \gg \Lambda_{\QCD}$, bottom-quark production is a perturbative
process and can be calculated order by order. Thus, in a four-flavour scheme
(4FS), where one does not consider $\Pb$-quarks as partons in the
proton, the lowest-order QCD production processes are gluon--gluon
fusion and quark--antiquark annihilation, $\Pg\Pg \to \PQb\PAQb \PH$ and
$\Pq\bar{\Pq}\to \PQb\PAQb \PH$, respectively. However, the inclusive cross
section for $\Pg\Pg \to (\PQb\PAQb )\PH$ develops logarithms of the form
$\ln(\muF/\Mb)$, which arise from the splitting of gluons into
nearly collinear $\PQb\PAQb $ pairs. The large factorisation scale
$\muF \approx \MH/4$ corresponds to the upper limit of the collinear
region up to which factorisation is
valid~\cite{Rainwater:2002hm,Plehn:2002vy,Maltoni:2003pn}. For
Higgs-boson masses $\MH \gg 4 \Mb$, the logarithms become large
and spoil the convergence of the perturbative series. The
$\ln(\muF/\Mb)$ terms can be summed to all orders in perturbation
theory by introducing bottom parton densities. This defines the
so-called five-flavour scheme (5FS). The use of bottom distribution
functions is based on the approximation that the outgoing $\Pb$-quarks are
at small transverse momentum. In this scheme, the LO process for the
inclusive $(\PQb\PAQb )\PH$ cross section is bottom fusion, 
$\PQb\PAQb  \to \PH$.
\end{sloppypar}

If all orders in perturbation theory were taken into account, the four-
and five-flavour schemes would be identical, but the way of ordering the
perturbative expansion is different. At any finite order, the two
schemes include different parts of the all-order result, and the cross
section predictions do thus not match exactly. While this leads to an
ambiguity in the way the cross section is calculated, it also offers an
opportunity to test the importance of various higher-order terms and the
reliability of the theoretical prediction. The 4FS calculation is
available at NLO~\cite{Dittmaier:2003ej,Dawson:2003kb}, while the
5FS cross section has been calculated at NNLO
accuracy~\cite{Harlander:2003ai}.  Electroweak corrections to the 5FS
process have been found to be small~\cite{Dittmaier:2006cz} and will not
be considered in the numerical results presented here, except for the
effective MSSM couplings used for dressing the predictions obtained 
within the SM.

\subsubsubsection*{Total cross sections}

Within the \mhmaxx\ scenario the Higgs-boson production cross sections
develop a significant dependence on the $\mu$ parameter. The ratios of
the total cross sections for different $\mu$ values and the predictions
for $\mu=200 \UGeV$ are shown in \Fref{YRHXS_MSSM_neutral_fig6} for two
values of $\tanb=5,30$. While for moderate values of $\tanb$ the effect
of the $\mu$ variation between $-400 \UGeV$ and $800 \UGeV$ is less than
$10{-}15\%$, larger effects up to a factor of about $2$ arise for large
values of $\tanb$. These effects emerge dominantly due to the $\mu$
dependence of the $\Delta_{\PQb}$ corrections to the bottom Yukawa
couplings. While the results for the associated Higgs-boson
production with bottom quarks are valid up to NNLO, the corresponding
results for the gluon-fusion process rely on the approximation that the
genuine SUSY QCD corrections are dominated by the same $\Delta_{\PQb}$
corrections to the bottom Yukawa couplings which, however, up to now 
has not yet been confirmed by explicit calculations of these corrections.
\begin{figure}
\subfigure[]{\includegraphics[width=0.49\textwidth]{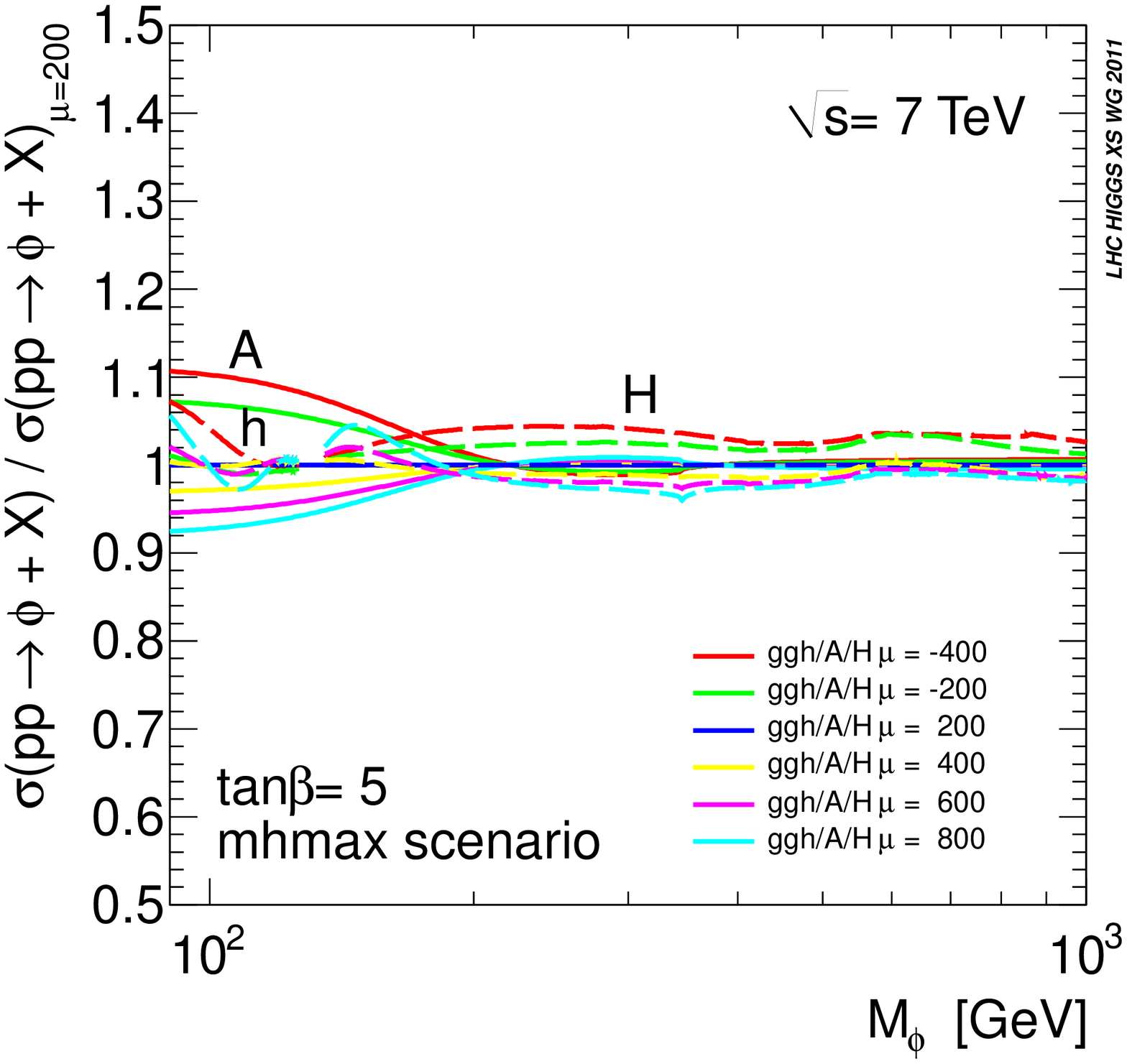}}
\subfigure[]{\includegraphics[width=0.49\textwidth]{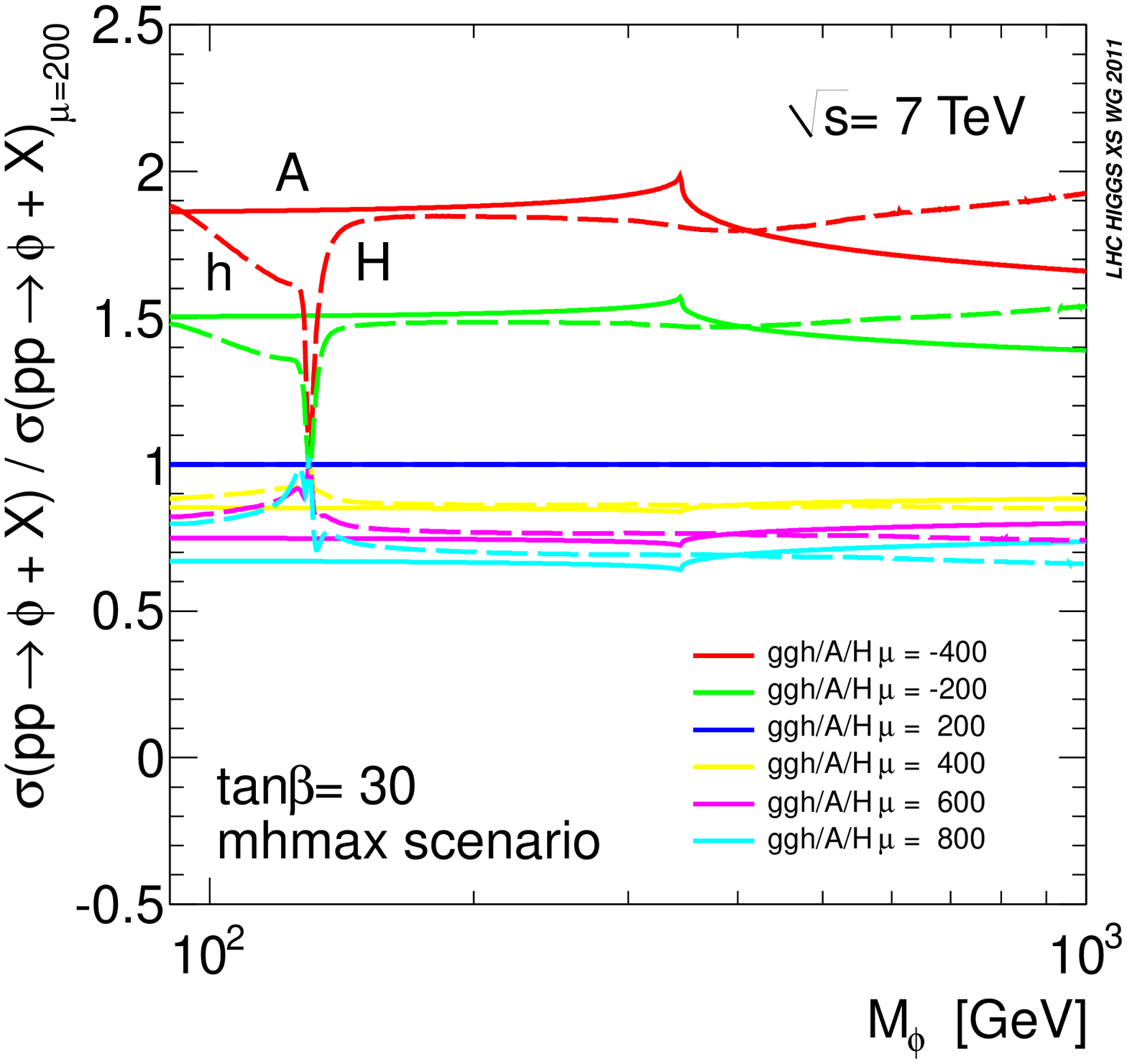}}
\subfigure[]{\includegraphics[width=0.49\textwidth]{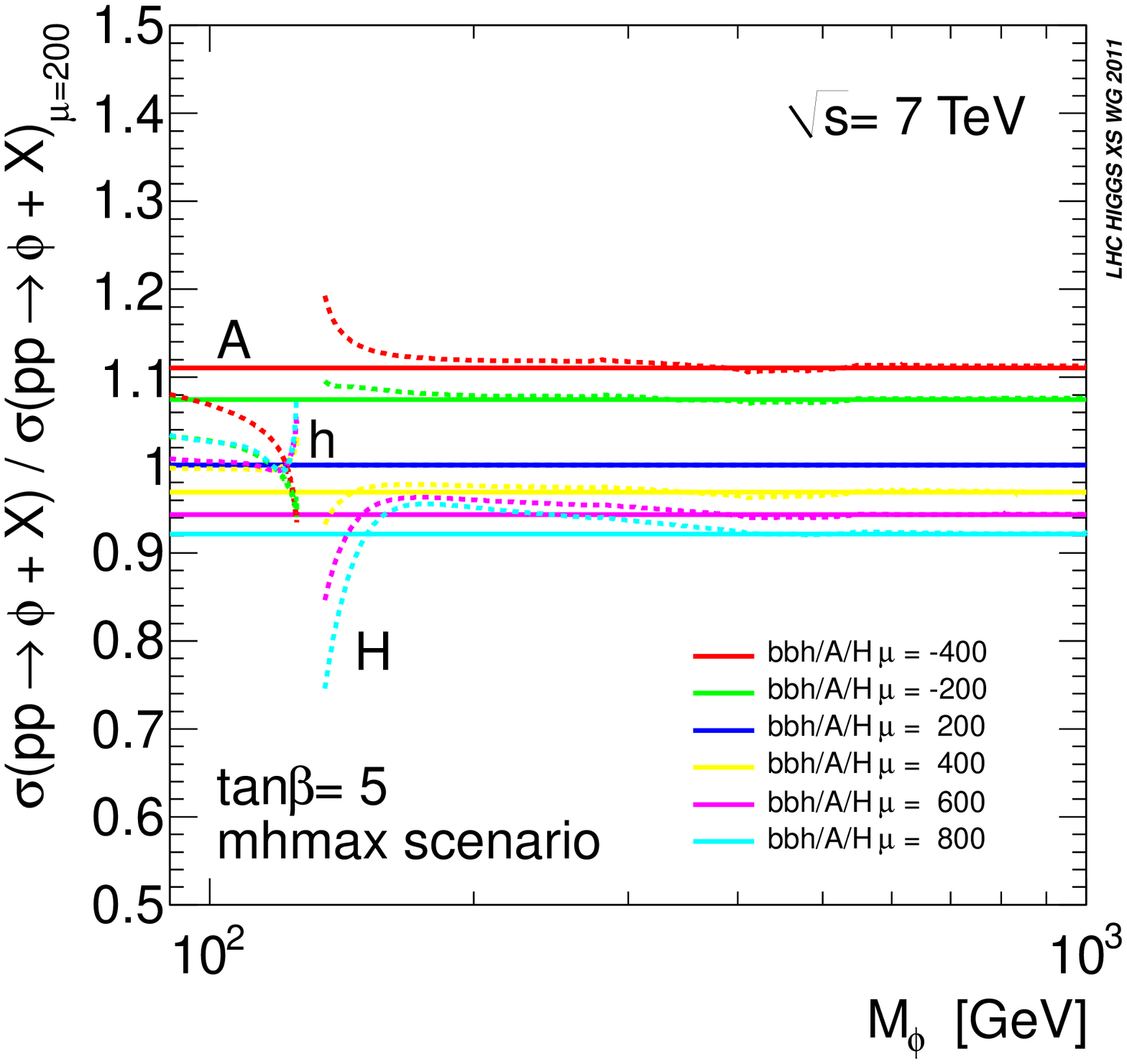}}
\subfigure[]{\includegraphics[width=0.49\textwidth]{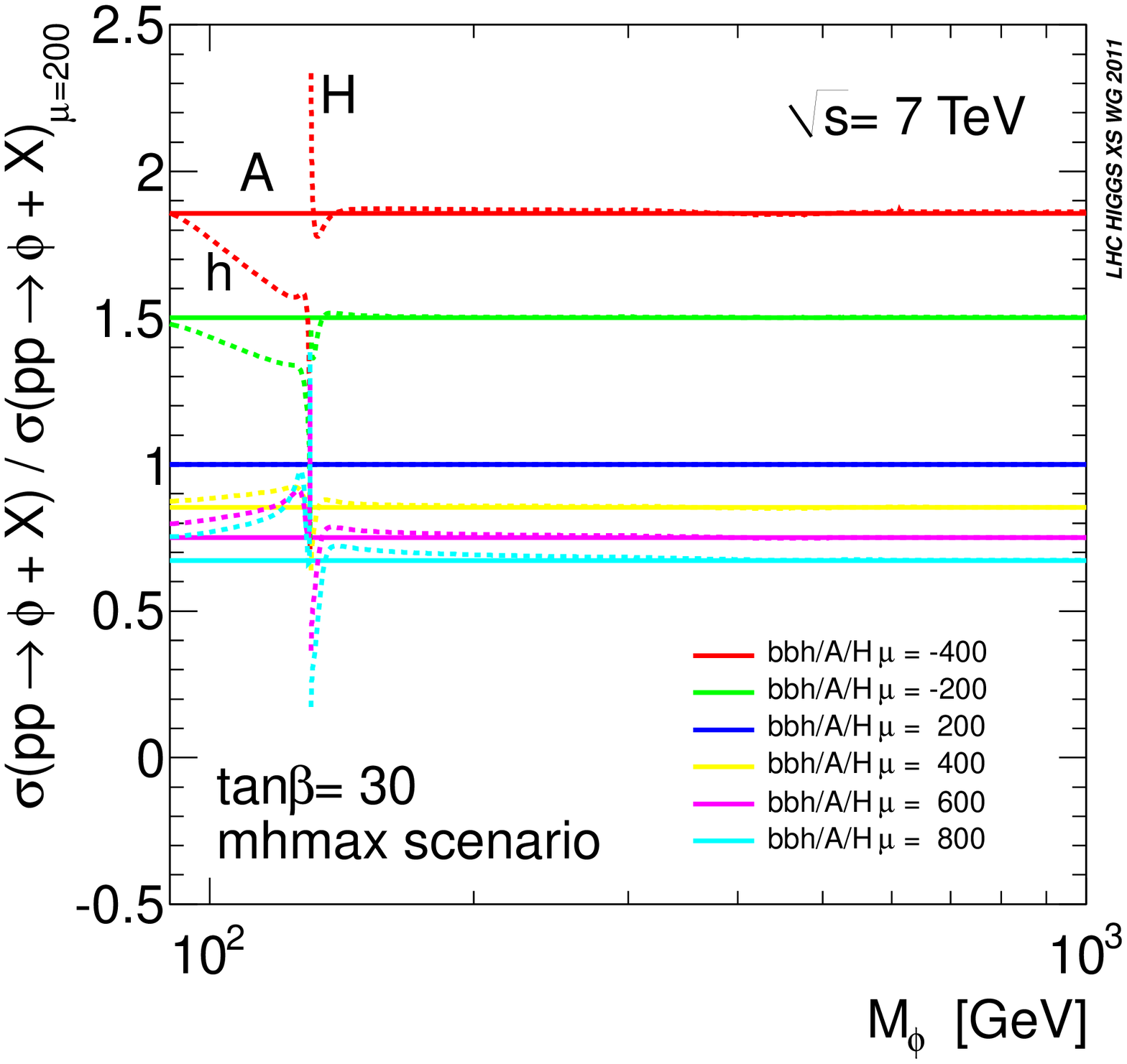}}
\caption{\label{YRHXS_MSSM_neutral_fig6} Ratios of the total MSSM
production cross sections via gluon fusion and Higgs radiation off
bottom quarks within the 5FS for $\sqrt{s}=7$\UTeV\ using NNLO and NLO
MSTW2008 PDFs \cite{Martin:2009iq,Martin:2009bu} for the \mhmaxx\
scenario with different $\mu$ values relative to the central prediction
with $\mu=200 \UGeV$; (a) gluon fusion for $\tanb=5$, (b) gluon fusion
for $\tanb=30$, (c) Higgs radiation off bottom quarks for $\tanb=5$, (d)
Higgs radiation off bottom quarks for $\tanb=30$.}
\end{figure}
In \Fref{YRHXS_MSSM_neutral_fig7} the central predictions for the
gluon-fusion processes $\Pg\Pg\to \PSh,\PSH,\PSA$ and neutral Higgs
radiation off bottom quarks within the 5FS are shown as a function of
the corresponding Higgs mass within the no-mixing scenario for two
values of $\tanb=5,30$.  It is clearly visible that Higgs-boson
radiation off bottom quarks plays the dominant role for $\tanb=30$ while
for $\tanb=5$ the gluon fusion is either dominant or competitive.
\begin{figure}
\subfigure[]{\includegraphics[width=0.49\textwidth]{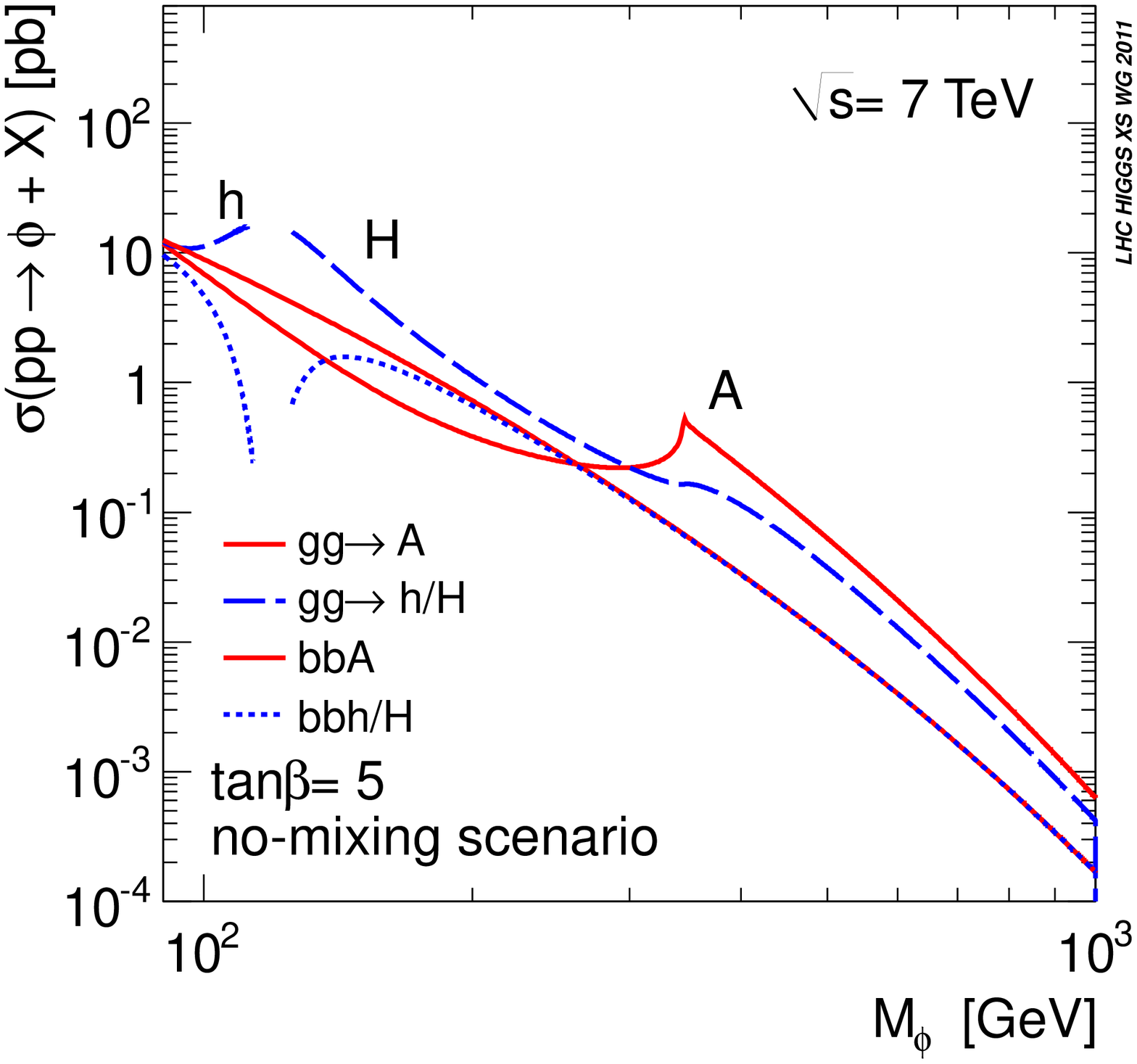}}
\subfigure[]{\includegraphics[width=0.49\textwidth]{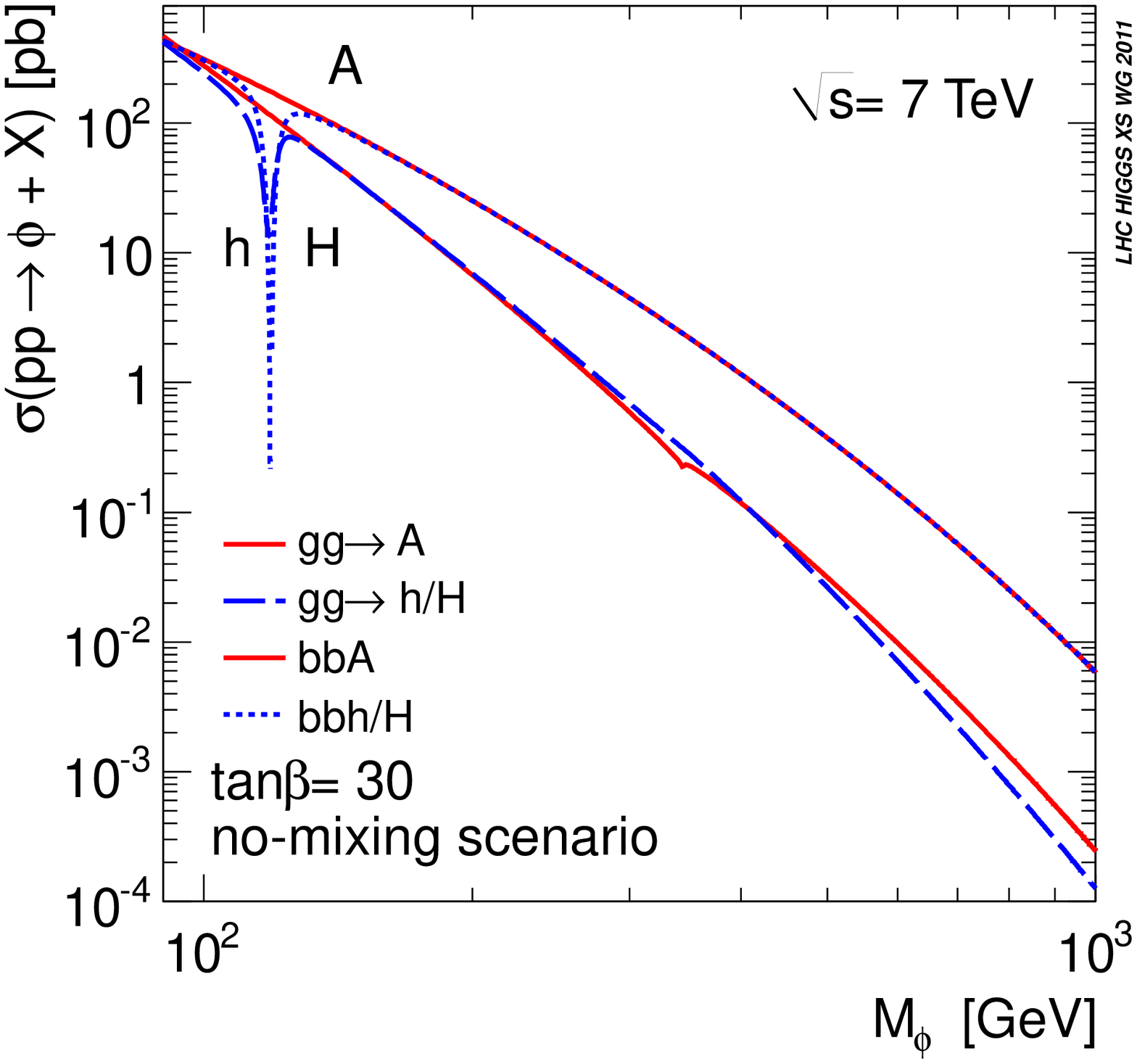}}
\caption{\label{YRHXS_MSSM_neutral_fig7} Central predictions for the
total MSSM production cross sections via gluon fusion and Higgs
radiation off
bottom quarks within the 5FS for $\sqrt{s}=7$\UTeV\ using NNLO and NLO
MSTW2008 PDFs \cite{Martin:2009iq,Martin:2009bu} for the no-mixing
scenario; (a) $\tanb=5$, (b) $\tanb=30$.}
\end{figure}

\begin{sloppypar}
The results for the total cross sections have been obtained from the grids
generated by \gghnnlo\ and \HIGLU\ for the gluon-fusion process and
\bbhnnlo~for $\PQb\PAQb\to \phi$ and rescaling the corresponding Yukawa
couplings by the MSSM factors calculated with \FeynHiggs.
\end{sloppypar}


%


\subsubsection{Santander matching}
\label{Santander-neutral}

A simple and pragmatic formula for the combination of the 
four- and five-flavour scheme calculations of 
bottom-quark associated Higgs-boson production
has been suggested in~\Bref{santander}. The matching formula originated
from discussions among the authors of \Bref{santander} at the {\it Higgs Days at Santander
    2009} and is therefore dubbed ``Santander matching''. We shall
briefly describe the matching scheme and provide matched
predictions for the inclusive cross section $\Pp\Pp \to (\PQb\PAQb )\PH+X$ at
the LHC operating at a centre-of-mass energy of $7\UTeV$ within the
Standard Model for illustrative purposes. 

The 4FS and 5FS calculations provide the unique description of the
cross section in the asymptotic limits $\MH/\Mb \to 1$ and
$\MH/\Mb \to \infty$, respectively. For phenomenologically
relevant Higgs-boson masses away from these asymptotic regions both
schemes are applicable and include different types of higher-order
contributions.  The matching suggested in \Bref{santander}
interpolates between the asymptotic limits of very light and very heavy
Higgs bosons. 

A comparison of the 4FS and 5FS calculations reveals that both are
in numerical agreement for moderate Higgs-boson masses (see Fig.\,23 of
\Bref{Dittmaier:2011ti}).  Once larger Higgs-boson masses are
considered, the effect of the collinear logarithms $\ln (\MH/\Mb)$ 
becomes more and more important and the two
approaches begin to differ. The two approaches are combined in
such a way that they are given variable weight, depending on the value
of the Higgs-boson mass. The difference between the two approaches is formally
logarithmic. Therefore, the dependence of their relative importance
on the Higgs-boson mass should be controlled by a logarithmic term. The
coefficients are determined such that
\begin{itemize}
\item[(a)] the 5FS gets $100\%$ weight in the limit $\MH/\Mb \to \infty$\,;
\item[(b)] the 4FS gets $100\%$ weight in the limit where the logarithms
are ``small''. There is obviously quite some arbitrariness in this
statement. In \Bref{santander} it is assumed that ``small'' means 
$\ln (\MH/\Mb) = 2$.
The consequence of this particular choice is that the 4FS and the 5FS
both get the same weight for Higgs-boson masses around 100\,GeV, consistent
with the observed agreement between the 4FS and the 5FS in this
region.%
\footnote{Note that one should use the {\it pole mass} for
$\Mb$ here rather than the running mass, since it is really the dynamical mass 
that rules the re-summed logarithms.}
\end{itemize}
This leads to the following formula
\begin{equation}
\begin{split}
\sigma^\text{matched}= \frac{\sigma^\text{4FS} +
  w\,\sigma^\text{5FS}}{1+w}\,,
\end{split}
\end{equation}
with the weight $w$ defined as 
\begin{equation}
\begin{split}
w = \ln\frac{\MH}{\Mb}  - 2\,,
\label{eq::tn}
\end{split}
\end{equation}
and $\sigma^\text{4FS}$ and $\sigma^\text{5FS}$ denote the total
inclusive cross section in the 4FS and the 5FS, respectively.
For $\Mb=4.75\UGeV$ and specific values of $\MH$, this leads to
\begin{equation}
\begin{split}
\sigma^\text{matched} \big|_{\MH=100\,\text{GeV}} 
&= 0.49\,\sigma^\text{4FS} + 0.51\,\sigma^\text{5FS}\,,\\
\sigma^\text{matched} \big|_{\MH=200\,\text{GeV}} 
&= 0.36\,\sigma^\text{4FS} + 0.64\,\sigma^\text{5FS}\,,\\
\sigma^\text{matched} \big|_{\MH=300\,\text{GeV}} 
&= 0.31\,\sigma^\text{4FS} + 0.69\,\sigma^\text{5FS}\,, \\
\sigma^\text{matched} \big|_{\MH=400\,\text{GeV}} 
&= 0.29\,\sigma^\text{4FS} + 0.71\,\sigma^\text{5FS}\,, \\
\sigma^\text{matched} \big|_{\MH=500\,\text{GeV}} 
&= 0.27\,\sigma^\text{4FS} + 0.73\,\sigma^\text{5FS}\, .
\end{split}
\end{equation}
A graphical representation of the weight factor $w$ is shown in
\fig{fig::mssm_xs4f5fhks}\,(a).

The theoretical uncertainties  in the 4FS and the 5FS calculations
should be added linearly, using the
weights $w$ defined in \eqn{eq::tn}.  This ensures that the combined
error is always larger than the minimum of the two individual
errors. Neglecting correlations and assuming equality of the
uncertainties in the 4FS and 5FS calculations would imply that the
matched uncertainty is reduced by a factor of $w/(1+w)$ with respect to
the common individual uncertainties.  This seems unreasonable. In the 
approach adopted in \Bref{santander} the matched uncertainty would 
be equal to the individual ones in this case. On the other hand,
taking the envelope of the 4FS and
5FS error bands seems overly conservative.

The estimates of the theoretical uncertainties in the 4FS and the 5FS 
calculations
are obtained through $\muF$, $\muR$, PDF, and
$\alphas$ variation
as described in \Bref{Dittmaier:2011ti}. They can be quite
asymmetric, which is why the combination should be done separately for
the upper and the lower uncertainty limits:
\begin{equation}
\begin{split}
\Delta\sigma_\pm = \frac{\Delta\sigma_\pm^\text{4FS}
  + w\,\Delta\sigma_\pm^\text{5FS}}{1+w}\,,
\end{split}
\end{equation}
where $\Delta\sigma_\pm^\text{4FS}$ and
$\Delta\sigma_\pm^\text{5FS}$ are the upper/lower uncertainty limits
of the 4FS and the 5FS, respectively.

We shall now discuss the numerical implications of the Santander matching
and provide matched
predictions for the inclusive cross section $\Pp\Pp \to (\PQb\PAQb )\PH+X$ 
at the LHC operating at a centre-of-mass energy of $7\UTeV$. The individual
numerical results for the 4FS and 5FS calculations have been
obtained in the context of the 
{\it LHC Higgs Cross Section Working Group} with input parameters as
described in \Bref{Dittmaier:2011ti}. Note that the cross section
predictions presented below correspond to Standard Model bottom Yukawa
couplings and a bottom-quark mass of $\Mb = 4.75 \UGeV$. SUSY
effects can be taken into account by simply rescaling the bottom Yukawa
coupling to the proper value~\cite{Dittmaier:2006cz,Dawson:2011pe}.

\fig{fig::mssm_xs4f5fhks}\,(b) shows the central values for the 4FS and
the 5FS cross section, as well as the matched result, as a function
of the Higgs-boson mass. The ratio of the central 4FS and the 5FS
predictions to the matched result is displayed in
\fig{fig::uncertainty}\,(b) (central dashed and dotted line): for
$\MH= 100 \UGeV$, the 4FS and the 5FS contribute with
approximately the same weight to the matched cross section, with
deviations between the individual 4FS and the 5FS predictions and
the matched one of less than $5\%$. With increasing Higgs-boson mass, the
4FS result deviates more and more from the matched cross section due
to its decreasing weight. At $\MH=500 \UGeV$, it agrees to less than
$20\%$ with the matched result, while the 5FS is still within $8\%$ of
the latter.

The corresponding theory error estimates are shown in
\refF{fig::uncertainty}. The absolute numbers are displayed in
panel\,(a), while in panel\,(b) they are shown relative to the central
value of the matched result. Up to $\MH\approx 300 \UGeV$, the
combined uncertainty band covers the central values of both the 4FS
and the 5FS. For larger Higgs-boson masses, the 4FS central value
is slightly outside this band.

%
\begin{figure}
    \begin{tabular}{cc} 
      \includegraphics[width=.49\textwidth]{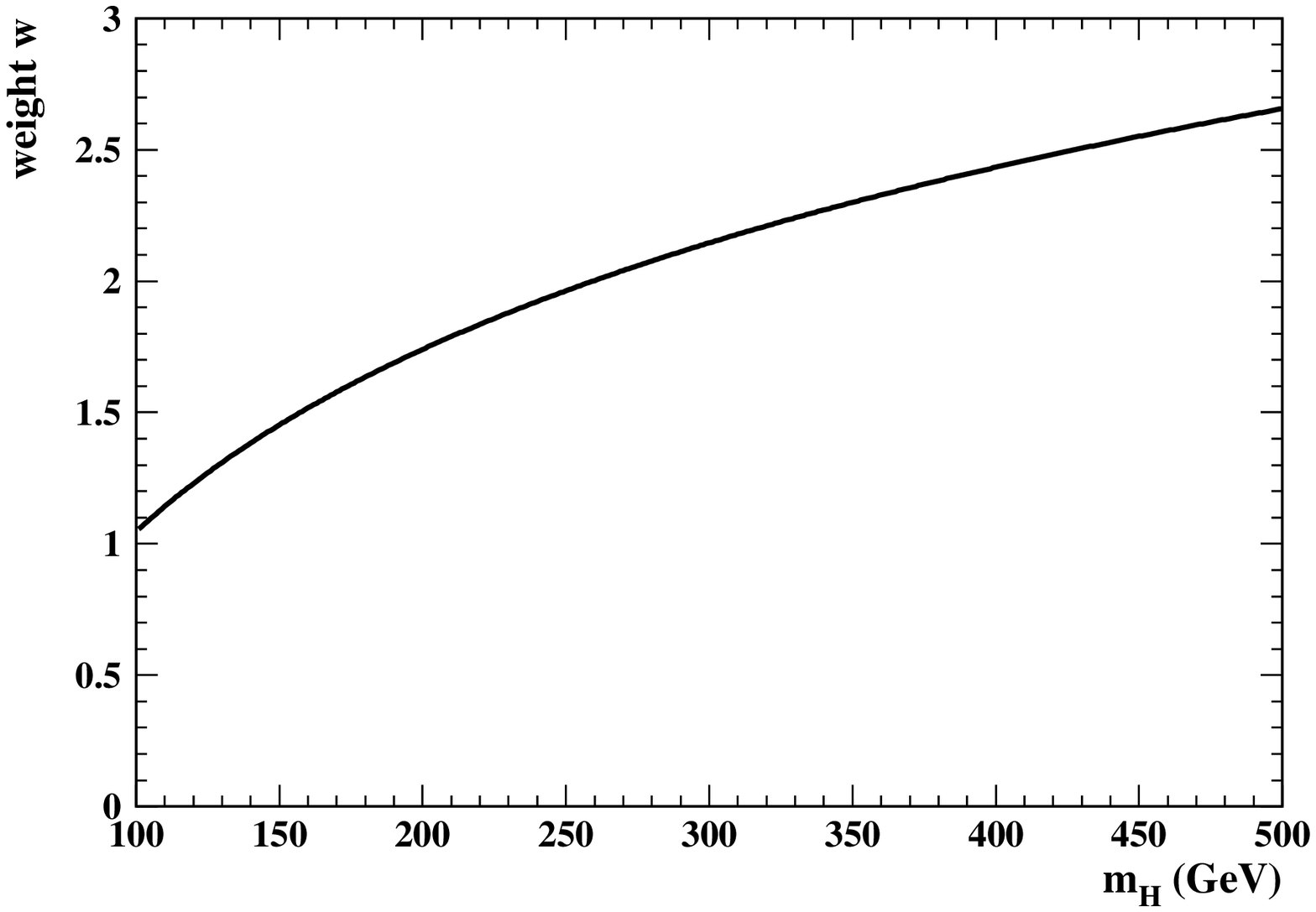} &
      \hspace*{-1em}
      \includegraphics[width=.49\textwidth]{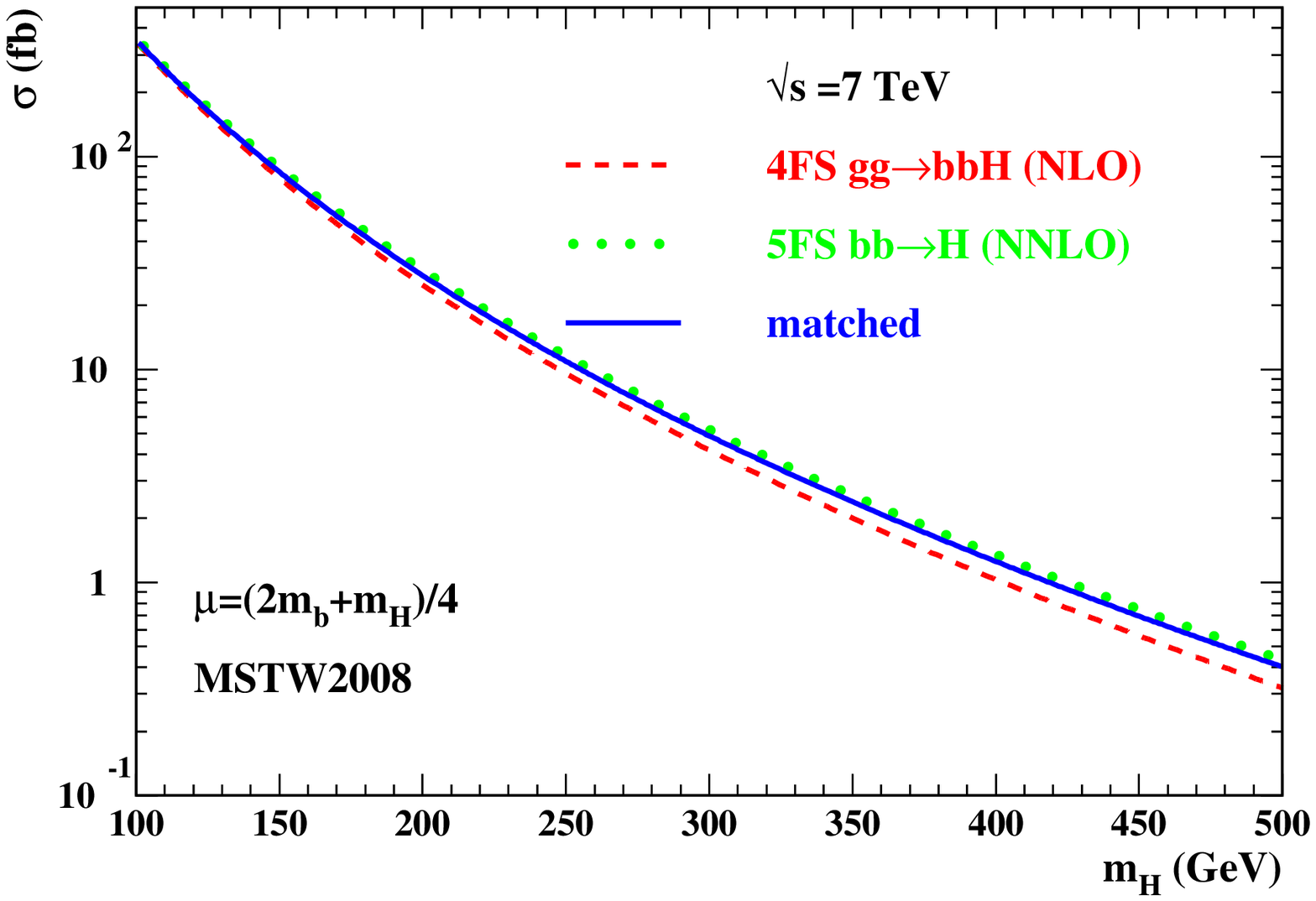} 
      \hspace*{-1em}
\\
      (a) & (b)
    \end{tabular}
      \caption{\label{fig::mssm_xs4f5fhks}\sloppy (a) Weight factor $w$,
        \eqn{eq::tn}, as a function of the Higgs-boson mass $\MH$.
        The bottom-quark pole mass has been set to $\Mb =
        4.75 \UGeV$. (b) Central values for the total inclusive cross
        section in the 4FS (red, dashed), the 5FS (green, dotted),
        and for the Santander-matched cross section (blue, solid).
        Here and in the following we use the MSTW2008\, PDF
        set~\cite{Martin:2009iq} (NLO for the 4FS, NNLO for
        the 5FS).}
\end{figure}
%

%
\begin{figure}
    \begin{tabular}{cc}
      \includegraphics[width=.49\textwidth]{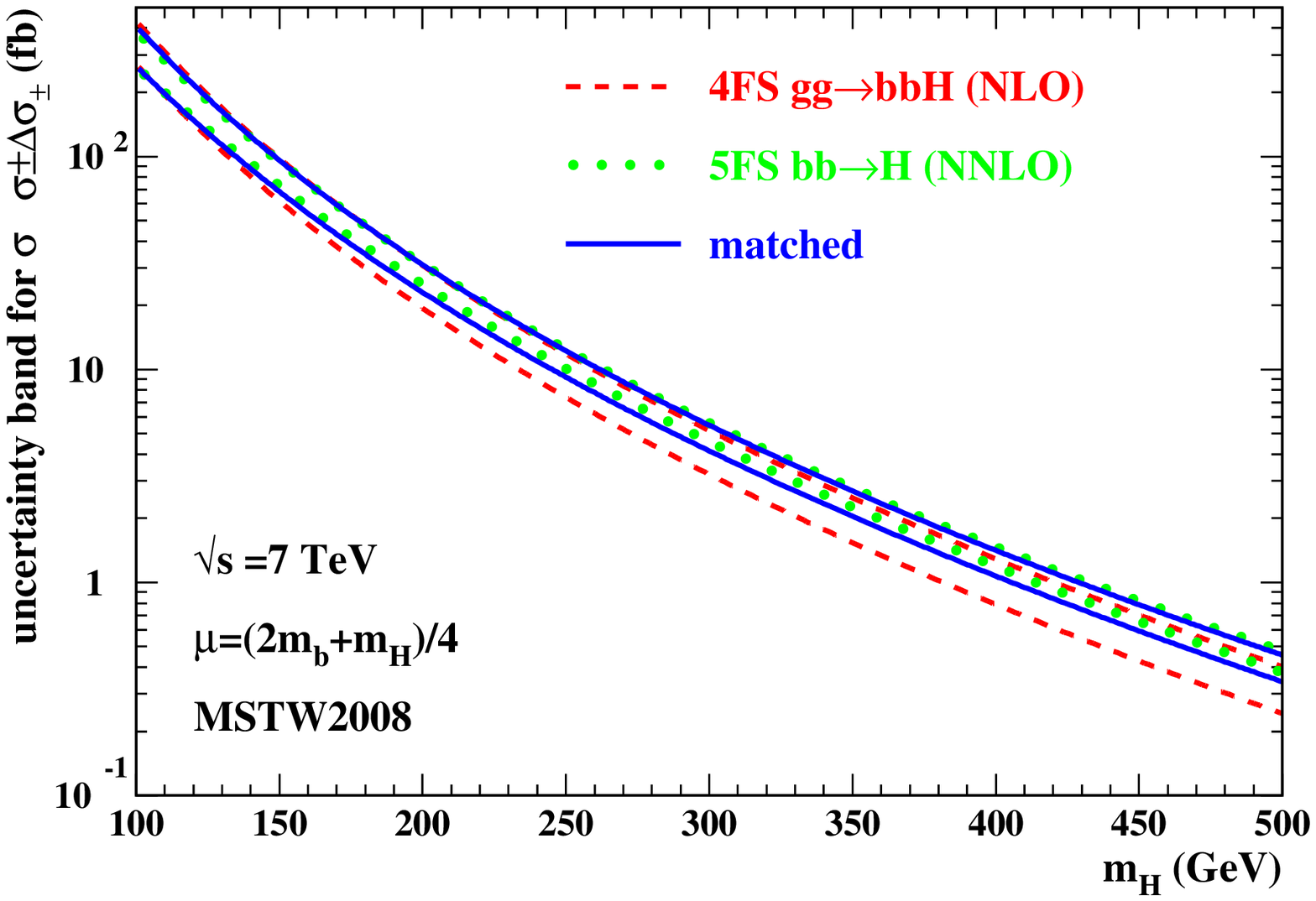} &
      \hspace*{-1em}
      \includegraphics[width=.49\textwidth]{%
        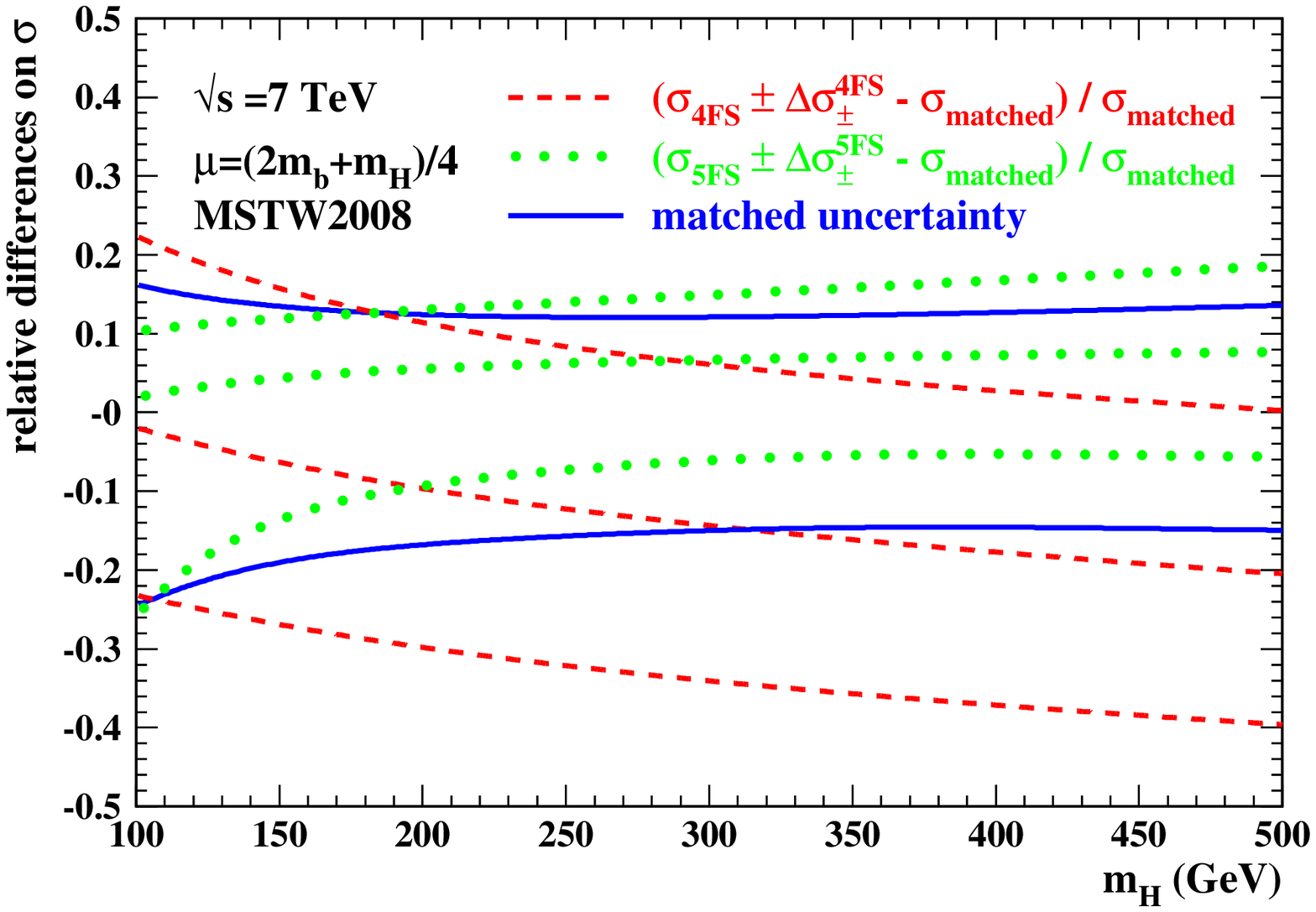}  
      \hspace*{-1em}
\\
      (a) & (b)
    \end{tabular}
      \caption{\label{fig::uncertainty}\sloppy (a) Theory uncertainty
        bands for the total inclusive cross section in the 4FS (red,
        dashed), the 5FS (green, dotted), and for the Santander-matched cross
        section (blue, solid).  (b) Uncertainty bands and central
        values, relative to the central value of the Santander-matched result
        (same line coding as panel (a)).}
\end{figure}
%


\subsection{Status of differential cross sections}

\subsubsection{Dressing SM predictions with effective couplings}
\label{sec:effcoup}


In making predictions for Higgs-boson production (or decay) processes in the
MSSM one has to face the fact that certain types of higher-order 
corrections have only been calculated in the SM case up to now, while
their counterpart for the case of the MSSM is not yet available. 
One the one hand, starting from dedicated MSSM calculations for Higgs cross
sections or decay widths treats higher-order corrections of SM type and
SUSY type on the same footing. On the other hand, this approach may be
lacking the known to be relevant (or at least the most up to date)
SM-type corrections. Consequently, 
it can be advantageous to start from SM-type processes including 
all the known higher-order corrections, 
and to dress suitable SM-calculation building blocks 
with appropriate MSSM coupling factors (effective couplings). 
At the same time, for internal consistency, the MSSM predictions for
Higgs-boson masses etc.\ have to be used.
While for the results for the MSSM Higgs production in gluon fusion we
discuss below the results of genuine MSSM calculations, for MSSM Higgs
production in association with bottom quarks the SM predictions are
dressed with appropriate effective couplings in the MSSM. For the 
weak-boson fusion channel we discuss the genuine MSSM result and compare
it to that that of the effective coupling approach. A similar comparison
for the
other Higgs production channels will be investigated in a later stage.


\subsubsection{Implementation of Higgs production via gluon fusion in
               the MSSM in {\sc POWHEG}}



The gluon-fusion mechanism, $\Pg\Pg \to \phi~~(\phi= \Ph,\PH,\PA)$, is
one of the most important processes for the production of the neutral Higgs
bosons of the MSSM. An
overview about the existing calculations is given in
\Sref{sec:mssm-calcsum}. 

The aim is to obtain results for Higgs-boson kinematic distributions,
but going beyond the approximation of restricting to the quark-loop
contributions (dressed with appropriate MSSM couplings). 
The results are obtained by combining all the
available NLO QCD information, \ie  including the superpartner
contributions, with the description of initial-state multiple-gluon
emission via a QCD parton shower (PS) Monte Carlo. The analysis relies
on the {\sc POWHEG} method~\cite{Nason:2004rx,Frixione:2007vw}, which
allows to systematically merge NLO calculations with vetoed PS,
avoiding double counting and preserving the NLO accuracy in the total
cross section. The merging procedure can be implemented using a
computer code framework called the {\sc POWHEG BOX}~\cite{Alioli:2010xd},
which allows to build an event generator that always assigns a
positive weight to the hardest event. The {\sc POWHEG} implementation
of the gluon-fusion Higgs production process in the SM was reported in
\Bref{Alioli:2008tz} (see also \Bref{Hamilton:2009za}) and
subsequent modifications to include the finite-quark-mass effects were
presented elsewhere~\cite{Bagnaschi:2011tu}.

We briefly summarise what is included in our {\sc POWHEG}
implementation of $\Pg\Pg \to \Ph,\PH$ in the MSSM (with more details in
\Bref{Bagnaschi:2011tu}). For the contributions of
diagrams with real-parton emission, as well as for the leading-order
(LO) virtual contributions, we use one-loop matrix elements with exact
dependence on the quark, squark, and Higgs masses from
\Bref{Bonciani:2007ex}. Concerning the NLO QCD virtual
contributions involving squarks, we use the results of
\Bref{Degrassi:2008zj} for the stop contributions, obtained in
the approximation of vanishing Higgs mass, and the results of
\Bref{Degrassi:2010eu} for the sbottom contributions, obtained
via an asymptotic expansion in the large supersymmetric masses that is
valid up to and including terms of ${\cal O}(\Mb^2/m_\phi^2)$, 
${\cal O}(\Mb/M_{\mathrm{SUSY}})$, and ${\cal O}(\MZ^2/M_{\mathrm{SUSY}}^2)$, 
with $M_{\mathrm{SUSY}}$ a generic superparticle mass.  For the remaining NLO QCD
contributions, arising from the two-loop diagrams with quarks and
gluons, we use the exact results of \Bref{Aglietti:2006tp}.  The
two-loop electroweak corrections, for which a complete calculation in
the MSSM is not available, are included in an approximate way by
properly rescaling the SM light-quark contributions given in
\Brefs{Aglietti:2004nj,Bonciani:2010ms}. This approximation is
motivated by the observation that, in the SM, the light-quark
contributions make up the dominant part of the electroweak corrections
when the Higgs mass is below the threshold for top-pair production.

As discussed above, it
is necessary to compute the entire spectrum of masses and couplings of
the model in a consistent way, starting from a given set of input
parameters. In our numerical analysis, we use the code 
{\sc SoftSusy}~\cite{Allanach:2001kg} to compute the MSSM spectrum and
the couplings 
of the Higgs bosons to quarks and squarks starting from a set of
running parameters expressed in the $\overline{\rm DR}$
renormalisation scheme.  However, it is in principle possible to
interface the code with other spectrum calculators (such as, \eg,
{\sc FeynHiggs}) that adopt different choices
of renormalisation conditions for the input parameters.

For illustrative purposes we present numerical results for the
production of the lightest CP-even Higgs boson, $\Ph$, at the LHC with
centre-of-mass energy of $7\UTeV$, in a representative region of the MSSM
parameter space. Events are generated with the described implementation of 
{\sc POWHEG}, then matched with the {\sc PYTHIA}
PS~\cite{Sjostrand:2006za}.  We compute the total inclusive cross
section 
for light Higgs production in gluon fusion, as well as the 
transverse-momentum distribution for a light Higgs boson produced in association
with a jet, and we compare them with the corresponding quantities
computed for a SM Higgs boson with the same mass.

The relevant MSSM Lagrangian parameters are chosen as:
\begin{equation}
\label{eq:MSSMinputs}
m_Q=m_U=m_D= 500 \UGeV,~~X_{\PQt} = 1250 \UGeV,
~~M_3=2\,M_2=4\,M_1= 400 \UGeV,~~\left|\mu\right| = 200 \UGeV,
\end{equation}
where $m_Q,\,m_U$ and $m_D$ are the soft-SUSY-breaking mass terms for
stop and sbottom squarks, $X_{\PQt} \equiv A_{\PQt} - \mu\cot\beta$ is the
left--right mixing term in the stop mass matrix (where $A_{\PQt}$ is the 
soft-SUSY-breaking Higgs--squark coupling and $\mu$ is the higgsino mass
parameter in the superpotential), and $M_i$ (for $i=1,2,3$) are the 
soft-SUSY-breaking gaugino masses. We consider the input parameters in
Eq.~(\ref{eq:MSSMinputs}) as expressed in the $\overline{\rm DR}$
renormalisation scheme, at a reference scale $Q= 500 \UGeV$. The choice
of $X_{\PQt}/m_Q$ resembles the choice in the \mhmaxx\ scenario.  We recall
that, in our conventions, the $\tan\beta$-dependent corrections to the
relation between the bottom mass and the bottom Yukawa
coupling~\cite{Hall:1993gn,Hempfling:1993kv,
Carena:1994bv,Pierce:1996zz,Carena:1999py,
Guasch:2003cv,Noth:2008tw,Noth:2010jy,Mihaila:2010mp} enhance the Higgs
couplings to bottom and sbottoms for $\mu<0$ and suppress them for
$\mu>0$.

We perform a scan on the parameters that determine the Higgs-boson
masses and mixing at tree level, $\MA$ and $\tan\beta$, varying them
in the ranges $90 \UGeV < \MA < 200 \UGeV$ and $2 <
\tan \beta < 50$. For each value of $\tan\beta$ we derive $A_{\PQt}$
from the condition on $X_{\PQt}$, then we fix the corresponding
Higgs--sbottom coupling as $A_{\PQb} = A_{\PQt}$. For each point in the parameter
space, we use {\sc SoftSusy} to compute the physical (\ie,
radiatively corrected) Higgs-boson masses $\Mh$ and $\MH$, and the
effective mixing angle $\alpha$ that diagonalises the 
radiatively-corrected mass matrix in the CP-even Higgs sector. We obtain from 
{\sc SoftSusy} also the MSSM running quark masses $\Mt(Q)$ and
$\Mb(Q)$, expressed in the $\overline{\rm DR}$ scheme at the
scale $Q=500 \UGeV$. The running quark masses are used both in the
calculation of the running stop and sbottom masses and mixing angles,
and in the calculation of the top and bottom contributions to the form
factors for Higgs-boson production (the latter are computed using the
$\overline{\rm DR}$ results presented in
Refs.~\cite{Degrassi:2008zj,Degrassi:2010eu}).

\begin{figure}
\begin{center}
\includegraphics[height=76mm,width=76mm]{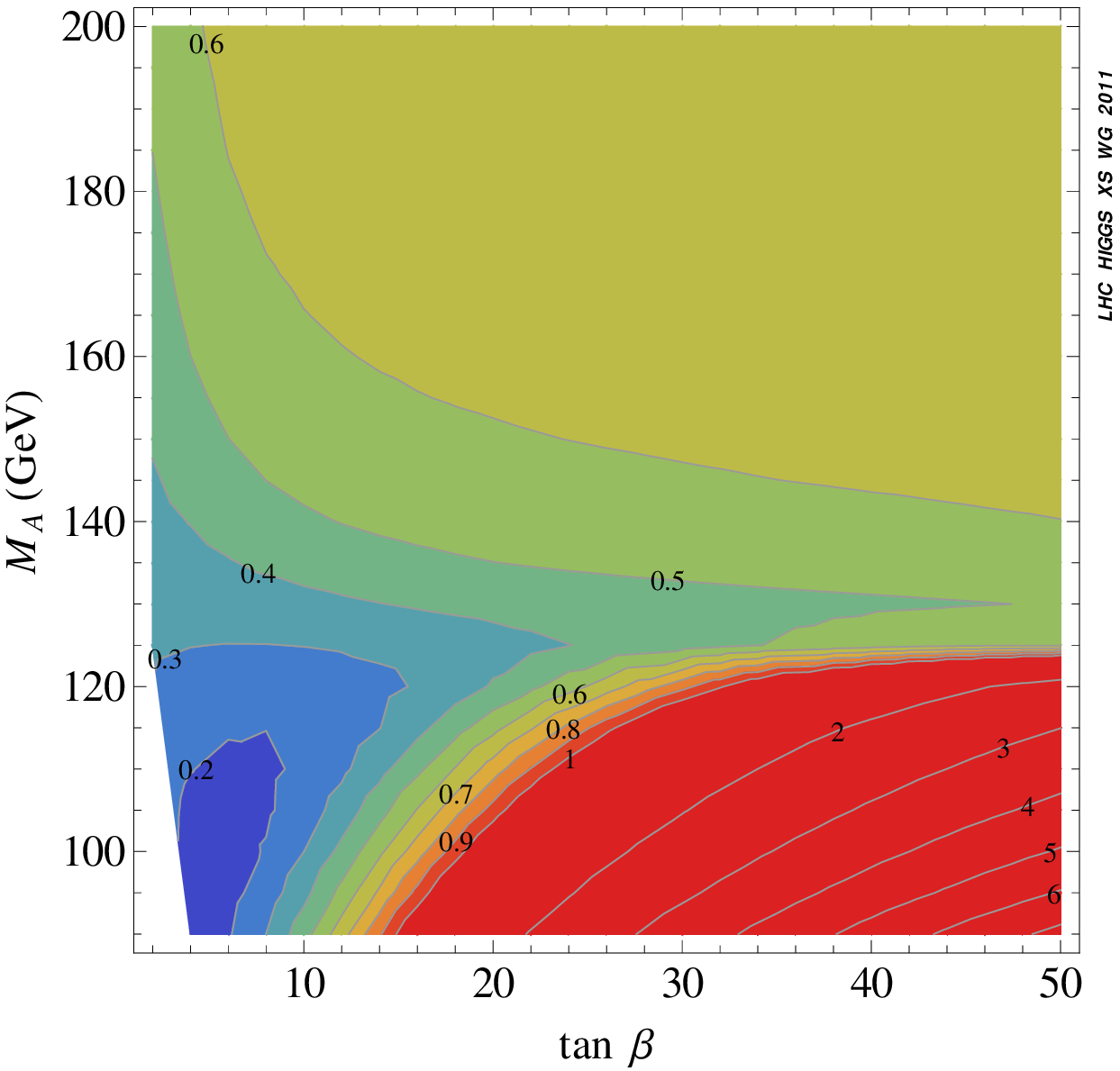}~~~~
\includegraphics[height=76mm,width=76mm]{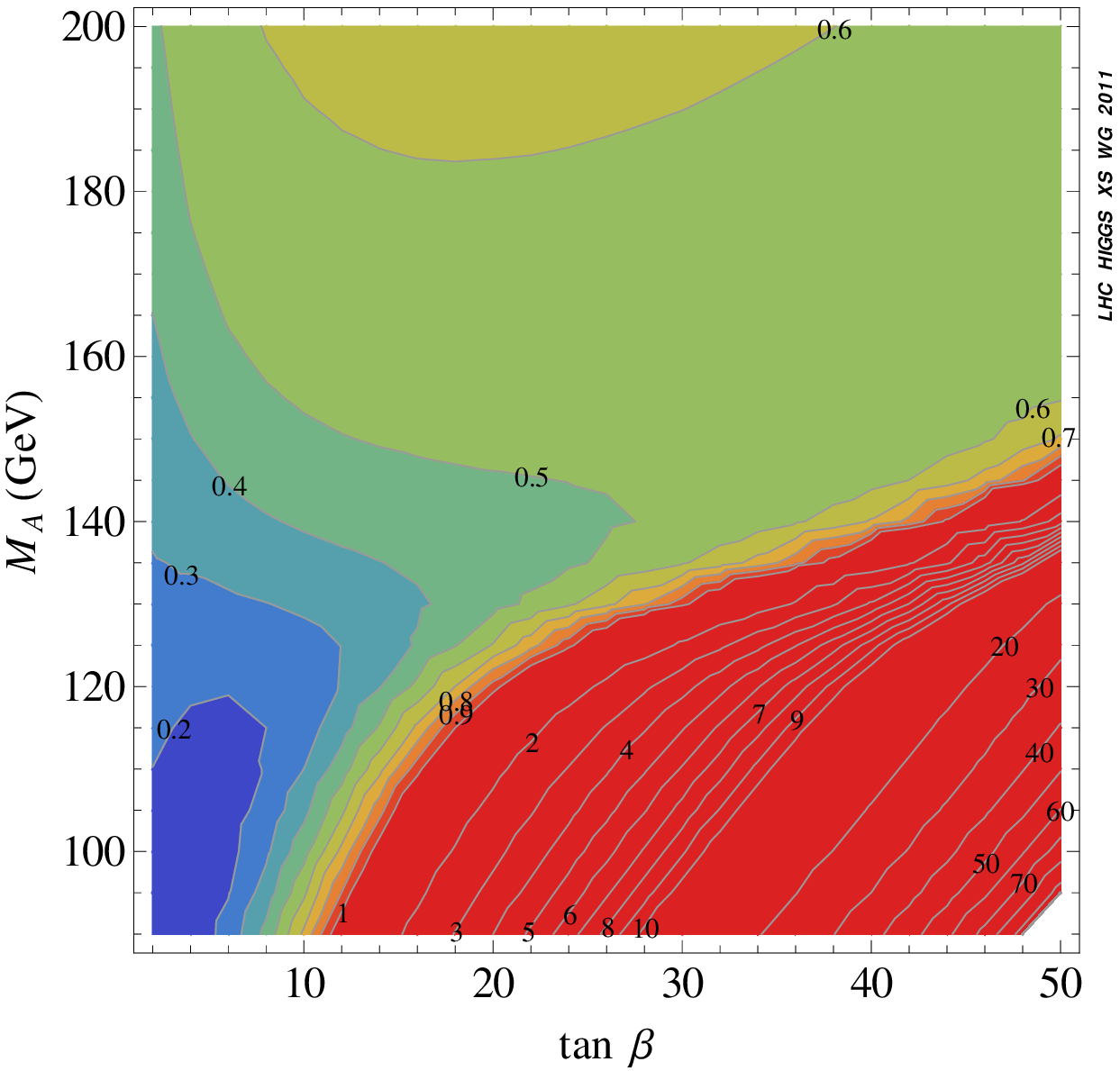}
\caption{\small Ratio of the total cross section for $\Ph$ production in
  the MSSM over the cross section for the production of a SM Higgs
  boson with the same mass. The plot on the left is for $\mu>0$ while
  the plot on the right is for $\mu<0$.
\label{fig:MSSMvsSMxsec}
} 
\end{center}
\end{figure}

In \refF{fig:MSSMvsSMxsec} we plot the ratio of the cross section
for the production of the lightest scalar $\Ph$ in the MSSM over the
cross section for the production of a SM Higgs boson with the same
mass. For a consistent comparison, we adopt the $\overline{\rm DR}$
scheme in both the MSSM and the SM calculations. The plot on the left
is obtained with $\mu>0$, while the plot on the right is obtained with
$\mu<0$. In order to interpret the plots, it is useful to recall that
for small values of $\MA$ it is the heaviest scalar $\PH$ that has
SM-like couplings to fermions, while the coupling of $\Ph$ to top
(bottom) quarks is suppressed (enhanced) by $\tan\beta$. In the
lower-left region of the plots, with small $\MA$ and moderate
$\tan\beta$, the enhancement of the bottom contribution does not
compensate for the suppression of the top contribution, and the MSSM
cross section is smaller than the corresponding SM cross section. On
the other hand, for sufficiently large $\tan\beta$ (in the lower-right
region of the plots) the enhancement of the bottom contribution
prevails, and the MSSM cross section becomes larger than the
corresponding SM cross section.  For $\mu<0$ the coupling of $\Ph$ to
bottom quarks is further enhanced by the $\tan\beta$-dependent
threshold corrections, and the ratio between the MSSM and SM
predictions can significantly exceed a factor of ten. We also note
that for sufficiently large $\MA$, \ie~when the couplings of $\Ph$ to
quarks approach their SM values, the MSSM cross section is smaller
than the SM cross section.

\begin{figure}
\begin{center}
\includegraphics[height=76mm,width=76mm]{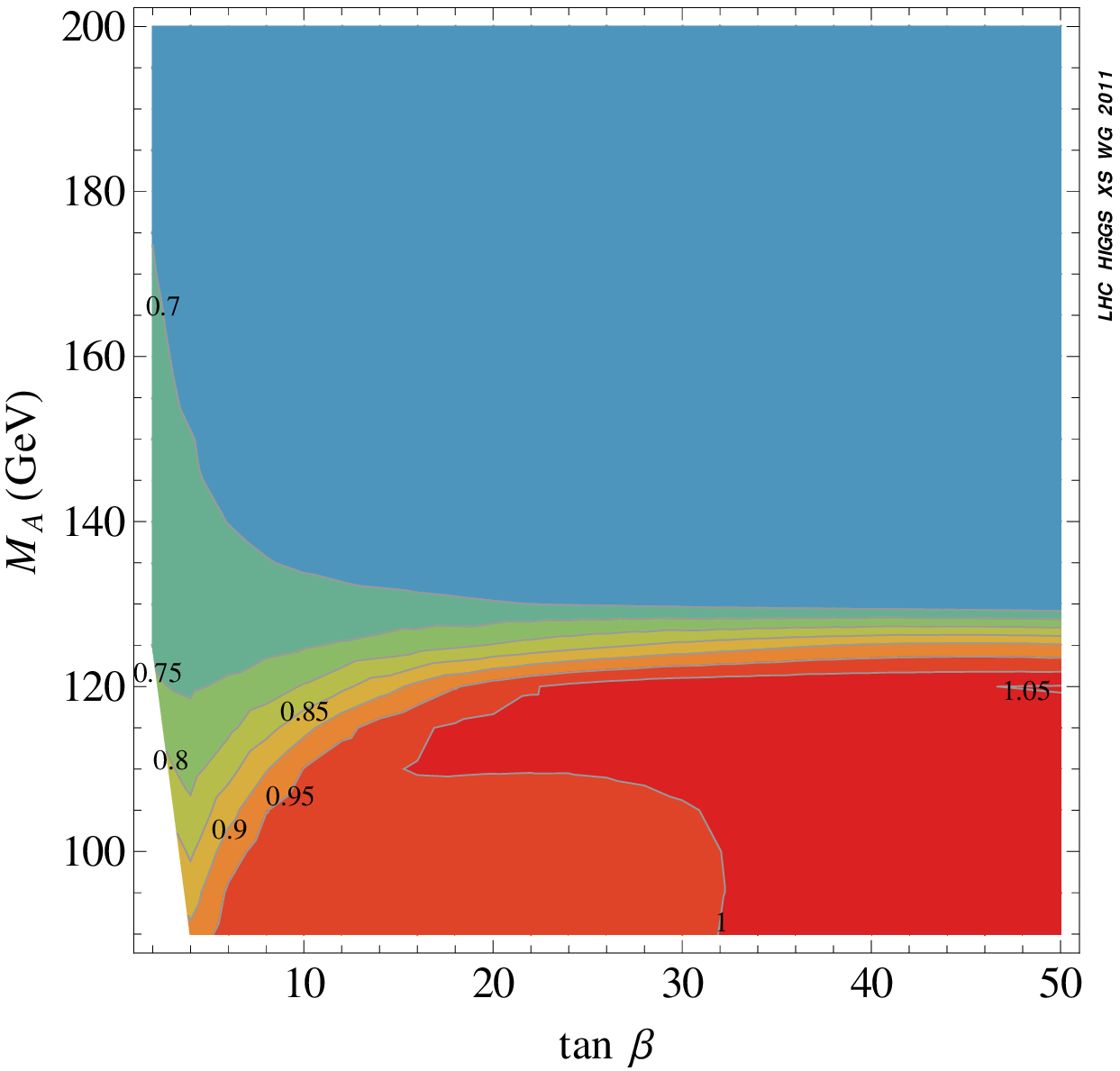}
~~~~
\includegraphics[height=76mm,width=76mm]{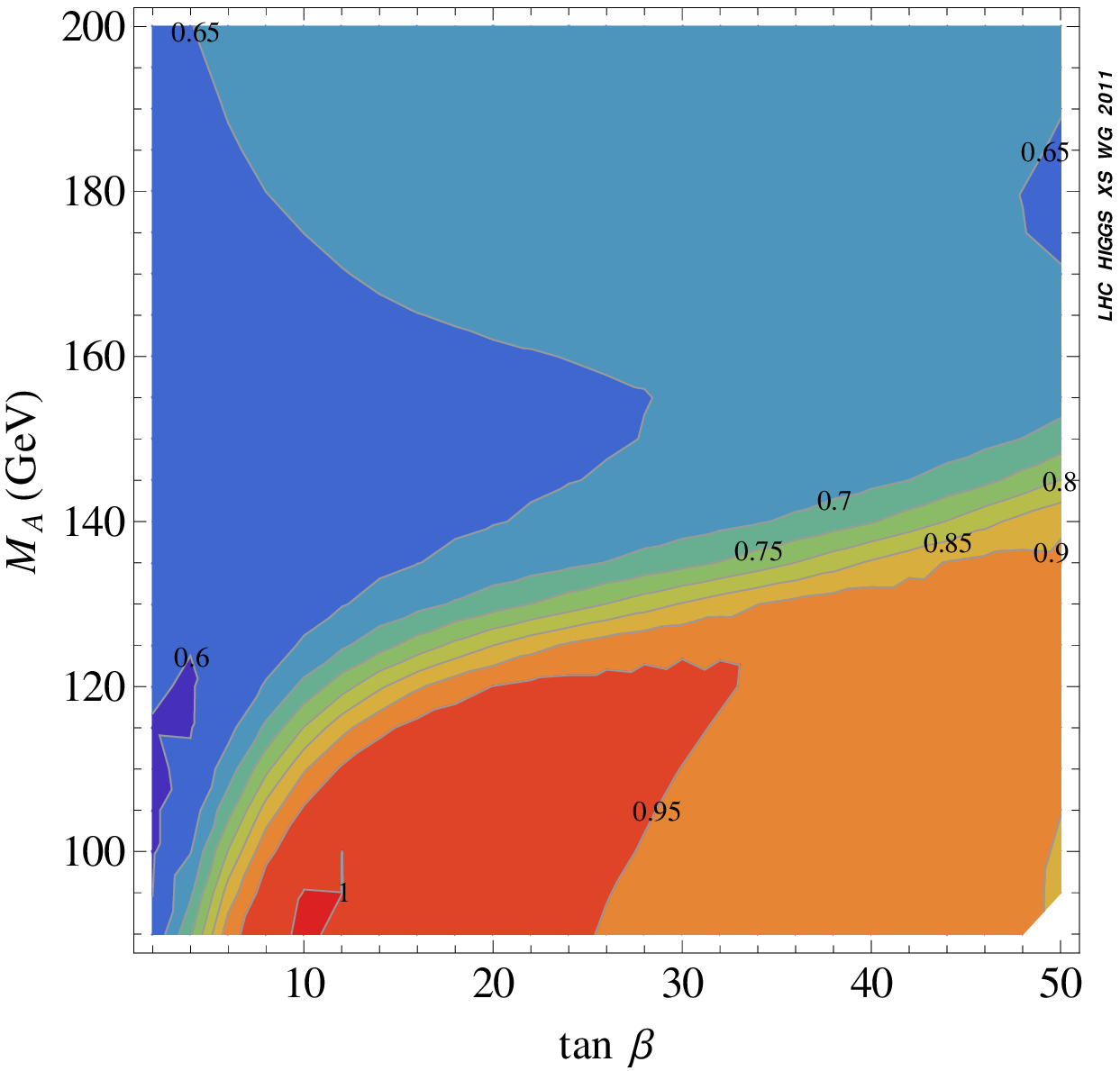}
\caption{\small Ratio of the full cross section for $\Ph$ production in
  the MSSM over the approximated cross section computed with only
  quarks running in the loops. The plot on the left is for $\mu>0$
  while the plot on the right is for $\mu<0$.
\label{fig:MSSMvsONLYQUARKS}
} 
\end{center}
\end{figure}

To assess the genuine effect of the squark contributions (as opposed to
the effect of the modifications in the Higgs--quark couplings), we plot
in \refF{fig:MSSMvsONLYQUARKS} the ratio of the full MSSM cross
section for the production of the lightest scalar $\Ph$ over the
approximated MSSM cross section computed with only quarks running in the
loops. As in \refF{fig:MSSMvsSMxsec}, the plot on the left is
obtained with $\mu>0$, while the plot on the right is obtained with
$\mu<0$. We observe that, in most of the considered region of the MSSM
parameter space, the squark contributions reduce the total cross
section. We identify three regions: $i)$ for sufficiently large
$\tan\beta$ and sufficiently small $\MA$ the squark contribution is
modest, ranging between $-10\%$ and $+5\%$; this region roughly
coincides with the one in which the total MSSM cross section is
dominated by the $\tan\beta$-enhanced bottom-quark contribution, and is
larger than the SM cross section; $ii)$ a transition region, where the
corrections rapidly become as large as $-30\%$; this region coincides
with the one in which the SM and MSSM cross sections are similar to each
other; $iii)$ for sufficiently large $\MA$ the squark correction is
almost constant, ranging between $-40\%$ and $-30\%$; this region
coincides with the one in which the MSSM cross section is smaller than
the corresponding SM cross section.

\begin{figure}
\begin{center}
\includegraphics[height=78mm,angle=-90]{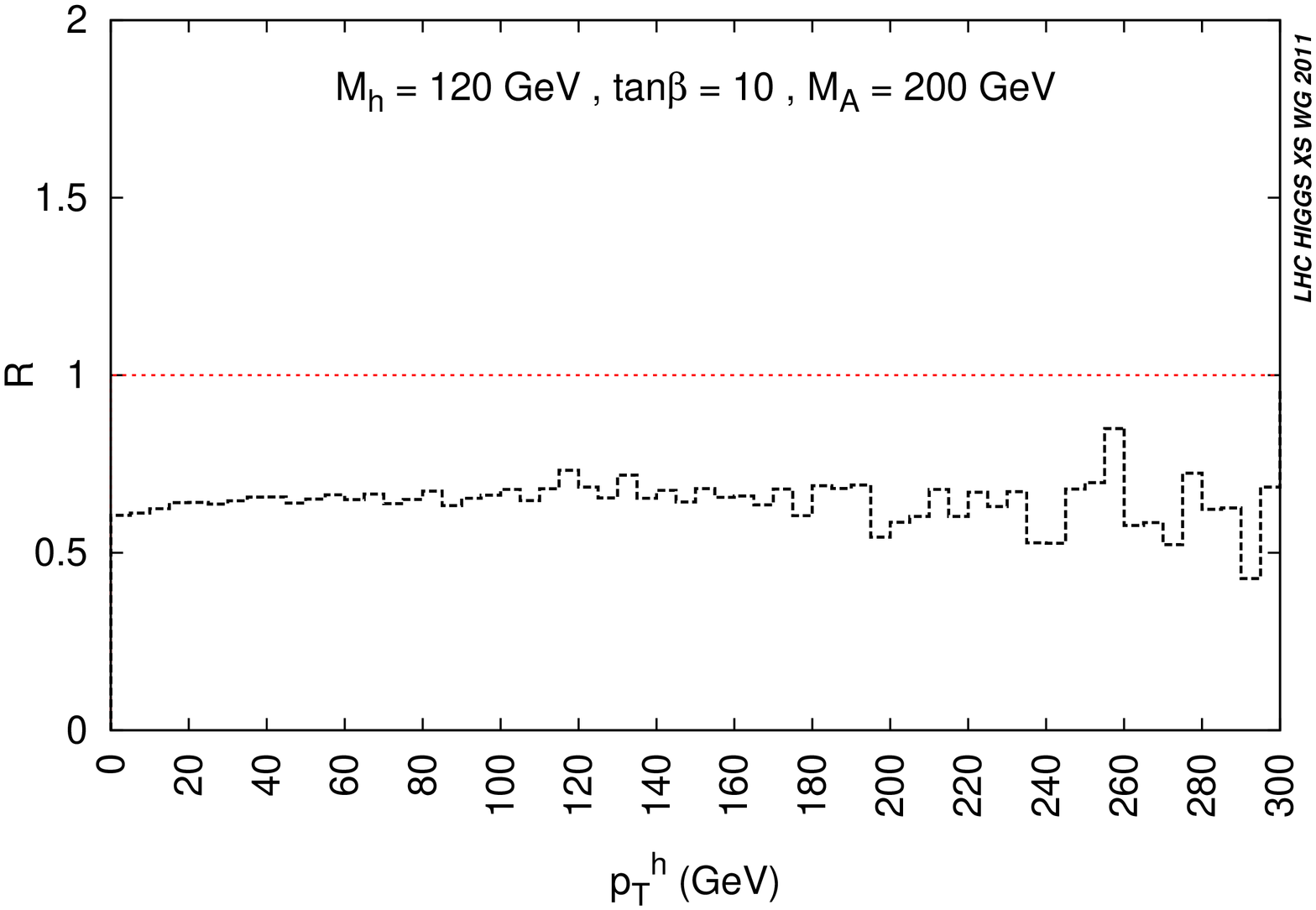}
~~
\includegraphics[height=78mm,angle=-90]{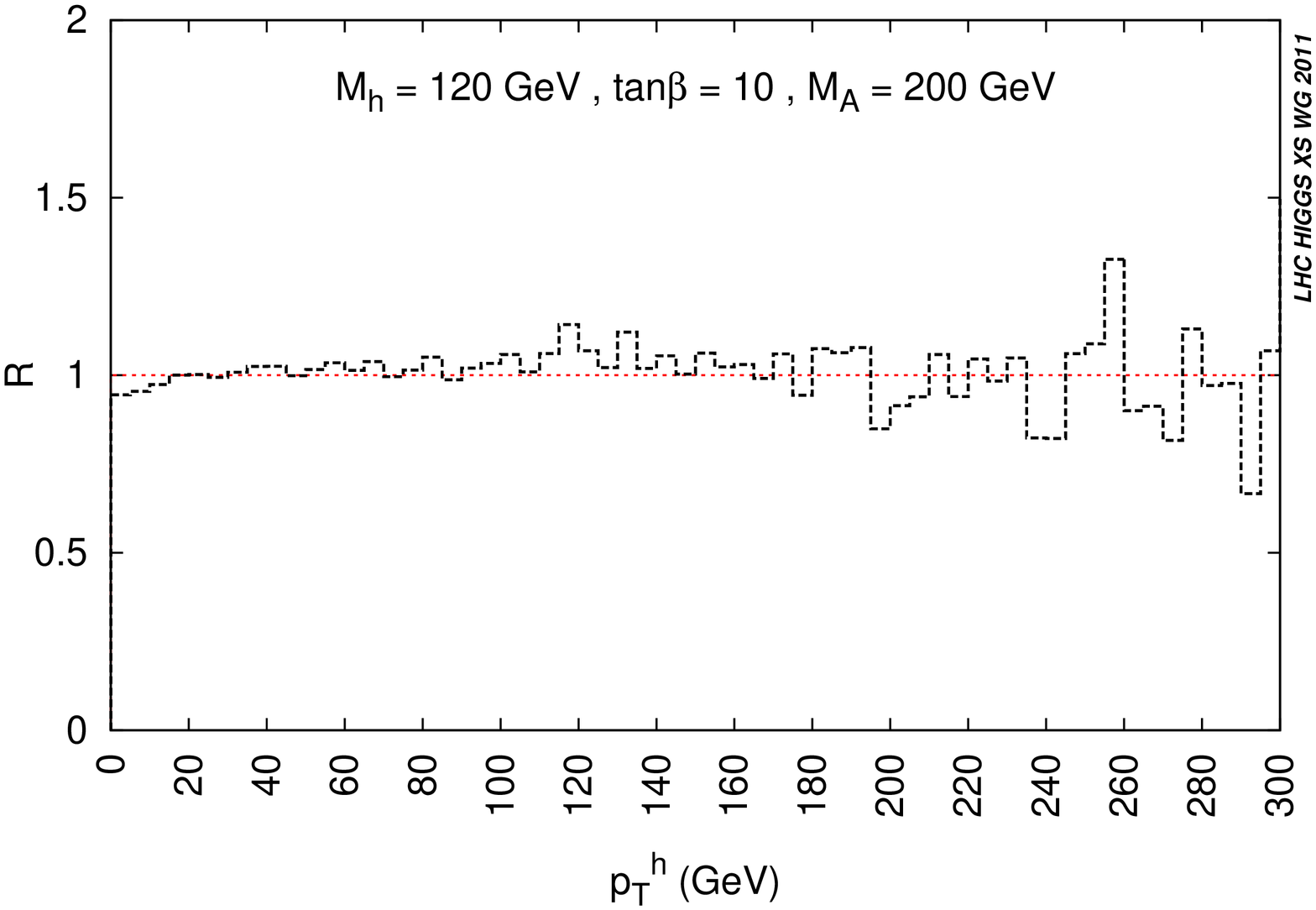}
\caption{\small Left: ratio of the transverse-momentum distribution of
  the lightest scalar $\Ph$ in the MSSM over the transverse-momentum
  distribution of a SM Higgs boson with the same mass. Right: ratio of
  the corresponding shapes.
\label{fig:tb10}
} 
\end{center}
\end{figure}

We now discuss the distribution of the transverse momentum $\pT^{\Ph}$ of
a light scalar $\Ph$ produced in association with a jet, considering two
distinct scenarios.  First, we take a point in the MSSM parameter
space ($\MA = 200 \UGeV$, $\tan \beta= 10$ and $\mu > 0$) in which the
coupling of $\Ph$ to the bottom quark is not particularly enhanced with
respect to the SM value, so that the bottom contribution to the cross
section is not particularly relevant. Because a light Higgs boson
cannot resolve the top and squark vertices, unless we consider very
large transverse momentum, we expect the form of the $\pT^{\Ph}$
distribution to be very similar to the one for a SM Higgs boson of
equal mass, the two distributions just differing by a scaling factor
related to the total cross section.  This is illustrated in the left
plot of \refF{fig:tb10}, where we show the ratio of the
transverse-momentum distribution for $\Ph$ over the distribution for a
SM Higgs boson of equal mass. In the right plot of
\refF{fig:tb10} we show the ratio of the corresponding shapes,
\ie\ the distributions normalised to the corresponding cross sections.
This ratio, as expected, is close to one in most of the $\pT^{\Ph}$ range.

\begin{figure}
\begin{center}
\includegraphics[height=78mm,angle=270]{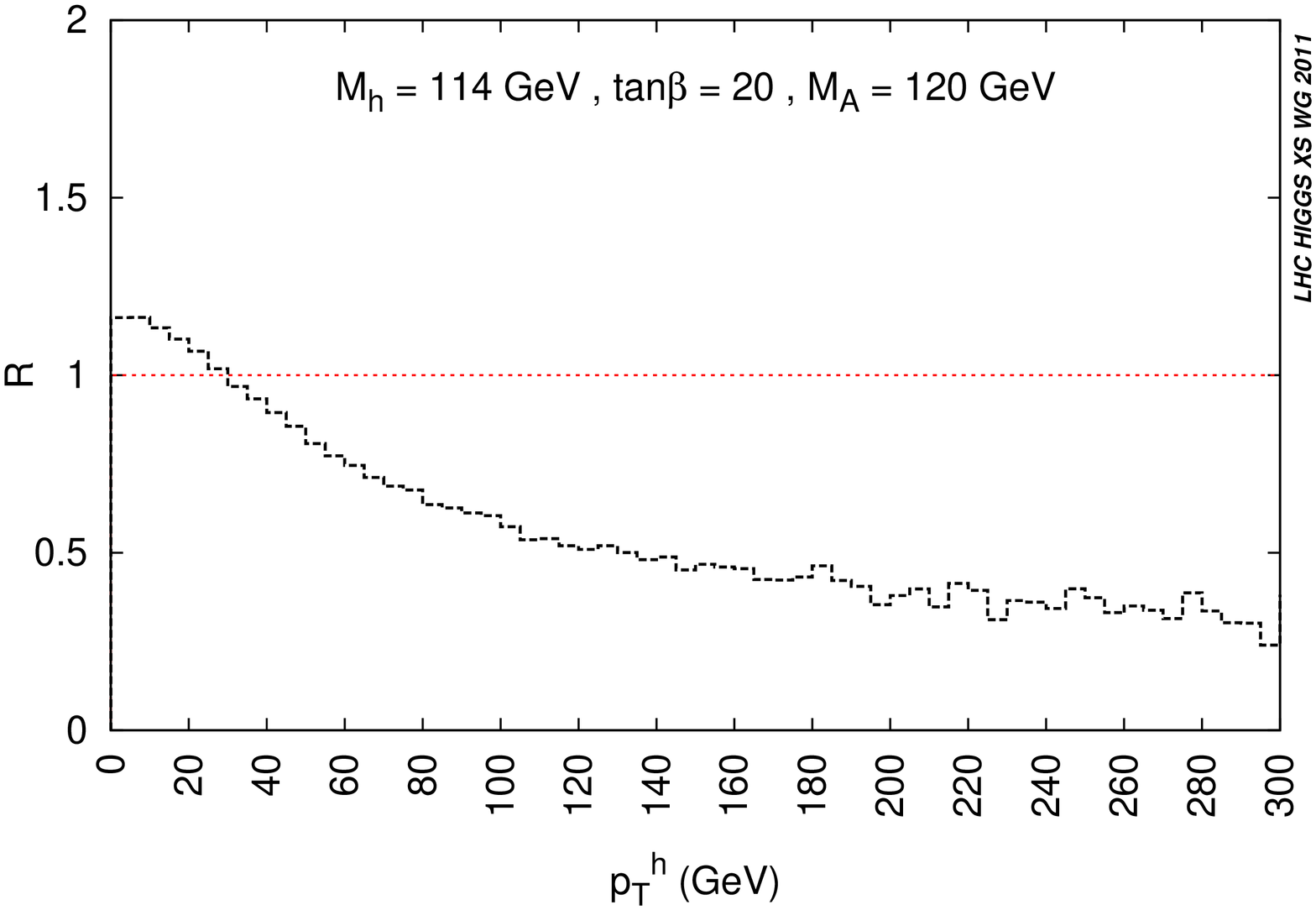}
~~
\includegraphics[height=78mm,angle=270]{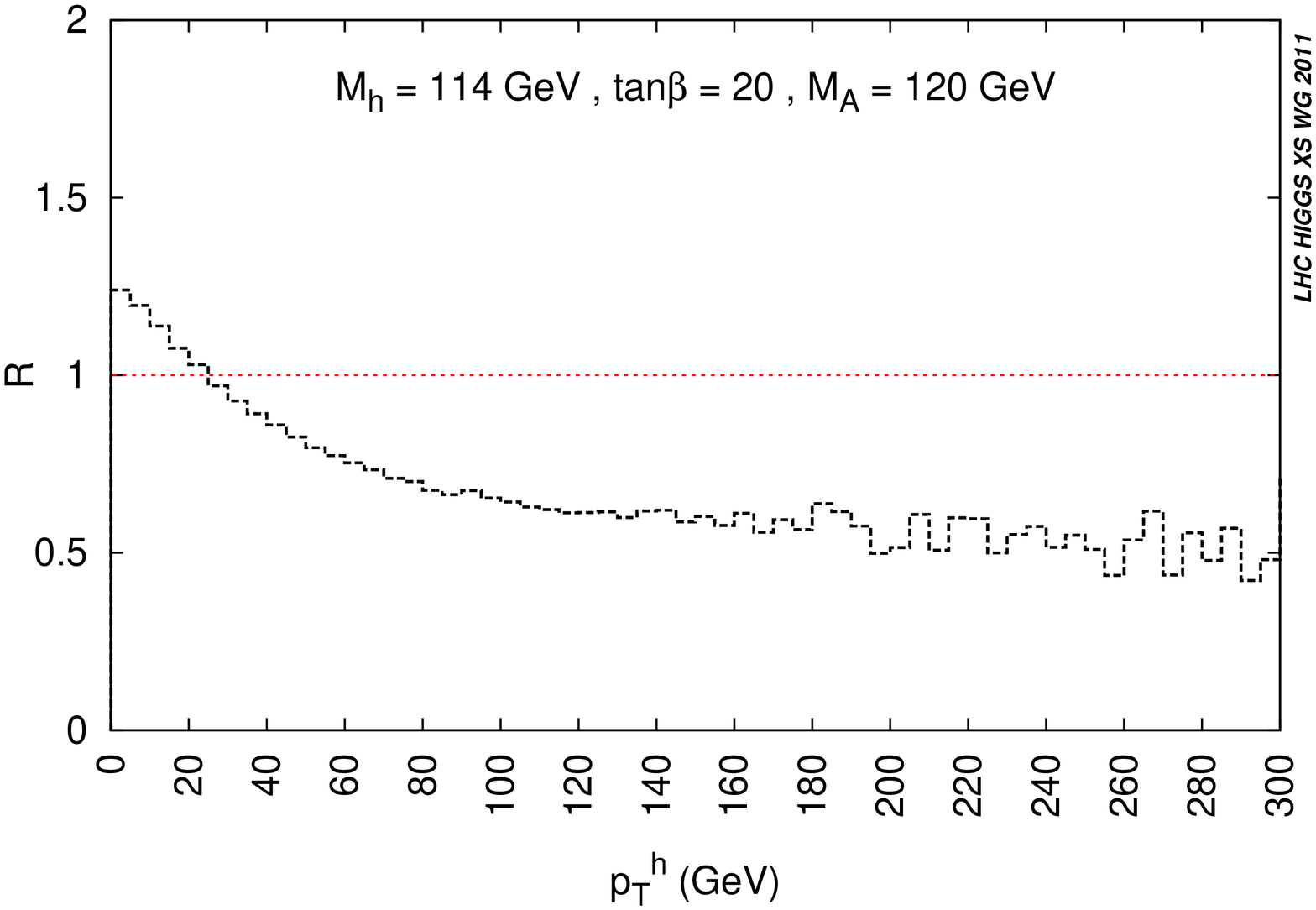}
\\
\includegraphics[height=78mm,angle=270]{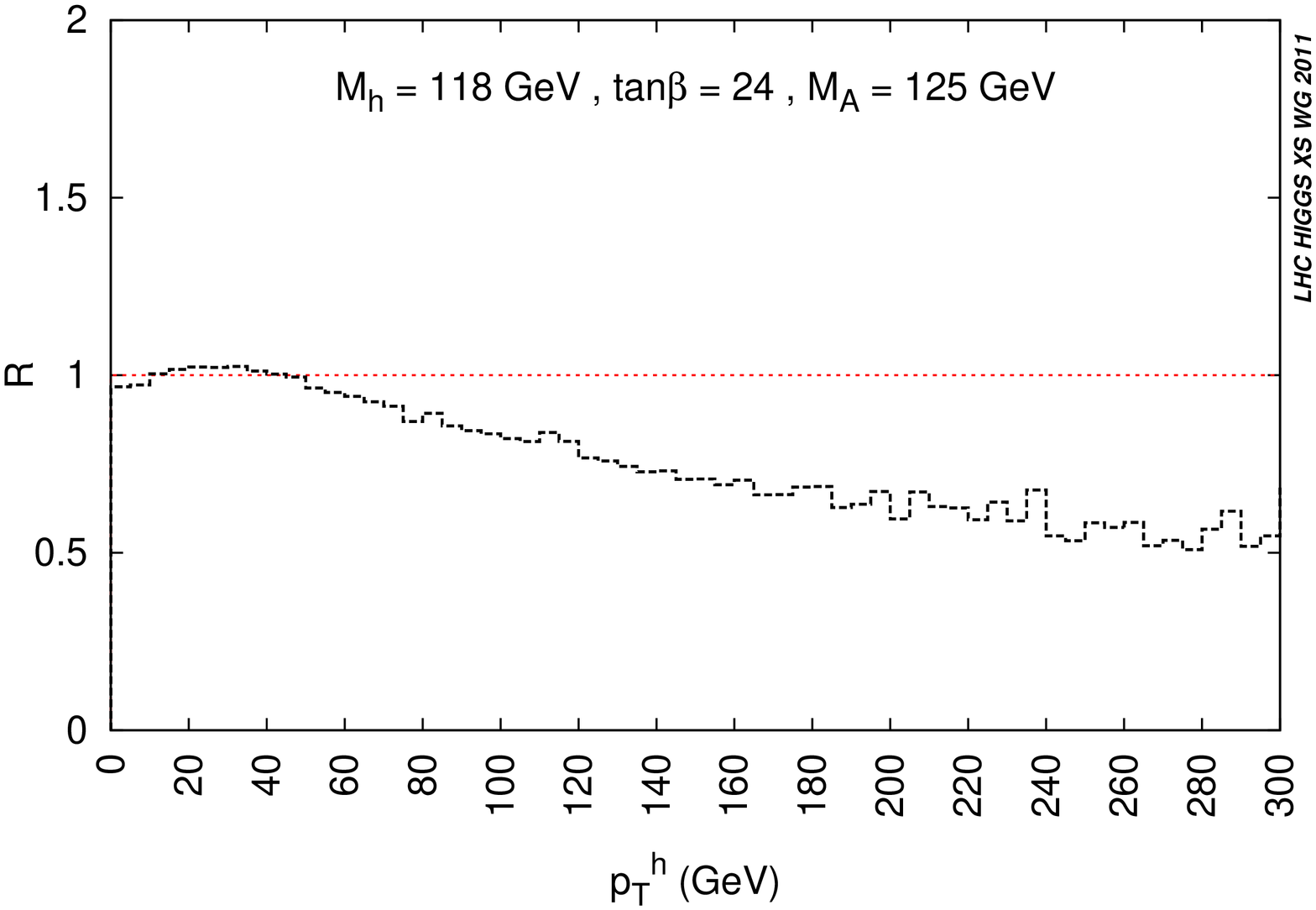}
~~
\includegraphics[height=78mm,angle=270]{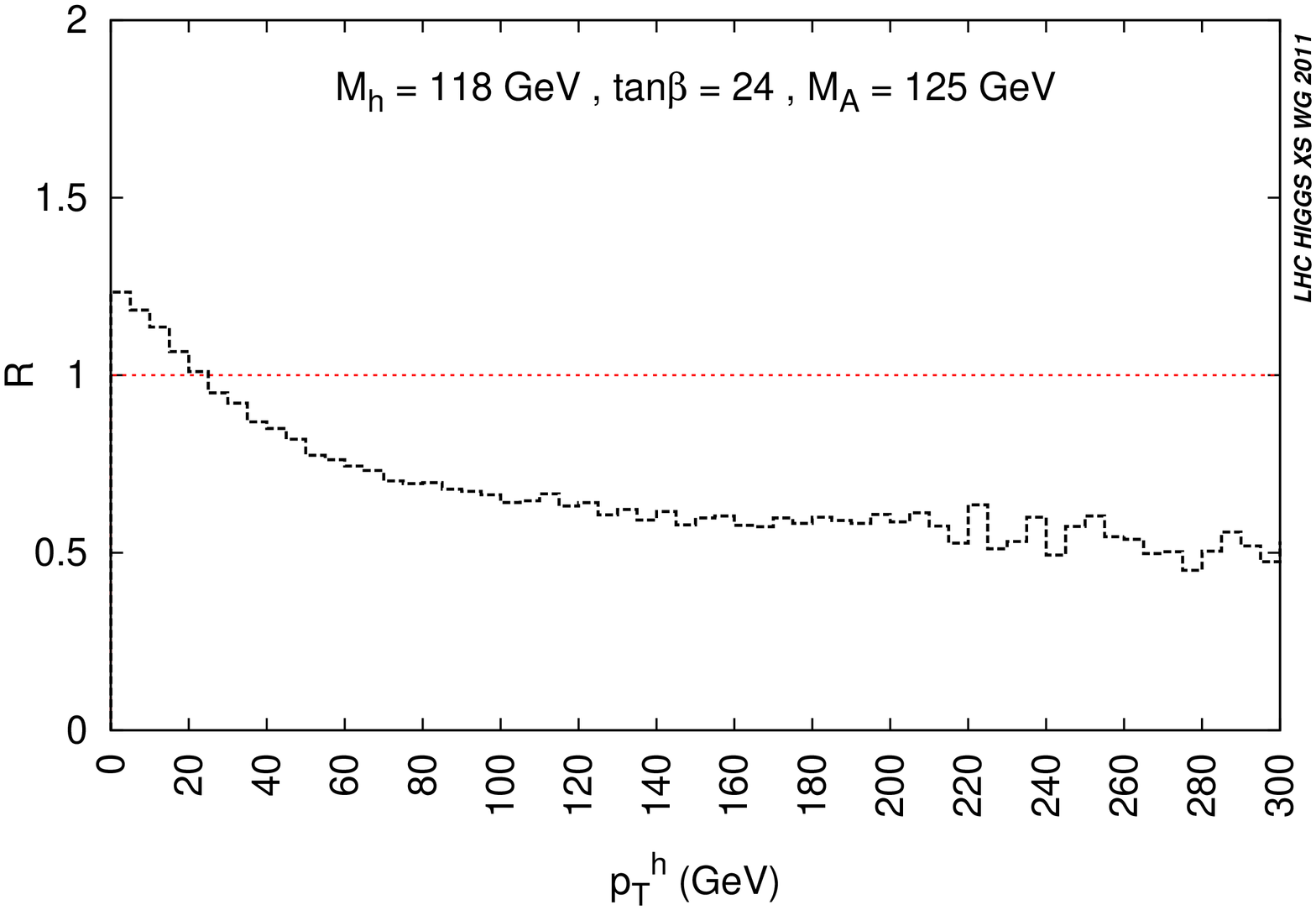}
\\
\includegraphics[height=78mm,angle=270]{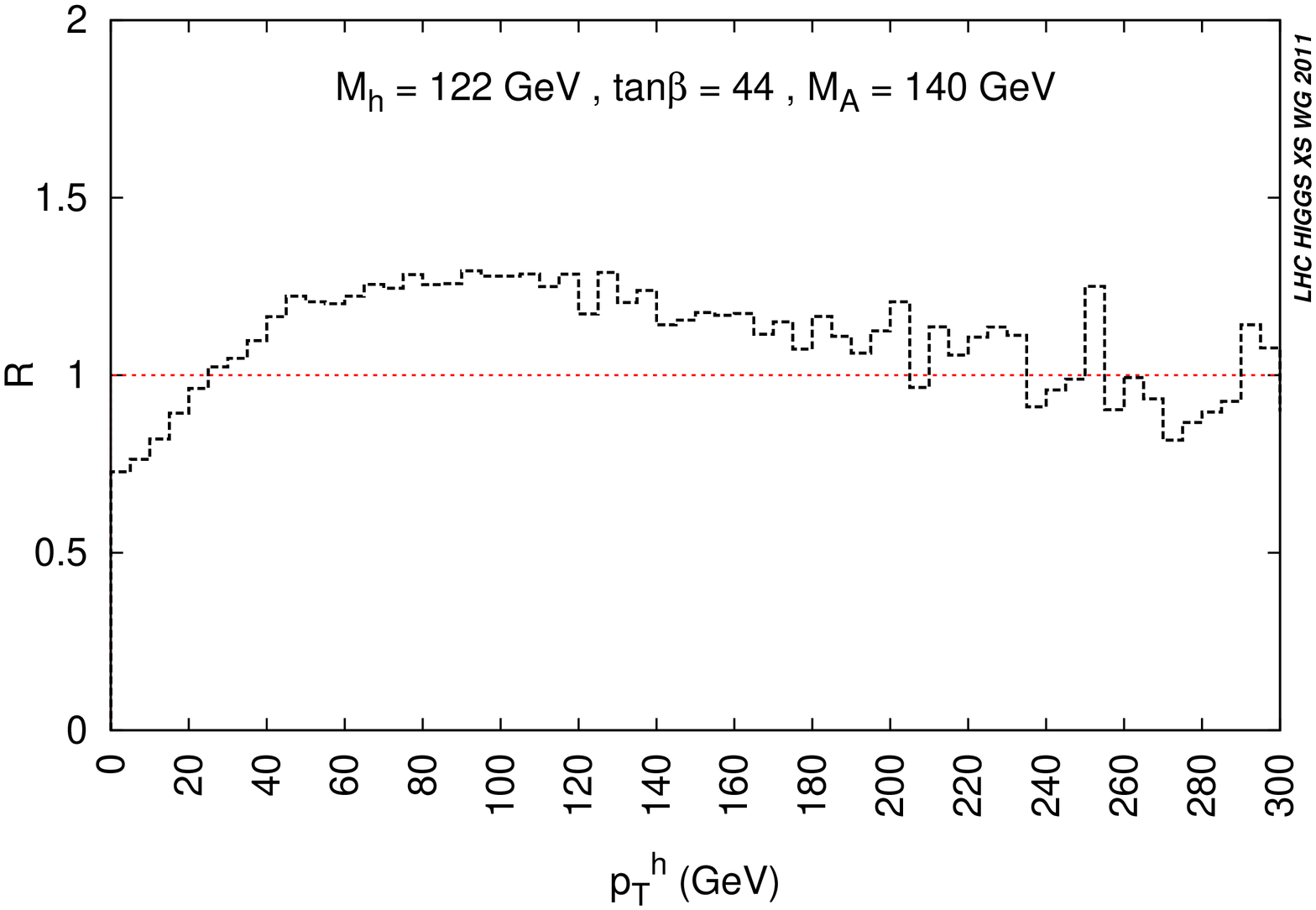}
~~
\includegraphics[height=78mm,angle=270]{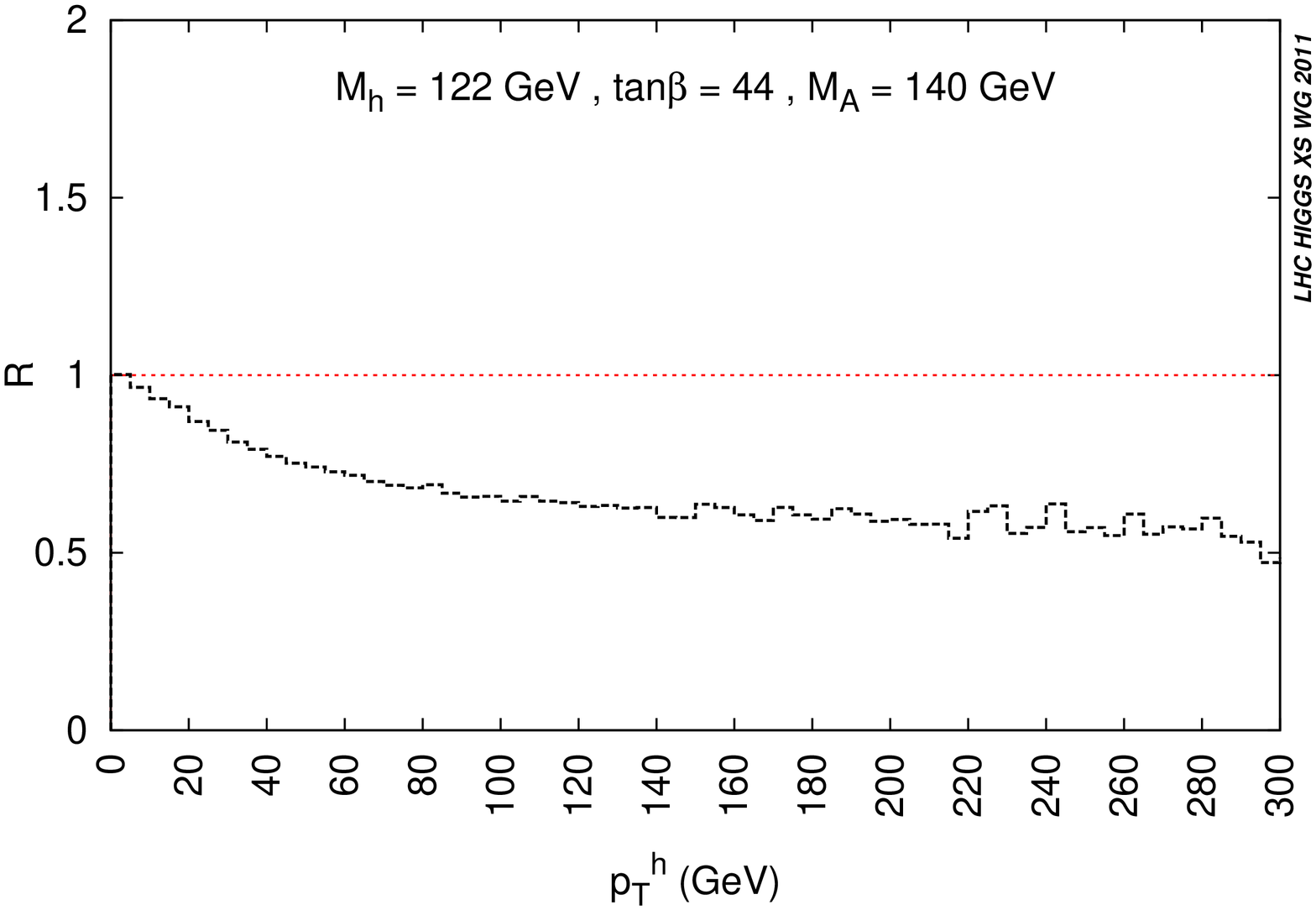}
\\
\caption{\small Left plots: ratio of the transverse momentum-distribution 
for the lightest scalar $\Ph$ in the MSSM over the
  distribution for a SM Higgs with the same mass. Right plots: ratio
  of the transverse-momentum distribution for $\Ph$ in the MSSM over the
  approximate distribution computed with only quarks running in the
  loops. The plots are obtained with $\mu<0$, and with three different
  choices of $\MA$ and $\tan\beta$ for which the MSSM and SM
  predictions for the total cross section agree within $5\%$.
\label{fig:distributions}
} 
\end{center}
\end{figure}

We then consider the opposite situation, namely when the coupling of
$\Ph$ to the bottom quark is significantly enhanced. In this situation
two tree-level channels, \ie~$\Pb \bar{\Pb} \to \Pg \Ph$ and 
$\Pb \Pg \to \Pb\Ph$,
can also contribute to the production mechanism and influence the
shape of the $\pT^{\Ph}$
distribution~\cite{Brein:2003df,Brein:2007da}. Leaving a study of the
effects of those additional channels to a future analysis, we will now
illustrate how the kinematic distribution of the Higgs boson can help
discriminate between the SM and the MSSM. To this purpose, we focus
on three points in the $(\MA, \tan\beta)$ plane characterised by the
fact that the total cross section for $\Ph$ production agrees, within
$5\%$, with the cross section for the production of a SM Higgs boson
of equal mass.  Thus, we are considering points around the curve
labeled ``1'' in \refF{fig:MSSMvsSMxsec} (we set $\mu<0$, as in
the right plot of that figure). In the left panels of
\refF{fig:distributions} we show the ratio of the transverse-momentum 
distribution for $\Ph$ over the distribution for a SM Higgs
boson of equal mass. We see that, because of the enhancement of the
bottom-quark contribution, the shape of the transverse momentum
distribution in the MSSM can differ significantly from the SM
case. Indeed, the region at small $\pT^{\Ph}$ can receive an enhancement
(at moderate $\tan\beta$) compared with the SM result, but also a
significant suppression (at large $\tan\beta$). The tail at large
$\pT^{\Ph}$ can receive a suppression down to a factor $4$ (at moderate
$\tan\beta$), but can also almost coincide with the SM result (at
large $\tan\beta$).

In the right panels of \refF{fig:distributions}, for the same
points in the $(\MA, \tan\beta)$ plane as in the left panels, we show
the ratio of the $\pT^{\Ph}$ distribution over the approximate
distribution computed with only quarks running in the loops.  Even
though the light Higgs boson cannot resolve the squark loops, we see
that the squark contributions affect the shape of the $\pT^{\Ph}$
distribution, because of the interference with the bottom
contribution. In particular, we observe that the squark contributions
may yield an enhancement of the distribution at small $\pT^{\Ph}$, and in
all cases they yield a negative correction for $\pT^{\Ph}>20 \UGeV$. The
negative correction becomes quite flat, at about $-40\%$, for
$\pT^{\Ph}>100 \UGeV$.

%


\subsubsection{Difference in kinematic acceptance between SM and MSSM
production of $\Pg\Pg\to \PH\to\PGt\PGt$} 

Current LHC MSSM $\PH\to\PGt\PGt$ analyses~\cite{Chatrchyan:2011nx,Aad:2011rv} 
use the Standard Model generation of $\Pg\Pg\to \PH$ events with the 
infinite-top-mass approximation as implemented in {\sc PYTHIA}~\cite{Sjostrand:2006za}
(used by CMS) or POWHEG~\cite{Alioli:2010xd} (used by ATLAS). Both ATLAS and
CMS collaborations have presented analyses in the \mhmaxx\
scenario~\cite{Carena:2002qg} as given in
Eq.~(\ref{YRHXS_MSSM_neutral_eq:mhmax}).
At large values of $\tan\beta$ in the \mhmaxx\ 
scenario $\Pg\Pg\to \PH$
production proceeds predominantly via a bottom-quark loop producing a
softer $\pT^{\PH}$ spectrum than that predicted in the Standard Model
production dominated by the top
loop~\cite{Langenegger:2006wu,Alwall:2011cy},%
\footnote{For 
$\tan\beta > 12$ and $\MA = 140 \UGeV$ in the \mhmaxx\ scenario 
the cross section of $\Pg\Pg \to \phi$ ($\phi = \Ph, \PH, \PA$)
production through only $\Pb$~loops~\cite{Baglio:2010ae} is less than $10\%$
different from the full MSSM $\Pg\Pg \to \phi$ production
cross section~\cite{Dittmaier:2011ti}.} 
as discussed in the previous section. As a result, the kinematic
acceptance of $\Pg\Pg\to \PH\to\PGt\PGt$ events may be overestimated in the
current LHC analyses. To determine the size of this effect,  
the {\sc PYTHIA} generated $\pT^{\PH}$ spectrum is re-weighted to
match the shape predicted when considering only the $\Pb$-loop
contribution to the $\Pg\Pg\to \PH$ process~\cite{Alwall:2011cy}. The
acceptance of $\PH\to\PGt\PGt$ events in the $\Pe{+}\PGt$-jet
final state in the CMS detector is then estimated before and after
re-weighting the events. 

Samples of $\Pg\Pg\to \PH\to\PGt\PGt$ events with $\MH = 140 \UGeV$ and
$\MH = 400 \UGeV$ are generated with {\sc PYTHIA} using 
the infinite-top-mass approximation, and the $\PGt$ leptons are decayed
with {\sc TAUOLA}. Using the $\pT^{\PH}$ spectra obtained with
$\Pb$~loops only~\cite{Alwall:2011cy}, the generated events are assigned
weights 
$w_{i} = N_{i}^{\Pb-{\rm loop}} / N_{i}^{\rm Pythia}$ in $i$ bins of 
$\pT^{\PH}$ where $N_{i}^{\rm Pythia}$ and $N_{i}^{\Pb-{\rm loop}}$ are the
normalised event rates expected from {\sc PYTHIA} and with only $\Pb$-loop
contribution, respectively. Events with $\pT^{\PH} < 240 \UGeV$ and 
$\pT^{\PH} < 300 \UGeV$ are considered for $\MH = 140 \UGeV$ and 
$\MH = 400 \UGeV$, respectively. \refF{fig:ggHAcceptancePtH}
shows the generated $\pT^{\PH}$ distributions for the two mass points
before and after applying the event weights.

\begin{figure}
\begin{center}
\includegraphics[width=0.45\textwidth]{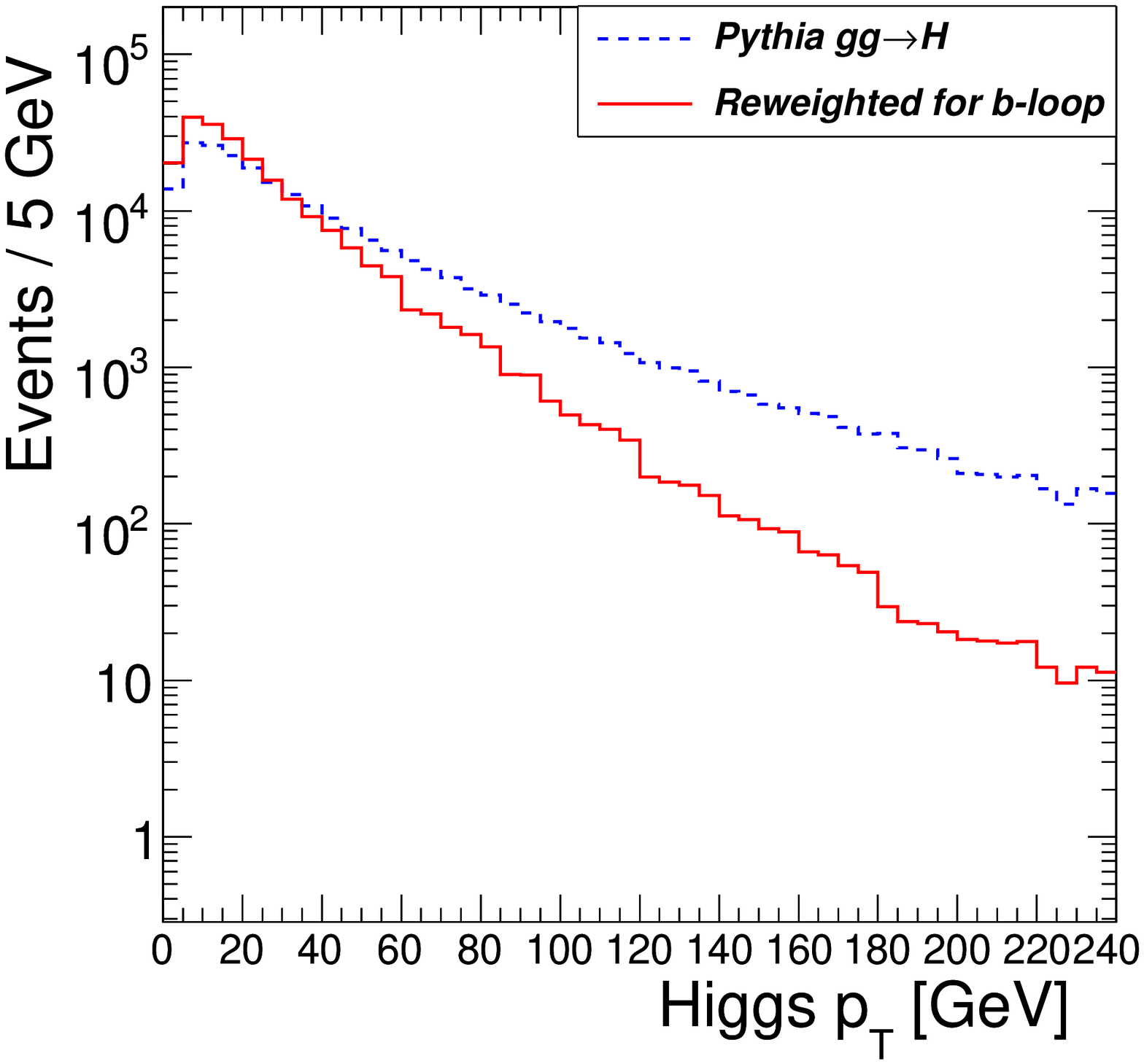} \hfill
\includegraphics[width=0.45\textwidth]{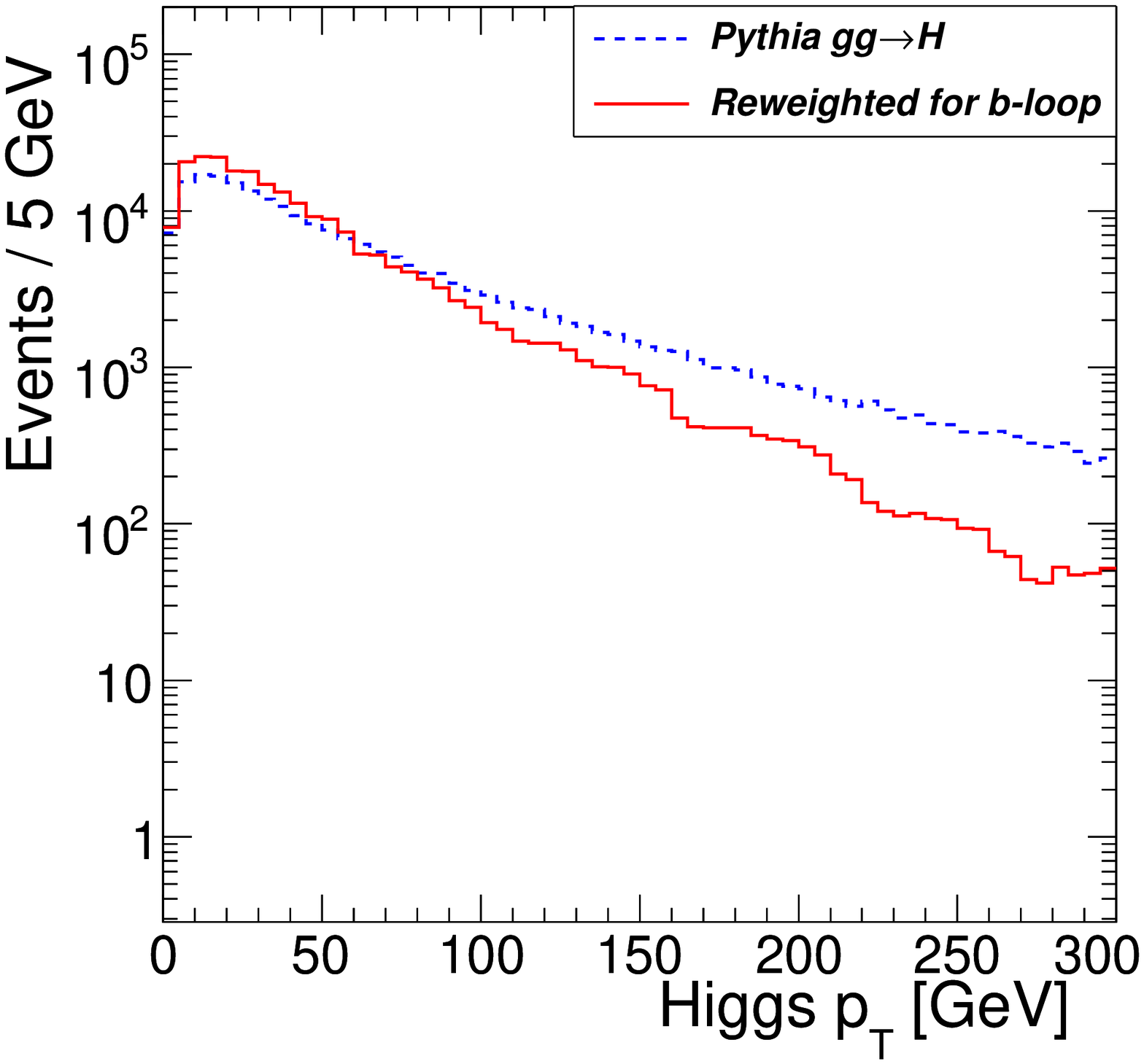}
\caption{Generated $\pT^{\PH}$ distributions for $\MH = 140 \UGeV$ (left) and 
$\MH = 400 \UGeV$ (right) for the $\Pg\Pg \to \PH$ process generated with {\sc PYTHIA} 
(dashed) and after re-weighting (solid) to correct for $\Pg\Pg \to \PH$ production
dominated by the $\Pb$-loop contribution.} 
\label{fig:ggHAcceptancePtH}
\end{center}
\end{figure}

\refF{fig:ggHAcceptancePtElecPtTau} shows the generated $\pT$
distributions of the visible $\PGt$ decay products: electrons and
$\PGt$-jets ($\PGt_{\Ph}$). The electrons and $\PGt$-jets are required
to pass kinematic selections at the generator level corresponding to the
kinematic selections of the CMS $\PH\to\PGt\PGt$
analysis~\cite{Chatrchyan:2011nx}. The electron is required to have
$\pT > 20 \UGeV$ and $|\eta| < 2.1$, and the $\PGt_{\Ph}$ is required
to have $\pT > 20 \UGeV$ and $|\eta| < 2.3$. Events containing a 
$\Pe{+}\PGt_{\Ph}$ pair separated by $\Delta R > 0.5$ are selected. The
acceptance, defined as the ratio of selected to generated events, is
calculated before and after applying the reweighting procedure and shown
in \refT{tab:ggHAcceptanceResults}. 

\begin{figure}
\begin{center}
\includegraphics[width=0.45\textwidth]{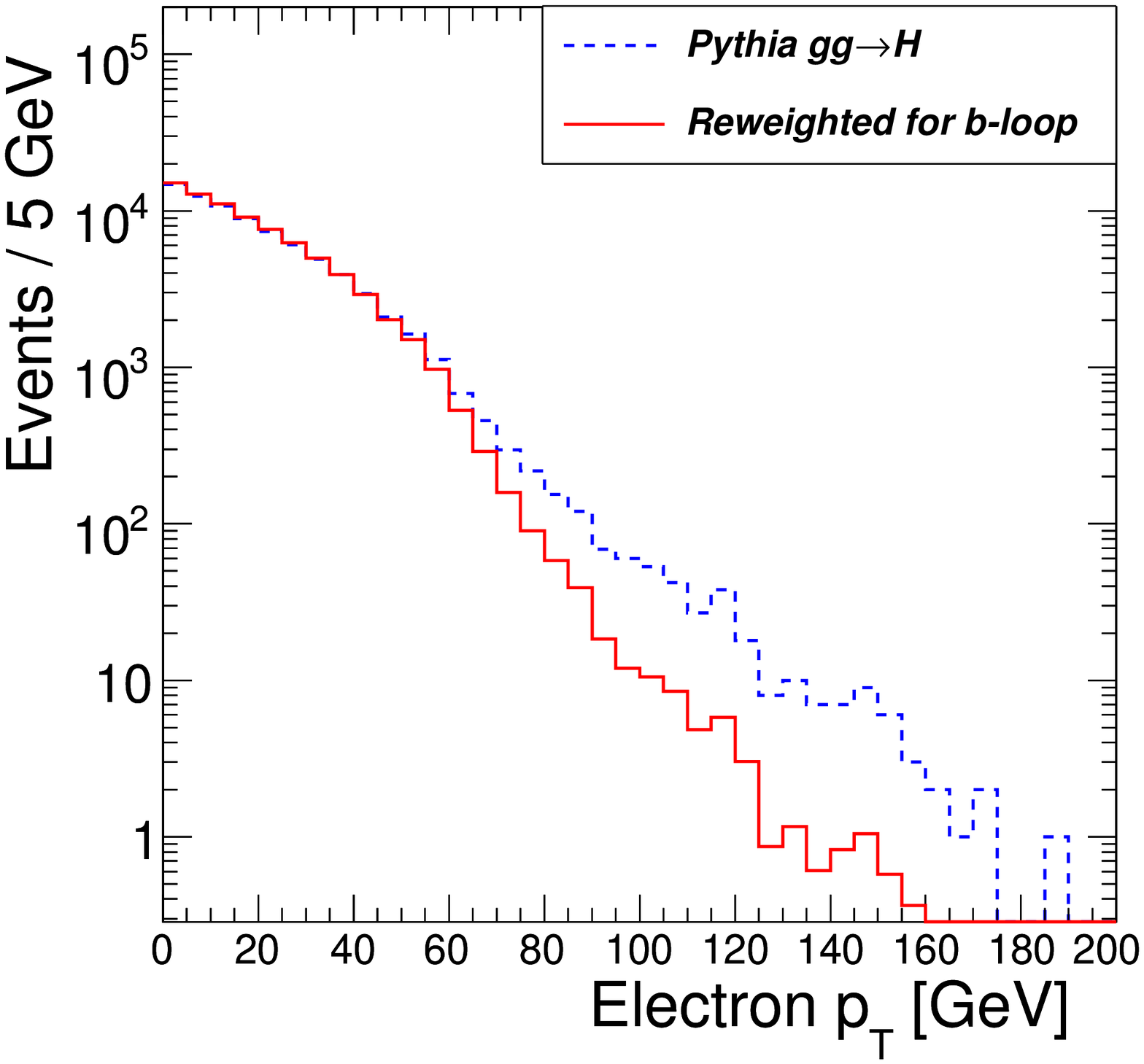} \hfill
\includegraphics[width=0.45\textwidth]{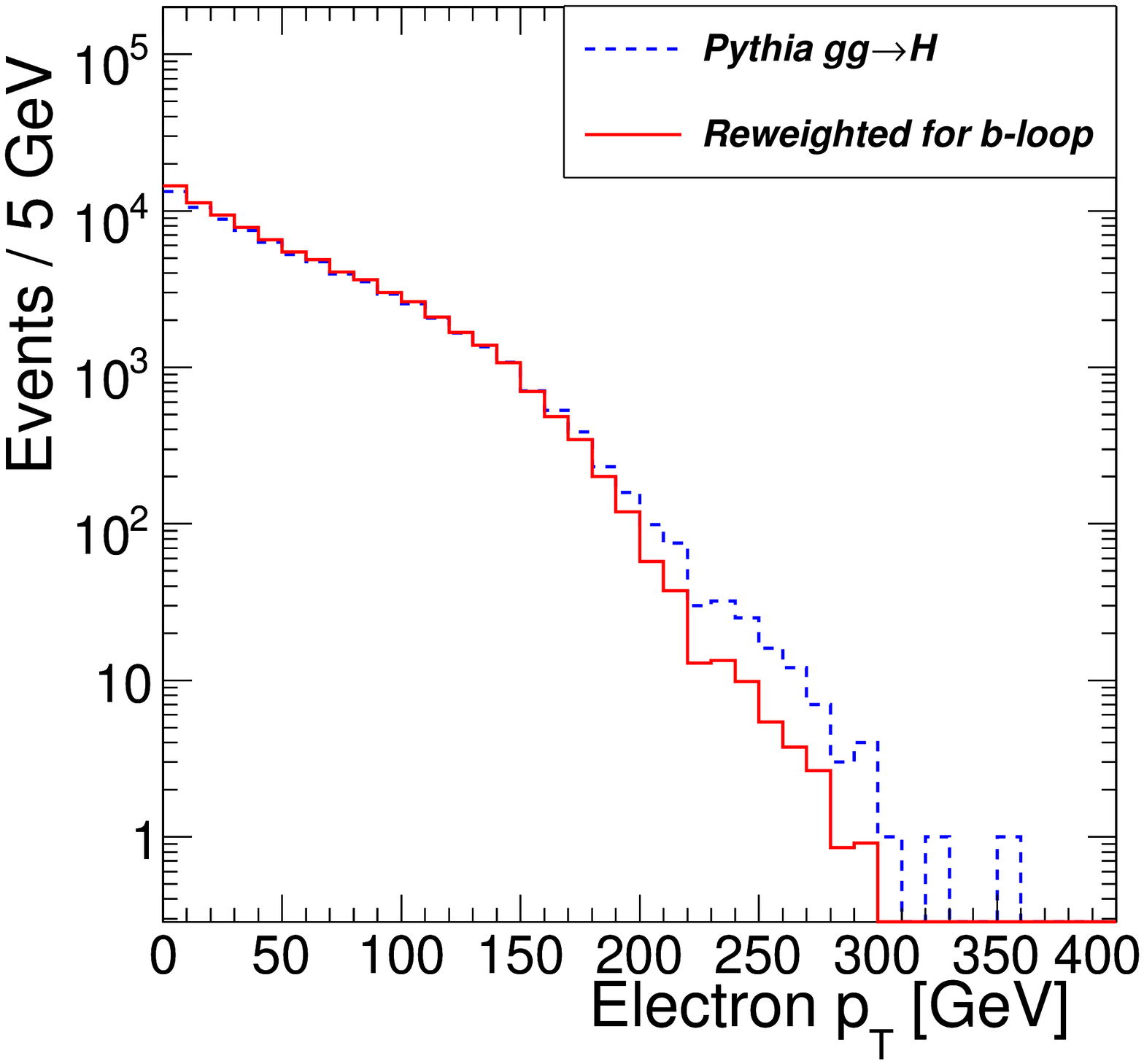} \hfill
\includegraphics[width=0.45\textwidth]{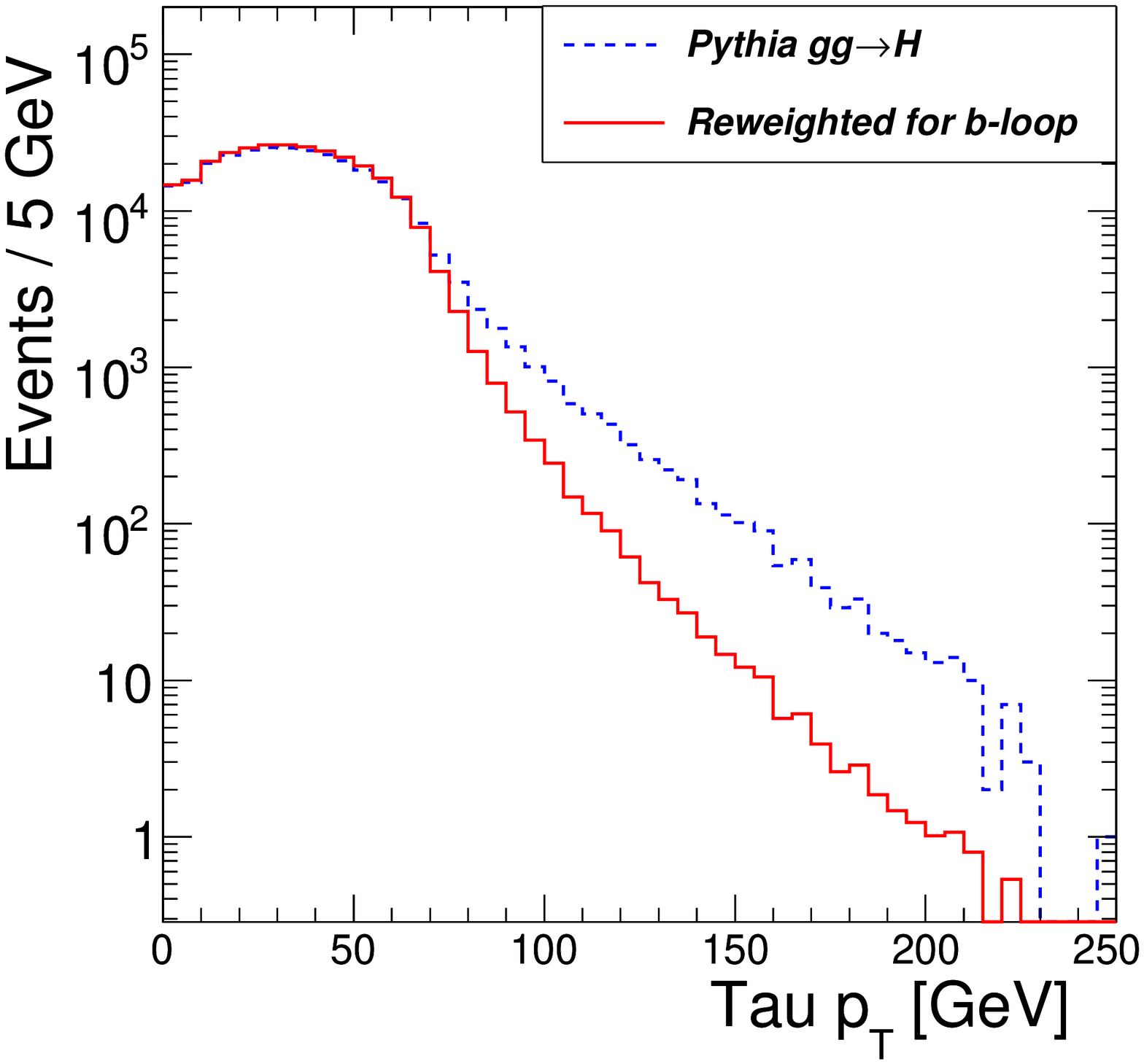} \hfill
\includegraphics[width=0.45\textwidth]{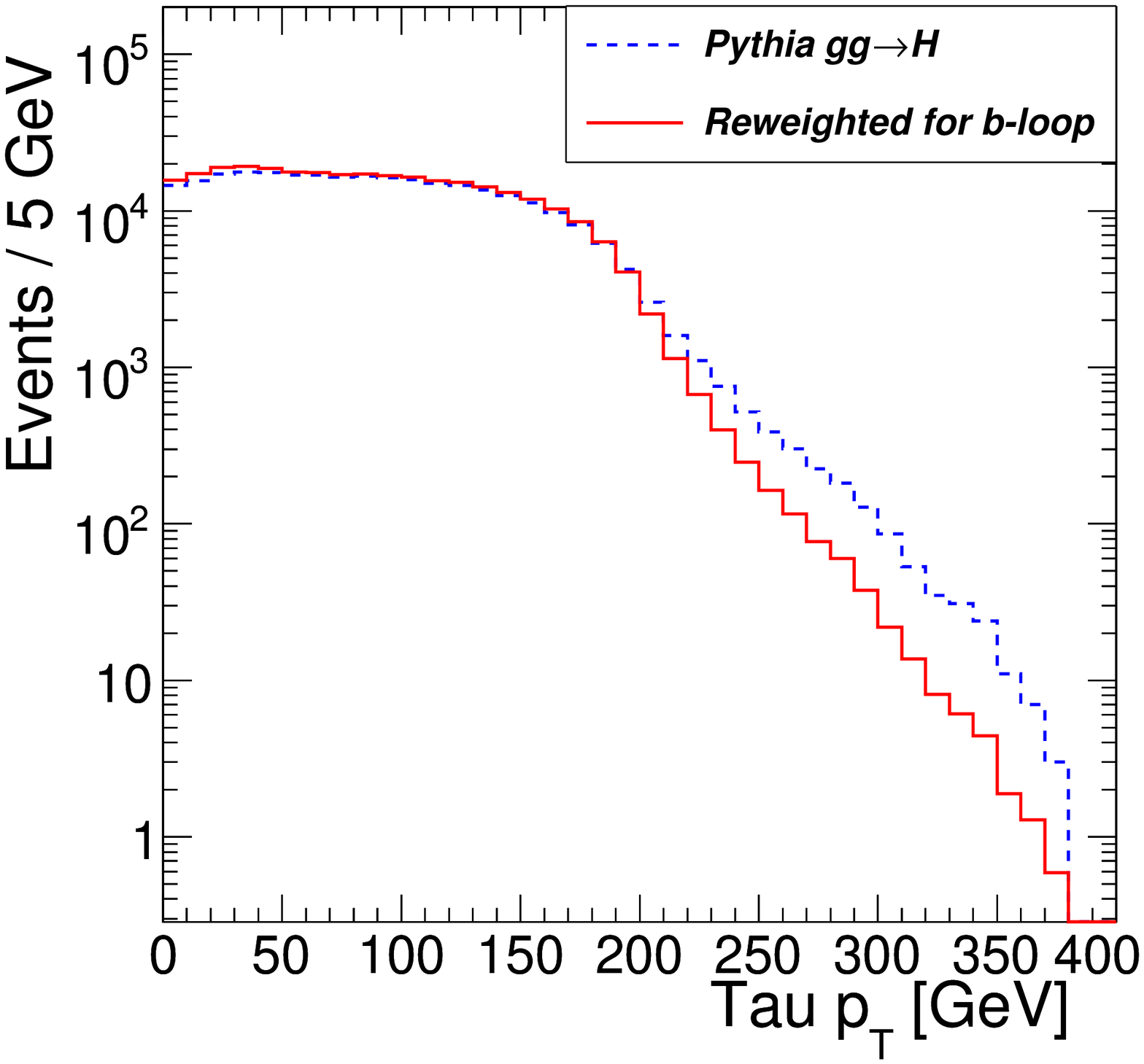}
\caption{Generated $\pT^{\Pe}$ and $\pT^{\PGt}$ distributions for 
$\MH = 140 \UGeV$ (left) and $\MH = 400 \UGeV$ (right) for the 
$\Pg\Pg \to \PH$ process 
generated with {\sc PYTHIA} (dashed) and after re-weighting (solid) to correct for 
$\Pg\Pg \to \PH$ production
dominated by the $\Pb$-loop contribution.} 
\label{fig:ggHAcceptancePtElecPtTau}
\end{center}
\end{figure}

\begin{table}
\begin{center}
\caption{The $\Pe{+}\PGt_{\Ph}$ acceptances before and after
re-weighting to correct for $\Pb$-loop contribution. 
\label{tab:ggHAcceptanceResults}}
\begin{tabular}{cccc}
\hline
$\MH$ [GeV]    & Acceptance & Acceptance & Correction factor \\ 
               & {\sc PYTHIA} $\Pg\Pg\to \PH$ & re-weighted for $\Pb$-loop & \\ 
\hline
   140 &  0.072 $\pm$ 0.001 & 0.070 $\pm$ 0.001 & 0.97 $\pm$ 0.01 \\
   400 &  0.149 $\pm$ 0.001 & 0.152 $\pm$ 0.001 & 1.02 $\pm$ 0.02 \\
\hline
\end{tabular}
\end{center}
\end{table}

It is found that in current analyses, the acceptance for the MSSM 
$\Pg\Pg\to \PH\to\PGt\PGt$ process with only $\PQb$~loops is smaller than for the SM 
$\Pg\Pg\to \PH\to\PGt\PGt$ process by approximately $3\%$ for 
$\MH = 140 \UGeV$, and consistent with the SM process for 
$\MH = 400 \UGeV$. The small size of the effect is due to the relatively
low $\pT$ thresholds on the visible $\PGt$ decay products used in the current analyses.



\subsubsection{Higher-order MSSM corrections for Higgs production in
vector-boson fusion}
\label{sec:mssm_vbfnlo}




\subsubsubsection{{\sc VBFNLO}}

Higgs production via (weak) vector-boson fusion (VBF) is an important SM
production channel at the LHC~\cite{Aad:2009wy,Ball:2007zza,Figy:2003nv}.  In
the MSSM, it is expected that at least one Higgs boson will have a significant
coupling to the weak bosons, meaning VBF should be an important production
channel in the MSSM as well. Since the latest release, {\sc VBFNLO}
\cite{Arnold:2011wj,Arnold:2008rz} has provided the ability to study production
of the three neutral Higgs bosons $\Ph, \PH$, and $\PA$ via vector-boson
fusion in the MSSM with real or complex parameters%
\footnote{At Born level in the MSSM with 
real parameters, the CP-odd Higgs boson $\PA$ is not produced.}.  

The MSSM implementation makes use of the anomalous Higgs couplings previously
available in {\sc VBFNLO}. These use the most general structure of the
coupling between a scalar particle and a pair of gauge bosons, which can be
written as \cite{Figy:2004pt}
\begin{eqnarray}
T^{\mu \nu} (q_{1}, q_{2}) &=& a_1(q_{1},q_{2}) g^{\mu \nu} + a_2(q_{1},q_{2})
\left[ (q_{1} q_{2}) \, g^{\mu \nu} - q_{1}^{\mu} q_{2}^{\nu} \right] +
%
 a_3(q_{1},q_{2}) \epsilon^{\mu \nu \rho \sigma} q_{1 \rho} q_{2 \sigma} ,
\label{eq:formfactors}
\end{eqnarray}
where $a_{1,2,3}$ are Lorentz-invariant formfactors and $q_{1,2}$ are the
momenta of the gauge bosons. Production of the MSSM Higgs bosons is implemented
in {\sc VBFNLO} by altering the value of the formfactor $a_{1}$ (which at
tree level is the only non-zero formfactor).  
The MSSM parameters can be specified either via a SLHA file or, if
{\sc VBFNLO} is linked to 
{\sc FeynHiggs}
\cite{Heinemeyer:1998yj,Heinemeyer:1998np,Degrassi:2002fi,Frank:2006yh}, by the
user.  

\subsubsubsection{Higgs mixing}
\label{sec:h_mix}

Mixing between the neutral Higgs bosons can be very significant
numerically,%
\footnote{In the MSSM with real parameters, mixing occurs between the
two neutral CP-even Higgs bosons, $\Ph$ and $\PH$, and if complex
parameters are allowed, mixing between all three neutral Higgs bosons,
$\Ph$, $\PH$, and $\PA$ must be considered.}
and needs to be taken into account in order to produce
phenomenologically relevant results.  There is some flexibility, however, in the
manner in which this mixing is included in the VBF process, 
and {\sc VBFNLO} provides several
options.  These are implemented by altering the formfactors of
Eq.~(\ref{eq:formfactors}) appropriately, and are as follows:

\begin{enumerate}
\item
Wave-function normalisation factors (the $Z$-factors as defined in
\Bref{Frank:2006yh}) are used which ensure the correct on-shell
properties of the produced Higgs boson. These wave-function
normalisation factors incorporate the potentially numerically large 
and process-independent Higgs propagator corrections. 
For this option, it is recommended that {\sc FeynHiggs} is used, which 
calculates the necessary $Z$-factors including all
one-loop corrections as well as the dominant two-loop corrections in the MSSM
with complex parameters.%
\footnote{If a SLHA file is used as input and this
option is chosen, {\sc VBFNLO} can calculate the $Z$-factors if linked to {\sc
LoopTools}~\cite{Hahn:1998yk,vanOldenborgh:1989wn}, but only in the MSSM with
real parameters, and only to one-loop level.} Since the higher-order
Higgs propagator corrections are contained in the $Z$-factors, 
the mixing angle $\alpha$ needs to be evaluated at lowest order. 

\item Since the Higgs mixing contributions are universal and appear in
the same way in higher-order diagrams as well, it seems natural to apply
them in a factorised way such that they multiply both the tree-level and
the loop-level contributions. Compared to the case where the $Z$-factors
are only applied at the tree level, this option can have a significant
numerical effect in rather extreme regions of the MSSM parameter space,
especially when CP-violating
effects are considered.
In the \mhmaxx\ scenario, however, including mixing at loop-level as well as
at tree level has very little effect, even in the non-decoupling regime. 

\item If the momentum dependence of the $Z$-factors and CP-violating
effects are neglected, the contribution of the $Z$-factors reduces to a 
correction to the Higgs mixing angle, resulting in an effective mixing angle,
$\alpha_{\mbox{\tiny{eff}}}$ (numerical values for
$\alpha_{\mbox{\tiny{eff}}}$ can either be taken from the SLHA file, or
from {\sc FeynHiggs}). It should be noted, however, that this simple
approximation is not sufficient to ensure the correct on-shell
properties of the produced Higgs boson.

\item The Higgs mixing can be treated like an additional
counterterm that is added at NLO.
\end{enumerate}


Various Higgs decays are included in {\sc VBFNLO}: 
$\PH \to \PGg\PGg$, 
$\PH \to \PGmp\PGmm$, 
$\PH \to \PGtp\PGtm$,
$\PH \to \PQb\PAQb $, 
$\PH \to \PW^{+}\PW^{-} \to \Pl^{+} \PGn_{\Pl} \Pl^{-}
\bar{\PGn}_{\Pl}$, 
$\PH \to \PZ\PZ \to \Pl^{+} \Pl^{-} \PGn_{\Pl} \PAGn_{\Pl}$ and 
$\PH \to \PZ\PZ \to \Pl^{+} \Pl^{-} \Pl^{+} \Pl^{-}$.  
When working in the MSSM, the branching ratios and
widths needed can either be read from the SLHA file or taken from the 
{\sc FeynHiggs} output.

\subsubsubsection{Electroweak corrections to VBF in the MSSM}

In the SM, the electroweak corrections to VBF have been found to be as
important numerically as the NLO QCD corrections $\mathcal{O}(-5\%)$
in the mass range $100{-}200\UGeV$ \cite{Ciccolini:2007jr,Ciccolini:2007ec,Figy:2010ct}.  In
the MSSM, these corrections~\cite{Hollik:2008xn,Rauch:2010mi,Figy:2010ct} 
are also potentially important and should be
considered for phenomenologically relevant studies.  In the decoupling
region of the MSSM parameter space, these corrections tend to be -- as
expected -- very similar to those in the SM for the light scalar Higgs
boson. In other areas of parameter space, however, the electroweak
corrections can differ significantly between the SM and the MSSM.

The complete SM-type and dominant SUSY NLO electroweak corrections to 
vector-boson fusion have been incorporated into the program {\sc VBFNLO}, as 
described in \Bref{Figy:2010ct},
supplementing the already existing SM NLO QCD corrections. 
In the SM, the full electroweak corrections to the $t$-channel VBF process
are included.  In the MSSM all SM-type boxes and pentagons, together
with all MSSM corrections to the vertex and self energy type diagrams
are incorporated. For the Higgs-boson vertex and self-energy corrections
this has been done 
using an effective $\PH\PV\PV$ vertex, with formfactors as in
Eq.~(\ref{eq:formfactors}), and an effective $\Pq\Pq\PV$ coupling is used
to include the loop corrections to that vertex.  To include the box and
pentagon diagrams, on the other hand, the full $2 \to 3$ matrix element
has been calculated.
An on-shell renormalisation scheme has been used, ensuring that the
renormalised mass parameters of the Higgs and gauge bosons
correspond to the physical masses.

The remaining SUSY electroweak
corrections -- \ie  the chargino and neutralino boxes and pentagons --
are sub-dominant in the
investigated \mhmaxx\ scenario~\cite{Hollik:2008xn,Rauch:2010mi}.  
Due to this, and the large CPU time
needed to calculate these corrections, they are not yet included in {\sc
VBFNLO}.  The SUSY QCD corrections were investigated in
Refs.~\cite{Djouadi:1999ht,Hollik:2008xn,Rauch:2010mi,Spira:1997dg} and
found to be small.

{\sc VBFNLO} offers several options for the parametrisation of the
electromagnetic coupling.  The choice of parametrisation has a significant
effect on the relative size of the electroweak corrections, as the charge
renormalisation constant, $\delta Z_{\Pe}$, must be altered to suit the LO
coupling.  


\subsubsubsection{VBF parameters and cuts}

The numerical results presented here were produced using {\sc VBFNLO} to
simulate VBF production of the light CP-even Higgs boson and have been
evaluated in the \mhmaxx\ scenario as described in
Eq.~(\ref{YRHXS_MSSM_neutral_eq:mhmax}). 

{\sc FeynHiggs-2.7.4} is used to calculate the MSSM parameters -- in
particular the Higgs boson masses, mixing, and widths.  Matching the setup of
the investigation of the vector-boson-fusion channel in the SM, described in
\refS{sec:VBF_setup}, the electroweak parameters used
here are as given in Appendix~A of 
\Bref{Dittmaier:2011ti}, the electromagnetic coupling is
defined via the Fermi constant $G_{\rm F}$, and the renormalisation and
factorisations scales are set to $\MW$.  The anti-$k_{T}$ algorithm is used
to construct the jets, and the cuts used are also as in
\refS{sec:VBF_setup}: 
\begin{equation}
 \Delta R = 0.5, \quad
  p_{\rm T_{j}} > 20 \UGeV, \quad
 |y_{j}| < 4.5, \quad
 |y_{j_{1}} -y_{j_{2}}| > 4, \quad
 m_{jj} > 600 \UGeV .
\end{equation}
The Higgs boson is generated on-shell.

For each parameter point the cross sections and distributions have been
calculated twice -- once in the MSSM, including NLO QCD and all MSSM
electroweak corrections except for the SUSY boxes and pentagons (\ie  those
containing charginos and neutralinos), and once in the SM with a Higgs mass
equal to that in the MSSM, including the complete SM QCD and electroweak
corrections.  The SM cross section can then be rescaled by a SUSY factor, in
order to assess the impact of the SUSY-type corrections.  Two SUSY factors
have been investigated, $S_{Z}$ and $S_{\alpha}$, which can be expressed as
follows,
\begin{eqnarray}
 S_{Z} &=& | Z_{\Ph\Ph} \sin(\beta-\alpha_{\mbox{\tiny{tree}}}) 
          + Z_{\Ph\PH} \cos(\beta-\alpha_{\mbox{\tiny{tree}}}) |^{2}, \nonumber \\
 S_{\alpha} &=& \sin^{2}(\beta-\alpha_{\mbox{\tiny{eff}}}).
\label{eqn:susyFAC}
\end{eqnarray}
Here $S_{Z}$ is an effective coupling that is composed from the full
wave-function-normalisation factors that ensure the correct on-shell
properties of the produced Higgs boson, see the description in
\refS{sec:h_mix}. $S_{\alpha}$, on the other hand,
is an approximate effective coupling arising from $S_{Z}$ in the limit
where the momentum dependence of the $Z$-factors is neglected.
For the full result within the MSSM, the wave-function normalisation
factors are used without further approximations.


\subsubsubsection{Total cross sections}

\refT{tab:mstw} shows the NLO cross sections (together with the
statistical errors from the Monte Carlo integration) in the MSSM and for
the rescaled SM with the MSTW2008NLO PDF set%
\footnote{The simulation has also been run using the CTEQ6.6 PDF set,
and very similar features were found. It should be noted that both PDF
sets are in principle not sufficient for processes where NLO electroweak
corrections are incorporated, since the PDF sets do not involve any 
QED contributions. However, the QED effects are expected to be small on the
one hand and on the other hand we intend to show the basic higher-order
corrections in this section, the size of which will not depend on the
choice of the PDFs significantly.}%
. As expected, in the decoupling regime (\ie  for large values of
$\MA$), the rescaling of the SM works well, giving adjusted cross
sections within $2\%$ of the true MSSM value.  The rescaling is worst
for high values of $\tan \beta$ and low values of $\MA$, in the
non-decoupling regime, where there is a factor of more than $3$ difference
between the rescaled SM cross sections and the true MSSM cross section,
implying that the SUSY corrections at this point in parameter space are
significant.  It should be noted, however, that in this region of
parameter space the leading-order cross section is extremely small. The
seemingly large corrections are directly related to the fact that the
cross section itself is heavily suppressed in this region. They
therefore do not endanger the reliability of the theoretical prediction
for Higgs-boson production in weak-boson fusion where it is
phenomenologically relevant.  In parameter regions where the 
leading-order cross section is heavily suppressed, one in general has to take
into account also the squared one-loop contribution, which may be of
comparable size as the product of the one-loop amplitude with the
tree-level contribution.  For the scenario under consideration, there is
not a large difference between the two rescaling factors described in
Eq.~(\ref{eqn:susyFAC}); $S_{Z}$ performs slightly better in the more
extreme regions of the scenario (\ie  large $\tan\beta$, small $\MA$),
whereas $S_{\alpha}$ (accidentally) provides a better description in the
decoupling region. 

In the SM, the electroweak corrections are of the order $-8\%$.  In the MSSM,
the electroweak corrections in the decoupling regime are slightly larger than
this, by approximately $0.5{-}1\%$, and in the non-decoupling regime the
electroweak corrections become much larger -- reaching more than -- $70\%$ for 
$\tan \beta = 50$ and $\MA = 100\UGeV$ (as explained above, this large
correction is related to the fact that the leading-order cross section
is heavily suppressed in this parameter region).

\begin{table}
\newcommand{\lstrut}{{$\strut\atop\strut$}}
\begin{center}
\caption {Higgs NLO cross sections at $7\UTeV$ with VBF cuts and the
MSTW2008NLO PDF set, for the \mhmaxx\ scenario in the MSSM and for the
corresponding SM cross section rescaled by the SUSY factors $S_{\alpha}$
and $S_{Z}$, as defined in Eq.~(\ref{eqn:susyFAC}).
\label{tab:mstw}}
\vspace{0.2cm}
\small
\begin{tabular}{ccccc}
\hline
$\tan \beta$ & $\MA$ [GeV] & NLO, MSSM [fb] & NLO, SM$\times S_{\alpha}$ [fb] & NLO, SM$\times S_{Z}$ [fb]\\
\hline
\multirow{3}{*}{3} & 100 & 148.261 $\pm$ 0.381 & 152.605 $\pm$ 0.298 & 153.425 $\pm$ 0.299 \\
 & 200 & 249.060 $\pm$ 2.123 & 250.880 $\pm$ 1.548 & 252.419 $\pm$ 1.558 \\
 & 500 & 253.411 $\pm$ 0.813 & 255.963 $\pm$ 0.698 & 257.394 $\pm$ 0.701 \\
\hline
\multirow{3}{*}{15} & 100 & 13.059 $\pm$ 0.052 & 16.108 $\pm$ 0.024 & 16.066 $\pm$ 0.024 \\
 & 200 & 236.179 $\pm$ 1.399 & 237.359 $\pm$ 1.392 & 238.746 $\pm$ 1.400 \\
 & 500 & 235.097 $\pm$ 2.989 & 239.307 $\pm$ 0.899 & 240.668 $\pm$ 0.904 \\
\hline
\multirow{3}{*}{50} & 100 & 0.544 $\pm$ 0.003 & 1.835 $\pm$ 0.007 & 1.806 $\pm$ 0.007 \\
 & 200 & 237.742 $\pm$ 0.438 & 238.196 $\pm$ 0.597 & 239.551 $\pm$ 0.600 \\
 & 500 & 236.179 $\pm$ 1.266 & 236.496 $\pm$ 2.518 & 237.832 $\pm$ 2.532 \\
\hline
\end{tabular}
\end{center}
\end{table}


\subsubsubsection{Differential distributions}

A number of differential distributions were generated for the parameter points
above.  Here, we examine a selection of them for the two `extreme' points 
\begin{eqnarray}
 \tan \beta = 3, \MA = 500 \UGeV, \nonumber \\
 \tan \beta = 50, \MA = 100 \UGeV.\nonumber
\end{eqnarray}
Each plot compares the leading-order result in the MSSM (solid black) with the
NLO MSSM value (dotted green) and the SM NLO result rescaled by the SUSY
factors $S_{\alpha}$ (pink dash-dotted) and $S_{Z}$ (red short-dashed) using
the MSTW2008NLO PDF set. 

\refF{fig:az_angle} shows the azimuthal-angle distribution, $\phi_{jj}$
between the two tagging jets.  This distribution is of special interest in VBF
studies, as it provides an opportunity to study the structure of the
$\PH\PV\PV$ coupling.  
As with the total cross sections, for the low-$\tan\beta$,
high-$\MA$ case, the rescaling procedure works reasonably well. For the
extreme case of 
high-$\tan\beta$ and low-$\MA$ (shown on the right-hand side of
\refF{fig:az_angle}), where the cross section is heavily suppressed,
a simple rescaling of the SM cross section would not be sufficient to 
accurately describe the MSSM result. It should be noted, however, that even in
this extreme case the 
shapes of the distributions are very similar at LO and NLO, in both the SM and
the MSSM, which means that the higher-order corrections do not significantly
alter the structure of the $\PH\PV\PV$ coupling. 

\begin{figure}
\begin{center}
     \subfigure[]{
         \resizebox{0.45\hsize}{!}{\includegraphics*{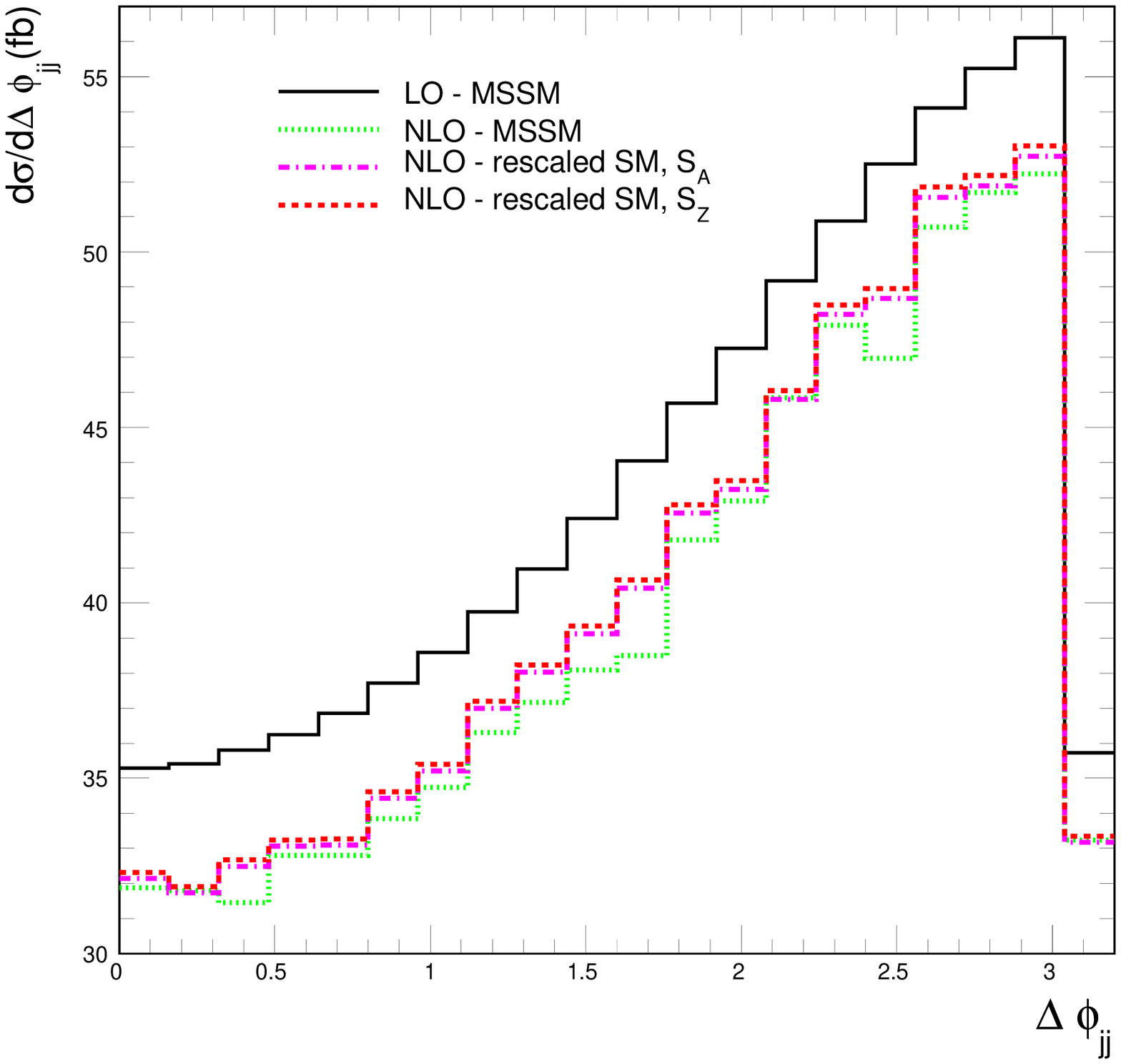}}
         } 
    \hspace{0.3cm}
     \subfigure[]{
         \resizebox{0.45\hsize}{!}{\includegraphics*{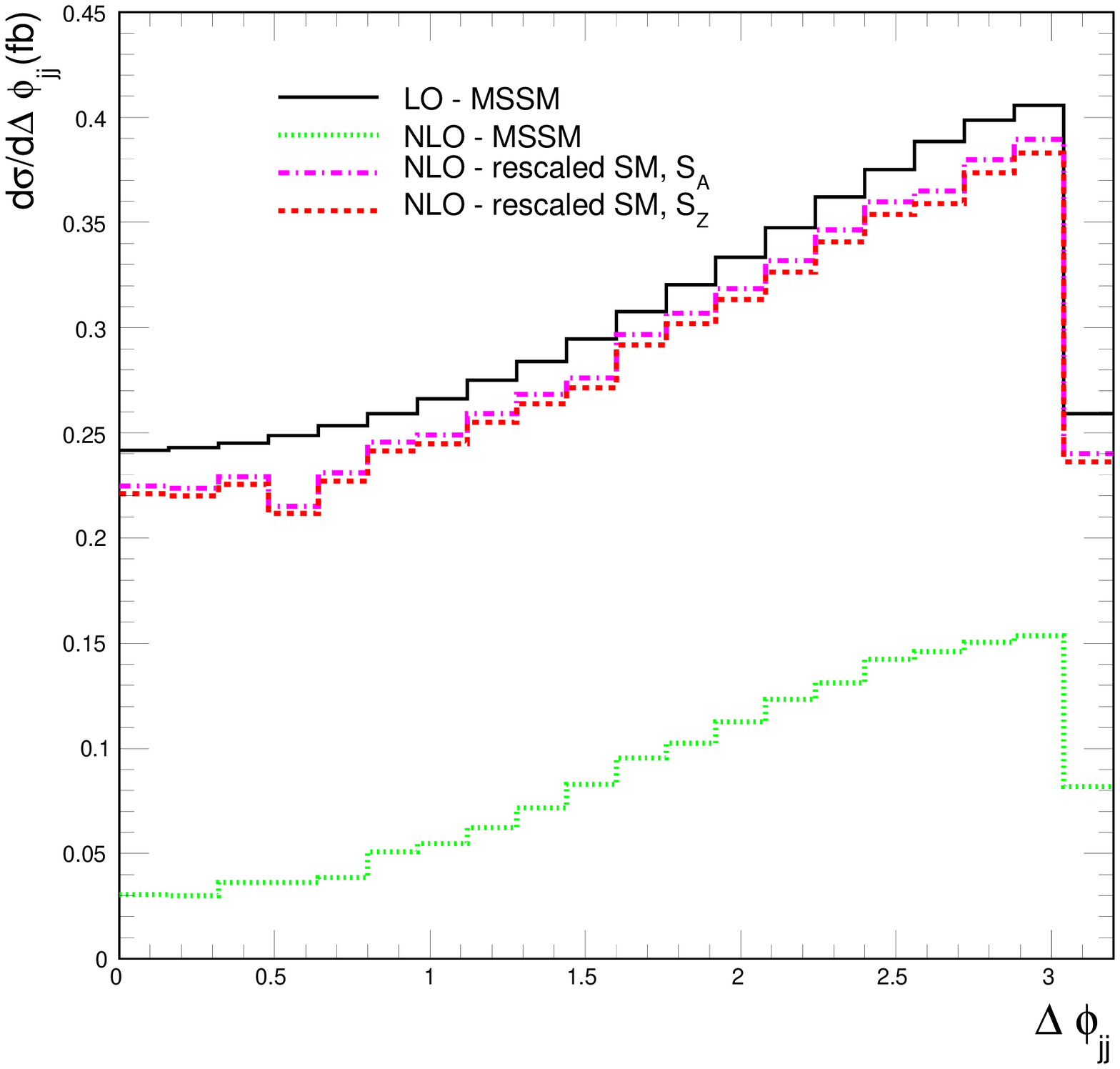}}
         } 
     \caption{Azimuthal-angle distributions for $\tan \beta = 3$, $\MA =
         500 \UGeV$ (left plot) and $\tan \beta = 50$, $\MA = 100 \UGeV$
         (right plot), comparing the MSSM results with the rescaled SM values
         with the MSTW2008NLO PDF set.}  
     \label{fig:az_angle}
\end{center}
\end{figure} 

\refF{fig:pt_j1} presents the differential distribution of the
transverse momentum of the leading jet (\ie  the jet with the highest
$\pT$).  As with the azimuthal-angle distribution, the SM rescaling is seen
to work well in the decoupling regime (left plot), but not in the
non-decoupling regime (right plot), where the relative effect of the 
MSSM electroweak corrections is much larger than in the SM, while the
cross section itself is heavily suppressed.  The shape, however, of 
the distribution remains essentially unchanged. The NLO corrections 
(and in particular the EW contribution) cause
an increasing reduction to the cross section at higher energy scales, as noted
in~\refS{sec:VBF_Differential_distributions}.  This feature is
enhanced in the MSSM.

\begin{figure}
\begin{center}
     \subfigure[]{
         \resizebox{0.45\hsize}{!}{\includegraphics*{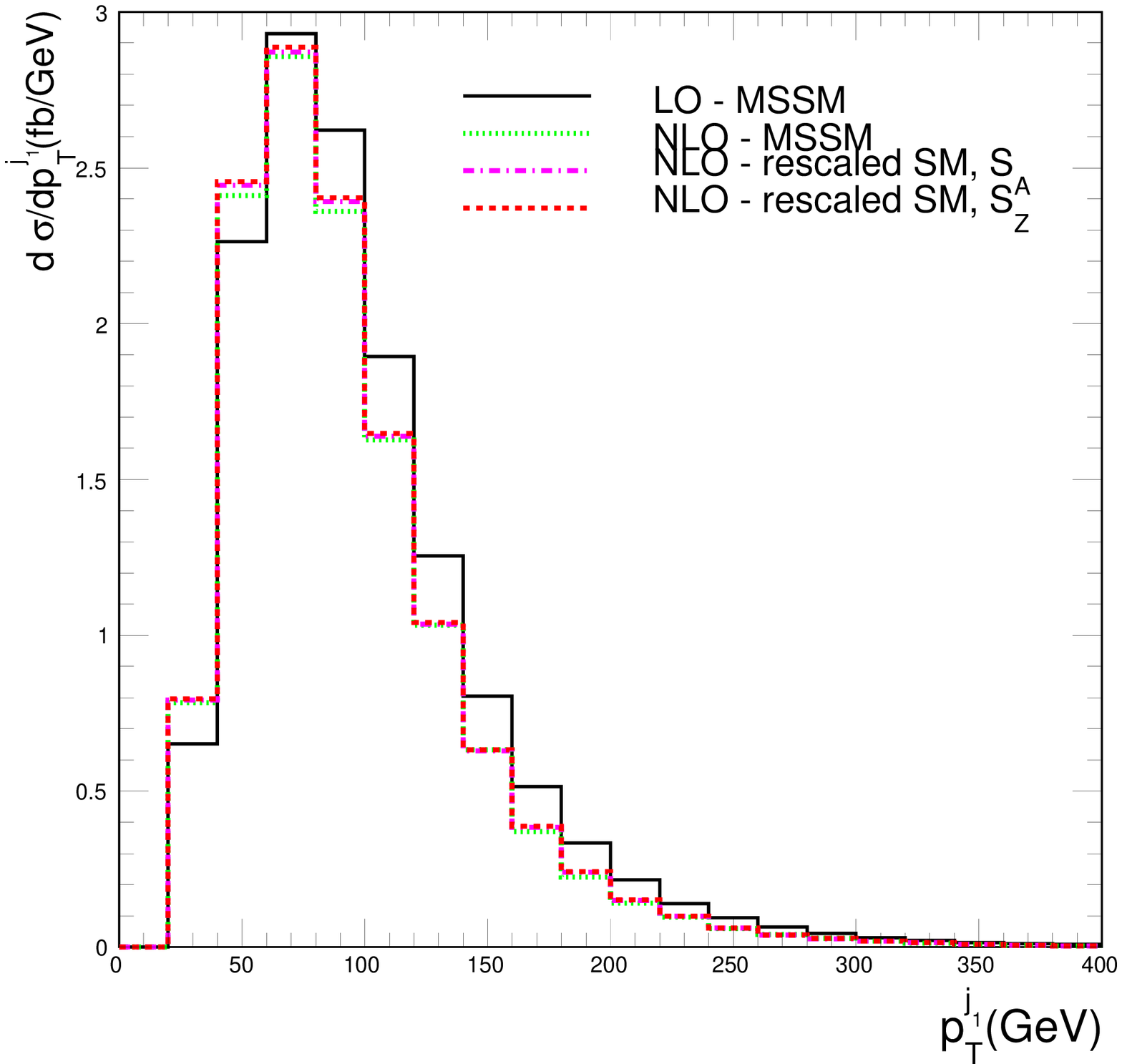}}
         } 
    \hspace{0.3cm}
     \subfigure[]{
         \resizebox{0.45\hsize}{!}{\includegraphics*{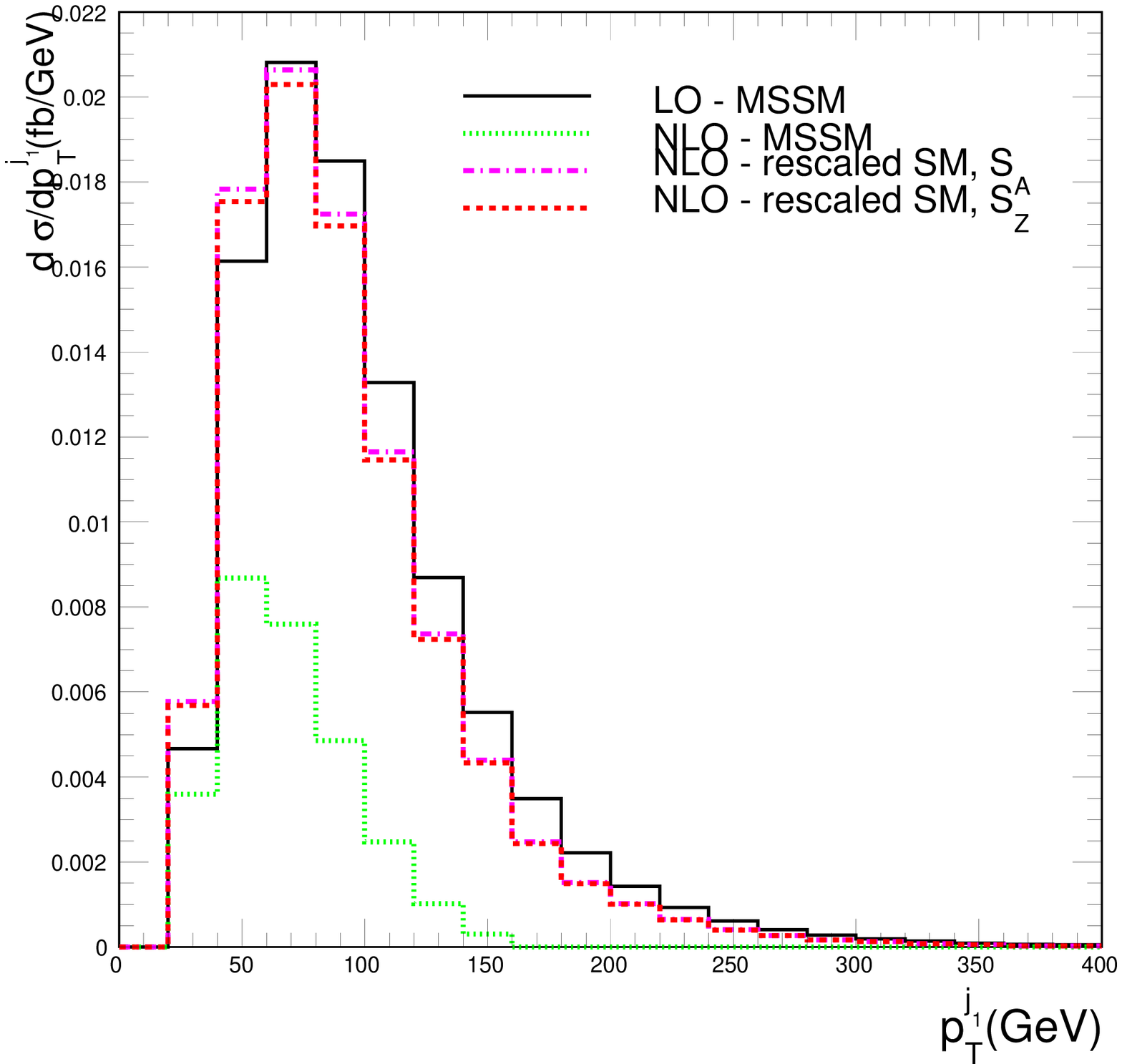}}
         } 
     \caption{Transverse momentum of the leading jet for $\tan \beta = 3$,
         $\MA = 500 \UGeV$ (left plot) and $\tan \beta = 50$, $\MA =
         100 \UGeV$ (right plot), comparing the MSSM results with the rescaled
         SM values with the MSTW2008NLO PDF set.}  
     \label{fig:pt_j1}
\end{center}
\end{figure} 

The Higgs transverse-momentum ($p_{\rm T,\PH}$) and rapidity ($y_{\PH}$) differential
distributions are shown in \refF{fig:pty_higgs}, which shows the same
general behaviour of the SM--MSSM comparison as the previous plots.  The NLO
corrections are much larger for the non-decoupling parameter point (right-hand
plots), resulting in poorer rescaling behaviour in the region where the
cross section itself is very small, but the shape of the
distributions is relatively unaffected.  As with the distribution of the
transverse momentum of the leading jet, $p_{\rm T,j1}$, the magnitude of the NLO
corrections increases with $\pT$, and increases at a greater rate in the
MSSM with $\tan \beta = 50$, $\MA = 100 \UGeV$ than in the corresponding
rescaled SM results.  The NLO corrections to the Higgs rapidity distribution
are relatively constant as a function of the Higgs boson's rapidity. For
high $\tan \beta$ and low $\MA$ a certain effect on the shape is
visible, but once again this behaviour appears in a parameter region
where the cross section itself is very small.

\begin{figure}
\begin{center}
     \subfigure[]{
         \resizebox{0.45\hsize}{!}{\includegraphics*{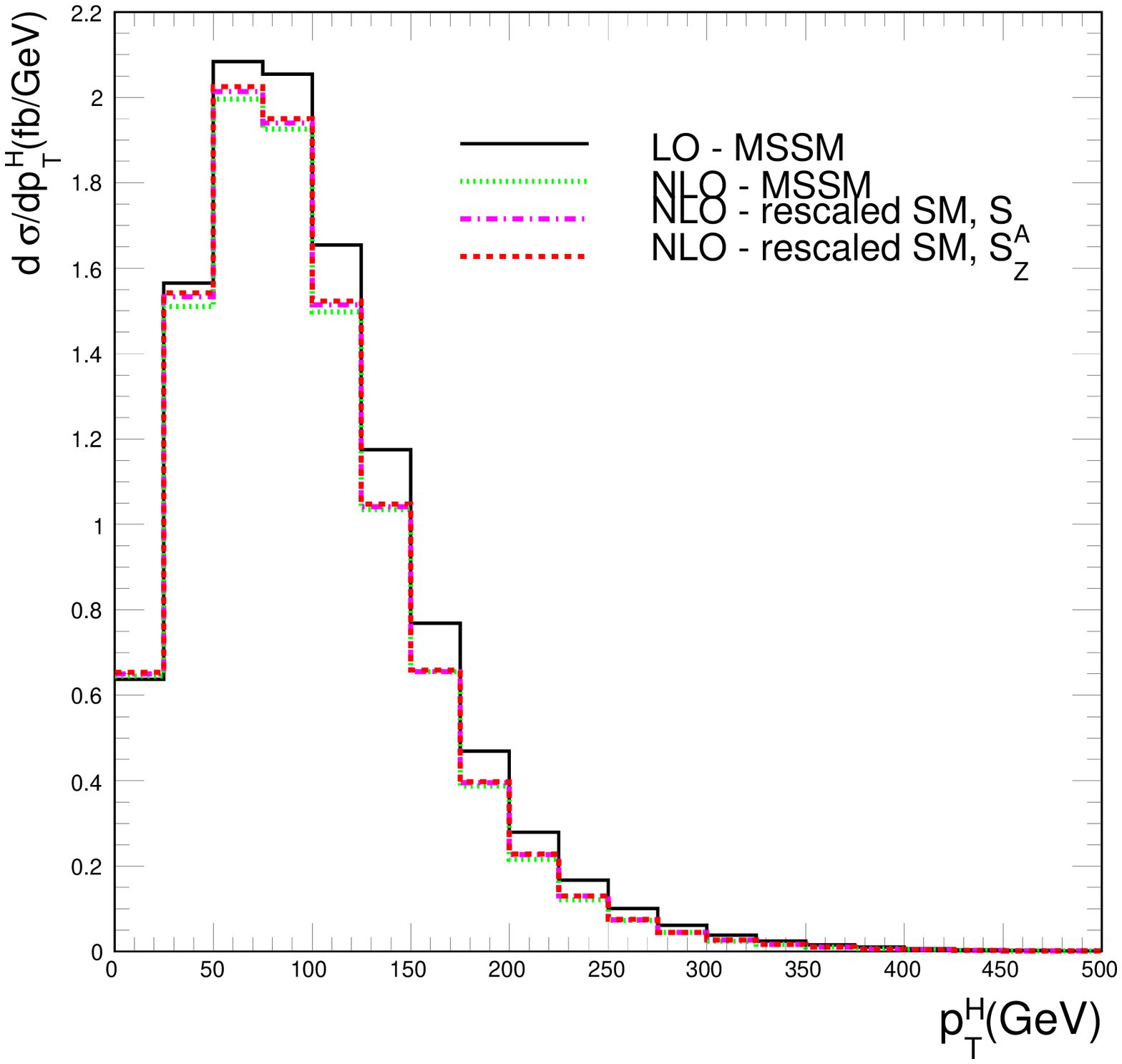}}
         } 
    \hspace{0.3cm}
     \subfigure[]{
         \resizebox{0.45\hsize}{!}{\includegraphics*{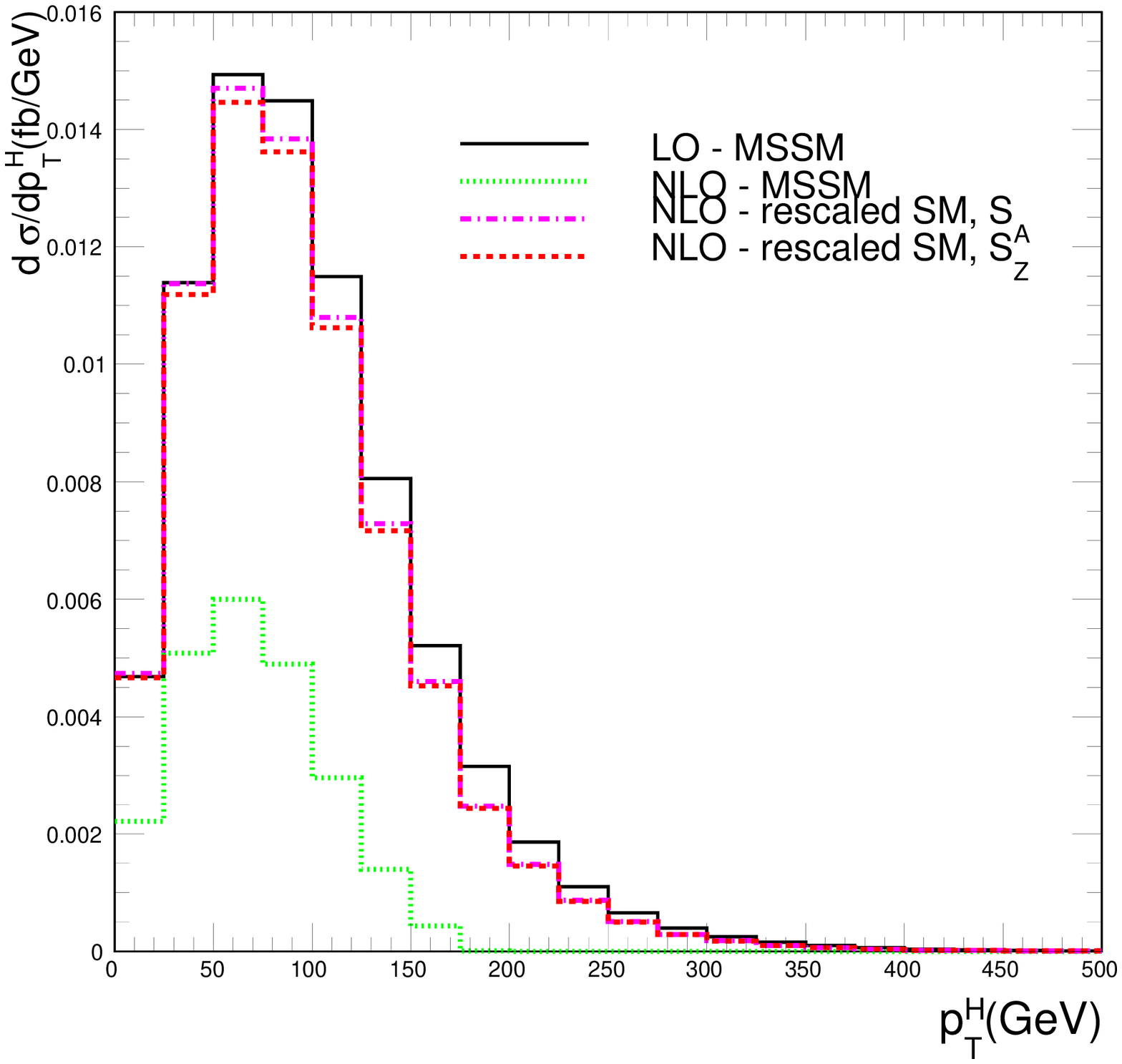}}
         } \\
     \subfigure[]{
         \resizebox{0.45\hsize}{!}{\includegraphics*{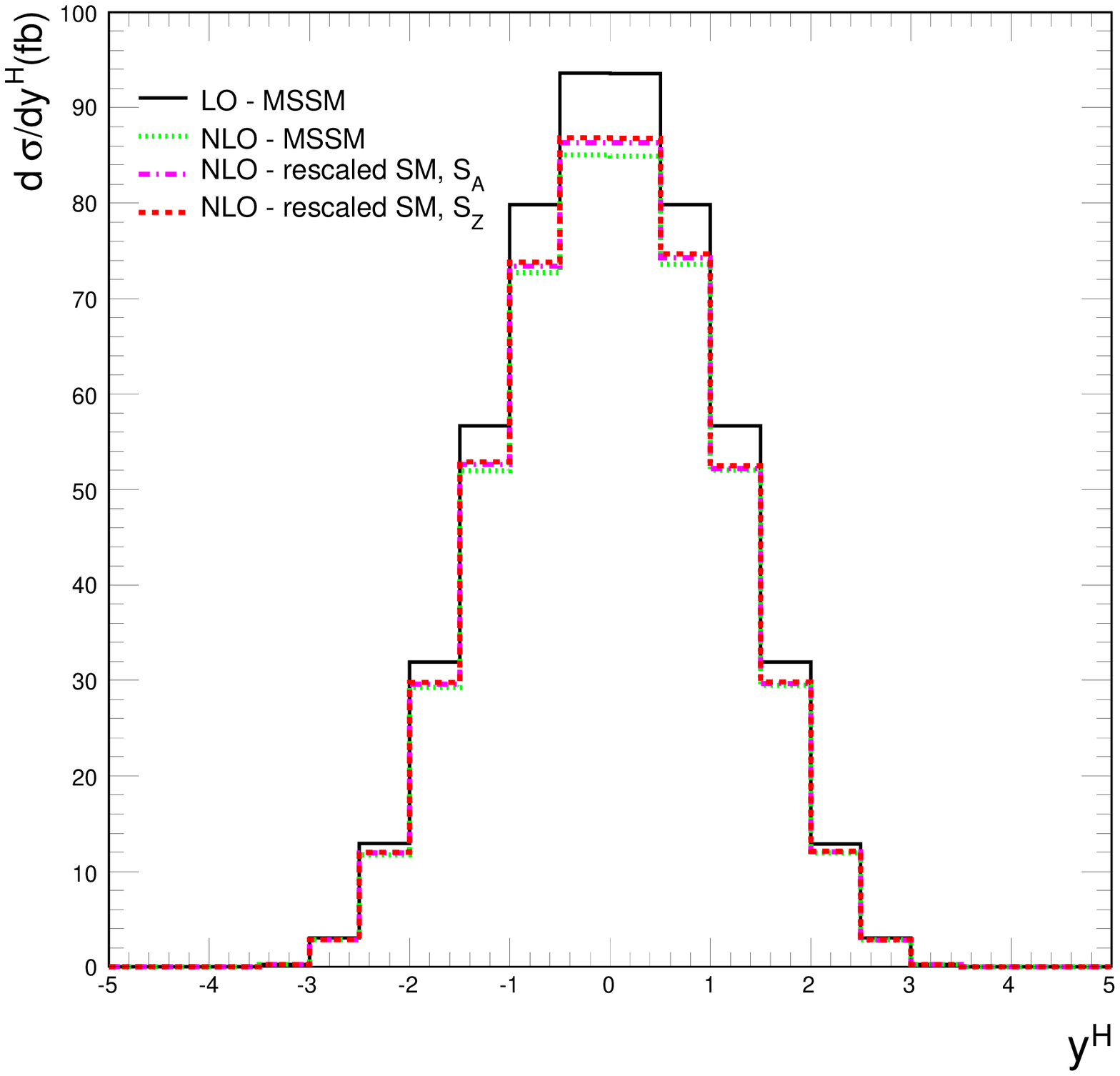}}
         } 
    \hspace{0.3cm}
     \subfigure[]{
         \resizebox{0.45\hsize}{!}{\includegraphics*{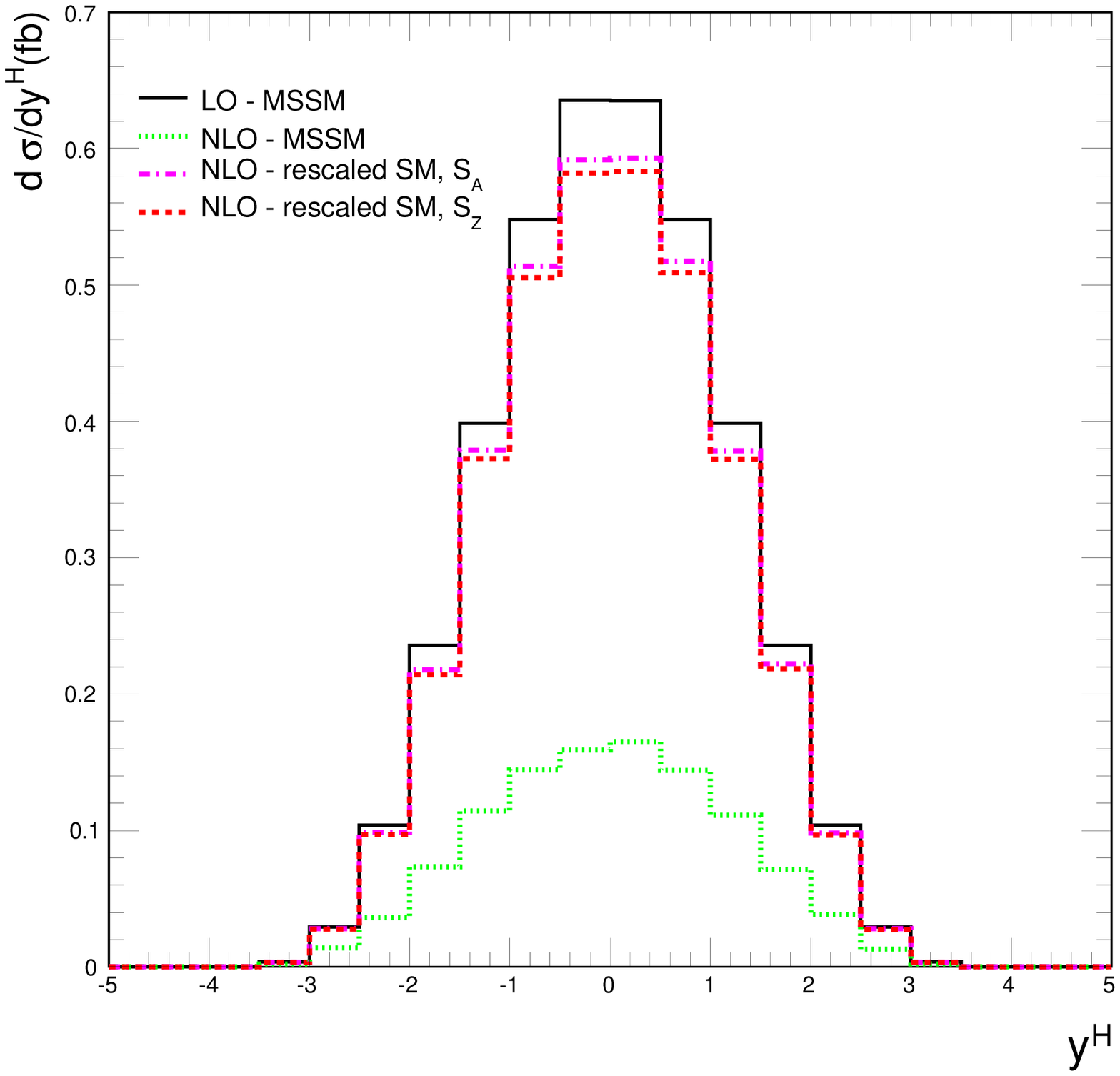}}
         } 
     \caption{Distribution of Higgs transverse momentum (upper plots) and
         rapidity (lower plots) for $\tan \beta = 3$, $\MA = 500 \UGeV$ (left
         plots) and $\tan \beta = 50$, $\MA = 100 \UGeV$ (right plots),
         comparing the MSSM results with the rescaled SM values with the
         MSTW2008NLO PDF set.}  
     \label{fig:pty_higgs}
\end{center}
\end{figure} 


\subsubsection{MSSM NLO corrections for Higgs radiation from bottom
quarks in the 5FS}

\label{sec:mssm_bbh}


Apart from the total inclusive cross section for Higgs production at the
LHC, it is important to have predictions for (a) kinematical
distributions of the Higgs boson and associated jets, and (b) the
fraction of events with zero, one, or more jets.  
%
In supersymmetric theories
the associated production of a Higgs boson with bottom quarks is the
dominant process in a large parameter region of the MSSM (see, e.g.,
\Bref{Dittmaier:2011ti}). The proper theory description of this
process has been a subject of discussion for quite some time. The result
is a rather satisfactory reconciliation of the two possible approaches,
the so-called four- and five-flavor scheme (4FS and 5FS), which led to
the ``Santander-matching'' procedure, see above.

Higher-order corrections in the 5FS are usually easier to calculate than
in the 4FS, because one deals with a $2\to 1$ rather than a $2\to 3$
process at LO. However, since the 5FS works in the collinear
approximation of the outgoing bottom quarks, effects from large
transverse momenta of the bottom quarks are taken into account only at
higher orders in this approach. In fact, NNLO plays a special role in
the 5FS, which becomes obvious by noticing that it is the first order
where the 5FS contains the LO Feynman diagram of the
4FS\cite{Harlander:2003ai}. Hence only then the 5FS includes two
outgoing bottom quarks at large transverse momentum.

With the total inclusive cross section under good theoretical control,
it is natural to study more differential quantities. Being a $2\to 1$
process, kinematical distributions in the 5FS are trivial
at LO, just like in gluon fusion: the $p_{\rm T}$ of the Higgs boson vanishes,
and the rapidity distribution is given by the boost of the partonic
relative to the hadronic system.

Non-trivial distributions require a jet in the final state.  The $p_{\rm
T}$
and $y$ distributions of the Higgs boson in the process $b\bar b\to
\phi+$jet were studied at NLO in \Bref{Harlander:2010cz} (here and
in what follows, $\phi\in\{\Ph,\PH,\PA\}$). Combining these results with the
NNLO inclusive total cross section\cite{Harlander:2003ai}, one may
obtain NNLO results with kinematical cuts.

As mentioned above, particularly interesting for experimental analyses
is the decomposition of the events into $\phi+n$-jet bins. The case
$n=0$ can be obtained at NNLO by calculating the case $n\geq 1$ at NLO
level, and subtracting it from the total inclusive cross
section\cite{Catani:2001cr,Harlander:2011fx}. For consistency, however,
both ingredients should be evaluated using NNLO PDFs and running of
$\alphas$. This is indicated by the superscript NLO$'$ in the following
equation:
\begin{equation}
\begin{split}
\sigma^\text{NNLO}_\text{jet-veto} \equiv \sigma^\text{NNLO}_{0\text{-jet}} = \sigma^\text{NNLO}_\text{tot} -
\sigma^{\text{NLO}'}_{\geq 1\text{-jet}}\,.
\label{eq::0jet}
\end{split}
\end{equation}
Note that this equation is understood without any flavor requirements
on the outgoing jet.


%
\begin{figure}
  \begin{center}
    \begin{tabular}{cc}
      \includegraphics[width=.45\textwidth]{%
        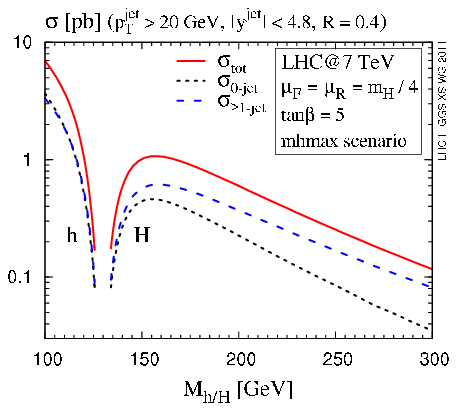} &
      \includegraphics[width=.45\textwidth]{%
        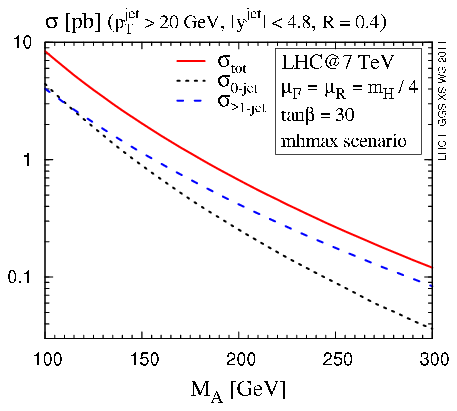} \\
      (a) & (b) \\
      \includegraphics[width=.45\textwidth]{%
        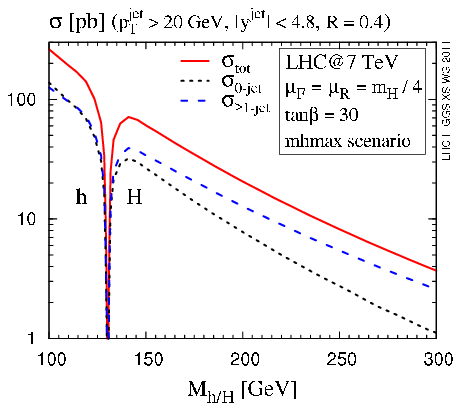} &
      \includegraphics[width=.45\textwidth]{%
        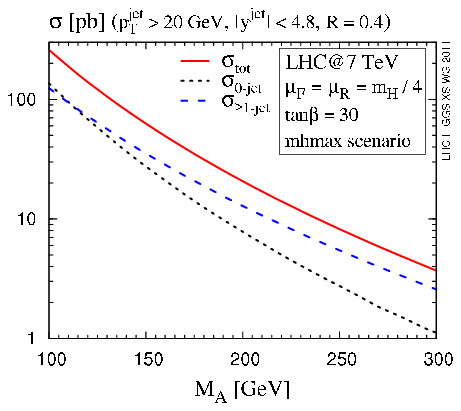} \\
      (c) & (d)
    \end{tabular}
    \caption[]{\label{fig::012jet}\sloppy
      Total inclusive (solid/red) and Higgs plus $n$-jet cross
      section for $n=0$ (dashed/blue) and $n\geq 1$ (dotted/black) in
      the \mhmaxx\ scenario for (a,b) $\tan\beta=5$ and (c,d)
      $\tan\beta=30$. Left and right column correspond to the CP-even
      and the CP-odd Higgs bosons, respectively. }
  \end{center}
\end{figure}
%

%
\begin{figure}
  \begin{center}
    \begin{tabular}{cc}
      \includegraphics[width=.45\textwidth]{%
        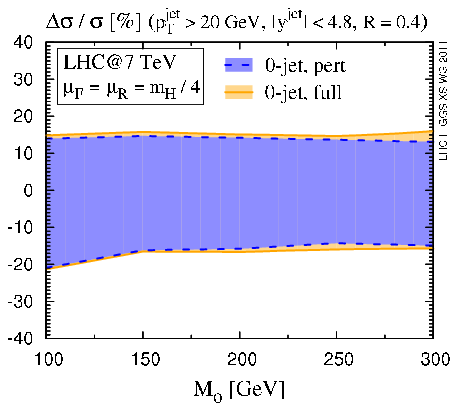} &
      \includegraphics[width=.45\textwidth]{%
        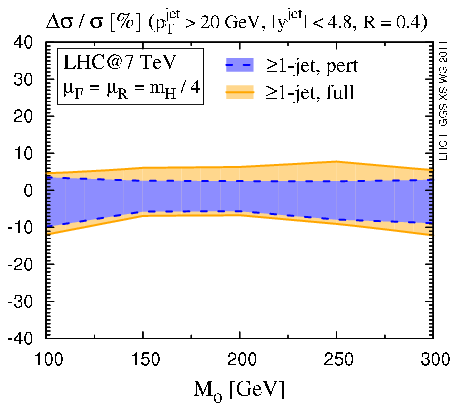} \\
      (a) & (b)
    \end{tabular}
    \caption[]{\label{fig::err012}\sloppy
      Relative perturbative and full error estimates for (a) the jet-vetoed
      cross section and (b) the inclusive $\phi+$jet cross section.
      The full error estimate is obtained by adding the perturbative and the
      PDF$+\alphas$-error quadratically.
    }
  \end{center}
\end{figure}
%

\begin{table}
\begin{center}
\caption[]{\label{tab::012jettb5}Numerical values for the total cross
  section $\sigma_\text{tot}^\text{NNLO}$, the $\phi+0$-jet rate
  $\sigma_\text{0-jet}^\text{NNLO}$, and the inclusive $\phi+$jet rate
  $\sigma_{\geq1\text{-jet}}^\text{NLO}$ for the
  \mhmaxx\ scenario for $\tan\beta=5$ ($\phi\in\{\Ph,\PH,\PA\}$).}
\renewcommand{\arraystretch}{1.2}
\vspace{.5em}
\begin{tabular}{c|ccc|ccc|ccc}
\hline\multicolumn{10}{c}{LHC @
  7\,TeV,\ \ $|y^\text{jet}|<4.8$,\ \ $p_{\rm T}^\text{jet} > 20 \UGeV$,\ \ 
$R = 0.4$;\ \ \mhmaxx$(\tan\beta=5)$} 
\\\hline
$\MA$&\multicolumn{3}{c|}{$\sigma_\text{tot}^\text{NNLO}$ [pb]}&\multicolumn{3}{c|}{$\sigma_{0\text{-jet}}^\text{NNLO}$ [pb]}&\multicolumn{3}{c}{$\sigma_{\geq1\text{-jet}}^\text{NLO}$ [pb]}\\ {}[GeV]&$\PA$&$\Ph$&$\PH$&$\PA$&$\Ph$&$\PH$&$\PA$&$\Ph$&$\PH$\\\hline
 $100$&$   8.42$&$   9.84$&$  0.392$&$   4.43$&$   5.38$&$  0.180$&$   4.03$&$   4.59$&$  0.214$\\
$110$&$   6.12$&$   7.10$&$  0.574$&$   3.09$&$   3.74$&$  0.261$&$   3.05$&$   3.40$&$  0.315$\\
$120$&$   4.54$&$   5.05$&$  0.788$&$   2.22$&$   2.59$&$  0.353$&$   2.35$&$   2.47$&$  0.436$\\
$130$&$   3.43$&$   3.54$&$  0.974$&$   1.62$&$   1.78$&$  0.429$&$   1.83$&$   1.77$&$  0.545$\\
$140$&$   2.63$&$   2.48$&$   1.07$&$   1.19$&$   1.23$&$  0.461$&$   1.44$&$   1.26$&$  0.606$\\
$150$&$   2.04$&$   1.78$&$   1.06$&$  0.894$&$  0.878$&$  0.446$&$   1.14$&$  0.913$&$  0.611$\\
$160$&$   1.60$&$   1.33$&$  0.974$&$  0.681$&$  0.652$&$  0.400$&$  0.924$&$  0.686$&$  0.573$\\
$170$&$   1.27$&$   1.03$&$  0.861$&$  0.522$&$  0.505$&$  0.344$&$  0.749$&$  0.536$&$  0.516$\\
$180$&$   1.02$&$  0.836$&$  0.744$&$  0.406$&$  0.407$&$  0.291$&$  0.613$&$  0.434$&$  0.453$\\
$190$&$  0.825$&$  0.698$&$  0.635$&$  0.320$&$  0.339$&$  0.242$&$  0.503$&$  0.364$&$  0.392$\\
$200$&$  0.673$&$  0.599$&$  0.539$&$  0.254$&$  0.290$&$  0.201$&$  0.417$&$  0.313$&$  0.337$\\
$210$&$  0.552$&$  0.526$&$  0.457$&$  0.203$&$  0.254$&$  0.165$&$  0.348$&$  0.275$&$  0.290$\\
$220$&$  0.457$&$  0.470$&$  0.387$&$  0.163$&$  0.227$&$  0.136$&$  0.292$&$  0.246$&$  0.249$\\
$230$&$  0.380$&$  0.427$&$  0.328$&$  0.133$&$  0.206$&$  0.113$&$  0.246$&$  0.224$&$  0.213$\\
$240$&$  0.318$&$  0.393$&$  0.278$&$  0.108$&$  0.190$&$ 0.0942$&$  0.208$&$  0.206$&$  0.184$\\
$250$&$  0.267$&$  0.366$&$  0.237$&$ 0.0894$&$  0.176$&$ 0.0785$&$  0.178$&$  0.192$&$  0.158$\\
$260$&$  0.226$&$  0.343$&$  0.203$&$ 0.0735$&$  0.165$&$ 0.0649$&$  0.152$&$  0.180$&$  0.136$\\
$270$&$  0.192$&$  0.324$&$  0.173$&$ 0.0600$&$  0.156$&$ 0.0541$&$  0.130$&$  0.170$&$  0.118$\\
$280$&$  0.163$&$  0.308$&$  0.149$&$ 0.0509$&$  0.148$&$ 0.0461$&$  0.112$&$  0.162$&$  0.102$\\
$290$&$  0.140$&$  0.294$&$  0.129$&$ 0.0426$&$  0.142$&$ 0.0388$&$ 0.0967$&$  0.155$&$ 0.0891$\\
$300$&$  0.120$&$  0.283$&$  0.111$&$ 0.0363$&$  0.136$&$ 0.0338$&$ 0.0836$&$  0.149$&$ 0.0774$\\
\hline
\end{tabular}
\end{center}
\end{table}

\begin{table}
\begin{center}
\caption[]{\label{tab::012jettb30}Same as Table\,\ref{tab::012jettb5},
  but for $\tan\beta=30$.}
\renewcommand{\arraystretch}{1.2}
\vspace{.5em}
\begin{tabular}{c|ccc|ccc|ccc}
\hline\multicolumn{10}{c}{LHC @
  7\,TeV,\ \ $|y^\text{jet}|<4.8$,\ \ $p_{\rm T}^\text{jet} > 20 \UGeV$,\ \ $R =
  0.4$;\ \ \mhmaxx$(\tan\beta=30)$} 
\\\hline
$\MA$&\multicolumn{3}{c|}{$\sigma_\text{tot}^\text{NNLO}$ [pb]}&\multicolumn{3}{c|}{$\sigma_{0\text{-jet}}^\text{NNLO}$ [pb]}&\multicolumn{3}{c}{$\sigma_{\geq1\text{-jet}}^\text{NLO}$ [pb]}\\ {}[GeV]&$\PA$&$\Ph$&$\PH$&$\PA$&$\Ph$&$\PH$&$\PA$&$\Ph$&$\PH$\\\hline
 $100$&$259$&$265$&$  0.776$&$136$&$140$&$  0.365$&$124$&$127$&$  0.415$\\
$110$&$188$&$195$&$   1.93$&$   94.9$&$   98.4$&$  0.906$&$   93.8$&$   97.0$&$   1.03$\\
$120$&$140$&$141$&$   7.19$&$   68.2$&$   68.9$&$   3.38$&$   72.3$&$   72.7$&$   3.85$\\
$130$&$105$&$   64.2$&$   46.0$&$   49.7$&$   30.6$&$   21.4$&$   56.3$&$   34.0$&$   24.8$\\
$140$&$   80.8$&$   11.1$&$   71.2$&$   36.6$&$   5.25$&$   32.2$&$   44.3$&$   5.92$&$   39.2$\\
$150$&$   62.7$&$   3.83$&$   59.8$&$   27.5$&$   1.81$&$   26.2$&$   35.2$&$   2.04$&$   33.6$\\
$160$&$   49.3$&$   2.03$&$   48.0$&$   20.9$&$  0.955$&$   20.4$&$   28.4$&$   1.08$&$   27.7$\\
$170$&$   39.1$&$   1.32$&$   38.4$&$   16.1$&$  0.621$&$   15.8$&$   23.0$&$  0.702$&$   22.6$\\
$180$&$   31.4$&$  0.961$&$   30.9$&$   12.5$&$  0.453$&$   12.3$&$   18.9$&$  0.513$&$   18.6$\\
$190$&$   25.4$&$  0.756$&$   25.1$&$   9.86$&$  0.356$&$   9.74$&$   15.5$&$  0.404$&$   15.3$\\
$200$&$   20.7$&$  0.625$&$   20.5$&$   7.81$&$  0.295$&$   7.74$&$   12.8$&$  0.334$&$   12.7$\\
$210$&$   17.0$&$  0.535$&$   16.9$&$   6.26$&$  0.252$&$   6.20$&$   10.7$&$  0.286$&$   10.6$\\
$220$&$   14.0$&$  0.471$&$   13.9$&$   5.01$&$  0.222$&$   4.97$&$   8.98$&$  0.252$&$   8.92$\\
$230$&$   11.7$&$  0.423$&$   11.6$&$   4.08$&$  0.200$&$   4.05$&$   7.56$&$  0.226$&$   7.51$\\
$240$&$   9.77$&$  0.386$&$   9.71$&$   3.33$&$  0.182$&$   3.31$&$   6.41$&$  0.206$&$   6.37$\\
$250$&$   8.22$&$  0.357$&$   8.17$&$   2.75$&$  0.168$&$   2.73$&$   5.46$&$  0.191$&$   5.43$\\
$260$&$   6.95$&$  0.334$&$   6.91$&$   2.26$&$  0.157$&$   2.25$&$   4.66$&$  0.178$&$   4.64$\\
$270$&$   5.90$&$  0.315$&$   5.87$&$   1.85$&$  0.148$&$   1.84$&$   3.99$&$  0.168$&$   3.97$\\
$280$&$   5.03$&$  0.299$&$   5.00$&$   1.57$&$  0.141$&$   1.56$&$   3.44$&$  0.160$&$   3.43$\\
$290$&$   4.30$&$  0.286$&$   4.28$&$   1.31$&$  0.135$&$   1.30$&$   2.97$&$  0.153$&$   2.96$\\
$300$&$   3.70$&$  0.274$&$   3.68$&$   1.12$&$  0.129$&$   1.11$&$   2.57$&$  0.146$&$   2.56$\\
\hline
\end{tabular}
\end{center}
\end{table}


As an exemplary case, we consider results for the jet parameters
(anti-$k_{\mathrm{T}}$\cite{Cacciari:2008gp})
\begin{equation}
\begin{split}
R = 0.4,\qquad
p_{\rm T}^\text{jet} > 20 \UGeV,\qquad
|\eta^\text{jet}| < 4.8.
\end{split}
\end{equation}
Fig.~\ref{fig::012jet} shows the contributions of the NNLO jet-vetoed
($\phi+0$-jet) and the NLO inclusive $\phi+$jet rate to the NNLO total cross
section in the \mhmaxx\ scenario for two different values of
$\tan\beta$.  The corresponding numbers are given in
Tables~\ref{tab::012jettb5} and \ref{tab::012jettb30}.  Note, however,
that the sum of the $\phi+n$-jet cross sections does not add up exactly
to the total rate, because they are evaluated at different perturbative
orders, and therefore with different sets of PDFs and $\alphas$
evolution. These numbers have been derived from the SM results of
\Bref{Harlander:2011fx}, reweighted by the MSSM bottom Yukawa
coupling with the help of {\tt FeynHiggs}.

Fig.\,\ref{fig::err012} displays the relative perturbative error
estimates,
obtained by varying the renormalisation scale $\mu_{\rm R}$ by a factor
of two around $\mu_0=M_\phi/4$ while keeping the factorisation scale
$\mu_{\rm F}$ fixed at $\mu_0$, and then doing the same with $\mu_{\rm
F}$ and $\mu_{\rm R}$ interchanged. The PDF$+\alphas$ uncertainties are
those from MSTW2008\cite{Martin:2009iq}. The full error estimate
was obtained by
quadratically adding the perturbative to the PDF$+\alphas$ error
estimate.  Note
that these plots are the same as in the SM; the precise numerical values
can therefore be taken from \Bref{Harlander:2011fx}.

The 5FS has also been used in order to evaluate associated $\phi+\Pb$
production through NLO\cite{Campbell:2002zm}, which is contained in the
recent $\phi+$jet calculation \cite{Harlander:2011fx}. Updated numbers
can be obtained quite easily with the help of the program {\tt
  MCFM}\cite{MCFMweb}. Here, however, we use the program described in
\Bref{Harlander:2011fx}. Employing again the NNLO result for the
total inclusive cross section, one may evaluate the $\PQb$-vetoed cross
section through NNLO, along the lines of Eq.\,(\ref{eq::0jet}):
\begin{equation}
\begin{split}
\sigma^\text{NNLO}_{\Pb\text{-jet veto}} \equiv \sigma^\text{NNLO}_{0\Pb} = \sigma^\text{NNLO}_\text{tot}
- \sigma^{\text{NLO}'}_{\geq 1\Pb\text{-jet}}\,.
\end{split}
\end{equation}
It should be recalled, however, that this quantity does not take into
account the finite $\Pb$-jet efficiency $\ep_{\Pb}$. This distinguishes it from
the Higgs cross section with zero $\Pb$-tags $\sigma_{0\Pb\text{-tag}}$, which
can be obtained by combining $\sigma_{0\Pb}$ with the rate
for having one or two $\Pb$ quarks in the final state, i.e.,
$\sigma_{1\Pb}$ and $\sigma_{2\Pb}$\cite{Harlander:2011fx}:
\begin{equation}
\begin{split}
\sigma^\text{NNLO}_{0\Pb\text{-tag}} &= 
\sigma^\text{NNLO}_{0\Pb} +
(1-\ep_{\PQb})\sigma^\text{NLO}_{1\Pb} +
(1-\ep_{\PQb})^2\sigma^\text{LO}_{2\Pb}.
\label{eq::0btag}
\end{split}
\end{equation}
In this case, we set the jet parameters to
\begin{equation}
\begin{split}
R=0.4,\qquad
p_{\rm T}^b > 20\,\text{GeV},\qquad
|\eta^b| < 2.5.
\end{split}
\end{equation}
The $\phi+0\Pb$-, $1\Pb$-, and $2\Pb$-contributions are shown, together with the
total cross section, in Fig.\,\ref{fig::012b}, for the same parameters
as in Fig.\,\ref{fig::012jet}. The numbers were again produced with the
help of the results from Ref.\,\cite{Harlander:2011fx}, reweighted by
the corresponding MSSM Yukawa coupling.


%
\begin{figure}
  \begin{center}
    \begin{tabular}{cc}
      \includegraphics[width=.45\textwidth]{%
        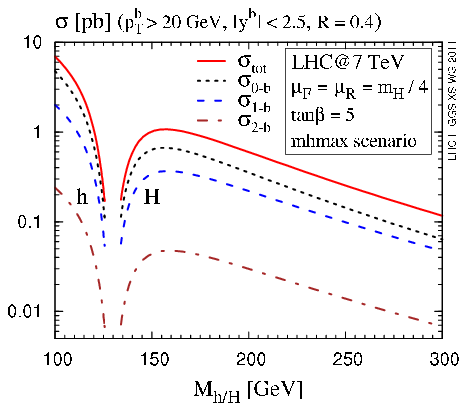} &
      \includegraphics[width=.45\textwidth]{%
        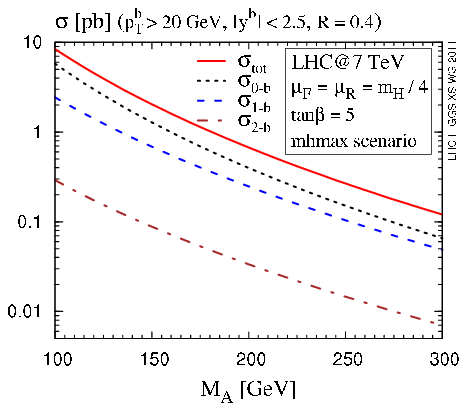} \\
      (a) & (b) \\
      \includegraphics[width=.45\textwidth]{%
        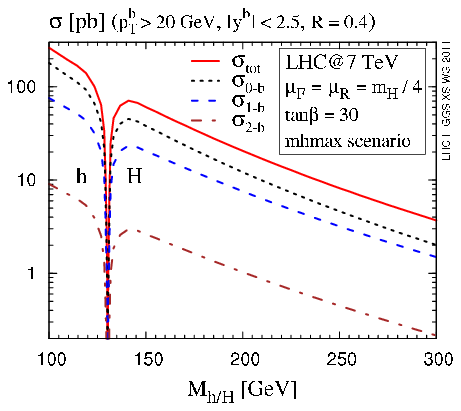} &
      \includegraphics[width=.45\textwidth]{%
        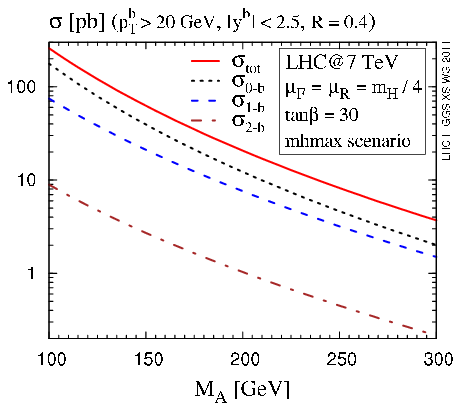} \\
      (c) & (d)
    \end{tabular}
    \caption[]{\label{fig::012b}\sloppy Total inclusive (solid/red)
      and Higgs plus $n\Pb$-jet cross section for $n=0$ (dashed/blue),
      $n=1$ (dotted/black), and $n=2$ (dash-dotted/brown) in the
      \mhmaxx\  scenario for (a,b) $\tan\beta=5$ and (c,d)
      $\tan\beta=30$. Left and right column correspond to the CP-even
      and the CP-odd Higgs bosons, respectively. }
  \end{center}
\end{figure}
%






%
\subsubsection{Comparison of b-jet acceptance and kinematic distributions in 
$\PQb\PAQb \rightarrow \PH$ production 
               between {\sc PYTHIA} and high-order calculations}

At large values of tan$\beta$ and $\MA$ the $\PQb\PAQb \rightarrow \phi$ process dominates the production of the
MSSM neutral Higgs bosons~\cite{Djouadi:2005gj}. The most recent SUSY $\phi \rightarrow \tau \tau$ analysis in CMS~\cite{CMSPAS:HIG-11-020} uses 
a combination of two analyses: with at least one b-tagged jet and with zero b-tagged jets. For the event generation
of $\Pp\Pp \rightarrow \PQb\PAQb\Ph$ process the {\sc PYTHIA} Monte Carlo (process 186) is used. The comparison of {\sc PYTHIA} with the 4FS calculations is shown 
in the earlier paper~\cite{Kinnunen:2004ji} for the $\pT$ distribution of b-jets where good agreement at the level of $10\%$ is found.
In this note we compare acceptance efficiency for b-jets, $\pT$, and $\eta$ distributions for b-jets and Higgs bosons as given by 
{\sc PYTHIA} with the high-order calculations in the 5FS described in
\Bref{Harlander:2011fx} and \refS{sec:mssm_bbh}.
The jets have been reconstructed from 
the final-state partons using the anti-$k_{\mathrm{T}}$ algorithm~\cite{Cacciari:2008gp} with parameter $R$=0.5. 
The jet is defined as a b-jet if at least one b quark is present in the list of the 
jet constituents. The kinematic acceptance cuts for b-jets used were the same as in the CMS analysis~\cite{CMSPAS:HIG-11-020}: 
$\pT> 20 \UGeV$, $| \eta | < $ 2.4. \refT{tab:bbH_acceptance_tab140.tex} and~\refT{tab:bbH_acceptance_tab400.tex} show the fraction 
of b-jets within the acceptance obtained from the high order calculations (second column) and from {\sc PYTHIA} generator (5-th column) 
for $\MH=140 \UGeV$ and $400 \UGeV$. The errors of the theoretical calculations due to the scale variation and PDF are shown in the third and fourth 
columns of \refT{tab:bbH_acceptance_tab140.tex} and \refT{tab:bbH_acceptance_tab400.tex}. The total theoretical cross section
for the Standard Model $\PQb\PAQb \rightarrow \PH$ process for $\MH=140 \UGeV$ is $108.85\Upb$ and for 
$\MH=400\UGeV$ it is $1.29\Upb$.
\begin{table}
\centering
\caption{The fraction of b-jets within the acceptance obtained from the high-order calculations in the 5FS (second column) and from {\sc PYTHIA} 
         generator (5-th column) for $\MH=140\UGeV$. The errors of the theoretical calculations due to the scale variation and PDF in \% are shown 
         in third and fourth columns}
\vspace{.5em}
\begin{tabular}{ccccc} 
\hline 
final state $i$  &  $\sigma_{i}^{\mathrm{th}}$/$\sigma _{\mathrm{tot}}^{\mathrm{th}}$ & scale error(\%) & PDF error (\%)    &  {\sc PYTHIA} $\Pp\Pp \rightarrow \PQb\PAQb \Ph$   \\ 
\hline 
 0 b-jet     &            0.6381                           & -14.4~~+8.8      & -4.6~~+3.6       &         0.621 \\
 one b-jet   &            0.3286                           & -6.9~~+4.4       & -3.2~~+5.0       &         0.322 \\
 two b-jets  &            0.0417                           & -33.1~~+59.0     & -3.0~~+2.3       &         0.057 \\
\hline 
\end{tabular}
\label{tab:bbH_acceptance_tab140.tex}
\end{table}

\begin{table}
\centering
\caption{The fraction of b-jets within the acceptance obtained from the high order calculations in the 5FS (second column) and from {\sc PYTHIA} 
         generator (5-th column) for $\MH=400\UGeV$. The errors of the theoretical calculations due to the scale variation and PDF are shown in 
          third and fourth columns}
\vspace{.5em}
\begin{tabular}{ccccc} 
\hline 
final state $i$  & $\sigma_{i}^{\mathrm{th}}$/$\sigma _{\mathrm{tot}}^{\mathrm{th}}$ & scale error (\%) & PDF error (\%)  &   {\sc PYTHIA} $\Pp\Pp \rightarrow \PQb\PAQb \Ph$   \\ 
\hline 
 0 b-jet     &            0.519                           & -7.2~~+8.9       &   -6.7~~+7.6    &         0.511 \\
 one b-jet   &            0.426                           & -8.9~~+5.8       &   -6.4~~+6.4    &         0.387 \\
 two b-jets  &            0.063                           & -28.4~~+45.5     &   -4.5~~+4.4    &         0.102 \\
\hline 
\end{tabular}
\label{tab:bbH_acceptance_tab400.tex}
\end{table}
Good agreement between {\sc PYTHIA} and the high order calculations in the 5FS is found for the fractions of zero b-jets and at least
one b-jet in the acceptance. The fraction of two b-jets predicted by the 5FS calculations is lower than given by {\sc PYTHIA}, but agrees
within the theoretical errors. 

We compare the differential distributions $\pT$ and rapidity ($y$) of the leading $\pT$ b-jet within the acceptance
between {\sc PYTHIA} and the NLO predictions. \refF{fig:bbHfig1} shows the distributions of $y^{\PQb}$ normalised on unity for $\MH=140\UGeV$ (left)
and $\MH=400\UGeV$ (right). One can see the difference between {\sc PYTHIA} and the NLO curves especially for the heavy Higgs boson. 
\begin{figure}
\begin{center}
\includegraphics[width=0.45\textwidth]{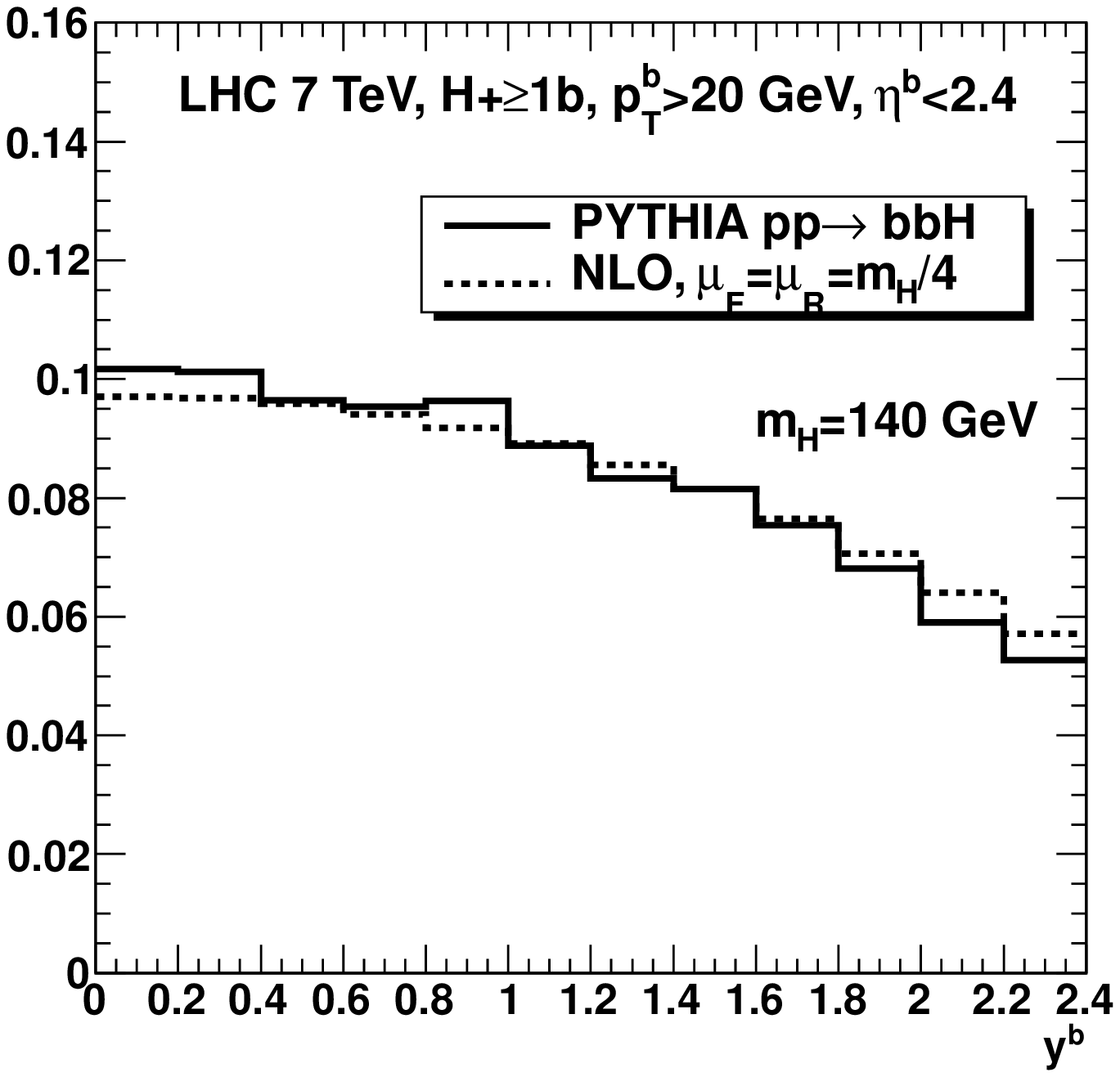} \hfill
\includegraphics[width=0.45\textwidth]{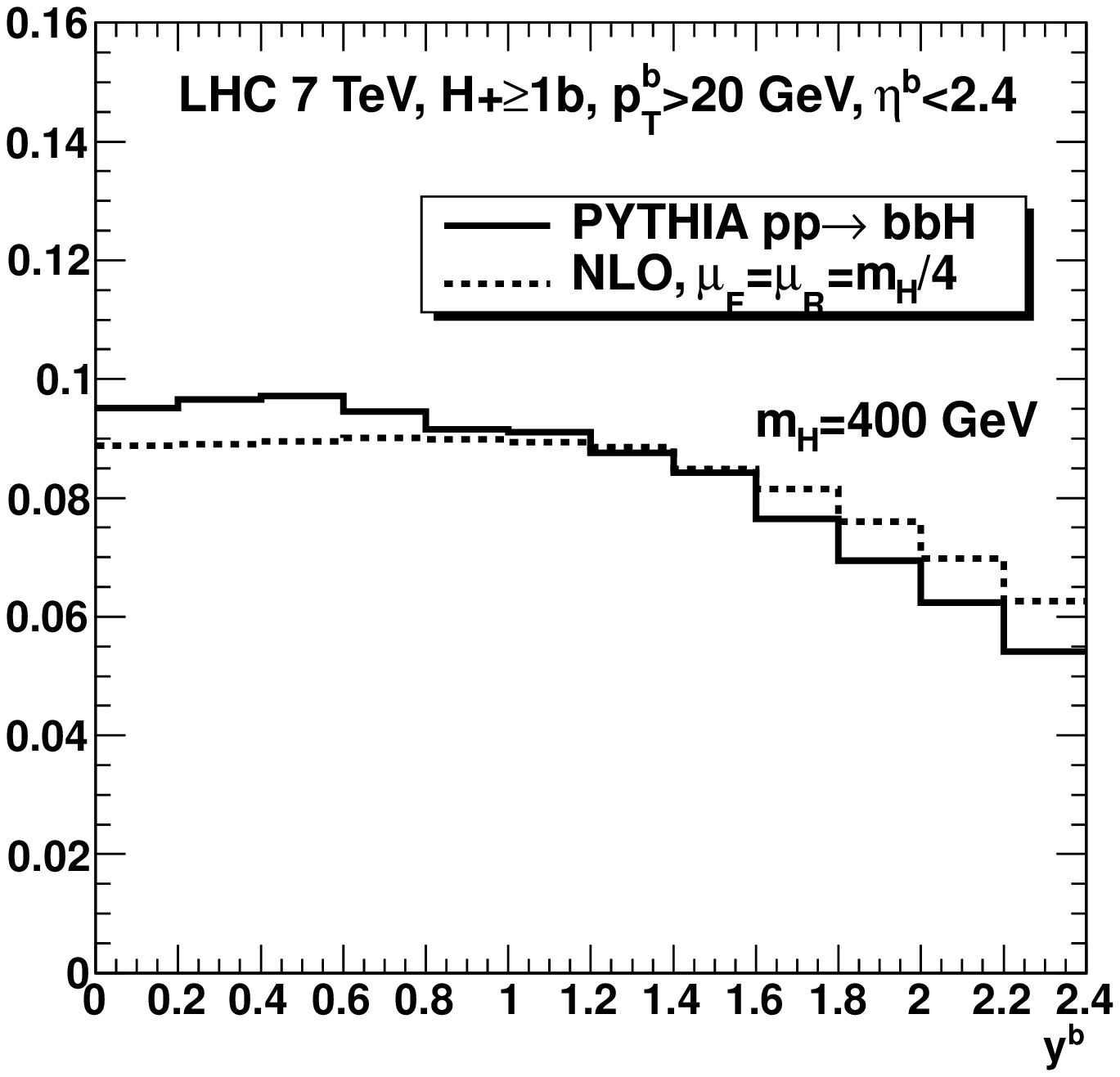}
\caption{The rapidity distributions of the leading $\pT$ b-jet normalised on unity for $\MH=140\UGeV$ (left) 
         and $\MH=400\UGeV$ (right) obtained with {\sc PYTHIA} (solid lines) and with NLO calculations (dashed lines).}
\label{fig:bbHfig1}
\end{center}
\end{figure}
\refF{fig:bbHfig2} shows the distributions of $\pT^{\PQb}$ normalised on unity for $\MH=140\UGeV$ (left)
and $\MH=400\UGeV$ (right). There is good agreement between {\sc PYTHIA} and the NLO predictions.
\begin{figure}
\begin{center}
\includegraphics[width=0.45\textwidth]{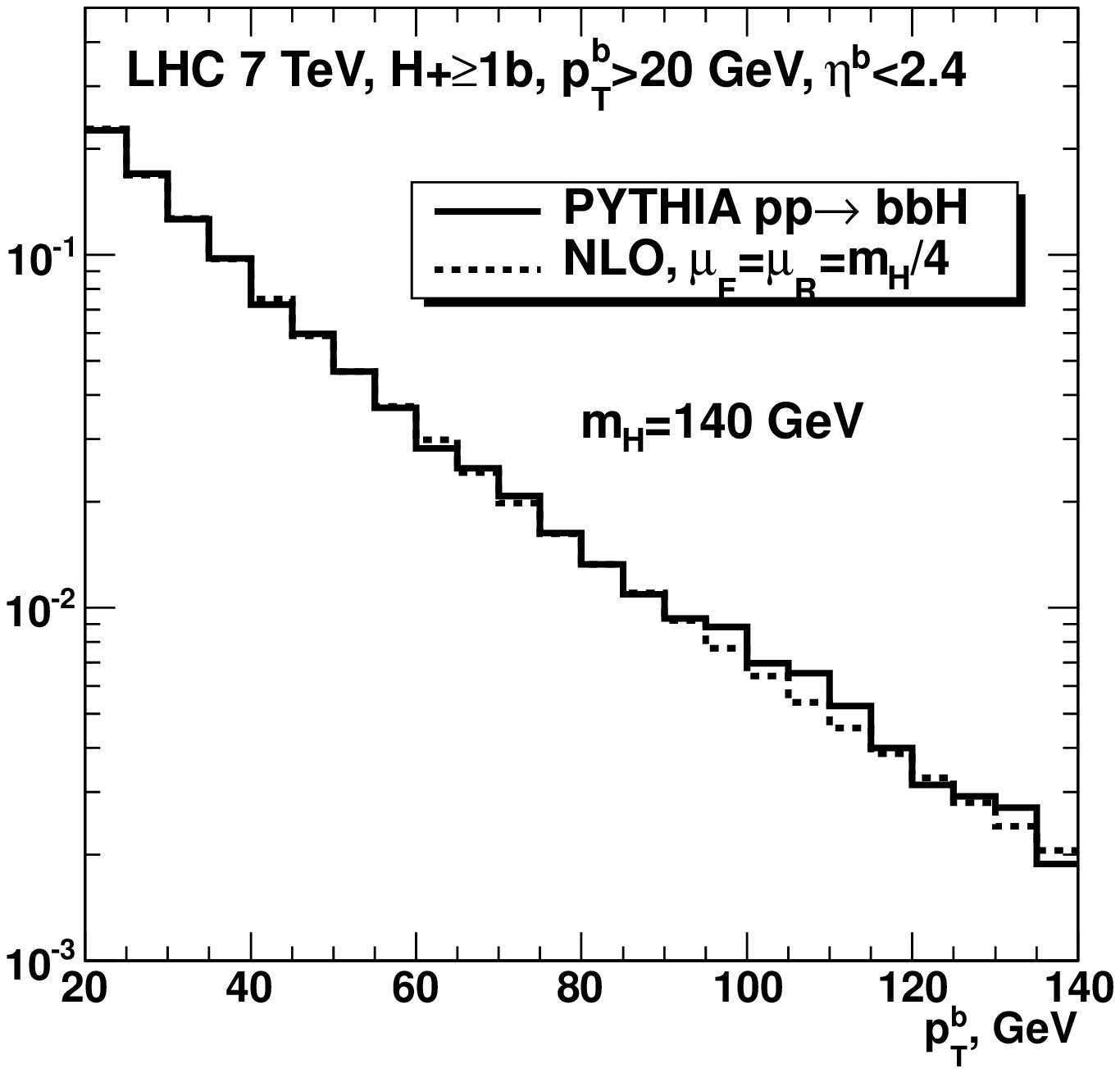} \hfill
\includegraphics[width=0.45\textwidth]{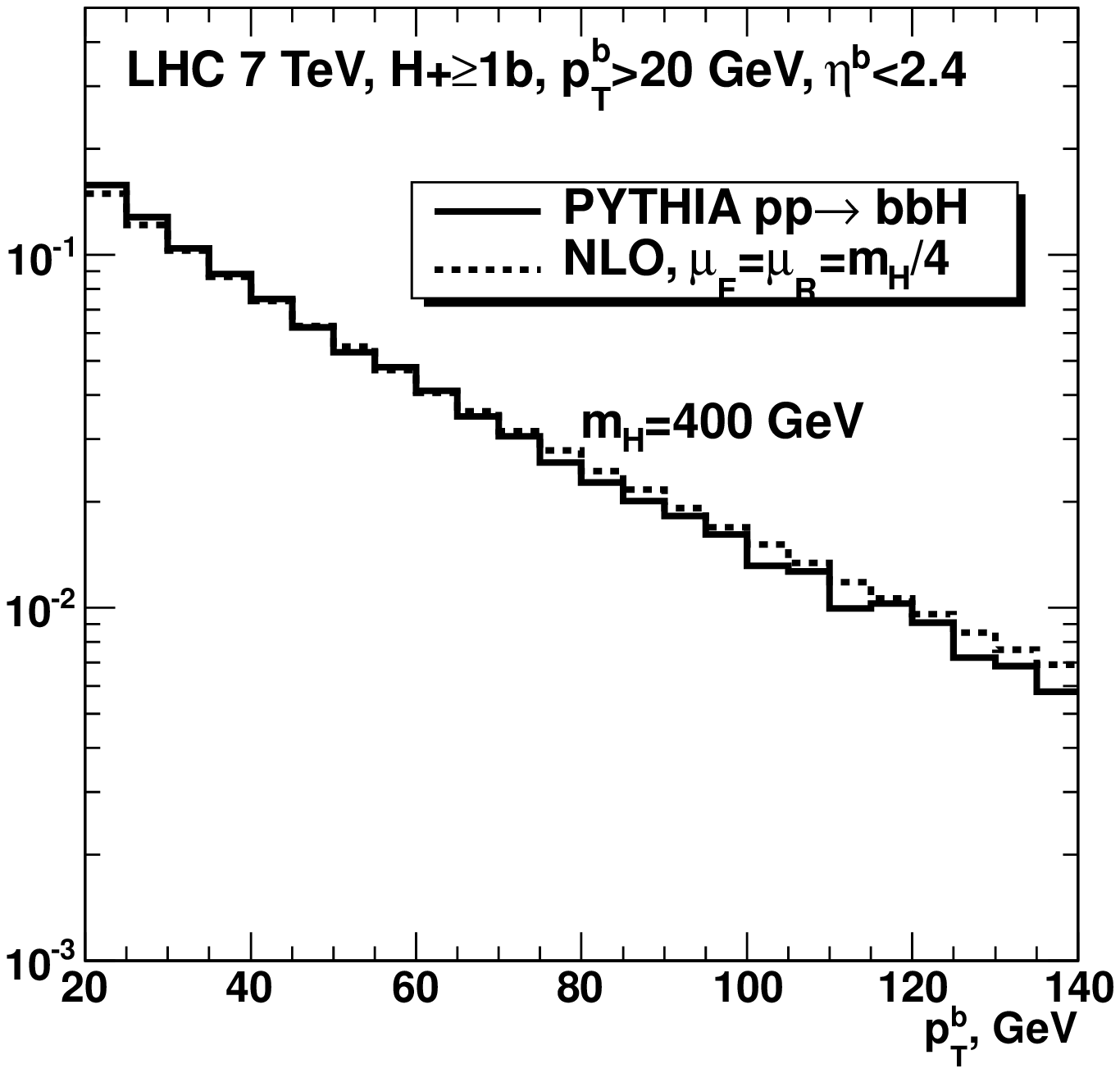}
\caption{The $\pT$ distributions of the leading $\pT$ b-jet normalised on unity for $\MH=140\UGeV$ (left) and 
$\MH=400\UGeV$ 
         (right) obtained with {\sc PYTHIA} (solid lines) and with NLO calculations (dashed lines).}
\label{fig:bbHfig2}
\end{center}
\end{figure}

We also compare the $\pT$ and rapidity distributions of the Higgs boson with at least one b-jet in the acceptance.

\clearpage

\newpage
\providecommand{\MHpm}{M_{\PSHpm}}
\providecommand{\lsim}
{\;\raisebox{-.3em}{$\stackrel{\displaystyle <}{\sim}$}\;}
\providecommand{\gsim}
{\;\raisebox{-.3em}{$\stackrel{\displaystyle >}{\sim}$}\;}
\providecommand{\mhmaxx}{{\ensuremath{m_{\rm h}^{\rm max}}}~}
\providecommand{\eqn}[1]{Eq.\,(\ref{#1})}
\providecommand{\fig}[1]{Fig.\,\ref{#1}}

\section{Charged-Higgs-boson production and decay\footnote{%
    M.~Flechl, S.~Heinemeyer, M.~Kr\"amer, S.~Lehti (eds.);
    M. Hauru, M. Spira and M. Ubiali.}}

Many extensions of the Standard Model, in particular supersymmetric
theories, require two Higgs doublets leading to five physical scalar
Higgs bosons, including two (mass-degenerate) charged particles $\PSHpm$.
The discovery of a charged Higgs boson would provide
unambiguous evidence for an extended Higgs sector beyond the Standard
Model. Searches at LEP have set a limit $\MHpm > 79.3$\UGeV\
on the mass of a charged Higgs boson in a general two-Higgs-doublet
model~\cite{Heister:2002ev}. 
One usually distinguishes a ``light charged Higgs'', $\MHpm < \Mt$, and a
``heavy charged Higgs'', $\MHpm > \Mt$. 
Within the MSSM, the charged Higgs boson
mass is constrained by the pseudoscalar Higgs mass and the $\PW$-boson
mass through $\MHpm^2 = \MA^2 + \MW^2$ at
tree level, with only moderate higher-order
corrections~\cite{Gunion:1988pc,Brignole:1991wp,Diaz:1991ki,Frank:2006yh}. A
mass limit on the MSSM charged Higgs boson can thus 
be derived from the limit on the pseudoscalar Higgs boson,
$\MA > 93.4$\UGeV~\cite{Schael:2006cr}, resulting in
$\MHpm \gsim 120$\UGeV. At the Tevatron, searches for light charged
Higgs bosons in top-quark decays $\PQt \to \PQb \PSHpm$~\cite{Aaltonen:2009ke,:2009zh} 
have placed some constraints
on the MSSM parameter space, but do not provide any further generic
bounds on $\MHpm$.

There are two main mechanisms for charged Higgs 
boson production at
the LHC:
\begin{displaymath}
\begin{array}{lcl}
  \mbox{top-quark decay:} & \PQt \to \PQb \PSHpm{\rm + X} &\; {\rm if}\;\;
  \MHpm \lsim m_{\rm t}\,, \\
  \mbox{associated production:} & \Pp\Pp \to \PQt \PQb\PSHpm{\rm + X}
  &\; {\rm if}\;\; \MHpm \gsim \Mt\,.
\end{array}
\end{displaymath}
The first process is dominant for the light charged Higgs, while the second
process dominates for a heavy charged Higgs boson.
Alternative production mechanisms like quark--antiquark annihilation
$\Pq\bar \Pq'\to \PSHpm$ and $\PSHpm+\mathrm{jet}$ production~\cite{Dittmaier:2007uw},
associated $\PSHpm \PWmp$ production~\cite{Eriksson:2006yt}, or
Higgs pair production~\cite{Krause:1997rc,Alves:2005kr,Brein:1999sy} have
suppressed rates, and it is not yet clear whether a signal could be
established in any of those channels.
Some of the above production processes may,
however, be enhanced in models with non-minimal flavour violation.
As the cross section for light-charged-Higgs-boson production are significantly 
higher than for heavy charged Higgs bosons, early searches at the LHC and also 
this section focuses mainly on those. All 
results shown below are for the \mhmaxx\ scenario of the MSSM.


\subsection{Light charged Higgs boson}

To estimate the cross section for events with charged Higgs bosons in 
top-quark pair production, the following ingredients are needed: The top-quark
pair production cross section, the  
branching ratio BR($\PQt \to \PQb \PSHp$) and the light-charged-Higgs-boson
decay branching ratios. 
Complete scans of the ($M_{\PSHpm},\tan\beta)$
plane for $\sqrt{s}=7 \UTeV$ are available in electronic format~\footnote{{\tt https://twiki.cern.ch/twiki/bin/view/LHCPhysics/MSSMCharged}}.

\subsubsection{Top-quark pair production cross section}

The $\ttbar$ production cross section at $\sqrt{s}=7\UTeV$ is predicted to be 
$165^{+4}_{-9}$(scale)$^{+7}_{-7}$(PDF)~pb by approximate NNLO
calculations~\cite{Moch:2008ai,Langenfeld:2009tc}  
recommended by the ATLAS Top Working Group~\cite{Aad:2011yb}. The scale
uncertainty is obtained as the ``envelope'' from a variation of the
renormalisation scale $\muR$ and the factorisation scale $\muF$ 
from $0.5$~to~$2$ times $\Mt$ (with $0.5 <   \muF/\muR <   2$). 
The PDF uncertainty, obtained using MSTW2008~\cite{Martin:2009iq}, is taken
at the $68\%$~C.L.; the two uncertainties should be added linearly.

\subsubsection{Top-quark decays via a charged Higgs boson}
\label{sec:chiggs_top_quark_decays}

The decay width calculation of the top quark to
a light charged Higgs boson is compared for two different programs, {\sc
FeynHiggs}, 
version\,2.8.5~\cite{Heinemeyer:1998yj,Heinemeyer:1998np,Degrassi:2002fi,Frank:2006yh}, and 
{\sc HDECAY}, version 4.43~\cite{Spira:1996if,hdecay2}. The
\mhmaxx\ benchmark scenario is used~\cite{Carena:2002qg},
which (in the on-shell scheme) is defined
in \eqn{YRHXS_MSSM_neutral_eq:mhmax}. 
We slightly deviate from the original definition and use $\MHpm$ instead of
$\MA$ as input parameter. Furthermore, also the $\mu$ parameter
is varied with values $\pm 1000, \pm 200$\UGeV~\cite{Carena:2005ek}.
The Standard Model parameters are taken as given in
\Bref{Dittmaier:2011ti}, Appendix~A.

\medskip
The {\sc FeynHiggs} calculation is based on the evaluation of
$\Gamma(\PQt \to \PWp \PQb)$ and $\Gamma(\PQt \to \PSHp \PQb)$. The former is
calculated at NLO according to \Bref{Campbell:2004ch}. The decay
to the charged Higgs boson and the bottom quark uses $\Mt(\Mt)$ and
$\Mb(\Mt)$ in the Yukawa coupling, where the
latter receives the additional correction factor $1/(1 +
\Delta_{\PQb})$. 
The numerical results presented
here are based on the evaluation of $\Delta_{\PQb}$ in
\Bref{Hofer:2009xb}.
Furthermore additional QCD corrections taken from
\Bref{Carena:1999py} are included, see also \Bref{Czarnecki:1992ig}.

The {\sc HDECAY} calculation is based on the evaluation of
$\Gamma(\PQt \to \PWp \PQb)$ and $\Gamma(\PQt \to \PSHp \PQb)$.
The decays were evaluated including the full NLO QCD corrections (including
bottom mass effects)~\cite{Campbell:2004ch} (and references therein).
The top and (kinematical) bottom masses are
taken as the pole masses while the bottom mass of the Yukawa coupling is taken
as running $\overline{{\rm MS}}$ mass at the scale of the top mass.
SUSY QCD and electroweak corrections are approximated via $\Delta_{\PQb}$ based 
on
\Brefs{Hall:1993gn,Hempfling:1993kv,Carena:1994bv,Pierce:1996zz,Carena:1999py,Guasch:2003cv,Noth:2008tw,Noth:2010jy,Mihaila:2010mp}.

\begin{figure}
  \centering
  \includegraphics[width=0.48\textwidth]{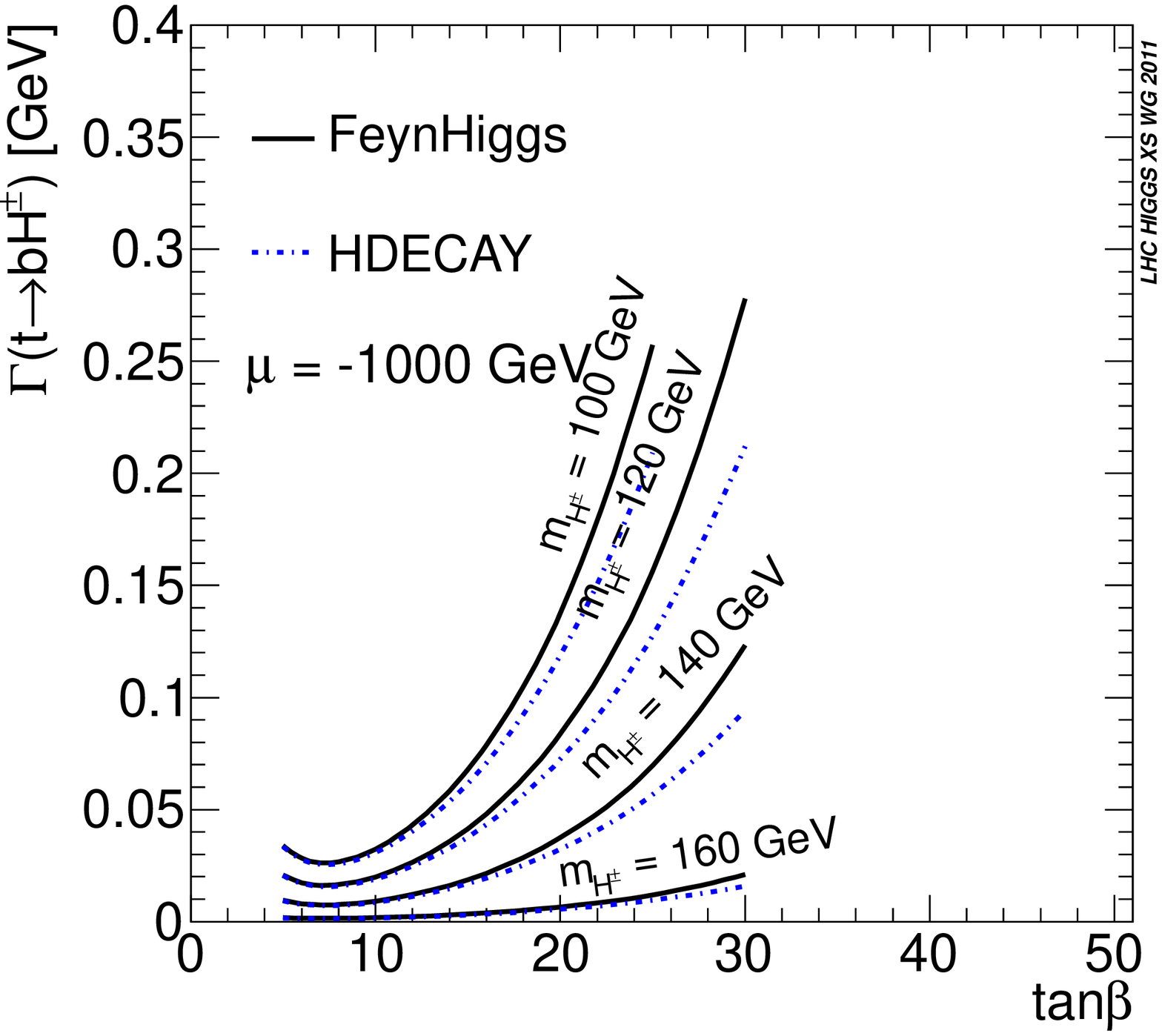} 
  \includegraphics[width=0.48\textwidth]{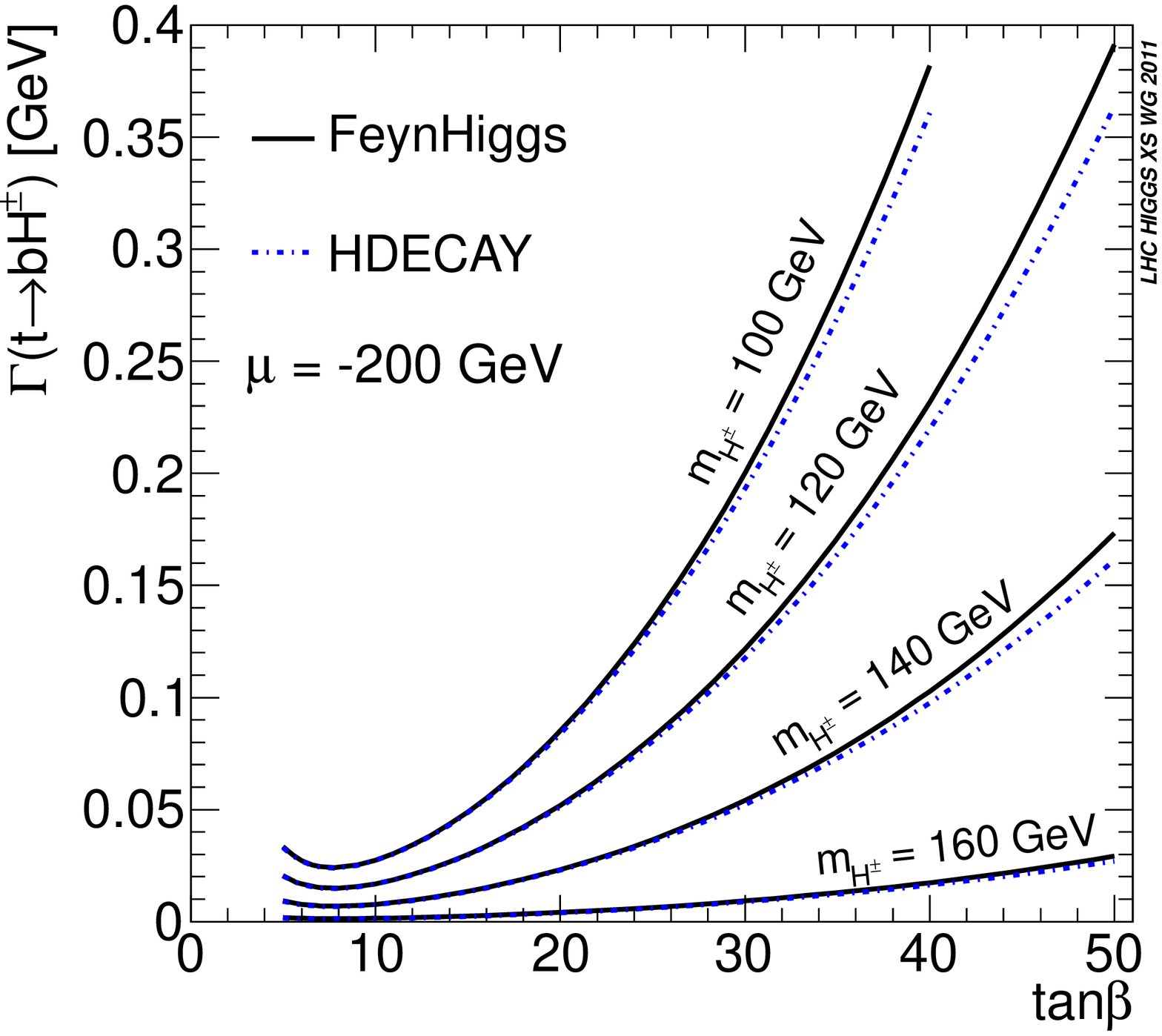} \\[-1.5em]
  \includegraphics[width=0.48\textwidth]{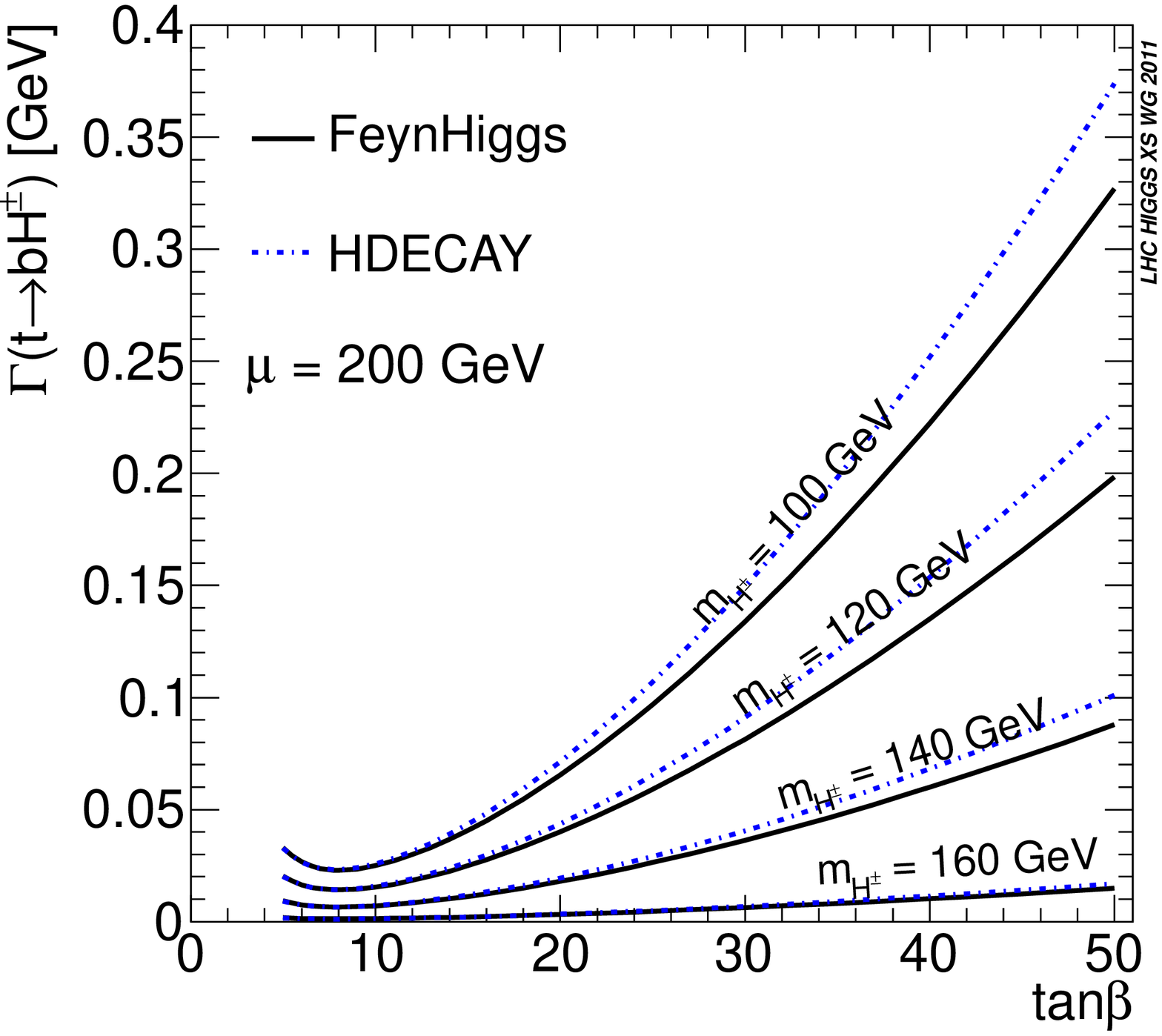} 
  \includegraphics[width=0.48\textwidth]{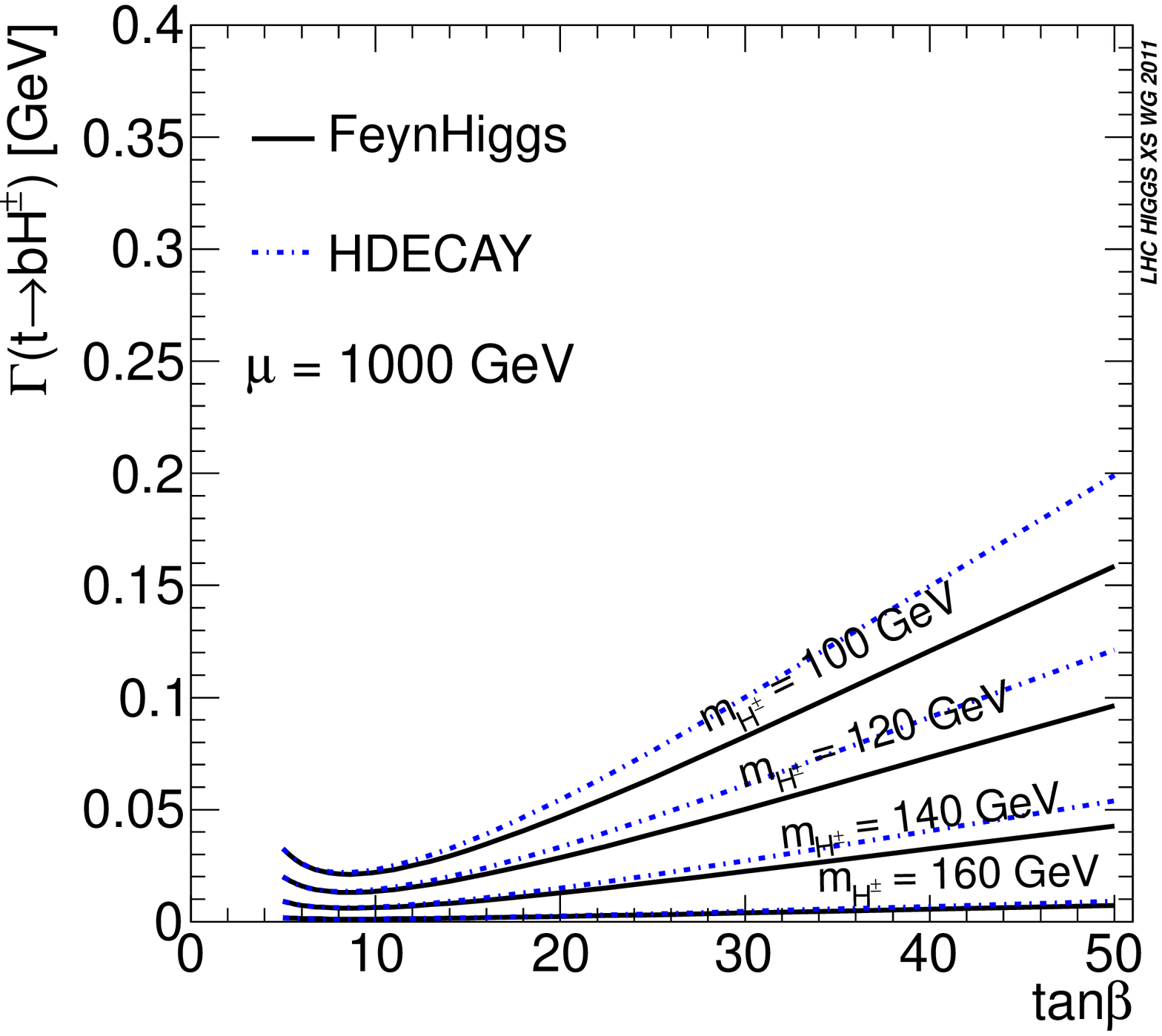} 
\vspace{-.5cm}
\caption{The decay width $\Gamma(\PQt \to \PQb \PSHpm)$ calculated
  with {\sc FeynHiggs} and {\sc HDECAY} as a function of $\tan\beta$
  for different values of $\mu$ and $\MHpm$.}
\label{fig:GammaTToBHp}
\end{figure}
%
\begin{figure}
  \centering
  \includegraphics[width=0.48\textwidth]{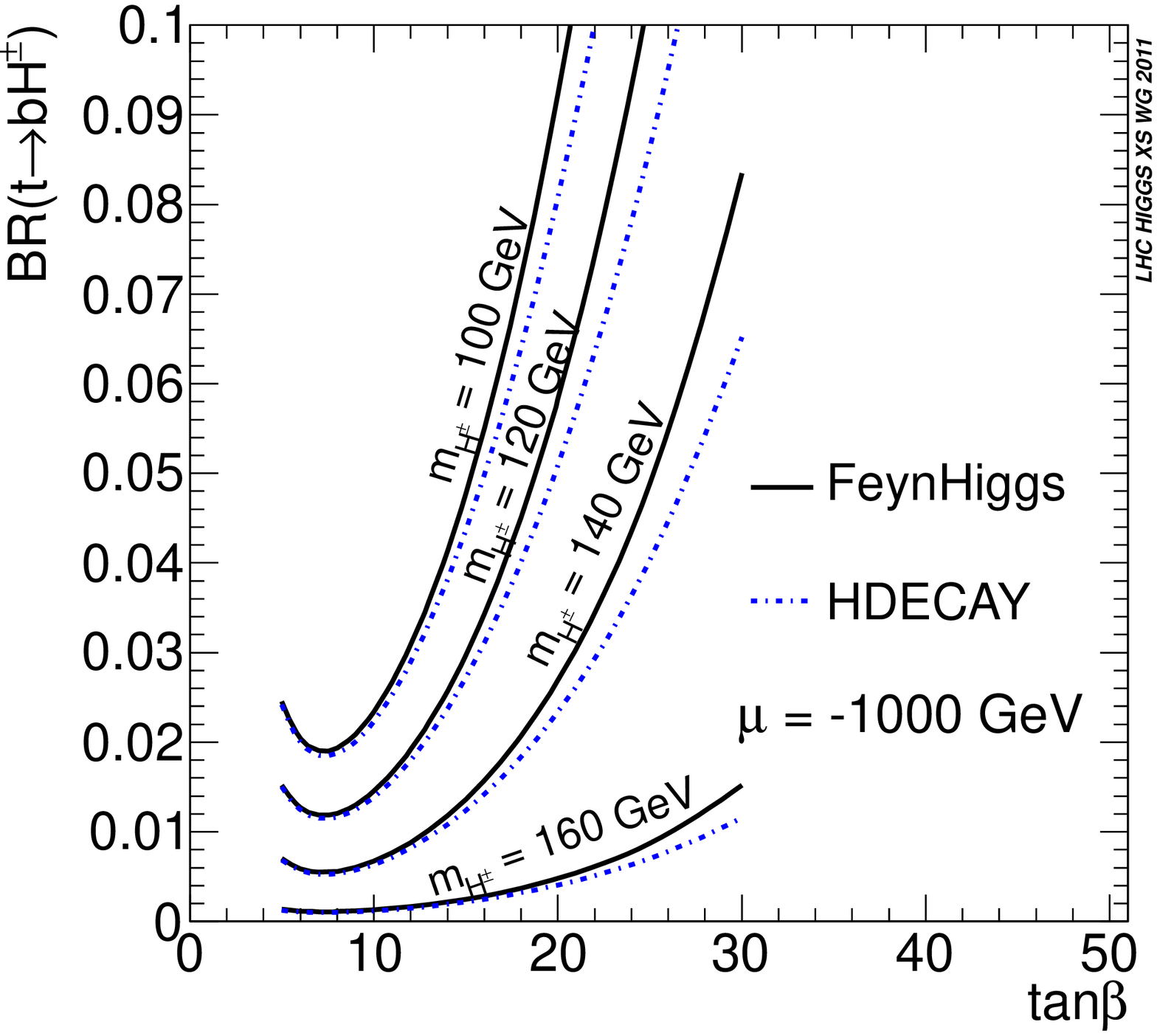} 
  \includegraphics[width=0.48\textwidth]{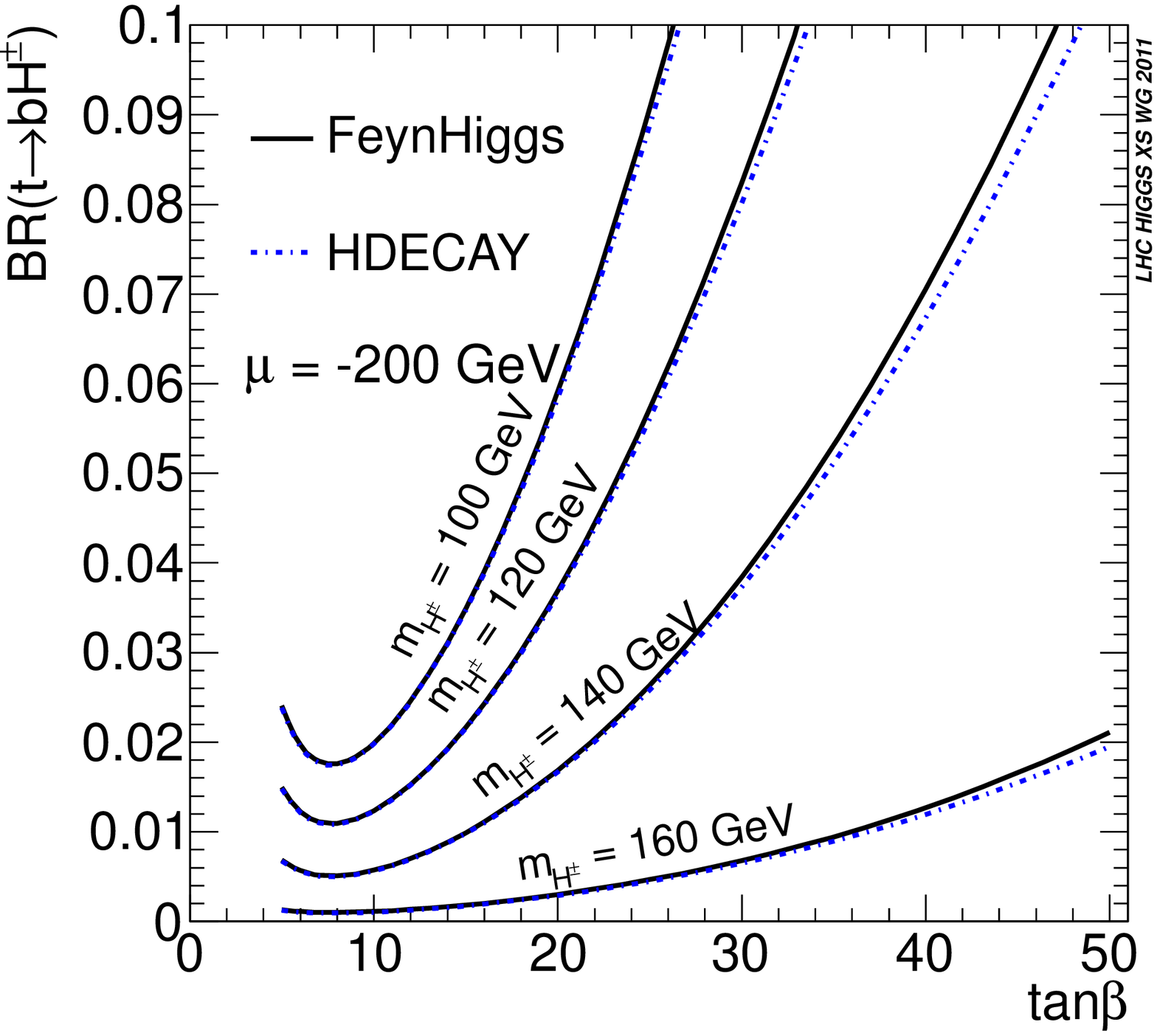} \\[-1.5em]
  \includegraphics[width=0.48\textwidth]{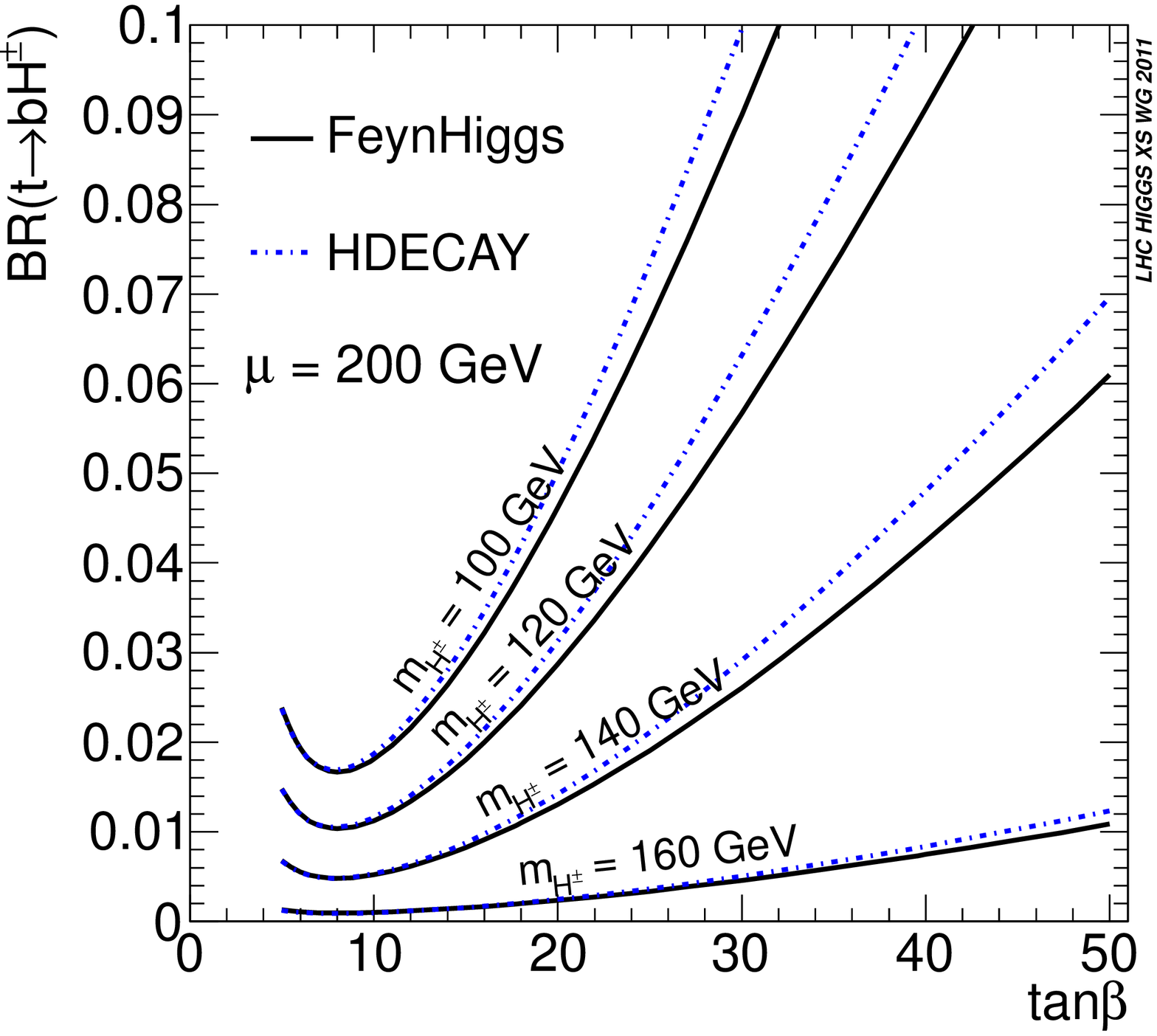} 
  \includegraphics[width=0.48\textwidth]{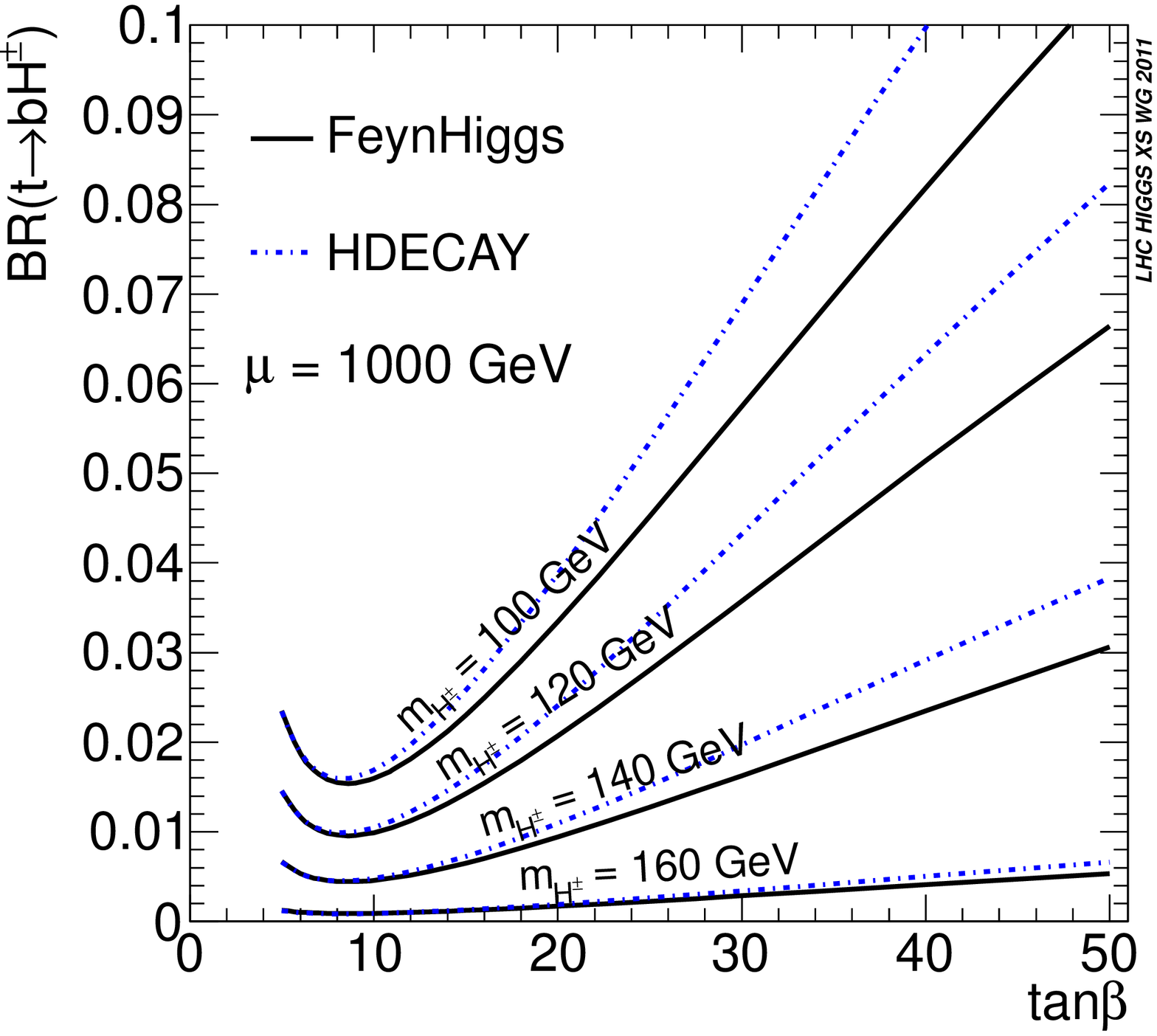} 
\vspace{-.6cm}
\caption{The branching fraction BR$(\PQt \to \PQb \PSHpm)$ calculated
  with {\sc FeynHiggs} and {\sc HDECAY} as a function of $\tan\beta$
  for different values of $\mu$ and $\MHpm$.}
\label{fig:BRTToBHp}
\vspace{.5em}
\includegraphics[height=6.8cm]{YRHXS2_chiggs/dGammatHpb_01.eps}
\hfill
\includegraphics[height=6.8cm]{YRHXS2_chiggs/sigmatt_BRtHpb_03B.eps}
\vspace*{-0.3cm}
\caption{
Left: Relative uncertainty in $\Gamma(\PQt \to \PQb \PSHpm)$ 
assuming a $3\%$ residual uncertainty in $\Delta_{\PQb}$.
Right: $\sigma_{\Pt\Pt} \cdot \mathrm{BR}(\PQt \to \PQb \PSHpm) \cdot \mathrm{BR}(\PQt \to \PQb \PWpm) \cdot 2$ including
scale and PDF uncertainties, uncertainties for missing electroweak and QCD
corrections, and $\Delta_{\PQb}$-induced uncertainties.
}
\label{fig:chiggs_gamma_unc}
\end{figure}
{\sc FeynHiggs} has been run with the selected set
of parameters. The {\sc FeynHiggs} output is then used to set the values
for the {\sc HDECAY} input parameters. The main result   
from the comparison is shown in \refFs{fig:GammaTToBHp}
and \ref{fig:BRTToBHp}. The decay width $\Gamma(\PQt \to \PWp \PQb)$ calculated
by both programs agrees very well,  
the difference being only of the order of $0.2\%$. The difference in 
$\Gamma(\PQt \to \PSHp \PQb)$ varies from negligible ($0.1\%$) to a maximum of
about $9\%$, the largest difference being at high values of $\tan\beta$.
The source of the differences is thought to come from differences in the
$\Delta_{\PQb}$ evaluation, largest at high values of $\tan\beta$, 
or from two-loop corrections to $\Delta_{\PQb}$.
A similarly good agreement is observed for the corresponding branching ratio. 

\refF{fig:chiggs_gamma_unc} (left) shows the relative uncertainty of the
total decay width $\Gamma(\PQt \to \PSHp \PQb)$, assuming a residual
uncertainty of the $\Delta_{\PQb}$ corrections of $3\%$ (which is reached after
the inclusion of the leading two-loop contributions to
$\Delta_{\PQb}$~\cite{Noth:2008tw,Noth:2010jy,Mihaila:2010mp}). 
Combining the calculations on $\ttbar$ production and decay, the 
cross section for the process $\Pp\Pp \to \ttbar \to \PW \PQb \PSHpm \PQb$ can
be predicted. The result for $\sigma_{\Pt\Pt} \cdot 
\mathrm{BR}(\PQt \to \PQb \PSHpm) \cdot \mathrm{BR}(\PQt \to \PQb \PWpm) \cdot 2$
at $\sqrt{s}=7 \UTeV$ is shown in
\refF{fig:chiggs_gamma_unc} (right), together with the dominating  
systematic uncertainties: PDF and scale uncertainties on the $\ttbar$ cross
section, $5\%$ for missing one-loop electroweak diagrams, $2\%$ for missing
two-loop QCD diagrams, and 
$\Delta_{\PQb}$-induced uncertainties. The theory uncertainties are added
linearly to the (quadratically) combined experimental uncertainties.


\subsubsection{Light-charged-Higgs-boson decay}

In the \mhmaxx\ scenario of the MSSM, the BR$(\PSHpm \to \PGt\PGn) \approx 1$
for all parameter values is still allowed by the LEP
experiments~\cite{Heister:2002ev}. 
(Only for very large values of $\MHpm$ the off-shell decay to $\Pt\PQb$ can
reach a level of up to $10\%$.)
The uncertainty on this assumption is less than $1\%$ and thus negligible
compared to other uncertainties.

The charged-Higgs-boson decay widths calculated with {\sc FeynHiggs} and 
{\sc HDECAY} in \mhmaxx\ benchmark scenario are compared
in the same manner as described in \refS{sec:chiggs_top_quark_decays}.
The total decay width of the charged Higgs boson is shown in
\refF{fig:GammaHp}. 
The decay channels
$\PH^{\pm} \to \PGt\PGn_{\PGt}$, $\PH^{\pm} \to \PA\PW$, $\PH^{\pm} \to \PQc\PQs$, 
$\PH^{\pm} \to \PH\PW$, $\PH^{\pm} \to \PGm\PGn_{\PGm}$, and $\PH^{\pm} \to \Pt\PQb$
available in both programs are studied. 
For $\PH^{\pm} \to \PGt\PGn_{\PGt}$, {\sc FeynHiggs} includes the
Higgs propagator corrections up to the two-loop level.
Concerning the latter, in {\sc HDECAY} these corrections are included in the
approximation of vanishing external momentum.
On the other hand, it 
includes the full NLO QCD corrections to charged-Higgs decays
into quarks, which are incorporated in {\sc FeynHiggs} only in the
approximation of a heavy charged Higgs boson.
The experimentally most interesting decay channel
$\PH^{\pm} \to \PGt\PGn_{\PGt}$ showed a good agreement, with {\sc HDECAY}
consistently predicting a $3.5\%$ larger decay width than {\sc FeynHiggs}, due
to the differences described above.
The result is shown in \refF{fig:GammaHpToTauNu}. A good agreement is also
found in the $\PH^{\pm} \to \PGm\PGn_{\PGm}$ channel, again with {\sc HDECAY}
predicting consistently a $\sim 3.5\%$ larger decay width than {\sc
FeynHiggs}. In the $\PH^{\pm} \to \PQc\PQs$ channel 
a notable discrepancy of $7{-}19\%$ is found. The result of the comparison in the 
$\PH^{\pm} \to \PQc\PQs$ channel is shown in \refF{fig:GammaHpToCS}.
%
The differences in $\PH^{\pm} \to \PQc\PQs$ may be attributed to 
the QCD corrections implemented in {\sc FeynHiggs}, which are valid only in
the limit of large charged-Higgs masses (in comparison to the quark masses),
whereas 
in {\sc HDECAY} they are more complete. 
This channel can only play a significant role for very low values of
$\tan\beta$ and is numerically negligible within the \mhmaxx\ scenario.
%
\begin{figure}
  \centering
\includegraphics[width=0.48\textwidth]{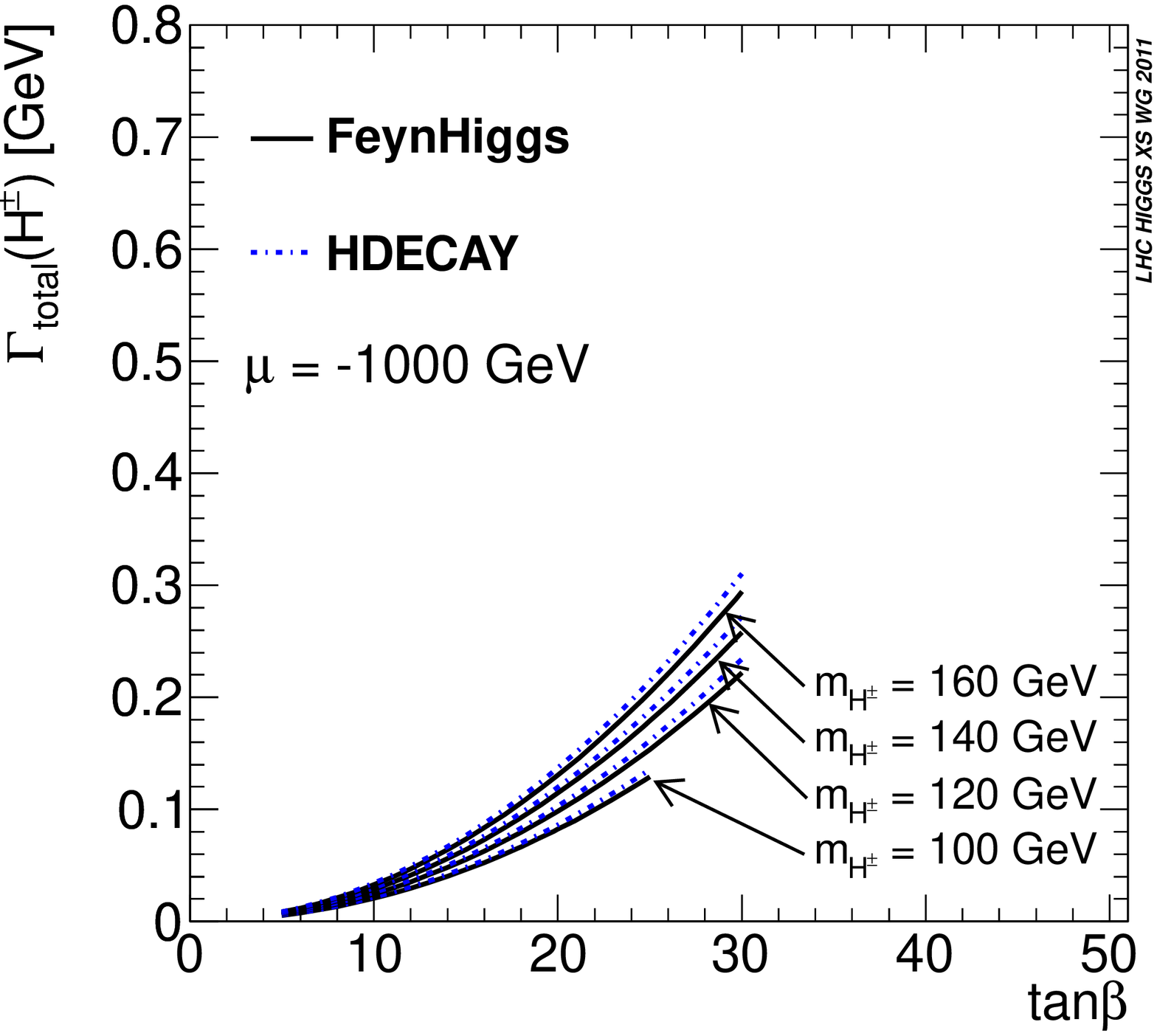} 
\includegraphics[width=0.48\textwidth]{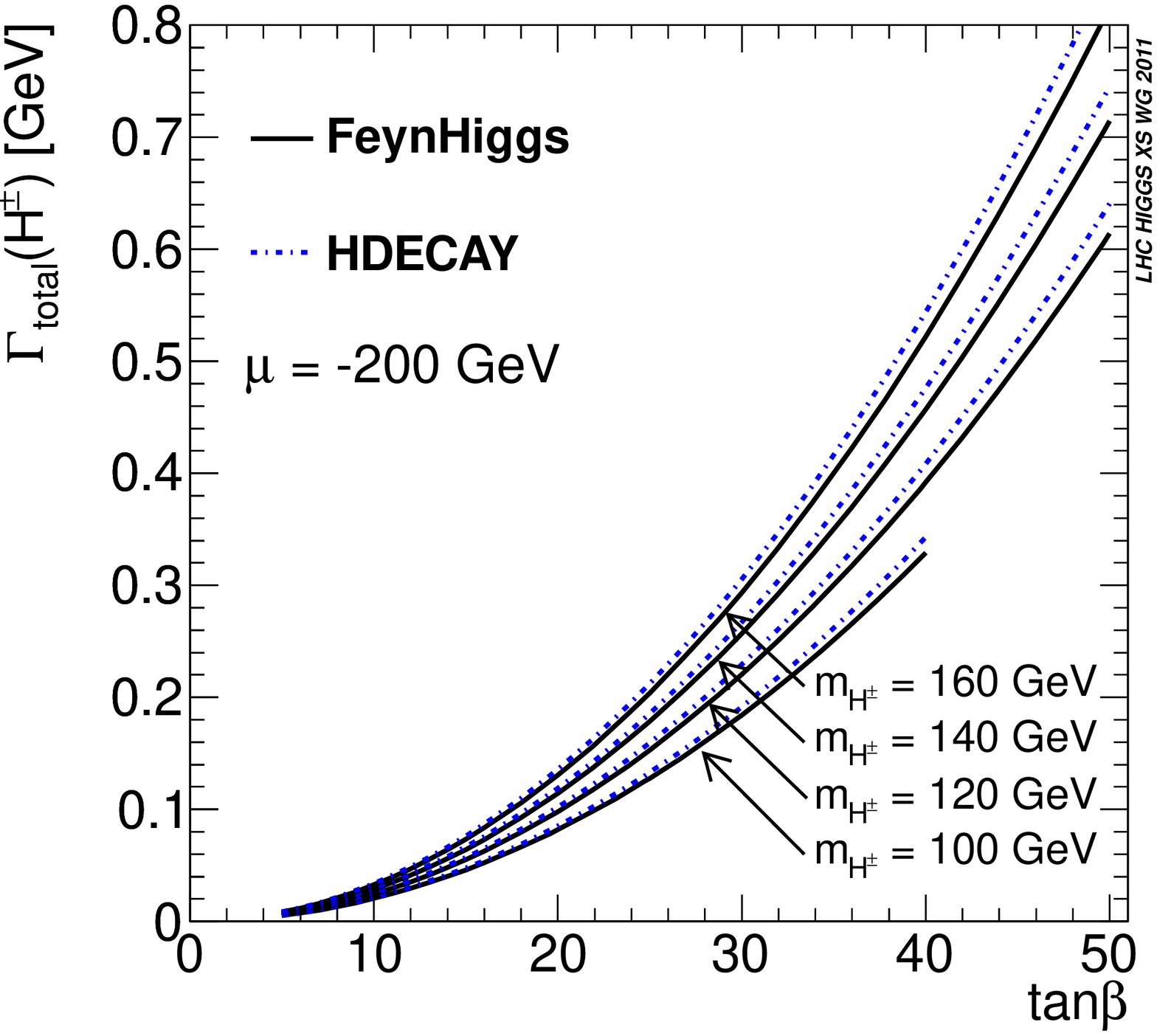} \\[-1.5em]
\includegraphics[width=0.48\textwidth]{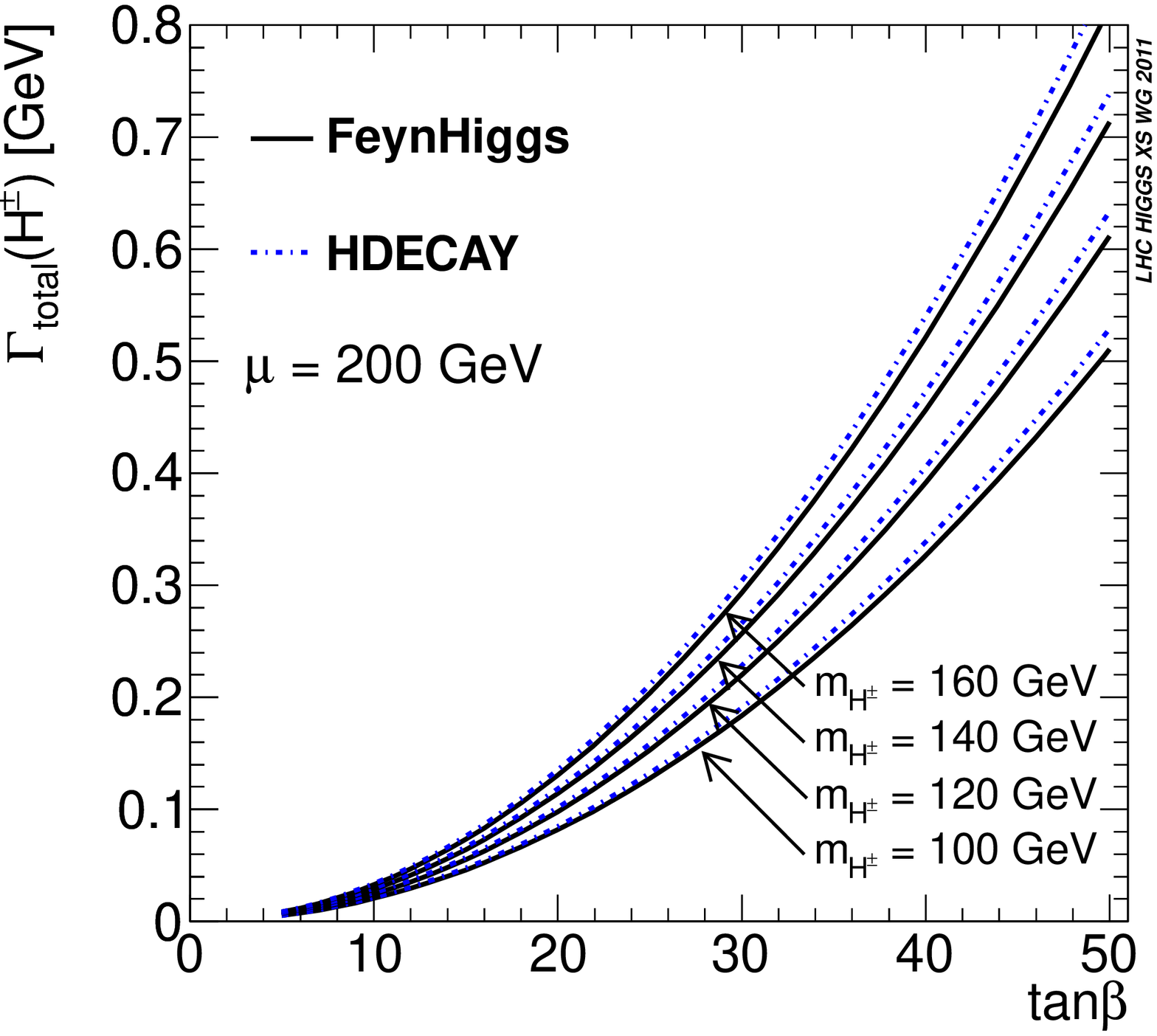} 
\includegraphics[width=0.48\textwidth]{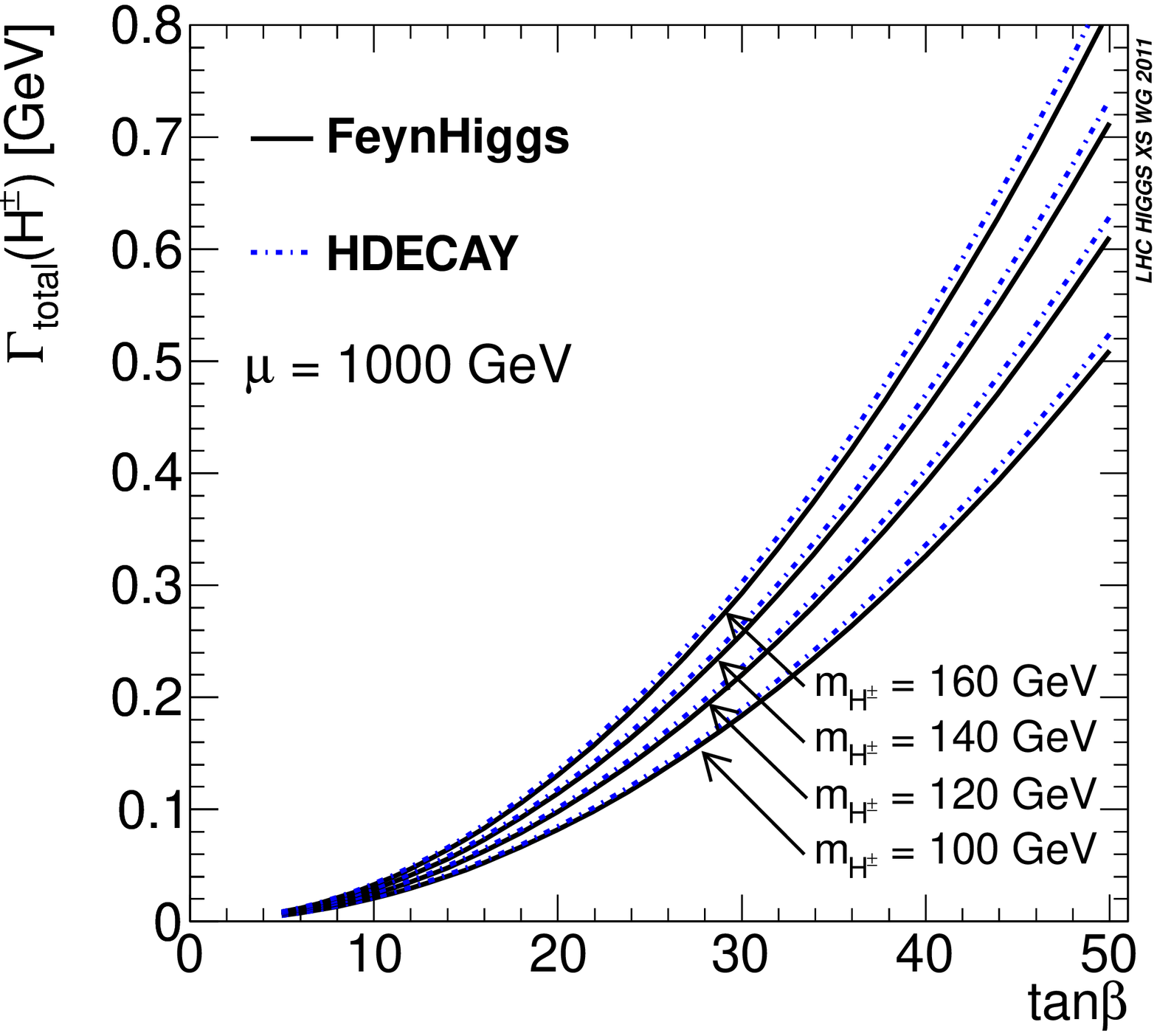}
\vspace{-.5cm}
\caption{The decay width of $\PSHpm$ calculated with {\sc FeynHiggs} and {\sc HDECAY} as a function
of $\tan\beta$ for different values of $\mu$ and $\MHpm$.}
\label{fig:GammaHp}
\end{figure}
\begin{figure}
  \centering
  \includegraphics[width=0.48\textwidth]{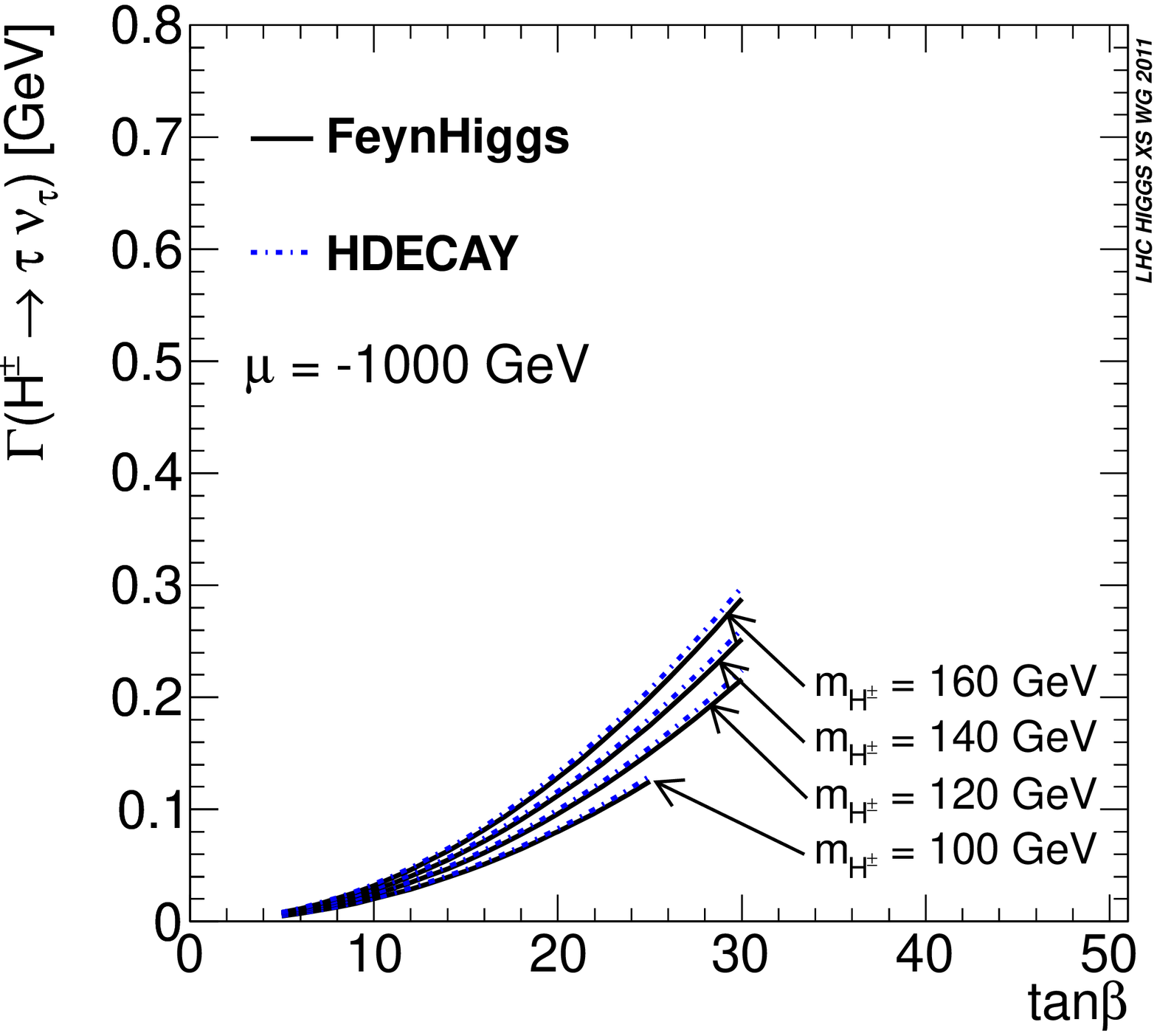} 
  \includegraphics[width=0.48\textwidth]{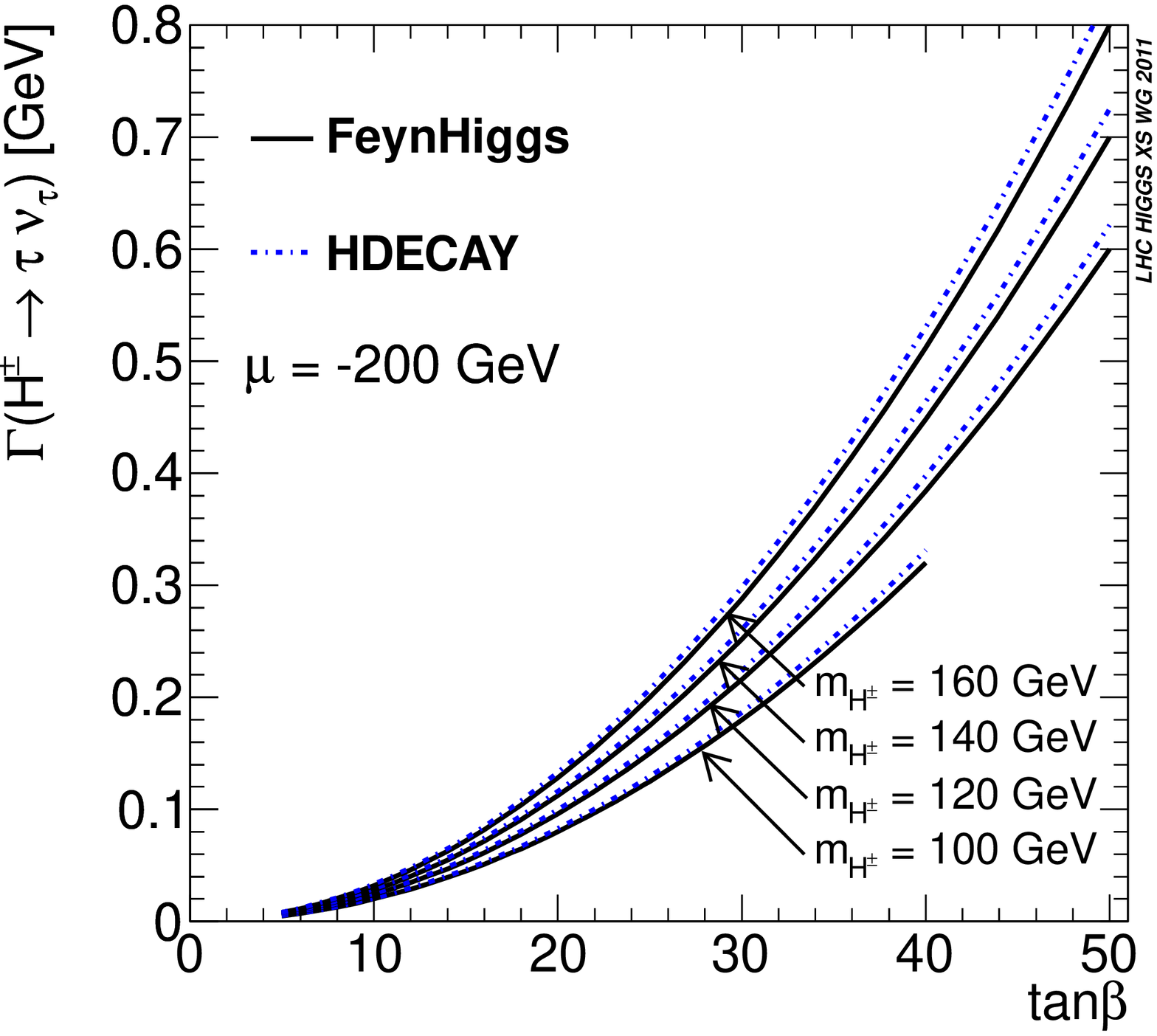} \\[-1.5em]
  \includegraphics[width=0.48\textwidth]{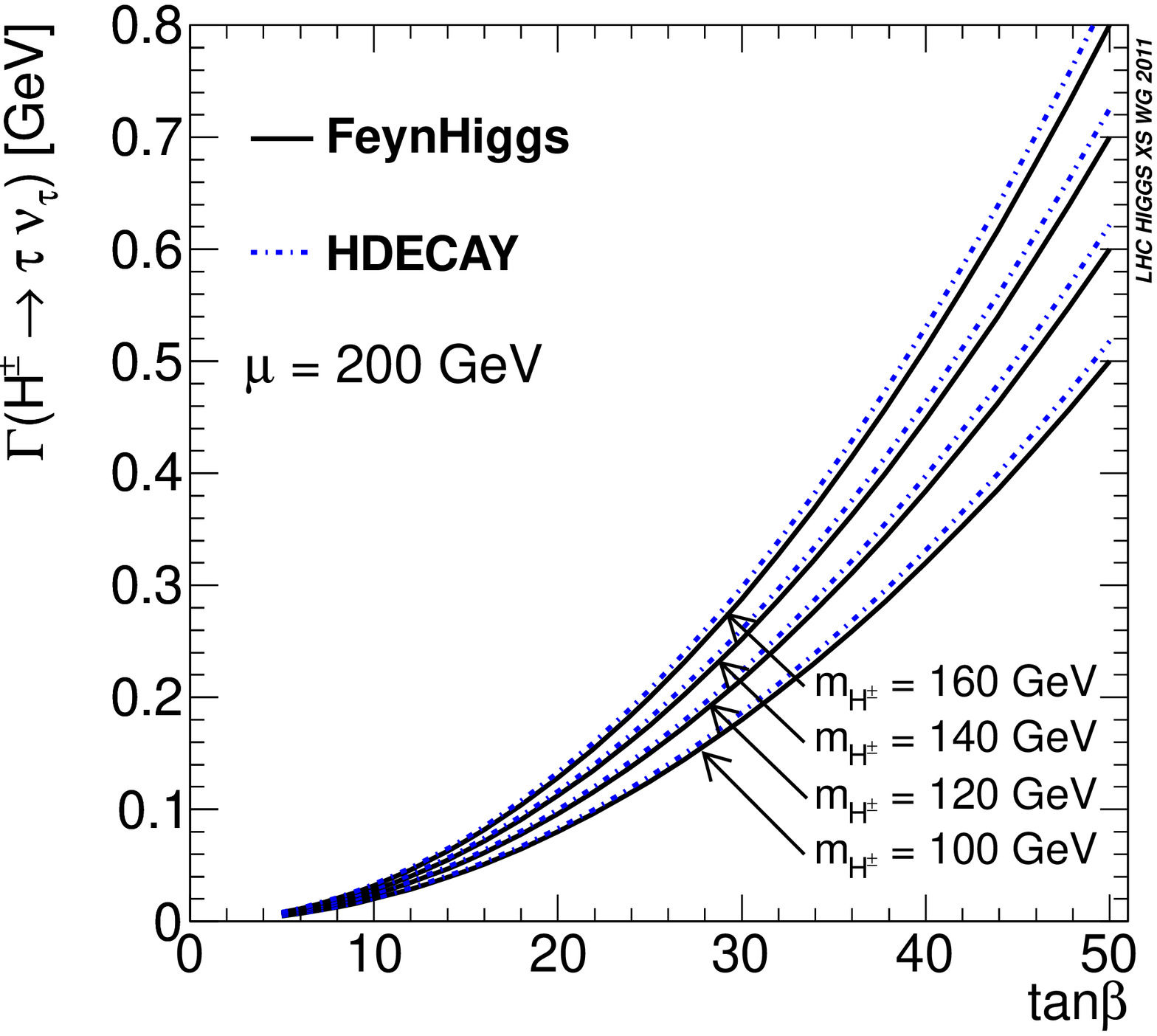} 
  \includegraphics[width=0.48\textwidth]{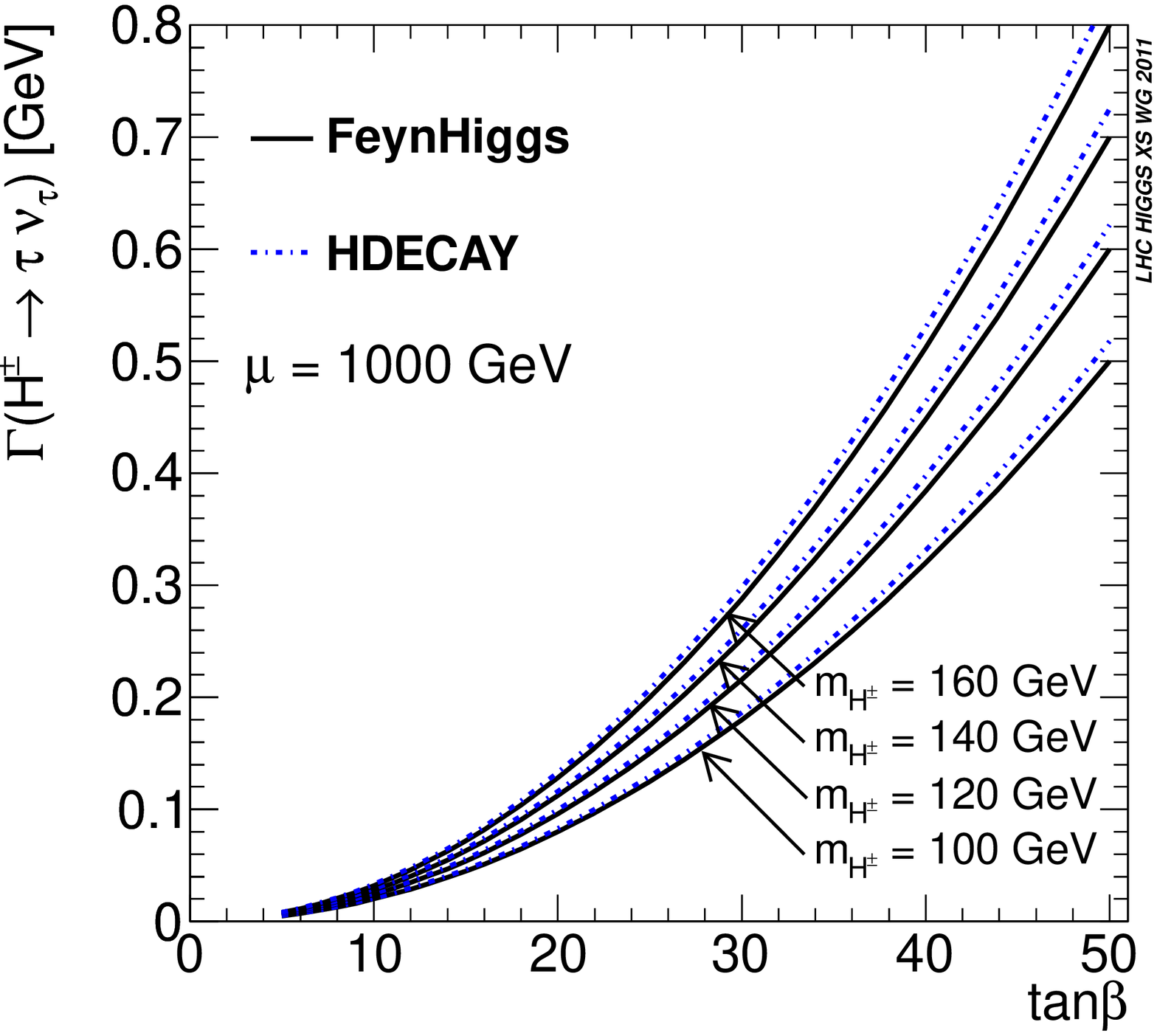} 
\vspace{-.5cm}
\caption{The decay width $\Gamma(\PH^{\pm} \to \PGt\PGn_{\PGt})$ calculated with {\sc FeynHiggs} and
{\sc HDECAY} as a function of $\tan\beta$ for different values of $\mu$ and $M_{\PH^{\pm}}$.}
\label{fig:GammaHpToTauNu}
\end{figure}
\begin{figure}
  \centering
  \includegraphics[width=0.48\textwidth]{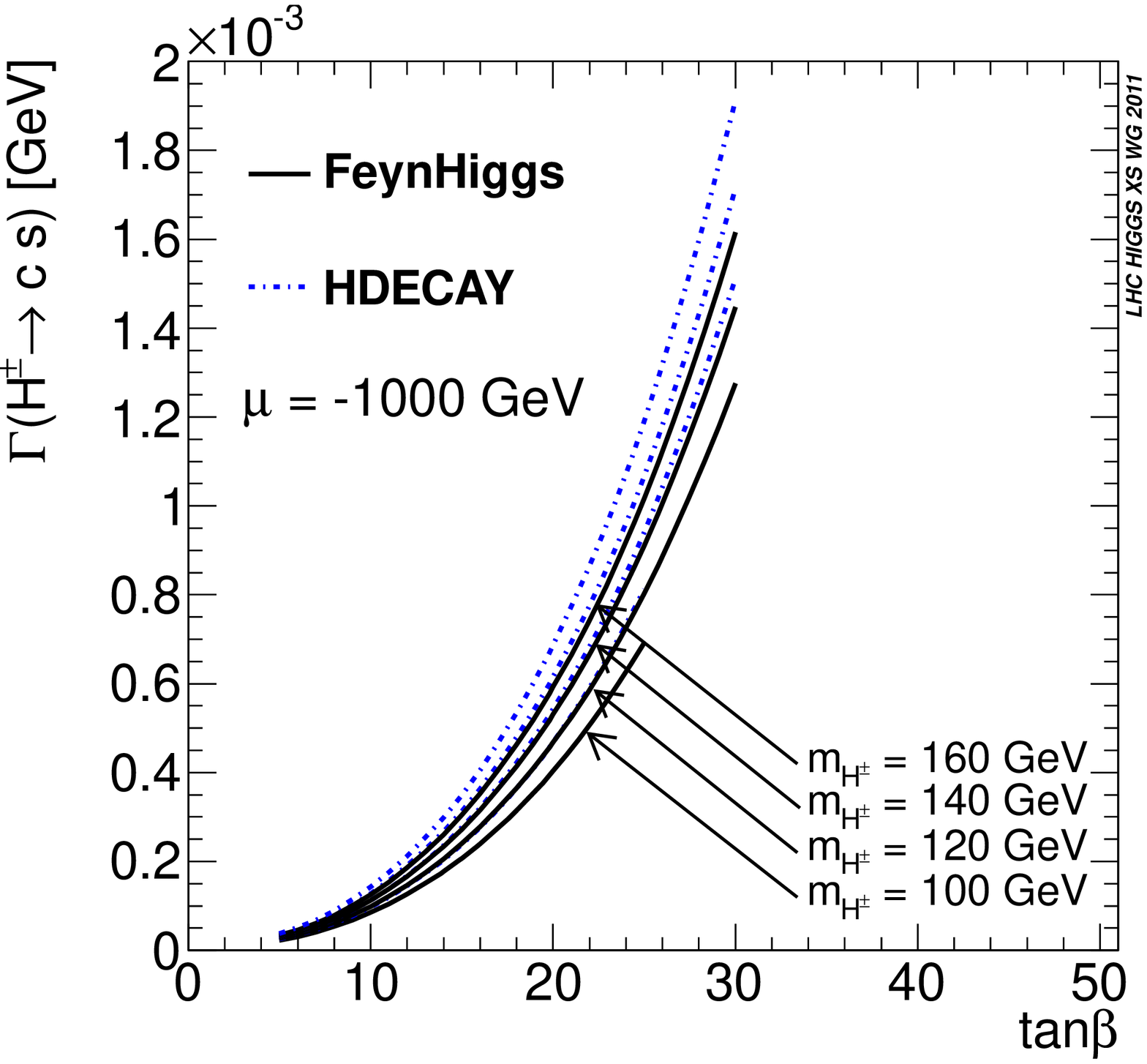} 
  \includegraphics[width=0.48\textwidth]{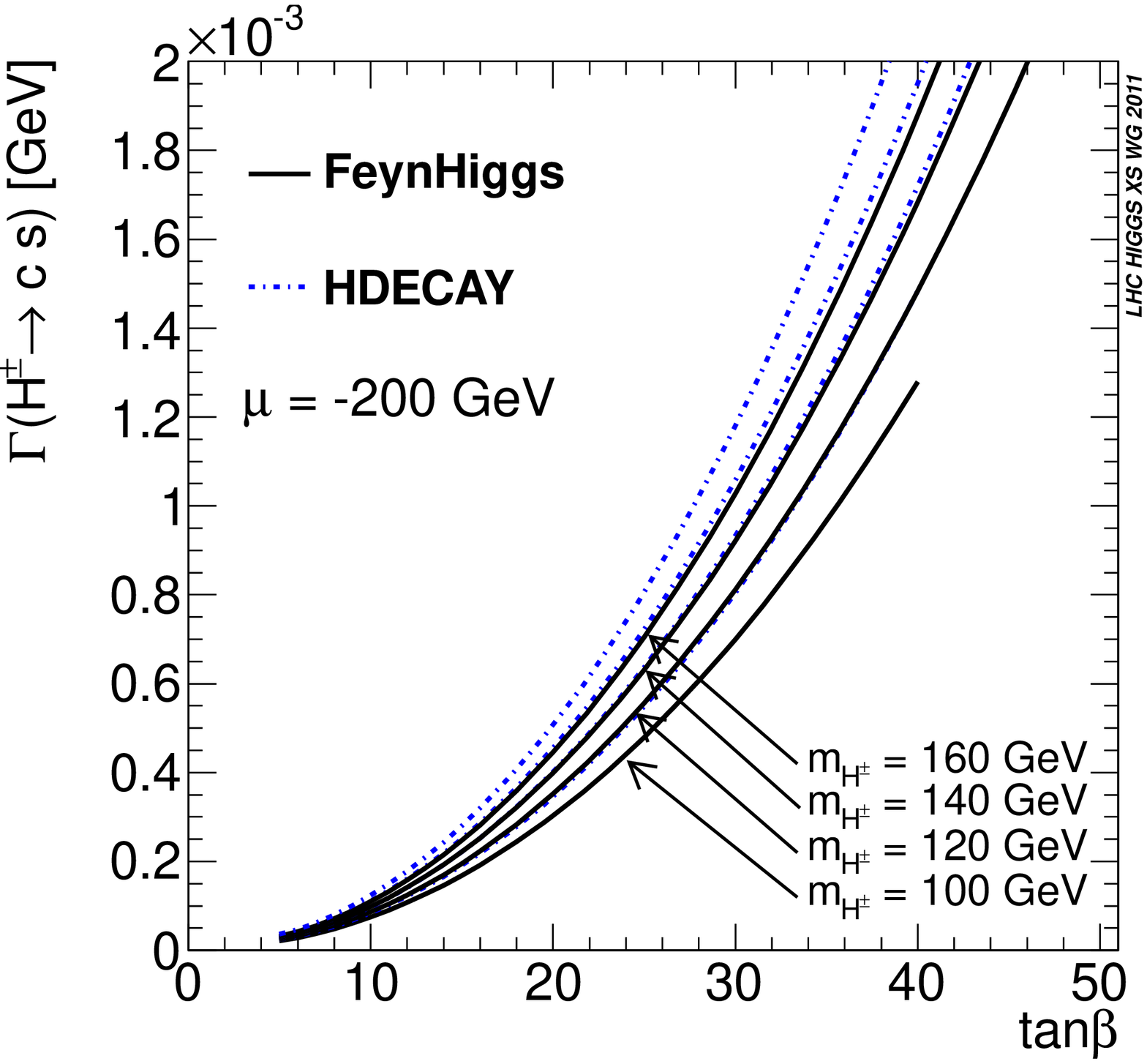} \\[-1.5em]
  \includegraphics[width=0.48\textwidth]{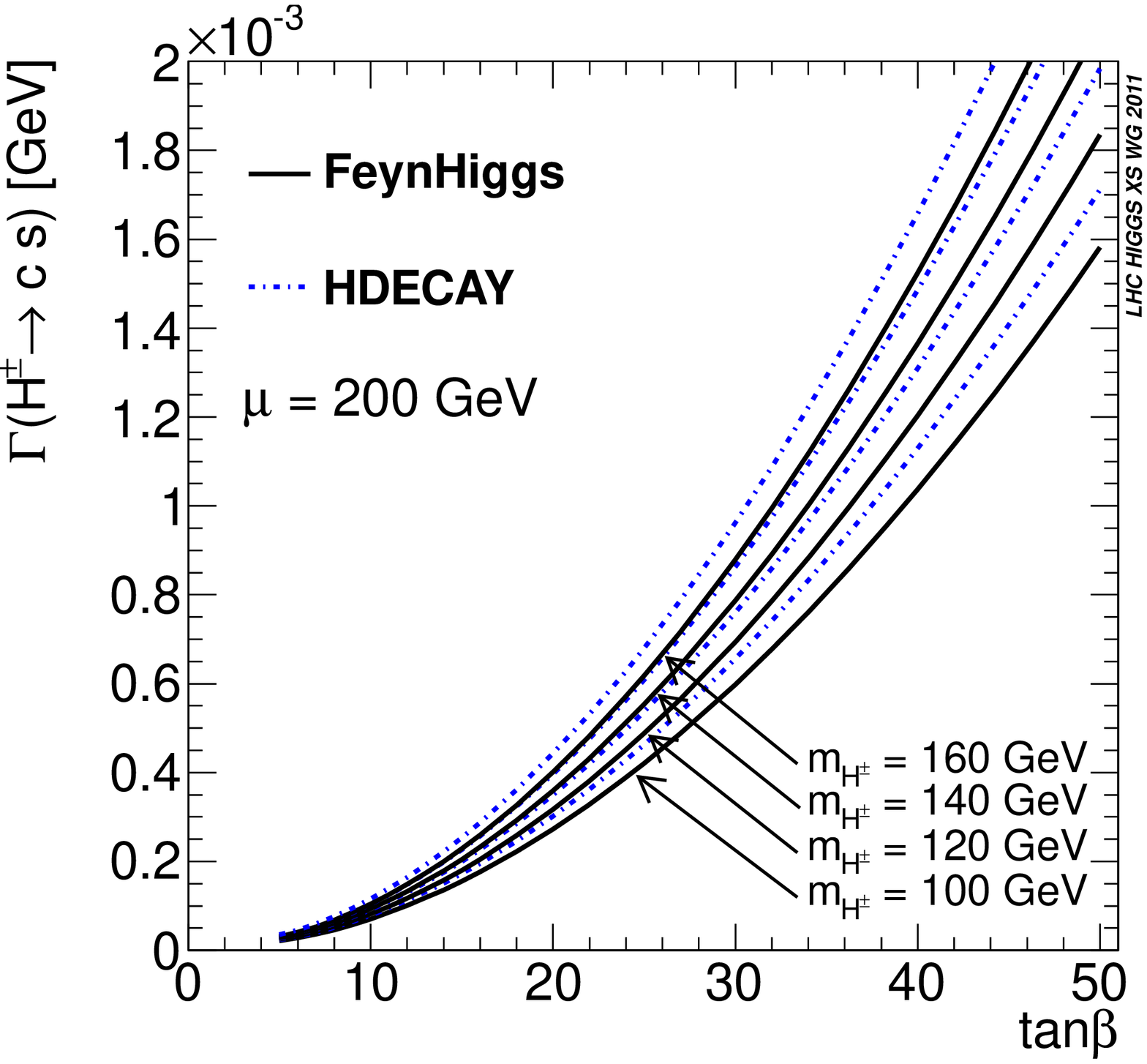} 
  \includegraphics[width=0.48\textwidth]{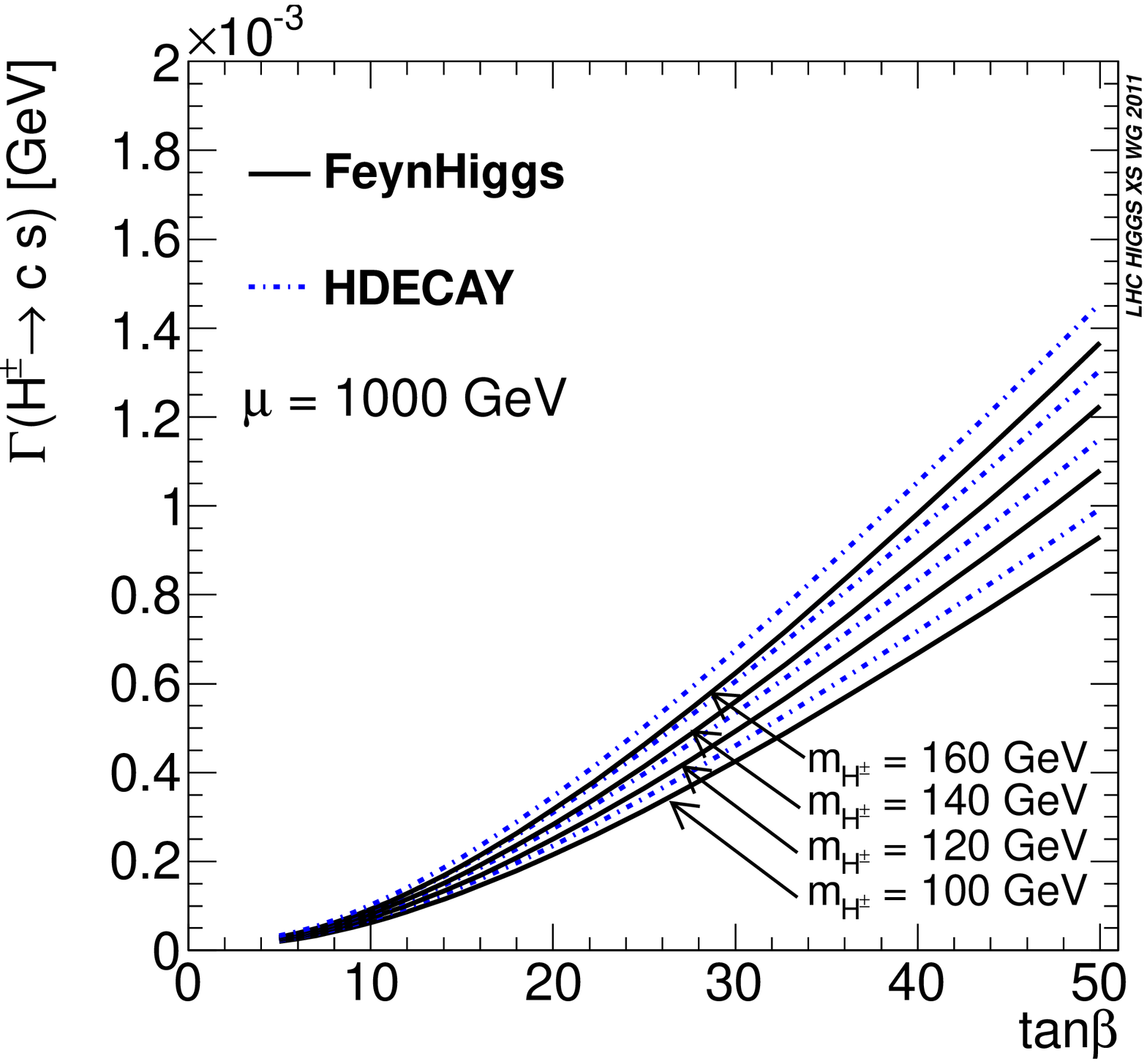} 
\vspace{-.5cm}
\caption{The decay width $\Gamma(\PH^{\pm} \to \PQc\PQs)$ calculated
  with {\sc FeynHiggs} and {\sc HDECAY} as a function of $\tan\beta$
  for different values of $\mu$ and $M_{\PH^{\pm}}$. A discrepancy
  of $7{-}19\%$ is observed.}
\label{fig:GammaHpToCS}
\end{figure}

\subsection{Heavy charged Higgs boson}
For heavy charged Higgs bosons,  $\MHpm \gsim \Mt$,
associated production $\Pp\Pp \to {\Pt}\PQb \PSHpm{\rm + X}$ is the
dominant production mode. 
Two different formalisms can be employed to calculate the cross
section for associated ${\Pt}\PQb\PSHpm$ production.  In the
four-flavour scheme (4FS) with no $\PQb$ quarks in the initial state,
the lowest-order QCD production processes are gluon--gluon fusion and
quark--antiquark annihilation, $\Pg\Pg \to {\Pt}\PQb\PSHpm$ and
$\Pq\bar \Pq \to {\Pt}\PQb\PSHpm$, respectively. Potentially
large logarithms $\propto \ln(\muF/\Mb)$, which arise from
the splitting of incoming gluons into nearly collinear $\PQb\bar \PQb$
pairs, can be summed to all orders in perturbation theory by
introducing bottom parton densities. This defines the five-flavour
scheme (5FS)~\cite{Barnett:1987jw}. The use of bottom distribution
functions is based on the approximation that the outgoing $\PQb$ quark
is at small transverse momentum and massless, and the virtual $\PQb$
quark is quasi on shell. In this scheme, the leading-order (LO)
process for the inclusive $\Pt\PQb\PSHpm$ cross section is
gluon--bottom fusion, $\Pg \PQb \to \Pt \PSHpm$.  The next-to-leading
order (NLO) cross section in the 5FS includes ${\cal
  O}(\alphas)$ corrections to $\Pg \PQb \to \Pt  \PSHpm$
and the tree-level processes $\Pg\Pg \to \Pt\PQb\PSHpm$ and $\Pq\bar
\Pq \to \Pt\PQb\PSHpm$. To all orders in perturbation theory the four-
and five-flavour schemes are identical, but the way of ordering the
perturbative expansion is different, and the results do not match
exactly at finite order. For the inclusive production of neutral Higgs 
bosons with bottom quarks, $\Pp\Pp \to \PQb\PAQb\PH{\rm + X}$, the
four- and five-flavour scheme calculations numerically agree within
their respective uncertainties, once higher-order QCD corrections are
taken into account~\cite{Dittmaier:2003ej, Campbell:2004pu,
  Dawson:2005vi, Buttar:2006zd}, see \refS{sec:MSSM} of this Report.
  
There has been considerable progress recently in improving the cross-section 
predictions for the associated production of charged Higgs
bosons with heavy quarks by calculating NLO SUSY QCD and electroweak
corrections in the four- and five-flavour schemes~\cite{Zhu:2001nt,
  Gao:2002is, Plehn:2002vy, Berger:2003sm, Kidonakis:2005hc,
  Peng:2006wv, Beccaria:2009my, Kidonakis:2010ux}, and the matching of
the NLO five-flavour scheme calculation with parton
showers~\cite{Weydert:2009vr}. 


\subsubsection{Santander matching}

A simple and pragmatic formula for the combination of the 
four- and five-flavour scheme calculations of 
bottom-quark associated Higgs-boson production
has been suggested in~\Bref{santander}. The matching formula originated
from discussions among the authors of \Bref{santander} at the 
{\it Higgs Days at Santander 2009} 
and is therefore dubbed ``Santander matching''. The matching
scheme has been described in some detail in
\refS{Santander-neutral}. Here we shall very briefly summarise
the main features of the scheme and provide matched
predictions for the inclusive cross section 
$\Pp\Pp \to {\Pt}\PQb \PSHpm{\rm + X}$
at the LHC operating at a centre-of-mass energy of $7\UTeV$. 

The 4FS and 5FS calculations provide the unique description of the
cross section in the asymptotic limits $\MH/\Mb \to 1$ and
$\MH/\Mb \to \infty$, respectively (here and in the following $\MH$ denotes a
generic Higgs boson mass). The two approaches are combined in
such a way that they are given variable weight, depending on the value
of the Higgs-boson mass. The difference between the two approaches is formally
logarithmic. Therefore, the dependence of their relative importance
on the Higgs-boson mass should be controlled by a logarithmic term.
This leads to the following formula \cite{santander}:
\begin{equation}
\begin{split}
\sigma^\text{matched}= \frac{\sigma^\text{4FS} +
  w\,\sigma^\text{5FS}}{1+w}\,,
\end{split}
\end{equation}
with the weight $w$ defined as 
\begin{equation}
\begin{split}
w = \ln\frac{\MH}{\Mb}  - 2\,,
\label{eq::t}
\end{split}
\end{equation}
and $\sigma^\text{4FS}$ and $\sigma^\text{5FS}$ denote the total
inclusive cross section in the 4FS and the 5FS, respectively.
For $\Mb=4.75\UGeV$ the weight factor increases from $w \approx 1.75$
at $\MH = 200\UGeV$ to $w\approx 2.65$ at $\MH = 500\UGeV$. 

The theoretical uncertainties  in the 4FS and the 5FS calculations
should be added linearly, using the
weights $w$ defined in \eqn{eq::t}.  This ensures that the combined
error is always larger than the minimum of the two individual
errors, see the discussion in \Bref{santander}. As the uncertainties
can be asymmetric, the combination should be done separately for
the upper and the lower uncertainty limits:
\begin{equation}
\begin{split}
\Delta\sigma_\pm = \frac{\Delta\sigma_\pm^\text{4FS}
  + w\,\Delta\sigma_\pm^\text{5FS}}{1+w}\,,
\end{split}
\label{eq::error}
\end{equation}
where $\Delta\sigma_\pm^\text{4FS}$ and
$\Delta\sigma_\pm^\text{5FS}$ are the upper/lower uncertainty limits
of the 4FS and the 5FS, respectively.

We shall now discuss the numerical implications of the Santander matching
and provide matched
predictions for the inclusive cross section $\Pp\Pp \to {\Pt}\PQb \PSHpm{\rm + X}$
at the LHC operating at a centre-of-mass energy of $7\UTeV$. The individual
numerical results for the 4FS and 5FS calculations have been
obtained 
with input parameters as
described in \Bref{Dittmaier:2011ti}. Note that the cross-section
predictions presented below correspond to NLO QCD results
for the production of heavy charged Higgs bosons in a
two-Higgs-doublet model. SUSY
effects can be taken into account by rescaling the bottom Yukawa
coupling to the proper value~\cite{Dittmaier:2006cz,Dawson:2011pe}.

Let us briefly discuss the choice of renormalisation and factorisation
scales and the estimate of the theory uncertainty in the
4FS and 5FS NLO calculations as presented in
\Bref{Dittmaier:2011ti}. In the 4FS both scales have been set to 
$\mu = (\Mt + \Mb + M_{\PSHm})/3$ as the default choice. The NLO scale 
uncertainty has been estimated from the variation of the scales by a
factor of three about the central scale, see also the discussion in  
\Bref{Dittmaier:2009np}. The residual NLO scale uncertainty in the 4FS 
is then approximately $\pm 30$\%. In the 5FS calculation the central
scale has been set to $\mu=(\Mt+M_{\PSHm})/4$ (see \Bref{Berger:2003sm}). 
A small residual NLO scale uncertainty of less then about $10\%$
is found when varying the scales by a factor three about the central
scale. We note that the scale uncertainty of the NLO 5FS calculation
is surprisingly small, and that the scale dependence of the NLO cross
section is monotonically rising when the scale is
decreased. Furthermore, the choice of the factorisation scale in the
5FS calculations is intricate and may depend both on the
kinematics of the process and on the Higgs-boson mass, see \eg 
\Brefs{Rainwater:2002hm,Boos:2003yi,Maltoni:2007tc}. Thus, a proper 
estimate of the theory uncertainty of the 5FS calculation may require 
further investigation. 

For the results presented below, however, we simply adopt the 4FS and
5FS cross section and theory uncertainty predictions as presented in 
\Bref{Dittmaier:2011ti}. As no estimate of the PDF error of the
5FS calculation has been given, we only consider the renormalisation
and factorisation scale
uncertainty. 

\refF{fig::xs4f5fhks}\,(a) shows the central values for the 4FS and
the 5FS cross section, as well as the matched result, as a function
of the Higgs-boson mass. The corresponding theory error estimates are shown in
\refF{fig::xs4f5fhks}\,(b). Taking the scale uncertainty into account, the 4FS and
5FS cross sections at NLO are consistent, even though the predictions
in the 5FS at our choice of the central scales are larger than those of
the 4FS by approximately $30\%$, almost independently of the Higgs-boson
mass. Qualitatively similar results have been obtained from a
comparison of 4FS and 5FS NLO calculations for single-top production
at the LHC~\cite{Campbell:2009ss}. We note that the lower end of the 
error band of the matched calculation approximately coincides with the 
central prediction of the 4FS, while the upper end of the band is very 
close to the upper band of the 5FS scale error. 

%
\begin{figure}
  \centerline{
    \begin{tabular}{cc} 
      \includegraphics[height=.4\textwidth, width=.49\textwidth]{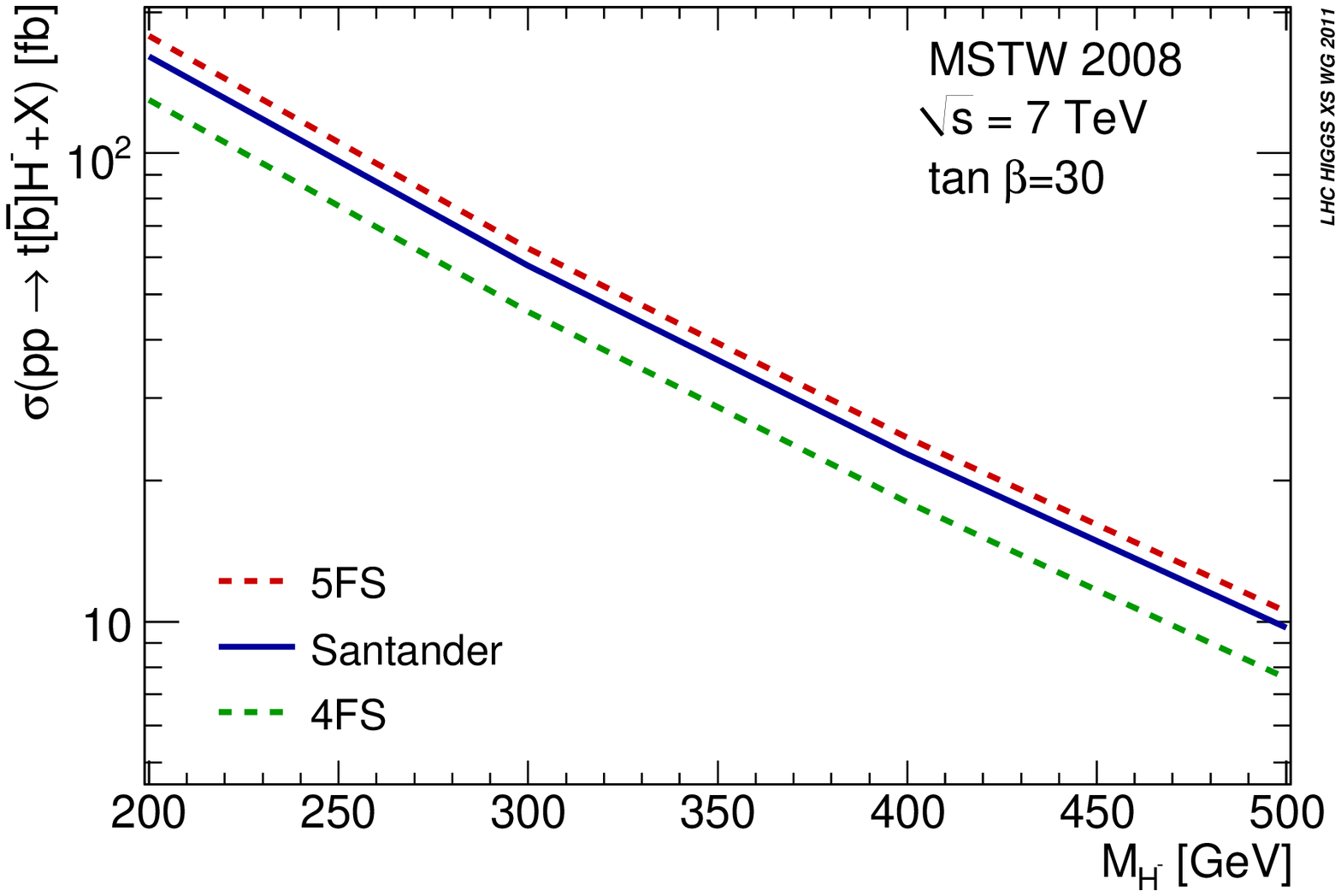} &
      \hspace*{-1em}
      \includegraphics[height=.4\textwidth, width=.49\textwidth]{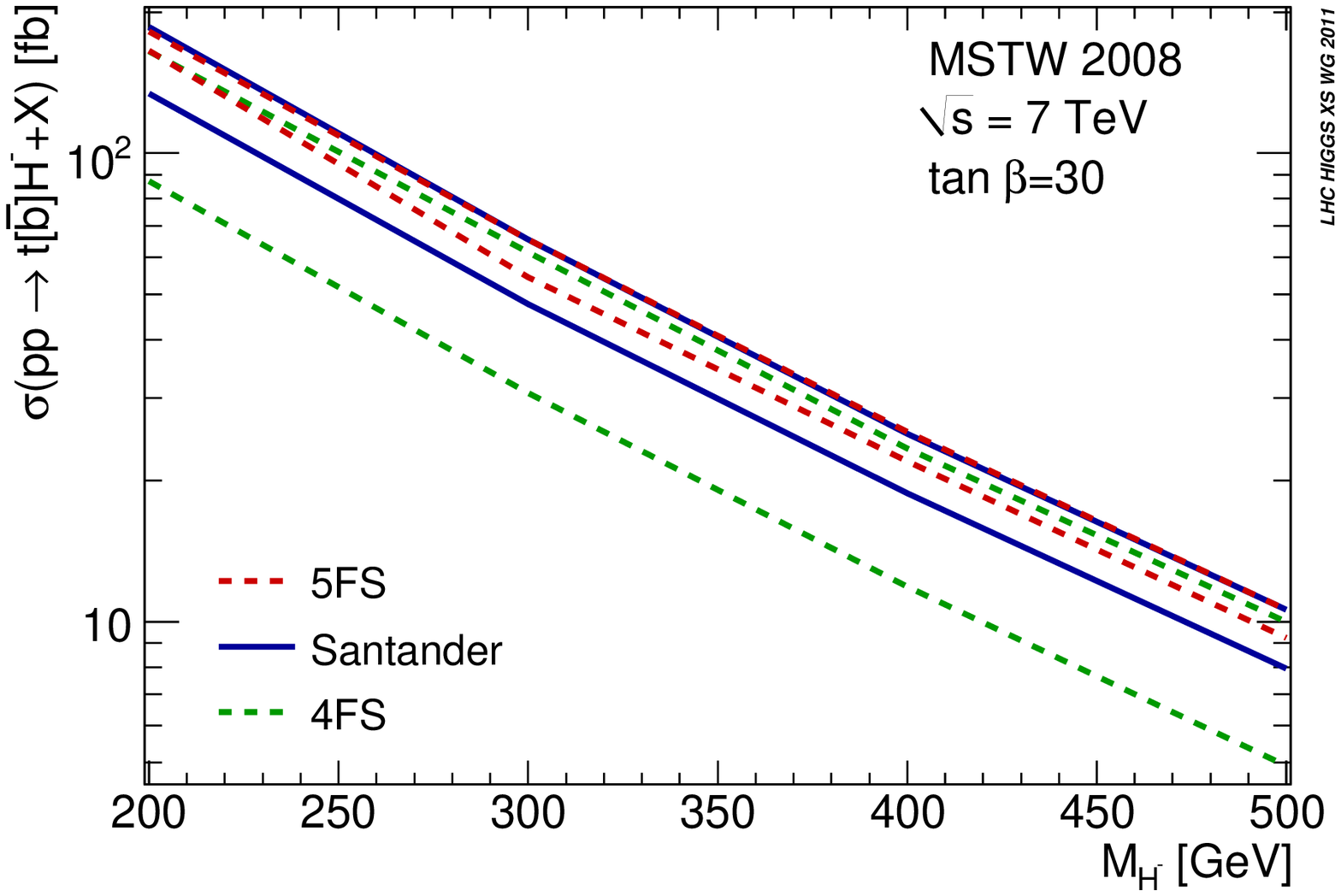} \\[-1em]
      (a) & (b)
    \end{tabular}
  }
      \caption{\label{fig::xs4f5fhks}\sloppy (a) Central values for the total inclusive cross
        section in the 5FS (red, dashed), the 4FS (green, dashed),
        and for the matched cross section (blue, solid). Here and in the
        following we use the MSTW2008\, PDF
        set~\cite{Martin:2009iq}. (b) Theory uncertainty
        bands for the total inclusive cross section in the 5FS (red,
        dashed), the 4FS (green, dashed), and for the matched cross
        section (blue, solid).}
\end{figure}
%

\subsubsection{Differential distributions}
Let us now turn to the transverse-momentum distributions
of the final-state particles. To illustrate the impact of the
higher-order QCD corrections on the shape of the distributions and to
analyse the difference between the 4FS and 5FS calculations, we
discuss results for the LHC operating at $14\UTeV$ energy, as presented 
in \Brefs{Dittmaier:2011ti,Weydert:2009vr}. In both calculations, the 
distributions have been evaluated for the
default scale choice $\mu = (m_{\rm t} + m_{\PQb} + M_{{\rm H}^-})/3$
and $\mu=(\Mt+M_{\PSHm})/4$, respectively. 

We first discuss the transverse-momentum distribution of the top and 
Higgs particles. As shown in \Brefs{Dittmaier:2011ti,Weydert:2009vr}, 
the impact of higher-order corrections, including the parton shower,
is small for the top and Higgs $\pT$. In
\refF{fig::pt4f5f}\,(a) and (b) we compare the NLO (4FS) and NLO +
parton-shower (5FS) results for the transverse-momentum distribution 
of the top and Higgs, respectively. 
%
\begin{figure}
  \centerline{
    \begin{tabular}{cc} 
      \includegraphics[height=.4\textwidth, width=.49\textwidth]{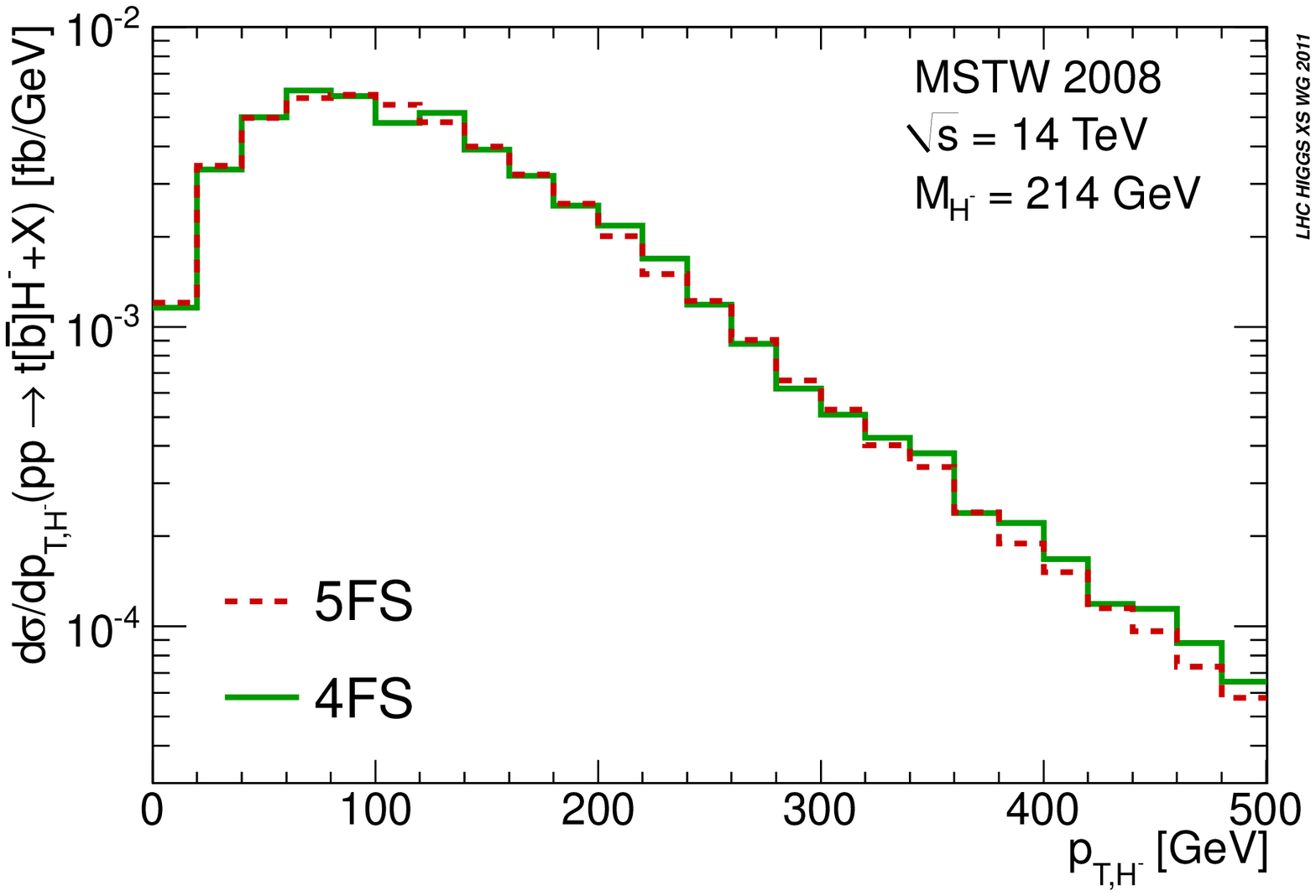} &
      \hspace*{-1em}
      \includegraphics[height=.4\textwidth, width=.49\textwidth]{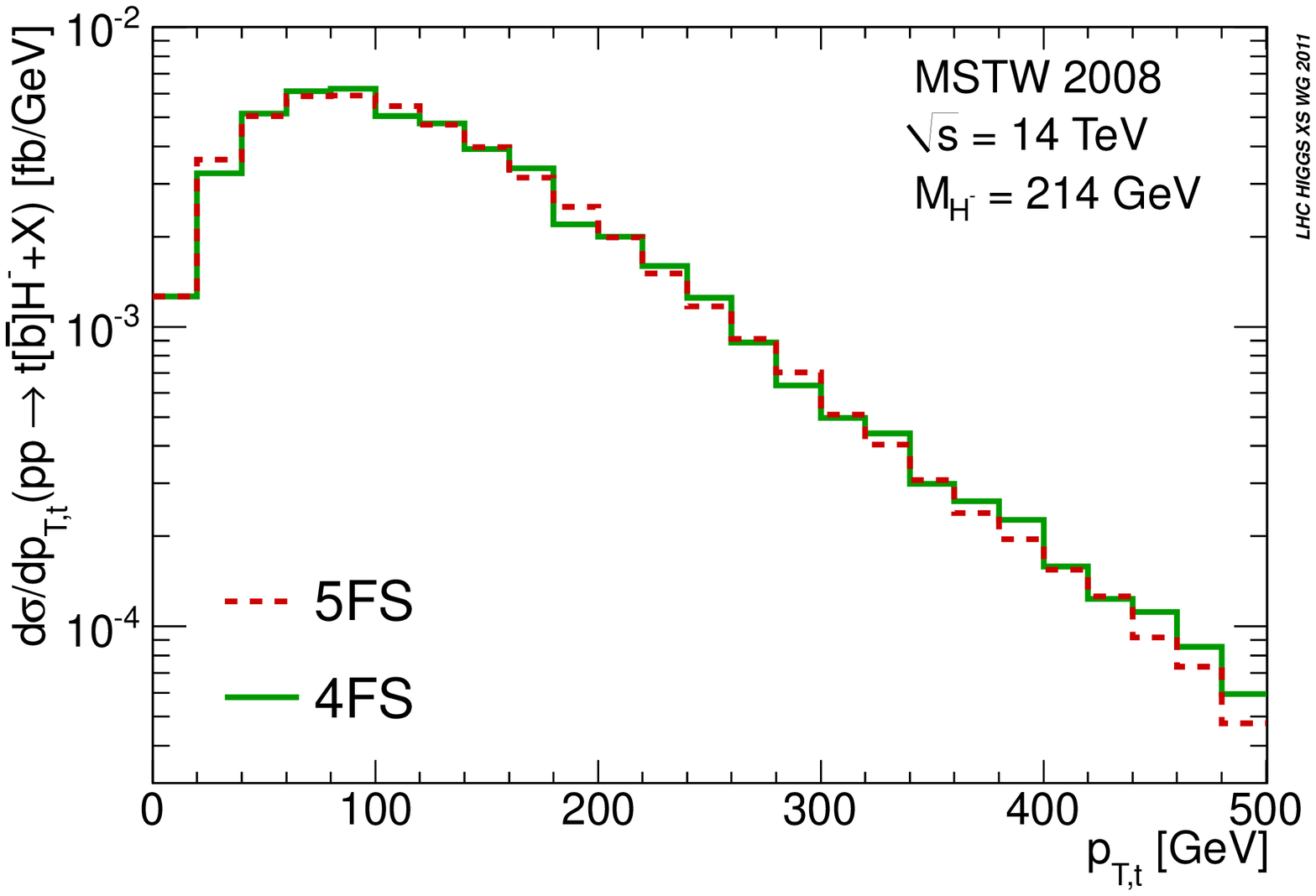} \\[-1em]
      (a) & (b)
    \end{tabular}
    }
      \caption{\label{fig::pt4f5f} \sloppy Normalised transverse-momentum distributions
      of the Higgs boson (a) and the top quark (b) for $\Pp\Pp \to {\rm t}\PAQb{\rm
    H}^-+X$ at the LHC ($14\UTeV$). Shown are results in the 4FS (NLO)
  and 5FS (NLO plus parton shower).}
\end{figure}
%
Note that we have normalised both distribution to one. It is evident
from  the comparison shown in \refF{fig::pt4f5f} that the shapes of the
Higgs and top transverse-momentum distribution in the 4FS and 5FS
agree very well.

The bottom $\pT$ distribution is described with different accuracy in the
4FS and 5FS calculations. While in the 4FS the kinematics of the process is
treated exactly already at LO, the 5FS is based on the approximation
that the bottom quark is produced at small transverse momentum. In the
5FS a finite bottom $\pT$ is thus only generated at NLO through the parton
process $\Pg\Pg \to {\Pt}\PQb\PSHpm$, which is the LO process of the
4FS.   We thus focus on the 
transverse-momentum distribution of the bottom quark as described within the 4FS
and compare the predictions at LO and at
NLO in \refF{fig::chiggsptb}, see \Bref{Dittmaier:2009np}.
%
\begin{figure}
  \centerline{
      \includegraphics[height=.4\textwidth, width=.49\textwidth]{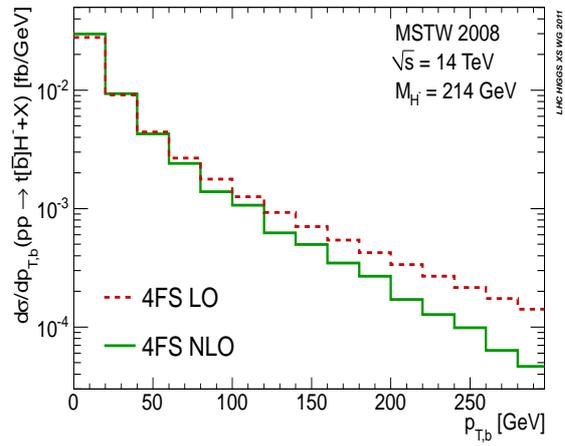} 
   } \vspace*{-1em}
      \caption{\label{fig::chiggsptb} \sloppy Transverse-momentum distributions
      of the bottom quark in the 4FS for $\Pp\Pp \to {\rm t}\PAQb{\rm
    H}^-+X$ at the LHC ($14\UTeV$). Shown are results at LO and NLO.}
\end{figure}
%
The bottom-quark $\pT$ distribution, which extends
to $p_{\mathrm{T,b}} \gg \Mb$, is significantly softened at NLO. The
large impact on the $p_{\mathrm{T,b}}$  
distribution is due to collinear gluon
radiation off bottom quarks that is enhanced by a factor $\alphas \ln(\Mb/p_{\mathrm{T,b}})$. The enhancement should
be significantly reduced if the bottom quarks are reconstructed from jets, since the application
of a jet algorithm treats the bottom--gluon system inclusively in the collinear cone, so that the
logarithmic enhancement cancels out.

\clearpage

\newpage
\newcommand{\FR}{\rm{\scriptscriptstyle{FR}}}
\newcommand{\myNLO}{\rm{\scriptscriptstyle{NLO}}}
\newcommand{\mySM}{\rm{\scriptscriptstyle{SM}}}
\newcommand{\myNF}{\rm{\scriptscriptstyle{NF}}}
\newcommand{\myF}{\rm{\scriptscriptstyle{F}}}
\newcommand{\myEW}{{\rm\scriptscriptstyle{E}\scriptscriptstyle{W}}}
\newcommand{\ssA}{{\scriptscriptstyle{A}}}
\newcommand{\ssQ}{{\scriptscriptstyle{Q}}}
\newcommand{\ssF}{{\scriptscriptstyle{F}}}
\newcommand{\ssD}{{\scriptscriptstyle{D}}}
\newcommand{\ssU}{{\scriptscriptstyle{U}}}
\newcommand{\ssL}{{\scriptscriptstyle{L}}}
\newcommand{\ssS}{{\scriptscriptstyle{S}}}
\newcommand{\ssH}{{\scriptscriptstyle{H}}}
\newcommand{\ssh}{{\scriptscriptstyle{h}}}
\newcommand{\ssW}{{\scriptscriptstyle{W}}}
\newcommand{\ssT}{{\scriptscriptstyle{T}}}
\newcommand{\ssZ}{{\scriptscriptstyle{Z}}}
\newcommand{\bqas}{\begin{eqnarray*}}
\newcommand{\eqas}{\end{eqnarray*}}
\newcommand{\nl}{\nonumber\\}
\def\mnew{\mpar{\hfil NEW \hfil}\ignorespaces}
\newcommand{\lpar}{\left(}                            
\newcommand{\rpar}{\right)} 
\newcommand{\lrbr}{\left[}
\newcommand{\rrbr}{\right]}
\newcommand{\lcbr}{\left\{}
\newcommand{\rcbr}{\right\}} 
\newcommand{\rbrak}[1]{\lrbr#1\rrbr}
\newcommand{\bq}{\begin{equation}}                    
\newcommand{\eq}{\end{equation}}
\newcommand{\bqa}{\arraycolsep 0.14em\begin{eqnarray}}
\newcommand{\eqa}{\end{eqnarray}}
\newcommand{\ba}[1]{\begin{array}{#1}}
\newcommand{\ea}{\end{array}}
\newcommand{\ben}{\begin{enumerate}}
\newcommand{\een}{\end{enumerate}}
\newcommand{\bei}{\begin{itemize}}
\newcommand{\eei}{\end{itemize}}
\renewcommand{\eqn}[1]{Eq.(\ref{#1})}
\newcommand{\eqns}[2]{Eqs.(\ref{#1})--(\ref{#2})}
\newcommand{\eqnss}[1]{Eqs.(\ref{#1})}
\newcommand{\eqnsc}[2]{Eqs.(\ref{#1}) and (\ref{#2})}
\newcommand{\eqnst}[3]{Eqs.(\ref{#1}), (\ref{#2}) and (\ref{#3})}
\newcommand{\eqnsf}[4]{Eqs.(\ref{#1}), (\ref{#2}), (\ref{#3}) and (\ref{#4})}
\newcommand{\eqnsv}[5]{Eqs.(\ref{#1}), (\ref{#2}), (\ref{#3}), (\ref{#4}) and (\ref{#5})}
\newcommand{\tbn}[1]{Tab.~\ref{#1}}
\newcommand{\tabn}[1]{Tab.~\ref{#1}}
\newcommand{\tbns}[2]{Tabs.~\ref{#1}--\ref{#2}}
\newcommand{\tabns}[2]{Tabs.~\ref{#1}--\ref{#2}}
\newcommand{\tbnsc}[2]{Tabs.~\ref{#1} and \ref{#2}}
\renewcommand{\fig}[1]{Fig.~\ref{#1}}
\newcommand{\figs}[2]{Figs.~\ref{#1}--\ref{#2}}
\newcommand{\sect}[1]{Section~\ref{#1}}
\newcommand{\sects}[2]{Section~\ref{#1} and \ref{#2}}
\newcommand{\sectm}[2]{Section~\ref{#1} -- \ref{#2}}
\newcommand{\subsect}[1]{Subsection~\ref{#1}}
\newcommand{\subsectm}[2]{Subsection~\ref{#1} -- \ref{#2}}
\newcommand{\appendx}[1]{Appendix~\ref{#1}}
\newcommand{\hsp}{\hspace{.5mm}}
\def\negs{\hspace{-0.26in}}
\def\negsh{\hspace{-0.13in}}
\newcommand{\barq}{\overline q}
\newcommand{\barb}{\overline b}
\newcommand{\bmid}{\Bigr|}
\newcommand{\Mbp}{\mathswitch {m_{\PQbpr}}}
\newcommand{\Mtp}{\mathswitch {m_{\PQtpr}}}
\newcommand{\Mlp}{\mathswitch {m_{\Pl'}}}
\newcommand{\Mnp}{\mathswitch {m_{\PGnl'}}}
\newcommand{\Mfp}{\mathswitch {m_{\Pf'}}}
\newcommand{\MGq}{\mathswitch {m_{\PQq}}}
\newcommand{\MGqs}{\mathswitch {m^2_{\PQq}}}
\newcommand{\MGl}{\mathswitch {m_{\Pl}}}
\newcommand{\MGf}{\mathswitch {m_{\Pf}}}
\newcommand{\MGfs}{\mathswitch {m^2_{\Pf}}}
\newcommand{\MGfq}{\mathswitch {m^4_{\Pf}}}

\newcommand{\MGqpr}{\mathswitch {m_{\ssQ}}}
\newcommand{\MGlpr}{\mathswitch {m_{\ssL}}}

\newcommand{\twol}{{\mbox{\scriptsize 2-loop}}}


\section{Predictions for Higgs production and decay with a 4th SM-like
  fermion generation\footnote{A.~Denner, S.~Dittmaier, A.~M\"uck, G.~Passarino, 
  M.~Spira, C.~Sturm, S.~Uccirati and M.M.~Weber.}}

\subsection{General setup}
We study the extension of the SM that includes a 4th generation of
heavy fermions, consisting of an up- and a down-type quark
$(\PQtpr,\PQbpr)$, a charged lepton $(\Pl')$, and a massive neutrino
$(\PGnl')$.  The 4th-generation fermions all have identical gauge
couplings as their SM copies and equivalent Yukawa couplings
proportional to their masses, but do not mix with the other three SM
generations.  The masses of the hypothetical new fermions in this
study are
\begin{align}
  \Mbp &= \Mlp= \Mnp = 600\UGeV, \nonumber \\
  \Mtp &= \Mbp + \left[1 + \frac{1}{5} \ln\biggl(\frac{\MH}{115\UGeV}\biggr)\right] \, 50 \UGeV,
\label{eq:masses}
\end{align}
where the relation among them is used to escape current exclusion
limits from electroweak (EW) precision data (see
\Brefs{Kribs:2007nz,Hashimoto:2010at}).  In the following we call the
Standard Model with $3$ generations ``SM3'' and the Standard Model
with a 4th generation of fermions ``SM4''.  Owing to screening (see
\refS{Hprod}), leading-order (LO) or next-to-leading-order (NLO) QCD
predictions typically depend only weakly on the precise values of
masses of the heavy fermions.  This is completely different for NLO EW
corrections, which are enhanced by powers of the masses of the heavy
fermions and induce a strong dependence of the results on these
masses.

Part of the results shown in the following have already been anticipated in
\Bref{Denner:2011vt}.

\subsection{Higgs production via gluon fusion \label{Hprod}}
So far, the experimental analysis has concentrated on models with
ultra-heavy 4th-generation fermions, excluding the possibility that
the Higgs boson decays to heavy neutrinos. Furthermore, the 2-loop
EW corrections have been included under the assumption that
they are dominated by light fermions.  At the moment the experimental
strategy consists in computing the cross-section ratio $R =
\sigma({\rm SM4})/\sigma({\rm SM3})$ with {\sc
  HIGLU}~\cite{Spira:1996if} while NLO EW radiative
corrections are switched off.

In this section we concentrate on full 2-loop EW corrections to
Higgs-boson production (through $\Pg\Pg$-fusion) at LHC in SM4 and
refer to the work of \Brefs{Spira:1995rr,Anastasiou:2011qw} for the
inclusion of QCD corrections.  The naive expectation is that light
fermions dominate the low-Higgs-boson-mass regime and, therefore, EW
corrections can be well approximated by the
ones~\cite{Actis:2008ts,Actis:2008ug} in SM3.  It is worth noting that
the leading behaviour of the EW corrections for high values of masses
in the 4th generation has been known for a long
time~\cite{Djouadi:1994ge,Djouadi:1997rj} (see also
\Bref{Fugel:2004ug}) showing an enhancement of radiative corrections.

To avoid misunderstandings we define the following terminology: for a
given amplitude $A$, in the limit $\MGf \to \infty$ we distinguish {\em
  decoupling} for $A \sim 1/\MGfs$ (or higher negative powers), {\em
  screening} for $A \to\,$ constant (or $A \sim \ln \MGfs$), and {\em
  enhancement} for $A \sim \MGfs$ (or higher positive powers).  To discuss
decoupling we need few definitions: SM3 is the usual SM with one
$\PQt{-}\PQb$ doublet; SM4 is the extension of SM3 with a new family
of heavy fermions, $\PQtpr{-}\PQbpr$ and $\Pl'{-}\PGnl'$.  All
relevant formulae for the asymptotic limit can be found in
\Brefs{Djouadi:1994ge,Djouadi:1997rj}.  Considering only EW
corrections, the amplitude for $\Pg\Pg$-fusion reads%
\footnote{Here we neglect the contributions of bottom-quark loops
  which amount to up to about $3\%$ within SM4 and $5{-}10\%$ in
SM3. The bottom-quark contributions are, however, included in our
numerical results for the gluon-fusion cross section.}
\bqa
A_{\mathrm{SM3}} &=& A^{\onel}_{\PQt} + A^{\NLO}_3, \quad
A^{\NLO}_3 = A^{\twol}_{\PQt} + \delta^{\FR}_{\PQt}\,A^{\onel}_{\PQt}~,
\nl
A_{\mathrm{SM4}} &=& A^{\onel}_{\ssQ} + A^{\NLO}_4, \quad
A^{\NLO}_4 = A^{\twol}_{\ssQ} + \delta^{\FR}_{\ssQ + \ssL}\,A^{\onel}_{\ssQ},
\label{defA}
\eqa
where 
\bqa
A_{\ssQ} = A_{\PQt + \PQtpr + \PQbpr}, \qquad 
\delta^{\FR}_{\ssQ + \ssL} = \delta^{\FR}_{\PQt + \PQtpr + \PQbpr + \Pl' + \PGnl'}~.
\eqa
In \eqn{defA} $\delta^{\FR}$ gives the contribution from finite renormalisation, 
including Higgs-boson wave-function renormalisation (see Sect.~$3.4$ of
\Bref{Actis:2008ts} for technical details).

First we recall the standard argument for the asymptotic behaviour in LO
$\Pg\Pg$-fusion, extendible to NLO and next-to-next-to-leading (NNLO)
QCD corrections~\cite{Spira:1995rr}, and give a simple argument to
prove enhancement at NLO EW level.

Any Feynman diagram contributing to the Higgs--gluon--gluon vertex has
dimension one and involves the Yukawa coupling (proportional to
$\MGf/\MW$) of the fermion $f$ circulating in the loop.  Naively this
suggests an asymptotic scaling of the amplitude proportional to
$\MGf^2$ for large $\MGf$, i.e.\ enhancement.  However, the total
$\PH\Pg\Pg$ amplitude must be proportional to $T_{\mu\nu} = p_1\cdot
p_2\,\delta_{\mu\nu} - p_{2\mu} p_{1\nu}$ (where $p_{1,2}$ are the two
gluon momenta) because of gauge invariance, and $T$ has dimension two.
Thus, the scaling of the whole amplitude for large $\MGf$ is reduced
by two powers in $\MGf$, leading to a constant limit, i.e.\ screening,
if there is exactly one Yukawa coupling.
The part of the diagram that is not proportional to $T$ cancels
in the total amplitude because of gauge invariance (all higher powers
of $\MGf$ will go away, and this explains the presence of huge
cancellations in the total amplitude).  At LO there is only one Yukawa
coupling as in NLO and NNLO QCD where one adds only gluon lines, so
there is screening.  At EW NLO there are diagrams with three Yukawa
couplings, therefore giving the net $\MGfs$ behaviour predicted in
\Bref{Djouadi:1994ge}, so there is enhancement and at the 2-loop level
it goes at most with $\MGfs$. We define
\bq
\sigma_{\mySM 3}\lpar \Pg\Pg \to \PH\rpar = 
\sigma^{\LO}_{\mySM 3}\lpar \Pg\Pg \to \PH\rpar\,\Bigl( 1 + \delta^{(3)}_{\myEW}\Bigr),
\quad
\sigma_{\mySM 4}\lpar \Pg\Pg \to \PH\rpar = 
\sigma^{\LO}_{\mySM 4}\lpar \Pg\Pg \to \PH\rpar\,\Bigl( 1 + \delta^{(4)}_{\myEW}\Bigr)~.
\eq
Analysing the results of \Brefs{Djouadi:1994ge,Djouadi:1997rj}, valid
for a light Higgs boson, one can see that in SM3 there is enhancement
in the quark sector for $\Mt \gg \Mb$ ($\delta^{(3)}_{\myEW} \sim
\GF m^2_{\PQt}$, where $\GF$ is the Fermi coupling constant).
The full calculation of \Bref{Actis:2008ug} shows that the physical
value for the top-quark mass is not large enough to make this
quadratic behaviour relevant with respect to the contribution from
light fermions.  From \Bref{Djouadi:1994ge} we can also understand
that a hypothetical SM3 with degenerate $\PQt{-}\PQb$ ($\Mt = \Mb =
\MGq$) would generate an enhancement in the small-Higgs-mass region
with the opposite sign ($\delta^{(3)}_{\myEW} \sim -\GF\MGqs$).
Moving to SM4, Eq.~(62) of \Bref{Djouadi:1997rj} shows that the
enhanced terms coming from finite renormalisation exactly cancel the
similar contribution from 2-loop diagrams for $\Mtp= \Mbp$, so that
for mass-degenerate $\PQtpr{-}\PQbpr$ (and $\Mt \gg \Mb$) we observe
screening in the 4th-generation quark sector.  This accidental
cancellation follows from the $3$ of colour $\mathrm{SU}(3)$ and from
the fact that we have $3$ heavy quarks contributing to LO almost with
the same rate (no enhancement at LO).  However, the same is not true
for $\Pl'{-}\PGnl'$ (there are no 2-loop diagrams with leptons in
$\Pg\Pg$-fusion), so we observe enhancement in the leptonic sector
of SM4, which actually dominates the behaviour at small values of
$\MH$.  To summarise the asymptotics at NLO EW:
\begin{itemize}
\item SM3 with heavy--light quark doublet ($\Mt \gg \Mb$): (positive) enhancement;
\item SM3 with a hypothetical heavy--heavy (mass-degenerate, $\Mt=\Mb$)
quark doublet: (negative) enhancement;
\item SM4 with heavy--light or heavy--heavy (mass-degenerate) quark doublets
in the 4th generation (and $\Mt \gg \Mb$):
enhancement in heavy--light, screening in heavy--heavy;
\item SM4 including heavy--heavy $\Pl'{-}\PGnl'$ doublet: enhancement.
\end{itemize}
We have verified that our (complete) results confirm the asymptotic
estimates of \Brefs{Djouadi:1994ge,Djouadi:1997rj}.

In order to prove that light-fermion dominance in SM3 below $300\UGeV$
is a numerical accident due to the fact that the top quark is not
heavy enough, we have computed $\delta^{(3)}_{\myEW}$ for a top quark of
$800\UGeV$ at $\MH= 100\UGeV$ and found top-quark dominance.  A
similar effect of the top quark is also present in SM4; if, for
instance, we fix all heavy-fermion masses $\Mfp$ to $600\UGeV$, we find
$\delta^{(4)}_{\myEW} = 12.1\%(29.1\%)$ at $\Mt = 172.5\UGeV(600\UGeV)$.

Moving to SM4, the LO $\Pg\Pg$-fusion cross section for a light Higgs
boson is about nine times the one of SM3, because three instead of one
heavy fermions propagate in the loop \cite{Georgi:1977gs}.
If one assumes that also NLO SM4
is dominated by light-fermion corrections, i.e., that EW
corrections are the same for SM3 and SM4, one expects
$\delta^{(4)}_{\myEW} \approx \delta^{(3)}_{\myEW}/3$ for very heavy
fermions.  According to \Bref{Djouadi:1997rj} this would be true
provided that no heavy leptons are included.

Our complete result (see \Bref{Passarino:2011kv}) is shown in
\refF{SM4_common_1} where the $\PQtpr{-}\PQbpr$ and the $\Pl'{-}\PGnl'$
doublets are included.  
Numbers for the relative EW corrections are listed in
\refT{tab:EWNLO}.
Electroweak NLO corrections due to the 4th generation are positive and
large for a light Higgs boson, positive but relatively small around
the $\PQt\PAQt$ threshold, and start to become negative around
$450\UGeV$.  
%
Increasing further the value of the Higgs-boson mass, the NLO effects
tend to become huge and negative ($\delta^{(4)}_{\myEW} < -100\%$)
showing minima around the heavy-quark thresholds. Above those
thresholds $\delta^{(4)}_{\myEW}$ starts to grow again and becomes
positive around $\MH= 1750\UGeV$ (not shown).
\begin{figure}
\begin{center}
\includegraphics[height=8cm]{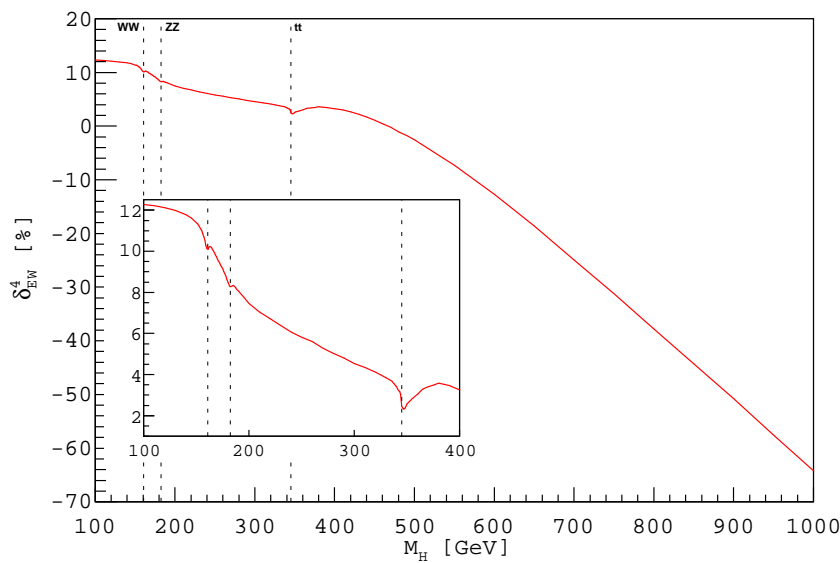}
\end{center}
\vspace{-0.6cm}
\caption[]{\label{SM4_common_1}
Relative corrections in SM4 ($\PQtpr{-}\PQbpr$ and $\Pl'{-}\PGnl'$ doublets) 
due to 2-loop EW corrections to $\Pg \Pg \to \PH$. 
The masses of the 4th-generation fermions are chosen according to
\Eref{eq:masses}. In the inset a blow-up of the small-$\MH$
region is shown.
}
\end{figure}
\begin{table}
\begin{center}
\caption[]{\label{tab:EWNLO}{NLO EW corrections to
$\Pg\Pg\to \PH+X$ cross sections in SM4 for the low-mass region:
$\delta^{(4)}_{\myEW}\, \pm \,\Delta$, where $\Delta$ is the numerical integration error.}}
\vspace{0.5cm}
\tabcolsep 4pt
\begin{tabular}{cc@{\hspace{6pt}}c|cc@{\hspace{6pt}}c|cc@{\hspace{6pt}}c}
\hline
$\MH$ [GeV] & $\delta^{(4)}_{\rm EW}$ [\%] & $\Delta$ [\%] &
$\MH$ [GeV] & $\delta^{(4)}_{\rm EW}$ [\%] & $\Delta$ [\%] &
$\MH$ [GeV] & $\delta^{(4)}_{\rm EW}$ [\%] & $\Delta$ [\%] \\
\hline
100 & 12.283 & 0.005 &  181 &  8.351 & 0.006 &  346 &   2.361 & 0.048  \\
110 & 12.212 & 0.005 &  182 &  8.278 & 0.006 &  347 &   2.327 & 0.048  \\
120 & 12.118 & 0.005 &  183 &  8.286 & 0.008 &  348 &   2.356 & 0.048  \\
130 & 11.982 & 0.005 &  184 &  8.320 & 0.009 &  349 &   2.470 & 0.048  \\
140 & 11.780 & 0.005 &  185 &  8.336 & 0.009 &  350 &   2.570 & 0.019  \\
145 & 11.620 & 0.005 &  186 &  8.300 & 0.009 &  351 &   2.629 & 0.019  \\
150 & 11.394 & 0.005 &  187 &  8.246 & 0.007 &  353 &   2.721 & 0.020  \\
151 & 11.333 & 0.005 &  188 &  8.176 & 0.009 &  355 &   2.834 & 0.019  \\
152 & 11.265 & 0.005 &  189 &  8.107 & 0.010 &  360 &   3.049 & 0.048  \\
153 & 11.191 & 0.004 &  190 &  8.057 & 0.010 &  365 &   3.300 & 0.048  \\
154 & 11.099 & 0.004 &  195 &  7.757 & 0.010 &  370 &   3.405 & 0.048  \\
155 & 10.994 & 0.004 &  200 &  7.465 & 0.010 &  375 &   3.492 & 0.048  \\
156 & 10.862 & 0.004 &  210 &  7.044 & 0.010 &  380 &   3.589 & 0.049  \\
157 & 10.705 & 0.004 &  220 &  6.720 & 0.018 &  385 &   3.521 & 0.049  \\
158 & 10.510 & 0.004 &  230 &  6.399 & 0.019 &  390 &   3.463 & 0.048  \\
159 & 10.290 & 0.004 &  240 &  6.065 & 0.020 &  400 &   3.260 & 0.048  \\
160 & 10.112 & 0.004 &  250 &  5.815 & 0.019 &  410 &   3.020 & 0.047  \\
161 & 10.104 & 0.004 &  260 &  5.600 & 0.019 &  420 &   2.680 & 0.042  \\
162 & 10.202 & 0.004 &  270 &  5.273 & 0.020 &  430 &   2.211 & 0.037  \\
163 & 10.238 & 0.004 &  280 &  5.038 & 0.020 &  440 &   1.710 & 0.039  \\
164 & 10.197 & 0.004 &  290 &  4.795 & 0.019 &  450 &   1.139 & 0.039  \\
165 & 10.115 & 0.004 &  300 &  4.541 & 0.020 &  460 &   0.444 & 0.040  \\
166 & 10.019 & 0.005 &  310 &  4.349 & 0.019 &  470 &  -0.223 & 0.042  \\
167 &  9.919 & 0.005 &  320 &  4.106 & 0.020 &  480 &  -1.060 & 0.047  \\
168 &  9.822 & 0.005 &  325 &  3.978 & 0.020 &  490 &  -1.740 & 0.049  \\
169 &  9.726 & 0.005 &  330 &  3.847 & 0.020 &  500 &  -2.542 & 0.036  \\
170 &  9.632 & 0.005 &  335 &  3.688 & 0.020 &  550 &  -7.281 & 0.048  \\
171 &  9.531 & 0.005 &  336 &  3.663 & 0.049 &  600 & -12.708 & 0.049  \\
172 &  9.432 & 0.005 &  337 &  3.589 & 0.049 &  650 & -18.593 & 0.049  \\
173 &  9.334 & 0.005 &  338 &  3.539 & 0.048 &  700 & -24.904 & 0.049  \\
174 &  9.234 & 0.005 &  339 &  3.463 & 0.047 &  750 & -31.227 & 0.049  \\
175 &  9.134 & 0.005 &  340 &  3.417 & 0.049 &  800 & -37.790 & 0.049  \\
176 &  9.027 & 0.005 &  341 &  3.317 & 0.049 &  850 & -44.322 & 0.049  \\
177 &  8.907 & 0.005 &  342 &  3.223 & 0.049 &  900 & -50.768 & 0.049  \\
178 &  8.786 & 0.005 &  343 &  3.169 & 0.049 &  950 & -57.516 & 0.049  \\
179 &  8.639 & 0.005 &  344 &  3.048 & 0.049 & 1000 & -64.187 & 0.045  \\
180 &  8.485 & 0.008 &  345 &  2.449 & 0.049 &      &         &        \\
\hline
\end{tabular}

\end{center}
\end{table}

Having computed the EW corrections $\delta^{(4)}_{\myEW}$ we should
discuss some aspects of their inclusion in the production cross
section $\sigma\lpar \Pg\Pg \to \PH + X\rpar$, i.e.\  their interplay
with QCD corrections and the remaining theoretical uncertainty. The
most accepted choice is given by
\bq
\sigma^{\myF} = \sigma^{\LO}
\,\Bigl( 1 + \delta_{\QCD}\Bigr)\,\,\Bigl( 1 + \delta_{\myEW}\Bigr)~,
\eq
which assumes complete factorisation of QCD and EW corrections. The
latter is based on the work of \Bref{Anastasiou:2008tj} where it is
shown that, at zero Higgs momentum, exact factorisation is violated
but with a negligible numerical impact; the result of
\Bref{Anastasiou:2008tj} can be understood in terms of soft-gluon
dominance.

The residual part beyond the soft-gluon-dominated part contributes up
to $5{-}10\%$ to the total inclusive cross section (for Higgs
masses up to $1\UTeV$).  Thus, with EW corrections we are talking about
non-factorizable effects which are at most $5\%$ in SM4.

EW corrections become $- 100\%$ just before the heavy-quark thresholds
making the use of the perturbative approach questionable. At EW NNLO
there are diagrams with five Yukawa couplings; according to our
argument on decoupling the enhancement at 3~loops goes as the 4th
power of the heavy-fermion mass, unless some accidental screening
occurs. Here we have no solid argument to estimate the remaining
uncertainty in the high-mass region and prefer to state that SM4 is in
a fully non-perturbative regime which should be approached with
extreme caution.  For the low-mass region we can do no more than make
educated guesses; therefore, assuming a leading 3-loop behaviour of
$\Mfp^4$, we estimate the remaining uncertainty to be of the order of
$(\alpha/\pi)^2(\Mfp/\MW)^4$ and thus $1{-}2\,\%$ in the interval $\MH
= 100{-}150\UGeV$ and almost negligible in the intermediate region,
$150{-}600\UGeV$.
\subsection{NLO corrections to $\PH\to 4\Pf$ in SM4}
\label{se:H4f}
\newcommand{\pfourf}{{\sc Prophecy4f} }

\begin{sloppypar}
The results of the $\PH\to 4\Pf$ decay channels have been
  obtained using the Monte Carlo generator
  \pfourf \cite{Prophecy4f,Bredenstein:2006rh,Bredenstein:2007ec}
  which has been extended to support the SM4.  \pfourf can
  calculate the EW and QCD NLO corrections to the partial widths for
  all 4f final states, i.e.\ leptonic, semi-leptonic, and hadronic
  final states.  Since the vector bosons are kept off shell, the
  results are valid for Higgs masses below, near, and above the
  on-shell gauge-boson production thresholds. Moreover, all
  interferences between WW and ZZ intermediate states are included at
  LO and NLO.
\end{sloppypar}

The additional corrections in SM4 arise from 4th-generation fermion
loops in the HWW/HZZ vertices, the gauge-boson self-energies, and the
renormalisation constants.  For the large 4th-generation masses of
${\cal O}(600 \UGeV)$ considered here, the 4th-generation Yukawa
couplings are large, and the total corrections are dominated by the
4th-generation corrections.  Numerically the NLO corrections amount to
about $-85\%$ and depend only weakly on the Higgs-boson mass for not
too large $\MH$, since the heavy-fermion masses in the scenario of
(\ref{eq:masses}) have only a weak Higgs mass dependence.  The
corrections from the 4th generation are taken into account at NLO with
their full mass dependence, but their behaviour for large masses can be
approximated well by the dominant corrections in the heavy-fermion
limit. In this limit the leading contribution can be absorbed into
effective HWW/HZZ interactions in the $\GF$ renormalisation scheme via
the Lagrangian
\beq
{\cal L}_{HVV} = \sqrt{\sqrt{2} \GF} H
\left[ 2 \MW^2 W^\dagger_\mu W^\mu (1 + \delta_{\PW}^{\mathrm{tot}}) + \MZ^2 Z_\mu Z^\mu (1 + \delta_{\PZ}^{\mathrm{tot}}) \right],
\eeq
where $W,Z,H$ denote the fields for the $\PW$, $\PZ$, and Higgs
bosons.  The higher-order corrections are contained in the factors
$\delta_V^{\mathrm{tot}}$ whose expansion up to 2-loop order is given
by
\beq
\delta_V^{\mathrm{tot}(1)} = \delta_u^{(1)} + \delta_V^{(1)}, \qquad
\delta_V^{\mathrm{tot}(2)} = \delta_u^{(2)} + \delta_V^{(2)} + \delta_u^{(1)} \delta_V^{(1)}.
\eeq
The 1-loop expressions for a single $\mathrm{SU}(2)$ doublet of heavy
fermions with masses $m_A$, $m_B$ are \cite{Chanowitz:1978uj}
\beq
\delta_u^{(1)} = N_c X_A \left[ \frac{7}{6} (1 + x) + \frac{x}{1-x} \ln x\right], \qquad
\delta_V^{(1)} = - 2 N_c X_A (1+x),
\label{eq:delta1}
\eeq
where $x = m_B^2 / m_A^2$,  $X_A = \GF m_A^2/(8 \sqrt{2} \pi^2)$, and
$N_c=3$ or $1$ for quarks or leptons, respectively.
The results for the 2-loop corrections $\delta_V^{\mathrm{tot}(2)}$ can
be found in \Bref{Kniehl:1995ra} for the QCD corrections of ${\cal
  O}(\alphas \GF \Mfp^2)$ and in \Bref{Djouadi:1997rj} for the EW
corrections of ${\cal O}(\GF^2 \Mfp^4)$.  The corrected partial
decay width is then given by
\beq
\Gamma_{\mathrm{NLO}} \;\approx\; \Gamma_{\mathrm{LO}} \left[1 + \delta_\Gamma^{(1)} +
\delta_\Gamma^{(2)}\right] \;=\;
\Gamma_{\mathrm{LO}} \left[1 + 2 \delta_V^{\mathrm{tot}(1)} +
(\delta_V^{\mathrm{tot}(1)})^2 + 2 \delta_V^{\mathrm{tot}(2)}\right].
\eeq

The size of the two-loop corrections $\delta_\Gamma^{(2)}$ is about
$+15\%$ for $\Mfp=600\GeV$,
again depending only very weakly on the Higgs mass. Due to
the large 1-loop corrections \pfourf includes the 2-loop QCD and EW
corrections in the heavy-fermion limit in addition to the exact 1-loop
corrections. Although the asymptotic two-loop corrections are not
directly applicable for a heavy Higgs boson, they can be viewed
as a qualitative estimate of the 2-loop effects. One should
keep in mind that for a Higgs boson heavier than about $600\GeV$ many
more uncertainties arise owing to the breakdown of perturbation theory.

The leading 2-loop terms can be taken as an estimate of the error
from unknown higher-order corrections. This implies an error of $15\%$
relative to the LO for the generic 4th-generation mass scale
$\Mfp=600\UGeV$ on the partial width for all $\PH\to 4\Pf$ decay
channels.  Assuming a scaling law of this error proportional to
$X_A^2$, the uncertainty is estimated to about $100X_A^2$ relative to
the LO prediction. However, since the correction grows large and
negative, the relative uncertainty on the corrected width gets
enhanced to $100X_A^2/(1-23X_A+100X_A^2)$, where the linear term in
$X_A$ parametrises the 1-loop correction and we have assumed the
mass splitting given in \refE{eq:masses}. For $\Mfp=600\UGeV$ this
results roughly in an uncertainty of $50\%$ on the corrected $\PH\to
4\Pf$ decay widths.

\subsection{$\PH\to \Pf\bar{\Pf}$}
The decay widths for $\PH\to \Pf\bar{\Pf}$ are calculated with {\sc
  HDECAY} \cite{Djouadi:1997yw,Spira:1997dg,hdecay2} which includes
the approximate NLO and NNLO EW corrections for the decay channels
into SM3 fermion pairs in the heavy SM4 fermion limit according to
\Bref{Djouadi:1997rj} and mixed NNLO EW/QCD corrections according to
\Bref{Kniehl:1995ra}. These corrections originate from the
wave-function renormalisation of the Higgs boson and are thus
universal for all fermion species. The leading 1-loop part is given by
$\delta_u^{(1)}$ of \Eref{eq:delta1}.  Numerically the EW 1-loop
correction to the partial decay widths into fermion pairs amounts to
about $+40\%$, while the 2-loop correction contributes an additional
$+20\%$.  The corrections are assumed to factorise from whatever is
included in {\sc HDECAY}, since the approximate expressions emerge as
corrections to the effective Lagrangian after integrating out the
heavy-fermion species.  Thus, {\sc HDECAY} multiplies the relative SM4
corrections with the full corrected SM3 result including QCD and
approximate EW corrections.  The scale of the strong coupling
$\alphas$ has been identified with the average mass of the heavy
quarks $\PQtpr, \PQbpr$ of the 4th generation. Since for the scenario
of \Eref{eq:masses}, the full SM4 2-loop corrections amount to about
$15{-}20\%$ relative to the LO result, and the NLO QCD and EW
corrections amount to about $+60\%$, the remaining
theoretical uncertainties can be estimated to be about $10\%$ for the
full partial decay widths into fermion pairs for the SM4 part, while
the uncertainties of the SM3 EW and QCD parts are negligible with
respect to that. The theoretical uncertainties scale with the
heavy-fermion masses as $0.1\times(\Mfp/600\GeV)^4$.

\subsection{$\PH\to \Pg\Pg, \PGg\PGg, \PGg\PZ$}
For the decay modes $\PH\to \Pg\Pg, \PGg\PGg, \PGg\PZ$, {\sc HDECAY}
\cite{Djouadi:1997yw,Spira:1997dg,hdecay2} is used as well.

For $\PH\to \Pg\Pg$, {\sc HDECAY} includes the NNNLO QCD corrections of
the SM3 in the limit of a heavy top quark
\cite{Inami:1982xt,Djouadi:1991tka,Spira:1995rr,Chetyrkin:1997iv,Baikov:2006ch},
applied to the results including the heavy-quark loops. 
While at NNLO the exact QCD corrections in SM4
\cite{Anastasiou:2011qw} are included, at NNNLO the relative SM3
corrections are added to the relative NNLO corrections and multiplied
by the LO result including the additional quark loops. Since the
failure of such an approximation is less than $1\%$ at NNLO, we assume
that at NNNLO it is negligible, i.e.\ much smaller than the
residual QCD scale uncertainty of about $3\%$.  In addition the full NLO EW
corrections of \refS{Hprod} have been included in factorised
form, since the dominant part of the QCD corrections emerges from the
gluonic contributions on top of the corrections to the effective
Lagrangian in the limit of heavy quarks.
Taking besides the scale uncertainty also the missing quark-mass
dependence at NLO and beyond into account, the total theoretical
uncertainties can be estimated to about $5\%$.

{\sc HDECAY} \cite{Djouadi:1997yw,Spira:1997dg,hdecay2} includes the
full NLO QCD corrections to the decay mode $\PH\to \PGg\PGg$ supplemented
by the additional contributions of the 4th-generation quarks and charged
leptons according to \Brefs{Djouadi:1990aj,Djouadi:1993ji,Spira:1995rr}.

The introduction of EW NLO corrections (2-loop level) to the decay 
$\PH \to \PGg\PGg$ requires particular attention. We write the amplitude as
\bq
A = A_{\LO} + X_{\PW}\,A_{\NLO} + X_{\PW}^2\,A_{\NNLO} + \dots,
\qquad X_{\PW}= \frac{\GF\MW^2}{8 \sqrt{2} \pi^2}.
\eq
Part of the problem in including the NLO EW term is related to the
fact that the cancellation between the $\PW$ and the fermion loops is
stronger in the SM4 than in the SM3 so that the LO result is
suppressed more, by about a factor of $2$ at the level of the
amplitude and thus a factor of $4$ at the level of the decay width.
Furthermore, the NLO corrections are strongly enhanced for ultra-heavy
fermions in the fourth generation;  assuming the mass scenario
of \eqn{eq:masses} for the heavy fermions and a Higgs mass of $100\UGeV$ we
get a correction to the decay width (interference of LO with NLO) of
$- 318\%$; clearly it does not make sense and one should always
remember that a badly behaving series should not be used to derive limits
on the parameters, \ie on the heavy-fermion masses.

Extending the same techniques used for $\PH \to \Pg\Pg$ in
\Bref{Passarino:2011kv}, we have computed the exact amplitude for $\PH
\to \PGg\PGg$ up to NLO; assuming for
simplicity $\Mbp = \Mtp = \MGqpr$ and $\Mlp = \Mnp = \MGlpr$ (but the
calculation gives results for any general setup), the amplitude can be
written as follows:
\bq
A = A_{\LO}\,\biggl[1  + X_{\PW}\,\Bigl( C_{\ssQ}\,\frac{\MGqpr^2}{\MW^2} +
C_{\ssL}\,\frac{\MGlpr^2}{\MW^2} + R \Bigr)\biggr],
\eq
where $C_{\ssQ,\ssL}$ and $R$ depend on masses. In the asymptotic region, 
$\MH < 2\,\MW \muchless \MGqpr, \MGlpr$ we require $R$ to be a constant and
parametrise the $C\,$-functions as
\bq
C_{\ssQ} = -\,\frac{192}{5}\,\lpar 1 + c_{\ssQ}\,\tau \rpar,
\qquad
C_{\ssL} = -\,\frac{32}{3}\,\lpar 1 + c_{\ssL}\,\tau \rpar,
\eq
where $C_{\ssQ,\ssL}$ are constant and $\tau^2 = \MH^2/(2\,\MW)^2$.
Note that for $\tau = R = 0$ this is the leading 2-loop behaviour
predicted in \Bref{Djouadi:1997rj} (see also \Bref{Fugel:2004ug} for
the top-dependent contribution which we hide here in $R$). By
performing a fit to our exact result we obtain a good agreement in the
asymptotic region, showing that the additional corrections
proportional to $\tau$ play a relevant role. For instance, with
fermions of the fourth generation heavier than $300\UGeV$ we have
fit/exact$\,- 1$ less than $5\%$ in the window $\MH =
[80{-}130]\UGeV$.

The major part of the NLO corrections emerges from an effective
Lagrangian in the heavy-particle limit, therefore we should consider
them as correction to the effective Feynman rules and thus to the
amplitude.  As a result, we define NLO EW corrections to $\PH \to
\PGg\PGg$ as follows:
\bq
\bmid A\bmid^2 = \bmid A_{\LO}\bmid^2\,\lpar 1 + \delta_{\myEW}^{(4)}\rpar =
\bmid A_{\LO} + X_{\PW}\,A_{\NLO} \bmid^2.
\label{EWnewd}
\eq
This choice is also partially based on the fact that, due to the
smallness of the LO amplitude, the interference between LO and NNLO is
expected to be suppressed as compared to the squared NLO amplitude.
Using \Eref{EWnewd}, the corrections for $\MH = 100\UGeV$ and $\MGf =
600\UGeV$ amount to $- 64.5\%$ as compared to the $- 318\%$ in the
conventional (unrealistic) approach.

To estimate the missing quartic corrections (in the heavy-fermion
masses) we assume a leading behaviour of $\MGqpr^4, \MGlpr^4$ for NNLO
EW, \ie no accidental cancellations.  For a conservative estimate of
the uncertainty we use
\bq
\bmid A \bmid^2 = \bmid A_{\LO} + X_{\PW}\,A_{\NLO} \bmid^2
\pm 2\,X_{\PW}^2\,\bmid A_{\LO}\bmid^2\,\bmid C_{\ssQ} + C_{\ssL}\bmid\,
\frac{\Mfp^4}{\MW^4},
\label{EWnewderr}
\eq
where we put $\MGqpr = \MGlpr = \Mfp$ in the last term; given our
setup the difference between $\Mtp$ and $\Mbp$ is irrelevant in
estimating the uncertainty.  Our educated guess for the error estimate
is to use the absolute value of the NLO leading coefficient as the
unknown coefficient in the NNLO one.

\begin{sloppypar}
The effect of including NLO EW corrections is thus better discussed in
terms of $|A_{\LO} + X_{\PW} A_{\NLO}|^2$; this quantity decreases
with increasing values of $\MH$ and approaches zero around $\MH =
150\GeV$ after which the effect is reversed.  The reason for this
behaviour is due to the smallness of $A_{\LO}$ due to cancellations
between bosonic and fermionic loops, while $A_{\NLO}$ is relatively
large, also due to absence of the accidental cancellations as the ones
observed in the $\Pg\Pg$ channel. Around $\MH = 150\UGeV$ the
credibility of our estimate for the effect of the NNLO corrections
becomes more questionable.
\end{sloppypar}

It is worth noting that for $\PH \to VV$ (see \refS{se:H4f}) the
situation is different. There is no accidentally small LO (there
SM3=SM4 in LO) and the square of $A_{\NLO}$ is taken into account by
the leading NNLO term taken from \Bref{Djouadi:1997rj}, giving the
$15\%$ at NNLO, which serves as our error estimate.

The decay mode $\PH\to \PGg\PZ$ is treated at LO only, since the NLO QCD
corrections within the SM3 are known to be small \cite{Spira:1991tj} and
can thus safely be neglected. The EW corrections in the SM3 as
well as the SM4 are unknown. This implies a theoretical uncertainty of
the order of $100\%$ in the intermediate Higgs-mass range within the SM4,
since large cancellations between the $\PW$ and fermion loops emerge at
LO similar to the decay mode $\PH\to \PGg\PGg$.

\subsection{Numerical results}
The results for the Higgs-boson production cross section via gluon
fusion have been obtained by including the NLO QCD corrections with
full quark-mass dependence \cite{Spira:1995rr} and the NNLO QCD
corrections in the limit of heavy quarks
\cite{Anastasiou:2011qw,Anastasiou:2010bt}.  The full EW corrections
\cite{Passarino:2011kv} have been included in factorised from as
discussed in \refS{Hprod}. We used the MSTW2008NNLO parton
density functions \cite{Martin:2009iq} with the strong coupling
normalised to $\alphas(\MZ)=0.11707$ at NNLO. The results for the
cross sections are displayed in \refT{tab:ggf-sm4} for a
centre-of-mass energy of $\sqrt{s}=7 \UTeV$. The renormalisation and
factorisation scales have been chosen as
$\mu_\mathrm{R}=\mu_\mathrm{F}=\MH/2$. These numbers are larger by
factors of $4{-}9$ than the corresponding SM3 cross sections. The QCD
uncertainties are about the same as in the SM3 case, while the
additional uncertainties due to the EW corrections have been discussed
in \refS{Hprod}.

The results for the Higgs branching fractions have been obtained in a
similar way as those for the results in SM3 in
\Brefs{Dittmaier:2011ti,Denner:2011mq}. While the partial widths for
$\PH\to\PW\PW/\PZ\PZ$ have been computed with {\sc Prophecy4f}, all
other partial widths have been calculated with {\sc  HDECAY}%
\footnote{Note that the EW corrections to the decay width of
  $H\to\gamma\gamma$ are omitted in the results presented in
  \refTs{tab:br-sm4-vv1}--\ref{tab:br-sm4-vv2}. This does practically
  not influence the other branching ratios.}. Then, the branching
ratios and the total width have been calculated from these partial
widths. The full results of the Higgs branching fractions for Higgs
masses up to $1 \UTeV$ are shown in
\refTs{tab:br-sm4-ff1}--\ref{tab:br-sm4-ff2} for the 2-fermion final
states and in \refTs{tab:br-sm4-vv1}--\ref{tab:br-sm4-vv2} for the
2-gauge-boson final states.  In the latter table also the total Higgs
width is given. \refTs{tab:br-sm4-comb1}--\ref{tab:br-sm4-comb2} list
the branching fractions for the $\Pep\Pem\Pep\Pem$ and
$\Pep\Pem\PGmp\PGmm$ final states as well as several combined
channels. Apart from the sum of all 4-fermion final states ($\PH \to
4\Pf$) the results for all-leptonic final states $\PH \to 4\Pl$ with
$\Pl = \Pe,\PGm,\PGt,\PGne,\PGnGm,\PGnGt$, the results for
all-hadronic final states $\PH \to 4\Pq$ with $\Pq =
\PQu,\PQd,\PQc,\PQs,\PQb$ and the semi-leptonic final states $\PH \to
2\Pl2\Pq$ are shown. To compare with the pure SM3,
\refF{fig:br-sm4-ratio} shows the ratios between the SM4 and SM3
branching fractions for the most important channels.

\begin{table}
\caption{SM4 Higgs-boson production cross section via gluon fusion
including NNLO QCD and NLO EW corrections using MSTW2008NNLO PDFs for
$\sqrt{s}=7 \UTeV$.}
\label{tab:ggf-sm4}
\tabcolsep 3.5pt
\begin{tabular}{cc|cc|cc|cc|cc}
\hline
$\MH$ [GeV] & $\sigma$ [pb] &
$\MH$ [GeV] & $\sigma$ [pb] &
$\MH$ [GeV] & $\sigma$ [pb] &
$\MH$ [GeV] & $\sigma$ [pb] &
$\MH$ [GeV] & $\sigma$ [pb] \\
\hline
$100 $&$ 244  $&$ 166 $&$ 77.5 $&$ 188 $&$ 56.7 $&$ 339 $&$ 13.5 $&$ 400  $&$ 9.59   $\\
$110 $&$ 199  $&$ 167 $&$ 76.3 $&$ 189 $&$ 56.0 $&$ 340 $&$ 13.5 $&$ 410  $&$ 8.80   $\\
$120 $&$ 165  $&$ 168 $&$ 75.2 $&$ 190 $&$ 55.2 $&$ 341 $&$ 13.4 $&$ 420  $&$ 8.04   $\\
$130 $&$ 138  $&$ 169 $&$ 74.1 $&$ 195 $&$ 51.8 $&$ 342 $&$ 13.3 $&$ 430  $&$ 7.33   $\\
$140 $&$ 117  $&$ 170 $&$ 73.0 $&$ 200 $&$ 48.6 $&$ 343 $&$ 13.3 $&$ 440  $&$ 6.67   $\\
$145 $&$ 107  $&$ 171 $&$ 71.9 $&$ 210 $&$ 43.0 $&$ 344 $&$ 13.2 $&$ 450  $&$ 6.05   $\\
$150 $&$ 99.2 $&$ 172 $&$ 70.9 $&$ 220 $&$ 38.2 $&$ 345 $&$ 13.1 $&$ 460  $&$ 5.49   $\\
\hline
\end{tabular}

\end{table}

\begin{table}
\vspace{-\headsep}
\caption{SM4 Higgs branching fractions for 2-fermion decay channels in
  the low- and intermediate-mass region.}
\label{tab:br-sm4-ff1}
\renewcommand{\arraystretch}{0.9}
\begin{tabular}{ccccccc}
\hline $\vphantom{\bar{\bar{\bar{\PQb}}}}$%
$\MH$ [GeV]&   $\Hbb$ & $\Htautau$ & $\Hmumu$ &   $\Hss$ &   $\Hcc$ &   $\Htt$ \\ \hline
  100 & $5.70\cdot 10^{-1}$ & $5.98\cdot 10^{-2}$ & $2.08\cdot 10^{-4}$ & $2.44\cdot 10^{-4}$ & $2.88\cdot 10^{-2}$ & $0.00$ \\
  110 & $5.30\cdot 10^{-1}$ & $5.67\cdot 10^{-2}$ & $1.97\cdot 10^{-4}$ & $2.26\cdot 10^{-4}$ & $2.68\cdot 10^{-2}$ & $0.00$ \\
  120 & $4.87\cdot 10^{-1}$ & $5.29\cdot 10^{-2}$ & $1.84\cdot 10^{-4}$ & $2.08\cdot 10^{-4}$ & $2.46\cdot 10^{-2}$ & $0.00$ \\
  130 & $4.36\cdot 10^{-1}$ & $4.82\cdot 10^{-2}$ & $1.67\cdot 10^{-4}$ & $1.86\cdot 10^{-4}$ & $2.20\cdot 10^{-2}$ & $0.00$ \\
  140 & $3.72\cdot 10^{-1}$ & $4.17\cdot 10^{-2}$ & $1.45\cdot 10^{-4}$ & $1.59\cdot 10^{-4}$ & $1.88\cdot 10^{-2}$ & $0.00$ \\
  145 & $3.33\cdot 10^{-1}$ & $3.75\cdot 10^{-2}$ & $1.30\cdot 10^{-4}$ & $1.42\cdot 10^{-4}$ & $1.68\cdot 10^{-2}$ & $0.00$ \\
  150 & $2.83\cdot 10^{-1}$ & $3.20\cdot 10^{-2}$ & $1.11\cdot 10^{-4}$ & $1.21\cdot 10^{-4}$ & $1.42\cdot 10^{-2}$ & $0.00$ \\
\hline\end{tabular}

\end{table}

\begin{table}
\vspace{-\headsep}
\caption{SM4 Higgs branching fractions for 2-fermion decay channels in
  the high-mass region.}
\label{tab:br-sm4-ff2}
\renewcommand{\arraystretch}{0.9}
\begin{tabular}{ccccccc}
\hline $\vphantom{\bar{\bar{\bar{\PQb}}}}$%
$\MH$ [GeV]&   $\Hbb$ & $\Htautau$ & $\Hmumu$ &   $\Hss$ &   $\Hcc$ &   $\Htt$ \\ \hline
  230 & $6.26\cdot 10^{-3}$ & $7.69\cdot 10^{-4}$ & $2.66\cdot 10^{-6}$ & $2.67\cdot 10^{-6}$ & $3.15\cdot 10^{-4}$ & $0.00$ \\
  240 & $5.42\cdot 10^{-3}$ & $6.71\cdot 10^{-4}$ & $2.33\cdot 10^{-6}$ & $2.31\cdot 10^{-6}$ & $2.73\cdot 10^{-4}$ & $0.00$ \\
  250 & $4.73\cdot 10^{-3}$ & $5.89\cdot 10^{-4}$ & $2.04\cdot 10^{-6}$ & $2.01\cdot 10^{-6}$ & $2.38\cdot 10^{-4}$ & $0.00$ \\
  260 & $4.17\cdot 10^{-3}$ & $5.24\cdot 10^{-4}$ & $1.82\cdot 10^{-6}$ & $1.78\cdot 10^{-6}$ & $2.10\cdot 10^{-4}$ & $2.40\cdot 10^{-7}$ \\
  270 & $3.71\cdot 10^{-3}$ & $4.69\cdot 10^{-4}$ & $1.62\cdot 10^{-6}$ & $1.58\cdot 10^{-6}$ & $1.86\cdot 10^{-4}$ & $1.07\cdot 10^{-5}$ \\
  280 & $3.32\cdot 10^{-3}$ & $4.23\cdot 10^{-4}$ & $1.47\cdot 10^{-6}$ & $1.41\cdot 10^{-6}$ & $1.67\cdot 10^{-4}$ & $5.13\cdot 10^{-5}$ \\
  290 & $2.99\cdot 10^{-3}$ & $3.82\cdot 10^{-4}$ & $1.33\cdot 10^{-6}$ & $1.27\cdot 10^{-6}$ & $1.50\cdot 10^{-4}$ & $1.45\cdot 10^{-4}$ \\
\hline\end{tabular}

\end{table}

\begin{table}
\vspace{-\headsep}
\caption{SM4 Higgs branching fractions for 2-gauge-boson decay
  channels and total Higgs width in the low- and intermediate-mass region.}
\label{tab:br-sm4-vv1}
\renewcommand{\arraystretch}{0.9}
\renewcommand{\arraystretch}{0.9}
\begin{tabular}{ccccccc}
\hline$\MH$ [GeV]&   $\Hgg$ & $\Hgaga$ &  $\HZga$ &   $\HWW$ &   $\HZZ$ & $\Gamma_{\PH}$ [GeV] \\ \hline
  100 & $3.39\cdot 10^{-1}$ & $1.31\cdot 10^{-4}$ & $1.70\cdot 10^{-5}$ & $1.67\cdot 10^{-3}$ & $1.35\cdot 10^{-4}$ & $5.52\cdot 10^{-3}$ \\
  110 & $3.78\cdot 10^{-1}$ & $1.72\cdot 10^{-4}$ & $1.35\cdot 10^{-4}$ & $7.20\cdot 10^{-3}$ & $5.83\cdot 10^{-4}$ & $6.41\cdot 10^{-3}$ \\
  120 & $4.10\cdot 10^{-1}$ & $2.26\cdot 10^{-4}$ & $4.06\cdot 10^{-4}$ & $2.27\cdot 10^{-2}$ & $2.37\cdot 10^{-3}$ & $7.49\cdot 10^{-3}$ \\
  130 & $4.28\cdot 10^{-1}$ & $2.95\cdot 10^{-4}$ & $8.51\cdot 10^{-4}$ & $5.77\cdot 10^{-2}$ & $7.12\cdot 10^{-3}$ & $8.92\cdot 10^{-3}$ \\
  140 & $4.20\cdot 10^{-1}$ & $3.81\cdot 10^{-4}$ & $1.46\cdot 10^{-3}$ & $1.29\cdot 10^{-1}$ & $1.68\cdot 10^{-2}$ & $1.11\cdot 10^{-2}$ \\
  145 & $4.01\cdot 10^{-1}$ & $4.29\cdot 10^{-4}$ & $1.80\cdot 10^{-3}$ & $1.86\cdot 10^{-1}$ & $2.36\cdot 10^{-2}$ & $1.28\cdot 10^{-2}$ \\
  150 & $3.63\cdot 10^{-1}$ & $4.74\cdot 10^{-4}$ & $2.13\cdot 10^{-3}$ & $2.75\cdot 10^{-1}$ & $3.09\cdot 10^{-2}$ & $1.55\cdot 10^{-2}$ \\
\hline\end{tabular}

\end{table}

\begin{table}
\vspace{-\headsep}
\caption{SM4 Higgs branching fractions for 2-gauge-boson decay
  channels and total Higgs width in the high-mass region.}
\label{tab:br-sm4-vv2}
\renewcommand{\arraystretch}{0.9}
\renewcommand{\arraystretch}{0.9}
\begin{tabular}{ccccccc}
\hline$\MH$ [GeV]&   $\Hgg$ & $\Hgaga$ &  $\HZga$ &   $\HWW$ &   $\HZZ$ & $\Gamma_{\PH}$ [GeV] \\ \hline
  230 & $1.81\cdot 10^{-2}$ & $5.65\cdot 10^{-5}$ & $3.05\cdot 10^{-4}$ & $6.95\cdot 10^{-1}$ & $2.79\cdot 10^{-1}$ & $9.94\cdot 10^{-1}$ \\
  240 & $1.71\cdot 10^{-2}$ & $5.20\cdot 10^{-5}$ & $2.75\cdot 10^{-4}$ & $6.92\cdot 10^{-1}$ & $2.84\cdot 10^{-1}$ & $1.19$ \\
  250 & $1.62\cdot 10^{-2}$ & $4.80\cdot 10^{-5}$ & $2.48\cdot 10^{-4}$ & $6.93\cdot 10^{-1}$ & $2.86\cdot 10^{-1}$ & $1.41$ \\
  260 & $1.56\cdot 10^{-2}$ & $4.49\cdot 10^{-5}$ & $2.26\cdot 10^{-4}$ & $6.91\cdot 10^{-1}$ & $2.90\cdot 10^{-1}$ & $1.65$ \\
  270 & $1.50\cdot 10^{-2}$ & $4.22\cdot 10^{-5}$ & $2.05\cdot 10^{-4}$ & $6.88\cdot 10^{-1}$ & $2.93\cdot 10^{-1}$ & $1.92$ \\
  280 & $1.46\cdot 10^{-2}$ & $4.00\cdot 10^{-5}$ & $1.88\cdot 10^{-4}$ & $6.88\cdot 10^{-1}$ & $2.95\cdot 10^{-1}$ & $2.21$ \\
  290 & $1.42\cdot 10^{-2}$ & $3.81\cdot 10^{-5}$ & $1.72\cdot 10^{-4}$ & $6.86\cdot 10^{-1}$ & $2.97\cdot 10^{-1}$ & $2.53$ \\
\hline\end{tabular}

\end{table}

\begin{table}
\vspace{-\headsep}
\caption{SM4 Higgs branching fractions for 4-fermion final states in
  the low- and intermediate-mass
  region with $\Pl = \Pe,\PGm,\PGt,\PGne,\PGnGm,\,\PGnGt$ and $\Pq = \PQu,\PQd,\PQc,\PQs,\PQb$.}
\label{tab:br-sm4-comb1}
\renewcommand{\arraystretch}{0.9}
\renewcommand{\arraystretch}{0.88}
\begin{tabular}{ccccccc}
\hline$\MH$ [GeV]&   $\PH \to 4\Pe$ & $\PH \to 2\Pe 2\PGm$ &   $\PH \to 4\Pl$ &   $\PH \to 4\Pq$ & $\PH \to 2\Pl 2\Pq$ &   $\PH \to 4\Pf$ \\ \hline 
   100 & $2.26\cdot 10^{-7}$ & $3.31\cdot 10^{-7}$ & $1.67\cdot 10^{-4}$ & $7.48\cdot 10^{-4}$ & $7.85\cdot 10^{-4}$ & $1.70\cdot 10^{-3}$ \\
   110 & $7.81\cdot 10^{-7}$ & $1.27\cdot 10^{-6}$ & $7.32\cdot 10^{-4}$ & $3.53\cdot 10^{-3}$ & $3.36\cdot 10^{-3}$ & $7.62\cdot 10^{-3}$ \\
   120 & $2.76\cdot 10^{-6}$ & $4.99\cdot 10^{-6}$ & $2.36\cdot 10^{-3}$ & $1.17\cdot 10^{-2}$ & $1.08\cdot 10^{-2}$ & $2.48\cdot 10^{-2}$ \\
   130 & $8.00\cdot 10^{-6}$ & $1.48\cdot 10^{-5}$ & $6.12\cdot 10^{-3}$ & $3.05\cdot 10^{-2}$ & $2.78\cdot 10^{-2}$ & $6.44\cdot 10^{-2}$ \\
   140 & $1.83\cdot 10^{-5}$ & $3.51\cdot 10^{-5}$ & $1.38\cdot 10^{-2}$ & $6.88\cdot 10^{-2}$ & $6.24\cdot 10^{-2}$ & $1.45\cdot 10^{-1}$ \\
   145 & $2.54\cdot 10^{-5}$ & $4.89\cdot 10^{-5}$ & $1.99\cdot 10^{-2}$ & $9.98\cdot 10^{-2}$ & $8.99\cdot 10^{-2}$ & $2.10\cdot 10^{-1}$ \\
   150 & $3.33\cdot 10^{-5}$ & $6.42\cdot 10^{-5}$ & $2.91\cdot 10^{-2}$ & $1.45\cdot 10^{-1}$ & $1.31\cdot 10^{-1}$ & $3.05\cdot 10^{-1}$ \\
\hline\end{tabular}

\end{table}

\begin{table}
\vspace{-\headsep}
\caption{SM4 Higgs branching fractions for 4-fermion final states in
  the high-mass
  region with $\Pl = \Pe,\PGm,\PGt,\PGne,\PGnGm,\,\PGnGt$ and $\Pq = \PQu,\PQd,\PQc,\PQs,\PQb$.}
\label{tab:br-sm4-comb2}
\renewcommand{\arraystretch}{0.88}
\begin{tabular}{ccccccc}
\hline$\MH$ [GeV]&   $\PH \to 4\Pe$ & $\PH \to 2\Pe 2\PGm$ &   $\PH \to 4\Pl$ &   $\PH \to 4\Pq$ & $\PH \to 2\Pl 2\Pq$ &   $\PH \to 4\Pf$ \\ \hline 
   230 & $2.83\cdot 10^{-4}$ & $5.70\cdot 10^{-4}$ & $9.04\cdot 10^{-2}$ & $4.70\cdot 10^{-1}$ & $4.14\cdot 10^{-1}$ & $9.74\cdot 10^{-1}$ \\
   240 & $2.88\cdot 10^{-4}$ & $5.80\cdot 10^{-4}$ & $9.04\cdot 10^{-2}$ & $4.71\cdot 10^{-1}$ & $4.15\cdot 10^{-1}$ & $9.76\cdot 10^{-1}$ \\
   250 & $2.90\cdot 10^{-4}$ & $5.86\cdot 10^{-4}$ & $9.05\cdot 10^{-2}$ & $4.71\cdot 10^{-1}$ & $4.16\cdot 10^{-1}$ & $9.78\cdot 10^{-1}$ \\
   260 & $2.93\cdot 10^{-4}$ & $5.92\cdot 10^{-4}$ & $9.05\cdot 10^{-2}$ & $4.72\cdot 10^{-1}$ & $4.16\cdot 10^{-1}$ & $9.79\cdot 10^{-1}$ \\
   270 & $2.95\cdot 10^{-4}$ & $5.97\cdot 10^{-4}$ & $9.05\cdot 10^{-2}$ & $4.73\cdot 10^{-1}$ & $4.17\cdot 10^{-1}$ & $9.80\cdot 10^{-1}$ \\
   280 & $2.98\cdot 10^{-4}$ & $6.02\cdot 10^{-4}$ & $9.06\cdot 10^{-2}$ & $4.73\cdot 10^{-1}$ & $4.17\cdot 10^{-1}$ & $9.81\cdot 10^{-1}$ \\
   290 & $2.99\cdot 10^{-4}$ & $6.08\cdot 10^{-4}$ & $9.06\cdot 10^{-2}$ & $4.74\cdot 10^{-1}$ & $4.18\cdot 10^{-1}$ & $9.82\cdot 10^{-1}$ \\
\hline\end{tabular}

\renewcommand{\arraystretch}{0.9}
\end{table}

\begin{figure}
\centerline{\includegraphics[height=9cm]{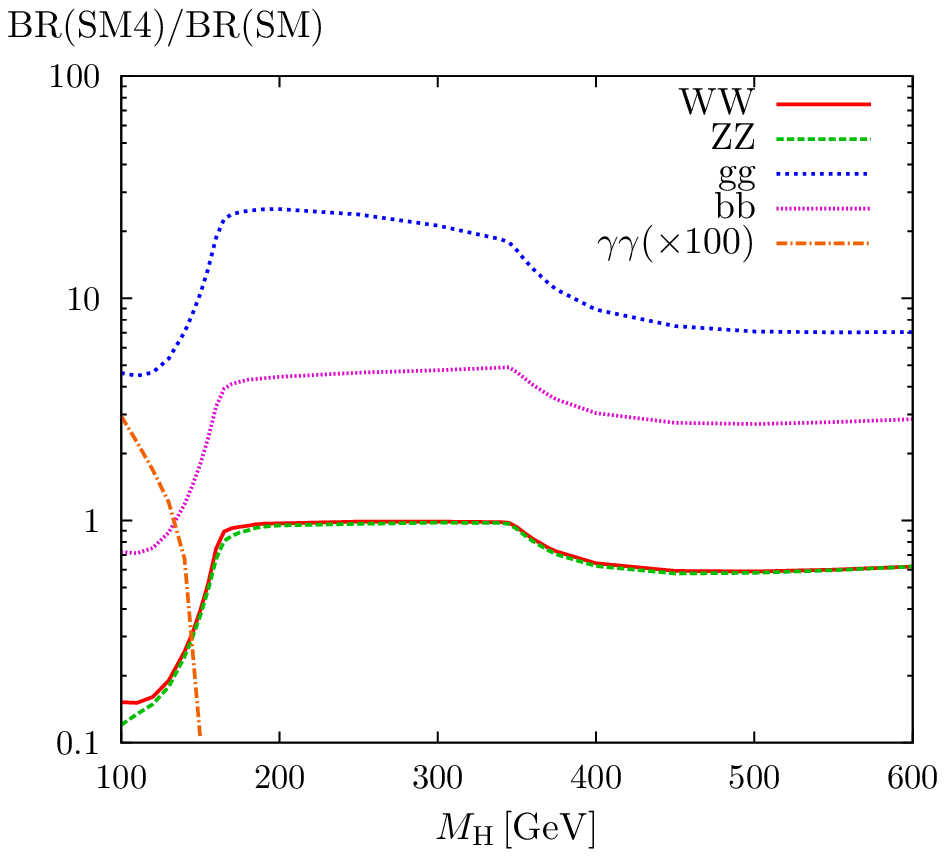}}
\vspace*{-.5em}
\caption{Ratio of branching fractions in SM4 with respect to SM3 for
  $\PW\PW$, $\PZ\PZ$, $\Pg\Pg$, $\PQb\bar{\PQb}$, and $\PGg\PGg$ decay
  channels ($\PGg\PGg$ ratio multiplied with $100$).}
\label{fig:br-sm4-ratio}
\end{figure}

Results for NLO EW corrections to the $\PH \to \PGg\PGg$ decay width are
shown in \refT{tab:EWNLOGG} for the window $[90{-}150]\GeV$.  The
effect on the branching fraction is shown in \refT{tab:EWNLOBRGG}.
The branching ratio for $\PH \to \PGg\PGg$ is strongly reduced in SM4.
\begin{table}
\begin{center}
\caption[]{\label{tab:EWNLOGG}{NLO EW corrections to the $\PH
    \to \PGg\PGg$ decay width according to \eqn{EWnewd} and estimate for the
    missing higher order (h.o.)  corrections from \eqn{EWnewderr}.}}
\vspace{1em}
\begin{tabular}{ccc}
\hline 
\rule[-1ex]{0ex}{3.5ex}%
$\MH\,$[GeV] & 
$\delta_{\myEW}^{(4)}\,[\%]$  & 
missing h.o.$\,[\%]$          \\
\hline
\rule[-0ex]{0ex}{2.5ex}%
$90   $&$    -53.9  $&$   \pm 13.9 $ \\
$95   $&$    -59.2  $&$   \pm 13.9 $ \\
$100  $&$    -64.5  $&$   \pm 13.9 $ \\
$105  $&$    -69.5  $&$   \pm 13.9 $ \\
$110  $&$    -74.4  $&$   \pm 13.9 $ \\
$115  $&$    -79.0  $&$   \pm 13.9 $ \\
$120  $&$    -83.3  $&$   \pm 13.9 $ \\
$125  $&$    -87.3  $&$    {+}13.9\,{-}12.7 $ \\
$130  $&$    -90.8  $&$    {+}13.9\,{-}9.2  $  \\
$135  $&$    -94.0  $&$    {+}13.9\,{-}6.0  $  \\
$140  $&$    -96.6  $&$    {+}13.9\,{-}3.4  $  \\
$145  $&$    -98.5  $&$    {+}13.9\,{-}1.5  $  \\
$150  $&$    -99.7  $&$    {+}13.9\,{-}0.3  $  
\rule[-1ex]{0ex}{2.5ex}\\
\hline
\end{tabular}
\end{center}
\end{table}
\begin{table}
\begin{center}
\caption[]{\label{tab:EWNLOBRGG}{Higgs branching fractions for $\PGg\PGg$ decay channel without and with NLO EW corrections.}}  \vspace{1em}
\begin{tabular}{ccc}
\hline 
\rule[-1ex]{0ex}{3.5ex}%
$\MH\,$[GeV] & 
w/o NLO EW       & 
w/ NLO EW           \\
\hline
\rule[0ex]{0ex}{2.5ex}%
$100$ & $1.31\,\cdot\,10^{-4}$  & $4.65\,\cdot\,10^{-5}$ \\    
$110$ & $1.72\,\cdot\,10^{-4}$  & $4.40\,\cdot\,10^{-5}$ \\    
$120$ & $2.26\,\cdot\,10^{-4}$  & $3.77\,\cdot\,10^{-5}$ \\    
$130$ & $2.95\,\cdot\,10^{-4}$  & $2.71\,\cdot\,10^{-5}$ \\    
$140$ & $3.81\,\cdot\,10^{-4}$  & $1.30\,\cdot\,10^{-5}$ \\    
$150$ & $4.74\,\cdot\,10^{-4}$  & $1.42\,\cdot\,10^{-6}$%
\rule[-1ex]{0ex}{2.5ex}\\
\hline
\end{tabular}
\end{center}
\end{table}

\clearpage


\clearpage


\newpage
\DeclareRobustCommand{\PV}{\HepParticle{V}{}{}\Xspace}
\DeclareRobustCommand{\PX}{\HepParticle{X}{}{}\Xspace}
\DeclareRobustCommand{\Pf}{\HepParticle{f}{}{}\Xspace}
\DeclareRobustCommand{\PF}{\HepParticle{F}{}{}\Xspace}
\newcommand{\myFR}{\rm{\scriptscriptstyle{FR}}}
\newcommand{\OS}{\mathrm{OS}}
\newcommand{\CPP}{\mathrm{CPP}}
\newcommand{\mUV}{\mathrm{UV}}
\newcommand{\mBW}{\mathrm{BW}}
\newcommand{\OFS}{\mathrm{OFS}}
\newcommand{\ssR}{{\mathrm{R}}}
\newcommand{\ssP}{{\mathrm{P}}}
\newcommand{\ssV}{{\mathrm{V}}}
\newcommand{\ssM}{{\mathrm{M}}}
\newcommand{\sH}{\mathrm{H}}
\newcommand{\HH}{\mathrm{HH}}
\newcommand{\VV}{\mathrm{VV}}
\newcommand{\ssWW}{{\scriptscriptstyle{WW}}}
\newcommand{\hatp}{{\hat p}}
\newcommand{\intrp}{\int_0^{\infty}}
\newcommand{\intMB}{\int_{-i\,\infty}^{+i\,\infty}}
\newcommand{\ord}[1]{{\cal O}\lpar#1\rpar}
\newcommand{\oD}{{\overline \Delta}}
\newcommand{\mv}{\mathswitch {M_{\PV}}}
\newcommand{\mbw}{\mathswitch {m_{\PW}}}
\newcommand{\mbws}{\mathswitch {m^2_{\PW}}}
\newcommand{\mhs}{\mathswitch {M^2_{\PH}}}
\newcommand{\mBh}{\mathswitch {{\overline M}_{\PH}}}
\newcommand{\mBhs}{\mathswitch {{\overline M}^2_{\PH}}}
\newcommand{\mhq}{\mathswitch {M^4_{\PH}}}
\newcommand{\mws}{\mathswitch {M^2_{\PW}}}
\newcommand{\mzs}{\mathswitch {M^2_{\PZ}}}
\newcommand{\cph}{\mathswitch {s_{\PH}}}
\newcommand{\cpw}{\mathswitch {s_{\PW}}}
\newcommand{\cpz}{\mathswitch {s_{\PZ}}}
\newcommand{\cpt}{\mathswitch {s_{\PQt}}}
\newcommand{\muh}{\mathswitch {\mu_{\PH}}}
\newcommand{\muhs}{\mathswitch {\mu^2_{\PH}}}
\newcommand{\gh}{\mathswitch {\gamma_{\PH}}}
\newcommand{\GOS}{\mathswitch {\Gamma_{\OS}}}
\newcommand{\GOL}{\mathswitch {\overline\Gamma}}
\newcommand{\muol}{\mathswitch {\overline\mu}}
\newcommand{\GOSu}{\mathswitch {\Gamma^{\OS}}}
\newcommand{\shat}{\mathswitch {\hat s}}
\providecommand{\muR}{\mathswitch {\mu_{\ssR}}}
\providecommand{\muF}{\mathswitch {\mu_{\ssF}}}
\newcommand{\muRs}{\mathswitch {\mu^2_{\ssR}}}
\newcommand{\muFs}{\mathswitch {\mu^2_{\ssF}}}
\newcommand{\tot}{{\mbox{\scriptsize tot}}}
\newcommand{\myprod}{{\mbox{\scriptsize prod}}}
\newcommand{\prop}{{\mbox{\scriptsize prop}}}
\newcommand{\dec}{{\mbox{\scriptsize dec}}}
\newcommand{\bos}{{\mbox{\scriptsize bos}}}
\newcommand{\peak}{{\mbox{\scriptsize peak}}}
\newcommand{\ren}{{\mbox{\scriptsize ren}}}
\newcommand{\mdiv}{{\mbox{\scriptsize div}}}
\newcommand{\ac}{{\mbox{\scriptsize all}}}
\newcommand{\HO}{{\mathrm{HO}}}
\newcommand{\eG}[1]{\mathswitch \Gamma\lpar #1 \rpar}
\newcommand{\eGs}[1]{\mathswitch \Gamma^2\lpar #1 \rpar}
\renewcommand{\ep}{\mathswitch \varepsilon}
\newcommand{\epb}{\mathswitch {\overline\varepsilon}}
\section{The (heavy) Higgs-boson lineshape\footnote{G.~Passarino (ed.), S.~Goria and D.~Rosco.}}
\label{sec:po}

\subsection{Introduction \label{PO:intro}}
In this section we consider the problem of a consistent definition of the Higgs-boson lineshape 
at LHC and address the question of optimal presentation of LHC data.

Some preliminary observations are needed: the Higgs system of the Standard Model has almost 
a non-perturbative behaviour, even at a low scale. Let us consider the traditional on-shell 
approach and the values for the on-shell decay width of the Higgs boson.

For $\MH = 140\UGeV$ the total width is $8.12\times 10^{-3}\UGeV$. The process
$\PH \to \PAQb \PQb$ has a partial with of $2.55\times 10^{-3}\UGeV$, other two-body
decays are almost negligible while $\PH \to 4\,\Pf$ has $4.64\times 10^{-3}\UGeV$, well below
the $\PW\PW$ threshold. Therefore the four-body decay, even below threshold, is more important
than the two-body ones. 
What are the corresponding implications? Since decay widths are related to the imaginary parts
of loop diagrams, the above statement is equivalent to say that, in terms of imaginary parts 
of the $\PH$ self-energy, three-loop diagrams are as important as one-loop diagrams.

We realise that most of the people just want to use some well-defined recipe
without having to dig any deeper; however, there is no alternative to a complete description
of LHC processes which has to include the complete matrix elements for all processes;
splitting the whole $S\,$-matrix element into components is just conventional wisdom.
However, the precise tone and degree of formality must be dictated by gauge invariance.

The framework discussed in this section is general enough and can be used for all
processes and for all kinematical regions; however, here the main focus will be on the
description of the Higgs-boson lineshape in the heavy-Higgs region, typically above
$600\UGeV$. The general argument is well documented in the literature and, for a complete 
description of all technical details, we refer to 
\Brefs{Actis:2008uh,Passarino:2010qk,Goria:2011wa}.
Part of the results discussed in this section and concerning a correct implementation 
of the Higgs-boson lineshape have been summarised in a talk at the BNL meeting of the 
HXSWG.\footnote{http://www.bnl.gov/hcs/}

One might wonder why considering a SM Higgs boson in such a high-mass range. There are 
are classic constraints on the Higgs-boson mass coming from unitarity, triviality, and
vacuum stability, precision electroweak data and absence of fine-tuning. However,
the search for a SM Higgs boson over a mass range from $80\UGeV$ to $1\UTeV$
is clearly indicated as a priority in many experimental papers, \eg\ 
{\it ATLAS: letter of intent for a general-purpose pp experiment at the large hadron collider 
at CERN}\footnote{http://atlas.web.cern.ch/Atlas/internal/tdr.html}. 

Coming back to the framework that we are introducing here, there is another important issue:
when working in the on-shell scheme one finds that the two-loop corrections to the on-shell 
Higgs width exceed the one-loop corrections if the on-shell Higgs mass is larger
than $900\UGeV$, as discussed in \Bref{Ghinculov:1996py}. This fact simply tells you
that perturbation theory diverges badly, starting from approximately $1\UTeV$.

The current recipe to handle theoretical uncertainty for heavy-Higgs-boson lineshape at
LHC is to assign a conservative estimate of $150\%\,\MH^3$ with \MH\ in 
{}\UTeV{}~\footnote{https://twiki.cern.ch/twiki/bin/view/LHCPhysics/HeavyHiggs}. 
This is equivalent to assume a $\pm 32\%$ at $\MH= 600\UGeV$ and
$\pm 77\%$ at $\MH= 800\UGeV$.

Recently the problem of going beyond the zero-width approximation has received 
new boost from the work of \Bref{Anastasiou:2011pi}: the program {\sc iHixs} allows
the study of the Higgs-boson lineshape for a finite width of the Higgs boson and computes the
cross section sampling over a Breit--Wigner distribution.

To start our discussion we consider the process $i j \to \PH ( \to \PF )  + \PX$ where 
$i,j\in\,$partons and $\PF$ is a generic final state (\eg $\PF= \PGg \PGg, 4\,\Pf$, \etc). For 
the sake of simplicity we neglect, for a moment, folding the partonic process with parton
distribution functions (PDFs). Since the Higgs boson is a scalar resonance we can split the 
whole process into three parts: {\em production}, {\em propagation}, and {\em decay}.
\subsection{Propagation\label{PO:prop}}
In quantum field theory (QFT) amplitudes are made out of propagators and vertices:
the Higgs (Dyson-resummed) propagator reads as follows:
\bq
\Delta_{\PH}(s) = \Bigl[ s - \mhs + 
S_{\HH}\lpar s,\Mt^2,\MH^2,\MW^2,\MZ^2\rpar \Bigr]^{-1},
\eq
where $\Mt$ and $M_i$ are a renormalised masses and $S_{\HH}$ is the renormalised Higgs 
self-energy (to all orders but with one-particle-irreducible diagrams). The first 
argument of the self-energy is the external momentum squared, the 
remaining ones are squared (renormalised) masses in the loops. 
We define complex poles for unstable particles as the (complex) solutions of the following system:
\bqa
\cph - \mhs + S_{\HH}\lpar \cph,\Mt^2,\MH^2,\MW^2,\MZ^2\rpar &=& 0,
\nl
\cpw - \mws + S_{\WW}\lpar \cpw,\Mt^2,\MH^2,\MW^2,\MZ^2\rpar &=& 0,
\eqa
\etc To lowest-order accuracy the Higgs propagator is rewritten as
\bq
\Delta^{-1}_{\PH} =  s - \cph.
\label{CMSprop}
\eq
The complex pole describing an unstable particle is conventionally parametrised as
\bq
s_i = \mu^2_i - i\,\mu_i\,\gamma_i,
\label{CPpar}
\eq
where $\mu_i$ is an input parameter (similar to the on-shell mass) while $\gamma_i$ can be 
computed (as the on-shell total width), say within the Standard Model (SM). 
Note that the pole of $\Delta$ fully embodies the propagation properties of a particle. 
We know that even in ordinary quantum mechanics the resonances are described as the 
complex energy poles in the scattering amplitude. 
The general formalism for describing unstable particles in QFT was developed long ago,
see \Brefs{Veltman:1963th,Jacob:1961zz,Valent:1974bd,Lukierski:1978ke,Bollini:1993yp};
for an implementation in gauge theories we refer to the work of
\Brefs{Grassi:2001bz,Grassi:2000dz,Kniehl:1998vy,Kniehl:1998fn}.

Consider the complex-mass scheme (CMS) introduced in \Bref{Denner:2005fg}
and extended at two-loop level in \Bref{Actis:2006rc}: here, at lowest level of accuracy, 
we use
\bq
\cph - \mhs + S_{\HH}\lpar \cph,\cpt,\cph,\cpw,\cpz\rpar = 0,
\label{exact}
\eq
where now $S_{\HH}$ is computed at one-loop level and $\cpw$ \etc are the experimental 
complex poles, derived from the corresponding on-shell masses and widths. For the $\PW$ 
and $\PZ$ bosons the input-parameter set (IPS) is defined in terms of pseudo-observables; 
at first, on-shell quantities are derived by fitting the experimental lineshapes with
\bq
\Sigma_{\VV}(s) = \frac{N}{(s - M^2_{\OS})^2 + 
s^2\,\Gamma^2_{\OS}/M^2_{\OS}}, \qquad V = \PW, \PZ,
\eq
where $N$ is an irrelevant (for our purposes) normalisation constant.
Secondly, we define pseu\-do-ob\-ser\-vab\-les 
\bq
M_{\ssP} = M_{\OS}\,\cos\psi, \qquad  
\Gamma_{\ssP} = \GOS\,\sin\psi, \qquad
\psi = \arctan \frac{\GOS}{M_{\OS}},
\eq
which are inserted in the IPS.

Renormalisation with complex poles should not be confused with a simple
recipe for the replacement of running widths with constant widths; there are 
many more ingredients in the scheme.
It is worth noting that perturbation theory based on $\cph$ instead of the on-shell mass has
much better convergence properties; indeed, as shown in \Bref{Ghinculov:1996py}, the two-loop 
corrections to the imaginary part of $\cph$ become as large as the one-loop ones {\em only}
when $\muh = 1.74\UTeV$. This suggests that the complex pole scheme is preferable also
from the point of view of describing the heavy-$\PH$-boson production at LHC. 

There is a substantial difference between $\PW,\PZ$ complex poles and the Higgs 
complex pole. In the first case $\PW,\PZ$ bosons decay predominantly into two (massless) 
fermions while for the Higgs boson below the $\PW\PW$ threshold the decay into 
four fermions is even larger than the decay into a $\PAQb\PQb$ pair. 
Therefore we cannot use for the Higgs boson the well-known result, valid for $\PW,\PZ$,
\bq
\Im S_{\VV}(s) \approx \frac{\GOSu_{\ssV}}{M^{\OS}_{\ssV}}\,s.
\eq
As a consequence of this fact we have
\bq
\mu^2_{\ssV} = M^2_{\ssV\,\OS} - \Gamma^2_{\ssV\,\OS} + \HO,
\quad
\gamma_{\ssV} = \GOSu_{\ssV}\,\Bigl[ 1 - \frac{1}{2}\,
\Bigl(\frac{\GOSu_{\ssV}}{M^{\OS}_{\ssV}}\Bigr)^2\Bigr] + \HO,
\label{cpV}
\eq
where the perturbative expansion is well under control since 
$\GOSu_{\ssV}/M^{\OS}_{\ssV} \muchless 1$. Here ``HO'' stands for further
higher-order contributions.
For the Higgs boson we have a different expansion,
\bqa
\gh &=& \Gamma_{\PH}\,\bigl[ 1 + \sum_n a_n\,\Gamma^n_{\PH}\bigr] +
\sum_{\ssV=\PW,\PZ}\,\Gamma_{\ssV}\,\frac{\mv}{\MH}
\sum_{n,l} b^{\ssV}_{nl}\,\Gamma^n_{\ssV}\,\Gamma^l_{\PH},
\nl
a_1 &=& - \bigl[ \Im\,S^h_{\HH\,,\,\OS}\bigr]^2,
\quad
a_2 = \frac{1}{2}\,\Bigl[ 
\frac{1}{\MH}\,\Im\,S^h_{\HH\,,\,\OS}
-\,\MH\,\Im\,S^{hh}_{\HH\,,\,\OS} \Bigr],
\nl
b^{\ssV}_{10} &=& -\,\Re\,S^v_{\HH\,,\,\OS}
+ \Re\,S^v_{\HH\,,\,\OS}\,\Re\,S^h_{\HH\,,\,\OS}
- \Im\,S^v_{\HH\,,\,\OS}\,\Im\,S^h_{\HH\,,\,\OS},
\nl
b^{\ssV}_{11} &=& \frac{1}{2\,\MH}\,\Im\,S^v_{\HH\,,\,\OS} -
\MH\,\Im\,S^{hv}_{\HH\,,\,\OS},
\quad
b^{\ssV}_{20} = - \frac{1}{2}\,\frac{\mv}{\MH}\,\Im\,S^{vv}_{\HH\,,\,\OS}.
\label{HO}
\eqa
Here we have $M_i = M^{\OS}_i$, $\Gamma_i = \GOSu_i$ (on-shell quantities) and
\bqa
S^a_{\HH\,,\,\OS} &=& \frac{\partial}{\partial M^2_a}\,
S_{\HH}\lpar \mhs,\Mt^2,\MH^2,\MW^2,\MZ^2\rpar,
\nl
S^{ab}_{\HH\,,\,\OS} &=& \frac{\partial^2}{\partial M^2_a \partial M^2_b}\,
S_{\HH}\lpar \mhs,\Mt^2,\MH^2,\MW^2,\MZ^2\rpar,
\eqa
where only the first few terms in the perturbative expansion are given.
In \eqn{HO} we used the following definition of the on-shell width
\bq
\GOSu_{\PH}\,M^{\OS}_{\PH} = \Im\,S_{\HH\,,\,\OS} + \HO.
\eq
Only the complex pole is gauge-parameter independent to all orders of perturbation 
theory while on-shell quantities are ill-defined beyond lowest order. Indeed, in the 
$R_{\xi}$ gauge, at lowest order, one has the following expression for the 
bosonic part:
\bq
\Im\,S_{\HH\,,\,\bos} = \frac{g^2}{4\,\mws}\,s^2\,\left[
\biggl( 1 - \frac{\mhq}{s^2}\biggr)\,\biggl( 1 - 4\,\xi_{\PW}\,
\frac{\mws}{s}\biggr)^{1/2}\,\theta\lpar s - 4\,\xi_{\PW} \mws\rpar
+ \frac{1}{2}\,\lpar \PW \to \PZ\rpar \right],
\eq
where $\xi_{\PV}$ ($\PV = \PW,\PZ$) are gauge parameters. Note that \eqn{HO} (the expansion) 
involves derivatives.

A technical remark: suppose that we use $\muh$ as a {\em free} input parameter and derive 
$\gh$ in the SM from the equation
\bq
\muh\,\gh =  \Im\,S_{\HH}\lpar \cph,\cpt,\cph,\cpw,\cpz\rpar.
\label{geq}
\eq
Since the bare $\PW,\PZ$ masses are replaced with complex poles also in couplings (to 
preserve gauge invariance) it follows that $\gh$ (solution of \eqn{geq}) is 
renormalisation scale dependent; while counterterms can be introduced to make the
self-energy ultraviolet finite, the $\muR$ dependence drops only in subtracted 
quantities, \ie after final renormalisation, something that would require knowledge of
the experimental Higgs complex pole. Following our intuition we fix the scale
at $\muh$.

After evaluating the coefficients of \eqn{HO} (\eg in the $\xi = 1$ gauge) it is easily 
seen that the expanded solution of \eqn{exact} is not a good approximation to the exact one, 
especially for high values of $\muh$. For instance, for $\muh = 500\UGeV$ we have an exact 
solution $\gh= 58.85\UGeV$ and an expanded one of $62.87\UGeV$ with an on-shell with of $68.0\UGeV$.
Therefore, we will use an exact, numerical solution for $\gh$; details on the
structure of the Higgs self-energy to all orders of perturbation theory can be found
in \Brefs{Actis:2006rb,Actis:2006rc}.

To compute $\gh$ at the same level of accuracy to which $\GOSu_{\PH}$ is known 
(see \Bref{Dittmaier:2011ti}) would require, at least, a three-loop calculation (the first 
instance where we have a four-fermion cut of the Higgs self-energy). 

Complex poles for unstable $\PW, \PZ, \PH$, and $\PQt$ also tell us that it is 
very difficult for a heavy Higgs boson to come out right within the SM;
these complex poles are solutions of a (coupled) system of equations
but for $\PW, \PZ$, and (partially) $\PQt$ we can compare with the corresponding
experimental quantities. Results are shown in \refT{tab:HTO_1} and clearly indicate
mismatches between predicted and experimental $\PW, \PQt$ complex poles.
Indeed, as soon as $\muh$ increases, it becomes more and more difficult to find complex
$\PW, \PQt$, and $\PZ$ poles with an imaginary part compatible with measurements.
\begin{table}
\begin{center}
\caption[]{\label{tab:HTO_1}{The Higgs-boson complex pole at fixed values of
the $\PW, \PQt$ complex poles compared with the complete solution for $\cph, \cpw$, and
$\cpt$.}}
\begin{tabular}{cccc}
\hline 
$\muh\,$[GeV] & $\gamma_{\PW}\,$ [GeV] fixed & $\gamma_{\PQt}\,$ [GeV] fixed &
$\gamma_{\PH}\,$ [GeV] derived \\ 
$200$ & $2.093$ & $1.481$ & $1.264$ \\
$250$ &         &         & $3.369$ \\
$300$ &         &         & $7.721$ \\ 
\hline
$\muh\,$[GeV] & $\gamma_{\PW}\,$ [GeV] derived & $\gamma_{\PQt}\,$ [GeV] derived &
$\gamma_{\PH}\,$ [GeV] derived \\ 
$200$ & $1.932$ & $1.171$ & $1.262$ \\
$250$ & $1.822$ & $0.923$ & $3.364$ \\
$300$ & $1.738$ & $0.689$ & $7.711$ \\
\hline
\end{tabular}
\end{center}
\end{table}

This simple fact raises the following question: What is the physical meaning of 
a heavy-Higgs-boson search? We have the usual and well-known considerations:
a Higgs boson above $600\UGeV$ requires new physics at $1\UTeV$, argument
based on partial-wave unitarity (which should not be taken quantitatively nor too 
literally); violation of unitarity bound possibly implies the presence of $J=0,1$ resonances,
but there is no way to predict their masses, simply scaling the $\PGp{-}\PGp$ system 
gives resonances in the $1\UTeV$ ballpark. 
Generally speaking, it would be a good idea to address this search as {\em search} for 
$J=0,1$ heavy resonances decaying into $\PV\PV \to 4\,\Pf$. 

In a model-independent approach both $\muh$ and $\gh$ should be kept free in order to perform 
a $2$-dimensional scan of the Higgs-boson lineshape. For the high-mass region this remains the 
recommended strategy. Once the fits are performed it will be left to theorists to struggle 
with the SM interpretation of the results.

To summarise, we have addressed the following question: What is the common sense 
definition of mass and width of the Higgs boson?
We have several options,
\bq
\cph = \muhs - i\,\muh\,\gh,
\quad
\cph = \lpar \mu'_{\PH} - \frac{i}{2}\,\gamma'_{\PH}\rpar^2, 
\quad
\cph = \frac{\mBhs - i\,{\GOL}_{\PH}\,\mBh}
{1 + {\GOL}^2_{\PH}/\mBhs}.
\eq
We may ask which one is correct, approximate or closer to the experimental peak. Here
we have to distinguish: for $\gh \muchless \muh$, $\mBh$ is a good approximation 
to the on-shell mass and it is closer to the experimental peak; for instance, for the $\PZ$ 
boson ${\overline M}_{\PZ}$ is equivalent to the mass measured at LEP.
However, in the high-mass scenario, where $\gh \sim \muh$, the situation 
changes and ${\overline O}_{\PH} \not= O^{\OS}_{\PH}$ for any observable (or pseudo-observable). 
Therefore, the message is: Do not use the computed on-shell width to estimate $\mBh$.
\subsection{Production and decay\label{PO:prodec}}
Before describing production and decay of an Higgs boson we underline the general structure
of any process containing a Higgs-boson intermediate state. The corresponding amplitude is given by
\bq
A(s) = \frac{f(s)}{s - \cph} + N(s),
\eq
where $N(s)$ denotes the part of the amplitude which is non-resonant.
Signal and background are defined as
\bq
A(s) = S(s) + B(s),
\qquad
S(s)= \frac{f(\cph)}{s - \cph}, \quad
B(s)= \frac{f(s) - f(\cph)}{s - \cph} + N(s).
\label{split}
\eq
As a first step we will show how to write $f(s)$ in a way such that pseudo-observables make 
their appearance. Consider the process $i j \to \PH \to \PF$ where $i,j\,\in\,$partons;
the complete cross section will be written as follows:
\bq
\sigma_{i j \to \PH \to \PF}(s) =
\frac{1}{2\,s}\,\int\,d\Phi_{i j \to \PF}\,
\Bigl[ \sum_{s,c} \bmid A_{i j \to \PH} \bmid^2 \Bigr]\,
\frac{1}{\bmid s - \cph\bmid^2}\,
\Bigl[ \sum_{s,c} \bmid A_{\PH \to \PF} \bmid^2 \Bigr],
\eq
where $\sum_{s,c}$ is over spin and colours (averaging on the initial state). 
Note that the background (\eg $\Pg\Pg \to 4\,\Pf$) has not been included and, strictly 
speaking and for reasons of gauge invariance, one should consider only the residue of the 
Higgs-resonant amplitude at the complex pole, as described in \eqn{split}. For the 
moment we will argue that the dominant corrections are the QCD ones where we have
no problem of gauge-parameter dependence. 
If we decide to keep the Higgs boson off-shell also in the resonant part of
the amplitude (interference signal/background remains unaddressed) then we can write
\bq
\sum_{s,c}\,\int\,d\Phi_{i j \to \PH}\,\bmid A_{i j \to \PH} \bmid^2 = s\,{\overline A}_{ij}(s).
\eq
For instance, we have
\bq
{\overline A}_{\Pg\Pg}(s) = \frac{\alphas^2}{\pi^2}\,\frac{G_{\ssF}\,s}{288\,\sqrt{2}}\,
\bmid \sum_q\,f\lpar \tau_q\rpar \bmid^2\,\lpar 1 + \delta_{\QCD}\rpar,
\eq
where $\tau_q = 4\,m^2_q/s$, $f(\tau_q)$ is defined in Eq.(3) of \Bref{Spira:1995rr} and where 
$\delta_{\QCD}$ gives the QCD corrections to $\Pg\Pg \to \PH$ up to NNLO + NLL order. 
Furthermore, we define
\bq
\Gamma_{\PH \to \PF} =
\frac{1}{2\,\sqrt{s}}\,\int\,d\Phi_{\PH \to \PF}\,\sum_{s,c} \bmid A_{\PH \to \PF} \bmid^2, 
\eq
which gives the partial decay width of a Higgs boson of virtuality $s$ into a final 
state $\PF$, and
\bq
\sigma_{i j \to \PH} = \frac{{\overline A}_{ij}(s)}{s},
\eq
which gives the production cross section of a Higgs boson of virtuality $s$.
We can write the final result in terms of pseudo-observables
\bq
\sigma\lpar i j \to \PH \to \PF \rpar = \frac{1}{\pi}\,
\sigma_{i j \to \PH}\,\frac{s^2}{\bmid s - \cph\bmid^2}\,
\frac{\Gamma_{\PH \to \PF}}{\sqrt{s}}.
\eq
It is also convenient to rewrite the result as 
\bq
\sigma\lpar i j \to \PH \to \PF \rpar = \frac{1}{\pi}
\sigma_{i j \to \PH}\,\frac{s^2}{\bmid s - \cph\bmid^2}\,
\frac{\Gamma^{\tot}_{\PH}}{\sqrt{s}}\,\hbox{BR}\lpar \PH \to \PF\rpar,
\label{QFT}
\eq
where we have introduced a sum over all final states,
\bq 
\Gamma^{\tot}_{\PH} = \sum_{f\in\PF}\,\Gamma_{\PH \to \PF}.
\eq
Note that we have written the phase-space integral for $i(p_1) + j(p_2) \to \PF$
as
\bqa
\int\,d\Phi_{i j \to \PF} &=&
\int\,\biggl(\prod_f\,d^4 p_f\,\delta^+(p^2_f)\biggr)\,
\delta^4( p_1 + p_2 - \sum_f p_f)
\nl
{}&=& \int\,d^4k\,\delta^4( k - p_1 - p_2)\,
\int\,\biggl(\prod_f\,d^4p_f\,\delta^+( p^2_f)\biggr)\,\delta^4( k - \sum_f p_f),
\eqa
where we assume that all initial and final states (\eg $\PGg \PGg$, $4\,\Pf$, \etc) 
are massless. It is worth noting that the introduction of complex poles does not 
imply complex kinematics. According to \eqn{split} only the residue of the propagator 
at the complex pole becomes complex, not any element of the phase-space integral. 

{\underline{Definition:}}
We define an off-shell production cross section (for all channels) as follows:
\bq
\sigma^\prop_{i j \to \ac}(s) = \frac{1}{\pi}\,
\sigma_{i j \to \PH}\,\frac{s^2}{\bmid s - \cph\bmid^2}\,
\frac{\Gamma^{\tot}_{\PH}}{\sqrt{s}}.
\label{sigmaPR}
\eq

When the cross section $i j \to \PH$ refers to an off-shell Higgs boson
the choice of the QCD scales should be made according to the virtuality 
and not to a fixed value. Therefore, always referring to \refF{HLS:complete},
for the PDFs and $\sigma_{i j \to \PH  + \PX}$ one should select 
$\muFs = \muRs = z\,s/4$ ($z\,s$ being the invariant mass of the detectable final state). 
Indeed, beyond LO one must not choose the invariant mass of the incoming partons for the 
renormalisation and factorisation scales, but an infrared-safe quantity fixed from the 
detectable final state. The argument is based on minimisation of the universal logarithms 
(DGLAP) and not the process-dependent ones.

The off-shell Higgs-boson production is currently computed according to the replacement
\bq
\sigma_{\OS}(\muhs)\,\delta( z\,s - \muhs)
\;\longrightarrow\;
\sigma_{\OFS}(z\,s)\,\hbox{BW}(z\,s),
\label{sigmaBW}
\eq
(\eg see \Bref{Alioli:2008tz}) at least at lowest QCD order, where the so-called modified 
Breit--Wigner distribution is defined by
\bq 
\hbox{BW}(s) = \frac{1}{\pi}\,\frac{s\,\GOSu_{\PH}/\muh}
{\lpar s - \muhs\rpar^2 + \lpar s\,\GOSu_{\PH}/\muh\rpar^2},
\label{BWdef}
\eq
where now $\muh = M^{\OS}_{\PH}$.
This ad-hoc Breit--Wigner cannot be derived from QFT and also is not normalizable
in $[0\,,\,+\infty]$. 
Note that this Breit--Wigner for a running width comes from the 
substitution of $\Gamma \to \Gamma(s) = \Gamma\,s/M^2$ in the Breit--Wigner for a 
fixed width $\Gamma$. This substitution is not justifiable.
Its practical purpose is to enforce a {\em physical} behaviour for low
virtualities of the Higgs boson, but the usage cannot be justified nor recommended.
For instance, if one considers $\PV\PV$ scattering and uses this distribution in the 
$s\,$-channel Higgs exchange, the behaviour for large values of $s$ spoils unitarity 
cancellation with the contact diagram. 
It is worth noting that the alternative replacement
\bq
\sigma_{\OS}(\muhs)\,\delta( z\,s - \muhs)
\;\longrightarrow\;
\sigma_{\OFS}(z\,s)\,\hbox{BW}(z\,s),
\quad
\hbox{BW}(z\,s) = 
\frac{1}{\pi}\,\frac{\muh\,\GOSu_{\PH}}
{\lpar z\,s - \muhs\rpar^2 + \lpar \muh\,\GOSu_{\PH}\rpar^2}, 
\label{BWFW}
\eq
has additional problems at the low-energy tail of the resonance due to the $\Pg\Pg$ luminosity,
creating an artificial increase of the lineshape at low virtualities.

Another important issue is that $\gh$, which appears in the
imaginary part of the inverse Dyson-resummed propagator, is not
the on-shell width, since they differ by higher-order terms and their
relations becomes non-perturbative when the on-shell width becomes of the
same order of the on-shell mass (typically, for on-shell masses above $800\,$GeV).

The complex-mass scheme can be translated into a more familiar language by introducing the
Bar-scheme. Using \eqn{CMSprop} with the parametrisation of \eqn{CPpar}
we perform the well-known transformation
\bq
\mBhs = \muhs + \gamma^2_{\PH} 
\qquad
\mu_{\sH}\,\gh = \mBh\,{\GOL}_{\PH}.
\label{Bars}
\eq
A remarkable identity follows, defining the Bar-scheme:
\bq
\frac{1}{s - \cph} =
\Bigl( 1 + i\,\frac{{\GOL}_{\PH}}{\mBh}\Bigr)\,
\Bigr( s - \mBhs + 
i\,\frac{{\GOL}_{\PH}}{\mBh}\,s \Bigr)^{-1},
\label{barid}
\eq
showing that the Bar-scheme is equivalent to introducing a running width in the propagator
with parameters that are not the on-shell ones. Special attention goes to the
numerator in \eqn{barid} which is essential in providing the right asymptotic
behaviour when $s \to \infty$, as needed for cancellations with contact terms in
$\PV\PV$ scattering. 
A sample of numerical results is shown in \refT{tab:HTO_2} where we compare the 
Higgs-boson complex pole to the corresponding quantities in the Bar-scheme.
\begin{table}
\begin{center}
\caption[]{\label{tab:HTO_2}{Higgs-boson complex pole; 
$\GOSu_{\PH}$ is the on-shell width, $\gh$ is defined in \eqn{CPpar} and the Bar-scheme
in \eqn{Bars}.}}
\begin{tabular}{ccccc}
\hline 
$\muh$[GeV]                 & 
$\GOSu_{\PH}$[GeV]  (YR)    &
$\gh\,$[GeV]                &
${\overline M}_{\PH}$[GeV]  &
$\GOL_{\PH}$[GeV]  \\
\hline 
$200$ & $1.43$   & $1.26$   & $200$    & $1.26$ \\
$400$ & $29.2$   & $24.28$  & $400.7$  & $24.24$ \\
$600$ & $123$    & $102.17$ & $608.6$  & $100.72$ \\
$700$ & $199$    & $159.54$ & $717.95$ & $155.55$ \\
$800$ & $304$    & $228.44$ & $831.98$ & $219.66$ \\
$900$ & $449$    & $307.63$ & $951.12$ & $291.09$ \\
\hline
\end{tabular}
\end{center}
\end{table}

The use of the complex pole is recommended even if the accuracy at which
its imaginary part can be computed is not of the same quality as the NLO
accuracy of the on-shell width. Nevertheless the use of a solid prediction (from a 
theoretical point of view) should always be preferred to the introduction of ill-defined
quantities (gauge-parameter dependent).

\subsubsection{Schemes\label{PO:schem}}
We are now in a position to give a more detailed description of the strategy behind
\eqn{split}. Consider the complete amplitude for a given process, \eg the one in
\refF{HLS:complete}; let $\lpar \shat= z s,\,\dots\rpar$ the full list of
Mandelstam invariants characterizing the process, then
\bq
A\lpar \shat,\,\dots\rpar \;=\; V_{\myprod}\lpar \shat,\,\dots\rpar\,
\Delta_{\prop}(\shat)\,V_{\dec}(\shat) \;+\; N\lpar \shat,\,\dots\rpar.
\eq
Here $V_{\myprod}$ denotes the amplitude for production, \eg $\Pg\Pg \to \PH(\shat)  + \PX$,
$\Delta_{\prop}(\shat)$ is the propagation function, $V_{\dec}(\shat)$ is the
amplitude for decay, \eg $\PH(\shat) \to 4\,\Pf$.
If no attempt is made to split $A(s)$, no ambiguity arises but, usually, the two components 
are known at different orders. Ho to define the signal? The following schemes are available:
\begin{description}
\setlength{\itemsep}{1.0em}
\item[ONBW]
\bq
S\lpar \shat,\,\dots\rpar = 
V_{\myprod}\lpar \muhs,\,\dots\rpar\,\Delta_{\prop}(\shat)\,V_{\dec}(\muhs),
\qquad
\Delta_{\prop}(\shat) = \hbox{Breit--Wigner},
\eq
in general violates gauge invariance, neglects the Higgs off-shellness and introduces the
ad hoc Breit--Wigner of \eqn{BWdef}.
\item[OFFBW]
\bq
S\lpar \shat,\,\dots\rpar = 
V_{\myprod}\lpar \shat,\,\dots\rpar\,\Delta_{\prop}(\shat)\,V_{\dec}(\shat),
\qquad
\Delta_{\prop}(\shat) = \hbox{Breit--Wigner},
\label{offbw}
\eq
in general violates gauge invariance and introduces the ad hoc Breit--Wigner.
\item[ONP]
\bq
S\lpar \shat,\,\dots\rpar = 
V_{\myprod}\lpar \muhs,\,\dots\rpar\,\Delta_{\prop}(\shat)\,V_{\dec}(\muhs),
\qquad
\Delta_{\prop}(\shat) = \hbox{propagator},
\eq
in general violates gauge invariance and neglects the Higgs off-shellness.
\item[OFFP]
\bq
S\lpar \shat,\,\dots\rpar = 
V_{\myprod}\lpar \shat,\,\dots\rpar\,\Delta_{\prop}(\shat)\,V_{\dec}(\shat),
\qquad
\Delta_{\prop}(\shat) = \hbox{propagator},
\label{offp}
\eq
in general violates gauge invariance.
\item[CPP]
\bq
S\lpar \shat,\,\dots\rpar = 
V_{\myprod}\lpar \cph,\,\dots\rpar\,\Delta_{\prop}(\shat)\,V_{\dec}(\cph),
\qquad
\Delta_{\prop}(\shat) = \hbox{propagator}.
\label{cpp}
\eq
Only the pole, the residue, and the reminder of the amplitude are gauge
invariant. Furthermore the CPP-scheme allows to identify POs like the production
cross section and any partial decay width by putting in one-to-one correspondence 
robust theoretical quantities and experimental data.
\end{description}
From the list above it follows that only the CPP-scheme should be used, once the
signal/background interference becomes available. However, the largest part of the available 
calculations is not yet equipped with Feynman integrals on the second Riemann sheet; furthermore
the largest corrections in the production are from QCD and we could argue that gauge
invariance is not an issue over there, so that the OFFP-scheme remains, at the moment, the 
most pragmatic alternative.

Implementation of the CPP-scheme requires some care in the presence of four-leg
processes (or more). The natural  choice is to perform analytical continuation
from real kinematics to complex invariants; consider, for instance, the process
$\Pg \Pg \to \PH j$ where the external $\PH$ leg is continued from real on-shell
mass to $\cph$. For the three invariants we use the following continuation: 
\bq
t = -\,\frac{s}{2}\,\Bigl[ 1 - \frac{\cph}{s} - 
\bigl( 1 - 4\,\frac{\cph}{s} \bigr)^{1/2}\,\cos\theta \Bigr],
\quad
u = -\,\frac{s}{2}\,\Bigl[ 1 - \frac{\cph}{s} + 
\bigl( 1 - 4\,\frac{\cph}{s} \bigr)^{1/2}\,\cos\theta \Bigr],
\eq
where $\theta$ is the scattering angle and $s = 4\,E^2$, where $2\,E$ is the centre-of-mass
energy.
To summarise, our proposal aims to support the use of \eqn{QFT} for producing
accurate predictions for the Higgs lineshape in the SM; \eqn{QFT} can be adapted easily for 
a large class of extensions of the SM. 
Similarly \eqn{sigmaPR} should be used to define the Higgs-boson production cross section
instead of $\sigma_{\OS}$ or $\sigma_{\OFS}$ of \eqn{sigmaBW}.

To repeat an obvious argument the zero-width approximation for the Higgs boson is usually
reported in comparing experimental studies and theoretical predictions. The approximation
has very low quality in the high-mass region where the Higgs boson has a non negligible
width. An integration over some distribution is a more accurate estimate of the signal
cross section and we claim that this distribution should be given by the complex propagator
and not by some ad hoc Breit--Wigner.

As we have already mentioned, most of this section is devoted to studying the Higgs-boson
lineshape in the high-mass region. However, nothing in the formalism is peculiar to that
application and the formalism itself forms the basis for extracting pseudo-observables from 
experimental data. This is a timely contribution to the relevant literature.

\subsubsection{Production cross section}
Before presenting detailed numerical results and comparisons we give the complete definition of
the production cross section,
\bq
\sigma^{\myprod} = 
\sum_{i,j}\,\int \hbox{PDF}\,\otimes\,\sigma^{\myprod}_{i j \to \ac} = 
\sum_{i,j}\,\int_{z_0}^1 dz \int_z^1 \frac{dv}{v}\,{\cal L}_{ij}(v)
\sigma^\prop_{ij \to \ac}(z s, v s, \muR, \muF),
\label{PDFprod_1}
\eq
where $z_0$ is a lower bound on the invariant mass of the $\PH$ decay products,
the luminosity is defined by
\bq
{\cal L}_{ij}(v) = \int_v^1 \frac{dx}{x}\,
f_i\lpar x,\muF\rpar\,f_j\lpar \frac{v}{x},\muF\rpar,
\label{PDFprod_2}
\eq
where $f_i$ is a parton distribution function and
\bq
\sigma^\prop_{ij \to \ac}(z s, v s, \muR, \muF) = \frac{1}{\pi}\,
\sigma_{ij \to \PH  + \PX}(z s, v s, \muR, \muF)\,\frac{ v z s^2}{\bmid z s - \cph\bmid^2}\,
\frac{\Gamma^{\tot}_{\PH}(z s)}{\sqrt{z s}}.
\label{PDFprod_3}
\eq
Therefore, $\sigma_{ij \to \PH  + \PX}(z s, v s, \muR)$ is the cross section for two
partons of invariant mass $v s$ ($z < v < 1$) to produce a final state containing 
a $\PH$ of virtuality $z s$ plus jets ($\PX$); it is made of several terms
\bq
\sum_{ij}\,\sigma_{ij \to \PH  + \PX}(z s, v s, \muR, \muF) = 
\sigma_{\Pg\Pg \to \PH}\,\delta\Bigl(1 - \frac{z}{v}\Bigr) + \sigma_{\Pg\Pg \to \PH\Pg} + 
\sigma_{\PQq\Pg \to \PH\PQq} + \sigma_{\PAQq\PQq \to \PH\Pg} +
\mbox{NNLO}.  
\label{PDFprod_4}
\eq
As a technical remark the complete phase-space integral for the process
$\hatp_i + \hatp_j \to p_k + \{f\}$ ($\hatp_i = x_i\,p_i$ \etc) is written as
\bqa
\int\,d\Phi_{ij \to \PF} &=& 
\int\,d\Phi_{\myprod}\,\int\,d\Phi_{\dec} =
\int d^4 p_k\,\delta^+(p^2_k)\,\biggl(\prod_{l=1,n} d^4 q_l\,\delta^+(q^2_l)\biggr)\,
\delta^4\biggl(\hatp_i + \hatp_j - p_k - \sum_l q_l\biggr)
\nl
{}&=&
\int d^4k\, d^4Q\,\delta^+(p^2_k)\,\delta^4 \biggl(\hatp_i + \hatp_j - p_k - Q\biggr)\, 
\int \biggl(\prod_{l=1,n} d^4q_l\,\delta^+(q^2_l)\biggr)\,\delta^4 \biggl(Q - \sum_l q_l\biggr),
\nonumber\\
\eqa
where $\int\,d\Phi_{\dec}$ is the phase-space for the process $Q \to \{f\}$ and
\bqa
\int\,d\Phi_{\myprod} &=& s\,\int dz\,\int d^4k\, d^4Q\,\delta^+(p^2_k)\,
\delta\lpar Q^2 - z s\rpar\,\theta(Q_0)\,
\delta^4 \lpar \hatp_i + \hatp_j - p_k - Q\rpar
\nl
{}&=&
s^2\,\int dz\, dv\, d{\hat t}\,\int d^4k\, d^4Q\,\delta^+(p^2_k)\,
\delta\lpar Q^2 - z s\rpar\,\theta(Q_0)\,
\delta^4 \lpar \hatp_i + \hatp_j - p_k - Q\rpar
\nl
{}&\times&
\delta\lpar (\hatp_i + \hatp_j)^2 - v s\rpar\,
\delta\lpar (\hatp_i + Q)^2 - {\hat t}\rpar.
\eqa
\eqnsc{PDFprod_1}{PDFprod_3} follow after folding with PDFs of argument $x_i$ and $x_j$, after 
using $x_i = x$, $x_j = v/x$ and after integration over ${\hat t}$. At NNLO there is an 
additional parton in the final state and five invariants are needed to describe the partonic 
process, plus the $\PH$ virtuality. However, one should remember that at NNLO use is made 
of the effective-theory approximation where the Higgs--gluon interaction is described by a 
local operator. Our variables $z, v$ are related to {\sc POWHEG} 
parametrisation~\cite{Alioli:2008tz}, $Y, \xi$, by $Y= \ln(x/\sqrt{z})$ and $v= z/(1-\xi)$. 
\subsection{Numerical results\label{PO:numer}}
In the following we will present numerical results obtained with the program {\sc HTO} 
(G.~Passarino, unpublished) that allows for the study of the Higgs-boson lineshape using 
complex poles. {\sc HTO} is a FORTRAN $95$ program that contains a translation of
the subroutine {\tt HIGGSNNLO} written by M.~Grazzini for computing
the total (on-shell) cross section for Higgs-boson production at NLO and NNLO.

The following acronyms will be used:
\begin{description}
\item[FW] Breit--Wigner Fixed Width (\eqn{BWFW}),
\item[RW] Breit--Wigner Running Width (\eqn{BWdef}),
\item[OS] parameters in the On-Shell scheme,
\item[Bar] parameters in Bar-scheme (\eqn{Bars}),
\item[FS] QCD renormalisation (factorisation) scales fixed,
\item[RS] QCD renormalisation (factorisation) scales running.
\end{description}
All results in this section refer to $\sqrt{s} = 7\UTeV$ and are based on the MSTW2008 
PDF sets~\cite{Martin:2009iq}. For complex $\PW,\PZ$ poles we use \eqn{cpV} with
HXSWG standard input.

The integrand in \eqn{PDFprod_1} has several peaks; the most 
evident is in the propagator and in {\sc HTO} a change of variable is performed,
\bqa
z s &=& \frac{1}{1+(\muh/\gh)^2}\,\Bigl[ \muhs + \muh\,\gh\,
\tan\lpar \muh \gh\,z'\rpar \Bigr],
\nl
z' &=& \frac{s}{\muh \gh}\,\Bigl[ \lpar a_m + a_{\ssM}\rpar\,\zeta - a_m \Bigr],
\quad a_m = \arctan\frac{\muhs-z_0 s}{\muh \gh},
\quad a_{\ssM} = \arctan\frac{s-\muhs}{\muh \gh}.
\eqa
Similarly, another change is performed,
\bq
v = \exp\{ \lpar 1 - u^2 \rpar\,\ln z\}, 
\qquad
x = \exp\{ \lpar 1 - y^2 \rpar\,\ln v\}. 
\eq
In comparing the OFFBW-scheme with the OFFP-scheme one should realise that there are
two sources of difference, the functional form of the distributions and the different
numerical values of the parameters in the distributions. To understand the impact of
the functional form we have performed a comparison where (unrealistically)
$\gh = \GOSu_{\PH}$ is used in the Higgs propagator; results are shown 
in \refT{tab:HTO_3}.
\begin{table}
\begin{center}
\caption[]{\label{tab:HTO_3}
{The re-weighting factor $w$ for the (total) integrated Higgs lineshape,
$\sigma^{\prop}/\sigma^{\mBW}$ using $\gh = \GOSu_{\PH}$
in \eqn{sigmaPR}.}}
\begin{tabular}{cccccc}
\hline 
$\muh\;$[GeV]               & 
$\GOSu_{\PH}\;$[GeV]        &
$\sigma^{\OS}\;$[pb]        &
$\sigma^{\mBW}\;$[pb]       &
$\sigma^{\prop}\;$[pb]      &
$w$                        \\
\hline
$ 200 $ & $ 1.43  $ & $ 5.249  $ & $ 5.674  $ & $ 5.459  $ & $ 0.962 $    \\
$ 300 $ & $ 8.43  $ & $ 2.418  $ & $ 2.724  $ & $ 2.585  $ & $ 0.949 $    \\
$ 400 $ & $ 29.2  $ & $ 2.035  $ & $ 1.998  $ & $ 1.927  $ & $ 0.964 $    \\
$ 500 $ & $ 68.0  $ & $ 0.8497 $ & $ 0.8108 $ & $ 0.7827 $ & $ 0.965 $    \\
$ 600 $ & $ 123.0 $ & $ 0.3275 $ & $ 0.3231 $ & $ 0.3073 $ & $ 0.951 $    \\
\hline
\end{tabular}
\end{center}
\end{table}

In \refT{tab:HTO_4} we compare the on-shell production cross section as given
in \Bref{Dittmaier:2011ti} with the off-shell cross section in the OFFBW-scheme
(\eqn{offbw}) and in the OFFP-scheme (\eqn{offp}).

An important question regarding the numerical impact of a calculation where the
on-shell Higgs boson is left off-shell and convoluted with some distribution is
how much of the effect survives the inclusion of theoretical uncertainties. In the
OFFBW-scheme we can compare with the results of \Bref{Anastasiou:2011pi}, at least
for values of the Higgs-boson mass below $300\UGeV$. We compare with Tables~2--3
of \Bref{Dittmaier:2011ti} at $300\UGeV$ and obtain the results shown in
\refT{tab:HTO_5} where only the QCD scale uncertainty is included. At these values of the 
Higgs-boson mass on-shell production and off-shell production sampled over a Breit--Wigner 
are compatible within the errors. The OFFP-scheme gives a slightly higher central
value of $2.86\Upb$.
The comparison shows drastically different results for high values of the mass
where the difference in central values is much bigger than the theoretical
uncertainty.
\begin{table}
\begin{center}
\caption[]{\label{tab:HTO_4}{Comparison of the on-shell production cross section as given
in \Bref{Dittmaier:2011ti} with the off-shell cross section in the OFFBW-scheme
(\eqn{offbw}) and in the OFFP-scheme (\eqn{offp}).}}
\begin{tabular}{cccccc}
\hline 
$\muh$[GeV]               & 
$\GOSu_{\PH}$[GeV]        &
$\gh$[GeV]                &
$\sigma^{\OS}$[pb]        &
$\sigma^{\mBW}$[pb]       &
$\sigma^{\prop}$[pb]      \\
\hline
$ 500 $ & $ 68.0  $ & $ 60.2 $ & $ 0.8497  $ & $ 0.8239  $ & $ 0.9367 $    \\
$ 550 $ & $ 93.1  $ & $ 82.8 $ & $ 0.5259  $ & $ 0.5161  $ & $ 0.5912 $    \\
$ 600 $ & $ 123   $ & $ 109  $ & $ 0.3275  $ & $ 0.3287  $ & $ 0.3784 $    \\
$ 650 $ & $ 158   $ & $ 139  $ & $ 0.2064  $ & $ 0.2154  $ & $ 0.2482 $    \\
$ 700 $ & $ 199   $ & $ 174  $ & $ 0.1320  $ & $ 0.1456  $ & $ 0.1677 $    \\
$ 750 $ & $ 248   $ & $ 205  $ & $ 0.0859  $ & $ 0.1013  $ & $ 0.1171 $    \\
$ 800 $ & $ 304   $ & $ 245  $ & $ 0.0567  $ & $ 0.0733  $ & $ 0.0850 $    \\
$ 850 $ & $ 371   $ & $ 277  $ & $ 0.0379  $ & $ 0.0545  $ & $ 0.0643 $    \\
$ 900 $ & $ 449   $ & $ 331  $ & $ 0.0256  $ & $ 0.0417  $ & $ 0.0509 $    \\
\hline
\end{tabular}
\end{center}
\end{table}
\begin{table}
\begin{center}
\caption[]{\label{tab:HTO_5}{The production cross section in pb at $\muh=300\UGeV$.
Results from {\sc HTO} are computed with running QCD scales.}}
\begin{tabular}{cccc}
\hline 
Tab.~2 of \Bref{Dittmaier:2011ti}  &
Tab.~3 of \Bref{Dittmaier:2011ti}  &
Tab.~5 of \Bref{Anastasiou:2011pi} &
{\sc HTO} RS-option\\
$2.42^{+0.14}_{-0.15}$ &
$2.45^{+0.16}_{-0.22}$ &
$2.57^{+0.15}_{-0.22}$ &
$2.81^{+0.25}_{-0.23}$ \\
\hline
\end{tabular}
\end{center}
\end{table}

In \refT{tab:HTO_5b} we show a more detailed comparison of the production cross section in the
OFFBW-scheme between {\sc iHixs} of \Bref{Anastasiou:2011pi} and our calculation.
$\Delta$ is the percentage error due to QCD scale uncertainties, and $\delta$ is the percentage ratio
{\sc HTO}/{\sc iHixs}.

\begin{table}
\begin{center}
\caption[]{\label{tab:HTO_5b}{Comparison of the production cross section in the
OFFBW-scheme between {\sc iHixs} (Table~5 of \Bref{Anastasiou:2011pi}) and our calculation.
$\Delta$ is the percentage error due to QCD scale uncertainties, and $\delta$ is the percentage ratio
{\sc HTO}/{\sc iHixs}.}}
\begin{tabular}{cccccc}
\hline 
$\muh$[GeV]              &
$\sigma_{\mathrm{iHixs}}$[pb] &
$\Delta_{\mathrm{iHixs}}[\%]$   &
$\sigma_{\mathrm{HTO}}$[pb]   &
$\Delta_{\mathrm{HTO}}[\%]$     &
$\delta[\%]$ \\
\hline 
$200 $&$  5.57 $&$   {+}7.19 \; {-}9.06 $&$ 5.63 $&$   {+}9.12 \; {-}9.30 $&$ 1.08$ \\
$220 $&$  4.54 $&$   {+}6.92 \; {-}8.99 $&$ 4.63 $&$   {+}8.93 \; {-}8.85 $&$ 1.98$ \\
$240 $&$  3.80 $&$   {+}6.68 \; {-}8.91 $&$ 3.91 $&$   {+}8.76 \; {-}8.51 $&$ 2.89$ \\
$260 $&$  3.25 $&$   {+}6.44 \; {-}8.84 $&$ 3.37 $&$   {+}8.61 \; {-}8.22 $&$ 3.69$ \\
$280 $&$  2.85 $&$   {+}6.18 \; {-}8.74 $&$ 2.97 $&$   {+}8.49 \; {-}7.98 $&$ 4.21$ \\
$300 $&$  2.57 $&$   {+}5.89 \; {-}8.58 $&$ 2.69 $&$   {+}8.36 \; {-}7.75 $&$ 4.67$ \\
\hline
\end{tabular}
\end{center}
\end{table}

Another quantity that is useful in describing the lineshape is obtained by introducing
\bq
M^2_{\peak} = \bigl( M^{\OS}_{\PH}\bigr)^2 + \bigl( \Gamma^{\OS}_{\PH}\bigr)^2,
\label{peak}
\eq
an by considering the following windows in the invariant mass, 
\bq
\Bigl[ M_{\peak} + \frac{n}{2}\,\Gamma^{\OS}_{\PH}\,,\,
       M_{\peak} + \frac{n+1}{2}\,\Gamma^{\OS}_{\PH} \Bigr],
\qquad n= 0,\pm 1, \dots.
\label{windows}
\eq
In \refT{tab:HTO_6} we present the ratio OFFP/OFFBW for the invariant mass distribution 
in the windows of \eqn{windows}.
\begin{table}
\begin{center}
\caption[]{\label{tab:HTO_6}{Scaling factor OFFP/OFFBW schemes (\eqnsc{offp}{offbw})
for the invariant-mass windows of \eqn{windows}.}}
\begin{tabular}{ccccccccc}
\hline 
$\muh$[GeV] & $n=-4$ & $n=-3$ & $n=-2$ & $n=-1$ & $n=0$ & $n=1$ & $n=2$ & $n=3$ \\
\hline 
$500$ & $1.24$ & $1.13$ & $1.08$ & $1.23$ & $1.32$ & $1.11$ & $0.97$ & $0.88$ \\
$600$ & $1.39$ & $1.20$ & $1.08$ & $1.27$ & $1.40$ & $1.12$ & $0.94$ & $0.84$ \\
$650$ & $1.50$ & $1.24$ & $1.08$ & $1.30$ & $1.44$ & $1.13$ & $0.93$ & $0.81$ \\
$700$ & $1.62$ & $1.28$ & $1.09$ & $1.34$ & $1.49$ & $1.13$ & $0.92$ & $0.79$ \\
$750$ & $1.78$ & $1.34$ & $1.10$ & $1.41$ & $1.57$ & $1.14$ & $0.90$ & $0.77$ \\
$800$ & $1.99$ & $1.41$ & $1.12$ & $1.49$ & $1.64$ & $1.14$ & $0.89$ & $0.75$ \\
\hline
\end{tabular}
\end{center}
\end{table}

In \refT{tab:HTO_7} we include uncertainties in the comparison; both OFFBW-scheme and 
OFFP-scheme are used with the RS option, \ie running QCD scales instead of a fixed one.  
Note that for $\sigma(\mu)$ we define a central value $\sigma_c = \sigma(\muol)$ 
and a scale error as $[ \sigma^-\,,\,\sigma^+ ]$, where
\bq
\sigma^- = \min_{\mu \in [\muol/2,2 \muol]}\,\sigma(\mu),
\qquad 
\sigma^+ = \max_{\mu \in [\muol/2,2 \muol]}\,\sigma(\mu),
\label{error}
\eq
and where $\mu = \muR = \muF$ and ${\muol}$ is the reference scale, static or
dynamic.
\begin{table}
\begin{center}
\caption[]{\label{tab:HTO_7}{Production cross section with errors due to QCD scale
variation and PDF uncertainty. First entry is the on-shell cross section of
Table~2 of \Bref{Dittmaier:2011ti}, second entry is OFFBW (\eqn{offbw}), last entry 
is OFFP \eqn{offp}.}}
\begin{tabular}{cccc}
\hline 
$\muh$[GeV]               & 
$\sigma$[pb]              &
Scale\,[\%]                 &
PDF\,[\%]                   \\
\hline
$600$ & $0.336$  & $+\,6.1\;-\,5.2$   & $+\,6.2\;-\,5.3$ \\ 
      & $0.329$  & $+\,11.2\;-\,11.3$ & $+\,4.7\;-\,4.8$ \\ 
      & $0.378$  & $+\,9.7\;-\,10.0$  & $+\,5.0\;-\,3.9$ \\ 
$650$ & $0.212$  & $+\,6.2\;-\,5.2$   & $+\,6.5\;-\,5.5$ \\ 
      & $0.215$  & $+\,12.4\;-\,12.2$ & $+\,5.1\;-\,5.3$ \\ 
      & $0.248$  & $+\,10.2\;-\,10.0$ & $+\,5.3\;-\,4.2$ \\ 
$700$ & $0.136$  & $+\,6.3\;-\,5.3$   & $+\,6.9\;-\,5.8$ \\ 
      & $0.146$  & $+\,13.9\;-\,13.3$ & $+\,5.6\;-\,5.8$ \\ 
      & $0.168$  & $+\,11.0\;-\,11.1$ & $+\,5.5\;-\,4.6$ \\ 
$750$ & $0.0889$ & $+\,6.4\;-\,5.4$   & $+\,7.2\;-\,6.1$ \\ 
      & $0.101$  & $+\,15.8\;-\,14.5$ & $+\,6.1\;-\,6.3$ \\ 
      & $0.117$  & $+\,12.2\;-\,11.9$ & $+\,5.8\;-\,5.1$ \\ 
$800$ & $0.0588$ & $+\,6.5\;-\,5.4$   & $+\,7.6\;-\,6.3$ \\ 
      & $0.0733$ & $+\,18.0\;-\,16.0$ & $+\,6.7\;-\,6.9$ \\ 
      & $0.0850$ & $+\,13.7\;-\,13.0$ & $+\,6.1\;-\,5.5$ \\ 
$850$ & $0.0394$ & $+\,6.5\;-\,5.5$   & $+\,8.0\;-\,6.6$ \\ 
      & $0.0545$ & $+\,20.4\;-\,17.6$ & $+\,7.2\;-\,7.5$ \\ 
      & $0.0643$ & $+\,15.7\;-\,14.3$ & $+\,6.5\;-\,6.0$ \\ 
$900$ & $0.0267$ & $+\,6.7\;-\,5.6$   & $+\,8.3\;-\,6.9$ \\ 
      & $0.0417$ & $+\,22.8\;-\,19.1$ & $+\,7.7\;-\,8.0$ \\ 
      & $0.0509$ & $+\,18.1\;-\,16.0$ & $+\,7.0\;-\,6.6$ \\ 
\hline
\end{tabular}
\end{center}
\end{table}

In \refT{tab:HTO_8} we compare results from Table~5 of \Bref{Anastasiou:2011pi}
with our OFFBW results with two options: 1) $\muRs, \muFs$ are fixed, 2) they run with $z/4$. 
For the three values of $\muh$ reported we find differences of $2.0\%, 3.7\%$, and
$4.7\%$, compatible with the scale uncertainty. Furthermore, we use 
\Bref{Dittmaier:2011ti} for input parameters which differ slightly from the one used
in \Bref{Anastasiou:2011pi}. Note that also the functional form ot the Breit--Wigner
is different since their default value is not the one in \eqn{BWdef} but
\bq 
\hbox{BW}(s) = \frac{1}{\pi}\,\frac{\sqrt{s}\,\Gamma_{\PH}(\sqrt{s})}
{\lpar  s - \muhs\rpar^2 + \muhs\,\Gamma^2_{\PH}(\muh)},
\label{XBWdef}
\eq
where $\Gamma_{\PH}(\sqrt{s})$ id the decay width of a Higgs boson at rest with mass
$\sqrt{s}$.
\begin{table}
\begin{center}
\caption[]{\label{tab:HTO_8}{Comparison of results from Table~5 of \Bref{Anastasiou:2011pi}
with our OFFBW results (\eqn{offbw}) with two options: 1) $\muRs, \muFs$ are fixed, 2) they run 
with $z/4$.}}
\begin{tabular}{cccccccc}
\hline 
&  {\sc iXis} & & {\sc HTO} &&&& \\
$\muh$[GeV] &  $\sigma$[pb]   & Scale\,[\%] &
                 $\sigma_1$[pb] & Scale\,[\%] &
                 $\sigma_2$[pb] & Scale\,[\%] \\
\hline
$220$ & $4.54$ & $+\,6.9\;-\,9.0$ & $4.63$ & $+\,8.9\,-\,8.9$ & $4.74$ & $+\,7.3\;-\,8.7$ \\
$260$ & $3.25$ & $+\,6.4\;-\,8.8$ & $3.37$ & $+\,8.6\,-\,8.2$ & $3.49$ & $+\,7.9\;-\,8.6$ \\
$300$ & $2.57$ & $+\,5.9\;-\,8.6$ & $2.69$ & $+\,8.4\,-\,7.8$ & $2.81$ & $+\,8.4\;-\,8.5$ \\
\hline
\end{tabular}
\end{center}
\end{table}

In \refT{tab:HTO_9} we use the OFFP-scheme of \eqn{offp} and look for the effect of
running QCD scales in the $\PH$ invariant-mass distribution. Differences are of the order of
$2{-}3\%$ apart from the high-mass side of the distribution for very high $\PH$ masses. 
\begin{table}
\begin{center}
\caption[]{\label{tab:HTO_9}{Scaling factor in the OFFP-scheme (\eqn{offp})
running/fixed QCD scales for the invariant-mass distribution.
Here $[x,y] = [M_{\peak} + x\,\Gamma^{\OS}_{\PH}\,,\,
M_{\peak} + y\,\Gamma^{\OS}_{\PH}]$.}}
\begin{tabular}{ccccccc}
\hline 
\vphantom{$\Big[$}
$\muh$[GeV] & 
$[-1\,,\,-\frac{1}{2}]$             &
$[-\frac{1}{2}\,,\,-\frac{1}{4}]$   &
$[-\frac{1}{4}\,,\,0]$              &
$[0\,,\,\frac{1}{4}]$               &
$[\frac{1}{4}\,,\,\frac{1}{2}]$     &
$[\frac{1}{2}\,,\,1]$               \\
\hline 
$600$ & $1.023$ & $1.021$ & $1.019$ & $1.018$ & $1.017$ & $1.016$  \\
$700$ & $1.027$ & $1.023$ & $1.021$ & $1.019$ & $1.018$ & $1.019$  \\
$800$ & $1.030$ & $1.024$ & $1.021$ & $1.021$ & $1.021$ & $1.136$  \\
\hline
\end{tabular}
\end{center}
\end{table}
As we have already mentioned, the only scheme respecting gauge invariance that allows us for
a proper definition of pseudo-observables is the CPP-scheme of \eqn{cpp}. It requires
analytical continuation of the Feynman integrals into the second Riemann sheet.
Once again, the definition of POs is conventional, but should put in one-to-one
correspondence well-defined theoretical predictions with derived experimental data.
In \refT{tab:HTO_10} we give a simple example by considering the process $\Pg\Pg \to \PH$ 
at lowest order and compare the traditional on-shell production cross section, see
the l.h.s. of \eqn{sigmaBW}, with the production cross section as defined in the CPP-scheme. 
Therefore we only consider $\Pg\Pg \to \PH$ at LO (\ie $v = z$) and put $\PH$ on its real 
mass shell or on the complex one.
Below $300\UGeV$ there is no visible difference, at $300\UGeV$ the two results start to
differ with an increasing gap up to $600\UGeV$ after which there is a plateau of about $25\%$ 
for higher values of $\muh$.
\begin{table}
\begin{center}
\caption[]{\label{tab:HTO_10}{$\gh$ and $\GOSu_{\PH}$ as a function of $\muh$. We also report
the ratio between the LO cross sections for $\Pg\Pg \to \PH$ in the CPP-scheme and in the
OS-scheme.}} 
\begin{tabular}{cccc}
\hline 
$\muh$[GeV]  & $\gh$[GeV]  & $\GOSu_{\PH}$[GeV] &
$\sigma_{\CPP}(\cph)\,/\,\sigma_{\OS}(M^{\OS}_{\PH})$ \\
\hline
$   300 $&$    7.58  $&$     8.43 $&$      0.999 $ \\
$   350 $&$    14.93 $&$    15.20 $&$      1.086 $ \\
$   400 $&$    26.66 $&$    29.20 $&$      1.155 $ \\
$   450 $&$    41.18 $&$    46.90 $&$      1.183 $ \\
$   500 $&$    58.85 $&$    68.00 $&$      1.204 $ \\
$   550 $&$    79.80 $&$    93.10 $&$      1.221 $ \\
$   600 $&$   104.16 $&$   123.00 $&$      1.236 $ \\
$   650 $&$   131.98 $&$   158.00 $&$      1.248 $ \\
$   700 $&$   163.26 $&$   199.00 $&$      1.256 $ \\
$   750 $&$   197.96 $&$   248.00 $&$      1.262 $ \\
$   800 $&$   235.93 $&$   304.00 $&$      1.264 $ \\
$   850 $&$   277.00 $&$   371.00 $&$      1.263 $ \\
$   900 $&$   320.96 $&$   449.00 $&$      1.258 $ \\
$   950 $&$   367.57 $&$   540.00 $&$      1.252 $ \\
$  1000 $&$   416.57 $&$   647.00 $&$      1.242 $ \\
\hline
\end{tabular}
\end{center}
\end{table}

The production cross section in \eqn{PDFprod_4} is made of one term corresponding to
$\Pg\Pg \to \PH$ which we denote by $\sigma^0$ and terms with at least one additional jet,
$\sigma^j$. The invariant mass of the two partons in the initial state is $v s$ with
$z < v < 1$, $z s$ being the Higgs virtuality. In \refT{tab:HTO_11} we introduce a cut
such that
\bq
z < v < \min\{ v_c\,z\,,\,1\}
\eq
and study the dependence of the result on $v_c$ by considering $\sigma^t = \sigma^0 + \sigma^j$ 
and $\delta= \sigma^t_c/\sigma^t$. The results show that a cut $z < v < 1.01\,z$ gives
already $85\%$ of the total answer.
\begin{table}
\begin{center}
\caption[]{\label{tab:HTO_11}{The effect of introducing a cut on the invariant mass
of the initial-state partons.}} 
\begin{tabular}{cccccc}
\hline 
$\muh$[GeV]        & 
$v_c$                &
$\sigma^0$[pb]     &
$\sigma^j$[pb]     &
$\sigma^t$[pb]     &
$\delta[\%]$         \\
\hline
$600$ & no cut  & $0.2677$ & $0.1107$ &  $0.3784$  &      \\ 
      & $1.001$ &          & $0.0127$ &  $0.2804$  & $74$ \\
      & $1.01$  &          & $0.0473$ &  $0.3149$  & $83$ \\ 
      & $1.05$  &          & $0.0918$ &  $0.3595$  & $95$ \\
$700$ & no cut  & $0.1208$ & $0.0469$ &  $0.1677$  &      \\    
      & $1.001$ &          & $0.0054$ &  $0.1262$  & $75$ \\
      & $1.01$  &          & $0.0201$ &  $0.1409$  & $84$ \\ 
      & $1.05$  &          & $0.0391$ &  $0.1599$  & $95$ \\
$800$ & no cut  & $0.0632$ & $0.0218$ &  $0.0850$  &      \\
      & $1.001$ &          & $0.0025$ &  $0.0657$  & $77$ \\ 
      & $1.01$  &          & $0.0094$ &  $0.0726$  & $85$ \\
      & $1.05$  &          & $0.0182$ &  $0.0814$  & $96$ \\ 
\hline
\end{tabular}
\end{center}
\end{table}

Although a final decision will only be taken in January 2012 it looks more and more likely 
that LHC will run at $8\UTeV$ in 2012; in \refT{tab:HTO_12} we have shown a comparison
between production cross sections compute at $7\UTeV$ and $8\UTeV$.
\begin{table}
\begin{center}
\caption[]{\label{tab:HTO_12}{Comparison of the production cross sections
in the OFFP-scheme (\eqn{offp}) at $7\UTeV$ and $8\UTeV$.}} 
\begin{tabular}{cccc}
\hline 
$\muh$[GeV]                 & 
$\sigma$[pb] $7\UTeV$       &
$\sigma$[pb] $8\UTeV$       &
ratio                         \\
\hline
$600$  & $0.378$  & $0.587$  & $1.55$ \\
$650$  & $0.248$  & $0.392$  & $1.58$ \\
$700$  & $0.168$  & $0.271$  & $1.67$ \\
$750$  & $0.117$  & $0.194$  & $1.67$ \\
$800$  & $0.0850$ & $0.145$  & $1.77$ \\
$850$  & $0.0643$ & $0.113$  & $1.77$ \\
$900$  & $0.0509$ & $0.0922$ & $1.87$ \\
\hline
\end{tabular}
\end{center}
\end{table}

On the left-hand side of \refF{fig:HTO_12} we compare the production cross section as 
computed with the OFFP-scheme of \eqn{offp} or with the OFFBW-scheme of \eqn{offbw}
for $\muh = 600\UGeV$. For the latter we use Breit--Wigner parameters in the OS-scheme 
(red curve) and in the Bar-scheme of \eqn{Bars} (blue curve). Deviations from the OFFP-scheme are 
maximal in the OS-scheme and much less pronounced in the Bar-scheme.

On the right-hand side of \refF{fig:HTO_12} we show the effect of using dynamical QCD 
scales for the $\Pg\Pg \to \PH + \PX \to \PH + \PGg\PGg$ cross section at $\muh= 400\UGeV$. 

On the left-hand side of \refF{fig:HTO_35} we consider the on-shell production cross section 
$\sigma^{\OS}( \Pp\Pp \to \PH )$ (black curve) which includes convolution with PDFs.
The blue curve gives the off-shell production cross section sampled over the (complex) Higgs
propagator while the red curves is sampled over a Breit--Wigner distribution. The
observed effect is substantial even in the low-mass region. 

On the right-hand side of \refF{fig:HTO_35} we show the differential $K$~factor for the process
$\Pp\Pp \to (\PH \to 4\,\Pe)  + \PX$, comparing the fixed-scale option,
$\muR = \muF = \MH/2$, and the running-scale option, $\muR = \muF = M(4\,\Pe)/2$.
For running QCD scales the $K$~factor is practically constant over a wide range of the Higgs
virtuality.

On the left-hand side of \refF{fig:HTO_67} we show the normalised invariant-mass distribution 
in the OFFP-scheme with running QCD scales for $600\UGeV$ (black), $700\UGeV$ (blue), 
$800\UGeV$ (red) in the windows $M_{\peak} \pm 2\,\GOS$, where $M_{\peak}$ is defined in \eqn{peak}.

Finally, on the right-hand side of \refF{fig:HTO_67} we show the normalised invariant-mass 
distribution in the OFFP-scheme (blue) and OFFBW-scheme (red) with running QCD scales 
at $800\UGeV$ in the window $M_{\peak} \pm 2\,\GOS$.
The invariant-mass distribution in the OFFP-scheme with running QCD scales for $800\UGeV$ in 
the window $M_{\peak} \pm 2\,\GOS$ is shown in \refF{fig:HTO_10}. The blue line refers 
to $8\UTeV$, the red one to $7\UTeV$.

\subsection{QCD scale error\label{PO_scerr}}
The conventional theoretical uncertainty associated with QCD scale variation is defined in 
\eqn{error}. Let us consider the production cross section at $800\UGeV$ in the OFFP-scheme with
running or fixed QCD scales, we obtain
\bq
\mbox{RS}\quad 0.08497\,{}^{+13.7\%}_{-13.0\%},
\qquad
\mbox{FS}\quad 0.07185\,{}^{+6.7\%}_{-8.5\%}.
\eq
The uncertainty with running scales is about twice the one where the scales are kept fixed.
One might wonder which is the major source and we have done the following; in estimating
the conventional uncertainty in the RS-option we keep $\muFs = \muol^2 = z s/4$ and 
vary $\muR$ between $\muol/2$ and $2\,\muol$. The corresponding result is
\bq
\mbox{RS}\quad 0.08497\,{}^{+3.9\%}_{-4.4\%}, \qquad\quad \muR\quad\mbox{only}.
\eq
Therefore, the main source of uncertainty comes from varying the factorisation scale.
However, in the fixed-scale option we obtain
\bq
\mbox{FS}\quad 0.07185\,{}^{+3.6\%}_{-4.4\%}, \qquad\quad \muR\quad\mbox{only}
\eq
and $\muF\,$-variation is less dominant. In any case the question of which variation is
dominating depends on the mass range; indeed, in the ONBW-scheme with fixed scales at
$140\UGeV$ we have
\bqa
\mbox{ONBW}\quad &12.27&\,{}^{+11.1\%}_{-10.1\%}, \qquad\quad \muR\quad\mbox{and}\quad \muF,
\nl
\mbox{ONBW}\quad &12.27&\,{}^{+8.7\%}_{-9.6\%}, \qquad\quad \muR\quad\mbox{only}
\eqa
showing $\muR$ dominance.

In order to compare this conventional definition of uncertainty with the work of 
\Bref{Cacciari:2011ze} we recall their definition of $p\%\,$ credible interval; given a series
\bq
\sigma = \sum_{n=l}^k\,c_n\,\alphas^n,
\eq
the $p\%\,$ credible interval $\sigma \pm \Delta\sigma^p$ is defined as
\bqa
\Delta\sigma^p &=& \alpha_s^{k+1}\,\max\lpar\{c\}\rpar\,\frac{n_c+1}{n_c}\,p\%,
\qquad p\% < \frac{n_c}{n_c+1},
\nl
\Delta\sigma^p &=& \alpha_s^{k+1}\,\max\lpar\{c\}\rpar\,
\Bigl[(n_c+1)\,(1-p\%)\Bigr]^{-1/n_c},
\qquad p\% > \frac{n_c}{n_c+1},
\eqa
with $n_c= k-l+1$. Using this definition we find that the $68\%(90\%)$ credible interval for 
$\muR$ uncertainty in the production cross section at $800\UGeV$ is $3.1\%(5.3\%)$ in substantial 
agreement with the conventional method. For ONBW-scheme at $140\UGeV$ the $90\%$
credible interval for $\muR$ uncertainty is $8.0\%$. It could be interesting to extend the 
method of \Bref{Cacciari:2011ze} to cover $\muF$ uncertainty.
\subsection{Concluding remarks\label{PO:sconc}}
In the last two decades many theoretical studies lead to improved estimates of the
SM Higgs-boson total cross section at hadron 
colliders~\cite{Spira:1995rr,Kramer:1996iq,Harlander:2001is,Anastasiou:2002yz,Ravindran:2002dc,Catani:2003zt} and of Higgs-boson partial 
decay widths~\cite{Prophecy4f}. Less work has been devoted to studying the 
Higgs-boson invariant-mass distribution (Higgs-boson lineshape).

In this section we made an attempt to resolve the problem by comparing different
theoretical inputs to the off-shellness of the Higgs boson. 
There is no question at all that the zero-width approximation should be avoided, especially
in the high-mass region where the on-shell width becomes of the same order as the
on-shell mass.

The propagation of the off-shell $\PH$ is usually parametrised in terms of some Breit--Wigner 
distribution (several variants are reported in the literature, \eqn{BWdef} and
\eqn{XBWdef}); we have shown evidence that only the Dyson-resummed propagator should be used, 
leading to the introduction of the $\PH$ complex pole, a gauge-invariant property
of the $S$-matrix.

Finally, when one accepts the idea that the Higgs boson is an off-shell intermediate
state the question of the most appropriate choice of QCD scales arises. We have shown
the effect of including dynamical choices for $\muR, \muF$ instead of a static choice.

For many purposes the well-known and convenient machinery of the on-shell approach can
be employed, but one should become aware of its limitations and potential pitfalls.
Although the basic comparison between different schemes was presented in this section, at
this time we are repeating that differences should not be taken as the source of
additional theoretical uncertainty; as a consequence, our recommendation is to use the 
OFFP-scheme described in \refS{PO:schem} (see \eqn{offp}) with running QCD scales. 
\clearpage

\begin{figure}
 \includegraphics[width=0.49\textwidth]{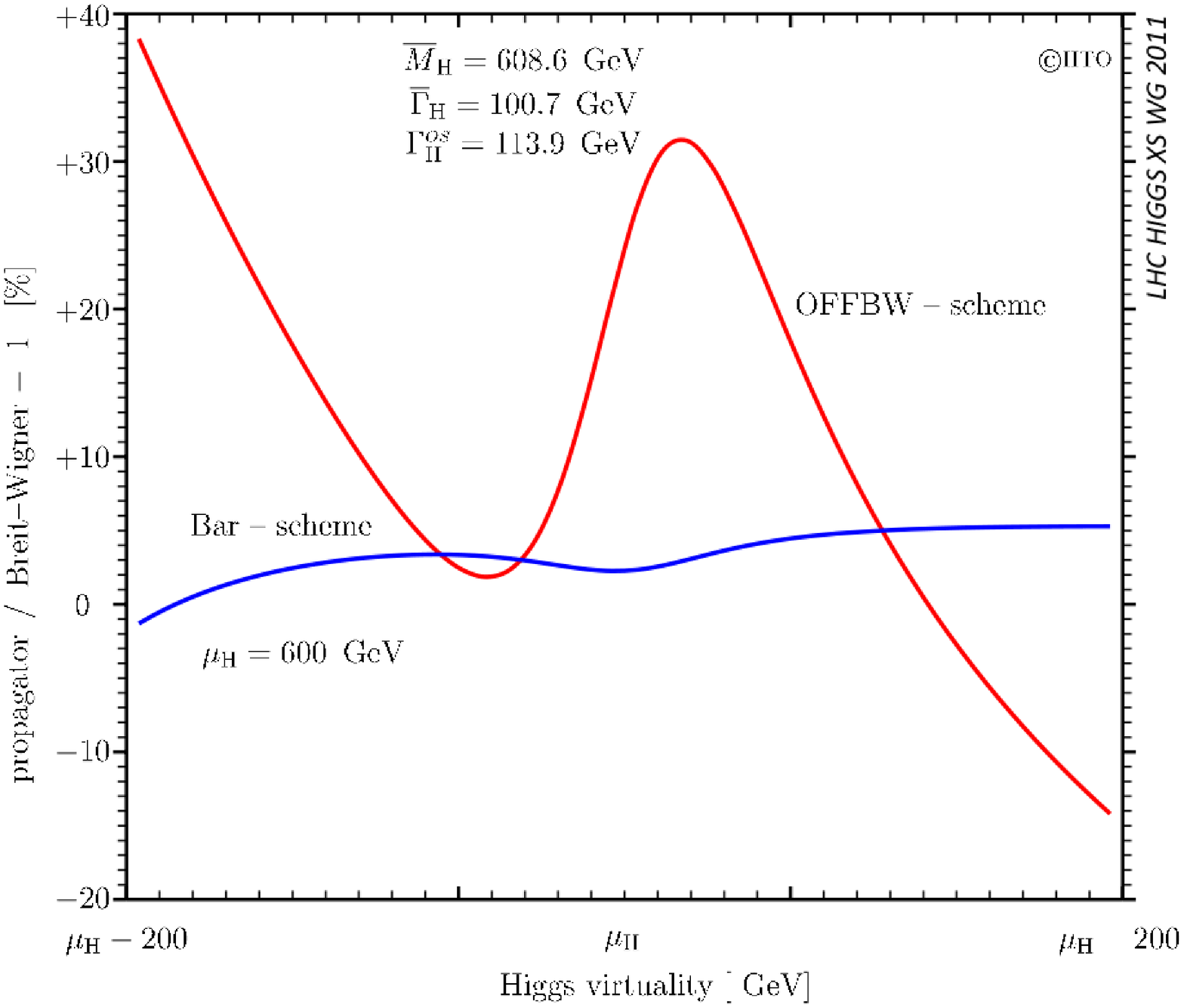}
 \includegraphics[width=0.49\textwidth]{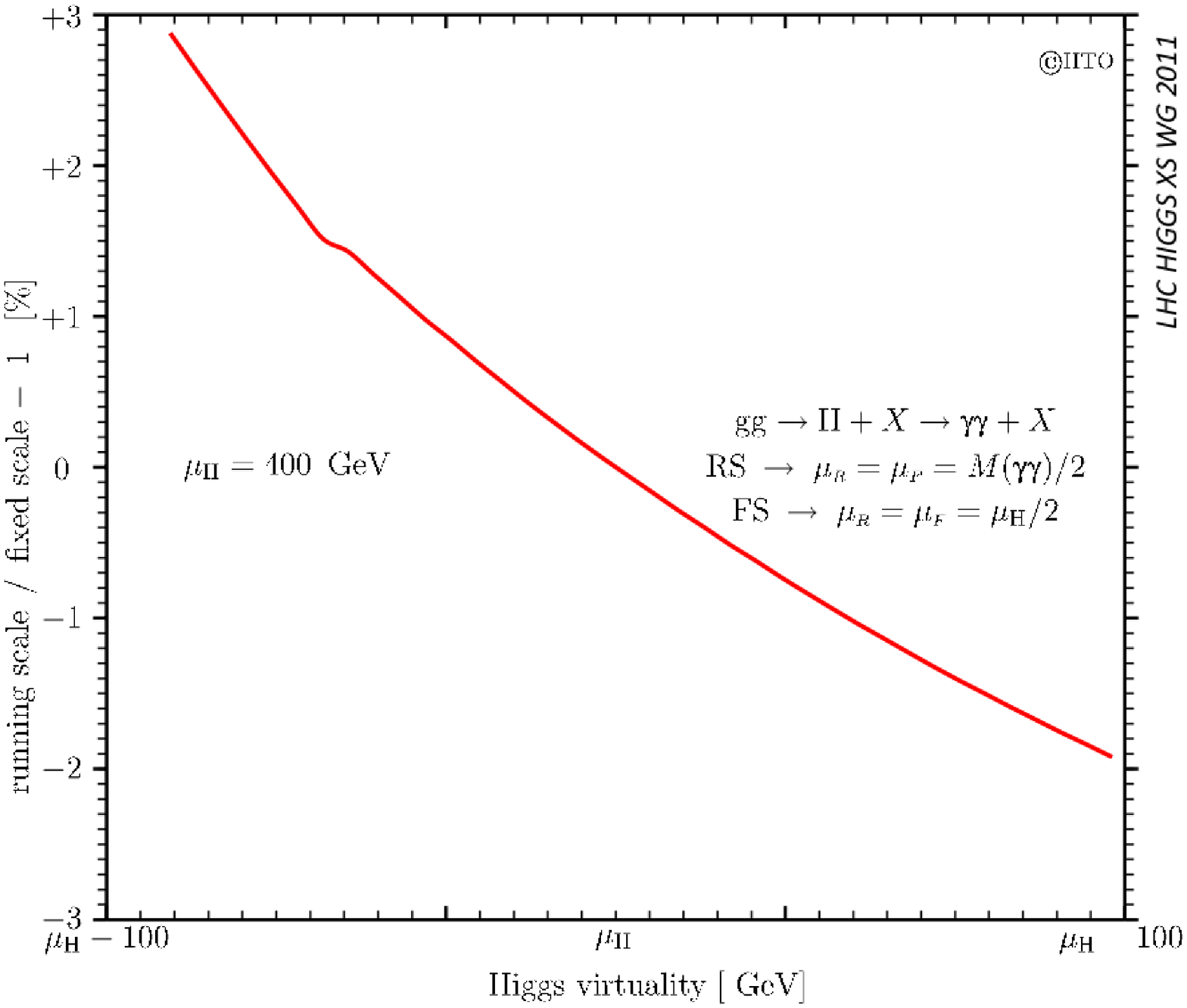}
  \caption{
In the left figure we show a comparison of production cross section as computed with the 
OFFP-scheme of \eqn{offp} or with the OFFBW-scheme of \eqn{offbw}. The red curve
gives Breit--Wigner parameters in the OS-scheme and the blue one in the Bar-scheme of
\eqn{Bars}.
In the right figure we show the effect of using dynamical QCD scales for the production 
cross section of \eqns{PDFprod_1}{PDFprod_3}.} 
\label{fig:HTO_12}
\end{figure}

\begin{figure}
  \includegraphics[width=0.49\textwidth]{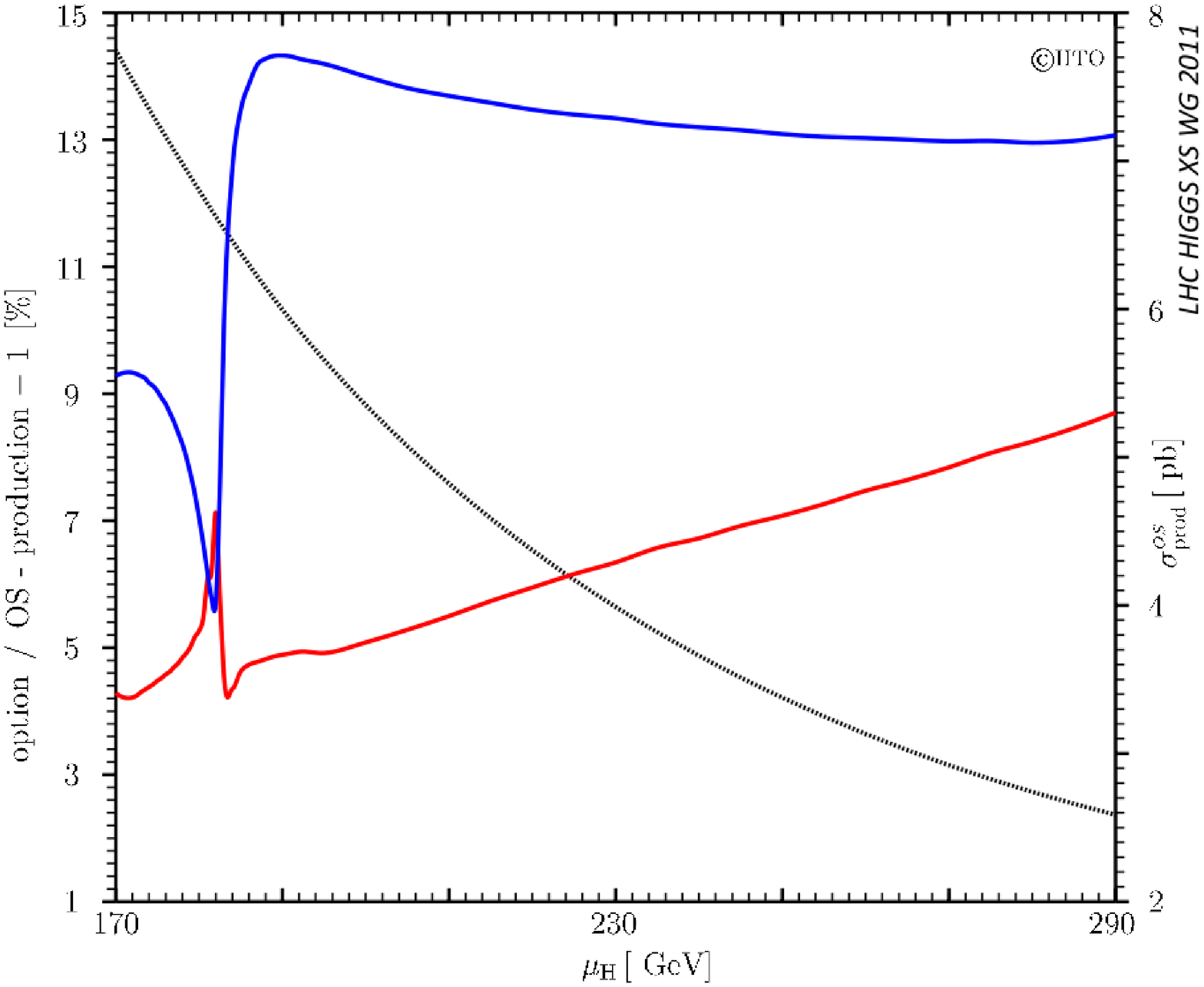}
\includegraphics[width=0.49\textwidth]{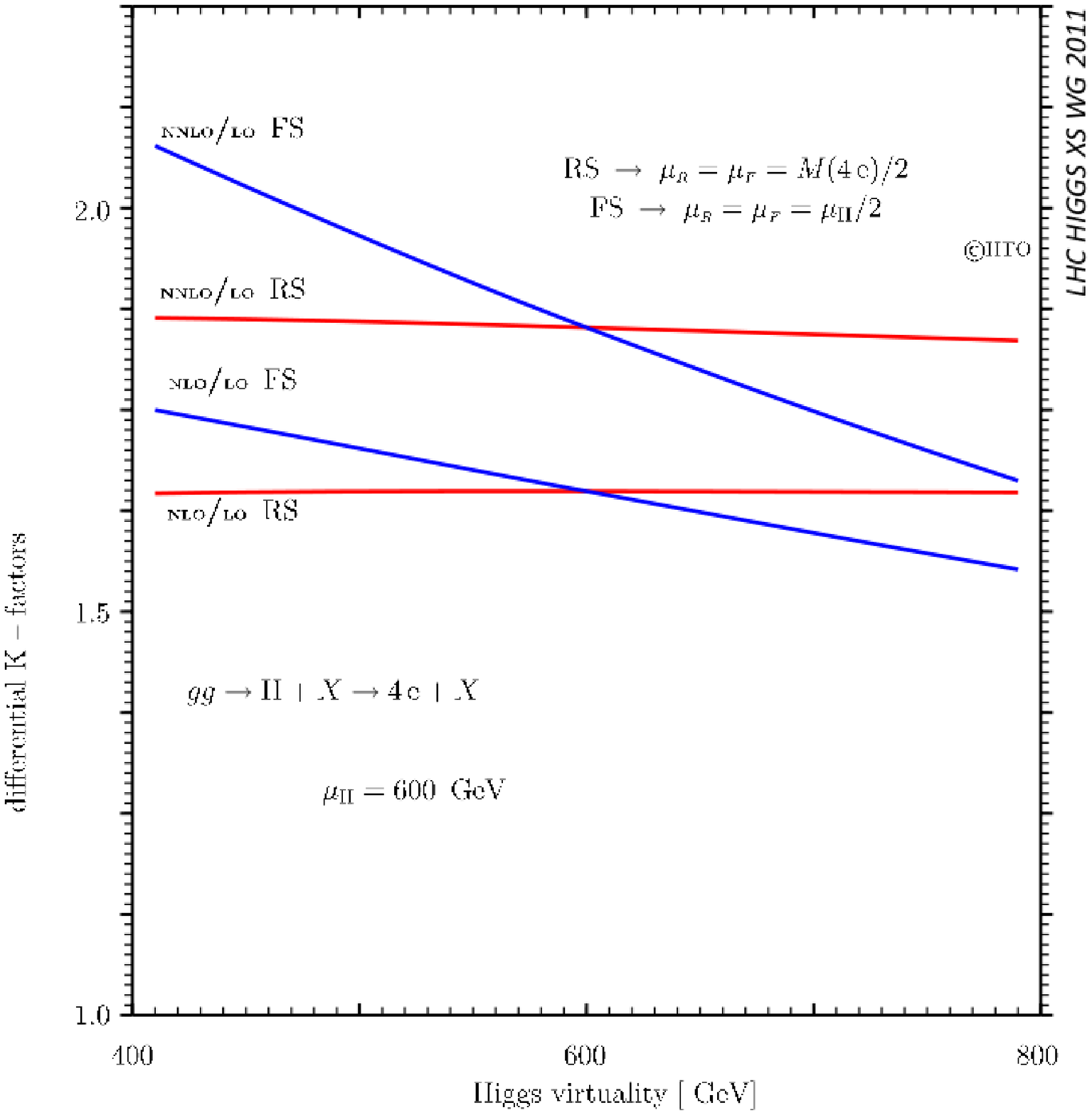}
  \caption{
In the left figure the blue curve gives the off-shell production cross section sampled over 
the (complex) Higgs propagator while the red curves is sampled over a Breit--Wigner distribution
The right figure showes differential $K$~factor for the process $\Pp\Pp \to (\PH \to 4\,\Pe)  
+ \PX$, comparing the fixed scale option, $\muR = \muF = \MH/2$, and the running scale option,
$\muR = \muF = M(4\,\Pe)/2$.}
\label{fig:HTO_35}
\end{figure}


\begin{figure}
  \includegraphics[width=0.49\textwidth]{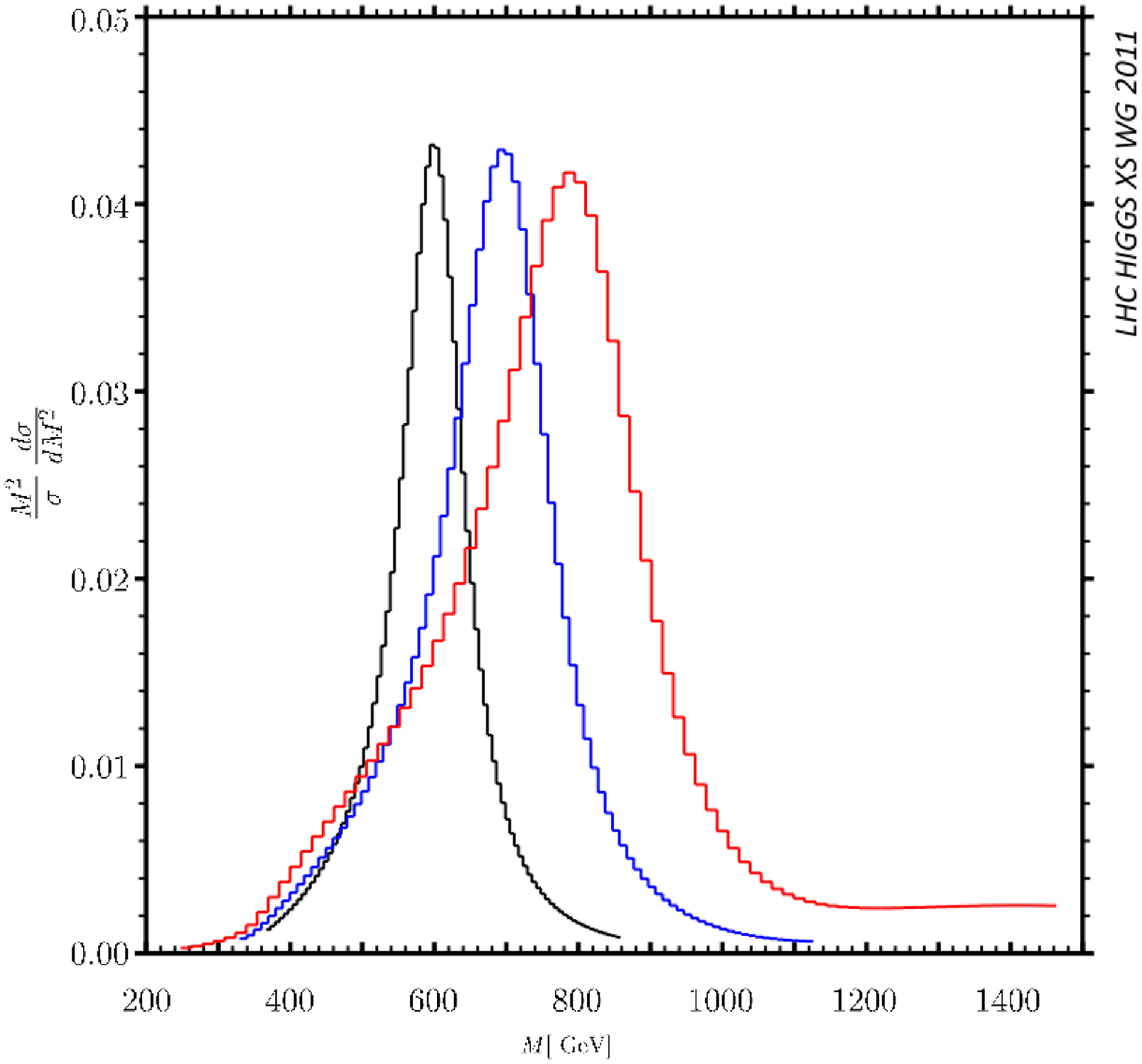}
 \includegraphics[width=0.49\textwidth]{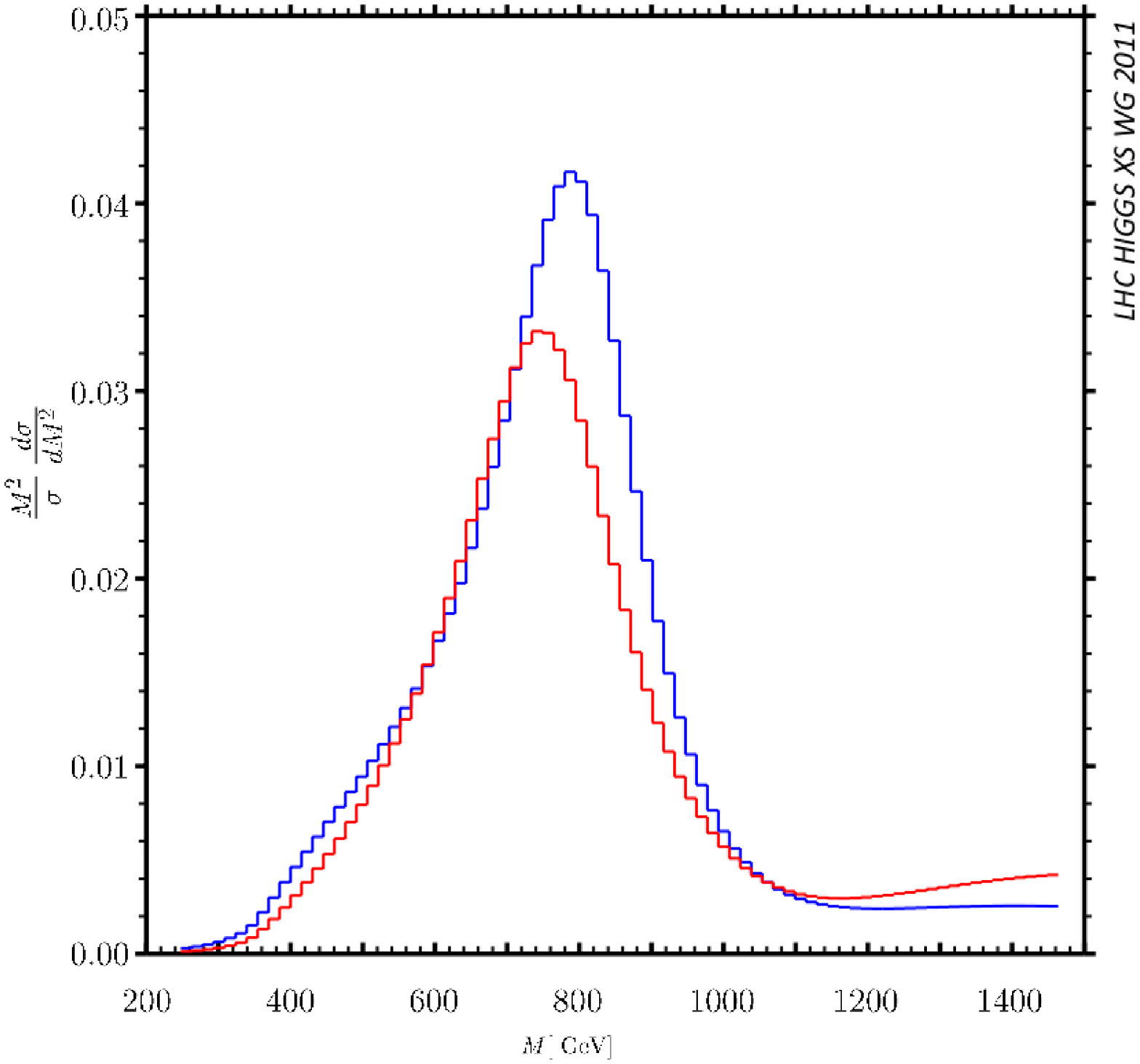}
  \caption{
The normalised invariant-mass distribution in the OFFP-scheme with
running QCD scales (left) for $600\UGeV$ (black), $700\UGeV$ (blue), $800\UGeV$ (red) in
the windows $M_{\peak} \pm 2\,\GOS$.
The normalised invariant-mass distribution in the OFFP-scheme (blue) and OFFBW-scheme (red)
with running QCD scales (right) at $800\UGeV$ in the window $M_{\peak} \pm 2\,\GOS$.}
\label{fig:HTO_67}
\end{figure}

\begin{figure}
 \centering
 \includegraphics[width=0.7\textwidth]{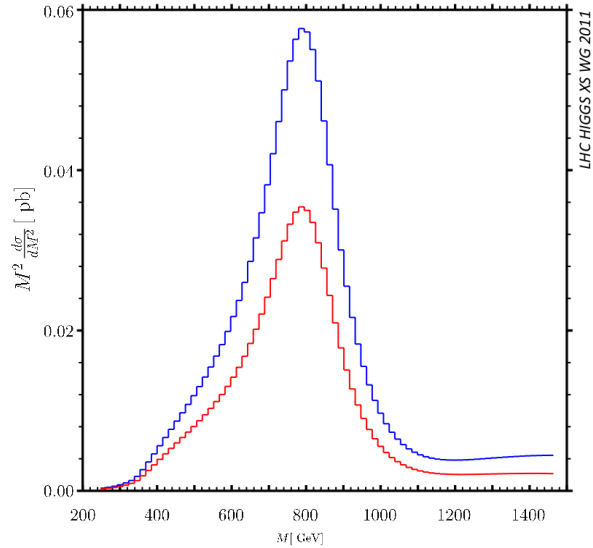}
  \caption{
The invariant-mass distribution in the OFFP-scheme with
running QCD scales for $800\UGeV$ in the window $M_{\peak} \pm 2\,\GOS$. The blue line refers 
to $8\UTeV$, the red one to $7\UTeV$.}
\label{fig:HTO_10}
\end{figure}

\clearpage

\begin{figure}
\vspace{2.cm}
\centering
\centerline{\includegraphics[bb=  0 0 595 842,width=16.cm]{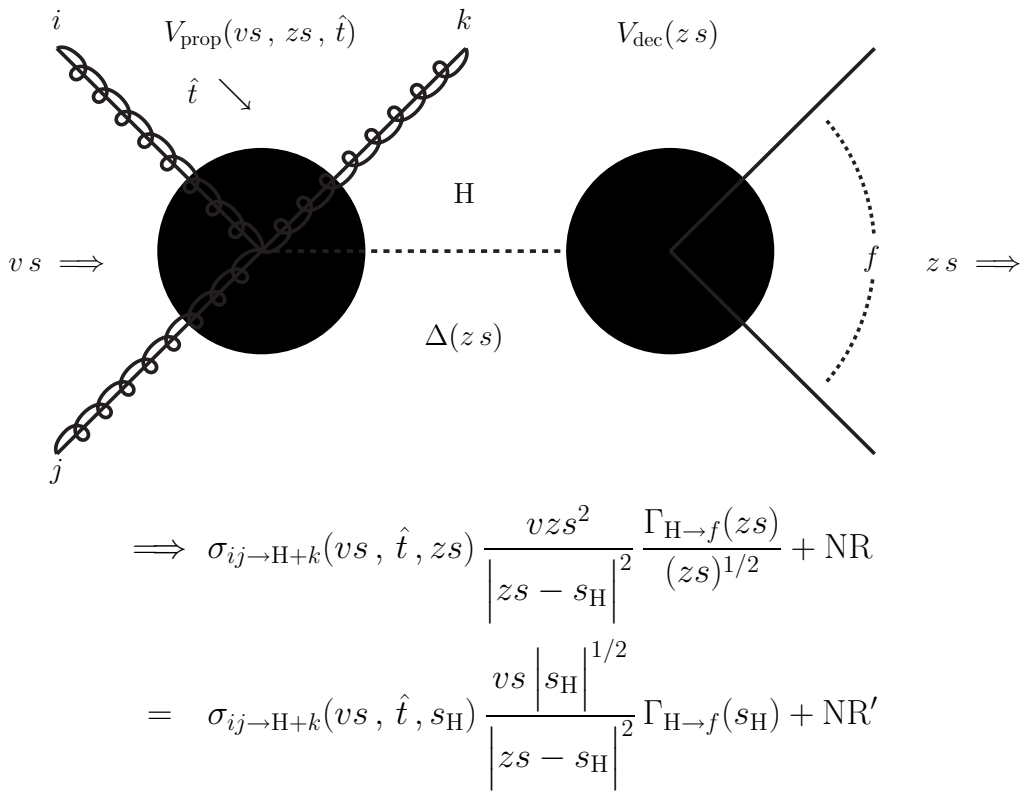}}
\vspace{-8.cm}
\caption[]{The resonant part of the process $i j \to \PH + k$ where $i,j$ and $k$
are partons ($\Pg$ or $\PQq$). We distinguish between production, propagation 
and decay, indicating the explicit dependence on Mandelstam invariants. If $s$ denotes 
the $\Pp\Pp$ invariant mass we distinguish between $v s$, the $i j$ invariant mass, $z s$, 
the Higgs-boson virtuality and ${\hat t}$ which is related to the Higgs-boson transverse 
momentum. Within NR we include the non-resonant contributions as well as their interference 
with the signal. NR includes those terms that are of higher order in the Laurent expansion 
of the signal around the complex pole. At NNLO QCD the parton $k$ is a pair $k_1{-}k_2$, and
more invariants are needed to characterise the production mechanism.}
\label{HLS:complete}
\end{figure}
\clearpage


\clearpage

\newpage
\section{Conclusions\footnote{S.~Dittmaier, C.~Mariotti, G.~Passarino
    and R.~Tanaka.}}

The present document is the result of the activities at the 
Higgs boson cross section workshops in 2011. 
The LHC Higgs cross section working group was created in January 2010 as a new joint effort 
between ATLAS, CMS, and the theory community.

In the previous Report~\cite{Dittmaier:2011ti} we presented a description of cross sections 
computed at next-to-next-to-leading order (NNLO) or NLO, for each of the production 
modes. 

Our goal in this second volume is to extend the previous studies to include differential 
cross sections. The motivation for such investigations is clear: Experimental analyses 
must impose cuts on the final state in order to extract the signal from background.  
A precise determination of the corresponding signal acceptance is needed. Hopefully, 
it is computed to the same order in the perturbative expansion as the inclusive cross section.  

Therefore, this Report seeks to answer important questions about the computation of cross 
sections that include acceptance cuts and differential distributions for all final 
states that are considered in the Higgs search at the LHC.

The possible distributions of interest in the myriad studies performed at the LHC cover a
broad spectrum. Our goal has been to identify and discuss the relevant issues that arise, and 
to give prescriptions for their solution. We must also stress that such prescriptions are 
not set in stone, but are subject to change if theoretical advances occur.  
If new more precise calculations become available, these new results will be incorporated into the 
relevant experimental investigations. It is our scientific mission and our raison d'\^{e}tre.
Updated results will be made available at the TWiki page: \\
https://twiki.cern.ch/twiki/bin/view/LHCPhysics/CrossSections.

Here is a summary of our guidelines: collecting tools with combined experiment/theory
expertise allowing controlled studies of different approaches;
stimulate authors to incorporate new features/algorithms into their codes;
understand differences between groups, which is even more important than continued
improvements within groups;
answer the general question `if a theoretical calculation is done, but it cannot be used by 
any experimentalists, does it make a sound ?'

If the Higgs boson exists in the simple form invoked by the Standard Model 
it should start to show up in the data soon, but probably not strongly enough to claim a discovery. 
If the elusive boson is a mirage, experiments should be able to rule it out completely. 

Alternative descriptions, such as some altogether different mechanism, are much more difficult 
to detect and, perhaps, less attractive. As we know from the history of science, this argument 
does not suffice to rule out the existence of such alternatives, and we should be ready to
cope with the future. But that's another story, for another time,
another Report. 
\begin{verse}
Sed fugit interea fugit irreparabile tempus, singula dum capti circumvectamur amore
\end{verse}
(But meanwhile it flees: time flees irretrievably, while we wander around, prisoners of our 
love of detail)

\clearpage

\begin{flushleft}
{\bf Acknowledgements}
\end{flushleft}

We are obliged to 
Ketevi Assamagan, 
Albert De Roeck, 
Eilam Gross, 
Andrey Korytov,
Sandra Kortner, 
Bill Murray, 
Aleandro Nisati, 
Christoph Paus,  
Gigi Rolandi and
Vivek Sharma
for their support and encouragement.


We would like to thank 
Lance Dixon,
Pietro Govoni,
Andrea Massironi and
Adrian Signer
for discussions. 


We are obliged to CERN, in particular to the IT Department and to the
Theory Unit for the support with logistics, especially to Elena Gianolio for
technical assistance.


S.Alekin, J. Bl\"umlein and S.~Moch have been supported on part by
Helmholtz Gemeinschaft under contract VH-HA-$101$ ({\it Alliance Physics at
the Tera\-scale}), by the
Deutsche Forschungsgemeinschaft in Sonderforschungsbereich/Transregio~$9$,
by the European Commission through contract PITN-GA-$2010-264564$ ({\it
LHCPhenoNet}), and by the
Russian Foundation for Basic Research under contract RFFI $08-02-91024$
CERN\_a.

A.~Bagnaschi, G.~Degrassi, P.~Slavich and A.~Vicini are
supported by the Research Executive Agency (REA) of the
European Union under the Grant Agreement number PITN-GA-$2010-264564$ (LHCPhenoNet). 

M.V.~Garzelli, A.~Kardos, and Z.~~Tr\'ocs\'any are supported in part by
   the LHCPhenoNet network PITN-GA-$2010-264564$,
   the Swiss National Science Foundation Joint Research Project SCOPES
   IZ$73Z0\_1/28079$, and the T\'AMOP $4.2.1./B-09/1/$KONV-$2010-0007$ project.

M.V.~Garzelli is supported in part by the MEC project FPA $2008-02984$.

S.~Heinemeyer is supported in part by CICYT (grant FPA $2010--22163-C02-01$) 
  and by the Spanish MICINN's Consolider-Ingenio $2010$ Program under grant MultiDark 
  CSD$2009-00064$.

J.~Huston would like to acknowledge National Science Foundation grant PHY-$1068318$

C.B. Jackson, H. Kim and J. Yu thank the U.S.~Department of Energy and the University 
of Texas at Arlington for the support.

P.~Jimenez-Delgado research is supported in part by the Swiss National Science 
Foundation (SNF) under contract $200020{-}138206$.

M.~Kr\"amer, M.~M\"uhlleitner and M.~Ubiali are supported by the DFG SFB/TR$9$
   ``Computational Particle Physics''.

P.~Nadolsky at SMU was supported in part by the U.S. DOE Early 
Career Research Award DE-SC$0003870$ and by Lightner-Sams Foundation.

G.~Passarino is supported by Ministero dell'Istruzione, dell'Universit\`a e della Ricerca
   Protocollo $2008$H$8$F$9$RA${}_{}002$. 

F.~Petriello is supported by the US DOE under contract DE-AC$02-06CH11357$ and the grant 
   DE-FG$02-91ER40684$.

L.~Reina is partially supported by the US DOE grant DE-FG$02-97IR41022$.

J.~Rojo has been supported by a Marie Curie 
Intra--European Fellowship of the European Community's 7th Framework 
Programme under contract number PIEF-GA-$2010-272515$.

G.~Salam is supported in part by grant ANR-$09$-BLAN-$0060$ from the French Agence Nationale 
   de la Recherche.

M.~Sch\"onherr's is supported by the Research Executive Agency (REA) of the European Union 
under the Grant Agreement number PITN-GA-$2010-264564$ (LHCPhenoNet).

F.~Siegert is supported by the German Research Foundation (DFG) via grant DI 784/2-1.

I.~Stewart is supported by the Office of Nuclear Physics of the U.S. Department of Energy 
   under the grant DE-FG$02-94ER40818$.

R.~S.~Thorne is supported partly by the London Centre for Terauniverse 
Studies (LCTS), using funding from the European Research Council via the 
Advanced Investigator Grant $267352$, and travel to LHC-Higgs Working Group 
meetings by an IPPP (Durham) Research Associateship. 

T.~Vickey is supported by the Oxford Oppenheimer Fund, the
Royal Society of the United Kingdom and the National Research
Foundation of the Republic of South Africa.

W.~Waalewijn is supported by the US DOE grant DE-FG$02-90ER40546$.

D.~Wackeroth is partially supported by the National Science
   Foundation under grants NSF-PHY-$0547564$ and NSF-PHY-$0757691$.

M. Warsinsky acknowledges the support of the Initiative and Networking Fund 
of the Helmholtz Association, contract HA-$101$ (Physics at the Terascale).




\clearpage

\bibliographystyle{atlasnote}
\bibliography{YRHXS2}

\clearpage
\appendix
\addcontentsline{toc}{section}{\appendixname}

\section{SM Higgs branching ratios}
\label{brappendix}

In this appendix we complete the listing of the branching fractions of the Standard Model 
Higgs boson discussed in \refS{se:BRs}.

\begin{table}\small \scriptsize
\renewcommand{\arraystretch}{1.4}
\setlength{\tabcolsep}{3ex}
\caption{SM Higgs branching ratios to two fermions and their total uncertainties (expressed in percentage). Very-low mass range.}
\label{BR-2fa}
\begin{center}
\begin{tabular}{lcccc}
\hline
$\MH$ [$\UGeV$] & $\Hbb$ & $\Htautau$ &$\Hmumu$ & $\Hcc$ \\
\hline
90.0   & 8.10\zehomi{1}$^{+1.6\%}_{-1.6\%}$      & 8.33\zehomi{2}$^{+7.0\%}_{-6.8\%}$      & 2.89\zehomi{4}$^{+7.3\%}_{-7.1\%}$   & 4.10\zehomi{2}$^{+12.1\%}_{-12.1\%}$     \\
95.0   & 8.02\zehomi{1}$^{+1.7\%}_{-1.7\%}$      & 8.32\zehomi{2}$^{+6.9\%}_{-6.8\%}$      & 2.89\zehomi{4}$^{+7.3\%}_{-7.1\%}$   & 4.05\zehomi{2}$^{+12.2\%}_{-12.1\%}$     \\
100.0  & 7.89\zehomi{1}$^{+1.8\%}_{-1.8\%}$      & 8.28\zehomi{2}$^{+6.9\%}_{-6.7\%}$      & 2.88\zehomi{4}$^{+7.2\%}_{-7.0\%}$   & 3.99\zehomi{2}$^{+12.2\%}_{-12.2\%}$     \\
105.0  & 7.71\zehomi{1}$^{+1.9\%}_{-2.0\%}$      & 8.17\zehomi{2}$^{+6.8\%}_{-6.7\%}$      & 2.84\zehomi{4}$^{+7.1\%}_{-7.0\%}$   & 3.90\zehomi{2}$^{+12.2\%}_{-12.2\%}$     \\
110.0  & 7.44\zehomi{1}$^{+2.1\%}_{-2.2\%}$      & 7.95\zehomi{2}$^{+6.6\%}_{-6.5\%}$      & 2.76\zehomi{4}$^{+7.0\%}_{-6.8\%}$   & 3.75\zehomi{2}$^{+12.2\%}_{-12.2\%}$     \\
110.5  & 7.40\zehomi{1}$^{+2.1\%}_{-2.2\%}$      & 7.91\zehomi{2}$^{+6.6\%}_{-6.5\%}$      & 2.75\zehomi{4}$^{+6.9\%}_{-6.8\%}$   & 3.74\zehomi{2}$^{+12.2\%}_{-12.2\%}$     \\
111.0  & 7.37\zehomi{1}$^{+2.2\%}_{-2.2\%}$      & 7.89\zehomi{2}$^{+6.6\%}_{-6.5\%}$      & 2.74\zehomi{4}$^{+6.9\%}_{-6.8\%}$   & 3.72\zehomi{2}$^{+12.2\%}_{-12.2\%}$     \\
\hline
\end{tabular}
\end{center}
\end{table}

\begin{table}\scriptsize
\renewcommand{\arraystretch}{1.4}
\setlength{\tabcolsep}{3ex}
\caption{SM Higgs branching ratios to two fermions and their total uncertainties (expressed in percentage). Low- and
intermediate-mass range.}
\label{BR-2fb}
\begin{center}
\begin{tabular}{lcccc}
\hline
$\MH$ [$\UGeV$] & $\Hbb$ & $\Htautau$ &$\Hmumu$ & $\Hcc$ \\
\hline
130.5  & 4.85\zehomi{1}$^{+3.8\%}_{-3.8\%}$    & 5.36\zehomi{2}$^{+5.2\%}_{-5.2\%}$    & 1.86\zehomi{4}$^{+5.5\%}_{-5.4\%}$    & 2.45\zehomi{2}$^{+12.2\%}_{-12.1\%}$     \\
131.0  & 4.76\zehomi{1}$^{+3.8\%}_{-3.9\%}$    & 5.26\zehomi{2}$^{+5.2\%}_{-5.1\%}$    & 1.82\zehomi{4}$^{+5.4\%}_{-5.3\%}$    & 2.41\zehomi{2}$^{+12.2\%}_{-12.1\%}$     \\
131.5  & 4.67\zehomi{1}$^{+3.9\%}_{-3.9\%}$    & 5.17\zehomi{2}$^{+5.1\%}_{-5.1\%}$    & 1.80\zehomi{4}$^{+5.4\%}_{-5.3\%}$    & 2.36\zehomi{2}$^{+12.2\%}_{-12.1\%}$     \\
132.0  & 4.58\zehomi{1}$^{+3.9\%}_{-4.0\%}$    & 5.07\zehomi{2}$^{+5.1\%}_{-5.0\%}$    & 1.77\zehomi{4}$^{+5.3\%}_{-5.2\%}$    & 2.32\zehomi{2}$^{+12.2\%}_{-12.1\%}$     \\
132.5  & 4.49\zehomi{1}$^{+4.0\%}_{-4.0\%}$    & 4.97\zehomi{2}$^{+5.0\%}_{-5.0\%}$    & 1.72\zehomi{4}$^{+5.2\%}_{-5.2\%}$    & 2.27\zehomi{2}$^{+12.2\%}_{-12.1\%}$     \\
133.0  & 4.40\zehomi{1}$^{+4.0\%}_{-4.1\%}$    & 4.88\zehomi{2}$^{+5.0\%}_{-4.9\%}$    & 1.69\zehomi{4}$^{+5.2\%}_{-5.1\%}$    & 2.23\zehomi{2}$^{+12.2\%}_{-12.1\%}$     \\
133.5  & 4.31\zehomi{1}$^{+4.1\%}_{-4.2\%}$    & 4.79\zehomi{2}$^{+4.9\%}_{-4.9\%}$    & 1.67\zehomi{4}$^{+5.1\%}_{-5.1\%}$    & 2.18\zehomi{2}$^{+12.2\%}_{-12.1\%}$     \\
\hline
\end{tabular}\end{center}
\end{table}

\begin{table}\small\scriptsize
\renewcommand{\arraystretch}{1.4}
\setlength{\tabcolsep}{3ex}
\caption{SM Higgs branching ratios to two fermions and their total uncertainties (expressed in percentage). Intermediate-mass range.}
\label{BR-2fc}
\begin{center}
\begin{tabular}{lcccc}
\hline
$\MH$ [$\UGeV$] & $\Hbb$ & $\Htautau$ &$\Hmumu$ & $\Hcc$ \\
\hline
172.0  & 7.10\zehomi{3}$^{+4.5\%}_{-4.5\%}$     & 8.25\zehomi{4}$^{+2.6\%}_{-2.6\%}$      & 2.86\zehomi{6}$^{+2.6\%}_{-2.6\%}$     & 3.57\zehomi{4}$^{+9.9\%}_{-9.9\%}$       \\
174.0  & 6.45\zehomi{3}$^{+4.5\%}_{-4.5\%}$     & 7.51\zehomi{4}$^{+2.6\%}_{-2.6\%}$      & 2.60\zehomi{6}$^{+2.6\%}_{-2.6\%}$     & 3.25\zehomi{4}$^{+10.0\%}_{-9.9\%}$       \\
176.0  & 5.91\zehomi{3}$^{+4.6\%}_{-4.5\%}$     & 6.89\zehomi{4}$^{+2.6\%}_{-2.6\%}$      & 2.39\zehomi{6}$^{+2.7\%}_{-2.6\%}$     & 2.97\zehomi{4}$^{+10.0\%}_{-9.9\%}$       \\
178.0  & 5.44\zehomi{3}$^{+4.6\%}_{-4.5\%}$     & 6.36\zehomi{4}$^{+2.6\%}_{-2.6\%}$      & 2.20\zehomi{6}$^{+2.7\%}_{-2.6\%}$     & 2.74\zehomi{4}$^{+10.0\%}_{-9.9\%}$       \\
180.0  & 5.01\zehomi{3}$^{+4.6\%}_{-4.5\%}$     & 5.87\zehomi{4}$^{+2.6\%}_{-2.6\%}$      & 2.04\zehomi{6}$^{+2.7\%}_{-2.6\%}$     & 2.52\zehomi{4}$^{+10.0\%}_{-9.9\%}$       \\
182.0  & 4.55\zehomi{3}$^{+4.6\%}_{-4.5\%}$     & 5.34\zehomi{4}$^{+2.6\%}_{-2.6\%}$      & 1.85\zehomi{6}$^{+2.7\%}_{-2.6\%}$     & 2.29\zehomi{4}$^{+10.0\%}_{-9.9\%}$       \\
184.0  & 4.09\zehomi{3}$^{+4.6\%}_{-4.5\%}$     & 4.81\zehomi{4}$^{+2.6\%}_{-2.6\%}$      & 1.67\zehomi{6}$^{+2.7\%}_{-2.6\%}$     & 2.06\zehomi{4}$^{+10.0\%}_{-9.9\%}$       \\
\hline
\end{tabular}\end{center}
\end{table}

\begin{table}\small\scriptsize
\renewcommand{\arraystretch}{1.3}
\setlength{\tabcolsep}{1.8ex}
\caption{SM Higgs branching ratios to two fermions and their total uncertainties (expressed in percentage). High-mass range.}
\label{BR-2fd}
\begin{center}\begin{tabular}{lccccc}
\hline
$\MH$ [$\UGeV$] & $\Hbb$ & $\Htautau$ &$\Hmumu$ & $\Hcc$ & $\Htt$\\
\hline
262.0  & 8.74\zehomi{4}$^{+4.7\%}_{-4.6\%}$   & 1.10\zehomi{4}$^{+2.6\%}_{-2.6\%}$      & 3.81\zehomi{7}$^{+2.6\%}_{-2.6\%}$    & 4.39\zehomi{5}$^{+10.1\%}_{-10.0\%}$    & 1.89\zehomi{7}$^{+342.0\%}_{-102.1\%}$     \\
264.0  & 8.52\zehomi{4}$^{+4.7\%}_{-4.6\%}$   & 1.07\zehomi{4}$^{+2.6\%}_{-2.6\%}$      & 3.72\zehomi{7}$^{+2.6\%}_{-2.6\%}$    & 4.28\zehomi{5}$^{+10.1\%}_{-10.0\%}$    & 4.49\zehomi{7}$^{+276.1\%}_{-91.5\%}$      \\
266.0  & 8.31\zehomi{4}$^{+4.7\%}_{-4.6\%}$   & 1.05\zehomi{4}$^{+2.6\%}_{-2.6\%}$      & 3.63\zehomi{7}$^{+2.6\%}_{-2.6\%}$    & 4.18\zehomi{5}$^{+10.1\%}_{-10.0\%}$    & 8.63\zehomi{7}$^{+210.1\%}_{-80.9\%}$      \\
268.0  & 8.11\zehomi{4}$^{+4.7\%}_{-4.6\%}$   & 1.02\zehomi{4}$^{+2.6\%}_{-2.6\%}$      & 3.55\zehomi{7}$^{+2.6\%}_{-2.6\%}$    & 4.08\zehomi{5}$^{+10.1\%}_{-10.0\%}$    & 1.46\zehomi{6}$^{+144.1\%}_{-70.3\%}$      \\
270.0  & 7.92\zehomi{4}$^{+4.7\%}_{-4.6\%}$   & 1.00\zehomi{4}$^{+2.6\%}_{-2.6\%}$      & 3.47\zehomi{7}$^{+2.6\%}_{-2.6\%}$    & 3.98\zehomi{5}$^{+10.1\%}_{-10.0\%}$    & 2.29\zehomi{6}$^{+78.1\%}_{-59.7\%}$      \\
272.0  & 7.73\zehomi{4}$^{+4.7\%}_{-4.6\%}$   & 9.80\zehomi{5}$^{+2.6\%}_{-2.6\%}$      & 3.40\zehomi{7}$^{+2.6\%}_{-2.6\%}$    & 3.89\zehomi{5}$^{+10.1\%}_{-10.0\%}$    & 3.37\zehomi{6}$^{+72.3\%}_{-56.6\%}$      \\
274.0  & 7.55\zehomi{4}$^{+4.7\%}_{-4.6\%}$   & 9.58\zehomi{5}$^{+2.6\%}_{-2.6\%}$      & 3.32\zehomi{7}$^{+2.6\%}_{-2.6\%}$    & 3.80\zehomi{5}$^{+10.1\%}_{-10.0\%}$    & 4.72\zehomi{6}$^{+66.5\%}_{-53.6\%}$      \\
\hline
\end{tabular}\end{center}
\end{table}

\begin{table}\small\scriptsize
\renewcommand{\arraystretch}{1.3}
\setlength{\tabcolsep}{2ex}
\caption{SM Higgs branching ratios to two fermions and their total uncertainties (expressed in percentage). Very-high-mass range.}
\label{BR-2fe}
\begin{center}\begin{tabular}{lccccc}
\hline
$\MH$ [$\UGeV$] & $\Hbb$ & $\Htautau$ &$\Hmumu$ & $\Hcc$ & $\Htt$\\
\hline
550.0  & 8.64\zehomi{5}$^{+7.7\%}_{-5.7\%}$       & 1.23\zehomi{5}$^{+5.5\%}_{-3.5\%}$       & 4.27\zehomi{8}$^{+5.5\%}_{-3.5\%}$       & 4.33\zehomi{6}$^{+13.1\%}_{-11.1\%}$             & 1.83\zehomi{1}$^{+5.4\%}_{-10.9\%}$      \\
560.0  & 8.28\zehomi{5}$^{+7.8\%}_{-5.8\%}$       & 1.19\zehomi{5}$^{+5.6\%}_{-3.7\%}$       & 4.10\zehomi{8}$^{+5.6\%}_{-3.6\%}$       & 4.15\zehomi{6}$^{+13.2\%}_{-11.2\%}$             & 1.80\zehomi{1}$^{+5.6\%}_{-11.3\%}$      \\
570.0  & 7.95\zehomi{5}$^{+7.8\%}_{-5.9\%}$       & 1.14\zehomi{5}$^{+5.8\%}_{-3.7\%}$       & 3.95\zehomi{8}$^{+5.7\%}_{-3.8\%}$       & 3.98\zehomi{6}$^{+13.3\%}_{-11.3\%}$             & 1.78\zehomi{1}$^{+5.7\%}_{-11.6\%}$      \\
580.0  & 7.64\zehomi{5}$^{+7.9\%}_{-6.0\%}$       & 1.10\zehomi{5}$^{+5.8\%}_{-4.0\%}$       & 3.80\zehomi{8}$^{+5.8\%}_{-4.0\%}$       & 3.83\zehomi{6}$^{+13.4\%}_{-11.4\%}$             & 1.75\zehomi{1}$^{+5.9\%}_{-11.9\%}$      \\
590.0  & 7.34\zehomi{5}$^{+8.1\%}_{-6.1\%}$       & 1.06\zehomi{5}$^{+6.0\%}_{-4.0\%}$       & 3.67\zehomi{8}$^{+6.0\%}_{-4.0\%}$       & 3.68\zehomi{6}$^{+13.5\%}_{-11.5\%}$             & 1.72\zehomi{1}$^{+6.2\%}_{-12.2\%}$      \\
600.0  & 7.06\zehomi{5}$^{+8.2\%}_{-6.3\%}$       & 1.02\zehomi{5}$^{+6.2\%}_{-4.2\%}$       & 3.54\zehomi{8}$^{+6.1\%}_{-4.2\%}$       & 3.54\zehomi{6}$^{+13.7\%}_{-11.6\%}$             & 1.69\zehomi{1}$^{+6.3\%}_{-12.4\%}$      \\
610.0  & 6.81\zehomi{5}$^{+8.3\%}_{-6.5\%}$       & 9.87\zehomi{6}$^{+6.2\%}_{-4.4\%}$       & 3.42\zehomi{8}$^{+6.2\%}_{-4.4\%}$       & 3.41\zehomi{6}$^{+13.8\%}_{-11.8\%}$             & 1.66\zehomi{1}$^{+6.4\%}_{-12.8\%}$      \\
\hline
\end{tabular}\end{center}
\end{table}

\begin{table}\small\scriptsize
\renewcommand{\arraystretch}{1.3}
\setlength{\tabcolsep}{1ex}
\caption{SM Higgs branching ratios to two gauge bosons and Higgs total width together with their total uncertainties (expressed in percentage). Very-low-mass range.}
\label{BR-2gba}
\begin{center}\begin{tabular}{lccccc|c}
\hline
$\MH$ [$\UGeV$] & $\Hgg$ & $\Hgaga$ &$\HZga$ & $\HWW$ & $\HZZ$ & $\Gamma_{\PH}$ [$\UGeV$]\\
\hline
90.0   & 6.12\zehomi{2}$^{+11.8\%}_{-11.4\%}$      & 1.22\zehomi{3}$^{+6.2\%}_{-6.2\%}$    & 0.00\zehoze $^{+0.0\%}_{-0.0\%}$      & 2.07\zehomi{3}$^{+5.8\%}_{-5.7\%}$      & 4.17\zehomi{4}$^{+5.8\%}_{-5.6\%}$              & 2.22\zehomi{3}$^{+5.2\%}_{-5.2\%}$       \\
95.0   & 6.74\zehomi{2}$^{+11.7\%}_{-11.3\%}$      & 1.39\zehomi{3}$^{+6.2\%}_{-6.1\%}$    & 4.48\zehomi{6}$^{+10.2\%}_{-10.1\%}$    & 4.67\zehomi{3}$^{+5.8\%}_{-5.6\%}$      & 6.65\zehomi{4}$^{+5.7\%}_{-5.6\%}$              & 2.35\zehomi{3}$^{+5.2\%}_{-5.1\%}$       \\
100.0  & 7.36\zehomi{2}$^{+11.6\%}_{-11.2\%}$      & 1.58\zehomi{3}$^{+6.2\%}_{-6.1\%}$    & 4.93\zehomi{5}$^{+10.2\%}_{-10.1\%}$    & 1.09\zehomi{2}$^{+5.7\%}_{-5.6\%}$      & 1.12\zehomi{3}$^{+5.7\%}_{-5.5\%}$              & 2.49\zehomi{3}$^{+5.1\%}_{-5.1\%}$       \\
105.0  & 7.95\zehomi{2}$^{+11.4\%}_{-11.1\%}$      & 1.77\zehomi{3}$^{+6.1\%}_{-5.9\%}$    & 1.71\zehomi{4}$^{+10.2\%}_{-9.9\%}$     & 2.40\zehomi{2}$^{+5.6\%}_{-5.5\%}$      & 2.13\zehomi{3}$^{+5.6\%}_{-5.5\%}$              & 2.65\zehomi{3}$^{+5.1\%}_{-5.0\%}$       \\
110.0  & 8.45\zehomi{2}$^{+11.2\%}_{-10.9\%}$      & 1.95\zehomi{3}$^{+5.9\%}_{-5.8\%}$    & 3.91\zehomi{4}$^{+9.9\%}_{-9.8\%}$      & 4.77\zehomi{2}$^{+5.4\%}_{-5.3\%}$      & 4.34\zehomi{3}$^{+5.4\%}_{-5.3\%}$              & 2.85\zehomi{3}$^{+4.9\%}_{-4.9\%}$       \\
110.5  & 8.49\zehomi{2}$^{+11.2\%}_{-10.8\%}$      & 1.97\zehomi{3}$^{+5.9\%}_{-5.8\%}$    & 4.18\zehomi{4}$^{+9.9\%}_{-9.8\%}$      & 5.09\zehomi{2}$^{+5.4\%}_{-5.3\%}$      & 4.67\zehomi{3}$^{+5.4\%}_{-5.3\%}$              & 2.87\zehomi{3}$^{+4.9\%}_{-4.9\%}$       \\
\hline
\end{tabular}\end{center}
\end{table}

\begin{table}\small\scriptsize
\renewcommand{\arraystretch}{1.4}
\setlength{\tabcolsep}{1ex}
\caption{SM Higgs branching ratios to two gauge bosons and Higgs total width together with their total uncertainties (expressed in percentage). Low- and intermediate.mass range.}
\label{BR-2gbb}
\begin{center}\begin{tabular}{lccccc|c}
\hline
$\MH$ [$\UGeV$] & $\Hgg$ & $\Hgaga$ &$\HZga$ & $\HWW$ & $\HZZ$ & $\Gamma_{\PH}$ [$\UGeV$]\\
\hline
130.5  & 7.90\zehomi{2}$^{+9.8\%}_{-9.5\%}$       & 2.24\zehomi{3}$^{+4.5\%}_{-4.3\%}$       & 1.99\zehomi{3}$^{+8.4\%}_{-8.4\%}$      & 3.12\zehomi{1}$^{+3.6\%}_{-3.5\%}$     & 4.13\zehomi{2}$^{+3.6\%}_{-3.5\%}$     & 5.01\zehomi{3}$^{+3.4\%}_{-3.4\%}$       \\
131.0  & 7.82\zehomi{2}$^{+9.7\%}_{-9.5\%}$       & 2.23\zehomi{3}$^{+4.4\%}_{-4.3\%}$       & 2.01\zehomi{3}$^{+8.3\%}_{-8.3\%}$      & 3.22\zehomi{1}$^{+3.5\%}_{-3.5\%}$     & 4.28\zehomi{2}$^{+3.5\%}_{-3.5\%}$     & 5.12\zehomi{3}$^{+3.4\%}_{-3.3\%}$       \\
131.5  & 7.73\zehomi{2}$^{+9.7\%}_{-9.4\%}$       & 2.22\zehomi{3}$^{+4.4\%}_{-4.2\%}$       & 2.06\zehomi{3}$^{+8.3\%}_{-8.3\%}$      & 3.31\zehomi{1}$^{+3.5\%}_{-3.4\%}$     & 4.42\zehomi{2}$^{+3.5\%}_{-3.4\%}$     & 5.23\zehomi{3}$^{+3.3\%}_{-3.3\%}$       \\
132.0  & 7.65\zehomi{2}$^{+9.6\%}_{-9.4\%}$       & 2.21\zehomi{3}$^{+4.3\%}_{-4.2\%}$       & 2.09\zehomi{3}$^{+8.2\%}_{-8.2\%}$      & 3.41\zehomi{1}$^{+3.4\%}_{-3.4\%}$     & 4.57\zehomi{2}$^{+3.4\%}_{-3.4\%}$     & 5.35\zehomi{3}$^{+3.2\%}_{-3.2\%}$       \\
132.5  & 7.56\zehomi{2}$^{+9.6\%}_{-9.4\%}$       & 2.20\zehomi{3}$^{+4.2\%}_{-4.1\%}$       & 2.13\zehomi{3}$^{+8.2\%}_{-8.2\%}$      & 3.51\zehomi{1}$^{+3.3\%}_{-3.3\%}$     & 4.72\zehomi{2}$^{+3.3\%}_{-3.3\%}$     & 5.47\zehomi{3}$^{+3.2\%}_{-3.2\%}$       \\
133.0  & 7.47\zehomi{2}$^{+9.5\%}_{-9.3\%}$       & 2.19\zehomi{3}$^{+4.2\%}_{-4.1\%}$       & 2.16\zehomi{3}$^{+8.1\%}_{-8.1\%}$      & 3.61\zehomi{1}$^{+3.3\%}_{-3.2\%}$     & 4.87\zehomi{2}$^{+3.3\%}_{-3.2\%}$     & 5.60\zehomi{3}$^{+3.1\%}_{-3.1\%}$       \\
133.5  & 7.37\zehomi{2}$^{+9.5\%}_{-9.3\%}$       & 2.17\zehomi{3}$^{+4.1\%}_{-4.0\%}$       & 2.19\zehomi{3}$^{+8.1\%}_{-8.1\%}$      & 3.70\zehomi{1}$^{+3.2\%}_{-3.2\%}$     & 5.02\zehomi{2}$^{+3.2\%}_{-3.2\%}$     & 5.74\zehomi{3}$^{+3.1\%}_{-3.1\%}$       \\
\hline
\end{tabular}\end{center}
\end{table}

\begin{table}\small\scriptsize
\renewcommand{\arraystretch}{1.4}
\setlength{\tabcolsep}{1ex}
\caption{SM Higgs branching ratios to two gauge bosons and Higgs total width together with their total uncertainties (expressed in percentage). Intermediate-mass range.}
\label{BR-2gbc}

\begin{center}\begin{tabular}{lccccc|c}
\hline
$\MH$ [$\UGeV$] & $\Hgg$ & $\Hgaga$ &$\HZga$ & $\HWW$ & $\HZZ$ & $\Gamma_{\PH}$ [$\UGeV$]\\
\hline
172.0  & 2.01\zehomi{3}$^{+7.7\%}_{-7.6\%}$    & 1.43\zehomi{4}$^{+1.6\%}_{-1.6\%}$    & 3.70\zehomi{4}$^{+5.6\%}_{-5.6\%}$    & 9.63\zehomi{1}$^{+0.0\%}_{-0.0\%}$       & 2.61\zehomi{2}$^{+0.1\%}_{-0.1\%}$      & 4.29\zehomi{1}$^{+0.6\%}_{-0.6\%}$       \\
174.0  & 1.88\zehomi{3}$^{+7.7\%}_{-7.6\%}$    & 1.32\zehomi{4}$^{+1.6\%}_{-1.6\%}$    & 3.48\zehomi{4}$^{+5.6\%}_{-5.5\%}$    & 9.60\zehomi{1}$^{+0.0\%}_{-0.0\%}$       & 2.98\zehomi{2}$^{+0.1\%}_{-0.1\%}$      & 4.77\zehomi{1}$^{+0.6\%}_{-0.6\%}$       \\
176.0  & 1.76\zehomi{3}$^{+7.7\%}_{-7.6\%}$    & 1.22\zehomi{4}$^{+1.6\%}_{-1.6\%}$    & 3.29\zehomi{4}$^{+5.6\%}_{-5.5\%}$    & 9.55\zehomi{1}$^{+0.0\%}_{-0.0\%}$       & 3.54\zehomi{2}$^{+0.1\%}_{-0.1\%}$      & 5.25\zehomi{1}$^{+0.6\%}_{-0.6\%}$       \\
178.0  & 1.65\zehomi{3}$^{+7.7\%}_{-7.6\%}$    & 1.14\zehomi{4}$^{+1.6\%}_{-1.6\%}$    & 3.12\zehomi{4}$^{+5.6\%}_{-5.5\%}$    & 9.47\zehomi{1}$^{+0.0\%}_{-0.0\%}$       & 4.44\zehomi{2}$^{+0.1\%}_{-0.1\%}$      & 5.75\zehomi{1}$^{+0.6\%}_{-0.6\%}$       \\
180.0  & 1.56\zehomi{3}$^{+7.7\%}_{-7.6\%}$    & 1.05\zehomi{4}$^{+1.6\%}_{-1.6\%}$    & 2.96\zehomi{4}$^{+5.6\%}_{-5.5\%}$    & 9.32\zehomi{1}$^{+0.0\%}_{-0.0\%}$       & 6.02\zehomi{2}$^{+0.1\%}_{-0.1\%}$      & 6.31\zehomi{1}$^{+0.6\%}_{-0.6\%}$       \\
182.0  & 1.44\zehomi{3}$^{+7.7\%}_{-7.6\%}$    & 9.69\zehomi{5}$^{+1.6\%}_{-1.6\%}$    & 2.76\zehomi{4}$^{+5.6\%}_{-5.5\%}$    & 9.03\zehomi{1}$^{+0.0\%}_{-0.0\%}$       & 9.00\zehomi{2}$^{+0.1\%}_{-0.1\%}$      & 7.00\zehomi{1}$^{+0.6\%}_{-0.6\%}$       \\
184.0  & 1.32\zehomi{3}$^{+7.7\%}_{-7.6\%}$    & 8.81\zehomi{5}$^{+1.6\%}_{-1.6\%}$    & 2.54\zehomi{4}$^{+5.6\%}_{-5.5\%}$    & 8.62\zehomi{1}$^{+0.0\%}_{-0.0\%}$       & 1.31\zehomi{1}$^{+0.0\%}_{-0.0\%}$      & 7.88\zehomi{1}$^{+0.6\%}_{-0.6\%}$       \\
\hline
\end{tabular}\end{center}
\end{table}

\begin{table}\small\scriptsize
\renewcommand{\arraystretch}{1.3}
\setlength{\tabcolsep}{0.5ex}
\caption{SM Higgs branching ratios to two gauge bosons and Higgs total width together with their total uncertainties (expressed in percentage). High-mass range.}
\label{BR-2gbd}
\begin{center}\begin{tabular}{lccccc|c}
\hline
$\MH$ [$\UGeV$] & $\Hgg$ & $\Hgaga$ &$\HZga$ & $\HWW$ & $\HZZ$ & $\Gamma_{\PH}$ [$\UGeV$]\\
\hline
262.0  & 6.64\zehomi{4}$^{+7.7\%}_{-7.5\%}$        & 2.11\zehomi{5}$^{+1.7\%}_{-1.7\%}$     & 8.39\zehomi{5}$^{+5.6\%}_{-5.6\%}$              & 6.98\zehomi{1}$^{+0.0\%}_{-0.0\%}$              & 3.00\zehomi{1}$^{+0.0\%}_{-0.0\%}$              & 4.91\zehoze $^{+0.5\%}_{-0.5\%}$       \\
264.0  & 6.61\zehomi{4}$^{+7.7\%}_{-7.5\%}$        & 2.06\zehomi{5}$^{+1.7\%}_{-1.7\%}$     & 8.21\zehomi{5}$^{+5.6\%}_{-5.6\%}$              & 6.98\zehomi{1}$^{+0.0\%}_{-0.0\%}$              & 3.00\zehomi{1}$^{+0.0\%}_{-0.0\%}$              & 5.07\zehoze $^{+0.5\%}_{-0.5\%}$       \\
266.0  & 6.59\zehomi{4}$^{+7.7\%}_{-7.6\%}$        & 2.01\zehomi{5}$^{+1.7\%}_{-1.7\%}$     & 8.04\zehomi{5}$^{+5.6\%}_{-5.6\%}$              & 6.98\zehomi{1}$^{+0.0\%}_{-0.0\%}$              & 3.01\zehomi{1}$^{+0.0\%}_{-0.0\%}$              & 5.23\zehoze $^{+0.5\%}_{-0.5\%}$       \\
268.0  & 6.57\zehomi{4}$^{+7.7\%}_{-7.6\%}$        & 1.96\zehomi{5}$^{+1.7\%}_{-1.7\%}$     & 7.88\zehomi{5}$^{+5.6\%}_{-5.6\%}$              & 6.97\zehomi{1}$^{+0.0\%}_{-0.0\%}$              & 3.01\zehomi{1}$^{+0.0\%}_{-0.0\%}$              & 5.39\zehoze $^{+0.5\%}_{-0.5\%}$       \\
270.0  & 6.55\zehomi{4}$^{+7.7\%}_{-7.6\%}$        & 1.91\zehomi{5}$^{+1.7\%}_{-1.7\%}$     & 7.72\zehomi{5}$^{+5.6\%}_{-5.6\%}$              & 6.97\zehomi{1}$^{+0.0\%}_{-0.0\%}$              & 3.02\zehomi{1}$^{+0.0\%}_{-0.0\%}$              & 5.55\zehoze $^{+0.5\%}_{-0.5\%}$       \\
272.0  & 6.54\zehomi{4}$^{+7.7\%}_{-7.6\%}$        & 1.87\zehomi{5}$^{+1.7\%}_{-1.7\%}$     & 7.56\zehomi{5}$^{+5.6\%}_{-5.6\%}$              & 6.96\zehomi{1}$^{+0.0\%}_{-0.0\%}$              & 3.02\zehomi{1}$^{+0.0\%}_{-0.0\%}$              & 5.72\zehoze $^{+0.5\%}_{-0.5\%}$       \\
\hline
\end{tabular}\end{center}
\end{table}

\begin{table}\small\scriptsize
\renewcommand{\arraystretch}{1.3}
\setlength{\tabcolsep}{1ex}
\caption{SM Higgs branching ratios to two gauge bosons and Higgs total width together with their total uncertainties (expressed in percentage). Very-high-mass range.}
\label{BR-2gbe}

\begin{center}\begin{tabular}{lccccc|c}
\hline
$\MH$ [$\UGeV$] & $\Hgg$ & $\Hgaga$ &$\HZga$ & $\HWW$ & $\HZZ$ & $\Gamma_{\PH}$ [$\UGeV$]\\
\hline
550.0  & 5.05\zehomi{4}$^{+10.4\%}_{-8.4\%}$       & 9.90\zehomi{8}$^{+6.1\%} _{-2.7\%}$      & 5.30\zehomi{6}$^{+8.6\%}_{-6.6\%}$      & 5.50\zehomi{1}$^{+2.4\%}_{-1.2\%}$    & 2.66\zehomi{1}$^{+2.5\%}_{-1.2\%}$      & 9.30\zehopl{1}$^{+1.6\%}_{-3.5\%}$       \\
560.0  & 4.88\zehomi{4}$^{+10.5\%}_{-8.6\%}$       & 8.10\zehomi{8}$^{+8.7\%} _{-3.2\%}$      & 4.95\zehomi{6}$^{+8.6\%}_{-6.6\%}$      & 5.52\zehomi{1}$^{+2.5\%}_{-1.2\%}$    & 2.67\zehomi{1}$^{+2.5\%}_{-1.2\%}$      & 9.86\zehopl{1}$^{+1.7\%}_{-3.6\%}$       \\
570.0  & 4.71\zehomi{4}$^{+10.6\%}_{-8.7\%}$       & 6.83\zehomi{8}$^{+11.5\%}_{-7.2\%}$      & 4.64\zehomi{6}$^{+8.6\%}_{-6.7\%}$      & 5.53\zehomi{1}$^{+2.5\%}_{-1.2\%}$    & 2.68\zehomi{1}$^{+2.5\%}_{-1.3\%}$      & 1.04\zehopl{2}$^{+1.9\%}_{-3.7\%}$       \\
580.0  & 4.55\zehomi{4}$^{+10.7\%}_{-8.9\%}$       & 5.91\zehomi{8}$^{+15.2\%}_{-10.3\%}$     & 4.35\zehomi{6}$^{+8.7\%}_{-6.9\%}$      & 5.55\zehomi{1}$^{+2.5\%}_{-1.2\%}$    & 2.70\zehomi{1}$^{+2.5\%}_{-1.3\%}$      & 1.10\zehopl{2}$^{+2.0\%}_{-3.8\%}$       \\
590.0  & 4.39\zehomi{4}$^{+10.9\%}_{-8.9\%}$       & 5.33\zehomi{8}$^{+17.8\%}_{-13.5\%}$     & 4.08\zehomi{6}$^{+8.9\%}_{-6.9\%}$      & 5.57\zehomi{1}$^{+2.5\%}_{-1.3\%}$    & 2.71\zehomi{1}$^{+2.5\%}_{-1.3\%}$      & 1.16\zehopl{2}$^{+2.2\%}_{-3.9\%}$       \\
600.0  & 4.24\zehomi{4}$^{+11.1\%}_{-9.1\%}$       & 5.03\zehomi{8}$^{+20.7\%}_{-16.0\%}$     & 3.84\zehomi{6}$^{+9.2\%}_{-7.1\%}$      & 5.58\zehomi{1}$^{+2.5\%}_{-1.3\%}$    & 2.72\zehomi{1}$^{+2.5\%}_{-1.3\%}$      & 1.23\zehopl{2}$^{+2.3\%}_{-4.0\%}$       \\
610.0  & 4.10\zehomi{4}$^{+11.0\%}_{-9.3\%}$       & 4.86\zehomi{8}$^{+22.0\%}_{-17.2\%}$     & 3.61\zehomi{6}$^{+9.3\%}_{-7.5\%}$      & 5.60\zehomi{1}$^{+2.5\%}_{-1.3\%}$    & 2.74\zehomi{1}$^{+2.6\%}_{-1.3\%}$      & 1.29\zehopl{2}$^{+2.5\%}_{-4.2\%}$       \\
\hline
\end{tabular}\end{center}
\end{table}


\begin{table}\small\scriptsize
\renewcommand{\arraystretch}{1.3}
\setlength{\tabcolsep}{1.5ex}
\caption{SM Higgs branching ratios for the different $\PH \to 4\Pl$ and
$\PH \to 2\Pl2\PGn$ final states and their total uncertainties (expressed in percentage). Very-low-mass range.}
\label{BR-4f1a}
\begin{center}\begin{tabular}{lccccccccc}
\hline
$\MH$  & $\PH \to 4\Pl$ & $\PH \to 4\Pl$ & $\PH \to \Pe\Pe\Pe\Pe$ & $\PH \to \Pe\Pe\PGm\PGm$ &
$\PH \to 2\Pl2\PGn$ & $\PH \to 2\Pl2\PGn$ & $\PH \to \Pe\PGn \Pe \PGn$ & $\PH \to \Pe \PGn \PGm \PGn$ &$\Delta$BR  \\
$[\UGeV]$ & $\Pl=\Pe,\PGm,\PGt$ & $\Pl=\Pe,\PGm$ & & & $\Pl=\Pe,\PGm,\PGt$ & $\Pl=\Pe,\PGm$ & & & [\%]\\ 
 & & & & & $\PGn={}$any & $\PGn={}$any& & & \\ 
\hline
90.0 & 4.89\zehomi{6} & 2.33\zehomi{6} & 7.01\zehomi{7} & 9.30\zehomi{7} & 2.10\zehomi{4} & 9.12\zehomi{5} & 1.75\zehomi{5} & 2.43\zehomi{5}  & $\pm$5.8\\
95.0 & 7.73\zehomi{6} & 3.67\zehomi{6} & 1.10\zehomi{6} & 1.48\zehomi{6} & 4.80\zehomi{4} & 2.10\zehomi{4} & 4.42\zehomi{5} & 5.49\zehomi{5}  & $\pm$5.8\\
100.0 & 1.28\zehomi{5} & 6.06\zehomi{6} & 1.78\zehomi{6} & 2.49\zehomi{6} & 1.14\zehomi{3} & 5.03\zehomi{4} & 1.13\zehomi{4} & 1.29\zehomi{4} & $\pm$5.7\\
105.0 & 2.37\zehomi{5} & 1.11\zehomi{5} & 3.18\zehomi{6} & 4.74\zehomi{6} & 2.53\zehomi{3} & 1.12\zehomi{3} & 2.59\zehomi{4} & 2.83\zehomi{4} & $\pm$5.6\\
110.0 & 4.72\zehomi{5} & 2.18\zehomi{5} & 6.04\zehomi{6} & 9.69\zehomi{6} & 5.08\zehomi{3} & 2.26\zehomi{3} & 5.32\zehomi{4} & 5.61\zehomi{4} & $\pm$5.4\\
\hline
\end{tabular}\end{center}
\end{table}

\begin{table}\small\scriptsize
\renewcommand{\arraystretch}{1.3}
\setlength{\tabcolsep}{1.5ex}
\caption{SM Higgs branching ratios for the different $\PH \to 4\Pl$ and
$\PH \to 2\Pl2\PGn$ final states and their total uncertainties (expressed in percentage). Low- and intermediate-mass range.}
\label{BR-4f1b}

\begin{center}\begin{tabular}{lccccccccc}
\hline
$\MH$  & $\PH \to 4\Pl$ & $\PH \to 4\Pl$ & $\PH \to \Pe\Pe\Pe\Pe$ & $\PH \to \Pe\Pe\PGm\PGm$ &
$\PH \to 2\Pl2\PGn$ & $\PH \to 2\Pl2\PGn$ & $\PH \to\Pe\PGn\Pe  \PGn$ & $\PH \to\Pe\PGn \PGm \PGn$ &$\Delta$BR  \\
$[\UGeV]$ & $\Pl=\Pe,\PGm,\PGt$ & $\Pl=\Pe,\PGm$ & & & $\Pl=\Pe,\PGm,\PGt$ & $\Pl=\Pe,\PGm$ & & & [\%]\\ 
 & & & & & $\PGn={}$any & $\PGn={}$any& & & \\ 
\hline
130.5 & 4.29\zehomi{4} & 1.93\zehomi{4} & 5.02\zehomi{5} & 9.29\zehomi{5} & 3.43\zehomi{2} & 1.55\zehomi{2} & 3.71\zehomi{3} & 3.67\zehomi{3} & $\pm$3.6\\
131.0 & 4.44\zehomi{4} & 2.00\zehomi{4} & 5.19\zehomi{5} & 9.62\zehomi{5} & 3.54\zehomi{2} & 1.60\zehomi{2} & 3.83\zehomi{3} & 3.79\zehomi{3} & $\pm$3.5\\
131.5 & 4.59\zehomi{4} & 2.07\zehomi{4} & 5.37\zehomi{5} & 9.95\zehomi{5} & 3.64\zehomi{2} & 1.65\zehomi{2} & 3.94\zehomi{3} & 3.90\zehomi{3} & $\pm$3.5\\
132.0 & 4.75\zehomi{4} & 2.14\zehomi{4} & 5.54\zehomi{5} & 1.03\zehomi{4} & 3.75\zehomi{2} & 1.70\zehomi{2} & 4.06\zehomi{3} & 4.01\zehomi{3} & $\pm$3.4\\
132.5 & 4.90\zehomi{4} & 2.21\zehomi{4} & 5.71\zehomi{5} & 1.06\zehomi{4} & 3.86\zehomi{2} & 1.75\zehomi{2} & 4.18\zehomi{3} & 4.13\zehomi{3} & $\pm$3.3\\
\hline
\end{tabular}\end{center}
\end{table}

\begin{table}\small\scriptsize
\setlength{\tabcolsep}{1.5ex}
\renewcommand{\arraystretch}{1.3}
\caption{SM Higgs branching ratios for the different $\PH \to 4\Pl$ and
$\PH \to 2\Pl2\PGn$ final states and their total uncertainties (expressed in percentage). Intermediate-mass range.}
\label{BR-4f1c}
\begin{center}\begin{tabular}{lccccccccc}
\hline
$\MH$  & $\PH \to 4\Pl$ & $\PH \to 4\Pl$ & $\PH \to \Pe\Pe\Pe\Pe$ & $\PH \to \Pe\Pe\PGm\PGm$ &
$\PH \to 2\Pl2\PGn$ & $\PH \to 2\Pl2\PGn$ & $\PH \to\Pe\PGn\Pe \PGn$ & $\PH \to\Pe\PGn \PGm \PGn$ &$\Delta$BR  \\
$[\UGeV]$ & $\Pl=\Pe,\PGm,\PGt$ & $\Pl=\Pe,\PGm$ & & & $\Pl=\Pe,\PGm,\PGt$ & $\Pl=\Pe,\PGm$ & & & [\%]\\ 
 & & & & & $\PGn={}$any & $\PGn={}$any& & & \\ 
\hline
172.0 & 2.66\zehomi{4} & 1.19\zehomi{4} & 3.00\zehomi{5} & 5.87\zehomi{5} & 1.03\zehomi{1} & 4.61\zehomi{2} & 1.15\zehomi{2} & 1.14\zehomi{2} & $\pm$0.0\\
174.0 & 3.04\zehomi{4} & 1.36\zehomi{4} & 3.42\zehomi{5} & 6.71\zehomi{5} & 1.03\zehomi{1} & 4.61\zehomi{2} & 1.14\zehomi{2} & 1.13\zehomi{2} & $\pm$0.0\\
176.0 & 3.61\zehomi{4} & 1.61\zehomi{4} & 4.06\zehomi{5} & 7.98\zehomi{5} & 1.03\zehomi{1} & 4.60\zehomi{2} & 1.14\zehomi{2} & 1.13\zehomi{2} & $\pm$0.0\\
178.0 & 4.52\zehomi{4} & 2.01\zehomi{4} & 5.07\zehomi{5} & 9.99\zehomi{5} & 1.02\zehomi{1} & 4.58\zehomi{2} & 1.13\zehomi{2} & 1.12\zehomi{2} & $\pm$0.0\\
180.0 & 6.12\zehomi{4} & 2.73\zehomi{4} & 6.85\zehomi{5} & 1.36\zehomi{4} & 1.01\zehomi{1} & 4.55\zehomi{2} & 1.12\zehomi{2} & 1.10\zehomi{2} & $\pm$0.0\\
\hline
\end{tabular}\end{center}
\end{table}

\begin{table}\small\scriptsize
\renewcommand{\arraystretch}{1.3}
\setlength{\tabcolsep}{1.5ex}
\caption{SM Higgs branching ratios for the different $\PH \to 4\Pl$ and
$\PH \to 2\Pl2\PGn$ final states and their total uncertainties (expressed in percentage). High-mass range.}
\label{BR-4f1d}
\begin{center}\begin{tabular}{lccccccccc}
\hline
$\MH$  & $\PH \to 4\Pl$ & $\PH \to 4\Pl$ & $\PH \to \Pe\Pe\Pe\Pe$ & $\PH \to \Pe\Pe\PGm\PGm$ &
$\PH \to 2\Pl2\PGn$ & $\PH \to 2\Pl2\PGn$ & $\PH \to\Pe\PGn\Pe \PGn$ & $\PH \to\Pe\PGn \PGm \PGn$ &$\Delta$BR  \\
$[\UGeV]$ & $\Pl=\Pe,\PGm,\PGt$ & $\Pl=\Pe,\PGm$ & & & $\Pl=\Pe,\PGm,\PGt$ & $\Pl=\Pe,\PGm$ & & & [\%]\\ 
 & & & & & $\PGn={}$any & $\PGn={}$any& & & \\ 
\hline
262.0 & 3.04\zehomi{3} & 1.35\zehomi{3} & 3.38\zehomi{4} & 6.75\zehomi{4} & 8.61\zehomi{2} & 4.10\zehomi{2} & 9.57\zehomi{3} & 8.22\zehomi{3} & $\pm$0.0\\
264.0 & 3.04\zehomi{3} & 1.35\zehomi{3} & 3.38\zehomi{4} & 6.76\zehomi{4} & 8.61\zehomi{2} & 4.10\zehomi{2} & 9.57\zehomi{3} & 8.22\zehomi{3} & $\pm$0.0\\
266.0 & 3.05\zehomi{3} & 1.35\zehomi{3} & 3.39\zehomi{4} & 6.77\zehomi{4} & 8.61\zehomi{2} & 4.09\zehomi{2} & 9.56\zehomi{3} & 8.21\zehomi{3} & $\pm$0.0\\
268.0 & 3.05\zehomi{3} & 1.36\zehomi{3} & 3.39\zehomi{4} & 6.78\zehomi{4} & 8.60\zehomi{2} & 4.09\zehomi{2} & 9.56\zehomi{3} & 8.21\zehomi{3} & $\pm$0.0\\
270.0 & 3.05\zehomi{3} & 1.36\zehomi{3} & 3.40\zehomi{4} & 6.79\zehomi{4} & 8.60\zehomi{2} & 4.09\zehomi{2} & 9.56\zehomi{3} & 8.21\zehomi{3} & $\pm$0.0\\
\hline
\end{tabular}\end{center}
\end{table}

\begin{table}\small\scriptsize
\renewcommand{\arraystretch}{1.3}
\setlength{\tabcolsep}{1.5ex}
\caption{SM Higgs branching ratios for the different $\PH \to 4\Pl$ and
$\PH \to 2\Pl2\PGn$ final states and their total uncertainties (expressed in percentage). Very-high-mass range.}
\label{BR-4f1e}
\begin{center}\begin{tabular}{lccccccccc}
\hline
$\MH$  & $\PH \to 4\Pl$ & $\PH \to 4\Pl$ & $\PH \to \Pe\Pe\Pe\Pe$ & $\PH \to \Pe\Pe\PGm\PGm$ &
$\PH \to 2\Pl2\PGn$ & $\PH \to 2\Pl2\PGn$ & $\PH \to\Pe\PGn\Pe \PGn$ & $\PH \to\Pe\PGn \PGm \PGn$ &$\Delta$BR  \\
$[\UGeV]$ & $\Pl=\Pe,\PGm,\PGt$ & $\Pl=\Pe,\PGm$ & & & $\Pl=\Pe,\PGm,\PGt$ & $\Pl=\Pe,\PGm$ & & & [\%]\\ 
 & & & & & $\PGn={}$any & $\PGn={}$any& & & \\ 
\hline
550.0 & 2.70\zehomi{3} & 1.20\zehomi{3} & 3.00\zehomi{4} & 6.00\zehomi{4} & 6.92\zehomi{2} & 3.32\zehomi{2} & 7.69\zehomi{3} & 6.50\zehomi{3} & $\pm$2.4\\
560.0 & 2.71\zehomi{3} & 1.21\zehomi{3} & 3.01\zehomi{4} & 6.03\zehomi{4} & 6.94\zehomi{2} & 3.33\zehomi{2} & 7.71\zehomi{3} & 6.51\zehomi{3} & $\pm$2.5\\
570.0 & 2.73\zehomi{3} & 1.21\zehomi{3} & 3.03\zehomi{4} & 6.06\zehomi{4} & 6.96\zehomi{2} & 3.34\zehomi{2} & 7.74\zehomi{3} & 6.53\zehomi{3} & $\pm$2.5\\
580.0 & 2.74\zehomi{3} & 1.22\zehomi{3} & 3.04\zehomi{4} & 6.09\zehomi{4} & 6.99\zehomi{2} & 3.35\zehomi{2} & 7.76\zehomi{3} & 6.55\zehomi{3} & $\pm$2.5\\
590.0 & 2.75\zehomi{3} & 1.22\zehomi{3} & 3.06\zehomi{4} & 6.12\zehomi{4} & 7.01\zehomi{2} & 3.36\zehomi{2} & 7.79\zehomi{3} & 6.57\zehomi{3} & $\pm$2.5\\
\hline
\end{tabular}\end{center}
\end{table}

\begin{table}\small\scriptsize
\renewcommand{\arraystretch}{1.3}
\setlength{\tabcolsep}{3ex}
\caption{SM Higgs branching ratios for the different four-fermion final states 
and their total uncertainties (expressed in percentage). Very-low-mass range.}
\label{BR-4f2a}
\begin{center}\begin{tabular}{lccccccc}
\hline
$\MH$  & $\PH \to 2\Pl2\Pq$ & $\PH \to 2\Pl2\Pq$ & $\PH \to \Pl\PGn_l \Pq\Pq$ & $\PH \to \PGn\PGn \Pq\Pq$ &
$\PH \to 4\Pq$ & $\PH \to 4\Pf$ &$\Delta$BR  \\
$[\UGeV]$ & $\Pl=\Pe,\PGm,\PGt$& $\Pl=\Pe,\PGm$
& $\Pl=\Pe,\PGm$ & $\PGn={}$any& $\Pq=\PQu\PQd\PQc\PQs\PQb$& $f={}$any
 &[\%]\\ 
& $\Pq=\PQu\PQd\PQc\PQs\PQb$ & $\Pq=\PQu\PQd\PQc\PQs\PQb$ &
$\Pq=\PQu\PQd\PQc\PQs\PQb$ & $\Pq=\PQu\PQd\PQc\PQs\PQb$ & & fermion &\\ 
\hline

90.0 & 5.87\zehomi{5} & 3.91\zehomi{5} & 3.03\zehomi{4} & 1.18\zehomi{4} & 1.05\zehomi{3} & 2.37\zehomi{3}  & $\pm$5.8 \\
95.0 & 9.31\zehomi{5} & 6.21\zehomi{5} & 6.85\zehomi{4} & 1.87\zehomi{4} & 2.32\zehomi{3} & 5.17\zehomi{3}  & $\pm$5.8 \\
100.0 & 1.57\zehomi{4} & 1.04\zehomi{4} & 1.61\zehomi{3} & 3.14\zehomi{4} & 5.34\zehomi{3} & 1.18\zehomi{2} & $\pm$5.7 \\
105.0 & 2.97\zehomi{4} & 1.98\zehomi{4} & 3.52\zehomi{3} & 5.95\zehomi{4} & 1.17\zehomi{2} & 2.58\zehomi{2} & $\pm$5.6 \\
110.0 & 6.06\zehomi{4} & 4.04\zehomi{4} & 6.99\zehomi{3} & 1.21\zehomi{3} & 2.34\zehomi{2} & 5.15\zehomi{2} & $\pm$5.4 \\
\hline
\end{tabular}\end{center}
\end{table}

\clearpage

\begin{table}\small\scriptsize
\renewcommand{\arraystretch}{1.3}
\setlength{\tabcolsep}{3ex}
\caption{SM Higgs branching ratios for the different four-fermion final states 
and their total uncertainties (expressed in percentage). Low- and intermediate-mass range.}
\label{BR-4f2b}
\begin{center}\begin{tabular}{lccccccc}
\hline
$\MH$  & $\PH \to 2\Pl2\Pq$ & $\PH \to 2\Pl2\Pq$ & $\PH \to \Pl\PGn_l \Pq\Pq$ & $\PH \to \PGn\PGn \Pq\Pq$ &
$\PH \to 4\Pq$ & $\PH \to 4\Pf$ &$\Delta$BR  \\
$[\UGeV]$ & $\Pl=\Pe,\PGm,\PGt$& $\Pl=\Pe,\PGm$
& $\Pl=\Pe,\PGm$ & $\PGn={}$any& $\Pq=\PQu\PQd\PQc\PQs\PQb$& $f={}$any
 &[\%]\\ 
& $\Pq=\PQu\PQd\PQc\PQs\PQb$ & $\Pq=\PQu\PQd\PQc\PQs\PQb$ &
$\Pq=\PQu\PQd\PQc\PQs\PQb$ & $\Pq=\PQu\PQd\PQc\PQs\PQb$ & & fermion &\\ 
\hline
130.5 & 5.79\zehomi{3} & 3.86\zehomi{3} & 4.57\zehomi{2} & 1.15\zehomi{2} & 1.61\zehomi{1} & 3.52\zehomi{1} & $\pm$3.6 \\
131.0 & 6.00\zehomi{3} & 4.00\zehomi{3} & 4.71\zehomi{2} & 1.20\zehomi{2} & 1.66\zehomi{1} & 3.63\zehomi{1} & $\pm$3.5 \\
131.5 & 6.21\zehomi{3} & 4.14\zehomi{3} & 4.85\zehomi{2} & 1.24\zehomi{2} & 1.71\zehomi{1} & 3.74\zehomi{1} & $\pm$3.5 \\
132.0 & 6.41\zehomi{3} & 4.28\zehomi{3} & 4.99\zehomi{2} & 1.28\zehomi{2} & 1.76\zehomi{1} & 3.85\zehomi{1} & $\pm$3.4 \\
132.5 & 6.62\zehomi{3} & 4.42\zehomi{3} & 5.14\zehomi{2} & 1.32\zehomi{2} & 1.81\zehomi{1} & 3.96\zehomi{1} & $\pm$3.3 \\
\hline
\end{tabular}\end{center}
\end{table}

\begin{table}\small\scriptsize
\renewcommand{\arraystretch}{1.3}
\setlength{\tabcolsep}{3ex}
\caption{SM Higgs branching ratios for the different four-fermion final states and their total uncertainties (expressed in percentage). Intermediate-mass range.}
\label{BR-4f2c}
\begin{center}\begin{tabular}{lccccccc}
\hline
$\MH$  & $\PH \to 2\Pl2\Pq$ & $\PH \to 2\Pl2\Pq$ & $\PH \to \Pl\PGn_l \Pq\Pq$ & $\PH \to \PGn\PGn \Pq\Pq$ &
$\PH \to 4\Pq$ & $\PH \to 4\Pf$ &$\Delta$BR  \\
$[\UGeV]$ & $\Pl=\Pe,\PGm,\PGt$& $\Pl=\Pe,\PGm$
& $\Pl=\Pe,\PGm$ & $\PGn={}$any& $\Pq=\PQu\PQd\PQc\PQs\PQb$& $f={}$any
 &[\%]\\ 
& $\Pq=\PQu\PQd\PQc\PQs\PQb$ & $\Pq=\PQu\PQd\PQc\PQs\PQb$ &
$\Pq=\PQu\PQd\PQc\PQs\PQb$ & $\Pq=\PQu\PQd\PQc\PQs\PQb$ & & fermion &\\ 
\hline
172.0 & 3.67\zehomi{3} & 2.44\zehomi{3} & 1.41\zehomi{1} & 7.30\zehomi{3} & 4.50\zehomi{1} & 9.89\zehomi{1} & $\pm$0.0 \\
174.0 & 4.19\zehomi{3} & 2.79\zehomi{3} & 1.41\zehomi{1} & 8.34\zehomi{3} & 4.51\zehomi{1} & 9.90\zehomi{1} & $\pm$0.0 \\
176.0 & 4.98\zehomi{3} & 3.32\zehomi{3} & 1.40\zehomi{1} & 9.92\zehomi{3} & 4.51\zehomi{1} & 9.91\zehomi{1} & $\pm$0.0 \\
178.0 & 6.24\zehomi{3} & 4.16\zehomi{3} & 1.39\zehomi{1} & 1.24\zehomi{2} & 4.52\zehomi{1} & 9.91\zehomi{1} & $\pm$0.0 \\
180.0 & 8.47\zehomi{3} & 5.65\zehomi{3} & 1.37\zehomi{1} & 1.69\zehomi{2} & 4.53\zehomi{1} & 9.92\zehomi{1} & $\pm$0.0 \\
\hline
\end{tabular}\end{center}
\end{table}

\begin{table}\small\scriptsize
\renewcommand{\arraystretch}{1.3}
\setlength{\tabcolsep}{3ex}
\caption{SM Higgs branching ratios for the different four-fermion final states and their total uncertainties (expressed in percentage). High-mass range.}
\label{BR-4f2d}
\begin{center}\begin{tabular}{lccccccc}
\hline
$\MH$  & $\PH \to 2\Pl2\Pq$ & $\PH \to 2\Pl2\Pq$ & $\PH \to \Pl\PGn_l \Pq\Pq$ & $\PH \to \PGn\PGn \Pq\Pq$ &
$\PH \to 4\Pq$ & $\PH \to 4\Pf$ &$\Delta$BR  \\
$[\UGeV]$ & $\Pl=\Pe,\PGm,\PGt$& $\Pl=\Pe,\PGm$
& $\Pl=\Pe,\PGm$ & $\PGn={}$any& $\Pq=\PQu\PQd\PQc\PQs\PQb$& $f={}$any
 &[\%]\\ 
& $\Pq=\PQu\PQd\PQc\PQs\PQb$ & $\Pq=\PQu\PQd\PQc\PQs\PQb$ &
$\Pq=\PQu\PQd\PQc\PQs\PQb$ & $\Pq=\PQu\PQd\PQc\PQs\PQb$ & & fermion &\\ 
\hline
262.0 & 4.22\zehomi{2} & 2.81\zehomi{2} & 1.02\zehomi{1} & 8.41\zehomi{2} & 4.64\zehomi{1} & 9.98\zehomi{1} & $\pm$0.0 \\
264.0 & 4.23\zehomi{2} & 2.82\zehomi{2} & 1.02\zehomi{1} & 8.42\zehomi{2} & 4.64\zehomi{1} & 9.98\zehomi{1} & $\pm$0.0 \\
266.0 & 4.23\zehomi{2} & 2.82\zehomi{2} & 1.02\zehomi{1} & 8.43\zehomi{2} & 4.64\zehomi{1} & 9.98\zehomi{1} & $\pm$0.0 \\
268.0 & 4.24\zehomi{2} & 2.83\zehomi{2} & 1.02\zehomi{1} & 8.44\zehomi{2} & 4.64\zehomi{1} & 9.98\zehomi{1} & $\pm$0.0 \\
270.0 & 4.24\zehomi{2} & 2.83\zehomi{2} & 1.02\zehomi{1} & 8.45\zehomi{2} & 4.64\zehomi{1} & 9.98\zehomi{1} & $\pm$0.0 \\
\hline
\end{tabular}\end{center}
\end{table}

\begin{table}\small\scriptsize
\renewcommand{\arraystretch}{1.3}
\setlength{\tabcolsep}{3ex}
\caption{SM Higgs branching ratios for the different four-fermion final states and their total uncertainties (expressed in percentage). Very-high-mass range.}
\label{BR-4f2e}
\centerline{
\begin{tabular}{lccccccc}
\hline
$\MH$  & $\PH \to 2\Pl2\Pq$ & $\PH \to 2\Pl2\Pq$ & $\PH \to \Pl\PGn_l \Pq\Pq$ & $\PH \to \PGn\PGn \Pq\Pq$ &
$\PH \to 4\Pq$ & $\PH \to 4\Pf$ &$\Delta$BR  \\
$[\UGeV]$ & $\Pl=\Pe,\PGm,\PGt$& $\Pl=\Pe,\PGm$
& $\Pl=\Pe,\PGm$ & $\PGn={}$any& $\Pq=\PQu\PQd\PQc\PQs\PQb$& $f={}$any
 &[\%]\\ 
& $\Pq=\PQu\PQd\PQc\PQs\PQb$ & $\Pq=\PQu\PQd\PQc\PQs\PQb$ &
$\Pq=\PQu\PQd\PQc\PQs\PQb$ & $\Pq=\PQu\PQd\PQc\PQs\PQb$ & & fermion &\\ 
\hline
550.0 & 3.75\zehomi{2} & 2.50\zehomi{2} & 8.06\zehomi{2} & 7.45\zehomi{2} & 3.80\zehomi{1} & 8.16\zehomi{1} & $\pm$2.4 \\
560.0 & 3.76\zehomi{2} & 2.51\zehomi{2} & 8.09\zehomi{2} & 7.49\zehomi{2} & 3.81\zehomi{1} & 8.19\zehomi{1} & $\pm$2.5 \\
570.0 & 3.78\zehomi{2} & 2.52\zehomi{2} & 8.11\zehomi{2} & 7.52\zehomi{2} & 3.82\zehomi{1} & 8.22\zehomi{1} & $\pm$2.5 \\
580.0 & 3.80\zehomi{2} & 2.53\zehomi{2} & 8.13\zehomi{2} & 7.56\zehomi{2} & 3.84\zehomi{1} & 8.25\zehomi{1} & $\pm$2.5 \\
590.0 & 3.82\zehomi{2} & 2.54\zehomi{2} & 8.16\zehomi{2} & 7.60\zehomi{2} & 3.85\zehomi{1} & 8.28\zehomi{1} & $\pm$2.5 \\
\hline
\end{tabular} }
\end{table}

\clearpage


\section{Weights for \POWHEG{} by {\sc HqT} program}
\label{app:hqt4powheg}

Table~\ref{tabZZ_1} completes the studies on $\PH\to\PZ\PZ$ in \refS{se:ZZ}
by the explicit factor for the {\sc POWHEG}/{\sc HqT} reweighting of the
Higgs-$\pT$ spectrum for several Higgs-boson masses.
\begin{table}
\renewcommand{\arraystretch}{0.91}
\setlength{\tabcolsep}{5pt}
\vspace{-\headsep}
\caption{Relative weights {\sc HqT} vs.\ {\sc POWHEG} for several Higgs-boson masses and $\pT$ values.}
\label{tabZZ_1}
\vspace{-0.3cm}
\begin{center}
{\tiny%
\begin{tabular}{lccccccccccccccccccccccccccc}
\hline
 & \multicolumn{19}{c}{$\MH$ [ GeV ]} \\ 
$\pT$ [ GeV ] & 115 & 120 & 130 & 140 & 150 & 160 & 170 & 180 & 190 & 200 & 220 & 250 & 300 & 350 & 400 & 450 & 500 & 550 & 600 \\
\hline
1.0  & 0.81 & 0.74 & 0.91 & 0.94 & 1.01 & 1.02 & 0.92 & 0.80 & 0.82 & 0.87 & 0.83 & 0.68 & 0.60 & 0.47 & 0.50 & 0.36 & 0.23 & 0.16 & 0.14 \\
5.0  & 1.27 & 1.26 & 1.25 & 1.25 & 1.22 & 1.24 & 1.23 & 1.22 & 1.21 & 1.22 & 1.18 & 1.18 & 1.14 & 1.12 & 1.10 & 1.06 & 1.03 & 1.00 & 0.97 \\
10.0  & 1.22 & 1.21 & 1.21 & 1.21 & 1.20 & 1.20 & 1.19 & 1.19 & 1.18 & 1.18 & 1.16 & 1.16 & 1.13 & 1.11 & 1.09 & 1.06 & 1.03 & 1.01 & 0.98 \\
15.0  & 1.16 & 1.16 & 1.16 & 1.16 & 1.16 & 1.16 & 1.16 & 1.16 & 1.16 & 1.15 & 1.14 & 1.14 & 1.12 & 1.10 & 1.09 & 1.06 & 1.04 & 1.01 & 0.99 \\
20.0  & 1.11 & 1.11 & 1.12 & 1.12 & 1.12 & 1.13 & 1.13 & 1.13 & 1.13 & 1.13 & 1.12 & 1.12 & 1.11 & 1.09 & 1.09 & 1.06 & 1.04 & 1.02 & 1.00 \\
25.0  & 1.06 & 1.07 & 1.08 & 1.08 & 1.09 & 1.09 & 1.10 & 1.10 & 1.10 & 1.10 & 1.10 & 1.10 & 1.09 & 1.09 & 1.08 & 1.06 & 1.04 & 1.03 & 1.01 \\
\hline
\end{tabular}}\end{center}
\end{table}

\section{2D K-factors for gluon-fusion Higgs signal production and prompt diphoton backgrounds}
\label{app:2dKfactorsSignalAndBackground}

\refTs{tab2dHNNLO}--\ref{tab2dPrompt} show the
doubly-differential $K$-factors for the $\Pg\Pg\to\PH\to\PGg\PGg$ discussed in
\refS{se:Kfac_ggHgaga}.
\begin{table}
\begin{small}
\begin{center}
\caption{Relative weights {\sc HNNLO} vs {\sc POWHEG} for a Higgs-boson mass $\MH=110$ GeV.}
\label{tab2dHNNLO}
\begin{tabular}{ccccccccccccccccccccc}
\hline
 $q_{T}$ [GeV] / $|Y_{\gamma\gamma}|$ & 0.00 & 0.25 & 0.50 & 0.75 & 1.00 & 1.25 & 1.50 & 1.75 & 2.00 & 2.25 \\
\hline
0.00 & 1.62 & 1.74 & 1.58 & 1.75 & 1.80 & 1.79 & 2.06 & 2.06 & 2.06 & 2.06 \\ 
4.00 & 1.62 & 1.74 & 1.58 & 1.75 & 1.80 & 1.79 & 2.06 & 2.06 & 2.06 & 2.06 \\ 
8.00 & 1.62 & 1.74 & 1.58 & 1.75 & 1.80 & 1.79 & 2.06 & 2.06 & 2.06 & 2.06 \\ 
12.00 & 1.62 & 1.74 & 1.58 & 1.75 & 1.80 & 1.79 & 2.06 & 2.06 & 2.06 & 2.06 \\ 
16.00 & 1.39 & 1.39 & 1.31 & 1.31 & 1.38 & 1.38 & 1.44 & 1.44 & 1.34 & 1.34 \\ 
20.00 & 1.23 & 1.23 & 1.29 & 1.29 & 1.27 & 1.27 & 1.11 & 1.11 & 1.34 & 1.34 \\ 
24.00 & 1.04 & 1.04 & 1.16 & 1.16 & 1.26 & 1.26 & 1.19 & 1.19 & 1.11 & 1.11 \\
\hline
\end{tabular}
\end{center}
\end{small}
\end{table}

\begin{table}
\begin{small}
\begin{center}
\caption{Relative weights {\sc HNNLO} vs {\sc POWHEG} for a Higgs-boson mass $\MH=120\UGeV$.}
\begin{tabular}{ccccccccccccccccccccc}
\hline
 $q_{T}$ [GeV] / $|Y_{\gamma\gamma}|$ & 0.00 & 0.25 & 0.50 & 0.75 & 1.00 & 1.25 & 1.50 & 1.75 & 2.00 & 2.25 \\
\hline
0.00 & 1.51 & 1.53 & 1.68 & 1.96 & 1.90 & 1.77 & 1.75 & 1.75 & 1.75 & 1.75 \\ 
4.00 & 1.51 & 1.53 & 1.68 & 1.96 & 1.90 & 1.77 & 1.75 & 1.75 & 1.75 & 1.75 \\ 
8.00 & 1.51 & 1.53 & 1.68 & 1.96 & 1.90 & 1.77 & 1.75 & 1.75 & 1.75 & 1.75 \\ 
12.00 & 1.51 & 1.53 & 1.68 & 1.96 & 1.90 & 1.77 & 1.75 & 1.75 & 1.75 & 1.75 \\ 
16.00 & 1.48 & 1.48 & 1.43 & 1.43 & 1.40 & 1.40 & 1.49 & 1.49 & 1.30 & 1.30 \\ 
20.00 & 1.24 & 1.24 & 1.31 & 1.31 & 1.25 & 1.25 & 1.18 & 1.18 & 1.30 & 1.30 \\ 
24.00 & 1.17 & 1.17 & 1.24 & 1.24 & 1.26 & 1.26 & 1.28 & 1.28 & 1.29 & 1.29 \\
\hline
\end{tabular}
\end{center}
\end{small}
\end{table}

\begin{table}
\begin{small}
\begin{center}
\caption{Relative weights {\sc HNNLO} vs {\sc POWHEG} for a Higgs-boson mass $\MH=130\UGeV$.}
\begin{tabular}{ccccccccccccccccccccc}
\hline
 $q_{T}$ [GeV] / $|Y_{\gamma\gamma}|$ & 0.00 & 0.25 & 0.50 & 0.75 & 1.00 & 1.25 & 1.50 & 1.75 & 2.00 & 2.25 \\
\hline
0.00 & 1.60 & 1.66 & 1.61 & 1.61 & 1.75 & 1.80 & 1.81 & 1.81 & 1.81 & 1.81 \\ 
4.00 & 1.60 & 1.66 & 1.61 & 1.61 & 1.75 & 1.80 & 1.81 & 1.81 & 1.81 & 1.81 \\ 
8.00 & 1.60 & 1.66 & 1.61 & 1.61 & 1.75 & 1.80 & 1.81 & 1.81 & 1.81 & 1.81 \\ 
12.00 & 1.60 & 1.66 & 1.61 & 1.61 & 1.75 & 1.80 & 1.81 & 1.81 & 1.81 & 1.81 \\ 
16.00 & 1.40 & 1.40 & 1.57 & 1.57 & 1.47 & 1.47 & 1.36 & 1.36 & 1.28 & 1.28 \\ 
20.00 & 1.31 & 1.31 & 1.31 & 1.31 & 1.09 & 1.09 & 1.17 & 1.17 & 1.28 & 1.28 \\ 
24.00 & 1.21 & 1.21 & 1.19 & 1.19 & 1.51 & 1.51 & 1.31 & 1.31 & 1.13 & 1.13 \\
\hline
\end{tabular}
\end{center}
\end{small}
\end{table}

\begin{table}
\begin{small}
\begin{center}
\caption{Relative weights {\sc HNNLO} vs {\sc POWHEG} for a Higgs-boson mass $\MH=140\UGeV$.}
\begin{tabular}{ccccccccccccccccccccc}
\hline
 $q_{T}$ [GeV] / $|Y_{\gamma\gamma}|$ & 0.00 & 0.25 & 0.50 & 0.75 & 1.00 & 1.25 & 1.50 & 1.75 & 2.00 & 2.25 \\
\hline
0.00 & 1.63 & 1.53 & 1.69 & 1.59 & 1.98 & 1.72 & 1.76 & 1.76 & 1.76 & 1.76 \\ 
4.00 & 1.63 & 1.53 & 1.69 & 1.59 & 1.98 & 1.72 & 1.76 & 1.76 & 1.76 & 1.76 \\ 
8.00 & 1.63 & 1.53 & 1.69 & 1.59 & 1.98 & 1.72 & 1.76 & 1.76 & 1.76 & 1.76 \\ 
12.00 & 1.63 & 1.53 & 1.69 & 1.59 & 1.98 & 1.72 & 1.76 & 1.76 & 1.76 & 1.76 \\ 
16.00 & 1.51 & 1.51 & 1.44 & 1.44 & 1.49 & 1.49 & 1.41 & 1.41 & 1.43 & 1.43 \\ 
20.00 & 1.32 & 1.32 & 1.28 & 1.28 & 1.16 & 1.16 & 1.21 & 1.21 & 1.43 & 1.43 \\ 
24.00 & 1.11 & 1.11 & 1.19 & 1.19 & 1.30 & 1.30 & 1.28 & 1.28 & 0.99 & 0.99 \\
\hline
\end{tabular}
\end{center}
\end{small}
\end{table}

\begin{table}
\setlength{\tabcolsep}{1.5ex}
\begin{center}
\begin{scriptsize}
\caption{Relative weights {\sc Diphox} + {\sc Gamma2MC} vs
{\sc Madgraph} $\gamma\gamma$+jets + {\sc Pythia} box.}
\label{tab2dPrompt}
\begin{tabular}{cccccccccccccc}
\hline
 $q_{\mathrm{T}}$ [GeV] / $M_{\gamma\gamma}$ [GeV] & 100 & 105 & 110 & 115 & 120 & 125 & 130 & 135 & 140 & 145 & 150 & 155 \\
 \hline
0.00 & 2.12 & 1.66 & 1.69 & 1.96 & 2.16 & 1.61 & 1.79 & 2.25 & 2.12 & 1.36 & 1.69 & 1.69 \\ 
4.00 & 1.45 & 1.26 & 1.21 & 1.33 & 1.53 & 1.20 & 1.20 & 1.48 & 1.33 & 0.95 & 1.69 & 1.69 \\ 
8.00 & 1.58 & 1.56 & 1.52 & 1.56 & 1.64 & 1.59 & 1.52 & 1.54 & 1.47 & 1.68 & 1.37 & 1.37 \\ 
12.00 & 1.12 & 1.19 & 1.17 & 1.14 & 1.17 & 1.06 & 1.30 & 1.19 & 1.22 & 1.18 & 1.37 & 1.37 \\ 
16.00 & 0.95 & 0.94 & 0.94 & 1.01 & 0.95 & 1.00 & 1.05 & 1.03 & 1.00 & 1.04 & 1.05 & 1.05 \\ 
20.00 & 0.90 & 0.93 & 0.90 & 0.96 & 0.94 & 0.93 & 0.91 & 0.96 & 0.91 & 0.98 & 1.05 & 1.05 \\ 
24.00 & 0.83 & 0.83 & 0.86 & 0.83 & 0.86 & 0.79 & 0.85 & 0.87 & 0.94 & 0.93 & 0.85 & 0.85 \\
\hline
\end{tabular}
\end{scriptsize}
\end{center}
\end{table}


\end{document}